\documentclass[12pt]{book}%
\usepackage{amsfonts}
\usepackage{amsmath}
\usepackage{fullpage}
\usepackage[round]{natbib}
\usepackage{amssymb}
\usepackage{makeidx}
\usepackage{hyperref}
\usepackage{fancyhdr}
\usepackage{graphicx}
\usepackage[calcwidth]{titlesec}
\usepackage[font={footnotesize},labelfont={bf}]{caption}
\usepackage{setspace}%
\providecommand{\U}[1]{\protect\rule{.1in}{.1in}}
\newtheorem{theorem}{Theorem}[section]

\newtheorem{corollary}{Corollary}[section]
\newtheorem{criterion}{Criterion}[section]
\newtheorem{definition}{Definition}[section]
\newtheorem{lemma}{Lemma}[section]
\newtheorem{notation}{Notation}[section]

\newtheorem{proposition}{Proposition}[section]
\newtheorem{remark}{Remark}[section]
\newtheorem{property}{Property}[section]

\newtheorem{exercise}{Exercise}[section]
\newenvironment{proof}[1][Proof]{\noindent\textbf{#1.} }{\ \rule{0.5em}{0.5em}}
\titleformat{\section}[hang]{\bfseries}
{\Large\thesection}{12pt}{\Large}[{\titlerule[0.5pt]}]
\titleformat{\chapter}[display]
{\normalfont\Large\filcenter}
{\titlerule[1pt]
\vspace{1pt}
\titlerule
\vspace{1pc}
\bfseries\LARGE\MakeUppercase{\chaptertitlename} \thechapter}
{1pc}
{\titlerule
\vspace{1pc}
\Huge\bfseries}
\makeindex
\begin{document}

\title{\textbf{{\Huge From Classical to Quantum Shannon Theory}}}
\author{Mark M. Wilde\\Hearne Institute for Theoretical Physics\\Department of Physics and Astronomy\\Center for Computation and Technology\\Louisiana State University\\Baton Rouge, Louisiana 70803, USA}
\maketitle

\noindent \textbf{Copyright Notice:}

\bigskip

\noindent The Work, ``Quantum Information Theory, 2nd edition'' is to be published by Cambridge University Press.

\bigskip

\noindent \copyright\ in the Work, Mark M.~Wilde, 2019

\bigskip

\noindent Cambridge University Press's catalogue entry for the Work can be found at \verb+www.cambridge.org+

\bigskip

\noindent NB: The copy of the Work, as displayed on this web site, is a draft, pre-publication copy only. The final, published version of the Work can be purchased through Cambridge University Press and other standard distribution channels. This draft copy is made available for personal use only and must not be sold or re-distributed.

\tableofcontents

\makeatother


\chapter*{How To Use This Book}

\section*{For Students}

Prerequisites for understanding the content in this book are a solid
background in probability theory and linear algebra. If you are new to
information theory, then there should be enough background in this book to get
you up to speed (Chapters~\ref{chap:classical-shannon-theory},
\ref{chap:info-entropy}, \ref{chap:additivity}, and
\ref{chap:classical-typicality}). However, classics on information theory such
as \cite{book1991cover} and \cite{M03} could be helpful as a reference. If you
are new to quantum mechanics, then there should be enough material in this
book (Part~II) to give you the background necessary for understanding quantum
Shannon theory. The book of \cite{book2000mikeandike} (sometimes
affectionately known as \textquotedblleft Mike and Ike\textquotedblright) has
become the standard starting point for students in quantum information science
and might be helpful as well. Some of the content in that book is available in
the dissertation of \cite{N98}. If you are familiar with Shannon's information
theory (at the level of \cite{book1991cover}, for example), then the present
book should be a helpful entry point into the field of quantum Shannon theory.
We build on intuition developed classically to help in establishing schemes
for communication over quantum channels. If you are familiar with quantum
mechanics, it might still be worthwhile to review Part II because some content
there might not be part of a standard course on quantum mechanics.

The aim of this book is to develop \textquotedblleft from the ground
up\textquotedblright\ many of the major, exciting, pre- and post-millennium
developments in the general area of study known as quantum Shannon theory. As
such, we spend a significant amount of time on quantum mechanics for quantum
information theory (Part~II), we give a careful study of the important unit
protocols of teleportation, super-dense coding, and entanglement distribution
(Part~III), and we develop many of the tools necessary for understanding
information transmission or compression (Part~IV). Parts~V and VI are the
culmination of this book, where all of the tools developed come into play for
understanding many of the important results in quantum Shannon theory.

\section*{For Instructors}

This book could be useful for self-learning or as a reference, but one of the
main goals is for it to be employed as an instructional aid for the classroom.
To aid instructors in designing a course to suit their own needs, a draft, pre-publication copy of this book is
available under a Creative Commons Attribution-NonCommercial-ShareAlike
license. This means that you can modify and redistribute this draft, pre-publication copy as you wish,
as long as you attribute the author, you do not use it for
commercial purposes, and you share a modification or derivative work under the
same license (see
\begin{quotation}
\texttt{http://creativecommons.org/licenses/by-nc-sa/3.0/}
\end{quotation}
for a readable summary of the terms of the license). These requirements can be
waived if you obtain permission from the present author. By releasing the draft, pre-publication copy of the book
under this license, I expect and encourage instructors to modify it for
their own needs. This will allow for the addition of new exercises, new
developments in the theory, and the latest open problems. It might also be a
helpful starting point for a book on a related topic, such as network quantum
Shannon theory.

I used an earlier version of this book in a one-semester course on quantum
Shannon theory at McGill University during Winter semester 2011 (in many parts
of the USA, this semester is typically called \textquotedblleft Spring
semester\textquotedblright). We almost went through the entire book, but it
might also be possible to spread the content over two semesters instead. Here
is the order in which we proceeded:

\begin{enumerate}
\item Introduction in Part I.

\item Quantum mechanics in Part II.

\item Unit protocols in Part III.

\item Chapter~\ref{chap:distance-measures} on distance measures,
Chapter~\ref{chap:info-entropy} on classical information and entropy, and
Chapter~\ref{chap:q-info-entropy} on quantum information and entropy.

\item The first part of Chapter~\ref{chap:classical-typicality}\ on classical
typicality and Shannon compression.

\item The first part of Chapter~\ref{chap:quantum-typicality}\ on quantum typicality.

\item Chapter~\ref{chap:schumach}\ on Schumacher compression.

\item Back to Chapters~\ref{chap:classical-typicality}\ and
\ref{chap:quantum-typicality} for the method of types.

\item Chapter~\ref{chap:ent-conc}\ on entanglement concentration.

\item Chapter~\ref{chap:classical-comm-HSW}\ on classical communication.

\item Chapter~\ref{chap:EA-classical}\ on entanglement-assisted classical communication.

\item The final explosion of results in Chapter~\ref{chap:coh-comm-noisy}%
\ (one of which is a route to proving the achievability part of the quantum
capacity theorem).
\end{enumerate}

The above order is just a particular order that suited the needs for the class
at McGill, but other orders are of course possible. One could sacrifice the
last part of Part III\ on the unit resource capacity region if there is no
desire to cover the quantum dynamic capacity theorem. One could also focus on
going from classical communication to private classical communication to
quantum communication in order to develop some more intuition behind the
quantum capacity theorem. I later did this when teaching the course at LSU\ in
Fall 2013. But just recently in Fall 2015, I went back to the ordering above
while including lectures devoted to the CHSH\ game and the new results in
Chapter~\ref{chap:mono-rel-ent}.

\section*{Other Sources}

There are many other sources to obtain a background in quantum Shannon theory.
The standard reference has become the book of \cite{book2000mikeandike}, but
it does not feature any of the post-millennium results in quantum Shannon
theory. Other excellent books that cover some aspects of quantum Shannon
theory are \citep{H06book,H02book,H12,Wat15}. Patrick Hayden has had a
significant hand as a collaborative guide for many PhD and Masters' theses in
quantum Shannon theory, during his time as a postdoctoral fellow at the
California Institute of Technology and as a professor at McGill University.
These include the theses of \cite{Yard05a}, \cite{A06}, \cite{Savov08,Savov12}%
, \cite{D10}, and \cite{D11}. All of these theses are excellent references.
Hayden also had a strong influence over the present author during the
development of the first edition of this book.

\chapter*{Preface to the Second Edition}

It has now been some years since I completed the first draft of the first
edition of this book. In this time, I have learned much from many
collaborators and I\ am grateful to them. During the past few years, Mario
Berta, Nilanjana Datta, Saikat Guha, and Andreas Winter have
strongly shaped my thinking about quantum information theory, and Mario and
Nilanjana in particular have influenced my technical writing style,
which is reflected in the new edition of the book. Also, the chance to work
with them and others has led me to new research directions in quantum
information theory that I never would have imagined on my own.

I am also thankful to Todd Brun, Paul Cuff, Ludovico Lami, Ciara Morgan, and
Giannicola Scarpa for using the book as the main text in their graduate
courses on quantum information theory and for feedback. One can try as much as
possible to avoid typos in a book, but inevitably, they seem to show up in
unexpected places. I am grateful to many people for pointing out typos or
errors and for suggesting how to fix them, including\ Todd Brun, Giulio
Chiribella, Paul Cuff, Dawei (David) Ding, Will Matthews, Milan Mosonyi, David
Reeb, and Marco Tomamichel. I also thank Corsin Pfister for helpful
discussions about unique linear extensions of quantum physical evolutions. I am grateful to David Tranah and the editorial staff at Cambridge University Press for their help with publishing the second edition.

So what's new in the second edition? Suffice it to say that every page of the
book has been rewritten and there are over 100 pages of new material! I
formulated many thoughts about revising during Fall 2013 while teaching a
graduate course on quantum information at LSU, and I then formulated many more
thoughts and made the actual changes during Fall 2015 (when teaching it
again). In that regard, I am thankful to both the Department of Physics and
Astronomy and the Center for Computation and Technology at LSU\ for providing
a great environment and support. I also thank the graduate students at
LSU\ who gave feedback during and after lectures. There are many little
changes throughout that will probably go unnoticed. For example, I have come
to prefer writing a quantum state shared between Alice and Bob as $\rho_{AB}$
rather than $\rho^{AB}$ (i.e., with system labels as subscripts rather than
superscripts). Admittedly, several collaborators influenced me here, but there
are a few good reasons for this convention:\ the phrase \textquotedblleft
state of a quantum system\textquotedblright\ suggests that the state $\rho$
should be \textquotedblleft resting on\textquotedblright\ the systems $AB$,
the often used partial trace $\operatorname{Tr}_{A}\{\rho_{AB}\}$ looks better
than $\operatorname{Tr}_{A}\{\rho^{AB}\}$, and the notation $\rho_{AB}$ is
more consistent with the standard notation $p_{X}$ for a probability
distribution corresponding to a random variable $X$. OK, that's perhaps minor.
Major changes include the addition of many new exercises, a detailed
discussion of Bell's theorem, the CHSH\ game, and Tsirelson's theorem, the
axiomatic approach to quantum channels, a proof of the Choi--Kraus theorem, a
definition of unital and adjoint maps, a discussion of states, channels, and
measurements all as quantum channels, the equivalence of purifications, the
adjoint map in terms of isometric extension, the definition of the diamond
norm and its interpretation, how a measurement achieves the fidelity, how the
Hilbert--Schmidt distance is not monotone with respect to channels, more
detailed definitions of classical and quantum relative entropies, new
continuity bounds for classical and quantum entropies, refinements of
classical entropy inequalities, streamlined proofs of data processing
inequalities using relative entropy, the equivalence of quantum entropy
inequalities like strong subadditivity and monotonicity of relative entropy,
Chapter~\ref{chap:mono-rel-ent}\ on recoverability, modified proofs of
additivity of channel information quantities, sequential decoding for
classical communication, simpler proofs of the Schumacher compression theorem,
a complete rewrite of Chapter~\ref{chap:ent-conc}, alternate proofs for the
achievability part of the HSW\ theorem, a proof for the classical capacity of
the erasure channel, simpler converse proofs for the entanglement-assisted
capacity theorem, a revised proof of the trade-off coding resource inequality,
a revised proof of the hashing bound, a simplified converse proof of the
quantum dynamic capacity theorem, a completely revised discussion of the
importance of the quantum dynamic capacity formula, and the addition of many
new references that have been influential in recent years. Minor changes
include improved presentations of many theorems and definitions throughout.

I am most grateful to my family for all of their support and encouragement
throughout my life, including my mother, father, sister, and brother and all
of my surrounding family members. I am still indebted to my wife Christabelle
and her family for warmth and love. Christabelle has been an unending source
of support and love for me. I dedicate this second edition to my nephews David
and Matthew.

\bigskip\bigskip\noindent Mark M.~Wilde\newline Baton Rouge, Louisiana,
USA\newline December 2015

\chapter*{Preface to the First Edition}

I began working on this book in the summer of 2008 in Los Angeles, with much
time to spare in the final months of dissertation writing. I had a strong
determination to review quantum Shannon theory, a beautiful area of quantum
information science that Igor Devetak had taught me three years earlier at USC
in fall 2005. I was carefully studying a manuscript entitled \textquotedblleft
Principles of Quantum Information Theory,\textquotedblright\ a text that Igor
had initiated in collaboration with Patrick Hayden and Andreas Winter. I read
this manuscript many times, and many parts of it I understood well, though
other parts I did not.

After a few weeks of reading and rereading, I decided \textquotedblleft if I
can write it out myself from scratch, perhaps I would then understand
it!\textquotedblright, and thus began the writing of the chapters on the
packing lemma, the covering lemma, and quantum typicality. I knew that Igor's
(now former) students Min-Hsiu Hsieh and Zhicheng Luo knew the topic well
because they had already written several quality research papers with him, so
I requested if they could meet with me weekly for an hour to review the
fundamentals. They kindly agreed and helped me quite a bit in understanding
the packing and covering techniques.

Not much later, after graduating, I began collaborating with Min-Hsiu on a
research project that Igor had suggested to the both of us: \textquotedblleft
find the triple trade-off capacity formulas of a quantum
channel.\textquotedblright\ This was perhaps the best starting point for me to
learn quantum Shannon theory because proving this theorem required an
understanding of most everything that had already been accomplished in the
area. After a month of effort, I continued to work with Min-Hsiu on this
project while joining Andreas Winter's Singapore group for a two-month visit.
As I learned more, I added more to the notes, and they continued to grow.

After landing a job in the DC area for January 2009, I realized that I had
almost enough material for teaching a course, and so I contacted local
universities in the area to see if they would be interested. Can Korman,
formerly chair of the Electrical Engineering Department at George Washington
University, was excited about the possibility. His enthusiasm was enough to
keep me going on the notes, and so I continued to refine and add to them in my
spare time in preparing for teaching. Unfortunately (or perhaps fortunately?),
the course ended up being canceled. This was disheartening to me, but in the
mean time, I had contacted Patrick Hayden to see if he would be interested in
having me join his group at McGill University for postdoctoral studies.
Patrick Hayden and David Avis then offered me a postdoctoral fellowship, and I
moved to Montr\'{e}al in October 2009.

After joining, I learned a lot by collaborating and discussing with Patrick
and his group members. Patrick offered me the opportunity to teach his
graduate class on quantum Shannon theory while he was away on sabbatical, and
this encouraged me further to persist with the notes.

I am grateful to everyone mentioned above for encouraging and supporting me
during this project, and I am also grateful to everyone who provided feedback
during the course of writing up. In this regard, I am especially grateful to
Dave Touchette for detailed feedback on all of the chapters in the book.
Dave's careful reading and spotting of errors has immensely improved the
quality of the book. I am grateful to my father, Gregory E. Wilde, Sr., for
feedback on earlier chapters and for advice and love throughout. I thank Ivan
Savov for encouraging me, for feedback, and for believing that this is an
important scholarly work. I also thank Constance Caramanolis, Raza-Ali Kazmi,
John M. Schanck, Bilal Shaw, and Anna Vershynina for valuable feedback. I am
grateful to Min-Hsiu Hsieh for the many research topics we have worked on
together that have enhanced my knowledge of quantum Shannon theory. I thank
Michael Nielsen and Victor Shoup for advice on Creative Commons licensing and
Kurt Jacobs for advice on book publishing. I am grateful to Sarah Payne and
David Tranah of Cambridge University Press for their extensive feedback on the
manuscript and their outstanding support throughout the publication process. I
acknowledge funding from the MDEIE (Quebec) PSR-SIIRI international
collaboration grant.

I am indebted to my mentors who took me on as a student during my career. Todd
Brun was a wonderful PhD supervisor---helpful, friendly, and encouraging of
creativity and original pursuit. Igor Devetak taught me quantum Shannon theory
in fall 2005 and helped me once per week during his office hours. He also
invited me to join Todd's and his group, and more recently, Igor provided much
encouragement and \textquotedblleft big-picture\textquotedblright\ feedback
during the writing of this book. Bart Kosko shaped me as a scholar during my
early years at USC and provided helpful advice regarding the book project.
Patrick Hayden has been an immense bedrock of support at McGill. His knowledge
of quantum information and many other areas is unsurpassed, and he has been
kind, inviting, and helpful during my time at McGill. I am also grateful to
Patrick for giving me the opportunity to teach at McGill and for advice
throughout the development of this book.

I thank my mother, father, sister, and brother and all of my surrounding
family members for being a source of love and support. Finally, I am indebted
to my wife Christabelle and her family for warmth and love. I dedicate this
book to the memory of my grandparents Joseph and Rose McMahon, and Norbert Jay
and Mary Wilde. \textit{Lux aeterna luceat eis, Domine.}

\bigskip\bigskip\noindent Mark M.~Wilde\newline Montreal, Quebec,
Canada\newline June 2011

\part{Introduction}

\chapter{Concepts in Quantum Shannon Theory}

\label{chap:intro-concepts}In these first few chapters, our aim is to
establish a firm grounding so that we can address some fundamental questions
regarding information transmission over quantum channels. This area of study
has become known as ``quantum Shannon theory'' in the broader quantum
information community, in order to distinguish this topic from other areas of
study in quantum information science. In this text, we will use the terms
``quantum Shannon theory'' and ``quantum information theory'' somewhat
interchangeably. We will begin by briefly overviewing several fundamental
aspects of the quantum theory. Our study of the quantum theory, in this
chapter and future ones, will be at an abstract level, without giving
preference to any particular physical system such as a spin-$\tfrac12$ particle or a
photon. This approach will be more beneficial for the purposes of our study,
but, here and there, we will make some reference to actual physical systems to
ground us in reality.

You may be wondering, what is \textit{quantum Shannon theory} and why do we
name this area of study as such? In short, quantum Shannon theory is the study
of the ultimate capability of noisy physical systems, governed by the laws of
quantum mechanics, to preserve information and correlations. Quantum
information theorists have chosen the name \textit{quantum Shannon theory} to
honor Claude Shannon, who single-handedly founded the field of classical
information theory, with a groundbreaking paper \citep{bell1948shannon}. In
particular, the name refers to the asymptotic theory of quantum information,
which is the main topic of study in this book. Information theorists since
Shannon have dubbed him the \textquotedblleft Einstein of the information
age.\textquotedblright\footnote{It is worthwhile to look up \textquotedblleft
Claude Shannon---Father of the Information Age\textquotedblright\ on YouTube
and watch several reknowned information theorists speak with awe about
\textquotedblleft the founding father\textquotedblright\ of information
theory.}\ The name \textit{quantum Shannon theory} is fit to capture this area
of study because we often use quantum versions of Shannon's ideas to prove
some of the main theorems in quantum Shannon theory.

We prefer the name \textquotedblleft quantum Shannon theory\textquotedblright%
\ over such names as \textquotedblleft quantum information
science\textquotedblright\ or just \textquotedblleft quantum
information.\textquotedblright\ These other names are too broad, encompassing
subjects as diverse as quantum computation, quantum algorithms, quantum
complexity theory, quantum communication complexity, entanglement theory,
quantum key distribution, quantum error correction, and even the experimental
implementation of quantum protocols. Quantum Shannon theory does overlap with
some of the aforementioned subjects, such as quantum computation, entanglement
theory, quantum key distribution, and quantum error correction, but the name
\textquotedblleft quantum Shannon theory\textquotedblright\ should evoke a
certain paradigm for quantum communication with which the reader will become
intimately familiar after some exposure to the topics in this book. For
example, it is necessary for us to discuss \textit{quantum gates} (a topic in
quantum computing) because quantum Shannon-theoretic protocols exploit them to
achieve certain information-processing tasks. Also, in
Chapter~\ref{chap:private-cap}, we are interested in the ultimate limitation
on the ability of a noisy quantum communication channel to transmit private
information (information that is secret from any third party besides the
intended receiver). This topic connects quantum Shannon theory with quantum
key distribution because the private information capacity of a noisy quantum
channel is strongly related to the task of using the quantum channel to
distribute a secret key. As a final connection, one of the most important
theorems of quantum Shannon theory is
\index{quantum capacity theorem}
the \textit{quantum capacity theorem}. This theorem determines the ultimate
rate at which a sender can reliably transmit quantum information over a
quantum channel to a receiver. The result provided by the quantum
capacity\ theorem is closely related to the theory of quantum error
correction, but the mathematical techniques used in quantum Shannon theory and
in quantum error correction are so different that these subjects merit
different courses of study.

Quantum Shannon theory intersects two of the great sciences of the twentieth
century:\ the quantum theory and information theory. It was really only a
matter of time before physicists, mathematicians, computer scientists, and
engineers began to consider the convergence of the two subjects because the
quantum theory was essentially established by 1926 and information theory by
1948. This convergence has sparked what we may call the \textquotedblleft
quantum information revolution\textquotedblright\ or what some refer to as the
\textquotedblleft second quantum revolution\textquotedblright%
\ \citep{dowling2003}\ (with the first revolution being the discovery of the quantum theory).

The fundamental components of the quantum theory are a set of postulates that
govern phenomena on the scale of atoms. Uncertainty is at the heart of the
quantum theory---\textquotedblleft quantum uncertainty\textquotedblright\ or
\textquotedblleft Heisenberg\ uncertainty\textquotedblright\ is not due to our
lack or loss of information or due to imprecise measurement capability, but
rather, it is a fundamental uncertainty inherent in nature itself. The
discovery of the quantum theory came about as a total shock to the physics
community, shaking the foundations of scientific knowledge. Perhaps it is for
this reason that every introductory quantum mechanics course delves into its
history in detail and celebrates the founding fathers of the quantum theory.
In this book, we do not discuss the history of the quantum theory in much
detail and instead refer to several great introductory books for these details
\citep{book1989bohm,book1994sakurai,Grif95a,Feynman98}. Physicists such as
Planck, Einstein, Bohr, de Broglie, Born, Heisenberg, Schr\"{o}dinger, Pauli,
Dirac, and von Neumann contributed to the foundations of the quantum theory in
the 1920s and 1930s. We introduce the quantum theory by \textit{briefly}
commenting on its history and major underlying concepts.

Information theory is the second great foundational science for quantum
Shannon theory. In some sense, it could be viewed as merely an application of
probability theory. Its aim is to quantify the ultimate compressibility of
information and the ultimate ability for a sender to transmit information
reliably to a receiver. It relies upon probability theory because
\textquotedblleft classical\textquotedblright\ uncertainty, arising from our
lack of total information about any given scenario, is ubiquitous throughout
all information-processing tasks. The uncertainty in classical information
theory is the kind that is present in the flipping of a coin or the shuffle of
a deck of cards, the uncertainty due to imprecise knowledge. \textquotedblleft
Quantum\textquotedblright\ uncertainty is inherent in nature itself and is
perhaps not as intuitive as the uncertainty that classical information theory
measures. We later expand further on these differing kinds of uncertainty, and
Chapter~\ref{chap:noisy-quantum-theory} shows how a theory of quantum
information captures both kinds of uncertainty within one
formalism.\footnote{Von Neumann established the density operator formalism in
his 1932 book on the quantum theory. This mathematical framework captures both
kinds of uncertainty \citep{book1996vonNeumann}.}

The history of classical information theory began with Claude Shannon.
Shannon's contribution is heralded as one of the single greatest contributions
to modern science because he established the field in his seminal paper
\citep{bell1948shannon}. In this paper, he coined the essential terminology,
and he stated and justified the main mathematical definitions and the two
fundamental theorems of information theory. Many successors have contributed
to information theory, but most, if not all, of the follow-up contributions
employ Shannon's line of thinking in some form. In quantum Shannon theory, we
will notice that many of Shannon's original ideas are present, though they
take a particular \textquotedblleft quantum\textquotedblright\ form.

One of the major assumptions in both classical information theory and quantum
Shannon theory is that local computation is free but communication is
expensive. In particular, for the classical case, we assume that each party
has unbounded computation available. For the quantum case, we assume that each
party has a fault-tolerant quantum computer available at his or her local
station and the power of each quantum computer is unbounded. We also assume
that both communication and a shared resource are expensive, and for this
reason, we keep track of these resources in a \textit{resource count}.
Sometimes, however, we might say that classical communication is free in order
to simplify a scenario. A simplification like this one can lead to greater
insights that might not be possible without making such an assumption.

We should first study and understand the postulates of the quantum theory in
order to study quantum Shannon theory properly. Your heart may sink when you
learn that the Nobel Prize-winning physicist Richard Feynman is famously
quoted as saying, \textquotedblleft I think I can safely say that nobody
understands quantum mechanics.\textquotedblright\ We should take the liberty
of clarifying Feynman's statement. Of course, Feynman does not intend to
suggest that no one knows how to work with the quantum theory. Many well-abled
physicists are employed to spend their days exploiting the laws of the quantum
theory to do fantastic things, such as the trapping of ions in a vacuum or
applying the quantum tunneling effect in a transistor to process a single
electron. I am hoping that you will give me the license to interpret Feynman's
statement. I think he means that it is very difficult for us to understand the
quantum theory intuitively because we do not experience the phenomena that it
predicts. If we were the size of atoms and we experienced the laws of quantum
theory on a daily basis, then perhaps the quantum theory would be as intuitive
to us as Newton's law of universal gravitation.\footnote{Of course, Newton's
law of universal gravitation was a revolutionary breakthrough because the
phenomenon of gravity is not entirely intuitive when a student first learns
it. However, we do experience the gravitational law in our daily lives, and I would
argue that this phenomenon is much more intuitive than, say, the phenomenon of
quantum entanglement.} Thus, in this sense, I would agree with
Feynman---nobody can really understand the quantum theory because it is not
part of our everyday experiences. Nevertheless, our aim in this book is to
work with the laws of quantum theory so that we may begin to gather insights
about what the theory predicts. Only by exposure to and practice with its
postulates can we really gain an intuition for its predictions. It is best to
imagine that the world in our everyday life does incorporate the postulates of
quantum mechanics, because, indeed, as many, many experiments have confirmed,
it does!

We delve into the history of the convergence of the quantum theory and
information theory in some detail in this introductory chapter because this
convergence does have an interesting history and is relevant to the topic of
this book. The purpose of this historical review is not only to become
familiar with the field itself but also to glimpse into the minds of the
founders of the field so that we may see the types of questions that are
important to think about when tackling new, unsolved
problems.\footnote{Another way to discover good questions is to attend parties
that well-established professors hold. The story goes that Oxford physicist
David Deutsch attended a 1981 party at the Austin, Texas house of reknowned
physicist John Archibald Wheeler, in which many attendees discussed the
foundations of computing \citep{ieeespec2001}. Deutsch claims that he could
immediately see that the quantum theory would give an improvement for
computation. A few years later, in 1985, he published an algorithm that was the
first instance of a quantum speed-up over the fastest classical algorithm
\citep{prsla1985deutsch}.} Many of the most important results come about from
asking simple, yet profound, questions and exploring the possibilities.

We first briefly review the history and the fundamental concepts of the
quantum theory before delving into the convergence of the quantum theory and
information theory. We build on these discussions by introducing some of the
initial fundamental contributions to quantum Shannon theory. The final part of
this chapter ends by posing some of the questions to which quantum Shannon
theory provides answers.

\section{Overview of the Quantum Theory}

\subsection{Brief History of the Quantum Theory}

A physicist living around 1890 would have been well pleased with the progress
of physics, but perhaps frustrated at the seeming lack of open research
problems. It seemed as though the Newtonian laws of mechanics, Maxwell's
theory of electromagnetism, and Boltzmann's theory of statistical mechanics
explained most natural phenomena. In fact, Max Planck, one of the founding
fathers of the quantum theory, was searching for an area of study in 1874\ and
his advisor gave him the following guidance:

\begin{quote}
\textquotedblleft In this field [of physics], almost everything is already
discovered, and all that remains is to fill a few holes.\textquotedblright
\end{quote}

\subsubsection{Two Clouds}

Fortunately, Planck did not heed this advice and instead began his physics
studies. Not everyone agreed with Planck's former advisor. Lord Kelvin stated
in his famous April 1900 lecture that \textquotedblleft two
clouds\textquotedblright\ surrounded the \textquotedblleft beauty and
clearness of theory\textquotedblright\ \citep{kelvin1901}. The first cloud was
the failure of Michelson and Morley to detect a change in the speed of light
as predicted by an \textquotedblleft ether theory,\textquotedblright\ and the
second cloud was the ultraviolet catastrophe, the classical prediction that a
blackbody emits radiation with an infinite intensity at high ultraviolet
frequencies. Also in 1900, Planck started the quantum revolution that began to
clear the second cloud. He assumed that light comes in discrete bundles of
energy and used this idea to produce a formula that correctly predicts the
spectrum of blackbody radiation \citep{planck1901}. A great cartoon lampoon of
the ultraviolet catastrophe shows Planck calmly sitting fireside with a
classical physicist whose face is burning to bits because of the intense
ultraviolet radiation that his classical theory predicts the fire is emitting
\citep{book2004mcevoy}. A few years later, \cite{einstein1905} contributed a
paper that helped to further clear the second cloud (he also cleared the first
cloud with his other 1905 paper on special relativity). He assumed that Planck
was right and showed that the postulate that light arrives in
\textquotedblleft quanta\textquotedblright\ (now known as the photon theory)
provides a simple explanation for the photoelectric effect, the phenomenon in
which electromagnetic radiation beyond a certain threshold frequency impinging
on a metallic surface induces a current in that metal.

These two explanations of Planck and Einstein fueled a theoretical revolution
in physics that some now call the first quantum revolution
\citep{dowling2003}. Some years later, \cite{debroglie1924} postulated that
every element of matter, whether an atom, electron, or photon, has both
particle-like behavior and wave-like behavior. Just two years later,
\cite{schrodinger1926} used the de Broglie idea to formulate a wave equation,
now known as Schr\"{o}dinger's equation, that governs the evolution of a
closed quantum-mechanical system. His formalism later became known as wave
mechanics and was popular among physicists because it appealed to notions with
which they were already familiar. Meanwhile, \cite{1925heisenberg} formulated
an \textquotedblleft alternate\textquotedblright\ quantum theory called matrix
mechanics. His theory used matrices and linear algebra, mathematics with which
many physicists at the time were not readily familiar. For this reason,
Schr\"{o}dinger's wave mechanics was more popular than Heisenberg's matrix
mechanics. In 1930, Paul Dirac published a textbook (now in its fourth edition
and reprinted 16 times) that unified the formalisms of Schr\"{o}dinger and
Heisenberg, showing that they were actually equivalent
\citep{citeulike:1280736}. In a later edition, he introduced the now
ubiquitous \textquotedblleft Dirac notation\textquotedblright\ for quantum
theory that we will employ in this book.

After the publication of Dirac's textbook, the quantum theory then stood on
firm mathematical grounding and the basic theory had been established. We thus
end our historical overview at this point and move on to the fundamental
concepts of the quantum theory.

\subsection{Fundamental Concepts of the Quantum Theory}

Quantum theory, as applied in quantum information theory, really has only a
few important concepts. We review each of these aspects of quantum theory
briefly in this section. Some of these phenomena are uniquely
\textquotedblleft quantum\textquotedblright\ but others do occur in the
classical theory. In short, these concepts are as follows:\footnote{I have
used Todd A. Brun's list from his lecture notes \citep{brun_lectures}.}

\begin{enumerate}
\item indeterminism,
\index{indeterminism}%

\item interference,
\index{quantum interference}%

\item uncertainty,

\item superposition,
\index{superposition}%

\item entanglement.
\end{enumerate}

The quantum theory is \textit{indeterministic} because the theory makes
predictions about probabilities of events only. This aspect of quantum theory
is in contrast with a deterministic classical theory such as that predicted by
the Newtonian laws. In the Newtonian system, it is possible to predict, with
certainty, the trajectories of all objects involved in an interaction if one
knows only the initial positions and velocities of all the objects. This
deterministic view of reality even led some to believe in determinism from a
philosophical point of view. For instance, the mathematician Pierre-Simon
Laplace once stated that a supreme intellect, colloquially known as
``Laplace's demon,'' could predict all future events from present and past events:

\begin{quote}
\textquotedblleft We may regard the present state of the universe as the
effect of its past and the cause of its future. An intellect which at a
certain moment would know all forces that set nature in motion, and all
positions of all items of which nature is composed, if this intellect were
also vast enough to submit these data to analysis, it would embrace in a
single formula the movements of the greatest bodies of the universe and those
of the tiniest atom; for such an intellect nothing would be uncertain and the
future just like the past would be present before its eyes.\textquotedblright
\end{quote}

The application of Laplace's statement to atoms is fundamentally incorrect,
but we can forgive him because the quantum theory had not yet been established
in his time. Many have extrapolated from Laplace's statement to argue the
invalidity of human free will. We leave such debates to
philosophers.\footnote{John Archibald Wheeler may disagree with this approach.
He once said, \textquotedblleft Philosophy is too important to be left to the
philosophers\textquotedblright\ \citep{PT09}.}

In reality, we never can possess full information about the positions and
velocities of every object in any given physical system. Incorporating
probability theory then allows us to make predictions about the probabilities
of events and, with some modifications, the classical theory becomes an
indeterministic theory. Thus, indeterminism is not a unique aspect of the
quantum theory but merely a feature of it. But this feature is so crucial to
the quantum theory that we list it among the fundamental concepts.

\textit{Interference}
\index{quantum interference}
is another feature of the quantum theory. It is also present in any classical
wave theory---constructive interference occurs when the crest of one wave
meets the crest of another, producing a stronger wave, while destructive
interference occurs when the crest of one wave meets the trough of another,
canceling out each other. In any classical wave theory, a wave occurs as a
result of many particles in a particular medium coherently displacing one
another, as in an ocean surface wave or a sound pressure wave, or as a result
of coherent oscillating electric and magnetic fields, as in an electromagnetic
wave. The strange aspect of interference in the quantum theory is that even a
single \textquotedblleft particle\textquotedblright\ such as an electron can
exhibit wave-like features, as in the famous double slit experiment (see,
e.g., \cite{elegantuniverse}\ for a history of these experiments). This
quantum interference is what contributes wave--particle duality to every
fundamental component of matter.

\textit{Uncertainty} is at the heart of the quantum theory. Uncertainty in the
quantum theory is fundamentally different from uncertainty in the classical
theory (discussed in the former paragraph about an indeterministic classical
theory). The archetypal example of uncertainty in the quantum theory occurs
for a single particle. This particle has two complementary variables: its
position and its momentum. The uncertainty principle states that it is
impossible to know both the particle's position and momentum to arbitrary
accuracy. This principle even calls into question the meaning of the word
\textquotedblleft know\textquotedblright\ in the previous sentence in the
context of quantum theory. We might say that we can only know that which we
measure, and thus, we can only know the position of a particle after
performing a precise measurement that determines it. If we follow with a
precise measurement of its momentum, we lose all information about the
position of the particle after learning its momentum. In quantum information
science, the BB84 protocol for quantum key distribution exploits the
uncertainty principle and statistical analysis to determine the presence of an
eavesdropper on a quantum communication channel by encoding information into
two complementary variables \citep{bb84}.

The \textit{superposition}
\index{superposition}%
principle states that a quantum particle can be in a linear combination state,
or \textit{superposed state}, of any two other allowable states. This
principle is a result of the linearity of quantum theory. Schrodinger's wave
equation is a linear differential equation, meaning that the linear
combination $\alpha\psi+\beta\phi$ is a solution of the equation if $\psi$ and
$\phi$ are both solutions of the equation. We say that the solution
$\alpha\psi+\beta\phi$ is a coherent superposition of the two solutions. The
superposition principle has dramatic consequences for the interpretation of
the quantum theory---it gives rise to the notion that a particle can somehow
\textquotedblleft be in one location and another\textquotedblright\ at the
same time. There are different interpretations of the meaning of the
superposition principle, but we do not highlight them here. We merely choose
to use the technical language that the particle is in a superposition of both
locations. The loss of a superposition can occur through the interaction of a
particle with its environment. Maintaining an arbitrary superposition of
quantum states is one of the central goals of a quantum communication protocol.

The last, and perhaps most striking, quantum feature that we highlight here is
\textit{entanglement}. There is no true classical analog of entanglement. The
closest analog of entanglement might be a secret key that two parties possess,
but even this analogy does not come close. Entanglement refers to the strong
quantum correlations that two or more quantum particles can possess. The
correlations in quantum entanglement are stronger than any classical
correlations in a precise, technical sense. \cite{S35} first coined the term
\textquotedblleft entanglement\textquotedblright\ after observing some of its
strange properties and consequences. Einstein, Podolsky, and Rosen then
presented an apparent paradox involving entanglement that raised concerns over
the completeness of the quantum theory \citep{epr1935}. That is, they
suggested that the seemingly strange properties of entanglement called the
uncertainty principle into question (and thus the completeness of the quantum
theory) and furthermore suggested that there might be some \textquotedblleft
local hidden-variable\textquotedblright\ theory that could explain the results
of experiments. It took about 30 years to resolve this paradox, but John Bell
did so by presenting a simple inequality, now known as a
\index{Bell inequality}
Bell inequality \citep{bell1964}. He showed that any two-particle classical
correlations that satisfy the assumptions of the \textquotedblleft local
hidden-variable theory\textquotedblright\ of Einstein, Podolsky, and Rosen
must be less than a certain amount. He then showed how the correlations of two
entangled quantum particles can violate this inequality, and thus,
entanglement has no explanation in terms of classical correlations but is
instead a uniquely quantum phenomenon. Experimentalists later verified that
two entangled quantum particles can violate Bell's inequality \citep{PhysRevLett.47.460}.

In quantum information science, the non-classical correlations in entanglement
play a fundamental role in many protocols. For example, entanglement is the
enabling resource in
\index{quantum teleportation}%
teleportation, a protocol that disembodies a quantum state in one location and
reproduces it in another. We will see many other examples of entanglement
throughout this book.

Entanglement theory concerns methods for quantifying the amount of
entanglement present not only in a two-particle state but also in a
multiparticle state. A large body of literature exists that investigates
entanglement theory~\citep{H42007}, but we only address aspects of it that are
relevant in our study of quantum Shannon theory.

The above five features capture the essence of the quantum theory, but we will
see more aspects of it as we progress through our overview in
Chapters~\ref{chap:noiseless-quantum-theory}, \ref{chap:noisy-quantum-theory},
and \ref{chap:purified-q-t}.

\section{The Emergence of Quantum Shannon Theory}

In the previous section, we discussed several unique quantum phenomena such as
superposition and entanglement, but it is not clear what kind of information
these unique quantum phenomena represent. Is it possible to find a convergence
of the quantum theory and Shannon's information theory, and if so, what is the convergence?

\subsection{The Shannon Information Bit}

A fundamental contribution of Shannon is the notion of a \textit{bit} as a
measure of information. Typically, when we think of a bit, we think of a
two-valued quantity that can be in the state \textquotedblleft
off\textquotedblright\ or the state \textquotedblleft on.\textquotedblright%
\ We represent this bit with a binary number that can be \textquotedblleft%
0\textquotedblright\ or \textquotedblleft1.\textquotedblright\ We also
associate a physical representation with a bit---this physical representation
can be whether a light switch is off or on, whether a transistor allows
current to flow or not, whether a large number of magnetic spins point in
one direction or another, the list going on and on. These are all physical
notions of a bit.

Shannon's notion of a bit is quite different from these physical notions, and
we motivate his notion with the example of a fair coin. Without flipping the
coin, we have no idea what the result of a coin flip will be---our best guess
at the result is to guess randomly. If someone else learns the result of a
random coin flip, we can ask this person the question:\ What was the result?
We then learn \textit{one bit of information}.

Though it may seem obvious, it is important to stress that we do not learn any
(or not as much) information if we do not ask the right question. This point
becomes even more important in the quantum case. Suppose that the coin is not
fair---without loss of generality, suppose the probability of
\textquotedblleft heads\textquotedblright\ is greater than the probability of
\textquotedblleft tails.\textquotedblright\ In this case, we would not be as
surprised to learn that the result of a coin flip is \textquotedblleft
heads.\textquotedblright\ We may say in this case that we would learn less than one
bit of information if we were to ask someone the result of the coin flip.

The Shannon binary entropy
\index{binary entropy}%
is a measure of information. Given a probability distribution $\left(
p,1-p\right)  $\ for a binary random variable, its Shannon binary entropy is%
\begin{equation}
h_{2}( p) \equiv-p\log p-( 1-p) \log( 1-p) , \label{eq-intro:bin-entropy}%
\end{equation}
where here and throughout the book (unless stated explicitly otherwise), the
logarithm is taken base two. The Shannon binary entropy measures information
in units of bits. We will discuss it in more detail in the next chapter and in
Chapter~\ref{chap:info-entropy}.

The Shannon bit, or Shannon binary entropy, is a measure of the surprise upon
learning the outcome of a random binary experiment. Thus, the Shannon bit has
a completely different interpretation from that of the physical bit. The
outcome of the coin flip resides in a physical bit, but it is the information
associated with the random nature of the physical bit that we would like to
measure. It is this notion of a bit that is important in information theory.

\subsection{A Measure of Quantum Information}

\label{sec-intro:measure-q-info}The above section discusses Shannon's notion
of a bit as a measure of information. A natural question is whether there is
an analogous measure of quantum information, but before we can even ask that
question, we might first wonder: What is \textit{quantum information}? As in
the classical case, there is a \textit{physical} notion of quantum
information. A quantum state always resides \textquotedblleft
in\textquotedblright\ a physical system. Perhaps another way of stating this
idea is that every physical system is in some quantum state. The physical
notion of a quantum bit, or qubit for short (pronounced \textquotedblleft cue
$\cdot\ $bit\textquotedblright), is a two-level quantum system. Examples of
two-level quantum systems are the spin of the electron, the polarization of a
photon, or an atom with a ground state and an excited state. The physical
notion of a qubit is straightforward to understand once we have a grasp of the
quantum theory.

A more pressing question for us in this book is to understand an
\textit{informational} notion of a qubit, as in the Shannon sense. In the
classical case, we quantify information by the amount of knowledge we gain
after learning the answer to a probabilistic question. In the quantum world,
what knowledge can we have of a quantum state?

Sometimes we may know the exact quantum state of a physical system because we
prepared the quantum system in a certain way. For example, we may prepare an
electron in its \textquotedblleft spin-up in the $z$
direction\textquotedblright\ state,\ where $\left\vert \uparrow_{z}%
\right\rangle $ denotes this state. If we prepare the state in this way, we
know for certain that the state is indeed $\left\vert \uparrow_{z}%
\right\rangle $ and no other state. Thus, we do not gain any information, or
equivalently, there is no removal of uncertainty if someone else tells us that
the state is $\left\vert \uparrow_{z}\right\rangle $. We may say that this
state has zero qubits of quantum information, where the term \textquotedblleft
qubit\textquotedblright\ now refers to a measure of the quantum information of
a state.

In the quantum world, we also have the option of measuring this state in the
$x$ direction. The postulates of quantum theory, given in
Chapter~\ref{chap:noiseless-quantum-theory}, predict that the state will then
be $\left\vert \uparrow_{x}\right\rangle $ or $\left\vert \downarrow
_{x}\right\rangle $ with equal probability after measuring in the $x$
direction. One interpretation of this aspect of quantum theory is that the
system does not have any definite state in the $x$ direction: in fact there is
maximal uncertainty about its $x$ direction, if we know that the physical
system has a definite $z$ direction. This behavior is one manifestation of the
Heisenberg uncertainty principle. So before performing the measurement, we
have no knowledge of the resulting state and we gain one Shannon bit of
information after learning the result of the measurement. If we use Shannon's
notion of entropy and perform an $x$ measurement, this classical measure loses
some of its capability here to capture our knowledge of the state of the
system. It is inadequate to capture our knowledge of the state because we
actually prepared it ourselves and know with certainty that it is in the state
$\left\vert \uparrow_{z}\right\rangle $. With these different notions of
information gain, which one is the most appropriate for the quantum case?

It turns out that the first way of thinking is the one that is most useful for
quantifying quantum information. If someone tells us the definite quantum
state of a particular physical system and this state is indeed the true state,
then we have complete knowledge of the state and thus do not learn more
\textquotedblleft qubits\textquotedblright\ of quantum information from this
point onward. This line of thinking is perhaps similar in one sense to the
classical world, but different from the classical world, in the sense of the
case presented in the previous paragraph.

Now suppose that a friend (let us call him \textquotedblleft
Bob\textquotedblright) randomly prepares quantum states as a probabilistic
ensemble. Suppose Bob prepares $\left\vert \uparrow_{z}\right\rangle $ or
$\left\vert \downarrow_{z}\right\rangle $ with equal probability. With only
this probabilistic knowledge, we acquire one bit of information if Bob reveals
which state he prepared. We could also perform a quantum measurement on the
system to determine what state Bob prepared (we discuss quantum measurements
in detail in Chapter~\ref{chap:noiseless-quantum-theory}). One reasonable
measurement to perform is a measurement in the $z$ direction. The result of
the measurement determines which state Bob actually prepared because both
states in the ensembles are states with definite $z$ direction. The result of
this measurement thus gives us one bit of information---the same amount that
we would learn if Bob informed us which state he prepared. It seems that most
of this logic is similar to the classical case---i.e., the result of the
measurement only gave us one Shannon bit of information.

Another measurement to perform is a measurement in the $x$ direction. If the
actual state prepared is $\left\vert \uparrow_{z}\right\rangle $, then the
quantum theory predicts that the state becomes $\left\vert \uparrow
_{x}\right\rangle $ or $\left\vert \downarrow_{x}\right\rangle $ with equal
probability. Similarly, if the actual state prepared is $\left\vert
\downarrow_{z}\right\rangle $, then the quantum theory predicts that the state
again becomes $\left\vert \uparrow_{x}\right\rangle $ or $\left\vert
\downarrow_{x}\right\rangle $ with equal probability. Calculating
probabilities, the resulting state is $\left\vert \uparrow_{x}\right\rangle $
with probability 1/2 and $\left\vert \downarrow_{x}\right\rangle $ with
probability 1/2. So the Shannon bit content of learning the result is again
one bit, but we arrived at this conclusion in a much different fashion from
the scenario in which we measured in the $z$ direction. How can we quantify
the \textit{quantum information} of this ensemble? We claim for now that this
ensemble contains one \textit{qubit} of quantum information and this result
derives from either the measurement in the $z$ direction or the measurement in
the $x$ direction for this particular ensemble.

Let us consider one final example that perhaps gives more insight into how we
might quantify quantum information. Suppose Bob prepares $\left\vert
\uparrow_{z}\right\rangle $ or $\left\vert \uparrow_{x}\right\rangle $ with
equal probability. The first state is spin-up in the $z$ direction and the
second is spin-up in the $x$ direction. If Bob reveals which state he
prepared, then we learn one Shannon bit of information. But suppose now that
we would like to learn the prepared state on our own, without the help of our
friend Bob. One possibility is to perform a measurement in the $z$ direction.
If the state prepared is $\left\vert \uparrow_{z}\right\rangle $, then we
learn this result with probability 1/2. But if the state prepared is
$\left\vert \uparrow_{x}\right\rangle $, then the quantum theory predicts that
the state becomes $\left\vert \uparrow_{z}\right\rangle $ or $\left\vert
\downarrow_{z}\right\rangle $ with equal probability (while we learn what the
new state is). Thus, quantum theory predicts that the act of measuring this
ensemble inevitably disturbs the state some of the time. Also, there is no way
that we can learn with certainty whether the prepared state is $\left\vert
\uparrow_{z}\right\rangle $ or $\left\vert \uparrow_{x}\right\rangle $. Using
a measurement in the $z$ direction, the resulting state is $\left\vert
\uparrow_{z}\right\rangle $ with probability 3/4 and $\left\vert
\downarrow_{z}\right\rangle $ with probability 1/4. We learn less than one
Shannon bit of information from this ensemble because the probability
distribution becomes skewed when we perform this particular measurement.

The probabilities resulting from the measurement in the $z$ direction are the
same that would result from an ensemble where Bob prepares $\left\vert
\uparrow_{z}\right\rangle $ with probability 3/4 and $\left\vert
\downarrow_{z}\right\rangle $ with probability 1/4 and we perform a
measurement in the $z$ direction. The actual Shannon entropy of the
distribution $(3/4,1/4)$ is about 0.81 bits, confirming our intuition that we
learn approximately less than one bit. A similar, symmetric analysis holds to
show that we gain 0.81 bits of information when we perform a measurement in
the $x$ direction.

We have more knowledge of the system in question if we gain less information
from performing measurements on it. In the quantum theory, we learn less about
a system if we perform a measurement on it that does not disturb it too much.
Is there a measurement that we can perform in which we learn the least amount
of information? Recall that learning the least amount of information is ideal
because it has the interpretation that we require fewer questions on average
to learn the result of a random experiment. Indeed, it turns out that a
measurement in the $x+z$ direction reveals the least amount of information.
Avoiding details for now, this measurement returns a state that we label
$\left\vert \uparrow_{x+z}\right\rangle $ with probability $\cos^{2}(\pi/8)$
and a state $\left\vert \downarrow_{x+z}\right\rangle $ with probability
$\sin^{2}(\pi/8)$. This measurement has the desirable effect that it causes
the least amount of disturbance to the original states in the ensemble. The
entropy of the distribution resulting from the measurement is about 0.6 bits
and is less than the one bit that we learn if Bob reveals the state. The
entropy $\approx0.6$ is also the least amount of information among all
possible sharp measurements that we may perform on the ensemble. We claim that
this ensemble contains $\approx0.6$ \textit{qubits} of quantum information.

We can determine the ultimate compressibility of classical data with Shannon's
source coding theorem (we overview this technique in the next chapter). Is
there a similar way that we can determine the ultimate compressibility of
quantum information? This question was one of the early and profitable ones
for quantum Shannon theory and the answer is affirmative. The technique for
quantum compression is called Schumacher compression%
\index{Schumacher compression}%
, named after Benjamin Schumacher. Schumacher used ideas similar to that of
Shannon---he created the notion of a quantum information source that emits
random physical qubits, and he invoked the law of large numbers to show that
there is a so-called \textit{typical subspace}
\index{typical subspace}%
where most of the quantum information really resides. This line of thought is
similar to that which we will discuss in the overview of data compression in
the next chapter. The size of the typical subspace for most quantum
information sources is exponentially smaller than the size of the space in
which the emitted physical qubits resides. Thus, one can \textquotedblleft
quantum compress\textquotedblright\ the quantum information to this subspace
without losing much. Schumacher's quantum source coding theorem then
quantifies, in an operational sense, the amount of actual quantum information
that the ensemble contains. The amount of actual quantum information
corresponds to the number of qubits, in the informational sense, that the
ensemble contains. It is this measure that is equivalent to the
\textquotedblleft optimal measurement\textquotedblright\ one that we suggested
in the previous paragraph. We will study this idea in more detail later when
we introduce the quantum theory and a rigorous notion of a quantum information source.

Some of the techniques of quantum Shannon theory are the direct
\textit{quantum} analog of the techniques from classical information theory.
We use the law of large numbers and the notion of the typical subspace, but we
require generalizations of measures from the classical world to determine how
\textquotedblleft close\textquotedblright\ two different quantum states are.
One measure, the \textit{fidelity}%
\index{fidelity}%
, has the operational interpretation that it gives the probability that one
quantum state would pass a test for being another. The \textit{trace distance
}%
\index{trace distance}%
is another distance measure that is perhaps more similar to a classical
distance measure---its classical analog is a measure of the closeness of two
probability distributions. The techniques in quantum Shannon theory also
reside firmly in the quantum theory and have no true classical analog for some
cases. Some of the techniques will seem similar to those in the classical
world, but the answer to some of the fundamental questions in quantum Shannon
theory are rather different from some of the answers in the classical world.
It is the purpose of this book to explore the answers to the fundamental
questions of quantum Shannon theory, and we now begin to ask what kinds of
tasks we can perform.

\subsection{Operational Tasks in Quantum Shannon Theory}

Quantum Shannon theory has several resources that two parties can exploit in a
quantum information-processing task. Perhaps the most natural quantum resource
is a \textit{noiseless qubit channel}. We can think of this resource as some
medium through which a physical qubit can travel without being affected by any
noise. One example of a noiseless qubit channel could be the free space
through which a photon travels, where it ideally does not interact with any
other particles along the way to its destination.\footnote{We should be
careful to note here that this is not actually a perfect channel because even
empty space can be noisy in quantum mechanics, but nevertheless, it is a
simple physical example to imagine.}

A \textit{noiseless classical bit channel} is a special case of a noiseless
qubit channel because we can always encode classical information into quantum
states. For the example of a photon, we can say that horizontal polarization
corresponds to a \textquotedblleft0\textquotedblright\ and vertical
polarization corresponds to a \textquotedblleft1.\textquotedblright\ We refer
to the dynamic resource of a noiseless classical bit channel as a
\textit{cbit}, in order to distinguish it from the noiseless qubit channel.

Perhaps the most intriguing resource that two parties can share is noiseless
entanglement. Any entanglement resource is a \textit{static resource} because
it is one that they share. Examples of static resources in the classical world
are an information source that we would like to compress or a common secret
key that two parties may possess. We actually have a way of measuring
entanglement that we discuss later on, and for this reason, we can say that a
sender and receiver have bits of entanglement or \textit{ebits}.

Entanglement turns out to be a useful resource in many quantum communication
tasks. One example where it is useful is in the
\index{quantum teleportation}%
teleportation protocol, where a sender and receiver use one ebit and two
classical bits to transmit one qubit faithfully. This protocol is an example
of the extraordinary power of noiseless entanglement. The name
\textquotedblleft teleportation\textquotedblright\ is really appropriate for
this protocol because the physical qubit vanishes from the sender's station
and appears at the receiver's station after the receiver obtains the two
transmitted classical bits. We will see later on that a noiseless qubit
channel can generate the other two noiseless resources, but it is impossible
for each of the other two noiseless resources to generate the noiseless qubit
channel. In this sense, the noiseless qubit channel is the strongest of the
three unit resources.

The first quantum information-processing task that we have discussed is
Schumacher compression. The goal of this task is to use as few noiseless qubit
channels as possible in order to transmit the output of a quantum information
source reliably. After we understand Schumacher compression in a technical
sense, the main focus of this book is to determine what quantum
information-processing tasks a sender and receiver can accomplish with the use
of a noisy quantum channel. The first and perhaps simplest task is to
determine how much classical information a sender can transmit reliably to a
receiver, by using a noisy quantum channel a large number of times. This task
is known as HSW\ coding, named after its discoverers Holevo, Schumacher, and
Westmoreland. The HSW\ coding theorem
\index{HSW theorem}%
is one quantum generalization of Shannon's channel coding theorem (the latter
overviewed in the next chapter). We can also assume that a sender and receiver
share some amount of noiseless entanglement prior to communication. They can
then use this noiseless entanglement in addition to a large number of uses of
a noisy quantum channel. This task is known as \textit{entanglement-assisted
classical communication}
\index{entanglement-assisted!classical communication}%
over a noisy quantum channel. The capacity theorem corresponding to this task
again highlights one of the marvelous features of entanglement. It shows that
entanglement gives a boost to the amount of noiseless classical communication
we can generate using a noisy quantum channel---the classical capacity is
generally higher with entanglement assistance than without it.

One of the most important theorems for quantum Shannon theory
\index{quantum capacity theorem}%
is the \textit{quantum channel capacity theorem}. Any proof of a capacity
theorem consists of two parts: one part establishes a lower bound on the
capacity and the other part establishes an upper bound. If the two bounds
coincide, then we have a characterization of the capacity in terms of these
bounds. The lower bound on the quantum capacity is colloquially known as the
\index{LSD theorem}%
LSD\ coding theorem,\footnote{The LSD\ coding theorem does not refer to the
synthetic crystalline compound, lysergic acid diethylamide (which one may
potentially use as a hallucinogenic drug), but refers rather to
\cite{PhysRevA.55.1613}, \cite{capacity2002shor}, and \cite{ieee2005dev}, all
of whom gave separate proofs of the lower bound on the quantum capacity with
increasing standards of rigor.} and it gives a characterization of the highest
rate at which a sender can transmit quantum information reliably over a noisy
quantum channel so that a receiver can recover it perfectly. The rate is
generally lower than the classical capacity because it is more difficult to
keep quantum information intact. As we have said before, it is possible to
encode classical information into quantum states, but this classical encoding
is only a special case of a quantum state. In order to preserve quantum
information, we have to be able to preserve arbitrary quantum states, not
merely a classical encoding within a quantum state.

The pinnacle of this book is in Chapter~\ref{chap:quantum-capacity}\ where we
finally reach our study of the quantum capacity theorem. All efforts and
technical developments in preceding chapters have this goal in
mind.\footnote{One goal of this book is to unravel the mathematical machinery
behind Devetak's proof of the quantum channel coding theorem
\citep{ieee2005dev}.} Our first coding theorem in the dynamic setting is the
HSW\ coding theorem. A rigorous study of this coding theorem lays an important
foundation---an understanding of the structure of a code for reliable
communication over a noisy quantum channel. The method for the HSW\ coding
theorem applies to the \textquotedblleft entanglement-assisted classical
capacity theorem,\textquotedblright\ which is one building block for other
protocols in quantum Shannon theory. We then develop a more complex coding
structure for sending private classical information over a noisy quantum
channel. In \textit{private coding}, we are concerned with coding in such a
way that the intended receiver can learn the transmitted message perfectly,
but a third-party eavesdropper cannot learn anything about what the sender
transmits to the intended receiver. This study of the private classical
capacity may seem like a detour at first, but it is closely linked with our
ultimate aim. The coding structure developed for sending private information
proves to be indispensable for understanding the structure of a quantum code.
There are strong connections between the goals of keeping classical
information private and keeping quantum information coherent. In the private
coding scenario, the goal is to avoid leaking any information to an
eavesdropper so that she cannot learn anything about the transmission. In the
quantum coding scenario, we can think of quantum noise as resulting from the
environment learning about the transmitted quantum information and this act of
learning disturbs the quantum information. This effect is related to the
information--disturbance trade-off that is fundamental in quantum information
theory. If the environment learns something about the state being transmitted,
there is inevitably some sort of noisy disturbance that affects the quantum
state. Thus, we can see a correspondence between private coding and quantum
coding. In quantum coding, the goal is to avoid leaking any information to the
environment because the avoidance of such a leak implies that there is no
disturbance to the transmitted state. So the role of the environment in
quantum coding is similar to the role of the eavesdropper in private coding,
and the goal in both scenarios is to decouple either the environment or
eavesdropper from the picture. It is then no coincidence that private codes
and quantum codes have a similar structure. In fact, we can say that the
quantum code inherits its structure from that of the private
code.\footnote{There are other methods of formulating quantum codes using
random subspaces
\citep{capacity2002shor,qcap2008first,qcap2008fourth,qcap2008second}, but we
prefer the approach of Devetak because we learn about other aspects of quantum
Shannon theory, such as the private capacity, along the way to proving the
quantum capacity theorem.}

We also consider \textquotedblleft trade-off\textquotedblright\ problems in
addition to discussing the quantum capacity theorem.
Chapter~\ref{chap:coh-comm-noisy}\ is another high point of the book,
featuring a whole host of results that emerge by combining several of the
ideas from previous chapters. The most appealing aspect of this chapter is
that we can construct virtually all of the protocols in quantum Shannon theory
from just one idea in Chapter~\ref{chap:EA-classical}. Also,
Chapter~\ref{chap:coh-comm-noisy} provides partial answers to many practical
questions concerning information transmission over noisy quantum channels.
Some example questions are as follows:

\begin{itemize}
\item How much quantum and classical information can a noisy quantum channel transmit?

\item An entanglement-assisted noisy quantum channel can transmit more
classical information than an unassisted one, but how much entanglement is
really necessary?

\item Does noiseless classical communication help in transmitting quantum
information reliably over a noisy quantum channel?

\item How much entanglement can a noisy quantum channel generate when aided by
classical communication?

\item How much quantum information can a noisy quantum channel communicate
when aided by entanglement?
\end{itemize}

These are examples of trade-off problems because they involve a noisy quantum
channel and either the consumption or generation of a noiseless resource. For
every combination of the generation or consumption of a noiseless resource,
there is a corresponding coding theorem that states what rates are achievable
(and in some cases optimal). Some of these trade-off questions admit
interesting answers, but some of them do not. Our final aim in these trade-off
questions is to determine the full triple trade-off solution where we study
the optimal ways of combining all three unit resources (classical
communication, quantum communication, and entanglement) with a noisy quantum channel.

The coding theorems for a noisy quantum channel are just as important as (if not
more important than) Shannon's classical coding theorems because they determine
the ultimate capabilities of information processing in a world where the
postulates of quantum theory apply. It is thought that quantum theory is the
ultimate theory underpinning all physical phenomena, and any theory of gravity
will have to incorporate the quantum theory in some fashion. Thus, it is
reasonable that we should be focusing our efforts now on a full Shannon theory
of quantum information processing in order to determine the tasks that these
systems can accomplish. In many physical situations, some of the assumptions
of quantum Shannon theory may not be justified (such as an independent and
identically distributed quantum channel), but nevertheless, it provides an
ideal setting in which we can determine the capabilities of these physical systems.

\subsection{History of Quantum Shannon Theory}

We conclude this introductory chapter by giving a brief overview of the
problems that researchers were thinking about that ultimately led to the
development of quantum Shannon theory.

\textbf{The 1970s}---The first researchers in quantum information theory were
concerned with transmitting classical data by optical means. They were
ultimately led to a quantum formulation because they wanted to transmit
classical information by means of a coherent laser. \textit{Coherent states}
are special quantum states that a coherent laser ideally emits. Glauber
provided a full quantum-mechanical theory of coherent states in two seminal
papers \citep{PhysRev.130.2529,PhysRev.131.2766}, for which he shared the
Nobel Prize in 2005 \citep{nobel2005glauber}. The first researchers of quantum
information theory were Helstrom, Gordon, Stratonovich, and Holevo.
\cite{gordon1964} first conjectured an important bound for our ability to
access classical information from a quantum system and \cite{levitin1969}
stated it without proof. \cite{Holevo73,japan1973holevo} later provided a
proof that the bound holds. This important bound is now known as the Holevo
bound%
\index{Holevo bound}%
, and it is useful in proving converse theorems (theorems concerning
optimality) in quantum Shannon theory. The simplest (yet rough) statement of
the Holevo bound states that it is not possible to transmit more than one
classical bit of information using a noiseless qubit channel, while at the
same time being able to decode it reliably---i.e., we get \textit{one cbit per
qubit}. \cite{Hel76} developed a full theory of quantum detection and quantum
estimation and published a textbook that discusses this theory.
\cite{Fannes73} contributed a useful continuity property of the entropy that
is also useful in proving converse theorems in quantum Shannon theory. Wiesner
also used the uncertainty principle to devise a notion of \textquotedblleft
quantum money\textquotedblright\ in 1970, but unfortunately, his work was not
accepted upon its initial submission. This work was \textit{way} ahead of its
time, and it was only until much later that it was accepted
\citep{Wiesner:1983:78}. Wiesner's ideas paved the way for the BB84 protocol
for quantum key distribution. Fundamental entropy inequalities, such as the
strong subadditivity of quantum entropy \citep{LR73,PhysRevLett.30.434} and
the monotonicity of quantum relative entropy \citep{Lindblad1975}, were proved
during this time as well. These entropy inequalities generalize the Holevo
bound and are foundational for establishing optimality theorems in quantum
Shannon theory.

\cite{Park1970}  produced one of the simplest, yet most profound, results that
is crucial to quantum information science (\cite{nat1982} and \cite{D82} also proved this result
many years later). He proved the \textit{no-cloning theorem},
\index{no-cloning theorem}%
showing that the postulates of the quantum theory imply the impossibility of
universally cloning quantum states. Given an arbitrary unknown quantum state,
it is impossible to build a device that can copy this state. This result has
deep implications for the processing of quantum information and shows a strong
divide between information processing in the quantum world and that in the
classical world. We will prove this theorem in
Chapter~\ref{chap:noiseless-quantum-theory} and use it time and again in our
reasoning. For a history of the no-cloning theorem, see \cite{O18}.


\textbf{The 1980s}---The 1980s witnessed only a few advances in quantum
information theory because just a handful of researchers thought about the
possibilities of linking quantum theory with information-theoretic ideas. The
Nobel Prize-winning physicist Richard Feynman published an interesting 1982
article that was one of the first to discuss computing with quantum-mechanical
systems \citep{ijtp1982feynman}. His interest was in using a quantum computer
to simulate quantum-mechanical systems---he figured there should be a speed-up
over a classical simulation if we instead use one quantum system to simulate
another. This work is less quantum Shannon theory than it is quantum
computing, but it is still a landmark because Feynman began to think about
exploiting the actual quantum information in a physical system, rather than
just using quantum systems to process classical information as the researchers
in the 1970s suggested.

The work of Wiesner on conjugate coding inspired an IBM\ physicist named
Charles Bennett. \cite{bb84} published a groundbreaking paper that detailed
the first quantum communication protocol: the BB84 protocol. This protocol
shows how a sender and a receiver can exploit a quantum channel to establish a
secret key. The security of this protocol, roughly speaking, relies on the
uncertainty principle%
\index{quantum key distribution}%
. If any eavesdropper tries to learn about the random quantum data that they
use to establish the secret key, this act of learning inevitably disturbs the
transmitted quantum data and the two parties can discover this disturbance by
noticing the change in the statistics of random sample data. The secret key
generation capacity of a noisy quantum channel is inextricably linked to the
BB84 protocol, and we study this capacity problem in detail when we study the
ability of quantum channels to communicate private information. Interestingly,
the physics community largely ignored the BB84 paper when Bennett and Brassard
first published it, likely because they presented it at an engineering
conference and the merging of physics and information had not yet taken effect.

\textbf{The 1990s}---The 1990s were a time of much increased activity in
quantum information science, perhaps some of the most exciting years with many
seminal results. One of the first major results was from Ekert. He published a
different way for performing quantum key distribution, this time relying on
the strong correlations of entanglement \citep{Ekert:1991:661}. He was unaware
of the BB84 protocol when he was working on his entanglement-based quantum key
distribution. The physics community embraced this result and a short time later,
Ekert and Bennett and Brassard became aware of each other's respective works
\citep{bbe92}. Bennett, Brassard, and Mermin later showed a sense in which
these two seemingly different schemes are equivalent \citep{Bennett:1992:557}.
Bennett later developed the B92 protocol for quantum key distribution using
any two non-orthogonal quantum states \citep{Bennett:1992:3121}.

Two of the most profound results that later impacted quantum Shannon theory
appeared in the early 1990s. First, \cite{PhysRevLett.69.2881} devised the
super-dense coding protocol. This protocol consumes one noiseless ebit of
entanglement and one noiseless qubit channel to simulate two noiseless
classical bit channels. Let us compare this result to that of Holevo. Holevo's
bound states that we can reliably send only one classical bit per qubit, but
the super-dense coding protocol states that we can double this rate if we
consume entanglement as well. Thus, entanglement is the enabler in this
protocol that boosts the classical rate beyond that possible with a noiseless
qubit channel alone. The next year, Bennett and some other coauthors reversed
the operations in the super-dense coding protocol to devise a protocol that
has more profound implications. They devised the \textit{teleportation
protocol} \citep{PhysRevLett.70.1895}---this
\index{quantum teleportation}%
protocol consumes two classical bit channels and one ebit to transmit a qubit
from a sender to a receiver. Right now, without any technical development yet,
it may be unclear how the qubit gets from the sender to the receiver. The original
authors described it as the \textquotedblleft disembodied transport of a
quantum state.\textquotedblright\ Suffice it for now to say that it is the
unique properties of entanglement (in particular, the ebit) that enable this
disembodied transport to occur. Yet again, it is entanglement that is the
resource that enables this protocol, but let us be careful not to overstate
the role of entanglement. Entanglement alone does not suffice for implementing
quantum teleportation. These protocols show that it is the unique combination
of entanglement and quantum communication or entanglement and classical
communication that yields these results. These two noiseless protocols are
cornerstones of quantum Shannon theory, originally suggesting that there are
interesting ways of combining the resources of classical communication,
quantum communication, and entanglement to formulate uniquely quantum
protocols and leading the way to more exotic protocols that combine the
different noiseless resources with noisy resources. Simple questions
concerning these protocols lead to quantum Shannon-theoretic protocols. In
super-dense coding, how much classical information can Alice send if the
quantum channel becomes noisy? What if the entanglement is noisy? In
teleportation, how much quantum information can Alice send if the classical
channel is noisy? What if the entanglement is noisy? Researchers addressed
these questions quite a bit after the original super-dense coding and
teleportation protocols were available, and we discuss these important
questions in this book.

The year 1994 was a landmark for quantum information science.
\cite{Shor:1994:124} published his algorithm that factors a number in
polynomial time---this algorithm gives an exponential speed-up over the best
known classical algorithm. We cannot overstate the importance of this
algorithm for the field. Its major application is to break RSA\ encryption
\citep{rsa}\ because the security of that encryption algorithm relies on the
computational difficulty of factoring a large number. This breakthrough
generated wide interest in the idea of a quantum computer and started the
quest to build one and study its capabilities.

Initially, much skepticism met the idea of building a practical quantum
computer \citep{landauer1995,PhysRevA.51.992}. Some experts thought that it
would be impossible to overcome errors that inevitably occur during quantum
interactions, due to the coupling of a quantum system with its environment.
Shor met this challenge by devising the first quantum error-correcting code
\citep{PhysRevA.52.R2493}\ and a scheme for fault-tolerant quantum computation
\citep{10.1109/SFCS.1996.548464}. His paper on quantum error correction is the
one most relevant for quantum Shannon theory. At the end of this paper, he
posed the idea of the quantum capacity of a noisy quantum channel as the
highest rate at which a sender and receiver can maintain the fidelity of a
quantum state when it is sent over a large number of uses of the noisy
channel. This open problem set the main task for researchers interested in
quantum Shannon theory. A flurry of theoretical activity then ensued in
quantum error correction
\citep{PhysRevA.54.1098,PhysRevLett.77.793,PhysRevLett.77.198,PhysRevA.54.1862,thesis97gottesman,PhysRevLett.78.405,ieee1998calderbank}\ and
fault-tolerant quantum computation
\citep{258579,kitaev1997,preskill1998,KLZ98}. These two areas are now
important subfields within quantum information science, but we do not focus on
them in any detail in this book.

Schumacher published a critical paper in 1995 as well \citep{PhysRevA.51.2738}
(we discussed some of his contributions in the previous section). This paper
gave the first informational notion of a qubit, and it even established the
now ubiquitous term \textquotedblleft qubit.\textquotedblright\ He proved the
quantum analog of Shannon's source coding theorem, giving the ultimate
compressibility of quantum information. He used the notion of a typical
subspace as an analogy of Shannon's typical set. This notion of a typical
subspace proves to be one of the most crucial ideas for constructing codes in
quantum Shannon theory, just as the notion of a typical set is so crucial for
Shannon's information theory.

Not much later, several researchers began investigating the capacity of a
noisy quantum channel for sending classical information
\citep{PhysRevA.54.1869}. \cite{Hol98}\ and \cite{PhysRevA.56.131}%
\ independently proved that the Holevo information of a quantum channel is an
achievable rate for classical communication over it. They appealed to
Schumacher's notion of a typical subspace and constructed channel codes for
sending classical information. The proof looks somewhat similar to the proof
of Shannon's channel coding theorem (discussed in the next chapter) after
taking a few steps away from it. The proof of the converse theorem proceeds
somewhat analogously to that of Shannon's theorem, with the exception that one
of the steps uses Holevo's bound from 1973. In hindsight, it is perhaps
somewhat surprising that it took over 20 years between the appearance of the
proof of Holevo's bound (the main step in the converse proof) and the
appearance of a direct coding theorem for sending classical information.

The quantum capacity theorem is perhaps one of the most fundamental theorems
of quantum Shannon theory. Initial work by several researchers provided some
insight into the quantum capacity theorem
\citep{BBPSSW96EPP,BDSW96,PhysRevLett.78.3217,PhysRevLett.80.5695}, and a
series of papers established an upper bound on the quantum
capacity~\citep{PhysRevA.54.2614,PhysRevA.54.2629,BNS98,BKN98}. For the lower
bound, \cite{PhysRevA.55.1613} was the first to construct an idea for a proof,
but it turns out that his proof was more of a heuristic argument.
\cite{capacity2002shor} then followed with another proof of the lower bound,
and some of Shor's ideas appeared much later in a full publication
\citep{qcap2008fourth}. \cite{ieee2005dev}\ and \cite{1050633} independently
solved the private capacity theorem at approximately the same time (with the
publication of the CWY\ paper appearing a year after Devetak's arXiv post).
Devetak took the proof of the private capacity theorem a step further and
showed how to apply its techniques to construct a quantum code that achieves a
good lower bound on the quantum capacity, while also providing an alternate,
cleaner proof of the converse theorem \citep{ieee2005dev}. It is Devetak's
technique that we mainly explore in this book because it provides some insight
into the coding structure (however, we also explore a different technique via
the entanglement-assisted classical capacity theorem).

\textbf{The 2000s}---In recent years, we have had many advancements in quantum
Shannon theory (technically some of the above contributions were in the 2000s,
but we did not want to break the continuity of the history of the quantum
capacity theorem). One major result was the proof of the entanglement-assisted
classical capacity theorem---it is the noisy version of the super-dense coding
protocol where the quantum channel is noisy
\citep{PhysRevLett.83.3081,ieee2002bennett,Hol01a}. This theorem assumes that
Alice and Bob share unlimited entanglement and they exploit the entanglement
and the noisy quantum channel to send classical information.

A few fantastic results have arisen in recent years. Horodecki, Oppenheim, and
Winter showed the existence of a
\index{state merging}%
state-merging protocol \citep{Horodecki:2005:673,Horodecki:2007:107}. This
protocol gives the minimum rate at which Alice and Bob consume noiseless qubit
channels in order for Alice to send her share of a quantum state to Bob. This
rate is the conditional quantum entropy---the protocol thus gives an
operational interpretation to this entropic quantity. What was most
fascinating about this result is that the conditional quantum entropy can be
negative in quantum Shannon theory. Prior to their work, no one really
understood what it meant for the conditional quantum entropy to become
negative \citep{W78,HH94,CA97}, but this state-merging result gave a
compelling operational interpretation. A negative rate implies that Alice and
Bob gain the ability for future quantum communication, instead of consuming
quantum communication as when the rate is positive.

Another fantastic result came
\index{superactivation}%
from \citep{science2008smith}. Suppose we have two noisy quantum channels and
each of them individually has zero capacity to transmit quantum information.
One would expect intuitively that the \textquotedblleft joint quantum
capacity\textquotedblright\ (when using them together) would also have zero
ability to transmit quantum information. But this result is not generally the
case in the quantum world. It is possible for some particular noisy quantum
channels with no individual quantum capacity to have a non-zero joint quantum
capacity. It is not clear yet how we might practically take advantage of such
a \textquotedblleft superactivation\textquotedblright\ effect, but the result
is nonetheless fascinating, counterintuitive, and not yet fully understood.

The latter part of the 2000s saw the unification of quantum Shannon theory.
The resource inequality framework was the first step because it unified many
previously known results into one formalism \citep{DHW03,DHW05RI}. Devetak,
Harrow, and Winter provided a family tree for quantum Shannon theory and
showed how to relate the different protocols in the tree to one another. We
will go into the theory of resource inequalities in some detail throughout
this book because it provides a tremendous conceptual simplification when
considering coding theorems in quantum Shannon theory. In fact, the last
chapter of this book contains a concise summary of many of the major quantum
Shannon-theoretic protocols in the language of resource inequalities.
\cite{ADHW06FQSW} published a work showing a sense in which the mother
protocol of the family tree can generate the father protocol. We have seen
unification efforts in the form of triple trade-off coding theorems
\citep{PhysRevA.68.062319,HW08GFP,HW09}. These theorems give the optimal
combination of classical communication, quantum communication, entanglement,
and an asymptotic noisy resource for achieving a variety of quantum
information-processing tasks.

We have also witnessed the emergence of a study of network quantum Shannon
theory. Some authors have tackled the quantum broadcasting paradigm
\citep{GS07,GSE07,DH2006,YHD2006}, where one sender transmits to multiple
receivers. A multiple-access quantum channel has many senders and one
receiver. Some of the same authors (and others) have tackled multiple-access
communication
\citep{Winter01,Yard05a,PhysRevA.72.062312,YDH05,YHD05MQAC,itit2008hsieh,CH08}.
This network quantum Shannon theory should become increasingly important as we
get closer to the ultimate goal of a quantum Internet.

Quantum Shannon theory has now established itself as an important and distinct
field of study. The next few chapters discuss the concepts that will prepare
us for tackling some of the major results in quantum Shannon theory.

\chapter{Classical Shannon Theory}

\label{chap:classical-shannon-theory}We cannot overstate the importance of
Shannon's contribution to modern science. His introduction of the field of
information theory and his solutions to its two main theorems demonstrate that
his ideas on communication were far beyond the other prevailing ideas in this
domain around 1948.

In this chapter, our aim is to discuss Shannon's two main contributions in a
descriptive fashion. The goal of this high-level discussion is to build up the
intuition for the problem domain of information theory and to understand the
main concepts before we delve into the analogous quantum information-theoretic
ideas. We avoid going into deep technical detail in this chapter, leaving such
details for later chapters where we formally prove both classical and quantum
Shannon-theoretic coding theorems. We do use some mathematics from probability
theory (namely, the law of large numbers).

We will be delving into the technical details of this chapter's material in
later chapters (specifically, Chapters~\ref{chap:info-entropy},
\ref{chap:additivity}, and \ref{chap:classical-typicality}). Once you have
reached later chapters that develop some more technical details, it might be
helpful to turn back to this chapter to get an overall flavor for the
motivation of the development.

\section{Data Compression}%

\index{data compression}%
We first discuss the problem of data compression. Those who are familiar with
the Internet have used several popular data formats such as JPEG, MPEG, ZIP,
GIF, etc. All of these file formats have corresponding algorithms for
compressing the output of an information source. A first glance at the
compression problem might lead one to believe that it is not possible to
compress the output of the information source to an arbitrarily small size,
and Shannon proved that this is the case. This result is the content of
Shannon's first noiseless coding theorem.

\subsection{An Example of Data Compression}

We begin with a simple example that illustrates the concept of an information
source. We then develop a scheme for coding this source so that it requires
fewer bits to represent its output faithfully.

Suppose that Alice is a sender and Bob is a receiver. Suppose further that a
noiseless bit channel connects Alice to Bob---a noiseless bit channel is one
that transmits information perfectly from sender to receiver, e.g., Bob
receives \textquotedblleft0\textquotedblright\ if Alice transmits
\textquotedblleft0\textquotedblright\ and Bob receives \textquotedblleft%
1\textquotedblright\ if Alice transmits~\textquotedblleft1.\textquotedblright%
\ Alice and Bob would like to minimize the number of times that they use this
noiseless channel because it is expensive to use it.

Alice would like to use the noiseless channel to communicate information to
Bob. Suppose that an information source randomly chooses from four symbols $\{
a,b,c,d\} $ and selects them with a skewed probability distribution:%
\begin{align}
\Pr\{ a\}  &  =1/2,\label{eq-intro:simple-info-source-1}\\
\Pr\{ b\}  &  =1/8,\\
\Pr\{ c\}  &  =1/4,\\
\Pr\{ d\}  &  =1/8. \label{eq-intro:simple-info-source}%
\end{align}
So it is clear that the symbol $a$ is the most likely one, $c$ the next
likely, and both $b$ and $d$ are least likely. We make the additional
assumption that the information source chooses each symbol independently of
all previous ones and chooses each with the same probability distribution
above. After the information source makes a selection, it gives the symbol to
Alice for coding.

A noiseless bit channel accepts only bits as input---it does not accept the
symbols $a$, $b$, $c$, $d$ as input. So, Alice has to encode her information
into bits. Alice could use the following coding scheme:%
\begin{equation}
a\rightarrow00,\ \ b\rightarrow01,\ \ c\rightarrow10,\ \ d\rightarrow11,
\end{equation}
where each binary representation of a letter is a \textit{codeword}. How do we
measure the performance of a particular coding scheme? The expected length of
a codeword is one way to measure performance. For the above example, the
expected length is equal to two bits. This measure reveals a problem with the
above scheme---the scheme does not take advantage of the skewed nature of the
distribution of the information source because each codeword has the same length.

One might instead consider a scheme that uses shorter codewords for symbols
that are more likely and longer codewords for symbols that are less
likely.\footnote{Such coding schemes are common. Samuel F. B. Morse employed
this idea in his popular Morse code. Also, in the movie \textit{The Diving
Bell and the Butterfly}, a writer becomes paralyzed with \textquotedblleft
locked-in\textquotedblright\ syndrome so that he can only blink his left eye.
An assistant then develops a \textquotedblleft blinking code\textquotedblright%
\ where she reads a list of letters in French, beginning with the most
commonly used letter and ending with the least commonly used letter. The
writer blinks when she says the letter he wishes and they finish an entire
book using this coding scheme.} Then the expected length of a codeword with
such a scheme should be shorter than that in the former scheme. The following
coding scheme gives an improvement in the expected length of a codeword:%
\begin{equation}
a\rightarrow0,\ \ b\rightarrow110,\ \ c\rightarrow10,\ \ d\rightarrow111.
\label{eq-intro:good-source-code}%
\end{equation}
This scheme has the advantage that any coded sequence is uniquely decodable.
For example, suppose that Bob obtains the following sequence:%
\begin{equation}
0011010111010100010.
\end{equation}
Bob can parse the above sequence as%
\begin{equation}
0\ 0\ 110\ 10\ 111\ 0\ 10\ 10\ 0\ 0\ 10,
\end{equation}
and determine that Alice transmitted the message%
\begin{equation}
aabcdaccaac. \label{eq-intro:source-sequence}%
\end{equation}
We can calculate the expected length of this coding scheme as follows:%
\begin{equation}
\frac{1}{2}(1)+\frac{1}{8}\left(  3\right)  +\frac{1}{4}(2)+\frac{1}{8}\left(
3\right)  =\frac{7}{4}.
\end{equation}
This scheme is thus more efficient because its expected length is $7/4$ bits
as opposed to two bits. It is a \textit{variable-length code} because the
number of bits in each codeword depends on the source symbol.

\subsection{A Measure of Information}

The above scheme suggests a way to measure information. Consider the
probability distribution
in~\eqref{eq-intro:simple-info-source-1}--\eqref{eq-intro:simple-info-source}.
Would we be more surprised to learn that the information source produced the
symbol $a$ or to learn that it produced the symbol $d$? The answer is $d$
because the source is less likely to produce it. Let $X$ denote a random
variable with distribution given in
\eqref{eq-intro:simple-info-source-1}--\eqref{eq-intro:simple-info-source}.
One measure of the surprise of symbol $x\in\left\{  a,b,c,d\right\}  $ is%
\begin{equation}
i( x) \equiv\log\left(  \frac{1}{p_{X}( x) }\right)  =-\log\left(  p_{X}( x)
\right)  ,
\end{equation}
where the logarithm is base two---this convention implies the units of this
measure are bits. This measure of surprise has the desirable property that it
is higher for lower probability events and lower for higher probability
events. Here, we take after Shannon, and we
\index{information content}%
name $i( x) $ the \textit{information content} or \textit{surprisal}
\index{surprisal}
of the symbol $x$. Observe that the length of each codeword in the coding
scheme in \eqref{eq-intro:good-source-code}\ is equal to the information
content of its corresponding symbol.

The information content has another desirable property called
\textit{additivity}. Suppose that the information source produces two symbols
$x_{1}$ and $x_{2}$, with corresponding random variables $X_{1}$ and $X_{2}$.
The probability for this event is $p_{X_{1}X_{2}}(x_{1},x_{2})$ and the joint
distribution factors as $p_{X_{1}}(x_{1})p_{X_{2}}(x_{2})$ if we assume the
source is \textit{memoryless}---that it produces each symbol independently.
The information content of the two symbols $x_{1}$ and $x_{2}$ is additive
because%
\begin{align}
i(x_{1},x_{2})  &  =-\log\left(  p_{X_{1}X_{2}}(x_{1},x_{2})\right) \\
&  =-\log\left(  p_{X_{1}}(x_{1})p_{X_{2}}(x_{2})\right) \\
&  =-\log\left(  p_{X_{1}}(x_{1})\right)  -\log\left(  p_{X_{2}}(x_{2})\right)
\\
&  =i(x_{1})+i(x_{2}).
\end{align}
In general, additivity is a desirable property for any information measure. We
will return to the issue of additivity in many different contexts in this book
(especially in Chapter~\ref{chap:additivity}).

The expected information content of the information source is%
\begin{equation}
\sum_{x}p_{X}(x)i(x)=-\sum_{x}p_{X}(x)\log\left(  p_{X}(x)\right)  .
\label{eq-intro:entropy}%
\end{equation}
The above quantity is so important in information theory that we give it a
name: the \textit{entropy} of the
\index{Shannon entropy}%
information source. The reason for its importance is that the entropy and
variations of it appear as the answer to many questions in information theory.
For example, in the above coding scheme, the expected length of a codeword is
the entropy of the information source because%
\begin{align}
-\frac{1}{2}\log\frac{1}{2}  &  -\frac{1}{8}\log\frac{1}{8}-\frac{1}{4}%
\log\frac{1}{4}-\frac{1}{8}\log\frac{1}{8}\nonumber\\
&  =\frac{1}{2}(1)+\frac{1}{8}\left(  3\right)  +\frac{1}{4}(2)+\frac{1}%
{8}\left(  3\right) \\
&  =\frac{7}{4}.
\end{align}
It is no coincidence that we chose the particular coding scheme in
\eqref{eq-intro:good-source-code}. The effectiveness of the scheme in this
example is related to the structure of the information source---the number of
symbols is a power of two and the probability of each symbol is the reciprocal
of a power of two.

\subsection{Shannon's Source Coding Theorem}

\label{intro:compression}%
\index{Shannon compression}%
\index{data compression!theorem}%
The next question to ask is whether there is any other scheme that can achieve
a better compression rate than the scheme in
\eqref{eq-intro:good-source-code}. This question is the one that Shannon asked
in his first coding theorem. To answer this question, we consider a more
general information source and introduce a notion of Shannon, the idea of the
\textit{set of typical sequences}.

We can represent a more general information source with a random variable $X$
whose realizations $x$ are \textit{letters} in an \textit{alphabet}
$\mathcal{X}$. Let $p_{X}( x) $ be the probability mass function associated
with random variable $X$, so that the probability of realization $x$ is
$p_{X}( x) $. Let $H( X) $ denote the entropy of the information source:%
\begin{equation}
H( X) \equiv-\sum_{x\in\mathcal{X}}p_{X}( x) \log\left(  p_{X}( x) \right)  .
\end{equation}
The entropy $H( X) $ is also the entropy of the random variable $X$. Another
way of writing it is $H( p) $, but we use the more common notation $H( X) $
throughout this book.

The information content $i( X) $ of random variable $X$ is%
\begin{equation}
i( X) \equiv-\log\left(  p_{X}( X) \right)  ,
\label{eq-intro:random-info-content}%
\end{equation}
and is itself a random variable. There is nothing wrong mathematically here
with having random variable $X$ as the argument to the density function
$p_{X}$, though this expression may seem self-referential at a first glance.
This way of thinking turns out to be useful later. Again, the expected
information content of $X$ is equal to the entropy:%
\begin{equation}
\mathbb{E}_{X}\left\{  -\log\left(  p_{X}( X) \right)  \right\}  =H( X) .
\end{equation}

\begin{exercise}
\label{ex-intro:uniform-entropy}Show that the entropy of a uniform random
variable is equal to $\log\left\vert \mathcal{X}\right\vert $, where
$\left\vert \mathcal{X}\right\vert $ is the size of the variable's alphabet.
\end{exercise}

\begin{figure}[ptb]
\begin{center}
\includegraphics[
width=4.8456in
]{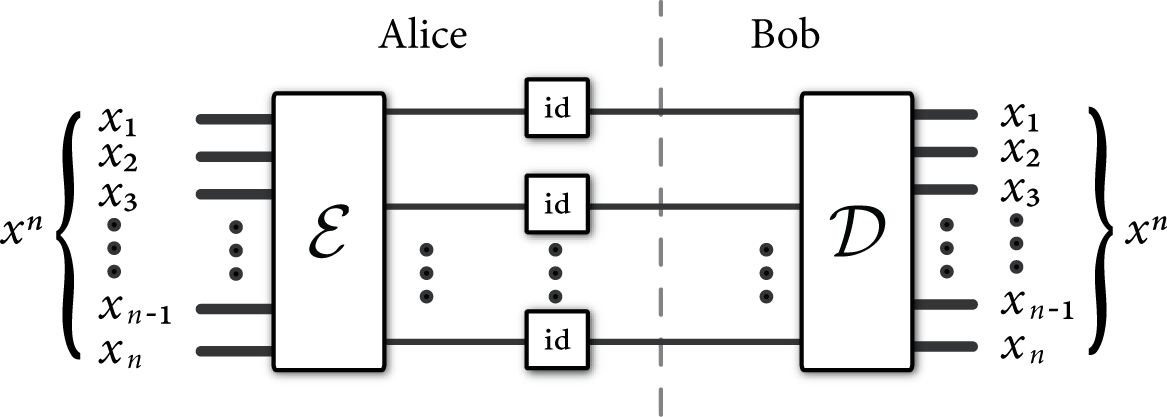}
\end{center}
\caption{This figure depicts Shannon's idea for a classical source code. The
information source emits a long sequence $x^{n}$ to Alice. She encodes this
sequence as a block with an encoder $\mathcal{E}$ and produces a codeword
whose length is less than that of the original sequence $x^{n}$ (indicated by
fewer lines coming out of the encoder $\mathcal{E}$). She transmits the
codeword over noiseless bit channels (each indicated by ``id'' which stands
for the identity bit channel) and Bob receives it. Bob decodes the transmitted
codeword with a decoder $\mathcal{D}$ and produces the original sequence that
Alice transmitted, only if their chosen code is good, in the sense that the
code has a small probability of error.}%
\label{fig-intro:classical-source-code}%
\end{figure}

We now turn to source coding the above information source. We \textit{could}
associate a binary codeword for each symbol $x$ as we did in the scheme in
\eqref{eq-intro:good-source-code}. But this scheme may lose some efficiency if
the size of our alphabet is not a power of two or if the probabilities are not
a reciprocal of a power of two as they are in our nice example. Shannon's
breakthrough idea was to let the source emit a large number of realizations
and then code the emitted data as a large block, instead of coding each symbol
as the above example does. This technique is called \textit{block coding}.
Shannon's other insight was to allow for a slight error in the compression
scheme, but to show that this error vanishes as the block size becomes
arbitrarily large. To make the block coding scheme more clear, Shannon
suggests to let the source emit the following sequence:%
\begin{equation}
x^{n}\equiv x_{1}x_{2}\cdots x_{n},
\end{equation}
where $n$ is a large number that denotes the size of the block of emitted data
and $x_{i}$, for all $i=1,\ldots,n$, denotes the $i$th emitted symbol. Let
$X^{n}$ denote the random variable associated with the sequence $x^{n}$, and
let $X_{i}$ be the random variable for the $i$th symbol $x_{i}$.
Figure~\ref{fig-intro:classical-source-code} depicts Shannon's idea for a
classical source code. 

An important assumption regarding this information source is that it is
independent and identically distributed (i.i.d.). The i.i.d.~assumption means
that each random variable $X_{i}$ has the same distribution as random variable
$X$, and we use the index $i$ merely to track to which symbol $x_{i}$ the
random variable $X_{i}$ corresponds. Under the i.i.d.~assumption, the
probability of any given emitted sequence $x^{n}$ factors as%
\begin{align}
p_{X^{n}}(x^{n})  &  =p_{X_{1},X_{2},\ldots,X_{n}}(x_{1},x_{2},\ldots,x_{n})\\
&  =p_{X_{1}}(x_{1})p_{X_{2}}(x_{2})\cdots p_{X_{n}}(x_{n})\\
&  =p_{X}(x_{1})p_{X}(x_{2})\cdots p_{X}(x_{n})\\
&  =\prod\limits_{i=1}^{n}p_{X}(x_{i}). \label{eq-intro:IID-factor}%
\end{align}
The above rule from probability theory results in a remarkable simplification
of the mathematics. Suppose that we now label the letters in the alphabet
$\mathcal{X}$ as $a_{1}$, \ldots, $a_{\left\vert \mathcal{X}\right\vert }$ in
order to distinguish the letters from the realizations. Let $N(a_{i}|x^{n})$
denote the number of occurrences of the letter $a_{i}$ in the sequence $x^{n}$
(where $i=1,\ldots,\left\vert \mathcal{X}\right\vert $). As an example,
consider the sequence in \eqref{eq-intro:source-sequence}. The quantities
$N(a_{i}|x^{n})$ for this example are%
\begin{align}
N(a|x^{n})  &  =5,\\
N(b|x^{n})  &  =1,\\
N(c|x^{n})  &  =4,\\
N(d|x^{n})  &  =1.
\end{align}
We can rewrite the result in \eqref{eq-intro:IID-factor} as%
\begin{equation}
p_{X^{n}}(x^{n})=\prod\limits_{i=1}^{n}p_{X}(x_{i})=\prod\limits_{i=1}%
^{\left\vert \mathcal{X}\right\vert }p_{X}(a_{i})^{N(a_{i}|x^{n})}.
\label{eq-intro:IID-factor-simpler}%
\end{equation}
Keep in mind that we are allowing the length $n$ of the emitted sequence to be
extremely large, so that it is much larger than the alphabet size $\left\vert
\mathcal{X}\right\vert $:%
\begin{equation}
n\gg\left\vert \mathcal{X}\right\vert .
\end{equation}
The formula on the right in \eqref{eq-intro:IID-factor-simpler} is much
simpler than the formula in \eqref{eq-intro:IID-factor} because it has fewer
iterations of multiplications. There is a sense in which the i.i.d.~assumption
allows us to permute the sequence $x^{n}$ as%
\begin{equation}
x^{n}\rightarrow\underbrace{a_{1}\cdots a_{1}}_{N(a_{1}|x^{n})}\underbrace
{a_{2}\cdots a_{2}}_{N(a_{2}|x^{n})}\cdots\underbrace{a_{\left\vert
\mathcal{X}\right\vert }\cdots a_{\left\vert \mathcal{X}\right\vert }%
}_{N(a_{\left\vert \mathcal{X}\right\vert }|x^{n})},
\end{equation}
because the probability calculation is invariant under this permutation. We
introduce the above way of thinking right now because it turns out to be
useful later when we develop some ideas in quantum Shannon theory
(specifically in Section~\ref{sec-ct:strong-cond-typ}). Thus, the formula on
the right in \eqref{eq-intro:IID-factor-simpler} characterizes the probability
of any given sequence $x^{n}$.

The above discussion applies to a particular sequence $x^{n}$\ that the
information source emits. Now, we would like to analyze the behavior of a
\textit{random sequence }$X^{n}$ that the source emits, and this distinction
between the realization $x^{n}$ and the random variable $X^{n}$ is important.
In particular, let us consider the sample average of the information content
of the random sequence $X^{n}$ (divide the information content of $X^{n}$ by
$n$ to get the sample average):%
\begin{equation}
-\frac{1}{n}\log\left(  p_{X^{n}}( X^{n}) \right)  .
\label{eq-intro:info-content-rand-seq}%
\end{equation}
It may seem strange at first glance that $X^{n}$, the argument of the
probability mass function $p_{X^{n}}$ is itself a random variable, but this
type of expression is perfectly well defined mathematically. (This
self-referencing type of expression is similar to
\eqref{eq-intro:random-info-content}, which we used to calculate the entropy.)
For reasons that will become clear shortly, we call the above quantity the
\textit{sample entropy}
\index{sample entropy}%
of the random sequence $X^{n}$.

Suppose now that we use the function $N( a_{i}|\bullet) $ to calculate the
number of appearances of the letter $a_{i}$ in the random sequence $X^{n}$. We
write the desired quantity as $N( a_{i}|X^{n}) $ and note that it is also a
random variable, whose random nature derives from that of $X^{n}$. We can
reduce the expression in \eqref{eq-intro:info-content-rand-seq}\ to the
following one with some algebra and the result
in~\eqref{eq-intro:IID-factor-simpler}:%
\begin{align}
-\frac{1}{n}\log\left(  p_{X^{n}}( X^{n}) \right)   &  =-\frac{1}{n}%
\log\left(  \prod\limits_{i=1}^{\left\vert \mathcal{X}\right\vert }p_{X}(
a_{i}) ^{N( a_{i}|X^{n}) }\right) \\
&  =-\frac{1}{n}\sum_{i=1}^{\left\vert \mathcal{X}\right\vert }\log\left(
p_{X}( a_{i}) ^{N( a_{i}|X^{n}) }\right) \\
&  =-\sum_{i=1}^{\left\vert \mathcal{X}\right\vert }\frac{N( a_{i}|X^{n}) }%
{n}\log\left(  p_{X}( a_{i}) \right)  .
\end{align}
We stress again that the above quantity is random.

Is there any way that we can determine the behavior of the above sample
entropy when $n$ becomes large?\ Probability theory gives us a way. The
expression $N(a_{i}|X^{n})/n$ represents an empirical distribution for the
letters $a_{i}$ in the alphabet $\mathcal{X}$. As $n$ becomes large, one form
of the law of large numbers%
\index{law of large numbers}
states that it is overwhelmingly likely that a random sequence has its
empirical distribution $N(a_{i}|X^{n})/n$ close to the true distribution
$p_{X}(a_{i})$, and conversely, it is highly unlikely that a random sequence
does not satisfy this property. Thus, a random emitted sequence $X^{n}$ is
highly likely to satisfy the following condition for all $\delta>0$ as $n$
becomes large:%
\begin{equation}
\lim_{n\rightarrow\infty}\Pr\left\{  \left\vert -\frac{1}{n}\log\left(
p_{X^{n}}(X^{n})\right)  -\sum_{i=1}^{\left\vert \mathcal{X}\right\vert }%
p_{X}(a_{i})\log\left(  \frac{1}{p_{X}(a_{i})}\right)  \right\vert \leq
\delta\right\}  =1.
\end{equation}
The quantity $-\sum_{i=1}^{\left\vert \mathcal{X}\right\vert }p_{X}(a_{i}%
)\log\left(  p_{X}(a_{i})\right)  $ is none other than the entropy $H(X)$ so
that the above expression is equivalent to the following one for all
$\delta>0$:%
\begin{equation}
\lim_{n\rightarrow\infty}\Pr\left\{  \left\vert -\frac{1}{n}\log\left(
p_{X^{n}}(X^{n})\right)  -H(X)\right\vert \leq\delta\right\}  =1.
\end{equation}
Another way of stating this property is as follows:

\begin{quote}
It is highly likely that the information source emits a sequence whose sample
entropy is close to the true entropy, and conversely, it is highly unlikely
that the information source emits a sequence that does not satisfy this
property.\footnote{Do not fall into the trap of thinking \textquotedblleft The
possible sequences that the source emits are typical
sequences.\textquotedblright\ That line of reasoning is quantitatively far
from the truth. In fact, what we can show is much different because the set of
typical sequences is much smaller than the set of all possible sequences.}
\end{quote}

\begin{figure}[ptb]
\begin{center}
\includegraphics[
width=4.8456in
]{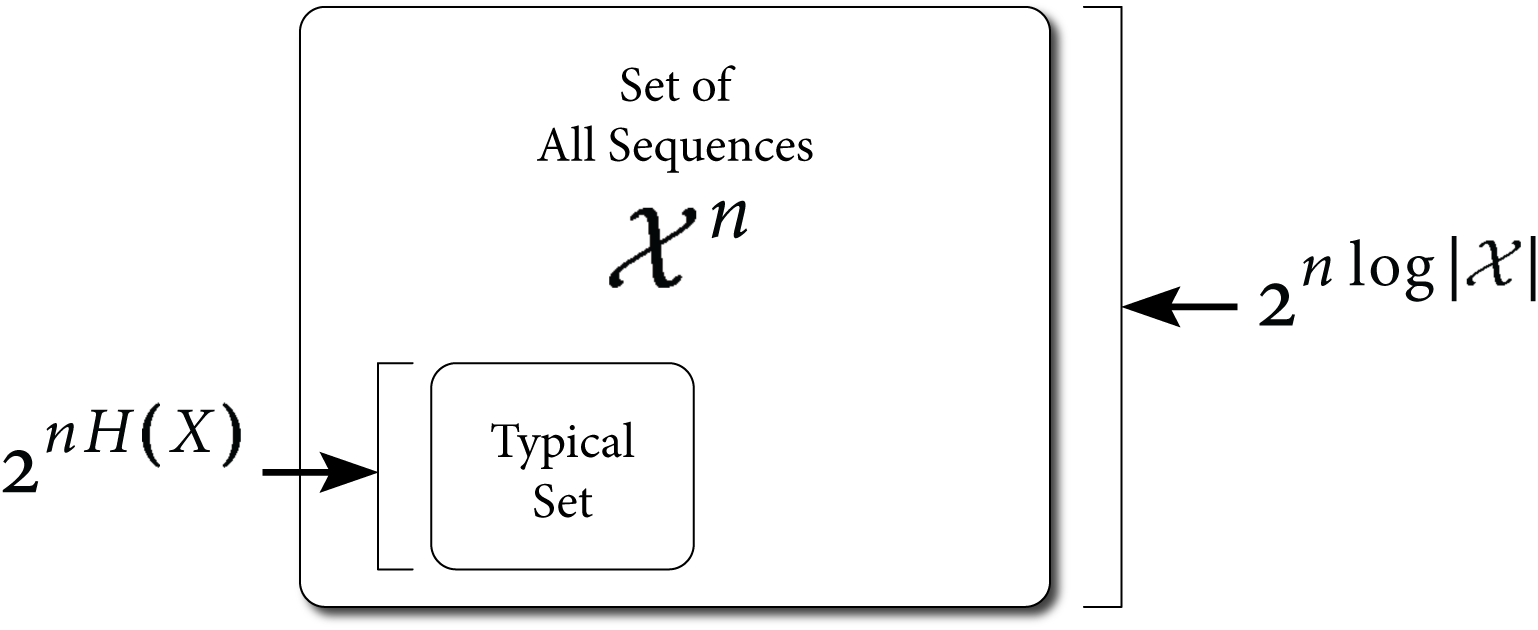}
\end{center}
\caption{This figure indicates that the typical set is much smaller
(exponentially smaller) than the set of all sequences. The typical set is
roughly the same size as the set of all sequences only when the entropy $H( X)
$ of the random variable $X$ is equal to $\log\left\vert \mathcal{X}%
\right\vert $---implying that the distribution of random variable $X$ is
uniform.}%
\label{fig-intro:typical-set}%
\end{figure}

Now we consider a particular realization $x^{n}$ of the random sequence
$X^{n}$. We name a particular sequence $x^{n}$ a \textit{typical sequence}
\index{typical sequence}%
if its sample entropy is close to the true entropy $H( X) $ and the set of all
typical sequences is the \textit{typical set}. Fortunately for data
compression, the set of typical sequences is not too large. In
Chapter~\ref{chap:classical-typicality} on typical sequences, we prove that
the size of this set is much smaller than the set of all sequences. We accept
it for now (and prove later) that the size of the typical set is$~\approx
2^{nH( X) }$, whereas the size of the set of all sequences is equal to
$\left\vert \mathcal{X}\right\vert ^{n}$. We can rewrite the size of the set
of all sequences as%
\begin{equation}
\left\vert \mathcal{X}\right\vert ^{n}=2^{n\log\left\vert \mathcal{X}%
\right\vert }.
\end{equation}
Comparing the size of the typical set to the size of the set of all sequences,
the typical set is exponentially smaller than the set of all sequences
whenever the random variable is not equal to the uniform random variable.
Figure~\ref{fig-intro:typical-set}\ illustrates this
concept.We summarize these two crucial properties of the typical set and
give another that we prove later:

\begin{property}
[Unit Probability]The probability that an emitted sequence is typical
approaches one as $n$ becomes large. Another way of stating this property is
that the typical set has almost all of the probability.
\end{property}

\begin{property}
[Exponentially Smaller Cardinality]The size of the typical set is
$\approx2^{nH( X) }$ and is exponentially smaller than the size $2^{n\log
\left\vert \mathcal{X}\right\vert }$ of the set of all sequences whenever
random variable $X$ is not uniform.
\end{property}

\begin{property}
[Equipartition]The probability of a \textit{particular} typical sequence is
roughly uniform$~\approx2^{-nH( X) }$. (The probability $2^{-nH( X) }$ is easy
to calculate if we accept that the typical set has all of the probability, its
size is $2^{nH( X) }$, and the distribution over typical sequences is uniform.)
\end{property}

These three
\index{asymptotic equipartition theorem}%
properties together are collectively known as the \textit{asymptotic
equipartition theorem}. The word \textquotedblleft
asymptotic\textquotedblright\ applies because the theorem exploits the
asymptotic limit when $n$ is large and the word \textquotedblleft
equipartition\textquotedblright\ refers to the third property above.

With the above notions of a typical set under our belt, a strategy for
compressing information should now be clear. The strategy is to compress only
the typical sequences that the source emits. We simply need to establish a
one-to-one encoding function that maps from the set of typical sequences (size
$2^{nH(X)}$) to the set of all binary strings of length $nH(X)$ (this set also
has size $2^{nH(X)}$). If the source emits an atypical sequence, we declare an
error. This coding scheme is reliable in the asymptotic limit because the
probability of an error\ event vanishes as $n$ becomes large, due to the unit
probability property in the asymptotic equipartition theorem. We measure the
rate of this block coding scheme as follows:%
\begin{equation}
\text{compression rate}\equiv\frac{\text{\#\ of noiseless channel bits}%
}{\text{\# of source symbols}}.
\end{equation}
For the case of Shannon compression, the number of noiseless channel bits is
equal to$~nH(X)$ and the number of source symbols is equal to$~n$. Thus, the
compression rate is equal to the entropy $H(X)$.

One may then wonder whether this rate of data compression is the best that we
can do---whether this rate is optimal (we could achieve a lower rate of
compression if it were not optimal). In fact, the above rate is the optimal
rate at which we can compress information, and this is the content of
Shannon's data compression theorem. We hold off on a formal proof of
optimality for now and delay it until we reach Chapter~\ref{chap:schumach}. We
just mention for now that this data compression protocol gives an
\textit{operational interpretation} to the Shannon entropy $H( X) $ because it
appears as the optimal rate of data compression.

The above discussion highlights the common approach in information theory for
establishing a coding theorem. Proving a coding theorem has two
parts---traditionally called the \textit{direct coding theorem} and the
\textit{converse theorem}. First, we give a coding scheme that can achieve a
given rate for an information-processing task. This first part includes a
direct construction of a coding scheme, hence the name \textit{direct coding
theorem}. The statement of the direct coding theorem for the above task is as follows:

\begin{quote}
If the rate of compression is greater than the entropy of the source, then
there exists a coding scheme that can achieve lossless data compression in the
sense that it is possible to make the probability of error for incorrectly
decoding arbitrarily small.
\end{quote}

The second task is to prove that the rate from the direct coding theorem is
optimal---that we cannot do any better than the suggested rate. We
traditionally call this part the converse theorem because it corresponds to
the converse of the above statement:

\begin{quote}
If there exists a coding scheme that can achieve lossless data compression
with arbitrarily small probability of decoding error, then the rate of
compression is greater than the entropy of the source.
\end{quote}

The techniques used in proving each part of the coding theorem are completely
different. For most coding theorems in information theory, we can prove the
direct coding theorem by appealing to the ideas of typical sequences and large
block sizes. That this technique gives a good coding scheme is directly
related to the asymptotic equipartition properties that govern the behavior of
random sequences of data as the length of the sequence becomes large. The
proof of a converse theorem relies on information inequalities that give tight
bounds on the entropic quantities appearing in the coding constructions. We
spend some time with information inequalities in
Chapter~\ref{chap:info-entropy}\ to build up our ability to prove converse theorems.

Sometimes, in the course of proving a direct coding theorem, one may think to
have found the optimal rate for a given information-processing task. Without a
matching converse theorem, it is not generally clear that the suggested rate
is optimal. So, always prove converse
theorems!\label{sec-intro:data-compression}

\section{Channel Capacity}

\label{sec-ccs:channel-cap}The next issue that we overview is the transmission
of information over a noisy classical channel. We begin with a standard
example---transmitting a single bit of information over a noisy bit-flip channel.

\subsection{An Example of an Error Correction Code}

We again have our protagonists, Alice and Bob, as respective sender and
receiver. This time, however, we assume that a noisy classical channel
connects them, so that information transfer is not reliable. Alice and Bob
realize that a noisy channel is not as expensive as a noiseless one, but it
still is expensive for them to use. For this reason, they would like to
maximize the amount of information that Alice can communicate reliably to Bob,
where reliable communication implies that there is a negligible probability of
error when transmitting this information.%
\begin{figure}
[ptb]
\begin{center}
\includegraphics[
width=2.0358in
]%
{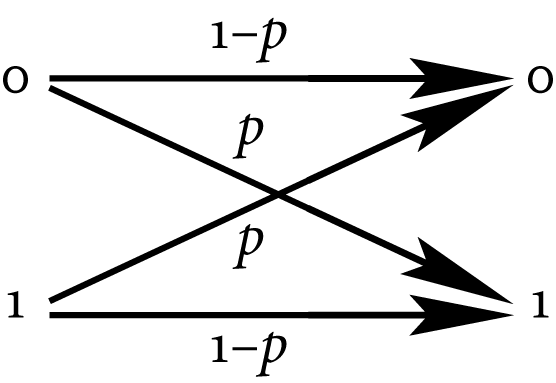}%
\caption{This figure depicts the action of the bit-flip channel. It preserves
the input bit with probability $1-p$ and flips it with probability $p$.}%
\label{fig-intro:bit-flip-channel}%
\end{center}
\end{figure}

The simplest example of a noisy classical channel is a bit-flip channel, with
the technical%
\index{binary symmetric channel}
name \textit{binary symmetric channel}. This channel flips the input bit with
probability $p$ and leaves it unchanged with probability $1-p$.
Figure~\ref{fig-intro:bit-flip-channel}\ depicts the action of the bit-flip
channel. Alice and Bob are allowed to use the channel multiple times, and in
so doing, we assume that the channel behaves independently from one use to the
next and behaves in the same random way as described above. For this reason,
we describe the multiple uses of the channel as i.i.d.~channels. This
assumption will be helpful when we go to the asymptotic regime of a large
number of uses of the channel.

Suppose that Alice and Bob just use the channel as is---Alice just sends plain
bits to Bob. This scheme works reliably only if the probability of bit-flip
error vanishes. So, Alice and Bob could invest their best efforts into
engineering the physical channel to make it reliable. But, generally, it is
not possible to engineer a classical channel this way for physical or
logistical reasons. For example, Alice and Bob may only have local computers
at their ends and may not have access to the physical channel because the
telephone company may control the channel.

Alice and Bob can employ a \textquotedblleft systems
engineering\textquotedblright\ solution to this problem rather than an
engineering of the physical channel. They can redundantly encode information
in a way such that Bob can have a higher probability of determining what Alice
is sending, effectively reducing the level of noise on the channel. A simple
example of this systems engineering solution is the three-bit majority vote
code.
\index{repetition code}%
Alice and Bob employ the following encoding:%
\begin{equation}
0\rightarrow000,\ \ \ \ \ \ 1\rightarrow111,
\end{equation}
where both \textquotedblleft000\textquotedblright\ and \textquotedblleft%
111\textquotedblright\ are \textit{codewords}. Alice transmits the codeword
\textquotedblleft000\textquotedblright\ with three independent uses of the
noisy channel if she really wants to communicate a \textquotedblleft%
0\textquotedblright\ to Bob and she transmits the codeword \textquotedblleft%
111\textquotedblright\ if she wants to send a \textquotedblleft%
1\textquotedblright\ to him. The \textit{physical} or \textit{channel} bits
are the actual bits that she transmits over the noisy channel, and the
\textit{logical} or \textit{information} bits are those that she intends for
Bob to receive. In our example, \textquotedblleft0\textquotedblright\ is a
logical bit and \textquotedblleft000\textquotedblright\ corresponds to the
physical bits.

The rate of this scheme is 1/3 because it encodes one information bit. The
term \textquotedblleft rate\textquotedblright\ is perhaps a misnomer for
coding scenarios that do not involve sending bits in a time sequence over a
channel. We may just as well use the majority vote code to store one bit in a
memory device that may be unreliable. Perhaps a more universal term is
\textit{efficiency}. Nevertheless, we follow convention and use the term
\textit{rate} throughout this book.

Of course, the noisy bit-flip channel does not always transmit these codewords
without error. So how does Bob decode in the case of error? He simply takes a
\textit{majority vote} to determine the transmitted message---he decodes as
\textquotedblleft0\textquotedblright\ if the number of zeros in the codeword
he receives is greater than the number of ones.%
\begin{table}[tbp] \centering
\begin{tabular}
[c]{c|c}\hline\hline
\textbf{Channel Output} & \textbf{Probability}\\\hline\hline
000 & $( 1-p) ^{3}$\\\hline
001, 010, 100 & $p( 1-p) ^{2}$\\\hline
011, 110, 101 & $p^{2}( 1-p) $\\\hline
111 & $p^{3}$\\\hline\hline
\end{tabular}
\caption{The first column gives the eight possible outputs of the noisy
bit-flip channel when Alice encodes a ``0'' with the majority vote code.
The second column gives the corresponding probability of Bob receiving the particular outputs.}\label{tbl-intro:majority-vote-probs}%
\end{table}%

We now analyze the performance of this simple \textquotedblleft systems
engineering\textquotedblright\ solution.
Table~\ref{tbl-intro:majority-vote-probs}\ enumerates the probability of
receiving every possible sequence of three bits, assuming that Alice transmits
a \textquotedblleft0\textquotedblright\ by encoding it as \textquotedblleft%
000.\textquotedblright\ The probability of no error is $( 1-p) ^{3}$, the
probability of a single-bit error is $3p( 1-p) ^{2}$, the probability of a
double-bit error is $3p^{2}( 1-p) $, and the probability of a total failure is
$p^{3}$. The majority vote solution can \textquotedblleft
correct\textquotedblright\ for no error and it corrects for all single-bit
errors, but it has no ability to correct for double-bit and triple-bit errors.
In fact, it actually incorrectly decodes these latter two scenarios by
\textquotedblleft correcting\textquotedblright\ \textquotedblleft%
011\textquotedblright, \textquotedblleft110\textquotedblright, or
\textquotedblleft101\textquotedblright\ to \textquotedblleft%
111\textquotedblright\ and decoding \textquotedblleft111\textquotedblright\ as
a \textquotedblleft1.\textquotedblright\ Thus, these latter two outcomes are
errors because the code has no ability to correct them. We can employ similar
arguments as above to the case in which Alice transmits a \textquotedblleft%
1\textquotedblright\ to Bob using the majority vote code.

When does this majority vote scheme perform better than no coding at all? It
is exactly when the probability of error with the majority vote code is less
than $p$, the probability of error with no coding. Letting $e$ denote the
event that an error occurs, the probability of error is equal to the following
quantity:%
\begin{equation}
\Pr\{ e\} =\Pr\{ e|0\} \Pr\{ 0\} +\Pr\{ e|1\} \Pr\{ 1\} .
\end{equation}
Our analysis above suggests that the conditional probabilities $\Pr\{ e|0\} $
and $\Pr\{ e|1\} $ are equal for the majority vote code because of the
symmetry in the noisy bit-flip channel. This result implies that the
probability of error is%
\begin{align}
\Pr\{ e\}  &  =3p^{2}( 1-p) +p^{3}\\
&  =3p^{2}-2p^{3}, \label{eq-intro:maj-vote-code-prob-error}%
\end{align}
because $\Pr\{ 0\} +\Pr\{ 1\} =1$. We consider the following inequality to
determine if the majority vote code reduces the probability of error:%
\begin{equation}
3p^{2}-2p^{3}<p.
\end{equation}
This inequality simplifies as%
\begin{align}
0  &  <2p^{3}-3p^{2}+p\\
\therefore0  &  <p\left(  2p-1\right)  \left(  p-1\right)  .
\end{align}
The only values of $p$ that satisfy the above inequality are $0<p<1/2$. Thus,
the majority vote code reduces the probability of error only when $0<p<1/2$,
i.e., when the noise on the channel is not too much. Too much noise has the
effect of causing the codewords to flip too often, throwing off Bob's decoder.

The majority vote code gives a way for Alice and Bob to reduce the probability
of error during their communication, but unfortunately, there is still a
non-zero probability for the noisy channel to disrupt their communication. Is
there any way that they can achieve reliable communication by reducing the
probability of error to zero?

One simple approach to achieve this goal is to exploit the majority vote idea
a second time. They can \textit{concatenate} two instances of the majority
vote code to produce a code with a larger number of physical bits.
Concatenation consists of using one code as an \textquotedblleft
inner\textquotedblright\ code and another as an \textquotedblleft
outer\textquotedblright\ code. There is no real need for us to distinguish
between the inner and outer code in this case because we use the same code for
both the inner and outer code. The concatenation scheme for our case first
encodes the message $i$, where $i\in\left\{  0,1\right\}  $, using the
majority vote code. Let us label the codewords as follows:%
\begin{equation}
\bar{0}\equiv000,\ \ \ \ \ \ \bar{1}\equiv111.
\end{equation}
For the second layer of the concatenation, we encode $\bar{0}$ and $\bar{1}$
with the majority vote code again:%
\begin{equation}
\bar{0}\rightarrow\bar{0}\bar{0}\bar{0},\ \ \ \ \ \ \bar{1}\rightarrow\bar
{1}\bar{1}\bar{1}.
\end{equation}
Thus, the overall encoding of the concatenated scheme is as follows:%
\begin{equation}
0\rightarrow000\ 000\ 000,\ \ \ \ \ \ 1\rightarrow111\ 111\ 111.
\end{equation}
The rate of the concatenated code is 1/9 and smaller than the original rate of
1/3. A simple application of the above performance analysis for the majority
vote code shows that this concatenation scheme reduces the probability of
error as follows:%
\begin{equation}
3[\Pr\{ e\} ]^{2} -2[\Pr\{ e\} ]^{3} =O( p^{4}) .
\end{equation}
The error probability $\Pr\{ e\} $ is in
\eqref{eq-intro:maj-vote-code-prob-error} and $O( p^{4}) $ indicates that the
leading order term of the left-hand side is the fourth power in $p$.

The concatenated scheme achieves a lower probability of error at the cost of
using more physical bits in the code. Recall that our goal is to achieve
reliable communication, where there is no probability of error. A first guess
for achieving reliable communication is to continue concatenating. If we
concatenate again, the probability of error reduces to $O( p^{6}) $, and the
rate drops to 1/27. We can continue indefinitely with concatenating to make
the probability of error arbitrarily small and achieve reliable communication,
but the problem is that the rate approaches zero as the probability of error
becomes arbitrarily small.

The above example seems to show that there is a trade-off between the rate of
the encoding scheme and the desired order of error probability. Is there a way
that we can code information for a noisy channel while maintaining a good rate
of communication?

\subsection{Shannon's Channel Coding Theorem}

Shannon's second breakthrough coding theorem provides an affirmative answer to
the above question. This answer came as a complete shock to communication
researchers in 1948. Furthermore, the techniques that Shannon used in
demonstrating this fact were rarely used by engineers at the time. We give a
broad overview of Shannon's main idea and techniques that he used to prove his
second important theorem---the noisy channel coding theorem%
\index{noisy channel coding theorem}%
.

\subsection{General Model for a Channel Code}

\label{sec-cst:channel-code}We first generalize some of the ideas in the above
example. We still have Alice trying to communicate with Bob, but this time,
she wants to be able to transmit a larger set of messages with asymptotically
perfect reliability, rather than merely sending \textquotedblleft%
0\textquotedblright\ or \textquotedblleft1.\textquotedblright\ Suppose that
she selects messages from a message set $\left[  M\right]  $ that consists of
$M$ messages:%
\begin{equation}
\left[  M\right]  \equiv\left\{  1,\ldots,M\right\}  .
\end{equation}
Suppose furthermore that Alice chooses a particular message $m$ with uniform
probability from the set $\left[  M\right]  $. This assumption of a uniform
distribution for Alice's messages indicates that we do not really care much
about the content of the actual message that she is transmitting. We just
assume total ignorance of her message because we only really care about her
ability to send any message reliably. The message set $\left[  M\right]  $
requires $\log(M)$ bits to represent it, where the logarithm is again base
two. This number becomes important when we calculate the rate of a channel code.

The next aspect of the model that we need to generalize is the noisy channel
that connects Alice to Bob. We used the bit-flip channel before, but this
channel is not general enough for our purposes. A simple way to extend the
channel model is to represent it as a conditional probability distribution
involving an input random variable $X$ and an output random variable$~Y$:%
\begin{equation}
\mathcal{N}:\ \ \ \ \ \ p_{Y|X}( y|x) .
\end{equation}
We use the symbol $\mathcal{N}$ to represent this more general channel model.
One assumption that we make about random variables $X$ and $Y$ is that they
are discrete, but the respective sizes of their outcome sets do not have to
match. The other assumption that we make concerning the noisy channel is that
it is i.i.d. Let $X^{n}\equiv X_{1}X_{2}\cdots X_{n}$ and $Y^{n}\equiv
Y_{1}Y_{2}\cdots Y_{n}$ be the random variables associated with respective
sequences $x^{n}\equiv x_{1}x_{2}\cdots x_{n}$ and $y^{n}\equiv y_{1}%
y_{2}\cdots y_{n}$. If Alice inputs the sequence $x^{n}$ to the $n$ inputs of
$n$ respective uses of the noisy channel, a possible output sequence may be
$y^{n}$. The i.i.d.~assumption allows us to factor the conditional probability
of the output sequence $y^{n}$:%
\begin{align}
p_{Y^{n}|X^{n}}( y^{n}|x^{n})  &  =p_{Y_{1}|X_{1}}( y_{1}|x_{1})
p_{Y_{2}|X_{2}}( y_{2}|x_{2}) \cdots p_{Y_{n}|X_{n}}( y_{n}|x_{n})\\
&  =p_{Y|X}( y_{1}|x_{1}) p_{Y|X}( y_{2}|x_{2}) \cdots p_{Y|X}( y_{n}|x_{n})\\
&  =\prod\limits_{i=1}^{n}p_{Y|X}( y_{i}|x_{i}) .
\end{align}
The technical name of this more general channel model is a \textit{discrete
memoryless channel}.

A coding scheme or \textit{code} translates all of Alice's messages into
codewords that can be input to $n$ i.i.d.~uses of the noisy channel. For
example, suppose that Alice selects a message $m$ to encode. We can write the
codeword corresponding to message $m$ as $x^{n}( m) $ because the input to the
channel is some codeword that depends on $m$.

\begin{figure}[ptb]
\begin{center}
\includegraphics[
width=4.8456in
]{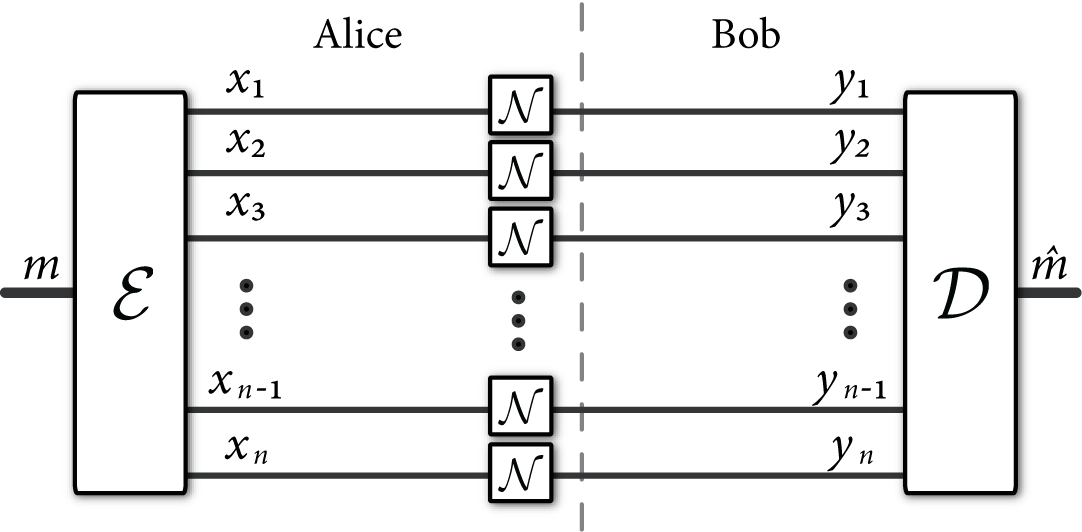}
\end{center}
\caption{This figure depicts Shannon's idea for a classical channel code.
Alice chooses a message $m$ from a message set $\left[  M\right]
\equiv\left\{  1,\ldots,M\right\}  $. She encodes the message $m$ with an
encoding operation $\mathcal{E}$. This encoding operation assigns a codeword
$x^{n}$ to the message $m$ and inputs the codeword $x^{n}$ to a large number
of i.i.d.\ uses of a noisy channel $\mathcal{N}$. The noisy channel randomly
corrupts the codeword $x^{n}$ to a sequence $y^{n}$. Bob receives the
corrupted sequence $y^{n}$ and performs a decoding operation $\mathcal{D}$ to
estimate the codeword $x^{n}$. This estimate of the codeword $x^{n}$ then
produces an estimate $\hat{m}$ of the message that Alice transmitted. A
reliable code has the property that Bob can decode each message $m\in\left[
M\right]  $ with a vanishing probability of error when the block length~$n$
becomes large.}%
\label{fig-intro:classical-channel-coding}%
\end{figure}The last part of the model involves Bob receiving the corrupted
codeword $y^{n}$ over the channel and determining a potential codeword $x^{n}$
with which it should be associated. We do not get into any details just yet
for this last decoding part---imagine for now that it operates similarly to
the majority vote code example.
Figure~\ref{fig-intro:classical-channel-coding}\ displays Shannon's model of
communication that we have described.

We calculate the $rate$ of a given coding scheme as follows:%
\begin{equation}
\operatorname{rate}\equiv\frac{\text{\# of message bits}}{\text{\#\ of channel
uses}}.
\end{equation}
In our model, the rate of a given coding scheme is%
\begin{equation}
R=\frac{1}{n}\log(M),
\end{equation}
where $\log(M)$ is the number of bits needed to represent any message in the
message set $\left[  M\right]  $ and $n$ is the number of channel uses. The
\textit{capacity} of a noisy channel is the highest rate at which it can
communicate information reliably.

We also need a way to determine the performance of any given code. Here, we
list several measures of performance. Let $\mathcal{C}\equiv\left\{
x^{n}(m)\right\}  _{m\in\left[  M\right]  }$ represent a code that Alice and
Bob choose, where $x^{n}(m)$ denotes each codeword corresponding to the
message $m$. Let $p_{e}(m,\mathcal{C})$ denote the probability of error when
Alice transmits a message $m\in\left[  M\right]  $ using the code
$\mathcal{C}$. We denote the average probability of error as%
\begin{equation}
\bar{p}_{e}(\mathcal{C})\equiv\frac{1}{M}\sum_{m=1}^{M}p_{e}(m,\mathcal{C}).
\end{equation}
The maximal probability of error is%
\begin{equation}
p_{e}^{\ast}(\mathcal{C})\equiv\max_{m\in\left[  M\right]  }p_{e}%
(m,\mathcal{C}).
\end{equation}
Our ultimate aim is to make the maximal probability of error $p_{e}^{\ast
}(\mathcal{C})$ arbitrarily small, but the average probability of error
$\bar{p}_{e}(\mathcal{C})$ is important in the analysis. These two performance
measures are related---the average probability of error is small if the
maximal probability of error is. Perhaps surprisingly, the maximal probability
is small for at least half of the messages if the average probability of error
is. We make this statement more quantitative in the following exercise.

\begin{exercise}
\label{ex-intro:expurgation}Let $\varepsilon\in[0,1/2]$ and let $p_{e}(
m,\mathcal{C}) $ denote the probability of error when Alice transmits a
message $m\in\left[  M\right]  $ using the code $\mathcal{C}$. Use Markov's
inequality
\index{Markov's inequality}%
to prove that the following upper bound on the average probability of error:%
\begin{equation}
\frac{1}{M}\sum_{m}p_{e}( m,\mathcal{C}) \leq\varepsilon
\end{equation}
implies the following upper bound for at least half of the messages $m$:%
\begin{equation}
p_{e}( m,\mathcal{C}) \leq2\varepsilon.
\end{equation}

\end{exercise}

You may have wondered why we use the random sequence $X^{n}$ to model the
inputs to the channel. We have already stated that Alice's message is a
uniform random variable, and the codewords in any coding scheme directly
depend on the message to be sent. For example, in the majority vote code, the
channel inputs are always \textquotedblleft000\textquotedblright\ whenever the
intended message is \textquotedblleft0\textquotedblright\ and similarly for
the channel inputs \textquotedblleft111\textquotedblright\ and the message
\textquotedblleft1\textquotedblright. So why is there a need to overcomplicate
things by modeling the channel inputs as the random variable $X^{n}$ when it
seems like each codeword is a deterministic function of the intended message?
We are not yet ready to answer this question but will return to it shortly.

We should also stress an important point before proceeding with Shannon's
ingenious scheme for proving the existence of reliable codes for a noisy
channel. In the above model, we described essentially two \textquotedblleft
layers of randomness\textquotedblright:

\begin{enumerate}
\item The first layer of randomness is the uniform random variable associated
with Alice's choice of a message.

\item The second layer of randomness is the noisy channel. The output of the
channel is a random variable because we cannot always predict the output of
the channel with certainty.
\end{enumerate}

It is not possible to \textquotedblleft play around\textquotedblright\ with
these two layers of randomness. The random variable associated with Alice's
message is fixed as a uniform random variable because we assume ignorance of
Alice's message. The conditional probability distribution of the noisy channel
is also fixed. We are assuming that Alice and Bob can learn the conditional
probability distribution associated with the noisy channel by estimating it.
Alternatively, we may assume that a third party has knowledge of the
conditional probability distribution and informs Alice and Bob of it in some
way. Regardless of how they obtain the knowledge of the distribution, we
assume that they both know it and that it is fixed.

\subsection{Proof Sketch of Shannon's Channel Coding Theorem}

\label{sec:random-code-idea}We are now ready to present an overview of
Shannon's technique for proving the existence of a code that can achieve the
capacity of a given noisy channel. Some of the methods that Shannon uses in
his outline of a proof are similar to those in the first coding theorem. We
again use the channel a large number of times so that the law of large numbers
from probability theory comes into play and allow for a small probability of
error that vanishes as the number of channel uses becomes large. If the notion
of typical sequences is so important in the first coding theorem, we might
suspect that it should be important in the noisy channel coding theorem as
well. The typical set captures a certain notion of efficiency because it is a
small set when compared to the set of all sequences, but it is the set that
has almost all of the probability. Thus, we should expect this efficiency to
come into play somehow in the channel coding theorem.

The aspect of Shannon's technique for proving the noisy channel coding theorem
that is different from the other ideas in the first theorem is the idea of
\textit{random coding}. Shannon's technique adds a \textit{third} layer of
randomness to the model given above (recall that the first two are Alice's
random message and the random nature of the noisy channel).

The third layer of randomness is to choose the codewords themselves in a
random fashion according to a random variable $X$, where we choose each letter
$x_{i}$ of a given codeword $x^{n}$ independently according to the
distribution $p_{X}( x_{i}) $. It is for this reason that we model the channel
inputs as a random variable. We can then write each codeword as a random
variable $X^{n}( m) $. The probability distribution for choosing a particular
codeword $x^{n}( m) $ is%
\begin{align}
\Pr\left\{  X^{n}( m) =x^{n}( m) \right\}   &  =p_{X_{1},X_{2},\ldots,X_{n}}(
x_{1}( m) ,x_{2}( m) ,\ldots,x_{n}( m) )\\
&  =p_{X}( x_{1}( m) ) p_{X}( x_{2}( m) ) \cdots p_{X}( x_{n}( m) )\\
&  =\prod\limits_{i=1}^{n}p_{X}( x_{i}( m) ) .
\end{align}
The important result to notice is that the probability for a given codeword
factors because we choose the code in an i.i.d.~fashion, and perhaps more
importantly, the distribution of each codeword has no explicit dependence on
the message $m$ with which it is associated. That is, the probability
distribution of the first codeword is exactly the same as the probability
distribution of all of the other codewords. The code $\mathcal{C}$\ itself
becomes a random variable in this scheme for choosing a code randomly. We now
let $\mathcal{C}$ refer to the random variable that represents a random code,
and we let $\mathcal{C}_{0}$ represent any particular deterministic code. The
probability of choosing a particular code$~\mathcal{C}_{0}=\left\{  x^{n}( m)
\right\}  _{m\in\left[  M\right]  }$ is%
\begin{equation}
p_{\mathcal{C}}( \mathcal{C}_{0}) =\prod\limits_{m=1}^{M}\prod\limits_{i=1}%
^{n}p_{X}( x_{i}( m) ) ,
\end{equation}
and this probability distribution again has no explicit dependence on each
message$~m$ in the code$~\mathcal{C}_{0}$.

Choosing the codewords in a random way allows for a dramatic simplification in
the mathematical analysis of the probability of error. One of Shannon's
breakthrough ideas was to analyze the \textit{expectation} of the average
probability of error, where the expectation is with respect to the random
code$~\mathcal{C}$, rather than analyzing the average probability of error
itself. The expectation of the average probability of error is%
\begin{equation}
\mathbb{E}_{\mathcal{C}}\left\{  \bar{p}_{e}( \mathcal{C}) \right\}  .
\end{equation}
This expectation is much simpler to analyze because of the random way that we
choose the code. Consider that%
\begin{equation}
\mathbb{E}_{\mathcal{C}}\left\{  \bar{p}_{e}( \mathcal{C}) \right\}
=\mathbb{E}_{\mathcal{C}}\left\{  \frac{1}{M}\sum_{m=1}^{M}p_{e}(
m,\mathcal{C}) \right\}  .
\end{equation}
Using linearity of the expectation, we can exchange the expectation with the
sum so that%
\begin{equation}
\mathbb{E}_{\mathcal{C}}\left\{  \bar{p}_{e}( \mathcal{C}) \right\}  =\frac
{1}{M}\sum_{m=1}^{M}\mathbb{E}_{\mathcal{C}}\left\{  p_{e}( m,\mathcal{C})
\right\}  .
\end{equation}
Now, the expectation of the probability of error for a particular message $m$
does not actually depend on the message $m$ because the distribution of each
random codeword $X^{n}( m) $ does not explicitly depend on $m$. This line of
reasoning leads to the dramatic simplification because $\mathbb{E}%
_{\mathcal{C}}\left\{  p_{e}( m,\mathcal{C}) \right\}  $ is then the same for
all messages. So we can then say that%
\begin{equation}
\mathbb{E}_{\mathcal{C}}\left\{  p_{e}( m,\mathcal{C}) \right\}
=\mathbb{E}_{\mathcal{C}}\left\{  p_{e}( 1,\mathcal{C}) \right\}  .
\end{equation}
(We could have equivalently chosen any message instead of the first.) We then
have that%
\begin{align}
\mathbb{E}_{\mathcal{C}}\left\{  \bar{p}_{e}( \mathcal{C}) \right\}   &
=\frac{1}{M}\sum_{m=1}^{M}\mathbb{E}_{\mathcal{C}}\left\{  p_{e}(
1,\mathcal{C}) \right\} \\
&  =\mathbb{E}_{\mathcal{C}}\left\{  p_{e}( 1,\mathcal{C}) \right\}  ,
\end{align}
where the last step follows because the quantity $\mathbb{E}_{\mathcal{C}%
}\left\{  p_{e}( 1,\mathcal{C}) \right\}  $ has no dependence on $m$. We now
only have to determine the expectation of the probability of error for
\textit{one message} instead of determining the expectation of the average
error probability of the whole set. This simplification follows because random
coding results in the equality of these two quantities.

Shannon then determined a way to obtain a bound on the expectation of the
average probability of error (we soon discuss this technique briefly) so that%
\begin{equation}
\mathbb{E}_{\mathcal{C}}\left\{  \bar{p}_{e}( \mathcal{C}) \right\}
\leq\varepsilon,
\end{equation}
where $\varepsilon$ is some number $\in(0,1)$ that we can make arbitrarily
small by letting the block size $n$ become arbitrarily large. If it is
possible to obtain a bound on the expectation of the average probability of
error, then surely there exists some deterministic code $\mathcal{C}_{0}$
whose average probability of error meets this same bound:%
\begin{equation}
\bar{p}_{e}( \mathcal{C}_{0}) \leq\varepsilon.
\end{equation}
If it were not so, then the original bound on the expectation would not be
possible. This step is the \textit{derandomization} step of Shannon's proof.
Ultimately, we require a deterministic code with a high rate and arbitrarily
small probability of error and this step shows the \textit{existence} of such
a code. The random coding technique is only useful for simplifying the
mathematics of the proof.

The last step of the proof is
\index{expurgation}%
the \textit{expurgation} step. It is an application of the result of
Exercise~\ref{ex-intro:expurgation}. Recall that our goal is to show the
existence of a high-rate code that has low maximal probability of error. But
so far we only have a bound on the average probability of error. In the
expurgation step, we simply throw out the half of the codewords with the worst
probability of error. Throwing out the worse half of the codewords reduces the
number of messages by a factor of two, but only has a negligible impact on the
rate of the code. Consider that the number of messages is $2^{nR}$ where $R$
is the rate of the code. Thus, the number of messages is $2^{n\left(
R-\frac{1}{n}\right)  }$ after throwing out the worse half of the codewords,
and the rate $R-\frac{1}{n}$ is asymptotically equal to the rate $R$. After
throwing out the worse half of the codewords, the result of
Exercise~\ref{ex-intro:expurgation} shows that the following bound then
applies to the maximal probability of error:%
\begin{equation}
p_{e}^{\ast}(\mathcal{C}_{0})\leq2\varepsilon.
\end{equation}
This last expurgation step ends the analysis of the probability of error.

We now discuss the size of the code that Alice and Bob employ. Recall that the
rate of the code is $R=\log(M)/n$. It is convenient to define the size $M$ of
the message set $\left[  M\right]  $ in terms of the rate $R$. When we do so,
the size of the message set is%
\begin{equation}
M=2^{nR}.
\end{equation}
What is peculiar about the message set size when defined this way is that it
grows exponentially with the number of channel uses. But recall that any given
code exploits $n$ channel uses to send $M$ messages. So when we take the limit
as the number of channel uses tends to infinity, we are implying that there
exists a sequence of codes whose message set size is $M=2^{nR}$ and number of
channel uses is $n$. We are focused on keeping the rate of the code constant
and use the limit $n\rightarrow\infty$ to make the probability of error vanish
for a certain fixed rate $R$.

What is the maximal rate at which Alice can communicate to Bob reliably?\ We
need to determine the number of distinguishable messages that Alice can
reliably send to Bob, and we require the notion of%
\index{conditional typicality}
\textit{conditional typicality} to do so. Consider that Alice chooses
codewords randomly according to random variable $X$ with probability
distribution $p_{X}( x) $. By the asymptotic equipartition theorem, it is
highly likely that each of the codewords that Alice chooses is a typical
sequence with sample entropy close to $H( X) $. In the coding scheme, Alice
transmits a particular codeword $x^{n}$ over the noisy channel and Bob
receives a random sequence $Y^{n}$. The random sequence $Y^{n}$ is a random
variable that depends on $x^{n}$ through the conditional probability
distribution $p_{Y|X}( y|x) $. We would like a way to determine the number of
possible output sequences that are likely to correspond to a particular input
sequence $x^{n}$. A useful entropic quantity for this situation is the
conditional entropy $H( Y|X) $, the technical details of which we leave for
Chapter~\ref{chap:info-entropy}.
\begin{figure}
[ptb]
\begin{center}
\includegraphics[
width=4.8456in
]%
{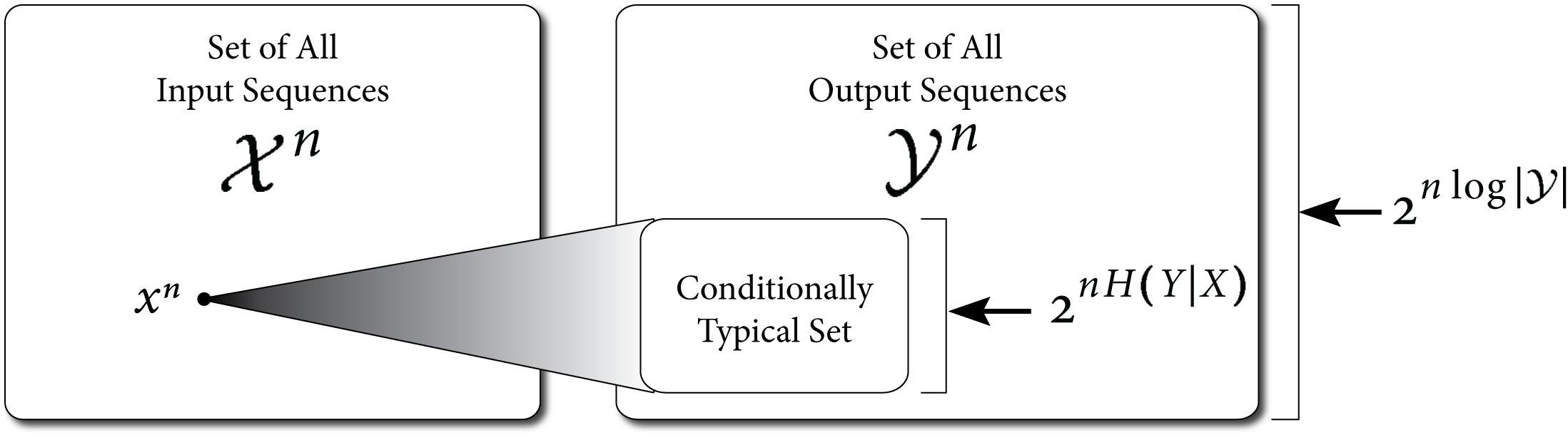}%
\caption{This figure depicts the notion of a conditionally typical set.
Associated to every input sequence $x^{n}$ is a conditionally typical set
consisting of the likely output sequences. The size of this conditionally
typical set is $\approx2^{nH\left(  Y|X\right)  }$. It is exponentially
smaller than the set of all output sequences whenever the conditional random
variable is not uniform.}%
\label{fig-intro:conditional-typical}%
\end{center}
\end{figure}
For now, just think of this conditional entropy as measuring the uncertainty
of a random variable $Y$ when one already knows the value of the random
variable $X$. The conditional entropy $H( Y|X) $ is always less than the
entropy $H( Y) $ unless $X$ and $Y$ are independent. This inequality holds
because knowledge of a correlated random variable $X$ does not increase the
uncertainty about $Y$. It turns out that there is a notion of conditional
typicality (depicted in Figure~\ref{fig-intro:conditional-typical}), similar
to the notion of typicality, and a similar asymptotic equipartition theorem
holds for conditionally typical sequences (more details in
Section~\ref{sec-ct:strong-cond-typ}). This theorem also has three important
properties. For each input sequence $x^{n}$, there is a corresponding
conditionally typical set with the following properties:

\begin{enumerate}
\item It has almost all of the probability---it is highly likely that a random
channel output sequence is conditionally typical given a particular input sequence.

\item Its size is $\approx2^{nH( Y|X) }.$

\item The probability of each conditionally typical sequence $y^{n}$, given
knowledge of the input sequence $x^{n}$, is $\approx2^{-nH( Y|X) }$.
\end{enumerate}

If we disregard knowledge of the input sequence used to generate an output
sequence, the probability distribution that generates the output sequences is%
\begin{equation}
p_{Y}( y) =\sum_{x}p_{Y|X}( y|x) p_{X}( x) .
\end{equation}
We can think that this probability distribution is the one that generates all
the possible output sequences. The likely output sequences are in an output
typical set of size$~2^{nH( Y) }$.%

\begin{figure}
[ptb]
\begin{center}
\includegraphics[
width=4.8456in
]%
{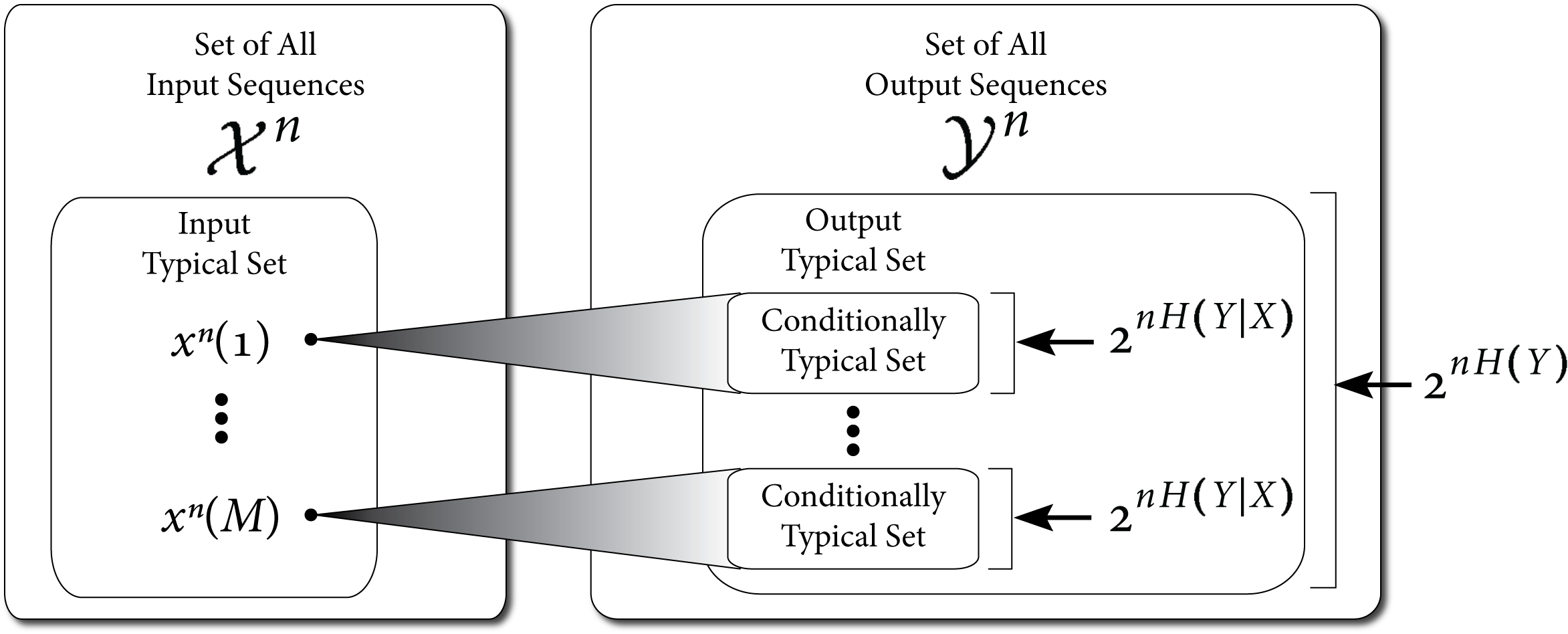}%
\caption{This figure depicts the packing argument that Shannon used. The
channel induces a conditionally typical set corresponding to each codeword
$x^{n}(i)$ where $i\in\left\{  1,\ldots,M\right\}  $. The size of each
conditionally typical output set is $2^{nH(Y|X)}$. The size of the typical set
of all output sequences is $2^{nH(Y)}$. These sizes suggest that we can divide
the output typical set into $M$ conditionally typical sets and be able to
distinguish $M\approx2^{nH(Y)}/2^{nH(Y|X)}$ messages without error.}%
\label{fig-intro:Shannon-packing}%
\end{center}
\end{figure}
We are now in a position to describe the structure of a random code and the
size of the message set. Alice generates $2^{nR}$ codewords according to the
distribution $p_{X}(x)$ and suppose for now that Bob has knowledge of the code
after Alice generates it. Suppose Alice sends one of the codewords over the
channel. Bob is ignorant of the transmitted codeword, so from his point of
view, the output sequences are generated according to the distribution
$p_{Y}(y)$. Bob then employs typical sequence decoding. He first determines if
the output sequence $y^{n}$ is in the typical output set of size $2^{nH(Y)}$.
If not, he declares an error. The probability of this type of error is small
by the asymptotic equipartition theorem. If the output sequence $y^{n}$ is in
the output typical set, he uses his knowledge of the code to determine a
conditionally typical set of size $2^{nH(Y|X)}$ to which the output sequence
belongs. If he decodes an output sequence $y^{n}$ to the wrong conditionally
typical set, then an error occurs. This last type of error suggests how they
might structure the code in order to prevent this type of error from
happening. If they structure the code so that the output conditionally typical
sets do not overlap too much, then Bob should be able to decode each output
sequence $y^{n}$ to a unique input sequence $x^{n}$ with high probability.
This line of reasoning suggests that they should divide the set of output
typical sequences into $M$ sets of conditionally typical output sets, each of
size $2^{nH(Y|X)}$. Thus, if they set the number of messages $M=2^{nR}$ as
follows:%
\begin{equation}
2^{nR}\approx\frac{2^{nH(Y)}}{2^{nH(Y|X)}}=2^{n\left(  H(Y)-H(Y|X)\right)  },
\end{equation}
then our intuition is that Bob should be able to decode correctly with high
probability. Such an argument is a \textquotedblleft packing\textquotedblright%
\ argument because it shows how to pack information into the space of all
output sequences. Figure~\ref{fig-intro:Shannon-packing}\ gives a visual
depiction of the packing argument. It turns out that this intuition is
correct---Alice can reliably send information to Bob if the quantity
$H(Y)-H(Y|X)$ bounds the rate $R$:%
\begin{equation}
R<H(Y)-H(Y|X).
\end{equation}
A rate less than $H(Y)-H(Y|X)$ ensures that we can make the expectation of the
average probability of error as small as we would like. We then employ the
derandomization and expurgation steps, discussed before, in order to show that
there exists a code whose maximal probability of error vanishes as the number
$n$ of channel uses tends to infinity.

The entropic quantity $H( Y) -H( Y|X) $ deserves special attention because it
is another important entropic quantity in information theory. It is the
\textit{mutual information}
\index{mutual information}%
between random variables $X$ and $Y$ and we denote it as%
\begin{equation}
I( X;Y) \equiv H( Y) -H( Y|X) .
\end{equation}
It is important because it arises as the limiting rate of reliable
communication.\ We will discuss its properties in more detail throughout this book.

There is one final step that we can take to strengthen the above coding
scheme. We mentioned before that there are three layers of randomness in the
coding construction:\ Alice's uniform choice of a message, the noisy channel,
and Shannon's random coding scheme. The first two layers of randomness we do
not have control over. But we actually do have control over the last layer of
randomness. Alice chooses the code according to the distribution $p_{X}( x) $.
She can choose the code according to any distribution that she would like. If
she chooses it according to $p_{X}( x) $, the resulting rate of the code is
the mutual information $I( X;Y) $. We will prove later on that the mutual
information $I( X;Y) $ is a concave function of the distribution $p_{X}( x) $
when the conditional distribution $p_{Y|X}( y|x) $ is fixed. Concavity implies
that there is a distribution $p_{X}^{\ast}( x) $ that maximizes the mutual
information. Thus, Alice should choose an optimal distribution $p_{X}^{\ast}(
x) $ when she randomly generates the code, and this choice gives the largest
possible rate of communication that they could have. This largest possible
rate is the \textit{capacity} of the channel and we denote it as%
\begin{equation}
C( \mathcal{N}) \equiv\max_{p_{X}( x) }I( X;Y) . \label{eq-intro:capacity}%
\end{equation}
Our discussion here is just an overview of Shannon's channel capacity theorem.
In Section~\ref{sec-ct:capacity}, we give a full proof of this theorem after
having developed some technical tools needed for a formal proof.

We clarify one more point. In the discussion of the operation of the code, we
mentioned that Alice and Bob both have knowledge of the code. Well, how can
Bob know the code if a noisy channel connects Alice to Bob? One solution to
this problem is to assume that Alice and Bob have unbounded computation on
their local ends. Thus, for a given code that uses the channel $n$ times, they
can both compute the above optimization problem and generate \textquotedblleft
test\textquotedblright\ codes randomly until they determine the best possible
code to employ for $n$ channel uses. They then both end up with the unique,
best possible code for $n$ uses of the given channel. This scheme might be
impractical, but nevertheless, it provides a justification for both of them to
have knowledge of the code that they use. Another solution to this problem is
simply to allow them to meet before going their separate ways in order to
coordinate on the choice of code.

We have said before that the capacity $C( \mathcal{N}) $ is the maximal rate
at which Alice and Bob can communicate. But in our discussion above, we did
not prove optimality---we only proved a direct coding theorem for the channel
capacity theorem. It took quite some time and effort to develop this elaborate
coding procedure---along the way, we repeatedly invoked one of the powerful
tools from probability theory, the law of large numbers. It perhaps seems
intuitive that typical sequence coding and decoding should lead to optimal
code constructions. Typical sequences exhibit some kind of asymptotic
efficiency by being the most likely to occur, but in the general case, their
cardinality is exponentially smaller than the set of all sequences. But is
this intuition about typical sequence coding correct? Is it possible that some
other scheme for coding might beat this elaborate scheme that Shannon devised?
\textit{Without a converse theorem that proves optimality, we would never
know!} If you recall from our previous discussion in
Section~\ref{sec-intro:data-compression} about coding theorems, we stressed
how important it is to prove a converse theorem that matches the rate that the
direct coding theorem suggests is optimal. For now, we delay the proof of the
converse theorem because the tools for proving it are much different from the
tools we described in this section. For now, accept that the formula in
\eqref{eq-intro:capacity} is indeed the optimal rate at which two parties can
communicate and we will prove this result in a later chapter.

We end the description of Shannon's channel coding theorem by summarizing the
statements of the direct coding theorem and the converse theorem. The
statement of the direct coding theorem is as follows:

\begin{quote}
If the rate of communication is less than the channel capacity, then it is
possible for Alice to communicate reliably to Bob, in the sense that a
sequence of codes exists whose maximal probability of error vanishes as the
number of channel uses tends to infinity.
\end{quote}

The statement of the converse theorem is as follows:

\begin{quote}
If a reliable sequence of codes exists, then the rate of this sequence of
codes is less than the channel capacity.
\end{quote}

Another way of stating the converse proves to be useful later on:

\begin{quote}
If the rate of a coding scheme is greater than the channel capacity, then a
reliable code does not exist, in the sense that the error probability of the
coding scheme is bounded away from zero.
\end{quote}

\section{Summary}

A general communication scenario involves one sender and one receiver. In the
classical setting, we discussed two information-processing tasks that they can
perform. The first task was data compression or source coding, and we assumed
that the sender and receiver are linked together by a noiseless classical bit
channel that they can use a large number of times. We can think of this
noiseless classical bit channel as a \textit{noiseless dynamic resource} that
the two parties share. The resource is dynamic because we assume that there is
some physical medium through which the physical carrier of information travels
in order to get from the sender to the receiver. It was our aim to count the
number of times they would have to use the noiseless resource in order to send
information reliably. The result of Shannon's source coding theorem is that
the entropy gives the minimum rate at which they have to use the noiseless
resource. The second task we discussed was channel coding and we assumed that
the sender and receiver are linked together by a noisy classical channel that
they can use a large number of times. This noisy classical channel is a
\textit{noisy dynamic resource} that they share. We can think of this
information-processing task as a \textit{simulation task}, where the goal is
to simulate a noiseless dynamic resource by using a noisy dynamic resource in
a redundant way. This redundancy is what allows Alice to communicate reliably
to Bob, and reliable communication implies that they have effectively
simulated a noiseless resource. We again had a resource count for this case,
where we counted $n$ as the number of times they use the noisy resource and
$nC$ is the number of noiseless bit channels they simulate (where $C$ is the
capacity of the channel). This notion of resource counting may not seem so
important for the classical case, but it becomes much more important for the
quantum case.

We now conclude our overview of Shannon's information theory. The main points
to take home from this overview are the ideas that Shannon employed for
constructing source and channel codes. We let the information source emit a
large sequence of data, or similarly, we use the channel a large number of
times so that we can invoke the law of large numbers from probability theory.
The result is that we can show vanishing error for both schemes by taking a
limit. In Chapter~\ref{chap:classical-typicality}, we develop the theory of
typical sequences in detail, proving many of the results taken for granted in
this overview.

In hindsight, Shannon's methods for proving the two coding theorems are merely
a \textit{tour de force} for one idea from probability theory: the law of
large numbers. Perhaps, this viewpoint undermines the contribution of Shannon,
until we recall that no one else had even come close to devising these methods
for data compression and channel coding. The theoretical development of
Shannon is one of the most important contributions to modern science because
his theorems determine the ultimate rate at which we can compress and
communicate classical information.

\part{The Quantum Theory}

\chapter{The Noiseless Quantum Theory}

\label{chap:noiseless-quantum-theory}The simplest quantum system is the
physical quantum bit or \textit{qubit}. The qubit is a two-level quantum
system---example qubit systems are the spin of an electron, the polarization
of a photon, or a two-level atom with a ground state and an excited state. We
do not worry too much about physical implementations in this chapter, but
instead focus on the mathematical postulates of the quantum theory and
operations that we can perform on qubits. From qubits we progress to a study
of physical \textit{qudits}. Qudits are quantum systems that have $d$ levels
and are an important generalization of qubits. Again, we do not discuss
physical realizations of qudits.

Noise can affect quantum systems, and we must understand methods of modeling
noise in the quantum theory because our ultimate aim is to construct schemes
for protecting quantum systems against the detrimental effects of noise. In
Chapter~\ref{chap:intro-concepts}, we remarked on the different types of noise
that occur in nature. The first, and perhaps more easily comprehensible type
of noise, is that which is due to our lack of information about a given
scenario. We observe this type of noise in a casino, with every shuffle of
cards or toss of dice. These events are random, and the random variables of
probability theory model them because the outcomes are unpredictable. This
noise is the same as that in all classical information-processing systems.

On the other hand, the quantum theory features a fundamentally different type
of noise. Quantum noise is inherent in nature and is not due to our lack of
information, but is due rather to nature itself. An example of this type of
noise is the \textquotedblleft Heisenberg noise\textquotedblright\ that
results from the uncertainty principle. If we know the momentum of a given
particle from performing a precise measurement of it, then we know absolutely
nothing about its position---a measurement of its position gives a random
result. Similarly, if we know the rectilinear polarization of a photon by
precisely measuring it, then a future measurement of its diagonal polarization
will give a random result. It is important to keep the distinction clear
between these two types of noise.

We explore the postulates of the quantum theory in this chapter, by paying
particular attention to qubits. These postulates apply to a closed quantum
system that is isolated from everything else in the universe. We label this
first chapter \textquotedblleft The Noiseless Quantum Theory\textquotedblright%
\ because closed quantum systems do not interact with their surroundings and
are thus not subject to corruption and information loss. Interaction with
surrounding systems can lead to loss of information in the sense of the
classical noise that we described above. Closed quantum systems do undergo a
certain type of quantum noise, such as that from the uncertainty principle and
the act of measurement, because they are subject to the postulates of the
quantum theory. The name \textquotedblleft noiseless quantum
theory\textquotedblright\ thus indicates the closed, ideal nature of the
quantum systems discussed.

This chapter introduces the four postulates of the quantum theory. The
mathematical tools of the quantum theory rely on the fundamentals of linear
algebra---vectors and matrices of complex numbers. It may seem strange at
first that we need to incorporate the machinery of linear algebra in order to
describe a physical system in the quantum theory, but it turns out that this
description uses the simplest set of mathematical tools to predict the
phenomena that a quantum system exhibits. A hallmark of the quantum theory is
that certain operations do not commute with one another, and matrices are the
simplest mathematical objects that capture this idea of non-commutativity.

\section{Overview}

\begin{figure}
[ptb]
\begin{center}
\includegraphics[
width=4.8456in
]%
{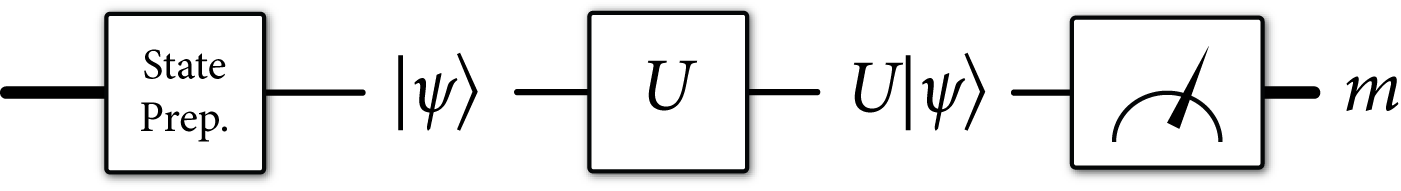}%
\caption{All of the steps in a typical noiseless quantum information
processing protocol. A classical control (depicted by the thick black line on
the left) initializes the state of a quantum system. The quantum system then
evolves according to some unitary operation (described in
Section~\ref{sec-noiseless-qt:evolution}). The final step is a measurement
that reads out some classical data~$m$ from the quantum system.}%
\label{fig-noiseless-qt:QIP}%
\end{center}
\end{figure}

We first briefly overview how information is processed with quantum systems.
This usually consists of three steps:\ state preparation, quantum operations,
and measurement. State preparation is the initialization of a quantum system
to some beginning state, depending on what operation we would like a quantum
system to execute. There could be some classical control device that
initializes the state of the quantum system. Observe that the input system for
this step is a classical system, and the output system is quantum. After
initializing the state of the quantum system, we perform some quantum
operations that evolve its state. This stage is where we can take advantage of
quantum effects for enhanced information-processing abilities. Both the input
and output systems of this step are quantum. Finally, we need some way of
reading out the result of the computation, and we can do so with a
measurement. The input system for this step is quantum, and the output is
classical. Figure~\ref{fig-noiseless-qt:QIP} depicts all of these steps. In a
quantum communication protocol, spatially separated parties may execute
different parts of these steps, and we are interested in keeping track of the
non-local resources needed to implement a communication protocol.
Section~\ref{sec-noiseless-qt:state-prep} describes quantum states (and thus
state preparation), Section~\ref{sec-noiseless-qt:evolution} describes the
noiseless evolution of quantum states, and
Section~\ref{sec-noiseless-qt:measurement}\ describes \textquotedblleft read
out\textquotedblright\ or measurement. For now, we assume that we can perform
all of these steps perfectly and later chapters discuss how to incorporate the
effects of noise.%

\section{Quantum Bits}

\label{sec-noiseless-qt:state-prep}The simplest quantum system is a two-state
system: a physical qubit. Let $\vert0\rangle$ denote one possible state of the
system. The left vertical bar and the right angle bracket indicate that we are
using the Dirac notation to represent this state. The Dirac notation has some
advantages for performing calculations in the quantum theory, and we highlight
some of these advantages as we progress through our development. Let
$\vert1\rangle$ denote another possible state of the qubit. We can encode a
classical bit or \textit{cbit} into a qubit with the following mapping:%
\begin{equation}
0\rightarrow\vert0\rangle,\ \ \ \ \ \ \ \ \ \ 1\rightarrow\vert1\rangle.
\end{equation}

So far, nothing in our description above distinguishes a classical bit from a
qubit, except for the funny vertical bar and angle bracket that we place
around the bit values. However, the quantum theory predicts that the above
states are not the only possible states of a qubit. Arbitrary
\textit{superpositions} (linear combinations) of the above two states are
possible as well because the quantum theory is a linear theory. Suffice it to
say that the linearity of the quantum theory results from the linearity of
Schr\"{o}dinger's equation that governs the evolution of quantum
systems.\footnote{We will not present Schr\"{o}dinger's equation in this book,
but instead focus on a \textquotedblleft quantum information\textquotedblright%
\ presentation of the quantum theory. Griffith's book on quantum mechanics
introduces the quantum theory from the Schr\"{o}dinger equation if you are
interested \citep{Grif95a}.} A general noiseless qubit can be in the following
state:%
\begin{equation}
\vert\psi\rangle\equiv\alpha\vert0\rangle+\beta\vert1\rangle,
\label{eq-qt:qubit}%
\end{equation}
where the coefficients $\alpha$ and $\beta$ are arbitrary complex numbers with
unit norm: $\left\vert \alpha\right\vert ^{2}+\left\vert \beta\right\vert
^{2}=1. $ The coefficients $\alpha$ and $\beta$ are
\index{probability amplitudes}%
\textit{probability amplitudes}---they are not probabilities themselves, but
they do allow us to calculate probabilities. The unit-norm constraint leads to
the \textit{Born rule} (the probabilistic interpretation) of the quantum
theory%
\index{Born rule}%
, and we speak more on this constraint and probability amplitudes when we
introduce the measurement postulate.

The possibility of superposition%
\index{superposition}
states indicates that we cannot represent the states $|0\rangle$ and
$|1\rangle$ with the Boolean algebra of the respective classical bits $0$ and
$1$ because Boolean algebra does not allow for superposition states. We
instead require the mathematics of \textit{linear algebra} to describe these
states. It is beneficial at first to define a vector representation of the
states $|0\rangle$ and $|1\rangle$:%
\begin{equation}
|0\rangle\equiv\left[
\begin{array}
[c]{c}%
1\\
0
\end{array}
\right]  ,\ \ \ \ \ \ \ \ \ |1\rangle\equiv\left[
\begin{array}
[c]{c}%
0\\
1
\end{array}
\right]  . \label{eq-qt:vec-rep-0}%
\end{equation}
The $|0\rangle$ and $|1\rangle$ states are called \textquotedblleft
kets\textquotedblright\ in the language of the Dirac notation, and it is best
at first to think of them merely as column vectors. The superposition state in
\eqref{eq-qt:qubit} then has a representation as the following two-dimensional
vector:%
\begin{equation}
|\psi\rangle=\alpha|0\rangle+\beta|1\rangle=\left[
\begin{array}
[c]{c}%
\alpha\\
\beta
\end{array}
\right]  .
\end{equation}
The representation of quantum states with vectors is helpful in understanding
some of the mathematics that underpins the theory, but it turns out to be much
more useful for our purposes to work directly with the Dirac notation. We give
the vector representation for now, but later on, we will exclusively employ
the Dirac notation.

\begin{figure}
[ptb]
\begin{center}
\includegraphics[
width=2.7752in
]%
{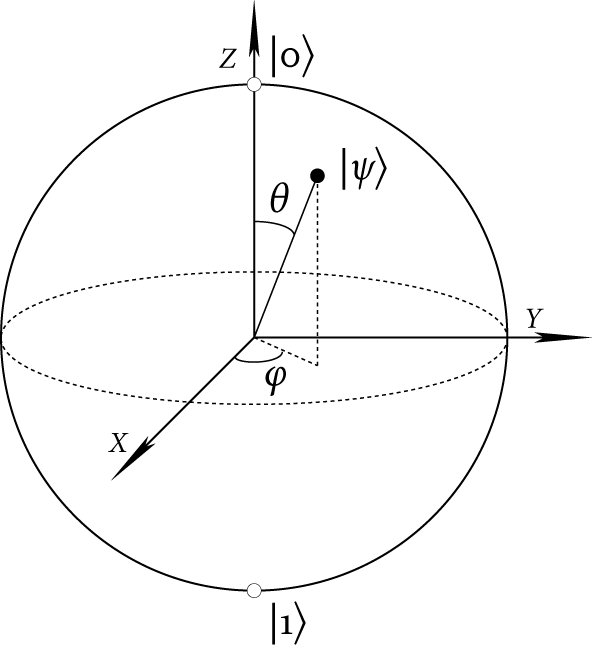}%
\caption{The Bloch sphere representation of a qubit. Any qubit $\vert
\psi\rangle$ admits a representation in terms of two angles $\theta$ and
$\varphi$ where $0\leq\theta\leq\pi$ and $0\leq\varphi<2\pi$. The state of any
qubit in terms of these angles is $\vert\psi\rangle=\cos( \theta/2)
\vert0\rangle+e^{i\varphi}\sin( \theta/2) \vert1\rangle$.}%
\label{fig-qt:bloch-sphere}%
\end{center}
\end{figure}

The \textit{Bloch sphere}%
\index{Bloch sphere}%
, depicted in Figure~\ref{fig-qt:bloch-sphere}, gives a valuable way to
visualize a qubit. Consider any two qubits that are equivalent up to a
differing global phase. For example,\ these two qubits could be%
\begin{equation}
|\psi_{0}\rangle\equiv|\psi\rangle,\ \ \ \ \ \ \ |\psi_{1}\rangle\equiv
e^{i\chi}|\psi\rangle, \label{eq-noiseless-qt:global-ph-irrelevant}%
\end{equation}
where $0\leq\chi<2\pi$. These two qubits are physically equivalent because
they give the same physical results when we measure them (more on this point
when we introduce the measurement postulate in
Section~\ref{sec-noiseless-qt:measurement}). Suppose that the probability
amplitudes $\alpha$ and $\beta$ have the following representations as complex
numbers:%
\begin{equation}
\alpha=r_{0}e^{i\varphi_{0}},\ \ \ \ \ \ \ \ \ \beta=r_{1}e^{i\varphi_{1}}.
\end{equation}
We can factor out the phase $e^{i\varphi_{0}}$ from both coefficients $\alpha$
and $\beta$, and we still have a state that is physically equivalent to the
state in \eqref{eq-qt:qubit}:%
\begin{equation}
|\psi\rangle\equiv r_{0}|0\rangle+r_{1}e^{i\left(  \varphi_{1}-\varphi
_{0}\right)  }|1\rangle,
\end{equation}
where we redefine $|\psi\rangle$ to represent the state because of the
equivalence mentioned in~\eqref{eq-noiseless-qt:global-ph-irrelevant}. Let
$\varphi\equiv\varphi_{1}-\varphi_{0}$, where $0\leq\varphi<2\pi$. Recall that
the unit-norm constraint requires $\left\vert r_{0}\right\vert ^{2}+\left\vert
r_{1}\right\vert ^{2}=1$. We can thus parametrize the values of $r_{0}$ and
$r_{1}$ in terms of one parameter $\theta$ so that%
\begin{equation}
r_{0}=\cos(\theta/2),\ \ \ \ \ \ \ \ \ r_{1}=\sin(\theta/2).
\end{equation}
The parameter $\theta$ varies between $0$ and $\pi$. This range of $\theta$
and the factor of two give a unique representation of the qubit. One may think
to have $\theta$ vary between $0$ and $2\pi$ and omit the factor of two, but
this parametrization would not uniquely characterize the qubit in terms of the
parameters $\theta$ and $\varphi$. The parametrization in terms of $\theta$
and $\varphi$ gives the Bloch sphere representation of the qubit in
\eqref{eq-qt:qubit}:%
\begin{equation}
|\psi\rangle\equiv\cos(\theta/2)|0\rangle+\sin(\theta/2)e^{i\varphi}|1\rangle.
\end{equation}

In linear algebra, column vectors are not the only type of vectors---row
vectors are useful as well. Is there an equivalent of a row vector in Dirac
notation? The Dirac notation provides an entity called a \textquotedblleft
bra,\textquotedblright\ that has a representation as a row vector. The bras
corresponding to the kets $\vert0\rangle$ and $\vert1\rangle$ are as follows:%
\begin{equation}
\langle0\vert\equiv\left[
\begin{array}
[c]{cc}%
1 & 0
\end{array}
\right]  ,\ \ \ \ \ \ \ \ \ \langle1\vert\equiv\left[
\begin{array}
[c]{cc}%
0 & 1
\end{array}
\right]  ,
\end{equation}
and are the matrix conjugate transpose of the kets $\vert0\rangle$ and
$\vert1\rangle$:%
\begin{equation}
\langle0\vert=\left(  \vert0\rangle\right)  ^{\dag}, \ \ \ \ \ \ \ \ \ \langle
1\vert=\left(  \vert1\rangle\right)  ^{\dag}.
\end{equation}
We require the conjugate transpose operation (as opposed to just the
transpose) because the mathematical representation of a general quantum state
can have complex entries.

The bras do not represent quantum states, but are helpful in calculating
probability amplitudes. For our example qubit in \eqref{eq-qt:qubit}, suppose
that we would like to determine the probability amplitude that the state is
$|0\rangle$. We can combine the state in \eqref{eq-qt:qubit} with the bra
$\langle0|$ as follows:%
\begin{align}
\langle0||\psi\rangle &  =\langle0|\left(  \alpha|0\rangle+\beta
|1\rangle\right) \\
&  =\alpha\langle0||0\rangle+\beta\langle0||1\rangle\\
&  =\alpha\left[
\begin{array}
[c]{cc}%
1 & 0
\end{array}
\right]  \left[
\begin{array}
[c]{c}%
1\\
0
\end{array}
\right]  +\beta\left[
\begin{array}
[c]{cc}%
1 & 0
\end{array}
\right]  \left[
\begin{array}
[c]{c}%
0\\
1
\end{array}
\right] \\
&  =\alpha\cdot1+\beta\cdot0\\
&  =\alpha. \label{eq-qt:braket}%
\end{align}
The above calculation may seem as if it is merely an exercise in linear
algebra, with a \textquotedblleft glorified\textquotedblright\ Dirac notation,
but it is a standard calculation in the quantum theory. A quantity like
$\langle0||\psi\rangle$ occurs so often in the quantum theory that we
abbreviate it as%
\begin{equation}
\left\langle 0|\psi\right\rangle \equiv\langle0||\psi\rangle,
\end{equation}
and the above notation is known as a \textquotedblleft
braket.\textquotedblright\footnote{It is for this (somewhat silly) reason that
Dirac decided to use the names \textquotedblleft bra\textquotedblright\ and
\textquotedblleft ket,\textquotedblright\ because putting them together gives
a \textquotedblleft braket.\textquotedblright\ The names in the notation may
be silly, but the notation itself has persisted over time because this way of
representing quantum states turns out to be useful. We will avoid the use of
the terms \textquotedblleft bra\textquotedblright\ and \textquotedblleft
ket\textquotedblright\ as much as we can, only resorting to these terms if
necessary.} The physical interpretation of the quantity $\left\langle
0|\psi\right\rangle $ is that it is the probability amplitude for being in the
state $|0\rangle$, and likewise, the quantity $\left\langle 1|\psi
\right\rangle $ is the probability amplitude for being in the state
$|1\rangle$. We can also determine that the amplitude $\left\langle
1|0\right\rangle $ (for the state $|0\rangle$ to be in the state $|1\rangle$)
and the amplitude $\left\langle 0|1\right\rangle $ are both equal to zero.
These two states are \textit{orthogonal states} because they have no overlap.
The amplitudes $\left\langle 0|0\right\rangle $ and $\left\langle
1|1\right\rangle $ are both equal to one by following a similar calculation.

Our next task may seem like a frivolous exercise, but we would like to
determine the amplitude for any state $\vert\psi\rangle$ to be in the state
$\vert\psi\rangle$, i.e., to be itself. Following the above method, this
amplitude is $\left\langle \psi|\psi\right\rangle $ and we calculate it as%
\begin{align}
\left\langle \psi|\psi\right\rangle  &  =\left(  \langle0\vert\alpha^{\ast
}+\langle1\vert\beta^{\ast}\right)  \left(  \alpha\vert0\rangle+\beta
\vert1\rangle\right) \label{eq-qt:unit-amplitude}\\
&  =\alpha^{\ast}\alpha\left\langle 0|0\right\rangle +\beta^{\ast}%
\alpha\left\langle 1|0\right\rangle +\alpha^{\ast}\beta\left\langle
0|1\right\rangle +\beta^{\ast}\beta\left\langle 1|1\right\rangle \\
&  =\left\vert \alpha\right\vert ^{2}+\left\vert \beta\right\vert ^{2}\\
&  =1, \label{eq-qt:unit-amplitude-1}%
\end{align}
where we have used the orthogonality relations of $\left\langle
0|0\right\rangle $, $\left\langle 1|0\right\rangle $, $\left\langle
0|1\right\rangle $, and $\left\langle 1|1\right\rangle $, and the unit-norm
constraint. We also write this in terms of the Euclidean norm of $\vert
\psi\rangle$ as
\begin{equation}
\Vert\vert\psi\rangle\Vert_{2} \equiv\sqrt{ \langle\psi\vert\psi\rangle} =1.
\label{eq-qt:Euclidean-norm}%
\end{equation}
We come back to the unit-norm constraint in our discussion of quantum
measurement, but for now, we have shown that any quantum state has a unit
amplitude for being itself.

The states $\vert0\rangle$ and $\vert1\rangle$ are a particular basis for a
qubit that we call the \textit{computational basis}. The computational basis
is the standard basis that we employ in quantum computation and communication,
but other bases are important as well. Consider that the following two vectors
form an orthonormal basis:%
\begin{equation}
\frac{1}{\sqrt{2}}\left[
\begin{array}
[c]{c}%
1\\
1
\end{array}
\right]  ,\ \ \ \ \ \ \ \ \frac{1}{\sqrt{2}}\left[
\begin{array}
[c]{c}%
1\\
-1
\end{array}
\right]  .
\end{equation}
The above alternate basis is so important in quantum information theory that
we define a Dirac notation shorthand for it, and we can also define the basis
in terms of the computational basis:%
\begin{equation}
\vert+\rangle\equiv\frac{\vert0\rangle+\vert1\rangle}{\sqrt{2}},
\ \ \ \ \ \ \ \ \ \vert-\rangle\equiv\frac{\vert0\rangle-\vert1\rangle}%
{\sqrt{2}}. \label{eq-qt:+}%
\end{equation}
The common names for this alternate basis are the \textquotedblleft$+$%
/$-$\textquotedblright\ basis, the Hadamard basis, or the diagonal basis. It
is preferable for us to use the Dirac notation, but we are using the vector
representation as an aid for now.

\begin{exercise}
\label{ex-qt:bloch-sphere}Determine the Bloch sphere angles $\theta$ and
$\varphi$ for the states $\vert+\rangle$ and $\vert-\rangle$.
\end{exercise}

What is the amplitude that the state in \eqref{eq-qt:qubit} is in the state
$\vert+\rangle$? What is the amplitude that it is in the state $\left\vert
-\right\rangle $? These are questions to which the quantum theory provides
simple answers. We employ the bra $\langle+\vert$ and calculate the amplitude
$\left\langle +|\psi\right\rangle $ as%
\begin{align}
\left\langle +|\psi\right\rangle  &  =\langle+\vert\left(  \alpha\vert
0\rangle+\beta\vert1\rangle\right) \\
&  =\alpha\left\langle +|0\right\rangle +\beta\left\langle +|1\right\rangle \\
&  =\frac{\alpha+\beta}{\sqrt{2}}. \label{eq-qt:+-amp}%
\end{align}
The result follows by employing the definition in \eqref{eq-qt:+}\ and doing
similar linear algebraic calculations as the example in \eqref{eq-qt:braket}.
We can also calculate the amplitude $\left\langle -|\psi\right\rangle $ as%
\begin{equation}
\left\langle -|\psi\right\rangle =\frac{\alpha-\beta}{\sqrt{2}}.
\label{eq-qt:--amp}%
\end{equation}
The above calculation follows from similar manipulations.

The $+$/$-$ basis is a \textit{complete} orthonormal basis, meaning that we
can represent any qubit state in terms of the two basis states $\left\vert
+\right\rangle $ and $\vert-\rangle$. Indeed, the above probability amplitude
calculations and the fact that the $+$/$-$ basis is complete imply that we can
represent the qubit in \eqref{eq-qt:qubit} as the following superposition
state:%
\begin{equation}
\vert\psi\rangle=\left(  \frac{\alpha+\beta}{\sqrt{2}}\right)  \vert
+\rangle+\left(  \frac{\alpha-\beta}{\sqrt{2}}\right)  \left\vert
-\right\rangle . \label{eq-qt:+-superposition}%
\end{equation}
The above representation is an alternate one if we would like to
\textquotedblleft see\textquotedblright\ the qubit state represented in the
$+$/$-$ basis. We can substitute the equalities in \eqref{eq-qt:+-amp} and
\eqref{eq-qt:--amp} to represent the state $\vert\psi\rangle$ as%
\begin{equation}
\vert\psi\rangle=\left\langle +|\psi\right\rangle \vert+\rangle+\left\langle
-|\psi\right\rangle \vert-\rangle.
\end{equation}
The amplitudes $\left\langle +|\psi\right\rangle $ and $\left\langle
-|\psi\right\rangle $ are both scalar quantities so that the above quantity is
equal to the following one:%
\begin{equation}
\vert\psi\rangle=\vert+\rangle\langle+|\psi\rangle+\vert-\rangle\langle
-|\psi\rangle.
\end{equation}
The order of the multiplication in the terms $\vert+\rangle\langle
+|\psi\rangle$ and $\vert-\rangle\langle-|\psi\rangle$ does not matter, i.e.,
the following equality holds:%
\begin{equation}
\vert+\rangle\left(  \left\langle +|\psi\right\rangle \right)  =\left(
\vert+\rangle\langle+\vert\right)  \vert\psi\rangle,
\end{equation}
and the same for $\vert-\rangle\left\langle -|\psi\right\rangle $. The
quantity on the left is a ket multiplied by an amplitude, whereas the quantity
on the right is a linear operator multiplying a ket, but linear algebra tells
us that these two quantities are equal. The operators $\vert+\rangle
\langle+\vert$ and $\vert-\rangle\langle-\vert$ are special operators---they
are rank-one projection operators, meaning that they project onto a
one-dimensional subspace. Using linearity, we have the following equality:%
\begin{equation}
\vert\psi\rangle=\left(  \vert+\rangle\langle+\vert+\vert-\rangle\langle
-\vert\right)  \vert\psi\rangle.
\end{equation}
The above equation indicates a seemingly trivial, but important point---the
operator $\vert+\rangle\langle+\vert+\vert-\rangle\langle-\vert$ is equal to
the identity operator and we can write%
\begin{equation}
I=\vert+\rangle\langle+\vert+\vert-\rangle\langle-\vert,
\end{equation}
where $I$ stands for the identity operator. This relation is known as the
\textit{completeness relation} or the \textit{resolution of the identity}.
Given any orthonormal basis, we can always construct a resolution of the
identity by summing over the rank-one projection operators formed from each of
the orthonormal basis states. For example, the computational basis states give
another way to form a resolution of the identity operator:%
\begin{equation}
I=\vert0\rangle\langle0\vert+\vert1\rangle\langle1\vert.
\end{equation}
This simple trick provides a way to find the representation of a quantum state
in any basis.

\section{Reversible Evolution}

\label{sec-noiseless-qt:evolution}Physical systems evolve as time progresses.
The application of a magnetic field to an electron can change its spin and
pulsing an atom with a laser can excite one of its electrons from a ground
state to an excited state. These are only a couple of ways in which physical
systems can change.

The Schr\"{o}dinger equation governs the evolution of a closed quantum system.
In this book, we will not even state the Schr\"{o}dinger equation, but we will
instead focus on an important implication of it. \textit{The evolution of a
closed quantum system is reversible if we do not learn anything about the
state of the system (that is, if we do not measure it)}. Reversibility implies
that we can determine the input state of an evolution given the output state
and knowledge of the evolution. An example of a single-qubit reversible
operation is a NOT\ gate:%
\begin{equation}
\vert0\rangle\rightarrow\vert1\rangle,\ \ \ \ \ \ \ \ \ \ \vert1\rangle
\rightarrow\vert0\rangle.
\end{equation}
In the classical world, we would say that the NOT\ gate merely flips the value
of the input classical bit. In the quantum world, the NOT gate flips the basis
states $\vert0\rangle$ and $\vert1\rangle$. The NOT\ gate is reversible
because we can simply apply the NOT\ gate again to recover the original input
state---the NOT\ gate is its own inverse.%

\begin{figure}
[ptb]
\begin{center}
\includegraphics[
width=3.039in
]%
{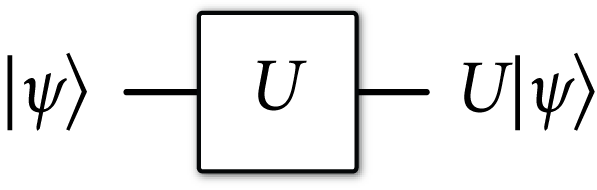}%
\caption{A quantum circuit diagram that depicts the evolution of a quantum
state $|\psi\rangle$ according to a unitary operator $U$.}%
\label{fig-qt:unitary}%
\end{center}
\end{figure}
In general, a closed quantum system evolves according to a unitary operator
$U$. Unitary evolution implies reversibility because a unitary operator always
possesses an inverse---its inverse is merely $U^{\dag}$, the conjugate
transpose. This property gives the relations:%
\begin{equation}
U^{\dag}U=UU^{\dag}=I.
\end{equation}
The unitary property also ensures that evolution preserves the unit-norm
constraint (an important requirement for a physical state that we discuss in
Section~\ref{sec-noiseless-qt:measurement}). Consider applying the unitary
operator $U$ to the example qubit state in \eqref{eq-qt:qubit}: $U|\psi
\rangle.$ Figure~\ref{fig-qt:unitary} depicts a quantum circuit diagram for
unitary evolution.

The bra that is dual to the above state is $\langle\psi\vert U^{\dag}$ (we
again apply the conjugate transpose operation to get the bra). We showed in
\eqref{eq-qt:unit-amplitude}--\eqref{eq-qt:unit-amplitude-1} that every
quantum state should have a unit amplitude for being itself. This relation
holds for the state $U\vert\psi\rangle$ because the operator $U$ is unitary:%
\begin{equation}
\langle\psi\vert U^{\dag}U\vert\psi\rangle=\langle\psi\vert I\vert\psi
\rangle=\left\langle \psi|\psi\right\rangle =1.
\end{equation}
The assumption that a vector always has a unit amplitude for being itself is
one of the crucial assumptions of the quantum theory, and the above reasoning
demonstrates that unitary evolution complements this assumption.

\begin{exercise}
A linear operator $T$ is norm preserving if $\Vert T \vert\psi\rangle\Vert_{2}
= \Vert\vert\psi\rangle\Vert_{2}$ holds for all quantum states $\vert
\psi\rangle$ (unit vectors), where the Euclidean norm is defined in
\eqref{eq-qt:Euclidean-norm}. Prove that an operator $T$ is an isometry (satisfying $T^\dag T = I$) if and
only if it is norm preserving. \textit{Hint: For showing the ``only-if'' part,
consider using the polarization identity:}
\begin{equation}
\langle\psi\vert\phi\rangle= \frac{1}{4} \left(  \Vert\vert\psi\rangle+
\vert\phi\rangle\Vert_{2}^{2} - \Vert\vert\psi\rangle- \vert\phi\rangle
\Vert_{2}^{2} + i \Vert\vert\psi\rangle+ i \vert\phi\rangle\Vert_{2}^{2} - i
\Vert\vert\psi\rangle- i \vert\phi\rangle\Vert_{2}^{2} \right)  .
\end{equation}

\end{exercise}

\subsection{Matrix Representations of Operators}

We now explore some properties of the NOT\ gate. Let $X$ denote the operator
corresponding to a NOT\ gate. The action of $X$ on the computational basis
states is as follows:%
\begin{equation}
X\vert i\rangle=\left\vert i\oplus1\right\rangle , \label{eq-qt:NOT}%
\end{equation}
where $i=\left\{  0,1\right\}  $ and $\oplus$ denotes binary addition. Suppose
the NOT\ gate acts on a superposition state:%
\begin{equation}
X\left(  \alpha\vert0\rangle+\beta\vert1\rangle\right)  .
\end{equation}
By linearity of the quantum theory, the $X$ operator distributes so that the
above expression is equal to the following one:%
\begin{equation}
\alpha X\vert0\rangle+\beta X\vert1\rangle=\alpha\vert1\rangle+\beta
\vert0\rangle.
\end{equation}
Indeed, the NOT gate $X$ merely flips the basis states of any quantum state
when represented in the computational basis.

We can determine a \textit{matrix representation}%
\index{matrix representation}
for the operator $X$ by using the bras $\langle0\vert$ and $\langle1\vert$.
Consider the relations in \eqref{eq-qt:NOT}. Let us combine the relations with
the bra $\langle0\vert$:%
\begin{equation}
\langle0\vert X\vert0\rangle=\left\langle 0|1\right\rangle
=0,\ \ \ \ \ \ \ \ \ \ \langle0\vert X\vert1\rangle=\left\langle
0|0\right\rangle =1.
\end{equation}
Likewise, we can combine with the bra $\langle1\vert$:%
\begin{equation}
\langle1\vert X\vert0\rangle=\left\langle 1|1\right\rangle
=1,\ \ \ \ \ \ \ \ \ \ \langle1\vert X\vert1\rangle=\left\langle
1|0\right\rangle =0.
\end{equation}
We can place these entries in a matrix to give a matrix representation of the
operator $X$:%
\begin{equation}
\left[
\begin{array}
[c]{cc}%
\langle0\vert X\vert0\rangle & \langle0\vert X\vert1\rangle\\
\langle1\vert X\vert0\rangle & \langle1\vert X\vert1\rangle
\end{array}
\right]  ,
\end{equation}
where we order the rows according to the bras and order the columns according
to the kets. We then say that%
\begin{equation}
X=\left[
\begin{array}
[c]{cc}%
0 & 1\\
1 & 0
\end{array}
\right]  ,
\end{equation}
and adopt the convention that the symbol $X$ refers to both the operator $X$
and its matrix representation (this is an abuse of notation, but it should be
clear from the context when $X$ refers to an operator and when it refers to
the matrix representation of the operator).

Let us now observe some uniquely quantum behavior. We would like to consider
the action of the NOT\ operator $X$ on the $+$/$-$ basis. First, let us
consider what happens if we operate on the $\vert+\rangle$ state with the $X$
operator. Recall that the state $\vert+\rangle=\left(  \vert0\rangle
+\vert1\rangle\right)  /\sqrt{2} $ so that%
\begin{equation}
X\vert+\rangle=X\left(  \frac{\vert0\rangle+\vert1\rangle}{\sqrt{2}}\right)
=\frac{X\vert0\rangle+X\vert1\rangle}{\sqrt{2}} =\frac{\vert1\rangle
+\vert0\rangle}{\sqrt{2}} =\vert+\rangle.
\end{equation}
The above development shows that the state $\vert+\rangle$ is a special state
with respect to the NOT\ operator $X$---it is an \textit{eigenstate} of $X$
with \textit{eigenvalue} one. An eigenstate of an operator is one that is
invariant under the action of the operator. The coefficient in front of the
eigenstate is the \textit{eigenvalue} corresponding to the eigenstate. Under a
unitary evolution, the coefficient in front of the eigenstate is just a
complex phase, but this global phase has no effect on the observations
resulting from a measurement of the state because two quantum states are
equivalent up to a differing global phase.

Now, let us consider the action of the NOT\ operator $X$ on the state
$\vert-\rangle$. Recall that $\vert-\rangle=\left(  \vert0\rangle
-\vert1\rangle\right)  /\sqrt{2} $. Calculating similarly, we get that%
\begin{equation}
X\vert-\rangle=X\left(  \frac{\vert0\rangle-\vert1\rangle}{\sqrt{2}}\right)
=\frac{X\vert0\rangle-X\vert1\rangle}{\sqrt{2}} =\frac{\vert1\rangle
-\vert0\rangle}{\sqrt{2}} =-\vert-\rangle.
\end{equation}
So the state $\vert-\rangle$ is also an eigenstate of the operator $X$, but
its eigenvalue is $-1$.

We can find a matrix representation of the $X$ operator in the $+$/$-$ basis
as well:%
\begin{equation}
\left[
\begin{array}
[c]{cc}%
\langle+\vert X\vert+\rangle & \langle+\vert X\left\vert -\right\rangle \\
\langle-\vert X\vert+\rangle & \langle-\vert X\vert-\rangle
\end{array}
\right]  =\left[
\begin{array}
[c]{cc}%
1 & 0\\
0 & -1
\end{array}
\right]  .
\end{equation}
This representation demonstrates that the $X$ operator is diagonal with
respect to the $+$/$-$ basis, and therefore, the $+$/$-$ basis is an
\textit{eigenbasis} for the $X$ operator. It is always handy to know the
eigenbasis of a unitary operator$~U$ because this eigenbasis gives the states
that are invariant under an evolution according to$~U$.

Let $Z$ denote the operator that flips states in the $+$/$-$ basis:%
\begin{equation}
Z\vert+\rangle\rightarrow\vert-\rangle,\ \ \ \ \ \ \ \ \ \ Z\vert
-\rangle\rightarrow\vert+\rangle.
\end{equation}
Using an analysis similar to that which we did for the $X$ operator, we can
find a matrix representation of the $Z$ operator in the $+$/$-$ basis:%
\begin{equation}
\left[
\begin{array}
[c]{cc}%
\langle+\vert Z\vert+\rangle & \langle+\vert Z\left\vert -\right\rangle \\
\langle-\vert Z\vert+\rangle & \langle-\vert Z\vert-\rangle
\end{array}
\right]  =\left[
\begin{array}
[c]{cc}%
0 & 1\\
1 & 0
\end{array}
\right]  .
\end{equation}
Interestingly, the matrix representation for the $Z$ operator in the $+$/$-$
basis is the same as that for the $X$ operator in the computational basis. For
this reason, we call the $Z$ operator the
\index{phase-flip operator}%
\textit{phase-flip }operator.\footnote{A more appropriate name might be the
\textquotedblleft bit flip in the $+$/$-$ basis operator,\textquotedblright%
\ but this name is too long, so we stick with the term \textquotedblleft phase
flip.\textquotedblright}

We expect the following steps to hold because the quantum theory is a linear
theory:%
\begin{align}
Z\left(  \frac{\vert+\rangle+\vert-\rangle}{\sqrt{2}}\right)   &
=\frac{Z\vert+\rangle+Z\vert-\rangle}{\sqrt{2}}=\frac{\left\vert
-\right\rangle +\vert+\rangle}{\sqrt{2}}=\frac{\vert+\rangle+\left\vert
-\right\rangle }{\sqrt{2}},\\
Z\left(  \frac{\vert+\rangle-\vert-\rangle}{\sqrt{2}}\right)   &
=\frac{Z\vert+\rangle-Z\vert-\rangle}{\sqrt{2}}=\frac{\left\vert
-\right\rangle -\vert+\rangle}{\sqrt{2}}=-\left(  \frac{\vert+\rangle
-\vert-\rangle}{\sqrt{2}}\right)  .
\end{align}
The above steps demonstrate that the states $\left(  \vert+\rangle+\left\vert
-\right\rangle \right)  /\sqrt{2} $ and $\left(  \vert+\rangle-\vert
-\rangle\right)  /\sqrt{2} $ are both eigenstates of the $Z$ operators. These
states are none other than the respective computational basis states
$\vert0\rangle$ and $\vert1\rangle$, by inspecting the definitions in
\eqref{eq-qt:+}. Thus, a matrix representation of the $Z$ operator in the
computational basis is%
\begin{equation}
\left[
\begin{array}
[c]{cc}%
\langle0\vert Z\vert0\rangle & \langle0\vert Z\vert1\rangle\\
\langle1\vert Z\vert0\rangle & \langle1\vert Z\vert1\rangle
\end{array}
\right]  =\left[
\begin{array}
[c]{cc}%
1 & 0\\
0 & -1
\end{array}
\right]  ,
\end{equation}
and is a diagonalization of the operator $Z$. So, the behavior of the $Z$
operator in the computational basis is the same as the behavior of the $X$
operator in the $+$/$-$ basis.

\subsection{Commutators and Anticommutators}

The \textit{commutator}%
\index{commutator}
$\left[  A,B\right]  $\ of two operators $A$ and $B$ is as follows:%
\begin{equation}
\left[  A,B\right]  \equiv AB-BA.
\end{equation}
Two operators commute if and only if their commutator is equal to zero.

The \textit{anticommutator}%
\index{anticommutator}
$\left\{  A,B\right\}  $ of two operators $A$ and $B$ is as follows:%
\begin{equation}
\left\{  A,B\right\}  \equiv AB+BA.
\end{equation}
We say that two operators \textit{anticommute} if their anticommutator is
equal to zero.

\begin{exercise}
Find a matrix representation for $\left[  X,Z\right]  $ in the basis $\left\{
\vert0\rangle,\vert1\rangle\right\}  $.
\end{exercise}

\subsection{The Pauli Matrices}

\label{sec-qt:Pauli-matrices}The convention%
\index{Pauli matrices}
in quantum theory is to take the computational basis as the \textit{standard
basis} for representing physical qubits. The standard matrix representation
for the above two operators is as follows when we choose the computational
basis as the standard basis:%
\begin{equation}
X\equiv%
\begin{bmatrix}
0 & 1\\
1 & 0
\end{bmatrix}
,\ \ \ Z\equiv%
\begin{bmatrix}
1 & 0\\
0 & -1
\end{bmatrix}
.
\end{equation}
The identity operator $I$ has the following representation in any basis:%
\begin{equation}
I\equiv%
\begin{bmatrix}
1 & 0\\
0 & 1
\end{bmatrix}
.
\end{equation}
Another operator, the $Y$ operator, is a useful one to consider as well. The
$Y$ operator has the following matrix representation in the computational
basis:%
\begin{equation}
Y\equiv%
\begin{bmatrix}
0 & -i\\
i & 0
\end{bmatrix}
.
\end{equation}
It is easy to check that $Y=iXZ$, and for this reason, we can think of the $Y$
operator as a combined bit and phase flip. The four matrices $I$, $X$, $Y$,
and $Z$ are special for the manipulation of physical qubits and are known as
the \textit{Pauli matrices}.

\begin{exercise}
Show that the Pauli matrices are all Hermitian, unitary, they square to the
identity, and their eigenvalues are $\pm1$.
\end{exercise}

\begin{exercise}
Represent the eigenstates of the $Y$ operator in the computational basis.
\end{exercise}

\begin{exercise}
Show that the Pauli matrices either commute or anticommute.
\end{exercise}

\begin{exercise}
\label{ex-qt:pauli-traces}Let us label the Pauli matrices as $\sigma_{0}\equiv
I$, $\sigma_{1}\equiv X$, $\sigma_{2}\equiv Y$, and $\sigma_{3}\equiv Z$. Show
that $\operatorname{Tr}\left\{  \sigma_{i}\sigma_{j}\right\}  =2\delta_{ij}$
for all $i,j\in\left\{  0,\ldots,3\right\}  $, where $\operatorname{Tr}$
denotes the trace of a matrix, defined as the sum of the entries along the
diagonal (see also Definition~\ref{def-nqt:trace}).
\end{exercise}

\subsection{Hadamard Gate}

Another important unitary operator is the transformation that takes the
computational basis to the $+$/$-$ basis. This transformation is the
\index{Hadamard gate}%
Hadamard transformation:%
\begin{equation}
\vert0\rangle\rightarrow\vert+\rangle, \ \ \ \ \ \ \ \ \ \vert1\rangle
\rightarrow\vert-\rangle. \label{eq-qt:hadamard-0}%
\end{equation}
Using the above relations, we can represent the Hadamard transformation as the
following operator:%
\begin{equation}
H\equiv\vert+\rangle\langle0\vert+\vert-\rangle\langle1\vert.
\label{eq-qt:hadamard-operator}%
\end{equation}
It is straightforward to check that the above operator implements the
transformation in~\eqref{eq-qt:hadamard-0}.

Now consider a generalization of the above construction. Suppose that one
orthonormal basis is $\left\{  \left\vert \psi_{i}\right\rangle \right\}
_{i\in\left\{  0,1\right\}  }$ and another is $\left\{  \left\vert \phi
_{i}\right\rangle \right\}  _{i\in\left\{  0,1\right\}  }$ where the index $i$
merely indexes the states in each orthonormal basis. Then the unitary operator
that takes states in the first basis to states in the second basis is%
\begin{equation}
\sum_{i=0,1}\vert\phi_{i}\rangle\langle\psi_{i}\vert.
\end{equation}

\begin{exercise}
Show that the Hadamard operator $H$ has the following matrix representation in
the computational basis:%
\begin{equation}
H=\frac{1}{\sqrt{2}}\left[
\begin{array}
[c]{cc}%
1 & 1\\
1 & -1
\end{array}
\right]  .
\end{equation}

\end{exercise}

\begin{exercise}
Show that the Hadamard operator is its own inverse by employing the above
matrix representation and by using its operator form in \eqref{eq-qt:hadamard-operator}.
\end{exercise}

\begin{exercise}
If the Hadamard gate is its own inverse, then it takes the states $\left\vert
+\right\rangle $ and $\vert-\rangle$ to the respective states $\vert0\rangle$
and $\vert1\rangle$ and we can represent it as the following operator:
$H=\vert0\rangle\langle+\vert+\vert1\rangle\langle-\vert. $ Show that
$\vert0\rangle\langle+\vert+\vert1\rangle\langle-\vert=\vert+\rangle
\langle0\vert+\vert-\rangle\langle1\vert. $
\end{exercise}

\begin{exercise}
Show that $HXH=Z$ and that $HZH=X$.
\end{exercise}

\subsection{Rotation Operators}%

\begin{figure}
[ptb]
\begin{center}
\includegraphics[
width=3.7101in
]%
{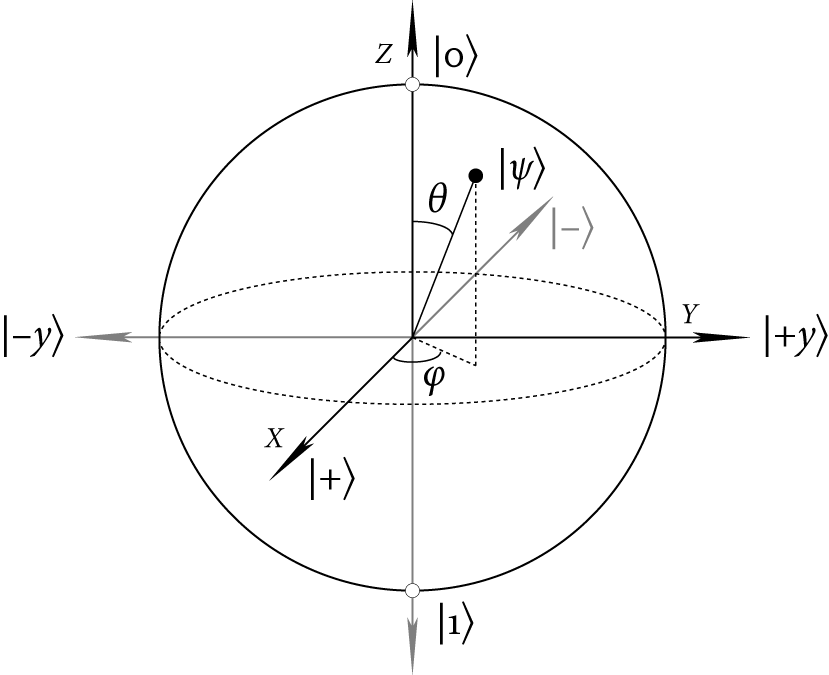}%
\caption{This figure provides more labels for states on the Bloch sphere. The
$Z$ axis has its points on the sphere as eigenstates of the Pauli $Z$
operator, the $X$ axis has eigenstates of the Pauli $X$ operator, and the $Y$
axis has eigenstates of the Pauli $Y$ operator. The rotation operators $R_{X}(
\phi) $, $R_{Y}( \phi) $, and $R_{Z}( \phi) $ rotate a state on the sphere by
an angle $\phi$ about the respective $X$, $Y$, and $Z$ axis.}%
\label{fig-qt:full-bloch-sphere}%
\end{center}
\end{figure}
We end this section on the evolution of quantum states by discussing
\textquotedblleft rotation evolutions\textquotedblright\ and by giving a more
complete picture of the Bloch sphere. The rotation operators $R_{X}( \phi) $,
$R_{Y}( \phi) $, $R_{Z}( \phi) $ are functions of the respective Pauli
operators $X$, $Y$, $Z$ where%
\begin{equation}
R_{X}( \phi) \equiv\exp\left\{  iX\phi/2\right\}  ,\ \ \ \ \ \ R_{Y}( \phi)
\equiv\exp\left\{  iY\phi/2\right\}  ,\ \ \ \ \ \ R_{Z}( \phi) \equiv
\exp\left\{  iZ\phi/2\right\}  ,
\end{equation}
and $\phi$ is some angle such that $0\leq\phi<2\pi$. How do we determine a
function of an operator? The standard way is to represent the operator in its
diagonal basis and apply the function to the non-zero eigenvalues of the
operator. For example, the diagonal representation of the $X$ operator is%
\begin{equation}
X=\vert+\rangle\langle+\vert-\vert-\rangle\langle-\vert.
\end{equation}
Applying the function $\exp\left\{  iX\phi/2\right\}  $ to the non-zero
eigenvalues of $X$ gives%
\begin{equation}
R_{X}( \phi) =\exp\left\{  i\phi/2\right\}  \vert+\rangle\langle+\vert
+\exp\left\{  -i\phi/2\right\}  \vert-\rangle\langle-\vert.
\end{equation}
This is a special case of the following more general convention that we follow
throughout this book:

\begin{definition}
[Function of a Hermitian operator]\label{def-qt:hermitian-op-function} Suppose
that a Hermitian operator $A$ has a spectral decomposition $A=\sum
_{i}a_{i}|i\rangle\langle i|$ for some orthonormal basis $\left\{
|i\rangle\right\}  $. Then the operator $f(A)$ for some function $f$ is
defined as follows:%
\begin{equation}
f(A)\equiv\sum_{i:a_{i}\in \operatorname{Dom}(f)}f(a_{i})|i\rangle\langle i|,
\end{equation}
where $\operatorname{Dom}(f)$ denotes the domain of the function $f$.
\end{definition}

\begin{exercise}
Show that the
\index{rotation operators}%
rotation operators $R_{X}( \phi) $, $R_{Y}( \phi) $, $R_{Z}( \phi) $ are equal
to the following expressions:%
\begin{align}
R_{X}( \phi)  &  =\cos( \phi/2) I+i\sin( \phi/2) X,\\
R_{Y}( \phi)  &  =\cos( \phi/2) I+i\sin( \phi/2) Y,\\
R_{Z}( \phi)  &  =\cos( \phi/2) I+i\sin( \phi/2) Z,
\end{align}
by using the facts that $\cos( \phi/2) =\frac{1}{2}\left(  e^{i\phi
/2}+e^{-i\phi/2}\right)  $ and $\sin( \phi/2) =\frac{1}{2i}\left(  e^{i\phi
/2}-e^{-i\phi/2}\right)  $ .
\end{exercise}

Figure~\ref{fig-qt:full-bloch-sphere} provides a more detailed picture of the
Bloch sphere since we have now established the Pauli operators and their
eigenstates. The computational basis states are the eigenstates of the $Z$
operator and are the north and south poles on the Bloch sphere. The $+$/$-$
basis states are the eigenstates of the $X$ operator and the calculation from
Exercise~\ref{ex-qt:bloch-sphere}\ shows that they are the \textquotedblleft
east and west poles\textquotedblright\ of the Bloch sphere. We leave it as
another exercise to show that the $Y$ eigenstates are the other poles along
the equator of the Bloch sphere.

\begin{exercise}
Determine the Bloch sphere angles $\theta$ and $\varphi$ for the eigenstates
of the Pauli $Y$ operator.
\end{exercise}

\section{Measurement}

\label{sec-noiseless-qt:measurement}Measurement is another type of evolution
that a quantum system can undergo. It is an evolution that allows us to
retrieve classical information from a quantum state and thus is the way that
we can \textquotedblleft read out\textquotedblright\ information. Suppose that
we would like to learn something about the quantum state $|\psi\rangle$ in
\eqref{eq-qt:qubit}. Nature prevents us from learning anything about the
probability amplitudes $\alpha$ and $\beta$ if we have only one quantum
measurement that we can perform on one copy of the state. Nature only allows
us to measure \textit{observables}. Observables%
\index{observable}
are physical variables such as the position or momentum of a particle. In the
quantum theory, we represent observables as Hermitian operators in part
because their eigenvalues are real numbers and every measuring device outputs
a real number. Examples of qubit observables that we can measure are the Pauli
operators $X$, $Y$, and $Z$.

\begin{figure}[ptb]
\begin{center}
\includegraphics[
width=3.4731in
]{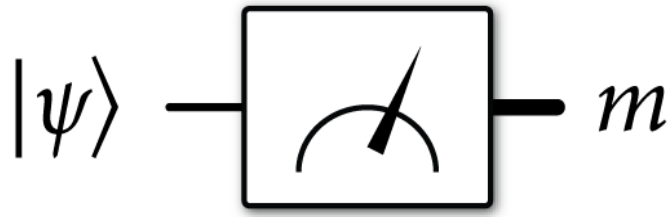}
\end{center}
\caption{This figure depicts our diagram of a quantum measurement. Thin lines
denote quantum information and thick lines denote classical information. The
result of the measurement is to output a classical variable $m$ according to a
probability distribution governed by the Born rule of the quantum theory.}%
\label{fig-qt:measurement}%
\end{figure}

Suppose that we measure the $Z$ operator. This measurement is called a
\textquotedblleft measurement in the computational basis\textquotedblright\ or
a \textquotedblleft measurement of the $Z$ observable\textquotedblright%
\ because we are measuring the eigenvalues of the $Z$ operator. The
measurement postulate of the quantum theory, also known as the \textit{Born
rule},
\index{Born rule}%
states that the system reduces to the state $\vert0\rangle$ with probability
$\left\vert \alpha\right\vert ^{2}$ and reduces to the state $\vert1\rangle$
with probability $\left\vert \beta\right\vert ^{2}$. That is, the resulting
probabilities are the squares of the probability amplitudes. After the
measurement, our measuring apparatus tells us whether the state reduced to
$\vert0\rangle$ or $\vert1\rangle$---it returns $+1$ if the resulting state is
$\vert0\rangle$ and returns $-1$ if the resulting state is $\vert1\rangle$.
These returned values are the eigenvalues of the $Z$ operator. The measurement
postulate is the aspect of the quantum theory that makes it probabilistic or
\textquotedblleft jumpy\textquotedblright\ and is part of the
\textquotedblleft strangeness\textquotedblright\ of the quantum theory.
Figure~\ref{fig-qt:measurement} depicts the notation for a measurement that we
will use in diagrams throughout this book.

What is the result if we measure the state $|\psi\rangle$ in the
$+$/$-$ basis? Consider that we can represent $|\psi\rangle$ as a
superposition of the $\vert+\rangle$ and $\left\vert -\right\rangle $ states,
as given in \eqref{eq-qt:+-superposition}. The measurement postulate then
states that a measurement of the $X$ operator gives the state $|+\rangle$ with
probability $\left\vert \alpha+\beta\right\vert ^{2}/2$ and the state
$\vert-\rangle$ with probability $\left\vert \alpha-\beta\right\vert ^{2}/2$.
Quantum interference is now coming into play because the amplitudes $\alpha$
and $\beta$ interfere with each other. So this effect plays an important role
in quantum information theory.

In some cases, the basis states $\vert0\rangle$ and $\vert1\rangle$ may not
represent the spin states of an electron, but may represent the
\textit{location} of an electron. So, a way to interpret this measurement
postulate is that the electron \textquotedblleft jumps into\textquotedblright%
\ one location or another depending on the outcome of the measurement. But
what is the state of the electron before the measurement? We will just say in
this book that it is in a superposed, indefinite, or unsharp state, rather
than trying to pin down a philosophical interpretation. Some might say that
the electron is in \textquotedblleft two different locations at the same
time.\textquotedblright

Also, we should stress that we cannot interpret the measurement postulate as
meaning that the state is in $\vert0\rangle$ or $\vert1\rangle$ with
respective probabilities $\left\vert \alpha\right\vert ^{2}$ and $\left\vert
\beta\right\vert ^{2}$ before the measurement occurs, because this latter
interpretation is physically different from what we described above and is
also completely classical. The superposition%
\index{superposition}
state $\alpha\vert0\rangle+\beta\vert1\rangle$ gives fundamentally different
behavior from the probabilistic description of a state that is in
$\vert0\rangle$ or $\vert1\rangle$ with respective probabilities $\left\vert
\alpha\right\vert ^{2}$ and $\left\vert \beta\right\vert ^{2}$. Suppose that
we have the two different descriptions of a state (superposition and
probabilistic) and measure the $Z$ operator. We get the same result for both
cases---the resulting state is $\vert0\rangle$ or $\vert1\rangle$ with
respective probabilities $\left\vert \alpha\right\vert ^{2}$ and~$\left\vert
\beta\right\vert ^{2}$.

But now suppose that we measure the $X$ operator. The superposed state gives
the result from before---we get the state $\vert+\rangle$ with probability
$\left\vert \alpha+\beta\right\vert ^{2}/2$ and the state $\left\vert
-\right\rangle $ with probability $\left\vert \alpha-\beta\right\vert ^{2}/2$.
The probabilistic description gives a\ much different result. Suppose that the
state is $\vert0\rangle$. We know that $\vert0\rangle$ is a uniform
superposition of $\vert+\rangle$ and $\vert-\rangle$:%
\begin{equation}
\vert0\rangle=\frac{\vert+\rangle+\vert-\rangle}{\sqrt{2}}.
\end{equation}
So the state collapses to $\vert+\rangle$ or $\vert-\rangle$ with equal
probability in this case. If the state is $\vert1\rangle$, then it collapses
again to $\vert+\rangle$ or $\vert-\rangle$ with equal probabilities. Summing
up these probabilities, it follows that a measurement of the $X$ operator
gives the state $\vert+\rangle$ with probability $\left(  \left\vert
\alpha\right\vert ^{2}+\left\vert \beta\right\vert ^{2}\right)  /2=1/2$ and
gives the state $\vert-\rangle$ with the same probability. These results are
fundamentally different from those in which the state is the superposition
state $\vert\psi\rangle$, and experiment after experiment has supported the
predictions of the quantum theory. Table~\ref{tbl-intro:classical-vs-quantum}%
\ summarizes the results that we just discussed.%

\begin{table}[tbp] \centering
\begin{tabular}
[c]{l|l|l}\hline\hline
\textbf{Quantum State} & \textbf{Probability of }$\vert+\rangle$ &
\textbf{Probability of }$\vert-\rangle$\\\hline\hline
Superposition state & $\left\vert \alpha+\beta\right\vert ^{2}/2$ &
$\left\vert \alpha-\beta\right\vert ^{2}/2$\\
Probabilistic description & $1/2$ & $1/2$\\\hline\hline
\end{tabular}
\caption{This table summarizes the differences in probabilities for a quantum
state in a superposition $\alpha \vert 0 \rangle + \beta \vert 1 \rangle $ and a classical state that
is a probabilistic mixture of $\vert 0 \rangle$ and $\vert 1 \rangle$.}\label{tbl-intro:classical-vs-quantum}%
\end{table}%

Now we consider a \textquotedblleft Stern--Gerlach\textquotedblright-like
argument to illustrate another example of fundamental quantum behavior
\citep{sterngerlach}. The Stern--Gerlach experiment was a crucial one for
determining the \textquotedblleft strange\textquotedblright\ behavior of
quantum spin states. Suppose that we prepare the state~$|0\rangle$. If we
measure this state in the $Z$ basis, the result is that we always obtain the
state$~|0\rangle$ because the prepared state is a definite $Z$ eigenstate.
Suppose now that we measure the $X$ operator. The state $|0\rangle$ is equal
to a uniform superposition of $\vert+\rangle$ and $|-\rangle$. The measurement
postulate then states that we get the state $|+\rangle$ or $\left\vert
-\right\rangle $ with equal probability after performing this measurement. If
we then measure the $Z$ operator again, the result is completely random. The
$Z$ measurement result is $|0\rangle$ or $|1\rangle$ with equal probability if
the result of the $X$ measurement is $|+\rangle$ and the same outcome occurs
if the result of the $X$ measurement is $\left\vert -\right\rangle $. This
argument demonstrates that the measurement of the $X$ operator
\textquotedblleft throws off\textquotedblright\ the measurement of the $Z$
operator. The Stern--Gerlach experiment was one of the earliest to validate
the predictions of the quantum theory.

\subsection{Probability, Expectation, and Variance of an Operator}

We have an alternate, more formal way of stating the measurement postulate
that turns out to be more useful for a general quantum system. Suppose that we
are measuring the$~Z\ $operator. The diagonal representation of this operator
is%
\begin{equation}
Z=|0\rangle\langle0|-|1\rangle\langle1|.
\end{equation}
Consider the Hermitian operator%
\begin{equation}
\Pi_{0}\equiv|0\rangle\langle0|. \label{eq-qt:Z-proj-0}%
\end{equation}
It is a projection operator because applying it twice has the same effect as
applying it once:$~\Pi_{0}^{2}=\Pi_{0}$. It projects onto the subspace spanned
by the single vector $|0\rangle$. A similar line of analysis applies to the
projection operator%
\begin{equation}
\Pi_{1}\equiv|1\rangle\langle1|. \label{eq-qt:Z-proj-1}%
\end{equation}
So we can represent the $Z$ operator as $\Pi_{0}-\Pi_{1}$. Performing a
measurement of the $Z$ operator is equivalent to asking the question:\ Is the
state $|0\rangle$ or $|1\rangle$? Consider the quantity $\langle\psi|\Pi
_{0}|\psi\rangle$:%
\begin{equation}
\langle\psi|\Pi_{0}|\psi\rangle=\left\langle \psi|0\right\rangle \left\langle
0|\psi\right\rangle =\alpha^{\ast}\alpha=\left\vert \alpha\right\vert ^{2}.
\end{equation}
A similar analysis demonstrates that%
\begin{equation}
\langle\psi|\Pi_{1}|\psi\rangle=\left\vert \beta\right\vert ^{2}.
\end{equation}
These two quantities then give the probability that the state reduces to
$|0\rangle$ or $|1\rangle$.

A more general way of expressing a measurement of the $Z$ basis is to say that
we have a set $\left\{  \Pi_{i}\right\}  _{i\in\left\{  0,1\right\}  }$\ of
measurement operators that determine the outcome probabilities. These
measurement operators also determine the state that results after the
measurement. If the measurement result is $+1$, then the resulting state is%
\begin{equation}
\frac{\Pi_{0}|\psi\rangle}{\sqrt{\langle\psi|\Pi_{0}|\psi\rangle}}=|0\rangle,
\end{equation}
where we implicitly ignore the irrelevant global phase factor $\frac{\alpha
}{\left\vert \alpha\right\vert }$. If the measurement result is $-1$, then the
resulting state is%
\begin{equation}
\frac{\Pi_{1}|\psi\rangle}{\sqrt{\langle\psi|\Pi_{1}|\psi\rangle}}=|1\rangle,
\end{equation}
where we again implicitly ignore the irrelevant global phase factor
$\frac{\beta}{\left\vert \beta\right\vert }$. Dividing by $\sqrt{\langle
\psi|\Pi_{i}|\psi\rangle}$ for $i=0,1$ ensures that the state resulting after
measurement corresponds to a physical state (a unit vector).

We can also measure any orthonormal basis in this way---this type of
projective measurement is called
\index{von Neumann measurement}%
a \textit{von Neumann measurement}. For any orthonormal basis $\left\{
\vert\phi_{i}\rangle\right\}  _{i\in\left\{  0,1\right\}  }$, the measurement
operators are $\left\{  \vert\phi_{i}\rangle\langle\phi_{i}\vert\right\}
_{i\in\left\{  0,1\right\}  }$, and the state reduces to $\vert\phi_{i}%
\rangle\langle\phi_{i}|\psi\rangle/\left\vert \left\langle \phi_{i}%
|\psi\right\rangle \right\vert $ with probability $\left\langle \psi|\phi
_{i}\right\rangle \left\langle \phi_{i}|\psi\right\rangle =\left\vert
\left\langle \phi_{i}|\psi\right\rangle \right\vert ^{2}$.

\begin{exercise}
Determine the set of measurement operators corresponding to a measurement of
the $X$ observable.
\end{exercise}

We might want to determine the expectation of the measurement result when
measuring the $Z$ operator. The probability of getting the $+1$ value
corresponding to the $|0\rangle$ state is $\left\vert \alpha\right\vert ^{2}$
and the probability of getting the $-1$ value corresponding to the $-1$
eigenstate is $\left\vert \beta\right\vert ^{2}$. Standard probability theory
then gives us a way to calculate the expected value of a measurement of the
$Z$ operator when the state is $|\psi\rangle$:%
\begin{equation}
\mathbb{E}\left[  Z\right]  =\left\vert \alpha\right\vert ^{2}\left(
1\right)  +\left\vert \beta\right\vert ^{2}\left(  -1\right)  =\left\vert
\alpha\right\vert ^{2}-\left\vert \beta\right\vert ^{2}.
\end{equation}
We can formulate an alternate way to write this expectation, by making use of
the Dirac notation:%
\begin{align}
\mathbb{E}\left[  Z\right]   &  =\left\vert \alpha\right\vert ^{2}\left(
1\right)  +\left\vert \beta\right\vert ^{2}\left(  -1\right) \\
&  =\langle\psi|\Pi_{0}|\psi\rangle+\langle\psi|\Pi_{1}|\psi\rangle\left(
-1\right) \\
&  =\langle\psi|\Pi_{0}-\Pi_{1}|\psi\rangle\\
&  =\langle\psi|Z|\psi\rangle.
\end{align}
It is common for physicists to denote the expectation as%
\begin{equation}
\left\langle Z\right\rangle \equiv\langle\psi|Z|\psi\rangle,
\end{equation}
when it is understood that the expectation is with respect to the state
$|\psi\rangle$. This type of expression is a general one and the next exercise
asks you to show that it works for the $X$ and $Y$ operators as well.

\begin{exercise}
Show that the expressions $\langle\psi\vert X\vert\psi\rangle$ and
$\langle\psi\vert Y\vert\psi\rangle$ give the respective expectations
$\mathbb{E}\left[  X\right]  $ and $\mathbb{E}\left[  Y\right]  $ when
measuring the state $\vert\psi\rangle$ in the respective $X$ and $Y$ basis.
\end{exercise}

We also might want to determine the variance of the measurement of the $Z$
operator. Standard probability theory again gives that%
\begin{equation}
\operatorname{Var}\left[  Z\right]  =\mathbb{E}\left[  Z^{2}\right]
-\mathbb{E}\left[  Z\right]  ^{2}.
\end{equation}
Physicists denote the standard deviation of the measurement of the $Z$
operator as%
\begin{equation}
\Delta Z\equiv\left\langle \left(  Z-\left\langle Z\right\rangle \right)
^{2}\right\rangle ^{1/2},
\end{equation}
and thus the variance is equal to $\left(  \Delta Z\right)  ^{2}$. Physicists
often refer to $\Delta Z$ as the uncertainty of the observable $Z$ when the
state is $\vert\psi\rangle$.

In order to calculate the variance Var$\left[  Z\right]  $, we really just
need the second moment $\mathbb{E}\left[  Z^{2}\right]  $ because we already
have the expectation $\mathbb{E}\left[  Z\right]  $:%
\begin{equation}
\mathbb{E}\left[  Z^{2}\right]  =\left\vert \alpha\right\vert ^{2}\left(
1\right)  ^{2}+\left\vert \beta\right\vert ^{2}\left(  -1\right)  ^{2}
=\left\vert \alpha\right\vert ^{2}+\left\vert \beta\right\vert ^{2}.
\end{equation}
We can again calculate this quantity with the Dirac notation. The quantity
$\langle\psi\vert Z^{2}\vert\psi\rangle$ is the same as $\mathbb{E}\left[
Z^{2}\right]  $ and the next exercise asks you for a proof.

\begin{exercise}
Show that $\mathbb{E}\left[  X^{2}\right]  =\langle\psi\vert X^{2}\vert
\psi\rangle$, $\mathbb{E}\left[  Y^{2}\right]  =\langle\psi\vert Y^{2}%
\vert\psi\rangle$, and $\mathbb{E}\left[  Z^{2}\right]  =\langle\psi\vert
Z^{2}\vert\psi\rangle$.
\end{exercise}

\subsection{The Uncertainty Principle}

\label{sec-qt:uncertainty-principle}The uncertainty principle%
\index{uncertainty principle}
is a fundamental feature of the quantum theory. In the case of qubits, one
instance of the uncertainty principle gives a lower bound on the product of
the uncertainty of the $Z$ operator and the uncertainty of the $X$ operator:%
\begin{equation}
\Delta Z\Delta X\geq\frac{1}{2}\left\vert \langle\psi|\left[  Z,X\right]
|\psi\rangle\right\vert . \label{eq-qt:uncertainty-principle}%
\end{equation}
We can prove this principle using the postulates of the quantum theory. Let us
define the operators $Z_{0}\equiv Z-\left\langle Z\right\rangle $ and
$X_{0}\equiv X-\left\langle X\right\rangle $. First, consider that%
\begin{equation}
\Delta Z\Delta X=\langle\psi|Z_{0}^{2}|\psi\rangle^{1/2}\langle\psi|X_{0}%
^{2}|\psi\rangle^{1/2}\geq\left\vert \langle\psi|Z_{0}X_{0}|\psi
\rangle\right\vert .
\end{equation}
The above step follows by applying the Cauchy--Schwarz inequality to the
vectors $X_{0}|\psi\rangle$ and $Z_{0}|\psi\rangle$. For any operator $A$, we
define its real part $\operatorname{Re}\left\{  A\right\}  $ as
$\operatorname{Re}\left\{  A\right\}  \equiv(A+A^{\dag})/2,$ and its imaginary
part $\operatorname{Im}\left\{  A\right\}  $\ as $\operatorname{Im}\left\{
A\right\}  \equiv(A-A^{\dag})/2i,$ so that $A=\operatorname{Re}\left\{
A\right\}  +i\operatorname{Im}\left\{  A\right\}  .$ So the real and imaginary
parts of the operator $Z_{0}X_{0}$ are%
\begin{align}
\operatorname{Re}\left\{  Z_{0}X_{0}\right\}   &  =\frac{Z_{0}X_{0}+X_{0}%
Z_{0}}{2}\equiv\frac{\left\{  Z_{0},X_{0}\right\}  }{2},\\
\operatorname{Im}\left\{  Z_{0}X_{0}\right\}   &  =\frac{Z_{0}X_{0}-X_{0}%
Z_{0}}{2i}\equiv\frac{\left[  Z_{0},X_{0}\right]  }{2i},
\label{eq-qt:anti-com}%
\end{align}
where $\left\{  Z_{0},X_{0}\right\}  $ is the anticommutator of $Z_{0}$ and
$X_{0}$ and $\left[  Z_{0},X_{0}\right]  $ is the commutator of the two
operators. We can then express the quantity $\left\vert \langle\psi|Z_{0}%
X_{0}|\psi\rangle\right\vert $ in terms of the real and imaginary parts of
$Z_{0}X_{0}$:%
\begin{align}
\left\vert \langle\psi|Z_{0}X_{0}|\psi\rangle\right\vert  &  =\left\vert
\langle\psi|\operatorname{Re}\left\{  Z_{0}X_{0}\right\}  |\psi\rangle
+i\langle\psi|\operatorname{Im}\left\{  Z_{0}X_{0}\right\}  |\psi
\rangle\right\vert \\
&  \geq\left\vert \langle\psi|\operatorname{Im}\left\{  Z_{0}X_{0}\right\}
|\psi\rangle\right\vert \\
&  =\left\vert \langle\psi|\left[  Z_{0},X_{0}\right]  |\psi\rangle\right\vert
/2\\
&  =\left\vert \langle\psi|\left[  Z,X\right]  |\psi\rangle\right\vert /2.
\end{align}
The first equality follows by substitution. The first inequality follows
because the magnitude of any complex number is greater than the magnitude of
its imaginary part. The second equality follows by substitution with
\eqref{eq-qt:anti-com}. Finally, the third equality follows from the result of
Exercise~\ref{ex-qt:commutator} below. We worked out the above derivation for
particular observables acting on qubit states, but note that it holds for
general observables and quantum states.

The commutator of the operators $Z$ and $X$ arises in the lower bound, and
thus, the non-commutativity of the operators $Z$ and $X$ is the fundamental
reason that there is an uncertainty principle for them. Also, there is no
uncertainty principle for any two operators that commute with each other.

It is worthwhile to interpret the uncertainty principle in
\eqref{eq-qt:uncertainty-principle}, which really receives an interpretation
after conducting a large number of independent experiments of two different
kinds. In the first kind of experiment, one prepares the state $\vert
\psi\rangle$ and measures the $Z$ observable. After repeating this experiment
independently many times, one can calculate an estimate of the standard
deviation $\Delta Z$, which becomes closer and closer to the true standard
deviation $\Delta Z$ as the number of independent experiments becomes large.
In the second kind of experiment, one prepares the state $\vert\psi\rangle$
and measures the $X$ observable. After repeating many times, one can calculate
an estimate of $\Delta X$. The uncertainty principle then states that the
product of the estimates (for a large number of independent experiments) is
bounded from below by the expectation of the commutator: $\frac{1}%
{2}\left\vert \langle\psi\vert\left[  X,Z\right]  \vert\psi\rangle\right\vert
$.

\begin{exercise}
\label{ex-qt:commutator}Show that $\left[  Z_{0},X_{0}\right]  =\left[
Z,X\right]  $ and that $\left[  Z,X\right]  = 2iY$.
\end{exercise}

\begin{exercise}
\label{ex-qt:lower-bound-unc-vanish}The uncertainty principle in
\eqref{eq-qt:uncertainty-principle} has the property that the lower bound has
a dependence on the state $\vert\psi\rangle$. Find a state $\vert\psi\rangle$
for which the lower bound on the uncertainty product $\Delta X\Delta Z$
vanishes.\footnote{Do not be alarmed by the result of this exercise! The usual
formulation of the uncertainty principle only gives a lower bound on the
uncertainty product. This lower bound never vanishes for the case of position
and momentum observables because the commutator of these two observables is
equal to the identity operator multiplied by $i$, but it can vanish for the
operators given in the exercise.}
\end{exercise}

\section{Composite Quantum Systems}%

\index{composite quantum systems}%
A single physical qubit is an interesting physical system that exhibits
uniquely quantum phenomena, but it is not particularly useful on its own (just
as a single classical bit is not particularly useful for classical
communication or computation). We can only perform interesting quantum
information-processing tasks when we combine qubits together. Therefore, we
should have a way to describe their behavior when they combine to form a
composite quantum system.

Consider two classical bits $c_{0}$ and $c_{1}$. In order to describe bit
operations on the pair of cbits, we write them as an ordered pair $\left(
c_{1},c_{0}\right)  $. The space of all possible bit values is the Cartesian
product $\mathbb{Z}_{2}\times\mathbb{Z}_{2}$ of two copies of the set
$\mathbb{Z}_{2}\equiv\left\{  0,1\right\}  $:%
\begin{equation}
\mathbb{Z}_{2}\times\mathbb{Z}_{2}\equiv\left\{  \left(  0,0\right)  ,\left(
0,1\right)  ,\left(  1,0\right)  ,(1,1)\right\}  .
\end{equation}
Typically, we make the abbreviation $c_{1}c_{0}\equiv\left(  c_{1}%
,c_{0}\right)  $ when representing cbit states.

We can represent the state of two cbits with particular states of qubits. For
example, we can represent the two-cbit state $00$ using the following mapping:%
\begin{equation}
00\rightarrow|0\rangle|0\rangle.
\end{equation}
Many times, we make the abbreviation $|00\rangle\equiv|0\rangle|0\rangle$ when
representing two-cbit states with qubits. Any two-cbit state $c_{1}c_{0}$ has
the following representation as a two-qubit state:%
\begin{equation}
c_{1}c_{0}\rightarrow\left\vert c_{1}c_{0}\right\rangle .
\end{equation}

The above qubit states are not the only possible states that can occur in the
quantum theory. By the superposition principle, any possible linear
combination of the set of two-cbit states is a possible two-qubit state:%
\begin{equation}
\left\vert \xi\right\rangle \equiv\alpha|00\rangle+\beta|01\rangle
+\gamma|10\rangle+\delta|11\rangle. \label{eq-qt:two-qubits}%
\end{equation}
The unit-norm condition $\left\vert \alpha\right\vert ^{2}+\left\vert
\beta\right\vert ^{2}+\left\vert \gamma\right\vert ^{2}+\left\vert
\delta\right\vert ^{2}=1$ again must hold for the two-qubit state to
correspond to a physical quantum state. It is now clear that the Cartesian
product is not sufficient for representing two-qubit quantum states because it
does not allow for linear combinations of states (just as the mathematics of
Boolean algebra is not sufficient to represent single-qubit states).

We again turn to linear algebra to determine a representation that suffices.
The \textit{tensor product}%
\index{tensor product}
is a mathematical operation that suffices to give a representation of
two-qubit quantum states. Suppose we have two two-dimensional vectors:%
\begin{equation}
\left[
\begin{array}
[c]{c}%
a_{1}\\
b_{1}%
\end{array}
\right]  ,\ \ \ \ \ \ \left[
\begin{array}
[c]{c}%
a_{2}\\
b_{2}%
\end{array}
\right]  .
\end{equation}
The tensor product of these two vectors is%
\begin{equation}
\left[
\begin{array}
[c]{c}%
a_{1}\\
b_{1}%
\end{array}
\right]  \otimes\left[
\begin{array}
[c]{c}%
a_{2}\\
b_{2}%
\end{array}
\right]  \equiv\left[
\begin{array}
[c]{c}%
a_{1}\left[
\begin{array}
[c]{c}%
a_{2}\\
b_{2}%
\end{array}
\right] \\
b_{1}\left[
\begin{array}
[c]{c}%
a_{2}\\
b_{2}%
\end{array}
\right]
\end{array}
\right]  =\left[
\begin{array}
[c]{c}%
a_{1}a_{2}\\
a_{1}b_{2}\\
b_{1}a_{2}\\
b_{1}b_{2}%
\end{array}
\right]  .
\end{equation}
One can understand this operation as taking the vector on the right and
stacking two copies of it together, while multiplying each copy by the
corresponding number in the first vector.

Recall, from \eqref{eq-qt:vec-rep-0}, the vector representation of the
single-qubit states $|0\rangle$ and $|1\rangle$. Using these vector
representations and the above definition of the tensor product, the two-qubit
basis states have the following vector representations:%
\begin{equation}
|00\rangle=\left[
\begin{array}
[c]{c}%
1\\
0\\
0\\
0
\end{array}
\right]  ,\ \ \ |01\rangle=\left[
\begin{array}
[c]{c}%
0\\
1\\
0\\
0
\end{array}
\right]  ,\ \ \ |10\rangle=\left[
\begin{array}
[c]{c}%
0\\
0\\
1\\
0
\end{array}
\right]  ,\ \ \ |11\rangle=\left[
\begin{array}
[c]{c}%
0\\
0\\
0\\
1
\end{array}
\right]  .
\end{equation}
A simple way to remember these representations is that the bits inside the ket
index the element equal to one in the vector. For example, the vector
representation of $|01\rangle$ has a one as its second element because $01$ is
the second index for the two-bit strings. The vector representation of the
superposition state in \eqref{eq-qt:two-qubits} is%
\begin{equation}
\left[
\begin{array}
[c]{c}%
\alpha\\
\beta\\
\gamma\\
\delta
\end{array}
\right]  .
\end{equation}

There are actually many different ways that we can write two-qubit states, and
we list all of these right now. Physicists have developed many shorthands, and
it is important to know each of them because they often appear in the
literature (this book even uses different notations depending on the context).
We may use any of the following two-qubit notations if the two qubits are
local to one party and only one party is involved in a protocol:%
\begin{align}
&  \alpha|0\rangle\otimes|0\rangle+\beta|0\rangle\otimes|1\rangle
+\gamma|1\rangle\otimes|0\rangle+\delta|1\rangle\otimes|1\rangle,\\
&  \alpha|0\rangle|0\rangle+\beta|0\rangle|1\rangle+\gamma|1\rangle
|0\rangle+\delta|1\rangle|1\rangle,\\
&  \alpha|00\rangle+\beta|01\rangle+\gamma|10\rangle+\delta|11\rangle.
\end{align}
We can put labels on the qubits if two or more parties, such as $A$ and $B$,
are involved%
\begin{align}
&  \alpha|0\rangle_{A}\otimes|0\rangle_{B}+\beta|0\rangle_{A}\otimes
|1\rangle_{B}+\gamma|1\rangle_{A}\otimes|0\rangle_{B}+\delta|1\rangle
_{A}\otimes|1\rangle_{B},\\
&  \alpha|0\rangle_{A}|0\rangle_{B}+\beta|0\rangle_{A}|1\rangle_{B}%
+\gamma|1\rangle_{A}|0\rangle_{B}+\delta|1\rangle_{A}|1\rangle_{B},\\
&  \alpha|00\rangle_{AB}+\beta|01\rangle_{AB}+\gamma|10\rangle_{AB}%
+\delta|11\rangle_{AB}.
\end{align}
This second scenario is different from the first scenario because two
spatially separated parties share the two-qubit state. If the state has
quantum correlations, then it can be valuable as a communication resource. We
go into more detail on this topic in Section~\ref{sec-qt:entanglement}, which
discusses \textit{entanglement}.

\subsection{Evolution of Composite Systems}

\label{sec-qt:tensor-product-ops-1}The postulate on unitary evolution extends
to the two-qubit scenario as well. First, let us establish that the tensor
product $A\otimes B$ of two operators $A$ and $B$ is%
\begin{align}
A\otimes B  &  \equiv\left[
\begin{array}
[c]{cc}%
a_{11} & a_{12}\\
a_{21} & a_{22}%
\end{array}
\right]  \otimes\left[
\begin{array}
[c]{cc}%
b_{11} & b_{12}\\
b_{21} & b_{22}%
\end{array}
\right] \label{eq-qt:operator-TP}\\
&  \equiv\left[
\begin{array}
[c]{cc}%
a_{11}\left[
\begin{array}
[c]{cc}%
b_{11} & b_{12}\\
b_{21} & b_{22}%
\end{array}
\right]  & a_{12}\left[
\begin{array}
[c]{cc}%
b_{11} & b_{12}\\
b_{21} & b_{22}%
\end{array}
\right] \\
a_{21}\left[
\begin{array}
[c]{cc}%
b_{11} & b_{12}\\
b_{21} & b_{22}%
\end{array}
\right]  & a_{22}\left[
\begin{array}
[c]{cc}%
b_{11} & b_{12}\\
b_{21} & b_{22}%
\end{array}
\right]
\end{array}
\right] \\
&  =\left[
\begin{array}
[c]{cccc}%
a_{11}b_{11} & a_{11}b_{12} & a_{12}b_{11} & a_{12}b_{12}\\
a_{11}b_{21} & a_{11}b_{22} & a_{12}b_{21} & a_{12}b_{22}\\
a_{21}b_{11} & a_{21}b_{12} & a_{22}b_{11} & a_{22}b_{12}\\
a_{21}b_{21} & a_{21}b_{22} & a_{22}b_{21} & a_{22}b_{22}%
\end{array}
\right]  .
\end{align}
The tensor-product operation for matrices is similar to what we did for
vectors, but now we are stacking copies of the matrix on the right both
vertically and horizontally, and multiplying each copy by the corresponding
number in the first matrix.

\begin{figure}[ptb]
\begin{center}
\includegraphics[
width=3.5in
]{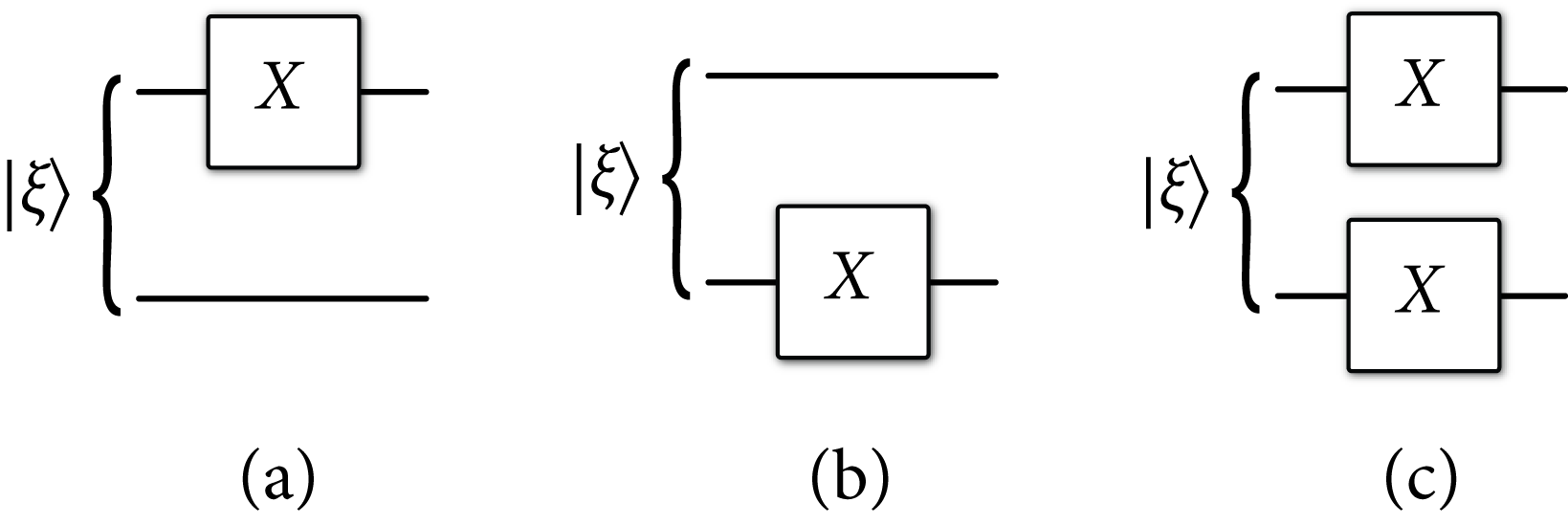}
\end{center}
\caption{This figure depicts circuits for the example two-qubit unitaries
$X_{1}I_{2}$, $I_{1}X_{2}$, and $X_{1}X_{2}$.}%
\label{fig-qt:two-qubit}%
\end{figure}

Consider the two-qubit state in \eqref{eq-qt:two-qubits}. We can perform a
NOT\ gate on the first qubit so that it changes to $\alpha|10\rangle
+\beta|11\rangle+\gamma|00\rangle+\delta|01\rangle.$ We can alternatively flip
its second qubit: $\alpha|01\rangle+\beta|00\rangle+\gamma|11\rangle
+\delta|10\rangle,$ or flip both at the same time: $\alpha|11\rangle
+\beta|10\rangle+\gamma|01\rangle+\delta|00\rangle.$
Figure~\ref{fig-qt:two-qubit} depicts quantum circuit representations of these
operations. These are all reversible operations because applying them again
gives the original state in \eqref{eq-qt:two-qubits}. In the first case, we
did nothing to the second qubit, and in the second case, we did nothing to the
first qubit. The identity operator acts on the qubits that have nothing happen
to them. 

Let us label the first qubit as \textquotedblleft$1$\textquotedblright\ and
the second qubit as \textquotedblleft$2$.\textquotedblright\ We can then label
the operator for the first operation as $X_{1}I_{2}$ because this operator
flips the first qubit and does nothing (applies the identity) to the second
qubit. We can also label the operators for the second and third operations
respectively as $I_{1}X_{2}$ and $X_{1}X_{2}$. The matrix representation of
the operator $X_{1}I_{2}$ is the tensor product of the matrix representation
of $X$ with the matrix representation of $I$---this relation similarly holds
for the operators $I_{1}X_{2}$ and $X_{1}X_{2}$. We show that it holds for the
operator $X_{1}I_{2}$ and ask you to verify the other two cases. We can use
the two-qubit computational basis to get a matrix representation for the
two-qubit operator~$X_{1}I_{2}$:%
\begin{multline}
\left[
\begin{array}
[c]{cccc}%
\langle00|X_{1}I_{2}|00\rangle & \langle00|X_{1}I_{2}|01\rangle &
\langle00|X_{1}I_{2}|10\rangle & \langle00|X_{1}I_{2}|11\rangle\\
\langle01|X_{1}I_{2}|00\rangle & \langle01|X_{1}I_{2}|01\rangle &
\langle01|X_{1}I_{2}|10\rangle & \langle01|X_{1}I_{2}|11\rangle\\
\langle10|X_{1}I_{2}|00\rangle & \langle10|X_{1}I_{2}|01\rangle &
\langle10|X_{1}I_{2}|10\rangle & \langle10|X_{1}I_{2}|11\rangle\\
\langle11|X_{1}I_{2}|00\rangle & \langle11|X_{1}I_{2}|01\rangle &
\langle11|X_{1}I_{2}|10\rangle & \langle11|X_{1}I_{2}|11\rangle
\end{array}
\right] \\
=\left[
\begin{array}
[c]{cccc}%
\left\langle 00|10\right\rangle  & \left\langle 00|11\right\rangle  &
\left\langle 00|00\right\rangle  & \left\langle 00|01\right\rangle \\
\left\langle 01|10\right\rangle  & \left\langle 01|11\right\rangle  &
\left\langle 01|00\right\rangle  & \left\langle 01|01\right\rangle \\
\left\langle 10|10\right\rangle  & \left\langle 10|11\right\rangle  &
\left\langle 10|00\right\rangle  & \left\langle 10|01\right\rangle \\
\left\langle 11|10\right\rangle  & \left\langle 11|11\right\rangle  &
\left\langle 11|00\right\rangle  & \left\langle 11|01\right\rangle
\end{array}
\right]  =\left[
\begin{array}
[c]{cccc}%
0 & 0 & 1 & 0\\
0 & 0 & 0 & 1\\
1 & 0 & 0 & 0\\
0 & 1 & 0 & 0
\end{array}
\right]  .
\end{multline}
This last matrix is equal to the tensor product $X\otimes I$ by inspecting the
definition of the tensor product for matrices in \eqref{eq-qt:operator-TP}.

\begin{exercise}
Show that the matrix representation of the operator $I_{1}X_{2}$ is equal to
the tensor product $I\otimes X$. Show the same for $X_{1}X_{2}$ and $X\otimes
X$.
\end{exercise}

\subsection{Probability Amplitudes for Composite Systems}

We relied on the orthogonality of the two-qubit computational basis states for
evaluating
\index{probability amplitudes}%
amplitudes such as $\left\langle 00|10\right\rangle $ or $\left\langle
00|00\right\rangle $ in the above matrix representation. It turns out that
there is another way to evaluate these amplitudes that relies only on the
orthogonality of the single-qubit computational basis states. Suppose that we
have four single-qubit states$~\vert\phi_{0}\rangle$, $\left\vert \phi
_{1}\right\rangle $, $|\psi_{0}\rangle$, $|\psi_{1}\rangle$, and we make the
following two-qubit states from them:%
\begin{equation}
\vert\phi_{0}\rangle\otimes|\psi_{0}\rangle,\ \ \ \ \ \ \ \ \ \vert\phi
_{1}\rangle\otimes|\psi_{1}\rangle.
\end{equation}
We may represent these states equally well as follows:%
\begin{equation}
\left\vert \phi_{0},\psi_{0}\right\rangle ,\ \ \ \ \ \ \ \ \ \ \left\vert
\phi_{1},\psi_{1}\right\rangle ,
\end{equation}
because the Dirac notation is versatile (virtually anything can go inside a
ket as long as its meaning is not ambiguous). The bra $\left\langle \phi
_{1},\psi_{1}\right\vert $ is dual to the ket $\left\vert \phi_{1},\psi
_{1}\right\rangle $, and we can use it to calculate the following amplitude:%
\begin{equation}
\left\langle \phi_{1},\psi_{1}|\phi_{0},\psi_{0}\right\rangle .
\end{equation}
This amplitude is equal to the multiplication of the single-qubit amplitudes:%
\begin{equation}
\left\langle \phi_{1},\psi_{1}|\phi_{0},\psi_{0}\right\rangle =\left\langle
\phi_{1}|\phi_{0}\right\rangle \left\langle \psi_{1}|\psi_{0}\right\rangle .
\label{eq-qt:amplitude-two-qubits}%
\end{equation}

\begin{exercise}
Verify that the amplitudes $\left\{  \left\langle ij|kl\right\rangle \right\}
_{i,j,k,l\in\left\{  0,1\right\}  }$ are respectively equal to the amplitudes
$\left\{  \left\langle i|k\right\rangle \left\langle j|l\right\rangle
\right\}  _{i,j,k,l\in\left\{  0,1\right\}  }$. By linearity, this exercise
justifies the relation in \eqref{eq-qt:amplitude-two-qubits} (at least for
two-qubit states).
\end{exercise}

\subsection{Controlled Gates}

An important two-qubit
\index{controlled gate}%
unitary evolution is the controlled-NOT\ (CNOT) gate. We consider its
classical version first. The classical gate acts on two cbits. It does nothing
if the first bit is equal to zero, and flips the second bit if the first bit
is equal to one:%
\begin{equation}
00\rightarrow00,\ \ \ \ \ 01\rightarrow01,\ \ \ \ \ 10\rightarrow
11,\ \ \ \ \ 11\rightarrow10.
\end{equation}
We turn this gate into a quantum gate\footnote{There are other terms for the
action of turning a classical operation into a quantum one. Some examples are
\textquotedblleft making it coherent,\textquotedblright\ \textquotedblleft
coherifying,\textquotedblright\ or the quantum gate is a \textquotedblleft
coherification\textquotedblright\ of the classical one. The term
\textquotedblleft coherify\textquotedblright\ is not a proper English word,
but we will use it regardless at certain points.} by demanding that it act in
the same way on the two-qubit computational basis states:%
\begin{equation}
|00\rangle\rightarrow|00\rangle,\ \ \ \ \ |01\rangle\rightarrow|01\rangle
,\ \ \ \ \ |10\rangle\rightarrow|11\rangle,\ \ \ \ \ |11\rangle\rightarrow
|10\rangle.
\end{equation}
By linearity, this behavior carries over to superposition states as well:%
\begin{equation}
\alpha|00\rangle+\beta|01\rangle+\gamma|10\rangle+\delta|11\rangle
\ \ \ \ \underrightarrow{\operatorname{CNOT}}\ \ \ \ \alpha|00\rangle
+\beta|01\rangle+\gamma|11\rangle+\delta|10\rangle.
\end{equation}

A useful operator representation of the CNOT\ gate is%
\begin{equation}
\operatorname{CNOT}\equiv\vert0\rangle\langle0\vert\otimes I+\vert
1\rangle\langle1\vert\otimes X. \label{eq-qt:CNOT}%
\end{equation}
The above representation truly captures the coherent quantum nature of the
CNOT gate. In the classical CNOT\ gate, we can say that it is a conditional
gate, in the sense that the gate applies to the second bit conditioned on the
value of the first bit. In the quantum CNOT\ gate, the second operation is
\textit{controlled} on the basis state of the first qubit (hence the choice of
the name \textquotedblleft controlled-NOT\textquotedblright). That is, the
gate acts on superpositions of quantum states and maintains these
superpositions, shuffling the probability amplitudes around while it does so.
The one case in which the gate has no effect is when the first qubit is
prepared in the state $\vert0 \rangle$ and the state of the second qubit is arbitrary.

\begin{figure}[ptb]
\begin{center}
\includegraphics[
width=2.6022in
]{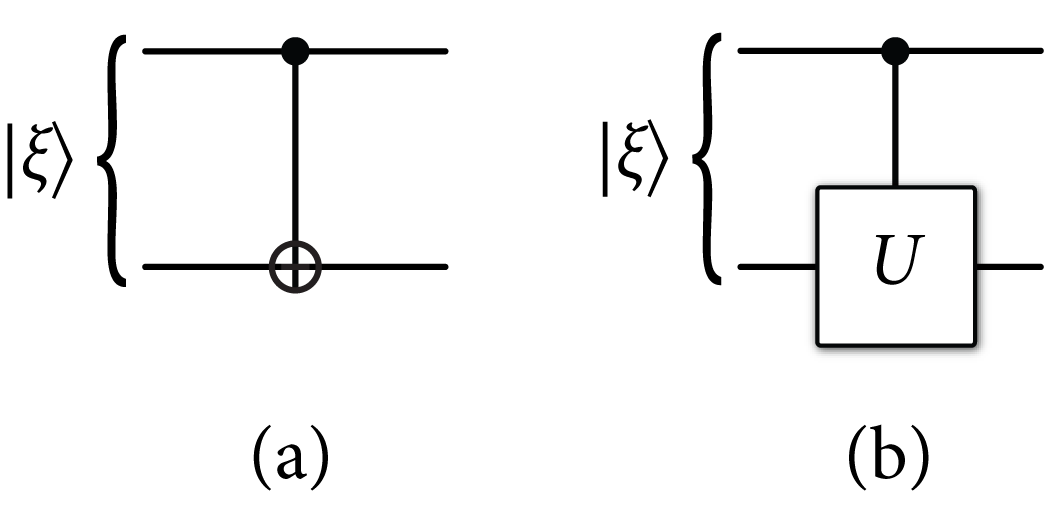}
\end{center}
\caption{Circuit diagrams that we use for (a)\ a CNOT\ gate and (b) a
controlled-$U$ gate.}%
\label{fig-qt:cnot}%
\end{figure}A controlled-$U$ gate is similar to the CNOT\ gate in
\eqref{eq-qt:CNOT}. It simply applies the unitary$~U$ (assumed to be a
single-qubit unitary) to the second qubit, controlled on the first qubit:%
\begin{equation}
\text{controlled-}U\equiv\vert0\rangle\langle0\vert\otimes I+\vert
1\rangle\langle1\vert\otimes U.
\end{equation}
The control qubit could alternatively be controlled with respect to any
orthonormal basis $\left\{  \vert\phi_{0}\rangle,\left\vert \phi
_{1}\right\rangle \right\}  $:%
\begin{equation}
\vert\phi_{0}\rangle\langle\phi_{0}\vert\otimes I+\vert\phi_{1}\rangle
\langle\phi_{1}\vert\otimes U.
\end{equation}
Figure~\ref{fig-qt:cnot}\ depicts the circuit diagrams for a
controlled-NOT\ and controlled-$U$ operation.

\begin{exercise}
Verify that the matrix representation of the $\operatorname{CNOT}$ gate in the
computational basis is%
\begin{equation}
\left[
\begin{array}
[c]{cccc}%
1 & 0 & 0 & 0\\
0 & 1 & 0 & 0\\
0 & 0 & 0 & 1\\
0 & 0 & 1 & 0
\end{array}
\right]  .
\end{equation}

\end{exercise}

\begin{exercise}
Consider applying Hadamards to the first and second qubits before and after a
$\operatorname{CNOT}$ acts on them. Show that this gate is equivalent to a
$\operatorname{CNOT}$ in the $+$/$-$ basis (recall that the $Z$ operator flips
the $+$/$-$ basis):%
\begin{equation}
H_{1}H_{2}\ \operatorname{CNOT}H_{1}H_{2}=\vert+\rangle\langle+\vert\otimes
I+\vert-\rangle\langle-\vert\otimes Z.
\end{equation}

\end{exercise}

\begin{exercise}
Show that two $\operatorname{CNOT}$ gates with the same control qubit commute.
\end{exercise}

\begin{exercise}
Show that two $\operatorname{CNOT}$ gates with the same target qubit commute.
\end{exercise}

\subsection{The No-Cloning Theorem}

\label{sec-qt:no-cloning}The no-cloning theorem%
\index{no-cloning theorem}
is one of the simplest results in the quantum theory, yet it has some of the
most profound consequences. It states that it is impossible to build a
\textit{universal copier} of quantum states. A universal copier would be a
device that could copy any arbitrary quantum state that is input to it. It may
be surprising at first to hear that copying quantum information is impossible
because copying classical information is ubiquitous.

We now give a simple proof of the no-cloning theorem. Suppose for a
contradiction that there is a two-qubit unitary operator $U$ acting as a
universal copier of quantum information. That is, if we input an arbitrary
state $|\psi\rangle=\alpha|0\rangle+\beta|1\rangle$ as the first qubit and
input an ancilla qubit $|0\rangle$ as the second qubit, such a device should
\textquotedblleft write\textquotedblright\ the first qubit to the second qubit
slot as follows:%
\begin{align}
U|\psi\rangle|0\rangle &  =|\psi\rangle|\psi\rangle\\
&  =\left(  \alpha|0\rangle+\beta|1\rangle\right)  \left(  \alpha
|0\rangle+\beta|1\rangle\right) \\
&  =\alpha^{2}|0\rangle|0\rangle+\alpha\beta|0\rangle|1\rangle+\alpha
\beta|1\rangle|0\rangle+\beta^{2}|1\rangle|1\rangle. \label{eq-qt:no-clone-1}%
\end{align}
The copier is universal, meaning that it copies an arbitrary state. In
particular, it also copies the states $|0\rangle$ and $|1\rangle$:
\begin{equation}
U|0\rangle|0\rangle=|0\rangle|0\rangle,\ \ \ \ \ \ \ \ \ U|1\rangle
|0\rangle=|1\rangle|1\rangle.
\end{equation}
Linearity of the quantum theory then implies that the unitary operator acts on
a superposition $\alpha|0\rangle+\beta|1\rangle$ as follows:%
\begin{equation}
U\left(  \alpha|0\rangle+\beta|1\rangle\right)  |0\rangle=\alpha
|0\rangle|0\rangle+\beta|1\rangle|1\rangle. \label{eq-qt:no-clone-2}%
\end{equation}
However, the consequence in \eqref{eq-qt:no-clone-1} contradicts the
consequence in \eqref{eq-qt:no-clone-2} because these two expressions do not
have to be equal for all $\alpha$ and $\beta$:%
\begin{equation}
\exists\alpha,\beta:\alpha^{2}|0\rangle|0\rangle+\alpha\beta|0\rangle
|1\rangle+\alpha\beta|1\rangle|0\rangle+\beta^{2}|1\rangle|1\rangle\neq
\alpha|0\rangle|0\rangle+\beta|1\rangle|1\rangle.
\label{eq-qt:no-cloning-contradiction}%
\end{equation}
Thus, linearity of the quantum theory contradicts the existence of a universal
quantum copier.

We would like to stress that this proof does not mean that it is impossible to
copy certain quantum states---it only implies the impossibility of a
\textit{universal} copier. Observe that \eqref{eq-qt:no-cloning-contradiction}
is satisfied for $\alpha= 1, \beta= 0$ or $\alpha=0, \beta=1$, so that we can
copy unknown classical states prepared in the basis $\vert0 \rangle, \vert1
\rangle$ (or any other orthonormal basis for that matter).

Another proof of the no-cloning theorem arrives at a contradiction by
exploiting unitarity of quantum evolutions. Let us again suppose that a
universal copier $U$\ exists. Consider two arbitrary states $|\psi\rangle$ and
$|\phi\rangle$. If a universal copier $U$ exists, then it performs the
following copying operation for both states:
\begin{equation}
U|\psi\rangle|0\rangle=|\psi\rangle|\psi\rangle,\ \ \ \ \ \ \ \ \ \ U|\phi
\rangle|0\rangle=\left\vert \phi\right\rangle |\phi\rangle.
\label{eq-qt:cloner}%
\end{equation}
Consider the probability amplitude $\langle\psi|\langle\psi||\phi\rangle
|\phi\rangle$:%
\begin{equation}
\langle\psi|\langle\psi||\phi\rangle|\phi\rangle=\left\langle \psi
|\phi\right\rangle \left\langle \psi|\phi\right\rangle =\left\langle \psi
|\phi\right\rangle ^{2}.
\end{equation}
The following relation for $\langle\psi|\langle\psi||\phi\rangle|\phi\rangle$
holds as well by using the results in \eqref{eq-qt:cloner} and the unitarity
property $U^{\dag}U=I$:%
\begin{align}
\langle\psi|\langle\psi||\phi\rangle|\phi\rangle &  =\langle\psi
|\langle0|U^{\dag}U|\phi\rangle|0\rangle\\
&  =\langle\psi|\langle0||\phi\rangle|0\rangle\\
&  =\left\langle \psi|\phi\right\rangle \left\langle 0|0\right\rangle \\
&  =\left\langle \psi|\phi\right\rangle .
\end{align}
As a consequence, we find that%
\begin{equation}
\langle\psi|\langle\psi||\phi\rangle|\phi\rangle=\left\langle \psi
|\phi\right\rangle ^{2}=\left\langle \psi|\phi\right\rangle ,
\end{equation}
by employing the above two results. The equality $\left\langle \psi
|\phi\right\rangle ^{2}=\left\langle \psi|\phi\right\rangle $ holds for
exactly two cases, $\left\langle \psi|\phi\right\rangle =1$ and $\left\langle
\psi|\phi\right\rangle =0$. The first case holds only when the two states\ are
the same state and the second case holds when the two states are orthogonal to
each other. Thus, it is impossible to copy quantum information in any other
case because we would contradict unitarity.

The no-cloning theorem has several applications in quantum information
processing. First, it underlies the security of the quantum key distribution
protocol because it ensures that an attacker cannot copy the quantum states
that two parties use to establish a secret key. It finds application in
quantum Shannon theory because we can use it to reason about the quantum
capacity of a certain quantum channel known as the erasure channel. We will
return to this point in Chapter~\ref{chap:quantum-capacity}.

\begin{exercise}
Suppose that two states $|\psi\rangle$ and $|\psi^{\perp}\rangle$ are
orthogonal: $\langle\psi|\psi^{\perp}\rangle=0.$ Construct a two-qubit unitary
that can copy the states, i.e., find a unitary $U$ that acts as follows:
$U|\psi\rangle|0\rangle=|\psi\rangle|\psi\rangle$, $U|\psi^{\perp}%
\rangle|0\rangle=|\psi^{\perp}\rangle|\psi^{\perp}\rangle.$
\end{exercise}

\begin{exercise}
[No-Deletion Theorem]Somewhat related to the no-cloning theorem, there is a
\index{no-deletion theorem}%
no-deletion theorem. Suppose that two copies of a quantum state $|\psi\rangle$
are available, and the goal is to delete one of these states by a unitary
interaction. That is, there should exist a universal quantum deleter $U$\ that
has the following action on the two copies of $|\psi\rangle$ and an ancilla
state $\left\vert A\right\rangle $, regardless of the input state
$|\psi\rangle$:%
\begin{equation}
U|\psi\rangle|\psi\rangle\left\vert A\right\rangle =|\psi\rangle
|0\rangle\left\vert A^{\prime}\right\rangle ,
\end{equation}
where $\left\vert A^{\prime}\right\rangle $ is another state. Show that this
is impossible.
\end{exercise}

\subsection{Measurement of Composite Systems}

The measurement postulate also extends to composite quantum systems. Suppose
again that we have the two-qubit quantum state in \eqref{eq-qt:two-qubits}. By
a straightforward analogy with the single-qubit case, we can determine the
following probability amplitudes:%
\begin{equation}
\left\langle 00|\xi\right\rangle =\alpha,\ \ \ \ \left\langle 01|\xi
\right\rangle =\beta,\ \ \ \ \left\langle 10|\xi\right\rangle =\gamma
,\ \ \ \ \left\langle 11|\xi\right\rangle =\delta.
\end{equation}
We can also define the following projection operators:%
\begin{equation}
\Pi_{00}\equiv|00\rangle\langle00|,\ \ \ \ \Pi_{01}\equiv|01\rangle
\langle01|,\ \ \ \ \Pi_{10}\equiv|10\rangle\langle10|,\ \ \ \ \Pi_{11}%
\equiv|11\rangle\langle11|,
\end{equation}
and apply the Born rule%
\index{Born rule}
to determine the probabilities for each result:%
\begin{equation}
\left\langle \xi\right\vert \Pi_{00}\left\vert \xi\right\rangle =\left\vert
\alpha\right\vert ^{2},\ \ \ \ \left\langle \xi\right\vert \Pi_{01}\left\vert
\xi\right\rangle =\left\vert \beta\right\vert ^{2},\ \ \ \ \left\langle
\xi\right\vert \Pi_{10}\left\vert \xi\right\rangle =\left\vert \gamma
\right\vert ^{2},\ \ \ \ \left\langle \xi\right\vert \Pi_{11}\left\vert
\xi\right\rangle =\left\vert \delta\right\vert ^{2}.
\end{equation}

Suppose that we wish to perform a measurement of the $Z$ operator on the first
qubit only. What is the set of projection operators that describes this
measurement? The answer is similar to what we found for the evolution of a
composite system. We apply the identity operator to the second qubit because
no measurement occurs on it. Thus, the set of measurement operators is%
\begin{equation}
\left\{  \Pi_{0}\otimes I,\Pi_{1}\otimes I\right\}  ,
\label{eq-qt:alice-projectors-comp}%
\end{equation}
where the definition of $\Pi_{0}$ and $\Pi_{1}$ is in
\eqref{eq-qt:Z-proj-0}--\eqref{eq-qt:Z-proj-1}. The state reduces to%
\begin{equation}
\frac{\left(  \Pi_{0}\otimes I\right)  \left\vert \xi\right\rangle }%
{\sqrt{\left\langle \xi\right\vert \left(  \Pi_{0}\otimes I\right)  \left\vert
\xi\right\rangle }}=\frac{\alpha|00\rangle+\beta|01\rangle}{\sqrt{\left\vert
\alpha\right\vert ^{2}+\left\vert \beta\right\vert ^{2}}},
\end{equation}
with probability $\left\langle \xi\right\vert \left(  \Pi_{0}\otimes I\right)
\left\vert \xi\right\rangle =\left\vert \alpha\right\vert ^{2}+\left\vert
\beta\right\vert ^{2}$, and reduces to%
\begin{equation}
\frac{\left(  \Pi_{1}\otimes I\right)  \left\vert \xi\right\rangle }%
{\sqrt{\left\langle \xi\right\vert \left(  \Pi_{1}\otimes I\right)  \left\vert
\xi\right\rangle }}=\frac{\gamma|10\rangle+\delta|11\rangle}{\sqrt{\left\vert
\gamma\right\vert ^{2}+\left\vert \delta\right\vert ^{2}}},
\end{equation}
with probability $\left\langle \xi\right\vert \left(  \Pi_{1}\otimes I\right)
\left\vert \xi\right\rangle =\left\vert \gamma\right\vert ^{2}+\left\vert
\delta\right\vert ^{2}$. Normalizing by $\sqrt{\left\langle \xi\right\vert
\left(  \Pi_{0}\otimes I\right)  \left\vert \xi\right\rangle }$ and
$\sqrt{\left\langle \xi\right\vert \left(  \Pi_{1}\otimes I\right)  \left\vert
\xi\right\rangle }$ again ensures that the resulting vector corresponds to a
physical state.

\section{Entanglement}

\label{sec-qt:entanglement}Composite quantum systems give rise to a uniquely
quantum phenomenon:\ \textit{entanglement}. Schr\"{o}dinger first observed
that two or more quantum systems can be entangled and coined the term after
noticing some of the bizarre consequences of this
phenomenon.\footnote{Schr\"{o}dinger actually used the German word
\textquotedblleft Verschr\"{a}nkung\textquotedblright\ to describe the
phenomenon, which literally translates as \textquotedblleft little parts that,
though far from one another, always keep the exact same distance from each
other.\textquotedblright\ The one-word English translation is
\textquotedblleft entanglement.\textquotedblright\ Einstein described the
\textquotedblleft Verschr\"{a}nkung\textquotedblright\ as a \textquotedblleft
spukhafte Fernwirkung,\textquotedblright\ most closely translated as
\textquotedblleft long-distance ghostly effect\textquotedblright\ or the more
commonly stated \textquotedblleft spooky action at a
distance.\textquotedblright}

We first consider a simple, unentangled state that two parties, Alice and Bob,
may share, in order to see how an unentangled state contrasts with an
entangled state. Suppose that they share the state%
\begin{equation}
|0\rangle_{A}|0\rangle_{B},
\end{equation}
where Alice has the qubit in system $A$ and Bob has the qubit in system $B$.
Alice can definitely say that her qubit is in the state $|0\rangle_{A}$ and
Bob can definitely say that his qubit is in the state $|0\rangle_{B}$. There
is nothing really too strange about this scenario.

Now, consider the composite quantum state $\left\vert \Phi^{+}\right\rangle
_{AB}$:%
\begin{equation}
\left\vert \Phi^{+}\right\rangle _{AB} \equiv\frac{1}{\sqrt{2}} \left(
\vert0\rangle_{A}\vert0\rangle_{B}+\vert1\rangle_{A}\vert1\rangle_{B}\right)
. \label{eq-qt:ebit}%
\end{equation}
Alice again has possession of the first qubit in system $A$ and Bob has
possession of the second qubit in system $B$. But now, it is not clear from
the above description how to determine the individual state of Alice or the
individual state of Bob. The above state is really a uniform superposition of
the joint state $\vert0\rangle_{A}\vert0\rangle_{B}$ and the joint state
$\vert1\rangle_{A}\vert1\rangle_{B}$, and it is not possible to describe
either Alice's or Bob's individual state in the noiseless quantum theory. We
also cannot describe the entangled state $\left\vert \Phi^{+}\right\rangle
_{AB}$ as a product state of the form $\vert\phi\rangle_{A}\vert\psi
\rangle_{B}$, for any states $\vert\phi\rangle_{A}$ or $\vert\psi\rangle_{B}$.
This leads to the following general definition:

\begin{definition}
[Pure-State Entanglement]A pure bipartite state $\vert\psi\rangle_{AB}$ is
entangled if it cannot be written as a product state $\vert\phi\rangle_{A}
\otimes\vert\varphi\rangle_{B}$ for any choices of states $\vert\phi
\rangle_{A}$ and $\vert\varphi\rangle_{B}$.
\end{definition}

\begin{exercise}
\label{ex-qt:+/-ebit}Show that the entangled state $\left\vert \Phi
^{+}\right\rangle _{AB}$ has the following representation in the $+$/$-$
basis:%
\begin{equation}
\left\vert \Phi^{+}\right\rangle _{AB}= \frac{1}{\sqrt{2}} \left(  \left\vert
+\right\rangle _{A}\vert+\rangle_{B}+\vert-\rangle_{A}\left\vert
-\right\rangle _{B}\right)  . \label{eq-qt:ebit-X-basis}%
\end{equation}

\end{exercise}

\begin{figure}
[ptb]
\begin{center}
\includegraphics[
width=1.8844in
]%
{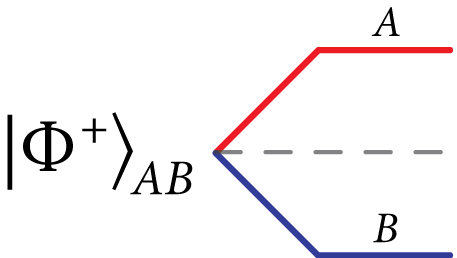}%
\caption{We use the above diagram to depict entanglement shared between two
parties $A$ and $B$. The diagram indicates that a source location creates the
entanglement and distributes one system to $A$ and the other system to $B$.
The standard unit of entanglement is the ebit $\left\vert \Phi^{+}%
\right\rangle _{AB}\equiv(\left\vert 00\right\rangle _{AB}+\left\vert
11\right\rangle _{AB})/\sqrt{2}$.}%
\label{fig-qt:shared-entanglement}%
\end{center}
\end{figure}

Figure~\ref{fig-qt:shared-entanglement} gives a graphical depiction of
entanglement. We use this depiction often throughout this book. Alice and Bob
must receive the entanglement in some way, and the diagram indicates that some
source distributes the entangled pair to them. It indicates that Alice and Bob
are spatially separated and they possess the entangled state after some time.
If they share the entangled state in \eqref{eq-qt:ebit}, we say that they
share one bit of entanglement, or one \textit{ebit}. The term
\textquotedblleft ebit\textquotedblright\ implies that there is some way to
quantify entanglement and we will make this notion clear in
Chapter~\ref{chap:ent-conc}.%

\subsection{Entanglement as a Resource}

In this book, we are interested in the use of entanglement as a resource. Much
of this book concerns the theory of quantum information-processing resources,
and we have a standard notation for the theory of resources. Let us represent
the resource of a shared ebit as
\begin{equation}
\left[  qq\right]  ,
\end{equation}
meaning that the ebit is a noiseless, quantum resource shared between two
parties. Square brackets indicate a noiseless resource, the letter $q$
indicates a quantum resource, and the two copies of the letter $q$ indicate a
two-party resource.

\label{sec-qt:common-randomness}Our first example of the use of entanglement
is its role in generating \textit{shared randomness}. We define one bit of
shared randomness as the following probability distribution for two binary
random variables $X_{A}$ and $X_{B}$:%
\begin{equation}
p_{X_{A},X_{B}}( x_{A},x_{B}) =\frac{1}{2}\delta( x_{A},x_{B}) ,
\label{eq-qt:common-randomness}%
\end{equation}
where $\delta$ is the Kronecker delta function. Suppose Alice possesses random
variable $X_{A}$ and Bob possesses random variable $X_{B}$. Thus, with
probability $1/2$, they either both have a zero or they both have a one. We
represent the resource of one bit of shared randomness as%
\begin{equation}
\left[  cc\right]  ,
\end{equation}
indicating that a bit of shared randomness is a noiseless, classical resource
shared between two parties.

Now suppose that Alice and Bob share an ebit and they decide that they will
each measure their qubits in the computational basis. Without loss of
generality, suppose that Alice performs a measurement first. Thus, Alice
performs a measurement of the $Z_{A}$ operator, meaning that she measures
$Z_{A}\otimes I_{B}$ (she cannot perform anything on Bob's qubit because they
are spatially separated). The projection operators for this measurement are
given in \eqref{eq-qt:alice-projectors-comp}, and they project the joint
state. Just before Alice looks at her measurement result, she does not know
the outcome, and we can describe the system as being in the following ensemble
of states:%
\begin{align}
&  |0\rangle_{A}|0\rangle_{B}\text{ with probability }\frac{1}{2},\\
&  |1\rangle_{A}|1\rangle_{B}\text{ with probability }\frac{1}{2}.
\end{align}
The interesting thing about the above ensemble is that Bob's result is already
determined even before he measures, just after Alice's measurement occurs.
Suppose that Alice knows the result of her measurement is $|0\rangle_{A}$.
When Bob measures his system, he obtains the state $|0\rangle_{B}$ with
probability one and \textit{Alice knows that he has measured this result}.
Additionally, Bob knows that Alice's state is $|0\rangle_{A}$ if he obtains
$|0\rangle_{B}$. The same results hold if Alice knows that the result of her
measurement is $|1\rangle_{A}$. Thus, this protocol is a method for them to
generate one bit of shared randomness as defined in \eqref{eq-qt:common-randomness}.

We can phrase the above protocol as the following \textit{resource
inequality}:%
\begin{equation}
\left[  qq\right]  \geq\left[  cc\right]  . \label{eq-qt:ent>comm-rand}%
\end{equation}
The interpretation of the above resource inequality is \textit{that there
exists a protocol which generates the resource on the right by consuming the
resource on the left and using only local operations}, and for this reason,
the resource on the left is a stronger resource than the one on the right. The
theory of resource inequalities plays a prominent role in this book and is a
useful shorthand for expressing quantum protocols.

A natural question is to wonder if there exists a protocol to generate
entanglement exclusively from shared randomness. It is not possible to do so
and one reason justifying this inequivalence of resources is another type of
inequality (different from the resource inequality mentioned above), called a
\index{Bell inequality}%
Bell's inequality. In short, Bell's theorem places an upper bound on the
correlations present in any two classical systems. Entanglement violates this
inequality, showing that it has no known classical equivalent. Thus,
entanglement is a strictly stronger resource than shared randomness and the
resource inequality in \eqref{eq-qt:ent>comm-rand} only holds in the given direction.

Shared randomness is a resource in classical information theory, and may be
useful in some scenarios, but it is actually a rather weak resource. Surely,
generating shared randomness is not the only use of entanglement. It turns out
that we can construct far more exotic protocols such as the teleportation
protocol or the super-dense coding protocol by combining the resource of
entanglement with other resources. We discuss these protocols in
Chapter~\ref{chap:three-noiseless}.

\begin{exercise}
Use the representation of the ebit in Exercise~\ref{ex-qt:+/-ebit} to show
that Alice and Bob can measure the $X$ operator to generate shared randomness.
This ability to obtain shared randomness by both parties measuring in either
the $Z$ or $X$ basis is the foundation for an entanglement-based secret key
distribution protocol.
\end{exercise}

\begin{exercise}
[Cloning Implies Signaling]Prove that if a universal quantum cloner were to
exist, then it would be possible for Alice to signal to Bob faster than the
speed of light by exploiting only the ebit state $\left\vert \Phi
^{+}\right\rangle _{AB}$ shared between them and no communication. That is,
show the existence of a protocol that would allow for this. (Hint: One
possibility is for Alice to measure the $X$ or $Z$ Pauli operator locally on
her share of the ebit, and then for Bob to exploit the universal quantum
cloner. Consider the representation of the ebit in \eqref{eq-qt:ebit} and
\eqref{eq-qt:ebit-X-basis}. Note that there could be a variety of answers to
this question because quantum theory becomes effectively nonlinear if we
assume the existence of a cloner!)
\end{exercise}

\subsection{Entanglement in the CHSH\ Game}

\label{sec-qt:CHSH-game}%

\index{CHSH game}%

\index{Bell's theorem}%
One of the simplest means for demonstrating the power of entanglement is with
a two-player game known as the CHSH\ game (after Clauser, Horne, Shimony, and
Holt), which is a particular variation of the original setup in Bell's
theorem. We first present the rules of the game, and then we find an upper
bound on the probability that players operating according to a classical
strategy can win. We finally leave it as an exercise to show that players
sharing a maximally entangled Bell state $\left\vert \Phi^{+}\right\rangle $
can have an approximately 10\% higher chance of winning the game using a
quantum strategy. This result, known as \textit{Bell's theorem}, represents
one of the most striking separations between classical and quantum physics.

\begin{figure}
[ptb]
\begin{center}
\includegraphics[
width=3.4411in
]%
{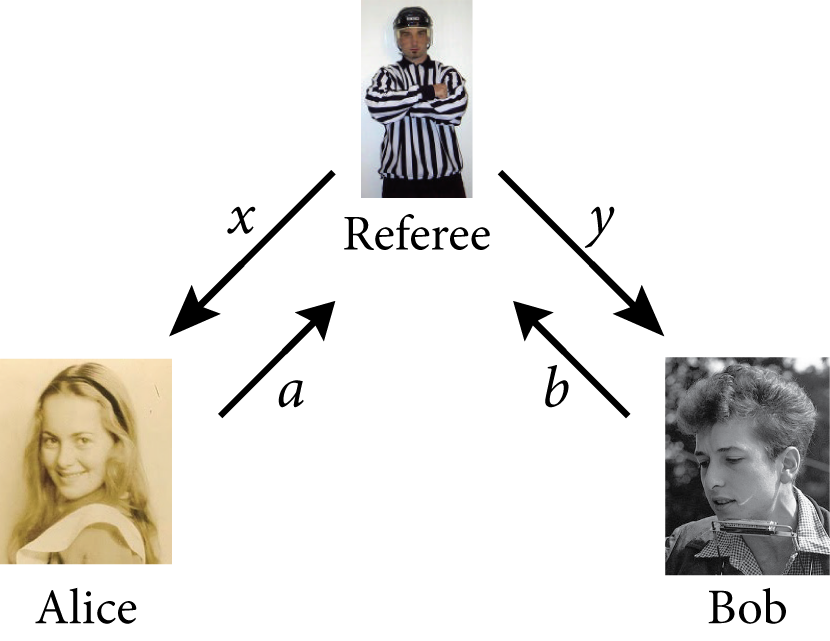}%
\caption{A depiction of the CHSH game. The referee distributes the bits $x$
and $y$ to Alice and Bob in the first round. In the second round, Alice and
Bob return the bits $a$ and $b$ to the referee.}%
\label{fig-qt:CHSH-game}%
\end{center}
\end{figure}

The players of the game are Alice and Bob, who are spatially separated from
each other from the time that the game starts until it is over. The game
begins with a referee selecting two bits $x$ and $y$ uniformly at random. The
referee then sends $x$ to Alice and $y$ to Bob. Alice and Bob are not allowed
to communicate with each other in any way at this point. Alice sends back to
the referee a bit $a$, and Bob sends back a bit $b$. Since they are spatially
separated, Alice's response bit $a$ cannot depend on Bob's input bit $y$, and
similarly, Bob's response bit $b$ cannot depend on Alice's input bit $x$.
After receiving the response bits $a$ and $b$, the referee determines if the
AND\ of $x$ and $y$ is equal to the exclusive OR of $a$ and $b$. If so, then
Alice and Bob win the game. That is, the winning condition is%
\begin{equation}
x\wedge y=a\oplus b. \label{eq-qt:chsh-predicate}%
\end{equation}
Figure~\ref{fig-qt:CHSH-game}\ depicts the CHSH\ game.%

We need to figure out an expression for the winning probability of the
CHSH\ game. Let $V(x,y,a,b)$ denote the following indicator function for
whether they win in a particular instance of the game:%
\begin{equation}
V(x,y,a,b)=\left\{
\begin{array}
[c]{cc}%
1 & \text{if }x\wedge y=a\oplus b\\
0 & \text{else}%
\end{array}
\right.  .
\end{equation}
There is a conditional probability distribution $p_{AB|XY}(a,b|x,y)$, which
corresponds to the particular strategy that Alice and Bob employ. Since the
inputs $x$ and $y$ are chosen uniformly at random and each take on two
possible values, the distribution $p_{XY}(x,y)$\ for $x$ and $y$ is as
follows:%
\begin{equation}
p_{XY}(x,y)=1/4.
\end{equation}
So an expression for the winning probability of the CHSH\ game is%
\begin{equation}
\frac{1}{4}\sum_{a,b,x,y}V(x,y,a,b)p_{AB|XY}(a,b|x,y).
\label{eq-qt:CHSH-winning-prob}%
\end{equation}

\begin{figure}[ptb]
\begin{center}
\includegraphics[
width=2.3865in
]{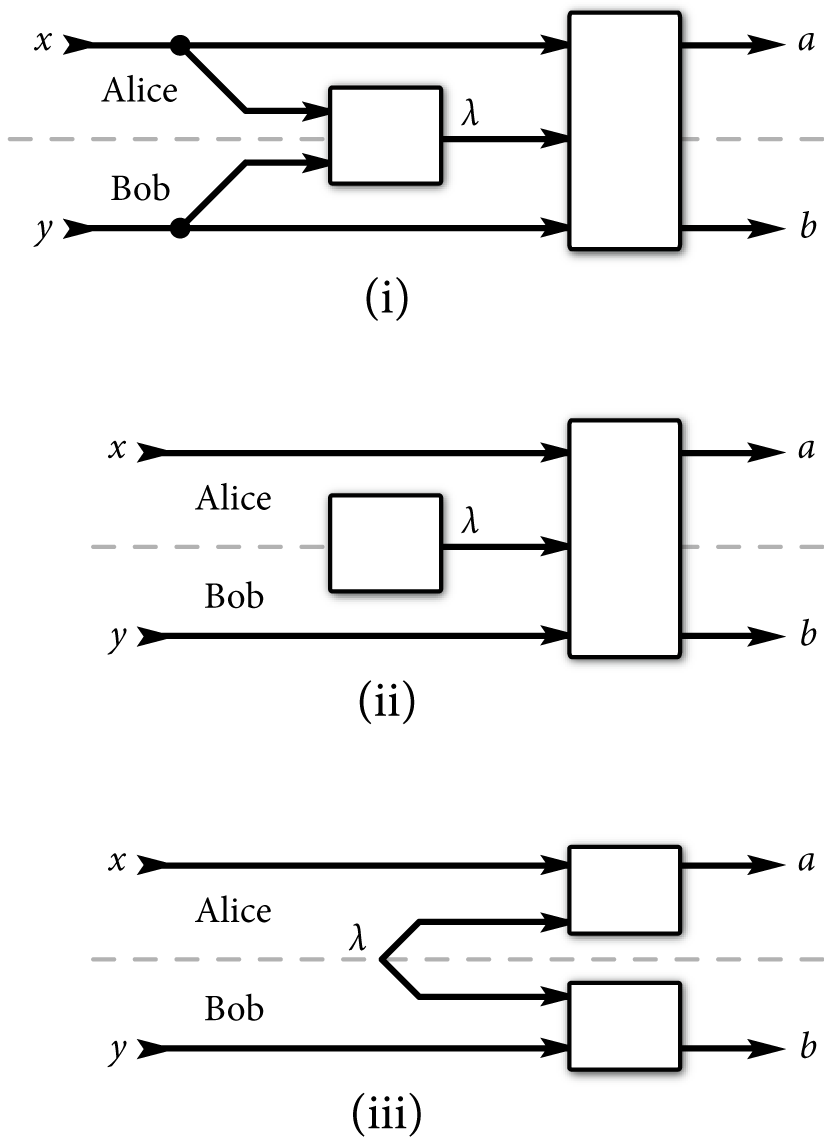}
\end{center}
\caption{Various reductions of a classical strategy in the CHSH\ game: (i) an
unconstrained strategy, (ii) strategy resulting from demanding that the
parameter $\lambda$ is independent of the input bits $x$ and $y$, and (iii)
further demanding that Alice and Bob's actions are independent and that they
do not have access to each other's input bits.}%
\label{fig-qt:chsh-reductions}%
\end{figure}In order to calculate this winning probability for a classical or
quantum strategy, we need to understand the distribution $p_{AB|XY}(a,b|x,y)$
further. In order to do so, we need a way for describing the strategy that
Alice and Bob employ. For this purpose, we will assume that there is a random
variable $\Lambda$ taking values $\lambda$, which describes either a classical
or quantum strategy, and its values could be all of the entries in a matrix
and even taking on continuous values. Using the law of total probability, we
can expand the conditional probability $p_{AB|XY}(a,b|x,y)$ as follows:%
\begin{equation}
p_{AB|XY}(a,b|x,y)=\int d\lambda\ p_{AB|\Lambda XY}(a,b|\lambda
,x,y)\ p_{\Lambda|XY}(\lambda|x,y),
\end{equation}
where $p_{\Lambda|XY}(\lambda|x,y)$ is a conditional probability distribution.
Decomposing the distribution $p_{AB|XY}(a,b|x,y)$ in this way leads to the
depiction of their strategy given in Figure~\ref{fig-qt:chsh-reductions}(i).

\subsubsection{Classical Strategies}

Let us suppose that they act according to a classical strategy. What is the
most general form of such a strategy?\ Looking at the picture in
Figure~\ref{fig-qt:chsh-reductions}(i), there are a few aspects of it which
are not consistent with our understanding of how the game works.

In a classical strategy, the random variable $\Lambda$ corresponds to
classical correlations that Alice and Bob can share \textit{before the game
begins}. They could meet beforehand and select a value $\lambda$\ of $\Lambda$
at random. According to the specification of the game, the input bits $x$ and
$y$ for Alice and Bob are chosen independently at random, and so the random
variable $\Lambda$ cannot depend on the bits $x$ and $y$. So the conditional
distribution $p_{\Lambda|XY}(\lambda|x,y)$ simplifies as follows:%
\begin{equation}
p_{\Lambda|XY}(\lambda|x,y)=p_{\Lambda}(\lambda),
\end{equation}
and Figure~\ref{fig-qt:chsh-reductions}(ii)\ reflects this constraint.

Next, Alice and Bob are spatially separated and acting independently, so that
the distribution $p_{AB|\Lambda XY}(a,b|\lambda,x,y)$ factors as follows:%
\begin{equation}
p_{AB|\Lambda XY}(a,b|\lambda,x,y)=p_{A|\Lambda XY}(a|\lambda
,x,y)\ p_{B|\Lambda XY}(b|\lambda,x,y).
\end{equation}
But we also said that Alice's strategy cannot depend on Bob's input bit $y$
and neither can Bob's strategy depend on Alice's input $x$, because they are
spatially separated. However, their strategies could depend on the random
variable $\Lambda$, which they are allowed to share before the game begins.
All of this implies that the conditional distribution describing their
strategy should factor as follows:%
\begin{equation}
p_{AB|\Lambda XY}(a,b|\lambda,x,y)=p_{A|\Lambda X}(a|\lambda,x)\ p_{B|\Lambda
Y}(b|\lambda,y),
\end{equation}
and Figure~\ref{fig-qt:chsh-reductions}(iii) reflects this change. Now
Figure~\ref{fig-qt:chsh-reductions}(iii)\ depicts the most general classical
strategy that Alice and Bob could employ if $\Lambda$ corresponds to a random
variable that Alice and Bob are both allowed to access before the game begins.

Putting everything together, the conditional distribution $p_{AB|XY}(a,b|x,y)$
for a classical strategy takes the following form:%
\begin{equation}
p_{AB|XY}(a,b|x,y)=\int d\lambda\ p_{A|\Lambda X}(a|\lambda,x)\ p_{B|\Lambda
Y}(b|\lambda,y)\ p_{\Lambda}(\lambda), \label{eq-qt:CHSH-classical-strat}%
\end{equation}
and we can now consider optimizing the winning probability in
\eqref{eq-qt:CHSH-winning-prob} with respect to all classical strategies.
Consider that any stochastic map $p_{A|\Lambda X}(a|\lambda,x)$ can be
simulated by applying a deterministic binary-valued function $f(a|\lambda
,x,n)$ to a local random variable $N$ taking values labeled by $n$. That is,
we can always find a random variable $N$ such that%
\begin{equation}
p_{A|\Lambda X}(a|\lambda,x)=\int dn\ f(a|\lambda,x,n)\ p_{N}(n).
\end{equation}
The same is true for the stochastic map $p_{B|\Lambda Y}(b|\lambda,y)$; i.e.,
there is a random variable $M$ such that%
\begin{equation}
p_{B|\Lambda Y}(b|\lambda,y)=\int dm\ g(b|\lambda,y,m)\ p_{M}(m),
\end{equation}
where $g$ is a deterministic binary-valued function. So this implies that%
\begin{align}
&  p_{AB|XY}(a,b|x,y)\nonumber\\
&  =\int d\lambda\ p_{A|\Lambda X}(a|\lambda,x)\ p_{B|\Lambda Y}%
(b|\lambda,y)\ p_{\Lambda}(\lambda)\\
&  =\int d\lambda\ \left[  \int dn\ f(a|\lambda,x,n)\ p_{N}(n)\right]
\ \left[  \int dm\ g(b|\lambda,y,m)\ p_{M}(m)\right]  \ p_{\Lambda}(\lambda)\\
&  =\int\int\int d\lambda\ dn\ dm\ f(a|\lambda,x,n)\ \ g(b|\lambda
,y,m)\ \ p_{\Lambda}(\lambda)\ p_{N}(n)\ p_{M}(m).
\end{align}
By inspecting the last line above, it is clear that we could then have the
shared random variable $\Lambda$ subsume the local random variables $N$ and
$M$, allowing us to write any conditional distribution $p_{AB|XY}(a,b|x,y)$
for a classical strategy as follows:%
\begin{equation}
p_{AB|XY}(a,b|x,y)=\int d\lambda\ f^{\prime}(a|\lambda,x)\ g^{\prime
}(b|\lambda,y)\ p_{\Lambda}(\lambda),
\end{equation}
where $f^{\prime}$ and $g^{\prime}$ are deterministic binary-valued functions
(related to $f$ and $g$). Substituting this expression into the winning
probability expression in \eqref{eq-qt:CHSH-winning-prob}, we find that%
\begin{align}
&  \frac{1}{4}\sum_{a,b,x,y}V(x,y,a,b)p_{AB|XY}(a,b|x,y)\nonumber\\
&  =\frac{1}{4}\sum_{a,b,x,y}V(x,y,a,b)\int d\lambda\ f^{\prime}%
(a|\lambda,x)\ g^{\prime}(b|\lambda,y)\ p_{\Lambda}(\lambda)\\
&  =\int d\lambda\ p_{\Lambda}(\lambda)\left[  \frac{1}{4}\sum_{a,b,x,y}%
V(x,y,a,b)\ f^{\prime}(a|\lambda,x)\ g^{\prime}(b|\lambda,y)\right] \\
&  \leq\frac{1}{4}\sum_{a,b,x,y}V(x,y,a,b)\ f^{\prime}(a|\lambda^{\ast
},x)\ g^{\prime}(b|\lambda^{\ast},y).
\end{align}
In the second equality, we just exchanged the integral over $\lambda$ with the
sum. In the inequality in the last step, we used the fact that the average is
always less than the maximum. That is, there is always a particular value
$\lambda^{\ast}$ that leads to the same or a higher winning probability than
when averaging over all values of $\lambda$. As a consequence of the above
development, we see that it suffices to consider deterministic strategies of
Alice and Bob when analyzing the winning probability.

Since we now know that deterministic strategies are optimal among all
classical strategies, let us focus on these. A deterministic strategy would
have Alice select a bit $a_{x}$ conditioned on the bit $x$ that she receives,
and similarly, Bob would select a bit $b_{y}$ conditioned on $y$. The
following table presents the winning conditions for the four different values
of $x$ and $y$ with this deterministic strategy:%
\begin{equation}%
\begin{tabular}
[c]{c|c|c|c}\hline\hline
$x$ & $y$ & $x\wedge y$ & $=a_{x}\oplus b_{y}$\\\hline\hline
0 & 0 & 0 & $=a_{0}\oplus b_{0}$\\\hline
0 & 1 & 0 & $=a_{0}\oplus b_{1}$\\\hline
1 & 0 & 0 & $=a_{1}\oplus b_{0}$\\\hline
1 & 1 & 1 & $=a_{1}\oplus b_{1}$\\\hline\hline
\end{tabular}
\ \ \ .
\end{equation}
However, we can observe that it is impossible for them to always win. If we
add the entries in the column $x\wedge y$, the binary sum is equal to one,
while if we add the entries in the column $=a_{x}\oplus b_{y}$, the binary sum
is equal to zero. Thus, it is impossible for all of these equations to be
satisfied. At most, only three out of four of them can be satisfied, so that
the maximal winning probability with a classical deterministic strategy
$p_{AB|XY}(a,b|x,y)$ is at most $3/4$:%
\begin{equation}
\frac{1}{4}\sum_{a,b,x,y}V(x,y,a,b)p_{AB|XY}(a,b|x,y)\leq\frac{3}{4}.
\end{equation}
We can then see that a strategy for them to achieve this upper bound is for
Alice and Bob always to return $a=0$ and $b=0$ no matter the values of $x$ and
$y$.

\subsubsection{Quantum Strategies}

What does a quantum strategy of Alice and Bob look like? Here the parameter
$\lambda$ can correspond to a shared quantum state $|\phi\rangle_{AB}$. Alice
and Bob perform local measurements depending on the values of the inputs $x$
and $y$ that they receive. We can write Alice's $x$-dependent measurement as
$\{\Pi_{a}^{(x)}\}$ where for each $x$, $\Pi_{a}^{(x)}$ is a projector and
$\sum_{a}\Pi_{a}^{(x)}=I$. Similarly, we can write Bob's $y$-dependent
measurement as $\{\Pi_{b}^{(y)}\}$. Then we instead employ the Born rule to
determine the conditional probability distribution $p_{AB|XY}(a,b|x,y)$:%
\begin{equation}
p_{AB|XY}(a,b|x,y)=\langle\phi|_{AB}\Pi_{a}^{(x)}\otimes\Pi_{b}^{(y)}%
|\phi\rangle_{AB},
\end{equation}
so that the winning probability with a particular quantum strategy is as
follows:%
\begin{equation}
\frac{1}{4}\sum_{a,b,x,y}V(x,y,a,b)\langle\phi|_{AB}\Pi_{a}^{(x)}\otimes
\Pi_{b}^{(y)}|\phi\rangle_{AB}.
\end{equation}

Interestingly, if Alice and Bob share a maximally entangled state, they can
achieve a higher winning probability than if they share classical correlations
only. This is one demonstration of the power of entanglement, and we leave it
as an exercise to prove that the following quantum strategy achieves a winning
probability of $\cos^{2}(\pi/8)\approx0.85$ in the CHSH game.

\begin{exercise}
Suppose that Alice and Bob share a maximally entangled state $\left\vert
\Phi^{+}\right\rangle $. Show that the following strategy has a winning
probability of $\cos^{2}(\pi/8)$. If Alice receives $x=0$ from the referee,
then she performs a measurement of Pauli $Z$ on her system and returns the
measurement outcome as ``a'' after identifying $a=0$ with the measurement
outcome $+1$ and $a=1$ with the measurement outcome $-1$. (The same convention
is applied to the following scenarios.) If she receives $x=1$, then she
performs a measurement of Pauli $X$ and returns the outcome as ``a.'' If Bob
receives $y=0$ from the referee, then he performs a measurement of $\left(
X+Z\right)  /\sqrt{2}$ on his system and returns the outcome as $b$. If Bob
receives $y=1$ from the referee, then he performs a measurement of $\left(
Z-X\right)  /\sqrt{2}$ and returns the outcome as $b$.
\end{exercise}

\subsubsection{Maximum Quantum Winning Probability}%

\index{Tsirelson's bound}%

Given that classical strategies cannot win with probability any larger than
$3/4$, it is natural to wonder if there is a bound on the winning probability
of a quantum strategy. It turns out that $\cos^{2}(\pi/8)$ is the maximum
probability with which Alice and Bob can win the CHSH game using a quantum
strategy, a result known as \textit{Tsirelson's bound}. To establish this
result, let us go back to the CHSH game. Conditioned on the inputs $x$ and $y$
being equal to 00, 01, or 10, we know that Alice and Bob win if they report
back the same results. The probability for this to happen with a given quantum
strategy is%
\begin{equation}
\langle\phi|_{AB}\Pi_{0}^{(x)}\otimes\Pi_{0}^{(y)}|\phi\rangle_{AB}%
+\langle\phi|_{AB}\Pi_{1}^{(x)}\otimes\Pi_{1}^{(y)}|\phi\rangle_{AB},
\end{equation}
and the probability for it not to happen is%
\begin{equation}
\langle\phi|_{AB}\Pi_{0}^{(x)}\otimes\Pi_{1}^{(y)}|\phi\rangle_{AB}%
+\langle\phi|_{AB}\Pi_{1}^{(x)}\otimes\Pi_{0}^{(y)}|\phi\rangle_{AB}.
\end{equation}
So, conditioned on $x$ and $y$ being equal to 00, 01, or 10, the probability
of winning minus the probability of losing is%
\begin{equation}
\langle\phi|_{AB}A^{(x)}\otimes B^{(y)}|\phi\rangle_{AB},
\end{equation}
where we define the \textit{observables} $A^{(x)}$ and $B^{(y)}$ as follows:%
\begin{align}
A^{(x)}  &  \equiv\Pi_{0}^{(x)}-\Pi_{1}^{(x)},\\
B^{(y)}  &  \equiv\Pi_{0}^{(y)}-\Pi_{1}^{(y)}.
\end{align}
If $x$ and $y$ are both equal to one, then Alice and Bob should report back
different results, and similar to the above, one can work out that the
probability of winning minus the probability of losing is equal to%
\begin{equation}
-\langle\phi|_{AB}A^{(1)}\otimes B^{(1)}|\phi\rangle_{AB}.
\end{equation}
Thus, when averaging over all values of the input bits, the probability of
winning minus the probability of losing is equal to%
\begin{equation}
\frac{1}{4}\langle\phi|_{AB}C_{AB}|\phi\rangle_{AB},
\label{eq-qt:win-lose-tsi}%
\end{equation}
where $C_{AB}$ is the CHSH\ operator, defined as%
\begin{equation}
C_{AB}\equiv A^{(0)}\otimes B^{(0)}+A^{(0)}\otimes B^{(1)}+A^{(1)}\otimes
B^{(0)}-A^{(1)}\otimes B^{(1)}.
\end{equation}
It is a simple exercise to check that%
\begin{equation}
C_{AB}^{2}=4I_{AB}-\left[  A^{(0)},A^{(1)}\right]  \otimes\left[
B^{(0)},B^{(1)}\right]  .
\end{equation}
The infinity norm $\left\Vert R\right\Vert _{\infty}$\ of an operator $R$ is
equal to its largest singular value. It obeys the following relations:%
\begin{align}
\left\Vert cR\right\Vert _{\infty}  &  =\left\vert c\right\vert \left\Vert
R\right\Vert _{\infty},\\
\left\Vert RS\right\Vert _{\infty}  &  \leq\left\Vert R\right\Vert _{\infty
}\left\Vert S\right\Vert _{\infty},\\
\left\Vert R+S\right\Vert _{\infty}  &  \leq\left\Vert R\right\Vert _{\infty
}+\left\Vert S\right\Vert _{\infty},
\end{align}
where $c\in\mathbb{C}$ and $S$ is another operator. Using these, we find that%
\begin{align}
\left\Vert C_{AB}^{2}\right\Vert _{\infty}  &  =\left\Vert 4I_{AB}-\left[
A^{(0)},A^{(1)}\right]  \otimes\left[  B^{(0)},B^{(1)}\right]  \right\Vert
_{\infty}\\
&  \leq4\left\Vert I_{AB}\right\Vert _{\infty}+\left\Vert \left[
A^{(0)},A^{(1)}\right]  \otimes\left[  B^{(0)},B^{(1)}\right]  \right\Vert
_{\infty}\\
&  =4+\left\Vert \left[  A^{(0)},A^{(1)}\right]  \right\Vert _{\infty
}\left\Vert \left[  B^{(0)},B^{(1)}\right]  \right\Vert _{\infty}\\
&  \leq4+2\cdot2=8,
\end{align}
implying that%
\begin{equation}
\left\Vert C_{AB}\right\Vert _{\infty}\leq\sqrt{8}=2\sqrt{2}.
\end{equation}
Given this and the expression in \eqref{eq-qt:win-lose-tsi}, the probability
of winning minus the probability of losing can never be larger than $\sqrt
{2}/2$ for any quantum strategy. Combined with the fact that the probability
of winning summed with the probability of losing is equal to one, we find that
the winning probability of any quantum strategy can never be larger than
$1/2+\sqrt{2}/4=\cos^{2}(\pi/8)$.

\subsection{The Bell States}%

\index{Bell states}%
There are other useful entangled states besides the standard ebit. Suppose
that Alice performs a $Z_{A}$ operation on her share of the ebit $\left\vert
\Phi^{+}\right\rangle _{AB}$. Then the resulting state is%
\begin{equation}
\left\vert \Phi^{-}\right\rangle _{AB}\equiv\frac{1}{\sqrt{2}}\left(
|00\rangle_{AB}-|11\rangle_{AB}\right)  . \label{eq-qt:bell1}%
\end{equation}
Similarly, if Alice performs an $X$ operator or a $Y$ operator, the global
state transforms to the following respective states (up to a global phase):%
\begin{align}
\left\vert \Psi^{+}\right\rangle _{AB}  &  \equiv\frac{1}{\sqrt{2}}\left(
|01\rangle_{AB}+|10\rangle_{AB}\right)  ,\label{eq-qt:bell2}\\
\left\vert \Psi^{-}\right\rangle _{AB}  &  \equiv\frac{1}{\sqrt{2}}\left(
|01\rangle_{AB}-|10\rangle_{AB}\right)  . \label{eq-qt:bell3}%
\end{align}
The states $\left\vert \Phi^{+}\right\rangle _{AB}$, $\left\vert \Phi
^{-}\right\rangle _{AB}$, $\left\vert \Psi^{+}\right\rangle _{AB}$, and
$\left\vert \Psi^{-}\right\rangle _{AB}$ are known as the \textit{Bell states
}and are the most important entangled states for a two-qubit system. They form
an orthonormal basis, called the \textit{Bell basis}, for a two-qubit space.
We can also label the Bell states as%
\begin{equation}
\left\vert \Phi^{zx}\right\rangle _{AB}\equiv Z_{A}^{z}X_{A}^{x}\left\vert
\Phi^{+}\right\rangle _{AB}, \label{eq-qt:Bell-state-bits}%
\end{equation}
where the two-bit binary number $zx$ indicates whether Alice applies $I_{A}$,
$Z_{A}$, $X_{A}$, or $Z_{A}X_{A}$. Then the states $\left\vert \Phi
^{00}\right\rangle _{AB}$, $\left\vert \Phi^{01}\right\rangle _{AB}$,
$\left\vert \Phi^{10}\right\rangle _{AB}$, and $\left\vert \Phi^{11}%
\right\rangle _{AB}$ are in correspondence with the respective states
$\left\vert \Phi^{+}\right\rangle _{AB}$, $\left\vert \Psi^{+}\right\rangle
_{AB}$, $\left\vert \Phi^{-}\right\rangle _{AB}$, and $\left\vert \Psi
^{-}\right\rangle _{AB}$.

\begin{exercise}
Show that the Bell states form an orthonormal basis:%
\begin{equation}
\left\langle \Phi^{z_{1}x_{1}}|\Phi^{z_{2}x_{2}}\right\rangle =\delta
_{z_{1},z_{2}}\delta_{x_{1},x_{2}}.
\end{equation}

\end{exercise}

\begin{exercise}
\label{ex-qt:bell-comp}Show that the following identities hold:%
\begin{align}
|00\rangle_{AB}  &  =\frac{1}{\sqrt{2}}\left(  \left\vert \Phi^{+}%
\right\rangle _{AB}+\left\vert \Phi^{-}\right\rangle _{AB}\right)  ,\\
|01\rangle_{AB}  &  =\frac{1}{\sqrt{2}}\left(  \left\vert \Psi^{+}%
\right\rangle _{AB}+\left\vert \Psi^{-}\right\rangle _{AB}\right)  ,\\
|10\rangle_{AB}  &  =\frac{1}{\sqrt{2}}\left(  \left\vert \Psi^{+}%
\right\rangle _{AB}-\left\vert \Psi^{-}\right\rangle _{AB}\right)  ,\\
|11\rangle_{AB}  &  =\frac{1}{\sqrt{2}}\left(  \left\vert \Phi^{+}%
\right\rangle _{AB}-\left\vert \Phi^{-}\right\rangle _{AB}\right)  .
\end{align}

\end{exercise}

\begin{exercise}
Show that the following identities hold by using the relation in
\eqref{eq-qt:Bell-state-bits}:%
\begin{align}
\left\vert \Phi^{+}\right\rangle _{AB}  &  =\frac{1}{\sqrt{2}}\left(
\left\vert ++\right\rangle _{AB}+\left\vert --\right\rangle _{AB}\right)  ,\\
\left\vert \Phi^{-}\right\rangle _{AB}  &  =\frac{1}{\sqrt{2}}\left(
\left\vert -+\right\rangle _{AB}+\left\vert +-\right\rangle _{AB}\right)  ,\\
\left\vert \Psi^{+}\right\rangle _{AB}  &  =\frac{1}{\sqrt{2}}\left(
\left\vert ++\right\rangle _{AB}-\left\vert --\right\rangle _{AB}\right)  ,\\
\left\vert \Psi^{-}\right\rangle _{AB}  &  =\frac{1}{\sqrt{2}}\left(
\left\vert -+\right\rangle _{AB}-\left\vert +-\right\rangle _{AB}\right)  .
\end{align}

\end{exercise}

Entanglement is one of the most useful resources in quantum computing, quantum
communication, and in the setting of quantum Shannon theory that we explore in
this book. Our goal in this book is merely to study entanglement as a
resource, but there are many other aspects of entanglement that one can study,
such as measures of entanglement, multiparty entanglement, and generalized
Bell's inequalities~\citep{H42007}.

\section{Summary and Extensions to Qudit States}

We now end our overview of the noiseless quantum theory by summarizing its
main postulates in terms of quantum states that are on $d$-dimensional
systems. Such states are called \textit{qudit states}, in analogy with the
name \textquotedblleft qubit\textquotedblright\ for two-dimensional quantum systems.

\subsection{Qudits}

A qudit
\index{qudit}%
state $\vert\psi\rangle$ is an arbitrary superposition of some set of
orthonormal basis states $\left\{  \vert j\rangle\right\}  _{j\in\left\{
0,\ldots,d-1\right\}  }$ for a $d$-dimensional quantum system:%
\begin{equation}
\vert\psi\rangle\equiv\sum_{j=0}^{d-1}\alpha_{j}\vert j\rangle.
\label{eq-qt:qudit-state}%
\end{equation}
The amplitudes $\alpha_{j}$ obey the normalization condition $\sum_{j=0}%
^{d-1}\left\vert \alpha_{j}\right\vert ^{2}=1$.

\subsection{Unitary Evolution}

\label{sec-nqt:generalized-Pauli}The first postulate of the quantum theory is
that we can perform a unitary (reversible) evolution $U$ on this state. The
resulting state is $U\vert\psi\rangle, $ meaning that we apply the operator
$U$ to the state $\vert\psi\rangle$.

One example of a unitary evolution is the cyclic shift operator $X( x) $ that
acts on the orthonormal states $\left\{  \vert j\rangle\right\}
_{j\in\left\{  0,\ldots,d-1\right\}  }$ as follows:%
\begin{equation}
X( x) \vert j\rangle=\left\vert x\oplus j\right\rangle , \label{eq-qt:X-op}%
\end{equation}
where $\oplus$ is a cyclic addition operator, meaning that the result of the
addition is $\left(  x+j\right)  \operatorname{mod}\left(  d\right)  $. Notice
that the $X$ Pauli operator has a similar behavior on the qubit computational
basis states because%
\begin{equation}
X\vert i\rangle=\left\vert i\oplus1\right\rangle ,
\end{equation}
for $i\in\left\{  0,1\right\}  $. Therefore, the operator $X( x) $ is a qudit
generalization of the $X$ Pauli operator.

\begin{exercise}
Show that the inverse of $X( x) $ is $X( -x) $.
\end{exercise}

\begin{exercise}
Show that the matrix representation $X( x) $ of the $X( x) $ operator, with
respect to the standard basis $\{\vert j \rangle\}$, is a matrix with elements
$\left[  X( x) \right]  _{i,j}=\delta_{i,j\oplus x}. $
\end{exercise}

Another example of a unitary evolution is the \textit{phase operator} $Z( z)
$. It applies a state-dependent phase to a basis state. It acts as follows on
the qudit computational basis states $\left\{  \vert j\rangle\right\}
_{j\in\left\{  0,\ldots,d-1\right\}  }$:%
\begin{equation}
Z( z) \vert j\rangle=\exp\left\{  i2\pi zj/d\right\}  \left\vert
j\right\rangle . \label{eq-qt:Z-op}%
\end{equation}
This operator is the qudit analog of the Pauli $Z$ operator. The $d^{2}$
operators $\left\{  X( x) Z( z) \right\}  _{x,z\in\left\{  0,\ldots
,d-1\right\}  }$ are known as the
\index{Heisenberg-Weyl operators}
\textit{Heisenberg--Weyl operators}.

\begin{exercise}
Show that $Z( 1) $ is equivalent to the Pauli $Z$ operator for the case that
the dimension $d=2$.
\end{exercise}

\begin{exercise}
Show that the inverse of $Z( z) $ is $Z( -z) $.
\end{exercise}

\begin{exercise}
Show that the matrix representation of the phase operator $Z( z) $, with
respect to the standard basis $\{\vert j \rangle\}$, is%
\begin{equation}
\left[  Z( z) \right]  _{j,k}=\exp\left\{  i2\pi zj/d\right\}  \delta_{j,k}.
\end{equation}
In particular, this result implies that the $Z( z) $ operator has a diagonal
matrix representation with respect to the qudit computational basis states
$\left\{  \vert j\rangle\right\}  _{j\in\left\{  0,\ldots,d-1\right\}  }$.
Thus, the qudit computational basis states $\left\{  \vert j\rangle\right\}
_{j\in\left\{  0,\ldots,d-1\right\}  }$ are eigenstates of the phase operator
$Z( z) $ (similar to the qubit computational basis states being eigenstates of
the Pauli $Z$ operator). The eigenvalue corresponding to the eigenstate $\vert
j\rangle$ is $\exp\left\{  i2\pi zj/d\right\}  $.
\end{exercise}

\begin{exercise}
Show that the eigenstates $|\widetilde{l}\rangle$\ of the cyclic shift
operator $X( 1) $ are the Fourier-transformed states $|\widetilde{l}\rangle$,
where%
\begin{equation}
|\widetilde{l}\rangle\equiv\frac{1}{\sqrt{d}}\sum_{j=0}^{d-1}\exp\left\{
i2\pi lj/d\right\}  \vert j\rangle, \label{eq-qt:X-eigenstates}%
\end{equation}
and $l$ is an integer in the set $\left\{  0,\ldots,d-1\right\}  $. Show that
the eigenvalue corresponding to the state $|\widetilde{l}\rangle$ is
$\exp\left\{  -i2\pi l/d\right\}  $. Conclude that these states are also
eigenstates of the operator $X( x) $, but the corresponding eigenvalues are
$\exp\left\{  -i2\pi lx/d\right\}  $.
\end{exercise}

\begin{exercise}
Show that the $+$/$-$ basis states are a special case of the states in
\eqref{eq-qt:X-eigenstates} when $d=2$.
\end{exercise}

\begin{exercise}
\label{ex-qt:fourier-gate}The Fourier transform%
\index{Fourier transform}
operator $F$ is a qudit analog of the Hadamard $H$. We define it to take $Z$
eigenstates to $X$ eigenstates: $F\equiv\sum_{j=0}^{d-1}|\widetilde{j}
\rangle\langle j\vert, $ where the states $|\widetilde{j} \rangle$ are defined
in \eqref{eq-qt:X-eigenstates}. It performs the following transformation on
the qudit computational basis states:%
\begin{equation}
\vert j\rangle\rightarrow\frac{1}{\sqrt{d}}\sum_{k=0}^{d-1}\exp\left\{  i2\pi
jk/d\right\}  \vert k\rangle.
\end{equation}
Show that the following relations hold for the Fourier transform operator $F$:
$FX( x) F^{\dag} =Z( x)$, $FZ( z) F^{\dag} =X( -z) $.
\end{exercise}

\begin{exercise}
Show that the commutation relations of the cyclic shift operator $X(x)$ and
the phase operator $Z(z)$ are as follows:%
\begin{multline}
X(x_{1})Z(z_{1})X(x_{2})Z(z_{2})=\\
\exp\left\{  2\pi i\left(  z_{1}x_{2}-x_{1}z_{2}\right)  /d\right\}
X(x_{2})Z(z_{2})X(x_{1})Z(z_{1}).
\end{multline}
You can get this result by first showing that%
\begin{equation}
X(x)Z(z)=\exp\left\{  -2\pi izx/d\right\}  Z(z)X(x).
\end{equation}

\end{exercise}

\subsection{Measurement of Qudits}

Measurement of qudits is similar to measurement of qubits. Suppose that we
have some state $\vert\psi\rangle$. Suppose further that we would like to
measure some Hermitian operator $A$ with the following diagonalization:%
\begin{equation}
A=\sum_{j}f( j) \Pi_{j},
\end{equation}
where $\Pi_{j}\Pi_{k}=\Pi_{j}\delta_{j,k}$, and $\sum_{j}\Pi_{j}=I$. A
measurement of the operator $A$ then returns the result $j$ with the following
probability:%
\begin{equation}
p( j) =\langle\psi\vert\Pi_{j}\vert\psi\rangle,
\end{equation}
and the resulting state is%
\begin{equation}
\frac{\Pi_{j}\vert\psi\rangle}{\sqrt{p( j) }}.
\end{equation}
The calculation of the expectation of the operator $A$ is similar to how we
calculate in the qubit case:%
\begin{equation}
\mathbb{E}\left[  A\right]  =\sum_{j}f( j) \langle\psi\vert\Pi_{j}\vert
\psi\rangle=\langle\psi\vert\sum_{j}f( j) \Pi_{j}\vert\psi\rangle=\langle
\psi\vert A\vert\psi\rangle.
\end{equation}

We give two quick examples of qudit operators that we might like to measure.
The operators $X(1)$ and $Z(1)$ are not completely analogous to the respective
Pauli $X$ and Pauli $Z$ operators because $X(1)$ and $Z(1)$ are not Hermitian.
Thus, we cannot directly measure these operators. Instead, we construct
operators that are essentially equivalent to \textquotedblleft measuring the
operators\textquotedblright\ $X(1)$ and $Z(1)$. Let us first consider the
$Z(1)$ operator. Its eigenstates are the qudit computational basis states
$\left\{  |j\rangle\right\}  _{j\in\left\{  0,\ldots,d-1\right\}  }$. We can
form the operator $M_{Z(1)}$ as
\begin{equation}
M_{Z(1)}\equiv\sum_{j=0}^{d-1}j|j\rangle\langle j|.
\label{eq-qt:position-observable}%
\end{equation}
Measuring this operator is equivalent to measuring in the qudit computational
basis. The expectation of this operator for a qudit $|\psi\rangle$ in the
state in\ \eqref{eq-qt:qudit-state} is%
\begin{align}
\mathbb{E}\left[  M_{Z(1)}\right]   &  =\langle\psi|M_{Z(1)}|\psi\rangle\\
&  =\sum_{j^{\prime}=0}^{d-1}\langle j^{\prime}\vert\alpha_{j^{\prime}}^{\ast
}\sum_{j=0}^{d-1}j|j\rangle\langle j|\sum_{j^{\prime\prime}=0}^{d-1}%
\alpha_{j^{\prime\prime}}\left\vert j^{\prime\prime}\right\rangle \\
&  =\sum_{j^{\prime},j,j^{\prime\prime}=0}^{d-1}j\alpha_{j^{\prime}}^{\ast
}\alpha_{j^{\prime\prime}}\left\langle j^{\prime}|j\right\rangle \left\langle
j|j^{\prime\prime}\right\rangle \\
&  =\sum_{j=0}^{d-1}j\left\vert \alpha_{j}\right\vert ^{2}.
\end{align}
Similarly, we can construct an operator $M_{X(1)}$ for \textquotedblleft
measuring the operator $X(1)$\textquotedblright\ by using the eigenstates
$|j\rangle_{X}$ of the $X(1)$ operator:%
\begin{equation}
M_{X(1)}\equiv\sum_{j=0}^{d-1}j|\widetilde{j}\rangle\langle\widetilde{j}|.
\end{equation}
We leave it as an exercise to determine the expectation when measuring the
$M_{X(1)}$ operator.

\begin{exercise}
Suppose the qudit is in the state $|\psi\rangle$
in\ \eqref{eq-qt:qudit-state}. Show that the expectation of the $M_{X\left(
1\right)  }$ operator is%
\begin{equation}
\mathbb{E}\left[  M_{X(1)}\right]  =\frac{1}{d}\sum_{j=0}^{d-1}j\left\vert
\sum_{j^{\prime}=0}^{d-1}\alpha_{j^{\prime}}\exp\left\{  -i2\pi j^{\prime
}j/d\right\}  \right\vert ^{2}.
\end{equation}
\textit{Hint}:\ First show that we can represent the state $|\psi\rangle$ in
the $X(1)$ eigenbasis as follows:%
\begin{equation}
|\psi\rangle=\sum_{l=0}^{d-1}\frac{1}{\sqrt{d}}\left(  \sum_{j=0}^{d-1}%
\alpha_{j}\exp\left\{  -i2\pi lj/d\right\}  \right)  |\widetilde{l}\rangle.
\end{equation}

\end{exercise}

\subsection{Composite Systems of Qudits}

We can define a system of multiple qudits again by employing the tensor
product. A general two-qudit state on systems $A$ and $B$ has the following
form:%
\begin{equation}
\left\vert \xi\right\rangle _{AB}\equiv\sum_{j,k=0}^{d-1}\alpha_{j,k}\vert
j\rangle_{A}\vert k\rangle_{B}.
\end{equation}
Evolution of two-qudit states is similar as before. Suppose Alice applies a
unitary $U_{A}$ to her qudit. The result is as follows:%
\begin{align}
\left(  U_{A}\otimes I_{B}\right)  \left\vert \xi\right\rangle _{AB}  &
=\left(  U_{A}\otimes I_{B}\right)  \sum_{j,k=0}^{d-1}\alpha_{j,k}\left\vert
j\right\rangle _{A}\vert k\rangle_{B}\\
&  =\sum_{j,k=0}^{d-1}\alpha_{j,k}\left(  U_{A}\vert j\rangle_{A}\right)
\vert k\rangle_{B},
\end{align}
which follows by linearity. Bob applying a local unitary $U_{B}$ has a similar
form. The application of some global unitary $U_{AB}$ results in the state
\begin{equation}
U_{AB}\left\vert \xi\right\rangle _{AB}.
\end{equation}

\subsubsection{The Qudit Bell States}

Two-qudit
\index{Bell states}%
states can be entangled as well. The maximally entangled qudit state is as
follows:%
\begin{equation}
\left\vert \Phi\right\rangle _{AB}\equiv\frac{1}{\sqrt{d}}\sum_{i=0}%
^{d-1}\vert i\rangle_{A}\vert i\rangle_{B}.
\label{eq-qt:max-ent-state-basis-1}%
\end{equation}
When Alice possesses the first qudit and Bob possesses the second qudit and
they are also separated in space, the above state is a resource known as an
\textit{edit} (pronounced \textquotedblleft ee $\cdot$ dit\textquotedblright).
It is useful in the qudit versions of the teleportation protocol and the
super-dense coding protocol discussed in Chapter~\ref{chap:three-noiseless}%
.\ Throughout the book, we often find it convenient to make use of the
unnormalized maximally entangled vector:%
\begin{equation}
\vert\Gamma\rangle_{AB}\equiv\sum_{i=0}^{d-1}\left\vert i\right\rangle
_{A}\vert i\rangle_{B}. \label{eq-qt:unnorm-max-ent}%
\end{equation}

Consider applying the operator $X( x) Z( z) $ to Alice's share of the
maximally entangled state $\left\vert \Phi\right\rangle _{AB}$. We use the
following notation:%
\begin{equation}
|\Phi^{x,z}\rangle_{AB}\equiv\left(  X_{A}( x) Z_{A}( z) \otimes I_{B}\right)
\left\vert \Phi\right\rangle _{AB}. \label{eq-qt:qudit-bell-states}%
\end{equation}
The $d^{2}$ states $\left\{  |\Phi^{x,z}\rangle_{AB}\right\}  _{x,z=0}^{d-1}$
are known as the qudit Bell states and are important in qudit quantum
protocols and in quantum Shannon theory.
Exercise~\ref{ex-qt:qudit-bell-states-ortho} asks you to verify that these
states form a complete, orthonormal basis. Thus, one can measure two qudits in
the qudit Bell basis. Similar to the qubit case, it is straightforward to see
that the qudit state can generate a \textit{dit} of shared randomness by
extending the arguments in Section~\ref{sec-qt:common-randomness}.

\begin{exercise}
\label{ex-qt:qudit-bell-states-ortho}Show that the set of states $\left\{
|\Phi^{x,z}\rangle_{AB}\right\}  _{x,z=0}^{d-1}$ forms a complete, orthonormal
basis:%
\begin{align}
\langle\Phi^{x_{1},z_{1}}|\Phi^{x_{2},z_{2}}\rangle &  =\delta_{x_{1},x_{2}%
}\delta_{z_{1},z_{2}},\\
\sum_{x,z=0}^{d-1}|\Phi^{x,z}\rangle\langle\Phi^{x,z}|_{AB}  &  =I_{AB}.
\end{align}

\end{exercise}

\begin{exercise}
[Transpose Trick]\label{ex-qt:bell-state-matrix-identity}Show that the
following \textquotedblleft transpose trick\textquotedblright\ or
\textquotedblleft ricochet\textquotedblright\ property holds for a maximally
entangled state $\left\vert \Phi\right\rangle _{AB}$ (as defined in
\eqref{eq-qt:max-ent-state-basis-1}) and any $d\times d$ matrix $M$:%
\begin{equation}
\left(  M_{A}\otimes I_{B}\right)  \left\vert \Phi\right\rangle _{AB}=\left(
I_{A}\otimes M_{B}^{T}\right)  \left\vert \Phi\right\rangle _{AB},
\end{equation}
where $M^{T}$ is the transpose of the operator $M$ with respect to the basis
$\left\{  \vert i\rangle_{B}\right\}  $ from
\eqref{eq-qt:max-ent-state-basis-1}. The implication is that some local action
of Alice on $\left\vert \Phi\right\rangle _{AB}$ is equivalent to Bob
performing the transpose of this action on his share of the state. Of course,
the same equality is true for the unnormalized maximally entangled vector
$\vert\Gamma\rangle_{AB}$ from \eqref{eq-qt:unnorm-max-ent}:%
\[
\left(  M_{A}\otimes I_{B}\right)  \vert\Gamma\rangle_{AB}=\left(
I_{A}\otimes M_{B}^{T}\right)  \vert\Gamma\rangle_{AB}.
\]

\end{exercise}

\section{Schmidt Decomposition}

The Schmidt decomposition%
\index{Schmidt decomposition}
is one of the most important tools for analyzing bipartite pure states in
quantum information theory, showing that it is possible to decompose any pure
bipartite state as a superposition of coordinated orthonormal states. It is a
consequence of the well known singular value decomposition theorem from linear
algebra. We state this result formally as the following theorem:

\begin{theorem}
[Schmidt decomposition]\label{thm-qt:schmidt}Suppose that we have a bipartite
pure state,%
\begin{equation}
|\psi\rangle_{AB}\in\mathcal{H}_{A}\otimes\mathcal{H}_{B},
\end{equation}
where $\mathcal{H}_{A}$ and $\mathcal{H}_{B}$ are finite-dimensional Hilbert
spaces, not necessarily of the same dimension, and $\left\Vert |\psi
\rangle_{AB}\right\Vert _{2}=1$. Then it is possible to express this state as
follows:%
\begin{equation}
|\psi\rangle_{AB}\equiv\sum_{i=0}^{d-1}\lambda_{i}\left\vert i\right\rangle
_{A}|i\rangle_{B},
\end{equation}
where the amplitudes $\lambda_{i}$ are real, strictly positive, and normalized
so that $\sum_{i}\lambda_{i}^{2}=1$, the states $\{|i\rangle_{A}\}$ form an
orthonormal basis for system $A$, and the states $\{|i\rangle_{B}\}$ form an
orthonormal basis for the system $B$. The vector $\left[  \lambda_{i}\right]
_{i\in\left\{  0,\ldots,d-1\right\}  }$ is called the vector of Schmidt
coefficients. The Schmidt rank $d$ of a bipartite state is equal to the number
of Schmidt coefficients $\lambda_{i}$\ in its Schmidt decomposition and
satisfies%
\begin{equation}
d\leq\min\left\{  \dim(\mathcal{H}_{A}),\dim(\mathcal{H}_{B})\right\}  .
\end{equation}

\end{theorem}

\begin{proof}
This is essentially a restatement of the singular value decomposition of a
matrix. Consider an arbitrary bipartite pure state $|\psi\rangle_{AB}%
\in\mathcal{H}_{A}\otimes\mathcal{H}_{B}$. Let $d_{A}\equiv\dim(\mathcal{H}%
_{A})$ and $d_{B}\equiv\dim(\mathcal{H}_{B})$. We can express $|\psi
\rangle_{AB}$ as follows:%
\begin{equation}
|\psi\rangle_{AB}=\sum_{j=0}^{d_{A}-1}\sum_{k=0}^{d_{B}-1}\alpha
_{j,k}|j\rangle_{A}|k\rangle_{B}, \label{eq-qt:schmidt-state}%
\end{equation}
for some amplitudes $\alpha_{j,k}$ and some orthonormal bases $\{\left\vert
j\right\rangle _{A}\}$ and $\{|k\rangle_{B}\}$ on the respective systems $A$
and $B$. Let us write the matrix formed by the coefficients $\alpha_{j,k}$ as
some $d_{A}\times d_{B}$ matrix $G$ where%
\begin{equation}
\left[  G\right]  _{j,k}=\alpha_{j,k}.
\end{equation}
Since every matrix has a singular value decomposition, we can write $G$ as%
\begin{equation}
G=U\Lambda V,
\end{equation}
where $U$ is a $d_{A}\times d_{A}$ unitary matrix, $V$ is a $d_{B}\times
d_{B}$ unitary matrix, and $\Lambda$ is a $d_{A}\times d_{B}$ matrix with
$d$\ real, strictly positive numbers $\lambda_{i}$ along the diagonal and
zeros elsewhere. Let us write the matrix elements of $U$ as $u_{j,i}$ and
those of $V$ as $v_{i,k}$. The above matrix equation is then equivalent to the
following set of equations:%
\begin{equation}
\alpha_{j,k}=\sum_{i=0}^{d-1}u_{j,i}\lambda_{i}v_{i,k}.
\end{equation}
Let us make this substitution into the expression for the state in
\eqref{eq-qt:schmidt-state}:%
\begin{equation}
|\psi\rangle_{AB}=\sum_{j=0}^{d_{A}-1}\sum_{k=0}^{d_{B}-1}\left(  \sum
_{i=0}^{d-1}u_{j,i}\lambda_{i}v_{i,k}\right)  \vert j\rangle_{A}|k\rangle_{B}.
\end{equation}
Readjusting some terms by exploiting the properties of the tensor product, we
find that%
\begin{align}
|\psi\rangle_{AB}  &  =\sum_{i=0}^{d-1}\lambda_{i}\left(  \sum_{j=0}^{d_{A}%
-1}u_{j,i}|j\rangle_{A}\right)  \otimes\left(  \sum_{k=0}^{d_{B}-1}%
v_{i,k}|k\rangle_{B}\right) \\
&  =\sum_{i=0}^{d-1}\lambda_{i}|i\rangle_{A}|i\rangle_{B},
\end{align}
where we define the orthonormal basis on the $A$ system as $\left\vert
i\right\rangle _{A}\equiv\sum_{j}u_{j,i}|j\rangle_{A}$ and we define the
orthonormal basis on the $B$ system as $|i\rangle_{B}\equiv\sum_{k}%
v_{i,k}|k\rangle_{B}$. This final step completes the proof of the theorem, but
Exercise~\ref{ex-qt:schmidt-ex} asks you to verify that the set of states
$\{|i\rangle_{A}\}$ form an orthonormal basis (the proof for the set of states
$\{|i\rangle_{B}\}$ is similar).
\end{proof}

The statement of Theorem~\ref{thm-qt:schmidt} is rather remarkable after
pausing to think about it further. For example, the Hilbert space
$\mathcal{H}_{A}$ of Alice could be a qubit Hilbert space of dimension two,
and the Hilbert space $\mathcal{H}_{B}$ of Bob could be of dimension one
billion (or some other large number). Then, in spite of this large dimension
for Bob's Hilbert space, if we know that the state of systems $A$ and $B$ is a
pure state, then it is always possible to find a two-dimensional subspace of
$\mathcal{H}_{B}$ which along with $\mathcal{H}_{A}$ suffices to represent the
state. So all those extra degrees of freedom are unnecessary in this example.
Often in quantum Shannon theory, we are optimizing certain functions over pure
states. In such cases, the Schmidt decomposition theorem is helpful in
limiting the size of the space we have to consider in such optimization problems.

\begin{remark}
\label{rem-qt:schmidt}The Schmidt decomposition applies not only to bipartite
systems but to any number of systems where we can make a bipartite cut of the
systems. For example, suppose that there is a state $|\phi\rangle_{ABCDE}$ on
systems $ABCDE$. We could say that $AB$ are part of one system and $CDE$ are
part of another system and write a Schmidt decomposition for this state as
follows:%
\begin{equation}
|\phi\rangle_{ABCDE}=\sum_{y}\sqrt{p_{Y}(y)}|y\rangle_{AB}|y\rangle_{CDE},
\end{equation}
where $\{|y\rangle_{AB}\}$ is an orthonormal basis for the joint system $AB$
and $\{|y\rangle_{CDE}\}$ is an orthonormal basis for the joint system $CDE$.
\end{remark}

\begin{exercise}
\label{ex-qt:schmidt-ex}Verify that the set of states $\{\left\vert
i\right\rangle _{A}\}$ from the proof of Theorem~\ref{thm-qt:schmidt} forms an
orthonormal basis by exploiting the unitarity of the matrix $U$.
\end{exercise}

\begin{exercise}
Prove that the Schmidt decomposition gives a way to identify if a pure state
is entangled or product. In particular, prove that a pure bipartite state is
entangled if and only if it has more than one Schmidt coefficient. First,
suppose that a pure bipartite state $|\phi\rangle_{AB}$\ has only one Schmidt
coefficient. Prove that its maximum overlap with a product state is equal to
one:%
\begin{equation}
\max_{|\varphi\rangle_{A},|\psi\rangle_{B}}\left\vert \langle\varphi
|_{A}\otimes\langle\psi|_{B}|\phi\rangle_{AB}\right\vert ^{2}=1.
\end{equation}
Now, suppose that there is more than one Schmidt coefficient for a state
$|\phi\rangle_{AB}$. Prove that this state's maximum overlap with a product
state is strictly less than one (and thus it cannot be written as a product
state):%
\begin{equation}
\max_{|\varphi\rangle_{A},|\psi\rangle_{B}}\left\vert \langle\varphi
|_{A}\otimes\langle\psi|_{B}|\phi\rangle_{AB}\right\vert ^{2}<1.
\end{equation}
(\textit{Hint: Use the Schmidt! Use the Schwarz! (as in Cauchy--Schwarz...) })
\end{exercise}

\section{History and Further Reading}

There are many great books on quantum mechanics that outline the mathematical
background. The books of \cite{book1989bohm}, \cite{book1994sakurai}, and
\cite{book2000mikeandike} are among these. The ideas for the resource
inequality formalism first appeared in the popular article \citep{B95} and
another of Bennett's papers \citep{Bennett04}. The no-deletion theorem is in
\citep{PB00}. The review article of the Horodecki family is a helpful
reference on the study of quantum entanglement~\citep{H42007}. Our
presentation of the CHSH game and its analysis follows the approach detailed
in \citep{S13}. The bound on the maximum quantum winning probability of the
CHSH game was established in \citep{T80}.

\chapter{The Noisy Quantum Theory}

\label{chap:noisy-quantum-theory}In general, we may not know for certain
whether we possess a particular quantum state. Instead, we may only have a
probabilistic description of an ensemble of quantum states. This chapter
re-establishes the foundations of the quantum theory so that they incorporate
a lack of complete information about a quantum system. The density operator
formalism is a powerful mathematical tool for describing this scenario. This
chapter also establishes how to model the noisy evolution of a quantum system,
and we explore models of noisy quantum channels that are generalizations of
the noisy classical channel discussed in Section~\ref{sec-cst:channel-code} of
Chapter~\ref{chap:classical-shannon-theory}.

You might have noticed that the development in the previous chapter relied on
the premise that the possessor of a quantum system has perfect knowledge of
the state of a given system. For instance, we assumed that Alice knows that
she possesses a qubit in the state $\vert\psi\rangle$ where%
\begin{equation}
\vert\psi\rangle=\alpha\vert0\rangle+\beta\vert1\rangle,
\end{equation}
for some $\alpha, \beta\in\mathbb{C}$ such that $\vert\alpha\vert^{2} +
\vert\beta\vert^{2} = 1$. Also, in some examples, we assumed that Alice and
Bob know that they share an ebit $\left\vert \Phi^{+}\right\rangle _{AB} $. We
even assumed perfect knowledge of a unitary evolution or a particular
measurement that a possessor of a quantum state may apply to it.

This assumption of perfect, definite knowledge of a quantum state is a
difficult one to justify in practice. In reality, it is challenging to
prepare, evolve, or measure a quantum state exactly as we wish. Slight errors
may occur in the preparation, evolution, or measurement due to imprecise
devices or to coupling with other degrees of freedom outside of the system
that we are controlling. An example of such imprecision can occur in the
coupling of two photons at a beamsplitter. We may not be able to tune the
reflectivity of the beamsplitter exactly or may not have the timing of the
arrival of the photons exactly set. The noiseless quantum theory as we
presented it in the previous section cannot handle such imprecisions.

In this chapter, we relax the assumption of perfect knowledge of the
preparation, evolution, or measurement of quantum states and develop a noisy
quantum theory that incorporates an imprecise knowledge of these states. The
noisy quantum theory fuses probability theory and the quantum theory into a
single formalism.

We proceed with the development of the noisy quantum theory in the following order:

\begin{enumerate}
\item We first present the density operator%
\index{density operator}
formalism, which gives a representation for a noisy, imprecise quantum state.

\item We then discuss a general form of measurements and the effect of them on
our description of a noisy quantum state.

\item We proceed to composite noisy systems, which admit a particular form,
and we discuss several possible states of composite noisy systems including
product states,
\index{separable states}%
separable states, classical--quantum states, entangled states, and arbitrary states.

\item Next, we consider the Kraus representation%
\index{Kraus representation}
of a quantum channel, which gives a way to describe noisy evolution, and we
discuss important examples of noisy quantum channels. We also stress how every
operation we have discussed so far, including preparations and measurements,
can be viewed as quantum channels.
\end{enumerate}

\section{Noisy Quantum States}

We generally may not have perfect knowledge of a prepared quantum state.
Suppose a third party, Bob, prepares a state for us and only gives us a
probabilistic description of it. That is, we might only know that Bob selects
the state $\vert\psi_{x}\rangle$ with a certain probability $p_{X}( x) $. Our
description of the state is then as an ensemble $\mathcal{E}$ of quantum
states where%
\begin{equation}
\mathcal{E}\equiv\left\{  p_{X}( x) ,\vert\psi_{x}\rangle\right\}
_{x\in\mathcal{X}}. \label{eq-qt:ensemble}%
\end{equation}
In the above, $X$ is a random variable with distribution $p_{X}( x) $. Each
realization$~x$ of random variable$~X$ belongs to an alphabet~$\mathcal{X}$.
Thus, the realization $x$ merely acts as an index, meaning that the quantum
state is $\vert\psi_{x}\rangle$ with probability $p_{X}( x) $. We also assume
that each state $\vert\psi_{x}\rangle$ is a $d$-dimensional\ qudit state.

A simple example is the following ensemble: $\left\{  \left\{  1/3,|1\rangle
\right\}  ,\left\{  2/3,|3\rangle\right\}  \right\}  $. The states $|1\rangle$
and $|3\rangle$ are in a four-dimensional Hilbert space with basis states
$\left\{  |0\rangle,|1\rangle,|2\rangle,|3\rangle\right\}  $. The
interpretation of this ensemble is that the state is$~|1\rangle$ with
probability$~1/3$ and the state is$~|3\rangle$ with probability$~2/3$.

\subsection{The Density Operator}

Suppose now that we have the ability to perform a perfect, projective
measurement of a system with ensemble description $\mathcal{E}$ in
\eqref{eq-qt:ensemble}. Let $\Pi_{j}$ be the elements of this projective
measurement so that $\sum_{j}\Pi_{j}=I$, and let $J$ be the random variable
corresponding to the measurement outcome $j$. Let us suppose at first, without
loss of generality, that the state in the ensemble is $|\psi_{x}\rangle$ for
some $x\in\mathcal{X}$. Then the Born rule%
\index{Born rule}
of the noiseless quantum theory states that the conditional probability
$p_{J|X}(j|x)$ of obtaining measurement result $j$ (given that the state is
$|\psi_{x}\rangle$) is equal to
\begin{equation}
p_{J|X}(j|x)=\langle\psi_{x}|\Pi_{j}|\psi_{x}\rangle,
\end{equation}
and the post-measurement state is $\Pi_{j}|\psi_{x}\rangle/\sqrt{p_{J|X}%
(j|x)}.$ However, we would also like to know the unconditional probability
$p_{J}(j)$ of obtaining measurement result $j$ for the ensemble description
$\mathcal{E}$. By the \textit{law of total probability}, the unconditional
probability $p_{J}(j)$ is%
\begin{align}
p_{J}(j)  &  =\sum_{x\in\mathcal{X}}p_{J|X}(j|x)p_{X}(x)\\
&  =\sum_{x\in\mathcal{X}}\langle\psi_{x}|\Pi_{j}|\psi_{x}\rangle p_{X}(x).
\label{eq-qt:density-op-develop}%
\end{align}

At this point, it is helpful for us to introduce the \textit{trace} of a
square operator, which will be used extensively throughout this book.

\begin{definition}
[Trace]\label{def-nqt:trace}The \textit{trace}
\index{trace}%
$\operatorname{Tr}\left\{  A\right\}  $\ of a square operator $A$ acting on a
Hilbert space~$\mathcal{H}$ is defined as follows:
\begin{equation}
\operatorname{Tr}\left\{  A\right\}  \equiv\sum_{i}\langle i\vert A\vert
i\rangle,
\end{equation}
where $\{\vert i\rangle\} $ is some complete, orthonormal basis for
$\mathcal{H}$.
\end{definition}

Observe that the trace operation is \textit{linear}. It is also independent of
which orthonormal basis we choose because
\begin{align}
\operatorname{Tr}\left\{  A\right\}   &  =\sum_{i}\langle i|A|i\rangle\\
&  =\sum_{i}\langle i|A\left(  \sum_{j}|\phi_{j}\rangle\langle\phi
_{j}|\right)  |i\rangle\\
&  =\sum_{i,j}\langle i|A|\phi_{j}\rangle\langle\phi_{j}|i\rangle\\
&  =\sum_{i,j}\langle\phi_{j}|i\rangle\langle i|A|\phi_{j}\rangle\\
&  =\sum_{j}\langle\phi_{j}|\left(  \sum_{i}|i\rangle\langle i|\right)
A|\phi_{j}\rangle\\
&  =\sum_{j}\langle\phi_{j}|A|\phi_{j}\rangle.
\end{align}
In the above, $\{|\phi_{j}\rangle\}$ is some other orthonormal basis for
$\mathcal{H}$ and we made use of the completeness relation: $I=\sum_{j}%
|\phi_{j}\rangle\langle\phi_{j}|=\sum_{i}|i\rangle\langle i|$.

\begin{exercise}
Prove that the trace
\index{trace!cyclicity}%
is cyclic. That is, for three operators $A$, $B$, and $C$, the following
relation holds $\operatorname{Tr}\{ABC \} = \operatorname{Tr}\{CAB \} =
\operatorname{Tr}\{BCA \} . $
\end{exercise}

Returning to \eqref{eq-qt:density-op-develop}, we can then show the following
useful property:%
\begin{align}
\langle\psi_{x}|\Pi_{j}|\psi_{x}\rangle &  =\langle\psi_{x}|\left(  \sum
_{i}|i\rangle\langle i|\right)  \Pi_{j}|\psi_{x}\rangle\\
&  =\sum_{i}\left\langle \psi_{x}|i\right\rangle \langle i|\Pi_{j}|\psi
_{x}\rangle\\
&  =\sum_{i}\langle i|\Pi_{j}|\psi_{x}\rangle\left\langle \psi_{x}%
|i\right\rangle \\
&  =\operatorname{Tr}\left\{  \Pi_{j}|\psi_{x}\rangle\langle\psi_{x}|\right\}
.
\end{align}
The first equality uses the completeness relation $\sum_{i}\vert i\rangle
\langle i|=I$. Thus, we continue with the development in
\eqref{eq-qt:density-op-develop} and show that%
\begin{align}
p_{J}(j)  &  =\sum_{x\in\mathcal{X}}\operatorname{Tr}\left\{  \Pi_{j}|\psi
_{x}\rangle\langle\psi_{x}|\right\}  p_{X}(x)\\
&  =\operatorname{Tr}\left\{  \Pi_{j}\sum_{x\in\mathcal{X}}p_{X}(x)|\psi
_{x}\rangle\langle\psi_{x}|\right\}  .
\end{align}
We can rewrite the last equation as follows:%
\begin{equation}
p_{J}(j)=\operatorname{Tr}\left\{  \Pi_{j}\rho\right\}  ,
\label{eq-nqt:new-born-rule}%
\end{equation}
introducing $\rho$ as the \textit{density operator} corresponding to the
ensemble $\mathcal{E}$:

\begin{definition}
[Density Operator]The
\index{density operator}%
\textit{density operator} $\rho$ corresponding to an ensemble $\mathcal{E}%
\equiv\left\{  p_{X}( x) ,\vert\psi_{x}\rangle\right\}  _{x\in\mathcal{X}}$ is
defined as
\begin{equation}
\rho\equiv\sum_{x\in\mathcal{X}}p_{X}( x) \vert\psi_{x}\rangle\langle\psi
_{x}\vert.
\end{equation}

\end{definition}

The operator $\rho$ as defined above is known as the \textit{density} operator
because it is the quantum generalization of a probability density function.
Throughout this book, we often use the symbols $\rho$, $\sigma$, $\tau$, $\pi
$, and $\omega$ to denote density operators.

We sometimes refer to the density operator as the \textit{expected density
operator} because there is a sense in which we are taking the expectation over
all of the states in the ensemble in order to obtain the density operator. We
can equivalently write the density operator as follows:%
\begin{equation}
\rho=\mathbb{E}_{X}\left\{  |\psi_{X}\rangle\langle\psi_{X}|\right\}  ,
\end{equation}
where the expectation is with respect to the random variable $X$. Note that we
are careful to use the notation $|\psi_{X}\rangle$ instead of the notation
$|\psi_{x}\rangle$ for the state inside of the expectation because the state
$|\psi_{X}\rangle$ is a random quantum state, random with respect to a
classical random variable $X$.

\begin{exercise}
Suppose the ensemble has a degenerate probability distribution, say $p_{X}( 0)
=1$ and $p_{X}( x) =0$ for all $x\neq0$. What is the density operator of this
degenerate ensemble?
\end{exercise}

\begin{exercise}
\label{ex-qt:alt-trace-max-ent}Prove the following equality:%
\begin{equation}
\operatorname{Tr}\left\{  A\right\}  =\langle\Gamma\vert_{RS}I_{R}\otimes
A_{S}\vert\Gamma\rangle_{RS},
\end{equation}
where $A$ is a square operator acting on a Hilbert space $\mathcal{H}_{S}$,
$I_{R}$ is the identity operator acting on a Hilbert space $\mathcal{H}_{R}$
isomorphic to $\mathcal{H}_{S}$ and $\vert\Gamma\rangle_{RS}$ is the
unnormalized maximally entangled vector from \eqref{eq-qt:unnorm-max-ent}.
This gives an alternate formula for the trace of a square operator $A$.
\end{exercise}

\begin{exercise}
\label{ex-nqt:trace-equal-eigs} Prove that $\operatorname{Tr}\{f(G^{\dag}G) \}
= \operatorname{Tr}\{f(G G^{\dag}) \}, $ where $G$ is \textbf{any} operator
(not necessarily Hermitian) and $f$ is any function. (Recall the convention
for a function of an operator given in
Definition~\ref{def-qt:hermitian-op-function}.)
\end{exercise}

\subsubsection{Properties of the Density Operator}

What are the properties that a given density operator corresponding to an
ensemble satisfies? Let us consider taking the trace of $\rho$:%
\begin{align}
\operatorname{Tr}\left\{  \rho\right\}   &  =\operatorname{Tr}\left\{
\sum_{x\in\mathcal{X}}p_{X}(x)|\psi_{x}\rangle\langle\psi_{x}|\right\} \\
&  =\sum_{x\in\mathcal{X}}p_{X}(x)\operatorname{Tr}\left\{  |\psi_{x}%
\rangle\langle\psi_{x}|\right\} \\
&  =\sum_{x\in\mathcal{X}}p_{X}(x)\left\langle \psi_{x}|\psi_{x}\right\rangle
\\
&  =\sum_{x\in\mathcal{X}}p_{X}(x)\\
&  =1.
\end{align}
The above development shows that every density operator corresponding to an
ensemble has \textit{unit trace}.

Let us consider taking the conjugate transpose of the density operator $\rho$:%
\begin{align}
\rho^{\dag}  &  =\left(  \sum_{x\in\mathcal{X}}p_{X}(x)|\psi_{x}\rangle
\langle\psi_{x}|\right)  ^{\dag}\\
&  =\sum_{x\in\mathcal{X}}p_{X}(x)\left(  |\psi_{x}\rangle\langle\psi
_{x}|\right)  ^{\dag}\\
&  =\sum_{x\in\mathcal{X}}p_{X}(x)|\psi_{x}\rangle\langle\psi_{x}|\\
&  =\rho.
\end{align}
Every density operator is thus a \textit{Hermitian} operator as well because
the conjugate transpose of $\rho$ is $\rho$.

Every density operator is furthermore \textit{positive semi-definite}%
\index{positive semi-definite}%
, meaning that
\begin{equation}
\langle\varphi|\rho|\varphi\rangle\geq0\ \ \ \ \forall\ |\varphi\rangle.
\end{equation}
We write $\rho\geq0$ to indicate that an operator is positive semi-definite. A
proof for non-negativity of any density operator $\rho$ is as follows:%
\begin{align}
\langle\varphi|\rho|\varphi\rangle &  =\langle\varphi|\left(  \sum
_{x\in\mathcal{X}}p_{X}(x)|\psi_{x}\rangle\langle\psi_{x}|\right)
|\varphi\rangle\\
&  =\sum_{x\in\mathcal{X}}p_{X}(x)\left\langle \varphi|\psi_{x}\right\rangle
\left\langle \psi_{x}|\varphi\right\rangle \\
&  =\sum_{x\in\mathcal{X}}p_{X}(x)\left\vert \left\langle \varphi|\psi
_{x}\right\rangle \right\vert ^{2}\geq0.
\end{align}
The inequality follows because each $p_{X}(x)$ is a probability and is
therefore non-negative.

\subsubsection{Ensembles and the Density Operator}

\label{sec-nqt:ensemble-density-unique}

Every ensemble has a unique density operator, but the opposite does not
necessarily hold: every density operator does not correspond to a unique
ensemble and could correspond to many ensembles. However, there are
restrictions on which ensembles can realize a given density operator and there
is a relation between them. We return to this question in
Section~\ref{sec-pqt:purif-equiv}, after we have developed more tools.

\begin{exercise}
Show that the ensembles $\left\{  \left\{  1/2,|0\rangle\right\}  ,\left\{
1/2,|1\rangle\right\}  \right\}  $ and $\left\{  \left\{  1/2,\left\vert
+\right\rangle \right\}  ,\left\{  1/2,\vert-\rangle\right\}  \right\}  $ have
the same density operator.
\end{exercise}

This last result has profound implications for the predictions of the quantum
theory because it is possible for two or more completely different ensembles
to have the same probabilities for measurement results. It also has important
implications for quantum Shannon theory as well.

By the spectral theorem, it follows that every density operator $\rho$ has a
spectral decomposition in terms of eigenstates $\left\{  |\phi_{x}%
\rangle\right\}  _{x\in\left\{  0,\ldots,d-1\right\}  }$ because every $\rho$
is Hermitian:%
\begin{equation}
\rho=\sum_{x=0}^{d-1}\lambda_{x}|\phi_{x}\rangle\langle\phi_{x}|,
\end{equation}
where the coefficients $\lambda_{x}$ are the eigenvalues.

\begin{exercise}
Show that the coefficients $\lambda_{x}$ are probabilities using the facts
that $\operatorname{Tr}\left\{  \rho\right\}  =1$ and $\rho\geq0$.
\end{exercise}

Thus, given any density operator $\rho$, we can define a \textquotedblleft
canonical\textquotedblright\ ensemble $\left\{  \lambda_{x},|\phi_{x}%
\rangle\right\}  $ corresponding to it. Note that this ensemble is not unique:
if $\lambda_{x}=\lambda_{x^{\prime}}$ for $x\neq x^{\prime}$, then the choice
of eigenvectors corresponding to these eigenvalues is not unique. The fact
that an ensemble can correspond to a density operator is so important for
quantum Shannon theory that we see this idea arise again and again throughout
this book. Any ensemble arising from the spectral theorem is the most
\textquotedblleft efficient\textquotedblright\ ensemble, in a sense, and we
will explore this idea more in Chapter~\ref{chap:schumach}\ on quantum data compression.

\subsubsection{Density Operator as the State}

We can also refer to the density operator as the \textit{state} of a given
quantum system because it is possible to use it to calculate probabilities for
any measurement performed on that system. We can make these calculations
without having an ensemble description---all we need is the density operator.
The noisy quantum theory also subsumes the noiseless quantum theory because
any state $\vert\psi\rangle$ has a corresponding density operator
$|\psi\rangle\langle\psi|$ in the noisy quantum theory, and all calculations
with this density operator in the noisy quantum theory give the same results
as using the state $\vert\psi\rangle$ in the noiseless quantum theory. For
these reasons, we will say that the \textit{state} of a given quantum system
is a density operator.

\begin{definition}
[Density Operator as the State]The state of a quantum system is given by a
density operator $\rho$, which is a positive semi-definite operator with trace
equal to one. Let $\mathcal{D}(\mathcal{H})$ denote the set of all density
operators acting on a Hilbert space $\mathcal{H}$.
\end{definition}

\noindent One of the most important states is the maximally mixed state $\pi$:

\begin{definition}
[Maximally Mixed State]%
\index{maximally mixed state}
The maximally mixed state $\pi$ is the density operator corresponding to a
uniform ensemble of orthogonal states $\left\{  \frac{1}{d},\vert
x\rangle\right\}  $, where $d$ is the dimensionality of the Hilbert space. The
maximally mixed state $\pi$ is then equal to
\begin{equation}
\pi\equiv\frac{1}{d}\sum_{x\in\mathcal{X}}\vert x\rangle\langle x\vert
=\frac{I}{d}. \label{eq-qt:maximally-mixed-state}%
\end{equation}

\end{definition}

\begin{exercise}
Show that $\pi$ is the density operator of the ensemble that chooses
$|0\rangle$, $|1\rangle$, $\vert+\rangle$, $\vert-\rangle$ with equal probability.
\end{exercise}

\begin{exercise}
[Convexity]Show that the set of density operators acting on a given Hilbert
space is a convex set. That is, if $\lambda\in\left[  0,1\right]  $ and $\rho$
and $\sigma$ are density operators, then $\lambda\rho+(1-\lambda)\sigma$ is a
density operator.
\end{exercise}

\begin{definition}
[Purity]The
\index{purity}
purity $P( \rho) $\ of a density operator $\rho$\ is equal to%
\begin{equation}
P( \rho) \equiv\operatorname{Tr}\left\{  \rho^{\dag}\rho\right\}
=\operatorname{Tr}\left\{  \rho^{2}\right\}  .
\end{equation}

\end{definition}

The purity is one particular measure of the noisiness of a quantum state. The
purity of a pure state is equal to one, and the purity of a mixed state is
strictly less than one, as the following exercise asks you to verify.

\begin{exercise}
Prove that the purity of a density operator $\rho$ is equal to one if and only
if $\rho$ is a pure state, such that it can be written as $\rho=\vert
\psi\rangle\langle\psi\vert$ for some unit vector $\psi$.
\end{exercise}

\subsubsection{The Density Operator on the Bloch Sphere}

Consider that the following pure%
\index{Bloch sphere}
qubit state%
\begin{equation}
|\psi\rangle\equiv\cos(\theta/2)|0\rangle+e^{i\varphi}\sin(\theta/2)|1\rangle
\end{equation}
has the following density operator representation:%
\begin{align}
|\psi\rangle\langle\psi|  &  =\left(  \cos(\theta/2)|0\rangle+e^{i\varphi}%
\sin(\theta/2)|1\rangle\right)  \left(  \cos(\theta/2)\langle0|+e^{-i\varphi
}\sin(\theta/2)\langle1|\right) \\
&  =\cos^{2}(\theta/2)|0\rangle\langle0|+e^{-i\varphi}\sin(\theta
/2)\cos(\theta/2)|0\rangle\langle1|\nonumber\\
&  \ \ \ \ \ \ \ +e^{i\varphi}\sin(\theta/2)\cos(\theta/2)|1\rangle
\langle0|+\sin^{2}(\theta/2)|1\rangle\langle1|.
\end{align}
The matrix representation, or \textit{density matrix}, of this density
operator with respect to the computational basis is as follows:%
\begin{equation}%
\begin{bmatrix}
\cos^{2}(\theta/2) & e^{-i\varphi}\sin(\theta/2)\cos(\theta/2)\\
e^{i\varphi}\sin(\theta/2)\cos(\theta/2) & \sin^{2}(\theta/2)
\end{bmatrix}
.
\end{equation}
Using trigonometric identities, it follows that the density matrix is equal to
the following matrix:%
\begin{equation}
\frac{1}{2}%
\begin{bmatrix}
1+\cos(\theta) & \sin(\theta)\left(  \cos\left(  \varphi\right)
-i\sin(\varphi)\right) \\
\sin(\theta)\left(  \cos(\varphi)+i\sin\left(  \varphi\right)  \right)  &
1-\cos(\theta)
\end{bmatrix}
.
\end{equation}
We can further exploit the Pauli matrices, defined in
Section~\ref{sec-qt:Pauli-matrices}, to represent the density matrix as
follows:%
\begin{equation}
\frac{1}{2}\left(  I+r_{x}X+r_{y}Y+r_{z}Z\right)  ,
\label{eq-qt:density-op-bloch}%
\end{equation}
where $r_{x}=\sin(\theta)\cos(\varphi)$, $r_{y}=\sin(\theta)\sin(\varphi)$,
and $r_{z}=\cos(\theta)$. The coefficients $r_{x}$, $r_{y}$, and $r_{z}$ are
none other than the Cartesian coordinate representation of the angles $\theta$
and $\varphi$, and they thus correspond to a unit vector.

More generally, the formula in \eqref{eq-qt:density-op-bloch} can represent an
arbitrary qubit density operator where the coefficients $r_{x}$, $r_{y}$, and
$r_{z}$ do not necessarily correspond to a unit vector, but rather a vector
$\mathbf{r}$\ such that $\left\Vert \mathbf{r}\right\Vert _{2}\leq1$. Consider
that the density matrix in \eqref{eq-qt:density-op-bloch} is as follows:%
\begin{equation}
\frac{1}{2}%
\begin{bmatrix}
1+r_{z} & r_{x}-ir_{y}\\
r_{x}+ir_{y} & 1-r_{z}%
\end{bmatrix}
. \label{eq-qt:general-density-op-bloch}%
\end{equation}
The above matrix corresponds to a valid density matrix because it has unit
trace, it is Hermitian, and it is non-negative (the next exercise asks you to
verify these facts). This alternate representation of the density matrix as a
vector in the Bloch sphere is useful for visualizing noisy qubit processes in
the noisy quantum theory.

\begin{exercise}
Show that the matrix in \eqref{eq-qt:general-density-op-bloch} has unit trace,
is Hermitian, and is non-negative for all $\mathbf{r}$\ such that $\left\Vert
\mathbf{r}\right\Vert _{2}\leq1$. It thus corresponds to any valid density matrix.
\end{exercise}

\begin{exercise}
Show that we can compute the Bloch sphere coordinates $r_{x}$, $r_{y}$, and
$r_{z}$ with the respective formulas $\operatorname{Tr}\left\{  X
\rho\right\}  $, $\operatorname{Tr}\left\{  Y \rho\right\}  $, and
$\operatorname{Tr}\left\{  Z \rho\right\}  $ using the representation in
\eqref{eq-qt:general-density-op-bloch} and the result of
Exercise~\ref{ex-qt:pauli-traces}.
\end{exercise}

\begin{exercise}
Show that the eigenvalues of a general qubit density operator with density
matrix representation in \eqref{eq-qt:general-density-op-bloch} are as
follows: $\frac{1}{2}\left(  1\pm\left\Vert \mathbf{r}\right\Vert _{2}\right)
. $
\end{exercise}

\begin{exercise}
Show that a mixture of pure states $|\psi_{j}\rangle$\ each with Bloch vector
$\mathbf{r}_{j}$ and probability $p( j) $ gives a density matrix with the
Bloch vector $\mathbf{r}$ where $\mathbf{r=}\sum_{j}p( j) \mathbf{r}_{j}. $
\end{exercise}

\subsection{An Ensemble of Ensembles}

\label{sec-qt:ensemble-of-ens}
\begin{figure}[ptb]
\begin{center}
\includegraphics[
width=4.4918in
]{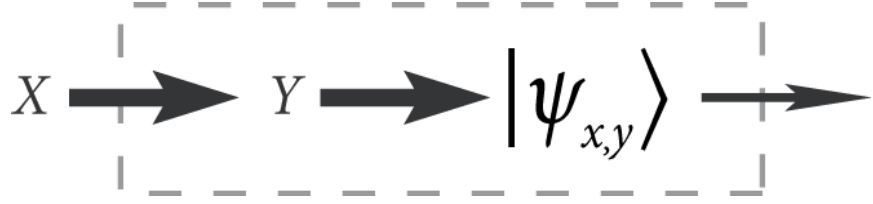}
\end{center}
\caption{The mixing process by which we can generate an \textquotedblleft
ensemble of ensembles.\textquotedblright\ First choose a realization $x$
according to distribution $p_{X}( x) $. Then choose a realization $y$
according to the conditional distribution $p_{Y|X}( y|x) $. Finally, choose a
state $\vert\psi_{x,y}\rangle$ according to the realizations $x$ and $y$. This
leads to an ensemble $\left\{  p_{X}( x) ,\rho_{x}\right\}  $ where $\rho
_{x}\equiv\sum_{y}p_{Y|X}( y|x) \vert\psi_{x,y}\rangle\langle\psi_{x,y}\vert
$.}%
\label{fig:ensemble-of-ensembles}%
\end{figure}

The most general ensemble that we can construct
is an \textit{ensemble of ensembles}, i.e., an ensemble $\mathcal{F}$ of
density operators where%
\begin{equation}
\mathcal{F}\equiv\left\{  p_{X}( x) ,\rho_{x}\right\}  .
\end{equation}
The ensemble $\mathcal{F}$ essentially has two layers of randomization. The
first layer is from the distribution $p_{X}( x) $. Each density operator
$\rho_{x}$ in $\mathcal{F}$\ arises from an ensemble $\left\{  p_{Y|X}( y|x)
,|\psi_{x,y}\rangle\right\}  $. The conditional distribution $p_{Y|X}( y|x)
$\ represents the second layer of randomization. Each$~\rho_{x}$ is a density
operator with respect to the above ensemble:%
\begin{equation}
\rho_{x}\equiv\sum_{y}p_{Y|X}( y|x) |\psi_{x,y}\rangle\langle\psi_{x,y}|.
\end{equation}
The ensemble $\mathcal{F}$ has its own density operator $\rho$ where%
\begin{align}
\rho &  \equiv\sum_{x,y}p_{Y|X}( y|x) p_{X}( x) |\psi_{x,y}\rangle\langle
\psi_{x,y}| =\sum_{x}p_{X}( x) \rho_{x}.
\end{align}
The density operator $\rho\,$\ is the density operator from the perspective of
someone who does not possess $x$. Figure~\ref{fig:ensemble-of-ensembles}
displays the process by which we can select the ensemble $\mathcal{F}%
$.

\subsection{Noiseless Evolution of an Ensemble}

Quantum states can evolve in a noiseless fashion either according to a unitary
operator or a measurement. In this section, we determine the noiseless
evolution of an ensemble and its corresponding density operator. We also show
how density operators evolve under a quantum measurement.

\subsubsection{Noiseless Unitary Evolution of a Noisy State}

We first consider noiseless evolution according to some unitary $U$. Suppose
we have the ensemble $\mathcal{E}$ in \eqref{eq-qt:ensemble} with density
operator $\rho$. Suppose without loss of generality that the state is
$|\psi_{x}\rangle$. Then the evolution postulate of the noiseless quantum
theory gives that the state after the unitary evolution is as follows:
$U|\psi_{x}\rangle$. This result implies that the evolution leads to a new
ensemble%
\begin{equation}
\mathcal{E}_{U}\equiv\left\{  p_{X}(x),U|\psi_{x}\rangle\right\}
_{x\in\mathcal{X}}.
\end{equation}
The density operator of the evolved ensemble is%
\begin{align}
\sum_{x\in\mathcal{X}}p_{X}(x)U|\psi_{x}\rangle\langle\psi_{x}|U^{\dag}  &
=U\left(  \sum_{x\in\mathcal{X}}p_{X}(x)|\psi_{x}\rangle\langle\psi
_{x}|\right)  U^{\dag}\\
&  =U\rho U^{\dag}.
\end{align}
Thus, the above relation shows that we can keep track of the evolution of the
density operator$~\rho$, rather than worrying about keeping track of the
evolution of every state in the ensemble$~\mathcal{E}$. It suffices to keep
track of only the density operator evolution because this operator is
sufficient to determine probabilities when performing any measurement on the system.

\subsubsection{Noiseless Measurement of a Noisy State}

In a similar fashion, we can analyze the result of a measurement on a system
with ensemble description $\mathcal{E}$ in \eqref{eq-qt:ensemble}. Suppose
that we perform a projective measurement with projection operators $\left\{
\Pi_{j}\right\}  _{j}$ where $\sum_{j}\Pi_{j}=I$. The main result of this
section is that two things happen after a measurement occurs. First, as shown
in the development preceding \eqref{eq-nqt:new-born-rule}, we receive the
outcome $j$ with probability $p_{J}(j) = \operatorname{Tr}\{\Pi_{j} \rho\}$.
Second, if the outcome of the measurement is $j$, then the state evolves as
follows:
\begin{equation}
\rho\longrightarrow\frac{\Pi_{j} \rho\Pi_{j}}{p_{J}(j)}.
\end{equation}

To see the above, let us suppose that the state in the ensemble $\mathcal{E}$
is $\vert\psi_{x}\rangle$. Then the noiseless quantum theory predicts that the
probability of obtaining outcome $j$ conditioned on the index $x$ is%
\begin{equation}
p_{J|X}( j|x) =\langle\psi_{x}\vert\Pi_{j}\vert\psi_{x}\rangle,
\end{equation}
and the resulting state is%
\begin{equation}
\frac{\Pi_{j}\vert\psi_{x}\rangle}{\sqrt{p_{J|X}( j|x) }}.
\end{equation}
Supposing that we receive outcome $j$, then we have a new ensemble:%
\begin{equation}
\mathcal{E}_{j}\equiv\left\{  p_{X|J}( x|j) , \frac{\Pi_{j}\vert\psi
_{x}\rangle}{\sqrt{p_{J|X}( j|x)} }\right\}  _{x\in\mathcal{X}}.
\end{equation}
The density operator for this ensemble is%
\begin{align}
&  \sum_{x\in\mathcal{X}}p_{X|J}( x|j) \frac{\Pi_{j}\vert\psi_{x}%
\rangle\langle\psi_{x}\vert\Pi_{j}}{p_{J|X}( j|x) }\nonumber\\
&  =\Pi_{j}\left(  \sum_{x\in\mathcal{X}}\frac{p_{X|J}( x|j) }{p_{J|X}( j|x)
}\vert\psi_{x}\rangle\langle\psi_{x}\vert\right)  \Pi_{j}\\
&  =\Pi_{j}\left(  \sum_{x\in\mathcal{X}}\frac{p_{J|X}( j|x) p_{X}( x)
}{p_{J|X}( j|x) p_{J}( j) }\vert\psi_{x}\rangle\langle\psi_{x}\vert\right)
\Pi_{j}\\
&  =\frac{\Pi_{j}\left(  \sum_{x\in\mathcal{X}}p_{X}( x) \vert\psi_{x}%
\rangle\langle\psi_{x}\vert\right)  \Pi_{j}}{p_{J}( j) }\\
&  =\frac{\Pi_{j}\rho\Pi_{j}}{p_{J}( j) }. \label{eq-qt:measurement-density}%
\end{align}
The second equality follows from applying the Bayes rule: $p_{X|J}( x|j) =
p_{J|X}( j|x) p_{X}( x)/ p_{J}( j) $.

\subsection{Probability Theory as a Special Case}

It may help to build some intuition for the noisy quantum theory by showing
how it contains probability theory as a special case. Indeed, we should expect
this containment of probability theory within the noisy quantum theory to hold
if the noisy quantum theory is making probabilistic predictions about the
physical world.

Let us again begin with an ensemble of quantum states, but this time, let us
pick the states in the ensemble to be special states, where they are all
orthogonal to one another. If the states in the ensemble are all orthogonal to
one another, then they are essentially classical states because there is a
measurement that distinguishes them from one another. So, let us pick the
ensemble to be $\left\{  p_{X}( x) ,\vert x\rangle\right\}  _{x\in\mathcal{X}%
}$ where the states $\left\{  \vert x\rangle\right\}  _{x\in\mathcal{X}}$ form
an orthonormal basis for a Hilbert space of dimension $\left\vert
\mathcal{X}\right\vert $. These states are classical because a measurement
with the following projection operators can distinguish them:%
\begin{equation}
\left\{  \vert x\rangle\langle x\vert\right\}  _{x\in\mathcal{X}}.
\end{equation}

The generalization of a probability distribution to the quantum world is the
density operator:%
\begin{equation}
p_{X}( x) \leftrightarrow\rho.
\end{equation}
The reason for this is that we can use the density operator to calculate
expectations and moments of observables. Furthermore, a probability
distribution can be encoded as a density operator that is diagonal with
respect to a known orthonormal basis, as follows:
\begin{equation}
\sum_{x\in\mathcal{X}} p_{X}(x) \vert x \rangle\langle x \vert.
\end{equation}

The generalization of a random variable is an observable. For example, let us
consider the following observable:%
\begin{equation}
X\equiv\sum_{x\in\mathcal{X}}x\vert x\rangle\langle x\vert,
\end{equation}
analogous to the observable in \eqref{eq-qt:position-observable}. We perform
the following calculation to determine the expectation of the observable $X$:%
\begin{equation}
\mathbb{E}_{\rho}\left[  X\right]  =\operatorname{Tr}\left\{  X\rho\right\}  .
\end{equation}
Explicitly calculating this quantity, we find that it is consistent with the
formula for the expectation of random variable $X$ with probability
distribution $p_{X}(x)$:%
\begin{align}
\operatorname{Tr}\left\{  X\rho\right\}   &  =\operatorname{Tr}\left\{
\sum_{x\in\mathcal{X}}x\vert x\rangle\langle x\vert\sum_{x^{\prime}%
\in\mathcal{X}}p_{X}( x^{\prime}) \vert x^{\prime}\rangle\langle x^{\prime
}\vert\right\} \\
&  =\sum_{x,x^{\prime}\in\mathcal{X}}x\ p_{X}( x^{\prime}) \left\vert
\left\langle x|x^{\prime}\right\rangle \right\vert ^{2}\\
&  =\sum_{x\in\mathcal{X}}x\ p_{X}( x) .
\end{align}

Another useful notion in probability theory is the notion of an indicator
random variable~$I_{A}( X) $. We define the indicator function%
\index{indicator function}
$I_{A}( x) $ for a set $A$ as follows:%
\begin{equation}
I_{A}( x) \equiv\left\{
\begin{array}
[c]{ccc}%
1 & : & x\in A\\
0 & : & x\notin A
\end{array}
\right.  .
\end{equation}
The expectation $\mathbb{E}\left[  I_{A}( X) \right]  $ of the indicator
random variable $I_{A}( X) $ is%
\begin{align}
\mathbb{E}\left[  I_{A}( X) \right]   &  =\sum_{x\in A}p_{X}( x) \equiv p_{X}(
A) ,
\end{align}
where $p_{X}( A) $ represents the probability of the set $A$. In the quantum
theory, we can define an indicator observable $I_{A}( X) $:%
\begin{equation}
I_{A}( X) \equiv\sum_{x\in A}\vert x\rangle\langle x\vert.
\end{equation}
It has eigenvalues equal to one for all eigenvectors with labels $x$\ in the
set $A$, and it has zero eigenvalues for those eigenvectors with labels
outside of $A$. It is straightforward to show that the expectation
$\operatorname{Tr}\left\{  I_{A}( X) \rho\right\}  $ of the indicator
observable $I_{A}( X) $ is $p_{X}( A) $.

You may have noticed that the indicator observable is also a projection
operator. So, according to the postulates of the quantum theory, we can
perform a measurement with elements:%
\begin{equation}
\left\{  I_{A}( X) ,I_{A^{c}}( X) \equiv I-I_{A}( X) \right\}  .
\end{equation}
The result of such a projective measurement is to project onto the subspace
given by $I_{A}( X) $ with probability $p_{X}( A) $ and to project onto the
complementary subspace given by $I_{A^{c}}( X) $ with probability $1-p_{X}( A)
$.

We highlight the connection between the noisy quantum theory and probability
theory with two more examples. First, suppose that we have two
\textit{disjoint} sets $A$ and $B$. Then the probability of their union is the
sum of the probabilities of the individual sets:%
\begin{equation}
\Pr\left\{  A\cup B\right\}  = \Pr\left\{  A\right\}  +\Pr\left\{  B\right\}
,
\end{equation}
and the probability of the complementary set $\left(  A\cup B\right)
^{c}=A^{c}\cap B^{c}$ is equal to $1-\Pr\left\{  A\right\}  -\Pr\left\{
B\right\}  $. We can perform the analogous calculation in the noisy quantum
theory. Let us consider two projection operators%
\begin{align}
\Pi( A)  &  \equiv\sum_{x\in A}\vert x\rangle\langle x\vert, \ \ \ \ \Pi( B)
\equiv\sum_{x\in B}\vert x\rangle\langle x\vert. \label{eq-qt:proj-A}%
\end{align}
The sum of these projection operators gives a projection onto the union set
$A\cup B$:%
\begin{equation}
\Pi\left(  A\cup B\right)  \equiv\sum_{x\in A\cup B}\vert x\rangle\langle
x\vert=\Pi( A) +\Pi( B) .
\end{equation}

\begin{exercise}
Show that $\operatorname{Tr}\left\{  \Pi(A\cup B)\rho\right\}  =\Pr
\{A\}+\Pr\{B\}$ whenever the projectors $\Pi(A)$ and $\Pi(B)$ satisfy
$\Pi(A)\Pi(B)=0$\ and the density operator $\rho$ is diagonal in the same
basis as $\Pi(A)$ and $\Pi(B)$.
\end{exercise}

We can also consider intersections of sets. Suppose that we have two sets $A$
and $B$. The intersection of these two sets consists of all the elements that
are common to both sets. There is an associated probability $\Pr\left\{  A\cap
B\right\}  $ with the intersection. We can again formulate this idea in the
noisy quantum theory. Consider the projection operators in
\eqref{eq-qt:proj-A}. The multiplication of these two projectors gives a
projector onto the intersection of the two spaces:%
\begin{equation}
\Pi\left(  A\cap B\right)  =\Pi( A) \Pi( B) .
\end{equation}

\begin{exercise}
Show that $\operatorname{Tr}\left\{  \Pi( A) \Pi( B) \rho\right\}
=\Pr\left\{  A\cap B\right\}  $ whenever the density operator $\rho$ is
diagonal in the same basis as $\Pi( A) $ and $\Pi( B) $.
\end{exercise}

Such ideas and connections to the classical world are crucial for
understanding quantum Shannon theory. Many times, we will be thinking about
unions of disjoint subspaces and it is helpful to make the analogy with a
union of disjoint sets. Also, in Chapter~\ref{chap:covering-lemma}\ on the
covering lemma, we will use projection operators to remove some of the support
of an operator, and this operation is analogous to taking intersections of sets.

Despite the fact that there is a strong connection for classical states, some
of this intuition breaks down by considering the non-orthogonality of quantum
states. For example, consider the case of the projectors $\Pi_{0}\equiv
\vert0\rangle\langle0\vert$ and $\Pi_{+}\equiv\vert+\rangle\langle+\vert$. The
two subspaces onto which these operators project do not intersect, yet we know
that the projectors have some overlap because their corresponding states are
non-orthogonal. One analogy of the intersection operation is to sandwich one
operator with another. For example, we can form the operators%
\begin{equation}
\Pi_{0}\Pi_{+}\Pi_{0},\ \ \ \ \ \ \ \ \ \ \Pi_{+}\Pi_{0}\Pi_{+}.
\end{equation}
If the two projectors were to commute, then this ordering would not matter,
and the resulting operator would be a projector onto the intersection of the
two subspaces. But this is not the case for our example here, and the
resulting operators are quite different.

\begin{exercise}
[Union Bound]\label{ex-nqt:union-bound}Prove a union bound for commuting
projectors $\Pi_{1}$ and $\Pi_{2}$ where $0\leq\Pi_{1},\Pi_{2}\leq I$ and for
an \textit{arbitrary} density operator $\rho$ (not necessarily diagonal in the
same basis as $\Pi_{1}$ and $\Pi_{2}$):%
\begin{equation}
\operatorname{Tr}\left\{  \left(  I-\Pi_{1}\Pi_{2}\right)  \rho\right\}
\leq\operatorname{Tr}\left\{  \left(  I-\Pi_{1}\right)  \rho\right\}
+\operatorname{Tr}\left\{  \left(  I-\Pi_{2}\right)  \rho\right\}  .
\end{equation}

\end{exercise}

\section{Measurement in the Noisy Quantum Theory}

\label{sec-nqt:more-gen-meas}We have described measurement in the quantum
theory using a set of projectors that form a resolution of the identity. For
example, the set $\left\{  \Pi_{j}\right\}  _{j}$ of projectors that satisfy
the condition $\sum_{j}\Pi_{j}=I$ form a valid projective quantum measurement.

There is an alternate description of quantum measurements that follows from
allowing the system of interest to interact unitarily with a probe system that
we measure after the interaction occurs. So suppose that the system of
interest is in a state $|\psi\rangle_{S}$ and that the probe is in a state
$|0\rangle_{P}$, so that the overall state before anything happens is as
follows:%
\begin{equation}
|\psi\rangle_{S}\otimes|0\rangle_{P}.
\end{equation}
Let $\{|0\rangle_{P},|1\rangle_{P},\ldots,|d-1\rangle_{P}\}$ be an orthonormal
basis for the probe system (assuming that it has dimension $d$). Now suppose
that the system and the probe interact according to a unitary $U_{SP}$, and
then we perform a measurement of the probe system, described by measurement
operators $\{|j\rangle\langle j|_{P}\}$. The probability to obtain outcome $j$
is%
\begin{equation}
p_{J}(j)=\left(  \langle\psi|_{S}\otimes\langle0|_{P}U_{SP}^{\dag}\right)
\left(  I_{S}\otimes|j\rangle\langle j|_{P}\right)  \left(  U_{SP}|\psi
\rangle_{S}\otimes|0\rangle_{P}\right)  , \label{eq-nqt:more-gen-meas-prob}%
\end{equation}
and the post-measurement state upon obtaining outcome $j$ is%
\begin{equation}
\frac{1}{\sqrt{p_{J}(j)}}\left(  I_{S}\otimes|j\rangle\langle j|_{P}\right)
\left(  U_{SP}|\psi\rangle_{S}\otimes|0\rangle_{P}\right)  .
\label{eq-nqt:more-gen-meas-state}%
\end{equation}

We can rewrite the expressions above in a different way. Let us expand the
unitary operator $U_{SP}$ in the orthonormal basis of the probe system $P$ as
follows:%
\begin{equation}
U_{SP}=\sum_{j,k}M_{S}^{j,k}\otimes|j\rangle\langle k|_{P},
\label{eq-nqt:U_SP-rep}%
\end{equation}
where $\{M_{S}^{j,k}\}$ is a set of operators. Up to a permutation of the $S$
and $P$ systems and using the mathematics of the tensor product (described in
Section~\ref{sec-qt:tensor-product-ops-1}), this is the same as writing the
unitary $U_{SP}$\ as follows:%
\begin{equation}%
\begin{bmatrix}
M_{S}^{0,0} & M_{S}^{0,1} & \cdots & M_{S}^{0,d-1}\\
M_{S}^{1,0} & M_{S}^{1,1} & \cdots & M_{S}^{1,d-1}\\
\vdots & \vdots & \ddots & \vdots\\
M_{S}^{d-1,0} & M_{S}^{d-1,1} & \cdots & M_{S}^{d-1,d-1}%
\end{bmatrix}
. \label{eq-nqt:matrix-rep-U_SP}%
\end{equation}
This set $\{M_{S}^{j,k}\}$ needs to satisfy some constraints corresponding to
the unitarity of $U_{SP}$. In particular, consider the following operator:%
\begin{equation}
\sum_{j}M_{S}^{j,0}\otimes|j\rangle\langle0|_{P}, \label{eq-nqt:U_SP-part}%
\end{equation}
which corresponds to the first column of operator-valued entries in $U_{SP}$,
as illustrated in \eqref{eq-nqt:matrix-rep-U_SP}. In what follows, we employ
the shorthand $M_{S}^{j}\equiv M_{S}^{j,0}$. From the fact that $U_{SP}^{\dag
}U_{SP}=I_{SP}=I_{S}\otimes I_{P}$, we deduce that the following equality must
hold%
\begin{align}
I_{S}\otimes|0\rangle\langle0|_{P}  &  =\left(  \sum_{j^{\prime}}%
M_{S}^{j^{\prime}\dag}\otimes|0\rangle\langle j^{\prime}|_{P}\right)  \left(
\sum_{j}M_{S}^{j}\otimes|j\rangle\langle0|_{P}\right) \\
&  =\sum_{j^{\prime},j}M_{S}^{j^{\prime}\dag}M_{S}^{j}\otimes|0\rangle
\left\langle j^{\prime}|j\right\rangle \langle0|_{P}\\
&  =\sum_{j}M_{S}^{j\dag}M_{S}^{j}\otimes|0\rangle\langle0|_{P},
\end{align}
where the last line follows from the fact that we chose an orthonormal basis
in the representation of $U_{SP}$ in \eqref{eq-nqt:U_SP-rep}. So the above
equality implies that the following condition holds%
\begin{equation}
\sum_{j}M_{S}^{j\dag}M_{S}^{j}=I_{S}.
\end{equation}

Plugging \eqref{eq-nqt:U_SP-rep} into \eqref{eq-nqt:more-gen-meas-prob} and
\eqref{eq-nqt:more-gen-meas-state}, a short calculation (similar to the above
one) reveals that they simplify as follows:%
\begin{align}
p_{J}(j)  &  =\langle\psi\vert M_{j}^{\dag}M_{j}\vert\psi\rangle,\\
\frac{1}{\sqrt{p_{J}(j)}}\left(  I_{S}\otimes|j\rangle\langle j|_{P}\right)
\left(  U_{SP}|\psi\rangle_{S}\otimes|0\rangle_{P}\right)   &  =\frac
{M_{j}\vert\psi\rangle_{S}\otimes|j\rangle_{P}}{\sqrt{p_{J}(j)}}.
\end{align}
Since the system and the probe are in a pure product state (and thus
independent of each other) after the measurement occurs, we can discard the
probe system and deduce that the post-measurement state of the system $S$ is
simply $M_{j}\vert\psi\rangle_{S}/\sqrt{p_{J}(j)}$.

Motivated by the above development, we allow for an alternate notion of
quantum measurement, saying that it consists of a set of measurement operators
$\left\{  M_{j}\right\}  _{j}$ that satisfy the following completeness
condition:%
\begin{equation}
\sum_{j}M_{j}^{\dag}M_{j}=I.
\end{equation}
Observe from the above development that this is the only constraint that the
operators $\{M_{j}\}$ need to satisfy. This constraint is a consequence of
unitarity, but can be viewed as a generalization of the completeness relation
for a set of projectors that constitute a projective quantum measurement.
Given a set of measurement operators of the above form, the probability for
obtaining outcome $j$ when measuring a state $\vert\psi\rangle$\ is%
\begin{equation}
p_{J}(j)\equiv\langle\psi\vert M_{j}^{\dag}M_{j}\vert\psi\rangle,
\end{equation}
and the post-measurement state when we receive outcome $j$ is%
\begin{equation}
\frac{M_{j}\vert\psi\rangle}{\sqrt{p_{J}(j)}}.
\end{equation}

Suppose that we instead have an ensemble $\left\{  p_{X}(x),|\psi_{x}%
\rangle\right\}  $ with density operator $\rho$. We can carry out an analysis
similar to that which led to \eqref{eq-qt:measurement-density} to conclude
that the probability $p_{J}(j)$ for obtaining outcome $j$ is%
\begin{equation}
p_{J}(j)\equiv\operatorname{Tr}\{M_{j}^{\dag}M_{j}\rho\},
\end{equation}
and the post-measurement state when we measure result $j$ is%
\begin{equation}
\frac{M_{j}\rho M_{j}^{\dag}}{p_{J}(j)}.
\end{equation}
The expression $p_{J}(j)=\operatorname{Tr}\{M_{j}^{\dag}M_{j}\rho\}$ is a
reformulation of the Born rule.

\subsection{POVM\ Formalism}

\label{sec-nqt:POVM}Sometimes, we simply may not care about the
post-measurement state of a quantum measurement, but instead we only care
about the probability for obtaining a particular outcome. For example, this
situation arises in the transmission of classical data over a quantum channel.
In this situation, we are merely concerned with minimizing the error
probabilities of the classical transmission. The receiver does not care about
the post-measurement state because he no longer needs it in the quantum
information-processing protocol. We can specify a measurement of this sort by
a POVM, defined as follows:

\begin{definition}
[POVM]A positive operator-valued measure%
\index{positive operator-valued measure}
(POVM) is a set $\left\{  \Lambda_{j}\right\}  _{j}$ of operators that satisfy
non-negativity and completeness:%
\begin{equation}
\forall j:\Lambda_{j}\geq0,\ \ \ \ \ \ \ \ \ \sum_{j}\Lambda_{j}=I.
\end{equation}

\end{definition}

\noindent The probability for obtaining outcome $j$ is%
\begin{equation}
\langle\psi|\Lambda_{j}|\psi\rangle,
\end{equation}
if the state is some pure state $|\psi\rangle$. The probability for obtaining
outcome $j$ is%
\begin{equation}
\operatorname{Tr}\left\{  \Lambda_{j}\rho\right\}  ,
\end{equation}
if the state is in a mixed state described by some density operator $\rho$.
This is another reformulation of the Born rule.

\begin{exercise}
Consider the following five \textquotedblleft Chrysler\textquotedblright%
\ states:%
\begin{equation}
|e_{k}\rangle\equiv\cos(2\pi k/5)|0\rangle+\sin(2\pi k/5)|1\rangle,
\end{equation}
where $k\in\left\{  0,\ldots,4\right\}  $. These states are the
\textquotedblleft Chrysler\textquotedblright\ states because they form a
pentagon on the $XZ$-plane of the Bloch sphere. Show that the following set of
operators forms a valid POVM: $\left\{  \tfrac{2}{5}|e_{k}\rangle\langle
e_{k}|\right\}  .$
\end{exercise}

\begin{exercise}
\label{ex-nqt:nayak}Suppose we have an ensemble $\left\{  p_{X}( x) ,\rho
_{x}\right\}  $ of density operators and a POVM with elements $\left\{
\Lambda_{x}\right\}  $ that should identify the states $\rho_{x}$ with high
probability, i.e., we would like $\operatorname{Tr}\left\{  \Lambda_{x}
\rho_{x}\right\}  $ to be as high as possible. The expected success
probability of the POVM is then%
\begin{equation}
\sum_{x}p_{X}( x) \operatorname{Tr}\left\{  \Lambda_{x} \rho_{x}\right\}  .
\end{equation}
Suppose that there exists some operator $\tau$ such that%
\begin{equation}
\tau\geq p_{X}( x) \rho_{x}\operatorname{,}%
\end{equation}
where the condition $\tau\geq p_{X}( x) \rho_{x}$ is the same as $\tau-p_{X}(
x) \rho_{x}\geq0$ (i.e., that the operator $\tau-p_{X}( x) \rho_{x}$ is a
positive semi-definite operator). Show that $\operatorname{Tr}\left\{
\tau\right\}  $ is an upper bound on the expected success probability of the
POVM. After doing so, consider the case of encoding $n$ bits into a
$d$-dimensional subspace. By choosing states uniformly at random (in the case
of the ensemble $\left\{  2^{-n},\rho_{i}\right\}  _{i\in\left\{  0,1\right\}
^{n}}$), show that the expected success probability is bounded above by
$d\ 2^{-n}$. Thus, it is not possible to store more than $n$ classical bits in
$n$ qubits and have a perfect success probability of retrieval.
\end{exercise}

\section{Composite Noisy Quantum Systems}

We are again interested in the behavior of two or more quantum systems when we
join them together. Some of the most exotic, truly \textquotedblleft
quantum\textquotedblright\ behavior occurs in joint quantum systems, and we
observe a marked departure from the classical world.

\subsection{Independent Ensembles}

Let us first suppose that we have two independent ensembles for quantum
systems $A$ and $B$. The first quantum system belongs to Alice and the second
quantum system belongs to Bob, and they may or may not be spatially separated.
Let $\{p_{X}(x),|\psi_{x}\rangle\}$ be the ensemble for the system $A$ and let
$\{p_{Y}(y),|\phi_{y}\rangle\}$ be the ensemble for the system $B$. Suppose
for now that the state on system $A$ is $|\psi_{x}\rangle$ for some $x$ and
the state on system $B$ is $|\phi_{y}\rangle$ for some $y$. Then, using the
composite system postulate of the noiseless quantum theory, the joint state
for a given $x$ and $y$ is $|\psi_{x}\rangle\otimes|\phi_{y}\rangle$. The
density operator for the joint quantum system is the expectation of the states
$|\psi_{x}\rangle\otimes|\phi_{y}\rangle$ with respect to the random variables
$X$ and $Y$ that describe the individual ensembles:%
\begin{equation}
\mathbb{E}_{X,Y}\left\{  \left(  |\psi_{X}\rangle\otimes|\phi_{Y}%
\rangle\right)  \left(  \langle\psi_{X}|\otimes\langle\phi_{Y}|\right)
\right\}  .
\end{equation}
The above expression is equal to the following one:%
\begin{equation}
\mathbb{E}_{X,Y}\left\{  |\psi_{X}\rangle\langle\psi_{X}|\otimes|\phi
_{Y}\rangle\langle\phi_{Y}|\right\}  ,
\end{equation}
because $\left(  |\psi_{x}\rangle\otimes|\phi_{y}\rangle\right)  \left(
\langle\psi_{x}|\otimes\langle\phi_{y}|\right)  =|\psi_{x}\rangle\langle
\psi_{x}|\otimes|\phi_{y}\rangle\langle\phi_{y}|$. We then explicitly write
out the expectation as a sum over probabilities:%
\begin{equation}
\sum_{x,y}p_{X}(x)p_{Y}(y)|\psi_{x}\rangle\langle\psi_{x}|\otimes|\phi
_{y}\rangle\langle\phi_{y}|.
\end{equation}
We can distribute the probabilities and the sum because the tensor product
obeys a distributive property:%
\begin{equation}
\sum_{x}p_{X}(x)|\psi_{x}\rangle\langle\psi_{x}|\otimes\sum_{y}p_{Y}%
(y)|\phi_{y}\rangle\langle\phi_{y}|.
\end{equation}
The density operator for this ensemble admits the following simple form:%
\begin{equation}
\rho\otimes\sigma, \label{eq-qt:product-state}%
\end{equation}
where $\rho=\sum_{x}p_{X}(x)|\psi_{x}\rangle\langle\psi_{x}|$ is the density
operator of the $X$ ensemble and $\sigma=\sum_{y}p_{Y}(y)|\phi_{y}%
\rangle\langle\phi_{y}|$ is the density operator of the $Y$ ensemble. We can
say that Alice's local density operator is $\rho$ and Bob's local density
operator is $\sigma$. The overall state is a tensor product of these two
density operators.

\begin{definition}
[Product State]%
\index{product state}
A density operator is a \textit{product state} if it is equal to a tensor
product of two or more density operators.
\end{definition}

We should expect the density operator to factor as it does above because we
assumed that the ensembles are independent. There is nothing much that
distinguishes this situation from the classical world, except for the fact
that the states in each respective ensemble may be non-orthogonal to other
states in the same ensemble. But even here, there is some equivalent
description of each ensemble in terms of an orthonormal basis so that there is
really not too much difference between this description and a joint
probability distribution that factors as two independent distributions.

\begin{exercise}
Show that the purity $P(\rho_{A})$ is equal to the following expression:%
\begin{equation}
P(\rho_{A})=\operatorname{Tr}\left\{  \left(  \rho_{A}\otimes\rho_{A^{\prime}%
}\right)  F_{AA^{\prime}}\right\}  .
\end{equation}
where system $A^{\prime}$ has a Hilbert space structure isomorphic to that of
system $A$ and $F_{AA^{\prime}}$ is the swap operator that has the following
action on kets in $A$ and $A^{\prime}$:%
\begin{equation}
\forall x,y\ \ \ \ \ F_{AA^{\prime}}|x\rangle_{A}|y\rangle_{A^{\prime}%
}=|y\rangle_{A}|x\rangle_{A^{\prime}}.
\end{equation}
(One can in fact show more generally that $\operatorname{Tr}\left\{
f(\rho_{A})\right\}  =\operatorname{Tr}\left\{  \left(  f(\rho_{A})\otimes
I_{A^{\prime}}\right)  F_{AA^{\prime}}\right\}  $ for any function $f$ on the
operators in system $A$.)
\end{exercise}

\subsection{Separable States}

Let us now consider two systems $A$ and $B$ whose corresponding ensembles are
correlated in a classical way. We describe this correlated ensemble as the
joint ensemble%
\begin{equation}
\left\{  p_{X}(x),|\psi_{x}\rangle\otimes|\phi_{x}\rangle\right\}  .
\end{equation}
It is straightforward to verify that the density operator of this correlated
ensemble has the following form:%
\begin{equation}
\mathbb{E}_{X}\left\{  \left(  |\psi_{X}\rangle\otimes|\phi_{X}\rangle\right)
\left(  \langle\psi_{X}|\otimes\left\langle \phi_{X}\right\vert \right)
\right\}  =\sum_{x}p_{X}(x)|\psi_{x}\rangle\langle\psi_{x}|\otimes|\phi
_{x}\rangle\langle\phi_{x}|. \label{eq-nqt:sep-state-motiv-1}%
\end{equation}
By ignoring Bob's system, Alice's local density operator is of the form%
\begin{equation}
\mathbb{E}_{X}\left\{  |\psi_{X}\rangle\langle\psi_{X}|\right\}  =\sum
_{x}p_{X}(x)|\psi_{x}\rangle\langle\psi_{x}|,
\end{equation}
and similarly, Bob's local density operator is%
\begin{equation}
\mathbb{E}_{X}\left\{  |\phi_{X}\rangle\langle\phi_{X}|\right\}  =\sum
_{x}p_{X}(x)|\phi_{x}\rangle\langle\phi_{x}|.
\end{equation}

States of the form in \eqref{eq-nqt:sep-state-motiv-1} can be generated by a
classical procedure. A third party generates a symbol $x$ according to the
probability distribution $p_{X}(x)$ and sends the symbol $x$ to both Alice and
Bob. Alice then prepares the state $\vert\psi_{x}\rangle$ and Bob prepares the
state $\vert\phi_{x}\rangle$. If they then discard the symbol $x$, the state
of their systems is given by \eqref{eq-nqt:sep-state-motiv-1}.

We can generalize this classical preparation procedure one step further, using
an idea similar to the \textquotedblleft ensemble of
ensembles\textquotedblright\ idea in Section~\ref{sec-qt:ensemble-of-ens}. Let
us suppose that we first generate a random variable $Z$ according to some
distribution $p_{Z}( z) $. We then generate two other ensembles, conditioned
on the value of the random variable $Z$. Let $\{p_{X|Z}( x|z) ,\left\vert
\psi_{x,z}\right\rangle \}$ be the first ensemble and let $\{p_{Y|Z}( y|z)
,|\phi_{y,z}\rangle\}$ be the second ensemble, where the random variables $X$
and $Y$ are independent when conditioned on $Z$. Let us label the density
operators of the first and second ensembles when conditioned on a particular
realization $z$ by $\rho_{z}$ and$~\sigma_{z}$, respectively. It is then
straightforward to verify that the density operator of an ensemble created
from this classical preparation procedure has the following form:%
\begin{equation}
\mathbb{E}_{X,Y,Z}\left\{  \left(  \left\vert \psi_{X,Z}\right\rangle
\otimes|\phi_{Y,Z}\rangle\right)  \left(  \left\langle \psi_{X,Z}\right\vert
\otimes\langle\phi_{Y,Z}|\right)  \right\}  =\sum_{z}p_{Z}( z) \rho_{z}%
\otimes\sigma_{z}. \label{eq-qt:separable-state}%
\end{equation}

\begin{exercise}
By ignoring Bob's system, we can determine Alice's local density operator.
Show that%
\begin{equation}
\mathbb{E}_{X,Y,Z}\left\{  \vert\psi_{X,Z}\rangle\langle\psi_{X,Z}%
\vert\right\}  =\sum_{z}p_{Z}( z) \rho_{z},
\label{eq-qt:local-state-separable}%
\end{equation}
so that the above expression is the density operator for Alice. It similarly
follows that the local density operator for Bob is%
\begin{equation}
\mathbb{E}_{X,Y,Z}\left\{  |\phi_{Y,Z}\rangle\langle\phi_{Y,Z}|\right\}
=\sum_{z}p_{Z}( z) \sigma_{z}.
\end{equation}

\end{exercise}

\begin{exercise}
\label{ex-nqt:separable-as-pure}Show that we can always write a state of the
form in \eqref{eq-qt:separable-state} as a convex combination of pure product
states:%
\begin{equation}
\sum_{w}p_{W}( w) \vert\phi_{w}\rangle\langle\phi_{w}\vert\otimes\vert\psi
_{w}\rangle\langle\psi_{w}\vert,
\end{equation}
by manipulating the general form in \eqref{eq-qt:separable-state}.
\end{exercise}

As a consequence of Exercise~\ref{ex-nqt:separable-as-pure}, we see that any
state of the form in \eqref{eq-qt:separable-state} can be written as a convex
combination of pure product states. Such states are called
\index{separable states}%
\textit{separable} states, defined formally as follows:

\begin{definition}
[Separable State]\label{def-nqt:separable-state}
\index{separable states}
A bipartite density operator $\sigma_{AB}$ is a separable state if it can be
written in the following form:
\begin{equation}
\sigma_{AB} = \sum_{x}p_{X}( x) \vert\psi_{x}\rangle\langle\psi_{x}\vert_{A}
\otimes\vert\phi_{x}\rangle\langle\phi_{x}\vert_{B}%
\end{equation}
for some probability distribution $p_{X}(x)$ and sets $\{ \vert\psi_{x}%
\rangle_{A}\}$ and $\{ \vert\phi_{x}\rangle_{B}\}$ of pure states.
\end{definition}

The term \textquotedblleft separable\textquotedblright\ implies that there is
no quantum entanglement in the above state, i.e., there is a completely
classical procedure that prepares the above state. In fact, this leads to the
definition of entanglement for a general bipartite density operator:

\begin{definition}
[Entangled State]\label{def-nqt:entangled-state}
\index{entangled state}
A bipartite density operator $\rho_{AB}$ is entangled if it is not separable.
\end{definition}

\begin{exercise}
[Convexity]Show that the set of separable states acting on a given
tensor-product Hilbert space is a convex set. That is, if $\lambda\in\left[
0,1\right]  $ and $\rho_{AB}$ and $\sigma_{AB}$ are separable states, then
$\lambda\rho_{AB}+(1-\lambda)\sigma_{AB}$ is a separable state.
\end{exercise}

\subsubsection{Separable States and the CHSH Game}%

\index{CHSH game}%

One motivation for Definitions~\ref{def-nqt:separable-state} and
\ref{def-nqt:entangled-state} was already given above: for a separable state,
there is a classical procedure that can be used to prepare it. Thus, for an
entangled state, there is no such procedure. That is, a non-classical
(quantum) interaction between the systems is necessary to prepare an entangled state.

Another related motivation is that separable states admit an explanation in
terms of a classical strategy for the CHSH game, discussed in
Section~\ref{sec-qt:CHSH-game}. Recall from \eqref{eq-qt:CHSH-classical-strat}
that classical strategies $p_{AB|XY}(a,b|x,y)$ are of the following form:
\begin{equation}
p_{AB|XY}(a,b|x,y) = \int d\lambda\ p_{\Lambda}(\lambda) \ p_{A|\Lambda
X}(a|\lambda,x) \ p_{B|\Lambda Y}(b|\lambda,y) .
\label{eq-nqt:classical-strat-CHSH-1}%
\end{equation}
If we allow for a continuous index $\lambda$ for a separable state, then we
can write such a state as follows:
\begin{equation}
\sigma_{AB} = \int d\lambda\ p_{\Lambda}( \lambda)\ \vert\psi_{\lambda}%
\rangle\langle\psi_{\lambda}\vert_{A} \otimes\vert\phi_{\lambda}\rangle
\langle\phi_{\lambda}\vert_{B} .
\end{equation}
Recall that in a general quantum strategy, there are measurements $\{\Pi
^{(x)}_{a} \}$ and $\{\Pi^{(y)}_{b} \}$, giving output bits $a$ and $b$ based
on the input bits $x$ and $y$ and leading to the following strategy:
\begin{align}
p_{AB|XY}(a,b|x,y)  &  = \operatorname{Tr} \{ (\Pi^{(x)}_{a} \otimes\Pi
^{(y)}_{b}) \sigma_{AB} \}\\
&  = \operatorname{Tr} \left\{  (\Pi^{(x)}_{a} \otimes\Pi^{(y)}_{b}) \left(
\int d\lambda\ p_{\Lambda}( \lambda)\ \vert\psi_{\lambda}\rangle\langle
\psi_{\lambda}\vert_{A} \otimes\vert\phi_{\lambda}\rangle\langle\phi_{\lambda
}\vert_{B} \right)  \right\} \\
&  = \int d\lambda\ p_{\Lambda}( \lambda)\ \operatorname{Tr} \left\{
\Pi^{(x)}_{a} \vert\psi_{\lambda}\rangle\langle\psi_{\lambda}\vert_{A}
\otimes\Pi^{(y)}_{b} \vert\phi_{\lambda}\rangle\langle\phi_{\lambda}\vert_{B}
\right\} \\
&  = \int d\lambda\ p_{\Lambda}( \lambda)\ \langle\psi_{\lambda}\vert_{A}
\Pi^{(x)}_{a} \vert\psi_{\lambda}\rangle_{A} \ \langle\phi_{\lambda}\vert_{B}
\Pi^{(y)}_{b} \vert\phi_{\lambda}\rangle_{B} .
\end{align}
By picking the probability distributions $p_{A|\Lambda X}(a|\lambda,x)$ and
$p_{B|\Lambda Y}(b|\lambda,y)$ in \eqref{eq-nqt:classical-strat-CHSH-1} as
follows:
\begin{align}
p_{A|\Lambda X}(a|\lambda,x)  &  = \langle\psi_{\lambda}\vert_{A} \Pi
^{(x)}_{a} \vert\psi_{\lambda}\rangle_{A},\\
p_{B|\Lambda Y}(b|\lambda,y)  &  = \langle\phi_{\lambda}\vert_{B} \Pi
^{(y)}_{b} \vert\phi_{\lambda}\rangle_{B} ,
\end{align}
we see that there is a classical strategy that can simulate any quantum
strategy which uses separable states in the CHSH game. Thus, the winning
probability of quantum strategies involving separable states are subject to
the classical bound of $3/4$ derived in Section~\ref{sec-qt:CHSH-game}. In
this sense, such strategies are effectively classical.

\subsection{Local Density Operators and Partial Trace}

\subsubsection{A First Example}

\label{sec-nqt:example-partial-trace}

Consider the entangled Bell state $\left\vert \Phi^{+}\right\rangle _{AB}$
shared on systems $A$ and $B$. In the above analyses, we determined a local
density operator description for both Alice and Bob. Now, we are curious if it
is possible to determine such a local density operator description for Alice
and Bob with respect to the state $\left\vert \Phi^{+}\right\rangle _{AB}$ or
more general ones.

As a first approach to this issue, recall that the density operator
description arises from its usefulness in determining the probabilities of the
outcomes of a particular measurement. We say that the density operator is
\textquotedblleft the state\textquotedblright\ of the system because it is a
mathematical representation that allows us to compute the probabilities
resulting from a physical measurement. So, if we would like to determine a
\textquotedblleft local density operator,\textquotedblright\ such a local
density operator should predict the result of a local measurement.

Let us consider a local POVM $\left\{  \Lambda^{j}\right\}  _{j}$ that Alice
can perform on her system. The global measurement operators for this local
measurement are $\{\Lambda_{A}^{j}\otimes I_{B}\}_{j}$ because nothing (the
identity) happens to Bob's system. The probability of obtaining outcome $j$
when performing this measurement on the state $\left\vert \Phi^{+}%
\right\rangle _{AB}$ is%
\begin{align}
\left\langle \Phi^{+}\right\vert _{AB}\Lambda_{A}^{j}\otimes I_{B}\left\vert
\Phi^{+}\right\rangle _{AB}  &  =\frac{1}{2}\sum_{k,l=0}^{1}\langle
kk|_{AB}\Lambda_{A}^{j}\otimes I_{B}\left\vert ll\right\rangle _{AB}%
\label{eq-qt:partial-trace-bell-1}\\
&  =\frac{1}{2}\sum_{k,l=0}^{1}\langle k|_{A}\Lambda_{A}^{j}|l\rangle
_{A}\left\langle k|l\right\rangle _{B}\\
&  =\frac{1}{2}\left(  \langle0|_{A}\Lambda_{A}^{j}|0\rangle_{A}+\langle
1|_{A}\Lambda_{A}^{j}|1\rangle_{A}\right) \\
&  =\frac{1}{2}\left(  \operatorname{Tr}\left\{  \Lambda_{A}^{j}%
|0\rangle\langle0|_{A}\right\}  +\operatorname{Tr}\left\{  \Lambda_{A}%
^{j}|1\rangle\langle1|_{A}\right\}  \right) \\
&  =\operatorname{Tr}\left\{  \Lambda_{A}^{j}\frac{1}{2}\left(  |0\rangle
\langle0|_{A}+|1\rangle\langle1|_{A}\right)  \right\} \\
&  =\operatorname{Tr}\left\{  \Lambda_{A}^{j}\pi_{A}\right\}  .
\label{eq-qt:partial-trace-bell}%
\end{align}
The above steps follow by applying the rules of taking the inner product with
respect to tensor product operators. The last line follows by recalling the
definition of the maximally mixed state $\pi$ in
\eqref{eq-qt:maximally-mixed-state}, where $\pi$ here is a qubit maximally
mixed state.

The above calculation demonstrates that we can predict the result of any local
\textquotedblleft Alice\textquotedblright\ measurement using the density
operator $\pi$. Therefore, it is reasonable to say that Alice's local density
operator is $\pi$, and we even go as far to say that her \textit{local state}
is $\pi$. A symmetric calculation shows that Bob's local state is also $\pi$.

This result concerning their local density operators may seem strange at
first. The following global state gives equivalent predictions for local
measurements:%
\begin{equation}
\pi_{A}\otimes\pi_{B}.
\end{equation}
Can we then conclude that an equivalent representation of the global state is
the above state? Absolutely not. The global state $\left\vert \Phi
^{+}\right\rangle _{AB}$ and the above state give drastically different
predictions for global measurements. Exercise~\ref{ex-qt:local-global}\ below
asks you to determine the probabilities for measuring the global operator
$Z_{A}\otimes Z_{B}$ when the global state is $\left\vert \Phi^{+}%
\right\rangle _{AB}$ or $\pi_{A}\otimes\pi_{B}$, and the result is that the
predictions are rather different.

\begin{exercise}
Show that the projection operators corresponding to a measurement of the
observable $Z_{A}\otimes Z_{B}$ are as follows:%
\begin{align}
\Pi_{\operatorname{even}}  &  \equiv\frac{1}{2}\left(  I_{A}\otimes
I_{B}+Z_{A}\otimes Z_{B}\right)  =|00\rangle\langle00|_{AB}+|11\rangle
\langle11|_{AB},\\
\Pi_{\operatorname{odd}}  &  \equiv\frac{1}{2}\left(  I_{A}\otimes I_{B}%
-Z_{A}\otimes Z_{B}\right)  =|01\rangle\langle01|_{AB}+|10\rangle
\langle10|_{AB}.
\end{align}
This measurement is a parity measurement, where the measurement operator
$\Pi_{\operatorname{even}}$ coherently measures even parity and the
measurement operator $\Pi_{\operatorname{odd}}$ measures odd parity.
\end{exercise}

\begin{exercise}
\label{ex-qt:local-global}Show that a parity measurement (defined in the
previous exercise) of the state $\left\vert \Phi^{+}\right\rangle _{AB}$
returns an even parity result with probability one, and a parity measurement
of the state $\pi_{A}\otimes\pi_{B}$ returns even or odd parity with equal
probability. Thus, despite the fact that these states have the same local
description, their global behavior is very different. Show that the same is
true for the phase parity measurement, given by%
\begin{align}
\Pi_{\operatorname{even}}^{X}  &  \equiv\frac{1}{2}\left(  I_{A}\otimes
I_{B}+X_{A}\otimes X_{B}\right)  ,\\
\Pi_{\operatorname{odd}}^{X}  &  \equiv\frac{1}{2}\left(  I_{A}\otimes
I_{B}-X_{A}\otimes X_{B}\right)  .
\end{align}

\end{exercise}

\begin{exercise}
Show that the maximally correlated state $\overline{\Phi}_{AB}$, where%
\begin{equation}
\overline{\Phi}_{AB}=\frac{1}{2}\left(  |00\rangle\langle00|_{AB}%
+|11\rangle\langle11|_{AB}\right)  ,
\end{equation}
gives results for local measurements that are the same as those for the
maximally entangled state$~\left\vert \Phi^{+}\right\rangle _{AB}$. Show that
the above parity measurements can distinguish these states.
\end{exercise}

\subsubsection{Partial Trace}

\label{sec-nqt:partial-trace} In general, we would like to determine a local
density operator that predicts the outcomes of all local measurements. The
general method for determining a local density operator is to employ%
\index{partial trace}
the \textit{partial trace operation}, which we motivate and define here, as a
generalization of the example discussed at the beginning of
Section~\ref{sec-nqt:example-partial-trace}.

Suppose that Alice and Bob share a bipartite state $\rho_{AB}$ and that Alice
performs a local measurement on her system, described by a POVM\ $\{\Lambda
_{A}^{j}\}$. Then the overall POVM\ on the joint system is $\{\Lambda_{A}%
^{j}\otimes I_{B}\}$ because we are assuming that Bob is not doing anything to
his system. According to the Born rule, the probability for Alice to receive
outcome $j$ after performing the measurement is given by the following
expression:%
\begin{equation}
p_{J}(j)=\operatorname{Tr}\{(\Lambda_{A}^{j}\otimes I_{B})\rho_{AB}\}.
\label{eq-nqt:local-global-partial-trace}%
\end{equation}
In order to evaluate the trace, we can choose any orthonormal basis that we
wish (see Definition~\ref{def-nqt:trace}\ and subsequent statements). Taking
$\{|k\rangle_{A}\}$ as an orthonormal basis for Alice's Hilbert space and
$\{|l\rangle_{B}\}$ as an orthonormal basis for Bob's Hilbert space, the set
$\{|k\rangle_{A}\otimes|l\rangle_{B}\}$ constitutes an orthonormal basis for
the tensor product of their Hilbert spaces. So we can evaluate
\eqref{eq-nqt:local-global-partial-trace} as follows:%
\begin{align}
\operatorname{Tr}\{(\Lambda_{A}^{j}\otimes I_{B})\rho_{AB}\}  &  =\sum
_{k,l}\left(  \langle k|_{A}\otimes\langle l|_{B}\right)  \left[  (\Lambda
_{A}^{j}\otimes I_{B})\rho_{AB}\right]  \left(  |k\rangle_{A}\otimes
|l\rangle_{B}\right) \\
&  =\sum_{k,l}\langle k|_{A}\left(  I_{A}\otimes\langle l|_{B}\right)  \left[
(\Lambda_{A}^{j}\otimes I_{B})\rho_{AB}\right]  \left(  I_{A}\otimes
|l\rangle_{B}\right)  |k\rangle_{A}\\
&  =\sum_{k,l}\langle k|_{A}\Lambda_{A}^{j}\left(  I_{A}\otimes\langle
l|_{B}\right)  \rho_{AB}\left(  I_{A}\otimes|l\rangle_{B}\right)
|k\rangle_{A}\\
&  =\sum_{k}\langle k|_{A}\Lambda_{A}^{j}\left[  \sum_{l}\left(  I_{A}%
\otimes\langle l|_{B}\right)  \rho_{AB}\left(  I_{A}\otimes|l\rangle
_{B}\right)  \right]  |k\rangle_{A} . \label{eq-nqt:partial-trace-dev-last}%
\end{align}
The first equality follows from the definition of the trace in
Definition~\ref{def-nqt:trace} and using the orthonormal basis $\{|k\rangle
_{A}\otimes|l\rangle_{B}\}$. The second equality follows because%
\begin{equation}
|k\rangle_{A}\otimes|l\rangle_{B}=\left(  I_{A}\otimes|l\rangle_{B}\right)
|k\rangle_{A}.
\end{equation}
The third equality follows because%
\begin{equation}
\left(  I_{A}\otimes\langle l|_{B}\right)  (\Lambda_{A}^{j}\otimes
I_{B})=\Lambda_{A}^{j}\left(  I_{A}\otimes\langle l|_{B}\right)  .
\end{equation}
The fourth equality follows by bringing the sum over $l$ inside. Using the
definition of the trace in Definition~\ref{def-nqt:trace} and the fact that
$\{|k\rangle_{A}\}$ is an orthonormal basis for Alice's Hilbert space, we can
rewrite \eqref{eq-nqt:partial-trace-dev-last} as%
\begin{equation}
\operatorname{Tr}\left\{  \Lambda_{A}^{j}\left[  \sum_{l}\left(  I_{A}%
\otimes\langle l|_{B}\right)  \rho_{AB}\left(  I_{A}\otimes|l\rangle
_{B}\right)  \right]  \right\}  . \label{eq-nqt:partial-trace-dev-last-1}%
\end{equation}
Our final step is to define the partial trace operation as follows:

\begin{definition}
[Partial Trace]\label{def-nqt:partial-trace}Let $X_{AB}$ be a square operator
acting on a tensor product Hilbert space $\mathcal{H}_{A}\otimes
\mathcal{H}_{B}$, and let $\{|l\rangle_{B}\}$ be an orthonormal basis for
$\mathcal{H}_{B}$. Then the partial trace over the Hilbert space
$\mathcal{H}_{B}$ is defined as follows:%
\begin{equation}
\operatorname{Tr}_{B}\{X_{AB}\}\equiv\sum_{l}\left(  I_{A}\otimes\langle
l|_{B}\right)  X_{AB}\left(  I_{A}\otimes|l\rangle_{B}\right)  .
\end{equation}
For simplicity, we often suppress the identity operators $I_{A}$ and write
this as follows:%
\begin{equation}
\operatorname{Tr}_{B}\{X_{AB}\}\equiv\sum_{l}\langle l|_{B}X_{AB}|l\rangle
_{B}.
\end{equation}

\end{definition}

For the same reason that the definition of the trace is invariant under the
choice of an orthonormal basis, the same is true for the partial trace
operation. We can also observe from the above definition that the partial
trace is a linear operation. Continuing with our development above, we can
define a local operator $\rho_{A}$, using the partial trace, as follows:%
\begin{equation}
\rho_{A}=\operatorname{Tr}_{B}\{\rho_{AB}\}.
\end{equation}
This then allows us to arrive at a rewriting of
\eqref{eq-nqt:partial-trace-dev-last-1} as $\operatorname{Tr}\{\Lambda_{A}%
^{j}\rho_{A}\}, $ which allows us to conclude that%
\begin{equation}
p_{J}(j)=\operatorname{Tr}\{(\Lambda_{A}^{j}\otimes I_{B})\rho_{AB}%
\}=\operatorname{Tr}\{\Lambda_{A}^{j}\rho_{A}\}.
\end{equation}
Thus, from the operator $\rho_{A}$, we can predict the outcomes of local
measurements that Alice performs on her system. Also important here is that
the global picture, in which we have a density operator $\rho_{AB}$ and a
measurement of the form $\{\Lambda_{A}^{j}\otimes I_{B}\}$, is consistent with
the local picture, in which the measurement is written as $\{\Lambda_{A}%
^{j}\}$ and the operator $\rho_{A}$ is used to calculate the probabilities
$p_{J}(j)$. The operator $\rho_{A}$ is itself a density operator, called the
\textit{local} or \textit{reduced density operator}, and the next exercise
asks you to verify that it is indeed a density operator.

\begin{exercise}
[Local Density Operator]Let
\index{density operator!local}%
$\rho_{AB}$ be a density operator acting on a bipartite Hilbert space. Prove
that $\rho_{A}=\operatorname{Tr}_{B}\{\rho_{AB}\}$ is a density operator,
meaning that it is positive semi-definite and has trace equal to one.
\end{exercise}

In conclusion, given a density operator $\rho_{AB}$ describing the joint state
held by Alice and Bob, we can always calculate a local density operator
$\rho_{A}$, which describes the local state of Alice if Bob's system is
inaccessible to her.

There is an alternate way of describing partial trace, of which it is helpful
to be aware. For a simple state of the form%
\begin{equation}
|x\rangle\langle x|_{A}\otimes|y\rangle\langle y|_{B},
\end{equation}
with $|x\rangle_{A}$ and $|y\rangle_{B}$ each unit vectors, the partial trace
has the following action:%
\begin{equation}
\operatorname{Tr}_{B}\left\{  |x\rangle\langle x|_{A}\otimes|y\rangle\langle
y|_{B}\right\}  =|x\rangle\langle x|_{A}\ \operatorname{Tr}\left\{
|y\rangle\langle y|_{B}\right\}  =|x\rangle\langle x|_{A},
\end{equation}
where we \textquotedblleft trace out\textquotedblright\ the second system to
determine the local density operator for the first. If the partial trace acts
on a tensor product of rank-one operators (not necessarily corresponding to a
state)%
\begin{equation}
\vert x_{1}\rangle\langle x_{2}\vert_{A} \otimes\vert y_{1}\rangle\langle
y_{2}\vert_{B} ,
\end{equation}
its action is as follows:%
\begin{align}
\operatorname{Tr}_{B}\left\{  \vert x_{1}\rangle\langle x_{2}\vert_{A}%
\otimes\vert y_{1}\rangle\langle y_{2}\vert_{B}\right\}   &  =\vert
x_{1}\rangle\langle x_{2}\vert_{A}\ \operatorname{Tr}\left\{  \vert
y_{1}\rangle\langle y_{2}\vert_{B}\right\} \\
&  =\vert x_{1}\rangle\langle x_{2}\vert_{A}\ \left\langle y_{2}%
|y_{1}\right\rangle .
\end{align}
In fact, an alternate way of defining the partial trace is as above and to
extend it by linearity.

\begin{exercise}
Show that the two notions of the partial trace operation are consistent. That
is, show that%
\begin{align}
\operatorname{Tr}_{B}\left\{  \vert x_{1}\rangle\langle x_{2}\vert_{A}%
\otimes\vert y_{1}\rangle\langle y_{2}\vert_{B}\right\}   &  =\sum_{i}\langle
i\vert_{B}\left(  \vert x_{1}\rangle\langle x_{2}\vert_{A}\otimes\vert
y_{1}\rangle\langle y_{2}\vert_{B}\right)  \vert i\rangle_{B}\\
&  =\vert x_{1}\rangle\langle x_{2}\vert_{A}\ \left\langle y_{2}%
|y_{1}\right\rangle ,
\end{align}
for some orthonormal basis $\{\vert i\rangle_{B}\}$ on Bob's system.
\end{exercise}

It can be helpful to see the alternate notion of partial trace worked out in
detail. The most general density operator on two systems $A$ and $B$ is some
operator $\rho_{AB}$ that is positive semi-definite with unit trace. We can
obtain the local density operator $\rho_{A}$ from $\rho_{AB}$ by tracing out
the $B$ system:%
\begin{equation}
\rho_{A}=\operatorname{Tr}_{B}\left\{  \rho_{AB}\right\}  .
\end{equation}
In more detail, let us expand an arbitrary density operator $\rho_{AB}$ with
an orthonormal basis $\{|i\rangle_{A}\otimes\vert j\rangle_{B}\}_{i,j}$ for
the bipartite (two-party) state:%
\begin{equation}
\rho_{AB}=\sum_{i,j,k,l}\lambda_{i,j,k,l}(|i\rangle_{A}\otimes|j\rangle
_{B})(\langle k|_{A}\otimes\langle l|_{B}).
\end{equation}
The coefficients $\lambda_{i,j,k,l}$ are the matrix elements of $\rho_{AB}$
with respect to the basis $\{|i\rangle_{A}\otimes\vert j\rangle_{B}\}_{i,j}$,
and they are subject to the constraint of non-negativity and unit trace for
$\rho_{AB}$. We can rewrite the above operator as%
\begin{equation}
\rho_{AB}=\sum_{i,j,k,l}\lambda_{i,j,k,l}|i\rangle\langle k|_{A}%
\otimes|j\rangle\langle l|_{B}.
\end{equation}
We can now evaluate the partial trace:%
\begin{align}
\rho_{A}  &  =\operatorname{Tr}_{B}\left\{  \sum_{i,j,k,l}\lambda
_{i,j,k,l}|i\rangle\langle k|_{A}\otimes|j\rangle\langle l|_{B}\right\} \\
&  =\sum_{i,j,k,l}\lambda_{i,j,k,l}\operatorname{Tr}_{B}\left\{  \vert
i\rangle\langle k|_{A}\otimes|j\rangle\langle l|_{B}\right\} \\
&  =\sum_{i,j,k,l}\lambda_{i,j,k,l}|i\rangle\langle k|_{A}\operatorname{Tr}%
\left\{  |j\rangle\langle l|_{B}\right\} \\
&  =\sum_{i,j,k,l}\lambda_{i,j,k,l}|i\rangle\langle k|_{A}\left\langle
j|l\right\rangle \\
&  =\sum_{i,j,k}\lambda_{i,j,k,j}|i\rangle\langle k|_{A}\\
&  =\sum_{i,k}\left(  \sum_{j}\lambda_{i,j,k,j}\right)  \vert i\rangle\langle
k|_{A}.
\end{align}
The second equality exploits the linearity of the partial trace operation. 

\begin{exercise}
Verify that the partial trace of a product state gives one of the density
operators in the product state:%
\begin{equation}
\operatorname{Tr}_{B}\left\{  \rho_{A}\otimes\sigma_{B}\right\}  =\rho_{A}.
\end{equation}
This result is consistent with the observation near \eqref{eq-qt:product-state}.
\end{exercise}

\begin{exercise}
Verify that the partial trace of a separable state gives the result in
\eqref{eq-qt:local-state-separable}:%
\begin{equation}
\operatorname{Tr}_{B}\left\{  \sum_{z}p_{Z}(z)\rho_{A}^{z}\otimes\sigma
_{B}^{z}\right\}  =\sum_{z}p_{Z}(z)\rho_{A}^{z}.
\end{equation}

\end{exercise}

\begin{exercise}
Consider the following density operator that embeds a joint probability
distribution $p_{X,Y}( x,y) $ in a bipartite quantum state:%
\begin{equation}
\rho=\sum_{x,y}p_{X,Y}( x,y) \vert x\rangle\langle x\vert\otimes\vert
y\rangle\langle y\vert,
\end{equation}
where the set of states $\left\{  \vert x\rangle\right\}  _{x}$ and $\left\{
\vert y\rangle\right\}  _{y}$ each form an orthonormal basis. Show that, in
this case, tracing out the second system is the same as taking the marginal
distribution $p_{X}( x) =\sum_{y}p_{X,Y}( x,y) $ of the joint distribution
$p_{X,Y}( x,y) $. That is, we are left with a density operator of the form%
\begin{equation}
\sum_{x}p_{X}( x) \vert x\rangle\langle x\vert.
\end{equation}
Keep in mind that the partial trace is a generalization of the marginalization
because it handles more exotic quantum states besides the above
\textquotedblleft classical\textquotedblright\ state.
\end{exercise}

\begin{exercise}
Show that the two partial traces in any order on a bipartite system are
equivalent to a full trace:%
\begin{equation}
\operatorname{Tr}\left\{  \rho_{AB}\right\}  =\operatorname{Tr}_{A}\left\{
\operatorname{Tr}_{B}\left\{  \rho_{AB}\right\}  \right\}  =\operatorname{Tr}%
_{B}\left\{  \operatorname{Tr}_{A}\left\{  \rho_{AB}\right\}  \right\}  .
\end{equation}

\end{exercise}

\begin{exercise}
Verify that Alice's local density operator does not change if Bob performs a
unitary operator or a measurement in which he does not inform her of the
measurement result.
\end{exercise}

\begin{exercise}
Prove that the partial trace operation obeys a cyclicity relation with respect
to operators that act exclusively on the system over which we trace. That is,
let $X_{AB}$ be a square operator acting on the tensor-product Hilbert space
$\mathcal{H}_{A}\otimes\mathcal{H}_{B}$, and let $Y_{B}$, $Z_{B}$ and $W_{B}$
be square operators acting on the Hilbert space $\mathcal{H}_{B}$. Prove that
\begin{align}
\operatorname{Tr}_{B}\{X_{AB}Y_{B}Z_{B}W_{B}\}  &  =\operatorname{Tr}%
_{B}\{W_{B}X_{AB}Y_{B}Z_{B}\}\\
&  =\operatorname{Tr}_{B}\{Z_{B}W_{B}X_{AB}Y_{B}\}\\
&  =\operatorname{Tr}_{B}\{Y_{B}Z_{B}W_{B}X_{AB}\}.
\end{align}
In the above, it is implicit that $Y_{B}=I_{A}\otimes Y_{B}$, etc.
\end{exercise}

\begin{exercise}
Recall that the purity of a density operator $\rho_{A}$ is equal to
$\operatorname{Tr}\left\{  \rho_{A}^{2}\right\}  $. Suppose that $\rho
_{A}=\operatorname{Tr}_{B}\left\{  \Phi_{AB}\right\}  $, where $\Phi_{AB}$ is
a maximally entangled state. Prove that the purity is equal to the inverse of
the dimension of the $A$ system.
\end{exercise}

\subsection{Classical--Quantum Ensemble}

\label{sec-nqt:classical-quantum}We end our overview of composite noisy
quantum systems by discussing one last type of joint ensemble:\ the
\textit{classical--quantum ensemble}. This ensemble is a generalization of the
\textquotedblleft ensemble of ensembles\textquotedblright\ from before.

Let us consider the following ensemble of density operators:%
\begin{equation}
\left\{  p_{X}( x) ,\rho_{A}^{x}\right\}  _{x\in\mathcal{X}}.
\label{eq-qt:ensemble-of-DOs}%
\end{equation}
The intuition here is that Alice prepares a quantum system in the state
$\rho_{A}^{x}$ with probability $p_{X}( x) $. She then passes this ensemble to
Bob, and it is Bob's task to learn about it. He can learn about the ensemble
if Alice prepares a large number of them in the same way.

There is generally a loss of the information in the random variable $X$ once
Alice has prepared this ensemble. It is easier for Bob to learn about the
distribution of the random variable $X$ if each density operator $\rho_{A}%
^{x}$ is a pure state $\vert x\rangle\langle x\vert$ where the states
$\left\{  \vert x\rangle\right\}  _{x\in\mathcal{X}}$ form an orthonormal
basis. The resulting density operator would be%
\begin{equation}
\rho_{A}=\sum_{x\in\mathcal{X}}p_{X}( x) \vert x\rangle\langle x\vert_{A}.
\end{equation}
Bob could then perform a measurement with measurement operators $\left\{
\vert x\rangle\langle x\vert\right\}  _{x\in\mathcal{X}}$, and learn about the
distribution $p_{X}( x) $ with a large number of measurements.

In the general case, the density operators $\left\{  \rho_{A}^{x}\right\}
_{x\in\mathcal{X}}$ do not correspond to pure states, much less orthonormal
ones, and it is more difficult for Bob to learn about random variable $X$. The
density operator of the ensemble is%
\begin{equation}
\rho_{A}=\sum_{x\in\mathcal{X}}p_{X}( x) \rho_{A}^{x},
\label{eq-qt:DO-ensemble-of-DOs}%
\end{equation}
and the information about the distribution of random variable $X$ becomes
\textquotedblleft mixed in\textquotedblright\ with the \textquotedblleft
mixedness\textquotedblright\ of the density operators $\rho_{x}$. There is
then no measurement that Bob can perform on $\rho$ that allows him to directly
learn about the probability distribution of random variable$~X$.

One solution to this issue is for Alice to prepare the following
classical--quantum ensemble:%
\begin{equation}
\left\{  p_{X}(x),|x\rangle\langle x|_{X}\otimes\rho_{A}^{x}\right\}
_{x\in\mathcal{X}}, \label{eq-nqt:cq-ensemble-1}%
\end{equation}
where we label the first system as $X$ and the second as $A$. She simply
correlates a state $|x\rangle$ with each density operator $\rho_{A}^{x}$,
where the states $\left\{  |x\rangle\right\}  _{x\in\mathcal{X}}$ form an
orthonormal basis. We call this ensemble a \textquotedblleft
classical--quantum\textquotedblright\ ensemble because the first system is
classical and the second system is quantum. This then leads to the notion of a
\textit{classical--quantum state}%
\index{classical-quantum state}
$\rho_{XA}$ defined as follows:

\begin{definition}
[Classical--Quantum State]\label{def-nqt:classical-quantum-state}The density
operator corresponding to a classical--quantum ensemble $\left\{
p_{X}(x),|x\rangle\langle x|_{X}\otimes\rho_{A}^{x}\right\}  _{x\in
\mathcal{X}}$, as discussed above, is called a classical--quantum state and
takes the following form:%
\begin{equation}
\rho_{XA}\equiv\sum_{x\in\mathcal{X}}p_{X}(x)|x\rangle\langle x|_{X}%
\otimes\rho_{A}^{x}.
\end{equation}
It is a particular kind of separable state of systems $X$ and $A$, in which
the individual states of the $X$ system are perfectly distinguishable and thus classical.
\end{definition}

The \textquotedblleft enlarged\textquotedblright\ ensemble in
\eqref{eq-nqt:cq-ensemble-1} lets Bob easily learn about random variable $X$
while at the same time he can learn about the ensemble that Alice prepares.
Bob can learn about the distribution of random variable $X$ by performing a
local measurement of the system $X$. He also can learn about the states
$\rho_{x}$ by performing a measurement on $A$ and combining the result of this
measurement with the result of the first measurement. The next exercises ask
you to verify these statements.

\begin{exercise}
Show that a local measurement of system $X$ reproduces the probability
distribution $p_{X}(x)$. Use local measurement operators $\left\{
|x\rangle\langle x|\right\}  _{x\in\mathcal{X}}$ to show that $p_{X}%
(x)=\operatorname{Tr}\left\{  \rho_{XA}\left(  |x\rangle\langle x|_{X}\otimes
I_{A}\right)  \right\}  $.
\end{exercise}

\begin{exercise}
Show that performing a measurement with measurement operators $\{\Lambda
_{A}^{j}\}$ on system $A$ is the same as performing a measurement of the
ensemble in \eqref{eq-qt:ensemble-of-DOs}. That is, show that
$\operatorname{Tr}\{\rho_{A}\Lambda_{A}^{j}\}=\operatorname{Tr}\{\rho
_{XA}(I_{X}\otimes\Lambda_{A}^{j})\}$, where $\rho_{A}$ is defined in \eqref{eq-qt:DO-ensemble-of-DOs}.
\end{exercise}

\begin{exercise}
[Lack of Convexity]Prove that the set of classical--quantum states is not a
convex set. That is, show that there exists a classical-quantum state
$\rho_{XA}$ and another $\sigma_{XA}$ and $\lambda\in\left[  0,1\right]  $,
such that $\lambda\rho_{XA}+(1-\lambda)\sigma_{XA}$ is not a
classical--quantum state.
\end{exercise}

\section{Quantum Evolutions}

\label{sec-nqt:noisy-evo}The evolution of a quantum state is never perfect. In
this section, we begin by discussing the most general approach to
understanding quantum evolutions:\ the \textit{axiomatic approach}. This
powerful approach starts with three physically reasonable axioms that should
hold for any quantum evolution and from there we deduce a set of mathematical
constraints that any quantum evolution should satisfy (this is known as the
\textit{Choi--Kraus theorem}). Throughout the book, we will refer to quantum
evolutions satisfying these constraints as \textit{quantum channels}. We then
show how noise resulting from the loss of information about a quantum system
or from lack of access to an environment system is equivalent to what we find
from the Choi--Kraus theorem. We finally discuss how every operation we have
discussed so far, including preparations and measurements, can be viewed as a
quantum channel, and we follow by giving several important examples of quantum channels.

\subsection{Axiomatic Approach to Quantum Evolutions}

\label{sec-nqt:axiomatic-approach}We now discuss a powerful approach to
understanding quantum physical evolutions called the \textit{axiomatic
approach}. Here we make three physically reasonable assumptions that any
quantum evolution should satisfy and then prove that these axioms imply
mathematical constraints on the form of any quantum physical evolution.

All of the constraints we impose are motivated by the reasonable requirement
for the output of the evolution to be a quantum state (density operator) if
the input to the evolution is a quantum state (density operator). An important
assumption to clarify at the outset is that we are viewing a quantum physical
evolution as a \textquotedblleft black box,\textquotedblright\ meaning that
Alice can prepare any state that she wishes before the evolution begins,
including pure states or mixed states. Critically, we even allow her to input
one share of an entangled state. This is a standard assumption in quantum
information theory, but one could certainly question whether this assumption
is reasonable. If we do accept this criterion as physically reasonable, then
the Choi--Kraus representation theorem for quantum evolutions follows as a consequence.

\begin{notation}
[Density Operators and Linear Operators]Let $\mathcal{D}(\mathcal{H})$ denote
the space of density operators acting on a Hilbert space $\mathcal{H}$, let
$\mathcal{L}(\mathcal{H})$ denote the space of square linear operators acting
on $\mathcal{H}$, and let $\mathcal{L}(\mathcal{H}_{A},\mathcal{H}_{B})$
denote the space of linear operators taking a Hilbert space $\mathcal{H}_{A}$
to a Hilbert space $\mathcal{H}_{B}$.
\end{notation}

Throughout this development, we let $\mathcal{N}$ denote a map which takes
density operators in $\mathcal{D}(\mathcal{H}_{A})$ to those in $\mathcal{D}%
(\mathcal{H}_{B})$. In general, the respective input and output Hilbert spaces
$\mathcal{H}_{A}$ and $\mathcal{H}_{B}$ need not be the same. Implicitly, we
have already stated a first physically reasonable requirement that we impose
on $\mathcal{N}$, namely, that\ $\mathcal{N}(\rho_{A})\in\mathcal{D}%
(\mathcal{H}_{B})$ if $\rho_{A}\in\mathcal{D}(\mathcal{H}_{A})$. Extending
this requirement, we demand that $\mathcal{N}$ should be \textit{convex
linear} when acting on $\mathcal{D}(\mathcal{H}_{A})$:%
\begin{equation}
\mathcal{N}(\lambda\rho_{A}+(1-\lambda)\sigma_{A})=\lambda\mathcal{N}(\rho
_{A})+(1-\lambda)\mathcal{N}(\sigma_{A}),
\label{eq-nqt:convex-linearity-channels}%
\end{equation}
where $\rho_{A},\sigma_{A}\in\mathcal{D}(\mathcal{H}_{A})$ and $\lambda
\in\left[  0,1\right]  $.

The physical interpretation of this convex-linearity requirement is in terms
of repeated experiments. Suppose a large number of experiments are conducted
in which identical quantum systems are prepared in the state $\rho_{A}$ for a
fraction $\lambda$ of the experiments and in the state $\sigma_{A}$ for the
other fraction $1-\lambda$ of the experiments. Suppose further that it is not
revealed which states are prepared for which experiments. Before you are
allowed to perform measurements on each system, the evolution $\mathcal{N}$ is
applied to each of the systems. The density operator characterizing the state
of each system for these experiments is then $\mathcal{N}(\lambda\rho
_{A}+(1-\lambda)\sigma_{A})$. You are then allowed to perform a measurement on
each system, which after a large number of experiments allow you to infer that
the density operator is $\mathcal{N}(\lambda\rho_{A}+(1-\lambda)\sigma_{A})$.
Now, in principle, it could have been revealed which fraction of the
experiments had the state $\rho_{A}$ prepared and which fraction had
$\sigma_{A}$ prepared. In this case, the density operator describing the
$\rho_{A}$ experiments would be $\mathcal{N}(\rho_{A})$ and that describing
the $\sigma_{A}$ experiments would be $\mathcal{N}(\sigma_{A})$. So, it is
reasonable to expect that the statistics observed in your measurement outcomes
in the first scenario would be consistent with those observed in the second
scenario, and this is the physical statement that the requirement \eqref{eq-nqt:convex-linearity-channels}\ makes.

Now, it is mathematically convenient to extend the domain and range of the
quantum channel to apply not only to density operators but to all linear
operators. To this end, it is possible to find a unique linear extension
$\widetilde{\mathcal{N}}$ of any quantum evolution $\mathcal{N}$ defined as
above (originally defined exclusively by its action on density operators and
satisfying convex linearity). See Appendix~\ref{app:unique-linear-ext}\ for a
full development of this idea. Thus, it is reasonable to associate this unique
linear extension $\widetilde{\mathcal{N}}$\ to the quantum physical evolution
$\mathcal{N}$ mathematically, and in what follows (and for the rest of the
book), we simply identify a physical evolution $\mathcal{N}$\ with its unique
linear extension $\widetilde{\mathcal{N}}$, and this is what we call a
\textit{quantum channel}. For these reasons, we now impose that any quantum
channel $\mathcal{N}$ is linear:

\begin{criterion}
[Linearity]\label{crit-nqt:linearity}A quantum channel $\mathcal{N}$ is a
linear map:%
\begin{equation}
\mathcal{N}(\alpha X_{A}+\beta Y_{A})=\alpha\mathcal{N}(X_{A})+\beta
\mathcal{N}(Y_{A}),
\end{equation}
where $X_{A},Y_{A}\in\mathcal{L}(\mathcal{H}_{A})$ and $\alpha,\beta
\in\mathbb{C}$.
\end{criterion}

We have already demanded that quantum physical evolutions should take density
operators to density operators. Combining with linearity (in particular, scale
invariance) implies that quantum channels should preserve the class of
positive semi-definite operators. That is, they should be positive maps, as
defined below:

\begin{definition}
[Positive Map]A
\index{positive map}%
linear map $\mathcal{M}:\mathcal{L}(\mathcal{H}_{A})\rightarrow\mathcal{L}%
(\mathcal{H}_{B})$ is positive if $\mathcal{M}(X_{A})$ is positive
semi-definite for all positive semi-definite $X_{A}\in\mathcal{L}%
(\mathcal{H}_{A})$.
\end{definition}

If we were dealing with classical systems, then positivity would be sufficient
to describe the class of physical evolutions. However, above we argued that we
are working in the \textquotedblleft black box\textquotedblright\ picture of
quantum physical evolutions, and here, in principle, we allow for Alice to
prepare the input system $A$ to be one share of an arbitrary two-party state
$\rho_{RA}\in\mathcal{D}(\mathcal{H}_{R}\otimes\mathcal{H}_{A})$, where $R$ is
a reference system of arbitrary size. So this means that the evolution
consisting of the identity acting on the reference system $R$ and the map
$\mathcal{N}$ acting on system $A$ should take $\rho_{RA}$ to a density
operator on systems $R$ and $B$. Let $\operatorname{id}_{R}\otimes
\mathcal{N}_{A\rightarrow B}$ denote this evolution, where $\operatorname{id}%
_{R}$ denotes the identity superoperator acting on the system $R$.

How do we describe the evolution $\operatorname{id}_{R}\otimes\mathcal{N}%
_{A\rightarrow B}$ mathematically? Let $X_{RA}$ be an arbitrary operator
acting on $\mathcal{H}_{R}\otimes\mathcal{H}_{A}$, and let $\{|i\rangle_{R}\}$
be an orthonormal basis for $\mathcal{H}_{R}$. Then we can expand $X_{RA}$
with respect to this basis as follows:%
\begin{equation}
X_{RA}=\sum_{i,j}|i\rangle\langle j|_{R}\otimes X_{A}^{i,j},
\end{equation}
and the action of $\operatorname{id}_{R}\otimes\mathcal{N}_{A\rightarrow B}$
on $X_{RA}$ (for linear $\mathcal{N}$) is defined as follows:%
\begin{align}
\left(  \operatorname{id}_{R}\otimes\mathcal{N}_{A\rightarrow B}\right)
\left(  X_{RA}\right)   &  =\left(  \operatorname{id}_{R}\otimes
\mathcal{N}_{A\rightarrow B}\right)  \left(  \sum_{i,j}|i\rangle\langle
j|_{R}\otimes X_{A}^{i,j}\right) \\
&  =\sum_{i,j}\left(  \operatorname{id}_{R}\otimes\mathcal{N}_{A\rightarrow
B}\right)  \left(  |i\rangle\langle j|_{R}\otimes X_{A}^{i,j}\right) \\
&  =\sum_{i,j}\operatorname{id}_{R}\left(  |i\rangle\langle j|_{R}\right)
\otimes\mathcal{N}_{A\rightarrow B}\left(  X_{A}^{i,j}\right) \\
&  =\sum_{i,j}|i\rangle\langle j|_{R}\otimes\mathcal{N}_{A\rightarrow
B}\left(  X_{A}^{i,j}\right)  . \label{eq-nqt:action-id-N}%
\end{align}
That is, the identity superoperator $\operatorname{id}_{R}$ has no effect on
the $R$ system. The above development leads to the notion of a linear map
being \textit{completely positive} and our next criterion for any quantum
physical evolution:

\begin{definition}
[Completely Positive Map]\label{def-nqt:completely-positive}A linear map%
\index{completely positive map}
$\mathcal{M}:\mathcal{L}(\mathcal{H}_{A})\rightarrow\mathcal{L}(\mathcal{H}%
_{B})$ is completely positive if $\operatorname{id}_{R}\otimes\mathcal{M}$ is
a positive map for a reference system $R$ of arbitrary size.
\end{definition}

\begin{criterion}
[Complete Positivity]\label{crit-nqt:CP}A quantum channel is a completely
positive map.
\end{criterion}

There is one last requirement that we impose for quantum physical evolutions,
known as \textit{trace preservation}. This requirement again stems from the
reasonable constraint that $\mathcal{N}$ should map density operators to
density operators. That is, it should be the case that $\operatorname{Tr}%
\{\rho_{A}\}=\operatorname{Tr}\{\mathcal{N}(\rho_{A})\}=1$ for all input
density operators $\rho_{A}$. However, now that we have argued for linearity of
every quantum physical evolution, trace preservation on density operators
combined with linearity implies that quantum channels are trace preserving on
the set of all operators. This is due to the fact that there are sets of
density operators that form a basis for $\mathcal{L}(\mathcal{H}_{A})$.
Indeed, one such basis of density operators is as follows:%
\begin{equation}
\rho_{A}^{x,y}=\left\{
\begin{array}
[c]{cc}%
|x\rangle\langle x|_{A} & \text{if }x=y\\
\frac{1}{2}\left(  |x\rangle_{A}+|y\rangle_{A}\right)  \left(  \langle
x|_{A}+\langle y|_{A}\right)  & \text{if }x<y\\
\frac{1}{2}\left(  |x\rangle_{A}+i|y\rangle_{A}\right)  \left(  \langle
x|_{A}-i\langle y|_{A}\right)  & \text{if }x>y
\end{array}
\right.  . \label{eq-nqt:density-op-basis}%
\end{equation}
Consider that for all $x,y$ such that $x<y$, the following holds%
\begin{align}
|x\rangle\langle y|_{A}  &  =\left(  \rho_{A}^{x,y}-\frac{1}{2}\rho_{A}%
^{x,x}-\frac{1}{2}\rho_{A}^{y,y}\right)  -i\left(  \rho_{A}^{y,x}-\frac{1}%
{2}\rho_{A}^{x,x}-\frac{1}{2}\rho_{A}^{y,y}\right)  ,\\
|y\rangle\langle x|_{A}  &  =\left(  \rho_{A}^{x,y}-\frac{1}{2}\rho_{A}%
^{x,x}-\frac{1}{2}\rho_{A}^{y,y}\right)  +i\left(  \rho_{A}^{y,x}-\frac{1}%
{2}\rho_{A}^{x,x}-\frac{1}{2}\rho_{A}^{y,y}\right)  ,
\end{align}
so that we can represent any operator $X_{A}$ as a linear combination of
density operators from the set $\left\{  \rho_{A}^{x,y}\right\}  $. This leads
to our final criterion for quantum channels:

\begin{criterion}
[Trace Preservation]\label{crit-nqt:TP}A quantum channel is trace preserving,%
\index{trace preserving map}
in the sense that $\operatorname{Tr}\{X_{A}\}=\operatorname{Tr}\{\mathcal{N}%
(X_{A})\}$ for all $X_{A}\in\mathcal{L}(\mathcal{H}_{A})$.
\end{criterion}

\begin{definition}
[Quantum Channel]A quantum channel is a linear, completely positive,
\index{completely positive trace-preserving map}%
trace preserving map, corresponding to a quantum physical evolution.
\end{definition}

Criteria~\ref{crit-nqt:linearity}, \ref{crit-nqt:CP}, and \ref{crit-nqt:TP}
detailed above lead naturally to the Choi--Kraus representation theorem, which
states that a map satisfies all three criteria if and only if it takes a
particular form according to a Choi--Kraus decomposition:

\begin{theorem}
[Choi--Kraus]\label{thm-nqt:choi-kraus-theorem}A map $\mathcal{N}%
:\mathcal{L}(\mathcal{H}_{A})\rightarrow\mathcal{L}(\mathcal{H}_{B})$
\index{Choi--Kraus theorem}%
(denoted also by $\mathcal{N}_{A\rightarrow B}$) is linear, completely
positive, and trace-preserving if and only if it has a Choi--Kraus
decomposition as follows:%
\begin{equation}
\mathcal{N}_{A\rightarrow B}(X_{A})=\sum_{l=0}^{d-1}V_{l}X_{A}V_{l}^{\dag},
\label{eq-nqt:kraus-decomp}%
\end{equation}
where $X_{A}\in\mathcal{L}(\mathcal{H}_{A})$, $V_{l}\in\mathcal{L}%
(\mathcal{H}_{A},\mathcal{H}_{B})$ for all $l\in\left\{  0,\ldots,d-1\right\}
$,%
\begin{equation}
\sum_{l=0}^{d-1}V_{l}^{\dag}V_{l}=I_{A}, \label{eq-nqt:trace-preserve-kraus}%
\end{equation}
and $d$ need not be any larger than $\dim(\mathcal{H}_{A})\dim(\mathcal{H}%
_{B})$.
\end{theorem}

Before we delve into a proof, it is helpful to give a sketch. There is an
easier part and a more challenging part of the proof. For the more challenging
part, a helpful tool is an operator called the Choi operator:

\begin{definition}
[Choi Operator]\label{def-nqt:choi-op}
\index{Choi operator}%
Let $\mathcal{H}_{R}$ and $\mathcal{H}_{A}$ be isomorphic Hilbert spaces, and
let $\{|i\rangle_{R}\}$ and $\{|i\rangle_{A}\}$ be orthonormal bases for
$\mathcal{H}_{R}$ and $\mathcal{H}_{A}$, respectively. Let $\mathcal{H}_{B}$
be some other Hilbert space, and let $\mathcal{N}:\mathcal{L}(\mathcal{H}%
_{A})\rightarrow\mathcal{L}(\mathcal{H}_{B})$ be a linear map (written also as
$\mathcal{N}_{A\rightarrow B}$). The Choi operator corresponding to
$\mathcal{N}_{A\rightarrow B}$ and the bases $\{|i\rangle_{R}\}$ and
$\{|i\rangle_{A}\}$ is defined as the following operator:%
\begin{equation}
\left(  \operatorname{id}_{R}\otimes\mathcal{N}_{A\rightarrow B}\right)
\left(  \vert\Gamma\rangle\langle\Gamma\vert_{RA}\right)  =\sum_{i,j=0}%
^{d_{A}-1}\vert i\rangle\langle j\vert_{R}\otimes\mathcal{N}_{A\rightarrow
B}(|i\rangle\langle j\vert_{A}), \label{eq:choi-matrix}%
\end{equation}
where $d_{A}\equiv\dim(\mathcal{H}_{A})$ and $\vert\Gamma\rangle_{RA}$ is an
unnormalized maximally entangled vector, as defined in
\eqref{eq-qt:unnorm-max-ent}:%
\begin{equation}
\vert\Gamma\rangle_{RA}\equiv\sum_{i=0}^{d_{A}-1}\left\vert i\right\rangle
_{R}\otimes|i\rangle_{A}.
\end{equation}
\index{Choi rank}%
The rank of the Choi operator is called the Choi rank.
\end{definition}

If $\mathcal{N}_{A\rightarrow B}$ is a completely positive map, then the Choi
operator is positive semi-definite. This follows as a direct consequence of
Definition~\ref{def-nqt:completely-positive} and the fact that $\vert
\Gamma\rangle\langle\Gamma\vert_{RA}$ is positive semi-definite. The converse
is true as well, and Exercise~\ref{ex-nqt:choi-CP}\ asks you to verify this.
The converse is in some sense a much more powerful statement.
Definition~\ref{def-nqt:completely-positive} suggests that we would have to
check a seemingly infinite number of cases in order to verify whether a given
linear map is completely positive, but the converse statement establishes that
we need to check only one condition: whether the Choi operator is positive semi-definite.

Why else is the Choi operator a useful tool? One other important reason is
that it encodes how a quantum channel acts on any possible input operator
$X_{A}$, and thus specifies the channel completely. Consider that we can
expand the Choi operator as a matrix of matrices (of total size $d_{A}%
d_{B}\times d_{A}d_{B}$) in the following way, by exploiting properties of the
tensor product:%
\begin{equation}%
\begin{bmatrix}
\mathcal{N}_{A\rightarrow B}(|0\rangle\langle0\vert_{A}) & \mathcal{N}%
_{A\rightarrow B}(|0\rangle\langle1\vert_{A}) & \cdots & \mathcal{N}%
_{A\rightarrow B}(|0\rangle\langle d_{A}-1\vert_{A})\\
\mathcal{N}_{A\rightarrow B}(|1\rangle\langle0\vert_{A}) & \mathcal{N}%
_{A\rightarrow B}(|1\rangle\langle1\vert_{A}) & \cdots & \mathcal{N}%
_{A\rightarrow B}(|1\rangle\langle d_{A}-1\vert_{A})\\
\vdots & \vdots & \ddots & \vdots\\
\mathcal{N}_{A\rightarrow B}(|d_{A}-1\rangle\langle0\vert_{A}) &
\mathcal{N}_{A\rightarrow B}(|d_{A}-1\rangle\langle1\vert_{A}) & \cdots &
\mathcal{N}_{A\rightarrow B}(|d_{A}-1\rangle\langle d_{A}-1\vert_{A})
\end{bmatrix}
.
\end{equation}
So if we would like to figure out how the channel $\mathcal{N}_{A\rightarrow
B}$ acts on an input operator $X_{A}$, we can first expand $X_{A}$ with
respect to the orthonormal basis $\{|i\rangle_{A}\}$ as $X_{A}=\sum
_{i,j}x^{i,j}|i\rangle\langle j|_{A}$ and then apply the channel, using
linearity:%
\begin{equation}
\mathcal{N}_{A\rightarrow B}(X_{A})=\mathcal{N}_{A\rightarrow B}\left(
\sum_{i,j}x^{i,j}|i\rangle\langle j|_{A}\right)  =\sum_{i,j}x^{i,j}%
\mathcal{N}_{A\rightarrow B}(|i\rangle\langle j|_{A}).
\label{eq-nqt:linear-action-of-channel}%
\end{equation}
So the procedure is to expand $X_{A}$ as above, multiple the $\left(
i,j\right)  $ coefficient $x^{i,j}$ with the $( i,j) $ entry in the Choi
operator, and then sum these operators over all indices $i$ and $j$.

\bigskip

\begin{proof}
[Proof of Theorem~\ref{thm-nqt:choi-kraus-theorem}]We first prove the easier
\textquotedblleft if-part\textquotedblright\ of the theorem. So let us suppose
that $\mathcal{N}_{A\rightarrow B}$ has the form in
\eqref{eq-nqt:kraus-decomp} and that the condition in
\eqref{eq-nqt:trace-preserve-kraus} holds as well. Then $\mathcal{N}%
_{A\rightarrow B}$ is clearly a linear map. It is completely positive because
$(\operatorname{id}_{R}\otimes\mathcal{N}_{A\rightarrow B})(X_{RA})\geq0$ if
$X_{RA}\geq0$ when $\mathcal{N}_{A\rightarrow B}$ has the form in
\eqref{eq-nqt:kraus-decomp}, and this holds for a reference system $R$ of
arbitrary size. That is, consider from \eqref{eq-nqt:action-id-N} that
$\{I_{R}\otimes V_{l}\}$ is a set of Kraus operators for the extended channel
$\operatorname{id}_{R}\otimes\mathcal{N}_{A\rightarrow B}$\ and thus%
\begin{equation}
(\operatorname{id}_{R}\otimes\mathcal{N}_{A\rightarrow B})(X_{RA}) =\sum
_{l=0}^{d-1}(I_{R}\otimes V_{l})X_{RA}(I_{R}\otimes V_{l}^{\dag}).
\end{equation}
We know that $(I_{R}\otimes V_{l})X_{RA}(I_{R}\otimes V_{l})^{\dag}\geq0$ for
all $l$ when $X_{RA}\geq0$, and the same is true for the sum. Trace
preservation follows because%
\begin{align}
\operatorname{Tr}\left\{  \mathcal{N}_{A\rightarrow B}(X_{A})\right\}   &
=\operatorname{Tr}\left\{  \sum_{l=0}^{d-1}V_{l}X_{A}V_{l}^{\dag}\right\} \\
&  =\operatorname{Tr}\left\{  \sum_{l=0}^{d-1}V_{l}^{\dag}V_{l}X_{A}\right\}
\\
&  =\operatorname{Tr}\left\{  X_{A}\right\}  ,
\end{align}
where the second line follows from linearity and cyclicity of trace and the
last line follows from the condition in \eqref{eq-nqt:trace-preserve-kraus}.

We now prove the more difficult \textquotedblleft only-if\textquotedblright%
\ part. Let $d_{A}\equiv\dim(\mathcal{H}_{A})$ and $d_{B}\equiv\dim
(\mathcal{H}_{B})$. Consider that we can diagonalize the Choi operator as
given in Definition~\ref{def-nqt:choi-op}, because it is positive
semi-definite:%
\begin{equation}
\mathcal{N}_{A\rightarrow B}\left(  \vert\Gamma\rangle\langle\Gamma\vert
_{RA}\right)  =\sum_{l=0}^{d-1}|\phi_{l}\rangle\langle\phi_{l}|_{RB},
\label{eq:decomposition-choi}%
\end{equation}
where $d\leq d_{A}d_{B}$ is the Choi rank of the map $\mathcal{N}%
_{A\rightarrow B}$. (This decomposition does not necessarily have to be such
that the vectors $\left\{  |\phi_{l}\rangle_{RB}\right\}  $ are orthonormal,
but keep in mind that there is always a choice such that $d\leq d_{A}d_{B}$.)
Consider by inspecting (\ref{eq:choi-matrix}) that%
\begin{equation}
\left(  \langle i|_{R}\otimes I_{B}\right)  \left(  \mathcal{N}_{A\rightarrow
B}\left(  \vert\Gamma\rangle\langle\Gamma\vert_{RA}\right)  \right)  \left(
\vert j\rangle_{R}\otimes I_{B}\right)  =\mathcal{N}_{A\rightarrow B}\left(
|i\rangle\langle j|\right)  .
\end{equation}

Now, consider that for any bipartite vector $|\phi\rangle_{RB}$, we can expand
it in terms of an orthonormal basis $\left\{  \vert j\rangle_{B}\right\}  $
and the basis $\left\{  \left\vert i\right\rangle _{R}\right\}  $ given above:%
\begin{equation}
|\phi\rangle_{RB}=\sum_{i=0}^{d_{A}-1}\sum_{j=0}^{d_{B}-1}\alpha_{ij}%
|i\rangle_{R}\otimes|j\rangle_{B}.
\end{equation}
Let $V_{A\rightarrow B}$ denote the following linear operator:%
\begin{equation}
V_{A\rightarrow B}\equiv\sum_{i=0}^{d_{A}-1}\sum_{j=0}^{d_{B}-1}\alpha
_{i,j}|j\rangle_{B}\langle i|_{A}, \label{eq-qt:V-lin-op}%
\end{equation}
where $\left\{  |i\rangle_{A}\right\}  $ is the orthonormal basis given above.
Then we see that%
\begin{align}
\left(  I_{R}\otimes V_{A\rightarrow B}\right)  \vert\Gamma\rangle_{RA}  &
=\sum_{i=0}^{d_{A}-1}\sum_{j=0}^{d_{B}-1}\alpha_{i,j}\left\vert j\right\rangle
_{B}\langle i|_{A}\sum_{k=0}^{d_{A}-1}|k\rangle_{R}\otimes|k\rangle
_{A}\label{eq-nqt:bipartite-vec-trick-1}\\
&  =\sum_{i=0}^{d_{A}-1}\sum_{j=0}^{d_{B}-1}\sum_{k=0}^{d_{A}-1}\alpha
_{i,j}|k\rangle_{R}\otimes|j\rangle_{B}\left\langle i|k\right\rangle _{A}\\
&  =\sum_{i=0}^{d_{A}-1}\sum_{j=0}^{d_{B}-1}\alpha_{ij}\left\vert
i\right\rangle _{R}\otimes|j\rangle_{B}\\
&  =|\phi\rangle_{RB}. \label{eq-nqt:bipartite-vec-trick-last}%
\end{align}
So this means that for all bipartite vectors $|\phi\rangle_{RB}$, we can find
a linear operator $V_{A\rightarrow B}$ such that $\left(  I_{R}\otimes
V_{A\rightarrow B}\right)  \vert\Gamma\rangle_{RA}=|\phi\rangle_{RB}$.
Consider also that%
\begin{align}
\langle i|_{R}|\phi\rangle_{RB}  &  =\langle i|_{R}\left(  I_{R}\otimes
V_{A\rightarrow B}\right)  \vert\Gamma\rangle_{RA}\\
&  =V_{A\rightarrow B}|i\rangle_{A}.
\end{align}
Applying this to our case of interest, for each $l$, we can write%
\begin{equation}
|\phi_{l}\rangle_{RB}=I_{R}\otimes\left(  V_{l}\right)  _{A\rightarrow B}%
\vert\Gamma\rangle_{RA},
\end{equation}
where $\left(  V_{l}\right)  _{A\rightarrow B}$ is some linear operator of the
form in (\ref{eq-qt:V-lin-op}). After making this observation, we realize that
it is possible to write%
\begin{align}
\mathcal{N}_{A\rightarrow B}\left(  |i\rangle\langle j|\right)   &  =\left(
\langle i|_{R}\otimes I_{B}\right)  \left(  \mathcal{N}_{A\rightarrow
B}\left(  \vert\Gamma\rangle\langle\Gamma\vert_{RA}\right)  \right)  \left(
|j\rangle_{R}\otimes I_{B}\right) \\
&  =\left(  \langle i|_{R}\otimes I_{B}\right)  \sum_{l=0}^{d-1}|\phi
_{l}\rangle\langle\phi_{l}|_{RB}\left(  |j\rangle_{R}\otimes I_{B}\right) \\
&  =\sum_{l=0}^{d-1}\left[  \left(  \langle i|_{R}\otimes I_{B}\right)
|\phi_{l}\rangle_{RB}\right]  \left[  \langle\phi_{l}|_{RB}\left(
|j\rangle_{R}\otimes I_{B}\right)  \right] \\
&  =\sum_{l=0}^{d-1}V_{l}|i\rangle\langle j|_{A}V_{l}^{\dag}.
\end{align}
By linearity of the map $\mathcal{N}_{A\rightarrow B}$, exploiting the above
result, and the development in \eqref{eq-nqt:linear-action-of-channel}, it
follows that the action of $\mathcal{N}_{A\rightarrow B}$ on any input
operator $X_{A}$ can be written as follows:%
\begin{equation}
\mathcal{N}_{A\rightarrow B}(X_{A})=\sum_{l=0}^{d-1}V_{l}X_{A}V_{l}^{\dag}.
\end{equation}

To prove the condition in \eqref{eq-nqt:trace-preserve-kraus}, let us begin by
exploiting the fact that the map $\mathcal{N}_{A\rightarrow B}$ is trace
preserving, so that%
\begin{equation}
\operatorname{Tr}\left\{  \mathcal{N}_{A\rightarrow B}\left(  \vert
i\rangle\langle j\vert_{A}\right)  \right\}  =\operatorname{Tr}\left\{
|i\rangle\langle j\vert_{A}\right\}  =\delta_{ij}.
\label{eq-qt:kraus-consistency}%
\end{equation}
for all operators $\left\{  |i\rangle\langle j\vert_{A}\right\}  _{i,j}$. But
consider also that%
\begin{align}
\operatorname{Tr}\left\{  \mathcal{N}_{A\rightarrow B}\left(  \vert
i\rangle\langle j\vert_{A}\right)  \right\}   &  =\operatorname{Tr}\left\{
\sum_{l}V_{l}\left(  |i\rangle\langle j\vert_{A}\right)  V_{l}^{\dag}\right\}
\\
&  =\operatorname{Tr}\left\{  \sum_{l}V_{l}^{\dag}V_{l}\left(  \vert
i\rangle\langle j\vert_{A}\right)  \right\} \\
&  =\langle j\vert_{A}\sum_{l}V_{l}^{\dag}V_{l}\left\vert i\right\rangle _{A}.
\end{align}
Thus, in order to have consistency with (\ref{eq-qt:kraus-consistency}), we
require that $\langle j\vert_{A}\sum_{l}V_{l}^{\dag}V_{l}\vert i\rangle
_{A}=\delta_{i,j}$, or equivalently, for \eqref{eq-nqt:trace-preserve-kraus}
to hold.
\end{proof}

\begin{remark}
If the decomposition in \eqref{eq:decomposition-choi} is a spectral
decomposition, then it follows that the Kraus operators $\left\{
V_{l}\right\}  $ are orthogonal with respect to the Hilbert--Schmidt inner
product:%
\begin{equation}
\operatorname{Tr}\left\{  V_{l}^{\dag}V_{k}\right\}  =\operatorname{Tr}%
\left\{  V_{l}^{\dag}V_{l}\right\}  \delta_{l,k}.
\end{equation}
This follows from the fact that%
\begin{align}
\delta_{l,k}\left\langle \phi_{l}|\phi_{l}\right\rangle  &  =\left\langle
\phi_{l}|\phi_{k}\right\rangle \\
&  =\langle\Gamma\vert_{RB}\left[  I_{R}\otimes\left(  V_{l}^{\dag}%
V_{k}\right)  _{B}\right]  \vert\Gamma\rangle_{RB}\\
&  =\operatorname{Tr}\left\{  V_{l}^{\dag}V_{k}\right\}  ,
\end{align}
where in the third line we have applied the result of
Exercise~\ref{ex-qt:alt-trace-max-ent}.
\end{remark}

\begin{exercise}
\label{ex-nqt:choi-CP}Prove that a linear map $\mathcal{N}$ is completely
positive if its corresponding Choi operator, as defined in
Definition~\ref{def-nqt:choi-op}, is a positive semi-definite operator. (Hint:
Use the fact that any positive semi-definite operator can be diagonalized, the
fact that $\operatorname{id}_{R}\otimes\mathcal{N}$ is linear, and use
something similar to \eqref{eq-nqt:bipartite-vec-trick-1}--\eqref{eq-nqt:bipartite-vec-trick-last}).
\end{exercise}

\subsection{Unique Specification of a Quantum Channel}

\label{sec-nqt:specify-channel-phi}We emphasize again that any linear map
$\mathcal{N}:\mathcal{L}(\mathcal{H}_{A})\rightarrow\mathcal{L}(\mathcal{H}%
_{B})$ is specified completely by its action $\mathcal{N}_{A\rightarrow
B}(|i\rangle\langle j\vert_{A})$\ on an operator of the form $|i\rangle\langle
j\vert_{A}$ where $\left\{  |i\rangle_{A}\right\}  $ is some orthonormal
basis. Thus, two linear maps $\mathcal{N}_{A\rightarrow B}$ and $\mathcal{M}%
_{A\rightarrow B}$ are equal if they have the same effect on all operators of
the form $|i\rangle\langle j\vert$:%
\begin{equation}
\mathcal{N}_{A\rightarrow B}=\mathcal{M}_{A\rightarrow B}\mathcal{\ \ \ \ }%
\Leftrightarrow\ \ \ \ \forall i,j\ \ \ \ \mathcal{N}_{A\rightarrow B}\left(
|i\rangle\langle j\vert_{A}\right)  =\mathcal{M}_{A\rightarrow B}\left(
|i\rangle\langle j\vert_{A}\right)  . \label{eq-qt:noisy-maps-equiv}%
\end{equation}

As a consequence, there is an interesting way to test whether two quantum
channels are equal to each other. Let us now consider a maximally entangled
qudit state $\left\vert \Phi\right\rangle _{RA}$ where%
\begin{equation}
\left\vert \Phi\right\rangle _{RA}=\frac{1}{\sqrt{d}}\sum_{i=0}^{d-1}%
|i\rangle_{R}|i\rangle_{A},
\end{equation}
and $d$ is the dimension of each system $R$ and $A$. The density operator
$\Phi_{RA}$ corresponding to $\left\vert \Phi\right\rangle _{RA}$ is as
follows:%
\begin{equation}
\Phi_{RA}=\frac{1}{d}\sum_{i,j=0}^{d-1}|i\rangle\langle j\vert_{R}%
\otimes|i\rangle\langle j\vert_{A}.
\end{equation}
Let us now send the $A$ system of $\Phi_{RA}$ through a quantum channel
$\mathcal{N}$:%
\begin{equation}
\left(  \operatorname{id}_{R}\otimes\mathcal{N}_{A\rightarrow B}\right)
\left(  \Phi_{RA}\right)  =\frac{1}{d}\sum_{i,j=0}^{d-1}|i\rangle\langle
j\vert_{R}\otimes\mathcal{N}_{A\rightarrow B}(\vert i\rangle\langle j\vert
_{A}). \label{eq-qt:max-ent-state-thru-N}%
\end{equation}

The resulting state completely characterizes the quantum channel $\mathcal{N}$
because the following map translates between the state in
\eqref{eq-qt:max-ent-state-thru-N} and the operators $\mathcal{N}%
_{A\rightarrow B}\left(  |i\rangle\langle j\vert_{A}\right)  $ in
\eqref{eq-qt:noisy-maps-equiv}:%
\begin{equation}
d\langle i\vert_{R}\left(  \operatorname{id}_{R}\otimes\mathcal{N}%
_{A\rightarrow B}\right)  \left(  \Phi_{RA}\right)  \vert j\rangle
_{R}=\mathcal{N}_{A\rightarrow B}\left(  |i\rangle\langle j\vert_{A}\right)  .
\end{equation}
Thus, we can completely characterize a quantum channel by determining the
quantum state resulting from sending one share of a maximally entangled state
through it, and the following condition is necessary and sufficient for any
two quantum channels to be equal:%
\begin{equation}
\mathcal{N}=\mathcal{M\ \ \ \ }\Leftrightarrow\ \ \ \ \left(
\operatorname{id}_{R}\otimes\mathcal{N}_{A\rightarrow B}\right)  \left(
\Phi_{RA}\right)  =\left(  \operatorname{id}_{R}\otimes\mathcal{M}%
_{A\rightarrow B}\right)  \left(  \Phi_{RA}\right)  .
\end{equation}
It is equivalent to the condition in \eqref{eq-qt:noisy-maps-equiv}.

\subsection{Serial Concatenation of Quantum Channels}

A quantum state may undergo not just one type of quantum evolution---it can of
course undergo one quantum channel followed by another quantum channel. Let
$\mathcal{N}:\mathcal{L}(\mathcal{H}_{A})\rightarrow\mathcal{L}(\mathcal{H}%
_{B})$ denote a first quantum channel and let $\mathcal{M}:\mathcal{L}%
(\mathcal{H}_{B})\rightarrow\mathcal{L}(\mathcal{H}_{C})$ denote a second
quantum channel. Suppose that the Kraus operators of $\mathcal{N}$ are
$\left\{  N_{k}\right\}  $ and the Kraus operators of $\mathcal{M}$ are
$\left\{  M_{k}\right\}  $. It is straightforward to define the serial
concatenation $\mathcal{M}_{B\rightarrow C}\circ\mathcal{N}_{A\rightarrow B}$
of these two quantum channels. Consider that the output of the first channel
is%
\begin{equation}
\mathcal{N}_{A\rightarrow B}(\rho_{A})\equiv\sum_{k}N_{k}\rho_{A}N_{k}^{\dag},
\end{equation}
for some input density operator $\rho_{A}\in\mathcal{D}(\mathcal{H}_{A})$. The
output of the serially concatenated channel $\mathcal{M}_{B\rightarrow C}%
\circ\mathcal{N}_{A\rightarrow B}$ is then%
\begin{equation}
\left(  \mathcal{M}_{B\rightarrow C}\circ\mathcal{N}_{A\rightarrow B}\right)
(\rho_{A})=\sum_{k}M_{k}\mathcal{N}_{A\rightarrow B}(\rho)M_{k}^{\dag}%
=\sum_{k,k^{\prime}}M_{k}N_{k^{\prime}}\rho_{A}N_{k^{\prime}}^{\dag}%
M_{k}^{\dag}.
\end{equation}
It is clear that the Kraus operators of the serially concatenated channel
$\mathcal{M}_{B\rightarrow C}\circ\mathcal{N}_{A\rightarrow B}$ are $\left\{
M_{k}N_{k^{\prime}}\right\}  _{k,k^{\prime}}$. Serial concatenation of
channels has an obvious generalization to a serial concatenation of more than
two channels.

\subsection{Parallel Concatenation of Quantum Channels}

We can also use two channels in parallel. That is, suppose that we send a
system $A$ through a channel $\mathcal{N}:\mathcal{L}(\mathcal{H}%
_{A})\rightarrow\mathcal{L}(\mathcal{H}_{C})$ and a system $B$ through a
channel $\mathcal{M}:\mathcal{L}(\mathcal{H}_{B})\rightarrow\mathcal{L}%
(\mathcal{H}_{D})$. Suppose further that the Kraus operators of $\mathcal{N}%
_{A\rightarrow C}$ are $\{N_{k}\}$ and those for $\mathcal{M}_{B\rightarrow
D}$ are $\{M_{k^{\prime}}\}$. Then the parallel concatenation of the two
channels is equal to the following serial concatenation:%
\begin{equation}
\mathcal{N}_{A\rightarrow C}\otimes\mathcal{M}_{B\rightarrow D}=(\mathcal{N}%
_{A\rightarrow C}\otimes\operatorname{id}_{D})(\operatorname{id}_{A}%
\otimes\mathcal{M}_{B\rightarrow D}),
\end{equation}
or equivalently%
\begin{equation}
\mathcal{N}_{A\rightarrow C}\otimes\mathcal{M}_{B\rightarrow D}%
=(\operatorname{id}_{C}\otimes\mathcal{M}_{B\rightarrow D})(\mathcal{N}%
_{A\rightarrow C}\otimes\operatorname{id}_{B}).
\end{equation}
Intuitively, if Alice is conducting a local action and Bob is as well, the
order in which they conduct their actions does not matter for determining the
final output state. We have already discussed that a set of Kraus operators
for $\mathcal{N}_{A\rightarrow C}\otimes\operatorname{id}_{D}$ is
$\{N_{k}\otimes I_{D}\}$ and a set for $\operatorname{id}_{A}\otimes
\mathcal{M}_{B\rightarrow D}$ is $\{I_{A}\otimes M_{k^{\prime}}\}$, so that it
is straightforward to verify that a set of Kraus operators for $\mathcal{N}%
_{A\rightarrow C}\otimes\mathcal{M}_{B\rightarrow D}$ is $\{N_{k}\otimes
M_{k^{\prime}}\}$. The parallel concatenated channel $\mathcal{N}%
_{A\rightarrow C}\otimes\mathcal{M}_{B\rightarrow D}$ thus has the following
action on an input density operator $\rho_{AB}\in\mathcal{D}(\mathcal{H}%
_{A}\otimes\mathcal{H}_{B})$:%
\begin{equation}
(\mathcal{N}_{A\rightarrow C}\otimes\mathcal{M}_{B\rightarrow D})(\rho
_{AB})=\sum_{k,k^{\prime}}\left(  N_{k}\otimes M_{k^{\prime}}\right)
(\rho_{AB})\left(  N_{k}\otimes M_{k^{\prime}}\right)  ^{\dag},
\end{equation}
Parallel concatenation of channels also has an obvious generalization to more
than two channels.

\subsection{Unital Maps and Adjoints of Quantum Channels}

\label{sec-nqt:adjoint-unital}Recall that the adjoint $G^{\dag}$\ of a linear
operator $G$\ is defined as the unique linear operator satisfying the
following set of equations:%
\begin{equation}
\left\langle y,Gx\right\rangle =\langle G^{\dag}y,x\rangle,
\label{eq-nqt:adjoint-vectors}%
\end{equation}
for all vectors $x$ and $y$, and with $\left\langle z,w\right\rangle =\sum
_{i}z_{i}^{\ast}w_{i}$ defined as the inner product between vectors $z$ and
$w$.

As an extension of this idea, we can define an inner product for operators:

\begin{definition}
[Hilbert--Schmidt Inner Product]The Hilbert--Schmidt inner product
\index{Hilbert--Schmidt inner product}%
between two operators $C,D\in\mathcal{L}(\mathcal{H})$ is defined as follows:%
\begin{equation}
\left\langle C,D\right\rangle \equiv\operatorname{Tr}\{C^{\dag}D\}.
\end{equation}

\end{definition}

This then allows us to define the
\index{adjoint}%
adjoint $\mathcal{N}^{\dag}$\ of a linear map $\mathcal{N}$ in a way similar
to \eqref{eq-nqt:adjoint-vectors}:

\begin{definition}
[Adjoint Map]\label{def-nqt:adjoint-map}Let $\mathcal{N}:\mathcal{L}%
(\mathcal{H}_{A})\rightarrow\mathcal{L}(\mathcal{H}_{B})$ be a linear map. The
adjoint $\mathcal{N}^{\dag}:\mathcal{L}(\mathcal{H}_{B})\rightarrow
\mathcal{L}(\mathcal{H}_{A})$\ of a linear map $\mathcal{N}$ is the unique
linear map satisfying the following set of equations:%
\begin{equation}
\left\langle Y,\mathcal{N}(X)\right\rangle =\langle\mathcal{N}^{\dag
}(Y),X\rangle,
\end{equation}
for all $X\in\mathcal{L}(\mathcal{H}_{A})$ and $Y\in\mathcal{L}(\mathcal{H}%
_{B})$.
\end{definition}

Another important class of linear maps are unital%
\index{unital map}
maps, defined as follows:

\begin{definition}
[Unital Map]\label{def-nqt:unital-map}A linear map $\mathcal{N}:\mathcal{L}%
(\mathcal{H}_{A})\rightarrow\mathcal{L}(\mathcal{H}_{B})$\ is unital if it
preserves the identity operator, in the sense that $\mathcal{N}(I_{A})=I_{B}$.
\end{definition}

Given the notion of an adjoint map, it is natural to inquire what is the
adjoint of a quantum channel, and furthermore, what is an interpretation of
it. So let us now suppose that $\mathcal{N}:\mathcal{L}(\mathcal{H}%
_{A})\rightarrow\mathcal{L}(\mathcal{H}_{B})$ is a quantum channel with a set
$\{V_{l}\}$\ of\ Kraus operators satisfying $\sum_{l}V_{l}^{\dag}V_{l}=I_{A}$.
Then we compute%
\begin{align}
\left\langle Y,\mathcal{N}(X)\right\rangle  &  =\operatorname{Tr}\left\{
Y^{\dag}\sum_{l}V_{l}XV_{l}^{\dag}\right\}  =\operatorname{Tr}\left\{
\sum_{l}V_{l}^{\dag}Y^{\dag}V_{l}X\right\} \\
&  =\operatorname{Tr}\left\{  \left(  \sum_{l}V_{l}^{\dag}YV_{l}\right)
^{\dag}X\right\}  =\left\langle \sum_{l}V_{l}^{\dag}YV_{l},X\right\rangle ,
\end{align}
where the second equality is from linearity and cyclicity of trace and the
last is from the definition of the Hilbert--Schmidt inner product. Thus, the
adjoint $\mathcal{N}^{\dag}$ of any quantum channel $\mathcal{N}$\ is given by%
\begin{equation}
\mathcal{N}^{\dag}(Y)=\sum_{l}V_{l}^{\dag}YV_{l}.
\label{eq-nqt:adjoint-map-kraus-ops}%
\end{equation}
The adjoint $\mathcal{N}^{\dag}$\ is completely positive, as one can verify by
applying Exercise~\ref{ex-nqt:choi-CP}. Furthermore, the adjoint
$\mathcal{N}^{\dag}$ is unital because%
\begin{equation}
\mathcal{N}^{\dag}(I_{B})=\sum_{l}V_{l}^{\dag}I_{B}V_{l}=\sum_{l}V_{l}^{\dag
}V_{l}=I_{A}.
\end{equation}
We summarize these results as follows:

\begin{proposition}
The adjoint $\mathcal{N}^{\dag}:\mathcal{L}(\mathcal{H}_{B})\rightarrow
\mathcal{L}(\mathcal{H}_{A})$ of a quantum channel $\mathcal{N}:\mathcal{L}%
(\mathcal{H}_{A})\rightarrow\mathcal{L}(\mathcal{H}_{B})$ is a completely
positive, unital map.
\end{proposition}

What is an interpretation of the adjoint of a quantum channel? It provides a
connection from the Schr\"{o}dinger picture of quantum physics, in which the
focus is on the evolution of states, to the Heisenberg picture, in which the
focus is on the evolution of observables or measurement operators. To see
this, let $\{\Lambda_{B}^{j}\}$ be a POVM, $\rho_{A}$ be a density operator,
and $\mathcal{N}:\mathcal{L}(\mathcal{H}_{A})\rightarrow\mathcal{L}%
(\mathcal{H}_{B})$ be a quantum channel. Suppose that we prepare the state
$\rho_{A}$, apply the channel $\mathcal{N}$, and then perform the measurement
$\{\Lambda_{B}^{j}\}$. The probability of getting outcome $j$ from the
measurement is given by the Born rule:%
\begin{equation}
p_{J}(j)=\operatorname{Tr}\{\Lambda_{B}^{j}\mathcal{N}(\rho_{A}%
)\}=\operatorname{Tr}\{\mathcal{N}^{\dag}(\Lambda_{B}^{j})\rho_{A}\},
\end{equation}
where the second equality follows because $\mathcal{N}^{\dag}$ is the adjoint
of $\mathcal{N}$. This latter expression is what corresponds to the Heisenberg
picture. Here, the interpretation is that each measurement operator
$\Lambda_{B}^{j}$\ \textquotedblleft evolves backwards\textquotedblright\ to
become $\mathcal{N}^{\dag}(\Lambda_{B}^{j})$ and then the measurement
$\{\mathcal{N}^{\dag}(\Lambda_{B}^{j})\}$ is performed on the state $\rho_{A}%
$. We should verify that the set $\{\mathcal{N}^{\dag}(\Lambda_{B}^{j})\}$
indeed constitutes a measurement. Consider that each $\mathcal{N}^{\dag
}(\Lambda_{B}^{j})$ is positive semi-definite, given that the adjoint is a
completely positive map, and that%
\begin{equation}
\sum_{j}\mathcal{N}^{\dag}(\Lambda_{B}^{j})=\mathcal{N}^{\dag}\left(  \sum
_{j}\Lambda_{B}^{j}\right)  =\mathcal{N}^{\dag}(I_{B})=I_{A},
\end{equation}
where the equalities are following because $\mathcal{N}^{\dag}$ is linear and
unital. The interpretation of the measurement $\{\mathcal{N}^{\dag}%
(\Lambda_{B}^{j})\}$ is that it is the physical procedure corresponding to
applying the channel $\mathcal{N}$ and then performing the measurement
$\{\Lambda_{B}^{j}\}$, which is of course a valid measurement procedure.

\section{Interpretations of Quantum Channels}

We now detail two interpretations of quantum channels that are consistent with
the Choi--Kraus theorem (Theorem~\ref{thm-nqt:choi-kraus-theorem}). The first
is that we can interpret the noise occurring in a quantum channel as the loss
of a measurement outcome, and the second is that we can interpret noise as
being due to a unitary interaction with an environment to which we do not have access.

\subsection{Noisy Evolution as the Loss of a Measurement Outcome}

\begin{figure}
[ptb]
\begin{center}
\includegraphics[
width=4.8456in
]%
{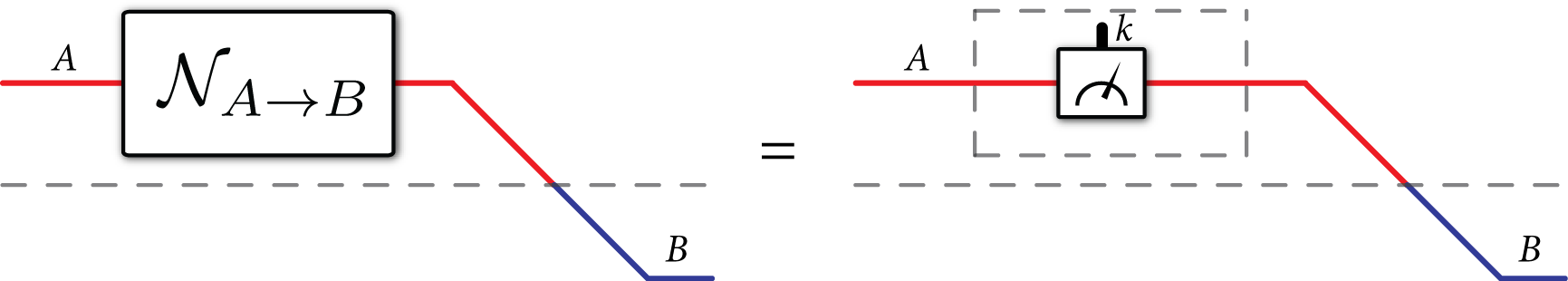}%
\caption{The diagram on the left depicts a quantum channel $\mathcal{N}%
_{A\rightarrow B}$ that takes a quantum system $A$ to a quantum system $B$.
This quantum channel has an interpretation in terms of the diagram on the
right, in which some third party performs a measurement on the input system
and does not inform the receiver of the measurement outcome.}%
\label{fig:noisy-quantum-channel}%
\end{center}
\end{figure}

We can interpret the noise resulting from a quantum channel as arising from
the loss of a measurement outcome (see Figure~\ref{fig:noisy-quantum-channel}%
). Suppose that the state of a system is described by a density operator
$\rho$ and that we then perform a measurement with a set $\left\{
M_{k}\right\}  $ of measurement operators for which $\sum_{k}M_{k}^{\dag}%
M_{k}=I$. The probability of obtaining outcome $k$ from the measurement is
given by the Born rule: $p_{K}(k)=\operatorname{Tr}\{M_{k}^{\dag}M_{k}\rho\}$,
and the post-measurement state is $M_{k}\rho M_{k}^{\dag}/p_{K}(k)$, as
discussed at the end of Section~\ref{sec-nqt:more-gen-meas}. Let us now
suppose that we lose track of the measurement outcome, or equivalently,
someone else measures the system and does not inform us of the measurement
outcome. The resulting ensemble description is then%
\begin{equation}
\left\{  p_{K}(k),M_{k}\rho M_{k}^{\dag}/p_{K}(k)\right\}  _{k}.
\end{equation}
The density operator corresponding to this ensemble is then%
\begin{equation}
\sum_{k}p_{K}(k)\frac{M_{k}\rho M_{k}^{\dag}}{p_{K}(k)}=\sum_{k}M_{k}\rho
M_{k}^{\dag}. \label{eq-qt:kraus-map}%
\end{equation}
We can thus write this evolution as a quantum channel $\mathcal{N}(\rho)$
where $\mathcal{N}(\rho)=\sum_{k}M_{k}\rho M_{k}^{\dag}$. The measurement
operators are playing the role of Kraus operators in this evolution.%

\subsection{Noisy Evolution from a Unitary Interaction}

There is another perspective on quantum noise that is helpful to consider. It
is equivalent to the perspective given in Chapter~\ref{chap:purified-q-t} when
we discuss isometric evolution. Suppose that a quantum system $A$ begins in
the state $\rho_{A}$ and that there is an environment system $E$ in a pure
state $|0\rangle_{E}$. So the initial state of the joint system $AE$ is
$\rho_{A}\otimes|0\rangle\langle0|_{E}.$ Suppose that these two systems
interact according to some unitary operator $U_{AE}$ acting on both systems
$A$ and $E$. If we only have access to the system $A$ after the interaction,
then we calculate the state $\sigma_{A}$\ of this system by taking the partial
trace over the environment $E$:%
\begin{equation}
\sigma_{A}=\operatorname{Tr}_{E}\left\{  U_{AE}\left(  \rho_{A}\otimes
|0\rangle\langle0|_{E}\right)  U_{AE}^{\dag}\right\}  .
\end{equation}
This evolution is equivalent to that of a completely positive,
trace-preserving map with Kraus operators $\{B_{i}\equiv\left(  I_{A}%
\otimes\langle i|_{E}\right)  U_{AE}\left(  I_{A}\otimes|0\rangle_{E}\right)
\}_{i}$. This follows because we can take the partial trace with respect to an
orthonormal basis $\{|i\rangle_{E}\}$\ for the environment:%
\begin{align}
&  \operatorname{Tr}_{E}\left\{  U_{AE}\left(  \rho_{A}\otimes|0\rangle
\langle0|_{E}\right)  U_{AE}^{\dag}\right\} \nonumber\\
&  =\sum_{i}\left(  I_{A}\otimes\langle i|_{E}\right)  U_{AE}\left(  \rho
_{A}\otimes|0\rangle\langle0|_{E}\right)  U_{AE}^{\dag}\left(  I_{A}%
\otimes|i\rangle_{E}\right) \\
&  =\sum_{i}\left(  I_{A}\otimes\langle i|_{E}\right)  U_{AE}\left(
I_{A}\otimes|0\rangle_{E}\right)  \left(  \rho_{A}\right)  \left(
I_{A}\otimes\langle0|_{E}\right)  U_{AE}^{\dag}\left(  I_{A}\otimes
|i\rangle_{E}\right) \\
&  =\sum_{i}B_{i}\rho_{A}B_{i}^{\dag}.
\end{align}
The first equality follows from Definition~\ref{def-nqt:partial-trace}\ for
partial trace. The second equality follows because%
\begin{equation}
\rho_{A}\otimes|0\rangle\langle0|_{E}=\left(  I_{A}\otimes|0\rangle
_{E}\right)  \left(  \rho_{A}\right)  \left(  I_{A}\otimes\langle
0|_{E}\right)  .
\end{equation}
That the operators $\left\{  B_{i}\right\}  $ are a legitimate set of Kraus
operators satisfying $\sum_{i}B_{i}^{\dag}B_{i}=I_{A}$ follows from the
unitarity of $U_{AE}$ and the orthonormality of the basis $\{\left\vert
i\right\rangle _{E}\}$:%
\begin{align}
\sum_{i}B_{i}^{\dag}B_{i}  &  =\sum_{i}\left(  I_{A}\otimes\langle
0|_{E}\right)  U_{AE}^{\dag}\left(  I_{A}\otimes|i\rangle_{E}\right)  \left(
I_{A}\otimes\langle i|_{E}\right)  U_{AE}\left(  I_{A}\otimes|0\rangle
_{E}\right) \\
&  =\left(  I_{A}\otimes\langle0|_{E}\right)  U_{AE}^{\dag}\left(
I_{A}\otimes\sum_{i}|i\rangle\langle i|_{E}\right)  U_{AE}\left(  I_{A}%
\otimes|0\rangle_{E}\right) \\
&  =\left(  I_{A}\otimes\langle0|_{E}\right)  U_{AE}^{\dag}U_{AE}\left(
I_{A}\otimes|0\rangle_{E}\right) \\
&  =\left(  I_{A}\otimes\langle0|_{E}\right)  I_{A}\otimes I_{E}\left(
I_{A}\otimes|0\rangle_{E}\right) \\
&  =I_{A}.
\end{align}

\section{Quantum Channels are All Encompassing}

In this section, we show how everything we have considered so far can be
viewed as a quantum channel. This includes physical evolutions as we have
discussed so far, but additionally (and perhaps surprisingly)\ density
operators, discarding of systems, and quantum measurements. From this
perspective, one could argue that that there really is just a single
underlying postulate of quantum physics, that everything we consider in the
theory is just a quantum channel of some sort.

\subsection{Preparation and Appending Channels}

The preparation of a system $A$ in a state $\rho_{A}\in\mathcal{D}%
(\mathcal{H}_{A})$ is a particular type of quantum channel, with trivial input
Hilbert space $\mathbb{C}$ and output Hilbert space $\mathcal{H}_{A}$. Let
$\rho_{A}=\sum_{x}p_{X}(x)|x\rangle\langle x|_{A}$ be a spectral decomposition
of $\rho_{A}$. Then the Kraus operators of this channel are $\{N_{x}%
\equiv\sqrt{p_{X}(x)}|x\rangle_{A}\}$, and we can easily verify that these are
legitimate Kraus operators by calculating%
\begin{equation}
\sum_{x}N_{x}^{\dag}N_{x}=\sum_{x}\left(  \sqrt{p_{X}(x)}\langle
x|_{A}\right)  \left(  \sqrt{p_{X}(x)}|x\rangle_{A}\right)  =\sum_{x}%
p_{X}(x)=1,
\end{equation}
so that the completeness relation holds, given that the number $1$ is the
identity for the trivial Hilbert space $\mathbb{C}$. Considering that the
number $1$ is also the only density operator in $\mathcal{D}(\mathbb{C})$, we
can view this channel as mapping the trivial density operator $1$ to a density
operator $\rho_{A}\in\mathcal{D}(\mathcal{H}_{A})$. It is thus a preparation channel.

\begin{definition}
[Preparation Channel]A preparation channel $\mathcal{P}_{A}\equiv
\mathcal{P}_{\mathbb{C}\rightarrow A}$\ prepares a quantum system $A$\ in a
given state $\rho_{A}\in\mathcal{D}(\mathcal{H}_{A})$.
\end{definition}

This leads to a related channel, called an appending channel:

\begin{definition}
[Appending Channel]An appending channel is the parallel concatenation of the
identity channel and a preparation channel.
\end{definition}

Thus, an appending channel has the following action on a system $B$ in the
state $\sigma_{B}$:%
\begin{equation}
\left(  \mathcal{P}_{A}\otimes\operatorname{id}_{B}\right)  (\sigma_{B}%
)=\rho_{A}\otimes\sigma_{B}.
\end{equation}
The Kraus operators of such an appending channel are then $\{\sqrt{p_{X}%
(x)}|x\rangle_{A}\otimes I_{B}\}$.

\subsection{Trace-out and Discarding Channels}

In some sense, the opposite of preparation is discarding. So suppose that we
completely discard the contents of a quantum system $A$. The channel that does
so is called a \textit{trace-out channel} $\operatorname{Tr}_{A}$, and its
action is to map any density operator $\rho_{A}\in\mathcal{D}(\mathcal{H}%
_{A})$ to the trivial density operator $1$. The Kraus operators of the
trace-out channel are $\{N_{x}\equiv\langle x|_{A}\}$, where $\{|x\rangle
_{A}\}$ is some orthonormal basis for the system $A$. These Kraus operators
satisfy the completeness relation because%
\begin{equation}
\sum_{x}N_{x}^{\dag}N_{x}=\sum_{x}|x\rangle\langle x|_{A}=I_{A}.
\end{equation}
This channel is in direct correspondence with the trace operation, given in
Definition~\ref{def-nqt:trace}.

Now suppose that we have two systems $A$ and $B$, and we would like to discard
system $A$ only. The channel that does so is a \textit{discarding channel},
which is the parallel concatenation of the trace-out channel
$\operatorname{Tr}_{A}$ and the identity channel $\operatorname{id}_{B}$. It
has the following action on a density operator $\rho_{AB}\in\mathcal{D}%
(\mathcal{H}_{A}\otimes\mathcal{H}_{B})$:%
\begin{equation}
\left(  \operatorname{Tr}_{A}\otimes\operatorname{id}_{B}\right)  (\rho
_{AB})=\sum_{x}\left(  \langle x|_{A}\otimes I_{B}\right)  \rho_{AB}\left(
|x\rangle_{A}\otimes I_{B}\right)  =\operatorname{Tr}_{A}\{\rho_{AB}\},
\end{equation}
where we have taken the Kraus operators of $\operatorname{Tr}_{A}%
\otimes\operatorname{id}_{B}$ to be $\{\langle x|_{A}\otimes I_{B}\}$.
Clearly, this channel is in direct correspondence with the partial trace
operation, given in Definition~\ref{def-nqt:partial-trace}.

\subsection{Unitary and Isometric Channels}

Unitary evolution is a special kind of quantum channel in which there is a
single Kraus operator $U\in\mathcal{L}(\mathcal{H})$, satisfying $UU^{\dag
}=U^{\dag}U=I_{\mathcal{H}}$. Unitary channels are thus completely positive,
trace-preserving, and unital. Let $\rho\in\mathcal{D}(\mathcal{H})$. Under the
action of a unitary channel $\mathcal{U}$, this state evolves as%
\begin{equation}
\mathcal{U}(\rho)=U\rho U^{\dag},
\end{equation}
where $\mathcal{U}(\rho)\in\mathcal{D}(\mathcal{H})$. Our convention
henceforth is to denote a unitary channel by $\mathcal{U}$ and a unitary
operator by~$U$.

There is a related, but more general kind of quantum channel called an
\textit{isometric} quantum
\index{isometric channel}%
channel. Before defining it, we need to define the notion of a linear isometry:

\begin{definition}
[Isometry]\label{def-pqt:isometry}%
\index{isometry}
Let $\mathcal{H}$ and $\mathcal{H}^{\prime}$ be Hilbert spaces such that
$\dim(\mathcal{H}) \leq\dim(\mathcal{H}^{\prime})$. An isometry $V$ is a
linear map from $\mathcal{H}$ to $\mathcal{H}^{\prime}$ such that $V^{\dag}V =
I_{\mathcal{H}}$. Equivalently, an isometry $V$ is a linear, norm-preserving
operator, in the sense that $\Vert\vert\psi\rangle\Vert_{2} = \Vert V
\vert\psi\rangle\Vert_{2}$ for all $\vert\psi\rangle\in\mathcal{H}$.
\end{definition}

An isometry is a generalization of a unitary, because it maps between spaces
of different dimensions and is thus generally rectangular and need not satisfy
$VV^{\dag}= I_{\mathcal{H}^{\prime}}$. Rather, it satisfies $VV^{\dag}=
\Pi_{\mathcal{H}^{\prime}}$, where $\Pi_{\mathcal{H}^{\prime}}$ is some
projection onto ${\mathcal{H}^{\prime}}$, because
\begin{equation}
(VV^{\dag})(VV^{\dag}) = V(V^{\dag}V)V^{\dag}= VI_{\mathcal{H}}V^{\dag}=
VV^{\dag}. \label{eq-pqt:isometry-projector}%
\end{equation}
In later chapters, we repeatedly use the notion of an isometry.

We can now define an isometric channel:

\begin{definition}
[Isometric Channel]A channel $\mathcal{V}:\mathcal{L}(\mathcal{H})
\to\mathcal{L}(\mathcal{H}^{\prime})$ is an isometric channel if there exists
a linear isometry $V: \mathcal{H} \to\mathcal{H}^{\prime}$ such that
\begin{equation}
\mathcal{V}(X) = V X V^{\dag},
\end{equation}
for $X \in\mathcal{L}(\mathcal{H})$.
\end{definition}

Isometric channels are completely positive and trace-preserving. Furthermore,
as in the case of unitary channels, there is just a single Kraus operator $V$
satisfying $V^{\dag}V = I_{\mathcal{H}}$.

\subsubsection{Reversing Unitary and Isometric Channels}

Suppose that we would like to reverse the action of a unitary channel
$\mathcal{U}$. It is easy to do so: the adjoint map $\mathcal{U}^{\dag}$ is a
unitary channel, and by performing it after $\mathcal{U}$, we get
\begin{equation}
(\mathcal{U}^{\dag}\circ\mathcal{U})(X) = U^{\dag}U X U^{\dag}U = X,
\end{equation}
for $X \in\mathcal{L}(\mathcal{H})$.

If we would like to reverse the action of an isometric channel $\mathcal{V}$,
we need to be a bit more careful. In this case, the adjoint map $\mathcal{V}%
^{\dag}$ is not a channel, because it is not trace-preserving. Consider that
\begin{align}
\operatorname{Tr}\{\mathcal{V}^{\dag}(Y)\}  &  =\operatorname{Tr}\{V^{\dag
}YV\}=\operatorname{Tr}\{VV^{\dag}Y\}\\
&  =\operatorname{Tr}\{\Pi_{\mathcal{H}^{\prime}}Y\}\leq\operatorname{Tr}%
\{Y\},
\end{align}
for $Y\in\mathcal{L}(\mathcal{H}^{\prime})$ and where the projection
$\Pi_{\mathcal{H}^{\prime}}\equiv VV^{\dag}$.

However, it is possible to construct a \textit{reversal channel} $\mathcal{R}$
for any isometric channel $\mathcal{V}$ in the following way:%
\begin{equation}
\mathcal{R}(Y)\equiv\mathcal{V}^{\dag}(Y)+\operatorname{Tr}\{\left(
I_{\mathcal{H}^{\prime}}-\Pi_{\mathcal{H}^{\prime}}\right)  Y\}\sigma,
\end{equation}
where $\sigma\in\mathcal{D}(\mathcal{H})$. One can verify that the map
$\mathcal{R}$ is completely positive, and it is trace-preserving because%
\begin{align}
\operatorname{Tr}\{\mathcal{R}(Y)\}  &  =\operatorname{Tr}\{\left[
\mathcal{V}^{\dag}(Y)+\operatorname{Tr}\{\left(  I_{\mathcal{H}^{\prime}}%
-\Pi_{\mathcal{H}^{\prime}}\right)  Y\}\sigma\right]  \}\\
&  =\operatorname{Tr}\{\mathcal{V}^{\dag}(Y)\}+\operatorname{Tr}\{\left(
I_{\mathcal{H}^{\prime}}-\Pi_{\mathcal{H}^{\prime}}\right)
Y\}\operatorname{Tr}\{\sigma\}\\
&  =\operatorname{Tr}\{\Pi_{\mathcal{H}^{\prime}}Y\}+\operatorname{Tr}%
\{\left(  I_{\mathcal{H}^{\prime}}-\Pi_{\mathcal{H}^{\prime}}\right)  Y\}\\
&  =\operatorname{Tr}\{Y\}.
\end{align}
Furthermore, it perfectly reverses the action of the isometric channel
$\mathcal{V}$ because%
\begin{align}
(\mathcal{R}\circ\mathcal{V})(X)  &  =\mathcal{V}^{\dag}(\mathcal{V}%
(X))+\operatorname{Tr}\{\left(  I_{\mathcal{H}^{\prime}}-\Pi_{\mathcal{H}%
^{\prime}}\right)  \mathcal{V}(X)\}\sigma\\
&  =V^{\dag}VXV^{\dag}V+\operatorname{Tr}\{\left(  I_{\mathcal{H}^{\prime}%
}-VV^{\dag}\right)  VXV^{\dag}\}\sigma\\
&  =X+\left[  \operatorname{Tr}\{VXV^{\dag}\}-\operatorname{Tr}\{VV^{\dag
}VXV^{\dag}\}\right]  \sigma\\
&  =X+\left[  \operatorname{Tr}\{V^{\dag}VX\}-\operatorname{Tr}\{V^{\dag
}VV^{\dag}VX\}\right]  \sigma\\
&  =X+\left[  \operatorname{Tr}\{X\}-\operatorname{Tr}\{X\}\right]  \sigma\\
&  =X,
\end{align}
for $X\in\mathcal{L}(\mathcal{H})$.

\subsection{Classical-to-Classical Channels}

\label{sec-nqt:c-to-c-channel}It is natural to expect that classical channels
are special
\index{classical-to-classical channel}%
cases of quantum channels, and indeed, this is the case. To see this, fix an
input probability distribution $p_{X}(x)$ and a classical channel
$p_{Y|X}(y|x)$. Fix an orthonormal basis $\{|x\rangle\}$ corresponding to the
input letters and an orthonormal basis $\{|y\rangle\}$\ corresponding to the
output letters. We can then encode the input probability distribution
$p_{X}(x)$ as a density operator $\rho$ of the following form:%
\begin{equation}
\rho=\sum_{x}p_{X}(x)|x\rangle\langle x|.
\end{equation}
Let $\mathcal{N}$\ be a quantum channel with the following Kraus operators%
\begin{equation}
\left\{  \sqrt{p_{Y|X}(y|x)}|y\rangle\langle x|\right\}  _{x,y}.
\end{equation}
(The fact that these are legitimate Kraus operators follows directly from the
fact that $p_{Y|X}(y|x)$ is a conditional probability distribution.)\ The
quantum channel then has the following action on the input $\rho$:%
\begin{align}
\mathcal{N}(\rho)  &  =\sum_{x,y}\sqrt{p_{Y|X}(y|x)}|y\rangle\langle x|\left(
\sum_{x^{\prime}}p_{X}(x^{\prime})\vert x^{\prime}\rangle\langle x^{\prime
}|\right)  \sqrt{p_{Y|X}(y|x)}|x\rangle\langle y|\\
&  =\sum_{x,y,x^{\prime}}p_{Y|X}(y|x)p_{X}(x^{\prime})\left\vert \left\langle
x^{\prime}|x\right\rangle \right\vert ^{2}|y\rangle\langle y|\\
&  =\sum_{x,y}p_{Y|X}(y|x)p_{X}(x)|y\rangle\langle y|\\
&  =\sum_{y}\left(  \sum_{x}p_{Y|X}(y|x)p_{X}(x)\right)  |y\rangle\langle y|.
\end{align}
Thus, the evolution is the same that a noisy classical channel $p_{Y|X}%
(y|x)$\ would enact on a probability distribution $p_{X}(x)$ by taking it to%
\begin{equation}
p_{Y}(y)=\sum_{x}p_{Y|X}(y|x)p_{X}(x)
\end{equation}
at the output.

Since a noiseless classical channel has $p_{Y|X}(y|x) = \delta_{x,y}$, we are
led to the following definition:

\begin{definition}
[Noiseless Classical Channel]\label{def-nqt:classical-channel-dephase}Let
$\{|x\rangle\}$ be an orthonormal basis for a Hilbert space $\mathcal{H}$. A
noiseless classical channel has the following action on a density operator
$\rho\in\mathcal{D}(\mathcal{H})$:
\begin{equation}
\rho\rightarrow\sum_{x}|x\rangle\langle x|\rho|x\rangle\langle x|.
\end{equation}
That is, it removes the off-diagonal elements of $\rho$ when represented as a
matrix with respect to the basis $\{|x\rangle\}$.
\end{definition}

\subsection{Classical-to-Quantum Channels}

\begin{figure}[ptb]
\begin{center}
\includegraphics[
width=3.8484in
]{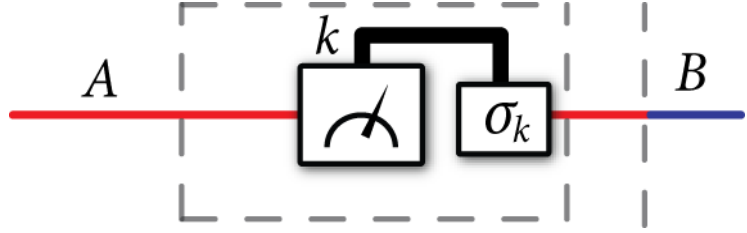}
\end{center}
\caption{This figure illustrates the internal workings of a classical--quantum
channel. It first measures the input state in some basis $\left\{
|k\rangle\right\}  $ and outputs a quantum state $\sigma_{k}$ conditioned on
the measurement outcome.}%
\label{fig-qt:cq-channel}%
\end{figure}\label{sec-nqt:-classical-to-quantum-channel}Classical-to-quantum
channels, or classical--quantum channels for short, are channels which take
classical systems to quantum systems. They thus go one step beyond both
classical-to-classical channels and preparation channels. More generally, they
make a given quantum system classical and then prepare a quantum state, as
discussed in the following definition:

\begin{definition}
[Classical--Quantum Channel]A classical--quantum channel first
\index{classical-to-quantum channel}%
measures the input state in a particular orthonormal basis and outputs a
density operator conditioned on the result of the measurement. Given an
orthonormal basis $\{|k\rangle_{A}\}$ and a set of states $\{\sigma_{B}^{k}%
\}$, each of which is in $\mathcal{D}(\mathcal{H}_{B})$, a classical--quantum
channel has the following action on an input density operator $\rho_{A}%
\in\mathcal{D}(\mathcal{H}_{A})$:%
\begin{equation}
\rho_{A}\rightarrow\sum_{k}\langle k|_{A}\rho_{A}\vert k\rangle_{A}\sigma
_{B}^{k}. \label{eq-nqt:cq-channel}%
\end{equation}

\end{definition}

Let us see how this comes about, using the definition above. The
classical--quantum channel first measures the input state $\rho_{A}$ in the
basis $\left\{  |k\rangle_{A}\right\}  $. Given that the result of the
measurement is $k$, the post measurement state is%
\begin{equation}
\frac{|k\rangle\langle k|\rho_{A}|k\rangle\langle k|}{\langle k|\rho
_{A}|k\rangle}.
\end{equation}
The channel then correlates a density operator $\sigma_{B}^{k}$ with the
post-measurement state$~k$:%
\begin{equation}
\frac{|k\rangle\langle k|\rho_{A}|k\rangle\langle k|}{\langle k|\rho
_{A}|k\rangle}\otimes\sigma_{B}^{k}.
\end{equation}
This action leads to an ensemble:%
\begin{equation}
\left\{  \langle k|\rho_{A}|k\rangle,\frac{|k\rangle\langle k|\rho
_{A}|k\rangle\langle k|}{\langle k|\rho_{A}|k\rangle}\otimes\sigma_{B}%
^{k}\right\}  ,
\end{equation}
and the density operator of the ensemble is%
\begin{equation}
\sum_{k}\langle k|\rho_{A}|k\rangle\frac{|k\rangle\langle k|\rho_{A}%
|k\rangle\langle k|}{\langle k|\rho_{A}|k\rangle}\otimes\sigma_{B}^{k}%
=\sum_{k}|k\rangle\langle k|\rho_{A}|k\rangle\langle k|\otimes\sigma_{B}^{k}.
\end{equation}
The channel then only outputs the system on the right (tracing out the first
system) so that the resulting channel is as given in \eqref{eq-nqt:cq-channel}.

\begin{exercise}
What is a set of Kraus operators for a classical--quantum channel?
\end{exercise}

\subsection{Quantum-to-Classical Channels (Measurement Channels)}

Quantum-to-classical, or quantum--classical channels for short, are
\index{quantum-to-classical channel}%
\index{measurement channel}%
in some sense the opposite of classical--quantum channels. They take a quantum
system to a classical one, and as such, they are in direct correspondence with
measurements. So sometimes they are referred to as measurement channels. They
also represent a way of generalizing classical channels different from
classical--quantum channels.

\begin{definition}
[Quantum--Classical Channels]Let $\{|x\rangle_{X}\}$ be an orthonormal basis
for a Hilbert space $\mathcal{H}_{X}$, and let $\{\Lambda_{A}^{x}\}$ be a
POVM\ acting on the system $A$. A quantum--classical channel has the following
action on an input density operator $\rho_{A}\in\mathcal{D}(\mathcal{H}_{A})$:%
\begin{equation}
\rho_{A}\rightarrow\sum_{x}\operatorname{Tr}\{\Lambda_{A}^{x}\rho
_{A}\}|x\rangle\langle x|_{X}. \label{eq-nqt:qc-channel}%
\end{equation}

\end{definition}

We should verify that this is indeed a quantum channel, by determining its
Kraus operators. Consider that the trace operation can be written as
$\operatorname{Tr}\{\cdot\}=\sum_{j}\langle j|_{A}\cdot|j\rangle_{A}$, where
$\{|j\rangle_{A}\}$ is some orthonormal basis for $\mathcal{H}_{A}$. Then we
can rewrite \eqref{eq-nqt:qc-channel} as%
\begin{align}
\sum_{x}\operatorname{Tr}\{\Lambda_{A}^{x}\rho_{A}\}|x\rangle\langle x|_{X}
&  =\sum_{x}\operatorname{Tr}\left\{  \sqrt{\Lambda_{A}^{x}}\rho_{A}%
\sqrt{\Lambda_{A}^{x}}\right\}  |x\rangle\langle x|_{X}\\
&  =\sum_{x,j}\langle j|_{A}\sqrt{\Lambda_{A}^{x}}\rho_{A}\sqrt{\Lambda
_{A}^{x}}|j\rangle_{A}|x\rangle\langle x|_{X}\\
&  =\sum_{x,j}|x\rangle_{X}\langle j|_{A}\sqrt{\Lambda_{A}^{x}}\rho_{A}%
\sqrt{\Lambda_{A}^{x}}|j\rangle_{A}\langle x|_{X}.
\end{align}
So this development reveals that a set of Kraus operators for the channel in
\eqref{eq-nqt:qc-channel} is $\{N_{x,j}\equiv|x\rangle_{X}\langle j|_{A}%
\sqrt{\Lambda_{A}^{x}}\}$. Let us verify the completeness relation for them:%
\begin{align}
\sum_{x,j}N_{x,j}^{\dag}N_{x,j}  &  =\sum_{x,j}\sqrt{\Lambda_{A}^{x}}%
|j\rangle_{A}\langle x|_{X}|x\rangle_{X}\langle j|_{A}\sqrt{\Lambda_{A}^{x}}\\
&  =\sum_{x,j}\sqrt{\Lambda_{A}^{x}}|j\rangle_{A}\langle j|_{A}\sqrt
{\Lambda_{A}^{x}}\\
&  =\sum_{x}\Lambda_{A}^{x}=I_{A},
\end{align}
where the last equality follows because $\{\Lambda_{A}^{x}\}$ is a POVM.

\subsection{Entanglement-Breaking Channels}%

\index{entanglement-breaking channel}%
\label{sec-nqt:entanglement-breaking}An important class of channels is the set
of
\index{entanglement-breaking channel}%
entanglement-breaking channels, and we will see that both quantum--classical
and classical--quantum channels are special cases of them.

\begin{definition}
[Entanglement-Breaking Channel]An
\index{entanglement-breaking channel}%
entanglement-breaking channel $\mathcal{N}^{\operatorname{EB}}:\mathcal{L}%
(\mathcal{H}_{A})\rightarrow\mathcal{L}(\mathcal{H}_{B})$ is defined by the
property that the channel $\operatorname{id}_{R}\otimes\mathcal{N}%
_{A\rightarrow B}^{\operatorname{EB}}$\ takes any state $\rho_{RA}$ to a
separable state, where $R$ is a reference system of arbitrary size.
\end{definition}

Fortunately, we do not need to check this property for all possible $\rho
_{RA}$. In fact, it suffices to check whether $\left(  \operatorname{id}%
_{R}\otimes\mathcal{N}_{A\rightarrow B}^{\operatorname{EB}}\right)  (\Phi
_{RA})$ is a separable state, where $\Phi_{RA}$ is a maximally entangled
state, as defined in \eqref{eq-qt:max-ent-state-basis-1}.

\begin{exercise}
Prove that a quantum channel $\mathcal{N}_{A\rightarrow B}$ is
\index{entanglement-breaking channel}%
entanglement-breaking if $\left(  \operatorname{id}_{R}\otimes\mathcal{N}%
_{A\rightarrow B}\right)  (\Phi_{RA})$ is a separable state, where $\Phi_{RA}$
is a maximally entangled state. (Hint:\ You can use a trick similar to that
which you used to solve Exercise~\ref{ex-nqt:choi-CP}. Alternatively, you can
inspect the proof of Theorem~\ref{thm-nqt:ent-break-unit-rank} below.)
\end{exercise}

\begin{exercise}
Show that both a classical--quantum channel and a quantum--classical channel
are
\index{entanglement-breaking channel}%
entanglement-breaking---i.e., if we input the $A$ system of a bipartite state
$\rho_{RA}$ to either of these channels, then the resulting state on systems
$RB$ is separable.
\end{exercise}

We can prove a more general structural theorem regarding entanglement-breaking
channels by exploiting its definition.

\begin{theorem}
\label{thm-nqt:ent-break-unit-rank}A channel is
\index{entanglement-breaking channel}%
entanglement-breaking if and only if it has a Kraus representation with Kraus
operators that are unit rank.
\end{theorem}

\begin{proof}
We first prove the \textquotedblleft if-part\textquotedblright\ of the
theorem. Suppose that the Kraus operators of a quantum channel $\mathcal{N}%
_{A\rightarrow B}$\ are%
\begin{equation}
\{N_{z}\equiv|\xi_{z}\rangle_{B}\langle\varphi_{z}|_{A}\}.
\label{eq-nqt:kraus-ops-ent-break}%
\end{equation}
Without loss of generality, we can take each $|\xi_{z}\rangle_{B}$ to be a
unit vector, simply by rescaling the corresponding $|\varphi_{z}\rangle_{A}$.
In order for this set to be a legitimate set of Kraus operators, the following
condition should hold%
\begin{equation}
I_{A}=\sum_{z}N_{z}^{\dag}N_{z}=\sum_{z}|\varphi_{z}\rangle_{A}\langle\xi
_{z}|_{B}|\xi_{z}\rangle_{B}\langle\varphi_{z}|_{A}=\sum_{z}|\varphi
_{z}\rangle\langle\varphi_{z}|_{A}. \label{eq-nqt:ent-break-meas}%
\end{equation}
Now consider when such a channel acts on one share of a general bipartite
state $\rho_{RA}\in\mathcal{D}(\mathcal{H}_{A}\otimes\mathcal{H}_{B})$:%
\begin{align}
(\operatorname{id}_{R}\otimes\mathcal{N}_{A\rightarrow B})(\rho_{RA})  &
=\sum_{z}\left(  I_{R}\otimes|\xi_{z}\rangle_{B}\langle\varphi_{z}%
|_{A}\right)  \rho_{RA}\left(  I_{R}\otimes|\varphi_{z}\rangle_{A}\langle
\xi_{z}|_{B}\right) \\
&  =\sum_{z}\operatorname{Tr}_{A}\{|\varphi_{z}\rangle\langle\varphi_{z}%
|_{A}\rho_{RA}\}\otimes|\xi_{z}\rangle\langle\xi_{z}|_{B}\\
&  =\sum_{z}p_{Z}(z)\rho_{R}^{z}\otimes|\xi_{z}\rangle\langle\xi_{z}|_{B},
\end{align}
where in the last line we define the state $\rho_{R}^{z}\equiv
\operatorname{Tr}_{A}\{|\varphi_{z}\rangle\langle\varphi_{z}|_{A}\rho
_{RA}\}/p_{Z}(z)$ and the probability distribution $p_{Z}$ from $p_{Z}%
(z)=\operatorname{Tr}\{|\varphi_{z}\rangle\langle\varphi_{z}|_{A}\rho_{RA}\}$
(the fact that $p_{Z}$ is a probability distribution follows from
\eqref{eq-nqt:ent-break-meas}). Consider now that the density operator in the
last line above is separable. Since $\rho_{RA}$ is arbitrary, the
\textquotedblleft if-part\textquotedblright\ of the theorem follows.

We now prove the \textquotedblleft only-if\textquotedblright\ part. Consider
that the output of an
\index{entanglement-breaking channel}%
entanglement-breaking channel $\mathcal{N}^{\operatorname{EB}}$\ acting on one
share of a maximally entangled state $\Phi_{RA}$ is as follows:%
\begin{equation}
\left(  \operatorname{id}_{R}\otimes\mathcal{N}_{A\rightarrow B}%
^{\operatorname{EB}}\right)  (\Phi_{RA})=\sum_{z}p_{Z}(z)|\phi_{z}%
\rangle\langle\phi_{z}|_{R}\otimes\vert\psi_{z}\rangle\langle\psi_{z}|_{B},
\end{equation}
where $p_{Z}$ is a probability distribution and $\{|\phi_{z}\rangle_{R}\}$ and
$\{|\psi_{z}\rangle_{B}\}$ are sets of pure states. This holds because the
output of a channel is a separable state (the channel \textquotedblleft
breaks\textquotedblright\ entanglement), and it is always possible to find a
representation of the separable state with pure states (see
Exercise~\ref{ex-nqt:separable-as-pure}). Now consider constructing a quantum
channel $\mathcal{M}$ with the following unit-rank Kraus operators:%
\begin{equation}
N_{z}\equiv\left\{  \sqrt{d\ p_{Z}(z)}|\psi_{z}\rangle_{B}\langle\phi
_{z}^{\ast}|_{A}\right\}  _{z},
\end{equation}
where $d$ is the Schmidt rank of the maximally entangled state $\Phi_{RA}%
$\ and$\ |\phi_{z}^{\ast}\rangle_{A}$ is the state $|\phi_{z}\rangle_{A}$ with
all of its elements conjugated with respect to the bases defined from
$\Phi_{RA}$. We should first verify that these Kraus operators form a valid
channel, by checking that $\sum_{z}N_{z}^{\dag}N_{z}=I_{A}$:%
\begin{align}
\sum_{z}N_{z}^{\dag}N_{z}  &  =\sum_{z}d\ p_{Z}(z)|\phi_{z}^{\ast}\rangle
_{A}\left\langle \psi_{z}|\psi_{z}\right\rangle _{B}\langle\phi_{z}^{\ast
}|_{A}\\
&  =d\ \sum_{z}p_{Z}(z)|\phi_{z}^{\ast}\rangle\langle\phi_{z}^{\ast}|_{A}.
\end{align}
Consider that%
\begin{align}
\operatorname{Tr}_{B}\left\{  \left(  \operatorname{id}_{R}\otimes
\mathcal{N}_{A\rightarrow B}^{\operatorname{EB}}\right)  (\Phi_{RA})\right\}
&  =\pi_{R}\\
&  =\operatorname{Tr}_{B}\left\{  \sum_{z}p_{Z}(z)|\phi_{z}\rangle\langle
\phi_{z}|_{R}\otimes\vert\psi_{z}\rangle\langle\psi_{z}|_{B}\right\} \\
&  =\sum_{z}p_{Z}(z)|\phi_{z}\rangle\langle\phi_{z}|_{R},
\end{align}
where $\pi_{R}$ is the maximally mixed state. Thus, it follows that
$\mathcal{M}$ is a valid quantum channel because%
\begin{align}
d\ \sum_{z}p_{Z}(z)|\phi_{z}\rangle\langle\phi_{z}|_{R}  &  =d\ \pi_{R}%
=I_{R}=\left(  I_{A}\right)  ^{\ast}\\
&  =d\ \sum_{z}p_{Z}(z)|\phi_{z}^{\ast}\rangle\langle\phi_{z}^{\ast}|_{A}%
=\sum_{z}N_{z}^{\dag}N_{z}.
\end{align}
Now let us consider the action of the channel$~\mathcal{M}$ on the maximally
entangled state:%
\begin{align}
&  \left(  \operatorname{id}_{R}\otimes\mathcal{M}_{A\rightarrow B}\right)
(\Phi_{RA})\\
&  =\frac{1}{d}\sum_{z,i,j}|i\rangle\langle j|_{R}\otimes\sqrt{d\ p_{Z}%
(z)}|\psi_{z}\rangle_{B}\langle\phi_{z}^{\ast}|_{A}|i\rangle\langle
j|_{A}|\phi_{z}^{\ast}\rangle_{A}\langle\psi_{z}|_{B}\sqrt{d\ p_{Z}(z)}\\
&  =\sum_{z,i,j}p_{Z}(z)\ |i\rangle\langle j|_{R}\otimes\left\langle \phi
_{z}^{\ast}|i\right\rangle \left\langle j|\phi_{z}^{\ast}\right\rangle
\ |\psi_{z}\rangle\langle\psi_{z}|_{B}\\
&  =\sum_{z,i,j}p_{Z}(z)\ |i\rangle\left\langle j|\phi_{z}^{\ast}\right\rangle
\left\langle \phi_{z}^{\ast}|i\right\rangle \langle j|_{R}\otimes\ |\psi
_{z}\rangle\langle\psi_{z}|_{B}\\
&  =\sum_{z}p_{Z}(z)\ |\phi_{z}\rangle\langle\phi_{z} |_{R}\otimes\ |\psi
_{z}\rangle\langle\psi_{z}|_{B}.
\end{align}
The last equality follows from recognizing $\sum_{i,j}\left\vert
i\right\rangle \langle j|\cdot|i\rangle\langle j\vert$ as the transpose
operation (with respect to the bases from $\Phi_{RA}$) and noting that the
transpose is equivalent to conjugation for a Hermitian operator $|\phi
_{z}\rangle\langle\phi_{z}|$. Finally, since the action of both $\mathcal{N}%
_{A\rightarrow B}^{\operatorname{EB}}$ and $\mathcal{M}_{A\rightarrow B}$ on
the maximally entangled state is the same, we can conclude that the two
channels are equal (see Section~\ref{sec-nqt:specify-channel-phi}). Thus,
$\mathcal{M}$ is a representation of the channel with unit-rank Kraus operators.
\end{proof}

The proof of the above theorem leads to the following important corollary:

\begin{corollary}
An
\index{entanglement-breaking channel}%
entanglement-breaking channel $\mathcal{N}_{A\rightarrow B}^{\operatorname{EB}%
}$ is a serial concatenation of a quantum--classical channel $\mathcal{M}%
_{A\rightarrow Z}$ with a classical--quantum channel $\mathcal{P}%
_{Z\rightarrow B}$, i.e., $\mathcal{N}_{A\rightarrow B}^{\operatorname{EB}%
}=\mathcal{P}_{Z\rightarrow B}\circ\mathcal{M}_{A\rightarrow Z}$. That is,
every entanglement-breaking channel can be written as a measurement followed
by a preparation.
\end{corollary}

\begin{proof}
Due to the above theorem, we can take the Kraus operators for an
entanglement-breaking channel $\mathcal{N}_{A\rightarrow B}^{\operatorname{EB}%
}$\ to be as in \eqref{eq-nqt:kraus-ops-ent-break}, with $\{|\xi_{z}%
\rangle_{B}\}$ a set of unit vectors and $\{|\varphi_{z}\rangle_{A}\}$
satisfying \eqref{eq-nqt:ent-break-meas}. Let $\{|z\rangle_{Z}\}$ be an
orthonormal basis for a Hilbert space $\mathcal{H}_{Z}$. Then take the
quantum--classical channel $\mathcal{M}_{A\rightarrow Z}$ to be%
\begin{equation}
\mathcal{M}_{A\rightarrow Z}(\rho_{A})=\sum_{z}\operatorname{Tr}\{|\varphi
_{z}\rangle\langle\varphi_{z}|_{A}\rho_{A}\}|z\rangle\langle z|_{Z}%
\end{equation}
and the classical--quantum channel $\mathcal{P}_{Z\rightarrow B}$\ to be%
\begin{equation}
\mathcal{P}_{Z\rightarrow B}(\sigma_{Z})=\sum_{z}\langle z|\sigma_{Z}%
|z\rangle_{Z}\ |\xi_{z}\rangle\langle\xi_{z}|_{B}.
\end{equation}
One can then verify that $\mathcal{N}_{A\rightarrow B}^{\operatorname{EB}%
}=\mathcal{P}_{Z\rightarrow B}\circ\mathcal{M}_{A\rightarrow Z}$.
\end{proof}

\subsection{Quantum Instruments}

\label{sec-pqt:instrument}The description of a quantum channel with Kraus
operators gives the most general evolution that a quantum state can undergo.
We may want to specialize this definition somewhat for another scenario.
Suppose that we would like to determine the most general evolution where the
input is a quantum system and the output consists of both a quantum system and
a classical system. Such a scenario may arise in a case where Alice is trying
to transmit both classical and quantum information, and Bob exploits a quantum
instrument to decode both kinds of information. A \textit{quantum instrument}
gives such an evolution with a hybrid output.

\begin{definition}
[Trace Non-Increasing Map]A linear map $\mathcal{M}$ is trace non-increasing
if $\operatorname{Tr}\{\mathcal{M}(X)\}\leq\operatorname{Tr}\{X\}$ for all
positive semi-definite $X\in\mathcal{L}(\mathcal{H})$, with $\mathcal{H}$ a
Hilbert space.
\end{definition}

\begin{definition}
[Quantum Instrument]A quantum instrument consists of a collection
$\{\mathcal{E}_{j}\}$\ of completely positive, trace non-increasing maps such
that the sum map $\sum_{j}\mathcal{E}_{j}$ is trace preserving. Let
$\{|j\rangle\}$\ be an orthonormal basis for a Hilbert space $\mathcal{H}_{J}%
$. The action of a quantum instrument on a density operator $\rho
\in\mathcal{D}(\mathcal{H})$ is the following quantum channel, which features
a quantum and classical output:%
\begin{equation}
\rho\rightarrow\sum_{j}\mathcal{E}_{j}(\rho)\otimes|j\rangle\langle j|_{J}.
\end{equation}

\end{definition}

\begin{figure}
[ptb]
\begin{center}
\includegraphics[
width=4.8456in
]%
{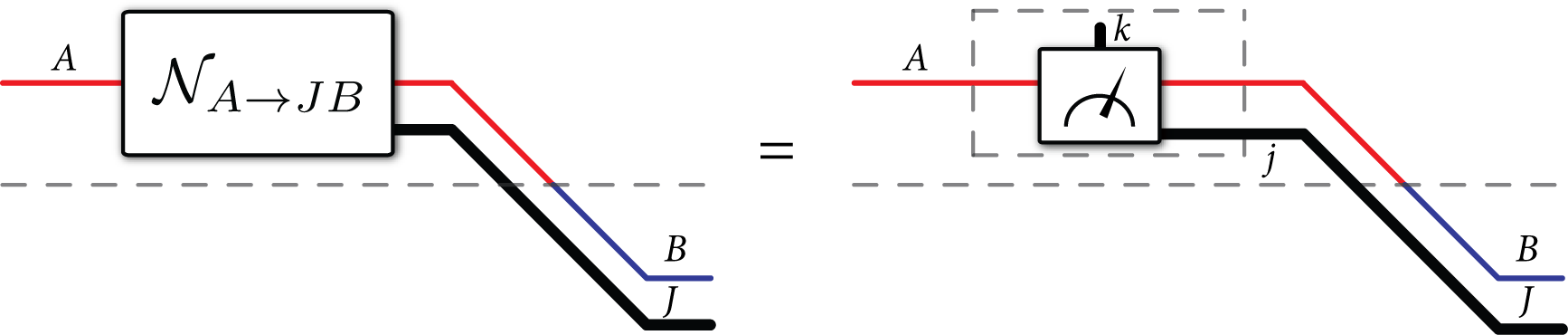}%
\caption{The figure on the left illustrates a quantum instrument, a general
noisy evolution that produces both a quantum and classical output. The figure
on the right illustrates the internal workings of a quantum instrument,
showing that it results from having only partial access to a measurement
outcome.}%
\label{fig-qt:instrument}%
\end{center}
\end{figure}

Let us see one way in which this definition comes about. Recall that we may
view a noisy quantum channel as arising from the forgetting of a measurement
outcome, as in \eqref{eq-qt:kraus-map}. Let us now suppose that some third
party performs a measurement with two outcomes $j$ and $k$, but does not give
us access to the measurement outcome $j$. Suppose that the measurement
operators for this two-outcome measurement are $\left\{  M_{j,k}\right\}
_{j,k}$. Let us first suppose that the third party performs the measurement on
a quantum system with density operator $\rho$ and gives us both of the
measurement outcomes. The post-measurement state in such a scenario is%
\begin{equation}
\frac{M_{j,k}\rho M_{j,k}^{\dag}}{p_{J,K}(j,k)},
\end{equation}
where the joint distribution of outcomes $j$ and $k$ is%
\begin{equation}
p_{J,K}(j,k)=\operatorname{Tr}\{M_{j,k}^{\dag}M_{j,k}\rho\}.
\end{equation}
We can calculate the marginal distributions $p_{J}(j)$ and $p_{K}(k)$
according to the law of total probability:%
\begin{align}
p_{J}(j)  &  =\sum_{k}p_{J,K}(j,k)=\sum_{k}\operatorname{Tr}\{M_{j,k}^{\dag
}M_{j,k}\rho\},\label{eq-qt-marginal-prob-instrument}\\
p_{K}(k)  &  =\sum_{j}p_{J,K}(j,k)=\sum_{j}\operatorname{Tr}\{M_{j,k}^{\dag
}M_{j,k}\rho\}.
\end{align}
Suppose the measuring device also places the classical outcomes in classical
registers $J$ and $K$, so that the post-measurement state is%
\begin{equation}
\frac{M_{j,k}\rho M_{j,k}^{\dag}}{p_{J,K}(j,k)}\otimes|j\rangle\langle
j|_{J}\otimes|k\rangle\langle k|_{K},
\end{equation}
where the sets $\left\{  |j\rangle\right\}  $ and $\left\{  |k\rangle\right\}
$ form respective orthonormal bases. Such an operation is possible physically,
and we could retrieve the classical information at some later point by
performing a complete projective measurement of the registers $J$ and $K$. If
we would like to determine the Kraus map for the overall quantum evolution, we
simply take the expectation over all measurement outcomes $j$ and $k$:%
\begin{multline}
\sum_{j,k}p_{J,K}(j,k)\left(  \frac{M_{j,k}\rho M_{j,k}^{\dag}}{p_{J,K}%
(j,k)}\right)  \otimes|j\rangle\langle j|_{J}\otimes|k\rangle\langle k|_{K}\\
=\sum_{j,k}M_{j,k}\rho M_{j,k}^{\dag}\otimes|j\rangle\langle j|_{J}%
\otimes|k\rangle\langle k|_{K}.
\end{multline}
Let us now suppose that we do not have access to the measurement result $k$.
This lack of access is equivalent to lacking access to classical register $K$.
To determine the resulting state, we should trace out the classical register
$K$. Our map then becomes%
\begin{equation}
\sum_{j,k}M_{j,k}\rho M_{j,k}^{\dag}\otimes|j\rangle\langle j|_{J}.
\end{equation}
The above map corresponds to a
\index{quantum instrument}%
quantum instrument, and is a general noisy quantum evolution that produces
both a quantum output and a classical output. Figure~\ref{fig-qt:instrument}%
\ depicts a quantum instrument.%

We can rewrite the above map more explicitly as follows:%
\begin{equation}
\sum_{j}\left(  \sum_{k}M_{j,k}\rho M_{j,k}^{\dag}\right)  \otimes
|j\rangle\langle j|_{J}=\sum_{j}\mathcal{E}_{j}( \rho) \otimes|j\rangle\langle
j|_{J},
\end{equation}
where we define%
\begin{equation}
\mathcal{E}_{j}(\rho)\equiv\sum_{k}M_{j,k}\rho M_{j,k}^{\dag}.
\end{equation}
Each $j$-dependent map $\mathcal{E}_{j}(\rho)$ is a completely positive
trace-non-increasing map because $\operatorname{Tr}\left\{  \mathcal{E}%
_{j}(\rho)\right\}  \leq1$. In fact, by examining the definition of
$\mathcal{E}_{j}(\rho)$ and comparing to
\eqref{eq-qt-marginal-prob-instrument}, it holds that%
\begin{equation}
\operatorname{Tr}\left\{  \mathcal{E}_{j}(\rho)\right\}  =p_{J}(j).
\end{equation}
It is important to note that the probability $p_{J}(j)$ is dependent on the
density operator $\rho$ that is input to the instrument. We can determine the
quantum output of the instrument by tracing over the classical register $J$.
The resulting quantum output is then%
\begin{equation}
\operatorname{Tr}_{J}\left\{  \sum_{j}\mathcal{E}_{j}(\rho)\otimes
|j\rangle\langle j|_{J}\right\}  =\sum_{j}\mathcal{E}_{j}(\rho).
\end{equation}
The above \textquotedblleft sum map\textquotedblright\ is a trace-preserving
map because%
\begin{equation}
\operatorname{Tr}\left\{  \sum_{j}\mathcal{E}_{j}(\rho)\right\}  =\sum
_{j}\operatorname{Tr}\left\{  \mathcal{E}_{j}( \rho) \right\}  =\sum_{j}%
p_{J}(j)=1,
\end{equation}
where the last equality follows because the marginal probabilities $p_{J}(j)$
sum to one. The above points that we have mentioned are the most salient for
the quantum instrument. We will exploit this type of evolution when we require
a device that outputs both a classical and quantum system.

We should stress that a quantum instrument is more general than applying a
mixture of CPTP\ maps to a quantum state. Suppose that we apply a mixture
$\left\{  \mathcal{N}_{j}\right\}  $\ of CPTP\ maps to a quantum state $\rho$,
chosen according to a distribution $p_{J}(j)$. The resulting expected state is
as follows:%
\begin{equation}
\sum_{j}p_{J}(j)|j\rangle\langle j|_{J}\otimes\mathcal{N}_{j}(\rho).
\end{equation}
The probabilities $p_{J}(j)$\ here are independent of the state $\rho$ that is
input to the mixture of CPTP\ maps, but this is not generally the case for a
quantum instrument. There, the probabilities $p_{J}( j) $ can depend on the
state $\rho$ that is input---it may be beneficial then to write these
probabilities as $p_{J}( j|\rho) $ because there is an implicit conditioning
on the state that is input to the instrument.

\section{Examples of Quantum Channels}

This section discusses some of the most important examples of quantum channels
that we will consider in this book. Throughout, we will be considering the
information-carrying ability of these various channels. They will provide some
useful, \textquotedblleft hands on\textquotedblright\ insight into quantum
Shannon theory.

\subsection{Noisy Evolution from a Random Unitary}

\label{sec-qt:noise-random-unitary}Perhaps the simplest example of a quantum
channel is the quantum bit-flip channel, which has the following action on a
qubit density operator $\rho$:%
\begin{equation}
pX\rho X^{\dag}+( 1-p) \rho. \label{eq-qt:bit-flip-channel}%
\end{equation}
The above density operator is more \textquotedblleft mixed\textquotedblright%
\ than the original density operator and we will make this statement more
precise in Chapter~\ref{chap:info-entropy}, when we study entropy. The
evolution $\rho\rightarrow pX\rho X^{\dag}+( 1-p) \rho$ is clearly a
legitimate quantum channel. Here the Kraus operators are $\{\sqrt{p}%
X,\sqrt{1-p}I\}$ and it is clear that they satisfy the completeness relation.

A generalization of the above discussion is to consider some ensemble of
unitaries (a random unitary) $\left\{  p(k),U_{k}\right\}  $ that we can apply
to a density operator $\rho$, resulting in the following output density
operator:%
\begin{equation}
\sum_{k}p(k)U_{k}\rho U_{k}^{\dag}.
\end{equation}

\subsection{Dephasing Channels}

We have already given the example of a noisy quantum bit-flip channel in
Section~\ref{sec-qt:noise-random-unitary}. Another important example is a bit
flip in the conjugate basis, or equivalently, a \textit{phase-flip channel}.
This channel acts as follows on any given density operator:%
\begin{equation}
\rho\rightarrow( 1-p) \rho+pZ\rho Z. \label{eq-qt:dephasing-channel}%
\end{equation}
It is also known as
\index{dephasing channel}%
a \textit{dephasing channel}.

For $p=1/2$, the action of the dephasing channel on a given quantum state is
equivalent to the action of measuring the qubit in the computational basis and
forgetting the result of the measurement. We make this idea more clear with an
example. First, suppose that we have a qubit%
\begin{equation}
|\psi\rangle=\alpha|0\rangle+\beta|1\rangle,
\end{equation}
and we measure it in the computational basis. Then the postulates of quantum
theory state that the qubit becomes $|0\rangle$ with probability $\left\vert
\alpha\right\vert ^{2}$ and it becomes $|1\rangle$ with probability
$\left\vert \beta\right\vert ^{2}$. Suppose that we forget the measurement
outcome, or alternatively, that we do not have access to it. Then our best
description of the qubit is with the following ensemble:%
\begin{equation}
\left\{  \left\{  \left\vert \alpha\right\vert ^{2},|0\rangle\right\}
,\left\{  \left\vert \beta\right\vert ^{2},|1\rangle\right\}  \right\}  .
\end{equation}
The density operator of this ensemble is%
\begin{equation}
\left\vert \alpha\right\vert ^{2}|0\rangle\langle0|+\left\vert \beta
\right\vert ^{2}|1\rangle\langle1|.
\end{equation}

Now let us check if the dephasing channel gives the same behavior as the
forgetful measurement above. We can consider the qubit as being an ensemble
$\left\{  1,|\psi\rangle\right\}  $, i.e., the state is certain to be
$|\psi\rangle$. The density operator of the ensemble is then $\rho$ where%
\begin{equation}
\rho=\left\vert \alpha\right\vert ^{2}|0\rangle\langle0|+\alpha\beta^{\ast
}|0\rangle\langle1|+\alpha^{\ast}\beta|1\rangle\langle0|+\left\vert
\beta\right\vert ^{2}|1\rangle\langle1|.
\end{equation}
If we act on the density operator $\rho$ with the dephasing channel with
$p=1/2$, then it preserves the density operator with probability $1/2$ and
phase flips the qubit with probability $1/2$:%
\begin{align}
&  \frac{1}{2}\rho+\frac{1}{2}Z\rho Z\nonumber\\
&  =\frac{1}{2}\left(  \left\vert \alpha\right\vert ^{2}|0\rangle
\langle0|+\alpha\beta^{\ast}|0\rangle\langle1|+\alpha^{\ast}\beta
|1\rangle\langle0|+\left\vert \beta\right\vert ^{2}|1\rangle\langle1|\right)
+\nonumber\\
&  \ \ \ \ \ \ \frac{1}{2}\left(  \left\vert \alpha\right\vert ^{2}%
|0\rangle\langle0|-\alpha\beta^{\ast}|0\rangle\langle1|-\alpha^{\ast}%
\beta|1\rangle\langle0|+\left\vert \beta\right\vert ^{2}|1\rangle
\langle1|\right) \\
&  =\left\vert \alpha\right\vert ^{2}|0\rangle\langle0|+\left\vert
\beta\right\vert ^{2}|1\rangle\langle1|.
\end{align}
The dephasing channel eliminates the off-diagonal terms of the density
operator when represented with respect to the computational basis. The
resulting density operator description is the same as what we found for the
forgetful measurement. It is also equivalent to a classical channel, as given
in Definition~\ref{def-nqt:classical-channel-dephase}.

\begin{exercise}
Verify that the action of the dephasing channel on the Bloch vector is%
\begin{multline}
\frac{1}{2}\left(  I+r_{x}X+r_{y}Y+r_{z}Z\right)  \rightarrow\\
\frac{1}{2}\left(  I+( 1-2p) r_{x}X+( 1-2p) r_{y}Y+r_{z}Z\right)  ,
\end{multline}
so that the channel preserves any component of the Bloch vector in the $Z$
direction, while shrinking any component in the $X$ or $Y$ direction.
\end{exercise}

\subsection{Pauli Channels}

A Pauli channel%
\index{Pauli channel}
is a generalization of the above dephasing channel and the bit-flip channel.
It simply applies a random Pauli operator according to a probability
distribution. The map for a qubit Pauli channel is%
\begin{equation}
\rho\rightarrow\sum_{i,j=0}^{1}p( i,j) Z^{i}X^{j}\rho X^{j}Z^{i}.
\end{equation}
The generalization of this channel to qudits is straightforward. We simply
replace the Pauli operators with the Heisenberg--Weyl operators. The Pauli
qudit channel is%
\begin{equation}
\rho\rightarrow\sum_{i,j=0}^{d-1}p( i,j) Z( i) X( j) \rho X^{\dag}( j)
Z^{\dag}( i) .
\end{equation}
These channels have been prominent in the study of quantum key distribution.

\begin{exercise}
We can write a Pauli channel as%
\begin{equation}
\rho\rightarrow p_{I}\rho+p_{X}X\rho X+p_{Y}Y\rho Y+p_{Z}Z\rho Z.
\end{equation}
Verify that the action of the Pauli channel on the Bloch vector is%
\begin{multline}
\left(  r_{x},r_{y},r_{z}\right)  \rightarrow\\
\left(  \left(  p_{I}+p_{X}-p_{Y}-p_{Z}\right)  r_{x},\ \left(  p_{I}%
+p_{Y}-p_{X}-p_{Z}\right)  r_{y},\ \left(  p_{I}+p_{Z}-p_{X}-p_{Y}\right)
r_{z}\right)  .
\end{multline}

\end{exercise}

\subsection{Depolarizing Channels}

\label{sec-nqt:depolarizing}The depolarizing channel%
\index{depolarizing channel}
is a \textquotedblleft worst-case scenario\textquotedblright\ channel. It
assumes that we completely lose the input qubit with some probability, i.e.,
it replaces the lost qubit with the maximally mixed state. The map for the
depolarizing channel is%
\begin{equation}
\rho\rightarrow( 1-p) \rho+p\pi, \label{eq-qt:depolarizing}%
\end{equation}
where $\pi$ is the maximally mixed state:\ $\pi=I/2$.

Most of the time, this channel is too pessimistic. Usually, we can learn
something about the physical nature of the channel by some estimation process.
We should only consider using the depolarizing channel as a model if we have
little to no information about the actual physical channel.

\begin{exercise}
[Pauli Twirl]Show that randomly applying the Pauli operators $I$, $X$, $Y$,
$Z$ with uniform probability to any density operator gives the maximally mixed
state:%
\begin{equation}
\frac{1}{4}\rho+\frac{1}{4}X\rho X+\frac{1}{4}Y\rho Y+\frac{1}{4}Z\rho Z=\pi.
\end{equation}
(Hint:\ Represent the density operator as $\rho=\left(  I+r_{x}X+r_{y}%
Y+r_{z}Z\right)  /2$ and apply the commutation rules of the Pauli operators.)
This is known as the \textquotedblleft twirling\textquotedblright\
\index{twirling}%
operation.
\end{exercise}

\begin{exercise}
Show that we can rewrite the depolarizing channel as the following Pauli
channel:%
\begin{equation}
\rho\rightarrow\left(  1-3p/4\right)  \rho+p\left(  \frac{1}{4}X\rho
X+\frac{1}{4}Y\rho Y+\frac{1}{4}Z\rho Z\right)  .
\end{equation}

\end{exercise}

\begin{exercise}
Show that the action of a depolarizing channel on the Bloch vector is%
\begin{equation}
\left(  r_{x},r_{y},r_{z}\right)  \rightarrow\left(  ( 1-p) r_{x},\ ( 1-p)
r_{y},\ ( 1-p) r_{z}\right)  .
\end{equation}
Thus, it uniformly shrinks the Bloch vector to become closer to the maximally
mixed state.
\end{exercise}

The generalization of the depolarizing channel to qudits is again
straightforward. It is the same as the map in \eqref{eq-qt:depolarizing}, with
the exception that the density operators $\rho$ and $\pi$ are qudit density operators.

\begin{exercise}
[Qudit Twirl]\label{ex-qt:uniformly-random-unitary}Show that randomly applying
the Heisenberg--Weyl operators%
\begin{equation}
\left\{  X( i) Z( j) \right\}  _{i,j\in\left\{  0,\ldots,d-1\right\}  }%
\end{equation}
with uniform probability to any qudit density operator gives the maximally
mixed state $\pi$:%
\begin{equation}
\frac{1}{d^{2}}\sum_{i,j=0}^{d-1}X( i) Z( j) \rho Z^{\dag}( j) X^{\dag}( i)
=\pi.
\end{equation}
(Hint: You can do the full calculation, or you can decompose this channel into
the composition of two completely dephasing channels where the first is a
dephasing in the computational basis and the next is a dephasing in the
conjugate basis).
\end{exercise}

\subsection{Amplitude Damping Channels}

The amplitude damping channel%
\index{amplitude damping channel}
is an approximation to a noisy evolution that occurs in many physical systems
ranging from optical systems to chains of spin-1/2 particles to spontaneous
emission of a photon from an atom.

In order to motivate this channel, we give a physical interpretation to our
computational basis states. Let us think of the $|0\rangle$ state as the
ground state of a two-level atom and let us think of the state $|1\rangle$ as
the excited state of the atom. Spontaneous emission is a process that tends to
decay the atom from its excited state to its ground state, even if the atom is
in a superposition of the ground and excited states. Let the parameter
$\gamma$ denote the probability of decay so that $0\leq\gamma\leq1$. One Kraus
operator that captures the decaying behavior is%
\begin{equation}
A_{0}=\sqrt{\gamma}|0\rangle\langle1|.
\end{equation}
The operator $A_{0}$ annihilates the ground state:%
\begin{equation}
A_{0}|0\rangle\langle0|A_{0}^{\dag}=0,
\end{equation}
and it decays the excited state to the ground state:%
\begin{equation}
A_{0}|1\rangle\langle1|A_{0}^{\dag}=\gamma|0\rangle\langle0|.
\end{equation}
The Kraus operator $A_{0}$ alone does not specify a physical map because
$A_{0}^{\dag}A_{0}=\gamma|1\rangle\langle1|$ (recall that the Kraus operators
of any channel should satisfy the condition $\sum_{k}A_{k}^{\dag}A_{k}=I$). We
can satisfy this condition by choosing another operator $A_{1}$ such that%
\begin{equation}
A_{1}^{\dag}A_{1}=I-A_{0}^{\dag}A_{0}=|0\rangle\langle0|+\left(
1-\gamma\right)  |1\rangle\langle1|.
\end{equation}
The following choice of $A_{1}$ satisfies the above condition:%
\begin{equation}
A_{1}\equiv|0\rangle\langle0|+\sqrt{1-\gamma}|1\rangle\langle1|.
\end{equation}
Thus, the operators $A_{0}$ and $A_{1}$ are valid Kraus operators for the
amplitude damping channel.

\begin{exercise}
Consider a single-qubit density operator with the following matrix
representation with respect to the computational basis:%
\begin{equation}
\rho=%
\begin{bmatrix}
1-p & \eta\\
\eta^{\ast} & p
\end{bmatrix}
,
\end{equation}
where $0\leq p\leq1$ and $\eta$ is some complex number. Show that applying the
amplitude damping channel with parameter $\gamma$ to a qubit with the above
density operator gives a density operator with the following matrix
representation:%
\begin{equation}%
\begin{bmatrix}
1-\left(  1-\gamma\right)  p & \sqrt{1-\gamma}\eta\\
\sqrt{1-\gamma}\eta^{\ast} & \left(  1-\gamma\right)  p
\end{bmatrix}
. \label{eq-nqt:amp-damp-output}%
\end{equation}

\end{exercise}

\begin{exercise}
Show that the amplitude damping channel obeys a composition rule. Consider an
amplitude damping channel $\mathcal{N}_{1}$ with transmission parameter
$\left(  1-\gamma_{1}\right)  $ and consider another amplitude damping channel
$\mathcal{N}_{2}$ with transmission parameter $\left(  1-\gamma_{2}\right)  $.
Show that the composition channel $\mathcal{N}_{2}\circ\mathcal{N}_{1}$ is an
amplitude damping channel with transmission parameter $\left(  1-\gamma
_{1}\right)  \left(  1-\gamma_{2}\right)  $. (Note that the transmission
parameter is equal to one minus the damping parameter.)
\end{exercise}

\subsection{Erasure Channels}

\label{sec-nqt:erasure}The erasure channel%
\index{erasure channel}
is another important channel in quantum Shannon theory. It admits a simple
model and is amenable to relatively straightforward analysis when we later
discuss its capacity. The erasure channel can serve as a simplified model of
photon loss in optical systems.

We first recall the classical definition of an erasure channel. A classical
erasure channel either transmits a bit with some probability $1-\varepsilon$
or replaces it with an erasure symbol $e$ with some probability $\varepsilon$.
The output alphabet contains one more symbol than the input alphabet, namely,
the erasure symbol $e$.

The generalization of the classical erasure channel to the quantum world is
straightforward. It implements the following map:%
\begin{equation}
\rho\rightarrow\left(  1-\varepsilon\right)  \rho+\varepsilon|e\rangle\langle
e|,
\end{equation}
where $|e\rangle$ is some state that is not in the input Hilbert space, and
thus is orthogonal to it. The output space of the erasure channel is larger
than its input space by one dimension. The interpretation of the quantum
erasure channel is similar to that for the classical erasure channel. It
transmits a qubit with probability $1-\varepsilon$ and \textquotedblleft
erases\textquotedblright\ it (replaces it with an orthogonal erasure state)
with probability $\varepsilon$.

\begin{exercise}
Show that the following operators are Kraus operators for the quantum
erasure channel: $\{\sqrt{1-\varepsilon}(|0\rangle_{B}\langle0|_{A}%
+|1\rangle_{B}\langle1|_{A}),\sqrt{\varepsilon}|e\rangle_{B}\langle
0|_{A},\sqrt{\varepsilon}|e\rangle_{B}\langle1|_{A}\}$.
\end{exercise}

At the receiving end of the channel, a simple measurement can determine
whether an erasure has occurred. We perform a measurement with measurement
operators $\left\{  \Pi_{\operatorname{in}},|e\rangle\langle e|\right\}  $,
where $\Pi_{\operatorname{in}}$ is the projector onto the input Hilbert space.
This measurement has the benefit of detecting no more information than
necessary. It merely detects whether an erasure occurs, and thus preserves the
quantum information at the input if an erasure does not occur.

\subsection{Conditional Quantum Channels}

\label{sec-nqt:conditional-quantum-encoder}We end this chapter by considering
one final type of evolution. A \textit{conditional quantum encoder}
$\mathcal{E}_{MA\rightarrow B}$, or
\index{conditional quantum channel}%
\textit{conditional quantum channel }, is a collection $\left\{
\mathcal{E}_{A\rightarrow B}^{m}\right\}  _{m}$\ of CPTP maps. Its inputs are
a classical system $M$ and a quantum system $A$ and its output is a quantum
system $B$. A conditional quantum encoder can function as an encoder of both
classical and quantum information.%

\begin{figure}
[ptb]
\begin{center}
\includegraphics[
width=4.8456in
]%
{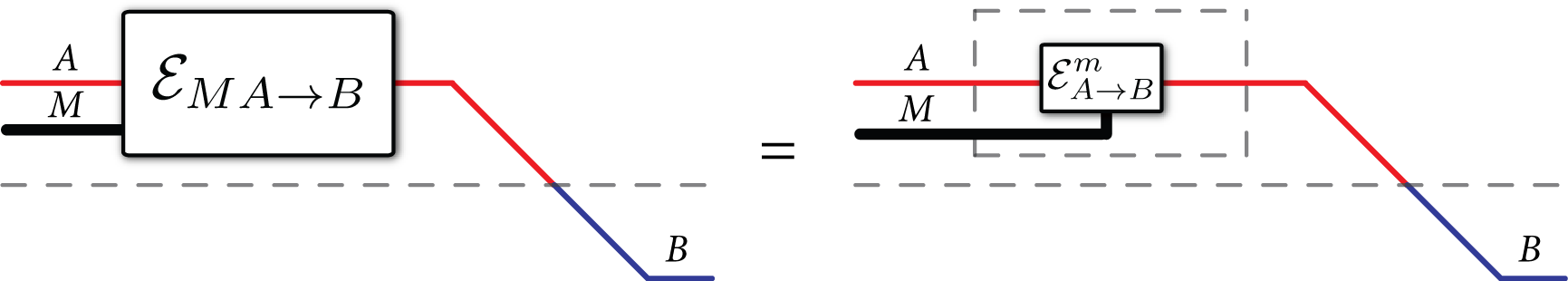}%
\caption{The figure on the left depicts a general operation, a conditional
quantum encoder, that takes a classical system to a quantum system. The figure
on the right depicts the inner workings of the conditional quantum encoder.}%
\label{fig-qt:cond-q-encoder}%
\end{center}
\end{figure}
A classical--quantum state $\rho_{MA}$, where%
\begin{equation}
\rho_{MA}\equiv\sum_{m}p( m) \vert m\rangle\langle m\vert_{M}\otimes\rho
_{A}^{m},
\end{equation}
can act as an input to a conditional quantum encoder $\mathcal{E}%
_{MA\rightarrow B}$. The action of the conditional quantum encoder
$\mathcal{E}_{MA\rightarrow B}$ on the classical--quantum state $\rho_{MA}$ is
as follows:%
\begin{equation}
\mathcal{E}_{MA\rightarrow B}\left(  \rho_{MA}\right)  =\operatorname{Tr}%
_{M}\left\{  \sum_{m}p( m) \vert m\rangle\langle m\vert_{M}\otimes
\mathcal{E}_{A\rightarrow B}^{m}\left(  \rho_{A}^{m}\right)  \right\}  .
\end{equation}
Figure~\ref{fig-qt:cond-q-encoder} depicts the behavior of the conditional
quantum encoder.

It is actually possible to write \textit{any} quantum channel as a conditional
quantum encoder when its input is a
\index{classical-quantum state}%
classical--quantum state. Indeed, consider any quantum channel $\mathcal{N}%
_{XA\rightarrow B}$ that has input systems $X$ and $A$ and output system $B$.
Suppose the Kraus decomposition of this channel is as follows:%
\begin{equation}
\mathcal{N}_{XA\rightarrow B}(\rho)\equiv\sum_{j}A_{j}\rho A_{j}^{\dag}.
\end{equation}
Suppose now that the input to the channel is the following classical--quantum
state:%
\begin{equation}
\sigma_{XA}\equiv\sum_{x}p_{X}(x)|x\rangle\langle x|_{X}\otimes\rho_{A}^{x}.
\end{equation}
Then the channel $\mathcal{N}_{XA\rightarrow B}$ acts as follows on the
classical--quantum state $\sigma_{XA}$:%
\begin{equation}
\mathcal{N}_{XA\rightarrow B}\left(  \sigma_{XA}\right)  =\sum_{j,x}%
A_{j}\left(  p_{X}(x)|x\rangle\langle x|_{X}\otimes\rho_{A}^{x}\right)
A_{j}^{\dag}. \label{eq-qt:cond-quantum-enc}%
\end{equation}
Consider that a classical--quantum state admits the following matrix
representation by exploiting the tensor product:%
\begin{align}
&  \sum_{x\in\mathcal{X}}p_{X}(x)|x\rangle\langle x|_{X}\otimes\rho_{A}^{x}\\
&  =%
\begin{bmatrix}
p_{X}(x_{1})\rho_{A}^{x_{1}} & 0 & \cdots & 0\\
0 & p_{X}(x_{2})\rho_{A}^{x_{2}} &  & \vdots\\
\vdots &  & \ddots & 0\\
0 & \cdots & 0 & p_{X}(x_{\left\vert \mathcal{X}\right\vert })\rho
_{A}^{x_{\left\vert \mathcal{X}\right\vert }}%
\end{bmatrix}
\\
&  =\bigoplus\limits_{x\in\mathcal{X}}p_{X}(x)\rho_{x}.
\end{align}
It is possible to specify a matrix representation for each Kraus operator
$A_{j}$ in terms of $\left\vert \mathcal{X}\right\vert $ block matrices:%
\begin{equation}
A_{j}=%
\begin{bmatrix}
A_{j,1} & A_{j,2} & \cdots & A_{j,\left\vert \mathcal{X}\right\vert }%
\end{bmatrix}
.
\end{equation}
Each operator $A_{j}\left(  p_{X}(x)|x\rangle\langle x|_{X}\otimes\rho_{A}%
^{x}\right)  A_{j}^{\dag}$ in the sum in \eqref{eq-qt:cond-quantum-enc} then
takes the following form:%
\begin{align}
&  A_{j}\left(  p_{X}(x)|x\rangle\langle x|_{X}\otimes\rho_{A}^{x}\right)
A_{j}^{\dag}\\
&  =%
\begin{bmatrix}
A_{j,1} & A_{j,2} & \cdots & A_{j,\left\vert \mathcal{X}\right\vert }%
\end{bmatrix}%
\begin{bmatrix}
p_{X}(x_{1})\rho_{A}^{x_{1}} & 0 & \cdots & 0\\
0 & \ddots &  & \vdots\\
\vdots &  & \ddots & 0\\
0 & \cdots & 0 & p_{X}(x_{\left\vert \mathcal{X}\right\vert })\rho
_{A}^{x_{\left\vert \mathcal{X}\right\vert }}%
\end{bmatrix}%
\begin{bmatrix}
A_{j,1}^{\dag}\\
A_{j,2}^{\dag}\\
\vdots\\
A_{j,\left\vert \mathcal{X}\right\vert }^{\dag}%
\end{bmatrix}
\\
&  =\sum_{x\in\left\vert \mathcal{X}\right\vert }p_{X}(x)A_{j,x}\rho_{A}%
^{x}A_{j,x}^{\dag}.
\end{align}
We can write the overall map as follows:%
\begin{align}
\mathcal{N}_{XA\rightarrow B}\left(  \sigma_{XA}\right)   &  =\sum_{j}%
\sum_{x\in\mathcal{X}}p_{X}(x)A_{j,x}\rho_{A}^{x}A_{j,x}^{\dag}\\
&  =\sum_{x\in\mathcal{X}}p_{X}(x)\sum_{j}A_{j,x}\rho_{A}^{x}A_{j,x}^{\dag}\\
&  =\sum_{x\in\mathcal{X}}p_{X}(x)\mathcal{N}_{A\rightarrow B}^{x}(\rho
_{A}^{x}),
\end{align}
where we define each map $\mathcal{N}_{A\rightarrow B}^{x}$\ as follows:%
\begin{equation}
\mathcal{N}_{A\rightarrow B}^{x}(\rho_{A}^{x})=\sum_{j}A_{j,x}\rho_{A}%
^{x}A_{j,x}^{\dag}.
\end{equation}
Thus, the action of any quantum channel on a classical--quantum state is the
same as the action of the conditional quantum encoder.

\begin{exercise}
Show that the condition $\sum_{j}A_{j}^{\dag}A_{j}=I$ implies the $\left\vert
\mathcal{X}\right\vert $\ conditions:%
\begin{equation}
\forall x\in\mathcal{X}:\sum_{j}A_{j,x}^{\dag}A_{j,x}=I.
\end{equation}

\end{exercise}

\section{Summary}

We give a brief summary of the main results in this chapter. We derived all of
these results from the noiseless quantum theory and an ensemble viewpoint. An
alternate viewpoint is to say that the density operator is the state of the
system and then give the postulates of quantum mechanics in terms of the
density operator. Regardless of which viewpoint you consider as more fundamental,
they are consistent with each other.

The density operator $\rho$\ for an ensemble $\left\{  p_{X}(x),|\psi
_{x}\rangle\right\}  $ is the following expectation:%
\begin{equation}
\rho=\sum_{x}p_{X}(x)|\psi_{x}\rangle\langle\psi_{x}|.
\end{equation}
The evolution of the density operator according to a unitary operator $U$ is%
\begin{equation}
\rho\rightarrow U\rho U^{\dag}.
\end{equation}
A measurement of the state according to a measurement $\left\{  M_{j}\right\}
$ where $\sum_{j}M_{j}^{\dag}M_{j}=I$ leads to the following post-measurement
state:%
\begin{equation}
\rho\rightarrow\frac{M_{j}\rho M_{j}^{\dag}}{p_{J}(j)},
\end{equation}
where the probability $p_{J}(j)$ for obtaining outcome $j$ is%
\begin{equation}
p_{J}(j)=\operatorname{Tr}\left\{  M_{j}^{\dag}M_{j}\rho\right\}  .
\end{equation}
The most general noisy evolution that a quantum state can undergo is according
to a completely positive, trace-preserving map $\mathcal{N}(\rho)$ that we can
write as follows:%
\begin{equation}
\mathcal{N}(\rho)=\sum_{j}A_{j}\rho A_{j}^{\dag},
\end{equation}
where $\sum_{j}A_{j}^{\dag}A_{j}=I$. A special case of this evolution is a
quantum instrument. A quantum instrument has a quantum input and a classical
and quantum output. The most general way to represent a quantum instrument is
as follows:%
\begin{equation}
\rho\rightarrow\sum_{j}\mathcal{E}_{j}(\rho)\otimes|j\rangle\langle j|_{J},
\end{equation}
where each map $\mathcal{E}_{j}$ is a completely positive,
trace-non-increasing map, where%
\begin{equation}
\mathcal{E}_{j}(\rho)=\sum_{k}A_{j,k}\rho A_{j,k}^{\dag},
\end{equation}
and $\sum_{j,k}A_{j,k}^{\dag}A_{j,k}=I$, so that the overall map is trace-preserving.

\section{History and Further Reading}

\cite{book2000mikeandike} have given an excellent introduction to noisy quantum
channels. \cite{W89} defined what it means for a multiparty quantum state to be entangled. \cite{HSR03} introduced
\index{entanglement-breaking channel}%
entanglement-breaking channels and proved several properties of them (e.g.,
the proof of Theorem~\ref{thm-nqt:ent-break-unit-rank}). \cite{DL70}
introduced the quantum instrument formalism, and \cite{O79} developed it
further. \cite{GBP97} introduced the quantum erasure channel and constructed
some simple quantum error-correcting codes for it. A discussion of the
conditional quantum channel appears in \cite{Yard05a}.

\chapter{The Purified Quantum Theory}

\label{chap:purified-q-t}The final chapter of our development of the quantum
theory gives perhaps the most powerful viewpoint, by providing a mathematical
tool, the purification theorem, which offers a completely different way of
thinking about noise in quantum systems. This theorem states that our lack of
information about a set of quantum states can be thought of as arising from
entanglement with another system to which we do not have access. The system to
which we do not have access is known as a \textit{purifying system}. In this
purified view of the quantum theory, noisy evolution arises from the
interaction of a quantum system with its environment. The interaction of a
quantum system with its environment leads to correlations between the quantum
system and its environment, and this interaction leads to a loss of
information because we cannot access the environment. The environment is thus
the purification of the output of the noisy quantum channel.

In Chapter~\ref{chap:noiseless-quantum-theory}, we introduced the noiseless
quantum theory. The noiseless quantum theory is a useful theory to learn so
that we can begin to grasp an intuition for some uniquely quantum behavior,
but it is an idealized model of quantum information processing. In
Chapter~\ref{chap:noisy-quantum-theory}, we introduced the noisy quantum
theory as a generalization of the noiseless quantum theory. The noisy quantum
theory can describe the behavior of imperfect quantum systems that are subject
to noise.

In this chapter, we actually show that we can view the noisy quantum theory
\textit{as a special case of the noiseless quantum theory}. This relation may
seem strange at first, but the purification theorem allows us to make this
connection. The quantum theory that we present in this chapter is a noiseless
quantum theory, but we name it \textit{the purified quantum theory}, in order
to distinguish it from the description of the noiseless quantum theory in
Chapter~\ref{chap:noiseless-quantum-theory}.

The purified quantum theory shows that it is possible to view noise as
resulting from entanglement of a system with another system. We have actually
seen a glimpse of this phenomenon in the previous chapter when we introduced
the notion of the local density operator, but we did not highlight it in
detail there. The example was the maximally entangled Bell state $\left\vert
\Phi^{+}\right\rangle _{AB}$. This state is a pure state on the two systems
$A$ and $B$, but the local density operator of Alice is the maximally mixed
state $\pi_{A}$. We saw that the local density operator is a mathematical
object that allows us to make all the predictions about any local measurement
or evolution. We also have seen that a density operator arises from an
ensemble, but there is also the reverse interpretation, that an ensemble
corresponds to a convex decomposition of any density operator. There is a
sense in which we can view this local density operator as arising from an
ensemble where we choose the states $\vert0\rangle$ and $\vert1\rangle$ with
equal probability $1/2$. The purification idea goes as far as to say that the
noisy ensemble for Alice with density operator $\pi_{A}$ arises from the
entanglement of her system with Bob's. We explore this idea in more detail in
this final chapter on the quantum theory.

\section{Purification}

\label{sec-pqt:purification}\begin{figure}[ptb]
\begin{center}
\includegraphics[
width=1.8248in
]{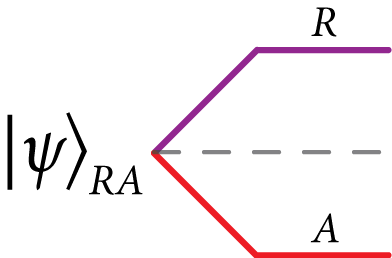}
\end{center}
\caption{This diagram depicts a purification $\vert\psi\rangle_{RA}$ of a
density operator $\rho_{A}$. The above diagram indicates that the reference
system $R$\ is generally entangled with the system $A$. An interpretation of
the purification theorem is that the noise inherent in a density operator
$\rho_{A}$ is due to entanglement with a reference system~$R$.}%
\label{fig:purification}%
\end{figure}

Suppose we are given a density operator $\rho_{A}$\ on a system $A$. Every
such density operator has a \textit{purification}, as defined below and
depicted in Figure~\ref{fig:purification}:

\begin{definition}
[Purification]A \textit{purification}%
\index{purification}
of a density operator $\rho_{A} \in\mathcal{D}(\mathcal{H}_{A})$ is a pure
bipartite state $\vert\psi\rangle_{RA} \in\mathcal{H}_{R} \otimes
\mathcal{H}_{A} $ on a \textit{reference} system $R$ and the original system
$A$, with the property that the reduced state on system $A$ is equal to
$\rho_{A}$:%
\begin{equation}
\rho_{A}=\operatorname{Tr}_{R}\left\{  \vert\psi\rangle\langle\psi\vert
_{RA}\right\}  .
\end{equation}

\end{definition}

Suppose that a spectral decomposition for the density operator $\rho_{A}$ is
as follows:%
\begin{equation}
\rho_{A}=\sum_{x}p_{X}(x)|x\rangle\langle x|_{A}.
\label{eq-qt:local-for-purify}%
\end{equation}
We claim that the following state $|\psi\rangle_{RA}$ is a purification of
$\rho_{A}$:%
\begin{equation}
|\psi\rangle_{RA}\equiv\sum_{x}\sqrt{p_{X}(x)}|x\rangle_{R}|x\rangle_{A},
\label{eq-pqt:standard-purification-1}%
\end{equation}
where the set $\{|x\rangle_{R}\}_{x}$\ of vectors is some set of orthonormal
vectors for the reference system $R$. The next exercise asks you to verify
this claim.

\begin{exercise}
Show that the state $\vert\psi\rangle_{RA}$, as defined in
\eqref{eq-pqt:standard-purification-1}, is a purification of the density
operator $\rho_{A}$, with a spectral decomposition as given in \eqref{eq-qt:local-for-purify}.
\end{exercise}

\begin{exercise}
[Canonical purification]\label{ex-pqt:canonical-pur}Let $\rho_{A}$ be a
density operator and let $\sqrt{\rho_{A}}$ be its unique positive
semi-definite square root (i.e., $\rho_{A}=\sqrt{\rho_{A}}\sqrt{\rho_{A}}$.)
We define the canonical purification of $\rho_{A}$ as follows:%
\begin{equation}
\left(  I_{R}\otimes\sqrt{\rho_{A}}\right)  \vert\Gamma\rangle_{RA},
\label{eq-pqt:canonical-pur}%
\end{equation}
where $\vert\Gamma\rangle_{RA}$ is the unnormalized maximally entangled vector
from \eqref{eq-qt:unnorm-max-ent}. Show that \eqref{eq-pqt:canonical-pur} is a
purification of $\rho_{A}$.
\end{exercise}

\subsection{Interpretation of Purifications}%

\index{purification!interpretation}%
The purification idea has an interesting physical interpretation: we can think
of the noisiness inherent in a particular quantum system as being due to
entanglement with some external reference system to which we do not have
access. That is, we can think that the density operator $\rho_{A}$ arises from
the entanglement of the system $A$ with the reference system $R$ and from our
lack of access to the system $R$.

Stated another way, the purification idea gives us a fundamentally different
way to interpret noise. The interpretation is that any noise on a local system
is due to entanglement with another system to which we do not have access.
This interpretation extends to the noise from a noisy quantum channel. We can
view this noise as arising from the interaction of the system that we possess
with an external environment over which we have no control.

The global state $\vert\psi\rangle_{RA}$ is a pure state, but a reduced state
$\rho_{A}$ is not a pure state in general because we trace over the reference
system to obtain it. A reduced state $\rho_{A}$ is pure if and only if the
global state $\vert\psi\rangle_{RA}$ is a pure product state.

\subsection{Equivalence of Purifications}

\label{sec-pqt:purif-equiv}%

\index{purification!equivalence}%
Theorem \ref{thm-pqt:purif-unitary-reference} below states that there is an
equivalence relation between all purifications of a given density operator
$\rho_{A}$. It is a consequence of the Schmidt decomposition
(Theorem~\ref{thm-qt:schmidt}). Before stating it, recall the definition of an
isometry from Definition~\ref{def-pqt:isometry}.

\begin{theorem}
\label{thm-pqt:purif-unitary-reference} All purifications of a density
operator are related by an isometry acting on the purifying system. That is,
let $\rho_{A}$ be a density operator, and let $|\psi\rangle_{R_{1}A}$ and
$|\varphi\rangle_{R_{2}A}$ be purifications of $\rho_{A}$, such that
$\dim(\mathcal{H}_{R_{1}})\leq\dim(\mathcal{H}_{R_{2}})$. Then there exists an
isometry $U_{R_{1}\rightarrow R_{2}}$ such that%
\begin{equation}
|\varphi\rangle_{R_{2}A}=\left(  U_{R_{1}\rightarrow R_{2}}\otimes
I_{A}\right)  |\psi\rangle_{R_{1}A}.
\end{equation}

\end{theorem}

\begin{proof}
Let us first suppose that the eigenvalues of $\rho_{A}$ are distinct, so that
a unique spectral decomposition of $\rho_{A}$ is as follows:%
\begin{equation}
\rho_{A}=\sum_{x}p_{X}(x)|x\rangle\langle x|_{A}.
\end{equation}
Then a Schmidt decomposition of $|\varphi\rangle_{R_{2}A}$ necessarily has the
form%
\begin{equation}
|\varphi\rangle_{R_{2}A}=\sum_{x}\sqrt{p_{X}(x)}|\varphi_{x}\rangle_{R_{2}%
}|x\rangle_{A},
\end{equation}
where $\{|\varphi_{x}\rangle_{R_{2}}\}$ is an orthonormal basis for the
$R_{2}$ system, and similarly, the Schmidt decomposition of $|\psi
\rangle_{R_{1}A}$ necessarily has the form%
\begin{equation}
|\psi\rangle_{R_{1}A}=\sum_{x}\sqrt{p_{X}(x)}|\psi_{x}\rangle_{R_{1}}%
|x\rangle_{A}.
\end{equation}
(If it were not the case then we could not have $\operatorname{Tr}_{R_{2}%
}\{|\varphi\rangle\langle\varphi|_{R_{2}A}\}=\operatorname{Tr}_{R_{1}}%
\{|\psi\rangle\langle\psi|_{R_{1}A}\}=\rho_{A}$, as given in the statement of
the theorem.) Given the above, we can take the isometry $U_{R_{1}\rightarrow
R_{2}}$ to be%
\begin{equation}
U_{R_{1}\rightarrow R_{2}}=\sum_{x}|\varphi_{x}\rangle_{R_{2}}\langle\psi
_{x}|_{R_{1}},
\end{equation}
which is an isometry because $U^{\dag}U=I_{R_{1}}$. If the eigenvalues of
$\rho_{A}$ are not distinct, then there is more freedom in the Schmidt
decompositions, but here we are free to choose them as above, and then the
development is the same.
\end{proof}

This theorem leads to a way of relating all convex decompositions of a given
density operator, addressing a question raised in
Section~\ref{sec-nqt:ensemble-density-unique}:

\begin{corollary}
Let two convex decompositions of a density operator $\rho$ be as follows:%
\begin{equation}
\rho=\sum_{x=1}^{d}p_{X}(x)|\psi_{x}\rangle\langle\psi_{x}|=\sum
_{y=1}^{d^{\prime}}p_{Y}(y)|\phi_{y}\rangle\langle\phi_{y}|,
\end{equation}
where $d^{\prime}\leq d$. Then there exists an isometry $U$ such that%
\begin{equation}
\sqrt{p_{X}(x)}|\psi_{x}\rangle=\sum_{y}U_{x,y}\sqrt{p_{Y}(y)}|\phi_{y}%
\rangle.
\end{equation}

\end{corollary}

\begin{proof}
Let $\{|x\rangle_{R}\}$ be an orthonormal basis for a purification system,
with a number of states equal to $\max\left\{  d,d^{\prime}\right\}  $. Then a
purification for the first decomposition is as follows:%
\begin{equation}
|\psi\rangle_{RA}\equiv\sum_{x}\sqrt{p_{X}(x)}|x\rangle_{R}\otimes|\psi
_{x}\rangle_{A},
\end{equation}
and a purification of the second decomposition is%
\begin{equation}
|\phi\rangle_{RA}\equiv\sum_{y}\sqrt{p_{Y}(y)}|y\rangle_{R}\otimes|\phi
_{y}\rangle_{A}.
\end{equation}
From Theorem~\ref{thm-pqt:purif-unitary-reference}, we know that there exists
an isometry $U_{R}$ such that $|\psi\rangle_{RA}=\left(  U_{R}\otimes
I_{A}\right)  |\phi\rangle_{RA}$. Then consider that%
\begin{align}
\sqrt{p_{X}(x)}|\psi_{x}\rangle_{A}  &  =\sum_{x^{\prime}}\sqrt{p_{X}%
(x^{\prime})}\langle x|_{R}|x^{\prime}\rangle_{R}\otimes|\psi_{x^{\prime}%
}\rangle_{A}=\left(  \langle x|_{R}\otimes I_{A}\right)  |\psi\rangle_{RA}\\
&  =\left(  \langle x|_{R}U_{R}\otimes I_{A}\right)  |\phi\rangle_{RA}%
=\sum_{y}\sqrt{p_{Y}(y)}\langle x|_{R}U_{R}|y\rangle_{R}|\phi_{y}\rangle_{A}\\
&  =\sum_{y}\sqrt{p_{Y}(y)}U_{x,y}|\phi_{y}\rangle_{A},
\end{align}
where in the last step we have defined $U_{x,y}=\langle x|_{R}U_{R}%
|y\rangle_{R}$.
\end{proof}

\begin{exercise}
Find a purification of the following classical--quantum state:%
\begin{equation}
\sum_{x}p_{X}(x)|x\rangle\langle x|_{X}\otimes\rho_{A}^{x}.
\end{equation}

\end{exercise}

\begin{exercise}
\label{ex-pqt:ensemble-purification}Let $\left\{  p_{X}(x),\rho_{A}%
^{x}\right\}  $ be an ensemble of density operators. Suppose that $\left\vert
\psi^{x}\right\rangle _{RA}$ is a purification of $\rho_{A}^{x}$. The expected
density operator of the ensemble is $\rho_{A}\equiv\sum_{x}p_{X}(x)\rho
_{A}^{x}. $ Find a purification of $\rho_{A}$.
\end{exercise}

\subsection{Extension of a Quantum State}

We can also define an \textit{extension}%
\index{extension}
of a quantum state $\rho_{A}$:

\begin{definition}
[Extension]An extension of a density operator $\rho_{A}\in\mathcal{D}%
(\mathcal{H}_{A})$ is a density operator $\Omega_{RA}\in\mathcal{D}%
(\mathcal{H}_{R}\otimes\mathcal{H}_{A})$ such that $\rho_{A}=\operatorname{Tr}%
_{R}\left\{  \Omega_{RA}\right\}  . $
\end{definition}

\noindent This notion can be useful, but keep in mind that we can always find
a purification $\vert\psi\rangle_{R^{\prime}RA}$ of the extension $\Omega
_{RA}$.

\section{Isometric Evolution}

A quantum channel admits a purification as well. We motivate this idea with a
simple example.

\subsection{Example: Isometric Extension of the Bit-Flip Channel}

Consider the bit-flip channel from \eqref{eq-qt:bit-flip-channel}---it applies
the identity operator with some probability $1-p$ and applies the bit-flip
Pauli operator $X$ with probability $p$. Suppose that we input a qubit system
$A$ in the state $\vert\psi\rangle$ to this channel. The ensemble
corresponding to the state at the output has the following form:%
\begin{equation}
\left\{  \left\{  1-p,\vert\psi\rangle\right\}  ,\left\{  p,X\vert\psi
\rangle\right\}  \right\}  ,
\end{equation}
and the density operator of the resulting state is%
\begin{equation}
( 1-p) \vert\psi\rangle\langle\psi\vert+pX\vert\psi\rangle\langle\psi\vert X.
\end{equation}
The following state is a purification of the above density operator (you
should quickly check that this relation holds):%
\begin{equation}
\sqrt{1-p}\vert\psi\rangle_{A}\vert0\rangle_{E}+\sqrt{p}X\vert\psi\rangle
_{A}\vert1\rangle_{E}.
\end{equation}
We label the original system as $A$ and label the purification system as $E$.
In this context, we can view the purification system as the environment of the channel.

There is another way for interpreting the dynamics of the above bit-flip
channel. Instead of determining the ensemble for the channel and then
purifying, we can say that the channel directly implements the following map
from the system $A$ to the larger joint system $AE$:%
\begin{equation}
\vert\psi\rangle_{A}\rightarrow\sqrt{1-p}\vert\psi\rangle_{A}\vert0\rangle
_{E}+\sqrt{p}X\vert\psi\rangle_{A}\vert1\rangle_{E}.
\label{eq-qt:isometric-example}%
\end{equation}
We see that any $p\in(0,1)$, i.e., any amount of noise in the channel, can
lead to entanglement of the input system with the environment $E$. We then
obtain the noisy dynamics of the channel by discarding (tracing out) the
environment system $E$.

\begin{exercise}
Find two input states for which the map in \eqref{eq-qt:isometric-example}
does not lead to entanglement between systems $A$ and $E$.
\end{exercise}

The map in \eqref{eq-qt:isometric-example} is an \textit{isometric extension}%
\index{isometric extension}
of the bit-flip channel. Let us label it as $U_{A\rightarrow AE}$ where the
notation indicates that the input system is $A$ and the output system is $AE$.
As discussed around Definition~\ref{def-pqt:isometry}, an isometry is similar
to a unitary operator but different because it maps states in one Hilbert
space (for an input system) to states in a larger Hilbert space (which could
be for a joint system). It generally does not admit a square matrix
representation, but instead admits a rectangular matrix representation. The
matrix representation of the isometric operation in
\eqref{eq-qt:isometric-example} consists of the following matrix elements:%
\begin{equation}
\left[
\begin{array}
[c]{cc}%
\langle0\vert_{A}\langle0\vert_{E}U_{A\rightarrow AE}\vert0\rangle_{A} &
\langle0\vert_{A}\langle0\vert_{E}U_{A\rightarrow AE}\vert1\rangle_{A}\\
\langle0\vert_{A}\langle1\vert_{E}U_{A\rightarrow AE}\vert0\rangle_{A} &
\langle0\vert_{A}\langle1\vert_{E}U_{A\rightarrow AE}\vert1\rangle_{A}\\
\langle1\vert_{A}\langle0\vert_{E}U_{A\rightarrow AE}\vert0\rangle_{A} &
\langle1\vert_{A}\langle0\vert_{E}U_{A\rightarrow AE}\vert1\rangle_{A}\\
\langle1\vert_{A}\langle1\vert_{E}U_{A\rightarrow AE}\vert0\rangle_{A} &
\langle1\vert_{A}\langle1\vert_{E}U_{A\rightarrow AE}\vert1\rangle_{A}%
\end{array}
\right]  =\left[
\begin{array}
[c]{cc}%
\sqrt{1-p} & 0\\
0 & \sqrt{p}\\
0 & \sqrt{1-p}\\
\sqrt{p} & 0
\end{array}
\right]  .
\end{equation}

There is no reason that we have to choose the environment states as we did in
\eqref{eq-qt:isometric-example}. We could have chosen the environment states
to be any orthonormal basis---isometric behavior only requires that the states
on the environment be distinguishable. This is related to the fact that all
purifications are related by an isometry acting on the purifying system (see
Theorem~\ref{thm-pqt:purif-unitary-reference}).

\subsubsection{An Isometry is Part of a Unitary on a Larger System}

We can view the dynamics in \eqref{eq-qt:isometric-example} as an interaction
between an initially pure environment and the qubit state $|\psi\rangle$. So,
an equivalent way to implement an isometric mapping is with a two-step
procedure. We first assume that the environment of the channel is in a pure
state$~|0\rangle_{E}$ before the interaction begins. The joint state of the
qubit $|\psi\rangle$ and the environment is%
\begin{equation}
|\psi\rangle_{A}|0\rangle_{E}.
\end{equation}
These two systems then interact according to a unitary operator $V_{AE}$. We
can specify two columns of the unitary operator (we make this more clear in a
bit) by means of the isometric mapping in \eqref{eq-qt:isometric-example}:%
\begin{equation}
V_{AE}|\psi\rangle_{A}|0\rangle_{E}=\sqrt{1-p}|\psi\rangle_{A}|0\rangle
_{E}+\sqrt{p}X|\psi\rangle_{A}|1\rangle_{E}. \label{eq-pqt:VAE_op}%
\end{equation}
In order to specify the full unitary $V_{AE}$, we must also specify how the
map behaves when the initial state of the qubit and the environment is%
\begin{equation}
|\psi\rangle_{A}|1\rangle_{E}.
\end{equation}
We choose the mapping to be as follows so that the overall interaction is
unitary:%
\begin{equation}
V_{AE}|\psi\rangle_{A}|1\rangle_{E}=\sqrt{p}|\psi\rangle_{A}|0\rangle
_{E}-\sqrt{1-p}X|\psi\rangle_{A}|1\rangle_{E}. \label{eq-pqt:VAE_op-1}%
\end{equation}

\begin{exercise}
Check that the operator $V_{AE}$, defined by \eqref{eq-pqt:VAE_op} and
\eqref{eq-pqt:VAE_op-1}, is unitary by determining its action on the
computational basis $\left\{  \vert0\rangle_{A}\vert0\rangle_{E},\vert
0\rangle_{A}\vert1\rangle_{E},\vert1\rangle_{A}\vert0\rangle_{E},\vert
1\rangle_{A}\vert1\rangle_{E}\right\}  $ and showing that all of the outputs
for each of these inputs form an orthonormal basis.
\end{exercise}

\begin{exercise}
Verify that the matrix representation of the full unitary operator $V_{AE}$,
defined by \eqref{eq-pqt:VAE_op} and \eqref{eq-pqt:VAE_op-1}, is%
\begin{equation}
\left[
\begin{array}
[c]{cccc}%
\sqrt{1-p} & \sqrt{p} & 0 & 0\\
0 & 0 & \sqrt{p} & -\sqrt{1-p}\\
0 & 0 & \sqrt{1-p} & \sqrt{p}\\
\sqrt{p} & -\sqrt{1-p} & 0 & 0
\end{array}
\right]  ,
\end{equation}
by considering the matrix elements $\langle i\vert_{A}\left\langle
j\right\vert _{E}V\vert k\rangle_{A}\left\vert l\right\rangle _{E}$.
\end{exercise}

\subsubsection{Complementary Channel}

We may not only be interested in the receiver's output of the quantum channel.
We may also be interested in determining the environment's output from the
channel. This idea becomes increasingly important as we proceed in our study
of quantum Shannon theory. We should consider all parties in a quantum
protocol, and the purified quantum theory allows us to do so. We consider the
environment as one of the parties in a quantum protocol because the
environment could also be receiving some quantum information from the sender.

We can obtain the environment's output from the quantum channel simply by
tracing out every system besides the environment. The map from the sender to
the environment is known as a
\index{complementary channel}%
\textit{complementary channel}. In our example of the isometric extension of
the bit-flip channel in \eqref{eq-qt:isometric-example}, we can check that the
environment receives the following output state if the channel input is
$|\psi\rangle_{A}$:%
\begin{align}
&  \operatorname{Tr}_{A}\left\{  \left(  \sqrt{1-p}|\psi\rangle_{A}%
|0\rangle_{E}+\sqrt{p}X|\psi\rangle_{A}|1\rangle_{E}\right)  \left(
\sqrt{1-p}\langle\psi|_{A}\langle0|_{E}+\sqrt{p}\langle\psi|_{A}X\langle
1|_{E}\right)  \right\} \nonumber\\
&  =\operatorname{Tr}_{A}\left\{  (1-p)|\psi\rangle\langle\psi|_{A}%
\otimes|0\rangle\langle0|_{E}+\sqrt{p(1-p)}X|\psi\rangle\langle\psi
|_{A}\otimes|1\rangle\langle0|_{E}\right\} \nonumber\\
&  \ \ \ \ \ \ \ \ \ \ \ \ +\operatorname{Tr}_{A}\left\{  \sqrt{p\left(
1-p\right)  }|\psi\rangle\langle\psi|_{A}X\otimes|0\rangle\langle
1|_{E}+pX|\psi\rangle\langle\psi|_{A}X\otimes|1\rangle\langle1|_{E}\right\} \\
&  =(1-p)|0\rangle\langle0|_{E}+\sqrt{p\left(  1-p\right)  }\langle\psi
|X|\psi\rangle|1\rangle\langle0|_{E}\nonumber\\
&  \ \ \ \ \ \ \ \ \ \ \ \ +\sqrt{p(1-p)}\langle\psi|X|\psi\rangle
|0\rangle\langle1|_{E}+p|1\rangle\langle1|_{E}\\
&  =(1-p)|0\rangle\langle0|_{E}+\sqrt{p\left(  1-p\right)  }\langle\psi
|X|\psi\rangle\left(  |1\rangle\langle0|_{E}+|0\rangle\langle1|_{E}\right)
+p|1\rangle\langle1|_{E}\\
&  =(1-p)|0\rangle\langle0|_{E}+\sqrt{p\left(  1-p\right)  }2\operatorname{Re}%
\left\{  \alpha^{\ast}\beta\right\}  \left(  |1\rangle\langle0|_{E}%
+|0\rangle\langle1|_{E}\right)  +p|1\rangle\langle1|_{E},
\end{align}
where in the last line we assume that the qubit $|\psi\rangle\equiv
\alpha|0\rangle+\beta|1\rangle$.

It is helpful to examine several cases of the above example. Consider the case
in which the noise parameter $p=0$ or $p=1$. In this case, the environment
receives one of the respective states $\vert0\rangle$ or $\vert1\rangle$.
Therefore, in these cases, the environment does not receive any of the quantum
information about the state $\vert\psi\rangle$ transmitted down the
channel---it does not learn anything about the probability amplitudes $\alpha$
or $\beta$. This viewpoint is a completely different way to see that the
channel is truly noiseless in these cases. A channel is noiseless if the
environment of the channel does not learn anything about the states that we
transmit through it, i.e., if the channel does not leak quantum information to
the environment. Now let us consider the case in which $p\in(0,1)$. As $p$
approaches $1/2$ from either above or below, the amplitude $\sqrt{p( 1-p) }$
of the off-diagonal terms is a monotonic function that reaches its peak at
$1/2$. Thus, at the peak $1/2$, the off-diagonal terms are the strongest,
implying that the environment is generally ``stealing'' much of the coherence
from the original quantum state~$\vert\psi\rangle$.

\begin{exercise}
Show that the receiver's output density operator for a bit-flip channel with
$p=1/2$ is the same as what the environment obtains.
\end{exercise}

\subsection{Isometric Extension of a Quantum Channel}

We now give a general definition for an isometric extension of a quantum channel:

\begin{definition}
[Isometric Extension]\label{def-pqt:isometric-ext}Let $\mathcal{H}_{A}$ and
$\mathcal{H}_{B}$ be Hilbert spaces, and let $\mathcal{N}:\mathcal{L}%
(\mathcal{H}_{A})\rightarrow\mathcal{L}(\mathcal{H}_{B})$ be a quantum
channel. Let $\mathcal{H}_{E}$ be a Hilbert space with dimension no smaller
than the Choi rank of the channel $\mathcal{N}$. An isometric extension
\index{isometric extension}
or Stinespring dilation
\index{Stinespring dilation}
$U:\mathcal{H}_{A}\rightarrow\mathcal{H}_{B}\otimes\mathcal{H}_{E}$ of the
channel $\mathcal{N}$ is a linear isometry such that
\begin{equation}
\operatorname{Tr}_{E}\{UX_{A}U^{\dag}\}=\mathcal{N}_{A\rightarrow B}(X_{A}),
\end{equation}
for $X_{A}\in\mathcal{L}(\mathcal{H}_{A})$. The fact that $U$ is an isometry
is equivalent to the following conditions:
\begin{equation}
U^{\dag}U=I_{A},\ \ \ \ \ \ \ \ \ UU^{\dag}=\Pi_{BE},
\end{equation}
where $\Pi_{BE}$ is a projection of the tensor-product Hilbert space
$\mathcal{H}_{B}\otimes\mathcal{H}_{E}$.
\end{definition}

\begin{notation}
We often write a channel $\mathcal{N} : \mathcal{L}(\mathcal{H}_{A})
\to\mathcal{L}(\mathcal{H}_{B})$ as $\mathcal{N}_{A \to B}$ in order to
indicate the input and output systems explicitly. Similarly, we often write an
isometric extension $U:\mathcal{H}_{A} \to\mathcal{H}_{B} \otimes
\mathcal{H}_{E}$ of $\mathcal{N}$ as $U_{A\rightarrow BE}^{\mathcal{N}}$ in
order to indicate its association with $\mathcal{N}$ explicitly, as well the
fact that it accepts an input system $A$ and has output systems $B$ and $E$.
The system $E$ is often referred to as an ``environment'' system. Finally,
there is a quantum channel $\mathcal{U}_{A\rightarrow BE}^{\mathcal{N}}$
associated to an isometric extension $U_{A\rightarrow BE}^{\mathcal{N}}$,
which is defined by
\begin{equation}
\mathcal{U}_{A\rightarrow BE}^{\mathcal{N}}(X_{A}) = U X_{A} U^{\dag},
\end{equation}
for $X_{A} \in\mathcal{L}(\mathcal{H}_{A})$. Note that $\mathcal{U}%
_{A\rightarrow BE}^{\mathcal{N}}$ is a quantum channel with a single Kraus
operator $U$ given that $U^{\dag}U = I_{A}$.
\end{notation}

\begin{figure}[ptb]
\begin{center}
\includegraphics[
width=2.3212in
]{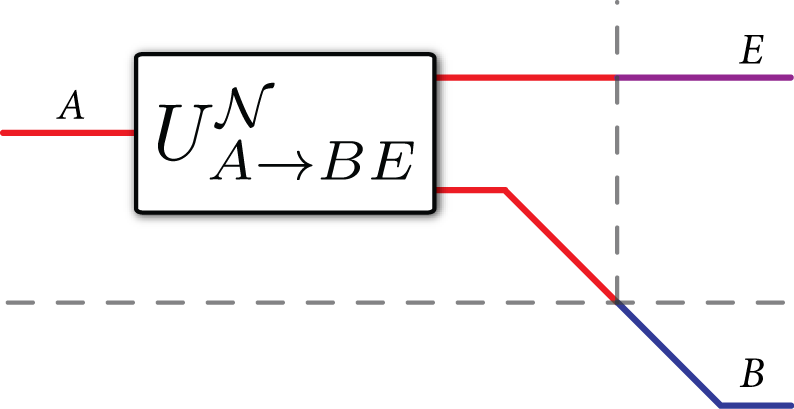}
\end{center}
\caption{This figure depicts an isometric extension $U_{A\rightarrow
BE}^{\mathcal{N}}$ of a quantum channel $\mathcal{N}_{A\rightarrow B}$. The
extension $U_{A\rightarrow BE}^{\mathcal{N}}$ includes the inaccessible
environment on system $E$ as a \textquotedblleft receiver\textquotedblright%
\ of quantum information. Ignoring the environment $E$ gives the quantum
channel $\mathcal{N}_{A\rightarrow B}$.}%
\label{fig:isometric-quantum-channel}%
\end{figure}

We can think of an isometric extension of a quantum channel as a purification
of that channel: the environment system $E$ is analogous to the purification
system from Section~\ref{sec-pqt:purification} because we trace over it to get
back the original channel. An isometric extension \textit{extends} the
original channel because it produces the evolution of the quantum channel
$\mathcal{N}_{A\rightarrow B}$ if we trace out the environment system $E$. It
also behaves as an \textit{isometry}---it is analogous to a rectangular matrix
that behaves somewhat like a unitary operator. The matrix representation of an
isometry is a rectangular matrix formed from selecting only a few of the
columns from a unitary matrix. The property $U^{\dag}U = I_{A}$ indicates that
the isometry behaves analogously to a unitary operator, because we can
determine an inverse operation simply by taking its conjugate transpose. The
property $UU^{\dag}= \Pi_{BE}$ distinguishes an isometric operation from a
unitary one. It states that the isometry takes states in the input system $A$
to a particular subspace of the joint system $BE$. The projector $\Pi_{BE}$
projects onto the subspace where the isometry takes input quantum states.
Figure~\ref{fig:isometric-quantum-channel}\ depicts a quantum circuit for an
isometric extension.

\subsubsection{Isometric Extension from Kraus Operators}

\label{sec-qt:iso-ext-kraus}It is possible to determine an isometric extension
of a quantum channel directly from a set of Kraus operators. Consider a
quantum channel $\mathcal{N}_{A\rightarrow B}$ with the following Kraus
representation:%
\begin{equation}
\mathcal{N}_{A\rightarrow B}( \rho_{A}) =\sum_{j}N_{j}\rho_{A}N_{j}^{\dag}.
\end{equation}
An isometric extension of the channel $\mathcal{N}_{A\rightarrow B}$ is the
following linear map:%
\begin{equation}
U_{A\rightarrow BE}^{\mathcal{N}}\equiv\sum_{j}N_{j}\otimes\left\vert
j\right\rangle _{E}. \label{eq-qt:channel-isometry}%
\end{equation}
It is straightforward to verify that the above map is an isometry:
\begin{align}
\left(  U^{\mathcal{N}}\right)  ^{\dag}U^{\mathcal{N}}  &  =\left(  \sum
_{k}N_{k}^{\dag}\otimes\langle k\vert_{E}\right)  \left(  \sum_{j}N_{j}%
\otimes\vert j\rangle_{E}\right) \\
&  =\sum_{k,j}N_{k}^{\dag}N_{j}\left\langle k|j\right\rangle \\
&  =\sum_{k}N_{k}^{\dag}N_{k}\\
&  =I_{A}.
\end{align}
The last equality follows from the completeness condition of the Kraus
operators. As a consequence, we get that $U^{\mathcal{N}}\left(
U^{\mathcal{N}}\right)  ^{\dag}$ is a projector on the joint system $BE$,
which follows by the same reasoning given in
\eqref{eq-pqt:isometry-projector}. Finally, we should verify that
$U^{\mathcal{N}}$ is an extension of $\mathcal{N}$. Applying the channel
$\mathcal{U}_{A\rightarrow BE}^{\mathcal{N}}$ to an arbitrary density operator
$\rho_{A}$ gives the following map:%
\begin{align}
\mathcal{U}_{A\rightarrow BE}^{\mathcal{N}}( \rho_{A})  &  \equiv
U^{\mathcal{N}}\rho_{A}\left(  U^{\mathcal{N}}\right)  ^{\dag}\\
&  =\left(  \sum_{j}N_{j}\otimes\vert j\rangle_{E}\right)  \rho_{A}\left(
\sum_{k}N_{k}^{\dag}\otimes\langle k\vert_{E}\right) \\
&  =\sum_{j,k}N_{j}\rho_{A}N_{k}^{\dag}\otimes\vert j\rangle\langle k\vert
_{E},
\end{align}
and tracing out the environment system gives back the original quantum channel
$\mathcal{N}_{A\rightarrow B}$:%
\begin{equation}
\operatorname{Tr}_{E}\left\{  \mathcal{U}_{A\rightarrow BE}^{\mathcal{N}%
}\left(  \rho_{A}\right)  \right\}  =\sum_{j}N_{j}\rho_{A}N_{j}^{\dag}
=\mathcal{N}_{A\rightarrow B}( \rho_{A}) .
\end{equation}

\begin{exercise}
Show that all isometric extensions of a quantum channel are equivalent up to
an isometry on the environment system (this is similar to the result of
Theorem~\ref{thm-pqt:purif-unitary-reference}).
\end{exercise}

\begin{exercise}
\label{ex:iso-extension-erasure}Show that an isometric extension of the
erasure channel is%
\begin{multline}
U_{A\rightarrow BE}^{\mathcal{N}}=\sqrt{1-\varepsilon}(|0\rangle_{B}%
\langle0|_{A}+|1\rangle_{B}\langle1|_{A})\otimes|e\rangle_{E}\\
+\sqrt{\varepsilon}|e\rangle_{B}\langle0|_{A}\otimes|0\rangle_{E}%
+\sqrt{\varepsilon}|e\rangle_{B}\langle1|_{A}\otimes|1\rangle_{E}\\
=\sqrt{1-\varepsilon}I_{A\rightarrow B}\otimes|e\rangle_{E}+\sqrt{\varepsilon
}I_{A\rightarrow E}\otimes|e\rangle_{B}.
\end{multline}

\end{exercise}

\begin{exercise}
Determine the resulting state when Alice inputs an arbitrary pure state
$\vert\psi\rangle$ into an isometric extension of the erasure channel. Verify
that Bob and Eve receive the same ensemble (they have the same local density
operator) when the erasure probability $\varepsilon=1/2$.
\end{exercise}

\begin{exercise}
Show that the matrix representation of an isometric extension of the erasure
channel is%
\begin{equation}
\left[
\begin{array}
[c]{cc}%
\langle0|_{B}\langle0|_{E}U_{A\rightarrow BE}^{\mathcal{N}}|0\rangle_{A} &
\langle0|_{B}\langle0|_{E}U_{A\rightarrow BE}^{\mathcal{N}}|1\rangle_{A}\\
\langle0|_{B}\langle1|_{E}U_{A\rightarrow BE}^{\mathcal{N}}|0\rangle_{A} &
\langle0|_{B}\langle1|_{E}U_{A\rightarrow BE}^{\mathcal{N}}|1\rangle_{A}\\
\langle0|_{B}\langle e|_{E}U_{A\rightarrow BE}^{\mathcal{N}}|0\rangle_{A} &
\langle0|_{B}\langle e|_{E}U_{A\rightarrow BE}^{\mathcal{N}}|1\rangle_{A}\\
\langle1|_{B}\langle0|_{E}U_{A\rightarrow BE}^{\mathcal{N}}|0\rangle_{A} &
\langle1|_{B}\langle0|_{E}U_{A\rightarrow BE}^{\mathcal{N}}|1\rangle_{A}\\
\langle1|_{B}\langle1|_{E}U_{A\rightarrow BE}^{\mathcal{N}}|0\rangle_{A} &
\langle1|_{B}\langle1|_{E}U_{A\rightarrow BE}^{\mathcal{N}}|1\rangle_{A}\\
\langle1|_{B}\langle e|_{E}U_{A\rightarrow BE}^{\mathcal{N}}|0\rangle_{A} &
\langle1|_{B}\langle e|_{E}U_{A\rightarrow BE}^{\mathcal{N}}|1\rangle_{A}\\
\langle e|_{B}\langle0|_{E}U_{A\rightarrow BE}^{\mathcal{N}}|0\rangle_{A} &
\langle e|_{B}\langle0|_{E}U_{A\rightarrow BE}^{\mathcal{N}}|1\rangle_{A}\\
\langle e|_{B}\langle1|_{E}U_{A\rightarrow BE}^{\mathcal{N}}|0\rangle_{A} &
\langle e|_{B}\langle1|_{E}U_{A\rightarrow BE}^{\mathcal{N}}|1\rangle_{A}\\
\langle e|_{B}\langle e|_{E}U_{A\rightarrow BE}^{\mathcal{N}}|0\rangle_{A} &
\langle e|_{B}\langle e|_{E}U_{A\rightarrow BE}^{\mathcal{N}}|1\rangle_{A}%
\end{array}
\right]  =\left[
\begin{array}
[c]{cc}%
0 & 0\\
0 & 0\\
\sqrt{1-\varepsilon} & 0\\
0 & 0\\
0 & 0\\
0 & \sqrt{1-\varepsilon}\\
\sqrt{\varepsilon} & 0\\
0 & \sqrt{\varepsilon}\\
0 & 0
\end{array}
\right]  .
\end{equation}

\end{exercise}

\begin{exercise}
Show that the matrix representation of an isometric extension $U_{A\rightarrow
BE}^{\mathcal{N}}$ of the amplitude damping channel is%
\begin{equation}
\left[
\begin{array}
[c]{cc}%
\langle0\vert_{B}\langle0\vert_{E}U_{A\rightarrow BE}^{\mathcal{N}}%
\vert0\rangle_{A} & \langle0\vert_{B}\langle0\vert_{E}U_{A\rightarrow
BE}^{\mathcal{N}}\vert1\rangle_{A}\\
\langle0\vert_{B}\langle1\vert_{E}U_{A\rightarrow BE}^{\mathcal{N}}%
\vert0\rangle_{A} & \langle0\vert_{B}\langle1\vert_{E}U_{A\rightarrow
BE}^{\mathcal{N}}\vert1\rangle_{A}\\
\langle1\vert_{B}\langle0\vert_{E}U_{A\rightarrow BE}^{\mathcal{N}}%
\vert0\rangle_{A} & \langle1\vert_{B}\langle0\vert_{E}U_{A\rightarrow
BE}^{\mathcal{N}}\vert1\rangle_{A}\\
\langle1\vert_{B}\langle1\vert_{E}U_{A\rightarrow BE}^{\mathcal{N}}%
\vert0\rangle_{A} & \langle1\vert_{B}\langle1\vert_{E}U_{A\rightarrow
BE}^{\mathcal{N}}\vert1\rangle_{A}%
\end{array}
\right]  =\left[
\begin{array}
[c]{cc}%
0 & \sqrt{\gamma}\\
1 & 0\\
0 & 0\\
0 & \sqrt{1-\gamma}%
\end{array}
\right]  .
\end{equation}

\end{exercise}

\begin{exercise}
Consider a full unitary $V_{AE\rightarrow BE}$ such that%
\begin{equation}
\operatorname{Tr}_{E}\left\{  V\left(  \rho_{A}\otimes|0\rangle\langle
0|_{E}\right)  V^{\dag}\right\}
\end{equation}
gives the amplitude damping channel. Show that a matrix representation of
$V$ is%
\begin{multline}
\left[
\begin{array}
[c]{cccc}%
\langle0|_{B}\langle0|_{E}V|0\rangle_{A}|0\rangle_{E} & \langle0|_{B}%
\langle0|_{E}V|0\rangle_{A}|1\rangle_{E} & \langle0|_{B}\langle0|_{E}%
V|1\rangle_{A}|0\rangle_{E} & \langle0|_{B}\langle0|_{E}V|1\rangle
_{A}|1\rangle_{E}\\
\langle0|_{B}\langle1|_{E}V|0\rangle_{A}|0\rangle_{E} & \langle0|_{B}%
\langle1|_{E}V|0\rangle_{A}|1\rangle_{E} & \langle0|_{B}\langle1|_{E}%
V|1\rangle_{A}|0\rangle_{E} & \langle0|_{B}\langle1|_{E}V|1\rangle
_{A}|1\rangle_{E}\\
\langle1|_{B}\langle0|_{E}V|0\rangle_{A}|0\rangle_{E} & \langle1|_{B}%
\langle0|_{E}V|0\rangle_{A}|1\rangle_{E} & \langle1|_{B}\langle0|_{E}%
V|1\rangle_{A}|0\rangle_{E} & \langle1|_{B}\langle0|_{E}V|1\rangle
_{A}|1\rangle_{E}\\
\langle1|_{B}\langle1|_{E}V|0\rangle_{A}|0\rangle_{E} & \langle1|_{B}%
\langle1|_{E}V|0\rangle_{A}|1\rangle_{E} & \langle1|_{B}\langle1|_{E}%
V|1\rangle_{A}|0\rangle_{E} & \langle1|_{B}\langle1|_{E}V|1\rangle
_{A}|1\rangle_{E}%
\end{array}
\right] \\
=\left[
\begin{array}
[c]{cccc}%
0 & -\sqrt{1-\gamma} & \sqrt{\gamma} & 0\\
1 & 0 & 0 & 0\\
0 & 0 & 0 & 1\\
0 & \sqrt{\gamma} & \sqrt{1-\gamma} & 0
\end{array}
\right]  .
\end{multline}

\end{exercise}

\begin{exercise}
Consider the full unitary operator for the amplitude damping channel from the
previous exercise. Show that the density operator%
\begin{equation}
\operatorname{Tr}_{B}\left\{  V\left(  \rho_{A}\otimes|0\rangle\langle
0|_{E}\right)  V^{\dag}\right\}
\end{equation}
that Eve receives has the following matrix representation:%
\begin{equation}%
\begin{bmatrix}
\gamma p & \sqrt{\gamma}\eta^{\ast}\\
\sqrt{\gamma}\eta & 1-\gamma p
\end{bmatrix}
\text{\ \ \ \ if \ \ \ \ } \rho_{A}=%
\begin{bmatrix}
1-p & \eta\\
\eta^{\ast} & p
\end{bmatrix}
.
\end{equation}
By comparing with \eqref{eq-nqt:amp-damp-output}, observe that the output to
Eve is the bit flip of the output of an amplitude damping channel with damping
parameter~$1-\gamma$.
\end{exercise}

\subsubsection{Complementary Channel}

In the purified quantum theory, it is useful to consider all parties that are
participating in a given protocol. One such party is the environment of the
channel, even if it is not necessarily an active participant in a protocol.
However, in a cryptographic setting, in some sense the environment is active,
and we associate it with an eavesdropper, thus personifying it as ``Eve.''

For any quantum channel $\mathcal{N}_{A\rightarrow B}$, there exists an
isometric extension $U_{A\rightarrow BE}^{\mathcal{N}}$ of that channel. The
\index{complementary channel}%
complementary channel $\mathcal{N}_{A\rightarrow E}^{c}$ is a quantum channel
from the sender to the environment, formally defined as follows:

\begin{definition}
[Complementary Channel]Let $\mathcal{N} : \mathcal{L}(\mathcal{H}_{A})
\to\mathcal{L}(\mathcal{H}_{B})$ be a quantum channel, and let $U:\mathcal{H}%
_{A} \to\mathcal{H}_{B} \otimes\mathcal{H}_{E}$ be an isometric extension of
the channel $\mathcal{N}$. The complementary channel $\mathcal{N}^{c} :
\mathcal{L}(\mathcal{H}_{A}) \to\mathcal{L}(\mathcal{H}_{E})$ of $\mathcal{N}%
$, associated with $U$, is defined as follows:
\begin{equation}
\mathcal{N}^{c}( X_{A}) =\operatorname{Tr}_{B}\left\{  U X_{A} U^{\dag
}\right\}  ,
\end{equation}
for $X_{A} \in\mathcal{L}(\mathcal{H}_{A})$.
\end{definition}

That is, we obtain a complementary channel by tracing out Bob's system $B$
from the output of an isometric extension. It captures the noise that Eve
\textquotedblleft sees\textquotedblright\ by having her system coupled to
Bob's system.

\begin{exercise}
Show that Eve's density operator (the output of a complementary channel) is of
the following form:%
\begin{equation}
\rho\rightarrow\sum_{i,j}\operatorname{Tr}\{ N_{i}\rho N_{j}^{\dag}\} \vert
i\rangle\langle j\vert,
\end{equation}
if we take an isometric extension of the channel to be of the form in \eqref{eq-qt:channel-isometry}.
\end{exercise}

The complementary channel is unique only up to an isometry acting on Eve's
system. It inherits this property from the fact that an isometric extension of
a quantum channel is unique only up to isometries acting on Eve's system. For
all practical purposes, this lack of uniqueness does not affect our study of
the noise that Eve sees because the measures of noise in
Chapter~\ref{chap:q-info-entropy}\ are invariant with respect to isometries
acting on Eve's system.

\subsection{Further Examples of Isometric Extensions}

\subsubsection{Generalized Dephasing Channels}

\label{sec-nqt:gen-deph}A generalized dephasing channel%
\index{dephasing channel!generalized}
is one that preserves states diagonal in some preferred orthonormal basis
$\left\{  |x\rangle\right\}  $, but it can add arbitrary phases to the
off-diagonal elements of a density operator represented in this basis. An
isometric extension of a generalized dephasing channel acts as follows on the
basis $\left\{  |x\rangle\right\}  $:%
\begin{equation}
U_{A\rightarrow BE}^{\mathcal{N}_{\operatorname{D}}}|x\rangle_{A}%
=|x\rangle_{B}|\varphi_{x}\rangle_{E},
\end{equation}
where $|\varphi_{x}\rangle_{E}$ is some state for the environment (these
states need not be mutually orthogonal). Thus, we can represent the isometry
as follows:%
\begin{equation}
U_{A\rightarrow BE}^{\mathcal{N}_{\operatorname{D}}}\equiv\sum_{x}%
|x\rangle_{B}|\varphi_{x}\rangle_{E}\langle x|_{A},
\end{equation}
and its action on a density operator $\rho$ is%
\begin{equation}
U^{\mathcal{N}_{\operatorname{D}}}\rho\left(  U^{\mathcal{N}_{\operatorname{D}%
}}\right)  ^{\dag}=\sum_{x,x^{\prime}}\langle x|\rho|x^{\prime}\rangle
\ \ |x\rangle\langle x^{\prime}|_{B}\otimes|\varphi_{x}\rangle\langle
\varphi_{x^{\prime}}|_{E}.
\end{equation}
Tracing out the environment gives the action of the channel $\mathcal{N}%
_{\operatorname{D}}$\ to the receiver%
\begin{equation}
\mathcal{N}_{\operatorname{D}}(\rho)=\sum_{x,x^{\prime}}\langle x|\rho
|x^{\prime}\rangle\langle\varphi_{x^{\prime}}|\varphi_{x}\rangle
\ \ |x\rangle\langle x^{\prime}|_{B},
\end{equation}
where we observe that this channel preserves the diagonal components $\left\{
|x\rangle\langle x|\right\}  $ of $\rho$, but it multiplies the $d\left(
d-1\right)  $ off-diagonal elements of $\rho$ by arbitrary phases, depending
on the $d\left(  d-1\right)  $ overlaps $\left\langle \varphi_{x^{\prime}%
}|\varphi_{x}\right\rangle $ of the environment states (where $x\neq
x^{\prime}$). Tracing out the receiver gives the action of the complementary
channel $\mathcal{N}_{\operatorname{D}}^{c}$ to the environment%
\begin{equation}
\mathcal{N}_{\operatorname{D}}^{c}(\rho)=\sum_{x}\langle x|\rho|x\rangle
\ |\varphi_{x}\rangle\langle\varphi_{x}|_{E}.
\end{equation}
Observe that the channel to the environment is
\index{entanglement-breaking channel}%
entanglement-breaking. That is, the action of the channel is the same as first
performing a complete projective measurement in the basis $\left\{
|x\rangle\right\}  $ and preparing a state $|\varphi_{x}\rangle_{E}$
conditioned on the outcome of the measurement (it is a classical--quantum
channel, as discussed in Section~\ref{sec-nqt:entanglement-breaking}).
Additionally, the receiver Bob can simulate the action of this channel to the
receiver by performing the same actions on the state that he receives.

\begin{exercise}
Explicitly show that the following qubit dephasing channel is a special case
of a generalized dephasing channel:%
\begin{equation}
\rho\rightarrow( 1-p) \rho+pZ\rho Z.
\end{equation}

\end{exercise}

\subsubsection{Quantum Hadamard Channels}

\label{sec-nqt:hadamard-channel}Quantum Hadamard channels%
\index{Hadamard channel}
are those whose complements are
\index{entanglement-breaking channel}%
entanglement-breaking, and so generalized dephasing channels are a subclass of
quantum Hadamard channels. We can write the output of a quantum Hadamard
channel as the Hadamard product (element-wise multiplication) of a
representation of the input density operator with another operator. To discuss
how this comes about, suppose that the complementary channel$~\mathcal{N}%
_{A\rightarrow E}^{c}$\ of a channel$~\mathcal{N}_{A\rightarrow B}$ is
entanglement-breaking. Then, using the fact that its Kraus operators
$\left\vert \xi_{i}\right\rangle _{E}\left\langle \zeta_{i}\right\vert _{A}$
are unit rank (see Theorem~\ref{thm-nqt:ent-break-unit-rank}) and the
construction in \eqref{eq-qt:channel-isometry} for an isometric extension, we
can write an isometric extension$~U^{\mathcal{N}^{c}}$\ for $\mathcal{N}^{c}$
as%
\begin{align}
U^{\mathcal{N}^{c}}\rho_{A}\left(  U^{\mathcal{N}^{c}}\right)  ^{\dag}  &
=\sum_{i,j}\left\vert \xi_{i}\right\rangle _{E}\left\langle \zeta
_{i}\right\vert _{A}\rho_{A}\left\vert \zeta_{j}\right\rangle _{A}\left\langle
\xi_{j}\right\vert _{E}\otimes\left\vert i\right\rangle _{B}\langle j|_{B}\\
&  =\sum_{i,j}\left\langle \zeta_{i}\right\vert _{A}\rho_{A}\left\vert
\zeta_{j}\right\rangle _{A}\left\vert \xi_{i}\right\rangle _{E}\left\langle
\xi_{j}\right\vert _{E}\otimes|i\rangle_{B}\langle j\vert_{B}.
\label{eq:iso-Hadamard}%
\end{align}
The sets $\{\left\vert \xi_{i}\right\rangle _{E}\}$ and $\{\left\vert
\zeta_{i}\right\rangle _{A}\}$ each do not necessarily consist of orthonormal
states, but the set $\{|i\rangle_{B}\}$ does because it is the environment of
the complementary channel. Tracing over the system $E$ gives the original
channel from system $A$ to $B$:%
\begin{equation}
\mathcal{N}_{A\rightarrow B}^{\operatorname{H}}(\rho_{A})=\sum_{i,j}%
\left\langle \zeta_{i}\right\vert _{A}\rho_{A}\left\vert \zeta_{j}%
\right\rangle _{A}\left\langle \xi_{j}|\xi_{i}\right\rangle _{E}|i\rangle
_{B}\langle j|_{B}. \label{eq:hadamard-product}%
\end{equation}
Let $\Sigma$ denote the matrix with elements$~\left[  \Sigma\right]
_{i,j}=\left\langle \zeta_{i}\right\vert _{A}\rho_{A}\left\vert \zeta
_{j}\right\rangle _{A}$, a representation of the input state $\rho$, and let
$\Gamma$ denote the matrix with elements$~\left[  \Gamma\right]
_{i,j}=\left\langle \xi_{i}|\xi_{j}\right\rangle _{E}$. Then, from
\eqref{eq:hadamard-product}, it is clear that the output of the channel is the
Hadamard product $\ast$\ of $\Sigma$ and $\Gamma^{\dag}$ with respect to the
basis $\{|i\rangle_{B}\}$:%
\begin{equation}
\mathcal{N}_{A\rightarrow B}^{\operatorname{H}}(\rho)=\Sigma\ast\Gamma^{\dag}.
\end{equation}
For this reason, such a channel is known as a Hadamard channel.

Hadamard channels are
\index{degradable channel}%
\textit{degradable}, as introduced in the following definition:

\begin{definition}
[Degradable Channel]%
\index{degradable channel}
Let $\mathcal{N}_{A\rightarrow B}$ be a quantum channel, and let
$\mathcal{N}_{A\rightarrow E}^{c}$ denote a complementary channel for
$\mathcal{N}_{A\rightarrow B}$. The channel $\mathcal{N}_{A\rightarrow B}$ is
degradable if there exists a degrading channel $\mathcal{D}_{B\rightarrow E}$
such that
\begin{equation}
\mathcal{D}_{B\rightarrow E}(\mathcal{N}_{A\rightarrow B}(X_{A}))=\mathcal{N}%
_{A\rightarrow E}^{c}(X_{A}),
\end{equation}
for all $X_{A}\in\mathcal{L}(\mathcal{H}_{A})$.
\end{definition}

To see that a quantum Hadamard channel is degradable, let Bob perform a
complete projective measurement of his state in the basis $\{|i\rangle_{B}\}$
and prepare the state $\left\vert \xi_{i}\right\rangle _{E}$ conditioned on
the outcome of the measurement. This procedure simulates the complementary
channel$~\mathcal{N}_{A\rightarrow E}^{c}$ and also implies that the degrading
channel~$\mathcal{D}_{B\rightarrow E}$ is
\index{entanglement-breaking channel}%
entanglement-breaking. To be more precise, the Kraus operators of the
degrading channel$~\mathcal{D}_{B\rightarrow E}$\ are $\{\left\vert \xi
_{i}\right\rangle _{E}\langle i|_{B}\}$ so that%
\begin{align}
\mathcal{D}_{B\rightarrow E}(\mathcal{N}_{A\rightarrow B}^{\operatorname{H}%
}(\sigma_{A}))  &  =\sum_{i}\left\vert \xi_{i}\right\rangle _{E}\langle
i|_{B}\mathcal{N}_{A\rightarrow B}(\sigma_{A})|i\rangle_{B}\left\langle
\xi_{i}\right\vert _{E}\\
&  =\sum_{i}\langle\zeta_{i}|_{A}\sigma_{A}\left\vert \zeta_{i}\right\rangle
_{A}\left\vert \xi_{i}\right\rangle \left\langle \xi_{i}\right\vert _{E},
\end{align}
demonstrating that this degrading channel simulates the complementary
channel$~\mathcal{N}_{A\rightarrow E}^{\operatorname{H}}$. Note that we can
view this degrading channel as the composition of two channels:\ a first
channel $\mathcal{D}_{B\rightarrow Y}^{1}$ performs the complete projective
measurement, leading to a classical variable $Y$, and a second channel
$\mathcal{D}_{Y\rightarrow E}^{2}$ performs the state preparation, conditioned
on the value of the classical variable $Y$. We can therefore write
$\mathcal{D}_{B\rightarrow E}=\mathcal{D}_{Y\rightarrow E}^{2}\circ
\mathcal{D}_{B\rightarrow Y}^{1}$. This particular form of the channel has
implications for its quantum capacity (see Chapter~\ref{chap:quantum-capacity}%
) and its more general capacities (see Chapter~\ref{chap:trade-off}). Observe
that a generalized dephasing channel from the previous section is a quantum
Hadamard channel because the channel to its environment is
\index{entanglement-breaking channel}%
entanglement-breaking.

\subsection{Isometric Extension and Adjoint of a Quantum Channel}

Recall the notion of an adjoint of a quantum channel from
Section~\ref{sec-nqt:adjoint-unital}. Here we show an alternate way of
representing an adjoint of a quantum channel using an isometric extension of it.

\begin{proposition}
\label{prop-pqt:adjoint-iso-ext}Let $\mathcal{N}:\mathcal{L}(\mathcal{H}%
_{A})\rightarrow\mathcal{L}(\mathcal{H}_{B})$ be a quantum channel and let
$U:\mathcal{H}_{A}\rightarrow\mathcal{H}_{B}\otimes\mathcal{H}_{E}$ be an
isometric extension of it. Then the adjoint map $\mathcal{N}^{\dag
}:\mathcal{L}(\mathcal{H}_{B})\rightarrow\mathcal{L}(\mathcal{H}_{A})$\ can be
written as follows:%
\begin{equation}
\mathcal{N}^{\dag}(Y_{B})=U^{\dag}(Y_{B}\otimes I_{E})U,
\label{eq-pqt:adjoint-isometric-ext}%
\end{equation}
for $Y_{B}\in\mathcal{L}(\mathcal{H}_{B})$.
\end{proposition}

\begin{proof}
We can see this by using the definition of the adjoint map
(Definition~\ref{def-nqt:adjoint-map}), the definition of an isometric
extension (Definition~\ref{def-pqt:isometric-ext}), and the definition of
partial trace (Definition~\ref{def-nqt:partial-trace}). Consider from the
definition of the adjoint map that $\mathcal{N}^{\dag}$ is such that%
\begin{equation}
\left\langle Y_{B},\mathcal{N}(X_{A})\right\rangle =\left\langle
\mathcal{N}^{\dag}(Y_{B}),X_{A}\right\rangle ,
\end{equation}
for all $X_{A}\in\mathcal{L}(\mathcal{H}_{A})$\ and $Y_{B}\in\mathcal{L}%
(\mathcal{H}_{B})$. Then%
\begin{align}
\left\langle Y_{B},\mathcal{N}(X_{A})\right\rangle  &  =\operatorname{Tr}%
\{Y_{B}^{\dag}\mathcal{N}(X_{A})\}\\
&  =\operatorname{Tr}\{Y_{B}^{\dag}\operatorname{Tr}_{E}\{UX_{A}U^{\dag}\}\}\\
&  =\operatorname{Tr}\{(Y_{B}^{\dag}\otimes I_{E})UX_{A}U^{\dag}\}\\
&  =\operatorname{Tr}\{U^{\dag}(Y_{B}^{\dag}\otimes I_{E})UX_{A}\}\\
&  =\operatorname{Tr}\{\left[  U^{\dag}(Y_{B}\otimes I_{E})U\right]  ^{\dag
}X_{A}\}\\
&  =\left\langle U^{\dag}\left(  Y_{B}\otimes I_{E}\right)  U,X_{A}%
\right\rangle .
\end{align}
The second equality is from the definition of an isometric extension. The
third equality follows by applying the definition of partial trace. The fourth
uses cyclicity of trace. Since we have shown that $\left\langle Y_{B}%
,\mathcal{N}(X_{A})\right\rangle =\left\langle U^{\dag}\left(  Y_{B}\otimes
I_{E}\right)  U,X_{A}\right\rangle $ for all $X_{A}\in\mathcal{L}%
(\mathcal{H}_{A})$\ and $Y_{B}\in\mathcal{L}(\mathcal{H}_{B})$, the statement
in \eqref{eq-pqt:adjoint-isometric-ext} follows.
\end{proof}

We can verify the formula in \eqref{eq-pqt:adjoint-isometric-ext} in a
different way. Suppose that we have a Kraus representation of the channel
$\mathcal{N}$ as follows:%
\begin{equation}
\mathcal{N}(X_{A})=\sum_{l}V_{l}X_{A}V_{l}^{\dag},
\end{equation}
where $V_{l}\in\mathcal{L}(\mathcal{H}_{A},\mathcal{H}_{B})$ for all $l$ and
$\sum_{l}V_{l}^{\dag}V_{l}=I_{A}$. An isometric extension $U$\ for this
channel is then as given in \eqref{eq-qt:channel-isometry}:%
\begin{equation}
U=\sum_{l}V_{l}\otimes|l\rangle_{E},
\end{equation}
where $\{|l\rangle_{E}\}$ is some orthonormal basis. We can then explicitly
compute the formula in \eqref{eq-pqt:adjoint-isometric-ext} as follows:%
\begin{align}
U^{\dag}(Y_{B}\otimes I_{E})U  &  =\left(  \sum_{l}V_{l}^{\dag}\otimes\langle
l|_{E}\right)  (Y_{B}\otimes I_{E})\left(  \sum_{l^{\prime}}V_{l^{\prime}%
}\otimes|l^{\prime}\rangle_{E}\right) \\
&  =\sum_{l,l^{\prime}}V_{l}^{\dag}Y_{B}V_{l^{\prime}}\langle l|l^{\prime
}\rangle_{E} =\sum_{l}V_{l}^{\dag}Y_{B}V_{l} =\mathcal{N}^{\dag}(Y_{B}),
\end{align}
where the last equality follows from what we calculated before in \eqref{eq-nqt:adjoint-map-kraus-ops}.

\section{Coherent Quantum Instrument}%

\index{quantum instrument!coherent}%
It is useful to consider an isometric extension of a quantum instrument (we
discussed quantum instruments in Section~\ref{sec-pqt:instrument}). This
viewpoint is important when we recall that a quantum instrument is the most
general map from a quantum system to a quantum system and a classical system.

Recall from Section~\ref{sec-pqt:instrument} that a quantum instrument acts as
follows on an input $\rho_{A}\in\mathcal{D}(\mathcal{H}_{A})$:%
\begin{equation}
\rho_{A}\rightarrow\sum_{j}\mathcal{E}_{A\rightarrow B}^{j}( \rho_{A})
\otimes\vert j\rangle\langle j\vert_{J}, \label{eq-pqt:instrument}%
\end{equation}
where each $\mathcal{E}_{A\rightarrow B}^{j}$ is a completely positive
trace-non-increasing (CPTNI) map that has the following form:%
\begin{equation}
\mathcal{E}_{A\rightarrow B}^{j}( \rho_{A}) =\sum_{k}M_{j,k}\rho_{A}
M_{j,k}^{\dag},
\end{equation}
such that $\sum_{k}M_{j,k}^{\dag}M_{j,k}\leq I$ for all $j$.

We now describe a particular coherent evolution that implements the above
transformation when we trace over certain degrees of freedom. A pure extension
of each CPTNI map $\mathcal{E}_{j}$ is as follows:%
\begin{equation}
U_{A\rightarrow BE}^{\mathcal{E}_{j}}\equiv\sum_{k}M_{j,k}\otimes|k\rangle
_{E},
\end{equation}
where the operator $M_{j,k}$ acts on the input system $A$ and the environment
system $E$ is large enough to accomodate all of the CPTNI maps $\mathcal{E}%
_{j}$. That is, if the first map $\mathcal{E}_{1}$ has states $\left\{
|1\rangle_{E},\ldots,\left\vert d_{1}\right\rangle _{E}\right\}  $, then the
second map $\mathcal{E}_{2}$ has states $\left\{  \left\vert d_{1}%
+1\right\rangle _{E},\ldots,\left\vert d_{1}+d_{2}\right\rangle _{E}\right\}
$ so that the states on $E$ are orthogonal for all the different maps
$\mathcal{E}_{j}$ that are part of the instrument. We can embed this pure
extension into the evolution in \eqref{eq-pqt:instrument} as follows:%
\begin{equation}
\rho_{A}\rightarrow\sum_{j}\mathcal{U}_{A\rightarrow BE}^{\mathcal{E}_{j}%
}(\rho_{A})\otimes|j\rangle\langle j|_{J},
\end{equation}
where $\mathcal{U}_{A\rightarrow BE}^{\mathcal{E}_{j}}(\rho_{A}%
)=U_{A\rightarrow BE}^{\mathcal{E}_{j}}(\rho_{A})(U_{A\rightarrow
BE}^{\mathcal{E}_{j}})^{\dag}$. This evolution is not quite fully coherent,
but a simple modification of it does make it fully coherent:%
\begin{equation}
\sum_{j}U_{A\rightarrow BE}^{\mathcal{E}_{j}}\otimes|j\rangle_{J}%
\otimes|j\rangle_{E_{J}}.
\end{equation}
The full action of the coherent instrument is then as follows:%
\begin{align}
\rho_{A}  &  \rightarrow\sum_{j,j^{\prime}}U_{A\rightarrow BE}^{\mathcal{E}%
_{j}}\rho_{A}\left(  U_{A\rightarrow BE}^{\mathcal{E}_{j^{\prime}}}\right)
^{\dag}\otimes|j\rangle\langle j^{\prime}\vert_{J}\otimes|j\rangle\langle
j^{\prime}\vert_{E_{J}}\\
&  =\sum_{j,k,j^{\prime},k^{\prime}}M_{j,k}\rho_{A}M_{j^{\prime},k^{\prime}%
}^{\dag}\otimes|k\rangle\langle k^{\prime}|_{E}\otimes|j\rangle\langle
j^{\prime}\vert_{J}\otimes|j\rangle\langle j^{\prime}\vert_{E_{J}}.
\end{align}
One can then check that tracing over the environmental degrees of freedom $E$
and $E_{J}$ reproduces the action of the quantum instrument in \eqref{eq-pqt:instrument}.

\section{Coherent Measurement}

\label{sec-pt:coherent-measurement}We end this chapter by discussing a
coherent measurement. This last section, combined with the notion of an
isometric extension of a quantum channel, shows that it is sufficient to
describe all of the quantum theory in the so-called \textquotedblleft
traditionalist\textquotedblright\ way by using only unitary evolutions and von
Neumann (complete projective) measurements.

Suppose that we have a set of measurement operators $\left\{  M_{j}\right\}
_{j}$ such that $\sum_{j}M_{j}^{\dag}M_{j}=I$. In the noisy quantum theory, we
found that the post-measurement state of a measurement on a quantum system $S$
with density operator $\rho$ is%
\begin{equation}
\frac{M_{j}\rho M_{j}^{\dag}}{p_{J}( j) }, \label{eq-qt:no-coherent-meas}%
\end{equation}
where the measurement outcome $j$ occurs with probability%
\begin{equation}
p_{J}( j) =\operatorname{Tr}\left\{  M_{j}^{\dag}M_{j}\rho\right\}  .
\end{equation}

We would like a way to perform the above measurement on system $S$ in a
\textit{coherent} fashion. The isometry in \eqref{eq-qt:channel-isometry}
gives a hint for how we can structure such a coherent measurement. We can
build the coherent measurement as the following isometry:%
\begin{equation}
U_{S\rightarrow SS^{\prime}}\equiv\sum_{j}M_{S}^{j}\otimes\left\vert
j\right\rangle _{S^{\prime}}.
\end{equation}
Appying this isometry to a density operator $\rho_{S}$ gives the following
state:
\begin{align}
\mathcal{U}_{S\rightarrow SS^{\prime}}(\rho_{S})  &  ={U}_{S\rightarrow
SS^{\prime}}\rho_{S}({U}_{S\rightarrow SS^{\prime}})^{\dag}\\
&  =\sum_{j,j^{\prime}}M_{S}^{j}\rho_{S}(M_{S}^{j^{\prime}})^{\dag}%
\otimes|j\rangle\langle j^{\prime}\vert_{S^{\prime}}.
\end{align}
We can then apply a complete projective measurement with projection operators
$\left\{  |j\rangle\langle j|\right\}  _{j}$ to the system $S^{\prime}$, which
gives the following post-measurement state:%
\begin{multline}
\frac{(I_{S}\otimes|j\rangle\langle j|_{S^{\prime}})(\mathcal{U}_{S\rightarrow
SS^{\prime}}(\rho_{S}))(I_{S}\otimes\vert j\rangle\langle j|_{S^{\prime}}%
)}{\operatorname{Tr}\left\{  (I_{S}\otimes|j\rangle\langle j|_{S^{\prime}%
})(U_{S\rightarrow SS^{\prime}}( \rho_{S}) )\right\}  }\\
=\frac{M_{S}^{j}\rho_{S}(M_{S}^{j})^{\dag}}{\operatorname{Tr}\left\{
(M_{S}^{j})^{\dag}M_{S}^{j}\rho_{S}\right\}  }\otimes|j\rangle\langle
j\vert_{S^{\prime}}.
\end{multline}
The result is then the same as that in \eqref{eq-qt:no-coherent-meas}. In
fact, this is the same as the way in which Section~\ref{sec-nqt:more-gen-meas}
motivated an alternate description of quantum measurements.

\begin{exercise}
\label{ex-pt:coherent-measurement}Suppose that there is a set of density
operators $\rho_{S}^{k}$ and a POVM\ $\left\{  \Lambda_{S}^{k}\right\}  $ that
identifies these states with high probability, in the sense that%
\begin{equation}
\forall k\ \ \ \ \operatorname{Tr}\left\{  \Lambda_{S}^{k}\rho_{S}%
^{k}\right\}  \geq1-\varepsilon,
\end{equation}
where $\varepsilon\in(0,1)$. Construct a coherent measurement $U_{S\rightarrow
SS^{\prime}}$ and show that the coherent measurement has a high probability of
success in the sense that%
\begin{equation}
\left\vert \left\langle \phi_{k}\right\vert _{RS}\langle k|_{S^{\prime}%
}U_{S\rightarrow SS^{\prime}}\left\vert \phi_{k}\right\rangle _{RS}\right\vert
\geq1-\varepsilon,
\end{equation}
where each $\left\vert \phi_{k}\right\rangle _{RS}$ is a purification of
$\rho_{k}$.
\end{exercise}

\section{History and Further Reading}

The purified view of quantum mechanics has long been part of quantum
information theory (e.g., see \cite{book2000mikeandike} or \cite{Yard05a}).
Early work of \cite{S55} showed that every linear CPTP map can be realized as
a linear isometry with an output on a larger Hilbert space and followed by a
partial trace. \cite{PhysRevA.71.032314} discussed some of the observations
about the amplitude damping channel that appear in our exercises.
\cite{cmp2005dev} introduced generalized dephasing channels in the context of
trade-off coding and they also introduced the notion of a degradable quantum
channel. \cite{KMNR07} studied the quantum Hadamard channels. Coherent
instruments and measurements appeared in
\citep{DW04,ieee2005dev,itit2008hsieh}\ as part of the decoder used in several
quantum coding theorems. We exploit them in
Chapters~\ref{chap:quantum-capacity} and~\ref{chap:trade-off}.

\part{Unit Quantum Protocols}

\chapter{Three Unit Quantum Protocols}

\label{chap:three-noiseless}This chapter begins our first exciting application
of the postulates of the quantum theory to quantum communication. We study the
fundamental, unit quantum communication protocols. These protocols involve a
single sender Alice and a single receiver Bob. The protocols are ideal and
noiseless because we assume that Alice and Bob can exploit perfect classical
communication, perfect quantum communication, and perfect entanglement. At the
end of this chapter, we suggest how to incorporate imperfections into these
protocols for later study.

Alice and Bob may wish to perform one of several quantum
information-processing tasks, such as the transmission of classical
information, quantum information, or entanglement. Several fundamental
protocols make use of these resources:

\begin{enumerate}
\item We will see that noiseless entanglement is an important resource in
quantum Shannon theory because it enables Alice and Bob to perform other
protocols that are not possible with classical resources only. We will present
a simple, idealized protocol for generating entanglement, named
\index{entanglement distribution}%
\textit{entanglement distribution}.

\item Alice may wish to communicate classical information to Bob. A trivial
method, named \textit{elementary coding}, is a simple way of doing so and we
discuss it briefly.

\item A more interesting technique for transmitting classical information is
\index{super-dense coding}%
\textit{super-dense coding}. It exploits a noiseless qubit channel and shared
entanglement to transmit more classical information than would be possible
with a noiseless qubit channel alone.

\item Finally, Alice may wish to transmit quantum information to Bob. A
trivial method for her to do so is to exploit a noiseless qubit channel.
However, it is useful to have other ways for transmitting quantum information
because such a resource is difficult to engineer in practice. An alternative,
surprising method for transmitting quantum information is
\index{quantum teleportation}%
\textit{quantum teleportation}. The teleportation protocol exploits classical
communication and shared entanglement to transmit quantum information.
\end{enumerate}

Each of these protocols is a fundamental unit protocol and provides a
foundation for asking further questions in quantum Shannon theory. In fact,
the discovery of these latter two protocols was the stimulus for much of the
original research in quantum Shannon theory. One could take each of these
protocols and ask about its performance if one or more of the resources
involved is noisy rather than noiseless. Later chapters of this book explore
many of these possibilities.

This chapter introduces the technique of \textit{resource counting}, which is
of practical importance because it quantifies the communication cost of
achieving a certain task. We include only non-local resources in a resource
count---non-local resources include classical or quantum communication or
shared entanglement.

It is important to minimize the use of certain resources, such as noiseless
entanglement or a noiseless qubit channel, in a given protocol because they
are expensive. Given a certain implementation of a quantum
information-processing task, we may wonder if there is a way of implementing
it that consumes fewer resources. A proof that a given protocol is the best
that we can hope to do is an optimality proof (also known as a converse proof,
as discussed in Section~\ref{intro:compression}). We argue, based on good
physical grounds, that the protocols in this chapter are the best
implementations of the desired quantum information-processing task.
Chapter~\ref{chap:trade-off} gives information-theoretic proofs of optimality.

\section{Non-Local Unit Resources}

We first briefly define what we mean by a noiseless qubit channel, a noiseless
classical bit channel, and noiseless entanglement. Each of these resources is
a \textit{non-local, unit resource}. A resource is \textit{non-local} if two
spatially separated parties share it or if one party uses it to communicate to
another. We say that a resource is \textit{unit} if it comes in some
\textquotedblleft gold standard\textquotedblright\ form, such as qubits,
classical bits, or entangled bits. It is important to establish these
definitions so that we can check whether a given protocol is truly simulating
one of these resources.

A noiseless qubit channel is any mechanism that implements the following map:%
\begin{equation}
\vert i\rangle_{A}\rightarrow\vert i\rangle_{B},
\end{equation}
extended linearly to arbitrary state vectors and where $i\in\left\{
0,1\right\}  $, $\{\vert0\rangle_{A},\vert1\rangle_{A}\}$ is some preferred
orthonormal basis on Alice's system, and $\{\vert0\rangle_{B},\vert
1\rangle_{B}\}$ is some preferred orthonormal basis on Bob's system. The bases
do not have to be the same, but it must be clear which basis each party is
using. The above map is linear so that it preserves arbitrary superposition
states (it preserves any qubit). For example, the map acts as follows on a
superposition state:%
\begin{equation}
\alpha\vert0\rangle_{A}+\beta\vert1\rangle_{A}\rightarrow\alpha\vert
0\rangle_{B}+\beta\vert1\rangle_{B}.
\end{equation}
We can also write it as the following isometry:%
\begin{equation}
\sum_{i=0}^{1}\vert i\rangle_{B}\langle i\vert_{A}.
\end{equation}
Any information-processing protocol that implements the above map simulates a
noiseless qubit channel. We label the communication resource of a noiseless
qubit channel as follows:%
\begin{equation}
\left[  q\rightarrow q\right]  ,
\end{equation}
where the notation indicates one forward use of a noiseless qubit channel.

A noiseless classical bit channel is any mechanism that implements the
following map:%
\begin{align}
|i\rangle\langle i|_{A}  &  \rightarrow|i\rangle\langle i|_{B},\\
|i\rangle\langle j|_{A}  &  \rightarrow0\ \ \ \text{for }i\neq j,
\end{align}
extended linearly to density operators and where $i,j\in\left\{  0,1\right\}
$ and the orthonormal bases are again arbitrary. This channel maintains the
diagonal elements of a density operator in the basis $\left\{  |0\rangle
_{A},\vert1\rangle_{A}\right\}  $, but it eliminates the off-diagonal
elements. We can write it as the following linear map acting on a density
operator $\rho_{A}$:%
\begin{equation}
\rho_{A}\rightarrow\sum_{i=0}^{1}|i\rangle_{B}\langle i|_{A}\rho_{A}%
|i\rangle_{A}\langle i|_{B}.
\end{equation}
The form above is consistent with
Definition~\ref{def-nqt:classical-channel-dephase} for noiseless classical
channels. This resource is weaker than a noiseless qubit channel because it
does not require Alice and Bob to maintain arbitrary superposition states---it
merely transfers classical information. Alice can use the above channel to
transmit classical information to Bob. She can prepare either of the classical
states $|0\rangle\langle0|$ or $\vert1\rangle\langle1|$, send it through the
classical channel, and Bob performs a computational basis measurement to
determine the message Alice transmits. We denote the communication resource of
a noiseless classical bit channel as follows:%
\begin{equation}
\left[  c\rightarrow c\right]  ,
\end{equation}
where the notation indicates one forward use of a noiseless classical bit channel.

We can study other ways of transmitting classical information. For example,
suppose that Alice flips a fair coin that chooses the state $|0\rangle_{A}$ or
$|1\rangle_{A}$ with equal probability. The resulting state is the following
density operator:%
\begin{equation}
\frac{1}{2}\left(  |0\rangle\langle0|_{A}+|1\rangle\langle1|_{A}\right)
\end{equation}
Suppose that she sends the above state through a noiseless classical channel.
The resulting density operator for Bob is as follows:%
\begin{equation}
\frac{1}{2}\left(  |0\rangle\langle0|_{B}+|1\rangle\langle1|_{B}\right)  .
\end{equation}

The above classical bit channel map does not preserve off-diagonal elements of
a density operator. Suppose instead that Alice prepares a superposition state%
\begin{equation}
\frac{\vert0\rangle_{A}+\vert1\rangle_{A}}{\sqrt{2}}.
\end{equation}
The density operator corresponding to this state is%
\begin{equation}
\frac{1}{2}\left(  \vert0\rangle\langle0\vert_{A}+\vert0\rangle\langle
1\vert_{A}+\vert1\rangle\langle0\vert_{A}+\vert1\rangle\langle1\vert
_{A}\right)  .
\end{equation}
Suppose Alice then transmits this state through the above classical channel.
The classical channel eliminates all the off-diagonal elements of the density
operator and the resulting state for Bob is as follows:%
\begin{equation}
\frac{1}{2}\left(  \vert0\rangle\langle0\vert_{B}+\vert1\rangle\langle
1\vert_{B}\right)  .
\end{equation}
Thus, it is impossible for a noiseless classical channel to simulate a
noiseless qubit channel because it cannot maintain arbitrary superposition
states. However, it is possible for a noiseless qubit channel to simulate a
noiseless classical bit channel, and we denote this fact with the following
\textit{resource inequality}:%
\begin{equation}
\left[  q\rightarrow q\right]  \geq\left[  c\rightarrow c\right]  .
\end{equation}
Noiseless quantum communication is therefore a stronger resource than
noiseless classical communication.

\begin{exercise}
\label{ex-3np:dephasing-classical}Show that the noisy dephasing channel in
\eqref{eq-qt:dephasing-channel} with $p=1/2$ is equal to a noiseless classical
bit channel.
\end{exercise}

The final resource that we consider is shared entanglement. The ebit is our
\textquotedblleft gold standard\textquotedblright\ resource for pure bipartite
(two-party) entanglement, and we will make this point more clear operationally
in Chapter~\ref{chap:ent-conc}. An ebit is the following state of two qubits:%
\begin{equation}
\left\vert \Phi^{+}\right\rangle _{AB}\equiv\frac{1}{\sqrt{2}}\left(
|00\rangle_{AB}+|11\rangle_{AB}\right)  ,
\end{equation}
where Alice possesses the first qubit and Bob possesses the second.

Below, we show how a noiseless qubit channel can generate a noiseless ebit
through a simple protocol named \textit{entanglement distribution}. However,
an ebit cannot simulate a noiseless qubit channel (for reasons which we
explain later). Therefore, noiseless quantum communication is the strongest of
all three resources, and entanglement and classical communication are in some
sense \textquotedblleft orthogonal\textquotedblright\ to one another because
neither can simulate the other.

\section{Protocols}

\subsection{Entanglement Distribution}

\label{sec-3np:ent-dist}The entanglement distribution%
\index{entanglement distribution}
protocol is the most basic of the three unit protocols. It exploits one use of
a noiseless qubit channel to establish one shared noiseless ebit. It consists
of the following two steps:

\begin{enumerate}
\item Alice prepares a Bell state locally in her laboratory. She prepares two
qubits in the state $|0\rangle_{A}|0\rangle_{A^{\prime}}$, where we label the
first qubit as $A$ and the second qubit as $A^{\prime}$. She performs a
Hadamard gate on qubit $A$ to produce the following state:%
\begin{equation}
\left(  \frac{|0\rangle_{A}+|1\rangle_{A}}{\sqrt{2}}\right)  |0\rangle
_{A^{\prime}}.
\end{equation}
She then performs a CNOT\ gate with qubit $A$ as the source qubit and qubit
$A^{\prime}$ as the target qubit. The state becomes the following Bell state:%
\begin{equation}
\left\vert \Phi^{+}\right\rangle _{AA^{\prime}}=\frac{|00\rangle_{AA^{\prime}%
}+|11\rangle_{AA^{\prime}}}{\sqrt{2}}.
\end{equation}

\item She sends qubit $A^{\prime}$ to Bob with one use of a noiseless qubit
channel. Alice and Bob then share the ebit $\left\vert \Phi^{+}\right\rangle
_{AB}$.
\end{enumerate}

\begin{figure}
[ptb]
\begin{center}
\includegraphics[
width=4.7608in
]%
{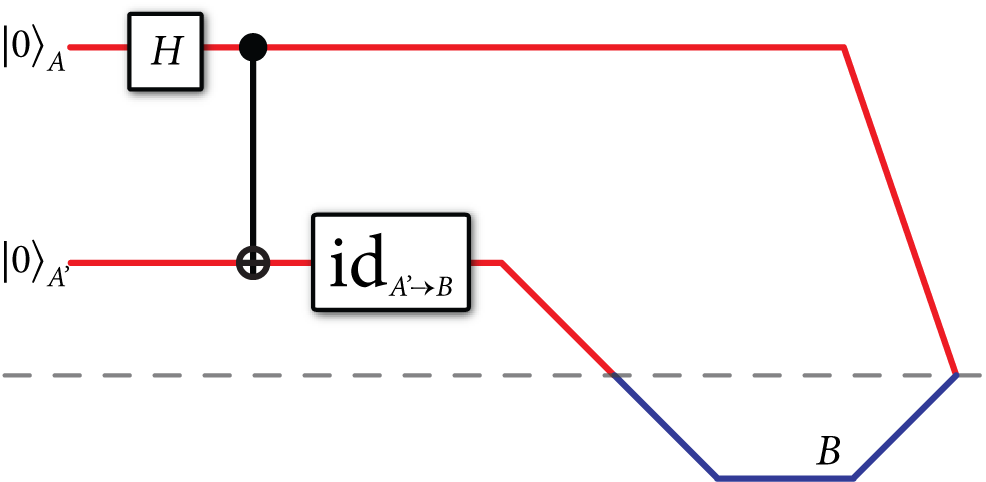}%
\caption{This figure depicts a protocol for entanglement distribution. Alice
performs local operations (the Hadamard and CNOT) and consumes one use of a
noiseless qubit channel to generate one noiseless ebit $\left\vert \Phi
^{+}\right\rangle _{AB}$\ shared with Bob.}%
\label{fig-3np:ent-dist}%
\end{center}
\end{figure}

\noindent Figure~\ref{fig-3np:ent-dist} depicts the entanglement distribution
protocol.%

The following resource inequality quantifies the non-local resources consumed
or generated in the above protocol:%
\begin{equation}
\left[  q\rightarrow q\right]  \geq\left[  qq\right]  ,
\label{eq-3np:ent-dist}%
\end{equation}
where $\left[  q\rightarrow q\right]  $ denotes one forward use of a noiseless
qubit channel and $\left[  qq\right]  $ denotes a shared, noiseless ebit. The
meaning of the resource inequality is that there exists a protocol that
consumes the resource on the left in order to generate the resource on the
right. The best analogy is to think of a resource inequality as a
\textquotedblleft chemical reaction\textquotedblright-like formula, where the
protocol is like a chemical reaction that transforms one resource into another.

There are several subtleties to notice about the above protocol and its
corresponding resource inequality:

\begin{enumerate}
\item We are careful with the language when describing the resource state. We
described the state $\left\vert \Phi^{+}\right\rangle $ as a Bell state%
\index{Bell states}
in the first step because it is a local state in Alice's laboratory. We only
used the term \textquotedblleft ebit\textquotedblright\
\index{ebit}%
to describe the state after the second step, when the state becomes a
non-local resource shared between Alice and Bob.

\item The resource count involves non-local resources only---we do not factor
any local operations, such as the Hadamard gate or the CNOT\ gate, into the
resource count. This line of thinking is different from the theory of
computation, where it is of utmost importance to minimize the number of steps
involved in a computation. In this book, we are developing a theory of quantum
communication and thus count non-local resources only.

\item We are assuming that it is possible to perform all local operations
perfectly. This line of thinking is another departure from practical concerns
that one might have in fault-tolerant quantum computation, the study of the
propagation of errors in quantum operations. Performing a CNOT\ gate is a
highly non-trivial task at the current stage of experimental development in
quantum computation, with most implementations being far from perfect.
Nevertheless, we proceed forward with this communication-theoretic line of thinking.
\end{enumerate}

The following exercises outline classical information-processing tasks that
are analogous to the task of entanglement distribution.

\begin{exercise}
Outline a protocol for \textit{shared randomness distribution}. Suppose that
Alice and Bob have available one use of a noiseless classical bit channel.
Give a method for them to implement the following resource inequality:%
\begin{equation}
\left[  c\rightarrow c\right]  \geq\left[  cc\right]  ,
\end{equation}
where $\left[  c\rightarrow c\right]  $ denotes one forward use of a noiseless
classical bit channel and $\left[  cc\right]  $ denotes a shared, non-local
bit of shared randomness.
\end{exercise}

\begin{exercise}
Consider three parties Alice, Bob, and Eve and suppose that a noiseless
private channel connects Alice to Bob. Privacy here implies that Eve does not
learn anything about the information that traverses the private
channel---Eve's probability distribution is independent of Alice and Bob's:%
\begin{equation}
p_{A,B,E}( a,b,e) =p_{A}( a) p_{B|A}( b|a) p_{E}( e) .
\end{equation}
For a noiseless private bit channel, $p_{B|A}( b|a) =\delta_{b,a}$. A
noiseless secret key corresponds to the following distribution:%
\begin{equation}
p_{A,B,E}( a,b,e) =\frac{1}{2}\delta_{b,a}p_{E}( e) ,
\end{equation}
where $\frac{1}{2}$ implies that the key is equal to \textquotedblleft%
0\textquotedblright\ or \textquotedblleft1\textquotedblright\ with equal
probability, $\delta_{b,a}$ implies a perfectly correlated secret key, and the
factoring of the distribution $p_{A,B,E}( a,b,e) $ implies the secrecy of the
key (Eve's information is independent of Alice and Bob's). The difference
between a noiseless private bit channel and a noiseless secret key is that the
private channel is a dynamic resource while the secret key is a shared, static
resource. Show that it is possible to upgrade the protocol for shared
randomness distribution to a protocol for \textit{secret key distribution,} if
Alice and Bob share a noiseless private bit channel. That is, show that they
can achieve the following resource inequality:%
\begin{equation}
\left[  c\rightarrow c\right]  _{\operatorname{priv}}\geq\left[  cc\right]
_{\operatorname{priv}},
\end{equation}
where $\left[  c\rightarrow c\right]  _{\operatorname{priv}}$ denotes one
forward use of a noiseless private bit channel and $\left[  cc\right]
_{\operatorname{priv}}$ denotes one bit of shared, noiseless secret key.
\end{exercise}

\subsubsection{Entanglement and Quantum Communication}

Can entanglement enable two parties to communicate quantum information?\ It is
natural to wonder if there is a protocol corresponding to the following
resource inequality:%
\begin{equation}
\left[  qq\right]  \overset{?}{\geq}\left[  q\rightarrow q\right]  .
\end{equation}
Unfortunately, it is physically impossible to construct a protocol that
implements the above resource inequality. The argument against such a protocol
arises from the theory of relativity. Specifically, the theory of relativity
prohibits information transfer or signaling at a speed greater than the speed
of light. Suppose that two parties share noiseless entanglement over a large
distance. That resource is a static resource, possessing only shared quantum
correlations. If a protocol were to exist that implements the above resource
inequality, it would imply that two parties could communicate quantum
information faster than the speed of light, because they would be exploiting
the entanglement for the instantaneous transfer of quantum information.

The entanglement distribution resource inequality is only \textquotedblleft
one-way,\textquotedblright\ as in \eqref{eq-3np:ent-dist}. Quantum
communication is therefore strictly stronger than shared entanglement when no
other non-local resources are available.

\subsection{Elementary Coding}

We can also send classical information using a noiseless qubit channel. A
simple protocol for doing so
\index{elementary coding}%
is \textit{elementary coding.} This protocol consists of the following steps:

\begin{enumerate}
\item Alice prepares either $\vert0\rangle$ or $\vert1\rangle$, depending on
the classical bit that she would like to send.

\item She transmits this state over the noiseless qubit channel, and Bob
receives the qubit.

\item Bob performs a measurement in the computational basis to determine the
classical bit that Alice transmitted.
\end{enumerate}

Elementary coding succeeds without error because Bob's measurement can always
distinguish the classical states $|0\rangle$ and $|1\rangle$. The following
resource inequality applies to elementary coding:%
\begin{equation}
\left[  q\rightarrow q\right]  \geq\left[  c\rightarrow c\right]  .
\end{equation}
Again, we are only counting non-local resources in the resource count---we do
not count the state preparation at the beginning or the measurement at the end.

If no other resources are available for consumption, the above resource
inequality is optimal---one cannot do better than to transmit one classical
bit of information per use of a noiseless qubit channel. This result may be a
bit frustrating at first, because it may seem that we could exploit the
continuous degrees of freedom in the probability amplitudes of a qubit state
for encoding more than one classical bit per qubit. Unfortunately, there is no
way that we can access the information in the continuous degrees of freedom
using any measurement scheme. The result of Exercise~\ref{ex-nqt:nayak}%
\ demonstrates the optimality of the above protocol, and it holds as well by
invoking the Holevo bound from Chapter~\ref{chap:q-info-entropy}.

\subsection{Quantum Super-Dense Coding}

\label{sec:dense-coding}We now outline a protocol named
\index{super-dense coding}%
\textit{super-dense coding}. It is named as such because it has the striking
property that noiseless entanglement can double the classical communication
ability of a noiseless qubit channel. It consists of three steps:

\begin{enumerate}
\item Suppose that Alice and Bob share an ebit $\left\vert \Phi^{+}%
\right\rangle _{AB}$. Alice applies one of four unitary operations $\left\{
I,X,Z,XZ\right\}  $ to her share of the above state. The state becomes one of
the following four
\index{Bell states}%
Bell states (up to a global phase), depending on the message that Alice
chooses:%
\begin{equation}
\left\vert \Phi^{+}\right\rangle _{AB}\operatorname{, \ \ \ \ \ \ \ }%
\left\vert \Phi^{-}\right\rangle _{AB}\operatorname{,\ \ \ \ \ \ \ \ }%
\left\vert \Psi^{+}\right\rangle _{AB},\ \ \ \ \ \ \ \ \left\vert \Psi
^{-}\right\rangle _{AB}.
\end{equation}
The definitions of these Bell states are in \eqref{eq-qt:bell1}--\eqref{eq-qt:bell3}.

\item She transmits her qubit to Bob with one use of a noiseless qubit channel.

\item Bob performs a Bell measurement%
\index{Bell measurement}
(a measurement in the basis $\{\left\vert \Phi^{+}\right\rangle _{AB}$,
$\left\vert \Phi^{-}\right\rangle _{AB}$, $\left\vert \Psi^{+}\right\rangle
_{AB}$, $\left\vert \Psi^{-}\right\rangle _{AB}\}$) to distinguish the four
states perfectly---he can distinguish the states because they are all
orthogonal to each other.
\end{enumerate}

\begin{figure}
[ptb]
\begin{center}
\includegraphics[
width=3.6876in
]%
{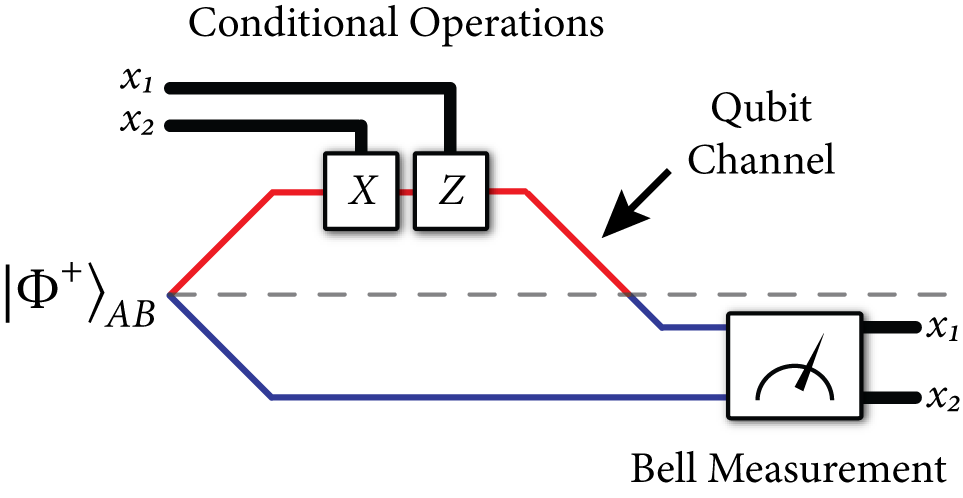}%
\caption{This figure depicts the dense coding protocol. Alice and Bob share an
ebit before the protocol begins. Alice would like to transmit two classical
bits $x_{1}x_{2}$ to Bob. She performs a Pauli rotation conditioned on her two
classical bits and sends her share of the ebit over a noiseless qubit channel.
Bob can then recover the two classical bits by performing a Bell measurement.}%
\label{fig:dense-coding}%
\end{center}
\end{figure}

Thus, Alice can transmit two classical bits (corresponding to the four
messages) if she shares a noiseless ebit with Bob and uses a noiseless qubit
channel. Figure~\ref{fig:dense-coding}\ depicts the protocol for quantum
super-dense coding.%

The super-dense coding protocol realizes the following resource inequality:%
\begin{equation}
\left[  qq\right]  +\left[  q\rightarrow q\right]  \geq2\left[  c\rightarrow
c\right]  .
\end{equation}
Notice again that the resource inequality counts the use of non-local
resources only---we do not count the local operations at the beginning of the
protocol or the Bell measurement at the end of the protocol.

Also, notice that we could have implemented two noiseless classical bit
channels with two instances of elementary coding:%
\begin{equation}
2\left[  q\rightarrow q\right]  \geq2\left[  c\rightarrow c\right]  .
\end{equation}
However, this method is not as powerful as the super-dense coding
protocol---in super-dense coding, we consume the weaker resource of an ebit to
help transmit two classical bits, instead of consuming the stronger resource
of an extra noiseless qubit channel.

The super-dense coding protocol also transmits the classical bits
\textit{privately}. Suppose a third party intercepts the qubit that Alice
transmits. There is no measurement that the third party can perform to
determine which message Alice transmits because the local density operator of
all of the Bell states is the same and equal to the maximally mixed state
$\pi_{A}$ (the information for the eavesdropper is the same irrespective of
each message that Alice transmits). The privacy of the protocol is due to
Alice and Bob sharing maximal entanglement. We exploit this aspect of the
super-dense coding protocol when we ``make it coherent'' in
Chapter~\ref{chap:coherent-communication}.

\subsection{Quantum Teleportation}

Perhaps the most striking protocol in noiseless quantum communication is the%
\index{quantum teleportation}
\textit{quantum teleportation protocol}. The protocol destroys the quantum
state of a qubit in one location and recreates it on a qubit at a distant
location, with the help of shared entanglement. Thus, the name
\textquotedblleft teleportation\textquotedblright\ corresponds well to the
mechanism that occurs.

The teleportation protocol is actually a flipped version of the super-dense
coding protocol, in the sense that Alice and Bob merely \textquotedblleft swap
their equipment.\textquotedblright\ The first step in understanding
teleportation is to perform a few algebraic steps using the tricks of the
tensor product and the Bell state substitutions from
Exercise~\ref{ex-qt:bell-comp}. Consider a qubit $|\psi\rangle_{A^{\prime}}%
$\ that Alice possesses, where%
\begin{equation}
|\psi\rangle_{A^{\prime}}\equiv\alpha|0\rangle_{A^{\prime}}+\beta
|1\rangle_{A^{\prime}}.
\end{equation}
Suppose she shares an ebit $\left\vert \Phi^{+}\right\rangle _{AB}$ with Bob.
The joint state of the systems $A^{\prime}$, $A$, and $B$ is as follows:%
\begin{equation}
|\psi\rangle_{A^{\prime}}\left\vert \Phi^{+}\right\rangle _{AB}.
\end{equation}
Let us first explicitly write out this state:%
\begin{equation}
|\psi\rangle_{A^{\prime}}\left\vert \Phi^{+}\right\rangle _{AB}=\left(
\alpha|0\rangle_{A^{\prime}}+\beta|1\rangle_{A^{\prime}}\right)  \left(
\frac{|00\rangle_{AB}+|11\rangle_{AB}}{\sqrt{2}}\right)  .
\end{equation}
Distributing terms gives the following equality:%
\begin{equation}
=\frac{1}{\sqrt{2}}\left[  \alpha|000\rangle_{A^{\prime}AB}+\beta
|100\rangle_{A^{\prime}AB}+\alpha|011\rangle_{A^{\prime}AB}+\beta
|111\rangle_{A^{\prime}AB}\right]  .
\end{equation}
We use the relations in Exercise~\ref{ex-qt:bell-comp} to rewrite the joint
system $A^{\prime}A$ in the Bell basis:%
\begin{equation}
=\frac{1}{2}\left[
\begin{array}
[c]{c}%
\alpha\left(  \left\vert \Phi^{+}\right\rangle _{A^{\prime}A}+\left\vert
\Phi^{-}\right\rangle _{A^{\prime}A}\right)  |0\rangle_{B}+\beta\left(
\left\vert \Psi^{+}\right\rangle _{A^{\prime}A}-\left\vert \Psi^{-}%
\right\rangle _{A^{\prime}A}\right)  |0\rangle_{B}\\
+\alpha\left(  \left\vert \Psi^{+}\right\rangle _{A^{\prime}A}+\left\vert
\Psi^{-}\right\rangle _{A^{\prime}A}\right)  |1\rangle_{B}+\beta\left(
\left\vert \Phi^{+}\right\rangle _{A^{\prime}A}-\left\vert \Phi^{-}%
\right\rangle _{A^{\prime}A}\right)  |1\rangle_{B}%
\end{array}
\right].
\end{equation}
Simplifying gives the following equality:%
\begin{equation}
=\frac{1}{2}\left[
\begin{array}
[c]{c}%
\left\vert \Phi^{+}\right\rangle _{A^{\prime}A}\left(  \alpha|0\rangle
_{B}+\beta|1\rangle_{B}\right)  +\left\vert \Phi^{-}\right\rangle _{A^{\prime
}A}\left(  \alpha|0\rangle_{B}-\beta|1\rangle_{B}\right) \\
+\left\vert \Psi^{+}\right\rangle _{A^{\prime}A}\left(  \alpha|1\rangle
_{B}+\beta|0\rangle_{B}\right)  +\left\vert \Psi^{-}\right\rangle _{A^{\prime
}A}\left(  \alpha|1\rangle_{B}-\beta|0\rangle_{B}\right)
\end{array}
\right]  .
\end{equation}
We can finally rewrite the state as four superposed terms, with a distinct
Pauli operator applied to Bob's system $B$ for each term in the superposition:%
\begin{equation}
=\frac{1}{2}\left[  \left\vert \Phi^{+}\right\rangle _{A^{\prime}A}%
|\psi\rangle_{B}+\left\vert \Phi^{-}\right\rangle _{A^{\prime}A}Z|\psi
\rangle_{B}+\left\vert \Psi^{+}\right\rangle _{A^{\prime}A}X|\psi\rangle
_{B}+\left\vert \Psi^{-}\right\rangle _{A^{\prime}A}XZ|\psi\rangle_{B}\right]
.
\end{equation}

\begin{figure}
[ptb]
\begin{center}
\includegraphics[
width=4.2393in
]%
{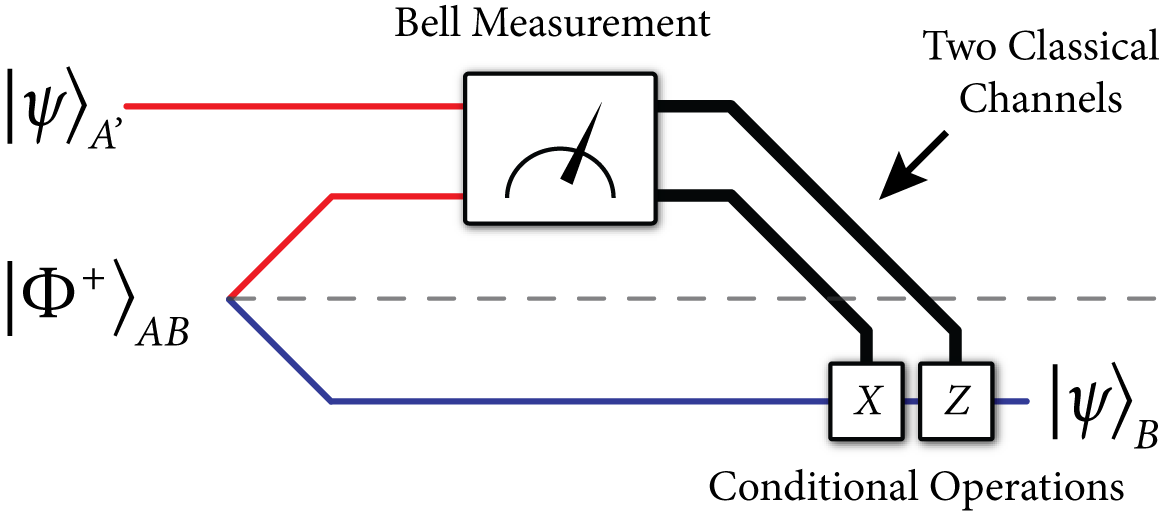}%
\caption{This figure depicts the teleportation protocol. Alice would like to
transmit an arbitrary quantum state $\vert\psi\rangle_{A^{\prime}}$ to Bob.
Alice and Bob share an ebit before the protocol begins. Alice can
\textquotedblleft teleport\textquotedblright\ her quantum state to Bob by
consuming the entanglement and two uses of a noiseless classical bit channel.}%
\label{fig:teleportation}%
\end{center}
\end{figure}
We now outline the three steps of the teleportation protocol (depicted in
Figure~\ref{fig:teleportation}):

\begin{enumerate}
\item Alice performs a Bell measurement on her systems $A^{\prime}A$. The
state collapses to one of the following four states with uniform probability:%
\begin{align}
&  \left\vert \Phi^{+}\right\rangle _{A^{\prime}A}|\psi\rangle_{B},\\
&  \left\vert \Phi^{-}\right\rangle _{A^{\prime}A}Z|\psi\rangle_{B},\\
&  \left\vert \Psi^{+}\right\rangle _{A^{\prime}A}X|\psi\rangle_{B},\\
&  \left\vert \Psi^{-}\right\rangle _{A^{\prime}A}XZ|\psi\rangle_{B}.
\end{align}
Notice that the state resulting from the measurement is a product state with
respect to the cut $A^{\prime}A\ |\ B$, regardless of the outcome of the
measurement. At this point, Alice knows whether Bob's state is $|\psi
\rangle_{B}$, $Z|\psi\rangle_{B}$, $X|\psi\rangle_{B}$, or $XZ|\psi\rangle
_{B}$ because she knows the result of the measurement. On the other hand, Bob
does not know anything about the state of his system $B$%
---Exercise~\ref{ex-qt:uniformly-random-unitary}\ states that his local
density operator is the maximally mixed state $\pi_{B}$\ just after Alice
performs the measurement. Thus, there is no teleportation of quantum
information at this point because Bob's local state is completely independent
of the original state $|\psi\rangle$. In other words, teleportation cannot be instantaneous.

\item Alice transmits two classical bits to Bob that indicate which of the
four measurement outcomes occurred. After Bob receives the classical
information, he is immediately certain which operation he needs to perform in
order to restore his state to Alice's original state $|\psi\rangle$. Notice
that he does not need to have knowledge of the state in order to restore
it---he only needs knowledge of the restoration operation.

\item Bob performs the restoration operation: one of the identity, a Pauli $X$
operator, a Pauli $Z$ operator, or the Pauli operator $ZX$, depending on the
classical information that he receives from Alice.
\end{enumerate}

Teleportation is an \textit{oblivious} protocol because Alice and Bob do not
require any knowledge of the quantum state being teleported in order to
perform it. We might also say that this feature of teleportation makes it
universal---it works independently of the input state.

You might think that the teleportation protocol violates the no-cloning
theorem because a \textquotedblleft copy\textquotedblright\ of the state
appears on Bob's system. But this violation does not occur at any point in the
protocol because the Bell measurement destroys the information about the state
of Alice's original information qubit while recreating it somewhere else.
Also, notice that the result of the Bell measurement is independent of the
particular probability amplitudes $\alpha$ and $\beta$ corresponding to the
state Alice wishes to teleport.

The teleportation protocol is not an instantaneous teleportation, as portrayed
in the television episodes of \textit{Star Trek}. There is no transfer of quantum
information instantaneously after the Bell measurement because Bob's local
description of the $B$ system is the maximally mixed state $\pi$. It is only
after he receives the classical bits to \textquotedblleft
telecorrect\textquotedblright\ his state that the transfer occurs. It must be
this way---otherwise, they would be able to communicate faster than the speed
of light, and superluminal communication is not allowed by the theory of relativity.

Finally, we can phrase the teleportation protocol as a resource inequality:%
\begin{equation}
\left[  qq\right]  +2\left[  c\rightarrow c\right]  \geq\left[  q\rightarrow
q\right]  .
\end{equation}
Again, we include only non-local resources in the resource count. The above
resource inequality is perhaps the most surprising of the three unit protocols
we have studied so far. It combines two resources, noiseless entanglement and
noiseless classical communication, that achieve noiseless quantum
communication even though they are both individually weaker than it. This
protocol and super-dense coding are two of the most fundamental protocols in
quantum communication theory because they sparked the notion that there are
clever ways of combining resources to generate other resources.

In Exercise~\ref{ex-3np:RSP} below, we discuss a variation of teleportation
called \textit{remote state preparation},%
\index{remote state preparation}
where Alice possesses a classical description of the state that she wishes to
teleport. With this knowledge, it is possible to reduce the amount of
classical communication necessary for teleportation.

\begin{exercise}
\label{ex-3np:RSP}\textit{Remote state preparation} is a variation of the
teleportation protocol. We consider a simple example of a remote state
preparation protocol. Suppose Alice possesses a classical description of a
state $|\psi\rangle\equiv\left(  |0\rangle+e^{i\phi}|1\rangle\right)
/\sqrt{2}$\ (on the equator of the Bloch sphere) and she shares an ebit
$\left\vert \Phi^{+}\right\rangle _{AB}$\ with Bob. Alice would like to
prepare the state $\vert\psi\rangle$ on Bob's system. Show that Alice can
prepare this state on Bob's system if she measures her system $A$\ in the
$\left\{  \left\vert \psi^{\ast}\right\rangle ,\left\vert \psi^{\perp\ast
}\right\rangle \right\}  $ basis, transmits one classical bit, and Bob
performs a recovery operation conditioned on the classical information. (Note
that $\left\vert \psi^{\ast}\right\rangle $ is the conjugate of the vector
$|\psi\rangle$).
\end{exercise}

\begin{exercise}
\textit{Third-party controlled teleportation} is another variation on the
teleportation protocol. Suppose that Alice, Bob, and Charlie possess a
GHZ\ state:%
\begin{equation}
|\Phi_{\operatorname{GHZ}}\rangle\equiv\frac{|000\rangle_{ABC}+|111\rangle
_{ABC}}{\sqrt{2}}.
\end{equation}
Alice would like to teleport an arbitrary qubit to Bob. She performs the usual
steps in the teleportation protocol. Give the final steps that Charlie should
perform and the information that he should transmit to Bob in order to
complete the teleportation protocol. (Hint: The resource inequality for the
protocol is as follows:%
\begin{equation}
\left[  qqq\right]  _{ABC}+2\left[  c\rightarrow c\right]  _{A\rightarrow
B}+\left[  c\rightarrow c\right]  _{C\rightarrow B}\geq\left[  q\rightarrow
q\right]  _{A\rightarrow B},
\end{equation}
where $\left[  qqq\right]  _{ABC}$ represents the resource of the GHZ\ state
shared between Alice, Bob, and Charlie, and the other resources are as before
with the directionality of communication indicated by the corresponding subscript.)
\end{exercise}

\begin{exercise}
\textit{Gate teleportation} is yet another variation of quantum teleportation
that is useful in fault-tolerant quantum computation. Suppose that Alice would
like to perform a single-qubit gate $U$\ on a qubit in state $|\psi\rangle$.
Suppose that the gate $U$ is difficult to perform, but that $U\sigma
_{i}U^{\dag}$, where $\sigma_{i}$ is one of the single-qubit Pauli operators,
is much less difficult to perform. A protocol for gate teleportation is as
follows. Alice and Bob first prepare the ebit $U_{B}\left\vert \Phi
^{+}\right\rangle _{AB}$. Alice performs a Bell measurement on her qubit
$|\psi\rangle_{A^{\prime}}$ and system $A$. She transmits two classical bits
to Bob and Bob performs one of the four corrective operations $U\sigma
_{i}U^{\dag}$ on his qubit. Show that this protocol works, i.e., that Bob's
final state is $U|\psi\rangle$.
\end{exercise}

\begin{exercise}
Show that it is possible to simulate a dephasing qubit channel by the
following technique. First, Alice prepares a maximally entangled Bell state
$\left\vert \Phi^{+}\right\rangle $. She sends one share of it to Bob through
a dephasing qubit channel. She and Bob perform the usual teleportation
protocol. Show that this procedure gives the same result as sending a qubit
through a dephasing channel. (Hint: This result holds because the dephasing
channel commutes with all Pauli operators.)
\end{exercise}

\begin{exercise}
Construct an \textit{entanglement swapping\ protocol}
\index{entanglement swapping}%
from the teleportation protocol. That is, suppose that Charlie and Alice
possess a bipartite state $\vert\psi\rangle_{CA}$. Show that if Alice
teleports her share of the state $\vert\psi\rangle_{CA}$ to Bob, then Charlie
and Bob share the state $\vert\psi\rangle_{CB}$. A special case of this
protocol is when the state $\vert\psi\rangle_{CA}$ is an ebit. Then the
protocol is equivalent to an entanglement swapping protocol.
\end{exercise}

\section{Optimality of the Three Unit Protocols}

\label{sec-3np:optimality}We now consider several arguments that may seem
somewhat trivial at first, but they are crucial for having a good theory of
quantum communication. We are always thinking about the optimality of certain
protocols---if there is a better, cheaper way to perform a given protocol,
then this would be advantageous. There are several questions that we can ask
about the above protocols:

\begin{enumerate}
\item In entanglement distribution, is one ebit per qubit the best that we can
do, or is it possible to generate more than one ebit with a single use of a
noiseless qubit channel?

\item In super-dense coding, is it possible to generate two noiseless
classical bit channels with less than one noiseless qubit channel or less than
one noiseless ebit? Is it possible to generate more than two classical bit
channels using the given resources?

\item In teleportation, is it possible to teleport more than one qubit using
the given resources? Is it possible to teleport using less than two classical
bits or less than one ebit?
\end{enumerate}

In this section, we answer all of these questions in the negative---all the
protocols as given are optimal protocols. Here, we begin to see the beauty of
the resource inequality formalism. It allows us to chain protocols together to
make new protocols. We exploit this idea in the forthcoming optimality arguments.

First, let us tackle the optimality of entanglement distribution. Is there a
protocol that implements any other resource inequality such as%
\begin{equation}
\left[  q\rightarrow q\right]  \geq E\left[  qq\right]  ,
\end{equation}
where the rate $E$ of entanglement generation is greater than one?

We show that such a resource inequality can never occur, i.e., it is optimal
for $E=1$. Suppose such a resource inequality with $E>1$ does exist. Under an
assumption of free forward classical communication, we can combine the above
resource inequality with teleportation to achieve the following resource
inequality:%
\begin{equation}
\left[  q\rightarrow q\right]  \geq E\left[  q\rightarrow q\right]  .
\end{equation}
We could then simply keep repeating this protocol to achieve an unbounded
amount of quantum communication, which is impossible. Thus, it must be that
$E=1$.

Next, we consider the optimality of super-dense coding. We again exploit a
proof by contradiction argument. Let us suppose that we have an unlimited
amount of entanglement available. Suppose that there exists some
\textquotedblleft super-duper\textquotedblright-dense coding protocol that
generates an amount of classical communication greater than that which
super-dense coding generates. That is, the classical communication output of
super-duper-dense coding is $2C$ where $C>1$, and its resource inequality is%
\begin{equation}
\left[  q\rightarrow q\right]  +\left[  qq\right]  \geq2C\left[  c\rightarrow
c\right]  .
\end{equation}
Then this super-duper-dense coding scheme (along with the infinite
entanglement) gives the following resource inequality:%
\begin{equation}
\left[  q\rightarrow q\right]  +\infty\left[  qq\right]  \geq2C\left[
c\rightarrow c\right]  +\infty\left[  qq\right]  .
\end{equation}
An infinite amount of entanglement is still available after executing the
super-duper-dense coding protocol because it consumes only a finite amount of
entanglement. We can then chain the above protocol with teleportation and
achieve the following resource inequality:%
\begin{equation}
2C\left[  c\rightarrow c\right]  +\infty\left[  qq\right]  \geq C\left[
q\rightarrow q\right]  +\infty\left[  qq\right]  .
\end{equation}
Overall, we have then shown a scheme that achieves the following resource
inequality:%
\begin{equation}
\left[  q\rightarrow q\right]  +\infty\left[  qq\right]  \geq C\left[
q\rightarrow q\right]  +\infty\left[  qq\right]  .
\end{equation}
We can continue with this protocol and perform it $k$ times so that we
implement the following resource inequality:%
\begin{equation}
\left[  q\rightarrow q\right]  +\infty\left[  qq\right]  \geq C^{k}\left[
q\rightarrow q\right]  +\infty\left[  qq\right]  .
\end{equation}
The result of this construction is that one noiseless qubit channel and an
infinite amount of entanglement can generate an infinite amount of quantum
communication. This result is impossible physically because entanglement does
not boost the capacity of a noiseless qubit channel. Also, the scheme is
exploiting just one noiseless qubit channel along with the entanglement to
generate an unbounded amount of quantum communication---it must be signaling
superluminally in order to do so. Thus, the rate of classical communication in
super-dense coding is optimal.

We leave the optimality arguments for teleportation as an exercise because
they are similar to those for the super-dense coding protocol. Note that it is
possible to prove optimality of these protocols without assumptions such as
free classical communication (for the case of entanglement distribution), and
we do so in Chapter~\ref{chap:unit-resource-cap}.

\begin{exercise}
Show that it is impossible for $C>1$ in the teleportation protocol where $C$
is with respect to the following resource inequality:%
\begin{equation}
2\left[  c\rightarrow c\right]  +\left[  qq\right]  \geq C\left[  q\rightarrow
q\right]  .
\end{equation}

\end{exercise}

\begin{exercise}
Show that the rates of the consumed resources in the teleportation and
super-dense coding protocols are optimal.
\end{exercise}

\section{Extensions for Quantum Shannon Theory}

\label{sec-3np:extensions-qst}The previous section sparked some good questions
that we might ask as a quantum Shannon theorist. We might also wonder what
types of communication rates are possible if some of the consumed resources
are noisy, rather than being perfect resources. We list some of these
questions below.

Let us first consider entanglement distribution. Suppose that the consumed
noiseless qubit channel in entanglement distribution is instead a noisy
quantum channel $\mathcal{N}$. The communication task is then known as
\textit{entanglement generation}. We can rephrase the communication task as
the following resource inequality:%
\begin{equation}
\left\langle \mathcal{N}\right\rangle \geq E\left[  qq\right]  .
\end{equation}
The meaning of the resource inequality is that we consume the resource of a
noisy quantum channel $\mathcal{N}$ in order to generate entanglement between
a sender and receiver at some rate $E$. We will make the definition of a
quantum Shannon-theoretic resource inequality more precise when we begin our
formal study of quantum Shannon theory, but the above definition should be
sufficient for now. The optimal rate of entanglement generation with the noisy
quantum channel $\mathcal{N}$ is known as the entanglement generation capacity
of $\mathcal{N}$. This task is intimately related to the quantum communication
capacity%
\index{quantum capacity theorem}
of $\mathcal{N}$, and we discuss the connection further in
Chapter~\ref{chap:quantum-capacity}.

Let us now turn to super-dense coding. Suppose that the consumed noiseless
qubit channel in super-dense coding is instead a noisy quantum channel
$\mathcal{N}$. The name for this task is then
\index{entanglement-assisted!classical communication}%
\textit{entanglement-assisted classical communication}. The following resource
inequality captures the corresponding communication task:%
\begin{equation}
\left\langle \mathcal{N}\right\rangle +E\left[  qq\right]  \geq C\left[
c\rightarrow c\right]  .
\end{equation}
The meaning of the resource inequality is that we consume a noisy quantum
channel $\mathcal{N}$ and noiseless entanglement at some rate $E$ to produce
noiseless classical communication at some rate$~C$. We will study this
protocol in depth in Chapter~\ref{chap:EA-classical}. We can also consider the
scenario in which the entanglement is no longer noiseless, but it is rather a
general bipartite state$~\rho_{AB}$ that Alice and Bob share. The task is then
known as noisy super-dense coding.\footnote{The name noisy super-dense coding
could just as well apply to the former task of entanglement-assisted classical
communication, but this terminology has \textquotedblleft
stuck\textquotedblright\ in the research literature\ for this specific quantum
information-processing task.} We study noisy super-dense coding in
Chapter~\ref{chap:coh-comm-noisy}. The corresponding resource inequality is as
follows (its meaning should be clear at this point):%
\begin{equation}
\langle\rho_{AB}\rangle+Q\left[  q\rightarrow q\right]  \geq C\left[
c\rightarrow c\right]  .
\end{equation}

We can ask the same questions for the teleportation protocol as well. Suppose
that the entanglement resource is instead a noisy bipartite state $\rho_{AB}$.
The task is then
\index{quantum teleportation!noisy}
\textit{noisy teleportation} and has the following resource inequality:%
\begin{equation}
\langle\rho_{AB}\rangle+C\left[  c\rightarrow c\right]  \geq Q\left[
q\rightarrow q\right]  .
\end{equation}

The questions presented in this section are some of the fundamental questions
in quantum Shannon theory. We arrived at these questions simply by replacing
the noiseless resources in the three fundamental noiseless protocols with
noisy ones. We will spend a significant amount of effort building up our
knowledge of quantum Shannon-theoretic tools that will be indispensable for
answering these questions.

\section{Three Unit Qudit Protocols}

We end this chapter by studying the qudit%
\index{qudit}
versions of the three unit protocols. It is useful to have these versions of
the protocols because we may want to process qudit systems with them.

The qudit resources are straightforward extensions of the qubit resources. A
noiseless qudit channel is the following map:%
\begin{equation}
|i\rangle_{A}\rightarrow|i\rangle_{B},
\end{equation}
where $\{|i\rangle_{A}\}_{i\in\left\{  0,\ldots,d-1\right\}  }$ is some
preferred orthonormal basis on Alice's system and $\{|i\rangle_{B}%
\}_{i\in\left\{  0,\ldots,d-1\right\}  }$ is some preferred basis on Bob's
system. We can also write the qudit channel map as the following isometry:%
\begin{equation}
I_{A\rightarrow B}\equiv\sum_{i=0}^{d-1}|i\rangle_{B}\langle i|_{A}.
\label{eq-3np:noiseless-qudit}%
\end{equation}
The map $I_{A\rightarrow B}$ preserves superposition states so that%
\begin{equation}
\sum_{i=0}^{d-1}\alpha_{i}|i\rangle_{A}\rightarrow\sum_{i=0}^{d-1}\alpha
_{i}|i\rangle_{B}.
\end{equation}
A noiseless classical dit channel or \textit{cdit} is the following map:%
\begin{align}
|i\rangle\langle i|_{A}  &  \rightarrow|i\rangle\langle i|_{B},\\
|i\rangle\langle j|_{A}  &  \rightarrow0\text{ for }i\neq j.
\end{align}
A noiseless maximally entangled qudit state or an \textit{edit} is as follows:%
\begin{equation}
\left\vert \Phi\right\rangle _{AB}\equiv\frac{1}{\sqrt{d}}\sum_{i=0}%
^{d-1}|i\rangle_{A}|i\rangle_{B}. \label{eq-3np:max-ent-state-1}%
\end{equation}

We quantify the \textquotedblleft dit\textquotedblright\ resources with bit
measures. For example, a noiseless qudit channel is the following resource:%
\begin{equation}
\log d\left[  q\rightarrow q\right]  ,
\end{equation}
where the logarithm is base two. Thus, one qudit channel can transmit $\log d$
qubits of quantum information so that the qubit remains our standard unit of
quantum information. We quantify the amount of information transmitted
according to the dimension of the space that is transmitted. For example,
suppose that a quantum system has eight levels. We can then encode three
qubits of quantum information in this eight-level system.

Likewise, a classical dit channel is the following resource:%
\begin{equation}
\log d\left[  c\rightarrow c\right]  ,
\end{equation}
so that a classical dit channel transmits $\log d$ classical bits. The
parameter $d$ here is the number of classical messages that the channel transmits.

Finally, an edit is the following resource:%
\begin{equation}
\log d\left[  qq\right]  .
\end{equation}
We quantify the amount of entanglement in a maximally entangled state by its
Schmidt rank (see Theorem~\ref{thm-qt:schmidt}). We measure entanglement in
units of ebits (we return to this issue in Chapter~\ref{chap:ent-conc}).

\subsection{Entanglement Distribution}

The extension of the entanglement distribution%
\index{entanglement distribution}
protocol to the qudit case is straightforward. Alice merely prepares the state
$\left\vert \Phi\right\rangle _{AA^{\prime}}$ in her laboratory and transmits
the system $A^{\prime}$ through a noiseless qudit channel. She can prepare the
state $\left\vert \Phi\right\rangle _{AA^{\prime}}$ with two gates:\ the qudit
analog of the Hadamard gate and the CNOT gate. The qudit analog of the
Hadamard gate is the Fourier gate $F$\ introduced in
Exercise~\ref{ex-qt:fourier-gate}\ where%
\begin{equation}
F:\left\vert l\right\rangle \rightarrow\frac{1}{\sqrt{d}}\sum_{j=0}^{d-1}%
\exp\left\{  \frac{2\pi ilj}{d}\right\}  \vert j\rangle,
\end{equation}
so that%
\begin{equation}
F\equiv\frac{1}{\sqrt{d}}\sum_{l,j=0}^{d-1}\exp\left\{  \frac{2\pi ilj}%
{d}\right\}  \vert j\rangle\langle l\vert.
\end{equation}
The qudit analog of the CNOT\ gate is the following controlled-shift gate:%
\begin{equation}
\operatorname{CNOT}_{d}\equiv\sum_{j=0}^{d-1}\vert j\rangle\langle
j\vert\otimes X( j) ,
\end{equation}
where $X( j) $ is the shift operator defined in \eqref{eq-qt:X-op}.

\begin{exercise}
Verify that Alice can prepare the maximally entangled qudit state $\left\vert
\Phi\right\rangle _{AA^{\prime}}$ locally by preparing $|0\rangle_{A}%
|0\rangle_{A^{\prime}}$, applying $F_{A}$ and $\operatorname{CNOT}_{d}$. Show
that%
\begin{equation}
\left\vert \Phi\right\rangle _{AA^{\prime}}=\operatorname{CNOT}_{d}\cdot
F_{A}|0\rangle_{A}|0\rangle_{A^{\prime}}.
\end{equation}

\end{exercise}

\noindent The resource inequality for this qudit entanglement distribution
protocol is as follows:%
\begin{equation}
\log d\left[  q\rightarrow q\right]  \geq\log d\left[  qq\right]  .
\end{equation}

\subsection{Quantum Super-Dense Coding}

The qudit version of the super-dense coding%
\index{super-dense coding}
protocol proceeds analogously to the qubit case, with some notable exceptions.
It still consists of three steps:

\begin{enumerate}
\item Alice and Bob begin with a maximally entangled state of the form in
\eqref{eq-3np:max-ent-state-1}. Alice applies one of $d^{2}$ unitary
operations in the set $\left\{  X( x) Z( z) \right\}  _{x,z=0}^{d-1}$ to her
qudit. The shared state then becomes one of the $d^{2}$ maximally entangled
qudit states in \eqref{eq-qt:qudit-bell-states}.

\item She sends her qudit to Bob with one use of a noiseless qudit channel.

\item Bob performs a measurement in the qudit Bell basis to determine the
message Alice sent. The result of Exercise~\ref{ex-qt:qudit-bell-states-ortho}%
\ is that these states are perfectly distinguishable with a measurement.
\end{enumerate}

\noindent This qudit super-dense coding protocol realizes the following
resource inequality:%
\begin{equation}
\log d\left[  qq\right]  +\log d\left[  q\rightarrow q\right]  \geq2\log
d\left[  c\rightarrow c\right]  .
\end{equation}

\subsection{Quantum Teleportation}

The operations in the qudit teleportation%
\index{quantum teleportation}
protocol are again similar to the qubit case. The protocol proceeds in three steps:

\begin{enumerate}
\item Alice possesses an arbitrary qudit $\vert\psi\rangle_{A^{\prime}}%
$\ where%
\begin{equation}
\vert\psi\rangle_{A^{\prime}}\equiv\sum_{i=0}^{d-1}\alpha_{i}\vert
i\rangle_{A^{\prime}}.
\end{equation}
Alice and Bob share a maximally entangled qudit state $\left\vert
\Phi\right\rangle _{AB}$ of the form in \eqref{eq-3np:max-ent-state-1}. The
joint state of Alice and Bob is then $\vert\psi\rangle_{A^{\prime}}\left\vert
\Phi\right\rangle _{AB}$. Alice performs a measurement in the basis
$\{|\Phi_{i,j}\rangle_{A^{\prime}A}\}_{i,j}$.

\item She transmits the measurement result $(i, j)$ to Bob with the use of two
classical dit channels.

\item Bob then applies the unitary transformation $Z_{B}( j) X_{B}( i) $ to
his state to \textquotedblleft telecorrect\textquotedblright\ it to Alice's
original qudit.
\end{enumerate}

We prove that this protocol works by analyzing the probability of the
measurement result and the post-measurement state on Bob's system. The
techniques that we employ here are different from those for the qubit case.

First, let us suppose that Alice would like to teleport the $A^{\prime}$
system of a state $|\psi\rangle_{RA^{\prime}}$\ that she shares with an
inaccessible reference system $R$. This way, our teleportation protocol
encompasses the most general setting in which Alice would like to teleport a
mixed state on $A^{\prime}$. Also, Alice shares the maximally entangled edit
state $\left\vert \Phi\right\rangle _{AB}$\ with Bob. Alice first performs a
measurement of the systems $A^{\prime}$ and $A$\ in the basis $\{|\Phi
_{i,j}\rangle_{A^{\prime}A}\}_{i,j}$ where%
\begin{equation}
|\Phi_{i,j}\rangle_{A^{\prime}A}=U_{A^{\prime}}^{ij}\left\vert \Phi
\right\rangle _{A^{\prime}A}, \label{eq-3np:Phi-states}%
\end{equation}
and%
\begin{equation}
U_{A^{\prime}}^{ij}\equiv Z_{A^{\prime}}(j)X_{A^{\prime}}(i).
\end{equation}
The measurement operators are thus%
\begin{equation}
|\Phi_{i,j}\rangle\langle\Phi_{i,j}|_{A^{\prime}A}.
\end{equation}
Then the unnormalized post-measurement state is%
\begin{equation}
|\Phi_{i,j}\rangle\langle\Phi_{i,j}|_{A^{\prime}A}\ |\psi\rangle_{RA^{\prime}%
}\left\vert \Phi\right\rangle _{AB},
\end{equation}
where here and in what follows, we have taken the common practice of omitting
tensor products with identity operators, instead leaving them implicit in
order to reduce clutter in the notation. We can rewrite this state as follows,
by exploiting the definition of $|\Phi_{i,j}\rangle_{A^{\prime}A}$ in
\eqref{eq-3np:Phi-states}:%
\begin{equation}
|\Phi_{i,j}\rangle\langle\Phi|_{A^{\prime}A}\ \left(  U_{A^{\prime}}%
^{ij}\right)  ^{\dag}\ |\psi\rangle_{RA^{\prime}}\left\vert \Phi\right\rangle
_{AB}.
\end{equation}
Recall the \textquotedblleft transpose trick\textquotedblright\ from
Exercise~\ref{ex-qt:bell-state-matrix-identity} that holds for any maximally
entangled state $\left\vert \Phi\right\rangle $. We can exploit this result to
show that the action of the unitary $\left(  U^{ij}\right)  ^{\dag}$ on the
$A^{\prime}$ system is the same as the action of the unitary $\left(
U^{ij}\right)  ^{\ast}$ on the $A$ system:%
\begin{equation}
|\Phi_{i,j}\rangle\langle\Phi|_{A^{\prime}A}\ \left(  U_{A}^{ij}\right)
^{\ast}\ |\psi\rangle_{RA^{\prime}}\left\vert \Phi\right\rangle _{AB}.
\end{equation}
Then the unitary $\left(  U_{A}^{ij}\right)  ^{\ast}$ commutes with the
systems $R$ and $A^{\prime}$:%
\begin{equation}
|\Phi_{i,j}\rangle\langle\Phi|_{A^{\prime}A}\ |\psi\rangle_{RA^{\prime}%
}\ \left(  U_{A}^{ij}\right)  ^{\ast}\left\vert \Phi\right\rangle _{AB}.
\end{equation}
We can again apply the transpose trick from
Exercise~\ref{ex-qt:bell-state-matrix-identity} to show that the state is
equal to%
\begin{equation}
|\Phi_{i,j}\rangle\langle\Phi|_{A^{\prime}A}\ |\psi\rangle_{RA^{\prime}%
}\ \left(  U_{B}^{ij}\right)  ^{\dag}\left\vert \Phi\right\rangle _{AB}.
\end{equation}
Then we can commute the unitary $\left(  U_{B}^{ij}\right)  ^{\dag}$ all the
way to the left, and we can switch the order of $|\psi\rangle_{RA^{\prime}}%
\ $and $\left\vert \Phi\right\rangle _{AB}$ without any problem because the
system labels are sufficient to track the states in these systems:%
\begin{equation}
\left(  U_{B}^{ij}\right)  ^{\dag}|\Phi_{i,j}\rangle\langle\Phi|_{A^{\prime}%
A}\ \left\vert \Phi\right\rangle _{AB}\ |\psi\rangle_{RA^{\prime}}.
\label{eq-3np:qudit-TP-almost-done}%
\end{equation}

Now let us consider the very special overlap $\langle\Phi|_{A^{\prime}%
A}\ \left\vert \Phi\right\rangle _{AB}$\ of the maximally entangled edit state
with itself on different systems:%
\begin{align}
\langle\Phi|_{A^{\prime}A}\ \left\vert \Phi\right\rangle _{AB}  &  =\left(
\frac{1}{\sqrt{d}}\sum_{i=0}^{d-1}\langle i\vert_{A^{\prime}}\langle
i\vert_{A}\right)  \left(  \frac{1}{\sqrt{d}}\sum_{j=0}^{d-1}\vert
j\rangle_{A}\vert j\rangle_{B}\right) \\
&  =\frac{1}{d}\sum_{i,j=0}^{d-1}\langle i\vert_{A^{\prime}}\langle i\vert
_{A}\vert j\rangle_{A}\vert j\rangle_{B} =\frac{1}{d}\sum_{i,j=0}^{d-1}\langle
i\vert_{A^{\prime}}\left\langle i|j\right\rangle _{A}\vert j\rangle_{B}\\
&  =\frac{1}{d}\sum_{i=0}^{d-1}\langle i\vert_{A^{\prime}}\vert i\rangle_{B}
=\frac{1}{d}\sum_{i=0}^{d-1}\vert i\rangle_{B}\langle i\vert_{A^{\prime}}\\
&  =\frac{1}{d}I_{A^{\prime}\rightarrow B}.
\end{align}
The first equality follows by definition. The second equality follows from
linearity and rearranging terms in the multiplication and summation. The third
and fourth equalities follow by realizing that $\langle i\vert_{A}\vert
j\rangle_{A}$ is an inner product and evaluating it for the orthonormal basis
$\left\{  \vert i\rangle_{A}\right\}  $. The fifth equality follows by
rearranging the bra and the ket. The final equality is our last important
realization: the operator $\sum_{i=0}^{d-1}\vert i\rangle_{B}\langle
i\vert_{A^{\prime}}$ is the noiseless qudit channel $I_{A^{\prime}\rightarrow
B}$\ that the teleportation protocol creates from the system $A^{\prime}$ to
$B$ (see the definition of a noiseless qudit channel in
\eqref{eq-3np:noiseless-qudit}). We might refer to this as the
\textquotedblleft teleportation map.\textquotedblright

We now apply the teleportation map to the state in
\eqref{eq-3np:qudit-TP-almost-done}:%
\begin{align}
\left(  U_{B}^{ij}\right)  ^{\dag}|\Phi_{i,j}\rangle\langle\Phi|_{A^{\prime}%
A}\ \left\vert \Phi\right\rangle _{AB}\ |\psi\rangle_{RA^{\prime}}  &
=\left(  U_{B}^{ij}\right)  ^{\dag}|\Phi_{i,j}\rangle_{A^{\prime}A}\frac{1}%
{d}I_{A^{\prime}\rightarrow B}\ |\psi\rangle_{RA^{\prime}}\\
&  =\frac{1}{d}\left(  U_{B}^{ij}\right)  ^{\dag}|\Phi_{i,j}\rangle
_{A^{\prime}A}|\psi\rangle_{RB}\\
&  =\frac{1}{d}|\Phi_{i,j}\rangle_{A^{\prime}A}\ \left(  U_{B}^{ij}\right)
^{\dag}|\psi\rangle_{RB}.
\end{align}
We can compute the probability of receiving outcome $i$ and $j$ from the
measurement when the input state is $|\psi\rangle_{RA^{\prime}}$. It is just
equal to the overlap of the above vector with itself:%
\begin{align}
p\left(  i,j|\psi\right)   &  =\left[  \frac{1}{d}\langle\Phi_{i,j}%
|_{A^{\prime}A}\ \langle\psi|_{RB}U_{B}^{ij}\right]  \left[  \frac{1}{d}%
|\Phi_{i,j}\rangle_{A^{\prime}A}\ \left(  U_{B}^{ij}\right)  ^{\dag}%
|\psi\rangle_{RB}\right] \\
&  =\frac{1}{d^{2}}\langle\Phi_{i,j}|_{A^{\prime}A}|\Phi_{i,j}\rangle
_{A^{\prime}A}\ \langle\psi|_{RB}U_{B}^{ij}\left(  U_{B}^{ij}\right)  ^{\dag
}|\psi\rangle_{RB}\\
&  =\frac{1}{d^{2}}\langle\Phi_{i,j}|_{A^{\prime}A}|\Phi_{i,j}\rangle
_{A^{\prime}A}\ \langle\psi|_{RB}|\psi\rangle_{RB} =\frac{1}{d^{2}}.
\end{align}
Thus, the probability of the outcome $(i,j)$ is completely random and
independent of the input state. We would expect this to be the case for a
universal teleportation protocol that operates independently of the input
state. Thus, after normalization, the state on Alice and Bob's system is%
\begin{equation}
|\Phi_{i,j}\rangle_{A^{\prime}A}\ \left(  U_{B}^{ij}\right)  ^{\dag}%
|\psi\rangle_{RB}.
\end{equation}

At this point, Bob does not know the result of the measurement. We obtain his
density operator by tracing over the systems $A^{\prime}$, $A$, and $R$ to
which he does not have access and taking the expectation over all the
measurement outcomes:%
\begin{multline}
\operatorname{Tr}_{A^{\prime}AR}\left\{  \frac{1}{d^{2}}\sum_{i,j=0}%
^{d-1}|\Phi_{i,j}\rangle\langle\Phi_{i,j}|_{A^{\prime}A}\ \left(  U_{B}%
^{ij}\right)  ^{\dag}\vert\psi\rangle\langle\psi\vert_{RB}U_{B}^{ij}\right\}
\\
=\frac{1}{d^{2}}\sum_{i,j=0}^{d-1}\ \left(  U_{B}^{ij}\right)  ^{\dag}\psi
_{B}U_{B}^{ij} =\pi_{B}.
\end{multline}
The first equality follows by evaluating the partial trace and by defining
$\psi_{B}\equiv\operatorname{Tr}_{R}\left\{  \vert\psi\rangle\langle\psi
\vert_{RB}\right\}  . $ The second equality follows because applying a
Heisenberg--Weyl operator uniformly at random completely randomizes a quantum
state to be the maximally mixed state (see
Exercise~\ref{ex-qt:uniformly-random-unitary}).

Now suppose that Alice sends the measurement results $i$ and $j$ over two uses
of a noiseless classical dit channel. Bob then knows that the state is%
\begin{equation}
\left(  U_{B}^{ij}\right)  ^{\dag}|\psi\rangle_{RB},
\end{equation}
and he can apply $U_{B}^{ij}$ to make the overall state become $|\psi
\rangle_{RB}$. This final step completes the teleportation process. The
resource inequality for the qudit teleportation protocol is as follows:%
\begin{equation}
\log d\left[  qq\right]  +2\log d\left[  c\rightarrow c\right]  \geq\log
d\left[  q\rightarrow q\right]  .
\end{equation}

\section{History and Further Reading}

This chapter presented the three important protocols that exploit the three
unit resources of classical communication, quantum communication, and
entanglement. We learned, perhaps surprisingly, that it is possible to combine
two resources together in interesting ways to simulate a different resource
(in both super-dense coding and teleportation). These combinations of
resources turn up quite a bit in quantum Shannon theory, and we see them in
their most basic form in this chapter.

\cite{PhysRevLett.69.2881} published the super-dense coding protocol, and
within a year, \cite{PhysRevLett.70.1895} realized that Alice and Bob could
teleport particles if they swap their operations with respect to the
super-dense coding protocol. These two protocols were the seeds of much later
work in quantum Shannon theory.

\chapter{Coherent Protocols}

\label{chap:coherent-communication}We introduced three protocols in the
previous chapter:\ entanglement distribution, teleportation, and super-dense
coding. The last two of these protocols, teleportation and super-dense coding,
are perhaps more interesting than entanglement distribution because they
demonstrate insightful ways for combining all three unit resources to achieve
an information-processing task.

It appears that teleportation and super-dense coding might be
\textquotedblleft inverse\textquotedblright\ protocols with respect to each
other because teleportation arises from super-dense coding when Alice and Bob
\textquotedblleft swap their equipment.\textquotedblright\ But there is a
fundamental asymmetry between these protocols when we consider their
respective resource inequalities. Recall that the resource inequality for
teleportation is%
\begin{equation}
2\left[  c\rightarrow c\right]  +\left[  qq\right]  \geq\left[  q\rightarrow
q\right]  , \label{eq-coh:tele}%
\end{equation}
while that for super-dense coding is%
\begin{equation}
\left[  q\rightarrow q\right]  +\left[  qq\right]  \geq2\left[  c\rightarrow
c\right]  . \label{eq-coh:dense}%
\end{equation}
The asymmetry in these protocols is that they are not \textit{dual under
resource reversal}. Two protocols are dual under resource reversal if the
resources that one consumes are the same that the other generates and vice
versa. Consider that the super-dense coding resource inequality in
\eqref{eq-coh:dense} generates two classical bit channels. Glancing at the
left-hand side of the teleportation resource inequality in
\eqref{eq-coh:tele}, we see that two classical bit channels generated from
super-dense coding are not sufficient to generate the noiseless qubit channel
on the right-hand side of \eqref{eq-coh:tele}---the protocol requires the
consumption of noiseless entanglement in addition to the consumption of the
two noiseless classical bit channels.

Is there a way for teleportation and super-dense coding to become dual under
resource reversal? One way is if we assume that \textit{entanglement is a free
resource}. This assumption is strong and we may have difficulty justifying it
from a practical standpoint because noiseless entanglement is extremely
fragile. It is also a powerful resource, as the teleportation and super-dense
coding protocols demonstrate. But in the theory of quantum communication, we
often make assumptions such as this one---such assumptions tend to give a
dramatic simplification of a problem. Continuing with our development, let us
assume that entanglement is a free resource and that we do not have to factor
it into the resource count. Under this assumption, the resource inequality for
teleportation becomes%
\begin{equation}
2\left[  c\rightarrow c\right]  \geq\left[  q\rightarrow q\right]  ,
\end{equation}
and that for super-dense coding becomes%
\begin{equation}
\left[  q\rightarrow q\right]  \geq2\left[  c\rightarrow c\right]  .
\end{equation}
Teleportation and super-dense coding are then dual under resource reversal
under the \textquotedblleft free-entanglement\textquotedblright\ assumption,
and we obtain the following \textit{resource equality}:%
\begin{equation}
\left[  q\rightarrow q\right]  =2\left[  c\rightarrow c\right]  .
\end{equation}

\begin{exercise}
Suppose that the quantum capacity of a quantum channel assisted by an
unlimited amount of entanglement is equal to some number $Q$. What is the
capacity of that entanglement-assisted channel for transmitting classical information?
\end{exercise}

\begin{exercise}
How can we obtain the following resource equality? (Hint: Assume that some
resource is free.)%
\begin{equation}
\left[  q\rightarrow q\right]  =\left[  qq\right]  .
\end{equation}
Which noiseless protocols did you use to show the above resource equality? The
above resource equality is a powerful statement: entanglement and quantum
communication are equivalent under the assumption that you have found.
\end{exercise}

\begin{exercise}
Suppose that the entanglement generation capacity of a quantum channel is
equal to some number $E$. What is the quantum capacity of that channel when
assisted by free, forward classical communication?
\end{exercise}

The above assumptions are useful for finding simple ways to make protocols
dual under resource reversal, and we will exploit them later in our proofs of
various capacity theorems in quantum Shannon theory. But it turns out that
there is a more clever way to make teleportation and super-dense coding dual
under resource reversal. In this chapter, we introduce a new resource---the
\textit{noiseless coherent bit channel}. This resource produces
\textquotedblleft coherent\textquotedblright\ versions of the teleportation
and super-dense coding protocols that are dual under resource reversal. The
payoff of this coherent communication technique is that we can exploit it to
simplify the proofs of various coding theorems of quantum Shannon theory. It
also leads to a deeper understanding of the relationship between the
teleportation and super-dense coding protocols from the previous chapter.

\section{Definition of Coherent Communication}

We begin by introducing the coherent bit channel%
\index{coherent bit channel}
as a classical channel that has ``quantum feedback'' (in a particular sense).
Recall from Exercise~\ref{ex-3np:dephasing-classical}\ that a classical bit
channel is equivalent to a dephasing channel that dephases in the
computational basis with dephasing parameter $p=1/2$. The CPTP\ map
corresponding to this completely dephasing channel is as follows:%
\begin{equation}
\mathcal{N}( \rho) =\frac{1}{2}\left(  \rho+Z\rho Z\right)  .
\end{equation}
An isometric extension $U_{A\rightarrow BE}^{\mathcal{N}}$ of the above
channel then follows by applying \eqref{eq-qt:channel-isometry}:%
\begin{equation}
U_{A\rightarrow BE}^{\mathcal{N}}=\frac{1}{\sqrt{2}}\left(  I_{A\rightarrow
B}\otimes\vert+\rangle_{E}+Z_{A\rightarrow B}\otimes\vert-\rangle_{E}\right)
,
\end{equation}
where we choose the orthonormal basis states of the environment $E$ to be
$\vert+\rangle$ and $\vert-\rangle$ (recall that we have unitary freedom in
the choice of the basis states for the environment). It is straightforward to
show that the isometry $U_{A\rightarrow BE}^{\mathcal{N}}$ is as follows by
expanding the operators $I$ and$~Z$ and the states $\vert+\rangle$ and
$\vert-\rangle$:%
\begin{equation}
U_{A\rightarrow BE}^{\mathcal{N}}=\vert0\rangle_{B}\langle0\vert_{A}%
\otimes\vert0\rangle_{E}+\vert1\rangle_{B}\langle1\vert_{A}\otimes
\vert1\rangle_{E}.
\end{equation}
Thus, a classical bit channel is equivalent to the following map, with its
action extended by linearity:%
\begin{equation}
\vert i\rangle_{A}\rightarrow\vert i\rangle_{B}\vert i\rangle_{E}:i\in\left\{
0,1\right\}  .
\end{equation}

\begin{figure}
[ptb]
\begin{center}
\includegraphics[
width=2.4094in
]%
{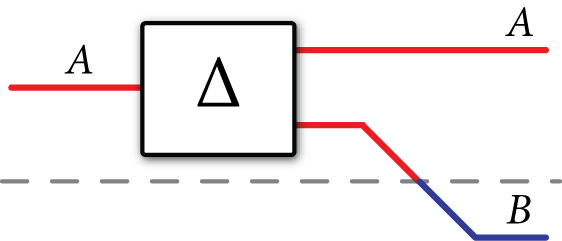}%
\caption{This figure depicts the operation of a coherent bit channel. It is
the \textquotedblleft coherification\textquotedblright\ of a classical bit
channel in which the sender $A$ has access to the environment's output.}%
\label{fig:coherent-channel}%
\end{center}
\end{figure}

A coherent bit channel is similar to the above classical bit channel map, with
the exception that we assume that Alice somehow regains control of the
environment of the channel:%
\begin{equation}
\vert i\rangle_{A}\rightarrow\vert i\rangle_{B}\vert i\rangle_{A}:i\in\left\{
0,1\right\}  . \label{eq-coh:cobit}%
\end{equation}
\textquotedblleft Coherence\textquotedblright\ in this context is also
synonymous with linearity---the maintenance and linear transformation of
superposed states. The coherent bit channel is similar to classical copying
because it copies the basis states while maintaining coherent superpositions.
We denote the resource of a coherent bit channel as follows:%
\begin{equation}
\left[  q\rightarrow qq\right]  .
\end{equation}
Figure~\ref{fig:coherent-channel}\ provides a visual depiction of the coherent
bit channel.%

\begin{exercise}
Show that the following resource inequality holds:%
\begin{equation}
\left[  q\rightarrow qq\right]  \geq\left[  c\rightarrow c\right]  .
\end{equation}
That is, devise a protocol that generates a noiseless classical bit channel
with one use of a noiseless coherent bit channel.
\end{exercise}

\section{Implementations of a Coherent Bit Channel}

\begin{figure}
[ptb]
\begin{center}
\includegraphics[
width=3.7542in
]%
{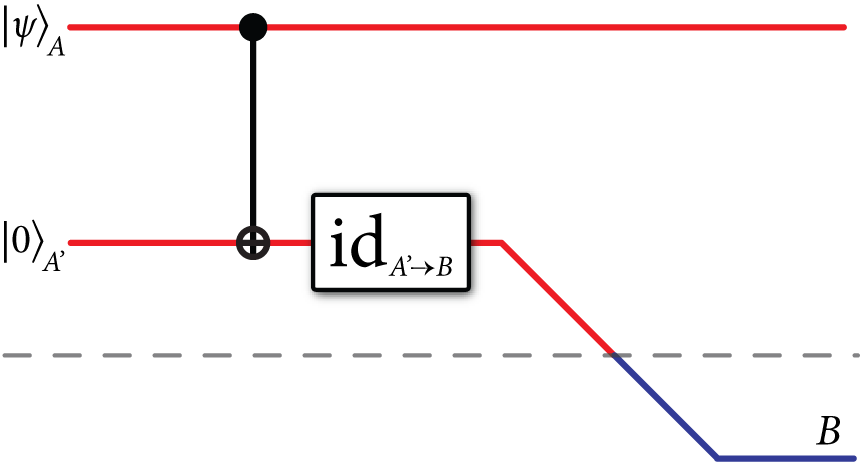}%
\caption{A simple protocol to implement a noiseless coherent channel with one
use of a noiseless qubit channel.}%
\label{fig-coh:naive-cobit}%
\end{center}
\end{figure}

How might we actually implement a coherent bit channel?\ The simplest way to
do so is with the aid of a local CNOT\ gate and a noiseless qubit channel. The
protocol proceeds as follows (Figure~\ref{fig-coh:naive-cobit}\ illustrates
the protocol):

\begin{enumerate}
\item Alice possesses an information qubit in the state $|\psi\rangle
_{A}\equiv\alpha|0\rangle_{A}+\beta|1\rangle_{A}$. She prepares an ancilla
qubit in the state $|0\rangle_{A^{\prime}}$.

\item Alice performs a local CNOT\ gate from qubit $A$ to qubit $A^{\prime}$.
The resulting state is%
\begin{equation}
\alpha\vert0\rangle_{A}\vert0\rangle_{A^{\prime}}+\beta\vert1\rangle_{A}%
\vert1\rangle_{A^{\prime}}.
\end{equation}

\item Alice transmits qubit $A^{\prime}$ to Bob with one use of a noiseless
qubit channel~$\operatorname{id}_{A^{\prime}\rightarrow B}$. The resulting
state is%
\begin{equation}
\alpha\vert0\rangle_{A}\vert0\rangle_{B}+\beta\vert1\rangle_{A}\vert
1\rangle_{B},
\end{equation}
and it is now clear that Alice and Bob have implemented a noiseless coherent
bit channel as defined in \eqref{eq-coh:cobit}.
\end{enumerate}

\noindent The above protocol realizes the following resource inequality:%
\begin{equation}
\left[  q\rightarrow q\right]  \geq\left[  q\rightarrow qq\right]  ,
\end{equation}
demonstrating that quantum communication generates coherent communication.

\begin{exercise}
Show that the following resource inequality holds:%
\begin{equation}
\left[  q\rightarrow qq\right]  \geq\left[  qq\right]  .
\end{equation}
That is, devise a protocol that generates a noiseless ebit with one use of a
noiseless coherent bit channel.
\end{exercise}

\begin{exercise}
Show that the following two resource inequalities cannot hold:%
\begin{align}
\left[  q\rightarrow qq\right]   &  \geq\left[  q\rightarrow q\right]  ,\\
\left[  qq\right]   &  \geq\left[  q\rightarrow qq\right]  .
\end{align}

\end{exercise}

We now have the following chain of resource inequalities:%
\begin{equation}
\left[  q\rightarrow q\right]  \geq\left[  q\rightarrow qq\right]  \geq\left[
qq\right]  .
\end{equation}
Thus, the power of the coherent bit channel lies in between that of a
noiseless qubit channel and a noiseless ebit.

\begin{exercise}
\label{ex-coh:coh-comm-ass-class}Another way to implement a noiseless coherent
bit channel is with a variation of teleportation that we name
\textquotedblleft coherent communication assisted by entanglement and
classical communication.\textquotedblright\ Suppose that Alice and Bob share
an ebit $\left\vert \Phi^{+}\right\rangle _{AB}$. Alice can append an ancilla
qubit $|0\rangle_{A^{\prime}}$ to this state and perform a local CNOT\ from
$A$ to $A^{\prime}$ to give the following state:%
\begin{equation}
|\Phi_{\operatorname{GHZ}}\rangle_{AA^{\prime}B}=\frac{1}{\sqrt{2}}\left(
|000\rangle_{AA^{\prime}B}+|111\rangle_{AA^{\prime}B}\right)  .
\end{equation}
Alice prepends an information qubit $|\psi\rangle_{A_{1}}\equiv\alpha
|0\rangle_{A_{1}}+\beta|1\rangle_{A_{1}}$ to the above state so that the
global state is as follows:%
\begin{equation}
|\psi\rangle_{A_{1}}|\Phi_{\operatorname{GHZ}}\rangle_{AA^{\prime}B}.
\end{equation}
Suppose Alice performs the usual teleportation operations on systems $A_{1}$,
$A$, and $A^{\prime}$. Give the steps that Alice and Bob should perform in
order to generate the state $\alpha|0\rangle_{A^{\prime}}|0\rangle_{B}%
+\beta|1\rangle_{A^{\prime}}|1\rangle_{B}$, thus implementing a noiseless
coherent bit channel. \textit{Hint: }The resource inequality for this protocol
is as follows:%
\begin{equation}
\left[  qq\right]  +\left[  c\rightarrow c\right]  \geq\left[  q\rightarrow
qq\right]  .
\end{equation}
This should be compared with the teleportation protocol, which corresponds to
$\left[  qq\right]  +2\left[  c\rightarrow c\right]  \geq\left[  q\rightarrow
q\right]  $.
\end{exercise}

\begin{exercise}
Determine a qudit version of coherent communication assisted by classical
communication and entanglement by modifying the steps in the above protocol.
\end{exercise}

\section{Coherent Dense Coding}

In the previous section, we introduced two protocols that implement a
noiseless coherent bit channel: the simple method in the previous section and
coherent communication assisted by classical communication and entanglement
(Exercise~\ref{ex-coh:coh-comm-ass-class}). We now introduce a different
method for implementing two coherent bit channels that makes more judicious
use of available resources. We name it
\index{super-dense coding!coherent}%
\textit{coherent super-dense coding }because it is a coherent version of the
super-dense coding protocol.%
\begin{figure}
[ptb]
\begin{center}
\includegraphics[
width=4.024in
]%
{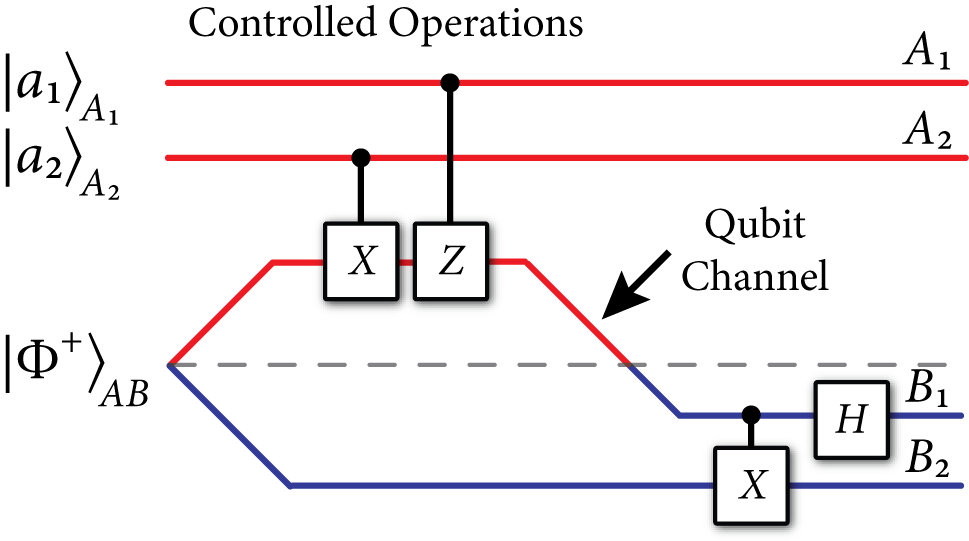}%
\caption{This figure depicts the protocol for coherent super-dense coding.}%
\label{fig:coherent-dense}%
\end{center}
\end{figure}

The protocol proceeds as follows (Figure~\ref{fig:coherent-dense}\ depicts the protocol):

\begin{enumerate}
\item Alice and Bob share one ebit in the state $\left\vert \Phi
^{+}\right\rangle _{AB}$ before the protocol begins.

\item Alice first prepares two qubits $A_{1}$ and $A_{2}$ in the state
$\left\vert a_{1}\right\rangle _{A_{1}}\left\vert a_{2}\right\rangle _{A_{2}}$
and prepends this state to the ebit. The global state is as follows:%
\begin{equation}
\left\vert a_{1}\right\rangle _{A_{1}}\left\vert a_{2}\right\rangle _{A_{2}%
}\left\vert \Phi^{+}\right\rangle _{AB},
\end{equation}
where $a_{1}$ and $a_{2}$ are binary-valued. This preparation step is
reminiscent of the super-dense coding protocol (recall that, in the
super-dense coding protocol, Alice has two classical bits she would like to communicate).

\item Alice performs a CNOT\ gate from register $A_{2}$ to register $A$ and
performs a controlled-$Z$ gate from register $A_{1}$ to register $A$. The
resulting state is as follows:%
\begin{equation}
\left\vert a_{1}\right\rangle _{A_{1}}\left\vert a_{2}\right\rangle _{A_{2}%
}Z_{A}^{a_{1}}X_{A}^{a_{2}}\left\vert \Phi^{+}\right\rangle _{AB}.
\end{equation}

\item Alice transmits the qubit in register $A$ to Bob. We rename this
register as $B_{1}$ and Bob's other register $B$ as $B_{2}$.

\item Bob performs a CNOT\ gate from his register $B_{1}$ to $B_{2}$ and
performs a Hadamard gate on $B_{1}$. The final state is as follows:%
\begin{equation}
\left\vert a_{1}\right\rangle _{A_{1}}\left\vert a_{2}\right\rangle _{A_{2}%
}\left\vert a_{1}\right\rangle _{B_{1}}\left\vert a_{2}\right\rangle _{B_{2}}.
\end{equation}

\end{enumerate}

The above protocol implements two coherent bit channels:\ one from $A_{1}$ to
$B_{1}$ and another from $A_{2}$ to $B_{2}$. You can check that the protocol
works for arbitrary superpositions of two-qubit states on $A_{1}$ and $A_{2}%
$---it is for this reason that this protocol implements two coherent bit
channels. The resource inequality corresponding to coherent super-dense coding
is%
\begin{equation}
\left[  qq\right]  +\left[  q\rightarrow q\right]  \geq2\left[  q\rightarrow
qq\right]  . \label{eq:coh-dense}%
\end{equation}

\begin{exercise}
Construct a \textit{qudit} version of coherent super-dense coding that
implements the following resource inequality:%
\begin{equation}
\log d\left[  qq\right]  +\log d\left[  q\rightarrow q\right]  \geq2\log
d\left[  q\rightarrow qq\right]  .
\end{equation}
(\textit{Hint}: The qudit generalization of a controlled-NOT gate is
$\sum_{i=0}^{d-1}\vert i\rangle\langle i\vert\otimes X( i) , $ where $X$ is
defined in \eqref{eq-qt:X-op}, of a controlled-$Z$ gate is $\sum_{j=0}%
^{d-1}\vert j\rangle\langle j\vert\otimes Z( j) , $ where $Z$ is defined in
\eqref{eq-qt:Z-op}, and of the Hadamard gate is the Fourier transform gate.)
\end{exercise}

\section{Coherent Teleportation}

\label{sec-coh:coh-tele}We now introduce a coherent version of the
teleportation protocol that we name%
\index{quantum teleportation!coherent}
\textit{coherent teleportation}. Let a $Z$ coherent bit channel $\Delta_{Z}$
be one that copies eigenstates of the $Z$ operator (this is as we defined a
coherent bit channel before). Let an $X$ coherent bit channel $\Delta_{X}$ be
one that copies eigenstates of the $X$ operator:%
\begin{align}
\Delta_{X}: \vert+\rangle_{A}  &  \rightarrow\vert+\rangle_{A}\vert
+\rangle_{B},\\
\vert-\rangle_{A}  &  \rightarrow\vert-\rangle_{A}\vert-\rangle_{B}.
\end{align}
It does not really matter which basis we use to define a coherent bit
channel---it just matters that it copies the orthogonal states of some basis.

\begin{exercise}
Show how to simulate an $X$ coherent bit channel using a $Z$ coherent bit
channel and local operations.
\end{exercise}

The protocol proceeds as follows (Figure~\ref{fig:coh-tele} depicts the
protocol):%
\begin{figure}
[ptb]
\begin{center}
\includegraphics[
width=4.9614in
]%
{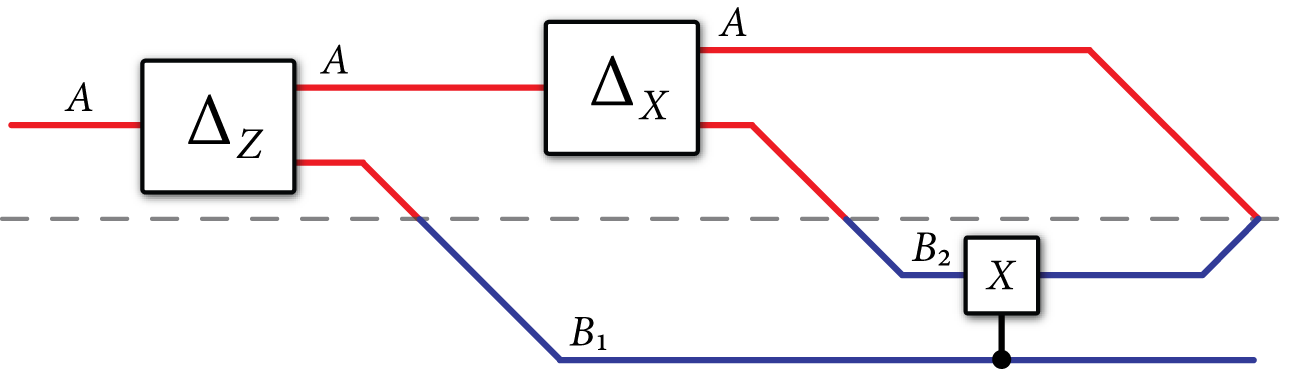}%
\caption{This figure depicts the protocol for coherent teleportation.}%
\label{fig:coh-tele}%
\end{center}
\end{figure}

\begin{enumerate}
\item Alice possesses an information qubit $\vert\psi\rangle_{A}$ where%
\begin{equation}
\vert\psi\rangle_{A}\equiv\alpha\vert0\rangle_{A}+\beta\vert1\rangle_{A}.
\end{equation}
She sends her qubit through a $Z$ coherent bit channel:%
\begin{equation}
\vert\psi\rangle_{A}\ \ \ \ \underrightarrow{\Delta_{Z}}\ \ \ \ \alpha
\vert0\rangle_{A}\vert0\rangle_{B_{1}}+\beta\vert1\rangle_{A}\vert
1\rangle_{B_{1}}\equiv|\tilde{\psi}\rangle_{AB_{1}}.
\end{equation}
Let us rewrite the above state $|\tilde{\psi}\rangle_{AB_{1}}$ as follows:%
\begin{align}
|\tilde{\psi}\rangle_{AB_{1}}  &  =\alpha\left(  \frac{\left\vert
+\right\rangle _{A}+\vert-\rangle_{A}}{\sqrt{2}}\right)  \vert0\rangle_{B_{1}%
}+\beta\left(  \frac{\vert+\rangle_{A}-\left\vert -\right\rangle _{A}}%
{\sqrt{2}}\right)  \vert1\rangle_{B_{1}}\\
&  =\frac{1}{\sqrt{2}}\left[  \vert+\rangle_{A}\left(  \alpha\vert
0\rangle_{B_{1}}+\beta\vert1\rangle_{B_{1}}\right)  +\vert-\rangle_{A}\left(
\alpha\vert0\rangle_{B_{1}}-\beta\vert1\rangle_{B_{1}}\right)  \right]  .
\end{align}

\item Alice sends her qubit $A$ through an $X$ coherent bit channel with
output systems $A$ and~$B_{2}$:%
\begin{multline}
|\tilde{\psi}\rangle_{AB_{1}}\ \ \ \ \underrightarrow{\Delta_{X}}%
\ \ \ \ \frac{1}{\sqrt{2}}\vert+\rangle_{A}\vert+\rangle_{B_{2}}\left(
\alpha\vert0\rangle_{B_{1}}+\beta\vert1\rangle_{B_{1}}\right) \\
+\frac{1}{\sqrt{2}}\vert-\rangle_{A}\vert-\rangle_{B_{2}}\left(  \alpha
\vert0\rangle_{B_{1}}-\beta\vert1\rangle_{B_{1}}\right)  .
\end{multline}

\item Bob then performs a CNOT\ gate from qubit $B_{1}$ to qubit $B_{2}$.
Consider that the action of the CNOT\ gate with the source qubit in the
computational basis and the target qubit in the +/$-$ basis is as follows:%
\begin{align}
|0\rangle|+\rangle &  \rightarrow|0\rangle|+\rangle,\\
|0\rangle|-\rangle &  \rightarrow|0\rangle|-\rangle,\\
|1\rangle|+\rangle &  \rightarrow|1\rangle|+\rangle,\\
|1\rangle|-\rangle &  \rightarrow-|1\rangle|-\rangle,
\end{align}
so that the last entry catches a phase of $\pi$ ($e^{i\pi}=-1$). Then this
CNOT\ gate brings the overall state to%
\begin{align}
&  \frac{1}{\sqrt{2}}\left[  |+\rangle_{A}\vert+\rangle_{B_{2}}\left(
\alpha|0\rangle_{B_{1}}+\beta|1\rangle_{B_{1}}\right)  +|-\rangle_{A}%
|-\rangle_{B_{2}}\left(  \alpha|0\rangle_{B_{1}}+\beta|1\rangle_{B_{1}%
}\right)  \right] \nonumber\\
&  =\frac{1}{\sqrt{2}}\left[  |+\rangle_{A}\vert+\rangle_{B_{2}}|\psi
\rangle_{B_{1}}+|-\rangle_{A}\left\vert -\right\rangle _{B_{2}}|\psi
\rangle_{B_{1}}\right] \\
&  =\frac{1}{\sqrt{2}}\left[  |+\rangle_{A}\vert+\rangle_{B_{2}}+|-\rangle
_{A}|-\rangle_{B_{2}}\right]  |\psi\rangle_{B_{1}}\\
&  =\left\vert \Phi^{+}\right\rangle _{AB_{2}}|\psi\rangle_{B_{1}}.
\end{align}
Thus, Alice teleports her information qubit to Bob, and both Alice and Bob
possess one ebit at the end of the protocol.
\end{enumerate}

\noindent The resource inequality for coherent teleportation is as follows:%
\begin{equation}
2[q\rightarrow qq]\geq\lbrack qq]+[q\rightarrow q]. \label{eq:coh-tele}%
\end{equation}

\begin{exercise}
Show how a cobit channel and an ebit can generate a GHZ\ state. That is,
demonstrate a protocol that realizes the following resource inequality:%
\begin{equation}
\left[  qq\right]  _{AB}+\left[  q\rightarrow qq\right]  _{BC}\geq\left[
qqq\right]  _{ABC}.
\end{equation}

\end{exercise}

\begin{exercise}
Outline the qudit version of the above coherent teleportation protocol. The
protocol should realize the following resource inequality:%
\begin{equation}
2\log d[q\rightarrow qq]\geq\log d[qq]+\log d[q\rightarrow q].
\end{equation}

\end{exercise}

\begin{exercise}
Outline a catalytic version of the coherent teleportation protocol by
modifying the original teleportation protocol. Let Alice possess an
information qubit $\vert\psi\rangle_{A^{\prime}}$ and let Alice and Bob share
an ebit $\left\vert \Phi^{+}\right\rangle _{AB}$. Replace the Bell measurement
with a controlled-NOT\ and Hadamard gate, replace the classical bit channels
with coherent bit channels, and replace Bob's conditional unitary operations
with controlled unitary operations. The resulting resource inequality should
be of the form:%
\begin{equation}
2\left[  q\rightarrow qq\right]  +\left[  qq\right]  \geq\left[  q\rightarrow
q\right]  +2\left[  qq\right]  .
\end{equation}
This protocol is catalytic in the sense that it gives the resource inequality
in \eqref{eq:coh-tele} when we cancel one ebit from each side.
\end{exercise}

\section{Coherent Communication Identity}

\label{sec-cc:coh-comm-identity}The fundamental result of this chapter is the
\index{coherent communication identity}%
\textit{coherent communication identity}:%
\begin{equation}
2[q\rightarrow qq]=[qq]+[q\rightarrow q]. \label{eq-coh:identity}%
\end{equation}
We obtain this identity by combining the resource inequality for coherent
super-dense coding in \eqref{eq:coh-dense} and the resource inequality for
coherent teleportation in \eqref{eq:coh-tele}. The coherent communication
identity demonstrates that coherent super-dense coding and coherent
teleportation are dual under resource reversal---the resources that coherent
teleportation consumes are the same as those that coherent super-dense coding
generates and vice versa.

The major application of the coherent communication identity is in noisy
quantum Shannon theory. We will find later that its application is in the
\textquotedblleft upgrading\textquotedblright\ of protocols that output
private classical information. Suppose that a protocol outputs private
classical bits. The super-dense coding protocol is one such example, as the
last paragraph of Section~\ref{sec:dense-coding}\ argues. Then it is possible
to upgrade the protocol by making it coherent, similar to the way in which we
made super-dense coding coherent by replacing conditional unitary operations
with controlled unitary operations.

We make this idea more precise with an example. The resource inequality for
entanglement-assisted classical coding (discussed in more detail in
Chapter~\ref{chap:EA-classical}) has the following form:%
\begin{equation}
\left\langle \mathcal{N}\right\rangle +E\left[  qq\right]  \geq C\left[
c\rightarrow c\right]  ,
\end{equation}
where $\mathcal{N}$ is a noisy quantum channel that connects Alice to Bob, $E$
is some rate of entanglement consumption, and $C$ is some rate of classical
communication. It is possible to upgrade the generated classical bits to
coherent bits, for reasons that are similar to those that we used in the
upgrading of super-dense coding. The resulting resource inequality has the
following form:%
\begin{equation}
\left\langle \mathcal{N}\right\rangle +E\left[  qq\right]  \geq C\left[
q\rightarrow qq\right]  .
\end{equation}
We can now employ the coherent communication identity in
\eqref{eq-coh:identity} and argue that any protocol that realizes the above
resource inequality can realize the following one:%
\begin{equation}
\left\langle \mathcal{N}\right\rangle +E\left[  qq\right]  \geq\frac{C}%
{2}\left[  q\rightarrow q\right]  +\frac{C}{2}\left[  qq\right]  ,
\end{equation}
merely by using the generated coherent bits in a coherent super-dense coding
protocol. We can then make a \textquotedblleft catalytic
argument\textquotedblright\ to cancel the ebits on both sides of the resource
inequality. The final resource inequality is as follows:%
\begin{equation}
\left\langle \mathcal{N}\right\rangle +\left(  E-\frac{C}{2}\right)  \left[
qq\right]  \geq\frac{C}{2}\left[  q\rightarrow q\right]  .
\end{equation}
The above resource inequality corresponds to a protocol for
\textit{entanglement-assisted quantum communication}, and it turns out to be
optimal for some channels\ as this protocol's converse theorem shows. This
optimality is due to the efficient translation of classical bits to coherent
bits and the application of the coherent communication identity.

\section{History and Further Reading}

\cite{prl2004harrow} introduced the idea of coherent communication. Later, the
idea of the coherent bit channel was generalized to the continuous-variable
case \citep{wilde:060303}. Coherent communication has many applications in
quantum Shannon theory which we will study in later chapters.

\chapter{Unit Resource Capacity Region}

\label{chap:unit-resource-cap}In Chapter~\ref{chap:three-noiseless}, we
presented the three unit protocols of teleportation, super-dense coding, and
entanglement distribution. We argued in Section~\ref{sec-3np:optimality} that
each of these protocols are individually optimal. For example, recall that the
entanglement distribution protocol is optimal because two parties cannot
generate more than one ebit from the use of one noiseless qubit channel.

In this chapter, we show that these three protocols are actually the most
important protocols---we do not need to consider any other protocols when the
noiseless resources of classical communication, quantum communication, and
entanglement are available. Combining these three protocols together is the
best that one can do with the unit resources.

In this sense, this chapter gives a good example of a converse proof of a
capacity theorem. We construct a three-dimensional region, known as the unit
resource achievable region, that the three unit protocols fill out. The
converse proof of this chapter employs physical arguments to show that the
unit resource achievable region is optimal, and we can then refer to it as
the
\index{unit resource capacity region}%
unit resource capacity region. We later exploit the development here when we
get to the study of trade-off capacities (see Chapter~\ref{chap:trade-off}).

\section{The Unit Resource Achievable Region}

Let us first recall the resource inequalities for the three unit protocols.
The resource inequality for teleportation is%
\begin{equation}
2[c\rightarrow c]+[qq]\geq\lbrack q\rightarrow q], \label{TP}%
\end{equation}
while that for super-dense coding is%
\begin{equation}
\lbrack q\rightarrow q]+[qq]\geq2[c\rightarrow c], \label{SD}%
\end{equation}
and that for entanglement distribution is as follows:%
\begin{equation}
\lbrack q\rightarrow q]\geq\lbrack qq]. \label{ED}%
\end{equation}
Each of the resources $\left[  q\rightarrow q\right]  $, $\left[  qq\right]
$, $\left[  c\rightarrow c\right]  $ is a \textit{unit resource}.

The above three unit protocols are sufficient to recover all other unit
protocols. For example, we can combine super-dense coding and entanglement
distribution to produce the following resource inequality:%
\begin{equation}
2[q\rightarrow q]+[qq]\geq2[c\rightarrow c]+\left[  qq\right]  .
\end{equation}
The above resource inequality is equivalent to the following one:%
\begin{equation}
\lbrack q\rightarrow q]\geq\lbrack c\rightarrow c], \label{QC}%
\end{equation}
after removing the entanglement from both sides and scaling by $1/2$ (we can
remove the entanglement here because it acts as a catalytic resource).

We can justify this scaling by considering a scenario in which we use the
above protocol $N$ times. For the first run of the protocol, we require one
ebit to get it started, but then every other run both consumes and generates
one ebit, giving%
\begin{equation}
2N[q\rightarrow q]+[qq]\geq2N[c\rightarrow c]+\left[  qq\right]  .
\end{equation}
Dividing by $N$ gives the rate of the task, and as $N$ becomes large, the use
of the initial ebit is negligible. We refer to (\ref{QC}) as \textquotedblleft
classical coding over a noiseless qubit channel.\textquotedblright

We can think of the above resource inequalities in a different way. Let us
consider a three-dimensional space with points of the form $(C,Q,E)$, where
$C$ corresponds to noiseless classical communication, $Q$ corresponds to
noiseless quantum communication, and $E$ corresponds to noiseless
entanglement. Each point in this space corresponds to a protocol involving the
unit resources. A coordinate of a point is negative if the point's
corresponding resource inequality consumes that coordinate's corresponding
resource, and a coordinate of a point is positive if the point's corresponding
resource inequality generates that coordinate's corresponding resource.%

\begin{figure}
[ptb]
\begin{center}
\includegraphics[
width=4.0413in
]%
{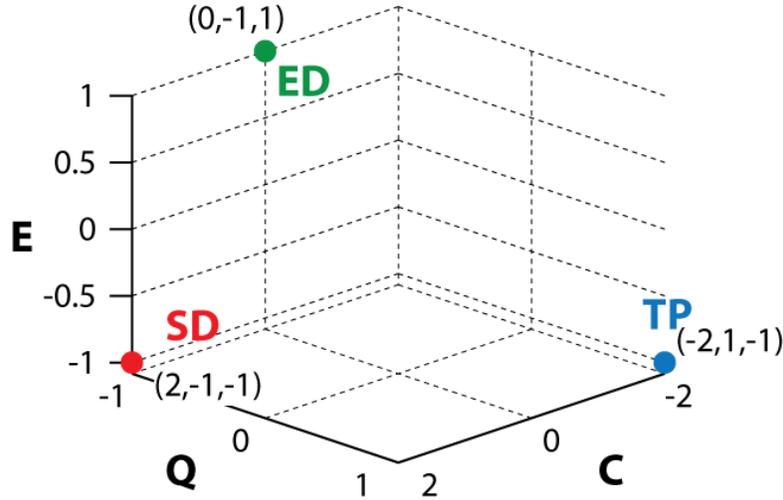}%
\caption{The three points corresponding to the three respective unit protocols
of entanglement distribution (ED), teleportation (TP), and super-dense coding
(SD).}%
\label{fig-ur:points}%
\end{center}
\end{figure}
For example, the point corresponding to the
\index{quantum teleportation}%
teleportation protocol is
\begin{equation}
x_{\operatorname{TP}}\equiv\left(  -2,1,-1\right)  ,
\end{equation}
because teleportation consumes two noiseless classical bit channels and one
ebit to generate one noiseless qubit channel. For similar reasons, the
respective points corresponding to super-dense coding and entanglement
distribution are as follows:%
\begin{equation}
x_{\operatorname{SD}} \equiv\left(  2,-1,-1\right)
,\ \ \ \ x_{\operatorname{ED}} \equiv\left(  0,-1,1\right)  .
\end{equation}
Figure~\ref{fig-ur:points}\ plots these three points in the three-dimensional
space of classical communication, quantum communication, and entanglement.

We can execute any of the three unit protocols just one time, or we can
execute any one of them $m$ times where $m$ is some positive integer.
Executing a protocol $m$ times then gives other points in the
three-dimensional space. That is, we can also achieve the points
$mx_{\operatorname{TP}}$, $mx_{\operatorname{SD}}$, and $mx_{\operatorname{ED}%
}$ for any positive $m$. This method allows us to fill up a certain portion of
the three-dimensional space and provides an interpretation for resource
inequalities with rational coefficients. Since the rational numbers are dense
in the real numbers, we can also allow for real number coefficients. This
becomes important later on when we consider combining the three unit protocols
in order to achieve certain rates of transmission (a communication rate can be
any real number). Thus, we can combine the protocols together to achieve any
point of the following form:%
\begin{equation}
\alpha x_{\operatorname{TP}}+\beta x_{\operatorname{SD}}+\gamma
x_{\operatorname{ED}},
\end{equation}
where $\alpha,\beta,\gamma\geq0$.

Let us further establish some notation. Let $L$ denote a line, $Q$ a quadrant,
and $O$ an octant in the three-dimensional space (it should be clear from the
context whether $Q$ refers to quantum communication or \textquotedblleft
quadrant\textquotedblright). For example, $L^{-00}$ denotes a line going in
the direction of negative classical communication:%
\begin{equation}
L^{-00}\equiv\left\{  \alpha\left(  -1,0,0\right)  :\alpha\geq0\right\}  .
\end{equation}
$Q^{0+-}$ denotes the quadrant where there is zero classical communication,
generation of quantum communication, and consumption of entanglement:%
\begin{equation}
Q^{0+-}\equiv\left\{  \alpha\left(  0,1,0\right)  +\beta\left(  0,0,-1\right)
:\alpha,\beta\geq0\right\}  .
\end{equation}
$O^{+-+}$ denotes the octant where there is generation of classical
communication, consumption of quantum communication, and generation of
entanglement:%
\begin{equation}
O^{+-+}\equiv\left\{
\begin{array}
[c]{c}%
\alpha\left(  1,0,0\right)  +\beta\left(  0,-1,0\right)  +\gamma\left(
0,0,1\right) \\
:\alpha,\beta,\gamma\geq0
\end{array}
\right\}  .
\end{equation}

It proves useful to have a \textquotedblleft set addition\textquotedblright%
\ operation between two regions $A$ and $B$ (known as the Minkowski sum):%
\begin{equation}
A+B\equiv\{a+b:a\in A,b\in B\}.
\end{equation}
The following relations hold%
\begin{align}
Q^{0+-}  &  =L^{0+0}+L^{00-},\\
O^{+-+}  &  =L^{+00}+L^{0-0}+L^{00+},
\end{align}
by using the above definition.

The following geometric objects lie in the $(C,Q,E)$ space:

\begin{enumerate}
\item The \textquotedblleft line of teleportation\textquotedblright%
\ $L_{\operatorname{TP}}$ is the following set of points:
\begin{equation}
L_{\operatorname{TP}}\equiv\left\{  \alpha\left(  -2,1,-1\right)  :\alpha
\geq0\right\}  . \label{eq:line-TP}%
\end{equation}

\item The \textquotedblleft line of super-dense coding\textquotedblright%
\ $L_{\operatorname{SD}}$ is the following set of points:
\begin{equation}
L_{\operatorname{SD}}\equiv\left\{  \beta\left(  2,-1,-1\right)  :\beta
\geq0\right\}  . \label{eq:line-SD}%
\end{equation}

\item The \textquotedblleft line of entanglement
distribution\textquotedblright\ $L_{\operatorname{ED}}$ is the following set
of points:
\begin{equation}
L_{\operatorname{ED}}\equiv\left\{  \gamma\left(  0,-1,1\right)  :\gamma
\geq0\right\}  . \label{eq:line-ED}%
\end{equation}

\end{enumerate}

\begin{definition}
Let $\widetilde{C}_{\emph{U}}$ denote the unit resource achievable region. It
consists of all linear combinations of the above protocols:%
\begin{equation}
\widetilde{C}_{\emph{U}}\equiv L_{\emph{TP}}+L_{\emph{SD}}+L_{\emph{ED}}.
\label{ut_C}%
\end{equation}

\end{definition}

\begin{figure}[ptb]
\begin{center}
\includegraphics[
width=3.7421in
]{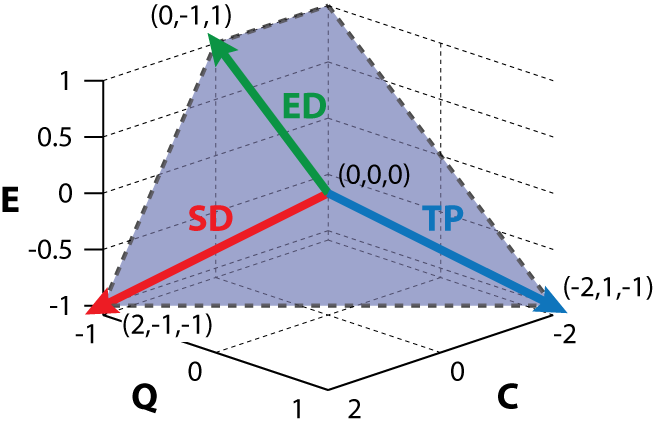}
\end{center}
\caption{This figure depicts the unit resource achievable region
$\widetilde{C}_{\operatorname{U}}$.}%
\label{fig-ur:unit}%
\end{figure}

The following matrix equation gives all achievable triples $(C,Q,E)$ in
$\widetilde{C}_{\operatorname{U}}$:%
\begin{equation}%
\begin{bmatrix}
C\\
Q\\
E
\end{bmatrix}
=%
\begin{bmatrix}
-2 & 2 & 0\\
1 & -1 & -1\\
-1 & -1 & 1
\end{bmatrix}%
\begin{bmatrix}
\alpha\\
\beta\\
\gamma
\end{bmatrix}
,
\end{equation}
where $\alpha,\beta,\gamma\geq0$. We can rewrite the above equation using the
matrix inverse:%
\begin{equation}%
\begin{bmatrix}
\alpha\\
\beta\\
\gamma
\end{bmatrix}
=%
\begin{bmatrix}
-1/2 & -1/2 & -1/2\\
0 & -1/2 & -1/2\\
-1/2 & -1 & 0
\end{bmatrix}%
\begin{bmatrix}
C\\
Q\\
E
\end{bmatrix}
,
\end{equation}
in order to express the coefficients $\alpha$, $\beta$, and $\gamma$ as a
function of the rate triples $(C,Q,E)$. The restriction of non-negativity of
$\alpha$, $\beta$, and $\gamma$ gives the following restriction on the
achievable rate triples $(C,Q,E)$:%
\begin{align}
C+Q+E  &  \leq0,\label{utriple1}\\
Q+E  &  \leq0,\label{utriple2}\\
C+2Q  &  \leq0. \label{utriple3}%
\end{align}
The above result implies that the achievable region $\widetilde{C}%
_{\operatorname{U}}$ in (\ref{ut_C}) is equivalent to all rate triples
satisfying \eqref{utriple1}--\eqref{utriple3}. Figure~\ref{fig-ur:unit}
displays the full unit resource achievable region.

\begin{definition}
The unit resource capacity region $C_{\emph{U}}$ is the closure of the set of
all points $(C,Q,E)$ in the $C,Q,E$ space, satisfying the following resource
inequality:%
\begin{equation}
0\geq C[c\rightarrow c]+Q[q\rightarrow q]+E[qq].
\label{eq:unit-capacity-region}%
\end{equation}

\end{definition}

The definition states that the unit resource capacity region consists of all
those points $\left(  C,Q,E\right)  $ that have corresponding protocols that
can implement them. The notation in the above definition may seem slightly
confusing at first glance until we recall that a resource with a negative rate
implicitly belongs on the left-hand side of the resource inequality.

Theorem~\ref{ut} below gives the optimal three-dimensional capacity region for
the three unit resources.

\begin{theorem}
\label{ut} The unit resource capacity region $C_{\emph{U}}$ is equal to the
unit resource achievable region $\widetilde{C}_{\emph{U}}$:%
\begin{equation}
C_{\emph{U}}=\widetilde{C}_{\emph{U}}.
\end{equation}

\end{theorem}

Proving the above theorem involves two steps:\ the \textit{direct coding
theorem} and the \textit{converse theorem}. For this case, the \textit{direct
coding theorem} establishes that the achievable region $\widetilde
{C}_{\operatorname{U}}$ is in the capacity region $C_{\operatorname{U}}$:%
\begin{equation}
\widetilde{C}_{\operatorname{U}}\subseteq C_{\operatorname{U}}.
\end{equation}
The \textit{converse theorem}, on the other hand, establishes optimality of
$\widetilde{C}_{\operatorname{U}}$:%
\begin{equation}
C_{\operatorname{U}}\subseteq\widetilde{C}_{\operatorname{U}}.
\end{equation}

\section{The Direct Coding Theorem}

The result of the direct coding theorem, that $\widetilde{C}_{\operatorname{U}%
}\subseteq C_{\operatorname{U}}$, is immediate from the definition in
(\ref{ut_C})\ of the unit resource achievable region $\widetilde
{C}_{\operatorname{U}}$, the definition in \eqref{eq:unit-capacity-region} of
the unit resource capacity region $C_{\operatorname{U}}$, and the theory of
resource inequalities. We can achieve points in the unit resource capacity
region simply by considering positive linear combinations of the three unit
protocols. The next section shows that the unit resource capacity region
consists of all and only those points in the unit resource achievable region.

\section{The Converse Theorem}

We employ the definition of $\widetilde{C}_{\operatorname{U}}$ in (\ref{ut_C})
and consider the eight octants of the $(C,Q,E)$ space individually in order to
prove the converse theorem (that $C_{\operatorname{U}}\subseteq\widetilde
{C}_{\operatorname{U}}$). Let $(\pm,\pm,\pm)$ denote labels for the eight
different octants.

It is possible to demonstrate the optimality of each of these three protocols
individually with a contradiction argument as we saw in
Chapter~\ref{chap:three-noiseless}. However, in the converse proof of
Theorem~\ref{ut}, we show that a mixed strategy combining these three unit
protocols is optimal.

We accept the following postulates and exploit them in order to prove the converse:

\begin{enumerate}
\item Entanglement alone cannot generate classical communication or quantum
communication or both.

\item Classical communication alone cannot generate entanglement or quantum
communication or both.

\item Holevo bound: One cannot generate more than one classical bit of
communication per use of a noiseless qubit channel alone.
\end{enumerate}

$\boldsymbol{(+,+,+)}$. This octant of $C_{\operatorname{U}}$ is empty because
a sender and receiver require some resources to implement classical
communication, quantum communication, and entanglement. (They cannot generate
a unit resource from nothing!)

$\boldsymbol{(+,+,-)}$. This octant of $C_{\operatorname{U}}$ is empty because
entanglement alone cannot generate either classical communication or quantum
communication or both.

$\boldsymbol{(+,-,+)}$. The task for this octant is to generate a noiseless
classical channel of $C$ bits and $E$ ebits of entanglement by consuming $|Q|$
qubits of quantum communication. We thus consider all points of the form
$\left(  C,Q,E\right)  $ where $C\geq0$, $Q\leq0$, and $E\geq0$. It suffices
to prove the following inequality:%
\begin{equation}
C+E\leq\left\vert Q\right\vert , \label{pnp}%
\end{equation}
because combining (\ref{pnp}) with $C\geq0$ and $E\geq0$ implies
\eqref{utriple1}--\eqref{utriple3}. The achievability of $(C,-|Q|,E)$ implies
the achievability of the point $(C+2E,-|Q|-E,0)$, because we can consume all
of the entanglement with super-dense coding (\ref{SD}):%
\begin{equation}
\left(  C+2E,-\left\vert Q\right\vert -E,0\right)  =\left(  C,-\left\vert
Q\right\vert ,E\right)  +\left(  2E,-E,-E\right)  .
\end{equation}
This new point implies that there is a protocol that consumes $\left\vert
Q\right\vert +E$ noiseless qubit channels to send $C+2E$ classical bits. The
following bound then applies%
\begin{equation}
C+2E\leq\left\vert Q\right\vert +E,
\end{equation}
because the Holevo bound (Exercise~\ref{ex-nqt:nayak}\ gives a simpler
statement of this bound)\ states that we can send only one classical bit per
qubit. The bound in (\ref{pnp}) then follows.

$\boldsymbol{(+,-,-)}$. The task for this octant is to simulate a classical
channel of size $C$ bits using $|Q|$ qubits of quantum communication and $|E|$
ebits of entanglement. We consider all points of the form $\left(
C,Q,E\right)  $ where $C\geq0$, $Q\leq0$, and $E\leq0$. It suffices to prove
the following inequalities:%
\begin{align}
C  &  \leq2|Q|,\label{pnn1}\\
C  &  \leq|Q|+|E|, \label{pnn2}%
\end{align}
because combining \eqref{pnn1}--\eqref{pnn2} with $C\geq0$ implies
\eqref{utriple1}--\eqref{utriple3}. The achievability of $(C,-|Q|,-|E|)$
implies the achievability of $(0,-|Q|+C/2,-|E|-C/2)$, because we can consume
all of the classical communication with teleportation (\ref{TP}):%
\begin{equation}
\left(  0,-|Q|+C/2,-|E|-C/2\right)  =\left(  C,-|Q|,-|E|\right)  +\left(
-C,C/2,-C/2\right)  .
\end{equation}
The following bound applies (quantum communication cannot be positive):%
\begin{equation}
-\left\vert Q\right\vert +C/2\leq0,
\end{equation}
because entanglement alone cannot generate quantum communication. The bound in
(\ref{pnn1}) then follows from the above bound. The achievability of
$(C,-|Q|,-|E|)$ implies the achievability of $(C,-|Q|-|E|,0)$ because we can
consume an extra $\left\vert E\right\vert $ qubit channels with entanglement
distribution (\ref{ED}):%
\begin{equation}
\left(  C,-|Q|-|E|,0\right)  =\left(  C,-|Q|,-|E|\right)  +\left(
0,-\left\vert E\right\vert ,\left\vert E\right\vert \right)  .
\end{equation}
The bound in (\ref{pnn2})\ then applies by the same Holevo bound argument as
in the previous octant.

$\boldsymbol{(-,+,+)}$. This octant of $C_{\operatorname{U}}$ is empty because
classical communication alone cannot generate either quantum communication or
entanglement or both.

$\boldsymbol{(-,+,-)}$. The task for this octant is to simulate a quantum
channel of size $Q$ qubits using $|E|$ ebits of entanglement and $|C|$ bits of
classical communication. We consider all points of the form $\left(
C,Q,E\right)  $ where $C\leq0$, $Q\geq0$, and $E\leq0$. It suffices to prove
the following inequalities:%
\begin{align}
Q  &  \leq\left\vert E\right\vert ,\label{npn1}\\
2Q  &  \leq\left\vert C\right\vert , \label{npn2}%
\end{align}
because combining them with $C\leq0$ implies
\eqref{utriple1}--\eqref{utriple3}. The achievability of the point
$(-|C|,Q,-|E|)$ implies the achievability of the point $(-|C|,0,Q-|E|)$,
because we can consume all of the quantum communication for entanglement
distribution (\ref{ED}):%
\begin{equation}
(-|C|,0,Q-|E|)=(-|C|,Q,-|E|)+\left(  0,-Q,Q\right)  .
\end{equation}
The following bound applies (entanglement cannot be positive):%
\begin{equation}
Q-\left\vert E\right\vert \leq0,
\end{equation}
because classical communication alone cannot generate entanglement. The bound
in \eqref{npn1} follows from the above bound. The achievability of the point
$(-|C|,Q,-|E|)$ implies the achievability of the point $(-|C|+2Q,0,-Q-|E|)$,
because we can consume all of the quantum communication for super-dense coding
\eqref{SD}:%
\begin{equation}
\left(  -|C|+2Q,0,-Q-|E|\right)  =\left(  -|C|,Q,-|E|\right)  +\left(
2Q,-Q,-Q\right)  .
\end{equation}
The following bound applies (classical communication cannot be positive):%
\begin{equation}
-\left\vert C\right\vert +2Q\leq0,
\end{equation}
because entanglement alone cannot create classical communication. The bound in
\eqref{npn2} follows from the above bound.

$\boldsymbol{(-,-,+)}$. The task for this octant is to create $E$ ebits of
entanglement using $|Q|$ qubits of quantum communication and $|C|$ bits of
classical communication. We consider all points of the form $\left(
C,Q,E\right)  $ where $C\leq0$, $Q\leq0$, and $E\geq0$. It suffices to prove
the following inequality:%
\begin{equation}
E\leq\left\vert Q\right\vert , \label{nnp}%
\end{equation}
because combining it with $Q\leq0$ and $C\leq0$ implies
\eqref{utriple1}--\eqref{utriple3}. The achievability of $(-|C|,-|Q|,E)$
implies the achievability of $(-|C|-2E,-|Q|+E,0)$, because we can consume all
of the entanglement with teleportation \eqref{TP}:%
\begin{equation}
\left(  -|C|-2E,-|Q|+E,0\right)  =\left(  -|C|,-|Q|,E\right)  +\left(
-2E,E,-E\right)  .
\end{equation}
The following bound applies (quantum communication cannot be positive):%
\begin{equation}
-\left\vert Q\right\vert +E\leq0,
\end{equation}
because classical communication alone cannot generate quantum communication.
The bound in \eqref{nnp} follows from the above bound.

$\boldsymbol{(-,-,-)}$. $\widetilde{C}_{\operatorname{U}}$ completely contains
this octant.

We have now proved that the set of inequalities in
\eqref{utriple1}--\eqref{utriple3} holds for all octants of the $\left(
C,Q,E\right)  $ space. The next exercises ask you to consider similar unit
resource achievable regions.

\begin{exercise}
Consider the resources of public classical communication:%
\begin{equation}
\left[  c\rightarrow c\right]  _{\operatorname{pub}},
\end{equation}
private classical communication:%
\begin{equation}
\left[  c\rightarrow c\right]  _{\operatorname{priv}},
\end{equation}
and shared secret key:%
\begin{equation}
\left[  cc\right]  _{\operatorname{priv}}.
\end{equation}
Public classical communication is equivalent to the following channel:%
\begin{equation}
\rho\rightarrow\sum_{i}\langle i\vert\rho\vert i\rangle\ \vert i\rangle\langle
i\vert_{B}\otimes\sigma_{E}^{i},
\end{equation}
so that an eavesdropper Eve obtains some correlations with the transmitted
state $\rho$. Private classical communication is equivalent to the following
channel:%
\begin{equation}
\rho\rightarrow\sum_{i}\langle i\vert\rho\vert i\rangle\ \vert i\rangle\langle
i\vert_{B}\otimes\sigma_{E},
\end{equation}
so that Eve's state is independent of the information that Bob receives.
Finally, a secret key is a state of the following form:%
\begin{equation}
\overline{\Phi}_{AB}\otimes\sigma_{E}\equiv\left(  \frac{1}{d}\sum_{i}\vert
i\rangle\langle i\vert_{A}\otimes\vert i\rangle\langle i\vert_{B}\right)
\otimes\sigma_{E},
\end{equation}
so that Alice and Bob share maximal classical correlation and Eve's state is
independent of it. There are three protocols that relate these three classical
resources. Secret key distribution is a protocol that consumes a noiseless
private channel to generate a noiseless secret key. It has the following
resource inequality:%
\begin{equation}
\left[  c\rightarrow c\right]  _{\operatorname{priv}}\geq\left[  cc\right]
_{\operatorname{priv}}.
\end{equation}
The one-time pad protocol exploits a shared secret key and a noiseless public
channel to generate a noiseless private channel (it simply XORs a bit of
secret key with the bit that the sender wants to transmit and this protocol is
provably unbreakable if the secret key is perfectly secret). It has the
following resource inequality:%
\begin{equation}
\left[  c\rightarrow c\right]  _{\operatorname{pub}}+\left[  cc\right]
_{\operatorname{priv}}\geq\left[  c\rightarrow c\right]  _{\operatorname{priv}%
}.
\end{equation}
Finally, private classical communication can simulate public classical
communication if we assume that Bob has a local register where he can place
information and he then gives this to Eve. It has the following resource
inequality:%
\begin{equation}
\left[  c\rightarrow c\right]  _{\operatorname{priv}}\geq\left[  c\rightarrow
c\right]  _{\operatorname{pub}}.
\end{equation}
Show that these three protocols fill out an optimal achievable region in the
space of public classical communication, private classical communication, and
secret key. Use the following three postulates to prove optimality: (1) public
classical communication alone cannot generate secret key or private classical
communication, (2) private key alone cannot generate public or private
classical communication, and (3) the net amount of public bit channel uses and
secret key bits generated cannot exceed the number of private bit channel uses consumed.
\end{exercise}

\begin{exercise}
Consider the resource of coherent communication from
Chapter~\ref{chap:coherent-communication}:%
\begin{equation}
\left[  q\rightarrow qq\right]  .
\end{equation}
Recall the coherent communication identity in \eqref{eq-coh:identity}:%
\begin{equation}
2\left[  q\rightarrow qq\right]  =\left[  q\rightarrow q\right]  +\left[
qq\right]  .
\end{equation}
Recall the other resource inequalities for coherent communication:%
\begin{equation}
\left[  q\rightarrow q\right]  \geq\left[  q\rightarrow qq\right]  \geq\left[
qq\right]  .
\end{equation}
Consider a space of points $\left(  C,Q,E\right)  $ where $C$ corresponds to
coherent communication, $Q$ to quantum communication, and $E$ to entanglement.
Determine the achievable region one obtains with the above resource
inequalities and another trivial resource inequality:%
\begin{equation}
\left[  qq\right]  \geq0.
\end{equation}
We interpret the above resource inequality as \textquotedblleft entanglement
consumption,\textquotedblright\ where Alice simply throws away entanglement.
\end{exercise}

\section{History and Further Reading}

The unit resource capacity region first appeared in \cite{HW09}\ in the
context of trade-off coding. The private unit resource capacity region later
appeared in \cite{WH10}.

\part{Tools of Quantum Shannon Theory}

\chapter{Distance Measures}

\label{chap:distance-measures}We discussed the major noiseless quantum
communication protocols such as teleportation, super-dense coding, their
coherent versions, and entanglement distribution in detail in
Chapters~\ref{chap:three-noiseless}, \ref{chap:coherent-communication}, and
\ref{chap:unit-resource-cap}. Each of these protocols relies on the assumption
that noiseless resources are available. For example, the entanglement
distribution protocol assumes that a noiseless qubit channel is available to
generate a noiseless ebit. This idealization allowed us to develop the main
principles of the protocols without having to think about more complicated
issues, but in practice, the protocols do not work as expected in the presence
of noise.

\begin{figure}
[ptb]
\begin{center}
\includegraphics[
width=4.8456in
]%
{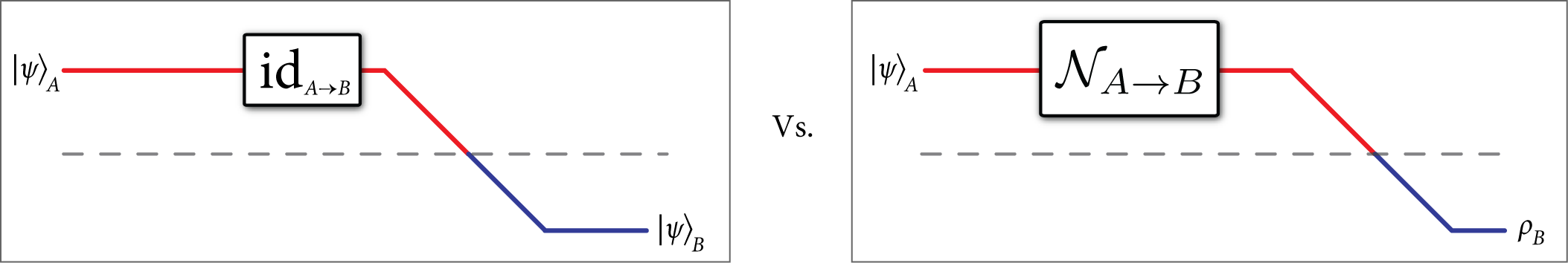}%
\caption{A distance measure quantifies how far the output of a given ideal
protocol (depicted on the left)\ is from an actual protocol that exploits a
noisy resource (depicted as the noisy quantum channel $\mathcal{N}%
_{A\rightarrow B}$ on the right).}%
\label{fig-dm:problem}%
\end{center}
\end{figure}

Given that quantum systems suffer noise in practice, we would like to have a
way to determine how well a protocol is performing. The simplest way to do so
is to compare the output of an ideal protocol to the output of the actual
protocol using a \textit{distance measure }of the two respective output
quantum states. That is, suppose that a quantum information-processing
protocol should ideally output some quantum state $|\psi\rangle$, but the
actual output of the protocol is a quantum state with density operator $\rho$.
Then a performance measure $P(\psi,\rho)$ should indicate how close the ideal
output is to the actual output. Figure~\ref{fig-dm:problem}\ depicts the
comparison of an ideal protocol with another protocol that is noisy.

This chapter introduces two distance measures that allow us to determine how
close two quantum states are to each other. The first distance measure that we
discuss is the \textit{trace distance} and the second is the \textit{fidelity}%
. (However, note that the fidelity is not a distance measure in the strict
mathematical sense---nevertheless, we exploit it as a \textquotedblleft
closeness\textquotedblright\ measure of quantum states because it admits an
intuitive operational interpretation.) These two measures are mostly
interchangeable, but we introduce both because it is often times more
convenient in a given situation to use one or the other.

Distance measures are particularly important in quantum Shannon theory because
they provide a way for us to determine how well a protocol is performing.
Recall that Shannon's method (outlined in
Chapter~\ref{chap:classical-shannon-theory}) for both the noiseless and noisy
coding theorem is to allow for a slight error in a protocol, but to show that
this error vanishes in the limit of large block length. In later chapters
where we prove quantum coding theorems, we borrow this technique of
demonstrating asymptotically small error, with either the trace distance or
the fidelity as the measure of performance.

\section{Trace Distance}

We first introduce the
\index{trace distance}%
trace distance. Our presentation is somewhat mathematical because we exploit
norms on linear operators in order to define it. Despite this mathematical
flavor, this section offers an intuitive operational interpretation of the
trace distance.

\subsection{Trace Norm}

\label{sec-dm:trace-norm}

\begin{definition}
[Trace Norm]The \textit{trace norm}
\index{trace norm}%
or \textit{Schatten 1-norm} $\left\Vert M\right\Vert _{1}$ of an operator
$M\in\mathcal{L}(\mathcal{H},\mathcal{H}^{\prime})$ is defined as%
\begin{equation}
\left\Vert M\right\Vert _{1}\equiv\operatorname{Tr}\left\{  |M|\right\}  ,
\end{equation}
where $|M|\equiv\sqrt{M^{\dag}M}$.
\end{definition}

\begin{proposition}
\label{prop-dm:trace-norm-sing-values}The trace norm of an operator
$M\in\mathcal{L}(\mathcal{H},\mathcal{H}^{\prime})$ is equal to the sum of its
singular values.
\end{proposition}

\begin{proof}
Recall from Definition~\ref{def-qt:hermitian-op-function} that any function
$f$ applied to a Hermitian operator $A$ is as follows:%
\begin{equation}
f(A)\equiv\sum_{i:\alpha_{i}\in \operatorname{Dom}(f)}f(\alpha_{i})|i\rangle\langle i|,
\end{equation}
where $\sum_{i}\alpha_{i}|i\rangle\langle i|$ is a spectral
decomposition of $A$. With these two definitions, it is straightforward to
show that the trace norm of $M$ is equal to the sum of its singular values.
Indeed, let $M=U\Sigma V$ be the singular value decomposition of $M$, where
$U$ and $V$ are unitary matrices and $\Sigma$ is a rectangular matrix with the
non-negative singular values along the diagonal. Then we can write%
\begin{equation}
M=\sum_{i=0}^{d-1}\sigma_{i}|u_{i}\rangle\langle v_{i}|,
\end{equation}
where $d$ is the rank of $M$, $\left\{  \sigma_{i}\right\}  $ are the strictly
positive singular values of $M$, $\left\{  |u_{i}\rangle\right\}  $ are the
orthonormal columns of $U$ in correspondence with the set $\left\{  \sigma
_{i}\right\}  $, and $\left\{  |v_{i}\rangle\right\}  $ are the orthonormal
rows of $V$ in correspondence with the set $\left\{  \sigma_{i}\right\}  $.
Then%
\begin{align}
M^{\dag}M  &  =\left[  \sum_{j=0}^{d-1}\sigma_{j}|v_{j}\rangle\langle
u_{j}|\right]  \left[  \sum_{i=0}^{d-1}\sigma_{i}|u_{i}\rangle\langle
v_{i}|\right] \\
&  =\sum_{i,j=0}^{d-1}\sigma_{j}\sigma_{i}|v_{j}\rangle\langle u_{j}%
||u_{i}\rangle\langle v_{i}|\\
&  =\sum_{i=0}^{d-1}\sigma_{i}^{2}|v_{i}\rangle\langle v_{i}|,
\end{align}
so that%
\begin{equation}
\sqrt{M^{\dag}M}=\sum_{i=0}^{d-1}\sqrt{\sigma_{i}^{2}}|v_{i}\rangle\langle
v_{i}|=\sum_{i=0}^{d-1}\sigma_{i}|v_{i}\rangle\langle v_{i}|,
\end{equation}
finally implying that%
\begin{equation}
\operatorname{Tr}\left\{  |M|\right\}  =\sum_{i=0}^{d-1}\sigma_{i}.
\end{equation}
This means also that%
\begin{equation}
\left\Vert M\right\Vert _{1}=\operatorname{Tr}\{\sqrt{MM^{\dag}}\},
\end{equation}
because the singular values of $MM^{\dag}$ and $M^{\dag}M$ are the same (this
is the key to Exercise~\ref{ex-nqt:trace-equal-eigs}). One can also easily
show that the trace norm of a Hermitian operator is equal to the absolute sum
of its eigenvalues.
\end{proof}

The trace norm is indeed a \textit{norm} because it satisfies the following
three properties: non-negative definiteness, homogeneity, and the triangle inequality.

\begin{property}
[Non-Negative Definiteness]The trace norm of an operator $M$ is non-negative
definite:%
\begin{equation}
\left\Vert M\right\Vert _{1}\geq0.
\end{equation}
The trace norm is equal to zero if and only if the operator $M$ is the zero
operator:%
\begin{equation}
\left\Vert M\right\Vert _{1}=0\ \ \ \Leftrightarrow\ \ \ M=0.
\end{equation}

\end{property}

\begin{property}
[Homogeneity]For any constant $c\in\mathbb{C}$,%
\begin{equation}
\left\Vert cM\right\Vert _{1}=\left\vert c\right\vert \left\Vert M\right\Vert
_{1}.
\end{equation}

\end{property}

\begin{property}
[Triangle Inequality]\label{prop-dm:triangle-ineq}For any two operators
$M,N\in\mathcal{L}(\mathcal{H},\mathcal{H}^{\prime})$, the following triangle
inequality holds:%
\begin{equation}
\left\Vert M+N\right\Vert _{1}\leq\left\Vert M\right\Vert _{1}+\left\Vert
N\right\Vert _{1}.
\end{equation}

\end{property}

Non-negative definiteness follows because the sum of the singular values of an
operator is non-negative, and the singular values are all equal to zero (and
thus the operator is equal to zero) if and only if the sum of the singular
values is equal to zero. Homogeneity follows directly from the fact that
$|cM|= |c| |M|$. We later give a proof of the triangle inequality (however,
for a special case only). Exercise~\ref{ex-dm:triangle-ineq-U} below asks you
to prove it for square operators.

Three other important properties of the trace norm are its invariance under
isometries, convexity, and a variational characterization. Each of the
properties below often arise as useful tools in quantum Shannon theory.

\begin{property}
[Isometric Invariance]The trace norm is invariant under multiplication by
isometries $U$ and $V$:%
\begin{equation}
\left\Vert UMV^{\dag}\right\Vert _{1}=\left\Vert M\right\Vert _{1}.
\end{equation}

\end{property}

\begin{property}
[Convexity]For any two operators $M,N\in\mathcal{L}(\mathcal{H},\mathcal{H}%
^{\prime})$ and $\lambda\in[0,1]$, the following inequality holds
\begin{equation}
\left\Vert \lambda M+(1-\lambda)N\right\Vert _{1}\leq\lambda\left\Vert
M\right\Vert _{1}+(1-\lambda)\left\Vert N\right\Vert _{1}.
\end{equation}

\end{property}

Isometric invariance holds because $M$ and $UMV^{\dag}$ have the same singular
values. Convexity follows directly from the triangle inequality and
homogeneity (thus, any norm is convex in this sense).

\begin{property}
[Variational characterization]\label{prop-dm:var-char-TD}For a square operator
$M\in\mathcal{L}(\mathcal{H})$, the following variational characterization of
the trace norm holds%
\begin{equation}
\left\Vert M\right\Vert _{1}=\max_{U} \left\vert \operatorname{Tr}\left\{
MU\right\}  \right\vert , \label{eq-dm:var-char-TD}%
\end{equation}
where the optimization is with respect to all unitary operators.
\end{property}

\begin{proof}
The above characterization follows by taking a singular value decomposition of
$M$ as $M=WDV$, with $W$ and $V$ unitaries and $D$ a diagonal matrix of
singular values. Applying the Cauchy--Schwarz inequality gives
\begin{align}
\left\vert \operatorname{Tr}\left\{  MU\right\}  \right\vert  &  =\left\vert
\operatorname{Tr}\left\{  WDVU\right\}  \right\vert =\left\vert
\operatorname{Tr}\left\{  \sqrt{D}\sqrt{D}VUW\right\}  \right\vert \\
&  \leq\sqrt{\operatorname{Tr}\left\{  \sqrt{D}\sqrt{D}\right\}  }%
\sqrt{\operatorname{Tr}\left\{  \left(  \sqrt{D}VUW\right)  ^{\dag}\sqrt
{D}VUW\right\}  }\\
&  =\operatorname{Tr}\left\{  D\right\}  =\left\Vert M\right\Vert _{1}.
\end{align}
The inequality is a consequence of the Cauchy--Schwarz inequality for the
Hilbert--Schmidt inner product:%
\begin{equation}
\left\vert \operatorname{Tr}\left\{  A^{\dag}B\right\}  \right\vert \leq
\sqrt{\operatorname{Tr}\left\{  A^{\dag}A\right\}  }\sqrt{\operatorname{Tr}%
\left\{  B^{\dag}B\right\}  }.\label{eq:Cauchy--Schwarz-Hilbert-Schmidt}%
\end{equation}
Equality holds by picking $U=V^{\dag}W^{\dag}$, from which we recover
(\ref{eq-dm:var-char-TD}).
\end{proof}

\begin{exercise}
\label{ex-dm:triangle-ineq-U} Prove that the triangle inequality
(Property~\ref{prop-dm:triangle-ineq}) holds for square operators
$M,N\in\mathcal{L}(\mathcal{H})$. (Hint: Use the characterization in
Property~\ref{prop-dm:var-char-TD}.)
\end{exercise}

\subsection{Trace Distance from the Trace Norm}

The trace norm induces a natural distance
\index{trace distance}%
measure, called the \textit{trace distance}.

\begin{definition}
[Trace Distance]\label{def-dm:trace-distance}Given any two operators
$M,N\in\mathcal{L}(\mathcal{H},\mathcal{H}^{\prime})$, the trace distance
between them is as follows:%
\begin{equation}
\left\Vert M-N\right\Vert _{1}.
\end{equation}

\end{definition}

The trace distance is especially useful as a measure of the distinguishability
of two quantum states with respective density operators $\rho$ and $\sigma$.
The following bounds apply to the trace distance between any two density
operators $\rho$ and $\sigma$:%
\begin{equation}
0\leq\left\Vert \rho-\sigma\right\Vert _{1}\leq2.
\label{eq-dm:bounds-trace-dist}%
\end{equation}
Sometimes it is useful to employ the \textit{normalized} trace distance
$\frac{1}{2}\left\Vert \rho-\sigma\right\Vert _{1}$, so that $\frac{1}%
{2}\left\Vert \rho-\sigma\right\Vert _{1}\in\left[  0,1\right]  $. The lower
bound in \eqref{eq-dm:bounds-trace-dist}\ applies when two quantum states are
equal---quantum states $\rho$ and $\sigma$ are equal to each other if and only
if their trace distance is zero. The physical implication of the trace
distance being equal to zero is that no measurement can distinguish $\rho$
from $\sigma$. The upper bound in \eqref{eq-dm:bounds-trace-dist}\ follows
from the triangle inequality:%
\begin{equation}
\left\Vert \rho-\sigma\right\Vert _{1}\leq\left\Vert \rho\right\Vert
_{1}+\left\Vert \sigma\right\Vert _{1}=2.
\end{equation}
The trace distance is maximum when $\rho$ and $\sigma$ have support on
orthogonal subspaces. Later, we will prove that this is the only case in which
this happens, after introducing the fidelity. The physical implication of
maximal trace distance is that there exists a measurement that can perfectly
distinguish $\rho$ from $\sigma$. We discuss these operational interpretations
of the trace distance in more detail in Section~\ref{sec-dm:op-int-trace}.

\begin{exercise}
Show that the trace distance between two qubit density operators $\rho$ and
$\sigma$\ is equal to the Euclidean distance between their respective Bloch
vectors $\overrightarrow{r}$ and $\overrightarrow{s}$, where%
\begin{equation}
\rho=\frac{1}{2}\left(  I+\overrightarrow{r}\cdot\overrightarrow{\sigma
}\right)  ,\ \ \ \ \ \ \sigma=\frac{1}{2}\left(  I+\overrightarrow{s}%
\cdot\overrightarrow{\sigma}\right)  .
\end{equation}
That is, show that $\left\Vert \rho-\sigma\right\Vert _{1}=\left\Vert
\overrightarrow{r}-\overrightarrow{s}\right\Vert _{2} $.
\end{exercise}

\begin{exercise}
Show that the trace distance obeys a telescoping property:%
\begin{equation}
\left\Vert \rho_{1}\otimes\rho_{2}-\sigma_{1}\otimes\sigma_{2}\right\Vert
_{1}\leq\left\Vert \rho_{1}-\sigma_{1}\right\Vert _{1}+\left\Vert \rho
_{2}-\sigma_{2}\right\Vert _{1},
\end{equation}
for any density operators $\rho_{1}$, $\rho_{2}$, $\sigma_{1}$, $\sigma_{2}$.
(Hint:\ First prove that $\left\Vert \rho\otimes\omega-\sigma\otimes
\omega\right\Vert _{1}=\left\Vert \rho-\sigma\right\Vert _{1}, $ for any
density operators $\rho$, $\sigma$, $\omega$.)
\end{exercise}

\begin{exercise}
\label{ex-dm:trace-inv-iso}Show that the trace distance is invariant with
respect to an isometric quantum channel, in the following sense:%
\begin{equation}
\left\Vert \rho-\sigma\right\Vert _{1}=\left\Vert U\rho U^{\dag}-U\sigma
U^{\dag}\right\Vert _{1}, \label{eq-dm:trace-inv-iso}%
\end{equation}
where $U$ is an isometry. The physical implication of
\eqref{eq-dm:trace-inv-iso} is that an isometric quantum channel applied to
both states does not increase or decrease the distinguishability of the two states.
\end{exercise}

\subsection{Trace Distance as a Probability Difference}

We now state and prove an important lemma that gives an alternative and useful
way for characterizing the trace distance. This particular characterization
finds application in many proofs of the lemmas that follow concerning trace distance.

\begin{lemma}
\label{lemma:trace-equiv}The normalized trace distance $\frac{1}{2}\left\Vert
\rho-\sigma\right\Vert _{1}$ between quantum states $\rho,\sigma\in
\mathcal{D}(\mathcal{H})$ is equal to the largest probability difference that
two states $\rho$ and $\sigma$ could give to the same measurement outcome
$\Lambda$:%
\begin{equation}
\frac{1}{2}\left\Vert \rho-\sigma\right\Vert _{1}=\max_{0\leq\Lambda\leq
I}\operatorname{Tr}\left\{  \Lambda\left(  \rho-\sigma\right)  \right\}  .
\end{equation}
The above maximization is with respect to all positive semi-definite operators
$\Lambda\in\mathcal{L}(\mathcal{H})$ that have their eigenvalues bounded from
above by one.
\end{lemma}

\begin{proof}
Consider that the difference operator $\rho-\sigma$ is Hermitian and so we can
diagonalize it as follows:%
\[
\rho-\sigma=\sum_{i}\lambda_{i}|i\rangle\langle i|,
\]
where $\{|i\rangle\}$ is an orthonormal basis of eigenvectors and
$\{\lambda_{i}\}$ is a set of real eigenvalues. Let us define%
\begin{equation}
P\equiv\sum_{i:\lambda_{i}\geq0}\lambda_{i}|i\rangle\langle
i|,\ \ \ \ \ \ \ \ \ \ Q\equiv\sum_{i:\lambda_{i}<0}\left\vert \lambda
_{i}\right\vert |i\rangle\langle i|, \label{eq-dm:P-Q-ops}%
\end{equation}
which implies that $P$ and $Q$ are positive semi-definite and that%
\begin{equation}
\rho-\sigma=P-Q.
\end{equation}
Consider also that $PQ=0$, and let $\Pi_{P}$ and $\Pi_{Q}$ denote the
projections onto the supports of $P$ and $Q$, respectively:%
\begin{equation}
\Pi_{P}\equiv\sum_{i:\lambda_{i}\geq0}|i\rangle\langle
i|,\ \ \ \ \ \ \ \ \ \ \Pi_{Q}\equiv\sum_{i:\lambda_{i}<0}|i\rangle\langle i|.
\label{eq-dm:P-Q-proj-ops}%
\end{equation}
Then it follows that%
\begin{align}
\Pi_{P}P\Pi_{P}  &  =P,\ \ \ \ \ \ \ \ \ \ \Pi_{Q}Q\Pi_{Q}=Q,\\
\Pi_{P}Q\Pi_{P}  &  =0,\ \ \ \ \ \ \ \ \ \ \Pi_{Q}P\Pi_{Q}=0.
\end{align}
The following property holds as well:%
\begin{equation}
\left\vert \rho-\sigma\right\vert =\left\vert P-Q\right\vert =P+Q.
\end{equation}
because the supports of $P$ and $Q$ are orthogonal and the absolute value of
the operator $P-Q$\ takes the absolute value of its eigenvalues. Therefore,%
\begin{equation}
\left\Vert \rho-\sigma\right\Vert _{1} =\operatorname{Tr}\left\{  \left\vert
\rho-\sigma\right\vert \right\}  =\operatorname{Tr}\left\{  P+Q\right\}
=\operatorname{Tr}\left\{  P\right\}  +\operatorname{Tr}\left\{  Q\right\}  .
\end{equation}
But%
\begin{align}
\operatorname{Tr}\left\{  P\right\}  -\operatorname{Tr}\left\{  Q\right\}   &
=\operatorname{Tr}\left\{  P-Q\right\}  =\operatorname{Tr}\left\{  \rho
-\sigma\right\} \\
&  =\operatorname{Tr}\left\{  \rho\right\}  -\operatorname{Tr}\left\{
\sigma\right\}  =0.
\end{align}
where the last equality follows because both quantum states have unit trace.
Therefore, $\operatorname{Tr}\left\{  P\right\}  =\operatorname{Tr}\left\{
Q\right\}  $ and%
\begin{equation}
\left\Vert \rho-\sigma\right\Vert _{1}=2\cdot\operatorname{Tr}\left\{
P\right\}  . \label{eq-dm:trace-distance-formula}%
\end{equation}
Consider then that%
\begin{align}
\operatorname{Tr}\left\{  \Pi_{P}\left(  \rho-\sigma\right)  \right\}   &
=\operatorname{Tr}\left\{  \Pi_{P}\left(  P-Q\right)  \right\}
=\operatorname{Tr}\left\{  \Pi_{P}P\right\} \\
&  =\operatorname{Tr}\left\{  P\right\}  =\frac{1}{2}\left\Vert \rho
-\sigma\right\Vert _{1}.
\end{align}
Now we prove that the operator $\Pi_{P}$ is the maximizing one. Let $\Lambda$
be any positive semi-definite operator with spectrum bounded above by one.
Then%
\begin{align}
\operatorname{Tr}\left\{  \Lambda\left(  \rho-\sigma\right)  \right\}   &
=\operatorname{Tr}\left\{  \Lambda\left(  P-Q\right)  \right\}  \leq
\operatorname{Tr}\left\{  \Lambda P\right\} \\
&  \leq\operatorname{Tr}\left\{  P\right\}  =\frac{1}{2}\left\Vert \rho
-\sigma\right\Vert _{1}.
\end{align}
The first inequality follows because $\Lambda$ and $Q$ are non-negative and so
$\operatorname{Tr}\{ \Lambda Q\} $ is non-negative. The second inequality
holds because $\Lambda\leq I$. The final equality follows from \eqref{eq-dm:trace-distance-formula}.
\end{proof}

\begin{exercise}
Let $\rho=|0\rangle\langle0|$ and $\sigma=|+\rangle\langle+|$. Compute $P$,
$Q$, $\Pi_{P}$, and $\Pi_{Q}$, as defined in \eqref{eq-dm:P-Q-ops} and
\eqref{eq-dm:P-Q-proj-ops}, for this choice of $\rho$ and $\sigma$. Compute
the trace distance $\left\Vert \rho-\sigma\right\Vert _{1}$.
\end{exercise}

\begin{exercise}
\label{ex-dm:trace-norm}Show that the trace norm of any Hermitian operator
$\omega$ is given by the following optimization:%
\begin{equation}
\left\Vert \omega\right\Vert _{1}=\max_{-I\leq\Lambda\leq I}\operatorname{Tr}%
\left\{  \Lambda\omega\right\}  .
\end{equation}

\end{exercise}

\subsection{Operational Interpretation of the Trace Distance}

\label{sec-dm:op-int-trace}We now provide an
\index{trace distance!operational interpretation}%
operational interpretation of the trace distance as the distinguishability of
two quantum states. The interpretation results from a hypothesis-testing%
\index{quantum hypothesis testing}
scenario. Suppose that Bob prepares one of two quantum states $\rho_{0}$ or
$\rho_{1}$ for Alice to distinguish. Suppose further that it is equally likely
\textit{a priori} for him to prepare either $\rho_{0}$ or $\rho_{1}$. Let $X$
denote the Bernoulli random variable assigned to the prior probabilities so
that $p_{X}(0)=p_{X}(1)=1/2$. Alice can perform a binary POVM\ with elements
$\Lambda\equiv\left\{  \Lambda_{0},\Lambda_{1}\right\}  $ to distinguish the
two states. That is, Alice guesses the state in question is $\rho_{0}$ if she
receives outcome \textquotedblleft0\textquotedblright\ from the measurement or
she guesses the state in question is $\rho_{1}$ if she receives outcome
\textquotedblleft1\textquotedblright\ from the measurement. Let $Y$ denote the
Bernoulli random variable assigned to the classical outcomes of her
measurement. The success probability $p_{\operatorname{succ}}(\Lambda)$ for
this hypothesis testing scenario is the sum of the probability of detecting
\textquotedblleft0\textquotedblright\ when the state is $\rho_{0}$ and the
probability of detecting \textquotedblleft1\textquotedblright\ when the state
is $\rho_{1}$:%
\begin{align}
p_{\operatorname{succ}}(\Lambda)  &  =p_{Y|X}(0|0)p_{X}(0)+p_{Y|X}%
(1|1)p_{X}(1)\\
&  =\operatorname{Tr}\left\{  \Lambda_{0}\rho_{0}\right\}  \frac{1}%
{2}+\operatorname{Tr}\left\{  \Lambda_{1}\rho_{1}\right\}  \frac{1}{2}.
\end{align}
We can simplify this expression using the completeness relation $\Lambda
_{0}+\Lambda_{1}=I$:%
\begin{align}
p_{\operatorname{succ}}(\Lambda)  &  =\frac{1}{2}\left(  \operatorname{Tr}%
\left\{  \Lambda_{0}\rho_{0}\right\}  +\operatorname{Tr}\left\{  \left(
I-\Lambda_{0}\right)  \rho_{1}\right\}  \right) \\
&  =\frac{1}{2}\left(  \operatorname{Tr}\left\{  \Lambda_{0}\rho_{0}\right\}
+\operatorname{Tr}\left\{  \rho_{1}\right\}  -\operatorname{Tr}\left\{
\Lambda_{0}\rho_{1}\right\}  \right) \\
&  =\frac{1}{2}\left(  \operatorname{Tr}\left\{  \Lambda_{0}\rho_{0}\right\}
+1-\operatorname{Tr}\left\{  \Lambda_{0}\rho_{1}\right\}  \right) \\
&  =\frac{1}{2}\left(  1+\operatorname{Tr}\left\{  \Lambda_{0}\left(  \rho
_{0}-\rho_{1}\right)  \right\}  \right)  .
\end{align}
Now Alice has freedom in choosing the POVM$\ \Lambda=\left\{  \Lambda
_{0},\Lambda_{1}\right\}  $ to distinguish the states $\rho_{0}$ and $\rho
_{1}$, and she would like to choose one that maximizes the success probability
$p_{\operatorname{succ}}(\Lambda)$. Thus, we can define the success
probability with respect to all measurements as follows:%
\begin{equation}
p_{\operatorname{succ}}\equiv\max_{\Lambda}p_{\operatorname{succ}}%
(\Lambda)=\max_{\Lambda}\frac{1}{2}\left(  1+\operatorname{Tr}\left\{
\Lambda_{0}\left(  \rho_{0}-\rho_{1}\right)  \right\}  \right)  .
\end{equation}
We can rewrite the above quantity in terms of the trace distance using its
characterization in Lemma~\ref{lemma:trace-equiv}\ because the expression
inside of the maximization involves only the operator$~\Lambda_{0}$:%
\begin{equation}
p_{\operatorname{succ}}=\frac{1}{2}\left(  1+\frac{1}{2}\left\Vert \rho
_{0}-\rho_{1}\right\Vert _{1}\right)  .
\end{equation}
Thus, the normalized trace distance has an operational interpretation that it
is linearly related to the maximum success probability in distinguishing two
quantum states $\rho_{0}$ and $\rho_{1}$ in a quantum hypothesis testing
experiment. From the above expression for the success probability, it is clear
that the states are indistinguishable when $\left\Vert \rho_{0}-\rho
_{1}\right\Vert _{1}$ is equal to zero. That is, it is just as good for Alice
to guess randomly what the state might be, and in this case, she can do no
better than to have $1/2$ probability of being correct. On the other hand, the
states are perfectly distinguishable when $\left\Vert \rho_{0}-\rho
_{1}\right\Vert _{1}$ is maximal and the measurement that distinguishes them
consists of two projectors: one projects onto the non-negative eigenspace of
$\rho_{0}-\rho_{1}$ and the other projects onto the negative eigenspace of
$\rho_{0}-\rho_{1}$. In this sense, we can say that the normalized trace
distance is the bias away from random guessing in a hypothesis testing experiment.

\begin{exercise}
Suppose that the prior probabilities in the above hypothesis-testing scenario
are not uniform but are rather equal to $p_{0}$ and $p_{1}$. Show that the
success probability is instead given by%
\begin{equation}
p_{\operatorname{succ}}=\frac{1}{2}\left(  1+\left\Vert p_{0}\rho_{0}%
-p_{1}\rho_{1}\right\Vert _{1}\right)  .
\end{equation}

\end{exercise}

\subsection{Trace Distance Lemmas}

We present several useful corollaries of Lemma~\ref{lemma:trace-equiv} and
their corresponding proofs. These corollaries include the triangle inequality,
measurement on close states, and monotonicity of trace distance. Each of these
corollaries finds application in many proofs in quantum Shannon theory.

\begin{lemma}
[Triangle Inequality]\label{lem:triangle-trace}The trace distance
\index{trace distance!triangle inequality}
obeys a triangle inequality. For any three quantum states $\rho$, $\sigma$,
$\tau\in\mathcal{D}(\mathcal{H})$, the following inequality holds:%
\begin{equation}
\left\Vert \rho-\sigma\right\Vert _{1}\leq\left\Vert \rho-\tau\right\Vert
_{1}+\left\Vert \tau-\sigma\right\Vert _{1}.
\end{equation}

\end{lemma}

\begin{proof}
Pick $\Pi$ as the maximizing operator for $\left\Vert \rho-\sigma\right\Vert
_{1}$ (according to Lemma~\ref{lemma:trace-equiv}) so that%
\begin{align}
\left\Vert \rho-\sigma\right\Vert _{1}  &  =2\cdot\operatorname{Tr}\left\{
\Pi\left(  \rho-\sigma\right)  \right\} \\
&  =2\cdot\operatorname{Tr}\left\{  \Pi\left(  \rho-\tau\right)  \right\}
+2\cdot\operatorname{Tr}\left\{  \Pi\left(  \tau-\sigma\right)  \right\} \\
&  \leq\left\Vert \rho-\tau\right\Vert _{1}+\left\Vert \tau-\sigma\right\Vert
_{1}.
\end{align}
The last inequality follows because the operator $\Pi$ maximizing $\left\Vert
\rho-\sigma\right\Vert _{1}$ in general is not the same operator that
maximizes both $\left\Vert \rho-\tau\right\Vert _{1}$ and $\left\Vert
\tau-\sigma\right\Vert _{1}$.
\end{proof}

\begin{corollary}
[Measurement on Close States]\label{lemma:trace-inequality}Suppose we have two
quantum states $\rho,\sigma\in\mathcal{D}(\mathcal{H})$ and an operator
$\Pi\in\mathcal{L}(\mathcal{H})$ such that $0\leq\Pi\leq I$. Then%
\begin{align}
\operatorname{Tr}\left\{  \Pi\rho\right\}   &  \geq\operatorname{Tr}\left\{
\Pi\sigma\right\}  - \frac{1}{2} \left\Vert \rho-\sigma\right\Vert _{1}\\
&  \geq\operatorname{Tr}\left\{  \Pi\sigma\right\}  - \left\Vert \rho
-\sigma\right\Vert _{1} . \label{eq-dm:lemma-trace-inequality}%
\end{align}

\end{corollary}

\begin{proof}
Consider the following arguments:%
\begin{align}
\frac{1}{2} \left\Vert \rho-\sigma\right\Vert _{1}  &  =\max_{0\leq\Lambda\leq
I}\left\{  \operatorname{Tr}\left\{  \Lambda\left(  \sigma-\rho\right)
\right\}  \right\} \\
&  \geq\operatorname{Tr}\left\{  \Pi\left(  \sigma-\rho\right)  \right\} \\
&  =\operatorname{Tr}\left\{  \Pi\sigma\right\}  -\operatorname{Tr}\left\{
\Pi\rho\right\}  .
\end{align}
The first equality follows from Lemma~\ref{lemma:trace-equiv}.
The first inequality follows because $\Lambda$ is the maximizing operator and
can only lead to a probability difference greater than that for another
operator $\Pi$ such that $0\leq\Pi\leq I$.
\end{proof}

The most common way that we employ Corollary~\ref{lemma:trace-inequality}\ in
quantum Shannon theory is in the following scenario. Suppose that a
measurement with operator $\Pi$\ succeeds with high probability on a quantum
state $\sigma$:%
\begin{equation}
\operatorname{Tr}\left\{  \Pi\sigma\right\}  \geq1-\varepsilon,
\label{eq-dm:high-prob}%
\end{equation}
where $\varepsilon$ is some small positive number. Suppose further that
another quantum state $\rho$ is $\varepsilon$-close in trace distance to
$\sigma$:%
\begin{equation}
\left\Vert \rho-\sigma\right\Vert _{1}\leq\varepsilon.
\label{eq-dm:close-state}%
\end{equation}
Then Corollary~\ref{lemma:trace-inequality} gives the intuitive result that
the measurement succeeds with high probability on the state $\rho$ that is
close to $\sigma$:%
\begin{equation}
\operatorname{Tr}\left\{  \Pi\rho\right\}  \geq1-2\varepsilon,
\end{equation}
by plugging \eqref{eq-dm:high-prob} and \eqref{eq-dm:close-state} into \eqref{eq-dm:lemma-trace-inequality}.

\begin{exercise}
\label{ex-dm:trace-ineq-herm}Prove that
\eqref{eq-dm:lemma-trace-inequality}\ holds for arbitrary Hermitian operators
$\rho$ and $\sigma$ by exploiting the result of
Exercise~\ref{ex-dm:trace-norm}.
\end{exercise}

\begin{figure}
[ptb]
\begin{center}
\includegraphics[
width=4.5455in
]%
{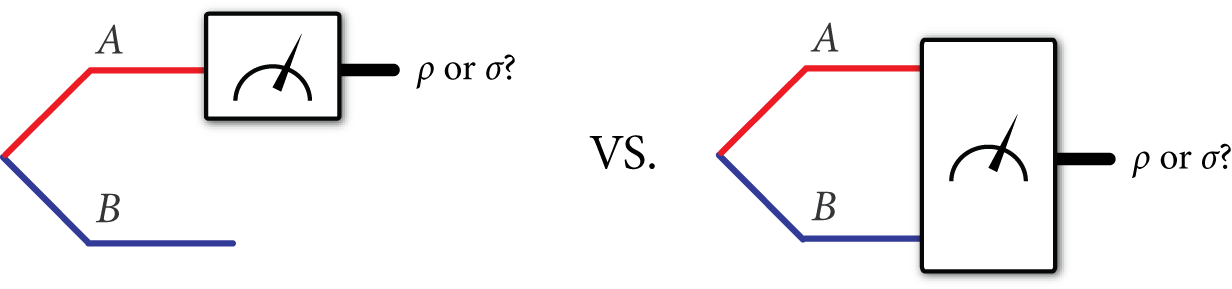}%
\caption{The task in this figure is for Bob to distinguish the state
$\rho_{AB}$ from the state $\sigma_{AB}$ using a binary-valued measurement.
Bob could perform an optimal measurement on system $A$ alone if he does not
have access to system $B$. If he has access to system $B$ as well, then he can
perform an optimal joint measurement on systems $A$ and $B$. We would expect
that he can distinguish the states more reliably if he performs a joint
measurement because there could be more information about the state available
in the other system $B$. Since the trace distance is a measure of
distinguishability, we would expect it to obey the following inequality:
$\left\Vert \rho_{A}-\sigma_{A}\right\Vert _{1}\leq\left\Vert \rho_{AB}%
-\sigma_{AB}\right\Vert _{1}$ (the states are less distinguishable if fewer
systems are available to be part of the distinguishability test).}%
\label{fig-dm:monotonicity-trace}%
\end{center}
\end{figure}

We next turn to the monotonicity of trace distance under the discarding of a
system. The interpretation of this corollary is that discarding of a system
does not increase distinguishability of two quantum states. That is, a global
measurement on the larger system might be able to distinguish the two states
better than a local measurement on an individual subsystem could. In fact, the
proof of monotonicity follows this intuition exactly, and
Figure~\ref{fig-dm:monotonicity-trace}\ depicts the intuition behind it.%

\begin{corollary}
[Monotonicity of Trace Distance]\label{lemma:trace-dist-monotone} Let
$\rho_{AB},\sigma_{AB}\in\mathcal{D}(\mathcal{H}_{A}\otimes\mathcal{H}_{B})$.
The trace distance%
\index{trace distance!monotonicity}
is monotone with respect to discarding of subsystems:%
\begin{equation}
\left\Vert \rho_{A}-\sigma_{A}\right\Vert _{1}\leq\left\Vert \rho_{AB}%
-\sigma_{AB}\right\Vert _{1}.
\end{equation}

\end{corollary}

\begin{proof}
Consider that%
\begin{equation}
\left\Vert \rho_{A}-\sigma_{A}\right\Vert _{1}=2\cdot\operatorname{Tr}\left\{
\Lambda_{A}\left(  \rho_{A}-\sigma_{A}\right)  \right\}  ,
\end{equation}
for some positive semi-definite operator $\Lambda_{A}\leq I_{A}$. Then%
\begin{align}
2\cdot\operatorname{Tr}\left\{  \Lambda_{A}\left(  \rho_{A}-\sigma_{A}\right)
\right\}   &  =2\cdot\operatorname{Tr}\left\{  \left(  \Lambda_{A}\otimes
I_{B}\right)  \left(  \rho_{AB}-\sigma_{AB}\right)  \right\} \\
&  \leq2\cdot\max_{0\leq\Lambda_{AB}\leq I}\operatorname{Tr}\left\{
\Lambda_{AB}\left(  \rho_{AB}-\sigma_{AB}\right)  \right\} \\
&  =\left\Vert \rho_{AB}-\sigma_{AB}\right\Vert _{1}.
\end{align}
The first equality follows because local predictions of the quantum theory
should coincide with its global predictions (as discussed in
Section~\ref{sec-nqt:partial-trace}). The inequality follows because the local
operator $\Lambda_{A}$ never gives a higher probability difference than a
maximization over all global operators. The last equality follows from the
characterization of the trace distance in Lemma~\ref{lemma:trace-equiv}.
\end{proof}

\begin{exercise}
[Monotonicity of Trace Distance]\label{ex-dm:mono-TD}Let $\rho,\sigma
\in\mathcal{D}(\mathcal{H}_{A})$ and $\mathcal{N}:\mathcal{L}(\mathcal{H}%
_{A})\rightarrow\mathcal{L}(\mathcal{H}_{B})$ be a quantum channel. Show that
the trace distance is monotone with respect to the action of the channel
$\mathcal{N}$:%
\begin{equation}
\left\Vert \mathcal{N}(\rho)-\mathcal{N}(\sigma)\right\Vert _{1}\leq\left\Vert
\rho-\sigma\right\Vert _{1}.
\end{equation}
(Hint:\ Use the result of Corollary~\ref{lemma:trace-dist-monotone} and
Exercise~\ref{ex-dm:trace-inv-iso}.)
\end{exercise}

The result of the previous exercise deserves an interpretation. It states that
a quantum channel $\mathcal{N}$ makes two quantum states $\rho$ and $\sigma$
less distinguishable from each other. That is, a noisy channel tends to
\textquotedblleft blur\textquotedblright\ two states to make them appear as if
they are more similar to each other than they are before the quantum channel acts.

\begin{exercise}
\label{thm-dm:meas-achieve-TD}Prove that a measurement achieves the trace
distance, in the following sense:%
\begin{equation}
\left\Vert \rho-\sigma\right\Vert _{1}=\max_{\left\{  \Lambda_{x}\right\}
}\sum_{x}\left\vert \operatorname{Tr}\{\Lambda_{x}\rho\}-\operatorname{Tr}%
\{\Lambda_{x}\sigma\}\right\vert ,
\end{equation}
where $\rho,\sigma\in\mathcal{D}(\mathcal{H})$ and the optimization is with
respect to all POVMs $\{\Lambda_{x}\}$. Hint:\ Use the result of
Exercise~\ref{ex-dm:mono-TD}\ to show the following bound for any choice of
POVM:%
\begin{equation}
\left\Vert \rho-\sigma\right\Vert _{1}\geq\sum_{x}\left\vert \operatorname{Tr}%
\{\Lambda_{x}\rho\}-\operatorname{Tr}\{\Lambda_{x}\sigma\}\right\vert .
\end{equation}
Next, use the developments in the proof of Lemma~\ref{lemma:trace-equiv}\ to
construct an optimal measurement that saturates this bound. (Further
hint:\ Consider the measurement $\{\Pi_{P},\Pi_{Q}\}$.)
\end{exercise}

\begin{exercise}
Show that the trace distance is \textit{strongly convex}. That is, for two
ensembles $\left\{  p_{X_{1}}(x),\rho_{x}\right\}  $ and $\left\{  p_{X_{2}%
}(x),\sigma_{x}\right\}  $ such that $\rho_{x},\sigma_{x}\in\mathcal{D}%
(\mathcal{H})$ for all $x$, the following inequality holds%
\begin{multline}
\left\Vert \sum_{x}p_{X_{1}}(x)\rho_{x}-\sum_{x}p_{X_{2}}(x)\sigma
_{x}\right\Vert _{1}\\
\leq\sum_{x}\left\vert p_{X_{1}}(x)-p_{X_{2}}(x)\right\vert +\sum_{x}p_{X_{1}%
}(x)\left\Vert \rho_{x}-\sigma_{x}\right\Vert _{1}.
\end{multline}

\end{exercise}

\subsection{Channel Distinguishability and the Diamond Norm}%

\index{diamond-norm distance}%
\index{channel distinguishability}
\label{sec-dm:diamond-norm}Given the operational interpretation of trace
distance in terms of the discrimination of quantum states (from
Section~\ref{sec-dm:op-int-trace}), a next natural question is to understand
how we can distinguish one quantum channel from another. That is, we would
like to understand how close two quantum channels are to each other in an
operational sense. For this purpose, there is a hypothesis testing scenario
which extends that from Section~\ref{sec-dm:op-int-trace}. In the protocol for
state discrimination from Section~\ref{sec-dm:op-int-trace}, there were really
just two steps:\ Bob prepares one of two states at random and sends the state
to Alice, who then performs a measurement in an attempt to figure out which
one Bob prepared.

When distinguishing channels, there is an extra degree of freedom:\ the
channel accepts an input quantum state which then gets transformed to an
output quantum state. This suggests that we should allow for an extra step in
a channel distinguishability scenario, in which Alice prepares a quantum
state. Let $\mathcal{N},\mathcal{M}:\mathcal{L}(\mathcal{H}_{A})\rightarrow
\mathcal{L}(\mathcal{H}_{B})$ be quantum channels. The augmented hypothesis
testing scenario consists of the following steps:

\begin{enumerate}
\item Alice prepares a state $\rho_{A}$ and sends it to Bob.

\item Bob flips a fair coin and based on the outcome, he acts on $\rho_{A}$
with either $\mathcal{N}$ or $\mathcal{M}$. Bob sends the output of the
channel to Alice.

\item Alice then performs a measurement to figure out which channel Bob applied.
\end{enumerate}

From our development in the previous section, we can immediately conclude that
the success probability in distinguishing the channels using such a protocol
is equal to%
\begin{equation}
\frac{1}{2}\left(  1+\frac{1}{2}\left\Vert \mathcal{N}_{A\rightarrow B}%
(\rho_{A})-\mathcal{M}_{A\rightarrow B}(\rho_{A})\right\Vert _{1}\right)  .
\end{equation}
However, it is clear that Alice could potentially increase the success
probability by maximizing this quantity with respect to her choice of the
input state. This leads to the following expression for the success
probability:%
\begin{equation}
\frac{1}{2}\left(  1+\frac{1}{2}\max_{\rho_{A}\in\mathcal{D}(\mathcal{H}_{A}%
)}\left\Vert \mathcal{N}_{A\rightarrow B}(\rho_{A})-\mathcal{M}_{A\rightarrow
B}(\rho_{A})\right\Vert _{1}\right)  .
\end{equation}

\begin{figure}
[ptb]
\begin{center}
\includegraphics[
width=4.1528in
]%
{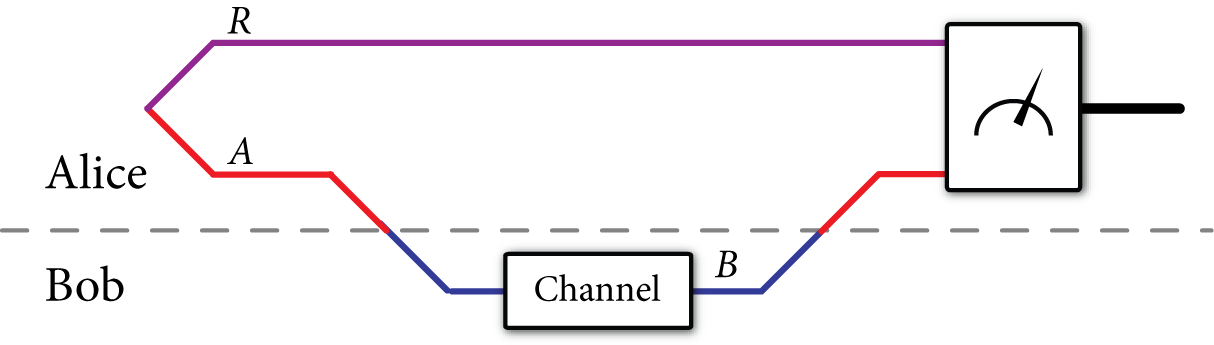}%
\caption{Protocol for Alice to distinguish one channel from another (described
in main text).}%
\label{fig-dm:diamond-norm}%
\end{center}
\end{figure}

This suggests that we should consider the quantity%
\begin{equation}
\max_{\rho_{A}\in\mathcal{D}(\mathcal{H}_{A})}\left\Vert \mathcal{N}%
_{A\rightarrow B}(\rho_{A})-\mathcal{M}_{A\rightarrow B}(\rho_{A})\right\Vert
_{1}%
\end{equation}
to be our measure of distinguishability between channels $\mathcal{N}$ and
$\mathcal{M}$. However, there is still a problem because the protocol for
distinguishing the channels is not as general as it could be. That is, it
excludes the possibility of Alice preparing an entangled state to distinguish
the channels. The most general protocol for distinguishing the channels
consists of the following steps (depicted in Figure~\ref{fig-dm:diamond-norm}):

\begin{enumerate}
\item Alice prepares a state $\rho_{RA}$ on systems $R$ and $A$ and sends
system $A$ to Bob. The reference system $R$ can have an arbitrarily large dimension.

\item Bob flips a fair coin and based on the outcome, he acts on the $A$
system with either $\mathcal{N}$ or $\mathcal{M}$, which produces an output
system $B$. Bob sends the system $B$ to Alice.

\item Alice then performs a measurement on systems $R$ and $B$ to figure out
which channel Bob applied.
\end{enumerate}

In such a protocol, we allow for the possibility of Alice preparing an
entangled state. It turns out that there can sometimes be a huge difference in
Alice's ability to figure out which channel was applied, if we allow or do not
allow for entangled states to be prepared (this depends on the channels).

By the same reasoning as before and allowing for an optimization over all
possible input states that Alice could prepare, Alice's success probability in
distinguishing the channels is as follows:%
\begin{equation}
\frac{1}{2}\left(  1+\frac{1}{2}\sup_{n}\max_{\rho_{R_{n}A}}\left\Vert \left(
\operatorname{id}_{R_{n}}\otimes\mathcal{N}_{A\rightarrow B}\right)
(\rho_{R_{n}A})-\left(  \operatorname{id}_{R_{n}}\otimes\mathcal{M}%
_{A\rightarrow B}\right)  (\rho_{R_{n}A})\right\Vert _{1}\right)  ,
\end{equation}
where $\rho_{R_{n}A}\in\mathcal{D}(\mathcal{H}_{R_{n}}\otimes\mathcal{H}_{A}%
)$. In the above formula, $n$ is a positive integer corresponding to the
dimension of the reference system $R_{n}$. A priori, we require a supremum
over this dimension size since we have not yet placed a bound on the dimension
needed for a reference system. The channel distance measure appearing in the
above formula is known as the diamond-norm distance between the channels:

\begin{definition}
[Diamond-Norm Distance]Let $\mathcal{N},\mathcal{M}:\mathcal{L}(\mathcal{H}%
_{A})\rightarrow\mathcal{L}(\mathcal{H}_{B})$ be quantum channels. The
diamond-norm distance is defined as%
\begin{equation}
\left\Vert \mathcal{N}-\mathcal{M}\right\Vert _{\diamondsuit}\equiv\sup
_{n}\max_{\rho_{R_{n}A}}\left\Vert \left(  \operatorname{id}_{R_{n}}%
\otimes\mathcal{N}_{A\rightarrow B}\right)  (\rho_{R_{n}A})-\left(
\operatorname{id}_{R_{n}}\otimes\mathcal{M}_{A\rightarrow B}\right)
(\rho_{R_{n}A})\right\Vert _{1},
\end{equation}
where $\rho_{R_{n}A}\in\mathcal{D}(\mathcal{H}_{R_{n}}\otimes\mathcal{H}_{A})$.
\end{definition}

Given the above definition, a natural question is whether we can place a bound
on the dimension of the reference system required. Indeed, this is possible,
as the following theorem states:

\begin{theorem}
Let $\mathcal{N},\mathcal{M}:\mathcal{L}(\mathcal{H}_{A})\rightarrow
\mathcal{L}(\mathcal{H}_{B})$ be quantum channels. Then%
\begin{equation}
\left\Vert \mathcal{N}-\mathcal{M}\right\Vert _{\diamondsuit}=\max
_{|\psi\rangle_{RA}}\left\Vert \left(  \operatorname{id}_{R}\otimes
\mathcal{N}_{A\rightarrow B}\right)  (|\psi\rangle\langle\psi|_{RA})-\left(
\operatorname{id}_{R}\otimes\mathcal{M}_{A\rightarrow B}\right)  (|\psi
\rangle\langle\psi|_{RA})\right\Vert _{1}, \label{eq-dm:diamond-norm-dim}%
\end{equation}
where the optimization is with respect to all $|\psi\rangle_{RA}\in
\mathcal{H}_{R}\otimes\mathcal{H}_{A}$ such that $\left\Vert |\psi\rangle
_{RA}\right\Vert _{2}=1$, with $\dim(\mathcal{H}_{R})=\dim(\mathcal{H}_{A})$.
\end{theorem}

\begin{proof}
This theorem follows as a consequence of the convexity of the trace norm and
the Schmidt decomposition. Indeed, let $\rho_{R_{n}A}$ be any density operator
for systems $R_{n}$ and $A$. Let $\sum_{x}p_{X}(x)|\psi^{x}\rangle\langle
\psi^{x}|_{R_{n}A}$ be a spectral decomposition of $\rho_{R_{n}A}$. From the
convexity of the trace norm, we find that%
\begin{align}
&  \left\Vert \left(  \operatorname{id}_{R_{n}}\otimes\mathcal{N}%
_{A\rightarrow B}\right)  (\rho_{R_{n}A})-\left(  \operatorname{id}_{R_{n}%
}\otimes\mathcal{M}_{A\rightarrow B}\right)  (\rho_{R_{n}A})\right\Vert
_{1}\nonumber\\
&  =\left\Vert \sum_{x}p_{X}(x)\left[  \left(  \operatorname{id}_{R_{n}%
}\otimes\mathcal{N}_{A\rightarrow B}\right)  (|\psi^{x}\rangle\langle\psi
^{x}|_{R_{n}A})-\left(  \operatorname{id}_{R_{n}}\otimes\mathcal{M}%
_{A\rightarrow B}\right)  (|\psi^{x}\rangle\langle\psi^{x}|_{R_{n}A})\right]
\right\Vert _{1}\\
&  \leq\sum_{x}p_{X}(x)\left\Vert \left[  \left(  \operatorname{id}_{R_{n}%
}\otimes\mathcal{N}_{A\rightarrow B}\right)  (|\psi^{x}\rangle\langle\psi
^{x}|_{R_{n}A})-\left(  \operatorname{id}_{R_{n}}\otimes\mathcal{M}%
_{A\rightarrow B}\right)  (|\psi^{x}\rangle\langle\psi^{x}|_{R_{n}A})\right]
\right\Vert _{1}\\
&  \leq\left\Vert \left[  \left(  \operatorname{id}_{R_{n}}\otimes
\mathcal{N}_{A\rightarrow B}\right)  (|\psi_{\ast}^{x}\rangle\langle\psi
_{\ast}^{x}|_{R_{n}A})-\left(  \operatorname{id}_{R_{n}}\otimes\mathcal{M}%
_{A\rightarrow B}\right)  (|\psi_{\ast}^{x}\rangle\langle\psi_{\ast}%
^{x}|_{R_{n}A})\right]  \right\Vert _{1},
\end{align}
where the last inequality follows because the average value is never larger
than the maximum value and we let $|\psi_{\ast}^{x}\rangle_{R_{n}A}$ denote
the state vector giving the maximum value. From the Schmidt decomposition
theorem (Theorem~\ref{thm-qt:schmidt}), the Schmidt rank of $|\psi_{\ast}%
^{x}\rangle_{R_{n}A}$ is no larger than $\dim(\mathcal{H}_{A})$, implying that
$|\psi_{\ast}^{x}\rangle_{R_{n}A}$ can be embedded in a tensor-product Hilbert
space $\mathcal{H}_{R}\otimes\mathcal{H}_{A}$ such that $\dim(\mathcal{H}%
_{R})=\dim(\mathcal{H}_{A})$. Since this bound holds for any density operator
$\rho_{R_{n}A}$, the statement of the theorem follows.
\end{proof}

As a consequence of the above theorem, we can take the result in
\eqref{eq-dm:diamond-norm-dim}\ to be the definition of the diamond-norm
distance. The main use of the diamond-norm distance is for comparing quantum
channels. For example, when studying the classical capacity of a quantum
channel (Chapter~\ref{chap:classical-comm-HSW}), we would like to compare how
well a given protocol simulates a noiseless classical channel, and the
diamond-norm distance gives a natural way to do so. The situation is similar
with the quantum capacity theorem (Chapter~\ref{chap:quantum-capacity}):\ here
we would like to compare how well a protocol simulates a noiseless quantum
channel, and we can quantify the performance using the diamond-norm distance.

\begin{exercise}
Suppose that Alice is restricted to use separable states on systems $R_{n}$
and $A$ to distinguish two quantum channels $\mathcal{N}$ and $\mathcal{M}$.
Show that the success probability in doing so is given by%
\begin{equation}
\frac{1}{2}\left(  1+\frac{1}{2}\max_{|\psi\rangle_{A}}\left\Vert
\mathcal{N}_{A\rightarrow B}(|\psi\rangle\langle\psi|_{A})-\mathcal{M}%
_{A\rightarrow B}(|\psi\rangle\langle\psi|_{A})\right\Vert _{1}\right)  ,
\end{equation}
where $|\psi\rangle\langle\psi|_{A}\in\mathcal{D}(\mathcal{H}_{A})$.
\end{exercise}

\begin{exercise}
Suppose that channels $\mathcal{N}$ and $\mathcal{M}$ are defined as follows:%
\begin{equation}
\mathcal{N}(X_{A})=\operatorname{Tr}\{X_{A}\}\rho_{B}%
,\ \ \ \ \ \ \ \ \ \ \mathcal{M}(X_{A})=\operatorname{Tr}\{X_{A}\}\sigma_{B},
\end{equation}
where $X_{A}\in\mathcal{L}(\mathcal{H}_{A})$ and $\rho_{B},\sigma_{B}%
\in\mathcal{D}(\mathcal{H}_{B})$. Show that%
\begin{equation}
\left\Vert \mathcal{N}-\mathcal{M}\right\Vert _{\diamondsuit}=\left\Vert
\rho_{B}-\sigma_{B}\right\Vert _{1}.
\end{equation}

\end{exercise}

\section{Fidelity}

\subsection{Pure-State Fidelity}

An alternate measure of the closeness of two quantum states is the
\index{fidelity}
\textit{fidelity}. We introduce its most simple form first. Suppose that we
input a particular pure state $|\psi\rangle$ to a quantum
information-processing protocol. Ideally, we may want the protocol to output
the same state that is input, but suppose that it instead outputs a pure state
$|\phi\rangle$. The pure-state fidelity $F(\psi,\phi)$\ is a measure of how
close the output state is to the input state.

\begin{definition}
[Pure-State Fidelity]Let $\vert\psi\rangle,\vert\phi\rangle\in\mathcal{H}$ be
pure states. The pure-state fidelity is the squared overlap of the states
$\vert\psi\rangle$ and $\vert\phi\rangle$:%
\begin{equation}
F( \psi,\phi) \equiv\left\vert \left\langle \psi|\phi\right\rangle \right\vert
^{2}. \label{eq-dm:pure-state-fidelity}%
\end{equation}

\end{definition}

The pure-state fidelity has the operational interpretation as the probability
that the output state $\vert\phi\rangle$\ would pass a test for being the same
as the input state $\vert\psi\rangle$, conducted by someone who knows the
input state (see Exercise~\ref{ex-dm:interp-fidelity}).

The pure-state fidelity is symmetric $F(\psi,\phi)=F(\phi,\psi)$, and it obeys
the following bounds:%
\begin{equation}
0\leq F(\psi,\phi)\leq1.
\end{equation}
It is equal to one if and only if the two states are the same, and it is equal
to zero if and only if the two states are orthogonal to each other. The
fidelity measure is \textit{not} a distance measure in the strict mathematical
sense because it is equal to one when two states are equal, whereas a distance
measure should be equal to zero when two states are equal.

\begin{exercise}
\label{ex-dist:fidelity-pure-states}Suppose that two pure quantum states
$|\psi\rangle,|\phi\rangle\in\mathcal{H}$ are as follows:%
\begin{equation}
|\psi\rangle\equiv\sum_{x}\sqrt{p(x)}|x\rangle,\ \ \ \ \ \ \ \ |\phi
\rangle\equiv\sum_{x}\sqrt{q(x)}|x\rangle,
\end{equation}
where $\left\{  |x\rangle\right\}  $ is some orthonormal basis for
$\mathcal{H}$. Show that the fidelity $F(\psi,\phi)$ between these two states
is equivalent to the \textit{Bhattacharyya overlap} (classical fidelity)
between the distributions $p(x)$ and$~q(x)$:%
\begin{equation}
F(\psi,\phi)=\left[  \sum_{x}\sqrt{p(x)q(x)}\right]  ^{2}.
\end{equation}

\end{exercise}

\subsection{Expected Fidelity}

Now let us suppose that the output of a given protocol is not a pure state,
but it is rather a mixed state with density operator $\rho$. In general, a
quantum information-processing protocol could be noisy and map the pure input
state $\vert\psi\rangle$ to a mixed state. We would like a way to compare
these two states.

\begin{definition}
[Expected Fidelity]The expected fidelity%
\index{fidelity!expected}
$F( \psi,\rho) $ between a pure state $\vert\psi\rangle\in\mathcal{H}$ and a
mixed state $\rho\in\mathcal{D}(\mathcal{H})$ is%
\begin{equation}
F( \psi,\rho) \equiv\langle\psi\vert\rho\vert\psi\rangle.
\label{eq-dm:expected-fidelity-def}%
\end{equation}

\end{definition}

We now justify the above definition of fidelity. Let us decompose $\rho$
according to a spectral decomposition $\rho=\sum_{x}p_{X}(x)|\phi_{x}%
\rangle\langle\phi_{x}|$. Recall that we can think of this output density
operator as arising from the ensemble $\left\{  p_{X}(x),|\phi_{x}%
\rangle\right\}  $. We generalize the pure-state fidelity from the previous
paragraph by defining it as the expected pure-state fidelity, where the
expectation is with respect to states in the ensemble:%
\begin{align}
F(\psi,\rho)  &  \equiv\mathbb{E}_{X}\left[  \left\vert \left\langle \psi
|\phi_{X}\right\rangle \right\vert ^{2}\right] \\
&  =\sum_{x}p_{X}(x)\left\vert \left\langle \psi|\phi_{x}\right\rangle
\right\vert ^{2}\\
&  =\sum_{x}p_{X}(x)\left\langle \psi|\phi_{x}\right\rangle \left\langle
\phi_{x}|\psi\right\rangle \\
&  =\langle\psi|\left(  \sum_{x}p_{X}(x)|\phi_{x}\rangle\langle\phi
_{x}|\right)  |\psi\rangle\\
&  =\langle\psi|\rho|\psi\rangle. \label{eq-dm:mixed-state-fid}%
\end{align}
The compact formula $F(\psi,\rho)=\langle\psi|\rho|\psi\rangle$ is a good way
to characterize the fidelity when the input state is pure and the output state
is mixed. We can see that the above fidelity measure is a generalization of
the pure-state fidelity in \eqref{eq-dm:pure-state-fidelity}. It obeys the
same bounds:%
\begin{equation}
0\leq F(\psi,\rho)\leq1,
\end{equation}
being equal to one if and only if the state $\rho$ is equal to $|\psi
\rangle\langle\psi|$ and equal to zero if and only if the support of $\rho$ is
orthogonal to $|\psi\rangle\langle\psi|$.

\begin{exercise}
\label{ex-dm:interp-fidelity}Given a state $\sigma\in\mathcal{D}(\mathcal{H}%
)$, we would like to see if it would pass a test for being close to a pure
state $\vert\varphi\rangle\in\mathcal{H} $. We can measure the POVM $\left\{
\vert\varphi\rangle\langle\varphi\vert,I-\vert\varphi\rangle\langle
\varphi\vert\right\}  $ with result $\varphi$ corresponding to a
\textquotedblleft pass\textquotedblright\ and the result $I-\varphi$
corresponding to a \textquotedblleft fail.\textquotedblright\ Show that the
fidelity is then equal to $\Pr\{\!\text{``pass''}\} $.
\end{exercise}

\begin{exercise}
\label{ex-dm:fidelity-triangle-like}Using the result of
Corollary~\ref{lemma:trace-inequality}, show that the following inequality
holds for a pure state $\vert\phi\rangle\in\mathcal{H} $ and mixed states
$\rho,\sigma\in\mathcal{D}(\mathcal{H})$:%
\begin{equation}
F( \phi, \rho) \leq F( \phi, \sigma) +\tfrac{1}{2}\left\Vert \rho
-\sigma\right\Vert _{1}.
\end{equation}

\end{exercise}

\subsection{Uhlmann Fidelity}

What is the most general form of the fidelity
\index{fidelity!Uhlmann}%
when both quantum states are mixed? We can borrow the above idea of the
pure-state fidelity that exploits the overlap between two pure states. Suppose
that we would like to determine the fidelity between two mixed states
$\rho_{A}$ and $\sigma_{A}$ that represent different states of some quantum
system $A$. Let $|\phi^{\rho}\rangle_{RA}$ and $\left\vert \phi^{\sigma
}\right\rangle _{RA}$ denote particular respective purifications of the mixed
states to some reference system $R$ (where for now we assume that the
reference system has the same dimension as the system $A$). We can define the
Uhlmann fidelity $F(\rho_{A},\sigma_{A})$\ between two mixed states $\rho_{A}$
and $\sigma_{A}$ as the maximum overlap between their respective
purifications, where the maximization is with respect to all purifications
$|\phi^{\rho}\rangle_{RA}$ and $\left\vert \phi^{\sigma}\right\rangle _{RA}$
of the respective states $\rho_{A}$ and$~\sigma_{A}$:%
\begin{equation}
F(\rho_{A},\sigma_{A})\equiv\max_{|\phi^{\rho}\rangle_{RA},\ \left\vert
\phi^{\sigma}\right\rangle _{RA}}\left\vert \langle\phi^{\rho}|\phi^{\sigma
}\rangle_{RA}\right\vert ^{2}.
\end{equation}
We can express the fidelity as a maximization over unitaries instead (recall
the result of Theorem~\ref{thm-pqt:purif-unitary-reference}\ that all
purifications are equivalent up to unitaries on the reference system):%
\begin{align}
F(\rho_{A},\sigma_{A})  &  =\max_{U^{\rho},U^{\sigma}}\left\vert \langle
\phi^{\rho}|_{RA}\left(  \left(  U_{R}^{\rho}\right)  ^{\dag}\otimes
I_{A}\right)  \left(  U_{R}^{\sigma}\otimes I_{A}\right)  \left\vert
\phi^{\sigma}\right\rangle _{RA}\right\vert ^{2}\\
&  =\max_{U^{\rho},U^{\sigma}}\left\vert \langle\phi^{\rho}|_{RA}\left(
U_{R}^{\rho}\right)  ^{\dag}U_{R}^{\sigma}\otimes I_{A}\left\vert \phi
^{\sigma}\right\rangle _{RA}\right\vert ^{2}.
\end{align}
It is unnecessary to maximize over two sets of unitaries because the product
$\left(  U_{R}^{\rho}\right)  ^{\dag}U_{R}^{\sigma}$ represents only a single
unitary. The final expression for the fidelity between two mixed states is
then defined as the Uhlmann fidelity.

\begin{definition}
[Uhlmann Fidelity]\label{def-dm:uhlmann}The Uhlmann fidelity $F( \rho
_{A},\sigma_{A}) $\ between two mixed states $\rho_{A}$ and $\sigma_{A}$ is
the maximum overlap between their respective purifications, where the
maximization is with respect to all unitaries $U$ acting on the purification
system $R$:%
\begin{equation}
F( \rho_{A},\sigma_{A}) =\max_{U}\left\vert \langle\phi^{\rho}|_{RA}%
U_{R}\otimes I_{A}\left\vert \phi^{\sigma}\right\rangle _{RA}\right\vert ^{2}.
\label{eq-dm:uhlmann-fidelity}%
\end{equation}

\end{definition}

We will find that this notion of fidelity generalizes both the pure-state
fidelity in \eqref{eq-dm:pure-state-fidelity} and the expected fidelity in
\eqref{eq-dm:mixed-state-fid}. This holds because the following formula for
the fidelity of two mixed states, characterized in terms of the Schatten
1-norm, is equivalent to the above Uhlmann characterization:%
\begin{equation}
F( \rho_{A},\sigma_{A}) =\left\Vert \sqrt{\rho_{A}}\sqrt{\sigma_{A}%
}\right\Vert _{1}^{2}. \label{eq-dm:fidelity-l1}%
\end{equation}
We state this result as Uhlmann's theorem.

\begin{theorem}
[Uhlmann's Theorem]\label{thm-dm:uhlmann-thm}The following two expressions for fidelity are equal:%
\begin{equation}
F( \rho_{A},\sigma_{A}) =\max_{U}\left\vert \langle\phi^{\rho}|_{RA}%
U_{R}\otimes I_{A}\left\vert \phi^{\sigma}\right\rangle _{RA}\right\vert
^{2}=\left\Vert \sqrt{\rho_{A}}\sqrt{\sigma_{A}}\right\Vert _{1}^{2}.
\label{eq-dm:uhlmann-char}%
\end{equation}

\end{theorem}

\begin{proof}
Let $|\phi^{\rho}\rangle_{RA}$ denote the canonical purification of $\rho_{A}$
(see Exercise~\ref{ex-pqt:canonical-pur}):%
\begin{equation}
|\phi^{\rho}\rangle_{RA}\equiv\left(  I_{R}\otimes\sqrt{\rho_{A}}\right)
\vert\Gamma\rangle_{RA}, \label{eq:uhlmann-pure-1}%
\end{equation}
where $\vert\Gamma\rangle_{RA}$ is the unnormalized maximally entangled
vector:%
\begin{equation}
\vert\Gamma\rangle_{RA}\equiv\sum_{i}|i\rangle_{R}|i\rangle_{A}.
\end{equation}
Therefore, the state $|\phi^{\rho}\rangle_{RA}$ is a particular purification
of $\rho$. Let $\left\vert \phi^{\sigma}\right\rangle _{RA}$ denote the
canonical purification of $\sigma_{A}$:%
\begin{equation}
\left\vert \phi^{\sigma}\right\rangle _{RA}\equiv\left(  I_{R}\otimes
\sqrt{\sigma_{A}}\right)  \vert\Gamma\rangle_{RA}. \label{eq:uhlmann-pure-2}%
\end{equation}
Consider that the overlap $\left\vert \langle\phi^{\rho}|U_{R}\otimes
I_{A}|\phi^{\sigma}\rangle\right\vert ^{2}$ is as follows:%
\begin{align}
\left\vert \langle\phi^{\rho}|U_{R}\otimes I_{A}|\phi^{\sigma}\rangle
\right\vert ^{2}  &  =\left\vert \langle\Gamma\vert_{RA}\left(  U_{R}%
\otimes\sqrt{\rho_{A}}\right)  \left(  I_{R}\otimes\sqrt{\sigma_{A}}\right)
\vert\Gamma\rangle_{RA}\right\vert ^{2}\\
&  =\left\vert \langle\Gamma\vert_{RA}\left(  U_{R}\otimes\sqrt{\rho_{A}}%
\sqrt{\sigma_{A}}\right)  \vert\Gamma\rangle_{RA}\right\vert ^{2}\\
&  =\left\vert \langle\Gamma\vert_{RA}\left(  I_{R}\otimes\sqrt{\rho_{A}}%
\sqrt{\sigma_{A}}U_{A}^{T}\right)  \left\vert \Gamma\right\rangle
_{RA}\right\vert ^{2}\\
&  =\left\vert \operatorname{Tr}\left\{  \sqrt{\rho_{A}}\sqrt{\sigma_{A}}%
U_{A}^{T}\right\}  \right\vert ^{2}.
\end{align}
The first equality follows by plugging in \eqref{eq:uhlmann-pure-1} and
\eqref{eq:uhlmann-pure-2}. The third equality follows from
Exercise~\ref{ex-qt:bell-state-matrix-identity}. The last equality follows
from Exercise~\ref{ex-qt:alt-trace-max-ent}. We can finally invoke
Property~\ref{prop-dm:var-char-TD} to establish that%
\begin{equation}
\max_{U_{A}}\left\vert \operatorname{Tr}\left\{  \sqrt{\rho_{A}}\sqrt
{\sigma_{A}}U_{A}^{T}\right\}  \right\vert ^{2}=\left\Vert \sqrt{\rho_{A}%
}\sqrt{\sigma_{A}}\right\Vert _{1}^{2},
\end{equation}
from which \eqref{eq-dm:uhlmann-char} follows.
\end{proof}

\begin{exercise}
Use the expression $\left\Vert \sqrt{\rho_{A}}\sqrt{\sigma_{A}}\right\Vert
_{1}^{2}$ for the fidelity and the\ Cauchy--Schwarz inequality for the
Hilbert--Schmidt inner product (from
\eqref{eq:Cauchy--Schwarz-Hilbert-Schmidt}) to prove that the quantum fidelity
between two density operators never exceeds one.
\end{exercise}

\begin{remark}
Note that we can define the fidelity function more generally for any two
positive semi-definite operators, which can sometimes be useful. That is, let
$P$ and $Q$ be positive semi-definite operators acting on the same Hilbert
space. Then we define%
\begin{equation}
F(P,Q)\equiv\left\Vert \sqrt{P}\sqrt{Q}\right\Vert _{1}^{2}.
\end{equation}
By applying the Cauchy--Schwarz inequality for the Hilbert--Schmidt inner
product (see \eqref{eq:Cauchy--Schwarz-Hilbert-Schmidt}) and the
characterization of the trace norm in Property~\ref{prop-dm:var-char-TD}, we
find that%
\begin{equation}
F(P,Q)\leq\operatorname{Tr}\{P\}\operatorname{Tr}\{Q\}.
\end{equation}

\end{remark}

\begin{remark}
Note that in the development above, we assumed that the dimension of the
reference system $R$ is equal to that of the system $A$. However, this is the
not the most general definition that we could have taken. We could have
defined fidelity as%
\begin{equation}
F(\rho_{A},\sigma_{A})=\sup_{\dim(\mathcal{H}_{R})}\max_{|\phi^{\rho}%
\rangle_{RA},\ \left\vert \phi^{\sigma}\right\rangle _{RA}}\left\vert
\langle\phi^{\rho}|_{RA}\left\vert \phi^{\sigma}\right\rangle _{RA}\right\vert
^{2},
\end{equation}
in which there is an extra optimization over the dimension of the reference
system $R$ in addition to the optimization over the purifications. However,
repeating an analysis similar to the above one would lead us to the conclusion
that%
\begin{equation}
F(\rho_{A},\sigma_{A})=\left\Vert \sqrt{\rho_{A}}\sqrt{\sigma_{A}}\right\Vert
_{1}^{2}%
\end{equation}
for this definition as well. Indeed, we have that%
\begin{equation}
\left\vert \langle\phi^{\rho}|_{RA}\left\vert \phi^{\sigma}\right\rangle
_{RA}\right\vert ^{2}=\left\vert \left[  \langle\Gamma\vert_{R^{\prime}A}%
\sqrt{\rho_{A}}\left(  V_{R^{\prime}\rightarrow R}\right)  ^{\dag}\right]
\left[  U_{R^{\prime}\rightarrow R}\sqrt{\sigma_{A}}\left\vert \Gamma
\right\rangle _{R^{\prime}A}\right]  \right\vert ^{2},
\end{equation}
where $R^{\prime}$ is a reference system with dimension equal to
$\dim(\mathcal{H}_{A})$ and $U_{R^{\prime}\rightarrow R}$ and $V_{R^{\prime
}\rightarrow R}$ are some isometries , given that all purifications are
related by an isometry acting on the reference system
(Theorem~\ref{thm-pqt:purif-unitary-reference}). Carrying through the same
analysis along with the characterization of the trace norm in
Exercise~\ref{ex-dm:trace-norm}\ then gives that $\left\vert \langle\phi
^{\rho}|_{RA}\left\vert \phi^{\sigma}\right\rangle _{RA}\right\vert
\leq\left\Vert \sqrt{\rho_{A}}\sqrt{\sigma_{A}}\right\Vert _{1}$, so that
having an arbitrarily large reference system does not help.
\end{remark}

\subsection{Properties of Fidelity}

We discuss some further properties of the fidelity that often prove useful.
Some of these properties are the counterpart of similar properties of the
trace distance. From the characterization of fidelity
in~\eqref{eq-dm:uhlmann-char}, we observe that it is symmetric in its
arguments:%
\begin{equation}
F( \rho,\sigma) =F( \sigma,\rho) .
\end{equation}
It obeys the following bounds:%
\begin{equation}
0\leq F( \rho,\sigma) \leq1. \label{eq-dm:bounds-fidelity}%
\end{equation}
The lower bound applies if and only if the respective supports of the two
states $\rho$ and $\sigma$ are orthogonal. To see this, suppose that the
supports of $\rho$ and $\sigma$ are orthogonal. This implies that $\sqrt{\rho
}\sqrt{\sigma} = 0$, so that $F( \rho,\sigma) = \Vert\sqrt{\rho}\sqrt{\sigma}
\Vert_{1}^{2} = 0 $. On the other hand, suppose that $F( \rho,\sigma)=0$. Then
by definition, this means that $\Vert\sqrt{\rho}\sqrt{\sigma} \Vert_{1}^{2} =
0$, and from non-negative definiteness of the trace norm, we find that
$\sqrt{\rho}\sqrt{\sigma} = 0$. This then implies that the supports of $\rho$
and $\sigma$ are orthogonal. The upper bound in \eqref{eq-dm:bounds-fidelity}
applies if and only if the two states $\rho$ and $\sigma$ are equal to each other.

\begin{exercise}
\label{ex-dm:fidelity-reduction-pure-mixed}Show that the definition of
fidelity in \eqref{eq-dm:fidelity-l1} reduces to
\eqref{eq-dm:pure-state-fidelity} when the two states are pure and to
\eqref{eq-dm:expected-fidelity-def} when one state is pure and the other is mixed.
\end{exercise}

\begin{property}
[Multiplicativity]Let
\index{fidelity!multiplicativity}%
$\rho_{1}, \sigma_{1} \in\mathcal{D}(\mathcal{H}_{1})$ and $\rho_{2},
\sigma_{2} \in\mathcal{D}(\mathcal{H}_{2})$. The fidelity is multiplicative
with respect to tensor products:%
\begin{equation}
F( \rho_{1}\otimes\rho_{2},\sigma_{1}\otimes\sigma_{2}) =F( \rho_{1}%
,\sigma_{1}) F( \rho_{2},\sigma_{2}) .
\end{equation}
This result holds by employing the definition of the fidelity in \eqref{eq-dm:fidelity-l1}.
\end{property}

The following monotonicity lemma is similar to the monotonicity lemma for
trace distance (Lemma~\ref{lemma:trace-dist-monotone}) and also bears the
similar interpretation that quantum states become more similar (less
distinguishable) under the discarding of subsystems.

\begin{lemma}
[Monotonicity]\label{lemma-dm:fidelity-monotone} Let $\rho_{AB}, \sigma_{AB}
\in\mathcal{D}(\mathcal{H}_{A} \otimes\mathcal{H}_{B})$. The fidelity
\index{fidelity!monotonicity}
is non-decreasing with respect to partial trace:
\begin{equation}
F( \rho_{AB},\sigma_{AB}) \leq F( \rho_{A},\sigma_{A}) ,
\end{equation}
where%
\begin{equation}
\rho_{A}=\operatorname{Tr}_{B}\{  \rho_{AB}\}  ,\ \ \ \ \ \ \sigma
_{A}=\operatorname{Tr}_{B}\{  \sigma_{AB}\}  .
\end{equation}

\end{lemma}

\begin{proof}
Consider a fixed purification $|\psi\rangle_{RAB}$ of $\rho_{A}$ and
$\rho_{AB}$ and a fixed purification $|\phi\rangle_{RAB}$ of $\sigma_{A}$ and
$\sigma_{AB}$. Then%
\begin{align}
\left\vert \langle\psi|_{RAB}U_{R}\otimes I_{A}\otimes I_{B}|\phi\rangle
_{RAB}\right\vert ^{2}  &  \leq\max_{U_{RB}}\left\vert \langle\psi
|_{RAB}U_{RB}\otimes I_{A}|\phi\rangle_{RAB}\right\vert ^{2}\\
&  =F(\rho_{A},\sigma_{A}),
\end{align}
where the first inequality follows because the maximization over unitaries
$U_{RB}$\ includes $U_{R}\otimes I_{A}$ and the equality is a consequence of
Uhlmann's theorem. Given that the inequality holds for all unitaries $U_{R}$,
we can conclude that%
\begin{equation}
F(\rho_{AB},\sigma_{AB})=\max_{U_{R}}\left\vert \langle\psi|_{RAB}U_{R}\otimes
I_{A}\otimes I_{B}|\phi\rangle_{RAB}\right\vert ^{2}\leq F(\rho_{A},\sigma
_{A}),
\end{equation}
where the equality is again a consequence of Uhlmann's theorem.
\end{proof}

\begin{property}
[Joint Concavity]\label{prop-dm:joint-concavity} Let $\rho_{x}, \sigma_{x}
\in\mathcal{D}(\mathcal{H})$ for all $x$ and let $p_{X}$ be a probability
distribution. The root fidelity%
\index{fidelity!joint concavity}
is jointly concave with respect to its input arguments:%
\begin{equation}
\sqrt{F}\left(  \sum_{x}p_{X}( x) \rho_{x},\sum_{x}p_{X}( x) \sigma
_{x}\right)  \geq\sum_{x}p_{X}( x) \sqrt{F}( \rho_{x},\sigma_{x}) .
\end{equation}

\end{property}

\begin{proof}
We prove joint concavity by exploiting the result of
Exercise~\ref{ex-pqt:ensemble-purification}. Suppose $|\phi^{\rho_{x}}%
\rangle_{RA}$ and $|\phi^{\sigma_{x}}\rangle_{RA}$ are respective Uhlmann
purifications of $\rho_{x}$ and $\sigma_{x}$ (these are purifications that
maximize the Uhlmann fidelity). Then%
\begin{equation}
F( \phi_{RA}^{\rho_{x}},\phi_{RA}^{\sigma_{x}}) =F( \rho_{x},\sigma_{x}) .
\end{equation}
Choose some orthonormal basis $\left\{  |x\rangle_{X}\right\}  $. Then%
\begin{equation}
|\phi^{\rho}\rangle\equiv\sum_{x}\sqrt{p_{X}(x)}|\phi^{\rho_{x}}\rangle
_{RA}|x\rangle_{X},\ \ \ \ \ \ |\phi^{\sigma}\rangle\equiv\sum_{x}\sqrt
{p_{X}(x)}|\phi^{\sigma_{x}}\rangle_{RA}|x\rangle_{X}%
\end{equation}
are respective purifications of $\sum_{x}p_{X}(x)\rho_{x}$ and $\sum_{x}%
p_{X}(x)\sigma_{x}$. The first inequality below holds by Uhlmann's theorem:%
\begin{align}
\sqrt{F}\left(  \sum_{x}p_{X}(x)\rho_{x},\sum_{x}p_{X}(x)\sigma_{x}\right)
&  \geq\left\vert \langle\phi^{\rho}|\phi^{\sigma}\rangle\right\vert \\
&  =\left\vert \sum_{x}p_{X}(x)\langle\phi^{\rho_{x}}|\phi^{\sigma_{x}}%
\rangle\right\vert \\
&  \geq\sum_{x}p_{X}(x)\left\vert \langle\phi^{\rho_{x}}|\phi^{\sigma_{x}%
}\rangle\right\vert \\
&  =\sum_{x}p_{X}(x)\sqrt{F}\left(  \rho_{x},\sigma_{x}\right)  ,
\end{align}
concluding the proof.
\end{proof}

\begin{property}
[Concavity]Let $\rho, \sigma, \tau\in\mathcal{D}(\mathcal{H})$ and $\lambda
\in\left[  0,1\right]  $. The fidelity is concave with respect to
\index{fidelity!concavity}%
one of its arguments:%
\begin{equation}
F( \lambda\rho+\left(  1-\lambda\right)  \tau,\sigma) \geq\lambda F(
\rho,\sigma) +\left(  1-\lambda\right)  F( \tau,\sigma) .
\end{equation}

\end{property}

\begin{proof}
Let $|\psi^{\sigma}\rangle_{RS}$ be a fixed purification of $\sigma_{S}$. Let
$|\psi^{\rho}\rangle_{RS}$ be a purification of $\rho_{S}$ such that%
\begin{equation}
\left\vert \left\langle \psi^{\sigma}|\psi^{\rho}\right\rangle \right\vert
^{2}=F(\rho,\sigma). \label{eq-dm:uhlmann-rho}%
\end{equation}
Similarly, let $|\psi^{\tau}\rangle_{RS}$ be a purification of $\tau_{S}$ such
that%
\begin{equation}
\left\vert \left\langle \psi^{\sigma}|\psi^{\tau}\right\rangle \right\vert
^{2}=F(\tau,\sigma). \label{eq-dm:uhlmann-tau}%
\end{equation}
Then consider that%
\begin{align}
&  \!\!\!\!\lambda F(\rho,\sigma)+\left(  1-\lambda\right)  F(\tau
,\sigma)\nonumber\\
&  =\lambda\left\vert \left\langle \psi^{\sigma}|\psi^{\rho}\right\rangle
\right\vert ^{2}+\left(  1-\lambda\right)  \left\vert \left\langle
\psi^{\sigma}|\psi^{\tau}\right\rangle \right\vert ^{2}\\
&  =\lambda\left\langle \psi^{\sigma}|\psi^{\rho}\right\rangle \left\langle
\psi^{\rho}|\psi^{\sigma}\right\rangle +\left(  1-\lambda\right)  \left\langle
\psi^{\sigma}|\psi^{\tau}\right\rangle \left\langle \psi^{\tau}|\psi^{\sigma
}\right\rangle \\
&  =\langle\psi^{\sigma}|_{RS}\left(  \lambda|\psi^{\rho}\rangle\langle
\psi^{\rho}|_{RS}+\left(  1-\lambda\right)  |\psi^{\tau}\rangle\langle
\psi^{\tau}|_{RS}\right)  |\psi^{\sigma}\rangle_{RS}\\
&  =F(|\psi^{\sigma}\rangle\langle\psi^{\sigma}|_{RS},\lambda|\psi^{\rho
}\rangle\langle\psi^{\rho}|_{RS}+\left(  1-\lambda\right)  |\psi^{\tau}%
\rangle\langle\psi^{\tau}|_{RS})\\
&  \leq F(\psi_{S}^{\sigma},\lambda\psi_{S}^{\rho}+\left(  1-\lambda\right)
\psi_{S}^{\tau})\\
&  =F(\lambda\rho+\left(  1-\lambda\right)  \tau,\sigma).
\end{align}
The first step is a rewriting using \eqref{eq-dm:uhlmann-rho} and
\eqref{eq-dm:uhlmann-tau}. The fourth equality is a consequence of
Exercise~\ref{ex-dm:fidelity-reduction-pure-mixed}. The inequality follows
from monotonicity of the fidelity with respect to partial trace
(Lemma~\ref{lemma-dm:fidelity-monotone}).
\end{proof}

\begin{exercise}
\label{ex-dm:Tr-fidelity-from-def}Let $\rho,\sigma\in\mathcal{D}(\mathcal{H}%
)$. Show that we can express the root fidelity as%
\begin{equation}
\sqrt{F}(\rho,\sigma)=\operatorname{Tr}\left\{  \sqrt{\rho^{1/2}\sigma
\rho^{1/2}}\right\}  =\operatorname{Tr}\left\{  \sqrt{\sigma^{1/2}\rho
\sigma^{1/2}}\right\}  ,
\end{equation}
using the definition in \eqref{eq-dm:fidelity-l1}.
\end{exercise}

\begin{exercise}
Let $\rho,\sigma\in\mathcal{D}(\mathcal{H})$. Show that the fidelity is
invariant with respect to an isometry $U\in\mathcal{L}(\mathcal{H}%
,\mathcal{H}^{\prime})$:%
\begin{equation}
F(\rho,\sigma)=F(U\rho U^{\dag},U\sigma U^{\dag}).
\end{equation}

\end{exercise}

\begin{exercise}
\label{ex-dm:mono-fidelity}Let $\rho,\sigma\in\mathcal{D}(\mathcal{H}_{A})$
and let $\mathcal{N}:\mathcal{L}(\mathcal{H}_{A})\rightarrow\mathcal{L}%
(\mathcal{H}_{B})$ be a quantum channel. Show that the fidelity is monotone
with respect to the channel $\mathcal{N}$:%
\begin{equation}
F(\rho,\sigma)\leq F(\mathcal{N}(\rho),\mathcal{N}(\sigma)).
\end{equation}

\end{exercise}

\begin{figure}
[ptb]
\begin{center}
\includegraphics[
width=4.0413in
]%
{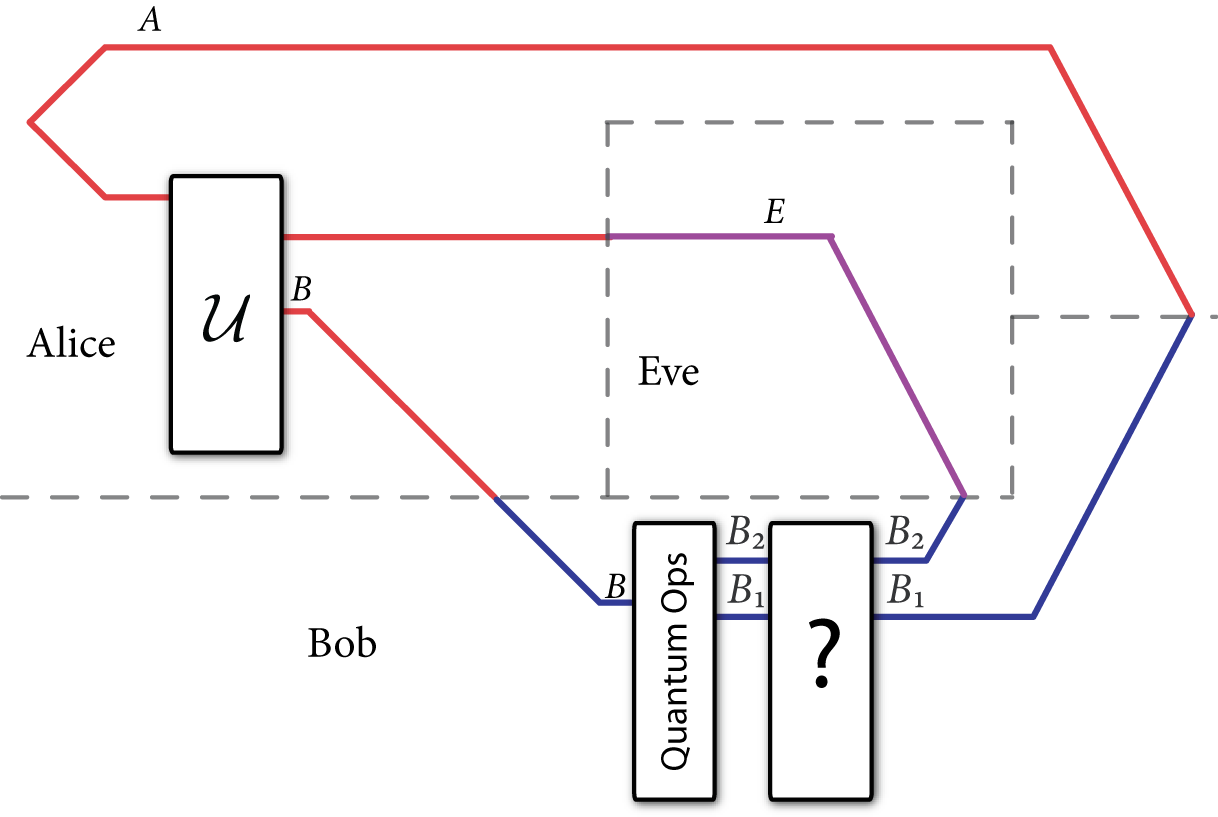}%
\caption{This figure depicts the protocol relevant to
Exercise~\ref{ex-dm:decouple}. Alice transmits one share of an entangled state
through a quantum channel with isometric extension channel $\mathcal{U}$. Bob
and Eve receive quantum systems as the output of the isometry. Bob performs
some quantum operations so that Alice, Bob, and Eve share the state in
\eqref{eq-dm:shared-state}. Exercise~\ref{ex-dm:decouple}\ asks you to
determine a decoupling unitary that Bob can perform to decouple his system
$B_{1}$ from Eve.}%
\label{fig-dm:exercise-decoupling}%
\end{center}
\end{figure}

\begin{exercise}
\label{ex-dm:decouple}Suppose that Alice uses a noisy quantum channel and a
sequence of quantum channels to generate the following state, shared with Bob
and Eve:%
\begin{equation}
\frac{1}{\sqrt{M}}\sum_{m}|m\rangle_{A}\left\vert m\right\rangle _{B_{1}%
}\left\vert \phi_{m}\right\rangle _{B_{2}E}, \label{eq-dm:shared-state}%
\end{equation}
where $\{|m\rangle_{A}\}$ and $\{|m\rangle_{B_{1}}\}$ are orthonormal bases
and $\{|\phi_{m}\rangle_{B_{2}E}\}$ is a set of states. Alice possesses the
system $A$, Bob possesses systems $B_{1}$ and $B_{2}$, and Eve possesses the
system $E$. Let $\phi_{E}^{m}$ denote the partial trace of $\left\vert
\phi_{m}\right\rangle _{B_{2}E}$ over Bob's system $B_{2}$ so that%
\begin{equation}
\phi_{E}^{m}\equiv\operatorname{Tr}_{B_{2}}\left\{  \vert\phi_{m}%
\rangle\langle\phi_{m}\vert_{B_{2}E}\right\}  .
\end{equation}
Suppose further that $F( \phi_{E}^{m},\theta_{E}) =1$, where $\theta_{E}$ is
some \textit{constant} density operator (independent of$~m$) for Eve's system
$E$. Determine a unitary that Bob can perform on his systems $B_{1}$ and
$B_{2}$ so that he \textit{decouples} Eve's system$~E$, in the sense that the
state after the decoupling unitary is as follows:%
\begin{equation}
\left(  \frac{1}{\sqrt{M}}\sum_{m}|m\rangle_{A}\left\vert m\right\rangle
_{B_{1}}\right)  \otimes\left\vert \phi_{\theta}\right\rangle _{B_{2}E},
\end{equation}
where $\left\vert \phi_{\theta}\right\rangle _{B_{2}E}$ is a purification of
the state $\theta_{E}$. The result is that Alice and Bob share maximal
entanglement between the respective systems $A$ and $B_{1}$ after Bob performs
the decoupling unitary. Figure~\ref{fig-dm:exercise-decoupling} displays the protocol.
\end{exercise}

\begin{exercise}
[Fidelity for Classical--Quantum States]\label{ex-dm:fid-cq-states}Show that
the root fidelity possesses the following property:%
\begin{equation}
\sqrt{F}\left(  \omega_{XB},\tau_{XB}\right)  =\sum_{x}\sqrt{p(x) q(x)}%
\sqrt{F}\left(  \omega_{x},\tau_{x}\right)  , \label{eq:fid-flags}%
\end{equation}
where%
\begin{equation}
\omega_{XB}\equiv\sum_{x}p(x)|x\rangle\langle x|_{X}\otimes\omega
_{x},\ \ \ \ \tau_{XB}\equiv\sum_{x}q(x)|x\rangle\langle x|_{X}\otimes\tau
_{x}, \label{eq:flag-states}%
\end{equation}
$p$ and $q$ are probability distributions, $\left\{  |x\rangle\right\}  $ is
some orthonormal basis, and $\omega_{x},\tau_{x}\in\mathcal{D}(\mathcal{H})$
for all~$x$.
\end{exercise}

\subsection{A Measurement Achieves the Fidelity}%

\index{fidelity!achieving measurement}%
There is a classical notion of fidelity for probability distributions, which
is sometimes called the classical fidelity or Bhattacharyya overlap. It is
defined as follows:

\begin{definition}
[Classical Fidelity]Let $p$ and $q$ be probability distributions defined over
a finite alphabet $\mathcal{X}$. The classical fidelity $F(p,q)$\ is defined
as follows:%
\begin{equation}
F(p,q)\equiv\left[  \sum_{x\in\mathcal{X}}\sqrt{p(x)q(x)}\right]  ^{2}.
\label{eq-dm:classical-fidelity}%
\end{equation}

\end{definition}

\begin{exercise}
\label{ex-dm:classical-fid-special-q-fid}Verify that the classical fidelity is
a special case of the quantum fidelity. That is, let $p$ and $q$ be
probability distributions defined over a finite alphabet $\mathcal{X}$, and
then place the entries of these distributions along the diagonal of commuting
matrices $\rho$ and $\sigma$, respectively. Show that $F(p,q)=F(\rho,\sigma)$.
\end{exercise}

Now suppose that we have two density operators $\rho,\sigma\in\mathcal{D}%
(\mathcal{H})$, and suppose further that we perform a POVM $\{\Lambda_{x}%
\}$\ on these states, leading to the following probability distributions:%
\begin{equation}
p(x)=\operatorname{Tr}\{\Lambda_{x}\rho
\},\ \ \ \ \ \ \ \ \ \ q(x)=\operatorname{Tr}\{\Lambda_{x}\sigma\}.
\end{equation}
We can then compute the classical fidelity of the distributions for the
measurement outcomes by using the formula in \eqref{eq-dm:classical-fidelity},
and this is a measure of distinguishability of the two quantum states, with
respect to a particular measurement. From the monotonicity of the quantum
fidelity with respect to quantum channels (Exercise~\ref{ex-dm:mono-fidelity}%
), it follows that the quantum fidelity $F(\rho,\sigma)$ never exceeds this
classical fidelity%
\begin{equation}
F(\rho,\sigma)\leq\left[  \sum_{x\in\mathcal{X}}\sqrt{\operatorname{Tr}%
\{\Lambda_{x}\rho\}\operatorname{Tr}\{\Lambda_{x}\sigma\}}\right]  ^{2}.
\label{eq-dm:fidelity-measurements-bound}%
\end{equation}
In particular, this bound follows from Exercise~\ref{ex-dm:mono-fidelity},
where the channel here is understood to be a measurement channel of the form
$\omega\rightarrow\sum_{x}\operatorname{Tr}\{\Lambda_{x}\omega\}|x\rangle
\langle x|$ and then we apply the result of
Exercise~\ref{ex-dm:classical-fid-special-q-fid}. What is perhaps surprising
is that there always exists a measurement that saturates the bound above,
leading to the following alternate characterization of fidelity:

\begin{theorem}
[Measurement Achieves Fidelity]\label{thm-dm:meas-achieve-fid}Let $\rho
,\sigma\in\mathcal{D}(\mathcal{H})$. Then%
\begin{equation}
F(\rho,\sigma)=\min_{\{\Lambda_{x}\}}\left[  \sum_{x\in\mathcal{X}}%
\sqrt{\operatorname{Tr}\{\Lambda_{x}\rho\}\operatorname{Tr}\{\Lambda_{x}%
\sigma\}}\right]  ^{2}, \label{eq-dm:fidelity-min-meas}%
\end{equation}
where the minimization is with respect to all POVMs.
\end{theorem}

\begin{proof}
As justified before the statement of the theorem, the bound in
\eqref{eq-dm:fidelity-measurements-bound} holds for any POVM. So here we
construct a specific POVM\ (known as the Fuchs-Caves measurement) that
saturates the bound. First consider the case in which $\sigma$ is positive
definite (and thus invertible). Consider the following operator (known as an
operator geometric mean of $\rho$ and $\sigma^{-1}$):%
\begin{equation}
M=\sigma^{-1/2}\left[  \sigma^{1/2}\rho\sigma^{1/2}\right]  ^{1/2}%
\sigma^{-1/2}.
\end{equation}
The operator $M$ is positive semi-definite, and thus has a spectral
decomposition:%
\begin{equation}
M=\sum_{y}\lambda_{y}|y\rangle\langle y|, \label{eq-dm:spec-dec-M-geo-mean}%
\end{equation}
with $\{\lambda_{y}\}$ a set of non-negative eigenvalues and $\{|y\rangle\}$ a
corresponding set of eigenvectors.

We will prove that the optimal measurement in \eqref{eq-dm:fidelity-min-meas}
is $\{|y\rangle\langle y|\}$. We begin by noting that a simple calculation
gives%
\begin{equation}
M\sigma M=\rho. \label{eq-dm:MsM-is-r}%
\end{equation}
So now consider the classical fidelity of the measurement $\{|y\rangle\langle
y|\}$:%
\begin{align}
\sum_{y}\sqrt{\operatorname{Tr}\{|y\rangle\langle y|\rho\}\operatorname{Tr}%
\{|y\rangle\langle y|\sigma\}}  &  =\sum_{y}\sqrt{\langle y|\rho
|y\rangle\ \langle y|\sigma|y\rangle}\\
&  =\sum_{y}\sqrt{\langle y|M\sigma M|y\rangle\ \langle y|\sigma|y\rangle}\\
&  =\sum_{y}\sqrt{\langle y|\lambda_{y}\sigma\lambda_{y}|y\rangle\ \langle
y|\sigma|y\rangle}\\
&  =\sum_{y}\lambda_{y}\langle y|\sigma|y\rangle .
\end{align}
The second equality follows from \eqref{eq-dm:MsM-is-r}. The third equality
follows because $M|y\rangle=\lambda_{y}|y\rangle$. Continuing, the last line
above is equal to%
\begin{equation}
\operatorname{Tr}\left\{  \sum_{y}\lambda_{y}|y\rangle\langle y|\sigma
\right\}  =\operatorname{Tr}\left\{  M\sigma\right\}  =\operatorname{Tr}%
\left\{  \left[  \sigma^{1/2}\rho\sigma^{1/2}\right]  ^{1/2}\right\}
=\sqrt{F}(\rho,\sigma).
\end{equation}
The last equality follows from Exercise~\ref{ex-dm:Tr-fidelity-from-def}.

For the case in which $\sigma$ is not invertible, we repeat the above
analysis, replacing $\rho$ with $\Pi_{\sigma}\rho\Pi_{\sigma}$, where
$\Pi_{\sigma}$ is the projection onto the support of $\sigma$. In this case,
the geometric mean operator $M$ has its support contained in the support of
$\sigma$, and one can find a spectral decomposition of $M$ as in
\eqref{eq-dm:spec-dec-M-geo-mean} so that%
\begin{equation}
\sqrt{F}(\Pi_{\sigma}\rho\Pi_{\sigma},\sigma)=\sum_{y}\sqrt{\operatorname{Tr}%
\{|y\rangle\langle y|\Pi_{\sigma}\rho\Pi_{\sigma}\}\operatorname{Tr}%
\{|y\rangle\langle y|\sigma\}}. \label{eq-dm:fid-meas-non-invertible}%
\end{equation}
Since the eigenvectors $\{|y\rangle\}$ do not necessarily span the whole
space, we can add additional orthonormal vectors all orthogonal to those in
$\{|y\rangle\}$, such that all of them taken together form a legitimate
measurement. Since both $\Pi_{\sigma}\rho\Pi_{\sigma}$ and $\sigma$ are
orthogonal to all of the new vectors, the probabilities for these measurement
outcomes are all equal to zero and thus they do not contribute anything to the
sum in \eqref{eq-dm:fid-meas-non-invertible}. Finally, we have that%
\begin{equation}
F(\Pi_{\sigma}\rho\Pi_{\sigma},\sigma)=F(\rho,\sigma)
\end{equation}
because $\sigma^{1/2}=\Pi_{\sigma}\sigma^{1/2}=\sigma^{1/2}\Pi_{\sigma}$, so
that%
\begin{equation}
\sqrt{F}(\rho,\sigma)=\operatorname{Tr}\left\{  \sqrt{\sigma^{1/2}\rho
\sigma^{1/2}}\right\}  =\operatorname{Tr}\left\{  \sqrt{\sigma^{1/2}%
\Pi_{\sigma}\rho\Pi_{\sigma}\sigma^{1/2}}\right\}  =\sqrt{F}(\Pi_{\sigma}%
\rho\Pi_{\sigma},\sigma),
\end{equation}
concluding the proof.
\end{proof}

\section{Relations between Trace Distance and Fidelity}

\label{sec-dm:fidelity-trace}In quantum Shannon theory, we are interested in
showing that a given quantum information-processing protocol approximates an
ideal protocol. We might do so by showing that the quantum output of the ideal
protocol, say $\rho$, is close to the quantum output of the actual protocol,
say $\sigma$. For example, we may be able to show that the fidelity between
$\rho$ and $\sigma$ is high:%
\begin{equation}
F( \rho,\sigma) \geq1-\varepsilon,
\end{equation}
where $\varepsilon$ is a small, positive real number that determines how well
$\rho$ approximates $\sigma$ according to the above fidelity criterion.
Typically, in a quantum Shannon-theoretic argument, we will take a limit to
show that it is possible to make $\varepsilon$ as small as we would like. As
the performance parameter $\varepsilon$ becomes vanishingly small, we expect
that $\rho$ and $\sigma$ are becoming approximately equal so that they are
identically equal when $\varepsilon$ vanishes in some limit.

We would naturally think that the trace distance should be small if the
fidelity is high because the trace distance vanishes when the fidelity is one
and vice versa (recall the conditions for saturation of the bounds in
\eqref{eq-dm:bounds-trace-dist} and \eqref{eq-dm:bounds-fidelity}). The next
theorem makes this intuition precise by establishing several relationships
between the trace distance and fidelity.

\begin{theorem}
[Relations between Fidelity and Trace Distance]%
\label{thm-dm:fidelity-trace-relation}The following bound applies to the trace
distance and the fidelity between two quantum states $\rho,\sigma
\in\mathcal{D}(\mathcal{H})$:%
\begin{equation}
1-\sqrt{F( \rho,\sigma) }\leq\frac{1}{2}\left\Vert \rho-\sigma\right\Vert
_{1}\leq\sqrt{1-F( \rho,\sigma) }.
\end{equation}

\end{theorem}

\begin{proof}
We first show that there is an exact relationship between fidelity and trace
distance for pure states. Let us pick two arbitrary pure states $|\psi
\rangle,|\phi\rangle\in\mathcal{H}$. We can write the state $|\phi\rangle$ in
terms of the state $|\psi\rangle$ and a vector $|\psi^{\perp}\rangle$
orthogonal to $|\psi\rangle$:%
\begin{equation}
|\phi\rangle=\cos(\theta)|\psi\rangle+\sin(\theta)|\psi^{\perp}\rangle.
\end{equation}
First, the fidelity between these two pure states is%
\begin{equation}
F(\psi,\phi)=\left\vert \left\langle \phi|\psi\right\rangle \right\vert
^{2}=\cos^{2}(\theta). \label{eq-dm:fidelity-pure}%
\end{equation}
Now let us determine the trace distance. The density operator $|\phi
\rangle\langle\phi|$ is as follows:%
\begin{align}
|\phi\rangle\langle\phi|  &  =\left(  \cos(\theta)|\psi\rangle+\sin
(\theta)|\psi^{\perp}\rangle\right)  \left(  \cos(\theta)\langle\psi
|+\sin(\theta)\langle\psi^{\perp}|\right) \\
&  =\cos^{2}(\theta)|\psi\rangle\langle\psi|+\sin(\theta)\cos(\theta
)|\psi^{\perp}\rangle\langle\psi|\nonumber\\
&  \ \ \ \ \ \ +\cos(\theta)\sin(\theta)|\psi\rangle\langle\psi^{\perp}%
|+\sin^{2}(\theta)|\psi^{\perp}\rangle\langle\psi^{\perp}|.
\end{align}
The matrix representation of the operator $|\psi\rangle\langle\psi
|-|\phi\rangle\langle\phi|$ with respect to the basis $\left\{  |\psi
\rangle,|\psi^{\perp}\rangle\right\}  $ is
\begin{equation}%
\begin{bmatrix}
1-\cos^{2}(\theta) & -\sin(\theta)\cos(\theta)\\
-\sin(\theta)\cos(\theta) & -\sin^{2}(\theta)
\end{bmatrix}
.
\end{equation}
It is straightforward to show that the eigenvalues of the above matrix are
$\left\vert \sin(\theta)\right\vert $ and $-\left\vert \sin(\theta)\right\vert
$ and it then follows that the trace distance between $|\psi\rangle$ and
$|\phi\rangle$ is the absolute sum of the eigenvalues:%
\begin{equation}
\left\Vert |\psi\rangle\langle\psi|-|\phi\rangle\langle\phi|\right\Vert
_{1}=2\left\vert \sin(\theta)\right\vert . \label{eq-dm:trace-pure}%
\end{equation}
Consider the following trigonometric relationship:%
\begin{equation}
\left(  \frac{2\left\vert \sin(\theta)\right\vert }{2}\right)  ^{2}=1-\cos
^{2}(\theta). \label{eq-dm:trace-fidelity-pure-relate}%
\end{equation}
Applying it gives the following relation between the fidelity and trace
distance for pure states:%
\begin{equation}
\left(  \frac{1}{2}\left\Vert |\psi\rangle\langle\psi|-|\phi\rangle\langle
\phi|\right\Vert _{1}\right)  ^{2}=1-F(\psi,\phi),
\end{equation}
by plugging \eqref{eq-dm:fidelity-pure} into the right-hand side\ of
\eqref{eq-dm:trace-fidelity-pure-relate} and \eqref{eq-dm:trace-pure} into the
left-hand side\ of \eqref{eq-dm:trace-fidelity-pure-relate}. Thus,%
\begin{equation}
\frac{1}{2}\left\Vert |\psi\rangle\langle\psi|-|\phi\rangle\langle
\phi|\right\Vert _{1}=\sqrt{1-F(\psi,\phi)}.
\end{equation}
To prove the upper bound for mixed states $\rho_{A}$ and $\sigma_{A}$, choose
purifications $|\phi^{\rho}\rangle_{RA}$ and $\left\vert \phi^{\sigma
}\right\rangle _{RA}$ of respective states $\rho_{A}$ and $\sigma_{A}$ such
that%
\begin{equation}
F(\rho_{A},\sigma_{A})=\left\vert \langle\phi^{\sigma}|\phi^{\rho}%
\rangle\right\vert ^{2}=F(\phi_{RA}^{\rho},\phi_{RA}^{\sigma}).
\end{equation}
(Recall that these purifications exist by Uhlmann's theorem.) Then%
\begin{align}
\frac{1}{2}\left\Vert \rho_{A}-\sigma_{A}\right\Vert _{1}  &  \leq\frac{1}%
{2}\left\Vert \phi_{RA}^{\rho}-\phi_{RA}^{\sigma}\right\Vert _{1}\\
&  =\sqrt{1-F(\phi_{RA}^{\rho},\phi_{RA}^{\sigma})}\\
&  =\sqrt{1-F(\rho_{A},\sigma_{A})},
\end{align}
where the first inequality follows by the monotonicity of the trace distance
under the discarding of systems (Lemma~\ref{lemma:trace-dist-monotone}).

To prove the lower bound for mixed states $\rho$ and $\sigma$, recall
Exercise~\ref{thm-dm:meas-achieve-TD}\ and
Theorem~\ref{thm-dm:meas-achieve-fid}. Exercise~\ref{thm-dm:meas-achieve-TD}
states that the trace distance is the maximum classical trace distance between
two probability distributions resulting from a POVM\ $\left\{  \Lambda
_{m}\right\}  $ acting on the states $\rho$ and $\sigma$:%
\begin{equation}
\left\Vert \rho-\sigma\right\Vert _{1}=\max_{\left\{  \Lambda_{m}\right\}
}\sum_{m}\left\vert p_{m}-q_{m}\right\vert ,
\end{equation}
where%
\begin{equation}
p_{m}\equiv\operatorname{Tr}\left\{  \Lambda_{m}\rho\right\}
,\ \ \ \ \ \ q_{m}\equiv\operatorname{Tr}\left\{  \Lambda_{m}\sigma\right\}  .
\end{equation}
Furthermore, Theorem~\ref{thm-dm:meas-achieve-fid} states that the quantum
fidelity is the minimum classical fidelity between two probability
distributions $p_{m}^{\prime}$ and $q_{m}^{\prime}$ resulting from a
measurement $\left\{  \Gamma_{m}\right\}  $ of the states $\rho$ and $\sigma$:%
\begin{equation}
F(\rho,\sigma)=\min_{\left\{  \Gamma_{m}\right\}  }\left(  \sum_{m}\sqrt
{p_{m}^{\prime}q_{m}^{\prime}}\right)  ^{2},
\end{equation}
where%
\begin{equation}
p_{m}^{\prime}\equiv\operatorname{Tr}\left\{  \Gamma_{m}\rho\right\}
,\ \ \ \ \ \ q_{m}^{\prime}\equiv\operatorname{Tr}\left\{  \Gamma_{m}%
\sigma\right\}  .
\end{equation}
We return to the proof. Suppose that the POVM\ $\{\Gamma_{m}\}$ achieves the
minimum classical fidelity and results in probability distributions
$p_{m}^{\prime}$ and $q_{m}^{\prime}$, so that%
\begin{equation}
F(\rho,\sigma)=\left(  \sum_{m}\sqrt{p_{m}^{\prime}q_{m}^{\prime}}\right)
^{2}.
\end{equation}
Consider that%
\begin{align}
\sum_{m}\left(  \sqrt{p_{m}^{\prime}}-\sqrt{q_{m}^{\prime}}\right)  ^{2}  &
=\sum_{m}p_{m}^{\prime}+q_{m}^{\prime}-2\sqrt{p_{m}^{\prime}q_{m}^{\prime}}\\
&  =2-2\sqrt{F(\rho,\sigma)}.
\end{align}
It also follows that%
\begin{align}
\sum_{m}\left(  \sqrt{p_{m}^{\prime}}-\sqrt{q_{m}^{\prime}}\right)  ^{2}  &
\leq\sum_{m}\left\vert \sqrt{p_{m}^{\prime}}-\sqrt{q_{m}^{\prime}}\right\vert
\left\vert \sqrt{p_{m}^{\prime}}+\sqrt{q_{m}^{\prime}}\right\vert \\
&  =\sum_{m}\left\vert p_{m}^{\prime}-q_{m}^{\prime}\right\vert \\
&  \leq\sum_{m}\left\vert p_{m}-q_{m}\right\vert \\
&  =\left\Vert \rho-\sigma\right\Vert _{1}.
\end{align}
The first inequality holds because $\left\vert \sqrt{p_{m}^{\prime}}%
-\sqrt{q_{m}^{\prime}}\right\vert \leq\left\vert \sqrt{p_{m}^{\prime}}%
+\sqrt{q_{m}^{\prime}}\right\vert $. The second inequality holds because the
distributions $p_{m}^{\prime}$ and $q_{m}^{\prime}$ minimizing the classical
fidelity in general have classical trace distance less than the distributions
$p_{m}$ and $q_{m}$ that maximize the classical trace distance. Thus, the
following inequality results%
\begin{equation}
2-2\sqrt{F(\rho,\sigma)}\leq\left\Vert \rho-\sigma\right\Vert _{1},
\end{equation}
and the lower bound in the statement of the theorem follows.
\end{proof}

Theorem~\ref{thm-dm:fidelity-trace-relation} allows us to complete our
understanding of the extreme values of trace distance and fidelity. We have
already argued that two states $\rho,\sigma\in\mathcal{D}(\mathcal{H})$ have
trace distance equal to zero if and only if $\rho= \sigma$.
Theorem~\ref{thm-dm:fidelity-trace-relation} allows us to conclude that
$F(\rho, \sigma) = 1$ if and only if $\rho= \sigma$. Similarly, we have argued
already that $F(\rho, \sigma) = 0$ if and only if the support of $\rho$ is
orthogonal to that of $\sigma$. Theorem~\ref{thm-dm:fidelity-trace-relation}
allows us to conclude that $\left\Vert \rho- \sigma\right\Vert _{1} = 2$ if
and only if the support of $\rho$ is orthogonal to that of $\sigma$.

The following two corollaries are simple consequences of
Theorem~\ref{thm-dm:fidelity-trace-relation}.

\begin{corollary}
\label{cor-dm:trace-imp-fid}Let $\rho,\sigma\in\mathcal{D}(\mathcal{H})$ and
fix $\varepsilon\in\lbrack0,1]$. Suppose that $\rho$ is $\varepsilon$-close to
$\sigma$ in trace distance:%
\begin{equation}
\left\Vert \rho-\sigma\right\Vert _{1}\leq\varepsilon.
\end{equation}
Then the fidelity between $\rho$ and $\sigma$ is greater than $1-\varepsilon$:%
\begin{equation}
F(\rho,\sigma)\geq1-\varepsilon.
\end{equation}

\end{corollary}

\begin{corollary}
\label{cor-dm:fid-imp-trace} Let $\rho,\sigma\in\mathcal{D}(\mathcal{H})$ and
fix $\varepsilon\in[0,1]$. Suppose the fidelity between $\rho$ and $\sigma$ is
greater than $1-\varepsilon$:%
\begin{equation}
F( \rho,\sigma) \geq1-\varepsilon.
\end{equation}
Then $\rho$ is $2\sqrt{\varepsilon}$-close to $\sigma$ in trace distance:%
\begin{equation}
\left\Vert \rho-\sigma\right\Vert _{1}\leq2\sqrt{\varepsilon}.
\end{equation}

\end{corollary}

\begin{exercise}
Let $\rho,\sigma\in\mathcal{D}(\mathcal{H})$. Prove the following lower bound
on the probability of error $p_{e}$\ in a quantum hypothesis test to
distinguish $\rho$ from $\sigma$:%
\begin{equation}
p_{e}\geq\frac{1}{2}\left(  1-\sqrt{1-F( \rho,\sigma) }\right)  .
\end{equation}
(Hint: Recall the development in Section~\ref{sec-dm:op-int-trace}.)
\end{exercise}

\section{Gentle Measurement}

The gentle measurement%
\index{gentle measurement}
and gentle operator lemmas are particular applications of
Theorem~\ref{thm-dm:fidelity-trace-relation}, and they concern the disturbance
of quantum states. We generally expect in quantum theory that certain
measurements might disturb the state which we are measuring. For example,
suppose a qubit is in the state $\vert0\rangle$. A measurement along the $X$
direction gives $+1$ and $-1$ with equal probability while drastically
disturbing the state to become either $\vert+\rangle$ or $\left\vert
-\right\rangle $, respectively. On the other hand, we might expect that the
measurement does not disturb the state by very much if one outcome is highly
likely. For example, suppose that we instead measure the qubit along the $Z$
direction. The measurement returns$~+1$ with unit probability while causing no
disturbance to the qubit. The \textquotedblleft gentle measurement
lemma\textquotedblright\ below quantitatively addresses the disturbance of
quantum states by demonstrating that a measurement with one outcome that is
highly likely causes only a little disturbance to the quantum state that we
measure (hence, the measurement is \textquotedblleft gentle\textquotedblright%
\ or \textquotedblleft tender\textquotedblright).

\begin{lemma}
[Gentle Measurement]\label{lem-dm:gentle-measurement}Consider a density
operator $\rho$ and a measurement\ operator$~\Lambda$ where $0\leq\Lambda\leq
I$. The measurement operator could be an element of a POVM. Suppose that the
measurement operator $\Lambda$\ has a high probability of detecting state
$\rho$:%
\begin{equation}
\operatorname{Tr}\left\{  \Lambda\rho\right\}  \geq1-\varepsilon,
\end{equation}
where $\varepsilon\in[0,1]$ (the probability of detection is high if
$\varepsilon$ is close to zero). Then the post-measurement state%
\begin{equation}
\rho^{\prime}\equiv\frac{\sqrt{\Lambda}\rho\sqrt{\Lambda}}{\operatorname{Tr}%
\left\{  \Lambda\rho\right\}  }%
\end{equation}
\ is $2\sqrt{\varepsilon}$-close to the original state $\rho$ in trace
distance:%
\begin{equation}
\left\Vert \rho-\rho^{\prime}\right\Vert _{1}\leq2\sqrt{\varepsilon}.
\end{equation}
Thus, the measurement does not disturb the state $\rho$ by much if
$\varepsilon$ is small.
\end{lemma}

\begin{proof}
Suppose first that $\rho$ is a pure state $|\psi\rangle\langle\psi|$. The
post-measurement state is then%
\begin{equation}
\frac{\sqrt{\Lambda}|\psi\rangle\langle\psi|\sqrt{\Lambda}}{\langle
\psi|\Lambda|\psi\rangle}.
\end{equation}
The fidelity between the original state $|\psi\rangle$ and the
post-measurement state above is as follows:%
\begin{align}
\langle\psi|\left(  \frac{\sqrt{\Lambda}|\psi\rangle\langle\psi|\sqrt{\Lambda
}}{\langle\psi|\Lambda|\psi\rangle}\right)  |\psi\rangle &  =\frac{\left\vert
\langle\psi|\sqrt{\Lambda}|\psi\rangle\right\vert ^{2}}{\langle\psi
|\Lambda|\psi\rangle} \geq\frac{\left\vert \langle\psi|\Lambda|\psi
\rangle\right\vert ^{2}}{\langle\psi|\Lambda|\psi\rangle}\\
&  =\langle\psi|\Lambda|\psi\rangle\geq1-\varepsilon.
\end{align}
The first inequality follows because $\sqrt{\Lambda}\geq\Lambda$ when
$\Lambda\leq I$. The second inequality follows from the hypothesis of the
lemma. Now let us consider when we have mixed states $\rho_{A}$ and $\rho
_{A}^{\prime}$. Suppose $|\psi\rangle_{RA}$ and $\left\vert \psi^{\prime
}\right\rangle _{RA}$ are respective purifications of $\rho_{A}$ and $\rho
_{A}^{\prime}$, where%
\begin{equation}
\left\vert \psi^{\prime}\right\rangle _{RA}\equiv\frac{I_{R}\otimes
\sqrt{\Lambda_{A}}|\psi\rangle_{RA}}{\sqrt{\langle\psi|I_{R}\otimes\Lambda
_{A}|\psi\rangle_{RA}}}.
\end{equation}
Then we can apply monotonicity of fidelity
(Lemma~\ref{lemma-dm:fidelity-monotone}) and the above result for pure states
to show that%
\begin{equation}
F(\rho_{A},\rho_{A}^{\prime})\geq F(\psi_{RA},\psi_{RA}^{\prime}%
)\geq1-\varepsilon.
\end{equation}
We obtain the bound on the trace distance $\left\Vert \rho_{A}-\rho
_{A}^{\prime}\right\Vert _{1}$ by exploiting
Corollary~\ref{cor-dm:fid-imp-trace}.
\end{proof}

The following is a variation on the gentle measurement lemma:

\begin{lemma}
[Gentle Operator]\label{lem-dm:gentle-operator}Consider a density operator
$\rho$ and a measurement\ operator $\Lambda$ where $0\leq\Lambda\leq I$. The
measurement operator could be an element of a POVM. Suppose that the
measurement operator $\Lambda$\ has a high probability of detecting state
$\rho$:%
\begin{equation}
\operatorname{Tr}\left\{  \Lambda\rho\right\}  \geq1-\varepsilon,
\label{eq-dm:gm-condition}%
\end{equation}
where $\varepsilon\in[0,1]$ (the probability is high if $\varepsilon$ is close
to zero). Then $\sqrt{\Lambda}\rho\sqrt{\Lambda}$ is $2\sqrt{\varepsilon}%
$-close to the original state $\rho$ in trace distance:%
\begin{equation}
\left\Vert \rho-\sqrt{\Lambda}\rho\sqrt{\Lambda}\right\Vert _{1}\leq
2\sqrt{\varepsilon}.
\end{equation}

\end{lemma}

\begin{proof}
Consider the following chain of inequalities:%
\begin{align}
&  \!\!\!\!\left\Vert \rho-\sqrt{\Lambda}\rho\sqrt{\Lambda}\right\Vert
_{1}\nonumber\\
&  =\left\Vert \left(  I-\sqrt{\Lambda}+\sqrt{\Lambda}\right)  \rho
-\sqrt{\Lambda}\rho\sqrt{\Lambda}\right\Vert _{1}\\
&  \leq\left\Vert \left(  I-\sqrt{\Lambda}\right)  \rho\right\Vert
_{1}+\left\Vert \sqrt{\Lambda}\rho\left(  I-\sqrt{\Lambda}\right)  \right\Vert
_{1}\\
&  =\operatorname{Tr}\left\vert \left(  I-\sqrt{\Lambda}\right)  \sqrt{\rho
}\cdot\sqrt{\rho}\right\vert +\operatorname{Tr}\left\vert \sqrt{\Lambda}%
\sqrt{\rho}\cdot\sqrt{\rho}\left(  I-\sqrt{\Lambda}\right)  \right\vert \\
&  \leq\sqrt{\operatorname{Tr}\left\{  \left(  I-\sqrt{\Lambda}\right)
^{2}\rho\right\}  \operatorname{Tr}\left\{  \rho\right\}  }+\sqrt
{\operatorname{Tr}\left\{  \Lambda\rho\right\}  \operatorname{Tr}\left\{
\rho\left(  I-\sqrt{\Lambda}\right)  ^{2}\right\}  }\\
&  \leq\sqrt{\operatorname{Tr}\left\{  \left(  I-\Lambda\right)  \rho\right\}
}+\sqrt{\operatorname{Tr}\left\{  \rho\left(  I-\Lambda\right)  \right\}  }\\
&  =2\sqrt{\operatorname{Tr}\left\{  \left(  I-\Lambda\right)  \rho\right\}  }
\leq2\sqrt{\varepsilon}.
\end{align}
The first inequality is a consequence of the triangle inequality. The second
equality follows from the definition of the trace norm and the fact that
$\rho$ is a positive semi-definite operator. The second inequality follows
from the Cauchy--Schwarz inequality for the Hilbert--Schmidt inner product
(see \eqref{eq:Cauchy--Schwarz-Hilbert-Schmidt}). The third inequality follows
because $\left(  1-\sqrt{x}\right)  ^{2}\leq1-x$ for $0\leq x\leq1$,
$\operatorname{Tr}\left\{  \rho\right\}  =1$, and $\operatorname{Tr}\left\{
\Lambda\rho\right\}  \leq1$. The final inequality follows from applying
\eqref{eq-dm:gm-condition} and because the square root function is monotone increasing.
\end{proof}

\begin{exercise}
Show that the gentle operator lemma holds for subnormalized positive
semi-definite operators $\rho$ (operators $\rho$ such that $\operatorname{Tr}%
\left\{  \rho\right\}  \leq1$).
\end{exercise}

Below is another variation on the gentle measurement lemma that applies to
ensembles of
\index{gentle measurement!for ensembles}%
quantum states.

\begin{lemma}
[Gentle Measurement for Ensembles]\label{lemma:GM-ensemble}Let $\left\{
p_{X}(x),\rho_{x}\right\}  $ be an ensemble with average density operator
$\overline{\rho}\equiv\sum_{x}p_{X}(x)\rho_{x}$. Given a positive
semi-definite operator $\Lambda$ with $\Lambda\leq I$ and $\operatorname{Tr}%
\left\{  \overline{\rho}\Lambda\right\}  \geq1-\varepsilon$ where
$\varepsilon\in[0,1]$, then%
\begin{equation}
\sum_{x}p_{X}(x)\left\Vert \rho_{x}-\sqrt{\Lambda}\rho_{x}\sqrt{\Lambda
}\right\Vert _{1}\leq2\sqrt{\varepsilon}.
\end{equation}

\end{lemma}

\begin{proof}
We can apply the same steps in the proof of the gentle operator lemma to get
the following inequality, holding for all $x$:%
\begin{equation}
\left\Vert \rho_{x}-\sqrt{\Lambda}\rho_{x}\sqrt{\Lambda}\right\Vert _{1}%
^{2}\leq4\left(  1-\operatorname{Tr}\left\{  \Lambda\rho_{x}\right\}  \right)
.
\end{equation}
Taking the expectation over both sides produces the following inequality:%
\begin{equation}
\sum_{x}p_{X}(x)\left\Vert \rho_{x}-\sqrt{\Lambda}\rho_{x}\sqrt{\Lambda
}\right\Vert _{1}^{2} \leq4\left(  1-\operatorname{Tr}\left\{  \Lambda
\rho\right\}  \right)  \leq4\varepsilon.
\end{equation}
Taking the square root of the above inequality gives the following one:%
\begin{equation}
\sqrt{\sum_{x}p_{X}(x)\left\Vert \rho_{x}-\sqrt{\Lambda}\rho_{x}\sqrt{\Lambda
}\right\Vert _{1}^{2}}\leq2\sqrt{\varepsilon}.
\end{equation}
Concavity of the square root then implies that
\begin{equation}
\sum_{x}p_{X}(x)\sqrt{\left\Vert \rho_{x}-\sqrt{\Lambda}\rho_{x}\sqrt{\Lambda
}\right\Vert _{1}^{2}}\leq2\sqrt{\varepsilon},
\end{equation}
concluding the proof.
\end{proof}

\begin{exercise}
[Coherent Gentle Measurement]Let $\left\{  \rho_{A}^{k}\right\}  $ be a
collection of density
\index{gentle measurement!coherent}%
operators and $\left\{  \Lambda_{A}^{k}\right\}  $ be a POVM\ such that for
all $k$:%
\begin{equation}
\operatorname{Tr}\left\{  \Lambda_{A}^{k}\rho_{A}^{k}\right\}  \geq
1-\varepsilon.
\end{equation}
Let $\left\vert \phi^{k}\right\rangle _{RA}$ be a purification of $\rho
_{A}^{k}$. Show that there exists a coherent gentle measurement $\mathcal{D}%
_{A\rightarrow AK}$ in the sense of Section~\ref{sec-pt:coherent-measurement}%
\ such that%
\begin{equation}
\left\Vert \mathcal{D}_{A\rightarrow AK}( \phi_{RA}^{k}) -\phi_{RA}^{k}%
\otimes|k\rangle\langle k|_{K}\right\Vert _{1}\leq2\sqrt{\varepsilon
(2-\varepsilon)}.
\end{equation}
(Hint:\ Use the result of Exercise~\ref{ex-pt:coherent-measurement}.)
\end{exercise}

\section{Fidelity of a Quantum Channel}

It is useful to have measures that determine how well a quantum channel
$\mathcal{N}$ preserves quantum information. We developed static distance
measures, such as the trace distance and the fidelity, in the previous
sections of this chapter. We would now like to exploit those measures in order
to define dynamic measures.

A \textquotedblleft first guess\textquotedblright\ measure of this sort is the
minimum fidelity $F_{\min}( \mathcal{N}) $, where%
\begin{equation}
F_{\min}( \mathcal{N}) \equiv\min_{\vert\psi\rangle}F( \psi,\mathcal{N}( \psi)
) .
\end{equation}
This measure seems like it may be a good one because we generally do not know
the state that Alice inputs to a noisy channel before transmitting to Bob.

It may seem somewhat strange that we chose to minimize over pure states in the
definition of the minimum fidelity. Are not mixed states the most general
states that occur in the quantum theory? It turns out that joint concavity of
the root fidelity (Property~\ref{prop-dm:joint-concavity}) and monotonicity of
the square function implies that we do not have to consider mixed states for
the minimum fidelity. Consider the following sequence of inequalities:%
\begin{align}
\sqrt{F}\left(  \rho,\mathcal{N}( \rho) \right)   &  =\sqrt{F}\left(  \sum
_{x}p_{X}( x) \vert x\rangle\langle x\vert,\mathcal{N}\left(  \sum_{x}p_{X}(
x) \vert x\rangle\langle x\vert\right)  \right) \\
&  =\sqrt{F}\left(  \sum_{x}p_{X}( x) \vert x\rangle\langle x\vert,\sum
_{x}p_{X}( x) \mathcal{N}( \vert x\rangle\langle x\vert) \right) \\
&  \geq\sum_{x}p_{X}( x) \sqrt{F}( \vert x\rangle\langle x\vert,\mathcal{N}(
\vert x\rangle\langle x\vert) )\\
&  \geq\sqrt{F}( \vert x_{\min}\rangle\langle x_{\min}\vert,\mathcal{N}( \vert
x_{\min}\rangle\langle x_{\min}\vert) ) .
\end{align}
The first equality follows by expanding the density operator $\rho$ with a
spectral decomposition. The second equality follows from linearity of the
quantum operation $\mathcal{N}$. The first inequality follows from joint
concavity of the root fidelity (Property~\ref{prop-dm:joint-concavity}), and
the last inequality follows because there exists some pure state $\left\vert
x_{\min}\right\rangle $ (one of the eigenstates of $\rho$) with fidelity never
larger than the expected fidelity in the previous line.

\subsection{Expected Fidelity of a Quantum Channel}

In general, the minimum fidelity is less useful than other measures of quantum
information preservation over a quantum channel. The difficulty with the
minimum fidelity is that it requires an optimization over the potentially
large space of input states. Since it could be somewhat difficult to
manipulate and compute in general, we introduce other ways to determine the
performance of a quantum channel.

We can simplify our notion of fidelity by instead restricting the states that
Alice sends and averaging the fidelity with respect to this set of states.
That is, suppose that Alice is transmitting states from an ensemble $\left\{
p_{X}( x) ,\rho_{x}\right\}  $ and we would like to determine how well a
quantum channel $\mathcal{N}$\ is preserving this source of quantum
information. Sending a particular state $\rho_{x}$ through a quantum channel
$\mathcal{N}$ produces the state $\mathcal{N}( \rho_{x}) $. The fidelity
between the transmitted state $\rho_{x}$ and the received state $\mathcal{N}(
\rho_{x}) $ is $F( \rho_{x},\mathcal{N}( \rho_{x}) ) $ as defined before. We
define the \textit{expected fidelity} of the ensemble as follows:%
\begin{equation}
\overline{F}( \mathcal{N}) \equiv\mathbb{E}_{X}\left[  F( \rho_{X}%
,\mathcal{N}( \rho_{X}) ) \right]  =\sum_{x}p_{X}( x) F( \rho_{x},\mathcal{N}(
\rho_{x}) ) .\label{eq-dm:expected-fidelity-1}%
\end{equation}
The expected fidelity indicates how well Alice is able to transmit the
ensemble on average to Bob. It again lies between zero and one, just as the
usual fidelity does.

A more general form of the expected fidelity is to consider the expected
performance for any quantum state instead of restricting ourselves to an
ensemble. That is, let us fix some quantum state $\vert\psi\rangle$ and apply
a random unitary $U$ to it, where we select the unitary according to the Haar
measure (this is the uniform distribution on unitaries). The state $U\vert
\psi\rangle$ represents a random quantum state and we can take the expectation
with respect to it in order to define the following more general notion of
expected fidelity:%
\begin{equation}
\overline{F}( \mathcal{N}) \equiv\mathbb{E}_{U}\left[  F( U\vert\psi
\rangle\langle\psi\vert U^{\dag},\mathcal{N}( U\vert\psi\rangle\langle
\psi\vert U^{\dag}) ) \right]  . \label{eq-dm:haar-expected-fidelity}%
\end{equation}
The above formula for the expected fidelity then becomes the following
integral over the Haar measure:%
\begin{equation}
\overline{F}( \mathcal{N}) =\int\langle\psi\vert U^{\dag}\mathcal{N}(
U\vert\psi\rangle\langle\psi\vert U^{\dag}) U\vert\psi\rangle\ dU.
\end{equation}

\subsection{Entanglement Fidelity}%

\begin{figure}
[ptb]
\begin{center}
\includegraphics[
width=4.8456in
]%
{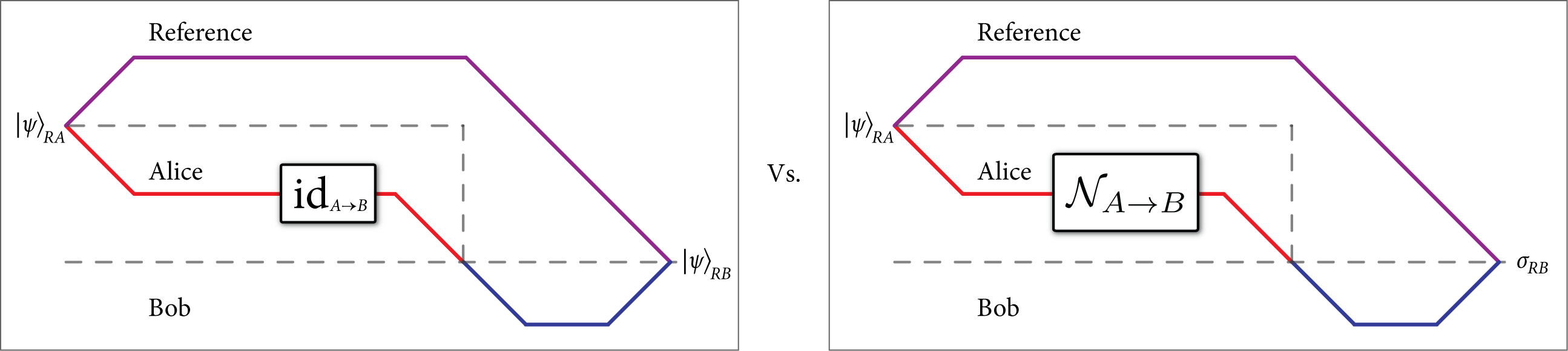}%
\caption{The entanglement fidelity compares the output of the ideal scenario
(depicted on the left) and the output of the noisy scenario (depicted on the
right).}%
\label{fig-dm:ent-fid}%
\end{center}
\end{figure}
We now consider a different measure of the ability of a quantum channel to
preserve quantum
\index{fidelity!entanglement fidelity}%
information. Suppose that Alice would like to transmit a quantum state with
density operator $\rho_{A}$. It admits a purification $\vert\psi\rangle_{RA}$
to a reference system $R$. Sending the $A$ system of $\vert\psi\rangle_{RA}$
through the identity channel id$_{A}$ gives back $\vert\psi\rangle_{RA}$.
Sending the $A$ system of $\vert\psi\rangle_{RA}$ through a quantum channel
$\mathcal{N}:\mathcal{L}(\mathcal{H}_{A}) \to\mathcal{L}(\mathcal{H}_{A})$
gives the state $\sigma_{RA}\equiv\left(  \operatorname{id}_{R}\otimes
\mathcal{N}_{A}\right)  ( \psi_{RA}) $. The entanglement fidelity is defined
as follows:

\begin{definition}
[Entanglement Fidelity]For $\rho$, $\mathcal{N}$, $\sigma$, and $|\psi\rangle$
as defined above, the entanglement fidelity is given by $F_{e}(\rho
,\mathcal{N})\equiv\langle\psi|\sigma|\psi\rangle$.
\end{definition}

It is a measure of how well the quantum channel $\mathcal{N}$ preserves
entanglement with another system. Figure~\ref{fig-dm:ent-fid} visually depicts
the two states that the entanglement fidelity compares.

The entanglement fidelity is invariant with respect to which purification of
the input that we pick. This follows simply because all purifications are
related by an isometry acting on the purifying system. That is, let
$|\psi\rangle_{R_{1}A}$ be one purification of $\rho_{A}$, let $|\varphi
\rangle_{R_{2}A}$ be a different one, and let $U_{R_{1}\rightarrow R_{2}}$ be
an isometry that relates them via $|\varphi\rangle_{R_{2}A}=U_{R_{1}%
\rightarrow R_{2}}|\psi\rangle_{R_{1}A}$. Then%
\begin{align}
&  \langle\varphi|_{R_{2}A}\left(  \operatorname{id}_{R_{2}}\otimes
\mathcal{N}_{A}\right)  \left(  \varphi_{R_{2}A}\right)  |\varphi
\rangle_{R_{2}A}\nonumber\\
&  =\langle\psi|_{R_{1}A}U_{R_{1}\rightarrow R_{2}}^{\dag}\left[  \left(
\operatorname{id}_{R_{2}}\otimes\mathcal{N}_{A}\right)  \left(  U_{R_{1}%
\rightarrow R_{2}}\psi_{R_{1}A}U_{R_{1}\rightarrow R_{2}}^{\dag}\right)
\right]  U_{R_{1}\rightarrow R_{2}}|\psi\rangle_{R_{1}A}\\
&  =\langle\psi|_{R_{1}A}U_{R_{1}\rightarrow R_{2}}^{\dag}U_{R_{1}\rightarrow
R_{2}}\left[  \left(  \operatorname{id}_{R_{1}}\otimes\mathcal{N}_{A}\right)
(\psi_{R_{1}A})\right]  U_{R_{1}\rightarrow R_{2}}^{\dag}U_{R_{1}\rightarrow
R_{2}}|\psi\rangle_{R_{1}A}\\
&  =\langle\psi|_{R_{1}A}\left[  \left(  \operatorname{id}_{R_{1}}%
\otimes\mathcal{N}_{A}\right)  (\psi_{R_{1}A})\right]  |\psi\rangle_{R_{1}A},
\end{align}
where the second equality follows because the isometry commutes with the
identity map id$_{R_{2}}$ and the last follows because $U_{R_{1}\rightarrow
R_{2}}$ is an isometry so that $U_{R_{1}\rightarrow R_{2}}^{\dag}%
U_{R_{1}\rightarrow R_{2}}=I_{R_{1}}$.

One of the benefits of considering the task of entanglement preservation is
that it implies the task of quantum communication. That is, if Alice can
devise a protocol that preserves the entanglement with another system, then
this same protocol will also be able to preserve quantum information that she transmits.

The following theorem gives a simple way to represent the entanglement
fidelity in terms of the Kraus operators of a given noisy quantum channel.

\begin{theorem}
\label{thm-dm:compact-e-fidelity} Let $\rho_{A} \in\mathcal{D}(\mathcal{H}%
_{A})$ and let $\mathcal{N}:\mathcal{L}(\mathcal{H}_{A}) \to\mathcal{L}%
(\mathcal{H}_{A})$ be a quantum channel. Suppose that $\left\{  K^{m}\right\}
$ is a set of Kraus operators for $\mathcal{N}$. Then the entanglement
fidelity $F_{e}( \rho,\mathcal{N}) $\ is equal to the following expression:%
\begin{equation}
F_{e}( \rho,\mathcal{N}) =\sum_{m}\left\vert \operatorname{Tr}\left\{
\rho_{A}K^{m}\right\}  \right\vert ^{2}.
\end{equation}

\end{theorem}

\begin{proof}
Given that the entanglement fidelity is invariant with respect to the choice
of purification, we can simply use the canonical purification $|\psi
\rangle_{RA}$ of $\rho_{A}$, i.e.,%
\begin{equation}
|\psi\rangle_{RA}=\left(  I_{R}\otimes\sqrt{\rho_{A}}\right)  \left\vert
\Gamma\right\rangle _{RA},
\end{equation}
where $\vert\Gamma\rangle_{RA}$ is the unnormalized maximally entangled vector
from \eqref{eq-qt:unnorm-max-ent}. We then find that%
\begin{align}
&  \!\!\!\!\!\langle\psi|_{RA}\left(  \operatorname{id}_{R}\otimes
\mathcal{N}_{A}\right)  (\psi_{RA})|\psi\rangle_{RA}\nonumber\\
&  =\langle\psi|_{RA}\sum_{m}\left(  I_{R}\otimes K_{A}^{m}\right)
|\psi\rangle\langle\psi|_{RA}\left(  I_{R}\otimes\left(  K_{A}^{m}\right)
^{\dag}\right)  |\psi\rangle_{RA}\\
&  =\sum_{m}\langle\psi|_{RA}\left(  I_{R}\otimes K_{A}^{m}\right)
|\psi\rangle\langle\psi|_{RA}\left(  I_{R}\otimes\left(  K_{A}^{m}\right)
^{\dag}\right)  |\psi\rangle_{RA}\\
&  =\sum_{m}\left\vert \langle\psi|_{RA}\left(  I_{R}\otimes K_{A}^{m}\right)
|\psi\rangle_{RA}\right\vert ^{2}.
\end{align}
Then consider that%
\begin{align}
&  \!\!\!\!\!\langle\psi|_{RA}\left(  I_{R}\otimes K_{A}^{m}\right)
|\psi\rangle_{RA}\nonumber\\
&  =\langle\Gamma\vert_{RA}\left(  I_{R}\otimes\sqrt{\rho_{A}}\right)  \left(
I_{R}\otimes K_{A}^{m}\right)  \left(  I_{R}\otimes\sqrt{\rho_{A}}\right)
\vert\Gamma\rangle_{RA}\\
&  =\langle\Gamma\vert_{RA}\left(  I_{R}\otimes\sqrt{\rho_{A}}K_{A}^{m}%
\sqrt{\rho_{A}}\right)  \vert\Gamma\rangle_{RA}\\
&  =\operatorname{Tr}\left\{  \sqrt{\rho_{A}}K_{A}^{m}\sqrt{\rho_{A}}\right\}
\\
&  =\operatorname{Tr}\left\{  \rho_{A}K_{A}^{m}\right\},
\end{align}
where we have used Exercise~\ref{ex-qt:alt-trace-max-ent}\ to establish the
third equality. So we find that%
\begin{equation}
\langle\psi|_{RA}\left(  \operatorname{id}_{R}\otimes\mathcal{N}_{A}\right)
(\psi_{RA})|\psi\rangle_{RA}=\sum_{m}\left\vert \operatorname{Tr}\left\{
\rho_{A}K^{m}\right\}  \right\vert ^{2},
\end{equation}
concluding the proof.
\end{proof}

\begin{exercise}
\label{ex-dm:convex-e-fidelity} Let $\rho_{1},\rho_{2} \in\mathcal{D}%
(\mathcal{H})$ and let $\mathcal{N}:\mathcal{L}(\mathcal{H}) \to
\mathcal{L}(\mathcal{H})$ be a quantum channel. Fix $\lambda\in[0,1]$. Show
that the entanglement fidelity is convex in the input state:%
\begin{equation}
F_{e}( \lambda\rho_{1}+\left(  1-\lambda\right)  \rho_{2},\mathcal{N})
\leq\lambda F_{e}( \rho_{1},\mathcal{N}) +\left(  1-\lambda\right)  F_{e}(
\rho_{2},\mathcal{N}) .
\end{equation}
(Hint: The result of Theorem~\ref{thm-dm:compact-e-fidelity} is useful here.)
\end{exercise}

\begin{exercise}
Prove that the entanglement fidelity does not depend upon the particular
choice of Kraus operators for a given channel. (Hint:\ Recall that there
always exists an isometry that relates two different Kraus representations of
a quantum channel, i.e., for a set $\left\{  K^{m}\right\}  $ of Kraus
operators and another set $\left\{  L^{n}\right\}  $, we have that%
\begin{equation}
K^{m}=\sum_{n}u_{mn}L^{n},
\end{equation}
where $u_{mn}$ are the entries of a unitary matrix.)
\end{exercise}

\subsection{Expected Fidelity and Entanglement Fidelity}

The entanglement fidelity and the expected fidelity provide seemingly
different methods for quantifying the ability of a noisy quantum channel to
preserve quantum information. Is there any way that we can show how they are related?

It turns out that they are indeed related. First, consider that the
entanglement fidelity is a lower bound on the channel's fidelity for
preserving the state $\rho$:%
\begin{equation}
F_{e}( \rho,\mathcal{N}) \leq F( \rho,\mathcal{N}( \rho) ) .
\label{eq-dm:e-fidelity-bound}%
\end{equation}
The above result follows simply from the monotonicity of fidelity under
partial trace (Lemma~\ref{lemma-dm:fidelity-monotone}). We can show that the
entanglement fidelity is always less than the expected fidelity in
\eqref{eq-dm:expected-fidelity-1} by combining convexity of entanglement
fidelity (Exercise~\ref{ex-dm:convex-e-fidelity}) and the bound in
\eqref{eq-dm:e-fidelity-bound}:%
\begin{align}
F_{e}\left(  \sum_{x}p_{X}( x) \rho_{x},\mathcal{N}\right)   &  \leq\sum
_{x}p_{X}( x) F_{e}( \rho_{x},\mathcal{N})\\
&  \leq\sum_{x}p_{X}( x) F( \rho_{x},\mathcal{N}( \rho_{x}) )\\
&  =\overline{F}( \mathcal{N}) .
\end{align}
Thus, any channel that preserves entanglement with some reference system
preserves the expected fidelity of an ensemble. In most cases, we only
consider the entanglement fidelity as the defining measure of performance of a
noisy quantum channel.

The relationship between entanglement fidelity and expected fidelity becomes
more exact (and more beautiful) in the case where we select a random quantum
state according to the Haar measure. It is possible to show that the expected
fidelity in \eqref{eq-dm:haar-expected-fidelity} relates to the entanglement
fidelity as follows:%
\begin{equation}
\overline{F}( \mathcal{N}) =\frac{dF_{e}( \pi,\mathcal{N}) +1}{d+1},
\label{eq-dm:relation-exp-ent-fid}%
\end{equation}
where $d$ is the dimension of the input system and $\pi$ is the maximally
mixed state with purification to the maximally entangled state.

\begin{exercise}
Prove that the relation in \eqref{eq-dm:relation-exp-ent-fid} holds for a
quantum depolarizing channel.
\end{exercise}

\section{The Hilbert--Schmidt Distance Measure}

\label{sec-dm:Hilbert-Schmidt-dist}One final distance measure that we develop
is the Hilbert--Schmidt distance%
\index{Hilbert--Schmidt distance}
measure. It is most similar to the familiar Euclidean distance measure of
vectors because an $\ell_{2}$-norm induces it. This distance measure does not
have an appealing operational interpretation like the trace distance and
fidelity do. Furthermore, it is neither generally increasing or decreasing
with respect to the action of quantum channels, and so one \textit{should not}
employ it as a distinguishability measure of quantum states. Nevertheless, it
can sometimes be helpful in calculations to exploit this distance measure and
to relate it to the trace distance via the bound in
Exercise~\ref{ex-dm:l2-vs-l1} below.

Let us define the Hilbert--Schmidt norm of an operator $M\in\mathcal{L}%
(\mathcal{H},\mathcal{H}^{\prime})$ as follows:%
\begin{equation}
\left\Vert M\right\Vert _{2}\equiv\sqrt{\operatorname{Tr}\left\{  M^{\dag
}M\right\}  }.
\end{equation}
It is straightforward to show that the above norm meets the three requirements
of a norm: non-negativity, homogeneity, and the triangle inequality. One can
compute this norm simply by summing the squares of the singular values of the
operator $M$ and taking the square root. The reasoning for this is the same as
that in the proof of Proposition~\ref{prop-dm:trace-norm-sing-values}.

The Hilbert--Schmidt norm induces the following Hilbert--Schmidt distance
measure:%
\begin{equation}
\left\Vert M-N\right\Vert _{2},
\end{equation}
where $M,N\in\mathcal{L}(\mathcal{H},\mathcal{H}^{\prime})$. We can then apply
this distance measure to quantum states $\rho$ and $\sigma$ simply by plugging
$\rho$ and $\sigma$ into the above formula in place of $M$ and $N$.

The Hilbert--Schmidt distance measure sometimes finds use in the proofs of
coding theorems in quantum Shannon theory because it is often easier to find
good bounds on it rather than on the trace distance. In some cases, we might
be taking expectations over ensembles of density operators and this
expectation often reduces to computing variances or covariances.

\begin{exercise}
\label{ex-dm:l2-vs-l1}Show that the following inequality holds for any
operator $X$%
\begin{equation}
\left\Vert X\right\Vert _{1}^{2}\leq d\left\Vert X\right\Vert _{2}^{2},
\label{eq-dm:l1-l2-bound}%
\end{equation}
where $d$ is the rank of $X$. (Hint: Use the Cauchy--Schwarz inequality for
the Hilbert--Schmidt inner product.)
\end{exercise}

\begin{exercise}
There are explicit counterexamples to the monotonicity of the Hilbert--Schmidt
distance. Let $A=|01\rangle\langle00|+|11\rangle\langle10|$ and $B=|01\rangle
\langle01|+|11\rangle\langle11|$ be Kraus operators for a channel
$\mathcal{N}(\omega)=A\omega A^{\dag}+B\omega B^{\dag}$, and consider the
states $\rho=|0\rangle\langle0|\otimes\pi$ and $\sigma=|1\rangle
\langle1|\otimes\pi$, where $\pi=I/2$. First verify that $\mathcal{N}$ is a
quantum channel and then show by explicit calculation that $\left\Vert
\rho-\sigma\right\Vert _{2}<\left\Vert \mathcal{N}(\rho)-\mathcal{N}%
(\sigma)\right\Vert _{2}$ for this example. On the other hand, let
$\rho=|0\rangle\langle0|$, $\sigma=|1\rangle\langle1|$, and $\mathcal{N}%
(\omega)=\frac{1}{2}\left(  \omega+X\omega X\right)  $. Show that $\left\Vert
\rho-\sigma\right\Vert _{2}>\left\Vert \mathcal{N}(\rho)-\mathcal{N}%
(\sigma)\right\Vert _{2}$ for this other example. Message:\ Do not use the
Hilbert--Schmidt distance as a measure of distinguishability!
\end{exercise}

\section{History and Further Reading}

\cite{Kholevo1972} introduced and studied a function of two quantum states $\rho$ and $\sigma$ that is very closely related to the Uhlmann fidelity from Definition~\ref{def-dm:uhlmann} and Theorem~\ref{thm-dm:uhlmann-thm}:
\begin{equation}
F_H(\rho,\sigma) \equiv \left[\operatorname{Tr} \{\sqrt{\rho} \sqrt{\sigma} \}\right]^2.
\end{equation}
This function is now known as the Holevo fidelity and was rediscovered as ``quantum affinity'' by \cite{LZ04}.
\cite{Kholevo1972} proved that $F_H(\rho,\sigma)$ is related to the normalized trace distance in the following way:
\begin{equation}
1-\sqrt{F_{H}(\rho,\sigma)}\leq\frac{1}{2}\left\Vert \rho-\sigma\right\Vert
_{1}\leq\sqrt{1-F_{H}(\rho,\sigma)}, \label{eq-TDF:holevo-ineq}
\end{equation}
which is very similar to the inequalities presented in Theorem~\ref{thm-dm:fidelity-trace-relation}.
In fact, by employing the inequality $F_H(\rho,\sigma) \leq F(\rho,\sigma)$, as well as \eqref{eq-TDF:holevo-ineq} and Theorem~\ref{thm-dm:fidelity-trace-relation}, we conclude the following generalization of Theorem~\ref{thm-dm:fidelity-trace-relation}:
\begin{equation}
1-\sqrt{F(\rho,\sigma)} \leq 1-\sqrt{F_{H}(\rho,\sigma)}\leq\frac{1}{2}\left\Vert \rho-\sigma\right\Vert
_{1}\leq\sqrt{1-F(\rho,\sigma)} \leq \sqrt{1-F_{H}(\rho,\sigma)}
\end{equation}
In spite of these interesting relations and various properties that the Holevo fidelity $F_H$ possesses, we have focused our attention on the Uhlmann fidelity in this chapter due to its widespread use in quantum information and other fields. For this reason, we refer to the Uhlmann fidelity simply as ``the fidelity.''

\cite{F96}\ and \cite{FG98} are a good starting point for learning more
regarding trace distance and fidelity. Other notable sources are
\cite{book2000mikeandike}, \cite{Yard05a}, and \cite{K07}. \cite{H69,Hel76}
demonstrated the operational interpretation of the trace distance in the
context of quantum hypothesis testing. \cite{U73} first proved the theorem
bearing his name, and \cite{J94} later presented a proof of this theorem for
the case of finite-dimensional quantum systems.
Theorem~\ref{thm-dm:meas-achieve-fid} is due to \cite{FC95}.
\cite{PhysRevA.54.2614} introduced the entanglement fidelity, and \cite{BKN98}
made further observations regarding it. \cite{Nielsen2002249} provided a
simple proof of the exact relation between entanglement fidelity and expected fidelity.

\cite{itit1999winter,thesis1999winter} originally proved the \textquotedblleft
gentle measurement\textquotedblright\ lemma. There, he used it to obtain a
variation of the direct part of the HSW\ coding theorem. Later, he used it to
prove achievable rates for the quantum multiple access
channel~\citep{Winter01}. \cite{ON07} subsequently improved this bound to
$2\sqrt{\varepsilon}$.

The counterexample to the monotonicity of the Hilbert--Schmidt distance is due
to \cite{Ozawa2000158}.

\chapter{Classical Information and Entropy}

\label{chap:info-entropy}All physical systems register bits of information,
whether it be an atom, an electrical current, the location of a billiard ball,
or a switch. Information can be classical, quantum, or a hybrid of both,
depending on the system. For example, an atom or an electron or a
superconducting system can register \textit{quantum }information because the
quantum theory applies to each of these systems, but we can safely argue that
the location of a billiard ball registers classical information only. These
atoms or electrons or superconducting systems can also register classical bits
because it is always possible for a quantum system to register classical bits.

The term \textit{information}, in the context of information theory, has a
precise meaning that is somewhat different from our prior \textquotedblleft
everyday\textquotedblright\ experience with it. Recall that the notion of the
physical bit refers to the physical representation of a bit, and the
information bit is a measure of how much we learn from the outcome of a random
experiment. Perhaps the word \textquotedblleft surprise\textquotedblright%
\ better captures the notion of information as it applies in the context of
information theory.

This chapter begins our formal study of classical information. Recall that
Chapter~\ref{chap:classical-shannon-theory}\ reviewed some of the major
operational tasks in classical information theory. Here, our approach is
somewhat different because our aim is to provide an intuitive understanding of
information measures, in terms of the parties who have access to the classical
systems. We define precise mathematical formulas that measure the amount of
information encoded in a single physical system or in multiple physical
systems. The advantage of developing this theory is that we can study
information in its own right without having to consider the details of the
physical system that registers it.

We first introduce the entropy in Section~\ref{sec-cie:entropy}\ as the
expected surprise of a random variable. We extend this basic notion of entropy
to develop other measures of information in Sections~\ref{sec-cie:cond-ent}%
--\ref{sec-cie:cond-mut-inf} that prove useful as intuitive informational
measures, but also, and perhaps more importantly, these measures are the
answers to operational tasks that one might wish to perform using noisy
resources. While introducing these quantities, we discuss and prove several
mathematical results concerning them that are important tools for the
practicing information theorist. These tools are useful both for proving
results and for increasing our understanding of the nature of information.
Section~\ref{sec-cie:info-ineq} introduces entropy inequalities that help us
to understand the limits on our ability to process information, and
Section~\ref{sec-cie:entropy-ineq-refinements}\ gives several refinements of
these entropy inequalities. Section~\ref{sec-cie:cl-info-quantum} ends the
chapter by applying the classical informational measures developed in the
forthcoming sections to the classical information that one can extract from a
quantum system.

\section{Entropy of a Random Variable}

\label{sec-cie:entropy}Consider a random variable $X$.
\index{Shannon entropy}%
Suppose that each realization $x$ of random variable $X$ belongs to an
alphabet~$\mathcal{X}$. Let $p_{X}(x)$ denote the probability density function
of $X$ so that $p_{X}(x)$ is the probability that realization $x$ occurs. The
\index{information content}%
information content $i(x)$ of a particular realization $x$\ is a measure of
the surprise that one has upon learning the outcome of a random experiment:%
\begin{equation}
i(x)\equiv-\log\left(  p_{X}(x)\right)  . \label{eq-ie:info-content}%
\end{equation}
The logarithm is base two and this choice implies that we measure surprise or
information in units of bits.

\begin{figure}
[ptb]
\begin{center}
\includegraphics[
width=3.5397in
]%
{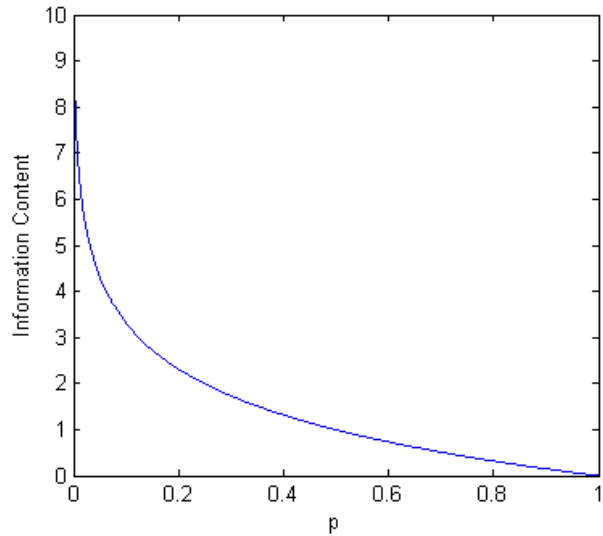}%
\caption{The information content or \textquotedblleft
surprise\textquotedblright\ in \eqref{eq-ie:info-content} as a function of a
probability $p$ ranging from $0$ to $1$. An event has a lower surprise if it
is more likely to occur and it has a higher surprise if it less likely to
occur.}%
\label{fig-ie:info-content}%
\end{center}
\end{figure}

Figure~\ref{fig-ie:info-content}\ plots the information content for values in
the unit interval. This measure of surprise behaves as we would hope---it is
higher for lower-probability events that surprise us more, and it is lower for
higher-probability events that do not surprise us as much. Inspection of the
figure reveals that the information content is non-negative for any
realization $x$.

The information content is also additive, due to the choice of the logarithm
function. Given two independent random experiments involving random variable
$X$ with respective realizations $x_{1}$ and $x_{2}$, we have that%
\begin{equation}
i( x_{1},x_{2}) =-\log\left(  p_{X,X}( x_{1},x_{2}) \right)  =-\log\left(
p_{X}( x_{1}) p_{X}( x_{2}) \right)  =i( x_{1}) +i( x_{2}) .
\end{equation}
Additivity is a property that we look for in measures of information (so much
so that we dedicate the whole of Chapter~\ref{chap:additivity}\ to this issue
for more general measures of information).%

The information content is a useful measure of surprise for particular
realizations of random variable $X$, but it does not capture a general notion
of the amount of surprise that a given random variable $X$ possesses. The
\index{Shannon entropy}%
entropy $H( X) $\ captures this general notion of the surprise of a random
variable $X$---it is the expected information content of random variable$~X$:%
\begin{equation}
H( X) \equiv\mathbb{E}_{X}\left\{  i( X) \right\}  .
\label{eq-cie:expected-info-content}%
\end{equation}
At a first glance, the above definition may seem strangely self-referential
because the argument of the probability density function $p_{X}( x) $ is
itself the random variable $X$, but this is well-defined mathematically.
Evaluating the above formula gives an expression which we take as the
definition for the entropy $H( X) $:

\begin{definition}
[Entropy]The entropy of a discrete random variable $X$ with probability
distribution $p_{X}(x)$ is
\begin{equation}
H( X) \equiv-\sum_{x}p_{X}( x) \log\left(  p_{X}( x) \right)  .
\end{equation}

\end{definition}

We adopt the convention that $0\cdot\log( 0) =0$ for realizations with zero
probability. The fact that $\lim_{\varepsilon\rightarrow0}\varepsilon\cdot
\log(1/\varepsilon)=0$ intuitively justifies this latter convention. (We can
interpret this convention as saying that the fact that the event has
probability zero is more important than or outweighs the fact that you would
be infinitely surprised if such an event would occur.)

\begin{figure}
[ptb]
\begin{center}
\includegraphics[
width=4.8456in
]%
{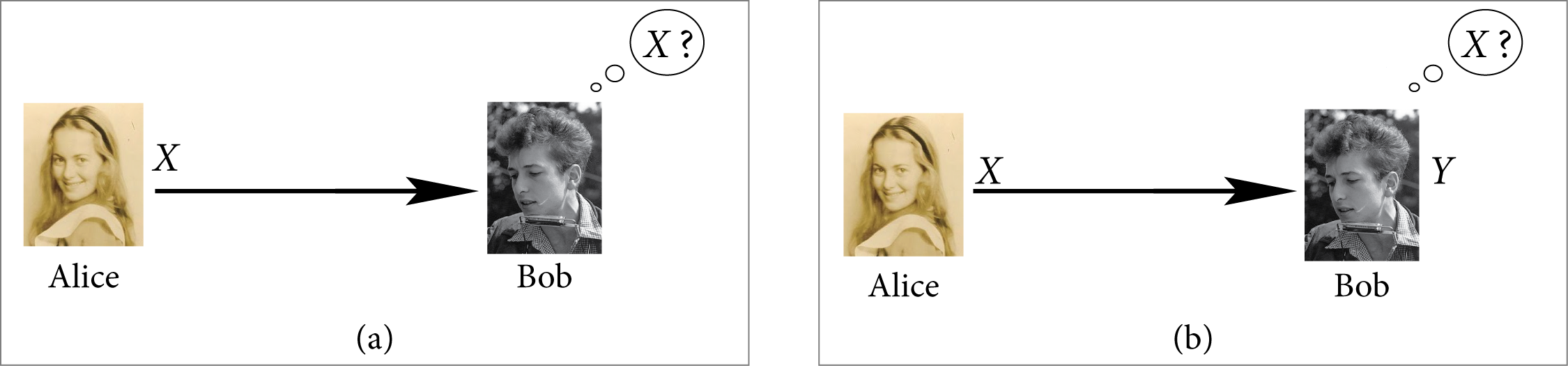}%
\caption{(a) The entropy $H( X) $ is the uncertainty that Bob has about random
variable $X$ before learning it. (b) The conditional entropy $H( X|Y) $ is the
uncertainty that Bob has about $X$ when he already possesses $Y$.}%
\label{fig-ie:alice-bob-entropy}%
\end{center}
\end{figure}

The entropy admits an intuitive interpretation. Suppose that Alice performs a
random experiment in her lab that selects a realization $x$ according to the
density $p_{X}( x) $ of random variable $X$. Suppose further that Bob has not
yet learned the outcome of the experiment. The interpretation of the entropy
$H( X) $\ is that it quantifies Bob's uncertainty about $X$ before learning
it---his expected information gain is $H( X) $ bits upon learning the outcome
of the random experiment. Shannon's noiseless coding theorem, described in
Chapter~\ref{chap:classical-shannon-theory}, makes this interpretation precise
by proving that Alice needs to send Bob bits at a rate $H( X) $ in order for
him to be able to decode a compressed message.
Figure~\ref{fig-ie:alice-bob-entropy}(a) depicts the interpretation of the
entropy $H( X) $, along with a similar interpretation for the conditional
entropy that we introduce in Section~\ref{sec-cie:cond-ent}.%

\subsection{The Binary Entropy Function}

A special case of the entropy occurs when the random variable $X$ is a
Bernoulli random variable with probability density $p_{X}( 0) =p$ and $p_{X}(
1) =1-p$. This Bernoulli random variable could correspond to the outcome of a
random coin flip. The entropy in this case is
\index{Shannon entropy!binary}%
known as the \textit{binary entropy function}:

\begin{definition}
[Binary Entropy]The binary entropy of $p\in[0,1]$ is
\begin{equation}
h_{2}( p) \equiv-p\log p-( 1-p) \log\left(  1-p\right)  .
\end{equation}

\end{definition}

\begin{figure}
[ptb]
\begin{center}
\includegraphics[
width=3.039in
]%
{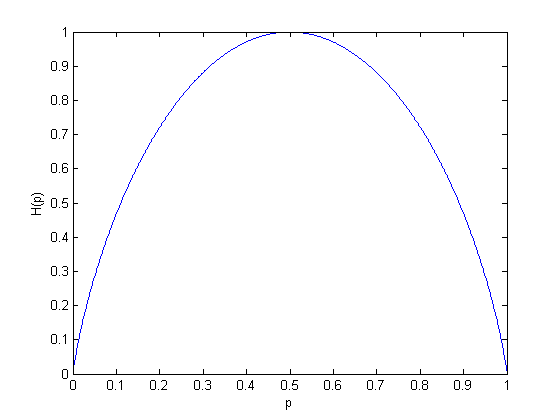}%
\caption{The binary entropy function $h_{2}(p)$ displayed as a function of the
parameter $p$.}%
\label{fig-ie:binary-entropy}%
\end{center}
\end{figure}

The binary entropy quantifies the number of bits that we learn from the
outcome of the coin flip. If the coin is unbiased ($p=1/2$), then we learn a
maximum of one bit ($h_{2}(p)=1$). If the coin is deterministic ($p=0$ or
$p=1$), then we do not learn anything from the outcome ($h_{2}(p)=0$).
Figure~\ref{fig-ie:binary-entropy}\ displays a plot of the binary entropy
function. The figure reveals that the binary entropy function $h_{2}(p)$ is a
concave function of the parameter $p$ and has its peak at $p=1/2$.%

\subsection{Mathematical Properties of Entropy}

\label{sec-ie:math-entropy}We now discuss five important mathematical
properties of the entropy $H( X) $.

\begin{property}
[Non-Negativity]The entropy $H(X)$ is non-negative for any discrete random
variable $X$ with probability density $p_{X}(x)$:%
\begin{equation}
H(X)\geq0.
\end{equation}

\end{property}

\begin{proof}%
\index{Shannon entropy!positivity}%
Non-negativity follows because entropy is the expected information content of
$i( X) $, and the information content itself is non-negative. It is perhaps
intuitive that the entropy should be non-negative because non-negativity
implies that we always learn some number of bits upon learning random variable
$X$ (if we already know beforehand what the outcome of a random experiment
will be, then we learn zero bits of information once we perform it). In a
classical sense, we can never learn a negative amount of information!
\end{proof}

\begin{property}
[Concavity]The entropy $H( X) $\ is concave in the probability density $p_{X}(
x) $.
\end{property}

\begin{proof}%
\index{Shannon entropy!concavity}%
We justify this result with a heuristic \textquotedblleft
mixing\textquotedblright\ argument for now, and provide a formal proof in
Section~\ref{sec-ie:class-inf-ineq}. Consider two random variables $X_{1}$ and
$X_{2}$ with two respective probability density functions $p_{X_{1}}( x) $ and
$p_{X_{2}}( x) $ whose realizations belong to the same alphabet. Consider a
Bernoulli random\ variable $B$\ with probabilities $q\in[0,1]$ and $1-q$
corresponding to its two respective realizations $b=1$ and $b=2$. Suppose that
we first generate a realization$~b$\ of random variable $B$ and then generate
a realization $x$ of random variable $X_{b}$. Random variable $X_{B}$ then
denotes a mixed version of the two random variables $X_{1}$ and $X_{2}$. The
probability density of $X_{B}$ is $p_{X_{B}}( x) =qp_{X_{1}}( x) +\left(
1-q\right)  p_{X_{2}}( x) $. Concavity of entropy is the following inequality:%
\begin{equation}
H( X_{B}) \geq qH( X_{1}) +\left(  1-q\right)  H( X_{2}) .
\end{equation}
Our heuristic argument is that this mixing process leads to more uncertainty
for the mixed random variable $X_{B}$ than the expected uncertainty over the
two individual random variables. We can think of this result as a physical
situation involving two gases. Two gases each have their own entropy, but the
entropy increases when we mix the two gases together. We later give a more
formal argument to justify concavity.
\end{proof}

\begin{property}
[Permutation Invariance]\label{prop-cie:inv-perm}The entropy is invariant with
respect to permutations of the realizations of random variable $X$.
\end{property}

\begin{proof}
That is, suppose that we apply some permutation $\pi$\ to realizations $x_{1}%
$, $x_{2}$, \ldots, $x_{\left\vert \mathcal{X}\right\vert }$ so that they
respectively become $\pi( x_{1}) $, $\pi( x_{2}) $, \ldots, $\pi(x_{\left\vert
\mathcal{X}\right\vert })$. Then the entropy is invariant under this shuffling
because it depends only on the probabilities of the realizations, not the
values of the realizations.
\end{proof}

\begin{property}
[Minimum Value]The entropy $H(X)$ vanishes if and only if $X$ is a
deterministic variable.
\end{property}

\begin{proof}
We would expect that the entropy of a \textit{deterministic }variable should
vanish, given the interpretation of entropy as the uncertainty of a random
experiment. This intuition holds true and it is the degenerate probability
density $p_{X}( x) =\delta_{x,x_{0}}$, where the realization $x_{0}$ has all
the probability and other realizations have vanishing probability, that gives
the minimum value of the entropy:\ $H( X) =0$ when $X$ has a degenerate
density. If $H(X) =0$, then this means that $p_{X}(x) \log[ 1/p_{X}(x)] = 0$
for all $x\in\mathcal{X}$, which in turn implies that either $p_{X}(x) = 0$ or
$p_{X}(x) = 1$ for all $x\in\mathcal{X}$. Since $p_{X}$ is required to be a
probability distribution, we can have that $p_{X}(x) = 1$ for just one
realization and $p_{X}(x) = 0$ for all others, so that $X$ is a deterministic
random variable.
\end{proof}

Sometimes, we may not have any prior information about the possible values of
a variable in a system, and we may decide that it is most appropriate to
describe them with a probability density function. How should we assign this
probability density if we do not have any prior information about the values?
Theorists and experimentalists often resort to a \textquotedblleft principle
of maximum entropy\textquotedblright\ or a \textquotedblleft principle of
maximal ignorance\textquotedblright---we should assign the probability density
to be the one that maximizes the entropy.

\begin{property}
[Maximum Value]\label{prop-cie:max-value-ent}The maximum value of the entropy
$H(X)$ for a random variable $X$ taking values in an alphabet $\mathcal{X}$ is
$\log\left\vert \mathcal{X}\right\vert $:%
\begin{equation}
H(X)\leq\log\left\vert \mathcal{X}\right\vert .
\end{equation}
The inequality is saturated if and only if $X$ is a uniform random variable on
$\mathcal{X}$.
\end{property}

\begin{proof}
First, note that the result of Exercise~\ref{ex-intro:uniform-entropy} is that
$\log\left\vert \mathcal{X}\right\vert $ is the entropy of the uniform random
variable on $\mathcal{X}$. Next, we can prove the above inequality with a
simple Lagrangian optimization by solving for the density $p_{X}(x)$ that
maximizes the entropy. Lagrangian optimization is well-suited for this task
because the entropy is concave in the probability density, and thus any local
maximum will be a global maximum. The Lagrangian $\mathcal{L}$ is as follows:%
\begin{equation}
\mathcal{L}\equiv H(X)+\lambda\left(  \sum_{x}p_{X}(x)-1\right)  ,
\end{equation}
where $H(X)$ is the quantity that we are maximizing, subject to the constraint
that the probability density $p_{X}(x)$ sums to one. The partial derivative
$\frac{\partial\mathcal{L}}{\partial p_{X}(x)}$ is as follows:%
\begin{equation}
\frac{\partial\mathcal{L}}{\partial p_{X}(x)} =-\log\left(  p_{X}(x)\right)
-1+\lambda.
\end{equation}
We set the partial derivative $\frac{\partial\mathcal{L}}{\partial p_{X}(x)}$
equal to zero to find the probability density that maximizes $\mathcal{L}$:%
\begin{align}
0  &  =-\log\left(  p_{X}(x)\right)  -1+\lambda\\
\Rightarrow p_{X}(x)  &  =2^{\lambda-1}.
\end{align}
The resulting probability density $p_{X}(x)$ depends only on a constant
$\lambda$, implying that it must be uniform $p_{X}(x)=1/\left\vert
\mathcal{X}\right\vert $. Thus, the uniform distribution\ maximizes the
entropy $H(X)$ when random variable $X$ is finite.
\end{proof}

\section{Conditional Entropy}

\label{sec-cie:cond-ent}Let us now suppose that
\index{Shannon entropy!conditional}%
Alice possesses random variable $X$ and Bob possesses some other random
variable $Y$. Random variables $X$ and $Y$ share correlations if they are not
statistically independent, and Bob then possesses \textquotedblleft side
information\textquotedblright\ about $X$ in the form of $Y$. Let $i(x|y)$
denote the conditional information content:%
\begin{equation}
i(x|y)\equiv-\log\left(  p_{X|Y}(x|y)\right)  .
\end{equation}
The entropy $H(X|Y=y)$\ of random variable $X$ conditioned on a particular
realization $y$\ of random variable $Y$ is the expected conditional
information content, where the expectation is with respect to$~X|Y=y$:%
\begin{align}
H(X|Y=y)  &  \equiv\mathbb{E}_{X|Y=y}\left\{  i(X|y)\right\} \\
&  =-\sum_{x}p_{X|Y}(x|y)\log\left(  p_{X|Y}(x|y)\right)  .
\end{align}
The relevant entropy that applies to the scenario where Bob possesses side
information is the conditional entropy $H(X|Y)$, defined as follows:

\begin{definition}
[Conditional Entropy]Let $X$ and $Y$ be discrete random variables with joint
probability distribution $p_{X,Y}(x,y)$. The conditional entropy $H(X|Y)$ is
the expected conditional information content, where the expectation is with
respect to both $X$ and~$Y$:%
\begin{align}
H( X|Y)  &  \equiv\mathbb{E}_{X,Y}\left\{  i( X|Y) \right\} \\
&  =\sum_{y}p_{Y}( y) H( X|Y=y)\label{eq-ie:class-cond-ent}\\
&  =-\sum_{y}p_{Y}( y) \sum_{x}p_{X|Y}( x|y) \log( p_{X|Y}( x|y) )\\
&  =-\sum_{x,y}p_{X,Y}( x,y) \log( p_{X|Y}( x|y) ).
\end{align}

\end{definition}

The conditional entropy $H( X|Y) $ as well deserves an interpretation. Suppose
that Alice possesses random variable $X$ and Bob possesses random variable
$Y$. The conditional entropy $H( X|Y) $\ is the amount of uncertainty that Bob
has about $X$ given that he already possesses$~Y$.
Figure~\ref{fig-ie:alice-bob-entropy}(b) depicts this interpretation.

The above interpretation of the conditional entropy $H( X|Y) $ immediately
suggests that it should be less than or equal to the entropy $H( X) $. That
is, having access to a side variable$~Y$ should only decrease our uncertainty
about another variable. We state this idea as the following theorem and give a
formal proof in Section~\ref{sec-ie:class-inf-ineq}.

\begin{theorem}
[Conditioning Does Not Increase Entropy]\label{thm-ie:cond-red-ent}The entropy
$H(X)$ is greater than or equal to the conditional entropy $H(X|Y)$:%
\begin{equation}
H(X)\geq H(X|Y),
\end{equation}
and equality occurs if and only if $X$ and $Y$ are independent random
variables. As a consequence of the fact that $H(X|Y)=\sum_{y}p_{Y}%
(y)H(X|Y=y)$, we see that the entropy is concave.
\end{theorem}

Non-negativity of conditional entropy follows from non-negativity of entropy
because conditional entropy is the expectation of the entropy $H( X|Y=y) $
with respect to the density $p_{Y}( y) $. It is again intuitive that
conditional entropy should be non-negative. Even if we have access to some
side information $Y$, we always learn some number of bits of information upon
learning the outcome of a random experiment involving $X$. Perhaps strangely,
we will see that \textit{quantum} conditional entropy can become negative,
defying our intuition of information in the classical sense given here.

\section{Joint Entropy}

\label{sec-cie:joint-ent}What if Bob knows neither $X$ nor $Y$?\ The natural
entropic quantity that describes his uncertainty is the
\index{Shannon entropy!joint}%
joint entropy$~H( X,Y) $. The joint entropy is merely the entropy of the joint
random variable$~\left(  X,Y\right)  $:

\begin{definition}
[Joint Entropy]Let $X$ and $Y$ be discrete random variables with joint
probability distribution $p_{X,Y}(x,y)$. The joint entropy $H(X,Y)$ is defined
as
\begin{align}
H( X,Y)  &  \equiv\mathbb{E}_{X,Y}\left\{  i( X,Y) \right\} \\
&  =-\sum_{x,y}p_{X,Y}( x,y) \log( p_{X,Y}( x,y) ) . \label{eq-cie:joint-ent}%
\end{align}

\end{definition}

The following exercise asks you to explore the relation between joint entropy
$H( X,Y) $, conditional entropy $H( Y|X) $, and marginal entropy $H( X) $. Its
proof follows by considering that the multiplicative probability relation
$p_{X,Y}( x,y) =p_{Y|X}( y|x) p_{X}( x) $\ of joint probability, conditional
probability, and marginal entropy becomes an additive relation under the
logarithms of the entropic definitions.

\begin{exercise}
\label{ex-ie:simple-ent-chain-rule}Verify that $H( X,Y) =H( X) +H( Y|X) =H( Y)
+H( X|Y) $.
\end{exercise}

\begin{exercise}
\label{ex-ie:ent-chain-rule}Extend the result of
Exercise~\ref{ex-ie:simple-ent-chain-rule} to prove the following chain rule
for entropy:%
\begin{equation}
H( X_{1},\ldots,X_{n}) =H( X_{1}) +H( X_{2}|X_{1}) +\cdots+H( X_{n}%
|X_{n-1},\ldots,X_{1}) .
\end{equation}

\end{exercise}

\begin{exercise}
\label{ex-ie:subadditivity}Prove that entropy is \textit{subadditive}:%
\begin{equation}
H( X_{1},\ldots,X_{n}) \leq\sum_{i=1}^{n}H( X_{i}) ,
\end{equation}
by exploiting Theorem~\ref{thm-ie:cond-red-ent}\ and the entropy chain rule in
Exercise~\ref{ex-ie:ent-chain-rule}.
\end{exercise}

\begin{exercise}
Prove that entropy is additive when the random variables $X_{1},\ldots,X_{n}$
are independent:%
\begin{equation}
H( X_{1},\ldots,X_{n}) =\sum_{i=1}^{n}H( X_{i}) .
\end{equation}

\end{exercise}

\section{Mutual Information}

\label{sec-cie:mutual-inf}We now introduce an
\index{mutual information}%
entropic measure of the common or mutual information that two parties possess.
Suppose that Alice possesses random variable $X$ and Bob possesses random
variable~$Y$.

\begin{definition}
[Mutual Information]Let $X$ and $Y$ be discrete random variables with joint
probability distribution $p_{X,Y}(x,y)$. The mutual information $I(X;Y)$ is
the marginal entropy $H( X) $ less the conditional entropy $H( X|Y) $:%
\begin{equation}
I( X;Y) \equiv H( X) -H( X|Y) . \label{eq-cie:mut-info-def}%
\end{equation}

\end{definition}

The mutual information quantifies the dependence or correlations of the two
random variables $X$ and $Y$. It measures how much knowing one random variable
reduces the uncertainty about the other random variable. In this sense, it is
the common information between the two random variables. Bob possesses $Y$ and
thus has an uncertainty $H( X|Y) $ about Alice's variable $X$. Knowledge of
$Y$ gives an information gain of $H( X|Y) $ bits about $X$ and then reduces
the overall uncertainty $H( X) $ about $X$, the uncertainty were he not to
have any side information at all about$~X$.

\begin{exercise}
Show that the mutual information is symmetric in its inputs:%
\begin{equation}
I( X;Y) =I( Y;X) ,
\end{equation}
implying additionally that%
\begin{equation}
I( X;Y) =H( Y) -H( Y|X) . \label{eq-ie:expand-MI}%
\end{equation}

\end{exercise}

We can also express the mutual information $I( X;Y) $ in terms of the
respective joint and marginal probability density functions $p_{X,Y}( x,y) $
and $p_{X}( x) $ and $p_{Y}( y) $:%
\begin{equation}
I( X;Y) =\sum_{x,y}p_{X,Y}( x,y) \log\left(  \frac{p_{X,Y}( x,y) }{p_{X}( x)
p_{Y}( y) }\right)  . \label{eq-ie:mut-info-dists}%
\end{equation}
The above expression leads to two insights regarding the mutual information
$I( X;Y) $. Two random variables $X$ and $Y$ possess zero bits of mutual
information if they are statistically independent (recall that the joint
density factors as $p_{X,Y}( x,y) =p_{X}( x) p_{Y}( y) $ when $X$ and $Y$ are
independent). That is, knowledge of $Y$ does not give any information about
$X$ when the random variables are statistically independent. Later, we show
that the converse statement is true as well. Also, two random variables
possess $H( X) $ bits of mutual information if they are perfectly correlated
in the sense that $Y=X$.

Theorem~\ref{thm-ie:mut-positive} below states that the mutual information $I(
X;Y) $ is non-negative for any random variables $X$ and $Y$---we provide a
formal proof in Section~\ref{sec-ie:class-inf-ineq}. However, this follows
naturally from the definition of mutual information in
\eqref{eq-cie:mut-info-def} and \textquotedblleft conditioning does not
increase entropy\textquotedblright\ (Theorem~\ref{thm-ie:cond-red-ent}).

\begin{theorem}
\label{thm-ie:mut-positive}The mutual information $I(X;Y)$ is non-negative for
any random variables$~X$ and$~Y$:%
\begin{equation}
I(X;Y)\geq0,
\end{equation}
and $I(X;Y)=0$ if and only if $X$ and $Y$ are independent random variables
(i.e., if $p_{X,Y}(x,y)=p_{X}(x)p_{Y}(y)$).
\end{theorem}

\section{Relative Entropy}

\label{sec-cie:relative-ent}The relative entropy%
\index{relative entropy}
is another important entropic quantity that quantifies how \textquotedblleft
far\textquotedblright\ one probability density function $p(x)$\ is from
another probability density function $q(x)$. It can be helpful to have a more
general definition in which we allow $q(x)$ to be a function taking
non-negative values. Before defining the relative entropy, we need the notion
of the support of a function.

\begin{definition}
[Support]Let $\mathcal{X}$ denote a finite set. The support of
\index{support!of a function}%
a function $f:\mathcal{X}\rightarrow\mathbb{R}$\ is equal to the subset of
$\mathcal{X}$ that takes non-zero values under $f$:%
\begin{equation}
\operatorname{supp}(f)\equiv\{x:f(x)\neq0\}.
\end{equation}

\end{definition}

\begin{definition}
[Relative Entropy]\label{def-cie:rel-ent}Let $p$ be a probability distribution
defined on the alphabet $\mathcal{X}$, and let $q:\mathcal{X}\rightarrow
\lbrack0,\mathbb{\infty)}$. The relative entropy $D(p\Vert q)$ is defined as
follows:
\begin{equation}
D(p\Vert q)\equiv\left\{
\begin{array}
[c]{cc}%
\sum_{x}p(x)\log\left(  p(x)/q(x)\right)  & \text{if }\operatorname{supp}%
(p)\subseteq\operatorname{supp}(q)\\
+\infty & \text{else}%
\end{array}
\right.  . \label{eq-ie:relative-entropy}%
\end{equation}

\end{definition}

According to the above definition, the relative entropy is equal to the
following expected log-likelihood ratio:%
\begin{equation}
D(p\Vert q)=\mathbb{E}_{X}\left\{  \log\left(  \frac{p(X)}{q(X)}\right)
\right\}  ,
\end{equation}
where $X$ is a random variable distributed according to $p$.

The above definition implies that the relative entropy is not symmetric under
interchange of $p(x)$ and $q(x)$. Thus, the relative entropy is not a distance
measure in the strict mathematical sense because it is not symmetric (nor does
it satisfy a triangle inequality).

The relative entropy has an interpretation in source coding, if we let $q(x)$
be a probability distribution. Suppose that an information source generates a
random variable $X$ according to the density $p(x)$. Suppose further that
Alice (the compressor) mistakenly assumes that the probability density of the
information source is instead $q(x)$ and codes according to this density. Then
the relative entropy quantifies the inefficiency that Alice incurs when she
codes according to the mistaken probability density---Alice requires
$H(X)+D\left(  p\Vert q\right)  $ bits on average to code (whereas she would
only require $H(X)$ bits on average to code if she used the true density
$p(x)$).

We might also see now that the mutual information $I(X;Y)$ is equal to the
relative entropy $D(p_{X,Y}\Vert p_{X}\otimes p_{Y})$ by comparing the
definition of relative entropy in \eqref{eq-ie:relative-entropy} and the
expression for the mutual information in \eqref{eq-ie:mut-info-dists} (by
$p_{X}\otimes p_{Y}$ we mean the product of the marginal distributions). In
this sense, the mutual information quantifies how far the two random variables
$X$ and $Y$ are from being independent because it calculates the
\textquotedblleft distance\textquotedblright\ of the joint density $p_{X,Y}%
$\ to the product $p_{X}\otimes p_{Y}$\ of the marginals.

Let $p_{X_{1}}$ and $p_{X_{2}}$ be two probability distributions defined over
the same alphabet. The relative entropy $D(p_{X_{1}}\Vert p_{X_{2}})$ admits a
pathological property. It can become infinite if the distribution $p_{X_{1}%
}(x_{1})$ does not have all of its support contained in the support of
$p_{X_{2}}(x_{2})$ (i.e., if there is some realization $x$ for which
$p_{X_{1}}(x)\neq0$ but $p_{X_{2}}(x)=0$). This can be somewhat bothersome if
we like the interpretation of relative entropy as a notion of distance. In an
extreme case, we would think that the distance between a deterministic binary
random variable $X_{2}$ where $\Pr\left\{  X_{2}=1\right\}  =1$ and one with
probabilities $\Pr\left\{  X_{1}=0\right\}  =\varepsilon$ and $\Pr\left\{
X_{1}=1\right\}  =1-\varepsilon$ should be on the order of $\varepsilon$ (this
is true for the classical trace distance). However, the relative entropy
$D(p_{X_{1}}\Vert p_{X_{2}})$\ in this case is infinite, in spite of our
intuition that these distributions are close. The interpretation in lossless
source coding is that it would require an infinite number of bits to code a
distribution $p_{X_{1}}$ losslessly if Alice mistakes it as $p_{X_{2}}$. Alice
thinks that the symbol $X_{2}=0$ never occurs, and in fact, she thinks that
the typical set consists of just one sequence of all ones and every other
sequence is atypical. But in reality, the typical set is quite a bit larger
than this, and it is only in the limit of an infinite number of bits that we
can say her compression is truly lossless.

\begin{exercise}
Verify that the definition of relative entropy in
Definition~\ref{def-cie:rel-ent}\ is consistent with the following limit:%
\begin{equation}
D(p\Vert q)=\lim_{\varepsilon\searrow0}D(p\Vert q+\varepsilon\mathbf{1}),
\end{equation}
where $\mathbf{1}$ denotes a vector of ones, so that the elements of
$q+\varepsilon\mathbf{1}$ are $q(x)+\varepsilon$.
\end{exercise}

\section{Conditional Mutual Information}

\label{sec-cie:cond-mut-inf}What is the common
\index{mutual information!conditional}%
information between two random variables $X$ and $Y$ when we have some side
information embodied in a random variable $Z$? The entropic quantity that
quantifies this common information is the conditional mutual information.

\begin{definition}
[Conditional Mutual Information]Let $X$, $Y$, and $Z$ be discrete random
variables. The conditional mutual information is defined as follows:%
\begin{align}
I( X;Y|Z)  &  \equiv H( Y|Z) -H( Y|X,Z)\\
&  =H( X|Z) -H( X|Y,Z)\\
&  =H( X|Z) +H( Y|Z) -H( X,Y|Z) .
\end{align}

\end{definition}

\begin{theorem}
[Strong Subadditivity]\label{thm-ie:class-ssa}The conditional
\index{strong subadditivity}%
mutual information $I(X;Y|Z)$ is non-negative:%
\begin{equation}
I(X;Y|Z)\geq0, \label{eq-ie:strong-subadd}%
\end{equation}
and the inequality is saturated if and only if $X-Z-Y$ is a Markov chain
(i.e., if $p_{X,Y|Z}(x,y|z)=p_{X|Z}(x|z)p_{Y|Z}(y|z)$).
\end{theorem}

\begin{proof}
The proof of the above theorem is a straightforward consequence of the
non-negativity of mutual information (Theorem~\ref{thm-ie:mut-positive}).
Consider the following equality:%
\begin{equation}
I(X;Y|Z)=\sum_{z}p_{Z}(z)I(X;Y|Z=z),
\end{equation}
where $I(X;Y|Z=z)$ is a mutual information with respect to the joint density
$p_{X,Y|Z}(x,y|z)$ and the marginal densities $p_{X|Z}(x|z)$ and
$p_{Y|Z}(y|z)$. Non-negativity of $I(X;Y|Z)$ then follows from non-negativity
of $p_{Z}(z)$ and $I(X;Y|Z=z)$. The saturation conditions then follow
immediately from those for mutual information given in
Theorem~\ref{thm-ie:mut-positive}\ (considering that the conditional mutual
information is a convex combination of mutual informations).
\end{proof}

The proof of the above classical version of strong subadditivity is perhaps
trivial in hindsight (it requires only a few arguments). The proof of the
quantum version of strong subaddivity is highly non-trivial on the other hand.
We discuss strong subadditivity of quantum entropy in the next chapter.

\begin{exercise}
The expression in \eqref{eq-ie:strong-subadd} represents the most compact way
to express the strong subadditivity of entropy. Show that the following
inequalities are equivalent ways of representing strong subadditivity:%
\begin{align}
H( XY|Z)  &  \leq H( X|Z) +H( Y|Z) ,\\
H( XYZ) +H( Z)  &  \leq H( XZ) +H( YZ) ,\\
H( X|YZ)  &  \leq H( X|Z) .
\end{align}

\end{exercise}

\begin{exercise}
\label{ex-ie:chain-rule-MI}Prove the following chain rule for mutual
information:%
\begin{multline}
I( X_{1},\ldots,X_{n};Y)\\
=I( X_{1};Y) +I( X_{2};Y|X_{1}) +\cdots+I( X_{n};Y|X_{1},\ldots,X_{n-1}) .
\end{multline}
The interpretation of the chain rule is that we can build up the correlations
between $X_{1},\ldots,X_{n}$ and $Y$ in $n$ steps: in a first step, we build
up the correlations between $X_{1}$ and $Y$, and now that $X_{1}$ is available
(and thus conditioned on), we build up the correlations between $X_{2}$ and
$Y$, etc.
\end{exercise}

\section{Entropy Inequalities}

\label{sec-cie:info-ineq}The entropic quantities introduced in the previous
sections each have bounds associated with them. These bounds are fundamental
limits on our ability to process and store information. We introduce several
bounds in this section: the non-negativity of relative entropy, two
data-processing inequalities, Fano's inequality, and a uniform bound for
continuity of entropy. Each of these inequalities plays an important role in
information theory, and we describe these roles in more detail in the
forthcoming subsections.

\subsection{Non-Negativity of Relative Entropy}

\label{sec-ie:class-inf-ineq}The relative entropy is always non-negative. This
seemingly innocuous result has several important implications---namely, the
maximal value of entropy, conditioning does not increase entropy
(Theorem~\ref{thm-ie:cond-red-ent}), non-negativity of mutual information
(Theorem~\ref{thm-ie:mut-positive}), and strong subadditivity
(Theorem~\ref{thm-ie:class-ssa})\ are straightforward corollaries of it. A
proof of this entropy inequality follows from the application of a simple
inequality:\ $\ln x\leq x-1$.

\begin{theorem}
[Non-Negativity of Relative Entropy]\label{thm-cie:rel-ent-positive}Let $p(x)$
be a probability%
\index{relative entropy!non-negativity}
distribution over the alphabet $\mathcal{X}$ and let $q:\mathcal{X}%
\rightarrow\lbrack0,1]$ be a function such that $\sum_{x}q(x)\leq1$. Then the
relative%
\index{relative entropy!positivity}
entropy $D(p\Vert q)$ is non-negative:%
\begin{equation}
D(p\Vert q)\geq0, \label{eq-cie:rel-ent-non-neg}%
\end{equation}
and $D(p\Vert q)=0$ if and only if $p=q$.
\end{theorem}

\begin{figure}[ptb]
\begin{center}
\includegraphics[
width=3.0in
]{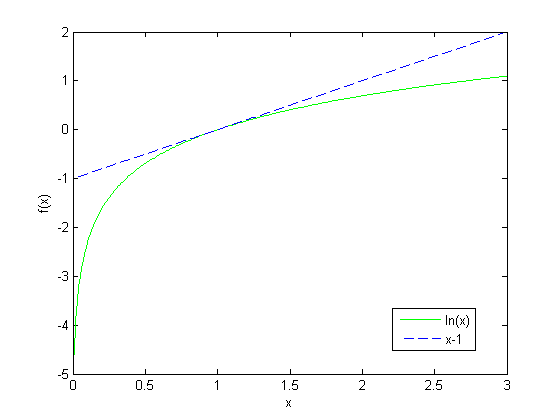}
\end{center}
\caption{A plot that compares the functions $\ln x$ and $x-1$, showing that
$\ln x\leq x-1$ for all positive$~x$.}%
\label{fig-ie:log-vs-x-1}%
\end{figure}

\begin{proof}
First, suppose that $\operatorname{supp}(p)\not \subseteq \operatorname{supp}%
(q)$. Then the relative entropy $D(p\Vert q)=+\infty$ and the inequality is
trivially satisfied.

Now, suppose that $\operatorname{supp}(p)\subseteq\operatorname{supp}(q)$. A
proof relies on the inequality $\ln x\leq x-1$ that holds for all $x\geq0$ and
saturates for this range if and only if $x=1$. (Brief justification:\ Let
$f(x)=x-1-\ln x$. Observe that $f(1)=0$, $f^{\prime}(1)=0$, $f^{\prime}(x)>0$
for $x>1$ and $f^{\prime}(x)<0$ for $x<1$. So $f(x)$ has a minimum at $x=1$
and is strictly increasing when $x>1$ and strictly decreasing when $x<1$. For
the saturation condition:\ Suppose that $f(x)=0$. Then $x=1$ is a solution of
the equation. Since the function is strictly decreasing when $x<1$ and
strictly increasing when $x>1$, $x=1$ is the only solution to $f(x)=0$%
.)\ Figure~\ref{fig-ie:log-vs-x-1}\ plots the functions $\ln x$ and $x-1$ to
compare them.

We first prove the inequality in \eqref{eq-cie:rel-ent-non-neg}. Consider the
following chain of inequalities:%
\begin{align}
D(p\Vert q)  &  =\sum_{x}p(x)\log\left(  \frac{p(x)}{q(x)}\right) \\
&  =-\frac{1}{\ln2}\sum_{x}p(x)\ln\left(  \frac{q(x)}{p(x)}\right) \\
&  \geq\frac{1}{\ln2}\sum_{x}p(x)\left(  1-\frac{q(x)}{p(x)}\right) \\
&  =\frac{1}{\ln2}\left(  \sum_{x}p(x)-\sum_{x}q(x)\right) \\
&  \geq0.
\end{align}
The sole inequality follows because $-\ln x\geq1-x$ (a simple rearrangement of
$\ln x\leq x-1$). The last inequality is a consequence of the assumption that
$\sum_{x}q(x)\leq1$.

Now suppose that $p=q$. It is then clear that $D(p\Vert q)=0$. Finally,
suppose that $D(p\Vert q)=0$. Then we necessarily have $\operatorname{supp}%
(p)\subseteq\operatorname{supp}(q)$, and the condition $D(p\Vert
q)=0$\ implies that both inequalities above are saturated. So, first we can
deduce that $q$ is a probability distribution since we assumed that $\sum
_{x}q(x)\leq1$ and the last inequality above is saturated. Next, the
inequality in the third line above is saturated, which implies that
$\ln\left(  q(x)/p(x)\right)  =1-q(x)/p(x)$ for all $x$ for which $p(x)>0$.
But this happens only when $q(x)/p(x)=1$ for all $x$ for which $p(x)>0$, which
allows us to conclude that $p=q$.
\end{proof}

We can now quickly prove several corollaries of the above theorem.

\begin{proof}
[Proofs of Property~\ref{prop-cie:max-value-ent},
Theorem~\ref{thm-ie:mut-positive}, Theorem~\ref{thm-ie:cond-red-ent}]Recall in
Section~\ref{sec-ie:math-entropy}\ that we proved that the entropy
$H(X)$\ takes the maximal value $\log\left\vert \mathcal{X}\right\vert $,
where$~\left\vert \mathcal{X}\right\vert $ is the size of the alphabet of $X$.
The proof method involved Lagrange multipliers. Here, we can prove this result
simply by computing the relative entropy $D(p_{X}\Vert\{1/\left\vert
\mathcal{X}\right\vert \})$, where $p_{X}$ is the probability density of $X$
and $\{1/\left\vert \mathcal{X}\right\vert \}$ is the uniform density, and
applying the non-negativity of relative entropy:%
\begin{align}
0  &  \leq D(p_{X}\Vert\{1/\left\vert \mathcal{X}\right\vert \})\\
&  =\sum_{x}p_{X}(x)\log\left(  \frac{p_{X}(x)}{\frac{1}{\left\vert
\mathcal{X}\right\vert }}\right) \\
&  =-H(X)+\sum_{x}p_{X}(x)\log\left\vert \mathcal{X}\right\vert \\
&  =-H(X)+\log\left\vert \mathcal{X}\right\vert .
\end{align}
It then follows that $H(X)\leq\log\left\vert \mathcal{X}\right\vert $ by
combining the first line with the last. Non-negativity of mutual information
(Theorem~\ref{thm-ie:mut-positive}) follows by recalling that
$I(X;Y)=D(p_{X,Y}\Vert p_{X}\otimes p_{Y})$ and applying the non-negativity of
relative entropy. The equality conditions follow from those for equality of
$D(p\Vert q)=0$. Conditioning does not increase entropy
(Theorem~\ref{thm-ie:cond-red-ent}) follows by noting that
$I(X;Y)=H(X)-H(X|Y)$ and applying Theorem~\ref{thm-ie:mut-positive}.
\end{proof}

\subsection{Data-Processing Inequality}

Another important inequality in classical information theory is the
\index{data processing inequality}
\textit{data-processing inequality}. There are at least two variations of it.
The first one states that correlations between random variables can only
decrease after we process one variable according to some stochastic function
that depends only on that variable. The next one states that the relative
entropy cannot increase if a channel is applied to both of its arguments.
These data-processing inequalities find application in the converse proof of a
coding theorem (the proof of the optimality of a communication rate).

\subsubsection{Mutual Information Data-Processing Inequality}

\begin{figure}[ptb]
\begin{center}
\includegraphics[
width=4in
]{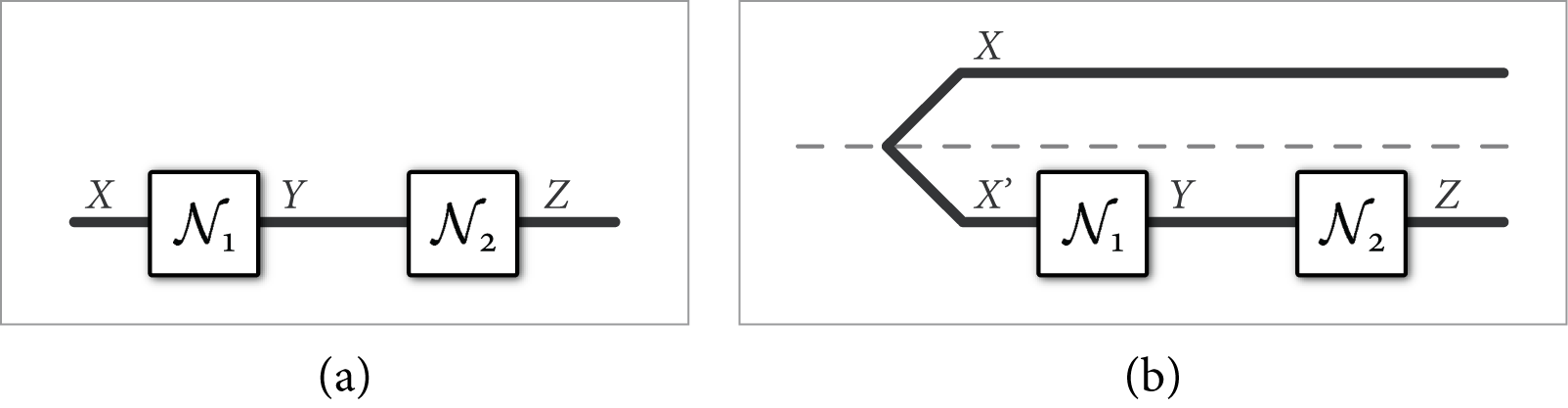}
\end{center}
\caption{Two slightly different depictions of the scenario in the
data-processing inequality. (a) The map $\mathcal{N}_{1}$ processes random
variable $X$ to produce some random variable $Y$, and the map $\mathcal{N}%
_{2}$ processes the random variable $Y$ to produce the random variable $Z$.
The inequality $I(X;Y)\geq I(X;Z)$ applies here because correlations can only
decrease after data processing. (b) This depiction of data processing helps us
to build intuition for data processing in the quantum world. The protocol
begins with two perfectly correlated random variables $X$ and $X^{\prime}%
$---perfect correlation implies that $p_{X,X^{\prime}}(x,x^{\prime}%
)=p_{X}(x)\delta_{x,x^{\prime}}$ and further that $H(X)=I(X;X^{\prime})$. We
process random variable $X^{\prime}$ with a stochastic map $\mathcal{N}_{1}$
to produce a random variable $Y$, and then further process $Y$ according to
the stochastic map $\mathcal{N}_{2}$ to produce random variable $Z$. By the
data-processing inequality, the following chain of inequalities holds:
$I(X;X^{\prime})\geq I(X;Y)\geq I(X;Z)$.}%
\label{fig-ie:classical-data-processing}%
\end{figure}%

\index{data processing inequality!mutual information}%
We detail the scenario that applies for the first data-processing inequality.
Suppose that we initially have two random variables $X$ and $Y$. We might say
that random variable $Y$ arises from random variable $X$ by processing $X$
according to a stochastic map $\mathcal{N}_{1}\equiv p_{Y|X}(y|x)$. That is,
the two random variables arise by first picking $X$ according to the density
$p_{X}(x)$ and then processing $X$ according to the stochastic map
$\mathcal{N}_{1}$. The mutual information $I(X;Y)$ quantifies the correlations
between these two random variables. Suppose then that we process$~Y$ according
to some other stochastic map $\mathcal{N}_{2}\equiv p_{Z|Y}(z|y)$ to produce a
random variable $Z$ (note that the map can also be deterministic because the
set of stochastic maps subsumes the set of deterministic maps). Then the first
data-processing inequality states that the correlations between $X$ and $Z$
must be less than the correlations between $X$ and~$Y$:%
\begin{equation}
I(X;Y)\geq I(X;Z),
\end{equation}
because data processing according to any stochastic map $\mathcal{N}_{2}$
cannot increase  correlations. Figure~\ref{fig-ie:classical-data-processing}%
(a) depicts the scenario described above.
Figure~\ref{fig-ie:classical-data-processing}(b) depicts a slightly different
scenario for data processing that helps build intuition for the forthcoming
notion of quantum data-processing in
Section~\ref{sec-ie:quantum-data-processing} of the next chapter.
Theorem~\ref{thm-ie:data-process} below states the classical data-processing inequality.

The scenario described in the above paragraph contains a major assumption: the
stochastic map $p_{Z|Y}(z|y)$ that produces random variable $Z$ depends on
random variable $Y$ only---it has no dependence on $X$, meaning that%
\begin{equation}
p_{Z|Y,X}(z|y,x)=p_{Z|Y}(z|y).
\end{equation}
This assumption is called the Markovian assumption and is the crucial
assumption in the proof of the data-processing inequality. We say that the
three random variables $X$, $Y$, and $Z$ form a \textit{Markov chain} and use
the notation $X\rightarrow Y\rightarrow Z$ to indicate this stochastic relationship.

\begin{theorem}
[Data-Processing Inequality]\label{thm-ie:data-process}Suppose three random
variables $X$, $Y$, and $Z$ form a Markov chain: $X\rightarrow Y\rightarrow
Z$. Then the following data-processing inequality holds
\begin{equation}
I( X;Y) \geq I( X;Z) .
\end{equation}

\end{theorem}

\begin{proof}
The Markov condition $X\rightarrow Y\rightarrow Z$ implies that random
variables $X$ and $Z$ are conditionally independent through $Y$ because%
\begin{align}
p_{X,Z|Y}(x,z|y)  &  =p_{Z|Y,X}(z|y,x)p_{X|Y}(x|y)\\
&  =p_{Z|Y}(z|y)p_{X|Y}(x|y).
\end{align}
We prove the data-processing inequality by manipulating the mutual information
$I(X;YZ)$. Consider the following equalities:%
\begin{equation}
I(X;YZ)=I(X;Y)+I(X;Z|Y)=I(X;Y). \label{eq-ie:markov-MI-expand}%
\end{equation}
The first equality follows from the chain rule for mutual information
(Exercise~\ref{ex-ie:chain-rule-MI}). The second equality follows because the
conditional mutual information $I(X;Z|Y)$ vanishes for a Markov chain
$X\rightarrow Y\rightarrow Z$---i.e., $X$ and $Z$ are conditionally
independent through $Y$ (recall Theorem~\ref{thm-ie:class-ssa}). We can also
expand the mutual information $I(X;YZ)$ in another way to obtain%
\begin{equation}
I(X;YZ)=I(X;Z)+I(X;Y|Z).
\end{equation}
Then the following equality holds for a Markov chain $X\rightarrow
Y\rightarrow Z$ by exploiting \eqref{eq-ie:markov-MI-expand}:%
\begin{equation}
I(X;Y)=I(X;Z)+I(X;Y|Z).
\end{equation}
The inequality in Theorem~\ref{thm-ie:data-process}\ follows because
$I(X;Y|Z)$ is non-negative for any random variables $X$, $Y$, and $Z$ (recall
Theorem~\ref{thm-ie:class-ssa}).
\end{proof}

By inspecting the above proof, we find the following:

\begin{corollary}
The following inequality holds for a Markov chain $X\rightarrow Y\rightarrow
Z$:%
\begin{equation}
I(X;Y)\geq I(X;Y|Z).
\end{equation}

\end{corollary}

\subsubsection{Relative Entropy Data-Processing Inequality}%

\index{data processing inequality!relative entropy}%
Another kind of data-processing inequality holds for the relative entropy,
known as monotonicity of relative entropy. This also is a consequence of the
non-negativity of relative entropy in Theorem~\ref{thm-cie:rel-ent-positive}.

\begin{corollary}
[Monotonicity of Relative Entropy]\label{cor-cie:mono-rel-ent}Let $p$ be a
probability distribution on an alphabet $\mathcal{X}$ and let $q:\mathcal{X}%
\rightarrow\lbrack0,\infty)$. Let $N(y|x)$ be a conditional probability
distribution (i.e., a classical channel). Then the relative entropy does not
increase after the channel $N(y|x)$ acts on $p$ and $q$:%
\begin{equation}
D(p\Vert q)\geq D(Np\Vert Nq), \label{eq-cie:rel-ent-mono}%
\end{equation}
where $Np$ is a probability distribution with elements $(Np)(y)\equiv\sum
_{x}N(y|x)p(x)$ and $Nq$ is a vector with elements $(Nq)(y)=\sum
_{x}N(y|x)q(x)$. Let $R$ be the channel defined by the following set of
equations:%
\begin{equation}
R(x|y)(Nq)(y)=N(y|x)q(x).
\end{equation}
The inequality in \eqref{eq-cie:rel-ent-mono}\ is saturated (i.e., $D(p\Vert
q)=D(Np\Vert Nq)$) if and only if $RNp=p$, where $RNp$ is a probability
distribution with elements $(RNp)(x)=\sum_{y,x^{\prime}}R(x|y)N(y|x^{\prime
})p(x^{\prime})$.
\end{corollary}

\begin{proof}
First, if $p$ and $q$ are such that $\operatorname{supp}(p)\not \subseteq
\operatorname{supp}(q)$, then the inequality is trivially true because
$D(p\Vert q)=+\infty$ in this case. So let us suppose that
$\operatorname{supp}(p)\subseteq\operatorname{supp}(q)$, which implies that
$\operatorname{supp}(Np)\subseteq\operatorname{supp}(Nq)$. Our first step is
to rewrite the terms in the inequality. To this end, consider the following
algebraic manipulations:%
\begin{align}
D(Np\Vert Nq)  &  =\sum_{y}(Np)(y)\log\left(  \frac{(Np)(y)}{(Nq)(y)}\right)
\\
&  =\sum_{y,x}N(y|x)p(x)\log\left(  \frac{(Np)(y)}{(Nq)(y)}\right) \\
&  =\sum_{x}p(x)\left[  \sum_{y}N(y|x)\log\left(  \frac{(Np)(y)}%
{(Nq)(y)}\right)  \right] \\
&  =\sum_{x}p(x)\log\exp\left[  \sum_{y}N(y|x)\log\left(  \frac{(Np)(y)}%
{(Nq)(y)}\right)  \right]  .
\end{align}
This implies that%
\begin{equation}
D(p\Vert q)-D(Np\Vert Nq)=D(p\Vert r), \label{eq-cie:rel-ent-mono-rewrite}%
\end{equation}
where%
\begin{equation}
r(x)\equiv q(x)\exp\left[  \sum_{y}N(y|x)\log\left(  \frac{(Np)(y)}%
{(Nq)(y)}\right)  \right]  .
\end{equation}
Now consider that%
\begin{align}
\sum_{x}r(x)  &  =\sum_{x}q(x)\exp\left[  \sum_{y}N(y|x)\log\left(
\frac{(Np)(y)}{(Nq)(y)}\right)  \right] \\
&  \leq\sum_{x}q(x)\sum_{y}N(y|x)\exp\left[  \log\left(  \frac{(Np)(y)}%
{(Nq)(y)}\right)  \right] \\
&  =\sum_{x}q(x)\sum_{y}N(y|x)\left(  \frac{(Np)(y)}{(Nq)(y)}\right) \\
&  =\sum_{y}\left[  \sum_{x}q(x)N(y|x)\right]  \frac{(Np)(y)}{(Nq)(y)}\\
&  =\sum_{y}(Np)(y) =1.
\end{align}
The inequality in the second line follows from convexity of the exponential
function. Since $r$ is a vector such that $\sum_{x}r(x)\leq1$, we can conclude
from Theorem~\ref{thm-cie:rel-ent-positive}\ that $D(p\Vert r)\geq0$, which by
\eqref{eq-cie:rel-ent-mono-rewrite}\ is the same as \eqref{eq-cie:rel-ent-mono}.

We now comment on the saturation conditions. First, suppose that $RNp=p$. By
monotonicity of relative entropy (what was just proved) under the application
of the channel $R$, we find that%
\begin{equation}
D(Np\Vert Nq)\geq D(RNp\Vert RNq)=D(p\Vert q),
\end{equation}
where the equality follows from the assumption that $RNp=p$, and the fact that
$RNq=q$. This last statement follows because%
\begin{equation}
(RNq)(x)=\sum_{y}R(x|y)(Nq)(y)=\sum_{y}N(y|x)q(x)=q(x).
\end{equation}
The other implication $D(p\Vert q)=D(Np\Vert Nq)\Rightarrow RNp=p$ is a
consequence of a later development, Theorem~\ref{thm-cie:refine-mono-rel-ent}.
\end{proof}

\subsection{Fano's Inequality}%

\index{Fano's inequality}%
Another entropy inequality that we consider is Fano's inequality. This
inequality also finds application in the converse proof of a coding theorem.

\begin{figure}
[ptb]
\begin{center}
\includegraphics[
width=1.8049in
]%
{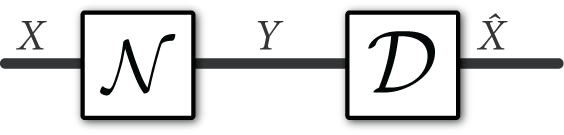}%
\caption{The classical communication scenario relevant in Fano's inequality.
Alice transmits a random variable $X$ over a noisy channel $\mathcal{N}$,
producing a random variable $Y$. Bob receives $Y$ and processes it according
to some decoding map $\mathcal{D}$ to produce his best estimate $\hat{X}$ of
$Y$.}%
\label{fig-ie:fano}%
\end{center}
\end{figure}

Fano's inequality applies to a general classical communication scenario.
Suppose Alice possesses some random variable $X$ that she transmits to Bob
over a noisy communication channel. Let $p_{Y|X}( y|x) $ denote the stochastic
map corresponding to the noisy communication channel. Bob receives a random
variable $Y$ from the channel and processes it in some way to produce his best
estimate $\hat{X}$ of the original random variable $X$.
Figure~\ref{fig-ie:fano}\ depicts this scenario.

The natural performance metric of this communication scenario is the
probability of error $p_{e}\equiv\Pr\{\hat{X}\neq X\}$---a low probability of
error corresponds to good performance. On the other hand, consider the
conditional entropy $H( X|Y) $. We interpreted it before as the uncertainty
about $X$ from the perspective of someone who already knows $Y$. If the
channel is noiseless ($p_{Y|X}( y|x) =\delta_{y,x}$), then there is no
uncertainty about $X$ because $Y$ is identical to $X$:%
\begin{equation}
H( X|Y) =0.
\end{equation}
As the channel becomes noisier, the conditional entropy $H( X|Y) $ increases
away from zero. In this sense, the conditional entropy $H( X|Y) $\ quantifies
the information about $X$ that is lost in the channel noise. We then might
naturally expect there to be a relationship between the probability of error
$p_{e}$\ and the conditional entropy $H( X|Y) $: the amount of information
lost in the channel should be low if the probability of error is low. Fano's
inequality provides a quantitative bound corresponding to this idea.%

\begin{theorem}
[Fano's Inequality]\label{thm-cie:fano}Suppose that Alice sends a random
variable $X$ through a noisy channel to produce random variable $Y$ and
further processing of $Y$ gives an estimate $\hat{X}$ of $X$. Thus,
$X\rightarrow Y\rightarrow\hat{X}$ forms a Markov chain. Let $p_{e}\equiv
\Pr\{\hat{X}\neq X\}$ denote the probability of error. Then the following
function of the error probability $p_{e}$ bounds the information lost in the
channel noise:%
\begin{equation}
H(X|Y)\leq H(X|\hat{X})\leq h_{2}(p_{e})+p_{e}\log\left(  \left\vert
\mathcal{X}\right\vert -1\right)  ,
\end{equation}
where $h_{2}(p_{e})$ is the binary entropy function. In particular, note that%
\begin{equation}
\lim_{p_{e}\rightarrow0}h_{2}(p_{e})+p_{e}\log\left(  \left\vert
\mathcal{X}\right\vert -1\right)  =0.
\end{equation}
The bound is sharp in the sense that there is a choice of random variables $X$
and $Y$ that saturates it.
\end{theorem}

\begin{proof}
Let $E$ denote an indicator random variable that indicates whether an error
occurs:%
\begin{equation}
E=\left\{
\begin{array}
[c]{ccc}%
0 & : & X=\hat{X}\\
1 & : & X\neq\hat{X}%
\end{array}
\right.  .
\end{equation}
Consider the entropy%
\begin{equation}
H(EX|\hat{X})=H(X|\hat{X})+H(E|X\hat{X}).
\end{equation}
The entropy $H(E|X\hat{X})$ on the right-hand side\ vanishes because there is
no uncertainty about the indicator random variable $E$ if we know both $X$ and
$\hat{X}$. Thus,%
\begin{equation}
H(EX|\hat{X})=H(X|\hat{X}). \label{eq-ie:fano-1}%
\end{equation}
Also, the data-processing inequality applies to the Markov chain $X\rightarrow
Y\rightarrow\hat{X}$:%
\begin{equation}
I(X;Y)\geq I(X;\hat{X}),
\end{equation}
and implies the following inequality:%
\begin{equation}
H(X|\hat{X})\geq H(X|Y). \label{eq-ie:fano-2}%
\end{equation}
Consider the following chain of inequalities:%
\begin{align}
H(EX|\hat{X})  &  =H(E|\hat{X})+H(X|E\hat{X})\\
&  \leq H(E)+H(X|E\hat{X})\\
&  =h_{2}(p_{e})+p_{e}H(X|\hat{X},E=1)\nonumber\\
&  \ \ \ \ \ \ \ +\left(  1-p_{e}\right)  H(X|\hat{X},E=0)\\
&  \leq h_{2}(p_{e})+p_{e}\log\left(  \left\vert \mathcal{X}\right\vert
-1\right)  . \label{eq-ie:fano-3}%
\end{align}
The first equality follows by expanding the entropy $H(EX|\hat{X})$. The first
inequality follows because conditioning reduces entropy. The second equality
follows by explicitly expanding the conditional entropy $H(X|E\hat{X})$ in
terms of the two possibilities of the error random variable $E$. The last
inequality follows from two facts: there is no uncertainty\ about $X$ when
there is no error (when $E=0$) and $\hat{X}$ is available, and the uncertainty
about $X$ when there is an error (when $E=1$) and $\hat{X}$ is available is
less than the uncertainty of a uniform distribution $\frac{1}{\left\vert
\mathcal{X}\right\vert -1}$\ for all of the other possibilities. Fano's
inequality follows from putting together \eqref{eq-ie:fano-2},
\eqref{eq-ie:fano-1}, and \eqref{eq-ie:fano-3}:%
\begin{equation}
H(X|Y)\leq H(X|\hat{X})=H(EX|\hat{X})\leq h_{2}(p_{e})+p_{e}\log\left(
\left\vert \mathcal{X}\right\vert -1\right)  .
\end{equation}

To establish the statement of saturation, let $X$ be a uniform random variable
and let $X^{\prime}$ be a copy of it (so that the joint distribution is
$p_{X,X^{\prime}}(x,x^{\prime}) = \delta_{x,x^{\prime}}/|\mathcal{X}|$ where
$\mathcal{X}$ is the alphabet). Let $p_{Y|X^{\prime}}$ denote a symmetric
channel such that the input $x^{\prime}$ is transmitted faithfully with
probability $1-\varepsilon$ and becomes one of the other $|\mathcal{X}|-1$
letters with probability $\varepsilon/ (|\mathcal{X}|-1)$, where
$\varepsilon\in[0,1]$. Observe that $Y$ is a uniform random variable, given
that $p_{Y|X^{\prime}}$ is a symmetric channel. Then $\operatorname{Pr}\{X\neq
Y\} = \varepsilon$, and we find that
\begin{equation}
H(X|Y) = H(Y|X) + H(X) - H(Y) = H(Y|X) = h_{2}(\varepsilon)+\varepsilon
\log\left(  \left\vert \mathcal{X}\right\vert -1\right) ,
\end{equation}
concluding the proof.
\end{proof}

\subsection{Continuity of Entropy}%

\index{Shannon entropy!continuity}%
That the entropy is a continuous function follows from the fact that the
function $-x\log x$ is continuous. However, it can be useful in applications
to have explicit continuity bounds. Before we give such bounds, we should
establish how we measure distance between probability distributions. A natural
way for doing so is to use the classical trace distance, defined as follows:

\begin{definition}
[Classical Trace Distance]Let $p,q:\mathcal{X}\rightarrow\mathbb{R}$, where
$\mathcal{X}$ is a finite alphabet. The classical trace distance between $p$
and $q$ is then%
\begin{equation}
\left\Vert p-q\right\Vert _{1}\equiv\sum_{x}\left\vert p(x)-q(x)\right\vert .
\end{equation}

\end{definition}

The classical trace distance is a special case of the trace distance from
Definition~\ref{def-dm:trace-distance}, in which we place the entries of $p$
and $q$ along the diagonal of some square matrices. If $p$ and $q$ are
probability distributions, then the operational meaning of the classical trace
distance is the same as it was in the fully quantum case:\ it is the bias in
the probability with which one could successfully distinguish $p$ and $q$ by
means of any binary hypothesis test. The optimal test is given in the
following lemma:

\begin{lemma}
\label{lem-cie:trace-dist-test-A}Let $p$ and $q$ be probability distributions
on a finite alphabet $\mathcal{X}$. Let $A\equiv\left\{  x:p(x)\geq
q(x)\right\}  $. Then%
\begin{equation}
\frac{1}{2}\left\Vert p-q\right\Vert _{1}=p(A)-q(A),
\end{equation}
where $p(A)\equiv\sum_{x\in A}p(x)$ and $q(A)\equiv\sum_{x\in A}q(x)$.
\end{lemma}

\begin{proof}
A proof for this lemma is very similar to that for
Lemma~\ref{lemma:trace-equiv}. It is brief and so we provide it here. Consider
that%
\begin{align}
0  &  =\sum_{x}\left[  p(x)-q(x)\right] \\
&  =\sum_{x\in A}\left[  p(x)-q(x)\right]  +\sum_{x\in A^{c}}\left[
p(x)-q(x)\right]  ,
\end{align}
which implies that%
\begin{equation}
\sum_{x\in A}\left[  p(x)-q(x)\right]  =\sum_{x\in A^{c}}\left[
q(x)-p(x)\right]  . \label{eq-cie:classical-TD-set-A}%
\end{equation}
Now consider that%
\begin{align}
\left\Vert p-q\right\Vert _{1}  &  =\sum_{x}\left\vert p(x)-q(x)\right\vert \\
&  =\sum_{x\in A}\left[  p(x)-q(x)\right]  +\sum_{x\in A^{c}}\left[
q(x)-p(x)\right] \\
&  =2\sum_{x\in A}\left[  p(x)-q(x)\right]  ,
\end{align}
where the last line follows from \eqref{eq-cie:classical-TD-set-A}.
\end{proof}

One can interpret the set $A$ and its complement $A^{c}$ as a binary
hypothesis test that one could perform after receiving a sample $x$ from
either the distribution $p$ or $q$ (without knowing which distribution was
used to generate the sample). If the sample $x$ is in $A$, then we would
decide that $p$ generated $x$, and otherwise, we would decide that $q$
generated $x$. The success probability for distinguishing $p$ from $q$ is then%
\begin{align}
\frac{1}{2}\left[  \sum_{x\in A}p(x)+\sum_{x\in A^{c}}q(x)\right]   &
=\frac{1}{2}\left[  1+\sum_{x\in A}\left[  p(x)-q(x)\right]  \right] \\
&  =\frac{1}{2}\left[  1+\frac{1}{2}\left\Vert p-q\right\Vert _{1}\right]  .
\end{align}
It turns out that this test is the optimal test, which follows from the same
reasoning given in Section~\ref{sec-dm:op-int-trace}. Thus, we can interpret
the classical trace distance as an important measure of distinguishability
between probability distributions, with an operational interpretation as given above.

We can now state an important entropy continuity bound.

\begin{theorem}
[Zhang--Audenaert]\label{thm-cie:cl-ent-continuity}Let $X$ and $Y$ be discrete
random variables taking values in a finite alphabet $\mathcal{A}$. Let $p_{X}$
and $p_{Y}$ denote their distributions, respectively. Then the following bound
holds%
\begin{equation}
\left\vert H(X)-H(Y)\right\vert \leq T\log(\left\vert \mathcal{A}\right\vert
-1)+h_{2}(T),
\end{equation}
where $T\equiv\frac{1}{2}\left\Vert p_{X}-p_{Y}\right\Vert _{1}$. Furthermore,
this bound is optimal, meaning that there exists a pair of random variables
saturating the bound for every $T\in\left[  0,1\right]  $ and alphabet size
$\left\vert \mathcal{A}\right\vert $.
\end{theorem}

In order to prove this theorem, we need to establish the notion of a maximal
coupling of two random variables.

\begin{definition}
[Coupling]A coupling of a pair $(X,Y)$ of two random variables is a pair
$(\hat{X},\hat{Y})$ of two other random variables that have the same marginal
distributions as those of $X$ and $Y$.
\end{definition}

\begin{definition}
[Maximal Coupling]A coupling $(\hat{X},\hat{Y})$\ of a pair of two random
variables $(X,Y)$ is maximal if $\Pr\{\hat{X}=\hat{Y}\}$ takes on its maximum
value, with respect to all couplings of $X$ and $Y$.
\end{definition}

We can now relate the classical trace distance to a maximal coupling:

\begin{lemma}
Let $X$ and $Y$ be discrete random variables taking values in a finite
alphabet~$\mathcal{A}$. Let $p_{X}$ and $p_{Y}$ denote their distributions,
respectively. If $(\hat{X},\hat{Y})$ is a maximal coupling of $X$ and $Y$,
then the following equalities hold%
\begin{align}
\Pr\{\hat{X}  &  =\hat{Y}\}=\sum_{u\in\mathcal{A}}\min\{p_{X}(u),p_{Y}%
(u)\},\label{eq-cie:TV-coupling-eq}\\
\Pr\{\hat{X}  &  \neq\hat{Y}\}=\frac{1}{2}\left\Vert p_{X}-p_{Y}\right\Vert
_{1}. \label{eq-cie:TV-coupling-neq}%
\end{align}

\end{lemma}

\begin{proof}
Let $\mathcal{B}=\left\{  u\in\mathcal{A}:p_{X}(u)<p_{Y}(u)\right\}  $. Let
$\mathcal{B}^{c}=\mathcal{A}\backslash\mathcal{B}$. Then for every coupling
$(\hat{X},\hat{Y})$ of $(X,Y)$, the following holds%
\begin{align}
\Pr\{\hat{X}=\hat{Y}\}  &  =\Pr\{\hat{X}=\hat{Y}\wedge\hat{Y}\in
\mathcal{B}\}+\Pr\{\hat{X}=\hat{Y}\wedge\hat{Y}\in\mathcal{B}^{c}\}\\
&  \leq\Pr\{\hat{X}\in\mathcal{B}\}+\Pr\{\hat{Y}\in\mathcal{B}^{c}\}\\
&  =\Pr\{X\in\mathcal{B}\}+\Pr\{Y\in\mathcal{B}^{c}\}\\
&  =\sum_{u\in\mathcal{B}}p_{X}(u)+\sum_{u\in\mathcal{B}^{c}}p_{Y}(u)\\
&  =\sum_{u\in\mathcal{B}}\min\{p_{X}(u),p_{Y}(u)\}+\sum_{u\in\mathcal{B}^{c}%
}\min\{p_{X}(u),p_{Y}(u)\}\\
&  =\sum_{u\in\mathcal{A}}\min\{p_{X}(u),p_{Y}(u)\}.
\label{eq-cie:last-step-coupling-bnd}%
\end{align}

We now establish a construction of a coupling that achieves the bound above,
so that it is maximal. Let $q\equiv\sum_{u\in\mathcal{A}}\min\{p_{X}%
(u),p_{Y}(u)\}$. Let $U$, $V$, $W$, and $J$ be independent discrete random
variables, with $p_{J}(0)=1-q$ and $p_{J}(1)=q$. Let $U$, $V$, and $W$ have
the following probability distributions:%
\begin{align}
p_{U}(u)  &  =\frac{1}{q}\left[  \min\{p_{X}(u),p_{Y}(u)\}\right]  ,\\
p_{V}(v)  &  =\frac{1}{1-q}\left[  p_{X}(v)-\min\{p_{X}(v),p_{Y}(v)\}\right]
,\\
p_{W}(w)  &  =\frac{1}{1-q}\left[  p_{Y}(w)-\min\{p_{X}(w),p_{Y}(w)\}\right]
,
\end{align}
where $u,v,w\in\mathcal{A}$. If $J=1$, then let $\hat{X}=\hat{Y}=U$, and if
$J=0$, then let $\hat{X}=V$ and $\hat{Y}=W$. So we find that for all
$x,y\in\mathcal{A}$%
\begin{align}
p_{\hat{X}}(x)  &  =q\ p_{X|J}(x|1)+\left(  1-q\right)  p_{X|J}(x|0)\\
&  =q\ p_{U}(x)+\left(  1-q\right)  p_{V}(x)\\
&  =p_{X}(x).
\end{align}
By similar reasoning, it follows that $p_{\hat{Y}}(y)=p_{Y}(y)$, so that we
can conclude that the constructed $(\hat{X},\hat{Y})$ is a coupling of
$(X,Y)$. It is maximal because%
\begin{equation}
\Pr\{\hat{X}=\hat{Y}\}\geq\Pr\left\{  J=1\right\}  =q,
\end{equation}
and we can conclude from \eqref{eq-cie:last-step-coupling-bnd} that $\Pr
\{\hat{X}=\hat{Y}\}=q$. The equality in \eqref{eq-cie:TV-coupling-neq} follows
from \eqref{eq-cie:TV-coupling-eq} and the following equality $\min\left\{
a,b\right\}  =\frac{1}{2}[a+b-\left\vert a-b\right\vert ]$, which holds for
all $a,b\in\mathbb{R}$.
\end{proof}

We can finally give a proof of Theorem~\ref{thm-cie:cl-ent-continuity}, by
invoking the above results and Fano's inequality (Theorem~\ref{thm-cie:fano}).

\begin{proof}
[Proof of Theorem~\ref{thm-cie:cl-ent-continuity}]Let $(\hat{X},\hat{Y})$ be a
maximal coupling of $(X,Y)$. Then
\begin{align}
\left\vert H(X)-H(Y)\right\vert  &  =\left\vert H(\hat{X})-H(\hat
{Y})\right\vert \\
&  =\left\vert H(\hat{X})-H(\hat{X}\hat{Y})+H(\hat{X}\hat{Y})-H(\hat
{Y})\right\vert \\
&  =\left\vert H(\hat{X}|\hat{Y})-H(\hat{Y}|\hat{X})\right\vert \\
&  \leq\max\left\{  H(\hat{X}|\hat{Y}),H(\hat{Y}|\hat{X})\right\} \\
&  \leq\Pr\{\hat{X}\neq\hat{Y}\}\log\left(  \left\vert \mathcal{A}\right\vert
-1\right)  +h_{2}(\Pr\{\hat{X}\neq\hat{Y}\})\\
&  =T\log\left(  \left\vert \mathcal{A}\right\vert -1\right)  +h_{2}(T).
\end{align}
The first equality follows because $(\hat{X},\hat{Y})$ is a maximal coupling
of $(X,Y)$. The second inequality is an application of Fano's inequality
(Theorem~\ref{thm-cie:fano}). The last equality follows from \eqref{eq-cie:TV-coupling-neq}.

The bound given in the statement of the theorem is optimal. An example of a
pair of probability distributions $p_{X}$ and $p_{Y}$\ saturating the bound is%
\begin{align}
p_{X}  &  =%
\begin{bmatrix}
1 & 0 & \cdots & 0
\end{bmatrix}
,\\
p_{Y}  &  =%
\begin{bmatrix}
1-\varepsilon & \varepsilon/\left(  \left\vert \mathcal{A}\right\vert
-1\right)  & \cdots & \varepsilon/\left(  \left\vert \mathcal{A}\right\vert
-1\right)
\end{bmatrix}
,
\end{align}
where $\varepsilon\in\left[  0,1\right]  $. The normalized classical trace
distance between $p_{X}$ and $p_{Y}$ is $\frac{1}{2}\left\Vert p_{X}%
-p_{Y}\right\Vert _{1}=\varepsilon$, while an explicit calculation reveals
that%
\begin{equation}
\left\vert H(X)-H(Y)\right\vert =H(Y)=\varepsilon\log\left(  \left\vert
\mathcal{A}\right\vert -1\right)  +h_{2}(\varepsilon),
\end{equation}
concluding the proof.
\end{proof}

\section{Near Saturation of Entropy Inequalities}

\label{sec-cie:entropy-ineq-refinements}The entropy inequalities discussed in
the previous section are foundational in classical information theory. In
fact, one can prove the converse parts of many classical capacity theorems by
making use of these entropy inequalities. Thus, it seems worthwhile to study
them in more detail and to explore the possibility of more refined statements.
For example, in each of these entropy inequalities, there is a condition for
when it is saturated:\ the non-negativity of relative entropy is saturated if
and only if $p=q$ and the non-negativity of mutual information is saturated if
and only if random variables $X$ and $Y$ are independent. It is thus a natural
question to consider what kind of statement we can make when these entropy
inequalities are nearly saturated.

\subsection{Pinsker Inequality}

One of the main tools for refining entropy inequalities in the classical case
is the Pinsker inequality, which provides a relation between the relative
entropy and the classical trace distance.

\begin{theorem}
[Pinsker Inequality]\label{thm-cie:pinsker}Let $p$ be a probability
distribution
\index{Pinsker inequality}
on a finite alphabet $\mathcal{X}$ and let $q:\mathcal{X}\rightarrow
\lbrack0,1]$ be such that $\sum_{x}q(x)\leq1$. Then%
\begin{equation}
D(p\Vert q)\geq\frac{1}{2\ln2}\left\Vert p-q\right\Vert _{1}^{2}.
\end{equation}

\end{theorem}

The utility of the Pinsker inequality is that it relates one measure of
distinguishability to another, and thus allows us to make precise statements
about the near saturation of an entropy inequality. For example,
Theorem~\ref{thm-cie:rel-ent-positive}\ states that $D(p\Vert q)\geq0$ for $p$
and $q$ as given above. The Pinsker inequality is a strong refinement of this
entropy inequality: it says more than just $D(p\Vert q)\geq0$, considering
that $\left\Vert p-q\right\Vert _{1}\geq0$. Letting $q$ be subnormalized as
above gives us a little more flexibility when applying the Pinsker inequality.

Before proving it, we establish the following lemma:

\begin{lemma}
\label{lem-cie:calc-lemma-pinsker}Let $a,b\in\left[  0,1\right]  $. Then
$a\ln\left(  \frac{a}{b}\right)  +\left(  1-a\right)  \ln\left(  \frac
{1-a}{1-b}\right)  \geq2\left(  a-b\right)  ^{2}$.
\end{lemma}

\begin{proof}
This bound follows by elementary calculus. First, if $b=0$ or $b=1$, then the
bound holds trivially. So suppose that $b\in\left(  0,1\right)  $. Let us
suppose furthermore (for now)\ that $a\geq b$. Consider the following
function:%
\begin{equation}
g(a,b)\equiv a\ln\left(  \frac{a}{b}\right)  +\left(  1-a\right)  \ln\left(
\frac{1-a}{1-b}\right)  -2\left(  a-b\right)  ^{2},
\end{equation}
so that $g(a,b)$ corresponds to the difference of the left-hand side\ and the
right-hand side\ in the statement of the lemma. Then%
\begin{align}
\frac{\partial g(a,b)}{\partial b}  &  =-\frac{a}{b}+\frac{1-a}{1-b}-4\left(
b-a\right) \\
&  =-\frac{a\left(  1-b\right)  }{b\left(  1-b\right)  }+\frac{b\left(
1-a\right)  }{b\left(  1-b\right)  }-4\left(  b-a\right) \\
&  =\frac{b-a}{b\left(  1-b\right)  }-4\left(  b-a\right) \\
&  =\frac{\left(  b-a\right)  \left(  4b^{2}-4b+1\right)  }{b\left(
1-b\right)  }\\
&  =\frac{\left(  b-a\right)  \left(  2b-1\right)  ^{2}}{b\left(  1-b\right)
}\\
&  \leq0,
\end{align}
with the last step holding from the assumption that $a\geq b$ and $b\in\left(
0,1\right)  $. Also, observe that both $\partial g(a,b)/\partial b=0$ and
$g(a,b)=0$ when $a=b$. Thus, the function $g(a,b)$ is decreasing in $b$ for
every $a$ whenever $b\leq a$ and reaches a minimum when $b=a$. So
$g(a,b)\geq0$ whenever $a\geq b$. The lemma holds in general by applying it to
$a^{\prime}\equiv1-a$ and $b^{\prime}\equiv1-b$ so that $b^{\prime}\geq
a^{\prime}$.
\end{proof}

\bigskip

\begin{proof}
[Proof of Pinsker's inequality (Theorem~\ref{thm-cie:pinsker})]The main idea
is to exploit the monotonicity of relative entropy and the lemma above. First
let us suppose that $q$ is a probability distribution. Let $\mathcal{A}$
denote a classical \textquotedblleft coarse-graining\textquotedblright%
\ channel that outputs $1$ if $x\in A$ and $0$ if $x\in A^{c}$, where $A$ is
the set defined in Lemma~\ref{lem-cie:trace-dist-test-A} (i.e., $A\equiv
\left\{  x:p(x)\geq q(x)\right\}  $). Then%
\begin{align}
D(p\Vert q)  &  \geq D(\mathcal{A}(p)\Vert\mathcal{A}(q))\\
&  =p(A)\log\left(  \frac{p(A)}{q(A)}\right)  +\left(  1-p(A)\right)
\log\left(  \frac{1-p(A)}{1-q(A)}\right) \\
&  \geq\frac{2}{\ln2}\left(  p(A)-q(A)\right)  ^{2}\\
&  =\frac{2}{\ln2}\left(  \frac{1}{2}\left\Vert p-q\right\Vert _{1}\right)
^{2}\\
&  =\frac{1}{2\ln2}\left\Vert p-q\right\Vert _{1}^{2}.
\end{align}
The first inequality follows from monotonicity of the relative entropy
(Corollary~\ref{cor-cie:mono-rel-ent}), with the quantities $p(A)$ and $q(A)$
defined in Lemma~\ref{lem-cie:trace-dist-test-A}. The second inequality is an
application of Lemma~\ref{lem-cie:calc-lemma-pinsker}, where we gain the
factor $\ln2$ by converting from natural logarithm to binary logarithm. The
second equality follows from Lemma~\ref{lem-cie:trace-dist-test-A}.

If $q$ is not a probability distribution (so that $\sum_{x}q(x)<1$), then we
can add an extra letter to $\mathcal{X}$ to make $q$ a probability
distribution. That is, we define a probability distribution $q^{\prime}$\ to
have its first $\left\vert \mathcal{X}\right\vert $ entries to be those from
$q$ and its last entry to be $1-\sum_{x}q(x)$. We also define an augmented
distribution $p^{\prime}$ to have its first $\left\vert \mathcal{X}\right\vert
$ entries to be those from $p$ and its last entry to be $0$. Then $0\cdot
\log\left(  0/\left[  1-\sum_{x}q(x)\right]  \right)  =0$, so that%
\begin{align}
D(p\Vert q)  &  =D(p^{\prime}\Vert q^{\prime})\\
&  \geq\frac{1}{2\ln2}\left\Vert p^{\prime}-q^{\prime}\right\Vert _{1}^{2}\\
&  =\frac{1}{2\ln2}\left(  \left\Vert p-q\right\Vert _{1}+\left[  1-\sum
_{x}q(x)\right]  \right)  ^{2}\\
&  \geq\frac{1}{2\ln2}\left\Vert p-q\right\Vert _{1}^{2},
\end{align}
concluding the proof.
\end{proof}

\subsection{Refinements of Entropy Inequalities}

The first refinement of an entropy inequality that we give is regarding the
non-negativity of mutual information and its maximal value:

\begin{theorem}
\label{thm-cie:mut-info-to-prod}Let $X$ and $Y$ be discrete random variables
taking values in alphabets $\mathcal{X}$ and $\mathcal{Y}$, respectively, let
$p_{XY}$ denote their joint distribution, and let $p_{X}\otimes p_{Y}$ denote
the product of their marginal distributions. Then%
\begin{equation}
I(X;Y)\geq\frac{2}{\ln2}\Delta^{2}, \label{eq-cie:mut-info-pinsker}%
\end{equation}
and%
\begin{equation}
I(X;Y)\leq\Delta\log(\min\{\left\vert \mathcal{X}\right\vert ,\left\vert
\mathcal{Y}\right\vert \}-1)+h_{2}(\Delta), \label{eq-cie:mutual-info-fannes}%
\end{equation}
where $\Delta=\frac{1}{2}\left\Vert p_{XY}-p_{X}\otimes p_{Y}\right\Vert _{1}$.
\end{theorem}

\begin{proof}
The first inequality is a direct application of the Pinsker inequality
(Theorem~\ref{thm-cie:pinsker}), noting that%
\begin{equation}
I(X;Y)=D(p_{XY}\Vert p_{X}\otimes p_{Y}).
\end{equation}
To prove the second inequality, let $\hat{X}$ and $\hat{Y}$ denote a pair of
random variables with joint distribution $p_{X}\otimes p_{Y}$. Then%
\begin{align}
I(X;Y)  &  =\left\vert I(X;Y)-I(\hat{X};\hat{Y})\right\vert \\
&  =\left\vert H(X|Y)-H(\hat{X}|\hat{Y})\right\vert \\
&  =\left\vert \sum_{y}p_{Y}(y)\left[  H(X|Y=y)-H(\hat{X}|\hat{Y}=y)\right]
\right\vert .
\end{align}
The first equality follows because $I(\hat{X};\hat{Y})=0$. The second equality
follows because $I(X;Y)=H(X)-H(X|Y)$, $I(\hat{X};\hat{Y})=H(\hat{X})-H(\hat
{X}|\hat{Y})$, and $H(X)=H(\hat{X})$. The third equality follows from the
definition of conditional entropy. Continuing,%
\begin{align}
&  \leq\sum_{y}p_{Y}(y)\left\vert H(X|Y=y)-H(\hat{X}|\hat{Y}=y)\right\vert \\
&  \leq\sum_{y}p_{Y}(y)\left[  \Delta(y)\log(\left\vert \mathcal{X}\right\vert
-1)+h_{2}(\Delta(y))\right] \\
&  \leq\Delta\log(\left\vert \mathcal{X}\right\vert -1)+h_{2}(\Delta).
\end{align}
The first inequality is a consequence of the triangle inequality. The second
inequality follows from an application of
Theorem~\ref{thm-cie:cl-ent-continuity}, and from defining%
\begin{equation}
\Delta(y)=\frac{1}{2}\sum_{x}\left\vert p_{X|Y}(x|y)-p_{X}(x)\right\vert .
\end{equation}
The final inequality follows because%
\begin{align}
\Delta &  =\frac{1}{2}\sum_{x,y}\left\vert p_{XY}(x,y)-p_{X}(x)p_{Y}%
(y)\right\vert \\
&  =\sum_{y}p_{Y}(y)\left[  \frac{1}{2}\sum_{x}\left\vert p_{X|Y}%
(x|y)-p_{X}(x)\right\vert \right] \\
&  =\sum_{y}p_{Y}(y)\Delta(y),
\end{align}
and from the fact that the binary entropy is concave. We obtain the other
bound%
\begin{equation}
I(X;Y)\leq\Delta\log(\left\vert \mathcal{Y}\right\vert -1)+h_{2}(\Delta),
\end{equation}
by expanding the mutual information as $I(X;Y)=|I(X;Y)-I(\hat{X};\hat
{Y})|=|H(Y|X)-H(\hat{Y}|\hat{X})|$ and proceeding in a similar fashion.
\end{proof}

We can make a similar kind of statement for the conditional mutual information:

\begin{theorem}
\label{thm-cie:cond-mut-info-to-Markov}Let $X$, $Y$, and $Z$ be discrete
random variables taking values in alphabets $\mathcal{X}$, $\mathcal{Y}$, and
$\mathcal{Z}$, respectively, let $p_{XYZ}$ denote their joint distribution,
and let $p_{X|Z}p_{Y|Z}p_{Z}$ denote another distribution which corresponds to
the Markov chain $X-Z-Y$. Then%
\begin{equation}
I(X;Y|Z)\geq\frac{2}{\ln2}\Delta^{2},
\end{equation}
and%
\begin{equation}
I(X;Y|Z)\leq\Delta\log(\min\{\left\vert \mathcal{X}\right\vert ,\left\vert
\mathcal{Y}\right\vert \}-1)+h_{2}(\Delta),
\end{equation}
where $\Delta=\frac{1}{2}\left\Vert p_{XYZ}-p_{X|Z}p_{Y|Z}p_{Z}\right\Vert
_{1}$.
\end{theorem}

\begin{proof}
A proof for the first inequality is similar to the proof of
\eqref{eq-cie:mut-info-pinsker}. One can write $I(X;Y|Z)=\sum_{z}%
p_{Z}(z)I(X;Y|Z=z)$, apply the Pinsker inequality
(Theorem~\ref{thm-cie:pinsker}), and then convexity of the square function and
the trace norm. Alternatively, one can directly compute that
$I(X;Y|Z)=D(p_{XYZ}\Vert p_{X|Z}p_{Y|Z}p_{Z})$ and then apply the Pinsker
inequality (Theorem~\ref{thm-cie:pinsker}).

A proof for the second inequality is similar to that for
\eqref{eq-cie:mutual-info-fannes}. Let $\hat{X}$, $\hat{Y}$, and $\hat{Z}$
denote a triple of random variables with joint distribution $p_{X|Z}%
p_{Y|Z}p_{Z}$. Then%
\begin{align}
I(X;Y|Z)  &  =\left\vert I(X;Y|Z)-I(\hat{X};\hat{Y}|\hat{Z})\right\vert \\
&  =\left\vert H(Y|Z)-H(Y|XZ)-\left[  H(\hat{Y}|\hat{Z})-H(\hat{Y}|\hat{X}%
\hat{Z})\right]  \right\vert \\
&  =\left\vert H(\hat{Y}|\hat{X}\hat{Z})-H(Y|XZ)\right\vert ,
\end{align}
where the first equality follows because $I(\hat{X};\hat{Y}|\hat{Z})=0$, the
second by expanding the conditional mutual informations, and the third because
$H(Y|Z)=H(\hat{Y}|\hat{Z})$. The rest of the steps proceed as in the proof of
\eqref{eq-cie:mutual-info-fannes}, and we end up with%
\begin{equation}
I(X;Y|Z)\leq\Delta\log(\left\vert \mathcal{Y}\right\vert -1)+h_{2}(\Delta).
\end{equation}
A proof for $I(X;Y|Z)\leq\Delta\log(\left\vert \mathcal{X}\right\vert
-1)+h_{2}(\Delta)$ follows by expanding the conditional mutual information
$I(X;Y|Z)$ in the other way, as $I(X;Y|Z)=H(X|Z)-H(X|YZ)$.
\end{proof}

\bigskip

Finally, the following theorem gives a strong refinement of the monotonicity
of relative entropy (Corollary~\ref{cor-cie:mono-rel-ent}). An important
implication is that if the relative entropy decrease $D(p\Vert q)-D(Np\Vert
Nq)$ is not too large under the action of a classical channel $N$, then one
can perform a recovery channel $R$, satisfying
\eqref{eq-cie:recovery-channel-spec}, such that the recovered distribution
$RNp$ is close to the original distribution $p$.

\begin{theorem}
[Refined Monotonicity of Relative Entropy]\label{thm-cie:refine-mono-rel-ent}%
Let $p$ be a probability distribution on a finite alphabet $\mathcal{X}$ and
let $q:\mathcal{X}\rightarrow\lbrack0,\infty)$ be a function such that
$\operatorname{supp}(p)\subseteq\operatorname{supp}(q)$. Let $N(y|x)$ be a
conditional probability distribution (classical channel). Then the following
refinement of the monotonicity of relative entropy holds:%
\begin{align}
D(p\Vert q)-D(Np\Vert Nq)  &  \geq D(p\Vert RNp)
\end{align}
where $Np$ is a probability distribution with elements $(Np)(y)\equiv\sum
_{x}N(y|x)p(x)$, $Nq$ is a vector with elements $(Nq)(y)=\sum_{x}N(y|x)q(x)$,
and the recovery channel $R(x|y)$ (a conditional probability distribution) is
defined by the following set of equations:%
\begin{equation}
R(x|y)(Nq)(y)=N(y|x)q(x), \label{eq-cie:recovery-channel-spec}%
\end{equation}
which correspond to the Bayes theorem if $q$ is a probability distribution.
Also, $RNp$ is a probability distribution with elements $(RNp)(x)=\sum
_{y,x^{\prime}}R(x|y)N(y|x^{\prime})p(x^{\prime})$.
\end{theorem}

\begin{proof}
Consider that%
\begin{align}
(RNp)(x)  &  =\sum_{y}\frac{N(y|x)q(x)}{(Nq)(y)}\left[  (Np)(y)\right] \\
&  =q(x)\sum_{y}\frac{N(y|x)(Np)(y)}{(Nq)(y)}.
\end{align}
Thus,%
\[
D(p\Vert RNp)=\sum_{x}p(x)\log\left(  \frac{p(x)}{q(x)\sum_{y}\frac
{N(y|x)(Np)(y)}{(Nq)(y)}}\right)  ,
\]
and by definition we have that%
\begin{align}
D(p\Vert q)  &  =\sum_{x}p(x)\log\left(  \frac{p(x)}{q(x)}\right)  ,\\
D(Np\Vert Nq)  &  =\sum_{y}(Np)(y)\log\left(  \frac{(Np)(y)}{(Nq)(y)}\right)
.
\end{align}
Then%
\begin{align}
D(p\Vert q)-D(p\Vert RNp)  &  =\sum_{x}p(x)\log\left(  \sum_{y}\frac
{N(y|x)(Np)(y)}{(Nq)(y)}\right) \\
&  \geq\sum_{x}p(x)\sum_{y}N(y|x)\log\left(  \frac{(Np)(y)}{(Nq)(y)}\right) \\
&  =\sum_{y}\left[  \sum_{x}N(y|x)p(x)\right]  \log\left(  \frac
{(Np)(y)}{(Nq)(y)}\right) \\
&  =\sum_{y}(Np)(y)\log\left(  \frac{(Np)(y)}{(Nq)(y)}\right) \\
&  =D(Np\Vert Nq).
\end{align}
The sole inequality is a consequence of concavity of the logarithm.
\end{proof}

\section{Classical Information from Quantum Systems}

\label{sec-cie:cl-info-quantum}We can always process classical information by
employing a quantum system as the carrier of information. The inputs and the
outputs to a quantum protocol can both be classical. For example, we can
prepare a quantum state according to some random variable $X$---the ensemble
$\left\{  p_{X}( x) ,\rho_{x}\right\}  $ captures this idea. We can retrieve
classical information from a quantum state in the form of some random variable
$Y$ by performing a measurement---the POVM\ $\left\{  \Lambda_{y}\right\}  $
captures this notion (recall that we employ the POVM\ formalism from
Section~\ref{sec-nqt:POVM}\ if we do not care about the state after the
measurement). Suppose that Alice prepares a quantum state according to the
ensemble $\left\{  p_{X}( x) ,\rho_{x}\right\}  $ and Bob measures the state
according to the POVM\ $\left\{  \Lambda_{y}\right\}  $. Recall that the
following formula gives the conditional probability $p_{Y|X}( y|x) $:%
\begin{equation}
p_{Y|X}( y|x) =\operatorname{Tr}\left\{  \Lambda_{y}\rho_{x}\right\}  .
\label{eq-ie:quantum-cond-density}%
\end{equation}

Is there any benefit to processing classical information using quantum
systems? Later, in Chapter~\ref{chap:classical-comm-HSW}, we see that there
indeed is an enhanced performance because we can achieve higher communication
rates in general by processing classical data using quantum resources. For
now, we extend our notions of entropy in a straightforward way to include the
above ideas.

\subsection{Shannon Entropy of a POVM}

\label{sec-cie:shannon-ent-POVM}The first notion that we can extend is the
Shannon entropy, by determining the Shannon entropy of a POVM. Suppose that
Alice prepares a quantum state $\rho$ (there is no classical index here). Bob
can then perform a particular POVM\ $\left\{  \Lambda_{x}\right\}  $\ to learn
about the quantum system. Let $X$ denote the random variable corresponding to
the classical output of the POVM. The probability density function $p_{X}(x)$
of random variable $X$ is then%
\begin{equation}
p_{X}(x)=\operatorname{Tr}\left\{  \Lambda_{x}\rho\right\}  .
\end{equation}
The Shannon entropy $H(X)$\ of the POVM $\left\{  \Lambda_{x}\right\}  $\ is%
\begin{align}
H(X)  &  =-\sum_{x}p_{X}(x)\log\left(  p_{X}(x)\right) \\
&  =-\sum_{x}\operatorname{Tr}\left\{  \Lambda_{x}\rho\right\}  \log\left(
\operatorname{Tr}\left\{  \Lambda_{x}\rho\right\}  \right)  .
\end{align}
In the next chapter, we prove that the minimum Shannon entropy with respect to
all rank-one POVMs is equal to a quantity known as the quantum entropy of the
density operator$~\rho$.

\subsection{Accessible Information}

\label{sec-cie:acc-info}Let us consider the scenario introduced at the
beginning of this section, in which Alice prepares an ensemble $\mathcal{E}%
\equiv\left\{  p_{X}(x),\rho_{x}\right\}  $ and Bob performs a POVM $\left\{
\Lambda_{y}\right\}  $. Suppose now that Bob is actually trying to retrieve as
much information as possible about the random variable $X$. The quantity that
governs how much information he can learn about random variable $X$ if he
possesses random variable $Y$ is the mutual information $I(X;Y)$. But here,
Bob can actually choose which measurement he would like to perform, and it
would be good for him to perform the measurement that maximizes his
information about $X$. The resulting quantity is known as the
\index{accessible information}%
accessible information $I_{\operatorname{acc}}(\mathcal{E})$ of the ensemble
$\mathcal{E}$ (because it is the information that Bob can access about random
variable $X$):%
\begin{equation}
I_{\operatorname{acc}}(\mathcal{E})\equiv\max_{\left\{  \Lambda_{y}\right\}
}I(X;Y),
\end{equation}
where the marginal density $p_{X}(x)$ is that from the ensemble and the
conditional density $p_{Y|X}(y|x)$ is given in
\eqref{eq-ie:quantum-cond-density}. In the next chapter, we show how to obtain
a natural bound on this quantity, called%
\index{Holevo bound}
the \textit{Holevo bound}. The bound arises from a quantum generalization of
the data-processing inequality.

\subsection{Classical Mutual Information of a Bipartite State}

A final quantity that we introduce is the classical mutual information
$I_{c}(\rho_{AB})$ of a bipartite state $\rho_{AB}$. Suppose that Alice and
Bob possess some bipartite state $\rho_{AB}$ and would like to extract maximal
classical correlation from it. That is, they each retrieve a random variable
by performing respective local POVMs $\left\{  \Lambda_{A}^{x}\right\}  $ and
$\left\{  \Lambda_{B}^{y}\right\}  $ on their shares of the bipartite state
$\rho_{AB}$. These measurements produce respective random variables $X$ and
$Y$, and they would like $X$ and $Y$ to be as correlated as possible. A good
measure of their resulting classical correlations obtainable from local
quantum information processing is as follows:%
\begin{equation}
I_{c}(\rho_{AB})\equiv\max_{\left\{  \Lambda_{A}^{x}\right\}  ,\left\{
\Lambda_{B}^{y}\right\}  }I(X;Y), \label{eq-ie:mut-info-bi-part-state}%
\end{equation}
where the joint distribution%
\begin{equation}
p_{X,Y}(x,y)\equiv\operatorname{Tr}\left\{  \left(  \Lambda_{A}^{x}%
\otimes\Lambda_{B}^{y}\right)  \rho_{AB}\right\}  .
\end{equation}

Suppose that the state $\rho_{AB}$ is classical, that is, it has the form%
\begin{equation}
\rho_{AB}=\sum_{x,y}p_{X,Y}(x,y)|x\rangle\langle x|_{A}\otimes|y\rangle\langle
y|_{B},
\end{equation}
where the states $|x\rangle_{A}$ form an orthonormal basis and so do the
states $|y\rangle_{B}$. Then, the optimal measurement in this case is for
Alice to perform a complete projective measurement in the basis $|x\rangle
_{A}$ and inform Bob to perform a similar measurement in the basis
$|y\rangle_{B}$. The amount of correlation they extract is then equal to
$I(X;Y)$.

\begin{exercise}
Prove that it suffices to consider maximizing with respect to rank-one POVMs
when computing \eqref{eq-ie:mut-info-bi-part-state}. (Hint:\ Consider refining
the POVM\ $\left\{  \Lambda_{x}\right\}  $ as the rank-one POVM$\ \left\{
|\phi_{x,z}\rangle\langle\phi_{x,z}|\right\}  $, where we spectrally decompose
$\Lambda_{x}$ as $\sum_{z}|\phi_{x,z}\rangle\langle\phi_{x,z}|$, and then
exploit the data-processing inequality.)
\end{exercise}

\section{History and Further Reading}

\cite{book1991cover} have given an excellent introduction to entropy and information
theory (some of the material in this chapter is similar to material appearing
in that book). \cite{M03} also gives a good introduction. E.~T.~Jaynes was an
advocate of the principle of maximum entropy, proclaiming its utility in
several sources~\citep{J57,J57a,J03}. A good exposition of Fano's inequality
appears on Scholarpedia~\citep{F08}. Theorem~\ref{thm-cie:cl-ent-continuity}
(continuity of entropy bound) is due to \cite{Z07}\ and \cite{A07}. The
particular proof that we give in this chapter is Zhang's \citep{Z07}, but we
used the presentation of \cite{Sason13}. The Pinsker inequality was first
proved by \cite{P60}, with subsequent enhancements by \cite{C67,K69,K67}.
Theorem~\ref{thm-cie:mut-info-to-prod}\ was realized in private communication
with Andreas Winter (August 2015). The refinement of the monotonicity of
relative entropy in Theorem~\ref{thm-cie:refine-mono-rel-ent}\ was established
by \cite{LW14}.

\chapter{Quantum Information and Entropy}

\label{chap:q-info-entropy}In this chapter, we discuss several information
measures that are important for quantifying the amount of information and
correlations that are present in quantum systems. The first fundamental
measure that we introduce is the
\index{von Neumann entropy}%
von Neumann entropy (or simply \textit{quantum entropy}). It is the quantum
generalization of the Shannon entropy, but it captures both classical and
quantum uncertainty in a quantum state.\footnote{We should point out the irony
in the historical development of classical and quantum entropy. The von
Neumann entropy has seen much widespread use in modern quantum information
theory, and perhaps this would make one think that von Neumann discovered this
quantity much after Shannon. But in fact, the reverse is true. Von Neumann
first discovered what is now known as the von Neumann entropy and applied it
to questions in statistical physics. Much later, Shannon determined an
information-theoretic formula and asked von Neumann what he should call it.
Von Neumann told him to call it the entropy for two reasons:\ 1) it was a
special case of the von Neumann entropy and 2) he would always have the
advantage in a debate because von Neumann claimed that no one at the time
really understood entropy.} The quantum entropy gives meaning to the notion of
an \textit{information qubit}. This notion is different from that of the
physical qubit, which is the description of a quantum state of an electron or
a photon. The information qubit is the fundamental quantum informational unit
of measure, determining how much quantum information is present in a quantum system.

The initial definitions here are analogous to the classical definitions of
entropy, but we soon discover a radical departure from the intuitive classical
notions from the previous chapter: the conditional quantum entropy can be
negative for certain quantum states. In the classical world, this negativity
simply does not occur, but it takes on a special meaning in quantum information
theory. Pure quantum states that are entangled have stronger-than-classical
correlations and are examples of states that have negative conditional
entropy. The negative of the conditional quantum entropy is so important in
quantum information theory that we even have a special name for it: the
\index{coherent information}%
coherent information. We discover that the coherent information obeys a
quantum data-processing inequality, placing it on a firm footing as a
particular informational measure of quantum correlations.

We then define several other quantum information measures, such as quantum
mutual information, that bear similar definitions as in the classical world,
but with Shannon entropies replaced with quantum entropies. This replacement
may seem to make quantum entropy somewhat trivial on the surface, but a simple
calculation reveals that a maximally entangled state on two qubits registers
\textit{two bits} of quantum mutual information (recall that the largest the
mutual information can be in the classical world is \textit{one bit} for the
case of two maximally correlated bits). We then discuss several entropy
inequalities that play an important role in quantum information
processing:\ the monotonicity of quantum relative entropy, strong
subadditivity, the quantum data-processing inequalities, and continuity of
quantum entropy.

\section{Quantum Entropy}

We might expect a measure of the entropy of a quantum system to be vastly
different from the classical measure of entropy from the previous chapter
because a quantum system possesses not only classical uncertainty but also
quantum uncertainty that arises from the uncertainty principle. But recall
that the density operator captures both types of uncertainty and allows us to
determine probabilities for the outcomes of any measurement on a given system.
Thus, a quantum measure of uncertainty should be a direct function of the
density operator, just as the classical measure of uncertainty is a direct
function of a probability density function. It turns out that this function
has a strikingly similar form to the classical entropy, as we see below.

\begin{definition}
[Quantum Entropy]\label{def:quantum-entropy}%
\index{von Neumann entropy}%
Suppose that Alice prepares some quantum system $A$ in a state $\rho_{A}%
\in\mathcal{D}(\mathcal{H}_{A})$. Then the entropy $H(A)_{\rho}$\ of the state
is defined as follows:%
\begin{equation}
H(A)_{\rho}\equiv-\operatorname{Tr}\left\{  \rho_{A}\log\rho_{A}\right\}  .
\end{equation}

\end{definition}

The entropy of a quantum system is also known as the \textit{von Neumann
entropy} or the \textit{quantum entropy} but we often simply refer to it as
the \textit{entropy}. We can denote it by $H(A)_{\rho}$ or $H(\rho_{A})$ to
show the explicit dependence on the density operator $\rho_{A}$. The quantum
entropy has a special relation to the eigenvalues of the density operator, as
the following exercise asks you to verify.

\begin{exercise}
\label{ex-qie:eigen-von-neumann}Consider a density operator $\rho_{A}$ with
the following spectral decomposition:%
\begin{equation}
\rho_{A}=\sum_{x}p_{X}(x)|x\rangle\langle x|_{A}.
\end{equation}
Show that the quantum entropy $H(A)_{\rho}$\ is the same as the Shannon
entropy $H(X)$\ of a random variable $X$ with probability distribution
$p_{X}(x)$.
\end{exercise}

In our definition of quantum entropy, we use the same notation $H$ as in the
classical case to denote the entropy of a quantum system. It should be clear
from the context whether we are referring to the entropy of a quantum or
classical system.

The quantum entropy admits an intuitive interpretation. Suppose that Alice
generates a quantum state $|\psi_{y}\rangle$ in her lab according to some
probability density $p_{Y}(y)$, corresponding to a random variable $Y$.
Suppose further that Bob has not yet received the state from Alice and does
not know which one she sent. The expected density operator from Bob's point of
view is then%
\begin{equation}
\sigma=\mathbb{E}_{Y}\left\{  |\psi_{Y}\rangle\langle\psi_{Y}|\right\}
=\sum_{y}p_{Y}(y)|\psi_{y}\rangle\langle\psi_{y}|.
\end{equation}
The interpretation of the entropy $H(\sigma)$\ is that it quantifies Bob's
uncertainty about the state Alice sent---his expected information gain is
$H(\sigma)$ qubits upon receiving and measuring the state that Alice sends.
Schumacher's noiseless quantum coding theorem, described in
Chapter~\ref{chap:schumach}, gives an alternative operational interpretation
of the quantum entropy by proving that Alice needs to send Bob qubits at a
rate $H(\sigma)$ in order for him to be able to decode a compressed quantum state.

The above interpretation of quantum entropy seems qualitatively similar to the
interpretation of classical entropy. However, there is a significant
quantitative difference that illuminates the difference between Shannon
entropy and quantum entropy. Let us consider an example. Suppose that Alice
generates a sequence $|\psi_{1}\rangle\left\vert \psi_{2}\right\rangle
\cdots\left\vert \psi_{n}\right\rangle $\ of quantum states according to the
following \textquotedblleft BB84\textquotedblright\ ensemble:%
\begin{equation}
\left\{  \left\{  1/4,|0\rangle\right\}  ,\left\{  1/4,|1\rangle\right\}
,\left\{  1/4,|+\rangle\right\}  ,\left\{  1/4,|-\rangle\right\}  \right\}  .
\end{equation}
Suppose that Alice and Bob share a noiseless classical channel. If she employs
Shannon's classical noiseless coding protocol, she should transmit classical
data to Bob at a rate of two classical channel uses per source state
$\left\vert \psi_{i}\right\rangle $ in order for him to reliably recover the
classical data needed to reproduce the sequence of states that Alice
transmitted (the Shannon entropy of the uniform distribution $1/4$ is $2$ bits).

Now let us consider computing the quantum entropy of the above ensemble.
First, we determine the expected density operator of Alice's ensemble:%
\begin{equation}
\frac{1}{4}\left(  |0\rangle\langle0|+|1\rangle\langle1|+|+\rangle
\langle+|+|-\rangle\langle-|\right)  =\pi,
\end{equation}
where $\pi$ is the maximally mixed state. The quantum entropy of the above
density operator is one qubit because the eigenvalues of $\pi$ are both equal
to $1/2$. Suppose now that a noiseless quantum channel connects Alice to
Bob---this is a channel that can preserve quantum coherence without any
interaction with an environment. Then Alice only needs to send qubits at a
rate of one channel use per source symbol if she employs a protocol known as
Schumacher compression%
\index{Schumacher compression}
(we discuss this protocol in detail in Chapter~\ref{chap:schumach}). Bob can
then reliably decode the qubits that Alice sent. The protocol also causes a
slight disturbance to the state, which however vanishes in the limit of many
invocations of the source. The above departure from classical information
theory holds in general---Exercise~\ref{ex-qie:shannon-vs-von-neumann}\ of
this chapter asks you to prove that the Shannon entropy of any ensemble is
never less than the quantum entropy of its expected density operator.

\subsection{Mathematical Properties of Quantum Entropy}

We now discuss several mathematical properties of the quantum
entropy:\ non-negativity, its minimum value, its maximum value, its invariance
with respect to isometries, and concavity. The first three of these properties
follow from the analogous properties in the classical world because the
quantum entropy of a density operator is the Shannon entropy of its
eigenvalues (see Exercise~\ref{ex-qie:eigen-von-neumann}). We state them
formally below:

\begin{property}
[Non-Negativity]The quantum entropy%
\index{von Neumann entropy!positivity}
$H(\rho)$ is non-negative for any density operator $\rho$:%
\begin{equation}
H(\rho)\geq0.
\end{equation}

\end{property}

\begin{proof}
This follows from non-negativity of Shannon entropy.
\end{proof}

\begin{property}
[Minimum Value]The minimum value of the quantum entropy is zero, and it occurs
when the density operator is a pure state.
\end{property}

\begin{proof}
The minimum value occurs when the eigenvalues of a density operator are
distributed with all the probability mass on one eigenvector and zero on the
others, so that the density operator is rank one and corresponds to a pure state.
\end{proof}

Why should the entropy of a pure quantum state vanish?\ It seems that there is
quantum uncertainty inherent in the state itself and that a measure of quantum
uncertainty should capture this fact. This last observation only makes sense
if we do not know anything about the state that is prepared. But if we know
exactly how it is prepared, we can perform a special quantum measurement to
verify this, and we do not learn anything from this measurement because the
outcome of it is always certain. For example, suppose that Alice prepares the
state $|\phi\rangle$ and Bob knows that she does so. He can then perform the
following measurement $\left\{  |\phi\rangle\langle\phi|,I-|\phi\rangle
\langle\phi|\right\}  $ to verify that she prepared this state. He always
receives the first outcome from the measurement and thus never gains any
information from it. Thus, in this sense it is reasonable that the entropy of
a pure state vanishes.

\begin{property}
[Maximum Value]\label{prop:max-val-von-ent}The maximum value of the quantum
entropy is $\log d$ where $d$ is the dimension of the system, and it occurs
for the maximally mixed state.
\end{property}

\begin{proof}
A proof of the above property is the same as that for the classical case.
\end{proof}

\begin{property}
[Concavity]\label{prop-qie:concavity}%
\index{von Neumann entropy!concavity}%
Let $\rho_{x}\in\mathcal{D}(\mathcal{H})$ and let $p_{X}(x)$ be a probability
distribution. The entropy is concave in the density operator:%
\begin{equation}
H(\rho)\geq\sum_{x}p_{X}(x)H(\rho_{x}),
\end{equation}
where $\rho\equiv\sum_{x}p_{X}(x)\rho_{x}$.
\end{property}

The physical interpretation of concavity is as before for classical entropy:
entropy can never decrease under a mixing operation. This inequality is a
fundamental property of the entropy, and we prove it after developing some
important entropic tools (see Exercise~\ref{ex-qie:concavity-entropy}).

\begin{property}
[Isometric Invariance]Let $\rho\in\mathcal{D}(\mathcal{H})$ and $U:\mathcal{H}%
\rightarrow\mathcal{H}^{\prime}$ be an isometry. The entropy of a density
operator is invariant with respect to isometries, in the following sense:%
\begin{equation}
H(\rho)=H(U\rho U^{\dag}).
\end{equation}

\end{property}

\begin{proof}
Isometric invariance of entropy follows by observing that the eigenvalues of a
density operator are invariant with respect to an isometry:%
\begin{align}
U\rho U^{\dag}  &  =U\sum_{x}p_{X}(x)|x\rangle\langle x|U^{\dag}\\
&  =\sum_{x}p_{X}(x)|\phi_{x}\rangle\langle\phi_{x}|,
\end{align}
where $\left\{  |\phi_{x}\rangle\right\}  $ is some orthonormal basis such
that $U|x\rangle=|\phi_{x}\rangle$. The above property follows because the
entropy is a function of the eigenvalues of a density operator.
\end{proof}

A unitary or isometric operator is a quantum generalization of a permutation
in this context (recall Property~\ref{prop-cie:inv-perm}\ of the classical entropy).

\subsection{Alternate Characterization of Quantum Entropy}

There is an interesting alternate characterization of the quantum entropy of a
state $\rho$ as the minimum Shannon entropy when a rank-one POVM is performed
on it (we discussed this briefly in Section~\ref{sec-cie:shannon-ent-POVM}).
In this sense, there is some optimal measurement to perform on $\rho$ such
that its entropy is equal to the quantum entropy, and this optimal measurement
is the \textquotedblleft right question to ask\textquotedblright\ (as we
discussed very early on in Section~\ref{sec-intro:measure-q-info}).

\begin{theorem}
Let $\rho\in\mathcal{D}(\mathcal{H})$. The quantum entropy $H(\rho)$ has the
following characterization:%
\begin{equation}
H(\rho)=\min_{\left\{  \Lambda_{y}\right\}  }\left[  -\sum_{y}%
\operatorname{Tr}\left\{  \Lambda_{y}\rho\right\}  \log\left(
\operatorname{Tr}\left\{  \Lambda_{y}\rho\right\}  \right)  \right]  ,
\end{equation}
where the minimum is restricted to be with respect to rank-one POVMs (those
with $\Lambda_{y}=|\phi_{y}\rangle\langle\phi_{y}|$ for some vectors
$|\phi_{y}\rangle$ such that $\operatorname{Tr}\left\{  |\phi_{y}%
\rangle\langle\phi_{y}|\right\}  \leq1$ and $\sum_{y}|\phi_{y}\rangle
\langle\phi_{y}|=I$).
\end{theorem}

\begin{proof}
In order to prove the above result, we should first realize that a complete
projective measurement in the eigenbasis of $\rho$ should achieve the minimum.
That is, if $\rho=\sum_{x}p_{X}(x)|x\rangle\langle x|$, we should expect that
the measurement $\left\{  |x\rangle\langle x|\right\}  $ achieves the minimum.
In this case, the Shannon entropy of the measurement is equal to the Shannon
entropy of $p_{X}(x)$, as discussed in Exercise~\ref{ex-qie:eigen-von-neumann}.

We now prove that any other rank-one POVM\ has a higher entropy than that
given by this measurement. Consider that the distribution of the measurement
outcomes for $\left\{  |\phi_{y}\rangle\langle\phi_{y}|\right\}  $ is equal to%
\begin{equation}
\operatorname{Tr}\left\{  |\phi_{y}\rangle\langle\phi_{y}|\rho\right\}
=\sum_{x}\left\vert \left\langle \phi_{y}|x\right\rangle \right\vert ^{2}%
p_{X}(x),
\end{equation}
so that we can think of $\left\vert \left\langle \phi_{y}|x\right\rangle
\right\vert ^{2}$ as a conditional probability distribution. Introducing
$\eta(p)\equiv-p\log p$, which is a concave function, we can write the quantum
entropy as%
\begin{align}
H(\rho)  &  =\sum_{x}\eta(p_{X}(x))\\
&  =\sum_{x}\eta(p_{X}(x))+\eta(p_{X}(x_{0})),
\end{align}
where $x_{0}$ is a symbol added to the alphabet of $x$ such that $p_{X}%
(x_{0})=0$. Let us denote the enlarged alphabet with the symbols $x^{\prime}$
so that $H(\rho)=\sum_{x^{\prime}}\eta(p_{X}(x^{\prime}))$. We know that
$\sum_{y}\left\vert \left\langle \phi_{y}|x\right\rangle \right\vert ^{2}=1$
from the fact that the set $\left\{  |\phi_{y}\rangle\langle\phi_{y}|\right\}
$ forms a POVM\ and $|x\rangle$ is a normalized state. We also know that
$\sum_{x}\left\vert \left\langle \phi_{y}|x\right\rangle \right\vert ^{2}%
\leq1$ because $\operatorname{Tr}\left\{  |\phi_{y}\rangle\langle\phi
_{y}|\right\}  \leq1$ for a rank-one POVM. Thinking of $\left\vert
\left\langle \phi_{y}|x\right\rangle \right\vert ^{2}$ as a distribution over
$x$, we can add a symbol $x_{0}$ with probability $1-\left\langle \phi
_{y}|\phi_{y}\right\rangle $ so that it makes a normalized distribution. Let
us call this distribution $p(x^{\prime}|y)$. We then have that%
\begin{align}
H(\rho)  &  =\sum_{x}\eta(p_{X}(x))\\
&  =\sum_{x,y}\left\vert \left\langle \phi_{y}|x\right\rangle \right\vert
^{2}\eta(p_{X}(x))\\
&  =\sum_{x^{\prime},y}p(x^{\prime}|y)\eta(p_{X}(x^{\prime}))\\
&  =\sum_{y}\left(  \sum_{x^{\prime}}p(x^{\prime}|y)\eta(p_{X}(x^{\prime
}))\right) \\
&  \leq\sum_{y}\eta\left(  \sum_{x^{\prime}}p(x^{\prime}|y)p_{X}(x^{\prime
})\right) \\
&  =\sum_{y}\eta\left(  \operatorname{Tr}\left\{  |\phi_{y}\rangle\langle
\phi_{y}|\rho\right\}  \right)  .
\end{align}
The third equality follows from the fact that $p_{X}(x_{0})=0$ for the added
symbol $x_{0}$. The only inequality follows from concavity of $\eta$. The last
expression is equal to the Shannon entropy of the POVM\ $\left\{  |\phi
_{y}\rangle\langle\phi_{y}|\right\}  $ when performed on the state $\rho$.
\end{proof}

\section{Joint Quantum Entropy}

The joint quantum entropy
\index{von Neumann entropy!joint}%
$H(AB)_{\rho}$\ of the density operator $\rho_{AB}\in\mathcal{D}%
(\mathcal{H}_{A}\otimes\mathcal{H}_{B})$\ for a bipartite system $AB$ follows
naturally from the definition of quantum entropy:%
\begin{equation}
H(AB)_{\rho}\equiv-\operatorname{Tr}\left\{  \rho_{AB}\log\rho_{AB}\right\}  .
\end{equation}
Now suppose that $\rho_{ABC}$ is a tripartite state, i.e., in $\mathcal{D}%
(\mathcal{H}_{A}\otimes\mathcal{H}_{B}\otimes\mathcal{H}_{C})$. Then the
entropy $H(AB)_{\rho}$ in this case is defined as above, where $\rho
_{AB}=\operatorname{Tr}_{C}\{\rho_{ABC}\}$. This is a convention that we take
throughout this book. We introduce a few of the properties of joint quantum
entropy in the subsections below.

\subsection{Marginal Entropies of a Pure Bipartite State}

The five properties of quantum entropy in the previous section may give you
the impression that the nature of quantum information is not too different
from that of classical information. We proved all these properties for the
classical case, and their proofs for the quantum case seem similar. The first
three even resort to the proofs in the classical case!

Theorem~\ref{thm-ie:marginal-entropies-pure-state} below is where we observe
our first radical departure from the classical world. It states that the
marginal entropies of a pure bipartite state are equal, while the entropy of
the overall state is equal to zero. Recall that the joint entropy $H(X,Y)$ of
two random variables $X$ and $Y$ is never less than either of the marginal
entropies $H(X)$ or $H(Y)$:%
\begin{equation}
H(X,Y)\geq H(X),\ \ \ \ \ \ \ \ \ \ H(X,Y)\geq H(Y).
\end{equation}
The above inequalities follow from the non-negativity of classical conditional
entropy. But in the quantum world, these inequalities do not always have to
hold, and the following theorem demonstrates that they do not hold for an
arbitrary pure bipartite quantum state with Schmidt rank greater than one (see
Theorem~\ref{thm-qt:schmidt} for a definition of Schmidt rank). The fact that
the joint quantum entropy can be less than the marginal quantum entropy is one
of the most fundamental differences between classical and quantum information.

\begin{theorem}
\label{thm-ie:marginal-entropies-pure-state}The marginal entropies
$H(A)_{\phi}$ and $H(B)_{\phi}$ of a pure bipartite state $|\phi\rangle_{AB}$
are equal:%
\begin{equation}
H(A)_{\phi}=H(B)_{\phi},
\end{equation}
while the joint entropy $H(AB)_{\phi}$ vanishes:%
\begin{equation}
H(AB)_{\phi}=0.
\end{equation}

\end{theorem}

\begin{proof}
The crucial ingredient for a proof of this theorem is the Schmidt
decomposition (Theorem~\ref{thm-qt:schmidt}). Recall that any bipartite state
$|\phi\rangle_{AB}$ admits a Schmidt decomposition of the following form:%
\begin{equation}
|\phi\rangle_{AB}=\sum_{i}\sqrt{\lambda_{i}}\left\vert i\right\rangle
_{A}|i\rangle_{B},
\end{equation}
where $\lambda_{i} >0$ for all $i$, $\sum_{i} \lambda_{i} = 1$, $\{|i\rangle
_{A}\}$ is some orthonormal set of vectors on system $A$, and $\{|i\rangle
_{B}\}$ is some orthonormal set on system $B$. Recall that the Schmidt rank is
equal to the number of non-zero coefficients $\lambda_{i}$. Then the
respective marginal states $\rho_{A}$\ and $\rho_{B}$ on systems $A$ and $B$
are as follows:%
\begin{equation}
\rho_{A}=\sum_{i}\lambda_{i}|i\rangle\langle i|_{A},\ \ \ \ \ \ \rho_{B}%
=\sum_{i}\lambda_{i}|i\rangle\langle i|_{B}.
\end{equation}
Thus, the marginal states admit a spectral decomposition with the same
eigenvalues. The theorem follows because the quantum entropy depends only on
the eigenvalues of a given spectral decomposition.
\end{proof}

The theorem applies not only to two systems $A$ and $B$, but it also applies
to any number of systems if we make a bipartite cut of the systems. For
example, if the state is $\vert\phi\rangle_{ABCDE}$, then the following
equalities (and others from different combinations) hold by applying
Theorem~\ref{thm-ie:marginal-entropies-pure-state} and
Remark~\ref{rem-qt:schmidt}:%
\begin{align}
H( A) _{\phi}  &  =H( BCDE) _{\phi},\\
H( AB) _{\phi}  &  =H( CDE) _{\phi},\\
H( ABC) _{\phi}  &  =H( DE) _{\phi},\\
H( ABCD) _{\phi}  &  =H( E) _{\phi}.
\end{align}

The closest analogy in the classical world to the above property is when we
copy a random variable $X$. That is, suppose that $X$ has a distribution
$p_{X}(x)$ and $\hat{X}$ is some copy of it so that the distribution of the
joint random variable $X\hat{X}$\ is $p_{X}(x)\delta_{x,\hat{x}}$. Then the
marginal entropies $H(X)$ and $H(\hat{X})$ are both equal. But observe that
the joint entropy $H(X\hat{X})$ is also equal to $H(X)$ and this is where the
analogy breaks down. That is, there is not a strong classical analogy of the
notion of purification.

\subsection{Additivity}

\begin{property}
[Additivity]Let $\rho_{A}\in\mathcal{D}(\mathcal{H}_{A})$ and $\sigma_{B}%
\in\mathcal{D}(\mathcal{H}_{B})$. The quantum entropy is additive
\index{von Neumann entropy!additivity}%
for tensor-product states:%
\begin{equation}
H(\rho_{A}\otimes\sigma_{B})=H(\rho_{A})+H(\sigma_{B}). \label{eq-qie:add-ent}%
\end{equation}

\end{property}

One can verify this property simply by diagonalizing both density operators
and resorting to the additivity of the joint Shannon entropies of the eigenvalues.

Additivity is an intuitive property that we would like to hold for any measure
of information. For example, suppose that Alice generates a large sequence
$\left\vert \psi_{x_{1}}\right\rangle \left\vert \psi_{x_{2}}\right\rangle
\cdots\left\vert \psi_{x_{n}}\right\rangle $\ of quantum states according to
the ensemble $\left\{  p_{X}(x),|\psi_{x}\rangle\right\}  $. She may be aware
of the classical indices $x_{1}x_{2}\cdots x_{n}$, but a third party to whom
she sends the quantum sequence may not be aware of these values. The
description of the state to this third party is then $\rho\otimes\cdots
\otimes\rho$, where $\rho\equiv\mathbb{E}_{X}\left\{  |\psi_{X}\rangle
\langle\psi_{X}|\right\}  $, and the quantum entropy of this $n$-fold tensor
product state is $H(\rho\otimes\cdots\otimes\rho)=nH(\rho)$, by applying
\eqref{eq-qie:add-ent} inductively.

\subsection{Joint Quantum Entropy of a Classical--Quantum State}

Recall from Definition~\ref{def-nqt:classical-quantum-state} that a
classical--quantum state is a bipartite state in which a classical system and
a quantum system are classically correlated. An example of such a state is as
follows:%
\begin{equation}
\rho_{XB}\equiv\sum_{x}p_{X}( x) \vert x\rangle\langle x\vert_{X}\otimes
\rho_{B}^{x}. \label{eq-qie:cq-state}%
\end{equation}
The joint quantum entropy of this state takes on a special form that appears
similar to entropies in the classical world.

\begin{theorem}
\label{thm-qie:joint-ent-cq-state}The joint entropy $H(XB)_{\rho}$ of a
classical--quantum state, as given in \eqref{eq-qie:cq-state}, is as follows:%
\begin{equation}
H(XB)_{\rho}=H(X)+\sum_{x}p_{X}(x)H(\rho_{B}^{x}),
\end{equation}
where $H(X)$ is the entropy of a random variable $X$ with distribution
$p_{X}(x)$.
\end{theorem}

\begin{proof}
Consider that $H(XB)_{\rho}=-\operatorname{Tr}\left\{  \rho_{XB}\log\rho
_{XB}\right\}  . $ So we need to evaluate the operator $\log\rho_{XB}$, and we
can find a simplified form for it because $\rho_{XB}$ is a classical-quantum
state:%
\begin{align}
\log\rho_{XB}  &  =\log\left[  \sum_{x}p_{X}(x)|x\rangle\langle x|_{X}%
\otimes\rho_{B}^{x}\right] \\
&  =\log\left[  \sum_{x}|x\rangle\langle x|_{X}\otimes p_{X}(x)\rho_{B}%
^{x}\right] \\
&  =\sum_{x}|x\rangle\langle x|_{X}\otimes\log\left[  p_{X}(x)\rho_{B}%
^{x}\right]  .
\end{align}
Then%
\begin{align}
&  \!\!\!\!\!-\operatorname{Tr}\left\{  \rho_{XB}\log\rho_{XB}\right\}
\nonumber\\
&  =-\operatorname{Tr}\left\{  \left[  \sum_{x}p_{X}(x)|x\rangle\langle
x|_{X}\otimes\rho_{B}^{x}\right]  \left[  \sum_{x^{\prime}}|x^{\prime}%
\rangle\langle x^{\prime}|_{X}\otimes\log\left[  p_{X}(x^{\prime})\rho
_{B}^{x^{\prime}}\right]  \right]  \right\} \\
&  =-\operatorname{Tr}\left\{  \sum_{x}p_{X}(x)|x\rangle\langle x|_{X}%
\otimes\left(  \rho_{B}^{x}\log\left[  p_{X}(x)\rho_{B}^{x}\right]  \right)
\right\} \\
&  =-\sum_{x}p_{X}(x)\operatorname{Tr}\left\{  \rho_{B}^{x}\log\left[
p_{X}(x)\rho_{B}^{x}\right]  \right\}  . \label{eq-qie:ent-cq-state}%
\end{align}
Consider that%
\begin{equation}
\log\left[  p_{X}(x)\rho_{B}^{x}\right]  =\log\left(  p_{X}(x)\right)
I+\log\rho_{B}^{x},
\end{equation}
which implies that \eqref{eq-qie:ent-cq-state} is equal to%
\begin{align}
&  -\sum_{x}p_{X}(x)\left[  \operatorname{Tr}\left\{  \rho_{B}^{x}\log\left[
p_{X}(x)\right]  \right\}  +\operatorname{Tr}\left\{  \rho_{B}^{x}\log\rho
_{B}^{x}\right\}  \right] \\
&  =-\sum_{x}p_{X}(x)\left[  \log\left[  p_{X}(x)\right]  +\operatorname{Tr}%
\left\{  \rho_{B}^{x}\log\rho_{B}^{x}\right\}  \right]  .
\end{align}
This last line is then equivalent to the statement of the theorem.
\end{proof}

\section{Potential yet Unsatisfactory Definitions of Conditional Quantum
Entropy}

The conditional quantum entropy may perhaps seem a bit difficult to define at
first because there is no formal notion of conditional probability in the
quantum theory. However, there are two senses which are perhaps closest to the
notion of conditional probability, but both of them do not lead to
satisfactory definitions of conditional quantum entropy. Nevertheless, it is
instructive for us to explore both of these notions briefly. The first arises
in the noisy quantum theory, and the second arises in the purified quantum theory.

We develop the first notion. Consider an arbitrary bipartite state $\rho_{AB}%
$. Suppose that Alice performs a complete projective measurement $\Pi
\equiv\left\{  |x\rangle\langle x|\right\}  $ of her system, where $\left\{
|x\rangle\right\}  $ is an orthonormal basis. This procedure leads to an
ensemble $\{p_{X}(x),|x\rangle\langle x|_{A}\otimes\rho_{B}^{x}\}$, where%
\begin{align}
\rho_{B}^{x}  &  \equiv\frac{1}{p_{X}(x)}\operatorname{Tr}_{A}\left\{  \left(
|x\rangle\langle x|_{A}\otimes I_{B}\right)  \rho_{AB}\left(  |x\rangle\langle
x|_{A}\otimes I_{B}\right)  \right\}  ,\\
p_{X}(x)  &  \equiv\operatorname{Tr}\left\{  \left(  |x\rangle\langle
x|_{A}\otimes I_{B}\right)  \rho_{AB}\right\}  .
\end{align}
One could then think of the density operators $\rho_{B}^{x}$\ as being
conditioned on the outcome of the measurement, and these density operators
describe the state of Bob given knowledge of the outcome of the measurement.

We could potentially define a conditional entropy as follows:%
\begin{equation}
H(B|A)_{\Pi}\equiv\sum_{x}p_{X}(x)H(\rho_{B}^{x}),
\end{equation}
in analogy with the definition of the classical entropy in
\eqref{eq-ie:class-cond-ent}. This approach might seem to lead to a useful
definition of conditional quantum entropy, but the problem with it is that the
entropy depends on the measurement chosen (the notation $H(B|A)_{\Pi}$
explicitly indicates this dependence). This problem does not occur in the
classical world because the probabilities for the outcomes of measurements do
not themselves depend on the measurement selected, unless we apply some coarse
graining to the outcomes. However, this dependence on measurement is a
fundamental aspect of the quantum theory.

We could then attempt to remove the dependence of the above definition on a
particular measurement $\Pi$\ by defining the conditional quantum entropy to
be the minimization of $H(B|A)_{\Pi}$ with respect to all possible
measurements. The intuition here is perhaps that entropy should be the minimal
amount of conditional uncertainty in a system after employing the best
possible measurement on the other. However, the removal of one problem leads
to another! This optimized conditional entropy is now difficult to compute as
the system grows larger, whereas in the classical world, the computation of
conditional entropy is simple if one knows the conditional probabilities. The
above idea is useful, but we leave it for now because there is a simpler
definition of conditional quantum entropy that plays a fundamental role in
quantum information theory.

The second notion of conditional probability is actually similar to the above
notion, though we present it in the purified viewpoint. Consider a tripartite
state $|\psi\rangle_{ABC}$ and a bipartite cut $A|BC$\ of the systems $A$,
$B$, and $C$. Theorem~\ref{thm-qt:schmidt}\ states that every bipartite state
admits a Schmidt decomposition, and the state $|\psi\rangle_{ABC}$ is no
exception. Thus, we can write a Schmidt decomposition for it as follows:%
\begin{equation}
|\psi\rangle_{ABC}=\sum_{x}\sqrt{p_{X}(x)}|x\rangle_{A}|\phi_{x}\rangle_{BC},
\end{equation}
where $p_{X}(x)$ is some probability density, $\left\{  |x\rangle\right\}  $
is an orthonormal basis for the system $A$, and $\left\{  |\phi_{x}%
\rangle\right\}  $ is an orthonormal basis for the systems $BC$. Each state
$|\phi_{x}\rangle_{BC}$ is a pure bipartite state, so we can again apply a
Schmidt decomposition to each of these states:%
\begin{equation}
|\phi_{x}\rangle_{BC}=\sum_{y}\sqrt{p_{Y|X}(y|x)}|y_{x}\rangle_{B}%
|y_{x}\rangle_{C},
\end{equation}
where $p_{Y|X}(y|x)$ is some conditional probability distribution depending on
the value of $x$, and $\{|y_{x}\rangle_{B}\}$ and $\{|y_{x}\rangle_{C}\}$ are
both orthonormal bases with dependence on the value $x$. Thus, the overall
state has the following form:%
\begin{equation}
|\psi\rangle_{ABC}=\sum_{x,y}\sqrt{p_{Y|X}(y|x)p_{X}(x)}|x\rangle_{A}%
|y_{x}\rangle_{B}|y_{x}\rangle_{C}.
\end{equation}
Suppose that Alice performs a complete projective measurement in the basis
$\{|x\rangle\langle x|_{A}\}$. The state on Bob and Charlie's systems is then
$|\psi_{x}\rangle_{BC}$, and each system on $B$ or $C$ has a marginal entropy
of $H(\sigma_{x})$ where $\sigma_{x}\equiv\sum_{y}p_{Y|X}(y|x)|y_{x}%
\rangle\langle y_{x}|$. We could potentially define the conditional quantum
entropy as%
\begin{equation}
\sum_{x}p_{X}(x)H(\sigma_{x}).
\end{equation}
This quantity does not depend on a measurement as before because we simply
choose the measurement from the Schmidt decomposition. But there are many
problems with the above notion of conditional quantum entropy: it is defined
only for pure quantum states, it is not clear how to apply it to a bipartite
quantum state, and the conditional entropy of Bob's system given Alice's and
that of Charlie's given Alice's is the same (which is perhaps the strangest of
all!). Thus this notion of conditional probability is not useful for a
definition of conditional quantum entropy.

\section{Conditional Quantum Entropy}%

\index{von Neumann entropy!conditional}%
The definition of conditional quantum entropy that has been most useful in
quantum information theory is the following simple one, inspired from the
relation between joint entropy and marginal entropy in
Exercise~\ref{ex-ie:simple-ent-chain-rule}.

\begin{definition}
[Conditional Quantum Entropy]\label{eq-ie:cond-quantum-entropy}Let $\rho
_{AB}\in\mathcal{D}(\mathcal{H}_{A}\otimes\mathcal{H}_{B})$. The conditional
quantum entropy $H(A|B)_{\rho}$ of $\rho_{AB}$ is equal to the difference of
the joint quantum entropy $H(AB)_{\rho}$ and the marginal entropy $H(B)_{\rho
}$:%
\begin{equation}
H(A|B)_{\rho}\equiv H(AB)_{\rho}-H(B)_{\rho}.
\end{equation}

\end{definition}

The above definition is the most natural one, both because it is
straightforward to compute for any bipartite state and because it obeys many
relations that the classical conditional entropy obeys (such as entropy chain
rules and conditioning reduces entropy). We explore many of these relations in
the forthcoming sections. For now, we state \textquotedblleft conditioning
cannot increase entropy\textquotedblright\ as the following theorem and tackle
its proof later on after developing a few more tools.

\begin{theorem}
[Conditioning Does Not Increase Entropy]\label{thm-qie:cond-reduce-ent}%
Consider a bipartite quantum state $\rho_{AB}$. Then the following inequality
applies to the marginal entropy $H( A) _{\rho}$ and the conditional quantum
entropy $H( A|B) _{\rho}$:%
\begin{equation}
H( A) _{\rho}\geq H( A|B) _{\rho}.
\end{equation}
We can interpret the above inequality as stating that conditioning cannot
increase entropy, even if the conditioning system is quantum.
\end{theorem}

\subsection{Conditional Entropy for Classical--Quantum States}

\label{sec-qie:cond-ent-cq}A classical--quantum state is an example of a state
for which conditional quantum entropy behaves as in the classical world.
Suppose that two parties share a classical--quantum state $\rho_{XB}$ of the
form in~\eqref{eq-qie:cq-state}. The system $X$ is classical and the system
$B$ is quantum, and the correlations between these systems are entirely
classical, determined by the probability distribution $p_{X}(x)$. Let us
calculate the conditional quantum entropy $H(B|X)_{\rho}$ for this state:%
\begin{align}
H(B|X)_{\rho}  &  =H(XB)_{\rho}-H(X)_{\rho}\\
&  =H(X)_{\rho}+\sum_{x}p_{X}(x)H(\rho_{B}^{x})-H(X)_{\rho}\\
&  =\sum_{x}p_{X}(x)H(\rho_{B}^{x}).
\end{align}
The first equality follows from Definition~\ref{eq-ie:cond-quantum-entropy}.
The second equality follows from Theorem~\ref{thm-qie:joint-ent-cq-state}, and
the final equality results from algebra.

The above form for conditional entropy is completely analogous with the
classical formula in \eqref{eq-ie:class-cond-ent} and holds whenever the
conditioning system is classical.

\subsection{Negative Conditional Quantum Entropy}

One of the properties of the conditional quantum entropy in
Definition~\ref{eq-ie:cond-quantum-entropy} that seems counterintuitive at
first sight is that it can be negative. This negativity holds for an ebit
$\left\vert \Phi^{+}\right\rangle _{AB}$ shared between Alice and Bob. The
marginal state on Bob's system is the maximally mixed state $\pi_{B}$. Thus,
the marginal entropy $H( B) $ is equal to one, but the joint entropy vanishes,
and so the conditional quantum entropy $H( A|B) =-1$.

What do we make of this result? Well, this is one of the fundamental
differences between the classical world and the quantum world, and perhaps is
the very essence of the departure from an informational standpoint. The
informational statement is that we can sometimes be more certain about the
joint state of a quantum system than we can be about any one of its individual
parts, and this is the reason that conditional quantum entropy can be
negative. This is in fact the same observation that Schr\"{o}dinger made
concerning entangled states~\citep{S35}:

\begin{quotation}
\textquotedblleft When two systems, of which we know the states by their
respective representatives, enter into temporary physical interaction due to
known forces between them, and when after a time of mutual influence the
systems separate again, then they can no longer be described in the same way
as before, viz.~by endowing each of them with a representative of its own. I
would not call that one but rather the characteristic trait of quantum
mechanics, the one that enforces its entire departure from classical lines of
thought. By the interaction the two representatives [the quantum states] have
become entangled. Another way of expressing the peculiar situation is: the
best possible knowledge of a whole does not necessarily include the best
possible knowledge of all its parts, even though they may be entirely separate
and therefore virtually capable of being `best possibly known,' i.e., of
possessing, each of them, a representative of its own. The lack of knowledge
is by no means due to the interaction being insufficiently known --- at least
not in the way that it could possibly be known more completely --- it is due
to the interaction itself.\textquotedblright
\end{quotation}

These explanations might aid somewhat in understanding a negative conditional
entropy, but the ultimate test for whether we truly understand an information
measure is if it is the answer to some operational task. The task which gives
an interpretation of the conditional quantum entropy is known as \textit{state
merging}. Suppose that Alice and Bob share $n$ copies of a bipartite state
$\rho_{AB}$ where $n$ is a large number and $A$ and $B$ are qubit systems. We
also allow them free access to a classical side channel, but we count the
number of times that they use a noiseless qubit channel. Alice would like to
send Bob qubits over a noiseless qubit channel so that he receives her share
of the state $\rho_{AB}$, i.e., so that he possesses all of the $A$ shares.
The naive approach would be for Alice simply to send her shares of the state
over the noiseless qubit channels, i.e., she would use the channel $n$ times
to send all $n$ shares. But the state-merging protocol allows her to do much
better, depending on the state $\rho_{AB}$. If the state $\rho_{AB}$ has
positive conditional quantum entropy, she needs to use the noiseless qubit
channel only $\approx nH(A|B)$ times (we will prove later that $H(A|B)\leq1$
for any bipartite state on qubit systems). However, if the conditional quantum
entropy is negative, she does not need to use the noiseless qubit channel at
all, and at the end of the protocol, Alice and Bob share $\approx nH(A|B)$
noiseless ebits! They can then use these ebits for future communication
purposes, such as a teleportation or super-dense coding protocol (see
Chapter~\ref{chap:three-noiseless}). Thus, a negative conditional quantum
entropy implies that Alice and Bob gain the potential for future quantum
communication, making clear in an operational sense what a negative
conditional quantum entropy means.\footnote{After Horodecki, Oppenheim, and
Winter published the state-merging%
\index{state merging}
protocol~\citep{Horodecki:2005:673}, the \textit{Bristol Evening Post}
featured a story about Andreas Winter with the amusing title \textquotedblleft
Scientist Knows Less Than Nothing,\textquotedblright\ as a reference to the
potential negativity of conditional quantum entropy. Of course, such a title
may seem a bit nonsensical to the layman, but it does grasp the idea that we
can know less about a part of a quantum system than we do about its whole.}
(We will cover this protocol in Chapter~\ref{chap:coh-comm-noisy}).

\begin{exercise}
\label{ex-ie:saturate-SSA}Let $\sigma_{ABC}=\rho_{AB}\otimes\tau_{C}$, where
$\rho_{AB}\in\mathcal{D}(\mathcal{H}_{A}\otimes\mathcal{H}_{B})$ and $\tau
_{C}\in\mathcal{D}(\mathcal{H}_{C})$. Show that $H(A|B)_{\rho}=H(A|BC)_{\sigma
}$.
\end{exercise}

\section{Coherent Information}

Negativity of the conditional quantum entropy is so important in quantum
information theory that we even have an information quantity and a special
notation to denote the negative of the conditional quantum entropy:

\begin{definition}
[Coherent Information]The
\index{coherent information}%
coherent information $I(A\rangle B)_{\rho}$\ of a bipartite state $\rho
_{AB}\in\mathcal{D}(\mathcal{H}_{A}\otimes\mathcal{H}_{B})$ is as follows:%
\begin{equation}
I(A\rangle B)_{\rho}\equiv H(B)_{\rho}-H(AB)_{\rho}.
\end{equation}

\end{definition}

You should immediately notice that this quantity is the negative of the
conditional quantum entropy in Definition~\ref{eq-ie:cond-quantum-entropy},
but it is perhaps more useful to think of the coherent information not merely
as the negative of the conditional quantum entropy, but as an information
quantity in its own right. This is why we employ a separate notation for it.
The \textquotedblleft$I$\textquotedblright\ is present because the coherent
information is an information quantity that measures quantum correlations,
much like the mutual information does in the classical case. For example, we
have already seen that the coherent information of an ebit is equal to one.
Thus, it is measuring the extent to which we know less about part of a system
than we do about its whole. Perhaps surprisingly, the coherent information
obeys a quantum data-processing inequality (discussed in
Section~\ref{sec-ie:quantum-data-processing}), which gives further support for
it having an \textquotedblleft$I$\textquotedblright\ present in its notation.
The Dirac symbol \textquotedblleft$\rangle$\textquotedblright\ is present to
indicate that this quantity is a quantum information quantity, having a good
meaning really only in the quantum world. The choice of \textquotedblleft%
$\rangle$\textquotedblright\ over \textquotedblleft$\langle$\textquotedblright%
\ also indicates a directionality from Alice to Bob, and this notation will
make more sense when we begin to discuss the coherent information of a quantum
channel in Chapter~\ref{chap:additivity}.

\begin{exercise}
\label{ex-qie:coh-info-neg}Calculate the coherent information $I(A\rangle
B)_{\Phi}$\ of the maximally entangled state%
\begin{equation}
\left\vert \Phi\right\rangle _{AB}\equiv\frac{1}{\sqrt{d}}\sum_{i=1}%
^{d}|i\rangle_{A}|i\rangle_{B}.
\end{equation}
Calculate the coherent information $I(A\rangle B)_{\overline{\Phi}}$ of the
maximally correlated state%
\begin{equation}
\overline{\Phi}_{AB}\equiv\frac{1}{d}\sum_{i=1}^{d}|i\rangle\langle
i|_{A}\otimes|i\rangle\langle i|_{B}.
\end{equation}

\end{exercise}

\begin{exercise}
\label{ex-qie:alternate-coh-info}Let $\rho_{AB}\in\mathcal{D}(\mathcal{H}%
_{A}\otimes\mathcal{H}_{B})$. Consider a purification $|\psi\rangle_{ABE}$ of
this state to some environment system $E$. Show that%
\begin{equation}
I(A\rangle B)_{\rho}=H(B)_{\psi}-H(E)_{\psi}.
\end{equation}
Thus, there is a sense in which the coherent information measures the
difference in the uncertainty of Bob and the uncertainty of the environment.
\end{exercise}

\begin{exercise}
[Duality of Conditional Entropy]\label{ex-qie:other-coh-info}Show that%
\index{von Neumann entropy!conditional!duality}
$-H(A|B)_{\rho}=I(A\rangle B)_{\rho}=H(A|E)_{\psi}$ for the purification in
the above exercise.
\end{exercise}

The coherent information can be both negative or positive depending on the
bipartite state for which we evaluate it, but it cannot be arbitrarily large
or arbitrarily small. The following theorem places a useful bound on its
absolute value.

\begin{theorem}
\label{thm-qie:bound-cond-ent}Let $\rho_{AB}\in\mathcal{D}(\mathcal{H}%
_{A}\otimes\mathcal{H}_{B})$. The following bound applies to the absolute
value of the conditional entropy $H(A|B)_{\rho}$:%
\begin{equation}
\left\vert H(A|B)_{\rho}\right\vert \leq\log\dim(\mathcal{H}_{A}).
\end{equation}
The bounds are saturated for $\rho_{AB}=\pi_{A}\otimes\sigma_{B}$, where
$\pi_{A}$ is the maximally mixed state and $\sigma_{B}\in\mathcal{D}%
(\mathcal{H}_{B})$, and for $\rho_{AB}=\Phi_{AB}$ (the maximally entangled state).
\end{theorem}

\begin{proof}
We first prove the inequality $H(A|B)_{\rho}\leq\log\dim(\mathcal{H}_{A})$:%
\begin{equation}
H(A|B)_{\rho} \leq H(A)_{\rho} \leq\log\dim(\mathcal{H}_{A}).
\end{equation}
The first inequality follows because conditioning reduces entropy
(Theorem~\ref{thm-qie:cond-reduce-ent}), and the second inequality follows
because the maximum value of the entropy $H(A)_{\rho}$ is $\log\dim
(\mathcal{H}_{A})$. We now prove the inequality $H(A|B)_{\rho}\geq-\log
\dim(\mathcal{H}_{A})$. Consider a purification $|\psi\rangle_{EAB}$\ of the
state $\rho_{AB}$. We then have that%
\begin{align}
H(A|B)_{\rho}  &  =-H(A|E)_{\psi}\\
&  \geq-H(A)_{\rho}\\
&  \geq-\log\dim(\mathcal{H}_{A}).
\end{align}
The first equality follows from Exercise~\ref{ex-qie:other-coh-info}. The
first and second inequalities follow by the same reasons as the inequalities
in the previous paragraph.
\end{proof}

\begin{exercise}
[Conditional Coherent Information]Consider a tripartite state $\rho_{ABC}$.
Show that%
\begin{equation}
I(A\rangle BC)_{\rho}=I(A\rangle B|C)_{\rho},
\end{equation}
where $I(A\rangle B|C)_{\rho}\equiv H(B|C)_{\rho}-H(AB|C)_{\rho}$ is the
\index{coherent information!conditional}%
conditional coherent information.
\end{exercise}

\begin{exercise}
[Conditional Coherent Information of a Classical--Quantum State]%
\label{ex-qie:cond-coh-info}Suppose we have a classical--quantum state
$\sigma_{XAB}$ where%
\begin{equation}
\sigma_{XAB}=\sum_{x}p_{X}(x)|x\rangle\langle x|_{X}\otimes\sigma_{AB}^{x},
\label{eq-qie:cqq-state}%
\end{equation}
$p_{X}$ is a probability distribution on a finite alphabet $\mathcal{X}$ and
$\sigma_{AB}^{x}\in\mathcal{D}(\mathcal{H}_{A}\otimes\mathcal{H}_{B})$ for all
$x\in\mathcal{X}$. Show that
\begin{equation}
I(A\rangle BX)_{\sigma}=\sum_{x}p_{X}(x)I(A\rangle B)_{\sigma^{x}}.
\end{equation}

\end{exercise}

\section{Quantum Mutual Information}

The standard informational measure of correlations in the classical world is
the mutual information, and such a quantity plays a prominent role in
measuring classical and quantum correlations in the quantum world as well.

\begin{definition}
[Quantum Mutual Information]The quantum mutual information%
\index{quantum mutual information}
of a bipartite state $\rho_{AB}\in\mathcal{D}(\mathcal{H}_{A}\otimes
\mathcal{H}_{B})$ is defined as follows:%
\begin{equation}
I(A;B)_{\rho}\equiv H(A)_{\rho}+H(B)_{\rho}-H(AB)_{\rho}.
\end{equation}

\end{definition}

The following relations hold for quantum mutual information, in analogy with
the classical case:%
\begin{align}
I(A;B)_{\rho}  &  =H(A)_{\rho}-H(A|B)_{\rho}\label{eq-ie:expand-quantum-MI}\\
&  =H(B)_{\rho}-H(B|A)_{\rho}.
\end{align}
These immediately lead to the following relations between quantum mutual
information and the coherent information:%
\begin{align}
I(A;B)_{\rho}  &  =H(A)_{\rho}+I(A\rangle B)_{\rho}\\
&  =H(B)_{\rho}+I(B\rangle A)_{\rho}.
\end{align}

The theorem below gives a fundamental lower bound on the quantum mutual
information---we merely state it for now and give a full proof later.

\begin{theorem}
[Non-Negativity of Quantum Mutual Information]\label{thm-ie:QMI-positive}
\index{quantum mutual information!non-negativity}%
The quantum mutual information $I( A;B) _{\rho}$ of any bipartite quantum
state $\rho_{AB}$ is non-negative:%
\begin{equation}
I( A;B) _{\rho}\geq0.
\end{equation}

\end{theorem}

\begin{exercise}
[Conditioning Does Not Increase Entropy]\label{ex-qie:cond-red-ent}Show that
non-negativity of quantum mutual information implies that conditioning does
not increase entropy (Theorem~\ref{thm-qie:cond-reduce-ent}).
\end{exercise}

\begin{exercise}
Calculate the quantum mutual information $I( A;B) _{\Phi}$ of the maximally
entangled state$~\Phi_{AB}$. Calculate the quantum mutual information $I( A;B)
_{\overline{\Phi}}$ of the maximally correlated state$~\overline{\Phi}_{AB}$.
\end{exercise}

\begin{exercise}
[Bound on Quantum Mutual Information]\label{ex-qie:dim-bound-MI}Let $\rho
_{AB}\in\mathcal{D}(\mathcal{H}_{A}\otimes\mathcal{H}_{B})$. Prove that the%
\index{quantum mutual information!dimension bound}
following bound applies to the quantum mutual information:%
\begin{equation}
I(A;B)_{\rho}\leq2\log\left[  \min\left\{  \dim(\mathcal{H}_{A}),\dim
(\mathcal{H}_{B})\right\}  \right]  .
\end{equation}
What is an example of a state that saturates the bound?
\end{exercise}

\begin{exercise}
Consider a pure state $|\psi\rangle_{RA}\in\mathcal{H}_{R}\otimes
\mathcal{H}_{A}$. Suppose that an isometry $U:\mathcal{H}_{A}\rightarrow
\mathcal{H}_{B}\otimes\mathcal{H}_{E}$ acts on the $A$ system of $|\psi
\rangle_{RA}$\ to produce the pure state $|\phi\rangle_{RBE}\in\mathcal{H}%
_{R}\otimes\mathcal{H}_{B}\otimes\mathcal{H}_{E}$. Show that%
\begin{equation}
I(R;B)_{\phi}+I(R;E)_{\phi}=I(R;A)_{\psi}.
\end{equation}

\end{exercise}

\begin{exercise}
Consider a tripartite pure state $|\psi\rangle_{SRA}\in\mathcal{H}_{S}%
\otimes\mathcal{H}_{R}\otimes\mathcal{H}_{A}$. Suppose that an isometry
$U:\mathcal{H}_{A}\rightarrow\mathcal{H}_{B}\otimes\mathcal{H}_{E}$ acts on
the $A$ system of $|\psi\rangle_{SRA}$ to produce the state $|\phi
\rangle_{SRBE}\in\mathcal{H}_{S}\otimes\mathcal{H}_{R}\otimes\mathcal{H}%
_{B}\otimes\mathcal{H}_{E}$. Show that%
\begin{equation}
I(R;A)_{\psi}+I(R;S)_{\psi}=I(R;B)_{\phi}+I(R;SE)_{\phi}.
\end{equation}

\end{exercise}

\begin{exercise}
[Entropy, Coherent Information, and Mutual Information]%
\label{ex-qie:entropy-games}Consider a pure state $|\phi\rangle_{ABE}$\ on
systems $ABE$. Using the Schmidt decomposition with respect to the bipartite
cut $A\ |\ BE$, we can write $|\phi\rangle_{ABE}$ as follows:%
\begin{equation}
|\phi\rangle_{ABE}=\sum_{x}\sqrt{p_{X}(x)}|x\rangle_{A}\otimes|\phi_{x}%
\rangle_{BE},
\end{equation}
for some orthonormal states $\{|x\rangle_{A}\}_{x\in\mathcal{X}}$ on system
$A$ and some orthonormal states $\{|\phi_{x}\rangle_{BE}\}$ on the joint
system $BE$. Prove the following relations:%
\begin{align}
I(A\rangle B)_{\phi}  &  =\frac{1}{2}I(A;B)_{\phi}-\frac{1}{2}I(A;E)_{\phi},\\
H(A)_{\phi}  &  =\frac{1}{2}I(A;B)_{\phi}+\frac{1}{2}I(A;E)_{\phi}.
\end{align}

\end{exercise}

\begin{exercise}
[Coherent Information and Private Information]We obtain a decohered version
$\overline{\phi}_{ABE}$\ of the state in Exercise~\ref{ex-qie:entropy-games}%
\ by measuring the $A$ system in the basis $\{|x\rangle_{A}\}_{x\in
\mathcal{X}}$. Let us now denote the $A$ system as the $X$ system because it
becomes a classical system after the measurement:%
\begin{equation}
\overline{\phi}_{XBE}=\sum_{x}p_{X}(x)|x\rangle\langle x|_{X}\otimes\phi
_{BE}^{x}.
\end{equation}
Prove the following relation:%
\begin{equation}
I(A\rangle B)_{\phi}=I(X;B)_{\overline{\phi}}-I(X;E)_{\overline{\phi}}.
\end{equation}
The quantity on the right-hand side\ is known as the private information,
because there is a sense in which it quantifies the classical information in
$X$ that is accessible to Bob while being private from Eve.
\end{exercise}

\begin{exercise}
[Additivity]\label{ex-qie:add-mutual-info}Let $\rho_{A_{1}B_{1}}\in
\mathcal{D}(\mathcal{H}_{A_{1}}\otimes\mathcal{H}_{B_{1}})$ and $\sigma
_{A_{2}B_{2}}\in\mathcal{D}(\mathcal{H}_{A_{2}}\otimes\mathcal{H}_{B_{2}})$.
Set $\omega_{A_{1}B_{1}A_{2}B_{2}}\equiv\rho_{A_{1}B_{1}}\otimes\sigma
_{A_{2}B_{2}}$. Prove that $I(A_{1}A_{2};B_{1}B_{2})_{\omega}=I(A_{1}%
;B_{1})_{\rho}+I(A_{2};B_{2})_{\sigma}$.
\end{exercise}

\subsection{Holevo Information}

Suppose that Alice prepares some classical ensemble $\mathcal{E}\equiv\left\{
p_{X}( x) ,\rho_{B}^{x}\right\}  $ and then hands this ensemble to Bob without
telling him the classical index $x$. The expected density operator of this
ensemble is%
\begin{equation}
\rho_{B}\equiv\mathbb{E}_{X}\left\{  \rho_{B}^{X}\right\}  =\sum_{x}p_{X}( x)
\rho_{B}^{x},
\end{equation}
and this density operator $\rho_{B}$\ characterizes the state from Bob's
perspective because he does not have knowledge of the classical index $x$. His
task is to determine the classical index $x$ by performing some measurement on
his system $B$. Recall from Section~\ref{sec-cie:acc-info} that the accessible
information quantifies Bob's information gain after performing some optimal
measurement $\left\{  \Lambda_{y}\right\}  $ on his system $B$:%
\begin{equation}
I_{\operatorname{acc}}( \mathcal{E}) =\max_{\left\{  \Lambda_{y}\right\}  }I(
X;Y) ,
\end{equation}
where $Y$ is a random variable corresponding to the outcome of the measurement.

What is the accessible information%
\index{accessible information}
of the ensemble?\ In general, this quantity is difficult to compute, but
another quantity, called the
\index{Holevo information}
Holevo information, provides a useful upper bound. The Holevo information
$\chi(\mathcal{E})$\ of the ensemble is%
\begin{equation}
\chi(\mathcal{E})\equiv H(\rho_{B})-\sum_{x}p_{X}(x)H(\rho_{B}^{x}).
\end{equation}
Exercise~\ref{ex-qie:holevo-bound}\ asks you to prove this upper bound after
we develop the quantum data-processing inequality for quantum mutual
information. The Holevo information characterizes the correlations between the
classical variable $X$ and the quantum system $B$.

\begin{exercise}
[Mutual Information of Classical--Quantum States]%
\label{ex-qie:holevo-cq-state}Consider the following classical--quantum state
representing the ensemble $\mathcal{E}$:%
\begin{equation}
\sigma_{XB}\equiv\sum_{x}p_{X}(x)|x\rangle\langle x|_{X}\otimes\rho_{B}^{x}.
\end{equation}
Show that the Holevo information $\chi(\mathcal{E})$ is equal to the mutual
information $I(X;B)_{\sigma}$:%
\begin{equation}
\chi(\mathcal{E})=I(X;B)_{\sigma}.
\end{equation}
In this sense, the quantum mutual information of a classical--quantum state is
most similar to the classical mutual information of Shannon.
\end{exercise}

\begin{exercise}
[Concavity of Quantum Entropy]\label{ex-qie:concavity-entropy}%
\index{von Neumann entropy!concavity}%
Prove the concavity of entropy (Property~\ref{prop-qie:concavity}) using
Theorem~\ref{thm-ie:QMI-positive}\ and the result of
Exercise~\ref{ex-qie:holevo-cq-state}.
\end{exercise}

\begin{exercise}
[Dimension Bound]Let $\sigma_{XB}\in\mathcal{D}(\mathcal{H}_{X}\otimes
\mathcal{H}_{B})$ be a classical--quantum state of the form:%
\begin{equation}
\sigma_{XB}=\sum_{x}p_{X}(x)|x\rangle\langle x|_{X}\otimes\sigma_{B}^{x}.
\end{equation}%
\index{Holevo information!dimension bound}
Prove that the following bound applies to the Holevo information:%
\begin{equation}
I(X;B)_{\sigma}\leq\log\left[  \min\left\{  \dim(\mathcal{H}_{X}%
),\dim(\mathcal{H}_{B})\right\}  \right]  .
\end{equation}
What is an example of a state that saturates the bound?
\end{exercise}

\section{Conditional Quantum Mutual Information}

\label{sec-qie:cond-MI}We define the conditional quantum mutual information
$I(A;B|C)_{\rho}$ of any tripartite state $\rho_{ABC}\in\mathcal{D}%
(\mathcal{H}_{A}\otimes\mathcal{H}_{B}\otimes\mathcal{H}_{C})$ similarly to
how we did in the classical case:%
\begin{equation}
I(A;B|C)_{\rho}\equiv H(A|C)_{\rho}+H(B|C)_{\rho}-H(AB|C)_{\rho}.
\end{equation}
In what follows, we sometimes abbreviate \textquotedblleft conditional quantum
mutual information\textquotedblright\ as CQMI.

One can exploit the above definition and the definition of quantum mutual
information to prove a chain rule for quantum mutual information.

\begin{property}
[Chain Rule for Quantum Mutual Information]\label{prop-qie:chain-CMI}The
quantum mutual information
\index{quantum mutual information!chain rule}
obeys a chain rule:%
\begin{equation}
I(A;BC)_{\rho}=I(A;B)_{\rho}+I(A;C|B)_{\rho}.
\end{equation}
The interpretation of the chain rule is that we can build up the correlations
between $A$ and $BC$ in two steps: in a first step, we build up the
correlations between $A$ and $B$, and now that $B$ is available (and thus
conditioned on), we build up the correlations between $A$ and $C$.
\end{property}

\begin{exercise}
\label{ex-qie:chain-rule-mut-info}Use the chain rule for quantum mutual
information to prove that%
\begin{equation}
I(A;BC)_{\rho}=I(AC;B)_{\rho}+I(A;C)_{\rho}-I(B;C)_{\rho}.
\end{equation}

\end{exercise}

\subsection{Non-Negativity of CQMI}

In the classical world, non-negativity of conditional mutual information
follows trivially from non-negativity of mutual information (recall
Theorem~\ref{thm-ie:class-ssa}). The proof of non-negativity of conditional
quantum mutual information is far from trivial in the quantum world, unless
the conditioning system is classical (see Exercise~\ref{ex-qie:trivial-SSA}).
It is a foundational result that non-negativity of this quantity holds because
so much of quantum information theory rests upon this theorem's shoulders (in
fact, we could say that this inequality is one of the \textquotedblleft
bedrocks\textquotedblright\ of quantum information theory). The list of its
corollaries includes the quantum data-processing inequality, the answers to
some additivity questions in quantum Shannon theory, the Holevo bound, and
others. The proof of Theorem~\ref{thm-ie:ssa-quantum}\ follows directly from
monotonicity of quantum relative entropy (Theorem~\ref{thm-qie:mono-rel-ent}),
which we prove in Chapter~\ref{chap:mono-rel-ent}. In fact, it is possible to
show that monotonicity of quantum relative entropy follows from strong
subadditivity as well, so that these two entropy inequalities are essentially
equivalent statements.

\begin{theorem}
[Non-Negativity of CQMI]\label{thm-ie:ssa-quantum}Let $\rho_{ABC}%
\in\mathcal{D}(\mathcal{H}_{A}\otimes\mathcal{H}_{B}\otimes\mathcal{H}_{C})$.
Then the conditional quantum mutual information
\index{quantum mutual information!conditional!non-negativity}
is non-negative:%
\begin{equation}
I(A;B|C)_{\rho}\geq0.
\end{equation}
This condition is equivalent to the strong subadditivity%
\index{strong subadditivity}
inequality in Exercise~\ref{ex-qie:strong-sub}, so we also refer to this
entropy inequality as strong subadditivity.
\end{theorem}

\begin{exercise}
[CQMI of Classical--Quantum States]\label{ex-qie:trivial-SSA}Consider a
classical--quantum state $\sigma_{XAB}$\ of the form in
\eqref{eq-qie:cqq-state}. Prove the following relation:%
\begin{equation}
I(A;B|X)_{\sigma}=\sum_{x}p_{X}(x)I(A;B)_{\sigma_{x}}.
\end{equation}
Conclude that non-negativity of conditional quantum mutual information is
trivial in this special case in which the conditioning system is classical,
simply by exploiting non-negativity of quantum mutual information
(Theorem~\ref{thm-ie:QMI-positive}).
\end{exercise}

\begin{exercise}
[Conditioning Does Not Increase Entropy]\label{ex-qie:SSA->cond-red-ent}Let
$\rho_{ABC}\in\mathcal{D}(\mathcal{H}_{A}\otimes\mathcal{H}_{B}\otimes
\mathcal{H}_{C})$. Show that Theorem~\ref{thm-ie:ssa-quantum}\ is equivalent
to the following stronger form of Theorem~\ref{thm-qie:cond-reduce-ent}:%
\begin{equation}
H(B|C)_{\rho}\geq H(B|AC)_{\rho}.
\end{equation}

\end{exercise}

\begin{exercise}
[Conditional Entropy \& Recoverability]Show that $H(B|C)_{\rho}=H(B|AC)_{\rho}
$ if there exists a recovery channel $\mathcal{R}_{C\rightarrow AC}$ such that
$\rho_{ABC}=\mathcal{R}_{C\rightarrow AC}(\rho_{BC})$ for $\rho_{ABC}%
\in\mathcal{D}(\mathcal{H}_{A}\otimes\mathcal{H}_{B}\otimes\mathcal{H}_{C})$.
(We will see later that this can be strengthened to \textquotedblleft if and
only if.\textquotedblright)
\end{exercise}

\begin{exercise}
[Concavity of Conditional Quantum Entropy]\label{ex-qie:concavity-cond-ent}%
Show that strong subadditivity implies that conditional entropy%
\index{von Neumann entropy!conditional!concavity}
is concave. That is, prove that%
\begin{equation}
\sum_{x}p_{X}(x)H(A|B)_{\rho^{x}}\leq H(A|B)_{\rho},
\end{equation}
where $p_{X}$ is a probability distribution on a finite alphabet $\mathcal{X}%
$, $\rho_{AB}^{x}\in\mathcal{D}(\mathcal{H}_{A}\otimes\mathcal{H}_{B})$ for
all $x\in\mathcal{X}$, and $\rho_{AB}\equiv\sum_{x}p_{X}(x)\rho_{AB}^{x}$.
\end{exercise}

\begin{exercise}
[Convexity of Coherent Information]Prove that coherent information%
\index{coherent information!convexity}
is convex:%
\begin{equation}
\sum_{x}p_{X}(x)I(A\rangle B)_{\rho^{x}}\geq I(A\rangle B)_{\rho},
\end{equation}
by exploiting the result of the above exercise.
\end{exercise}

\begin{exercise}
[Strong Subadditivity]\label{ex-qie:strong-sub}%
Theorem~\ref{thm-ie:ssa-quantum} also goes by the name of \textquotedblleft
strong subadditivity\textquotedblright\ because it is an example of a function
$\phi$\ that is strongly subadditive:%
\begin{equation}
\phi(E)+\phi(F)\geq\phi(E\cap F)+\phi(E\cup F). \label{eq-qie:SSA}%
\end{equation}
Let $\rho_{ABC}\in\mathcal{D}(\mathcal{H}_{A}\otimes\mathcal{H}_{B}%
\otimes\mathcal{H}_{C})$. Show that non-negativity of conditional quantum
mutual information is equivalent to the following strong subadditivity%
\index{strong subadditivity}
of quantum entropy:%
\begin{equation}
H(AC)_{\rho}+H(BC)_{\rho}\geq H(C)_{\rho}+H(ABC)_{\rho},
\end{equation}
where we think of $\phi$ in \eqref{eq-qie:SSA} as the entropy function $H$,
the argument $E$ in \eqref{eq-qie:SSA} as $AC$, and the argument $F$ in
\eqref{eq-qie:SSA} as $BC$.
\end{exercise}

\begin{exercise}
[Duality of CQMI]%
\index{quantum mutual information!conditional!duality}
Let $|\psi\rangle_{ABCD}\in\mathcal{H}_{A}\otimes\mathcal{H}_{B}%
\otimes\mathcal{H}_{C}\otimes\mathcal{H}_{D}$ be a pure state. Prove that%
\begin{equation}
I(A;B|C)_{\psi}=I(A;B|D)_{\psi}.
\end{equation}

\end{exercise}

\begin{exercise}
[Dimension Bound]\label{ex-qie:CMI-dim-bound}
\index{quantum mutual information!conditional!dimension bound}%
Let $\rho_{ABC}\in\mathcal{D}(\mathcal{H}_{A}\otimes\mathcal{H}_{B}%
\otimes\mathcal{H}_{C})$. Prove the following dimension bound:%
\begin{equation}
I(A;B|C)_{\rho}\leq2\log\left[  \min\left\{  \dim(\mathcal{H}_{A}%
),\dim(\mathcal{H}_{B})\right\}  \right]  .
\end{equation}
Let $\sigma_{XBC}\in\mathcal{D}(\mathcal{H}_{X}\otimes\mathcal{H}_{B}%
\otimes\mathcal{H}_{C})$ be a classical--quantum--quantum state of the form%
\begin{equation}
\sum_{x}p_{X}(x)|x\rangle\langle x|_{X}\otimes\sigma_{BC}^{x}.
\end{equation}
Prove that%
\begin{equation}
I(X;B|C)_{\sigma}\leq\log\dim(\mathcal{H}_{X}).
\end{equation}

\end{exercise}

\begin{exercise}
[Araki--Lieb triangle inequality]\label{ex-qie:araki-lieb}
\index{Araki--Lieb triangle inequality}%
Let $\rho_{AB}\in\mathcal{D}(\mathcal{H}_{A}\otimes\mathcal{H}_{B})$. Show
that%
\begin{equation}
\left\vert H(A)_{\rho}-H(B)_{\rho}\right\vert \leq H(AB)_{\rho}.
\end{equation}

\end{exercise}

\section{Quantum Relative Entropy}

The quantum relative entropy is one of the most important entropic quantities
in quantum information theory, mainly because we can reexpress many of the
entropies given in the previous sections in terms of it. This in turn allows
us to establish many properties of these quantities from the properties of
relative entropies. Its definition is a natural extension of that for the
classical relative entropy (see Definition~\ref{def-cie:rel-ent}). Before
defining it, we need the notion of the support of an operator:

\begin{definition}
[Kernel and Support]%
\index{support!of an operator}
\index{kernel}%
The kernel of an operator $A\in\mathcal{L}(\mathcal{H},\mathcal{H}^{\prime})$
is defined as%
\begin{equation}
\ker(A)\equiv\{|\psi\rangle\in\mathcal{H}:A|\psi\rangle=0\}.
\end{equation}
The support of $A$ is the subspace of $\mathcal{H}$\ orthogonal to its kernel:%
\begin{equation}
\operatorname{supp}(A)\equiv \ker(A)^\perp \equiv  \{|\psi\rangle\in\mathcal{H}: \langle \psi | \phi \rangle = 0 \quad \forall
|\phi\rangle \in \ker(A)\}.
\end{equation}
If $A$ is Hermitian and thus has a spectral decomposition as $A=\sum
_{i:a_{i}\neq0}a_{i}|i\rangle\langle i|$, then $\operatorname{supp}%
(A)=\operatorname{span}\{|i\rangle:a_{i}\neq0\}$. The projection onto the
support of $A$ is denoted by%
\begin{equation}
\Pi_{A}\equiv\sum_{i:a_{i}\neq0}|i\rangle\langle i|.
\end{equation}

\end{definition}

\begin{definition}
[Quantum Relative Entropy]\label{def-qie:q-rel-ent}The quantum relative
entropy%
\index{quantum relative entropy}
$D(\rho\Vert\sigma)$ between a density operator $\rho\in\mathcal{D}%
(\mathcal{H})$ and a positive semi-definite operator $\sigma\in\mathcal{L}%
(\mathcal{H})$ is defined as follows:%
\begin{equation}
D(\rho\Vert\sigma)\equiv\operatorname{Tr}\{  \rho\left[  \log\rho
-\log\sigma\right]  \}  , \label{eq-qie:rel-ent}%
\end{equation}
if the following support condition is satisfied%
\begin{equation}
\operatorname{supp}(\rho)\subseteq\operatorname{supp}(\sigma),
\label{eq-qie:supp-cond-rel-ent}%
\end{equation}
and it is defined to be equal to $+\infty$ otherwise.
\end{definition}

This definition is consistent with the classical definition in
Definition~\ref{def-cie:rel-ent}. However, we should note that there could be
several ways to generalize the classical definition to obtain a quantum
definition of relative entropy. For example, one could take%
\begin{equation}
D^{\prime}(\rho\Vert\sigma)=\operatorname{Tr}\left\{  \rho\log\left(
\rho^{1/2}\sigma^{-1}\rho^{1/2}\right)  \right\}  ,
\end{equation}
as a definition and it reduces to the classical definition in
Definition~\ref{def-cie:rel-ent} as well. In fact, it is easy to see that
there are an infinite number of quantum generalizations of the classical
definition of relative entropy. So how do we single out which definition is
the right one to use? The definition given in \eqref{eq-qie:rel-ent} is the
answer to a meaningful quantum information-processing task in the context of
quantum hypothesis testing (we do not elaborate on this further here but just
mention that it is known as the quantum Stein's lemma). Furthermore, this
definition generalizes the quantum entropic quantities we have given in this
chapter, which all in turn are the answers to meaningful quantum
information-processing tasks. For these reasons, we take the definition given
in \eqref{eq-qie:rel-ent} as \textit{the} quantum relative entropy. Recall
that it was this same line of reasoning that allowed us to single out the
entropy and the mutual information as meaningful measures of information in
the classical case.

Similar to the classical case, we can intuitively think of the quantum
relative entropy as a distance measure between quantum states. But it is not
strictly a distance measure in the mathematical sense because it is not
symmetric and it does not obey a triangle inequality.

The following proposition justifies why we take the definition of quantum
relative entropy to have the particular support conditions as given above:

\begin{proposition}
Let $\rho\in\mathcal{D}(\mathcal{H})$ and $\sigma\in\mathcal{L}(\mathcal{H})$
be positive semi-definite. The quantum relative entropy is consistent with the
following limit:%
\begin{equation}
D(\rho\Vert\sigma)=\lim_{\varepsilon\searrow0}D(\rho\Vert\sigma+\varepsilon
I).
\end{equation}

\end{proposition}

\begin{proof}
First observe that the operator $\sigma+\varepsilon I$ has support equal to
$\mathcal{H}$ for all $\varepsilon>0$, so that $D(\rho\Vert\sigma+\varepsilon
I)$ is finite for all $\varepsilon>0$. We will see that the limit is finite
and consistent with \eqref{eq-qie:rel-ent} if \eqref{eq-qie:supp-cond-rel-ent}
is satisfied, and otherwise the limit blows up to infinity.\ The idea in
proving this proposition is to represent both $\rho$ and $\sigma$ with respect to
the decomposition $\mathcal{H}=\operatorname{supp}(\sigma)\oplus\ker(\sigma)$.
Let $\Pi_{\sigma}$ denote the projection onto $\operatorname{supp}(\sigma)$
and let $\Pi_{\sigma}^{\perp}$ denote the projection onto $\ker(\sigma)$. So
we take%
\begin{equation}
\rho=%
\begin{bmatrix}
\rho_{00} & \rho_{01}\\
\rho_{10} & \rho_{11}%
\end{bmatrix}
,\ \ \ \ \ \ \ \ \ \ \sigma=%
\begin{bmatrix}
\sigma_{0} & 0\\
0 & 0
\end{bmatrix}
.
\end{equation}

First suppose that the support condition in \eqref{eq-qie:supp-cond-rel-ent}
is satisfied. Then this means that $\rho_{01}=\rho_{10}^{\dag}=0$
and$\ \rho_{11}=0$. Observe that%
\begin{equation}
D(\rho\Vert\sigma+\varepsilon I)=\operatorname{Tr}\left\{  \rho\log
\rho\right\}  -\operatorname{Tr}\left\{  \rho\log\left(  \sigma+\varepsilon
I\right)  \right\}  .
\end{equation}
The first term is finite for any $\rho$, so we should focus on the second term
exclusively, since this is where an issue could arise. Then%
\begin{align}
\operatorname{Tr}\left\{  \rho\log\left(  \sigma+\varepsilon I\right)
\right\}   &  =\operatorname{Tr}\left\{
\begin{bmatrix}
\rho_{00} & 0\\
0 & 0
\end{bmatrix}
\log%
\begin{bmatrix}
\sigma_{0}+\varepsilon\Pi_{\sigma} & 0\\
0 & \varepsilon\Pi_{\sigma}^{\perp}%
\end{bmatrix}
\right\} \nonumber\\
&  =\operatorname{Tr}\left\{
\begin{bmatrix}
\rho_{00} & 0\\
0 & 0
\end{bmatrix}%
\begin{bmatrix}
\log\left(  \sigma_{0}+\varepsilon\Pi_{\sigma}\right)  & 0\\
0 & \log\left(  \varepsilon\Pi_{\sigma}^{\perp}\right)
\end{bmatrix}
\right\} \\
&  =\operatorname{Tr}\left\{  \rho_{00}\log\left(  \sigma_{0}+\varepsilon
\Pi_{\sigma}\right)  \right\}  +\operatorname{Tr}\left\{  0\cdot\log\left(
\varepsilon\Pi_{\sigma}^{\perp}\right)  \right\} \\
&  =\operatorname{Tr}\left\{  \rho_{00}\log\left(  \sigma_{0}+\varepsilon
\Pi_{\sigma}\right)  \right\}  .
\end{align}
Taking the limit $\varepsilon\searrow0$, we get%
\begin{equation}
\lim_{\varepsilon\searrow0}\operatorname{Tr}\left\{  \rho_{00}\log\left(
\sigma_{0}+\varepsilon\Pi_{\sigma}\right)  \right\}  =\operatorname{Tr}%
\left\{  \rho_{00}\log\sigma_{0}\right\}  =\operatorname{Tr}\left\{  \rho
\log\sigma\right\}  .
\end{equation}
So we can conclude that%
\begin{equation}
\lim_{\varepsilon\searrow0}D(\rho\Vert\sigma+\varepsilon I)=\operatorname{Tr}%
\left\{  \rho\log\rho\right\}  -\operatorname{Tr}\left\{  \rho\log
\sigma\right\}  ,
\end{equation}
in this case.

Now suppose that the support condition in \eqref{eq-qie:supp-cond-rel-ent} is
not satisfied. Then $\rho_{11}\neq0$, and we find that%
\begin{align}
\operatorname{Tr}\left\{  \rho\log\left(  \sigma+\varepsilon I\right)
\right\}   &  =\operatorname{Tr}\left\{
\begin{bmatrix}
\rho_{00} & \rho_{01}\\
\rho_{10} & \rho_{11}%
\end{bmatrix}%
\begin{bmatrix}
\log\left(  \sigma_{0}+\varepsilon\Pi_{\sigma}\right)  & 0\\
0 & \log\left(  \varepsilon\Pi_{\sigma}^{\perp}\right)
\end{bmatrix}
\right\} \nonumber\\
&  =\operatorname{Tr}\left\{  \rho_{00}\log\left(  \sigma_{0}+\varepsilon
\Pi_{\sigma}\right)  \right\}  +\operatorname{Tr}\left\{  \rho_{11}\cdot
\log\left(  \varepsilon\Pi_{\sigma}^{\perp}\right)  \right\}  ,
\end{align}
and thus $\lim_{\varepsilon\searrow0}D(\rho\Vert\sigma+\varepsilon I)=+\infty
$, given that $\lim_{\varepsilon\searrow0}\left[  -\log\varepsilon\right]
=+\infty$.
\end{proof}

One of the most fundamental entropy inequalities in quantum information theory
is the monotonicity of quantum relative entropy. When the arguments to the
quantum relative entropy are quantum states, the physical interpretation of
this entropy inequality\ is that states become less distinguishable when noise
acts on them. We defer a proof of this theorem until
Chapter~\ref{chap:mono-rel-ent}, where we also establish a strengthening of it.

\begin{theorem}
[Monotonicity of Quantum Relative Entropy]\label{thm-qie:mono-rel-ent}Let
$\rho\in\mathcal{D}(\mathcal{H})$, $\sigma\in\mathcal{L}(\mathcal{H})$ be
positive semi-definite, and $\mathcal{N}:\mathcal{L}(\mathcal{H}%
)\rightarrow\mathcal{L}(\mathcal{H}^{\prime})$ be a quantum channel. The
quantum relative entropy%
\index{quantum relative entropy!monotonicity}
can only decrease or stay the same if we apply the same quantum channel
$\mathcal{N}$\ to $\rho$ and $\sigma$:%
\begin{equation}
D(\rho\Vert\sigma)\geq D(\mathcal{N}(\rho)\Vert\mathcal{N}(\sigma)).
\label{eq-qie:mono-rel-ent}%
\end{equation}

\end{theorem}

\noindent Theorem~\ref{thm-qie:mono-rel-ent}\ then implies non-negativity of
quantum relative entropy in certain cases.

\begin{theorem}
[Non-Negativity]\label{thm-qie:non-neg-rel-ent}Let $\rho\in\mathcal{D}%
(\mathcal{H})$, and let $\sigma\in\mathcal{L}(\mathcal{H})$ be positive
semi-definite and such that $\operatorname{Tr}\{\sigma\}\leq1$. Then the
\index{quantum relative entropy!non-negativity}%
quantum relative entropy $D(\rho\Vert\sigma)$ is non-negative:%
\begin{equation}
D(\rho\Vert\sigma)\geq0,
\end{equation}
and $D(\rho\Vert\sigma)=0$ if and only if $\rho=\sigma$.
\end{theorem}

\begin{proof}
The first part of the theorem follows from applying
Theorem~\ref{thm-qie:mono-rel-ent}, taking the quantum channel to be the
trace-out map. We then have that%
\begin{equation}
D(\rho\Vert\sigma) \geq D(\operatorname{Tr}\{\rho\}\Vert\operatorname{Tr}%
\{\sigma\}) =\operatorname{Tr}\{\rho\}\log\!\left(  \frac{\operatorname{Tr}%
\{\rho\}}{\operatorname{Tr}\{\sigma\}}\right)  \geq0.
\end{equation}
If $\rho=\sigma$, then the support condition in
\eqref{eq-qie:supp-cond-rel-ent} is satisfied and plugging into
\eqref{eq-qie:rel-ent}\ gives that $D(\rho\Vert\sigma)=0$. Now suppose that
$D(\rho\Vert\sigma)=0$. This means that the inequality above is saturated and
thus $\operatorname{Tr}\{\sigma\}=\operatorname{Tr}\{\rho\}=1$, so that
$\sigma$ is a density operator. Let $\mathcal{M}$\ be an arbitrary measurement
channel. From the monotonicity of quantum relative entropy
(Theorem~\ref{thm-qie:mono-rel-ent}), we can conclude that $D(\mathcal{M}%
(\rho)\Vert\mathcal{M}(\sigma))=0$. The equality condition for the
non-negativity of the classical relative entropy
(Theorem~\ref{thm-cie:rel-ent-positive})\ in turn implies that $\mathcal{M}%
(\rho)=\mathcal{M}(\sigma)$. Now since this equality holds for any possible
measurement channel, we can conclude that $\rho=\sigma$. (For example, we
could take $\mathcal{M}$ to be the optimal measurement for the trace distance,
which would allow us to conclude that $\max_{\mathcal{M}}\left\Vert
\mathcal{M}(\rho)-\mathcal{M}(\sigma)\right\Vert _{1}=\left\Vert \rho
-\sigma\right\Vert _{1}=0$, and hence $\rho=\sigma$.)
\end{proof}

\subsection{Deriving Other Entropies from Quantum Relative Entropy}

There is a sense in which the quantum relative entropy is a \textquotedblleft
parent quantity\textquotedblright\ for other entropies in quantum information
theory, such as the quantum entropy, the conditional quantum entropy, the
quantum mutual information, and the conditional quantum mutual information.
The following exercises explore these relations. The main tool needed to solve
some of them is the non-negativity of quantum relative entropy.

\begin{exercise}
[Operator Logarithm]\label{ex-qie:log-operators}Let $P_{A}\in\mathcal{L}%
(\mathcal{H}_{A})$ and $Q_{B}\in\mathcal{L}(\mathcal{H}_{B})$ be positive
semi-definite operators. Show that the following identity holds:%
\begin{equation}
\log\left(  P_{A}\otimes Q_{B}\right)  =\log\left(  P_{A}\right)  \otimes
I_{B}+I_{A}\otimes\log\left(  Q_{B}\right)  .
\end{equation}

\end{exercise}

\begin{exercise}
[Mutual Information and Relative Entropy]\label{ex-qie:rel-ent-mut-info}Let
$\rho_{AB}\in\mathcal{D}(\mathcal{H}_{A}\otimes\mathcal{H}_{B})$. Show that
the following identities hold:%
\begin{align}
I(A;B)_{\rho}  &  =D(\rho_{AB}\Vert\rho_{A}\otimes\rho_{B})\\
&  =\min_{\sigma_{B}}D(\rho_{AB}\Vert\rho_{A}\otimes\sigma_{B})\\
&  =\min_{\omega_{A}}D(\rho_{AB}\Vert\omega_{A}\otimes\rho_{B})\\
&  =\min_{\omega_{A},\sigma_{B}}D(\rho_{AB}\Vert\omega_{A}\otimes\sigma_{B}),
\end{align}
where the optimizations are with respect to $\omega_{A}\in\mathcal{D}%
(\mathcal{H}_{A})$ and $\sigma_{B}\in\mathcal{D}(\mathcal{H}_{B})$.
\end{exercise}

\begin{exercise}
[Conditional and Relative Entropy]\label{ex-qie:rel-ent-cond-ent}Let
$\rho_{AB}\in\mathcal{D}(\mathcal{H}_{A}\otimes\mathcal{H}_{B})$. Show that
the following identities hold:%
\begin{align}
I(A\rangle B)_{\rho}  &  =D(\rho_{AB}\Vert I_{A}\otimes\rho_{B})\\
&  =\min_{\sigma_{B}\in\mathcal{D}(\mathcal{H}_{B})}D(\rho_{AB}\Vert
I_{A}\otimes\sigma_{B}). \label{eq-qie:coh-info-rel-ent-min}%
\end{align}
Note that these are equivalent to%
\begin{align}
H(A|B)_{\rho}  &  =-D(\rho_{AB}\Vert I_{A}\otimes\rho_{B})\\
&  =-\min_{\sigma_{B}\in\mathcal{D}(\mathcal{H}_{B})}D(\rho_{AB}\Vert
I_{A}\otimes\sigma_{B}).
\end{align}

\end{exercise}

\begin{exercise}
[CQMI and Relative Entropy]Let $\rho_{ABC}\in\mathcal{D}(\mathcal{H}%
_{A}\otimes\mathcal{H}_{B}\otimes\mathcal{H}_{C})$. Let $\omega_{ABC}$ be the
following positive semi-definite operator:%
\begin{equation}
\omega_{ABC}\equiv2^{\left[  \log\rho_{AC}+\log\rho_{BC}-\log\rho_{C}\right]
},
\end{equation}
where identities are implicit if not written (e.g., $\rho_{BC}$ is a shorthand
for $I_{A}\otimes\rho_{BC}$). Show that%
\begin{equation}
I(A;B|C)_{\rho}=D(\rho_{ABC}\Vert\omega_{ABC}).
\end{equation}

\end{exercise}

\begin{exercise}
[Dimension Bound]\label{ex-qie:coh-info-dim-bound-part-two}Let $\rho_{ABC}%
\in\mathcal{D}(\mathcal{H}_{A}\otimes\mathcal{H}_{B}\otimes\mathcal{H}_{C})$.
Prove the following dimension bound:%
\begin{equation}
I(A\rangle BC)_{\rho}\leq I(AC\rangle B)_{\rho}+\log\dim(\mathcal{H}_{C}).
\end{equation}
(Hint:\ One way to do this is to use the formula in
\eqref{eq-qie:coh-info-rel-ent-min}. Another way is to use the chain rule and
previous dimension bounds.)
\end{exercise}

\begin{corollary}
[Subadditivity of Quantum Entropy]\label{cor-qie:subadd}The quantum entropy%
\index{von Neumann entropy!subadditivity}
is subadditive for a bipartite state $\rho_{AB}$:%
\begin{equation}
H(A)_{\rho}+H(B)_{\rho}\geq H(AB)_{\rho}.
\end{equation}

\end{corollary}

\begin{proof}
Subadditivity of entropy is equivalent to non-negativity of quantum mutual
information. We can prove non-negativity by exploiting the result of
Exercise~\ref{ex-qie:rel-ent-mut-info} and non-negativity of quantum relative
entropy (Theorem~\ref{thm-qie:non-neg-rel-ent}).
\end{proof}

\subsection{Mathematical Properties of Quantum Relative Entropy}

This section contains several auxiliary mathematical properties of quantum
relative entropy, including its isometric invariance, additivity for
tensor-product states, its form for classical--quantum states (these are left
as exercises). There are two other properties given which are commonly used in
relative entropy calculations.

\begin{exercise}
[Isometric Invariance]\label{ex-qie:rel-ent-unit-invar}
\index{quantum relative entropy!isometric invariance}%
Let $\rho\in\mathcal{D}(\mathcal{H})$ and $\sigma\in\mathcal{L}(\mathcal{H})$
be positive semi-definite. Show that the quantum relative entropy is invariant
with respect to an isometry $U:\mathcal{H}\rightarrow\mathcal{H}^{\prime}$:%
\begin{equation}
D(\rho\Vert\sigma)=D(U\rho U^{\dag}\Vert U\sigma U^{\dag}).
\end{equation}

\end{exercise}

\begin{exercise}
[Additivity of Quantum Relative Entropy]\label{ex-qie:add-q-rel-ent}
\index{quantum relative entropy!additivity}%
Let $\rho_{1}\in\mathcal{D}(\mathcal{H}_{1})$ and $\rho_{2}\in\mathcal{D}%
(\mathcal{H}_{1})$ be density operators, and let $\sigma_{1}\in\mathcal{L}%
(\mathcal{H}_{1})$ and $\sigma_{2}\in\mathcal{L}(\mathcal{H}_{2})$ be positive
semi-definite operators. Show that the quantum relative entropy is additive in
the following sense:%
\begin{equation}
D(\rho_{1}\otimes\rho_{2}\Vert\sigma_{1}\otimes\sigma_{2})=D(\rho_{1}%
\Vert\sigma_{1})+D(\rho_{2}\Vert\sigma_{2}).
\end{equation}
We can apply the above additivity relation inductively to conclude that%
\begin{equation}
D(\rho^{\otimes n}\Vert\sigma^{\otimes n})=nD(\rho\Vert\sigma),
\end{equation}
for $\rho\in\mathcal{D}(\mathcal{H})$ and $\sigma\in\mathcal{L}(\mathcal{H})$
positive semi-definite.
\end{exercise}

\begin{exercise}
[Relative Entropy of Classical--Quantum States]\label{ex-qie:cq-rel-ent}
\index{quantum relative entropy!for classical--quantum states}%
Show that the quantum relative entropy between classical--quantum states
$\rho_{XB}$ and $\sigma_{XB}$ is as follows:%
\begin{equation}
D(\rho_{XB}\Vert\sigma_{XB})=\sum_{x}p(x)D(\rho_{B}^{x}\Vert\sigma_{B}%
^{x}) + D(p\| q),\ \text{where}%
\end{equation}%
\begin{equation}
\rho_{XB}\equiv\sum_{x}p(x)|x\rangle\langle x|_{X}\otimes\rho_{B}%
^{x},\ \ \ \ \ \ \ \ \ \ \sigma_{XB}\equiv\sum_{x}q(x)|x\rangle\langle
x|_{X}\otimes\sigma_{B}^{x},
\end{equation}
with $p$ and $q$ probability distributions over a finite alphabet $\mathcal{X}$,
$\rho_{B}^{x}\in\mathcal{D}(\mathcal{H}_{B})$ for all $x\in\mathcal{X}$, and
$\sigma_{B}^{x}\in\mathcal{L}(\mathcal{H}_{B})$ positive semi-definite for all
$x\in\mathcal{X}$.
\end{exercise}

\begin{exercise}
\label{ex-qie:rel-ent-scaling}Let $a,b>0$, $\rho\in\mathcal{D}(\mathcal{H})$,
and $\sigma\in\mathcal{L}(\mathcal{H})$ be positive semi-definite. Show that%
\begin{equation}
D(a\rho\Vert b\sigma)=a\left[  D(\rho\Vert\sigma)+\log\left(  a/b\right)
\right]  .
\end{equation}
(Note that we only defined quantum relative entropy to have its first argument
equal to a density operator, but one could more generally allow for the first
argument to be positive semi-definite.)
\end{exercise}

\begin{proposition}
\label{prop-qie:rel-ent-s-s'}Let $\rho\in\mathcal{D}(\mathcal{H})$ and
$\sigma,\sigma^{\prime}\in\mathcal{L}(\mathcal{H})$ be positive semi-definite.
Suppose that $\sigma\leq\sigma^{\prime}$. Then
\begin{equation}
D(\rho\Vert\sigma^{\prime})\leq D(\rho\Vert\sigma).
\end{equation}

\end{proposition}

\begin{proof}
The assumption that $\sigma\leq\sigma^{\prime}$ is equivalent to
$\sigma^{\prime}-\sigma$ being positive semi-definite. Then the following
operator is positive semi-definite: $\sigma\otimes|0\rangle\langle
0|_{X}+\left(  \sigma^{\prime}-\sigma\right)  \otimes|1\rangle\langle1|_{X}$,
and as a consequence%
\begin{equation}
D(\rho\Vert\sigma)=D(\rho\otimes|0\rangle\langle0|_{X}\Vert\left[
\sigma\otimes|0\rangle\langle0|_{X}+\left(  \sigma^{\prime}-\sigma\right)
\otimes|1\rangle\langle1|_{X}\right]  ),
\end{equation}
which follows by a direct calculation (essentially the same reasoning as that
used to solve Exercise~\ref{ex-qie:cq-rel-ent}). By monotonicity of quantum
relative entropy (Theorem~\ref{thm-qie:mono-rel-ent}), the quantum relative
entropy does not increase after discarding the system $X$, so that%
\begin{multline}
D(\rho\otimes|0\rangle\langle0|_{X}\Vert\left[  \sigma\otimes|0\rangle
\langle0|_{X}+\left(  \sigma^{\prime}-\sigma\right)  \otimes|1\rangle
\langle1|_{X}\right]  )\\
\geq D(\rho\Vert\left[  \sigma+\left(  \sigma^{\prime}-\sigma\right)  \right]
)=D(\rho\Vert\sigma^{\prime}),
\end{multline}
concluding the proof.
\end{proof}

\section{Quantum Entropy Inequalities}

Monotonicity of quantum relative entropy has as its corollaries many of the
important entropy inequalities in quantum information theory (but keep in mind
that some of these also imply the monotonicity of quantum relative entropy).

\begin{corollary}
[Strong Subadditivity]\label{cor-qie:SSA}
\index{strong subadditivity}%
Let $\rho_{ABC}\in\mathcal{D}(\mathcal{H}_{A}\otimes\mathcal{H}_{B}%
\otimes\mathcal{H}_{C})$. The quantum entropy is strongly subadditive, in the
following sense:%
\begin{equation}
H(AC)_{\rho}+H(BC)_{\rho}\geq H(ABC)_{\rho}+H(C)_{\rho}.
\end{equation}

\end{corollary}

\begin{proof}
Consider from Exercise~\ref{ex-qie:strong-sub} that%
\begin{equation}
I(A;B|C)_{\rho}=H(AC)_{\rho}+H(BC)_{\rho}-H(ABC)_{\rho}-H(C)_{\rho},
\label{eq-qie:SSA-3}%
\end{equation}
so that%
\begin{equation}
I(A;B|C)_{\rho}=H(B|C)_{\rho}-H(B|AC)_{\rho}.
\end{equation}
From Exercise~\ref{ex-qie:rel-ent-cond-ent}, we know that%
\begin{align}
-H(B|AC)_{\rho}  &  =D(\rho_{ABC}\Vert I_{B}\otimes\rho_{AC}),\\
H(B|C)_{\rho}  &  =-D(\rho_{BC}\Vert I_{B}\otimes\rho_{C}).
\label{eq-qie:SSA-4}%
\end{align}
Then%
\begin{align}
D(\rho_{ABC}\Vert I_{B}\otimes\rho_{AC})  &  \geq D(\operatorname{Tr}%
_{A}\{\rho_{ABC}\}\Vert\operatorname{Tr}_{A}\{I_{B}\otimes\rho_{AC}%
\})\label{eq-qie:SSA-1}\\
&  =D(\rho_{BC}\Vert I_{B}\otimes\rho_{C}). \label{eq-qie:SSA-2}%
\end{align}
The inequality is a consequence of the monotonicity of quantum relative
entropy (Theorem~\ref{thm-qie:mono-rel-ent}), taking $\rho=\rho_{ABC}$,
$\sigma=I_{B}\otimes\rho_{AC}$, and $\mathcal{N}=\operatorname{Tr}_{A}$. By
\eqref{eq-qie:SSA-3}--\eqref{eq-qie:SSA-4}, the inequality in
\eqref{eq-qie:SSA-1}--\eqref{eq-qie:SSA-2} is equivalent to the inequality in
the statement of the corollary.
\end{proof}

\begin{corollary}
[Joint Convexity of Quantum Relative Entropy]Let $p_{X}$ be a probability
distribution over a finite alphabet $\mathcal{X}$, $\rho^{x}\in\mathcal{D}%
(\mathcal{H})$ for all $x\in\mathcal{X}$, and $\sigma^{x}\in\mathcal{L}%
(\mathcal{H})$ be positive semi-definite for all $x\in\mathcal{X}$. Set
$\overline{\rho}\equiv\sum_{x}p_{X}(x)\rho^{x}$ and $\overline{\sigma}%
\equiv\sum_{x}p_{X}(x)\sigma^{x}$. The quantum relative entropy%
\index{quantum relative entropy!joint convexity}
is jointly convex in its arguments:%
\begin{equation}
\sum_{x}p_{X}(x)D(\rho^{x}\Vert\sigma^{x})\geq D(\overline{\rho}\Vert
\overline{\sigma}).
\end{equation}

\end{corollary}

\begin{proof}
Consider classical--quantum states of the following form:%
\begin{align}
\rho_{XB}  &  \equiv\sum_{x}p_{X}(x)|x\rangle\langle x|_{X}\otimes\rho_{B}%
^{x},\\
\sigma_{XB}  &  \equiv\sum_{x}p_{X}(x)|x\rangle\langle x|_{X}\otimes\sigma
_{B}^{x}.
\end{align}
Observe that $\overline{\rho}=\rho_{B}$ and $\overline{\sigma}=\sigma_{B}$.
Then%
\begin{equation}
\sum_{x}p_{X}(x)D(\rho_{B}^{x}\Vert\sigma_{B}^{x})=D(\rho_{XB}\Vert\sigma
_{XB})\geq D(\rho_{B}\Vert\sigma_{B}).
\end{equation}
The equality follows from Exercise~\ref{ex-qie:cq-rel-ent}, and the inequality
follows from monotonicity of quantum relative entropy
(Theorem~\ref{thm-qie:mono-rel-ent}), where we take the channel to be the
partial trace over the system $X$.
\end{proof}

\begin{corollary}
[Unital Channels Increase Entropy]\label{cor-qie:unital-ch-inc-ent}Let
$\rho\in\mathcal{D}(\mathcal{H})$ and let $\mathcal{N}:\mathcal{L}%
(\mathcal{H})\rightarrow\mathcal{L}(\mathcal{H})$ be a unital quantum channel
(see Definition~\ref{def-nqt:unital-map}). Then%
\begin{equation}
H(\mathcal{N}(\rho))\geq H(\rho). \label{eq-qie:dephasing-inc-ent}%
\end{equation}

\end{corollary}

\begin{proof}
Consider that%
\begin{align}
H(\rho)  &  =-D(\rho\Vert I),\\
H(\mathcal{N}(\rho))  &  =-D(\mathcal{N}(\rho)\Vert I)=-D(\mathcal{N}%
(\rho)\Vert\mathcal{N}(I)),
\end{align}
where in the last equality, we have used that $\mathcal{N}$ is a unital
quantum channel. The inequality in \eqref{eq-qie:dephasing-inc-ent} is a
consequence of the monotonicity of quantum relative entropy
(Theorem~\ref{thm-qie:mono-rel-ent}) because $D(\rho\Vert I)\geq
D(\mathcal{N}(\rho)\Vert\mathcal{N}(I))$.
\end{proof}

A particular example of a unital channel occurs when we completely dephase a
density operator $\rho$ with respect to some dephasing basis $\left\{
|y\rangle\right\}  $. Let $\omega$ denote the dephased version of $\rho$:%
\begin{equation}
\omega\equiv\Delta_{Y}(\rho)=\sum_{y}|y\rangle\langle y|\rho|y\rangle\langle
y|.
\end{equation}
Then the entropy $H(\omega)$ of the completely dephased state is never smaller
than the entropy $H(\rho)$ of the original state. More generally, if we have a
set of projectors $\{\Pi_{x}\}$ satisfying $\sum_{x}\Pi_{x}=I$, then the
channel $\rho\rightarrow\sum_{x}\Pi_{x}\rho\Pi_{x}$ is unital, so that%
\begin{equation}
H\left(  \sum_{x}\Pi_{x}\rho\Pi_{x}\right)  \geq H(\rho).
\end{equation}

The quantum relative entropy itself is not a distance measure, but it actually
gives a useful upper bound on the trace distance between two quantum states.
This result is known as the quantum Pinsker inequality. Thus, in this sense,
we can think of the quantum relative entropy as being comparable to a distance
measure when it is small---if the quantum relative entropy between two quantum
states is small, then their trace distance is small as well. We can view the
quantum Pinsker inequality as a refinement of the statement that the quantum
relative entropy is non-negative (Theorem~\ref{thm-qie:non-neg-rel-ent}).

\begin{theorem}
[Quantum Pinsker Inequality]\label{thm-qie:rel-ent-trace}
\index{Pinsker inequality!quantum}%
Let $\rho\in\mathcal{D}(\mathcal{H})$ and let $\sigma\in\mathcal{L}%
(\mathcal{H})$ be positive semi-definite such that $\operatorname{Tr}%
\{\sigma\}\leq1$. Then%
\begin{equation}
D(\rho\Vert\sigma)\geq\frac{1}{2\ln2}\left\Vert \rho-\sigma\right\Vert
_{1}^{2}.
\end{equation}

\end{theorem}

\begin{proof}
This is a direct consequence of the classical Pinsker inequality
(Theorem~\ref{thm-cie:pinsker}) and the fact that a measurement achieves the
trace distance (see Exercise~\ref{thm-dm:meas-achieve-TD}). To get the
statement for subnormalized $\sigma$, we add an extra dimension to the Hilbert
space $\mathcal{H}$. Let $\rho^{\prime}\equiv\rho\oplus\left[  0\right]  $ and
$\sigma^{\prime}\equiv\sigma\oplus\left[  1-\operatorname{Tr}\{\sigma
\}\right]  $, so that $\sigma^{\prime}$ is a density operator. Let
$\mathcal{M}$ denote a measurement channel that achieves the trace distance
for $\rho^{\prime}$ and $\sigma^{\prime}$ (see
Exercise~\ref{thm-dm:meas-achieve-TD}). Then%
\begin{align}
D(\rho\Vert\sigma)  &  =D(\rho^{\prime}\Vert\sigma^{\prime})\\
&  \geq D(\mathcal{M}(\rho^{\prime})\Vert\mathcal{M}(\sigma^{\prime}))\\
&  \geq\frac{1}{2\ln2}\left\Vert \mathcal{M}(\rho^{\prime})-\mathcal{M}%
(\sigma^{\prime})\right\Vert _{1}^{2}\\
&  =\frac{1}{2\ln2}\left\Vert \rho^{\prime}-\sigma^{\prime}\right\Vert
_{1}^{2}\\
&  =\frac{1}{2\ln2}\left[  \left\Vert \rho-\sigma\right\Vert _{1}%
+(1-\operatorname{Tr}\{\sigma\})\right]  ^{2}\\
&  \geq\frac{1}{2\ln2}\left\Vert \rho-\sigma\right\Vert _{1}^{2}.
\end{align}
The first inequality is a consequence of the monotonicity of quantum relative
entropy (Theorem~\ref{thm-qie:mono-rel-ent}). The second inequality follows
from the classical Pinsker inequality (Theorem~\ref{thm-cie:pinsker}). The
second equality follows because we chose the measurement channel $\mathcal{M}$
to achieve the trace distance.
\end{proof}

\subsection{Equivalence of Quantum Entropy Inequalities}

We have already seen how the monotonicity of quantum relative entropy
(Theorem~\ref{thm-qie:mono-rel-ent}) implies many of the important entropy
inequalities in quantum information theory. However, what is less obvious is
that some of these other entropy inequalities also imply the monotonicity of
quantum relative entropy. Thus, we can say that together, these entropy
inequalities constitute a \textquotedblleft law\textquotedblright\ of quantum
information theory, saying either that information, correlations, or
distinguishability decrease under the action of a quantum channel or
conditional uncertainty increases under the action of a quantum channel on a
conditioning system. We formalize this equivalence in the following theorem:

\begin{theorem}
\label{thm-qie:equivalence-ent-ineq}The following statements are equivalent,
in the sense that one can prove the other statements as a consequence of one
of them:

\begin{enumerate}
\item The quantum relative entropy is monotone with respect to quantum
channels: $D(\rho\Vert\sigma)\geq D(\mathcal{N}(\rho)\Vert\mathcal{N}%
(\sigma))$, where $\rho,\sigma\in\mathcal{D}(\mathcal{H})$, and $\mathcal{N}%
:\mathcal{L}(\mathcal{H})\rightarrow\mathcal{L}(\mathcal{H}^{\prime})$ is a
quantum channel.

\item The quantum relative entropy is monotone with respect to partial trace:
$D(\rho_{AB}\Vert\sigma_{AB})\geq D(\rho_{B}\Vert\sigma_{B})$, where
$\rho_{AB},\sigma_{AB}\in\mathcal{D}(\mathcal{H}_{A}\otimes\mathcal{H}_{B})$.

\item The quantum relative entropy is jointly convex: $\sum_{x}p_{X}%
(x)D(\rho^{x}\Vert\sigma^{x})\geq D(\overline{\rho}\Vert\overline{\sigma})$,
where $p_{X}$ is a probability distribution over a finite alphabet
$\mathcal{X}$, $\rho^{x},\sigma^{x}\in\mathcal{D}(\mathcal{H})$ for all
$x\in\mathcal{X}$, $\rho\equiv\sum_{x}p_{X}(x)\rho^{x}$, and $\sigma\equiv
\sum_{x}p_{X}(x)\sigma^{x}$.

\item The conditional quantum mutual information is non-negative:
$I(A;B|C)_{\rho}\geq0$, where $\rho_{ABC}\in\mathcal{D}(\mathcal{H}_{A}%
\otimes\mathcal{H}_{B}\otimes\mathcal{H}_{C})$.

\item The conditional quantum entropy is concave: $H(A|B)_{\rho}\geq\sum
_{x}p_{X}(x)H(A|B)_{\rho^{x}}$, where $p_{X}$ is a probability distribution on
a finite alphabet $\mathcal{X}$, $\rho_{AB}^{x}\in\mathcal{D}(\mathcal{H}%
_{A}\otimes\mathcal{H}_{B})$ for all $x\in\mathcal{X}$, and $\rho_{AB}%
\equiv\sum_{x}p_{X}(x)\rho_{AB}^{x}$.
\end{enumerate}
\end{theorem}

\begin{proof}
We have already proved 2-4 starting from 1, and 5 from 4 as well (some of
these are left as exercises), so it remains to work our way back up to 1 from
the others. Consider that we can get 1 from 2 by using the Stinespring
dilation theorem. That is, let $U:\mathcal{H}\rightarrow\mathcal{H}^{\prime
}\otimes\mathcal{H}_{E}$ be an isometric extension of the channel
$\mathcal{N}$.\ Consider that%
\begin{align}
D(\rho\Vert\sigma)  &  =D(U\rho U^{\dag}\Vert U\sigma U^{\dag})\\
&  \geq D(\operatorname{Tr}_{E}\{U\rho U^{\dag}\}\Vert\operatorname{Tr}%
_{E}\{U\sigma U^{\dag}\})\\
&  =D(\mathcal{N}(\rho)\Vert\mathcal{N}(\sigma)).
\end{align}
The first equality follows from invariance of quantum relative entropy with
respect to isometries (Exercise~\ref{ex-qie:rel-ent-unit-invar}). The
inequality follows from monotonicity with respect to partial trace (by
assumption), and the last equality follows from the fact that $U$ is an
isometric extension of $\mathcal{N}$.

We can also get 1 from 3 by a related approach. Let $d=\dim(\mathcal{H}_{E})$
and $\left\{  V_{E}^{i}\right\}  $ be a Heisenberg--Weyl set of unitaries for
the environment system $E$. Consider that%
\begin{align}
D(\mathcal{N}(\rho)\Vert\mathcal{N}(\sigma))  &  =D(\mathcal{N}(\rho
)\otimes\pi_{E}\Vert\mathcal{N}(\sigma)\otimes\pi_{E})\\
&  =D\left(  \frac{1}{d^{2}}\sum_{i}V_{E}^{i}U\rho U^{\dag}\left(  V_{E}%
^{i}\right)  ^{\dag}\middle\Vert\frac{1}{d^{2}}\sum_{i}V_{E}^{i}U\sigma
U^{\dag}\left(  V_{E}^{i}\right)  ^{\dag}\right) \\
&  \leq\frac{1}{d^{2}}\sum_{i}D(V_{E}^{i}U\rho U^{\dag}\left(  V_{E}%
^{i}\right)  ^{\dag}\Vert V_{E}^{i}U\sigma U^{\dag}\left(  V_{E}^{i}\right)
^{\dag})\\
&  =D(\rho\Vert\sigma).
\end{align}
The first equality follows from additivity of quantum relative entropy
(Exercise~\ref{ex-qie:add-q-rel-ent}) and the fact that $D(\pi_{E}\Vert\pi
_{E})=0$. The second equality follows from the fact that a random application
of a Heisenberg--Weyl unitary is equivalent to a channel that traces out
system $E$ and replaces it with $\pi_{E}$ (see
Exercise~\ref{ex-qt:uniformly-random-unitary}). The inequality follows from
joint convexity (by assumption), and the last equality follows from invariance
of quantum relative entropy with respect to isometries
(Exercise~\ref{ex-qie:rel-ent-unit-invar}).

We now show that concavity of conditional entropy implies monotonicity of
quantum relative entropy, which has the most involved proof. Before doing so,
we need a somewhat advanced theorem from matrix analysis. Suppose that $f$ is
a differentiable function on an open neighborhood of the spectrum of some
self-adjoint operator $A$. Then its derivative $Df$ at $A$ is given by%
\begin{equation}
Df(A):K\rightarrow\sum_{\lambda,\eta}f^{\left[  1\right]  }(\lambda,\eta
)P_{A}(\lambda)KP_{A}(\eta),
\end{equation}
where $A=\sum_{\lambda}\lambda P_{A}(\lambda)$ is the spectral decomposition
of $A$, and $f^{\left[  1\right]  }$ is what is known as the first divided
difference function. In particular, if $x\longmapsto A(x)\in\mathcal{L}%
(\mathcal{H})$ (where $A(x)$ is positive semi-definite) is a differentiable
function on an open interval in $\mathbb{R}$, with derivative $A^{\prime}$,
then%
\begin{equation}
\frac{d}{dx}f(A(x))=\sum_{\lambda,\eta}f^{\left[  1\right]  }(\lambda
,\eta)P_{A(x)}(\lambda)A^{\prime}(x)P_{A(x)}(\eta), \label{eq:op-deriv}%
\end{equation}
so that%
\begin{equation}
\frac{d}{dx}\text{Tr}\left\{  f(A(x))\right\}  =\text{Tr}\left\{  f^{\prime
}(A(x))A^{\prime}(x)\right\}  . \label{eq-qie:like-chain-rule-diff}%
\end{equation}
Note how \eqref{eq-qie:like-chain-rule-diff}\ appears strikingly similar to
the usual chain rule. So if $A(x)=A+xB$, then%
\begin{equation}
\frac{d}{dx}\text{Tr}\left\{  f(A(x))\right\}  =\text{Tr}\left\{  f^{\prime
}(A(x))B\right\}  , \label{eq:critical-one}%
\end{equation}
which is the main formula that we need to proceed. In what follows, we will be
taking $A(x)=\sigma_{AB}+x\rho_{AB}$, where $\rho_{AB},\sigma_{AB}%
\in\mathcal{D}(\mathcal{H}_{A}\otimes\mathcal{H}_{B})$ and $x>0$. If we do not
have that $\operatorname{supp}(\rho_{AB})\subseteq\operatorname{supp}%
(\sigma_{AB})$, then we can take $\sigma_{AB}^{\prime}\equiv(1-\varepsilon
)\sigma_{AB}+\varepsilon\pi_{AB}$ for $\varepsilon\in(0,1)$ and $\pi_{AB}$ the
maximally mixed state. After doing so, we can take a limit as $\varepsilon
\rightarrow0$ at the end. We also make use of the standard fact that the
function $f:X\rightarrow X^{-1}$ is everywhere differentiable on the set of
invertible density operators, and at an invertible $X$, its derivative is
$f^{\prime}(X):Y\rightarrow-X^{-1}YX^{-1}$. Consider that the conditional
entropy is homogeneous, in the sense that%
\begin{equation}
H(A|B)_{xG}=xH(A|B)_{G},
\end{equation}
where $G_{AB}\in\mathcal{L}(\mathcal{H}_{A}\otimes\mathcal{H}_{B})$ is a
positive semi-definite operator. Let%
\begin{equation}
\xi_{YAB}\equiv\frac{1}{x+1}|0\rangle\langle0|_{Y}\otimes\sigma_{AB}+\frac
{x}{x+1}|1\rangle\langle1|_{Y}\otimes\rho_{AB}.
\end{equation}
Then it follows from homogeneity and concavity of conditional entropy (taking
5 true by assumption) that%
\begin{align}
H(A|B)_{\sigma+x\rho}  &  =(x+1)H(A|B)_{\xi}\\
&  \geq(x+1)\left[  \frac{1}{x+1}H(A|B)_{\sigma}+\frac{x}{x+1}H(A|B)_{\rho
}\right] \\
&  =H(A|B)_{\sigma}+xH(A|B)_{\rho}.
\end{align}
Manipulating the above inequality then gives%
\begin{equation}
\frac{H(A|B)_{\sigma+x\rho}-H(A|B)_{\sigma}}{x}\geq H(A|B)_{\rho},
\end{equation}
and taking the limit as $x\searrow0$ gives%
\begin{equation}
\lim_{x\searrow0}\frac{H(A|B)_{\sigma+x\rho}-H(A|B)_{\sigma}}{x}=\left.
\frac{d}{dx}H(A|B)_{\sigma+x\rho}\right\vert _{x=0}\geq H(A|B)_{\rho}.
\label{eq:lim}%
\end{equation}
We now evaluate the limit on the left-hand side. So we consider%
\begin{multline}
\frac{d}{dx}H(A|B)_{\sigma+x\rho}=\frac{d}{dx}\left[  -\operatorname{Tr}%
\left\{  \left(  \sigma_{AB}+x\rho_{AB}\right)  \log\left(  \sigma_{AB}%
+x\rho_{AB}\right)  \right\}  \right] \\
+\frac{d}{dx}\operatorname{Tr}\left\{  \left(  \sigma_{B}+x\rho_{B}\right)
\log\left(  \sigma_{B}+x\rho_{B}\right)  \right\}  .
\end{multline}
We evaluate this by using $\frac{d}{dy}\left[  g(y)\log g(y)\right]  =\left[
\log g(y)+1\right]  g^{\prime}(y)$ (up to a scale factor of $\ln2$ from using
the binary logarithm) and (\ref{eq:critical-one}) to find that%
\begin{equation}
\frac{d}{dx}\operatorname{Tr}\left\{  \left(  \sigma_{AB}+x\rho_{AB}\right)
\log\left(  \sigma_{AB}+x\rho_{AB}\right)  \right\}  =\operatorname{Tr}%
\left\{  \left[  \log\left(  \sigma_{AB}+x\rho_{AB}\right)  +I_{AB}\right]
\rho_{AB}\right\}  ,
\end{equation}
so that%
\begin{equation}
\frac{d}{dx}H(A|B)_{\sigma+x\rho}=-\operatorname{Tr}\left\{  \rho_{AB}%
\log\left(  \sigma_{AB}+x\rho_{AB}\right)  \right\}  +\operatorname{Tr}%
\left\{  \rho_{B}\log\left(  \sigma_{B}+x\rho_{B}\right)  \right\}  ,
\end{equation}
and thus%
\begin{equation}
\left.  \frac{d}{dx}H(A|B)_{\sigma+x\rho}\right\vert _{x=0}=-\operatorname{Tr}%
\left\{  \rho_{AB}\log\sigma_{AB}\right\}  +\operatorname{Tr}\left\{  \rho
_{B}\log\sigma_{B}\right\}  .
\end{equation}
Substituting back into the inequality~\eqref{eq:lim}, we find that%
\begin{equation}
-\operatorname{Tr}\left\{  \rho_{AB}\log\sigma_{AB}\right\}
+\operatorname{Tr}\left\{  \rho_{B}\log\sigma_{B}\right\}  \geq
-\operatorname{Tr}\left\{  \rho_{AB}\log\rho_{AB}\right\}  +\operatorname{Tr}%
\left\{  \rho_{B}\log\rho_{B}\right\}  ,
\end{equation}
which is equivalent to%
\begin{equation}
D(\rho_{AB}\Vert\sigma_{AB})\geq D(\rho_{B}\Vert\sigma_{B}).
\end{equation}
If the support condition $\operatorname{supp}(\rho_{AB})\subseteq
\operatorname{supp}(\sigma_{AB})$\ does not hold, then we can take
$\sigma_{AB}^{\prime}$ as mentioned above and all of the above development
holds. At the end, we can take the limit as $\varepsilon\rightarrow0$ to find
that $D(\rho_{AB}\Vert\sigma_{AB})=+\infty$, so that the inequality holds
trivially in this case.
\end{proof}

\subsection{Quantum Data Processing}

\label{sec-ie:quantum-data-processing}The quantum data-processing
inequalities
\index{quantum data processing inequality}
discussed below are similar in spirit to the classical data-processing
inequality. Recall that the classical data-processing inequality states that
processing classical data reduces classical correlations. The quantum
data-processing inequalities state that processing \textit{quantum }data
reduces \textit{quantum }correlations.

One variant applies to the following scenario. Suppose that Alice and Bob
share some bipartite state$~\rho_{AB}$. The coherent information $I(A\rangle
B)_{\rho}$ is one measure of the quantum correlations present in this state.
Bob then processes his system $B$ according to some quantum channel
$\mathcal{N}_{B\rightarrow B^{\prime}}$ to produce some quantum system
$B^{\prime}$ and let $\sigma_{AB^{\prime}}$ denote the resulting state. The
quantum data-processing inequality states that this step of quantum data
processing reduces quantum correlations, in the sense that%
\begin{equation}
I(A\rangle B)_{\rho}\geq I(A\rangle B^{\prime})_{\sigma}.
\end{equation}

\begin{theorem}
[Data Processing for Coherent Information]\label{thm-ie:quantum-data-process}%
\index{data processing inequality!coherent information}%
Let $\rho_{AB}\in\mathcal{D}(\mathcal{H}_{A}\otimes\mathcal{H}_{B})$ and let
$\mathcal{N}:\mathcal{L}(\mathcal{H}_{B})\rightarrow\mathcal{L}(\mathcal{H}%
_{B^{\prime}})$ be a quantum channel. Set $\sigma_{AB^{\prime}}\equiv
\mathcal{N}_{B\rightarrow B^{\prime}}(\rho_{AB})$. Then the following quantum
data-processing inequality holds%
\begin{equation}
I(A\rangle B)_{\rho}\geq I(A\rangle B^{\prime})_{\sigma}.
\end{equation}

\end{theorem}

\begin{proof}
This is a consequence of Exercise~\ref{ex-qie:rel-ent-cond-ent} and
Theorem~\ref{thm-qie:mono-rel-ent}. By Exercise~\ref{ex-qie:rel-ent-cond-ent},
we know that%
\begin{align}
I(A\rangle B)_{\rho}  &  =D(\rho_{AB}\Vert I_{A}\otimes\rho_{B}),\\
I(A\rangle B^{\prime})_{\sigma}  &  =D(\sigma_{AB^{\prime}}\Vert I_{A}%
\otimes\sigma_{B^{\prime}})\\
&  =D(\mathcal{N}_{B\rightarrow B^{\prime}}(\rho_{AB})\Vert I_{A}%
\otimes\mathcal{N}_{B\rightarrow B^{\prime}}(\rho_{B}))\\
&  =D(\mathcal{N}_{B\rightarrow B^{\prime}}(\rho_{AB})\Vert\mathcal{N}%
_{B\rightarrow B^{\prime}}(I_{A}\otimes\rho_{B})).
\end{align}
The statement then follows from the monotonicity of quantum relative entropy
by picking $\rho=\rho_{AB}$, $\sigma=I_{A}\otimes\rho_{B}$, and $\mathcal{N}%
=\operatorname{id}_{A}\otimes\mathcal{N}_{B\rightarrow B^{\prime}}$ in
Theorem~\ref{thm-qie:mono-rel-ent}.
\end{proof}

\begin{theorem}
[Data Processing for Mutual Information]\label{cor-qie:QDP}%
\index{data processing inequality!quantum mutual information}%
Let $\rho_{AB}\in\mathcal{D}(\mathcal{H}_{A}\otimes\mathcal{H}_{B})$,
$\mathcal{N}:\mathcal{L}(\mathcal{H}_{A})\rightarrow\mathcal{L}(\mathcal{H}%
_{A^{\prime}})$ be a quantum channel, and $\mathcal{M}:\mathcal{L}%
(\mathcal{H}_{B})\rightarrow\mathcal{L}(\mathcal{H}_{B^{\prime}})$ be a
quantum channel. Set $\sigma_{A^{\prime}B^{\prime}}\equiv(\mathcal{N}%
_{A\rightarrow A^{\prime}}\otimes\mathcal{M}_{B\rightarrow B^{\prime}}%
)(\rho_{AB})$. Then the following quantum data-processing inequality applies
to the quantum mutual information:%
\begin{equation}
I(A;B)_{\rho}\geq I(A^{\prime};B^{\prime})_{\sigma}.
\end{equation}

\end{theorem}

\begin{proof}
From Exercise~\ref{ex-qie:rel-ent-mut-info}, we know that%
\begin{align}
I(A;B)_{\rho}  &  =D(\rho_{AB}\Vert\rho_{A}\otimes\rho_{B}),\\
I(A^{\prime};B^{\prime})_{\sigma}  &  =D(\sigma_{A^{\prime}B^{\prime}}%
\Vert\sigma_{A^{\prime}}\otimes\sigma_{B^{\prime}})\\
&  =D((\mathcal{N}_{A\rightarrow A^{\prime}}\otimes\mathcal{M}_{B\rightarrow
B^{\prime}})(\rho_{AB})\Vert\mathcal{N}_{A\rightarrow A^{\prime}}(\rho
_{A})\otimes\mathcal{M}_{B\rightarrow B^{\prime}}(\rho_{B}))\\
&  =D((\mathcal{N}_{A\rightarrow A^{\prime}}\otimes\mathcal{M}_{B\rightarrow
B^{\prime}})(\rho_{AB})\Vert(\mathcal{N}_{A\rightarrow A^{\prime}}%
\otimes\mathcal{M}_{B\rightarrow B^{\prime}})(\rho_{A}\otimes\rho_{B})).
\end{align}
The statement then follows from the monotonicity of quantum relative entropy
by picking $\rho=\rho_{AB}$, $\sigma=\rho_{A}\otimes\rho_{B}$, and
$\mathcal{N}=\mathcal{N}_{A\rightarrow A^{\prime}}\otimes\mathcal{M}%
_{B\rightarrow B^{\prime}}$ in Theorem~\ref{thm-qie:mono-rel-ent}.
\end{proof}

\begin{exercise}
Let $\rho_{AB}\in\mathcal{D}(\mathcal{H}_{A}\otimes\mathcal{H}_{B})$,
$\mathcal{N}:\mathcal{L}(\mathcal{H}_{A})\rightarrow\mathcal{L}(\mathcal{H}%
_{A^{\prime}})$ be a unital quantum channel, and $\mathcal{M}:\mathcal{L}%
(\mathcal{H}_{B})\rightarrow\mathcal{L}(\mathcal{H}_{B^{\prime}})$ be a
quantum channel. Set $\sigma_{A^{\prime}B^{\prime}}\equiv(\mathcal{N}%
_{A\rightarrow A^{\prime}}\otimes\mathcal{M}_{B\rightarrow B^{\prime}}%
)(\rho_{AB})$.\ Prove that%
\begin{equation}
I(A\rangle B)_{\rho}\geq I(A^{\prime}\rangle B^{\prime})_{\sigma}.
\end{equation}

\end{exercise}

\begin{exercise}
[Holevo Bound]\label{ex-qie:holevo-bound}Use
\index{Holevo bound}%
the quantum data-processing inequality to show that the Holevo information
$\chi(\mathcal{E})$ is an upper bound on the accessible information
$I_{\operatorname{acc}}(\mathcal{E})$:%
\begin{equation}
I_{\operatorname{acc}}(\mathcal{E})\leq\chi(\mathcal{E}),
\end{equation}
where $\mathcal{E}$ is an ensemble of quantum states. (See
Section~\ref{sec-cie:acc-info}\ for a definition of accessible information.)
\end{exercise}

\begin{exercise}
[Shannon Entropy vs.~von Neumann Entropy]\label{ex-qie:shannon-vs-von-neumann}%
Consider an ensemble $\left\{  p_{X}(x),|\psi_{x}\rangle\right\}  $. The
expected density operator of the ensemble is%
\begin{equation}
\rho\equiv\sum_{x}p_{X}(x)|\psi_{x}\rangle\langle\psi_{x}|.
\end{equation}
Use the quantum data-processing inequality to show that the Shannon entropy
$H(X)$ is never less than the quantum entropy of the expected density operator
$\rho$:%
\begin{equation}
H(X)\geq H(\rho).
\end{equation}
(Hint:\ Begin with a classical shared randomness state $\sum_{x}%
p_{X}(x)|x\rangle\langle x|_{X}\otimes|x\rangle\langle x|_{X^{\prime}}$ and
apply a preparation channel to system $X^{\prime}$). Conclude that the Shannon
entropy of the ensemble is strictly greater than the quantum entropy whenever
the states in the ensemble are non-orthogonal.
\end{exercise}

\begin{exercise}
\label{ex-qie:cond-ent-c-q-non-neg}Use the idea in the above exercise to show
that the conditional entropy $H(X|B)_{\rho}$ is always non-negative whenever
the state $\rho_{XB}$ is a classical--quantum state:%
\begin{equation}
\rho_{XB}\equiv\sum_{x}p_{X}(x)|x\rangle\langle x|_{X}\otimes\rho_{B}^{x}.
\end{equation}
Additionally, show that $H(X|B)_{\rho}\geq0$ is equivalent to $H(B)_{\rho}\leq
H(X)_{\rho}+H(B|X)_{\rho}$ and $I(X;B)_{\rho}\leq H(X)_{\rho}$.
\end{exercise}

\begin{exercise}
[Separability and Negativity of Coherent Information]%
\label{ex-qie:sep-implies-coh-info-neg}Show that the following inequality
holds for any separable state $\rho_{AB}$:%
\begin{equation}
\max\left\{  I(A\rangle B)_{\rho},I(B\rangle A)_{\rho}\right\}  \leq0.
\end{equation}

\end{exercise}

\begin{exercise}
[Quantum Data Processing for CQMI]%
\index{data processing inequality!conditional quantum mutual information}%
Let $\rho_{ABC}\in\mathcal{D}(\mathcal{H}_{A}\otimes\mathcal{H}_{B}%
\otimes\mathcal{H}_{B})$, $\mathcal{N}:\mathcal{L}(\mathcal{H}_{A}%
)\rightarrow\mathcal{L}(\mathcal{H}_{A^{\prime}})$ be a quantum channel, and
$\mathcal{M}:\mathcal{L}(\mathcal{H}_{B})\rightarrow\mathcal{L}(\mathcal{H}%
_{B^{\prime}})$ be a quantum channel. Set $\sigma_{A^{\prime}B^{\prime}%
C}\equiv(\mathcal{N}_{A\rightarrow A^{\prime}}\otimes\mathcal{M}_{B\rightarrow
B^{\prime}})(\rho_{ABC})$. Prove that%
\begin{equation}
I(A;B|C)_{\rho}\geq I(A^{\prime};B^{\prime}|C)_{\sigma}.
\end{equation}

\end{exercise}

\subsection{Entropic Uncertainty Principle}

\label{sec-qie:ent-unc-princ}%
\index{uncertainty principle}%
The uncertainty principle reviewed in
Section~\ref{sec-qt:uncertainty-principle}\ aims to capture a fundamental
feature of quantum mechanics, namely, that there is an unavoidable uncertainty
in the measurement outcomes of incompatible (non-commuting) observables. This
uncertainty principle is a radical departure from classical intuitions, where
there it seems as if there should not be any obstacle to measuring
incompatible observables such as position and momentum.

However, the uncertainty principle that we reviewed before (the standard
version in most textbooks) suffers from a few deficiencies. First, the measure
of uncertainty used there is the standard deviation, which is not just a
function of the probabilities of measurement outcomes but also of the values
of the outcomes. Thus, the values of the outcomes may skew the uncertainty
measure (however, one could always relabel the values in order to avoid this
difficulty). More importantly however, from an information-theoretic
perspective, there is not a clear operational interpretation for the standard
deviation as there is for entropy. Second, the lower bound in
\eqref{eq-qt:uncertainty-principle} depends not only on the observables but
also on the state. In Exercise~\ref{ex-qt:lower-bound-unc-vanish}, we saw how
this lower bound can vanish for a state even when the distributions
corresponding to the measurement outcomes in fact do have uncertainty. So, it
would be ideal to separate this lower bound into two terms:\ one which depends
only on measurement incompatibility and another which depends only on the state.

Additionally, it might seem as if giving two parties access to a maximally
entangled state allows them to defy the uncertainty principle (and this is
what confounded Einstein, Podolsky, and Rosen after quantum mechanics had been
established). Indeed, suppose that Alice and Bob share a Bell state
$\left\vert \Phi^{+}\right\rangle =2^{-1/2}\left(  |00\rangle+|11\rangle
\right)  =2^{-1/2}\left(  \left\vert ++\right\rangle +\left\vert
--\right\rangle \right)  $. If Alice measures the Pauli $Z$ observable on her
system, then Bob can guess the outcome of her measurement with certainty.
Also, if Alice were instead to measure the Pauli $X$ observable on her system,
then Bob would also be able to guess the outcome of her measurement with
certainty, in spite of the fact that $Z$ and $X$ are incompatible observables.
So, a revision of the uncertainty principle is clearly needed to account for
this possibility, in the scenario where Bob shares a \textit{quantum memory}
correlated with Alice's system.

The \textit{uncertainty principle in the presence of quantum memory} is such a
revision that meets all of the desiderata stated above. It quantifies
uncertainty in terms of quantum entropy rather than with standard deviation,
and it also accounts for the scenario in which an observer has a quantum
memory correlated with the system being measured. So, suppose that Alice and
Bob share systems $A$ and $B$, respectively, that are in some state $\rho
_{AB}$. If Alice performs a measurement channel corresponding to a POVM
$\left\{  \Lambda_{A}^{x}\right\}  $ on her system $A$, then the
post-measurement state is as follows:%
\begin{equation}
\sigma_{XB}\equiv\sum_{x}|x\rangle\langle x|_{X}\otimes\operatorname{Tr}%
_{A}\left\{  \left(  \Lambda_{A}^{x}\otimes I_{B}\right)  \rho_{AB}\right\}  .
\label{eq-ie:EURQM-1st-state}%
\end{equation}
In the above classical--quantum state, the measurement outcomes $x$ are
encoded into orthonormal states $\left\{  |x\rangle\right\}  $ of the
classical register $X$, and the probability for obtaining outcome $x$ is
$\operatorname{Tr}\left\{  \left(  \Lambda_{A}^{x}\otimes I_{B}\right)
\rho_{AB}\right\}  $. We would like to quantify the uncertainty that Bob has
about the outcome of the measurement, and a natural quantity for doing so is
the conditional quantum entropy $H(X|B)_{\sigma}$. Similarly, starting from
the state $\rho_{AB}$, Alice could choose to perform some other measurement
channel corresponding to POVM\ $\left\{  \Gamma_{A}^{z}\right\}  $ on her
system $A$. In this case, the post-measurement state is as follows:%
\begin{equation}
\tau_{ZB}\equiv\sum_{z}|z\rangle\langle z|_{Z}\otimes\operatorname{Tr}%
_{A}\left\{  \left(  \Gamma_{A}^{z}\otimes I_{B}\right)  \rho_{AB}\right\}  ,
\label{eq-ie:EURQM-2nd-state}%
\end{equation}
with a similar interpretation as before. We can quantify Bob's uncertainty
about the measurement outcome $z$ in terms of the conditional quantum entropy
$H(Z|B)_{\tau}$. We define Bob's total uncertainty about the measurement
outcomes to be the sum of both entropies: $H(X|B)_{\sigma}+H(Z|B)_{\tau}$. We
will call this the \textit{uncertainty sum}, in analogy with the uncertainty
product in \eqref{eq-qt:uncertainty-principle}.

We stated above that it would be desirable to have a lower bound on the
uncertainty sum consisting of a measurement incompability term and a
state-dependent term. One way to quantify the incompatibility of the POVMs
$\left\{  \Lambda_{A}^{x}\right\}  $ and $\left\{  \Gamma_{A}^{z}\right\}  $
is in terms of the following quantity:%
\begin{equation}
c\equiv\max_{x,z}\left\Vert \sqrt{\Lambda_{A}^{x}}\sqrt{\Gamma_{A}^{z}%
}\right\Vert _{\infty}^{2}, \label{eq-ie:EURQM-meas-incomp}%
\end{equation}
where $\left\Vert \cdot\right\Vert _{\infty}$ is the infinity norm of an
operator (for the finite-dimensional case, $\left\Vert A\right\Vert _{\infty}$
is just the maximal eigenvalue of $\left\vert A\right\vert $). To grasp an
intuition for this incompatibility measure, suppose that $\left\{  \Lambda
_{A}^{x}\right\}  $ and $\left\{  \Gamma_{A}^{z}\right\}  $ are actually
complete projective measurements with one common element. In this case, it
follows that $c=1$, so that the measurements are regarded as maximally
compatible. On the other hand, if the measurements are of Pauli observables
$X$ and $Z$, these are maximally incompatible for a two-dimensional Hilbert
space and $c=1/2$. We now state the uncertainty principle in the presence of
quantum memory:

\begin{theorem}
[Uncertainty Principle with Quantum Memory]\label{thm-qie:EURQSI}Suppose that
Alice and Bob share a state $\rho_{AB}$ and that Alice performs either of the
POVMs $\left\{  \Lambda_{A}^{x}\right\}  $ or $\left\{  \Gamma_{A}%
^{z}\right\}  $ on her share of the state (with at least one of $\left\{
\Lambda_{A}^{x}\right\}  $ or $\left\{  \Gamma_{A}^{z}\right\}  $ being a
rank-one POVM). Then Bob's total uncertainty about the measurement outcomes
has the following lower bound:%
\begin{equation}
H(X|B)_{\sigma}+H(Z|B)_{\tau}\geq\log\left(  1/c\right)  +H(A|B)_{\rho},
\end{equation}
where the states $\sigma_{XB}$ and $\tau_{ZB}$ are defined in
\eqref{eq-ie:EURQM-1st-state} and \eqref{eq-ie:EURQM-2nd-state}, respectively,
and the measurement incompatibility is defined in \eqref{eq-ie:EURQM-meas-incomp}.
\end{theorem}

Interestingly, the lower bound given in the above theorem consists of both the
measurement incompatibility and the state-dependent term $H(A|B)_{\rho}$. As
we know from Exercise~\ref{ex-qie:sep-implies-coh-info-neg}, when the
conditional quantum entropy $H(A|B)_{\rho}$ becomes negative, this implies
that the state $\rho_{AB}$ is entangled (but not necessarily the converse).
Thus, a negative conditional entropy implies that the lower bound on the
uncertainty sum can become lower than $\log\left(  1/c\right)  $, and
furthermore, that it might be possible to reduce Bob's total uncertainty about
the measurement outcomes down to zero. Indeed, this is the case for the
example we mentioned before with measurements of Pauli $X$ and $Z$ on the
maximally entangled Bell state. One can verify for this case that
$\log(1/c)=1$ and $H(A|B)=-1$, so that this is consistent with the fact that
$H(X|B)_{\sigma}+H(Z|B)_{\tau}=0$ for this example. We now give a path to
proving the above theorem (leaving the final steps as an exercise).

\bigskip

\begin{proof}
We actually prove the following uncertainty relation instead:%
\begin{equation}
H(X|B)_{\sigma}+H(Z|E)_{\omega}\geq\log(1/c), \label{eq-ie:alt-unc-rel}%
\end{equation}
where $\omega_{ZE}$ is a classical--quantum state of the following form:%
\begin{equation}
\omega_{ZE}\equiv\sum_{z}|z\rangle\langle z|_{Z}\otimes\operatorname{Tr}%
_{AB}\left\{  \left(  \Gamma_{A}^{z}\otimes I_{BE}\right)  \phi_{ABE}^{\rho
}\right\}  ,
\end{equation}
and $\phi_{ABE}^{\rho}$ is a purification of $\rho_{AB}$. We leave it as an
exercise to demonstrate that the above uncertainty relation implies the one in
the statement of the theorem whenever $\Gamma_{A}^{z}$ is a rank-one POVM.
Consider the following isometric extensions of the measurement channels for
$\left\{  \Lambda_{A}^{x}\right\}  $ and $\left\{  \Gamma_{A}^{z}\right\}  $:%
\begin{align}
U_{A\rightarrow XX^{\prime}A}  &  \equiv\sum_{x}|x\rangle_{X}\otimes
|x\rangle_{X^{\prime}}\otimes\sqrt{\Lambda_{A}^{x}},\\
V_{A\rightarrow ZZ^{\prime}A}  &  \equiv\sum_{z}|z\rangle_{Z}\otimes
|z\rangle_{Z^{\prime}}\otimes\sqrt{\Gamma_{A}^{z}},
\end{align}
where $\left\{  |x\rangle\right\}  $ and $\left\{  |z\rangle\right\}  $ are
both orthonormal bases. Let $\omega_{ZZ^{\prime}ABE}$ denote the following
state:%
\begin{equation}
\left\vert \omega\right\rangle _{ZZ^{\prime}ABE}\equiv V_{A\rightarrow
ZZ^{\prime}A}\left\vert \phi^{\rho}\right\rangle _{ABE},
\end{equation}
so that $\omega_{ZE}=\ \operatorname{Tr}_{Z^{\prime}AB}\{\omega_{ZZ^{\prime
}ABE}\}$. Exercise~\ref{ex-qie:other-coh-info}\ establishes that%
\begin{equation}
H(Z|E)_{\omega}=-H(Z|Z^{\prime}AB)_{\omega},
\end{equation}
so that \eqref{eq-ie:alt-unc-rel} is equivalent to%
\begin{equation}
-H(Z|Z^{\prime}AB)_{\omega}\geq\log(1/c)-H(X|B)_{\sigma}.
\label{eq-qie:ent-unc-rewrite-1}%
\end{equation}
Recalling the result of Exercise~\ref{ex-qie:rel-ent-cond-ent}, we then have
that \eqref{eq-qie:ent-unc-rewrite-1} is equivalent to%
\begin{equation}
D(\omega_{ZZ^{\prime}AB}\Vert I_{Z}\otimes\omega_{Z^{\prime}AB})\geq
\log(1/c)+D(\sigma_{XB}\Vert I_{X}\otimes\sigma_{B}),
\label{eq-qie:ent-unc-rewrite-2}%
\end{equation}
where we observe that $\sigma_{B}=\omega_{B}$. So we aim to prove
\eqref{eq-qie:ent-unc-rewrite-2}. Consider the following chain of
inequalities:%
\begin{align}
&  D(\omega_{ZZ^{\prime}AB}\Vert I_{Z}\otimes\omega_{Z^{\prime}AB})\\
&  \geq D\left(  \omega_{ZZ^{\prime}AB}\middle\Vert VV^{\dag}\left(
I_{Z}\otimes\omega_{Z^{\prime}AB}\right)  VV^{\dag}\right) \\
&  =D\left(  \rho_{AB}\middle\Vert V^{\dag}\left(  I_{Z}\otimes\omega
_{Z^{\prime}AB}\right)  V\right) \\
&  =D\left(  U\rho_{AB}U^{\dag}\middle\Vert UV^{\dag}\left(  I_{Z}%
\otimes\omega_{Z^{\prime}AB}\right)  VU^{\dag}\right)  .
\label{eq-qie:ent-unc-proof-step}%
\end{align}
The first inequality follows from monotonicity of quantum relative entropy
under the channel $\rho\rightarrow\Pi\rho\Pi+\left(  I-\Pi\right)  \rho\left(
I-\Pi\right)  $, where the projector $\Pi\equiv VV^{\dag}$, and also from the
fact that $(I-\Pi)\omega_{ZZ^{\prime}AB}(I-\Pi)=0$. The first equality follows
from invariance of quantum relative entropy with respect to isometries
(Exercise~\ref{ex-qie:rel-ent-unit-invar})\ and the fact that $\omega
_{ZZ^{\prime}AB}=V\rho_{AB}V^{\dag}$. The second equality again follows from
invariance of quantum relative entropy with respect to isometries. Let us
define $\sigma_{XX^{\prime}ABE}$ as%
\begin{equation}
\left\vert \sigma\right\rangle _{XX^{\prime}ABE}\equiv U_{A\rightarrow
XX^{\prime}A}\left\vert \phi^{\rho}\right\rangle _{ABE}.
\end{equation}
We then have that \eqref{eq-qie:ent-unc-proof-step}\ is equal to%
\begin{equation}
D\left(  \sigma_{XX^{\prime}AB}\middle\Vert UV^{\dag}\left(  I_{Z}%
\otimes\omega_{Z^{\prime}AB}\right)  VU^{\dag}\right)  ,
\label{eq-qie:ent-unc-proof-step-2}%
\end{equation}
and explicitly evaluating $UV^{\dag}\left(  I_{Z}\otimes\omega_{Z^{\prime}%
AB}\right)  VU^{\dag}$ as%
\begin{align}
&  UV^{\dag}\left(  I_{Z}\otimes\omega_{Z^{\prime}AB}\right)  VU^{\dag}\\
&  =U\sum_{z^{\prime},z}\left\langle z^{\prime}|z\right\rangle _{Z}\left(
\langle z^{\prime}|_{Z^{\prime}}\otimes\sqrt{\Gamma_{A}^{z^{\prime}}}\right)
\omega_{Z^{\prime}AB}\left(  |z\rangle_{Z^{\prime}}\otimes\sqrt{\Gamma_{A}%
^{z}}\right)  U^{\dag}\\
&  =U\sum_{z}\left(  \langle z|_{Z^{\prime}}\otimes\sqrt{\Gamma_{A}^{z}%
}\right)  \omega_{Z^{\prime}AB}\left(  |z\rangle_{Z^{\prime}}\otimes
\sqrt{\Gamma_{A}^{z}}\right)  U^{\dag},
\end{align}
gives that \eqref{eq-qie:ent-unc-proof-step-2} is equal to%
\begin{equation}
D\left(  \sigma_{XX^{\prime}AB}\middle\Vert U\sum_{z}\left(  \langle
z|_{Z^{\prime}}\otimes\sqrt{\Gamma_{A}^{z}}\right)  \omega_{Z^{\prime}%
AB}\left(  |z\rangle_{Z^{\prime}}\otimes\sqrt{\Gamma_{A}^{z}}\right)  U^{\dag
}\right)  .
\end{equation}
We trace out the $X^{\prime}A$ systems and exploit monotonicity of quantum
relative entropy and cyclicity of trace to show that the above is not less
than%
\begin{equation}
D\left(  \sigma_{XB}\middle\Vert\sum_{z,x}|x\rangle\langle x|_{X}%
\otimes\operatorname{Tr}_{Z^{\prime}A}\left\{  \left(  |z\rangle\langle
z|_{Z^{\prime}}\otimes\sqrt{\Gamma_{A}^{z}}\Lambda_{A}^{x}\sqrt{\Gamma_{A}%
^{z}}\right)  \omega_{Z^{\prime}AB}\right\}  \right)  .
\label{eq-qie:ent-unc-proof-step-3}%
\end{equation}
Using the fact that $\sqrt{\Gamma_{A}^{z}}\Lambda_{A}^{x}\sqrt{\Gamma_{A}^{z}%
}=|\sqrt{\Gamma_{A}^{z}}\sqrt{\Lambda_{A}^{x}}|^{2}\leq cI$, it follows that%
\begin{equation}
\sum_{z,x}|x\rangle\langle x|_{X}\otimes\operatorname{Tr}_{Z^{\prime}%
A}\left\{  \left(  |z\rangle\langle z|_{Z^{\prime}}\otimes\left[  cI_{A}%
-\sqrt{\Gamma_{A}^{z}}\Lambda_{A}^{x}\sqrt{\Gamma_{A}^{z}}\right]  \right)
\omega_{Z^{\prime}AB}\right\}
\end{equation}
is a positive semi-definite operator, or equivalently, that%
\begin{equation}
\sum_{z,x}|x\rangle\langle x|_{X}\otimes\operatorname{Tr}_{Z^{\prime}%
A}\left\{  \left(  |z\rangle\langle z|_{Z^{\prime}}\otimes\sqrt{\Gamma_{A}%
^{z}}\Lambda_{A}^{x}\sqrt{\Gamma_{A}^{z}}\right)  \omega_{Z^{\prime}%
AB}\right\}  \leq c\ I_{X}\otimes\omega_{B}.
\end{equation}
Proposition~\ref{prop-qie:rel-ent-s-s'}\ and
Exercise~\ref{ex-qie:rel-ent-scaling}\ imply that
\eqref{eq-qie:ent-unc-proof-step-3}\ is not less than%
\begin{align}
D( \sigma_{XB}\Vert c\ I_{X}\otimes\omega_{B})  &  =\log(1/c)+D( \sigma
_{XB}\Vert I_{X}\otimes\omega_{B})\\
&  =\log(1/c)+D( \sigma_{XB}\Vert I_{X}\otimes\sigma_{B}) ,
\end{align}
which finally proves the inequality in \eqref{eq-ie:alt-unc-rel}. We now leave
it as an exercise to prove the statement of the theorem starting from the
inequality in \eqref{eq-ie:alt-unc-rel}.
\end{proof}

\begin{exercise}
Prove that \eqref{eq-ie:alt-unc-rel} implies Theorem~\ref{thm-qie:EURQSI}.
\end{exercise}

\begin{exercise}
Prove that Theorem~\ref{thm-qie:EURQSI} implies the following entropic
uncertainty relation for a state $\rho_{A}$ on a single system:%
\begin{equation}
H(X)+H(Z)\geq\log(1/c)+H(A)_{\rho},
\end{equation}
where $H(X)$ and $H(Z)$ are the Shannon entropies of the measurement outcomes.
\end{exercise}

\section{Continuity of Quantum Entropy}

Suppose that two density operators $\rho$ and $\sigma$ are close in trace
distance. We might then expect several properties to hold: the fidelity
between them should be close to one and we would suspect that their entropies
should be close. Theorem~\ref{thm-dm:fidelity-trace-relation}\ states that the
fidelity is close to one if the trace distance is small.%
\index{von Neumann entropy!continuity}%

An important theorem below, the Fannes--Audenaert inequality%
\index{Fannes-Audenaert Inequality}%
, states that quantum entropies are close as well. This theorem usually finds
application in a proof of a converse theorem in quantum Shannon theory.
Usually, the specification of any good protocol (in the sense of
asymptotically vanishing error) involves placing a bound on the trace distance
between the actual state resulting from a protocol and the ideal state that it
should produce. The Fannes--Audenaert inequality then allows us to translate
these statements of error into informational statements that bound the
asymptotic rates of communication in any good protocol.

\begin{theorem}
[Fannes--Audenaert Inequality]\label{thm-ie:fannes-actual}%
\index{Fannes-Audenaert Inequality}%
Let $\rho,\sigma\in\mathcal{D}(\mathcal{H})$ and suppose that $\frac{1}%
{2}\left\Vert \rho-\sigma\right\Vert _{1}\leq\varepsilon\in\left[  0,1\right]
$. Then the following inequality holds%
\begin{equation}
\left\vert H(\rho)-H(\sigma)\right\vert \leq\left\{
\begin{array}
[c]{cc}%
\varepsilon\log\left[  \dim(\mathcal{H})-1\right]  +h_{2}(\varepsilon) &
\text{if }\varepsilon\in\left[  0,1-1/\dim(\mathcal{H})\right] \\
\log\dim(\mathcal{H}) & \text{else}%
\end{array}
\right.  .
\end{equation}
Putting these together, a universal bound is%
\begin{equation}
\left\vert H(\rho)-H(\sigma)\right\vert \leq\varepsilon\log\dim(\mathcal{H}%
)+h_{2}(\varepsilon).
\end{equation}

\end{theorem}

\begin{proof}
A proof of this theorem follows by applying the classical result in
Theorem~\ref{thm-cie:cl-ent-continuity}. We first prove that%
\begin{equation}
H(\rho)-H(\sigma)\leq\varepsilon\log\left[  \dim(\mathcal{H})-1\right]
+h_{2}(\varepsilon)
\end{equation}
if $\varepsilon\in\left[  0,1-1/\dim(\mathcal{H})\right]  $. First note that
the function $f(\varepsilon)\equiv\varepsilon\log\left[  \dim(\mathcal{H}%
)-1\right]  +h_{2}(\varepsilon)$ is monotone non-decreasing on the interval
$\left[  0,1-1/\dim(\mathcal{H})\right]  $ because $f^{\prime}(\varepsilon
)=\log\left[  \dim(\mathcal{H})-1\right]  +\log\left(  \frac{1-\varepsilon
}{\varepsilon}\right)  \geq0$ for $\varepsilon\in\left[  0,1-1/\dim
(\mathcal{H})\right]  $. Let $\sigma=\sum_{y}p(y)|y\rangle\langle y|$ be a
spectral decomposition of $\sigma$. Let $\overline{\Delta}_{\sigma}$ denote
the following completely dephasing channel:%
\begin{equation}
\overline{\Delta}_{\sigma}(\omega)=\sum_{y}|y\rangle\langle y|\omega
|y\rangle\langle y|.
\end{equation}
Then $\overline{\Delta}_{\sigma}(\sigma)=\sigma$ and
Corollary~\ref{cor-qie:unital-ch-inc-ent}\ gives that $H(\rho)\leq
H(\overline{\Delta}_{\sigma}(\rho))$, so that%
\begin{equation}
H(\rho)-H(\sigma)\leq H(\overline{\Delta}_{\sigma}(\rho))-H(\overline{\Delta
}_{\sigma}(\sigma)).
\end{equation}
At the same time, we know from the monotonicity of trace distance with respect
to channels (Exercise~\ref{ex-dm:mono-TD}) that%
\begin{equation}
\left\Vert \rho-\sigma\right\Vert _{1}\geq\left\Vert \overline{\Delta}%
_{\sigma}(\rho)-\overline{\Delta}_{\sigma}(\sigma)\right\Vert _{1}.
\end{equation}
Putting everything together, we find that%
\begin{align}
H(\rho)-H(\sigma)  &  \leq H(\overline{\Delta}_{\sigma}(\rho))-H(\overline
{\Delta}_{\sigma}(\sigma))\\
&  \leq f\left(  \frac{1}{2}\left\Vert \overline{\Delta}_{\sigma}%
(\rho)-\overline{\Delta}_{\sigma}(\sigma)\right\Vert _{1}\right) \\
&  \leq f\left(  \frac{1}{2}\left\Vert \rho-\sigma\right\Vert _{1}\right)  ,
\end{align}
where the second inequality follows from
Theorem~\ref{thm-cie:cl-ent-continuity} and the last inequality follows
because $f$ is monotone non-decreasing on the interval $\left[  0,1-1/\dim
(\mathcal{H})\right]  $.

The other inequality $H(\sigma)-H(\rho)\leq\varepsilon\log\left[
\dim(\mathcal{H})-1\right]  +h_{2}(\varepsilon)$ follows by the same proof
method, but instead dephasing with respect to the eigenbasis of $\rho$. The
bound $\left\vert H(\rho)-H(\sigma)\right\vert \leq\log(\dim(\mathcal{H}))$
follows trivially from the fact that the entropy is non-negative and cannot
exceed $\log(\dim(\mathcal{H}))$.
\end{proof}

There is another variation of this theorem, which we state below. We do not
however give a full proof, but instead just argue for it. The proof sketch
makes use of an original insight by \cite{A07}.

\begin{theorem}
[Fannes--Audenaert Inequality]\label{thm-ie:fannes-actual copy(1)}%
\index{Fannes-Audenaert Inequality}%
Let $\rho,\sigma\in\mathcal{D}(\mathcal{H})$ and let $T\equiv\frac{1}%
{2}\left\Vert \rho-\sigma\right\Vert _{1}$. Then the following inequality
holds%
\begin{equation}
\left\vert H(\rho)-H(\sigma)\right\vert \leq T\log\left[  \dim(\mathcal{H}%
)-1\right]  +h_{2}(T).
\end{equation}
Furthermore, this bound is optimal because there exists a pair of states that
saturates it for all $T\in\left[  0,1\right]  $ and dimension $\dim
(\mathcal{H})$.
\end{theorem}

\begin{proof}
The following inequalities are known from the theory of matrix analysis
\cite[Inequality IV.62]{B97}:%
\begin{multline}
T_{0}\equiv\frac{1}{2}\left\Vert \operatorname{Eig}^{\downarrow}%
(\rho)-\operatorname{Eig}^{\downarrow}(\sigma)\right\Vert _{1}\leq T=\frac
{1}{2}\left\Vert \rho-\sigma\right\Vert _{1}\label{eq-qie:mat-ana-ineq-TD}\\
\leq T_{1}\equiv\frac{1}{2}\left\Vert \operatorname{Eig}^{\downarrow}%
(\rho)-\operatorname{Eig}^{\uparrow}(\sigma)\right\Vert _{1},
\end{multline}
where $\operatorname{Eig}^{\downarrow}(A)$ is the list of eigenvalues of
Hermitian $A$ in non-increasing order and $\operatorname{Eig}^{\uparrow}(A)$
is the list in non-decreasing order. Then applying the fact that the entropy
depends only on the eigenvalues and is invariant with respect to permutations
of them, we find that%
\begin{align}
\left\vert H(\rho)-H(\sigma)\right\vert  &  =\left\vert H(\operatorname{Eig}%
^{\downarrow}(\rho))-H(\operatorname{Eig}^{\downarrow}(\sigma))\right\vert \\
&  \leq T_{0}\log\left[  \dim(\mathcal{H})-1\right]  +h_{2}(T_{0}),
\end{align}
where the inequality follows from Theorem~\ref{thm-cie:cl-ent-continuity}.
Similarly, we find that%
\begin{align}
\left\vert H(\rho)-H(\sigma)\right\vert  &  =\left\vert H(\operatorname{Eig}%
^{\downarrow}(\rho))-H(\operatorname{Eig}^{\uparrow}(\sigma))\right\vert \\
&  \leq T_{1}\log\left[  \dim(\mathcal{H})-1\right]  +h_{2}(T_{1}).
\end{align}
We know from \eqref{eq-qie:mat-ana-ineq-TD}\ that $T=\lambda T_{0}%
+(1-\lambda)T_{1}$ for some $\lambda\in\left[  0,1\right]  $. Applying
concavity of the binary entropy, we find that%
\begin{multline}
\left\vert H(\rho)-H(\sigma)\right\vert \leq\lambda\left[  T_{0}\log\left[
\dim(\mathcal{H})-1\right]  +h_{2}(T_{0})\right] \\
+(1-\lambda)\left[  T_{1}\log\left[  \dim(\mathcal{H})-1\right]  +h_{2}%
(T_{1})\right] \\
\leq T\log\left[  \dim(\mathcal{H})-1\right]  +h_{2}(T).
\end{multline}
The inequality is optimal because choosing $\rho=|0\rangle\langle0|$ and
$\sigma=(1-\varepsilon)|0\rangle\langle0|+\varepsilon/(d-1)|1\rangle
\langle1|+\cdots+\varepsilon/(d-1)|d-1\rangle\langle d-1|$ saturates the bound
for all $\varepsilon\in\left[  0,1\right]  $ and for all dimensions $d$.
\end{proof}

An important theorem below, the Alicki--Fannes--Winter (AFW) inequality%
\index{Alicki-Fannes-Winter inequality}%
, states that conditional quantum entropies are close as well. This statement
does follow directly from the Fannes--Audenaert inequality, but the main
advantage of the AFW\ inequality is that the upper bound has a dependence only
on the dimension of the first system in the conditional entropy (no dependence
on the conditioning system). The AFW\ inequality also finds application in a
proof of a converse theorem in quantum Shannon theory.

\begin{theorem}
[AFW Inequality]\label{thm-qie:AFW-cont-ent}Let $\rho_{AB},\sigma_{AB}%
\in\mathcal{D}(\mathcal{H}_{A}\otimes\mathcal{H}_{B})$. Suppose that%
\begin{equation}
\frac{1}{2}\left\Vert \rho_{AB}-\sigma_{AB}\right\Vert _{1}\leq\varepsilon,
\end{equation}
for $\varepsilon\in\left[  0,1\right]  $. Then%
\begin{equation}
\left\vert H(A|B)_{\rho}-H(A|B)_{\sigma}\right\vert \leq2\varepsilon\log
\dim(\mathcal{H}_{A})+g_2(\varepsilon)  ,
\end{equation}
where $g_2(\varepsilon) \equiv (\varepsilon +1) \log_2 (\varepsilon + 1) - \varepsilon \log_2 \varepsilon$.
If $\rho_{XB}$ and $\sigma_{XB}$ are classical--quantum and have the following
form:%
\begin{align}
\rho_{XB}  &  =\sum_{x}p(x)|x\rangle\langle x|_{X}\otimes\rho_{B}^{x},\\
\sigma_{XB}  &  =\sum_{x}q(x)|x\rangle\langle x|_{X}\otimes\sigma_{B}^{x},
\end{align}
where $p$ and $q$ are probability distributions defined over a finite alphabet
$\mathcal{X}$, $\{|x\rangle\}$ is an orthonormal basis, and $\rho_{B}%
^{x},\sigma_{B}^{x}\in\mathcal{D}(\mathcal{H}_{B})$ for all $x\in\mathcal{X}$,
then%
\begin{align}
\left\vert H(X|B)_{\rho}-H(X|B)_{\sigma}\right\vert  &  \leq\varepsilon
\log\dim(\mathcal{H}_{X})+g_2(\varepsilon)  ,\\
\left\vert H(B|X)_{\rho}-H(B|X)_{\sigma}\right\vert  &  \leq\varepsilon
\log\dim(\mathcal{H}_{B})+g_2(\varepsilon)  .
\end{align}

\end{theorem}

\begin{proof}
The bounds trivially hold when $\varepsilon=0$, so henceforth we assume that
$\varepsilon\in(0,1]$. All of the upper bounds are monotone non-decreasing
with $\varepsilon$, so it suffices to assume that $\frac{1}{2}\left\Vert
\rho_{AB}-\sigma_{AB}\right\Vert _{1}=\varepsilon$. Let $\rho_{AB}-\sigma
_{AB}=P_{AB}-Q_{AB}$ be a decomposition of $\rho_{AB}-\sigma_{AB}$ into its
positive part $P_{AB}\geq0$ and its negative part $Q_{AB}\geq0$ (as in the
proof of Lemma~\ref{lemma:trace-equiv}). Let $\Delta_{AB}\equiv P_{AB}%
/\varepsilon$. Since $\operatorname{Tr}\{P_{AB}\}=\frac{1}{2}\left\Vert
\rho_{AB}-\sigma_{AB}\right\Vert _{1}$ (see the proof of
Lemma~\ref{lemma:trace-equiv}), it follows that $\Delta_{AB}$ is a density
operator. Now consider that%
\begin{align}
\rho_{AB}  &  =\sigma_{AB}+(\rho_{AB}-\sigma_{AB}) \label{eq-qie:Delta'-PSD-1}%
\\
&  =\sigma_{AB}+P_{AB}-Q_{AB}\\
&  \leq\sigma_{AB}+P_{AB}\\
&  =\sigma_{AB}+\varepsilon\Delta_{AB}\\
&  =\left(  1+\varepsilon\right)  \left(  \frac{1}{1+\varepsilon}\sigma
_{AB}+\frac{\varepsilon}{1+\varepsilon}\Delta_{AB}\right) \\
&  =\left(  1+\varepsilon\right)  \omega_{AB}, \label{eq-qie:Delta'-PSD-last}%
\end{align}
where we define $\omega_{AB}\equiv\frac{1}{1+\varepsilon}\sigma_{AB}%
+\frac{\varepsilon}{1+\varepsilon}\Delta_{AB}$. Now let $\Delta_{AB}^{\prime
}\equiv\frac{1}{\varepsilon}\left[  \left(  1+\varepsilon\right)  \omega
_{AB}-\rho_{AB}\right]  $. It follows from
\eqref{eq-qie:Delta'-PSD-1}--\eqref{eq-qie:Delta'-PSD-last} that $\Delta
_{AB}^{\prime}$ is positive semi-definite. Furthermore, one can check that
$\operatorname{Tr}\{\Delta_{AB}^{\prime}\}=1$, so that $\Delta_{AB}^{\prime}$
is a density operator. One can also quickly check that%
\begin{equation}
\omega_{AB}=\frac{1}{1+\varepsilon}\rho_{AB}+\frac{\varepsilon}{1+\varepsilon
}\Delta_{AB}^{\prime}=\frac{1}{1+\varepsilon}\sigma_{AB}+\frac{\varepsilon
}{1+\varepsilon}\Delta_{AB}.
\end{equation}
Now consider that%
\begin{align}
H(A|B)_{\omega}  &  =-D(\omega_{AB}\Vert I_{A}\otimes\omega_{B})\\
&  =H(\omega_{AB})+\operatorname{Tr}\{\omega_{AB}\log\omega_{B}\}\\
&  \leq \nonumber h_{2}\left(  \frac{\varepsilon}{1+\varepsilon}\right)  +\frac
{1}{1+\varepsilon}H(\rho_{AB})+\frac{\varepsilon}{1+\varepsilon}H(\Delta
_{AB}^{\prime})\\
&  \ \ \ +\frac{1}{1+\varepsilon}\operatorname{Tr}\{\rho_{AB}\log\omega
_{B}\}+\frac{\varepsilon}{1+\varepsilon}\operatorname{Tr}\{\Delta_{AB}%
^{\prime}\log\omega_{B}\}\\
&  =\nonumber h_{2}\left(  \frac{\varepsilon}{1+\varepsilon}\right)  -\frac
{1}{1+\varepsilon}D(\rho_{AB}\Vert I_{A}\otimes\omega_{B})\\
&  \ \ \ -\frac{\varepsilon}{1+\varepsilon}D(\Delta_{AB}^{\prime}\Vert
I_{A}\otimes\omega_{B})\\
&  \leq h_{2}\left(  \frac{\varepsilon}{1+\varepsilon}\right)  +\frac
{1}{1+\varepsilon}H(A|B)_{\rho}+\frac{\varepsilon}{1+\varepsilon
}H(A|B)_{\Delta^{\prime}}.
\end{align}
The first equality follows from Exercise~\ref{ex-qie:rel-ent-cond-ent}, and
the second equality follows from the definition of quantum relative entropy.
The first inequality follows because $H(AB)\leq H(Y)+H(AB|Y)$ for a
classical--quantum state on systems $Y$ and $AB$ (see
Exercise~\ref{ex-qie:cond-ent-c-q-non-neg}), here taking the state as%
\begin{equation}
\frac{1}{1+\varepsilon}|0\rangle\langle0|_{Y}\otimes\rho_{AB}+\frac
{\varepsilon}{1+\varepsilon}|1\rangle\langle1|_{Y}\otimes\Delta_{AB}^{\prime}.
\end{equation}
The third equality follows from algebra and the definition of quantum relative
entropy. The last inequality follows from
Exercise~\ref{ex-qie:rel-ent-cond-ent}. From concavity of the conditional
entropy (Exercise~\ref{ex-qie:concavity-cond-ent}), we have that%
\begin{equation}
H(A|B)_{\omega}\geq\frac{1}{1+\varepsilon}H(A|B)_{\sigma}+\frac{\varepsilon
}{1+\varepsilon}H(A|B)_{\Delta}.
\end{equation}
Putting together the upper and lower bounds on $H(A|B)_{\omega}$, and observing that $g_2(\varepsilon) = \left(  1+\varepsilon\right)
h_{2}\!\left(  \frac{\varepsilon}{1+\varepsilon}\right)$, we find that%
\begin{align}
H(A|B)_{\sigma}-H(A|B)_{\rho}  &  \leq g_2(\varepsilon) +\varepsilon\left[
H(A|B)_{\Delta^{\prime}}-H(A|B)_{\Delta}\right] \\
&  \leq g_2(\varepsilon)  +2\varepsilon\log\dim(\mathcal{H}_{A}),
\end{align}
where the second inequality follows from a dimension bound for the conditional
entropy (Theorem~\ref{thm-qie:bound-cond-ent}).

The statements for classical--quantum states follow because the density
operator $\Delta$ is classical--quantum in this case and we know that
$H(X|B)_{\Delta},H(B|X)_{\Delta}\geq0$ (see
Exercise~\ref{ex-qie:cond-ent-c-q-non-neg}).
\end{proof}

\begin{exercise}
[AFW for Coherent Information]Prove that%
\begin{equation}
\left\vert I(A\rangle B)_{\rho}-I(A\rangle B)_{\sigma}\right\vert
\leq2\varepsilon\log\dim(\mathcal{H}_{A})+g_2(\varepsilon)  ,
\end{equation}
with $\frac{1}{2}\left\Vert \rho_{AB}-\sigma_{AB}\right\Vert _{1}%
\leq\varepsilon\in\left[  0,1\right]  $.
\end{exercise}

\begin{exercise}
[AFW for Quantum Mutual Information]\label{ex:Alicki-Fannes-mut}Prove that%
\begin{equation}
\left\vert I(A;B)_{\rho}-I(A;B)_{\sigma}\right\vert \leq3\varepsilon\log
\dim(\mathcal{H}_{A})+2 g_2(\varepsilon)  ,
\end{equation}
for any $\rho_{AB}$ and $\sigma_{AB}$ with $\frac{1}{2}\left\Vert \rho
_{AB}-\sigma_{AB}\right\Vert _{1}\leq\varepsilon\in\left[  0,1\right]  $.
\end{exercise}

We can also use these results to get a refinement of the non-negativity of
mutual information and the dimension upper bounds on mutual information and
conditional mutual information. A refinement of the non-negativity of
conditional mutual information (strong subadditivity) will appear in
Chapter~\ref{chap:mono-rel-ent}. The refinements of mutual information are
quantified in terms of the trace distance between $\rho_{AB}$ and the product
of its marginals. The refinement of conditional mutual information is
quantified in terms of the trace distance between $\rho_{ABC}$ and a
\textquotedblleft recovered version\textquotedblright\ of $\rho_{BC}$, which
represents a quantum generalization of a Markov chain. So these results
represent quantum generalizations of the statements in
Theorems~\ref{thm-cie:mut-info-to-prod} and
\ref{thm-cie:cond-mut-info-to-Markov}.

\begin{theorem}
Let $\rho_{AB}\in\mathcal{D}(\mathcal{H}_{A}\otimes\mathcal{H}_{B})$ and let
$\Delta\equiv\frac{1}{2}\left\Vert \rho_{AB}-\rho_{A}\otimes\rho
_{B}\right\Vert _{1}$. Then%
\begin{align}
I(A;B)  &  \geq\frac{2}{\ln2}\Delta^{2},\\
I(A;B)_{\rho}  &  \leq2\Delta\log\left[  \min\left\{  \dim(\mathcal{H}%
_{A}),\dim(\mathcal{H}_{B})\right\}  \right]  +g_2(\Delta).
\end{align}

\end{theorem}

\begin{proof}
The first inequality is a direct application of
Exercise~\ref{ex-qie:rel-ent-mut-info}\ and the quantum Pinsker inequality
(Theorem~\ref{thm-qie:rel-ent-trace}). Let $\omega_{AB}\equiv\rho_{A}%
\otimes\rho_{B}$. The next inequality follows because%
\begin{align}
I(A;B)_{\rho}  &  =\left\vert I(A;B)_{\rho}-I(A;B)_{\omega}\right\vert \\
&  =\left\vert H(A)_{\rho}-H(A|B)_{\rho}-\left[  H(A)_{\omega}-H(A|B)_{\omega
}\right]  \right\vert \\
&  =\left\vert H(A|B)_{\omega}-H(A|B)_{\rho}\right\vert \\
&  \leq2\Delta\log\dim(\mathcal{H}_{A})+g_2(\Delta),
\end{align}
where in the last line we applied Theorem~\ref{thm-qie:AFW-cont-ent}. The
other inequality%
\begin{equation}
I(A;B)_{\rho}\leq2\Delta\log\dim(\mathcal{H}_{B})+g_2(\Delta)
\end{equation}
follows by expanding the mutual information in the other way.
\end{proof}

\begin{theorem}
\label{thm-qie:CQMI-distance-to-markov}Let $\rho_{ABC}\in\mathcal{D}%
(\mathcal{H}_{A}\otimes\mathcal{H}_{B}\otimes\mathcal{H}_{C})$ and let%
\begin{equation}
\Delta\equiv\frac{1}{2}\inf_{\mathcal{R}_{C\rightarrow AC}}\left\Vert
\rho_{ABC}-\mathcal{R}_{C\rightarrow AC}(\rho_{BC})\right\Vert _{1},
\label{eq-qie:distance-to-Markov}%
\end{equation}
where the optimization is with respect to channels $\mathcal{R}:\mathcal{L}%
(\mathcal{H}_{C})\rightarrow\mathcal{L}(\mathcal{H}_{A}\otimes\mathcal{H}%
_{C})$. Then%
\begin{equation}
I(A;B|C)_{\rho},\ I(A;B|C)_{\sigma}\leq2\Delta\log\dim(\mathcal{H}%
_{B})+g_2(\Delta),
\end{equation}
where $\sigma_{ABC}\equiv\mathcal{R}_{C\rightarrow AC}^{\ast}(\rho_{BC})$ with
$\mathcal{R}_{C\rightarrow AC}^{\ast}$ the optimal recovery channel in \eqref{eq-qie:distance-to-Markov}.
\end{theorem}

\begin{proof}
Let $\Delta_{\mathcal{R}}\equiv\frac{1}{2}\left\Vert \rho_{ABC}-\mathcal{R}%
_{C\rightarrow AC}(\rho_{BC})\right\Vert _{1}$ for some $\mathcal{R}%
_{C\rightarrow AC}$ and define $\omega_{ABC}\equiv\mathcal{R}_{C\rightarrow
AC}(\rho_{BC})$. Consider that%
\begin{align}
I(A;B|C)_{\rho}  &  =H(B|C)_{\rho}-H(B|AC)_{\rho}\\
&  \leq H(B|AC)_{\omega}-H(B|AC)_{\rho}\\
&  \leq2\Delta_{\mathcal{R}}\log\dim(\mathcal{H}_{B})+g_2(\Delta_{\mathcal{R}}).
\end{align}
The first inequality follows from quantum data processing
(Theorem~\ref{thm-ie:quantum-data-process}) and the second from
Theorem~\ref{thm-qie:AFW-cont-ent}. Since the inequality holds for all
recovery channels and the upper bound is monotone non-decreasing in
$\Delta_{\mathcal{R}}$, we can conclude the inequality in the statement of the
theorem. Now consider that%
\begin{align}
I(A;B|C)_{\sigma}  &  =H(B|C)_{\sigma}-H(B|AC)_{\sigma}\\
&  \leq H(B|C)_{\sigma}-H(B|C)_{\rho}\\
&  \leq2\Delta\log\dim(\mathcal{H}_{B})+g_2(\Delta).
\end{align}
The justifications for these inequalities are the same as those for the above
ones (we additionally need to use monotonicity of the trace distance with
respect to partial trace).
\end{proof}

\section{History and Further Reading}

\label{sec-qie:history-QE}The quantum entropy and its relatives, such as
conditional entropy and mutual information, are useful information measures
and suffice for our studies in this book. However, the quantum entropy is
certainly not the only information measure worthy of study. In recent years,
entropic measures such as the min- and max-entropy have emerged (and their
smoothed variants), and they are useful in developing a more general theory
that applies beyond the i.i.d.~setting that we study in this book. In fact,
one could view this theory as more fundamental than the theory presented in
this book, since the \textquotedblleft one-shot\textquotedblright\ results
often imply the i.i.d.~results studied in this book. Rather than developing
this theory in full, we point to several excellent references on the subject \citep{Renner2005,D09,DR09,KRS09,T15}.

The fact that the conditional entropy can be negative is discussed in
\cite{W78,HH94,CA97}. \cite{Holevo73} proved the important bound bearing his
name. Only later was it understood that this bound is a consequence of the
monotonicity of quantum relative entropy. \cite{LR73,PhysRevLett.30.434}
established the strong subadditivity of quantum entropy and concavity of
conditional quantum entropy by invoking an earlier result of \cite{L73}, and
\cite{LR73} also showed that the monotonicity of quantum relative entropy with
respect to partial trace is a consequence of the concavity of conditional
quantum entropy (the proof of this given in
Theorem~\ref{thm-qie:equivalence-ent-ineq}\ is due to them). \cite{araki1970}
proved the inequality in Exercise~\ref{ex-qie:araki-lieb}. \cite{U62}
established the modern definition of the quantum relative entropy. The fact
that it is non-negative for states is a result known as Klein's inequality
(see \cite{LR68} for attribution to Klein). \cite{Lindblad1975} established
the monotonicity of quantum relative entropy for separable Hilbert spaces
based on the results of \cite{LR73,PhysRevLett.30.434}, and \cite{U77}
extended Lindblad's result to more general settings.\ \cite{CCYZ11}\ proved
Proposition~\ref{prop-qie:rel-ent-s-s'}. \cite[Theorem~1.15]{OP93} proved the
quantum Pinsker inequality (stated here as Theorem~\ref{thm-qie:rel-ent-trace}%
).\ The coherent information first appeared in \cite{PhysRevA.54.2629}, where
they proved that it obeys a quantum data-processing inequality (this was the
first clue that the coherent information would be an important information
quantity for characterizing quantum capacity).

Entropic uncertainty relations have a long and interesting history. We do not
review this history here but instead point to the survey article
\citep{CBTW15}. There has been much interest in entropic uncertainty
relations, with perhaps the most notable advance being the entropic
uncertainty relation in the presence of quantum memory \citep{BCCRR10}. The
proof that we give for Theorem~\ref{thm-qie:EURQSI} is the same as that in
\citep{CCYZ11}, which in turn exploits ideas from \citep{TR11}.

\cite{Fannes73} proved his eponymous inequality, and \cite{A07} gave a
significant improvement of it. \cite{Winter15}\ proved\ the inequality in
Theorem~\ref{thm-qie:AFW-cont-ent}. Earlier, \cite{AF04} proved a weaker
version of the inequality in Theorem~\ref{thm-qie:AFW-cont-ent} (which however
has been extremely useful for many purposes in quantum information theory).
\cite{BSW14}\ proved Theorem~\ref{thm-qie:CQMI-distance-to-markov} (see also
\cite{FR14}).

\chapter{Quantum Entropy Inequalities and Recoverability}

\label{chap:mono-rel-ent}The quantum entropy inequalities discussed in the
previous chapter lie at the core of quantum Shannon theory and in fact underlie
some important principles of physics such as the uncertainty principle (see
Section~\ref{sec-qie:ent-unc-princ}). In fact, we will use these entropy
inequalities to prove the converse parts of every coding theorem appearing in
the last two parts of this book. Their prominence in both quantum Shannon
theory and other areas of physics motivates us to study them in more detail.
We delved into more depth in Chapter~\ref{chap:info-entropy}\ regarding many
of the classical entropy inequalities, and in the process, we established
necessary and sufficient conditions for the saturation of the inequalities
(Section~\ref{sec-ie:class-inf-ineq}), while also understanding the near
saturation of the entropy inequalities
(Section~\ref{sec-cie:entropy-ineq-refinements}). The aim of this chapter is
to carry out a similar program for all of the quantum entropy inequalities
presented in the previous chapter. The outcome will be a proof for the
monotonicity of quantum relative entropy (Theorem~\ref{thm-qie:mono-rel-ent}),
with the added benefit of an understanding of the saturation and near
saturation of this quantum entropy inequality.

\section{Recoverability Theorem}

The main theorem in this chapter can be summarized informally as follows: if
the decrease in quantum relative entropy between two quantum states after a
quantum channel acts is relatively small, then it is possible to perform a
recovery channel, such that we can perfectly recover one state while
approximately recovering the other. This can be interpreted as quantifying how
well one can reverse the action of a quantum channel. Throughout, we take
$\rho$, $\sigma$, and $\mathcal{N}$ as given in the following definition:

\begin{definition}
\label{def:rho-sig-N}Let $\rho\in\mathcal{D}(\mathcal{H})$ and let $\sigma
\in\mathcal{L}(\mathcal{H})$ be positive semi-definite, such that
$\operatorname{supp}(\rho)\subseteq\operatorname{supp}(\sigma)$. Let
$\mathcal{N}:\mathcal{L}(\mathcal{H})\rightarrow\mathcal{L}(\mathcal{H}%
^{\prime})$ be a quantum channel.
\end{definition}

\noindent The formal statement of the theorem is as follows:

\begin{theorem}
\label{thm-qeir:main-theorem}Given $\rho$, $\sigma$, and $\mathcal{N}$ as in
Definition~\ref{def:rho-sig-N}, there exists a recovery channel $\mathcal{R}%
_{\sigma,\mathcal{N}}:\mathcal{L}(\mathcal{H}^{\prime})\rightarrow
\mathcal{L}(\mathcal{H})$, depending only on $\sigma$ and $\mathcal{N}$, such
that%
\begin{align}
D(\rho\Vert\sigma)-D(\mathcal{N}(\rho)\Vert\mathcal{N}(\sigma))  &  \geq-\log
F(\rho,(\mathcal{R}_{\sigma,\mathcal{N}}\circ\mathcal{N})(\rho
)),\ \ \ \text{and}\label{eq-qeir:main-inequality}\\
(\mathcal{R}_{\sigma,\mathcal{N}}\circ\mathcal{N})(\sigma)  &  =\sigma.
\label{eq-qeir:perfect-recovery-sigma-prop}%
\end{align}

\end{theorem}

Given that the quantum fidelity $F$ takes values between zero and one, we can
immediately conclude that%
\begin{equation}
-\log F(\rho,(\mathcal{R}_{\sigma,\mathcal{N}}\circ\mathcal{N})(\rho))\geq0,
\end{equation}
so that the above theorem implies the monotonicity of quantum relative entropy
(Theorem~\ref{thm-qie:mono-rel-ent}) as a consequence. Furthermore, the
recovery channel satisfying \eqref{eq-qeir:main-inequality}\ has the property
that it perfectly recovers $\sigma$ from $\mathcal{N}(\sigma)$ (satisfying
\eqref{eq-qeir:perfect-recovery-sigma-prop}), a fact which we prove later and
which makes the inequality in \eqref{eq-qeir:main-inequality} non-trivial.

The proof given here for Theorem~\ref{thm-qeir:main-theorem} relies on the
method of complex interpolation and the notion of a R\'{e}nyi generalization
of a relative entropy difference.\ We review this background first before
going through the proof. One of the consequences of
Theorem~\ref{thm-qeir:main-theorem} is to provide physically meaningful
improvements to many quantum entropy inequalities discussed in the previous
chapter, such as strong subadditivity, joint convexity of quantum relative
entropy, and concavity of conditional quantum entropy. We explore these
consequences in Section~\ref{sec-qeir:corollaries}.

\section{Schatten Norms and Complex Interpolation}

The proof of Theorem~\ref{thm-qeir:main-theorem}\ given here requires a bit of
mathematical background before we can delve into it. So we first begin by
defining the Schatten norms and several of their properties. We then review
some essential results from complex analysis, that lead to a complex
interpolation theorem known as the Stein--Hirschman interpolation theorem.

\subsection{Schatten Norms and Duality}

An important technical tool in the proof given here is the Schatten $p$-norm
\index{Schatten norm}
of an operator~$A$, defined as%
\begin{equation}
\left\Vert A\right\Vert _{p}\equiv\left[  \operatorname{Tr}\left\{  \left\vert
A\right\vert ^{p}\right\}  \right]  ^{1/p},
\end{equation}
where $A\in\mathcal{L}(\mathcal{H})$,$\ |A|\equiv\sqrt{A^{\dag}A}$, and
$p\geq1$. We have already studied two special cases of this norm, which are
the trace norm when $p=1$ (Section~\ref{sec-dm:trace-norm}) and the
Hilbert--Schmidt norm when $p=2$ (Section~\ref{sec-dm:Hilbert-Schmidt-dist}).
One can show, along the same lines as the proof for
Proposition~\ref{prop-dm:trace-norm-sing-values}, that $\left\Vert
A\right\Vert _{p}$ is equal to the $p$-norm of the singular values of $A$.
That is, if $\sigma_{i}(A)$ is the vector of singular values of $A$, then%
\begin{equation}
\left\Vert A\right\Vert _{p}=\left[  \sum_{i}\sigma_{i}(A)^{p}\right]  ^{1/p}.
\label{eq-qeir:p-norm-sing-vals}%
\end{equation}
The convention is for $\left\Vert A\right\Vert _{\infty}$ to be defined as the
largest singular value of $A$ because $\left\Vert A\right\Vert _{p}$ converges
to this in the limit as $p\rightarrow\infty$. In the proof of
Theorem~\ref{thm-qeir:main-theorem}, we repeatedly use the fact that
$\left\Vert A\right\Vert _{p}$ is unitarily invariant. That is, $\left\Vert
A\right\Vert _{p}$ is invariant with respect to linear isometries, in the
sense that $\left\Vert A\right\Vert _{p}=\left\Vert UAV^{\dag}\right\Vert
_{p}$, where $U,V\in\mathcal{L}(\mathcal{H},\mathcal{H}^{\prime})$ are linear
isometries satisfying $U^{\dag}U=I_{\mathcal{H}}$ and $V^{\dag}%
V=I_{\mathcal{H}}$. Isometric invariance follows from
\eqref{eq-qeir:p-norm-sing-vals} and because these isometries do not change
the singular values of $A$. From these norms, one can define information
measures relating quantum states and channels, with the main one used here
known as a R\'{e}nyi generalization of a relative entropy difference.

Extending the Cauchy--Schwarz inequality is an important inequality known as
\index{H\"{o}lder inequality}
the H\"{o}lder inequality:%
\begin{equation}
\left\vert \left\langle A,B\right\rangle \right\vert =\left\vert
\operatorname{Tr}\{A^{\dag}B\}\right\vert \leq\left\Vert A\right\Vert
_{p}\left\Vert B\right\Vert _{q}, \label{eq-qeir:holder}%
\end{equation}
holding for $p,q\in\left[  1,\infty\right]  $ such that $\frac{1}{p}+\frac
{1}{q}=1$ and $A,B\in\mathcal{L}(\mathcal{H})$. When $p,q\in\left[
1,\infty\right]  $ and $\frac{1}{p}+\frac{1}{q}=1$, $p$ and $q$ are said to be
H\"{o}lder conjugates of each other. One can see that Cauchy--Schwarz is a
special case by picking $p=q=2$. Observe that equality is achieved in
\eqref{eq-qeir:holder}\ if $A$ and $B$ are such that $A^{\dag}=a\left\vert
B\right\vert ^{q/p}U^{\dag}$ for some constant $a\geq0$ and where $U$ is a
unitary such that $B=U\left\vert B\right\vert $ is a left polar decomposition
of $B$ (see Theorem~\ref{thm-app:polar}). The H\"{o}lder inequality along with
the sufficient equality condition is enough for us to conclude the following
variational expression for the $p$-norm in terms of its H\"{o}lder dual
$q$-norm:%
\begin{equation}
\left\Vert A\right\Vert _{p}=\max_{\left\Vert B\right\Vert _{q}\leq
1}\operatorname{Tr}\{A^{\dag}B\}.
\end{equation}
This expression can be very useful in calculations.

\begin{exercise}
Prove that $\left\Vert AB\right\Vert _{1}\leq\left\Vert A\right\Vert
_{p}\left\Vert B\right\Vert _{q}$ for $p,q\in\left[  1,\infty\right]  $ such
that $\frac{1}{p}+\frac{1}{q}=1$ and $A,B\in\mathcal{L}(\mathcal{H})$.
\end{exercise}

Throughout we adopt the convention from
Definition~\ref{def-qt:hermitian-op-function}\ and define $f(A)$ for a
function $f$ and a positive semi-definite operator $A$ as follows:
$f(A)\equiv\sum_{i:\lambda_{i}\in \operatorname{Dom}(f)}f(\lambda_{i})|i\rangle\langle i|$, where
$A=\sum_{i}\lambda_{i}|i\rangle\langle i|$ is a spectral decomposition of $A$.
We denote the support of $A$ by $\operatorname{supp}(A)$, and we let $\Pi_{A}$
denote the projection onto the support of $A$.

\subsection{Complex Analysis}

We now review a few concepts from complex analysis. We will not prove these
results in detail, but the purpose instead is to recall them, and the
interested reader can follow references to books on complex analysis for
details of proofs. The culmination of the development is the Stein--Hirschman
complex interpolation theorem (Theorem~\ref{thm-qeir:op-hirschman}).

The derivative of a complex-valued function $f:\mathbb{C}\rightarrow
\mathbb{C}$ at a point $z_{0}\in\mathbb{C}$\ is defined in the usual way as%
\begin{equation}
\left.  \frac{df(z)}{dz}\right\vert _{z=z_{0}}=\lim_{z\rightarrow z_{0}}%
\frac{f(z)-f(z_{0})}{z-z_{0}}.
\end{equation}
In order for this limit to exist, it must be the same for all possible
directions that one could take in the complex plane to approach $z_{0}$, and
this requirement demarcates a substantial difference between differentiability
of real functions and complex ones. Complex differentiability shares several
properties with
\index{holomorphic function}%
real differentiability:\ it is linear and obeys the product rule, the quotient
rule, and the chain rule. If $f$ is complex differentiable at every point
$z_{0}$ in an open set $U$, then we say that $f$ is \textit{holomorphic} on
$U$.

There is a connection between real differentiability and complex
\index{Cauchy--Riemann equations}
differentiability, given by the Cauchy--Riemann equations. Let
$f(x+iy)=u(x,y)+iv(x,y)$ where $x,y\in\mathbb{R}$ and $u,v:\mathbb{R}%
\rightarrow\mathbb{R}$. If $f$ is holomorphic, then $u$ and $v$ have first
partial derivatives with respect to $x$ and $y$ and satisfy the
Cauchy--Riemann equations:%
\begin{equation}
\frac{\partial u}{\partial x}=\frac{\partial v}{\partial y},\ \ \ \ \ \ \frac
{\partial u}{\partial y}=-\frac{\partial v}{\partial x}.
\end{equation}
The converse is not always true. However, if the first partial derivatives of
$u$ and $v$ are continuous and satisfy the Cauchy--Riemann equations, then $f$
is holomorphic. An important holomorphic function for our purposes is given in
the following exercise:

\begin{exercise}
\label{ex-qeir:x^z-holo}Verify that $f(z)=a^{z}=e^{\left[  \ln a\right]  z}$,
where $a>0$ and $z\in\mathbb{C}$, is a holomorphic function everywhere in the
complex plane. (Hint:\ Use that $e^{z}=e^{x}\left[  \cos(y)+i\sin(y)\right]  $
for $z=x+iy$ and $x,y\in\mathbb{R}$.)
\end{exercise}

Holomorphic functions have good closure properties. That is, the sums,
products, and compositions of holomorphic functions are holomorphic as well,
given that complex differentiation is linear and satisfies the product,
quotient, and chain rules. Note that the quotient of two holomorphic functions
is holomorphic wherever the denominator is not equal to zero.

The \textit{maximum modulus principle}
\index{maximum modulus principle}%
is an important principle that holomorphic functions obey. Formally, it is the
following statement:\ let $f:\mathbb{C}\rightarrow\mathbb{C}$ be a function
holomorphic on some connected, bounded open subset $U$ of $\mathbb{C}$. If
$z_{0}\in U$ is such that $\left\vert f(z_{0})\right\vert \geq\left\vert
f(z)\right\vert $ for all $z$ in a neighborhood of $z_{0}$, then the function
$f$ is constant on $U$. A consequence of this is that if $f$ is not constant
on a bounded, connected, open subset $U$ of $\mathbb{C}$, then it achieves its
maximum on the boundary of $U$.

The maximum modulus principle has an extension to an unbounded strip in
$\mathbb{C}$, which we call the \textit{maximum modulus principle on a strip}.
Let $S$ denote the standard strip in $\mathbb{C}$, $\overline{S}$ its closure,
and $\partial\overline{S}$ its boundary:%
\begin{align}
S  &  \equiv\left\{  z\in\mathbb{C}:0<\operatorname{Re}\{z\}<1\right\}  ,\\
\overline{S}  &  \equiv\left\{  z\in\mathbb{C}:0\leq\operatorname{Re}%
\{z\}\leq1\right\}  ,\\
\partial\overline{S}  &  \equiv\left\{  z\in\mathbb{C}:\operatorname{Re}%
\{z\}=0\vee\operatorname{Re}\{z\}=1\right\}  .
\end{align}
Let $f:\overline{S}\rightarrow\mathbb{C}$ be bounded on $\overline{S}$,
holomorphic on $S$, and continuous on $\partial\overline{S}$. Then the
supremum of $\left\vert f\right\vert $ is attained on $\partial\overline{S}$.
That is, $\sup_{z\in\overline{S}}\left\vert f(z)\right\vert =\sup
_{z\in\partial\overline{S}}\left\vert f(z)\right\vert $.

The maximum modulus principle on a strip implies a result known as the
\index{Hadamard three-lines theorem}
Hadamard three-lines theorem:

\begin{theorem}
[Hadamard Three-Lines]Let $f:\overline{S}\rightarrow\mathbb{C}$ be a function
that is bounded on $\overline{S}$, holomorphic on $S$, and continuous on the
boundary $\partial\overline{S}$. Let $\theta\in\left(  0,1\right)  $ and
$M(\theta)\equiv\sup_{t\in\mathbb{R}}\left\vert f(\theta+it)\right\vert $.
Then $\ln M(\theta)$ is a convex function on $\left[  0,1\right]  $, implying
that%
\begin{equation}
\ln M(\theta)\leq\left(  1-\theta\right)  \ln M(0)+\theta\ln M(1).
\end{equation}

\end{theorem}

\noindent There is a strengthening of the Hadamard three-lines theorem due to
Hirschman, which in fact implies the Hadamard three-lines theorem:

\begin{theorem}
[Hirschman]\label{thm-qeir:hirschman}%
\index{Hirschman's theorem}
Let $f(z):\overline{S}\rightarrow\mathbb{C}$ be a function that is bounded on
$\overline{S}$, holomorphic on $S$, and continuous on the boundary
$\partial\overline{S}$. Then for $\theta\in(0,1)$, the following bound holds%
\begin{equation}
\ln\left\vert f(\theta)\right\vert \leq\int_{-\infty}^{\infty}dt\ \left(
\alpha_{\theta}(t)\ln\left[  \left\vert f(it)\right\vert ^{1-\theta}\right]
+\beta_{\theta}(t)\ln\left[  \left\vert f(1+it)\right\vert ^{\theta}\right]
\right)  ,
\end{equation}
where%
\begin{align}
\alpha_{\theta}(t)  &  \equiv\frac{\sin(\pi\theta)}{2(1-\theta)\left[
\cosh(\pi t)-\cos(\pi\theta)\right]  },\label{eq-qeir:alpha-t}\\
\beta_{\theta}(t)  &  \equiv\frac{\sin(\pi\theta)}{2\theta\left[  \cosh(\pi
t)+\cos(\pi\theta)\right]  }. \label{eq-qeir:beta-t}%
\end{align}

\end{theorem}

For a fixed $\theta\in(0,1)$, we have that $\alpha_{\theta}(t),\beta_{\theta
}(t)\geq0$ for all $t\in\mathbb{R}$ and%
\begin{equation}
\int_{-\infty}^{\infty}dt\ \alpha_{\theta}(t)=\int_{-\infty}^{\infty}%
dt\ \beta_{\theta}(t)=1\ ,
\end{equation}
(see, e.g., \cite[Exercise~1.3.8]{G08}) so that $\alpha_{\theta}(t)$ and
$\beta_{\theta}(t)$ can be interpreted as probability density functions.
Furthermore, we have that%
\begin{equation}
\lim_{\theta\searrow0}\beta_{\theta}(t)=\frac{\pi}{2}\left[  \cosh(\pi
t)+1\right]  ^{-1}\equiv\beta_{0}(t)\ , \label{eq:dist-limit}%
\end{equation}
where $\beta_{0}$ is also a probability density function on $\mathbb{R}$. With
these observations, we can see that Hirschman's theorem implies the Hadamard
three-lines theorem, given that an expectation can never exceed a supremum.

\subsection{Complex Interpolation of Schatten Norms}

We can extend much of the development above to operator-valued functions,
which is needed to prove Theorem~\ref{thm-qeir:main-theorem}. Let
$G:\mathbb{C}\rightarrow\mathcal{L}(\mathcal{H})$ be an operator-valued
function.
\index{holomorphic function!operator-valued}%
We say that $G(z)$ is holomorphic if every function mapping $z$ to a matrix
entry is holomorphic. For our purposes in what follows, we are interested in
operator-valued functions of the form $A^{z}$, where $A$ is a positive
semi-definite operator. In this case, we apply the convention from
Definition~\ref{def-qt:hermitian-op-function}\ and take $A^{z}=\sum
_{i:\lambda_{i}\neq0}\lambda_{i}^{z}|i\rangle\langle i|$, where $A=\sum
_{i}\lambda_{i}|i\rangle\langle i|$ is an eigendecomposition of $A$ with
$\lambda_{i}\geq0$ for all $i$. Given the result of
Exercise~\ref{ex-qeir:x^z-holo}\ combined with the closure properties of
holomorphic functions mentioned above, we can conclude that $A^{z}$ is
holomorphic if $A$ is positive semi-definite.

We can now establish a version of the Hirschman theorem which applies to
operator-valued functions and allows for bounding their Schatten norms. This
is one of the main technical tools that we need to establish
Theorem~\ref{thm-qeir:main-theorem}.

\begin{theorem}
[Stein--Hirschman]\label{thm-qeir:op-hirschman}
\index{Stein--Hirschman interpolation theorem}%
Let $G:\overline{S}\rightarrow L(\mathcal{H})$ be an operator-valued function
that is bounded on $\overline{S}$, holomorphic on $S$, and continuous on the
boundary $\partial\overline{S}$. Let $\theta\in(0,1)$ and define $p_{\theta}$
by%
\begin{equation}
\frac{1}{p_{\theta}}=\frac{1-\theta}{p_{0}}+\frac{\theta}{p_{1}}\ ,
\end{equation}
where $p_{0},p_{1}\in\left[  1,\infty\right]  $. Then the following bound
holds%
\begin{equation}
\ln\left\Vert G(\theta)\right\Vert _{p_{\theta}}\leq\int_{-\infty}^{\infty
}dt\ \left(  \alpha_{\theta}(t)\ln\left[  \left\Vert G(it)\right\Vert _{p_{0}%
}^{1-\theta}\right]  +\beta_{\theta}(t)\ln\left[  \left\Vert
G(1+it)\right\Vert _{p_{1}}^{\theta}\right]  \right)  \ ,
\label{eq:oper-hirschman}%
\end{equation}
where $\alpha_{\theta}(t)$ and $\beta_{\theta}(t)$ are defined in \eqref{eq-qeir:alpha-t}--\eqref{eq-qeir:beta-t}.
\end{theorem}

\begin{proof}
For fixed $\theta\in(0,1)$, let $q_{\theta}$ be the H\"{o}lder conjugate of
$p_{\theta}$, defined by%
\begin{equation}
\frac{1}{p_{\theta}}+\frac{1}{q_{\theta}}=1\ .
\end{equation}
Similarly, let $q_{0}$ and $q_{1}$ be H\"{o}lder conjugates of $p_{0}$ and
$p_{1}$, respectively. From the sufficient equality condition for the
H\"{o}lder inequality, we can find an operator $X$ such that $\left\Vert
X\right\Vert _{q_{\theta}}=1$ and $\operatorname{Tr}\{XG(\theta)\}=\left\Vert
G(\theta)\right\Vert _{p_{\theta}}$. We can write the singular value
decomposition for $X$ in the form $X=UD^{1/q_{\theta}}V$ (implying
$\operatorname{Tr}\{D\}=1$). For $z\in S$, define%
\begin{equation}
X(z)\equiv UD^{\frac{1-z}{q_{0}}+\frac{z}{q_{1}}}V\ .
\end{equation}
As a consequence, $X(z)$ is bounded on $\overline{S}$, holomorphic on $S$, and
continuous on the boundary $\partial\overline{S}$. Also, observe that
$X(\theta)=X$. Then the following function satisfies the requirements needed
to apply Theorem~\ref{thm-qeir:hirschman}:%
\begin{equation}
g(z)\equiv\operatorname{Tr}\{X(z)G(z)\}\ .
\end{equation}
Indeed, we have that%
\begin{align}
\ln\left\Vert G(\theta)\right\Vert _{p_{\theta}}  &  =\ln\left\vert
g(\theta)\right\vert \\
&  \leq\int_{-\infty}^{\infty}dt\ \left(  \alpha_{\theta}(t)\ln\left[
\left\vert g(it)\right\vert ^{1-\theta}\right]  +\beta_{\theta}(t)\ln\left[
\left\vert g(1+it)\right\vert ^{\theta}\right]  \right)  .
\label{eq:app-hirschman}%
\end{align}
Now, from applying H\"{o}lder's inequality and the facts that $\left\Vert
X(it)\right\Vert _{q_{0}}=1=\left\Vert X(1+it)\right\Vert _{q_{1}}$, we find
that%
\begin{equation}
\left\vert g(it)\right\vert =\left\vert \operatorname{Tr}%
\{X(it)G(it)\}\right\vert \leq\left\Vert X(it)\right\Vert _{q_{0}}\left\Vert
G(it)\right\Vert _{p_{0}}=\left\Vert G(it)\right\Vert _{p_{0}},
\end{equation}
and
\begin{align}
\left\vert g(1+it)\right\vert  &  =\left\vert \operatorname{Tr}%
\{X(1+it)G(1+it)\}\right\vert \\
&  \leq\left\Vert X(1+it)\right\Vert _{q_{1}}\left\Vert G(1+it)\right\Vert
_{p_{1}}\\
&  =\left\Vert G(1+it)\right\Vert _{p_{1}}.
\end{align}
Bounding \eqref{eq:app-hirschman} from above using these inequalities then
gives \eqref{eq:oper-hirschman}.
\end{proof}

The theorem above is known as a complex interpolation theorem because it
allows us to obtain estimates on the \textquotedblleft
intermediate\textquotedblright\ norm in terms of other norms which might be
available. Furthermore, we are interpolating through the holomorphic family of
operators given by $G(z)$.

\section{Petz Recovery Map}%

\index{Petz recovery map}%
The channel appearing in the lower bound of
Theorem~\ref{thm-qeir:main-theorem}$\ $has an explicit form and is constructed
from a map known as the Petz recovery map, which we define as follows:

\begin{definition}
Let $\sigma\in\mathcal{L}(\mathcal{H})$ be positive semi-definite, and let
$\mathcal{N}:\mathcal{L}(\mathcal{H})\rightarrow\mathcal{L}(\mathcal{H}%
^{\prime})$ be a quantum channel. The Petz recovery map $\mathcal{P}%
_{\sigma,\mathcal{N}}:\mathcal{L}(\mathcal{H}^{\prime})\rightarrow
\mathcal{L}(\mathcal{H})$ is a completely positive, trace-non-increasing
linear map defined as follows for $Q\in\mathcal{L}(\mathcal{H}^{\prime})$:%
\begin{equation}
\mathcal{P}_{\sigma,\mathcal{N}}\left(  Q\right)  \equiv\sigma^{1/2}%
\mathcal{N}^{\dag}\left(  \left[  \mathcal{N}(\sigma)\right]  ^{-1/2}Q\left[
\mathcal{N}(\sigma)\right]  ^{-1/2}\right)  \sigma^{1/2}.
\label{eq:Petz-channel-Rel-ent}%
\end{equation}

\end{definition}

The Petz recovery map $\mathcal{P}_{\sigma,\mathcal{N}}$ is linear, and it is
completely positive because it is equal to a serial concatenation of three
completely positive maps: $Q\rightarrow\left[  \mathcal{N}(\sigma)\right]
^{-1/2}Q\left[  \mathcal{N}(\sigma)\right]  ^{-1/2}$, $Q\rightarrow
\mathcal{N}^{\dag}(Q)$, and $M\rightarrow\sigma^{1/2}M\sigma^{1/2}$ for
$M\in\mathcal{L}(\mathcal{H})$. It is trace-non-increasing because the
following holds for positive semi-definite $Q$:%
\begin{align}
\operatorname{Tr}\{\mathcal{P}_{\sigma,\mathcal{N}}\left(  Q\right)  \}  &
=\operatorname{Tr}\left\{  \sigma\mathcal{N}^{\dag}\left(  \left[
\mathcal{N}(\sigma)\right]  ^{-1/2}Q\left[  \mathcal{N}(\sigma)\right]
^{-1/2}\right)  \right\} \\
&  =\operatorname{Tr}\left\{  \mathcal{N}(\sigma)\left[  \mathcal{N}%
(\sigma)\right]  ^{-1/2}Q\left[  \mathcal{N}(\sigma)\right]  ^{-1/2}\right\}
\\
&  =\operatorname{Tr}\{\Pi_{\mathcal{N}(\sigma)}Q\}\leq\operatorname{Tr}\{Q\}.
\end{align}

An important special case of the Petz recovery map occurs when $\sigma$ and
$\mathcal{N}$ are effectively classical. That is, suppose that $\mathcal{N}$
is a classical-to-classical channel with Kraus operators $\{\sqrt
{N(y|x)}|y\rangle\langle x|\}$ (see Section~\ref{sec-nqt:c-to-c-channel}),
where $N(y|x)$ is a conditional probability distribution. Suppose further that
$\sigma=\sum_{x}q(x)|x\rangle\langle x|$, with $q(x)\geq0$ for all $x$. In
this case, one can check that the Petz recovery map is a
classical-to-classical channel with Kraus operators $\{\sqrt{R(x|y)}%
|x\rangle\langle y|\}$, where $R(x|y)$ is a conditional probability
distribution given by the Bayes theorem, satisfying%
\begin{equation}
R(x|y)(Nq)(y)=N(y|x)q(x),
\end{equation}
for all $x$ and $y$, where $(Nq)(y)=\sum_{x}N(y|x)q(x)$. We leave the details
of this calculation as an exercise for the reader and point out that this
recovery channel appears in the refinement of the monotonicity of classical
relative entropy from Theorem~\ref{thm-cie:refine-mono-rel-ent}.

We can also define a partial isometric map $\mathcal{U}_{\sigma,t}$ in the
following way:%
\begin{equation}
\mathcal{U}_{\sigma,t}(M)\equiv\sigma^{it}M\sigma^{-it}.
\label{eq-qeir:unitaries}%
\end{equation}
Since $\sigma^{it}\sigma^{-it}=\Pi_{\sigma}$, we can conclude that%
\begin{equation}
\mathcal{U}_{\sigma,t}(\sigma)=\sigma,
\label{eq-qeir:unitary-perfect-recovery}%
\end{equation}
so that this isometric map does not have any effect when $\sigma$ is input. In
the case that $\sigma$ is positive definite, $\mathcal{U}_{\sigma,t}$ is a
unitary channel. We can then define a rotated or \textquotedblleft
swiveled\textquotedblright\ Petz map, which plays an important role in the
construction of a recovery channel satisfying the lower bound in
Theorem~\ref{thm-qeir:main-theorem}.

\begin{definition}
[Rotated Petz Map]\label{def-qeir:rotated-petz}Let $\sigma\in\mathcal{L}%
(\mathcal{H})$ be positive semi-definite, and let $\mathcal{N}:\mathcal{L}%
(\mathcal{H})\rightarrow\mathcal{L}(\mathcal{H}^{\prime})$ be a quantum
channel.
\index{Petz recovery map!rotated}%
A rotated Petz map is defined as follows for $Q\in\mathcal{L}(\mathcal{H}%
^{\prime})$:%
\begin{equation}
\mathcal{R}_{\sigma,\mathcal{N}}^{t}\left(  Q\right)  \equiv(\mathcal{U}%
_{\sigma,-t}\circ\mathcal{P}_{\sigma,\mathcal{N}}\circ\mathcal{U}%
_{\mathcal{N}(\sigma),t})(Q).
\end{equation}

\end{definition}

\begin{proposition}
[Perfect Recovery]\label{prop-qeir:perfect-recovery}Let $\sigma\in
\mathcal{L}(\mathcal{H})$ be positive semi-definite, and let $\mathcal{N}%
:\mathcal{L}(\mathcal{H})\rightarrow\mathcal{L}(\mathcal{H}^{\prime})$ be a
quantum channel. A rotated Petz map $\mathcal{R}_{\sigma,\mathcal{N}}^{t}%
$\ perfectly recovers $\sigma$ from $\mathcal{N}(\sigma)$:%
\begin{equation}
\mathcal{R}_{\sigma,\mathcal{N}}^{t}(\mathcal{N}(\sigma))=\sigma.
\end{equation}

\end{proposition}

\begin{proof}
Consider that%
\begin{align}
\mathcal{P}_{\sigma,\mathcal{N}}(\mathcal{N}(\sigma))  &  =\sigma
^{1/2}\mathcal{N}^{\dag}\left(  \left[  \mathcal{N}(\sigma)\right]
^{-1/2}\mathcal{N}(\sigma)\left[  \mathcal{N}(\sigma)\right]  ^{-1/2}\right)
\sigma^{1/2}\\
&  =\sigma^{1/2}\mathcal{N}^{\dag}(\Pi_{\mathcal{N}(\sigma)})\sigma^{1/2}\\
&  \leq\sigma^{1/2}\mathcal{N}^{\dag}(I)\sigma^{1/2}=\sigma^{1/2}I\sigma
^{1/2}=\sigma.
\end{align}
The inequality follows because $\Pi_{\mathcal{N}(\sigma)}\leq I$ and
$\mathcal{N}^{\dag}$ is a completely positive map. The second to last equality
follows because the adjoint is unital.

Now we prove the other operator inequality $\mathcal{P}_{\sigma,\mathcal{N}%
}(\mathcal{N}(\sigma))\geq\sigma$, which will allow us to conclude that
$\mathcal{P}_{\sigma,\mathcal{N}}(\mathcal{N}(\sigma))=\sigma$. Let
$U:\mathcal{H}\rightarrow\mathcal{H}^{\prime}\otimes\mathcal{H}_{E}$ be an
isometric extension of the channel $\mathcal{N}$. From
Lemma~\ref{lem-app:support-1}, we know that $\operatorname{supp}\left(
U\sigma U^{\dag}\right)  \subseteq\operatorname{supp}\left(  \mathcal{N}%
(\sigma)\otimes I_{E}\right)  $, which implies that $\Pi_{U\sigma U^{\dag}%
}\leq\Pi_{\mathcal{N}(\sigma)\otimes I_{E}}=\Pi_{\mathcal{N}(\sigma)}\otimes
I_{E}$.\ Then for any vector $|\psi\rangle\in\mathcal{H}$, we have that%
\begin{align}
\langle\psi|\Pi_{\sigma}|\psi\rangle &  =\langle\psi|U^{\dag}\Pi_{U\sigma
U^{\dag}}U|\psi\rangle\\
&  \leq\langle\psi|U^{\dag}\left(  \Pi_{\mathcal{N}(\sigma)}\otimes
I_{E}\right)  U|\psi\rangle\\
&  =\operatorname{Tr}\{U|\psi\rangle\langle\psi|U^{\dag}\left(  \Pi
_{\mathcal{N}(\sigma)}\otimes I_{E}\right)  \}\\
&  =\operatorname{Tr}\{\mathcal{N}(|\psi\rangle\langle\psi|)\Pi_{\mathcal{N}%
(\sigma)}\}\\
&  =\operatorname{Tr}\{|\psi\rangle\langle\psi|\mathcal{N}^{\dag}%
(\Pi_{\mathcal{N}(\sigma)})\}\\
&  =\langle\psi|\mathcal{N}^{\dag}(\Pi_{\mathcal{N}(\sigma)})|\psi\rangle.
\end{align}
Since $|\psi\rangle$ is arbitrary, this establishes that $\Pi_{\sigma}%
\leq\mathcal{N}^{\dag}(\Pi_{\mathcal{N}(\sigma)})$. We can then use this
operator inequality to see that%
\begin{align}
\mathcal{P}_{\sigma,\mathcal{N}}(\mathcal{N}(\sigma))  &  =\sigma
^{1/2}\mathcal{N}^{\dag}(\Pi_{\mathcal{N}(\sigma)})\sigma^{1/2}\\
&  \geq\sigma^{1/2}\Pi_{\sigma}\sigma^{1/2}=\sigma.
\end{align}
Finally, we can conclude that%
\begin{align}
\mathcal{R}_{\sigma,\mathcal{N}}^{t}(\mathcal{N}(\sigma))  &  =(\mathcal{U}%
_{\sigma,-t}\circ\mathcal{P}_{\sigma,\mathcal{N}}\circ\mathcal{U}%
_{\mathcal{N}(\sigma),t})(\mathcal{N}(\sigma))\\
&  =(\mathcal{U}_{\sigma,-t}\circ\mathcal{P}_{\sigma,\mathcal{N}}%
)(\mathcal{N}(\sigma))\\
&  =\mathcal{U}_{\sigma,-t}(\sigma)=\sigma,
\end{align}
where the second and last equalities follow from \eqref{eq-qeir:unitary-perfect-recovery}.
\end{proof}

\begin{exercise}
\label{ex-qeir:rotated-Petz-TNI}Verify that a rotated Petz map satisfies the
following:%
\begin{equation}
\operatorname{Tr}\{\mathcal{R}_{\sigma,\mathcal{N}}^{t}(Q)\}=\operatorname{Tr}%
\{\Pi_{\mathcal{N}(\sigma)}Q\}\leq\operatorname{Tr}\{Q\},
\end{equation}
for positive semi-definite $Q\in\mathcal{L}(\mathcal{H}^{\prime})$, which
implies that it is trace non-increasing.
\end{exercise}

\begin{exercise}
Let $\omega$ be positive semi-definite and let $\left\langle A,B\right\rangle
_{\omega}\equiv\operatorname{Tr}\{A^{\dag}\omega^{1/2}B\omega^{1/2}\}$. The
adjoint of the Petz recovery map (with respect to the Hilbert--Schmidt inner
product) is equal to%
\begin{equation}
\mathcal{P}_{\sigma,\mathcal{N}}^{\dag}(M)=\left[  \mathcal{N}(\sigma)\right]
^{-1/2}\mathcal{N}\left(  \sigma^{1/2}M\sigma^{1/2}\right)  \left[
\mathcal{N}(\sigma)\right]  ^{-1/2},
\end{equation}
where $M\in\mathcal{L}(\mathcal{H})$. Show that $\mathcal{P}_{\sigma
,\mathcal{N}}^{\dag}$ is the unique linear map with domain
$\operatorname{supp}(\sigma)$ and range $\operatorname{supp}(\mathcal{N}%
(\sigma))$, which satisfies the following for all $M\in\mathcal{L}%
(\mathcal{H})$ and $Q\in\mathcal{L}(\mathcal{H}^{\prime})$:%
\begin{equation}
\left\langle M,\mathcal{N}^{\dag}(Q)\right\rangle _{\sigma}=\left\langle
\mathcal{P}_{\sigma,\mathcal{N}}^{\dag}(M),Q\right\rangle _{\mathcal{N}%
(\sigma)}.
\end{equation}
(Hint:\ Consider picking $M=|i\rangle\langle j|$ and $Q=|k\rangle\langle l|$.)
\end{exercise}

\section{R\'{e}nyi Information Measure}

Given $\rho$, $\sigma$, and $\mathcal{N}$ as in Definition~\ref{def:rho-sig-N}%
,
\index{R\'{e}nyi generalization of a relative entropy difference}
we define a R\'{e}nyi information measure $\widetilde{\Delta}_{\alpha}$ known
as a R\'{e}nyi generalization of a relative entropy difference:%
\begin{align}
\widetilde{\Delta}_{\alpha}(\rho,\sigma,\mathcal{N}) & \equiv\frac{1 }%
{\alpha-1}\ln\widetilde{Q}_{\alpha}(\rho,\sigma,\mathcal{N}%
),\label{eq:renyi-diff}\\
\widetilde{Q}_{\alpha}(\rho,\sigma,\mathcal{N})  & \equiv\left\Vert \left(
\left[  \mathcal{N}(\rho)\right]  ^{\frac{1-\alpha}{2\alpha}}\left[
\mathcal{N}(\sigma)\right]  ^{\frac{\alpha-1}{2\alpha}}\otimes I_{E}\right)
U\sigma^{\frac{1-\alpha}{2\alpha}}\rho^{1/2}\right\Vert ^{2\alpha}_{2\alpha},
\end{align}
where $\alpha\in(0,1)\cup(1,\infty)$ and $U:\mathcal{H}\rightarrow
\mathcal{H}^{\prime}\otimes\mathcal{H}_{E}$ is an isometric extension of the
channel $\mathcal{N}$. That is, $U$ is a linear isometry satisfying
$\operatorname{Tr}_{E}\{U(\cdot)U^{\dag}\}=\mathcal{N}(\cdot)$ and $U^{\dag
}U=I_{\mathcal{H}}$. Recall that all isometric extensions of a channel are
related by an isometry acting on the environment system~$E$, so that the
definition in \eqref{eq:renyi-diff} is invariant under any such choice. Recall
from Proposition~\ref{prop-pqt:adjoint-iso-ext} that the adjoint
$\mathcal{N}^{\dag}$\ of a channel is given in terms of an isometric
extension~$U$ as $\mathcal{N}^{\dag}(\cdot)=U^{\dag}\left(  (\cdot)\otimes
I_{E}\right)  U$.

The following lemma is one of the main reasons that we say that $\widetilde
{\Delta}_{\alpha}(\rho,\sigma,\mathcal{N})$ is a R\'{e}nyi generalization of a
relative entropy difference.

\begin{lemma}
The following limit holds for $\rho$, $\sigma$, and $\mathcal{N}$ as given in
Definition~\ref{def:rho-sig-N}:%
\begin{equation}
\frac{1}{\ln2}\lim_{\alpha\rightarrow1}\widetilde{\Delta}_{\alpha}(\rho
,\sigma,\mathcal{N})=D(\rho\Vert\sigma)-D(\mathcal{N}(\rho)\Vert
\mathcal{N}(\sigma)). \label{eq:rel-ent-diff-a-1}%
\end{equation}

\end{lemma}

\begin{proof}
Let $\Pi_{\omega}$ denote the projection onto the support of $\omega$. From
the condition $\operatorname{supp}(\rho)\subseteq\operatorname{supp}(\sigma)$,
it follows that $\operatorname{supp}(\mathcal{N}(\rho))\subseteq
\operatorname{supp}\left(  \mathcal{N}(\sigma)\right)  $ (see
Lemma~\ref{lem-app:support-2}). We can then conclude that%
\begin{equation}
\Pi_{\sigma}\Pi_{\rho}=\Pi_{\rho},\ \ \ \ \ \ \ \ \Pi_{\mathcal{N}(\rho)}%
\Pi_{\mathcal{N}(\sigma)}=\Pi_{\mathcal{N}(\rho)}. \label{eq:Pi-rho}%
\end{equation}
We also know that $\operatorname{supp}(U\rho U^{\dag})\subseteq
\operatorname{supp}\left(  \mathcal{N}(\rho)\otimes I_{E}\right)  $ (see
Lemma~\ref{lem-app:support-1}), so that%
\begin{equation}
\left(  \Pi_{\mathcal{N}(\rho)}\otimes I_{E}\right)  \Pi_{U\rho U^{\dag}}%
=\Pi_{U\rho U^{\dag}}. \label{eq:Pi-U-rho}%
\end{equation}
When $\alpha=1$, we find from the above facts that%
\begin{align}
\widetilde{Q}_{1}(\rho,\sigma,\mathcal{N})  &  =\left\Vert \left(
\Pi_{\mathcal{N}(\rho)}\Pi_{\mathcal{N}(\sigma)}\otimes I_{E}\right)
U\Pi_{\sigma}\rho^{1/2}\right\Vert _{2}^{2}\\
&  =\left\Vert \left(  \Pi_{\mathcal{N}(\rho)}\otimes I_{E}\right)  U\Pi
_{\rho}\rho^{1/2}\right\Vert _{2}^{2}\\
&  =\left\Vert \left(  \Pi_{\mathcal{N}(\rho)}\otimes I_{E}\right)  \Pi_{U\rho
U^{\dag}}U\rho^{1/2}\right\Vert _{2}^{2}\\
&  =\left\Vert \Pi_{U\rho U^{\dag}}U\rho^{1/2}\right\Vert _{2}^{2}=\left\Vert
\rho^{1/2}\right\Vert _{2}^{2}=1.
\end{align}
So from the definition of the derivative, this means that%
\begin{align}
\lim_{\alpha\rightarrow1}\widetilde{\Delta}_{\alpha}(\rho,\sigma,\mathcal{N})
&  =\lim_{\alpha\rightarrow1}\frac{\ln\widetilde{Q}_{\alpha}(\rho
,\sigma,\mathcal{N})-\ln\widetilde{Q}_{1}(\rho,\sigma,\mathcal{N})}{\alpha
-1}\\
&  =\left.  \frac{d}{d\alpha}\left[  \ln\widetilde{Q}_{\alpha}(\rho
,\sigma,\mathcal{N})\right]  \right\vert _{\alpha=1}\\
&  =\frac{1}{\widetilde{Q}_{1}(\rho,\sigma,\mathcal{N})}\left.  \frac
{d}{d\alpha}\left[  \widetilde{Q}_{\alpha}(\rho,\sigma,\mathcal{N})\right]
\right\vert _{\alpha=1}\\
&  =\left.  \frac{d}{d\alpha}\left[  \widetilde{Q}_{\alpha}(\rho
,\sigma,\mathcal{N})\right]  \right\vert _{\alpha=1}. \label{eq:limit-a-1-exp}%
\end{align}
Let $\alpha^{\prime}\equiv\frac{\alpha-1}{\alpha}$. Now consider that%
\begin{equation}
\widetilde{Q}_{\alpha}(\rho,\sigma,\mathcal{N})=\operatorname{Tr}\left\{
\left[  \rho^{1/2}\sigma^{-\alpha^{\prime}/2}\mathcal{N}^{\dag}\left(
\mathcal{N}(\sigma)^{\alpha^{\prime}/2}\mathcal{N}(\rho)^{-\alpha^{\prime}%
}\mathcal{N}(\sigma)^{\alpha^{\prime}/2}\right)  \sigma^{-\alpha^{\prime}%
/2}\rho^{1/2}\right]  ^{\alpha}\right\}  .
\end{equation}
Define the function%
\begin{equation}
\widetilde{Q}_{\alpha,\beta}(\rho,\sigma,\mathcal{N})\equiv\operatorname{Tr}%
\left\{  \left[  \rho^{1/2}\sigma^{-\alpha^{\prime}/2}\mathcal{N}^{\dag
}\left(  \mathcal{N}(\sigma)^{\alpha^{\prime}/2}\mathcal{N}(\rho
)^{-\alpha^{\prime}}\mathcal{N}(\sigma)^{\alpha^{\prime}/2}\right)
\sigma^{-\alpha^{\prime}/2}\rho^{1/2}\right]  ^{\beta}\right\}  ,
\end{equation}
and consider that%
\begin{align}
\left.  \frac{d}{d\alpha}\left[  \widetilde{Q}_{\alpha}(\rho,\sigma
,\mathcal{N})\right]  \right\vert _{\alpha=1}  &  =\left.  \frac{d}{d\alpha
}\widetilde{Q}_{\alpha,\alpha}(\rho,\sigma,\mathcal{N})\right\vert _{\alpha
=1}\\
&  =\left.  \frac{d}{d\alpha}\widetilde{Q}_{\alpha,1}(\rho,\sigma
,\mathcal{N})\right\vert _{\alpha=1}+\left.  \frac{d}{d\beta}\widetilde
{Q}_{1,\beta}(\rho,\sigma,\mathcal{N})\right\vert _{\beta=1}.
\label{eq:chain-break-up}%
\end{align}
We first compute $\widetilde{Q}_{1,\beta}(\rho,\sigma,\mathcal{N})$ as
follows:%
\begin{align}
\widetilde{Q}_{1,\beta}(\rho,\sigma,\mathcal{N})  &  =\operatorname{Tr}%
\left\{  \left[  \rho^{1/2}\Pi_{\sigma}\mathcal{N}^{\dag}(\Pi_{\mathcal{N}%
(\sigma)}\Pi_{\mathcal{N}(\rho)}\Pi_{\mathcal{N}(\sigma)})\Pi_{\sigma}%
\rho^{1/2}\right]  ^{\beta}\right\} \\
&  =\operatorname{Tr}\left\{  \left[  \rho^{1/2}\mathcal{N}^{\dag}%
(\Pi_{\mathcal{N}(\rho)})\rho^{1/2}\right]  ^{\beta}\right\} \\
&  =\operatorname{Tr}\left\{  \left[  \rho^{1/2}U^{\dag}\left(  \Pi
_{\mathcal{N}(\rho)}\otimes I_{E}\right)  U\rho^{1/2}\right]  ^{\beta}\right\}
\\
&  =\operatorname{Tr}\left\{  \left[  \left(  \Pi_{\mathcal{N}(\rho)}\otimes
I_{E}\right)  U\rho U^{\dag}\left(  \Pi_{\mathcal{N}(\rho)}\otimes
I_{E}\right)  \right]  ^{\beta}\right\} \\
&  =\operatorname{Tr}\left\{  \left[  U\rho U^{\dag}\right]  ^{\beta}\right\}
=\operatorname{Tr}\left\{  \rho^{\beta}\right\}  .
\end{align}
So then%
\begin{align}
\left.  \frac{d}{d\beta}\widetilde{Q}_{1,\beta}(\rho,\sigma,\mathcal{N}%
)\right\vert _{\beta=1}  &  =\left.  \frac{d}{d\beta}\operatorname{Tr}\left\{
\rho^{\beta}\right\}  \right\vert _{\beta=1} =\left.  \operatorname{Tr}%
\left\{  \rho^{\beta}\ln\rho\right\}  \right\vert _{\beta=1}\\
&  =\operatorname{Tr}\left\{  \rho\ln\rho\right\}  . \label{eq:Q_beta_1_term}%
\end{align}
Now we turn to the other term $\frac{d}{d\alpha}\widetilde{Q}_{\alpha,1}%
(\rho,\sigma,\mathcal{N})$. First consider that%
\begin{align}
\frac{d}{d\alpha}\left(  -\alpha^{\prime}\right)   &  =\frac{d}{d\alpha
}\left(  \frac{1-\alpha}{\alpha}\right)  =\frac{d}{d\alpha}\left(  \frac
{1}{\alpha}-1\right)  =-\frac{1}{\alpha^{2}},\\
\widetilde{Q}_{\alpha,1}(\rho,\sigma,\mathcal{N})  &  =\operatorname{Tr}%
\left\{  \rho\sigma^{-\alpha^{\prime}/2}\mathcal{N}^{\dag}\left(
\mathcal{N}(\sigma)^{\alpha^{\prime}/2}\mathcal{N}(\rho)^{-\alpha^{\prime}%
}\mathcal{N}(\sigma)^{\alpha^{\prime}/2}\right)  \sigma^{-\alpha^{\prime}%
/2}\right\}  .
\end{align}
Now we show that $\frac{d}{d\alpha}\widetilde{Q}_{\alpha,1}(\rho
,\sigma,\mathcal{N})$ is equal to%
\begin{multline}
\frac{d}{d\alpha}\operatorname{Tr}\left\{  \rho\sigma^{-\alpha^{\prime}%
/2}\mathcal{N}^{\dag}\left(  \mathcal{N}(\sigma)^{\alpha^{\prime}%
/2}\mathcal{N}(\rho)^{-\alpha^{\prime}}\mathcal{N}(\sigma)^{\alpha^{\prime}%
/2}\right)  \sigma^{-\alpha^{\prime}/2}\right\} \\
=\operatorname{Tr}\left\{  \rho\left[  \frac{d}{d\alpha}\sigma^{-\alpha
^{\prime}/2}\right]  \mathcal{N}^{\dag}\left(  \mathcal{N}(\sigma
)^{\alpha^{\prime}/2}\mathcal{N}(\rho)^{-\alpha^{\prime}}\mathcal{N}%
(\sigma)^{\alpha^{\prime}/2}\right)  \sigma^{-\alpha^{\prime}/2}\right\} \\
+\operatorname{Tr}\left\{  \rho\sigma^{-\alpha^{\prime}/2}\mathcal{N}^{\dag
}\left(  \left[  \frac{d}{d\alpha}\mathcal{N}(\sigma)^{\alpha^{\prime}%
/2}\right]  \mathcal{N}(\rho)^{-\alpha^{\prime}}\mathcal{N}(\sigma
)^{\alpha^{\prime}/2}\right)  \sigma^{-\alpha^{\prime}/2}\right\} \\
+\operatorname{Tr}\left\{  \rho\sigma^{-\alpha^{\prime}/2}\mathcal{N}^{\dag
}\left(  \mathcal{N}(\sigma)^{\alpha^{\prime}/2}\left[  \frac{d}{d\alpha
}\mathcal{N}(\rho)^{-\alpha^{\prime}}\right]  \mathcal{N}(\sigma
)^{\alpha^{\prime}/2}\right)  \sigma^{-\alpha^{\prime}/2}\right\} \\
+\operatorname{Tr}\left\{  \rho\sigma^{-\alpha^{\prime}/2}\mathcal{N}^{\dag
}\left(  \mathcal{N}(\sigma)^{\alpha^{\prime}/2}\mathcal{N}(\rho
)^{-\alpha^{\prime}}\left[  \frac{d}{d\alpha}\mathcal{N}(\sigma)^{\alpha
^{\prime}/2}\right]  \right)  \sigma^{-\alpha^{\prime}/2}\right\} \\
+\operatorname{Tr}\left\{  \rho\sigma^{-\alpha^{\prime}/2}\mathcal{N}^{\dag
}\left(  \mathcal{N}(\sigma)^{\alpha^{\prime}/2}\mathcal{N}(\rho
)^{-\alpha^{\prime}}\mathcal{N}(\sigma)^{\alpha^{\prime}/2}\right)  \left[
\frac{d}{d\alpha}\sigma^{-\alpha^{\prime}/2}\right]  \right\}
\end{multline}%
\begin{multline}
=\frac{1}{\alpha^{2}}\Bigg[-\frac{1}{2}\operatorname{Tr}\left\{  \rho\left[
\ln\sigma\right]  \sigma^{-\alpha^{\prime}/2}\mathcal{N}^{\dag}\left(
\mathcal{N}(\sigma)^{\alpha^{\prime}/2}\mathcal{N}(\rho)^{-\alpha^{\prime}%
}\mathcal{N}(\sigma)^{\alpha^{\prime}/2}\right)  \sigma^{-\alpha^{\prime}%
/2}\right\} \\
+\frac{1}{2}\operatorname{Tr}\left\{  \rho\sigma^{-\alpha^{\prime}%
/2}\mathcal{N}^{\dag}\left(  \left[  \ln\mathcal{N}(\sigma)\right]
\mathcal{N}(\sigma)^{\alpha^{\prime}/2}\mathcal{N}(\rho)^{-\alpha^{\prime}%
}\mathcal{N}(\sigma)^{\alpha^{\prime}/2}\right)  \sigma^{-\alpha^{\prime}%
/2}\right\} \\
-\operatorname{Tr}\left\{  \rho\sigma^{-\alpha^{\prime}/2}\mathcal{N}^{\dag
}\left(  \mathcal{N}(\sigma)^{\alpha^{\prime}/2}\left[  \ln\mathcal{N}%
(\rho)\right]  \mathcal{N}(\rho)^{-\alpha^{\prime}}\mathcal{N}(\sigma
)^{\alpha^{\prime}/2}\right)  \sigma^{-\alpha^{\prime}/2}\right\} \\
+\frac{1}{2}\operatorname{Tr}\left\{  \rho\sigma^{-\alpha^{\prime}%
/2}\mathcal{N}^{\dag}\left(  \mathcal{N}(\sigma)^{\alpha^{\prime}%
/2}\mathcal{N}(\rho)^{-\alpha^{\prime}}\mathcal{N}(\sigma)^{\alpha^{\prime}%
/2}\left[  \ln\mathcal{N}(\sigma)\right]  \right)  \sigma^{-\alpha^{\prime}%
/2}\right\} \\
-\frac{1}{2}\operatorname{Tr}\left\{  \rho\sigma^{-\alpha^{\prime}%
/2}\mathcal{N}^{\dag}\left(  \mathcal{N}(\sigma)^{\alpha^{\prime}%
/2}\mathcal{N}(\rho)^{-\alpha^{\prime}}\mathcal{N}(\sigma)^{\alpha^{\prime}%
/2}\right)  \sigma^{-\alpha^{\prime}/2}\left[  \ln\sigma\right]  \right\}
\Bigg].
\end{multline}
Taking the limit as $\alpha\rightarrow1$ gives%
\begin{multline}
\left.  \frac{d}{d\alpha}\widetilde{Q}_{\alpha,1}(\rho,\sigma,\mathcal{N}%
)\right\vert _{\alpha=1}=-\frac{1}{2}\operatorname{Tr}\left\{  \rho\left[
\ln\sigma\right]  \Pi_{\sigma}\mathcal{N}^{\dag}\left(  \Pi_{\mathcal{N}%
(\sigma)}\Pi_{\mathcal{N}(\rho)}\Pi_{\mathcal{N}(\sigma)}\right)  \Pi_{\sigma
}\right\} \\
+\frac{1}{2}\operatorname{Tr}\left\{  \rho\Pi_{\sigma}\mathcal{N}^{\dag
}\left(  \left[  \ln\mathcal{N}(\sigma)\right]  \Pi_{\mathcal{N}(\sigma)}%
\Pi_{\mathcal{N}(\rho)}\Pi_{\mathcal{N}(\sigma)}\right)  \Pi_{\sigma}\right\}
\\
-\operatorname{Tr}\left\{  \rho\Pi_{\sigma}\mathcal{N}^{\dag}\left(
\Pi_{\mathcal{N}(\sigma)}\left[  \ln\mathcal{N}(\rho)\right]  \Pi
_{\mathcal{N}(\rho)}\Pi_{\mathcal{N}(\sigma)}\right)  \Pi_{\sigma}\right\} \\
+\frac{1}{2}\operatorname{Tr}\left\{  \rho\Pi_{\sigma}\mathcal{N}^{\dag
}\left(  \Pi_{\mathcal{N}(\sigma)}\Pi_{\mathcal{N}(\rho)}\Pi_{\mathcal{N}%
(\sigma)}\left[  \ln\mathcal{N}(\sigma)\right]  \right)  \Pi_{\sigma}\right\}
\\
-\frac{1}{2}\operatorname{Tr}\left\{  \rho\Pi_{\sigma}\mathcal{N}^{\dag
}\left(  \Pi_{\mathcal{N}(\sigma)}\Pi_{\mathcal{N}(\rho)}\Pi_{\mathcal{N}%
(\sigma)}\right)  \left[  \ln\sigma\right]  \Pi_{\sigma}\right\} .
\end{multline}
We now simplify the first three terms and note that the last two are Hermitian
conjugates of the first two:%
\begin{align}
&  \!\!\!\!\!\! \operatorname{Tr}\left\{  \rho\left[  \ln\sigma\right]
\Pi_{\sigma}\mathcal{N}^{\dag}(\Pi_{\mathcal{N}(\sigma)}\Pi_{\mathcal{N}%
(\rho)}\Pi_{\mathcal{N}(\sigma)})\Pi_{\sigma}\right\} \nonumber\\
&  =\operatorname{Tr}\left\{  \rho\left[  \ln\sigma\right]  \mathcal{N}^{\dag
}(\Pi_{\mathcal{N}(\rho)})\right\} \\
&  =\operatorname{Tr}\left\{  \mathcal{N(}\rho\left[  \ln\sigma\right]
)\left(  \Pi_{\mathcal{N}(\rho)}\right)  \right\} \\
&  =\operatorname{Tr}\left\{  U\rho\left[  \ln\sigma\right]  U^{\dag}\left(
\Pi_{\mathcal{N}(\rho)}\otimes I_{E}\right)  \right\} \\
&  =\operatorname{Tr}\left\{  \Pi_{U\rho U^{\dag}}U\rho U^{\dag}U\left[
\ln\sigma\right]  U^{\dag}\left(  \Pi_{\mathcal{N}(\rho)}\otimes I_{E}\right)
\right\} \\
&  =\operatorname{Tr}\left\{  U\rho U^{\dag}U\left[  \ln\sigma\right]
U^{\dag}\right\} \\
&  =\operatorname{Tr}\left\{  \rho\left[  \ln\sigma\right]  \right\}  ,
\end{align}%
\begin{align}
&  \!\!\!\!\!\!\operatorname{Tr}\left\{  \rho\Pi_{\sigma}\mathcal{N}^{\dag
}(\left[  \ln\mathcal{N}(\sigma)\right]  \Pi_{\mathcal{N}(\sigma)}%
\Pi_{\mathcal{N}(\rho)}\Pi_{\mathcal{N}(\sigma)})\Pi_{\sigma}\right\}
\nonumber\\
&  =\operatorname{Tr}\left\{  \rho\mathcal{N}^{\dag}(\left[  \ln
\mathcal{N}(\sigma)\right]  \Pi_{\mathcal{N}(\rho)})\right\} \\
&  =\operatorname{Tr}\left\{  \mathcal{N}(\rho)\left[  \ln\mathcal{N}%
(\sigma)\right]  \Pi_{\mathcal{N}(\rho)}\right\} \\
&  =\operatorname{Tr}\left\{  \mathcal{N}(\rho)\left[  \ln\mathcal{N}%
(\sigma)\right]  \right\}  ,
\end{align}%
\begin{align}
&  \!\!\!\!\!\!\operatorname{Tr}\left\{  \rho\Pi_{\sigma}\mathcal{N}^{\dag
}(\Pi_{\mathcal{N}(\sigma)}\left[  \ln\mathcal{N}(\rho)\right]  \Pi
_{\mathcal{N}(\rho)}\Pi_{\mathcal{N}(\sigma)})\Pi_{\sigma}\right\} \nonumber\\
&  =\operatorname{Tr}\left\{  \rho\mathcal{N}^{\dag}(\left[  \ln
\mathcal{N}(\rho)\right]  \Pi_{\mathcal{N}(\rho)})\right\} \\
&  =\operatorname{Tr}\left\{  \mathcal{N}(\rho)\left(  \left[  \ln
\mathcal{N}(\rho)\right]  \Pi_{\mathcal{N}(\rho)}\right)  \right\} \\
&  =\operatorname{Tr}\left\{  \mathcal{N}(\rho)\left[  \ln\mathcal{N}%
(\rho)\right]  \right\}  .
\end{align}
This then implies that the following equality holds%
\begin{multline}
\left.  \frac{d}{d\alpha}\widetilde{Q}_{\alpha,1}(\rho,\sigma,\mathcal{N}%
)\right\vert _{\alpha=1}=-\operatorname{Tr}\left\{  \mathcal{N}\left(
\rho\left[  \ln\sigma\right]  \right)  \right\} \label{eq:last-deriv}\\
+\operatorname{Tr}\left\{  \mathcal{N}(\rho)\left[  \ln\mathcal{N}%
(\sigma)\right]  \right\}  -\operatorname{Tr}\left\{  \mathcal{N}(\rho)\left[
\ln\mathcal{N}(\rho)\right]  \right\}  .
\end{multline}
Putting together (\ref{eq:limit-a-1-exp}), (\ref{eq:chain-break-up}),
(\ref{eq:Q_beta_1_term}), and (\ref{eq:last-deriv}), we can then conclude the
statement of the theorem.
\end{proof}

For $\alpha=1/2$, observe that%
\begin{align}
\widetilde{\Delta}_{1/2}(\rho,\sigma,\mathcal{N})  &  =-\ln\left\Vert \left(
\left[  \mathcal{N}(\rho)\right]  ^{1/2}\left[  \mathcal{N}(\sigma)\right]
^{-1/2}\otimes I_{E}\right)  U\sigma^{1/2}\rho^{1/2}\right\Vert _{1}^{2}\\
&  =-\ln F( \rho,\mathcal{P}_{\sigma,\mathcal{N}}(\mathcal{N}(\rho))) .
\end{align}
where $F(\rho,\sigma)\equiv\left\Vert \sqrt{\rho}\sqrt{\sigma}\right\Vert
_{1}^{2}$ is the quantum fidelity. Thus, if $\widetilde{\Delta}_{\alpha}%
(\rho,\sigma,\mathcal{N})$ were monotone non-decreasing with respect to
$\alpha$, we could combine these observations to conclude that%
\begin{align}
D(\rho\Vert\sigma)-D(\mathcal{N}(\rho)\Vert\mathcal{N}(\sigma))  &  =\frac
{1}{\ln2}\widetilde{\Delta}_{1}(\rho,\sigma,\mathcal{N})\\
&  \overset{?}{\geq}\frac{1}{\ln2}\widetilde{\Delta}_{1/2}(\rho,\sigma
,\mathcal{N})\\
&  =-\log F\left(  \rho,\mathcal{P}_{\sigma,\mathcal{N}}(\mathcal{N}%
(\rho))\right)  .
\end{align}
If this were true, then we could conclude that
Theorem~\ref{thm-qeir:main-theorem}\ would be true with the recovery channel
taken to be the Petz recovery map. However, it is not known whether this is
true, and we will instead invoke the Stein--Hirschman theorem to conclude that
a convex combination of rotated Petz maps satisfies the bound stated in
Theorem~\ref{thm-qeir:main-theorem}.

\section{Proof of the Recoverability Theorem}

\label{sec:main-result}This section presents the proof of
Theorem~\ref{thm-qeir:main-theorem}. In fact, we prove a stronger statement,
which implies Theorem~\ref{thm-qeir:main-theorem} for a particular recovery
channel that we discuss below.

\begin{theorem}
\label{thm-qeir:rel-ent-better-main}Let $\rho$, $\sigma$, and $\mathcal{N}$ be
as given in Definition~\ref{def:rho-sig-N}. Then the following inequality
holds%
\begin{equation}
D(\rho\Vert\sigma)-D(\mathcal{N(}\rho)\Vert\mathcal{N(}\sigma))\geq
-\int_{-\infty}^{\infty}dt\ \beta_{0}(t)\ \log\left[  F\left(  \rho
,(\mathcal{R}_{\sigma,\mathcal{N}}^{t/2}\circ\mathcal{N})(\rho)\right)
\right]  , \label{eq:rel-ent-ineq}%
\end{equation}
where $\beta_{0}(t)=\frac{\pi}{2}\left[  \cosh(\pi t)+1\right]  ^{-1}$ is a
probability density function for $t\in\mathbb{R}$ and $\mathcal{R}%
_{\sigma,\mathcal{N}}^{t/2}$ is a rotated Petz recovery map from
Definition~\ref{def-qeir:rotated-petz}.
\end{theorem}

\begin{proof}
We can prove this result by employing Theorem~\ref{thm-qeir:op-hirschman}. We
first establish the inequality in (\ref{eq:rel-ent-ineq}). Let $U:\mathcal{H}%
\rightarrow\mathcal{H}^{\prime}\otimes\mathcal{H}_{E}$ be an isometric
extension of the channel $\mathcal{N}$. Pick%
\begin{equation}
G(z)\equiv\left(  \left[  \mathcal{N}(\rho)\right]  ^{z/2}\left[
\mathcal{N}(\sigma)\right]  ^{-z/2}\otimes I_{E}\right)  U\sigma^{z/2}%
\rho^{1/2},
\end{equation}
for $z\in\overline{S}$, $p_{0}=2$, $p_{1}=1$, and $\theta\in\left(
0,1\right)  $, which fixes $p_{\theta}=\frac{2}{1+\theta}$. The operator
valued-function $G(z)$ satisfies the conditions needed to apply
Theorem~\ref{thm-qeir:op-hirschman}. For the choices above, we find
\begin{align}
\left\Vert G(\theta)\right\Vert _{2/\left(  1+\theta\right)  }  &  =\left\Vert
\left(  \left[  \mathcal{N}(\rho)\right]  ^{\theta/2}\left[  \mathcal{N}%
(\sigma)\right]  ^{-\theta/2}\otimes I_{E}\right)  U\sigma^{\theta/2}%
\rho^{1/2}\right\Vert _{2/\left(  1+\theta\right)  },\\
\left\Vert G\left(  it\right)  \right\Vert _{2}  &  =\left\Vert \left(
\left[  \mathcal{N}(\rho)\right]  ^{it/2}\left[  \mathcal{N}(\sigma)\right]
^{-it/2}\otimes I_{E}\right)  U\sigma^{it}\rho^{1/2}\right\Vert _{2}%
\nonumber\\
&  \leq\left\Vert \rho^{1/2}\right\Vert _{2}\nonumber\\
&  =1,\\
\left\Vert G\left(  1+it\right)  \right\Vert _{1}  &  =\left\Vert \left(
\left[  \mathcal{N}(\rho)\right]  ^{(1+it)/2}\left[  \mathcal{N}%
(\sigma)\right]  ^{-\left(  1+it\right)  /2}\otimes I_{E}\right)
U\sigma^{\left(  1+it\right)  /2}\rho^{1/2}\right\Vert _{1}\nonumber\\
&  =\left\Vert \left(  \left[  \mathcal{N}(\rho)\right]  ^{\frac{it}{2}%
}\left[  \mathcal{N}(\rho)\right]  ^{\frac{1}{2}}\left[  \mathcal{N}%
(\sigma)\right]  ^{-\frac{it}{2}}\left[  \mathcal{N}(\sigma)\right]
^{-\frac{1}{2}}\otimes I_{E}\right)  U\sigma^{\frac{1}{2}}\sigma^{\frac{it}%
{2}}\rho^{\frac{1}{2}}\right\Vert _{1}\nonumber\\
&  =\left\Vert \left(  \left[  \mathcal{N}(\rho)\right]  ^{1/2}\left[
\mathcal{N}(\sigma)\right]  ^{-it/2}\left[  \mathcal{N}(\sigma)\right]
^{-1/2}\otimes I_{E}\right)  U\sigma^{1/2}\sigma^{it/2}\rho^{1/2}\right\Vert
_{1}\nonumber\\
&  =\sqrt{F}\left(  \rho,\left(  \mathcal{U}_{\sigma,-t/2}\circ\mathcal{P}%
_{\sigma,\mathcal{N}}\circ\mathcal{U}_{\mathcal{N}(\sigma),t/2}\right)
(\mathcal{N}(\rho))\right) \nonumber\\
&  =\sqrt{F}(\rho,(\mathcal{R}_{\sigma,\mathcal{N}}^{t/2}\circ\mathcal{N}%
)(\rho)). \label{eq:M1}%
\end{align}
Then we can apply Theorem~\ref{thm-qeir:op-hirschman} to conclude that%
\begin{multline}
\ln\left\Vert \left(  \left[  \mathcal{N(}\rho)\right]  ^{\theta/2}\left[
\mathcal{N(}\sigma)\right]  ^{-\theta/2}\otimes I_{E}\right)  U\sigma
^{\theta/2}\rho^{1/2}\right\Vert _{2/\left(  1+\theta\right)  }%
\label{eq:leads-to-improved-bnd}\\
\leq\int_{-\infty}^{\infty}dt\ \beta_{\theta}(t)\ln\left[  F\left(
\rho,(\mathcal{R}_{\sigma,\mathcal{N}}^{t/2}\circ\mathcal{N})(\rho)\right)
^{\theta/2}\right]  .
\end{multline}
This implies that%
\begin{multline}
-\frac{2}{\theta}\ln\left\Vert \left(  \left[  \mathcal{N(}\rho)\right]
^{\theta/2}\left[  \mathcal{N(}\sigma)\right]  ^{-\theta/2}\otimes
I_{E}\right)  U\sigma^{\theta/2}\rho^{1/2}\right\Vert _{2/\left(
1+\theta\right)  }\label{eq:critical-inequality}\\
\geq-\int_{-\infty}^{\infty}dt\ \beta_{\theta}(t)\ \ln\left[  F\left(
\rho,(\mathcal{R}_{\sigma,\mathcal{N}}^{t/2}\circ\mathcal{N})(\rho)\right)
\right]  .
\end{multline}
Letting $\theta=\left(  1-\alpha\right)  /\alpha$, we see that this is the
same as%
\begin{equation}
\widetilde{\Delta}_{\alpha}(\rho,\sigma,\mathcal{N})\geq-\int_{-\infty
}^{\infty}dt\ \beta_{\left(  1-\alpha\right)  /\alpha}(t)\ \ln\left[  F\left(
\rho,(\mathcal{R}_{\sigma,\mathcal{N}}^{t/2}\circ\mathcal{N})(\rho)\right)
\right]  . \label{eq:alpha-bound}%
\end{equation}
Since the inequality in \eqref{eq:critical-inequality} holds for all
$\theta\in\left(  0,1\right)  $ and thus \eqref{eq:alpha-bound} holds for all
$\alpha\in\left(  1/2,1\right)  $, we can take the limit as $\alpha\nearrow1$
and apply \eqref{eq:rel-ent-diff-a-1} and the dominated convergence theorem to
conclude that \eqref{eq:rel-ent-ineq} holds.
\end{proof}

With the theorem above in hand, Theorem~\ref{thm-qeir:main-theorem}\ follows
as a consequence by taking $\mathcal{R}_{\sigma,\mathcal{N}}$ to be the
following recovery channel:%
\begin{equation}
\mathcal{R}_{\sigma,\mathcal{N}}(Q)\equiv\int_{-\infty}^{\infty}dt\ \beta
_{0}(t)\ \mathcal{R}_{\sigma,\mathcal{N}}^{t/2}(Q)+\operatorname{Tr}%
\{(I-\Pi_{\mathcal{N(}\sigma)})Q\}\omega,
\label{eq-qeir:recovery-channel-beta}%
\end{equation}
where $Q\in\mathcal{L}(\mathcal{H}^{\prime})$ and $\omega\in\mathcal{D}%
(\mathcal{H})$. This is because%
\begin{multline}
-\int_{-\infty}^{\infty}dt\ \beta_{0}(t)\ \log\left[  F\left(  \rho
,(\mathcal{R}_{\sigma,\mathcal{N}}^{t/2}\circ\mathcal{N})(\rho)\right)
\right] \\
\geq-\log\left[  F\left(  \rho,\left(  \int_{-\infty}^{\infty}dt\ \beta
_{0}(t)\mathcal{R}_{\sigma,\mathcal{N}}^{t/2}\circ\mathcal{N}\right)
(\rho)\right)  \right] \\
\geq-\log\left[  F(\rho,(\mathcal{R}_{\sigma,\mathcal{N}}\circ\mathcal{N}%
)(\rho))\right]  ,
\end{multline}
where the first inequality is due to the concavity of both the logarithm and
the fidelity, and the second inequality follows from the assumption that
$\operatorname{supp}(\rho)\subseteq\operatorname{supp}(\sigma)$ and a
reasoning similar to that in the proof of
Proposition~\ref{prop-qie:rel-ent-s-s'} (one can also argue this last step
from the operator monotonicity of the square root function).

The extra term $\operatorname{Tr}\{(I-\Pi_{\mathcal{N(}\sigma)})Q\}\omega$ is
needed to ensure that $\mathcal{R}_{\sigma,\mathcal{N}}$ is trace-preserving
in addition to being completely positive. Trace preservation of $\mathcal{R}%
_{\sigma,\mathcal{N}}$ follows because%
\begin{align}
\operatorname{Tr}\{\mathcal{R}_{\sigma,\mathcal{N}}(Q)\}  &  =\int_{-\infty
}^{\infty}dt\ \beta_{0}(t)\ \operatorname{Tr}\{\mathcal{R}_{\sigma
,\mathcal{N}}^{t/2}(Q)\}+\operatorname{Tr}\{(I-\Pi_{\mathcal{N(}\sigma)})Q\}\\
&  =\int_{-\infty}^{\infty}dt\ \beta_{0}(t)\ \operatorname{Tr}\{\Pi
_{\mathcal{N(}\sigma)}Q\}+\operatorname{Tr}\{(I-\Pi_{\mathcal{N(}\sigma
)})Q\}\\
&  =\operatorname{Tr}\{\Pi_{\mathcal{N(}\sigma)}Q\}+\operatorname{Tr}%
\{(I-\Pi_{\mathcal{N(}\sigma)})Q\} =\operatorname{Tr}\{Q\},
\end{align}
where the second equality follows from Exercise~\ref{ex-qeir:rotated-Petz-TNI}%
. Observe that this recovery channel has the \textquotedblleft perfect
recovery of $\sigma$\textquotedblright\ property mentioned in
\eqref{eq-qeir:perfect-recovery-sigma-prop}, which follows from
Proposition~\ref{prop-qeir:perfect-recovery} and the particular form in \eqref{eq-qeir:recovery-channel-beta}.

As a corollary of Theorem~\ref{thm-qeir:rel-ent-better-main}, we obtain
equality conditions for the monotonicity of quantum relative entropy:

\begin{corollary}
[Equality Conditions]Let $\rho$, $\sigma$, and $\mathcal{N}$ be as given in
Definition~\ref{def:rho-sig-N}. Then%
\begin{equation}
D(\rho\Vert\sigma)=D(\mathcal{N}(\rho)\Vert\mathcal{N}(\sigma))
\end{equation}
if and only if all rotated Petz recovery maps perfectly recover $\rho$ from
$\mathcal{N}(\rho)$:%
\begin{equation}
\forall t\in\mathbb{R}:(\mathcal{R}_{\sigma,\mathcal{N}}^{t}\circ
\mathcal{N})(\rho)=\rho. \label{eq-qeir:perfect-recovery-rho}%
\end{equation}

\end{corollary}

\begin{proof}
Recall from Proposition~\ref{prop-qeir:perfect-recovery}\ that, independent of
the conditions in the statement of the corollary, we always have that
$(\mathcal{R}_{\sigma,\mathcal{N}}^{t}\circ\mathcal{N})(\sigma)=\sigma$ for
all $t\in\mathbb{R}$.

We start by proving the \textquotedblleft only if\textquotedblright\ part.
Suppose that $\forall t\in\mathbb{R}:(\mathcal{R}_{\sigma,\mathcal{N}}%
^{t}\circ\mathcal{N})(\rho)=\rho$. Then for a particular $t\in\mathbb{R}$, the
monotonicity of quantum relative entropy implies that%
\begin{align}
D(\rho\Vert\sigma)  &  \geq D(\mathcal{N(}\rho)\Vert\mathcal{N(}\sigma))\\
D(\mathcal{N(}\rho)\Vert\mathcal{N(}\sigma))  &  \geq D((\mathcal{R}%
_{\sigma,\mathcal{N}}^{t}\circ\mathcal{N})(\rho)\Vert(\mathcal{R}%
_{\sigma,\mathcal{N}}^{t}\circ\mathcal{N})(\sigma))\\
&  =D(\rho\Vert\sigma),
\end{align}
which in turn imply that $D(\rho\Vert\sigma)=D(\mathcal{N}(\rho)\Vert
\mathcal{N}(\sigma))$.

We now prove the \textquotedblleft if part\textquotedblright\ of the theorem.
Suppose that $D(\rho\Vert\sigma)=D(\mathcal{N}(\rho)\Vert\mathcal{N}(\sigma
))$. By Theorem~\ref{thm-qeir:rel-ent-better-main}, we can conclude that%
\begin{equation}
\int_{-\infty}^{\infty}dt\ \beta_{0}(t)\ \left[  -\log\left[  F\left(
\rho,(\mathcal{R}_{\sigma,\mathcal{N}}^{t/2}\circ\mathcal{N})(\rho)\right)
\right]  \right]  =0.
\end{equation}
Since $\beta_{0}(t)$ is a positive definite function for all $t\in\mathbb{R}$,
$-\log F\geq0$, the recovery maps $\mathcal{R}_{\sigma,\mathcal{N}}^{t/2}$ are
continuous in $t$ and so is the fidelity, we can conclude that%
\begin{equation}
-\log\left[  F\left(  \rho,(\mathcal{R}_{\sigma,\mathcal{N}}^{t/2}%
\circ\mathcal{N})(\rho)\right)  \right]  =0
\end{equation}
for all $t\in\mathbb{R}$, which is the same as $F(\rho,(\mathcal{R}%
_{\sigma,\mathcal{N}}^{t/2}\circ\mathcal{N})(\rho))=1$ for all $t\in
\mathbb{R}$. We can then conclude that \eqref{eq-qeir:perfect-recovery-rho}
holds because the fidelity between two states is equal to one if and only if
the states are the same.
\end{proof}

\section{Refinements of Quantum Entropy Inequalities}

\label{sec-qeir:corollaries}Theorem~\ref{thm-qeir:main-theorem} leads to a
strengthening of many quantum entropy inequalities, including strong
subadditivity of quantum entropy, concavity of conditional entropy, joint
convexity of relative entropy, non-negativity of quantum discord, and the
Holevo bound. We list these as corollaries and give brief proofs for them in
the following subsections.

\subsection{Strong Subadditivity}%

\index{strong subadditivity}%
Recall the conditional quantum mutual information of a tripartite state
$\rho_{ABC}$:%
\begin{equation}
I(A;B|C)_{\rho}\equiv H(AC)_{\rho}+H(BC)_{\rho}-H(C)_{\rho}-H(ABC)_{\rho}.
\end{equation}
Strong subadditivity is the statement that $I(A;B|C)_{\rho}\geq0$ for all
tripartite states $\rho_{ABC}$.

Corollary~\ref{thm-qeir:CMI} below gives an improvement of strong
subadditivity. It is a direct consequence of
Theorem~\ref{thm-qeir:main-theorem}\ after choosing%
\begin{equation}
\rho=\rho_{ABC},\ \ \ \ \sigma=\rho_{AC}\otimes I_{B},\ \ \ \ \mathcal{N}%
=\operatorname{Tr}_{A},
\end{equation}
so that%
\begin{equation}
\mathcal{N}(\rho)=\rho_{BC},\ \ \ \ \mathcal{N}(\sigma)=\rho_{C}\otimes
I_{B},\ \ \ \ \mathcal{N}^{\dag}(\cdot)=(\cdot)\otimes I_{A},
\end{equation}
and%
\begin{align}
D(\rho\Vert\sigma)-D(\mathcal{N}(\rho)\Vert\mathcal{N}(\sigma))  &
=D(\rho_{ABC}\Vert\rho_{AC}\otimes I_{B})-D(\rho_{BC}\Vert\rho_{C}\otimes
I_{B})\nonumber\\
&  =I(A;B|C)_{\rho},\\
\mathcal{P}_{\sigma,\mathcal{N}}(\cdot)  &  =\sigma^{1/2}\mathcal{N}^{\dag
}\left(  \left[  \mathcal{N}(\sigma)\right]  ^{-1/2}(\cdot)\left[
\mathcal{N}(\sigma)\right]  ^{-1/2}\right)  \sigma^{1/2}\nonumber\\
&  =\rho_{AC}^{1/2}\left[  \rho_{C}^{-1/2}(\cdot)\rho_{C}^{-1/2}\otimes
I_{A}\right]  \rho_{AC}^{1/2}. \label{eq:Petz-channel-CMI}%
\end{align}

\begin{corollary}
\label{thm-qeir:CMI}Let $\rho_{ABC}\in\mathcal{D}(\mathcal{H}_{A}%
\otimes\mathcal{H}_{B}\otimes\mathcal{H}_{C})$. Then the following inequality
holds%
\begin{equation}
I(A;B|C)_{\rho}\geq-\log\left[  F(\rho_{ABC},\mathcal{R}_{C\rightarrow
AC}(\rho_{BC}))\right]  , \label{eq:main-result}%
\end{equation}
where the recovery channel $\mathcal{R}_{C\rightarrow AC}=\int_{-\infty
}^{\infty}dt\ \beta_{0}(t)\ \mathcal{R}_{\rho_{AC},\operatorname{Tr}_{A}%
}^{t/2}$ perfectly recovers $\rho_{AC}$ from $\rho_{C}$.
\end{corollary}

\subsection{Concavity of Conditional Quantum Entropy}

Let $\mathcal{E}\equiv\left\{  p_{X}(x),\rho_{AB}^{x}\right\}  $ be an
ensemble of bipartite quantum states with expectation $\overline{\rho}%
_{AB}\equiv\sum_{x}p_{X}(x)\rho_{AB}^{x}$. Concavity of conditional entropy is
the statement that%
\begin{equation}
H(A|B)_{\overline{\rho}}\geq\sum_{x}p_{X}(x)H(A|B)_{\rho^{x}}.
\end{equation}

\index{von Neumann entropy!conditional!concavity}
Let $\omega_{XAB}$ denote the following classical-quantum state in which we
have encoded the ensemble$~\mathcal{E}$:%
\begin{equation}
\omega_{XAB}\equiv\sum_{x}p_{X}(x)|x\rangle\langle x|_{X}\otimes\rho_{AB}^{x}.
\end{equation}
We can rewrite%
\begin{align}
H(A|B)_{\overline{\rho}}-\sum_{x}p_{X}(x)H(A|B)_{\rho^{x}}  &  =H(A|B)_{\omega
}-H(A|BX)_{\omega}\\
&  =I(A;X|B)_{\omega}\\
&  =H(X|B)_{\omega}-H\left(  X|AB\right)  _{\omega}\\
&  =D(\omega_{XAB}\Vert I_{X}\otimes\omega_{AB})-D(\omega_{XB}\Vert
I_{X}\otimes\omega_{B}).
\end{align}
We can see the last line above as a relative entropy difference (as defined in
the right-hand side of \eqref{eq:rel-ent-diff-a-1}) by picking $\rho
=\omega_{XAB}$, $\sigma=I_{X}\otimes\omega_{AB}$, and $\mathcal{N}%
=\operatorname{Tr}_{A}$.

Applying Theorem~\ref{thm-qeir:main-theorem} and
Exercise~\ref{ex-dm:fid-cq-states}, we find the following improvement of
concavity of conditional entropy:

\begin{corollary}
Let an ensemble $\mathcal{E}$\ be as given above. Then the following
inequality holds%
\begin{equation}
H(A|B)_{\overline{\rho}}-\sum_{x}p_{X}(x)H(A|B)_{\rho^{x}}\geq-2\log\sum
_{x}p_{X}(x)\sqrt{F}(\rho_{AB}^{x},\mathcal{R}_{B\rightarrow AB}(\rho_{B}%
^{x})),
\end{equation}
where the recovery map $\mathcal{R}_{B\rightarrow AB}\equiv\int_{-\infty
}^{\infty}dt\ \beta_{0}(t)\ \mathcal{R}_{\overline{\rho}_{AB}%
,\operatorname{Tr}_{A}}^{t/2}$ perfectly recovers $\overline{\rho}_{AB}$ from
$\overline{\rho}_{B}$.
\end{corollary}

\subsection{Joint Convexity of Quantum Relative Entropy}

Let $\left\{  p_{X}(x),\rho_{x}\right\}  $ be an ensemble of density operators
and $\left\{  p_{X}(x),\sigma_{x}\right\}  $ be an ensemble of positive
semi-definite operators such that $\operatorname{supp}\left(  \rho_{x}\right)
\subseteq\operatorname{supp}(\sigma_{x})$ for all $x$ and with expectations
$\overline{\rho}\equiv\sum_{x}p_{X}(x)\rho_{x}$ and $\overline{\sigma}%
\equiv\sum_{x}p_{X}(x)\sigma_{x}$. Joint convexity of quantum relative
entropy
\index{quantum relative entropy!joint convexity}%
is the statement that distinguishability of these ensembles does not increase
under the loss of the classical label:%
\begin{equation}
\sum_{x}p_{X}(x)D(\rho_{x}\Vert\sigma_{x})\geq D(\overline{\rho}\Vert
\overline{\sigma}).
\end{equation}
By picking%
\begin{align}
\rho &  =\rho_{XB}\equiv\sum_{x}p_{X}(x)|x\rangle\langle x|_{X}\otimes\rho
_{x},\\
\sigma &  =\sigma_{XB}\equiv\sum_{x}p_{X}(x)|x\rangle\langle x|_{X}%
\otimes\sigma_{x},\\
\mathcal{N}  &  =\operatorname{Tr}_{X},
\end{align}
and applying Theorem~\ref{thm-qeir:main-theorem}, we arrive at the following
improvement of joint convexity of quantum relative entropy:

\begin{corollary}
\label{cor-qeir:JC-rel-ent}Let ensembles be as given above. Then the following
inequality holds%
\begin{equation}
\sum_{x}p_{X}(x)D(\rho_{x}\Vert\sigma_{x})-D(\overline{\rho}\Vert
\overline{\sigma})\geq-\log F(\rho_{XB},\mathcal{R}_{\sigma_{XB}%
,\operatorname{Tr}_{X}}(\overline{\rho})),
\end{equation}
where the recovery map $\mathcal{R}_{\sigma_{XB},\operatorname{Tr}_{X}}%
=\int_{-\infty}^{\infty}dt\ \beta_{0}(t)\ \mathcal{R}_{\sigma_{XB}%
,\operatorname{Tr}_{X}}^{t/2}$ perfectly recovers $\sigma_{XB}$ from
$\sigma_{B}$.
\end{corollary}

\subsection{Non-Negativity of Quantum Discord}

\label{sec-qeir:discord}Let $\rho_{AB}$ be a bipartite density operator and
let $\left\{  |\varphi_{x}\rangle\langle\varphi_{x}|_{A}\right\}  $ be a
rank-one quantum measurement on system $A$ (i.e., the vectors $|\varphi
_{x}\rangle_{A}$ satisfy $\sum_{x}|\varphi_{x}\rangle\langle\varphi_{x}%
|_{A}=I_{A}$). It suffices for us to consider rank-one measurements for our
\index{quantum discord}%
discussion here because every quantum measurement can be refined to have a
rank-one form, such that it delivers more classical information to the
experimentalist observing the apparatus. Then the (unoptimized) quantum
discord is defined to be the difference between the following mutual
informations:%
\begin{align}
I(A;B)_{\rho}  &  -I(X;B)_{\omega}, \mbox{   where }\\
\omega_{XB}  &  \equiv\mathcal{M}_{A\rightarrow X}(\rho_{AB}),\\
\mathcal{M}_{A\rightarrow X}(\cdot)  &  \equiv\sum_{x}\langle\varphi_{x}%
|_{A}(\cdot)|\varphi_{x}\rangle_{A}|x\rangle\langle x|_{X}.
\end{align}
The quantum channel $\mathcal{M}_{A\rightarrow X}$ is a measurement channel,
so that the state $\omega_{XB}$ is the classical-quantum state resulting from
the measurement. The set $\left\{  |x\rangle_{X}\right\}  $ is an orthonormal
basis so that $X$ is a classical system. The quantum discord is non-negative,
and by applying Theorem~\ref{thm-qeir:main-theorem},\ we find the following
improvement of this entropy inequality:

\begin{corollary}
\label{cor-qeir:discord}Let $\rho_{AB}$ and $\mathcal{M}_{A\rightarrow X}$ be
as given above. Then the following inequality holds%
\begin{equation}
I(A;B)_{\rho}-I(X;B)_{\omega}\geq-\log F(\rho_{AB},\mathcal{E}_{A}(\rho
_{AB})), \label{eq:discord-bound}%
\end{equation}
where%
\begin{equation}
\mathcal{E}_{A}\equiv\int_{-\infty}^{\infty}dt\ \beta_{0}(t)\ (\mathcal{U}%
_{\rho_{A},t}\circ\mathcal{P}_{\rho_{A},\mathcal{M}_{A\rightarrow X}}%
\circ\mathcal{M}_{A\rightarrow X})
\end{equation}
is an
\index{entanglement-breaking channel}%
entanglement-breaking map, $\mathcal{P}_{\rho_{A},\mathcal{M}_{A\rightarrow
X}}\circ\mathcal{M}_{A\rightarrow X}$ is an entanglement-breaking map:%
\begin{equation}
(\mathcal{P}_{\rho_{A},\mathcal{M}_{A\rightarrow X}}\circ\mathcal{M}%
_{A\rightarrow X})(\cdot)=\sum_{x}\langle\varphi_{x}|_{A}(\cdot)|\varphi
_{x}\rangle_{A}\frac{\rho_{A}^{1/2}|\varphi_{x}\rangle\langle\varphi_{x}%
|_{A}\rho_{A}^{1/2}}{\langle\varphi_{x}|_{A}\rho_{A}|\varphi_{x}\rangle_{A}},
\label{eq:EB-channel}%
\end{equation}
and the partial isometric map $\mathcal{U}_{\rho_{A},t}$ is defined from
\eqref{eq-qeir:unitaries}. The recovery map $\int_{-\infty}^{\infty}%
dt\ \beta_{0}(t)\ (\mathcal{U}_{\rho_{A},t}\circ\mathcal{P}_{\rho
_{A},\mathcal{M}_{A\rightarrow X}})$ perfectly recovers $\rho_{A}$ from
$\mathcal{M}_{A\rightarrow X}(\rho_{A})$.
\end{corollary}

\begin{proof}
We start with the rewriting%
\begin{equation}
I(A;B)_{\rho}-I(X;B)_{\omega}=D\left(  \rho_{AB}\Vert\rho_{A}\otimes
I_{B}\right)  -D\left(  \omega_{XB}\Vert\omega_{X}\otimes I_{B}\right)  ,
\end{equation}
and follow by picking $\rho=\rho_{AB}$, $\sigma=\rho_{A}\otimes I_{B}$, and
$\mathcal{N}=\mathcal{M}_{A\rightarrow X}$, and applying
Theorem~\ref{thm-qeir:main-theorem}. This then shows the corollary with a
recovery map of the form $\int_{-\infty}^{\infty}dt\ \beta_{0}(t)\ \mathcal{R}%
_{\rho_{A},\mathcal{M}_{A\rightarrow X}}^{t/2}$.

The channel $\mathcal{R}_{\rho_{A},\mathcal{M}_{A\rightarrow X}}^{t}%
\circ\mathcal{M}_{A\rightarrow X}$ is
\index{entanglement-breaking channel}%
entanglement-breaking because it consists of a measurement channel
$\mathcal{M}_{A\rightarrow X}$ followed by a preparation. We now work out the
form for the recovery map in \eqref{eq:discord-bound}. Consider that
$\mathcal{M}_{A\rightarrow X}\left(  \rho_{A}\right)  =\sum_{x}\langle
\varphi_{x}|_{A}\rho_{A}|\varphi_{x}\rangle_{A}|x\rangle\langle x|_{X}$, so
that%
\begin{multline}
\mathcal{U}_{\mathcal{M}_{A\rightarrow X}(\rho_{A}),-t}(\cdot)=\\
\left[  \sum_{x}\left[  \langle\varphi_{x}|_{A}\rho_{A}|\varphi_{x}\rangle
_{A}\right]  ^{-it}|x\rangle\langle x|_{X}\right]  (\cdot)\left[
\sum_{x^{\prime}}\left[  \left\langle \varphi_{x^{\prime}}\right\vert _{A}%
\rho_{A}|\varphi_{x^{\prime}}\rangle_{A}\right]  ^{it}\left\vert x^{\prime
}\right\rangle \langle x^{\prime}|_{X}\right]  .
\end{multline}
Thus, when composing $\mathcal{M}_{A\rightarrow X}$ with $\mathcal{U}%
_{\mathcal{M}_{A\rightarrow X}(\rho_{A}),-t}$, the phases cancel out to give
the following relation: $\mathcal{U}_{\mathcal{M}_{A\rightarrow X}(\rho
_{A}),-t}\left(  \mathcal{M}_{A\rightarrow X}(\cdot)\right)  =\mathcal{M}%
_{A\rightarrow X}(\cdot)$. One can then work out that%
\begin{align}
&  (\mathcal{P}_{\rho_{A},\mathcal{M}_{A\rightarrow X}}\circ\mathcal{M}%
_{A\rightarrow X})(\cdot)\nonumber\\
&  =\rho_{A}^{1/2}\mathcal{M}^{\dag}\left(  \left[  \mathcal{M}_{A\rightarrow
X}\left(  \rho_{A}\right)  \right]  ^{-1/2}\mathcal{M}_{A\rightarrow X}%
(\cdot)\left[  \mathcal{M}_{A\rightarrow X}(\rho_{A})\right]  ^{-1/2}\right)
\rho_{A}^{1/2}\\
&  =\sum_{x}\langle\varphi_{x}|_{A}(\cdot)|\varphi_{x}\rangle_{A}\frac
{\rho_{A}^{1/2}|\varphi_{x}\rangle\langle\varphi_{x}|_{A}\rho_{A}^{1/2}%
}{\langle\varphi_{x}|_{A}\rho_{A}|\varphi_{x}\rangle_{A}},
\end{align}
concluding the proof.
\end{proof}

\subsection{Holevo Bound}

The Holevo bound\ is a special case of the non-negativity
\index{Holevo bound}%
of quantum discord in which $\rho_{AB}$ is a quantum-classical state, which we
write explicitly as%
\begin{equation}
\rho_{AB}=\sum_{y}p_{Y}(y)\rho_{A}^{y}\otimes|y\rangle\langle y|_{Y},
\label{eq:cq-holevo}%
\end{equation}
where each $\rho_{A}^{y}$ is a density operator, so that $\rho_{A}=\sum
_{y}p_{Y}(y)\rho_{A}^{y}$. The Holevo bound states that the mutual information
of the state $\rho_{AB}$ in \eqref{eq:cq-holevo} is never smaller than the
mutual information after system $A$ is measured. By applying
Corollary~\ref{cor-qeir:discord} and \eqref{eq:fid-flags}, we find the
following improvement:

\begin{corollary}
[Holevo Bound]\label{cor-qeir:holevo} Let $\rho_{AB}$ be as in
\eqref{eq:cq-holevo},\ and let $\mathcal{M}_{A\rightarrow X}$ and $\omega
_{XB}$ be as in Section~\ref{sec-qeir:discord}, respectively. Then the
following inequality holds%
\begin{equation}
I(A;B)_{\rho}-I(X;B)_{\omega}\geq-2\log\sum_{y}p_{Y}(y)\sqrt{F}\left(
\rho_{A}^{y},\mathcal{E}_{A}\left(  \rho_{A}^{y}\right)  \right)  ,
\end{equation}
where $\mathcal{E}_{A}$ is an
\index{entanglement-breaking channel}%
entanglement-breaking map of the form in \eqref{eq:EB-channel} and the partial
isometric map $\mathcal{U}_{\rho_{A},t}$ is defined from \eqref{eq-qeir:unitaries}.
\end{corollary}

\section{History and Further Reading}

Section~\ref{sec-qie:history-QE}\ reviews the history of many quantum entropy
inequalities in quantum information. Here, we detail the history of the
refinements. \cite{Petz1986,Petz1988} considered the equality conditions for
monotonicity of quantum relative entropy, defined what is now known as the
Petz recovery map (there called \textquotedblleft transpose
channel\textquotedblright), and gave many such equality conditions, one of
which is that $D(\rho\Vert\sigma)=D(\mathcal{N}(\rho)\Vert\mathcal{N}%
(\sigma))$ if and only if the Petz recovery map perfectly recovers $\rho$ from
$\mathcal{N}(\rho)$: $(\mathcal{P}_{\sigma,\mathcal{N}}\circ\mathcal{N}%
)(\rho)=\rho$. Later, \cite{HJPW04} invoked Petz's result to elucidate the
structure of tripartite states that saturate the strong subadditivity of
quantum entropy with equality. \cite{Mosonyi2004}\ then invoked Petz's result
to establish the structure of triples $(\rho,\sigma,\mathcal{N})$ for which
the monotonicity of quantum relative entropy is saturated with equality (see
also \citep{M05}). The transpose channel was independently discovered many
years after Petz's work by \cite{BK02} in the context of approximate quantum
error correction.

The topic of the near saturation of quantum entropy inequalities is a more
recent development. \cite{BCY11} established a lower bound on conditional
mutual information related to how entangled the two unconditioned systems are
with respect to the 1-LOCC\ norm. Interest then turned to obtaining lower
bounds in terms of the trace norm. Much of the very recent work was inspired
by the posting of \cite{Winterconj} and a presentation of \cite{K13conj}. The
main conjecture of \cite{Winterconj} (hitherto unproven) is that
\begin{equation}
D(\rho\Vert\sigma)-D(\mathcal{N(}\rho)\Vert\mathcal{N(}\sigma))\geq
D(\rho\Vert(\mathcal{R}_{\sigma,\mathcal{N}}\circ\mathcal{N})(\rho)),
\label{eq-qeir:winter-conj}%
\end{equation}
for some recovery channel $\mathcal{R}_{\sigma,\mathcal{N}}$, depending only
on $\sigma$ and $\mathcal{N}$ and such that $(\mathcal{R}_{\sigma,\mathcal{N}%
}\circ\mathcal{N})(\sigma)=\sigma$. \cite{CL14} then established an
interesting lower bound on the monotonicity of quantum relative entropy with
respect to partial trace, which directly leads to a related lower bound on
conditional quantum mutual information, as pointed out by \cite{Z14} by making
use of the well known relation $I(A;B|C)_{\rho}=D(\rho_{ABC}\Vert I_{B}%
\otimes\rho_{AC})-D(\rho_{BC}\Vert I_{B}\otimes\rho_{C})$. Motivated by these
developments, \cite{BSW14} defined a R\'{e}nyi generalization of the
conditional mutual information. This notion and known properties of R\'{e}nyi
entropies led them to conjecture that $I(A;B|C)_{\rho}\geq-\log F(\rho
_{ABC},\mathcal{P}_{\rho_{AC},\operatorname{Tr}_{A}}(\rho_{BC}))$, which
hitherto remains unproven. Shortly thereafter, \cite{SBW14} put forward the
notion of a R\'{e}nyi generalization of a relative entropy difference, which
is one of the main tools used in the proof of
Theorem~\ref{thm-qeir:main-theorem}.

\cite{FR14} established the following:%
\begin{equation}
I(A;B|C)_{\rho}\geq-\log F(\rho_{ABC},\mathcal{R}_{C\rightarrow AC}(\rho
_{BC})), \label{eq-qeir:FR-bound}%
\end{equation}
where $\mathcal{R}_{C\rightarrow AC}$ is a recovery channel consisting of some
unitary acting on $C$, followed by the Petz recovery map $\mathcal{P}%
_{\rho_{AC},\operatorname{Tr}_{A}}$, and then some unitary acting on $AC$.
Since their argument made use of the probabilistic method, they could not give
further information about the aforementioned unitaries. Concurrently,
\cite{SW14} defined the \textquotedblleft fidelity of
recovery\textquotedblright%
\begin{equation}
F(A;B|C)_{\rho}\equiv\sup_{\mathcal{R}_{C\rightarrow AC}}F(\rho_{ABC}%
,\mathcal{R}_{C\rightarrow AC}(\rho_{BC}))
\end{equation}
as an information measure analogous to conditional mutual information and they
proved many of its properties. \cite{LW14}, based on earlier arguments in
\cite{Winterconj}, used the result of \cite{FR14} to establish a lower bound
on an entanglement measure known as squashed entanglement and further
elaborated on the conjecture in \eqref{eq-qeir:winter-conj}. \cite{BHOS14}
proved the bound $I(A;B|C)_{\rho}\geq-\log F(A;B|C)_{\rho}$ by invoking
quantum state redistribution, but their method gave less information about the
structure of the recovery channel. \cite{W14} showed how to apply the bound in
\eqref{eq-qeir:FR-bound} to give lower bounds on multipartite information
measures, which had an interpretation in terms of local recoverability.
\cite{BLW14} used the techniques of \cite{FR14} to establish a non-trivial
lower bound on a relative entropy difference. \cite{DW15} proved that
$\widetilde{\Delta}_{\alpha}(\rho,\sigma,\mathcal{N})\geq0$ for all $\alpha
\in(1/2,1)\cup(1,\infty)$ in addition to other related
statements.\ \cite{BT15} proved that the fidelity of recovery is
multiplicative with respect to tensor-product states, which then simplified
the proof of the bound $I(A;B|C)_{\rho}\geq-\log F(A;B|C)_{\rho}$.
\cite{SOR15} showed that the recovery map in
\eqref{eq-qeir:FR-bound}\ possesses a universal property in the sense that
$\mathcal{R}_{C\rightarrow AC}$ could be taken to depend only on the marginal
state $\rho_{AC}$. Their argument also led to the conclusion that the
unitaries could be taken to commute with $\rho_{C}$ and $\rho_{AC}$.

\cite{W15} invoked the Hadamard three-lines theorem and the notion of a
R\'{e}nyi generalization of a relative entropy difference to establish the
following bound:%
\begin{equation}
D(\rho\Vert\sigma)-D(\mathcal{N(}\rho)\Vert\mathcal{N(}\sigma))\geq
-\log\left[  \sup_{t\in\mathbb{R}}F(\rho,(\mathcal{R}_{\sigma,\mathcal{N}}%
^{t}\circ\mathcal{N})(\rho))\right]  ,
\end{equation}
where $(\mathcal{R}_{\sigma,\mathcal{N}}^{t}\circ\mathcal{N})(\sigma)=\sigma$.
Notice that the recovery map above is not universal, because the optimal $t$
could depend on $\rho$. \cite{W15} also established an upper bound on
$D(\rho\Vert\sigma)-D(\mathcal{N(}\rho)\Vert\mathcal{N(}\sigma))$, which in
some cases has an interpretation in terms of recoverability. \cite{DW15a}
subsequently defined \textquotedblleft swiveled R\'{e}nyi
entropies\textquotedblright\ in an attempt to address some of the open
questions raised in \citep{BSW14,SBW14}. An upshot of this work was to
establish further lower and upper bounds on $D(\rho\Vert\sigma)-D(\mathcal{N(}%
\rho)\Vert\mathcal{N(}\sigma))$.\ \cite{Sutter15} then showed the following
lower bound\ by a different method of proof, known as \textquotedblleft the
pinched Petz recovery\textquotedblright\ approach:%
\begin{equation}
D(\rho\Vert\sigma)-D(\mathcal{N(}\rho)\Vert\mathcal{N(}\sigma))\geq D_{M}%
(\rho\Vert(\mathcal{R}_{\rho,\sigma,\mathcal{N}}\circ\mathcal{N})(\rho)),
\end{equation}
where $(\mathcal{R}_{\rho,\sigma,\mathcal{N}}\circ\mathcal{N})(\sigma)=\sigma$
and $\mathcal{R}_{\rho,\sigma,\mathcal{N}}$ is some convex combination of
rotated Petz maps and $D_{M}$ is the \textquotedblleft measured relative
entropy\textquotedblright, which satisfies $D_{M}\geq-\log F$. \cite{Junge15}
invoked Hirschman's improvement \citep{H52}\ of the Hadamard three-lines
theorem and the method of proof from \citep{W15} to establish
Theorem~\ref{thm-qeir:main-theorem}, with the consequence of having an
explicit recovery channel $\mathcal{R}_{\sigma,\mathcal{N}}$\ which depends
only on $\sigma$ and $\mathcal{N}$\ and satisfies $(\mathcal{R}_{\sigma
,\mathcal{N}}\circ\mathcal{N})(\sigma)=\sigma$. This approach also directly
recovers the equality conditions from\ \cite{Petz1986,Petz1988}.

More information about the theory of complex interpolation is available in
\cite{BL76,RS75}. \cite{G08}\ provides an excellent review of Hirschman's
theorem. \cite{S56} proved the Stein--Hirschman interpolation theorem, which
is essential to the proof of Theorem~\ref{thm-qeir:main-theorem}. \cite{Z00}
and \cite{zurek} defined the quantum discord. \cite{J12}\ proved
Proposition~\ref{prop-qeir:perfect-recovery} for $t=0$.

\chapter{The Information of Quantum Channels}

\label{chap:additivity}We introduced several classical and quantum entropic
quantities in Chapters~\ref{chap:info-entropy} and \ref{chap:q-info-entropy}:
entropy, conditional entropy, joint entropy, mutual information, relative
entropy, and conditional mutual information. Each of these entropic quantities
is static, in the sense that each is evaluated with respect to random
variables or quantum systems that certain parties possess.

In this chapter, we introduce several dynamic entropic quantities for
channels, whether they be classical or quantum. We derive these measures by
exploiting the static measures from the two previous chapters. They come about
by sending one share of a state through a channel, computing a static measure
with respect to the input--output state, and then maximizing the static
measure with respect to all possible states that we can transmit through the
channel. This process then gives rise to a dynamic measure that quantifies the
ability of a channel to preserve correlations. For example, we could send one
share of a pure entangled state $\vert\phi\rangle_{AA^{\prime}}$ through a
quantum channel $\mathcal{N}_{A^{\prime}\rightarrow B}$---this transmission
gives rise to some bipartite state $\mathcal{N}_{A^{\prime}\rightarrow B}%
(\phi_{AA^{\prime}})$. We would then evaluate the mutual information of the
resulting state and maximize the mutual information with respect to all such
pure input states:%
\begin{equation}
\max_{\phi_{AA^{\prime}}}I( A;B) _{\omega},
\end{equation}
where $\omega_{AB}\equiv\mathcal{N}_{A^{\prime}\rightarrow B}(\phi
_{AA^{\prime}})$. The above quantity is a dynamic information measure of the
channel's ability to preserve
correlations---Section~\ref{sec-add:mut-info-channel}\ introduces this
quantity as the mutual information of the channel$~\mathcal{N}$.

For now, we simply think of the quantities in this chapter as measures of a
channel's ability to preserve correlations. Later, we show that these
quantities have explicit operational interpretations in terms of a channel's
ability to perform a certain task, such as the transmission of classical or
quantum information.\footnote{Giving operational interpretations to
information measures is in fact one of the main goals of this book!} Such an
operational interpretation gives meaning to an entropic measure---otherwise,
it is difficult to understand a measure in an information-theoretic sense
without having a specific operational task to which it corresponds.

Recall that the entropy obeys an additivity property for any two independent
random variables $X_{1}$ and $X_{2}$:%
\begin{equation}
H(X_{1},X_{2})=H(X_{1})+H(X_{2}). \label{eq-add:ent-add}%
\end{equation}
The above additivity property extends to a large sequence $X_{1},\ldots,X_{n}$
of independent and identically distributed random variables. That is, applying
\eqref{eq-add:ent-add} inductively shows that $nH(X)$ is equal to the entropy
of the sequence:%
\begin{equation}
H(X_{1},\ldots,X_{n})=\sum_{i=1}^{n}H(X_{i})=\sum_{i=1}^{n}H(X)=nH(X),
\end{equation}
where random variable $X$ has the same distribution as each of $X_{1}%
,\ldots,X_{n}$. Similarly, quantum entropy is additive for any two quantum
systems in a product state $\rho\otimes\sigma$:%
\begin{equation}
H(\rho\otimes\sigma)=H(\rho)+H(\sigma), \label{eq-add:von-ent-add}%
\end{equation}
and applying \eqref{eq-add:von-ent-add}\ inductively to a sequence of quantum
states gives the following similar simple formula: $H(\rho^{\otimes
n})=nH(\rho)$. Additivity is a desirable property and a natural expectation
that we have for any measure of information evaluated on independent systems.

In analogy with the static measures, we would like additivity to hold for the
dynamic information measures. Without additivity holding, we cannot really
make sense of a given measure because we would have to evaluate the measure on
a potentially infinite number of independent channel uses. This evaluation on
so many channel uses is generally an impossible optimization problem.
Additionally, the requirement to maximize over so many uses of the channel
does not identify a given measure as a unique measure of a channel's ability
to perform a certain task. There could generally be other measures that are
equal to the original one when we take the limit of many channel uses. Thus, a
measure does not have much substantive meaning if additivity does not hold.

We devote this chapter to the discussion of several dynamic measures.
Additivity holds in the general case for only three of the dynamic measures
presented here:\ the mutual information of a classical channel, the private
information of a classical wiretap channel, and the mutual information of a
quantum channel. For all other measures, there are known counterexamples of
channels for which additivity does not hold. In this chapter, we do not
discuss the counterexamples, but instead focus only on classes of channels for
which additivity does hold, in an effort to understand it in a technical
sense. The proof techniques for additivity exploit many of the ideas
introduced in the two previous chapters and give us a chance to practice with
what we have learned there on one of the most important problems in quantum
Shannon theory.

\section{Mutual Information of a Classical Channel}

\label{sec-cie:mut-inf-channel}Suppose that we would
\index{mutual information!of a classical channel}
like to determine how much information we can transmit through a classical
channel $\mathcal{N}$. Recall our simplified model of a classical channel
$\mathcal{N}$\ from Chapter~\ref{chap:classical-shannon-theory},\ in which
some conditional probability density $p_{Y|X}( y|x) $ models the effects of
noise. That is, we obtain some random variable $Y$ if we input a random
variable $X$ to the channel.

What is a good measure of the information throughput of this channel? The
mutual information is perhaps the best starting point. Suppose that random
variables $X$ and $Y$ are Bernoulli. If the classical channel is noiseless and
$X$ is completely random, the input and output random variables $X$ and $Y$
are perfectly correlated, the mutual information $I( X;Y) $\ is equal to one
bit, and the sender can transmit one bit per transmission as we would expect.
If the classical channel is completely noisy (in the sense that it prepares an
output that has a constant probability distribution irrespective of the
input), the input and output random variables are independent and the mutual
information is equal to zero bits. This observation matches our intuition that
the sender should not be able to transmit any information through this
completely noisy channel.

In the above model for a classical channel, the conditional probability
density $p_{Y|X}( y|x) $\ remains fixed, but we can \textquotedblleft play
around\textquotedblright\ with the input random variable $X$ by modifying its
probability density $p_{X}( x) $.\footnote{Recall the idea from
Section~\ref{sec:random-code-idea} where Alice and Bob actually choose a code
for the channel randomly according to the density $p_{X}( x) $.} Thus, we
still \textquotedblleft have room\textquotedblright\ for optimizing the mutual
information of the channel $\mathcal{N}$ by modifying this input density. This
gives us the following definition:

\begin{definition}
[Mutual Information of a Classical Channel]The mutual information
$I(\mathcal{N})$\ of a classical channel $\mathcal{N} \equiv p_{Y|X}$ is
defined as follows:%
\begin{equation}
I(\mathcal{N})\equiv\max_{p_{X}(x)}I(X;Y).
\label{eq-ie:mutual-info-classical-channel}%
\end{equation}

\end{definition}

\subsection{Regularized Mutual Information of a Classical Channel}

We now consider whether exploiting multiple uses of a classical channel
$\mathcal{N}$ and allowing for correlations between its inputs can increase
its mutual information. That is, suppose that we have two independent uses of
a classical channel $\mathcal{N}$ available. Let $X_{1}$ and $X_{2}$ denote
the random variables input to the respective first and second copies of the
channel, and let $Y_{1}$ and $Y_{2}$ denote the output random variables. Each
of the two uses of the channel is equivalent to the mapping $p_{Y|X}(y|x)$ so
that the channel uses are independent and identically distributed. Let
$\mathcal{N}\otimes\mathcal{N}$ denote the \textit{joint channel} that
corresponds to the mapping%
\begin{equation}
p_{Y_{1},Y_{2}|X_{1},X_{2}}(y_{1},y_{2}|x_{1},x_{2})=p_{Y_{1}|X_{1}}%
(y_{1}|x_{1})p_{Y_{2}|X_{2}}(y_{2}|x_{2}),
\end{equation}
where both $p_{Y_{1}|X_{1}}(y_{1}|x_{1})$ and $p_{Y_{2}|X_{2}}(y_{2}|x_{2})$
are equal to the mapping $p_{Y|X}(y|x)$. The mutual information of a classical
joint channel is as follows:%
\begin{equation}
I(\mathcal{N}\otimes\mathcal{N})\equiv\max_{p_{X_{1},X_{2}}(x_{1},x_{2}%
)}I(X_{1},X_{2};Y_{1},Y_{2}).
\end{equation}

\begin{figure}
[ptb]
\begin{center}
\includegraphics[
width=2.1759in
]%
{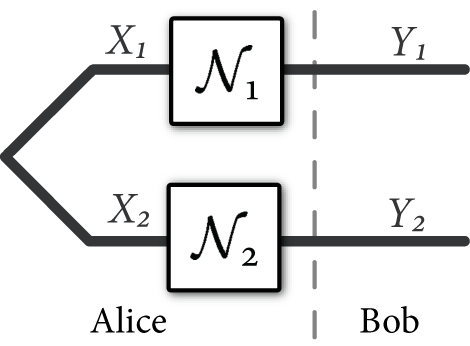}%
\caption{This figure displays the scenario for determining whether the mutual
information of two classical channels $\mathcal{N}_{1}$ and $\mathcal{N}_{2}%
$\ is additive. The question of additivity is equivalent to the possibility of
classical correlations being able to enhance the mutual information of two
classical channels. The result proved in
Theorem~\ref{thm-ie:additivity-classical}\ is that the mutual information is
additive for any two classical channels, so that classical correlations cannot
enhance it.}%
\label{fig-add:class-add}%
\end{center}
\end{figure}

We might think that we could increase the mutual information of this classical
channel by allowing for correlations between the inputs to the channels
through a correlated distribution $p_{X_{1},X_{2}}(x_{1},x_{2})$. That is,
there could be some superadditive effect if the mutual information of the
classical joint channel $\mathcal{N}\otimes\mathcal{N}$ is strictly greater
than two individual mutual informations:%
\begin{equation}
I(\mathcal{N}\otimes\mathcal{N})\overset{?}{>}2I(\mathcal{N}).
\label{eq-ie:superadditive-classical}%
\end{equation}
Figure~\ref{fig-add:class-add}\ displays the scenario corresponding to the
above question.%

In fact, we can take the above argument to its extreme, by defining the
regularized mutual information $I_{\operatorname{reg}}(\mathcal{N})$ of a
classical channel as follows:%
\begin{equation}
I_{\operatorname{reg}}(\mathcal{N})\equiv\lim_{n\rightarrow\infty}\frac{1}%
{n}I(\mathcal{N}^{\otimes n}). \label{eq-ie:reg-mutual-classical}%
\end{equation}
In the above definition, the quantity $I(\mathcal{N}^{\otimes n})$ is defined
as follows:%
\begin{equation}
I(\mathcal{N}^{\otimes n})\equiv\max_{p_{X^{n}}(x^{n})}I(X^{n};Y^{n}),
\end{equation}
and $\mathcal{N}^{\otimes n}$ denotes $n$ channels corresponding to the
following conditional probability distribution:%
\begin{equation}
p_{Y^{n}|X^{n}}(y^{n}|x^{n})=\prod\limits_{i=1}^{n}p_{Y_{i}|X_{i}}(y_{i}%
|x_{i}),
\end{equation}
where $X^{n}\equiv X_{1},X_{2},\ldots,X_{n}$, $x^{n}\equiv x_{1},x_{2}%
,\ldots,x_{n}$, and $Y^{n}\equiv Y_{1},Y_{2},\ldots,Y_{n}$. The potential
superadditive effect would have the following form after bootstrapping the
inequality in \eqref{eq-ie:superadditive-classical}\ to the regularization:%
\begin{equation}
I_{\operatorname{reg}}(\mathcal{N})\overset{?}{>}I(\mathcal{N}).
\end{equation}

\begin{exercise}
Determine a finite upper bound on $I_{\operatorname{reg}}(\mathcal{N})$. Thus,
this quantity is always finite.
\end{exercise}

The next section shows that the above strict inequalities do not hold for a
classical channel, implying that no such superadditive effect occurs for its
mutual information. In fact, the mutual information of a classical channel
obeys an additivity property that represents one of the cornerstones of our
understanding of classical information theory. This additivity property is the
statement that%
\begin{equation}
I(\mathcal{N}\otimes\mathcal{M})=I(\mathcal{N}) +I(\mathcal{M})
\end{equation}
for any two classical channels $\mathcal{N}$ and $\mathcal{M}$, and thus%
\begin{equation}
I_{\operatorname{reg}}(\mathcal{N})=I(\mathcal{N}),
\end{equation}
by an inductive argument. Thus, classical correlations between inputs do not
increase the mutual information of a classical channel.

We are stressing the importance of additivity in classical information theory
because recent research has demonstrated that superadditive effects can occur
in quantum Shannon theory (see Section~\ref{sec-cc:superadditivity}, for
example). These quantum results imply that our understanding of quantum
Shannon theory is not complete, but they also demonstrate the fascinating
possibility that quantum correlations can increase the information throughput
of a quantum channel.

\subsection{Additivity}

The mutual information of classical channels satisfies the important and
\index{mutual information!of a classical channel!additivity}
natural property of additivity. We prove the strongest form of additivity that
occurs for the mutual information of two different classical channels. Let
$\mathcal{N}_{1}$ and $\mathcal{N}_{2}$ denote two \textit{different
}classical channels corresponding to the respective mappings $p_{Y_{1}|X_{1}%
}(y_{1}|x_{1})$ and $p_{Y_{2}|X_{2}}(y_{2}|x_{2})$, and let $\mathcal{N}%
_{1}\otimes\mathcal{N}_{2}$ denote the joint channel that corresponds to the
mapping%
\begin{equation}
p_{Y_{1},Y_{2}|X_{1},X_{2}}(y_{1},y_{2}|x_{1},x_{2})=p_{Y_{1}|X_{1}}%
(y_{1}|x_{1})p_{Y_{2}|X_{2}}(y_{2}|x_{2}).
\end{equation}
The mutual information of the joint channel is then as follows:%
\begin{equation}
I(\mathcal{N}_{1}\otimes\mathcal{N}_{2})\equiv\max_{p_{X_{1},X_{2}}%
(x_{1},x_{2})}I(X_{1},X_{2};Y_{1},Y_{2}). \label{eq-ie:tandem-channel}%
\end{equation}
The following theorem states the additivity property.

\begin{theorem}
[Additivity of Mutual Information of Classical Channels]%
\label{thm-ie:additivity-classical}The mutual information of the classical
joint channel $\mathcal{N}_{1}\otimes\mathcal{N}_{2}$ is the sum of their
individual mutual informations:%
\begin{equation}
I(\mathcal{N}_{1}\otimes\mathcal{N}_{2})=I(\mathcal{N}_{1})+I(\mathcal{N}%
_{2}).
\end{equation}

\end{theorem}

\begin{proof}
We first prove the inequality $I(\mathcal{N}_{1}\otimes\mathcal{N}_{2})\geq
I(\mathcal{N}_{1})+I(\mathcal{N}_{2})$. This inequality is more trivial to
prove than the other direction. Let $p_{X_{1}}^{\ast}(x_{1})$ and $p_{X_{2}%
}^{\ast}(x_{2})$ denote distributions that achieve the respective maximums of
$I(\mathcal{N}_{1})$ and $I(\mathcal{N}_{2})$. The joint probability
distribution for all input and output random variables is then as follows:%
\begin{equation}
p_{X_{1},X_{2},Y_{1},Y_{2}}(x_{1},x_{2},y_{1},y_{2})=p_{X_{1}}^{\ast}%
(x_{1})p_{X_{2}}^{\ast}(x_{2})p_{Y_{1}|X_{1}}(y_{1}|x_{1})p_{Y_{2}|X_{2}%
}(y_{2}|x_{2}).
\end{equation}
Observe that $X_{1}$ and $Y_{1}$ are independent of $X_{2}$ and $Y_{2}$. Then
the following chain of inequalities holds:%
\begin{align}
I(\mathcal{N}_{1})+I(\mathcal{N}_{2})  &  =I(X_{1};Y_{1})+I(X_{2};Y_{2})\\
&  =I(X_{1},X_{2};Y_{1},Y_{2})\\
&  \leq I(\mathcal{N}_{1}\otimes\mathcal{N}_{2}).
\end{align}
The first equality follows by evaluating the mutual informations
$I(\mathcal{N}_{1})$ and $I(\mathcal{N}_{2})$ with respect to the maximizing
distributions $p_{X_{1}}^{\ast}(x_{1})$ and $p_{X_{2}}^{\ast}(x_{2})$. The
second equality follows because the mutual information is additive with
respect to the independent joint random variables $\left(  X_{1},Y_{1}\right)
$ and $\left(  X_{2},Y_{2}\right)  $. The final inequality follows because the
input distribution $p_{X_{1}}^{\ast}(x_{1})p_{X_{2}}^{\ast}(x_{2})$ is a
particular input distribution of the more general form $p_{X_{1},X_{2}}%
(x_{1},x_{2})$ needed in the maximization of the mutual information of the
joint channel $\mathcal{N}_{1}\otimes\mathcal{N}_{2}$.

We now prove the non-trivial inequality $I(\mathcal{N}_{1}\otimes
\mathcal{N}_{2})\leq I(\mathcal{N}_{1})+I(\mathcal{N}_{2})$. Let
$p_{X_{1},X_{2}}^{\ast}(x_{1},x_{2})$ denote a distribution that maximizes
$I(\mathcal{N}_{1}\otimes\mathcal{N}_{2})$, and let%
\begin{equation}
q_{X_{1}|X_{2}}(x_{1}|x_{2})\ \text{and}\ q_{X_{2}}(x_{2})
\end{equation}
be distributions such that%
\begin{equation}
p_{X_{1},X_{2}}^{\ast}(x_{1},x_{2})=q_{X_{1}|X_{2}}(x_{1}|x_{2})q_{X_{2}%
}(x_{2}).
\end{equation}
Recall that the conditional probability distribution for the joint channel
$\mathcal{N}_{1}\otimes\mathcal{N}_{2}$ is as follows:%
\begin{equation}
p_{Y_{1},Y_{2}|X_{1},X_{2}}(y_{1},y_{2}|x_{1},x_{2})=p_{Y_{1}|X_{1}}%
(y_{1}|x_{1})p_{Y_{2}|X_{2}}(y_{2}|x_{2}). \label{eq-ie:y1-x2-ind}%
\end{equation}
By summing over $y_{2}$, we observe that $Y_{1}$ and $X_{2}$ are independent when conditioned on $X_1$
because%
\begin{equation}
p_{Y_{1}|X_{1},X_{2}}(y_{1}|x_{1},x_{2})=p_{Y_{1}|X_{1}}(y_{1}|x_{1}).
\end{equation}
Also, the joint distribution $p_{X_{1},Y_{1},Y_{2}|X_{2}}(x_{1},y_{1}%
,y_{2}|x_{2})$ has the form%
\begin{equation}
p_{X_{1},Y_{1},Y_{2}|X_{2}}(x_{1},y_{1},y_{2}|x_{2})=p_{Y_{1}|X_{1}}%
(y_{1}|x_{1})q_{X_{1}|X_{2}}(x_{1}|x_{2})p_{Y_{2}|X_{2}}(y_{2}|x_{2}).
\label{eq-ie:y2-x1-y1-cond-ind}%
\end{equation}
Then $Y_{2}$ is conditionally independent of $X_{1}$ and $Y_{1}$ when
conditioning on $X_{2}$. Consider the following chain of inequalities:%
\begin{align}
I(\mathcal{N}_{1}\otimes\mathcal{N}_{2})  &  =I(X_{1},X_{2};Y_{1},Y_{2})\\
&  =H(Y_{1},Y_{2})-H(Y_{1},Y_{2}|X_{1},X_{2})\\
&  =H(Y_{1},Y_{2})-H(Y_{1}|X_{1},X_{2})-H(Y_{2}|Y_{1},X_{1},X_{2})\\
&  =H(Y_{1},Y_{2})-H(Y_{1}|X_{1})-H(Y_{2}|X_{2})\\
&  \leq H(Y_{1})+H(Y_{2})-H(Y_{1}|X_{1})-H(Y_{2}|X_{2})\\
&  =I(X_{1};Y_{1})+I(X_{2};Y_{2})\\
&  \leq I(\mathcal{N}_{1})+I(\mathcal{N}_{2}).
\end{align}
The first equality follows from the definition of $I(\mathcal{N}_{1}%
\otimes\mathcal{N}_{2})$ in \eqref{eq-ie:tandem-channel} and by evaluating the
mutual information with respect to the distributions $p_{X_{1},X_{2}}^{\ast
}(x_{1},x_{2})$, $p_{Y_{1}|X_{1}}(y_{1}|x_{1})$, and $p_{Y_{2}|X_{2}}%
(y_{2}|x_{2})$. The second equality follows by expanding the mutual
information $I(X_{1},X_{2};Y_{1},Y_{2})$. The third equality follows from the
entropy chain rule. The fourth equality follows because $H(Y_{1}|X_{1}%
,X_{2})=H(Y_{1}|X_{1})$ given that $Y_{1}$ is independent of $X_{2}$ when conditioned on $X_1$ as
pointed out in \eqref{eq-ie:y1-x2-ind}. Also, the equality follows because
$H(Y_{2}|Y_{1},X_{1},X_{2})=H(Y_{2}|X_{2})$ given that $Y_{2}$ is
conditionally independent of $X_{1}$ and $Y_{1}$ as pointed out in
\eqref{eq-ie:y2-x1-y1-cond-ind}. The first inequality follows from
subadditivity of entropy (Exercise~\ref{ex-ie:subadditivity}). The last
equality follows from the definition of mutual information, and the final
inequality follows because the marginal distributions for $X_{1}$ and $X_{2}$
can only achieve a mutual information less than or equal to the respective
maximizing marginal distributions for $I(\mathcal{N}_{1})$ and $I(\mathcal{N}%
_{2})$.
\end{proof}

A simple corollary of Theorem~\ref{thm-ie:additivity-classical} is that
correlations between input random variables cannot increase the mutual
information of a classical channel. The proof follows by a straightforward
induction argument. Thus, the \textit{single-letter} expression in
\eqref{eq-ie:mutual-info-classical-channel} for the mutual information of a
classical channel suffices for understanding the ability of a classical
channel to maintain correlations between its input and output.

\begin{corollary}
\label{cor-ie:class-mut=reg-class-mut}The regularized mutual information of a
classical channel is equal to its mutual information:%
\begin{equation}
I_{\operatorname{reg}}(\mathcal{N})=I(\mathcal{N}).
\end{equation}

\end{corollary}

\begin{proof}
We prove the result using induction on $n$, by showing that $I(\mathcal{N}%
^{\otimes n})=nI(\mathcal{N})$ for all $n$, implying that the limit in
\eqref{eq-ie:reg-mutual-classical} is not necessary. The base case for $n=1$
is trivial. Suppose the result holds for $n$: $I(\mathcal{N}^{\otimes
n})=nI(\mathcal{N})$. The following chain of equalities then proves the
inductive step:%
\begin{align}
I(\mathcal{N}^{\otimes n+1})  &  =I(\mathcal{N}\otimes\mathcal{N}^{\otimes
n})\\
&  =I(\mathcal{N})+I(\mathcal{N}^{\otimes n})\\
&  =I(\mathcal{N})+nI(\mathcal{N}).
\end{align}
The first equality follows because the channel $\mathcal{N}^{\otimes n+1}$ is
equivalent to the parallel concatenation of $\mathcal{N}$ and $\mathcal{N}%
^{\otimes n}$. The second critical equality follows from the application of
Theorem~\ref{thm-ie:additivity-classical} because the distributions of
$\mathcal{N}$ and $\mathcal{N}^{\otimes n}$ factor as in
\eqref{eq-ie:y1-x2-ind}. The final equality follows from the induction hypothesis.
\end{proof}

\subsection{The Problem with Regularization}

The main problem with a regularized information quantity is that it does not
uniquely characterize the capacity of a channel for a given
information-processing task, whether it be a classical or quantum one. This
motivates the task of proving additivity of an information quantity.

We illustrate this point now with an example. Let $I_{a}(\mathcal{N})$ denote
the following function of a classical channel $\mathcal{N}$:%
\begin{equation}
I_{a}(\mathcal{N})=\max_{p_{X}(x)}\left[  H(X)-aH(X|Y)\right]  ,
\end{equation}
where $a>1$. From the definitions we can conclude that%
\begin{equation}
I(\mathcal{N})\geq I_{a}(\mathcal{N}),
\end{equation}
and in fact that%
\begin{equation}
\frac{1}{n}I(\mathcal{N}^{\otimes n})\geq\frac{1}{n}I_{a}(\mathcal{N}^{n}),
\end{equation}
for any fixed positive integer $n$. However, something interesting happens
when we regularize the quantity $I_{a}(\mathcal{N})$, i.e., when we consider
the following limit:%
\begin{equation}
\lim_{n\rightarrow\infty}\frac{1}{n}I_{a}(\mathcal{N}^{\otimes n}%
)=\lim_{n\rightarrow\infty}\frac{1}{n}\max_{p_{X^{n}}(x^{n})}\left[
H(X^{n})-aH(X^{n}|Y^{n})\right]  .
\end{equation}
Let $n$ be a large fixed integer (large enough so that the law of large
numbers comes into play). That is, we can fix constants $\varepsilon\in(0,1)$
and $\delta>0$. From Shannon's channel coding theorem, we know that there
exists a code $\left\{  x^{n}(m)\right\}  _{m\in\mathcal{M}}$\ of rate
$\frac{1}{n}\log\left\vert \mathcal{M}\right\vert =I(\mathcal{N})-\delta$ and
length $n$ for the channel $\mathcal{N}$, such that the probability of error
when decoding is no larger than $\varepsilon$. So we can pick $p_{X^{n}}%
(x^{n})$ to be the uniform distribution over the codewords for this code, and
for this choice, we get%
\begin{equation}
H(X^{n})=\log\left\vert \mathcal{M}\right\vert =n\left[  I(\mathcal{N}%
)-\delta\right]  .
\end{equation}
Furthermore, we can apply Fano's inequality\ (Theorem~\ref{thm-cie:fano}) to
conclude that%
\begin{align}
H(X^{n}|Y^{n})  &  \leq h_{2}(\varepsilon)+\varepsilon\log\left[  \left\vert
\mathcal{M}\right\vert -1\right] \\
&  \leq h_{2}(\varepsilon)+\varepsilon n\left[  I(\mathcal{N})-\delta\right]
.
\end{align}
Putting these inequalities together implies that%
\begin{align}
\frac{1}{n}I_{a}(\mathcal{N}^{\otimes n})  &  =\frac{1}{n}\max_{p_{X^{n}%
}(x^{n})}\left[  H(X^{n})-aH(X^{n}|Y^{n})\right] \\
&  \geq\frac{1}{n}\left(  n\left[  I(\mathcal{N})-\delta\right]  -a\left[
h_{2}(\varepsilon)+\varepsilon n\left[  I(\mathcal{N})-\delta\right]  \right]
\right) \\
&  =\left(  1-a\varepsilon\right)  I(\mathcal{N})-a\delta\varepsilon-\frac
{1}{n}\left[  \delta-ah_{2}(\varepsilon)\right]  .
\end{align}
Taking the limit $n\rightarrow\infty$ then gives%
\begin{equation}
\lim_{n\rightarrow\infty}\frac{1}{n}I_{a}(\mathcal{N}^{\otimes n})\geq\left(
1-a\varepsilon\right)  I(\mathcal{N})-a\delta\varepsilon.
\end{equation}
However, as $n$ gets larger, we can make both $\varepsilon$ and $\delta$ go to
zero (in principle we could write $\varepsilon$ and $\delta$ as explicit
functions of $n$ that go to zero as $n\rightarrow\infty$). This establishes
that%
\begin{equation}
\lim_{n\rightarrow\infty}\frac{1}{n}I_{a}(\mathcal{N}^{\otimes n})\geq
I(\mathcal{N}).
\end{equation}
However, we have for every $n$ that%
\begin{align}
\frac{1}{n}\left[  H(X^{n})-aH(X^{n}|Y^{n})\right]   &  \leq\frac{1}{n}\left[
H(X^{n})-H(X^{n}|Y^{n})\right] \\
&  \leq\lim_{n\rightarrow\infty}\frac{1}{n}I(\mathcal{N}^{\otimes n})\\
&  =I(\mathcal{N}),
\end{align}
where the last equality follows from
Corollary~\ref{cor-ie:class-mut=reg-class-mut}. These two bounds lead us to
the conclusion that%
\begin{equation}
\lim_{n\rightarrow\infty}\frac{1}{n}I_{a}(\mathcal{N}^{\otimes n}%
)=I(\mathcal{N}).
\end{equation}
Thus, even though for a given channel $\mathcal{N}$ we can have a strict
inequality between $I_{a}(\mathcal{N})$ and $I(\mathcal{N})$ on the
single-copy level (and even on the multi-copy level for finite $n$), this
difference gets washed away or \textquotedblleft blurred\textquotedblright\ in
the limit $n\to\infty$ for any finite $a>1$. For this reason, we should always
be wary of a regularized characterization of capacity, knowing that it is
incomplete. It is only when we have additivity of an information quantity that
we can conclude that it fully characterizes capacity for a given
information-processing task. Unfortunately, regularization is a problem that
plagues much of quantum Shannon theory (except for a few tasks, such as
entanglement-assisted classical communication, discussed in
Chapter~\ref{chap:EA-classical}).

\subsection{Optimizing the Mutual Information of a Classical Channel}

The definition in \eqref{eq-ie:mutual-info-classical-channel} seems like a
suitable definition for the mutual information of a classical channel, but how
difficult is the maximization problem that it sets out?
Theorem~\ref{thm-ie:mut-concave-input} below states an important property of
the mutual information $I(X;Y)$ that allows us to answer this question.
Suppose that we fix the conditional density $p_{Y|X}(y|x)$, but can vary the
input density $p_{X}(x)$. Theorem~\ref{thm-ie:mut-concave-input} below states
that the mutual information $I(X;Y)$ is a concave function of the density
$p_{X}(x)$. In particular, this result implies that the channel mutual
information $I(\mathcal{N})$ has a global maximum, and the optimization
problem is therefore a straightforward computation that can exploit convex
optimization methods.

\begin{theorem}
\label{thm-ie:mut-concave-input}Suppose that we fix the conditional
probability density $p_{Y|X}(y|x)$. Then the mutual information $I(X;Y)$\ is
concave in the marginal density $p_{X}(x)$:%
\begin{equation}
\lambda I(X_{1};Y)+\left(  1-\lambda\right)  I(X_{2};Y)\leq I(Z;Y),
\end{equation}
where random variable $X_{1}$ has density $p_{X_{1}}(x)$, $X_{2}$ has density
$p_{X_{2}}(x)$, and $Z$ has density $\lambda p_{X_{1}}(x)+\left(
1-\lambda\right)  p_{X_{2}}(x)$ for $\lambda\in[0,1]$.
\end{theorem}

\begin{proof}
Let us fix the density $p_{Y|X}(y|x)$. The density $p_{Y}(y)$ is a linear
function of $p_{X}(x)$ because $p_{Y}(y)=\sum_{x}p_{Y|X}(y|x)p_{X}(x)$. Thus
$H(Y)$ is concave in $p_{X}(x)$. Recall that the conditional entropy
$H(Y|X)=\sum_{x}p_{X}(x)H(Y|X=x)$. The entropy $H(Y|X=x)$ is fixed when the
conditional probability density $p_{Y|X}(y|x)$ is fixed. Thus, $H(Y|X)$ is a
linear function of $p_{X}(x)$. These two results imply that the mutual
information $I(X;Y)$ is concave in the marginal density $p_{X}(x)$ when the
conditional density $p_{Y|X}(y|x)$ is fixed.
\end{proof}

\section{Private Information of a Wiretap Channel}

\label{sec-add:private-info-class}
\begin{figure}
[ptb]
\begin{center}
\includegraphics[
width=4.0421in
]%
{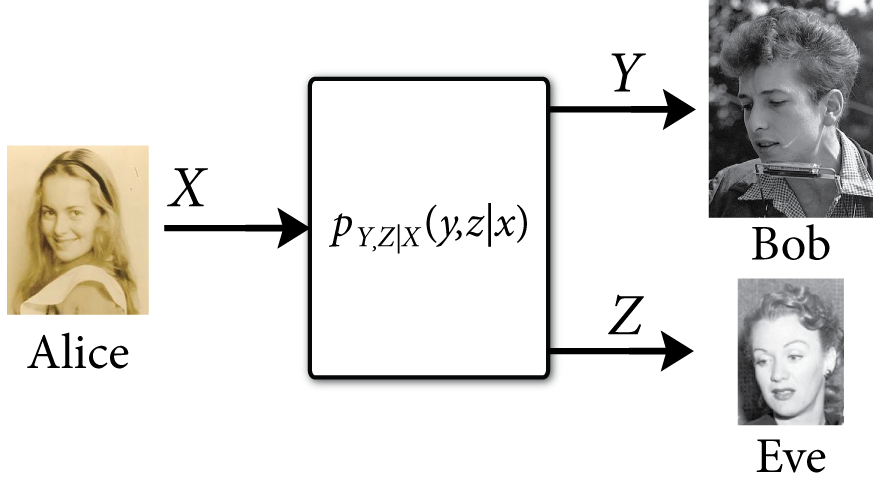}%
\caption{The setting for the classical wiretap channel.}%
\label{fig-add:wiretap}%
\end{center}
\end{figure}

Suppose now that we extend the above
two-user classical communication scenario%
\index{private information!of a wiretap channel}
to a three-user communication scenario, where the parties involved are Alice,
Bob, and Eve. Suppose that Alice would like to communicate to Bob while
keeping her messages private from Eve. The channel $\mathcal{N}$%
\ corresponding to this setting is called the wiretap channel, which has the
following conditional probability density:%
\begin{equation}
p_{Y,Z|X}(y,z|x).
\end{equation}
Alice has access to the input random variable $X$, Bob receives output random
variable $Y$, and Eve receives the random variable $Z$.
Figure~\ref{fig-add:wiretap}\ depicts this setting.%

We would like to establish a measure of information throughput for this
scenario. It might seem intuitive that it should be the amount of correlations
that Alice can establish with Bob, less the correlations that Eve receives:
$I(X;Y)-I(X;Z)$. However, a more general procedure would allow Alice to pick
an auxiliary random variable $U$ and then to pick $X$ given the value of $U$,
leading to the following information quantity:%
\begin{equation}
I(U;Y)-I(U;Z).
\end{equation}
But Alice can maximize over all such input distributions $p_{U,X}(u,x)$\ on
her end. This leads us to the following definition:

\begin{definition}
[Private Information of a Wiretap Channel]The private information
$P(\mathcal{N})$ of a classical wiretap channel $\mathcal{N}\equiv p_{Y,Z|X}%
$\ is defined as follows:%
\begin{equation}
P(\mathcal{N})\equiv\max_{p_{U,X}(u,x)}[I(U;Y)-I(U;Z)].
\end{equation}

\end{definition}

It is possible to provide an operational interpretation of the private
information, but we do not do that here. We instead focus on the additivity
properties of the private information $P(\mathcal{N})$.

One may wonder if the above quantity is non-negative, given that it is equal
to the difference of two mutual informations. Non-negativity does hold, and a
simple proof demonstrates this fact.

\begin{property}
The private
\index{private information!of a wiretap channel!positivity}
information $P( \mathcal{N}) $ of a wiretap channel is non-negative:%
\begin{equation}
P( \mathcal{N}) \geq0.
\end{equation}

\end{property}

\begin{proof}
We can choose the joint density $p_{U,X}(u,x)$ in the maximization of
$P(\mathcal{N})$ to be the degenerate distribution $p_{U,X}(u,x)=\delta
_{u,u_{0}}\delta_{x,x_{0}}$ for some values $u_{0}$ and $x_{0}$. Then both
mutual informations $I(U;Y)$ and $I(U;Z)$ vanish, and their difference
vanishes as well. The private information $P(\mathcal{N})$ can then only be
greater than or equal to zero because the above choice is a particular choice
of the density $p_{U,X}(u,x)$ and $P(\mathcal{N})$ requires a maximization
over all such distributions.
\end{proof}

\begin{exercise}
Show that adding an auxiliary random variable cannot increase the mutual
information of a classical channel. That is, let $\mathcal{N}\equiv p_{Y|X}$
and show that $I(\mathcal{N}) = \max_{p_{U,X}}I(U;Y)$. Explain why it is
possible for an auxiliary random variable to increase the private information
of a classical wiretap channel.
\end{exercise}

\subsection{Additivity of Private Information}

\begin{figure}[ptb]
\begin{center}
\includegraphics[
width=2.0625in
]{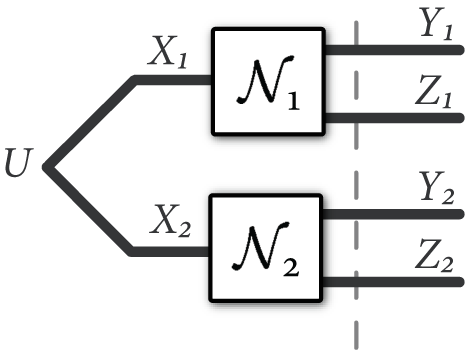}
\end{center}
\caption{This figure displays the scenario for determining whether the private
information of two classical wiretap channels $\mathcal{N}_{1}$ and
$\mathcal{N}_{2}$\ is additive. The question of additivity is equivalent to
the possibility of classical correlations being able to enhance the private
information of two classical wiretap channels. The result stated in
Theorem~\ref{thm-ie:additivity-classical copy(1)}\ is that the private
information is additive for two wiretap channels, so that classical
correlations cannot enhance the private information.}%
\label{fig-add:priv-add}%
\end{figure}

The private
\index{private information!of a wiretap channel!additivity for degraded channels}
information of general classical wiretap channels is additive. This result
follows essentially from an application of the chain rule for mutual
information. Figure~\ref{fig-add:priv-add}\ displays the scenario
corresponding to the analysis involved in determining whether the private
information is additive.

\begin{theorem}
[Additivity of Private Information]\label{thm-ie:additivity-classical copy(1)}%
Let $\mathcal{N}_{i}$\ be a classical wiretap channel with distribution
$p_{Y_{i},Z_{i}|X_{i}}$, for $i\in\{1,2\}$. The private information of the
classical joint channel $\mathcal{N}_{1}\otimes\mathcal{N}_{2}$ is the sum of
the individual private informations of the channels:%
\begin{equation}
P(\mathcal{N}_{1}\otimes\mathcal{N}_{2})=P(\mathcal{N}_{1})+P(\mathcal{N}%
_{2}).
\end{equation}

\end{theorem}

\begin{proof}
The inequality $P(\mathcal{N}_{1}\otimes\mathcal{N}_{2})\geq P(\mathcal{N}%
_{1})+P(\mathcal{N}_{2})$ is trivial and we leave it as an exercise for the
reader to complete.

We thus prove the following non-trivial inequality: $P(\mathcal{N}_{1}%
\otimes\mathcal{N}_{2})\leq P(\mathcal{N}_{1})+P(\mathcal{N}_{2})$. Let
$p_{U,X_{1},X_{2}}(u,x_{1},x_{2})$ be any probability distribution that we
could consider for the quantity $P(\mathcal{N}_{1}\otimes\mathcal{N}_{2})$,
where $U$ is an auxiliary random variable. The joint channel has the following
probability distribution:%
\begin{equation}
p_{Y_{1},Z_{1}|X_{1}}(y_{1},z_{1}|x_{1})p_{Y_{2},Z_{2}|X_{2}}(y_{2}%
,z_{2}|x_{2}).
\end{equation}
Consider the following chain of equalities:%
\begin{align}
&  I(U;Y_{1}Y_{2})-I(U;Z_{1}Z_{2})\nonumber\\
&  =I(U;Y_{1})+I(U;Y_{2}|Y_{1})-I(U;Z_{2})-I(U;Z_{1}|Z_{2})\\
&  =I(U;Y_{1}|Z_{2})+I(U;Y_{2}|Y_{1})-I(U;Z_{2}|Y_{1})-I(U;Z_{1}|Z_{2})\\
&  =I(U;Y_{1}|Z_{2})-I(U;Z_{1}|Z_{2})+I(U;Y_{2}|Y_{1})-I(U;Z_{2}|Y_{1}).
\end{align}
The first equality follows from an application of the chain rule for mutual
information. The second equality follows from the identity%
\begin{equation}
I(U;Y_{1})-I(U;Z_{2})=I(U;Y_{1}|Z_{2})-I(U;Z_{2}|Y_{1}),
\end{equation}
which can be verified using definitions. The third equality is a simple
rearrangement. We now focus on the term $I(U;Y_{1}|Z_{2})-I(U;Z_{1}|Z_{2})$.
Note that the probability distribution for these random variables can be
written as%
\begin{equation}
p_{Y_{1},Z_{1}|X_{1}}(y_{1},z_{1}|x_{1})p_{X_{1}|U}(x_{1}|u)p_{U|Z_{2}%
}(u|z_{2})p_{Z_{2}}(z_{2}),
\end{equation}
where $p_{U|Z_{2}}(u|z_{2})=\sum_{x_{2}}p_{U,X_{2}|Z_{2}}(u,x_{2}|z_{2})$.
(The fact that the distribution factors in this way is where the assumption of
independent wiretap channels comes into play.) Consider that%
\begin{align}
&  I(U;Y_{1}|Z_{2})-I(U;Z_{1}|Z_{2})\nonumber\\
&  =\sum_{z_{2}}p_{Z_{2}}(z_{2})\left[  I(U;Y_{1}|Z_{2}=z_{2})-I(U;Z_{1}%
|Z_{2}=z_{2})\right] \\
&  \leq\max_{z_{2}}\left[  I(U;Y_{1}|Z_{2}=z_{2})-I(U;Z_{1}|Z_{2}%
=z_{2})\right] \\
&  \leq P(\mathcal{N}_{1}).
\end{align}
The first equality follows because the conditional mutual informations can be
expanded as a convex combination of mutual informations. The last inequality
follows because $p_{U|Z_{2}}(u|z_{2})$ is a particular distribution for an
auxiliary random variable $U|Z_{2}=z_{2}$. By the same kind of reasoning, we
find that%
\begin{equation}
I(U;Y_{2}|Y_{1})-I(U;Z_{2}|Y_{1})\leq P(\mathcal{N}_{2}),
\end{equation}
and can then conclude that%
\begin{equation}
I(U;Y_{1}Y_{2})-I(U;Z_{1}Z_{2})\leq P(\mathcal{N}_{1})+P(\mathcal{N}_{2}).
\end{equation}
Since this inequality holds for any auxiliary random variable $U$, we can
finally conclude that $P(\mathcal{N}_{1}\otimes\mathcal{N}_{2})\leq
P(\mathcal{N}_{1})+P(\mathcal{N}_{2})$.
\end{proof}

\begin{exercise}
Show that the sum of the individual private informations can never be greater
than the private information of the classical joint channel:%
\begin{equation}
P(\mathcal{N}_{1}\otimes\mathcal{N}_{2})\geq P(\mathcal{N}_{1})+P(\mathcal{N}%
_{2}).
\end{equation}

\end{exercise}

\begin{exercise}
Show that the regularized private information of a wiretap channel
$\mathcal{N}$ is equal to its private information: $\lim_{n\rightarrow\infty
}\frac{1}{n}P(\mathcal{N}^{\otimes n})=P(\mathcal{N})$.
\end{exercise}

\subsection{Degraded Wiretap Channels}

The formula for the private information simplifies for a particular type of
wiretap channel, called a physically degraded wiretap channel. A wiretap
channel is physically degraded if $X$, $Y$, and $Z$ form the following Markov
chain:\ $X\rightarrow Y\rightarrow Z$. That is, the channel distribution
factors as $p_{Y,Z|X}(y,z|x)=p_{Z|Y}(z|y)p_{Y|X}(y|x)$, so that there is some
channel $p_{Z|Y}(z|y)$ that Bob can apply to his output to simulate the
channel $p_{Z|X}(z|x)$ to Eve:%
\begin{equation}
p_{Z|X}(z|x)=\sum_{y} p_{Z|Y}(z|y)p_{Y|X}(y|x).
\end{equation}
This condition allows us to apply the data-processing inequality to give the
following simplified formula for the private information of a degraded wiretap
channel, in which there is no need for an auxiliary random variable:

\begin{proposition}
The private information $P(\mathcal{N})$ of a degraded classical wiretap
channel $\mathcal{N}\equiv p_{Y,Z|X}$\ simplifies as follows:%
\begin{equation}
P(\mathcal{N})\equiv\max_{p_{X}(x)}[I(X;Y)-I(X;Z)].
\end{equation}

\end{proposition}

\begin{proof}
Consider that we always have%
\begin{equation}
\max_{p_{U,X}(u,x)}[I(U;Y)-I(U;Z)]\geq\max_{p_{X}(x)}[I(X;Y)-I(X;Z)],
\end{equation}
by picking the distribution on the left-hand side as $p_{U,X}(u,x)=p_{X}%
(x)\delta_{x,u}$. That is, we set $U$ to be a random variable with the same
alphabet size as the channel input and then just send $U$ directly into the
channel. To prove the other inequality, we need to use the assumption of
degradability, which implies that $U-X-Y-Z$ is a Markov chain. Consider that%
\begin{align}
I(U;Y)  &  =I(X;Y)+I(U;Y|X)-I(X;Y|U)\\
&  =I(X;Y)-I(X;Y|U).
\end{align}
The first equality follows from two applications of the chain rule for mutual
information. The second equality follows because $U-X-Y$ is a Markov chain, so
that $I(U;Y|X)=0$. By the same reasoning, we have that
$I(U;Z)=I(X;Z)-I(X;Z|U)$. Then%
\begin{align}
I(U;Y)-I(U;Z)  &  =I(X;Y)-I(X;Y|U)-\left[  I(X;Z)-I(X;Z|U)\right] \\
&  =I(X;Y)-I(X;Z)-\left[  I(X;Y|U)-I(X;Z|U)\right] \\
&  \leq I(X;Y)-I(X;Z)\\
&  \leq\max_{p_{X}(x)}[I(X;Y)-I(X;Z)].
\end{align}
The inequality follows from the assumption that the wiretap channel is
degraded, so that we can apply the data-processing inequality and conclude
that $I(X;Y|U)\geq I(X;Z|U)$. Since the inequality holds for any auxiliary
random variable $U$, we can conclude that%
\begin{equation}
\max_{p_{U,X}(u,x)}[I(U;Y)-I(U;Z)]\leq\max_{p_{X}(x)}[I(X;Y)-I(X;Z)],
\end{equation}
which completes the proof.
\end{proof}

An analogous notion of degradability exists in the quantum setting, and
Section~\ref{sec-add:coh-info}\ demonstrates that degradable quantum channels
have additive coherent information. The coherent information of a quantum
channel is a measure of how much quantum information a sender can transmit
through that channel to a receiver and thus is an important quantity to
consider for quantum data transmission.

\begin{exercise}
Show that the private information of a degraded wiretap channel can also be
written as $\max_{p_{X}(x)}I(X;Y|Z)$.
\end{exercise}

\section{Holevo Information of a Quantum Channel}

We now turn our attention to
\index{Holevo information!of a channel}%
the case of dynamic information measures for quantum channels, and we begin
with a measure of classical correlations. Suppose that Alice would like to
establish classical correlations with Bob, by means of a quantum channel.
Alice can prepare an ensemble $\left\{  p_{X}(x),\rho^{x}\right\}  $\ in her
laboratory, where the states $\rho^{x}$ are acceptable inputs to the quantum
channel. She keeps a copy of the classical index $x$ in some classical
register $X$. The expected density operator of this ensemble is the following
classical--quantum state:%
\begin{equation}
\rho_{XA}\equiv\sum_{x}p_{X}(x)|x\rangle\langle x|_{X}\otimes\rho_{A}^{x}.
\label{eq-add:cq-state-mixed}%
\end{equation}
Such a preparation is the most general way that Alice can correlate classical
data with a quantum state to input to the channel. Let $\rho_{XB}$ be the
state that arises from sending the $A$ system through the quantum channel
$\mathcal{N}_{A\rightarrow B}$:%
\begin{equation}
\rho_{XB}\equiv\sum_{x}p_{X}(x)|x\rangle\langle x|_{X}\otimes\mathcal{N}%
_{A\rightarrow B}(\rho_{A}^{x}). \label{eq-add:cq-state-holevo}%
\end{equation}

We would like to determine a measure of the ability of the quantum channel to
preserve classical correlations. We can appeal to ideas from the classical
case in Section~\ref{sec-cie:mut-inf-channel}, while incorporating the static
quantum measures from Chapter~\ref{chap:q-info-entropy}. A good measure of the
input--output classical correlations is the Holevo information of the above
classical--quantum state: $I(X;B)_{\rho}$. This measure corresponds to a
particular preparation that Alice chooses, but observe that she can prepare
the input ensemble in such a way as to achieve the highest possible
correlations. Maximizing the Holevo information over all possible preparations
gives a dynamic measure called the Holevo information of the channel.

\begin{definition}
[Holevo Information of a Quantum Channel]The Holevo information $\chi
(\mathcal{N})$ of a channel $\mathcal{N}$\ is a measure of the classical
correlations that Alice can establish with Bob:%
\begin{equation}
\chi(\mathcal{N})\equiv\max_{\rho_{XA}}I(X;B)_{\rho}, \label{eq-add:holevo}%
\end{equation}
where the maximization is with respect to all input classical--quantum states
of the form in \eqref{eq-add:cq-state-mixed} and $I(X;B)_{\rho}$ is evaluated
with respect to the state in \eqref{eq-add:cq-state-holevo}.
\end{definition}

\subsection{Additivity of the Holevo Information for Specific Channels}

\label{sec-add:holevo-additivity-ent-break}The Holevo information of a quantum
channel is generally not additive (by no means is this obvious!). The question
of additivity for this case is \textit{not} whether classical correlations can
enhance the Holevo information, but it is \textit{rather} whether quantum
correlations can enhance it. That is, Alice can choose an ensemble of the form
$\{p_{X}(x),\rho_{A_{1}A_{2}}^{x}\}$ for input to two uses of the quantum
channel. The conditional density operators $\rho_{A_{1}A_{2}}^{x}$ can be
entangled and these quantum correlations can potentially increase the Holevo information.

\begin{figure}[ptb]
\begin{center}
\includegraphics[
width=1.5349in
]{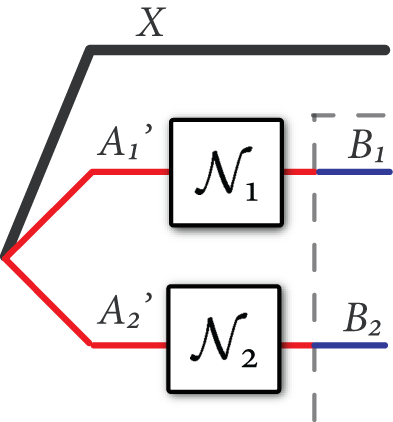}
\end{center}
\caption{The scenario for determining whether the Holevo information of two
quantum channels $\mathcal{N}_{1}$ and $\mathcal{N}_{2}$\ is additive. The
question of additivity is equivalent to the possibility of quantum
correlations being able to enhance the Holevo information of two quantum
channels. The result stated in Theorem~\ref{thm-add:holevo-EB-add}\ is that
the Holevo information is additive for the tensor product of an
entanglement-breaking channel and any other quantum channel, so that quantum
correlations cannot enhance the Holevo information in this case. This is
perhaps intuitive because an entanglement-breaking channel destroys quantum
correlations in the form of quantum entanglement.}%
\label{fig-add:hol-add}%
\end{figure}

The question of additivity of the Holevo information of a quantum channel was
a longstanding open conjecture in quantum information theory---many
researchers thought that quantum correlations would not enhance it and that
additivity would hold. But recent research has demonstrated a counterexample
to the additivity conjecture, and perhaps unsurprisingly in hindsight, this
counterexample exploits maximally entangled states to demonstrate
superadditivity (see Section~\ref{sec-cc:superadditivity}).
Figure~\ref{fig-add:hol-add}\ displays the scenario corresponding to the
question of additivity of the Holevo information.

Additivity of Holevo information may not hold for all quantum channels, but it
is possible to prove its additivity for certain classes of quantum channels.
One such class for which additivity holds is the class of
\index{entanglement-breaking channel}%
entanglement-breaking channels, and the proof of additivity is perhaps the
simplest for this case.

\begin{theorem}
[Additivity for Entanglement-Breaking Channels]\label{thm-add:holevo-EB-add}%
Suppose that a quantum channel $\mathcal{N}^{\operatorname{EB}}$ is%
\index{entanglement-breaking channel}
entanglement breaking and another channel $\mathcal{M}$ is arbitrary. Then the
Holevo information$~\chi(\mathcal{N}^{\operatorname{EB}}\otimes\mathcal{M})$
of the tensor-product channel $\mathcal{N}^{\operatorname{EB}}\otimes
\mathcal{M}$ is the sum of the individual Holevo informations$~\chi
(\mathcal{N}^{\operatorname{EB}})$ and$~\chi(\mathcal{M})$:%
\begin{equation}
\chi(\mathcal{N}^{\operatorname{EB}}\otimes\mathcal{M})=\chi(\mathcal{N}%
^{\operatorname{EB}})+\chi(\mathcal{M}).
\end{equation}

\end{theorem}

\begin{proof}
The trivial inequality $\chi(\mathcal{N}^{\operatorname{EB}}\otimes
\mathcal{M})\geq\chi(\mathcal{N}^{\operatorname{EB}})+\chi(\mathcal{M})$ holds
for any two quantum channels $\mathcal{N}^{\operatorname{EB}}$ and
$\mathcal{M}$ because we can choose the input ensemble on the left-hand
side\ to be a tensor product of the ones that individually maximize the terms
on the right-hand side.

We now prove the non-trivial inequality $\chi(\mathcal{N}^{\operatorname{EB}%
}\otimes\mathcal{M})\leq\chi(\mathcal{N}^{\operatorname{EB}})+\chi
(\mathcal{M})$ that holds when $\mathcal{N}^{\operatorname{EB}}$ is
entanglement breaking. Let $\rho_{XB_{1}B_{2}}$ be a state that maximizes the
Holevo information $\chi(\mathcal{N}^{\operatorname{EB}}\otimes\mathcal{M})$,
where%
\begin{align}
\rho_{XB_{1}B_{2}}  &  \equiv(\mathcal{N}_{A_{1}\rightarrow B_{1}%
}^{\operatorname{EB}}\otimes\mathcal{M})(\rho_{XA_{1}A_{2}}),\\
\rho_{XA_{1}A_{2}}  &  \equiv\sum_{x}p_{X}(x)|x\rangle\langle x|_{X}%
\otimes\rho_{A_{1}A_{2}}^{x}.
\end{align}
Let $\rho_{XB_{1}A_{2}}$ be the state after only the entanglement-breaking
channel $\mathcal{N}_{A_{1}\rightarrow B_{1}}^{\operatorname{EB}}$ acts. We
can write this state as follows:%
\begin{align}
\rho_{XB_{1}A_{2}}  &  \equiv\mathcal{N}_{A_{1}\rightarrow B_{1}%
}^{\operatorname{EB}}(\rho_{XA_{1}A_{2}}) =\sum_{x}p_{X}(x)|x\rangle\langle
x|_{X}\otimes\mathcal{N}_{A_{1}\rightarrow B_{1}}^{\operatorname{EB}}%
(\rho_{A_{1}A_{2}}^{x})\\
&  =\sum_{x}p_{X}(x)|x\rangle\langle x|_{X}\otimes\sum_{y}p_{Y|X}%
(y|x)\ \sigma_{B_{1}}^{x,y}\otimes\theta_{A_{2}}^{x,y}\\
&  =\sum_{x,y}p_{Y|X}(y|x)p_{X}(x)|x\rangle\langle x|_{X}\otimes\sigma_{B_{1}%
}^{x,y}\otimes\theta_{A_{2}}^{x,y}.
\end{align}
The third equality follows because the channel $\mathcal{N}^{\operatorname{EB}%
}$ breaks any entanglement in the state $\rho_{A_{1}A_{2}}^{x}$, leaving
behind a separable state $\sum_{y}p_{Y|X}(y|x)\ \sigma_{B_{1}}^{x,y}%
\otimes\theta_{A_{2}}^{x,y}$. Then the state $\rho_{XB_{1}B_{2}}$ has the form%
\begin{equation}
\rho_{XB_{1}B_{2}}=\sum_{x,y}p_{Y|X}(y|x)p_{X}(x)|x\rangle\langle
x|_{X}\otimes\sigma_{B_{1}}^{x,y}\otimes\mathcal{M}(\theta_{A_{2}}^{x,y}).
\end{equation}
Let $\omega_{XYB_{1}B_{2}}$ be an extension of $\rho_{XB_{1}B_{2}}$ where%
\begin{equation}
\omega_{XYB_{1}B_{2}}\equiv\sum_{x,y}p_{Y|X}(y|x)p_{X}(x)|x\rangle\langle
x|_{X}\otimes|y\rangle\langle y|_{Y}\otimes\sigma_{B_{1}}^{x,y}\otimes
\mathcal{M}(\theta_{A_{2}}^{x,y}), \label{eq-add:ent-break-state}%
\end{equation}
and $\operatorname{Tr}_{Y}\left\{  \omega_{XYB_{1}B_{2}}\right\}
=\rho_{XB_{1}B_{2}}$. Then the following chain of inequalities holds%
\begin{align}
\chi(\mathcal{N}^{\operatorname{EB}}\otimes\mathcal{M})  &  =I(X;B_{1}%
B_{2})_{\rho}\\
&  =I(X;B_{1})_{\rho}+I(X;B_{2}|B_{1})_{\rho}\\
&  \leq\chi(\mathcal{N}^{\operatorname{EB}})+I(X;B_{2}|B_{1})_{\rho}%
\end{align}
The first equality follows because we took $\rho_{XB_{1}B_{2}}$ to be a state
that maximizes the Holevo information $\chi(\mathcal{N}^{\operatorname{EB}%
}\otimes\mathcal{M})$ of the tensor-product channel $\mathcal{N}%
^{\operatorname{EB}}\otimes\mathcal{M}$. The second equality is an application
of the chain rule for conditional mutual information
(Property~\ref{prop-qie:chain-CMI}). The inequality follows because the Holevo
information $I(X;B_{1})_{\rho}$ is with respect to the following state:%
\begin{equation}
\rho_{XB_{1}}\equiv\sum_{x}p_{X}(x)|x\rangle\langle x|_{X}\otimes
\mathcal{N}_{A_{1}\rightarrow B_{1}}^{\operatorname{EB}}(\rho_{A_{1}}^{x}),
\end{equation}
whereas the Holevo information of the channel $\mathcal{N}_{A_{1}\rightarrow
B_{1}}^{\operatorname{EB}}$ is defined to be the maximal Holevo information
with respect to all input ensembles. Now let us focus on the term
$I(X;B_{2}|B_{1})_{\rho}$. Consider that%
\begin{align}
I(X;B_{2}|B_{1})_{\rho}  &  =I(X;B_{2}|B_{1})_{\omega}\\
&  \leq I(XB_{1};B_{2})_{\omega}\\
&  \leq I(XYB_{1};B_{2})_{\omega}\\
&  =I(XY;B_{2})_{\omega}+I(B_{1};B_{2}|XY)_{\omega}\\
&  =I(XY;B_{2})_{\omega}\\
&  \leq\chi(\mathcal{M}).
\end{align}
The first equality follows because the reduced state of $\omega_{XYB_{1}B_{2}%
}$ on systems $X$, $B_{1}$, and $B_{2}$ is equal to $\rho_{XB_{1}B_{2}}$. The
first inequality follows from the chain rule: $I(X;B_{2}|B_{1})=I(XB_{1}%
;B_{2})-I(B_{1};B_{2})\leq I(XB_{1};B_{2})$. The second inequality follows
from the quantum data-processing inequality. The second equality is again from
the chain rule for conditional mutual information. The third equality is the
crucial one that exploits the
\index{entanglement-breaking channel}%
entanglement-breaking property. It follows by examining
\eqref{eq-add:ent-break-state} and observing that the state $\omega
_{XYB_{1}B_{2}}$ on systems $B_{1}$ and $B_{2}$ is product when conditioned on
classical variables $X$ and $Y$, so that the conditional mutual information
between systems $B_{1}$ and $B_{2}$ given both $X$ and $Y$ is equal to zero.
The final inequality follows because $\omega_{XYB_{2}}$ is a particular state
of the form needed in the maximization of $\chi(\mathcal{M})$.
\end{proof}

\begin{corollary}
The regularized Holevo information of an
\index{entanglement-breaking channel}%
entanglement-breaking quantum channel$~\mathcal{N}^{\operatorname{EB}}$ is
equal to its Holevo information:%
\begin{equation}
\chi_{\operatorname{reg}}( \mathcal{N}^{\operatorname{EB}}) =\chi(
\mathcal{N}^{\operatorname{EB}}) .
\end{equation}

\end{corollary}

\begin{proof}
A proof of this property uses the same induction argument as in
Corollary~\ref{cor-ie:class-mut=reg-class-mut}\ and exploits the additivity
property in Theorem~\ref{thm-add:holevo-EB-add} above.
\end{proof}

\subsection{Optimizing the Holevo Information}

\subsubsection{Pure States are Sufficient}

The following theorem allows us to simplify the optimization problem that
\eqref{eq-add:holevo} sets out---we show that it is sufficient to consider
ensembles of pure states at the input.

\begin{theorem}
\label{thm-add:pure-states-suff-holevo}It is sufficient to maximize the Holevo
information with respect to pure states:%
\begin{equation}
\chi(\mathcal{N})=\max_{\rho_{XA}}I(X;B)_{\rho}=\max_{\tau_{XA}}I(X;B)_{\tau},
\end{equation}
where%
\begin{equation}
\tau_{XA}\equiv\sum_{x}p_{X}(x)|x\rangle\langle x|_{X}\otimes|\phi_{x}%
\rangle\langle\phi_{x}|_{A},
\end{equation}
and $\rho_{XB}$ and $\tau_{XB}$ are the states that result from sending the
$A$ system of $\rho_{XA}$ and $\tau_{XA}$\ through the quantum channel
$\mathcal{N}_{A\rightarrow B}$, respectively.
\end{theorem}

\begin{proof}
Suppose that $\rho_{XA}$ is any state of the form in
\eqref{eq-add:cq-state-mixed}. Consider a spectral decomposition of the states
$\rho_{A}^{x}$:%
\begin{equation}
\rho_{A}^{x}=\sum_{y}p_{Y|X}(y|x)\psi_{A}^{x,y},
\end{equation}
where the states $\psi_{A}^{x,y}$ are pure. Then let $\sigma_{XYA}$ denote the
following state:%
\begin{equation}
\sigma_{XYA}\equiv\sum_{x}p_{Y|X}(y|x)p_{X}(x)|x\rangle\langle x|_{X}%
\otimes|y\rangle\langle y|_{Y}\otimes\psi_{A}^{x,y},
\end{equation}
so that $\operatorname{Tr}_{Y}\{\sigma_{XYA}\}=\rho_{XA}$. Also, observe that
$\sigma_{XYA}$ is a state of the form $\tau_{XA}$ with $XY$ as the classical
system. Let $\sigma_{XYB}$ denote the state that results from sending the $A$
system through the quantum channel $\mathcal{N}_{A\rightarrow B}$. Then the
following relations hold:%
\begin{equation}
I(X;B)_{\rho}=I(X;B)_{\sigma}\leq I(XY;B)_{\sigma}.
\end{equation}
The equality follows because $\operatorname{Tr}_{Y}\{\sigma_{XYB}\}=\rho_{XB}$
and the inequality follows from the quantum data-processing inequality. It
then suffices to consider ensembles with only pure states because the state
$\sigma_{XYB}$ is a state of the form $\tau_{XB}$ with the combined system
$XY$ acting as the classical system.
\end{proof}

\subsubsection{Concavity in the Distribution and Convexity in the Signal
States}

We now show that the Holevo information%
\index{Holevo information!of a channel!concavity}
is concave as a function of the input distribution when the signal states are fixed.

\begin{theorem}
\label{thm-ie:holevo-concave-input}The Holevo information $I(X;B)$\ is concave
in the input distribution when the signal states are fixed, in the sense that%
\begin{equation}
\lambda I(X;B)_{\sigma}+\left(  1-\lambda\right)  I(X;B)_{\tau}\leq
I(X;B)_{\omega},
\end{equation}
where $\sigma_{XB}$ and $\tau_{XB}$ are of the form%
\begin{align}
\sigma_{XB}  &  \equiv\sum_{x}p_{X}(x)|x\rangle\langle x|_{X}\otimes
\mathcal{N}(\rho^{x}),\\
\tau_{XB}  &  \equiv\sum_{x}q_{X}(x)|x\rangle\langle x|_{X}\otimes
\mathcal{N}(\rho^{x}),
\end{align}
and $\omega_{XB}$\ is a mixture of the states $\sigma_{XB}$ and $\tau_{XB}$ of
the form%
\begin{equation}
\omega_{XB}\equiv\sum_{x}\left[  \lambda p_{X}(x)+\left(  1-\lambda\right)
q_{X}(x)\right]  |x\rangle\langle x|_{X}\otimes\mathcal{N}(\rho^{x}),
\end{equation}
where $0\leq\lambda\leq1$.
\end{theorem}

\begin{proof}
Let $\omega_{XUB}$ be the state%
\begin{equation}
\omega_{XUB}\equiv\sum_{x}\left[  p_{X}(x)|x\rangle\langle x|_{X}%
\otimes\lambda|0\rangle\langle0|_{U}+q_{X}(x)|x\rangle\langle x|_{X}%
\otimes\left(  1-\lambda\right)  |1\rangle\langle1|_{U}\right]  \otimes
\mathcal{N}(\rho^{x}).
\end{equation}
Observe that $\operatorname{Tr}_{U}\left\{  \omega_{XUB}\right\}  =\omega
_{XB}$. Then the statement of concavity is equivalent to $I(X;B|U)_{\omega
}\leq I(X;B)_{\omega}$. We can rewrite this as $H(B|U)_{\omega}%
-H(B|UX)_{\omega}\leq H(B)_{\omega}-H(B|X)_{\omega}$. Observe that
$H(B|UX)_{\omega}=H(B|X)_{\omega}$, i.e., one can calculate that both of these
conditional entropies are equal to%
\begin{equation}
\sum_{x}\left[  \lambda p_{X}(x)+\left(  1-\lambda\right)  q_{X}(x)\right]
H(\mathcal{N}(\rho^{x})).
\end{equation}
The statement of concavity then becomes $H(B|U)_{\omega}\leq H(B)_{\omega}$,
which follows from concavity of quantum entropy.
\end{proof}

The Holevo information%
\index{Holevo information!of a channel!concavity}
is convex as a function of the signal states when the input distribution is fixed.

\begin{theorem}
\label{thm-ie:holevo-convex-input-states}The Holevo information $I(X;B)$\ is
convex in the signal states when the input distribution is fixed, in the sense
that%
\begin{equation}
\lambda I(X;B)_{\sigma}+\left(  1-\lambda\right)  I(X;B)_{\tau}\geq
I(X;B)_{\omega},
\end{equation}
where $\sigma_{XB}$ and $\tau_{XB}$ are of the form%
\begin{align}
\sigma_{XB}  &  \equiv\sum_{x}p_{X}(x)|x\rangle\langle x|_{X}\otimes
\mathcal{N}(\sigma^{x}),\\
\tau_{XB}  &  \equiv\sum_{x}p_{X}(x)|x\rangle\langle x|_{X}\otimes
\mathcal{N}(\tau^{x}),
\end{align}
and $\omega_{XB}$\ is a mixture of the states $\sigma_{XB}$ and $\tau_{XB}$ of
the form%
\begin{equation}
\omega_{XB}\equiv\sum_{x}p_{X}(x)|x\rangle\langle x|_{X}\otimes\mathcal{N}(
\lambda\sigma^{x}+\left(  1-\lambda\right)  \tau^{x}) ,
\end{equation}
where $0\leq\lambda\leq1$.
\end{theorem}

\begin{proof}
Let $\omega_{XUB}$ be the state%
\begin{equation}
\omega_{XUB}\equiv\sum_{x}p_{X}(x)|x\rangle\langle x|_{X}\otimes\left[
\lambda|0\rangle\langle0|_{U}\otimes\mathcal{N}(\sigma^{x})+\left(
1-\lambda\right)  |1\rangle\langle1|_{U}\otimes\mathcal{N}(\tau^{x})\right]  .
\end{equation}
Observe that $\operatorname{Tr}_{U}\left\{  \omega_{XUB}\right\}  =\omega
_{XB}$. Then convexity in the input states is equivalent to the statement
$I(X;B|U)_{\omega}\geq I(X;B)_{\omega}$. Consider that $I(X;B|U)_{\omega
}=I(X;BU)_{\omega}-I(X;U)_{\omega}$, by the chain rule for the quantum mutual
information. Since the input distribution $p_{X}(x)$ is fixed, there are no
correlations between $X$ and the convexity variable $U$, so that
$I(X;U)_{\omega}=0$. Thus, the above inequality is equivalent to
$I(X;BU)_{\omega}\geq I(X;B)_{\omega}$, which follows from the quantum
data-processing inequality.
\end{proof}

In the above two theorems, we have shown that the Holevo information is either
concave or convex depending on whether the signal states or the input
distribution are fixed, respectively. Thus, the computation of the Holevo
information of a general quantum channel becomes difficult as the input
dimension of the channel grows larger, since a local maximum of the Holevo
information is not necessarily a global maximum. However, if the channel has a
classical input and a quantum output, the computation of the Holevo
information is straightforward because the only input parameter is the input
distribution, and we proved that the Holevo information is a concave function
of the input distribution.

\section{Mutual Information of a Quantum Channel}

\label{sec-add:mut-info-channel}We now consider a measure of the ability of a
quantum channel to preserve quantum
\index{quantum mutual information!of a channel}
correlations. The way that we arrive at this measure is similar to what we
have seen before. Alice prepares some pure quantum state $\phi_{AA^{\prime}}$
in her laboratory, and inputs the $A^{\prime}$ system to a quantum channel
$\mathcal{N}_{A^{\prime}\rightarrow B}$---this transmission gives rise to the
following bipartite state:%
\begin{equation}
\rho_{AB}=\mathcal{N}_{A^{\prime}\rightarrow B}(\phi_{AA^{\prime}}).
\end{equation}
The quantum mutual information $I(A;B)_{\rho}$ is a static measure of
correlations present in the state $\rho_{AB}$. To maximize the correlations
that the quantum channel can establish, Alice should maximize the quantum
mutual information $I(A;B)_{\rho}$ with respect to all possible pure states
that she can input to the channel $\mathcal{N}_{A^{\prime}\rightarrow B}$.
This procedure leads to the definition of the mutual information
$I(\mathcal{N})$\ of a quantum channel:%
\begin{equation}
I(\mathcal{N})\equiv\max_{\phi_{AA^{\prime}}}I(A;B)_{\rho}.
\label{eq-ie:mutual-info-q-channel}%
\end{equation}

The mutual information of a quantum channel corresponds to an important
operational task that is not particularly obvious from the above discussion.
Suppose that Alice and Bob share unlimited bipartite entanglement in whatever
form they wish, and suppose they have access to a large number of independent
uses of the channel $\mathcal{N}_{A^{\prime}\rightarrow B}$. Then the mutual
information of the channel corresponds to the maximal amount of classical
information that they can transmit in such a setting. This setting is the
noisy analog of the super-dense coding protocol from
Chapter~\ref{chap:three-noiseless} (recall the discussion in
Section~\ref{sec-3np:extensions-qst}). By teleportation, the maximal amount of
quantum information that they can transmit is half of the mutual information
of the channel. We discuss how to prove these statements rigorously in
Chapter~\ref{chap:EA-classical}.

\subsection{Additivity}

There might be little
\index{quantum mutual information!of a channel!additivity}
reason to expect that the quantum mutual information of a quantum channel is
additive, given that the Holevo information is not. But perhaps surprisingly,
additivity does hold for the mutual information of a quantum channel! This
result means that we completely understand this measure of information
throughput, and it also means that we understand the operational task to which
it corresponds (entanglement-assisted classical coding discussed in the
previous section).

\begin{figure}[ptb]
\begin{center}
\includegraphics[
width=2.0826in
]{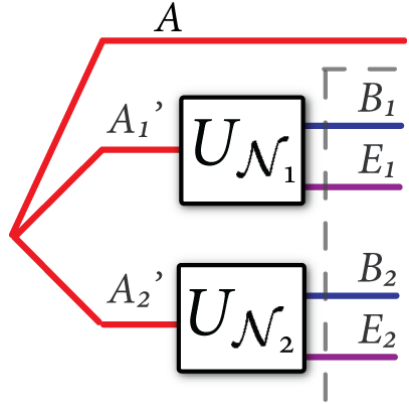}
\end{center}
\caption{This figure displays the scenario for determining whether the mutual
information of two quantum channels $\mathcal{N}_{1}$ and $\mathcal{N}_{2}%
$\ is additive. The question of additivity is equivalent to the possibility of
quantum correlations between channel inputs being able to enhance the mutual
information of two quantum channels. The result stated in
Theorem~\ref{thm-add:mut-info-additive}\ is that the mutual information is
additive for any two quantum channels, so that quantum correlations cannot
enhance it.}%
\label{fig-add:mut-info}%
\end{figure}

We might intuitively attempt to explain this phenomenon in terms of this
operational task---Alice and Bob already share unlimited entanglement between
their terminals and so entangled correlations at the input of the channel do
not lead to any superadditive effect as it does for the Holevo information.
This explanation is somewhat rough, but perhaps the additivity proof explains
best why additivity holds. The crucial tool in the proof is the chain rule for
mutual information and one application of subadditivity of entropy
(Corollary~\ref{cor-qie:subadd}). Figure~\ref{fig-add:mut-info}\ illustrates
the setting corresponding to the analysis for additivity of the mutual
information of a quantum channel.

\begin{theorem}
[Additivity of Q.~Mutual Information of Q.~Channels]%
\label{thm-add:mut-info-additive}Let $\mathcal{N}$ and $\mathcal{M}$ be any
quantum channels. Then the mutual information of the tensor-product channel
$\mathcal{N}\otimes\mathcal{M}$ is the sum of their individual mutual
informations:%
\begin{equation}
I(\mathcal{N}\otimes\mathcal{M})=I(\mathcal{N})+I(\mathcal{M}).
\end{equation}

\end{theorem}

\begin{proof}
We first prove the trivial inequality $I(\mathcal{N}\otimes\mathcal{M})\geq
I(\mathcal{N})+I(\mathcal{M})$. Let $\phi_{A_{1}A_{1}^{\prime}}$ and
$\psi_{A_{2}A_{2}^{\prime}}$ be states that maximize the respective mutual
informations $I(\mathcal{N})$ and $I(\mathcal{M})$. Let%
\begin{equation}
\rho_{A_{1}A_{2}B_{1}B_{2}}\equiv(\mathcal{N}_{A_{1}^{\prime}\rightarrow
B_{1}}\otimes\mathcal{M}_{A_{2}^{\prime}\rightarrow B_{2}})(\phi_{A_{1}%
A_{1}^{\prime}}\otimes\psi_{A_{2}A_{2}^{\prime}}).
\end{equation}
Observe that the state $\rho_{A_{1}A_{2}B_{1}B_{2}}$ is a particular state of
the form required in the maximization of $I(\mathcal{N}\otimes\mathcal{M})$,
by taking $A\equiv A_{1}A_{2}$. Then the following holds%
\begin{align}
I(\mathcal{N})+I(\mathcal{M})  &  =I(A_{1};B_{1})_{\mathcal{N}(\phi)}%
+I(A_{2};B_{2})_{\mathcal{M}(\psi)}\\
&  =I(A_{1}A_{2};B_{1}B_{2})_{\rho}\\
&  \leq I(\mathcal{N}\otimes\mathcal{M}).
\end{align}
The first equality follows by evaluating the mutual informations
$I(\mathcal{N})$ and $I(\mathcal{M})$ with respect to the maximizing states
$\phi_{A_{1}A_{1}^{\prime}}$ and $\psi_{A_{2}A_{2}^{\prime}}$. The second
equality follows from the fact that mutual information is additive with
respect to tensor-product states (Exercise~\ref{ex-qie:add-mutual-info}). The
final inequality follows because the input state $\phi_{A_{1}A_{1}^{\prime}%
}\otimes\psi_{A_{2}A_{2}^{\prime}}$ is a particular input state of the more
general form $\phi_{AA_{1}^{\prime}A_{2}^{\prime}}$ needed in the maximization
of the quantum mutual information of the tensor-product channel $\mathcal{N}%
\otimes\mathcal{M}$.

We now prove the non-trivial inequality $I(\mathcal{N}\otimes\mathcal{M})\leq
I(\mathcal{N})+I(\mathcal{M})$. Let $\phi_{AA_{1}^{\prime}A_{2}^{\prime}}$ be
a state that maximizes the mutual information $I(\mathcal{N}\otimes
\mathcal{M})$ and let%
\begin{equation}
\rho_{AB_{1}B_{2}}\equiv(\mathcal{N}_{A_{1}^{\prime}\rightarrow B_{1}}%
\otimes\mathcal{M}_{A_{2}^{\prime}\rightarrow B_{2}})(\phi_{AA_{1}^{\prime
}A_{2}^{\prime}}),
\end{equation}
Consider the following chain of inequalities:%
\begin{align}
I(\mathcal{N}\otimes\mathcal{M})  &  =I(A;B_{1}B_{2})_{\rho}\\
&  =I(A;B_{1})_{\rho}+I(AB_{1};B_{2})_{\rho}-I(B_{1};B_{2})_{\rho}\\
&  \leq I(A;B_{1})_{\rho}+I(AB_{1};B_{2})_{\rho}\\
&  \leq I(\mathcal{N})+I(\mathcal{M}).
\end{align}
The first equality follows from the definition of $I(\mathcal{N}%
\otimes\mathcal{M})$ in \eqref{eq-ie:mutual-info-q-channel} and evaluating
$I(A;B_{1}B_{2})$ with respect to the maximizing state $\phi$. The second
equality follows by expanding the quantum mutual information, using
Exercise~\ref{ex-qie:chain-rule-mut-info}. The first inequality follows
because $I(B_{1};B_{2})_{\rho}\geq0$. The last inequality follows because
$I(A;B_{1})_{\rho}\leq I(\mathcal{N})$ and $I(AB_{1};B_{2})_{\rho}\leq
I(\mathcal{M})$, where we have applied the result of
Exercise~\ref{ex-add:mut-info-pure-suffice} and the fact that the $A$ system
extends the $A_{1}$ system which is input to the channel $\mathcal{N}$ and the
fact that the $AB_{1}$ systems extend the $A_{2}$ system which is input to the
channel $\mathcal{M}$.
\end{proof}

\begin{corollary}
\label{cor-add:mutual-info-single-letter}The regularized mutual information of
a quantum channel $\mathcal{N}$ is equal to its mutual information:%
\begin{equation}
I_{\operatorname{reg}}(\mathcal{N})=I(\mathcal{N}).
\end{equation}

\end{corollary}

\begin{proof}
A proof of this property uses the same induction argument as in
Corollary~\ref{cor-ie:class-mut=reg-class-mut}\ and exploits the additivity
property in Theorem~\ref{thm-add:mut-info-additive} above.
\end{proof}

\begin{exercise}
[Alternate Mutual Information of a Quantum Channel]\label{ex-add:alt-mut-info}%
Let $\rho_{XAB}$ denote a state of the following form:%
\begin{equation}
\rho_{XAB}\equiv\sum_{x}p_{X}(x)|x\rangle\langle x|_{X}\otimes\mathcal{N}%
_{A^{\prime}\rightarrow B}(\phi_{AA^{\prime}}^{x}).
\end{equation}
Consider the following alternate definition of the mutual information of a
quantum channel:%
\begin{equation}
I_{\operatorname{alt}}(\mathcal{N})\equiv\max_{\rho_{XAB}}I(AX;B),
\end{equation}
where the maximization is with respect to states of the form $\rho_{XAB}$.
Show that%
\begin{equation}
I_{\operatorname{alt}}(\mathcal{N})=I(\mathcal{N}).
\end{equation}

\end{exercise}

\begin{exercise}
Compute the mutual information of a dephasing channel with dephasing
parameter$~p$.
\end{exercise}

\begin{exercise}
Compute the mutual information of an erasure channel with erasure
parameter$~\varepsilon$.
\end{exercise}

\begin{exercise}
[Pure States are Sufficient]\label{ex-add:mut-info-pure-suffice}Let
$\mathcal{N}_{A^{\prime}\rightarrow B}$ be a quantum channel. Show that it is
sufficient to consider pure states $\phi_{AA^{\prime}}$ for determining the
mutual information of a quantum channel. That is, one does not need to
consider mixed states $\rho_{AA^{\prime}}$ in the optimization task because%
\begin{equation}
\max_{\phi_{AA^{\prime}}}I(A;B)_{\mathcal{N}(\phi)}=\max_{\rho_{AA^{\prime}}%
}I(A;B)_{\mathcal{N}(\rho)}.
\end{equation}
(Hint: Consider purifying and using the quantum data-processing inequality.)
\end{exercise}

\subsection{Optimizing the Mutual Information of a Quantum Channel}

We now show that the mutual information of a quantum channel is concave as a
function of the input state. This result allows us to compute this quantity
with standard convex optimization techniques.

\begin{theorem}
\label{thm-ie:mutual-concave-input}The mutual information $I(A;B)$\ is concave
in the input state, in the sense that%
\begin{equation}
\sum_{x}p_{X}(x)I(A;B)_{\rho_{x}}\leq I(A;B)_{\sigma},
\end{equation}
where $\rho_{AB}^{x}\equiv\mathcal{N}_{A^{\prime}\rightarrow B}(\phi
_{AA^{\prime}}^{x})$, $\sigma_{A^{\prime}}\equiv\sum_{x}p_{X}(x)\rho
_{A^{\prime}}^{x}$, $\phi_{AA^{\prime}}$ is a purification of $\sigma
_{A^{\prime}}$, and $\sigma_{AB}\equiv\mathcal{N}_{A^{\prime}\rightarrow
B}(\phi_{AA^{\prime}})$.
\end{theorem}

\begin{proof}
Let $\rho_{XABE}$ be the following classical--quantum state:%
\begin{equation}
\rho_{XABE}\equiv\sum_{x}p_{X}(x)|x\rangle\langle x|_{X}\otimes\mathcal{U}%
_{A^{\prime}\rightarrow BE}^{\mathcal{N}}(\phi_{AA^{\prime}}^{x}),
\end{equation}
where $U_{A^{\prime}\rightarrow BE}^{\mathcal{N}}$ is an isometric extension
of the channel. Consider the following chain of inequalities:%
\begin{align}
\sum_{x}p_{X}(x)I(A;B)_{\rho_{x}}  &  =I(A;B|X)_{\rho}\\
&  =H(A|X)_{\rho}+H(B|X)_{\rho}-H(AB|X)_{\rho}\\
&  =H(BE|X)_{\rho}+H(B|X)_{\rho}-H(E|X)_{\rho}\\
&  =H(B|EX)_{\rho}+H(B|X)_{\rho}\\
&  \leq H(B|E)_{\rho}+H(B)_{\rho}\\
&  =H(B|E)_{\sigma}+H(B)_{\sigma}\\
&  =I(A;B)_{\sigma}.
\end{align}
The first equality follows because the conditioning system $X$ in
$I(A;B|X)_{\rho}$ is classical. The second equality follows by expanding the
quantum mutual information. The third equality follows because the state on
$ABE$ is pure when conditioned on $X$. The fourth equality follows from the
definition of conditional quantum entropy. The inequality follows from strong
subadditivity and concavity of quantum entropy. The equality follows by
inspecting the definition of the state $\sigma$, and the final equality
follows because the state is pure on systems $ABE$.
\end{proof}

\section{Coherent Information of a Quantum Channel}

\label{sec-add:coh-info}This section presents an alternative, important
measure of the ability of a
\index{coherent information!of a channel}%
quantum channel to preserve quantum correlations: the coherent information of
the channel. The way we arrive at this measure is similar to how we did for
the mutual information of a quantum channel. Alice prepares a pure state
$\phi_{AA^{\prime}}$ and inputs the $A^{\prime}$ system to a quantum channel
$\mathcal{N}_{A^{\prime}\rightarrow B}$. This transmission leads to a
bipartite state $\rho_{AB}$ where%
\begin{equation}
\rho_{AB}=\mathcal{N}_{A^{\prime}\rightarrow B}(\phi_{AA^{\prime}}).
\end{equation}
The coherent information of the state that arises from the channel is as
follows: $I(A\rangle B)_{\rho}=H(B)_{\rho}-H(AB)_{\rho}$, leading to our next definition.

\begin{definition}
[Coherent Information of a Quantum Channel]The coherent information
$Q(\mathcal{N})$ of a quantum channel is the maximum of the coherent
information with respect to all input pure states:%
\begin{equation}
Q(\mathcal{N})\equiv\max_{\phi_{AA^{\prime}}}I(A\rangle B)_{\rho}.
\end{equation}

\end{definition}

The coherent information of a quantum channel corresponds to an important
operational task (perhaps the most important for quantum information). It is a
good lower bound on the capacity for Alice to transmit quantum information to
Bob, but it is actually equal to such a quantum communication capacity of a
quantum channel in some special cases. We prove these results  in
Chapter~\ref{chap:quantum-capacity}.

\begin{exercise}
\label{ex-add:alt-coh-info}Let $I_{c}(\rho,\mathcal{N})$ denote the coherent
information of a channel $\mathcal{N}$ when state $\rho$ is its input:%
\begin{equation}
I_{c}(\rho,\mathcal{N})\equiv H(\mathcal{N}(\rho))-H(\mathcal{N}^{c}(\rho)),
\end{equation}
where $\mathcal{N}^{c}$ is a channel complementary to the original channel
$\mathcal{N}$. Show that%
\begin{equation}
Q(\mathcal{N})=\max_{\rho}I_{c}(\rho,\mathcal{N}).
\end{equation}
An equivalent way of writing the above expression on the right-hand side\ is%
\begin{equation}
\max_{\phi_{AA^{\prime}}}\left[  H(B)_{\psi}-H(E)_{\psi}\right]  ,
\end{equation}
where $|\psi\rangle_{ABE}\equiv U_{A^{\prime}\rightarrow BE}^{\mathcal{N}%
}|\phi\rangle_{AA^{\prime}}$ and $U_{A^{\prime}\rightarrow BE}^{\mathcal{N}}$
is an isometric extension of the channel $\mathcal{N}$.
\end{exercise}

The following property points out that the coherent information of a channel
is always non-negative, even though the coherent information of any given
state can sometimes be negative.

\begin{property}
[Non-Negativity of Channel Coherent Information]The coherent information $Q(
\mathcal{N}) $ of a quantum channel $\mathcal{N}$\ is non-negative:%
\index{coherent information!of a channel!positivity}
\begin{equation}
Q( \mathcal{N}) \geq0.
\end{equation}

\end{property}

\begin{proof}
We can choose the input state $\phi_{AA^{\prime}}$ to be a product state of
the form $\psi_{A}\otimes\varphi_{A^{\prime}}$. The coherent information of
this state vanishes:%
\begin{align}
I( A\rangle B) _{\psi\otimes\mathcal{N}(\varphi)}  &  =H( B) _{\mathcal{N}%
(\varphi)}-H( AB) _{\psi\otimes\mathcal{N}(\varphi)}\\
&  =H( B) _{\mathcal{N}(\varphi)}-H( A) _{\psi}-H( B) _{\mathcal{N}(\varphi
)}\\
&  =0.
\end{align}
The first equality follows by evaluating the coherent information for the
product state. The second equality follows because the state on $AB$ is
product. The last equality follows because the state on $A$ is pure.
Non-negativity then holds because the coherent information of a channel can
only be greater than or equal to this amount, given that it involves a
maximization over all input states and the above state is a particular input state.
\end{proof}

\subsection{Additivity of Coherent Information for Some Channels}

The coherent information
\index{coherent information!of a channel!additivity for a degradable channel}
of a quantum channel is generally not additive for arbitrary quantum channels.
One might potentially view this situation as unfortunate, but it implies that
quantum Shannon theory is a richer theory than its classical counterpart.
Attempts to understand why and how this quantity is not additive have led to
many breakthroughs (see Section~\ref{sec-q-cap:strangeness}).

\textit{Degradable} quantum channels%
\index{degradable channel}
form a special class of channels for which the coherent information is
additive. These channels have a property that is analogous to a property of
the degraded wiretap channels from Section~\ref{sec-add:private-info-class}.
To understand this property, recall that any quantum channel $\mathcal{N}%
_{A^{\prime}\rightarrow B}$ has a complementary channel $\mathcal{N}%
_{A^{\prime}\rightarrow E}^{c}$, realized by considering an isometric
extension of the channel and tracing over Bob's system.

\begin{definition}
[Degradable Quantum Channel]A quantum channel $\mathcal{N}_{A^{\prime
}\rightarrow B}$\ is degradable if there exists a degrading channel
$\mathcal{D}_{B\rightarrow E}$ such that%
\begin{equation}
\mathcal{N}_{A^{\prime}\rightarrow E}^{c}(\rho_{A^{\prime}})=\mathcal{D}%
_{B\rightarrow E}(\mathcal{N}_{A^{\prime}\rightarrow B}(\rho_{A^{\prime}})),
\end{equation}
for any input state $\rho_{A^{\prime}}$ and where $\mathcal{N}_{A^{\prime
}\rightarrow E}^{c}$ is a channel complementary to $\mathcal{N}_{A^{\prime
}\rightarrow B}$.
\end{definition}

The intuition behind a
\index{degradable channel}%
degradable quantum channel is that the channel from Alice to Eve is noisier
than the channel from Alice to Bob, in the sense that Bob can simulate the
channel to Eve by applying a degrading channel to his system. The picture to
consider for the analysis of additivity is the same as that in
Figure~\ref{fig-add:mut-info}.

There are also antidegradable channels, defined in the opposite way:

\begin{definition}
[Antidegradable Quantum Channel]%
\index{antidegradable channel}%
A quantum channel $\mathcal{N}_{A^{\prime}\rightarrow B}$\ is antidegradable
if there exists a degrading channel $\mathcal{D}_{E\rightarrow B}$ such that%
\begin{equation}
\mathcal{D}_{E\rightarrow B}(\mathcal{N}_{A^{\prime}\rightarrow E}^{c}%
(\rho_{A^{\prime}}))=\mathcal{N}_{A^{\prime}\rightarrow B}(\rho_{A^{\prime}}),
\end{equation}
for any input state $\rho_{A^{\prime}}$ and where $\mathcal{N}_{A^{\prime
}\rightarrow E}^{c}$ is a channel complementary to $\mathcal{N}_{A^{\prime
}\rightarrow B}$.
\end{definition}

\begin{theorem}
[Additivity for Degradable Quantum Channels]\label{thm:coh-info-additivity}%
\index{degradable channel}%
Let$~\mathcal{N}$ and $\mathcal{M}$ be any quantum channels that are
degradable. Then the coherent information of the tensor-product channel
$\mathcal{N}\otimes\mathcal{M}$ is the sum of their individual coherent
informations:%
\begin{equation}
Q(\mathcal{N}\otimes\mathcal{M})=Q(\mathcal{N})+Q(\mathcal{M}).
\end{equation}

\end{theorem}

\begin{proof}
We leave the proof of the inequality $Q(\mathcal{N}\otimes\mathcal{M})\geq
Q(\mathcal{N})+Q(\mathcal{M})$ as Exercise~\ref{ex-add:coh-info-superadd}%
\ below, and we prove the non-trivial inequality $Q(\mathcal{N}\otimes
\mathcal{M})\leq Q(\mathcal{N})+Q(\mathcal{M})$ that holds when quantum
channels $\mathcal{N}$ and $\mathcal{M}$ are degradable. Consider a pure state
$\phi_{AA_{1}^{\prime}A_{2}^{\prime}}$ that serves as the input to the two
quantum channels. Let $U_{A_{1}^{\prime}\rightarrow B_{1}E_{1}}^{\mathcal{N}}$
denote an isometric extension of the first channel and let $U_{A_{2}^{\prime
}\rightarrow B_{2}E_{2}}^{\mathcal{M}}$ denote an isometric extension of the
second channel. Let%
\begin{align}
\sigma_{AB_{1}E_{1}A_{2}^{\prime}}  &  \equiv U^{\mathcal{N}}\phi\left(
U^{\mathcal{N}}\right)  ^{\dag},\\
\theta_{AA_{1}^{\prime}B_{2}E_{2}}  &  \equiv U^{\mathcal{M}}\phi\left(
U^{\mathcal{M}}\right)  ^{\dag},\\
\rho_{AB_{1}E_{1}B_{2}E_{2}}  &  \equiv\left(  U^{\mathcal{N}}\otimes
U^{\mathcal{M}}\right)  \phi\left(  \left(  U^{\mathcal{N}}\right)  ^{\dag
}\otimes\left(  U^{\mathcal{M}}\right)  ^{\dag}\right)  .
\end{align}
We need to show that $Q(\mathcal{N}\otimes\mathcal{M})=Q(\mathcal{N}%
)+Q(\mathcal{M})$ when both channels are degradable. Furthermore, let
$\rho_{AB_{1}E_{1}B_{2}E_{2}}$ be a state that maximizes $Q(\mathcal{N}%
\otimes\mathcal{M})$. Consider the following chain of inequalities:%
\begin{align}
Q(\mathcal{N}\otimes\mathcal{M})  &  =I(A\rangle B_{1}B_{2})_{\rho}\\
&  =H(B_{1}B_{2})_{\rho}-H(AB_{1}B_{2})_{\rho}\\
&  =H(B_{1}B_{2})_{\rho}-H(E_{1}E_{2})_{\rho}\\
&  =H(B_{1})_{\rho}-H(E_{1})_{\rho}+H(B_{2})_{\rho}-H(E_{2})_{\rho}\nonumber\\
&  \ \ \ \ \ \ \ -\left[  I(B_{1};B_{2})_{\rho}-I(E_{1};E_{2})_{\rho}\right]
\\
&  \leq H(B_{1})_{\rho}-H(E_{1})_{\rho}+H(B_{2})_{\rho}-H(E_{2})_{\rho}\\
&  =H(B_{1})_{\sigma}-H(AA_{2}^{\prime}B_{1})_{\sigma}+H(B_{2})_{\theta
}-H(AA_{1}^{\prime}B_{2})_{\theta}\\
&  =I(AA_{2}^{\prime}\rangle B_{1})_{\sigma}+I(AA_{1}^{\prime}\rangle
B_{2})_{\theta}\\
&  \leq Q(\mathcal{N})+Q(\mathcal{M}).
\end{align}
The first equality follows from the definition of $Q(\mathcal{N}%
\otimes\mathcal{M})$ and because we set $\rho$ to be a state that maximizes
the tensor-product channel coherent information. The second equality follows
from the definition of coherent information, and the third equality follows
because the state $\rho$ is pure on systems $AB_{1}E_{1}B_{2}E_{2}$. The
fourth equality follows by expanding the entropies in the previous line. The
first inequality follows because there is a degrading channel from both
$B_{1}$ to $E_{1}$ and $B_{2}$ to $E_{2}$, allowing us to apply the quantum
data-processing inequality to get $I(B_{1};B_{2})_{\rho}\geq I(E_{1}%
;E_{2})_{\rho}$. The fifth equality follows because the entropies of $\rho$,
$\sigma$, and $\theta$ on the given reduced systems are equal and because the
state $\sigma$ on systems $AA_{2}^{\prime}B_{1}E_{1}$ is pure and the state
$\theta$ on systems $AA_{1}^{\prime}B_{2}E_{2}$ is pure. The last equality
follows from the definition of coherent information, and the final inequality
follows because the coherent informations are less than their respective
maximizations over all possible input states.
\end{proof}

\begin{corollary}
The regularized coherent information of a degradable%
\index{degradable channel}
quantum channel is equal to its coherent information: $Q_{\operatorname{reg}%
}(\mathcal{N})=Q(\mathcal{N})$.
\end{corollary}

\begin{proof}
A proof of this property uses the same induction argument as in
Corollary~\ref{cor-ie:class-mut=reg-class-mut}\ and exploits the additivity
property in Theorem~\ref{thm:coh-info-additivity} above.
\end{proof}

\begin{exercise}
\label{ex-add:erasure-degradable}%
\index{degradable channel}%
Consider the quantum erasure channel where the erasure parameter
$\varepsilon\in\left[  0,1/2\right]  $. Find the channel that degrades this
one, reproducing the channel from input to environment.
\end{exercise}

\begin{exercise}
[Superadditivity of Coherent Information]\label{ex-add:coh-info-superadd}Show
that the coherent information of the tensor-product channel $\mathcal{N}%
\otimes\mathcal{M}$ is never less than the sum of their individual coherent
informations: $Q(\mathcal{N}\otimes\mathcal{M})\geq Q(\mathcal{N}%
)+Q(\mathcal{M})$.
\end{exercise}

\begin{exercise}%
\index{degradable channel}%
Prove using monotonicity of relative entropy that the coherent information is
subadditive for a degradable channel: $Q(\mathcal{N})+Q(\mathcal{M})\geq
Q(\mathcal{N}\otimes\mathcal{M})$.
\end{exercise}

\begin{exercise}
Consider a quantity known as the reverse coherent information:%
\begin{equation}
Q_{\operatorname{rev}}(\mathcal{N})\equiv\max_{\phi_{AA^{\prime}}}I(B\rangle
A)_{\omega},
\end{equation}
where $\omega_{AB}\equiv\mathcal{N}_{A^{\prime}\rightarrow B}(\phi
_{AA^{\prime}})$. Show that the reverse coherent information is additive with
respect to any quantum channels $\mathcal{N}$ and $\mathcal{M}$:
$Q_{\operatorname{rev}}(\mathcal{N}\otimes\mathcal{M})=Q_{\operatorname{rev}%
}(\mathcal{N})+Q_{\operatorname{rev}}(\mathcal{M})$.
\end{exercise}

\begin{exercise}
\label{ex-add:antidegrad-zero-coh-info}Prove that the
\index{antidegradable channel}%
coherent information of an antidegradable channel is equal to zero. (Hint:
Consider using the identity from Exercise~\ref{ex-qie:entropy-games}.)
\end{exercise}

\begin{exercise}
Prove that an
\index{entanglement-breaking channel}%
entanglement-breaking channel is antidegradable.
\end{exercise}

\subsection{Optimizing the Coherent Information}

We would like to determine how difficult it is to maximize the coherent
information of a quantum channel. For general channels, this problem is
difficult, but it turns out to be straightforward for the class of degradable%
\index{degradable channel}
quantum channels. Theorem~\ref{thm-add:coh-deg-concave-input} below states an
important property of the coherent information $Q(\mathcal{N})$ of a
degradable quantum channel $\mathcal{N}$\ that allows us to answer this
question. The theorem states that the coherent information $Q(\mathcal{N})$ of
a degradable quantum channel is a concave function of the input density
operator $\rho_{A^{\prime}}$ over which we maximize it. In particular, this
result implies that a local maximum of the coherent information $Q(\mathcal{N}%
)$ is a global maximum since the set of density operators is convex, and the
optimization problem is therefore a straightforward computation that can
exploit convex optimization methods. The theorem below exploits the
characterization of the channel coherent information\ from
Exercise~\ref{ex-add:alt-coh-info}.

\begin{theorem}
\label{thm-add:coh-deg-concave-input}Suppose that a quantum channel
$\mathcal{N}$ is degradable.%
\index{degradable channel}
Then the coherent information $I_{c}\left(  \rho,\mathcal{N}\right)  $\ is
concave in the input density operator:%
\begin{equation}
\sum_{x}p_{X}(x)I_{c}(\rho_{x},\mathcal{N})\leq I_{c}\left(  \sum_{x}%
p_{X}(x)\rho_{x},\mathcal{N}\right)  ,
\end{equation}
where $p_{X}(x)$ is a probability density function and each $\rho_{x}$ is a
density operator.
\end{theorem}

\begin{proof}
Consider the following states:%
\begin{align}
\sigma_{XB}  &  \equiv\sum_{x}p_{X}(x)|x\rangle\langle x|_{X}\otimes
\mathcal{N}(\rho_{x}),\\
\theta_{XE}  &  \equiv\sum_{x}p_{X}(x)|x\rangle\langle x|_{X}\otimes\left(
\mathcal{T}\circ\mathcal{N}\right)  \left(  \rho_{x}\right)  ,
\end{align}
where $\mathcal{T}$ is a degrading channel for the channel $\mathcal{N}$, so
that $\mathcal{T}\circ\mathcal{N}=\mathcal{N}^{c}$. Then the following
statements hold:%
\begin{align}
I(X;B)_{\sigma}  &  \geq I(X;E)_{\theta}\\
\therefore\ \ \ H(B)_{\sigma}-H(B|X)_{\sigma}  &  \geq H(E)_{\theta
}-H(E|X)_{\theta}\\
\therefore\ \ \ H(B)_{\sigma}-H(E)_{\theta}  &  \geq H(B|X)_{\sigma
}-H(E|X)_{\theta}%
\end{align}%
\begin{multline}
\therefore\ \ \ H\left(  \mathcal{N}\left(  \sum_{x}p_{X}(x)\rho_{x}\right)
\right)  -H\left(  \mathcal{N}^{c}\left(  \sum_{x}p_{X}(x)\rho_{x}\right)
\right) \\
\geq\sum_{x}p_{X}(x)\left[  H\left(  \mathcal{N}(\rho_{x})\right)  -H\left(
\mathcal{N}^{c}(\rho_{x})\right)  \right]
\end{multline}%
\begin{equation}
\therefore\ \ \ I_{c}\left(  \sum_{x}p_{X}(x)\rho_{x},\mathcal{N}\right)
\geq\sum_{x}p_{X}(x)I_{c}\left(  \rho_{x},\mathcal{N}\right)  .
\end{equation}
The first statement is the crucial one and follows from the quantum
data-processing inequality and the fact that the map $\mathcal{T}$ degrades
Bob's state to Eve's state. The second and third statements follow from the
definition of quantum mutual information and rearranging entropies. The fourth
statement follows by plugging in the density operators into the entropies in
the previous statement. The final statement follows from the alternate
definition of coherent information in Exercise~\ref{ex-add:alt-coh-info}.
\end{proof}

\section{Private Information of a Quantum Channel}

The private information%
\index{private information!of a quantum channel}
of a quantum channel is the last information measure that we consider in this
chapter. Alice would like to establish classical correlations with Bob, but
does not want the environment of the channel to have access to these classical
correlations. The ensemble that she prepares is similar to the one we
considered for the Holevo information. The expected density operator of the
ensemble she prepares is a classical--quantum state of the form%
\begin{equation}
\rho_{XA^{\prime}}\equiv\sum_{x}p_{X}(x)|x\rangle\langle x|_{X}\otimes
\rho_{A^{\prime}}^{x}. \label{eq-ie:private-info-state}%
\end{equation}
Sending the $A^{\prime}$ system through an isometric extension $U_{A^{\prime
}\rightarrow BE}^{\mathcal{N}}$ of a quantum channel $\mathcal{N}$ leads to a
state $\rho_{XBE}$. A good measure of the private classical correlations that
she can establish with Bob is the difference of the classical correlations she
can establish with Bob, less the classical correlations that Eve can obtain:
$I(X;B)_{\rho}-I(X;E)_{\rho}$, leading to our next definition
(Chapter~\ref{chap:private-cap} discusses the operational task corresponding
to this information quantity).

\begin{definition}
[Private Information of a Quantum Channel]The private information
$P(\mathcal{N})$ of a quantum channel $\mathcal{N}$ is defined as follows:%
\begin{equation}
P(\mathcal{N})\equiv\max_{\rho_{XA^{\prime}}}I(X;B)_{\rho}-I(X;E)_{\rho},
\end{equation}
where the maximization is with respect to all states of the form in
\eqref{eq-ie:private-info-state} and the entropic quantities are evaluated
with respect to the state $\mathcal{U}^{\mathcal{N}}_{A^{\prime}\to BE}%
(\rho_{XA^{\prime}})$.
\end{definition}

\begin{property}
The private information $P( \mathcal{N}) $ of a quantum channel $\mathcal{N}$
is non-negative:%
\begin{equation}
P( \mathcal{N}) \geq0.
\end{equation}

\end{property}

\begin{proof}
We can choose the input state $\rho_{XA^{\prime}}$ to be a state of the form
$|0\rangle\langle0|_{X}\otimes\psi_{A^{\prime}}$, where $\psi_{A^{\prime}}$ is
pure. The private information of the output state vanishes%
\begin{equation}
I(X;B)_{|0\rangle\langle0|\otimes\mathcal{N}(\psi)}-I(X;E)_{|0\rangle
\langle0|\otimes\mathcal{N}^{c}(\psi)}=0.
\end{equation}
The equality follows just by evaluating both mutual informations for the above
state. The above property then holds because the private information of a
channel can only be greater than or equal to this amount, given that it
involves a maximization over all input states and the above state is a
particular input state.
\end{proof}

The regularized private information is as follows:%
\begin{equation}
P_{\operatorname{reg}}(\mathcal{N})=\lim_{n\rightarrow\infty}\frac{1}%
{n}P(\mathcal{N}^{\otimes n}).
\end{equation}

\subsection{Private Information and Coherent Information}

The private information of a quantum channel bears a\ special relationship to
that channel's coherent information. It is always at least as great as the
coherent information of the channel and is equal to it for certain channels.
The following theorem states the former inequality, and the next theorem
states the equivalence for
\index{degradable channel}%
degradable quantum channels.

\begin{theorem}
The private information $P(\mathcal{N})$ of any quantum channel $\mathcal{N}$
is never smaller than its coherent information $Q(\mathcal{N})$:%
\begin{equation}
Q(\mathcal{N})\leq P(\mathcal{N}).
\end{equation}

\end{theorem}

\begin{proof}
We can see this relation through a few steps. Consider a pure state
$\phi_{AA^{\prime}}$ that maximizes the coherent information $Q(\mathcal{N})$,
and let $\phi_{ABE}$ denote the state that arises from sending the $A^{\prime
}$ system through an isometric extension $U_{A^{\prime}\rightarrow
BE}^{\mathcal{N}}$ of the channel $\mathcal{N}$. Let $\phi_{A^{\prime}}$
denote the reduction of this state to the $A^{\prime}$ system. Suppose that it
admits the following spectral decomposition:%
\begin{equation}
\phi_{A^{\prime}}=\sum_{x}p_{X}(x)|\phi_{x}\rangle\langle\phi_{x}|_{A^{\prime
}}.
\end{equation}
We can create an augmented classical--quantum state that correlates a
classical variable with the index $x$:%
\begin{equation}
\sigma_{XA^{\prime}}\equiv\sum_{x}p_{X}(x)|x\rangle\langle x|_{X}\otimes
|\phi_{x}\rangle\langle\phi_{x}|_{A^{\prime}}.
\label{eq-ie:expand-to-private-state}%
\end{equation}
Let $\sigma_{XBE}$ denote the state that results from sending the $A^{\prime}$
system through an isometric extension $U_{A^{\prime}\rightarrow BE}%
^{\mathcal{N}}$ of the channel $\mathcal{N}$. Then the following chain of
inequalities holds:%
\begin{align}
Q(\mathcal{N})  &  =I(A\rangle B)_{\phi}\\
&  =H(B)_{\phi}-H(E)_{\phi}\\
&  =H(B)_{\sigma}-H(E)_{\sigma}\\
&  =H(B)_{\sigma}-H(B|X)_{\sigma}-H(E)_{\sigma}+H(B|X)_{\sigma}\\
&  =I(X;B)_{\sigma}-H(E)_{\sigma}+H(E|X)_{\sigma}\\
&  =I(X;B)_{\sigma}-I(X;E)_{\sigma}\\
&  \leq P(\mathcal{N}).
\end{align}
The first equality follows from evaluating the coherent information of the
state $\phi_{ABE}$ that maximizes the coherent information of the channel. The
second equality follows because the state $\phi_{ABE}$ is pure. The third
equality follows from the definition of $\sigma_{XBE}$ in
\eqref{eq-ie:expand-to-private-state} and its relation to $\phi_{ABE}$. The
fourth equality follows by adding and subtracting $H(B|X)_{\sigma}$, and the
next one follows from the definition of the mutual information $I(X;B)_{\sigma
}$ and the fact that the state of $\sigma_{XBE}$ on systems $B$ and $E$ is
pure when conditioned on $X$. The last equality follows from the definition of
the mutual information $I(X;E)_{\sigma}$. The final inequality follows because
the state $\sigma_{XBE}$ is a particular state of the form in
\eqref{eq-ie:private-info-state}, and $P(\mathcal{N})$ involves a maximization
over all states of that form.
\end{proof}

\begin{theorem}
\label{thm-ie:degradable-priv-coh}Suppose that a quantum channel $\mathcal{N}$
is degradable.%
\index{degradable channel}
Then its private information $P( \mathcal{N}) $ is equal to its coherent
information $Q( \mathcal{N}) $:%
\begin{equation}
P( \mathcal{N}) =Q( \mathcal{N}) .
\end{equation}

\end{theorem}

\begin{proof}
We prove the inequality $P(\mathcal{N})\leq Q(\mathcal{N})$ for degradable%
\index{degradable channel}
quantum channels because we have already proven that $Q(\mathcal{N})\leq
P(\mathcal{N})$ for any quantum channel $\mathcal{N}$. Consider a
classical--quantum state $\rho_{XBE}$ that arises from transmitting the
$A^{\prime}$ system of the state in \eqref{eq-ie:private-info-state} through
an isometric extension $U_{A^{\prime}\rightarrow BE}^{\mathcal{N}}$ of the
channel. Suppose further that this state maximizes $P(\mathcal{N})$. We can
take a spectral decomposition of each $\rho_{A^{\prime}}^{x}$ in the ensemble
to be as follows:%
\begin{equation}
\rho_{A^{\prime}}^{x}=\sum_{y}p_{Y|X}(y|x)\psi_{A^{\prime}}^{x,y},
\end{equation}
where each state $\psi_{A^{\prime}}^{x,y}$ is pure. We can construct the
following extension of the state $\rho_{XBE}$:%
\begin{equation}
\sigma_{XYBE}\equiv\sum_{x,y}p_{Y|X}(y|x)p_{X}(x)|x\rangle\langle
x|_{X}\otimes|y\rangle\langle y|_{Y}\otimes\mathcal{U}_{A^{\prime}\rightarrow
BE}^{\mathcal{N}}(\psi_{A^{\prime}}^{x,y}).
\end{equation}
Then the following chain of inequalities holds:%
\begin{align}
P(\mathcal{N})  &  =I(X;B)_{\rho}-I(X;E)_{\rho}\\
&  =I(X;B)_{\sigma}-I(X;E)_{\sigma}\\
&  =I(XY;B)_{\sigma}-I(Y;B|X)_{\sigma}-\left[  I(XY;E)_{\sigma}%
-I(Y;E|X)_{\sigma}\right] \\
&  =I(XY;B)_{\sigma}-I(XY;E)_{\sigma}-\left[  I(Y;B|X)_{\sigma}%
-I(Y;E|X)_{\sigma}\right]  .
\end{align}
The first equality follows from the definition of $P(\mathcal{N})$ and because
we set $\rho$ to be the state that maximizes it. The second equality follows
because $\rho_{XBE}=\operatorname{Tr}_{Y}\left\{  \sigma_{XYBE}\right\}  $.
The third equality follows from the chain rule for quantum mutual information.
The fourth equality follows from a rearrangement of entropies. Continuing,%
\begin{align}
&  \leq I(XY;B)_{\sigma}-I(XY;E)_{\sigma}\\
&  =H(B)_{\sigma}-H(B|XY)_{\sigma}-H(E)_{\sigma}+H(E|XY)_{\sigma}\\
&  =H(B)_{\sigma}-H(B|XY)_{\sigma}-H(E)_{\sigma}+H(B|XY)_{\sigma}\\
&  =H(B)_{\sigma}-H(E)_{\sigma}\\
&  \leq Q(\mathcal{N}).
\end{align}
The first inequality (the crucial one) follows because there is a degrading
channel from $B$ to $E$, so that the quantum data-processing inequality
implies that $I(Y;B|X)_{\sigma}\geq I(Y;E|X)_{\sigma}$. The second equality is
a rewriting of entropies, the third follows because the state of $\sigma$ on
systems $B$ and $E$ is pure when conditioned on classical systems $X$ and $Y$,
and the fourth follows by canceling entropies. The last inequality follows
because the entropy difference $H(B)_{\sigma}-H(E)_{\sigma}$ is less than the
maximum of that difference over all possible input states.
\end{proof}

\subsection{Additivity of Private Information of Degradable Channels}

\begin{figure}[ptb]
\begin{center}
\includegraphics[
width=2.0358in
]{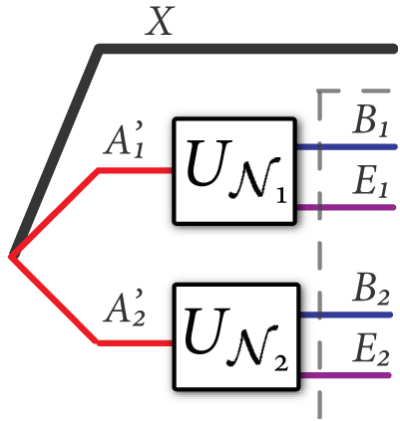}
\end{center}
\caption{This figure displays the scenario for determining whether the private
information of two quantum channels $\mathcal{N}_{1}$ and $\mathcal{N}_{2}%
$\ is additive. The question of additivity is equivalent to the possibility of
quantum correlations between channel inputs being able to enhance the private
information of two quantum channels. The result stated in
Theorem~\ref{thm:private-additivity-degradable}\ is that the private
information is additive for any two degradable quantum channels, so that
quantum correlations cannot enhance it in this case.}%
\label{fig-add:private-add}%
\end{figure}

The private information of
\index{private information!of a quantum channel!additivity for a degradable channel}
general quantum channels is not additive, but it is so in the case of%
\index{degradable channel}
degradable quantum channels. The method of proof is somewhat similar to that
in the proof of Theorem~\ref{thm:coh-info-additivity}, essentially exploiting
the degradability property. Figure~\ref{fig-add:private-add}\ illustrates the
setting to consider for additivity of the private
information.

\begin{theorem}
[Additivity for Degradable Quantum Channels]%
\label{thm:private-additivity-degradable}Let $\mathcal{N}$ and $\mathcal{M}$
be any quantum channels that are
\index{degradable channel}%
degradable. Then the private information of the tensor-product channel
$\mathcal{N}\otimes\mathcal{M}$ is the sum of their individual private
informations:%
\begin{equation}
P(\mathcal{N}\otimes\mathcal{M})=P(\mathcal{N})+P(\mathcal{M}).
\end{equation}
Furthermore, it holds that%
\begin{equation}
P(\mathcal{N}\otimes\mathcal{M})=Q(\mathcal{N}\otimes\mathcal{M}%
)=Q(\mathcal{N})+Q(\mathcal{M}).
\end{equation}

\end{theorem}

\begin{proof}
We first prove the more trivial inequality $P(\mathcal{N}\otimes
\mathcal{M})\geq P(\mathcal{N})+P(\mathcal{M})$. Let $\rho_{X_{1}A_{1}%
^{\prime}}$ and $\sigma_{X_{2}A_{2}^{\prime}}$ be states of the form in
\eqref{eq-ie:private-info-state} that maximize the respective private
informations $P(\mathcal{N})$ and $P(\mathcal{M})$. Let $\theta_{X_{1}%
X_{2}A_{1}^{\prime}A_{2}^{\prime}}$ be the tensor product of these two states:
$\theta=\rho\otimes\sigma$. Let $\rho_{X_{1}B_{1}E_{1}}$ and $\sigma
_{X_{2}B_{2}E_{2}}$ be the states that arise from sending $\rho_{X_{1}%
A_{1}^{\prime}}$ and $\sigma_{X_{2}A_{2}^{\prime}}$ through the respective
isometric extensions $U_{A_{1}^{\prime}\rightarrow B_{1}E_{1}}^{\mathcal{N}}$
and $U_{A_{2}^{\prime}\rightarrow B_{2}E_{2}}^{\mathcal{M}}$. Let
$\theta_{X_{1}X_{2}B_{1}B_{2}E_{1}E_{2}}$ be the state that arises from
sending $\theta_{X_{1}X_{2}A_{1}^{\prime}A_{2}^{\prime}}$ through the
tensor-product channel $\mathcal{U}_{A_{1}^{\prime}\rightarrow B_{1}E_{1}%
}^{\mathcal{N}}\otimes\mathcal{U}_{A_{2}^{\prime}\rightarrow B_{2}E_{2}%
}^{\mathcal{M}}$. Then%
\begin{align}
&  P(\mathcal{N})+P(\mathcal{M})\nonumber\\
&  =I(X_{1};B_{1})_{\rho}-I(X_{1};E_{1})_{\rho}+I(X_{2};B_{2})_{\sigma
}-I(X_{2};E_{2})_{\sigma}\\
&  =I(X_{1}X_{2};B_{1}B_{2})_{\theta}-I(X_{1}X_{2};E_{1}E_{2})_{\theta}\\
&  \leq P(\mathcal{N}\otimes\mathcal{M}).
\end{align}
The first equality follows from the definition of the private informations
$P(\mathcal{N})$ and $P(\mathcal{M})$ and by evaluating them on the respective
states $\rho_{X_{1}A_{1}^{\prime}}$ and $\sigma_{X_{2}A_{2}^{\prime}}$ that
maximize them. The second equality follows because the mutual information is
additive on tensor-product states (see Exercise~\ref{ex-qie:add-mutual-info}).
The final inequality follows because the state $\theta_{X_{1}X_{2}B_{1}%
B_{2}E_{1}E_{2}}$ is a particular state of the form needed in the maximization
of the private information of the tensor-product channel $\mathcal{N}%
\otimes\mathcal{M}$.

We now prove the inequality $P(\mathcal{N}\otimes\mathcal{M})\leq
P(\mathcal{N})+P(\mathcal{M})$. Let $\rho_{XA_{1}^{\prime}A_{2}^{\prime}}$ be
a state that maximizes $P(\mathcal{N}\otimes\mathcal{M})$ where%
\begin{equation}
\rho_{XA_{1}^{\prime}A_{2}^{\prime}}\equiv\sum_{x}p_{X}(x)|x\rangle\langle
x|_{X}\otimes\rho_{A_{1}^{\prime}A_{2}^{\prime}}^{x},
\end{equation}
and let $\rho_{XB_{1}B_{2}E_{1}E_{2}}$ be the state that arises from sending
$\rho_{XA_{1}^{\prime}A_{2}^{\prime}}$ through the tensor-product channel
$\mathcal{U}_{A_{1}^{\prime}\rightarrow B_{1}E_{1}}^{\mathcal{N}}%
\otimes\mathcal{U}_{A_{2}^{\prime}\rightarrow B_{2}E_{2}}^{\mathcal{M}}$.
Consider a spectral decomposition of each state $\rho_{A_{1}^{\prime}%
A_{2}^{\prime}}^{x}$:%
\begin{equation}
\rho_{A_{1}^{\prime}A_{2}^{\prime}}^{x}=\sum_{y}p_{Y|X}(y|x)\psi
_{A_{1}^{\prime}A_{2}^{\prime}}^{x,y},
\end{equation}
where each state $\psi_{A_{1}^{\prime}A_{2}^{\prime}}^{x,y}$ is pure. Let
$\sigma_{XYA_{1}^{\prime}A_{2}^{\prime}}$ be an extension of $\rho
_{XA_{1}^{\prime}A_{2}^{\prime}}$ where%
\begin{equation}
\sigma_{XYA_{1}^{\prime}A_{2}^{\prime}}\equiv\sum_{x,y}p_{Y|X}(y|x)p_{X}%
(x)|x\rangle\langle x|_{X}\otimes|y\rangle\langle y|_{Y}\otimes\psi
_{A_{1}^{\prime}A_{2}^{\prime}}^{x,y},
\end{equation}
and let $\sigma_{XYB_{1}E_{1}B_{2}E_{2}}$ be the state that arises from
sending $\sigma_{XYA_{1}^{\prime}A_{2}^{\prime}}$ through the tensor-product
channel $\mathcal{U}_{A_{1}^{\prime}\rightarrow B_{1}E_{1}}^{\mathcal{N}%
}\otimes\mathcal{U}_{A_{2}^{\prime}\rightarrow B_{2}E_{2}}^{\mathcal{M}}$.
Consider the following chain of inequalities:%
\begin{align}
&  P(\mathcal{N}\otimes\mathcal{M})\\
&  =I(X;B_{1}B_{2})_{\rho}-I(X;E_{1}E_{2})_{\rho}\\
&  =I(X;B_{1}B_{2})_{\sigma}-I(X;E_{1}E_{2})_{\sigma}\\
&  =I(XY;B_{1}B_{2})_{\sigma}-I(XY;E_{1}E_{2})_{\sigma}\nonumber\\
&  \ \ \ \ \ \ -\left[  I(Y;B_{1}B_{2}|X)_{\sigma}-I(Y;E_{1}E_{2}|X)_{\sigma
}\right] \\
&  \leq I(XY;B_{1}B_{2})_{\sigma}-I(XY;E_{1}E_{2})_{\sigma}\\
&  =H(B_{1}B_{2})_{\sigma}-H(B_{1}B_{2}|XY)_{\sigma}-H(E_{1}E_{2})_{\sigma
}+H(E_{1}E_{2}|XY)_{\sigma}\\
&  =H(B_{1}B_{2})_{\sigma}-H(B_{1}B_{2}|XY)_{\sigma}-H(E_{1}E_{2})_{\sigma
}+H(B_{1}B_{2}|XY)_{\sigma}\\
&  =H(B_{1}B_{2})_{\sigma}-H(E_{1}E_{2})_{\sigma}\\
&  =H(B_{1})_{\sigma}-H(E_{1})_{\sigma}+H(B_{2})_{\sigma}-H(E_{2})_{\sigma
}\nonumber\\
&  \ \ \ \ \ \ -\left[  I(B_{1};B_{2})_{\sigma}-I(E_{1};E_{2})_{\sigma}\right]
\\
&  \leq H(B_{1})_{\sigma}-H(E_{1})_{\sigma}+H(B_{2})_{\sigma}-H(E_{2}%
)_{\sigma}\\
&  \leq Q(\mathcal{N})+Q(\mathcal{M})\\
&  =P(\mathcal{N})+P(\mathcal{M}).
\end{align}
The first equality follows from the definition of $P(\mathcal{N}%
\otimes\mathcal{M})$ and from evaluating it on the state $\rho$ that maximizes
it. The second equality follows because the state $\sigma_{XYB_{1}E_{1}%
B_{2}E_{2}}$ is equal to the state $\rho_{XB_{1}E_{1}B_{2}E_{2}}$ after
tracing out the system $Y$. The third equality follows from the chain rule for
mutual information:\ $I(XY;B_{1}B_{2})=I(Y;B_{1}B_{2}|X)+I(X;B_{1}B_{2})$. It
holds that $I(Y;B_{1}B_{2}|X)_{\sigma}\geq I(Y;E_{1}E_{2}|X)_{\sigma}$ because
there is a degrading channel from $B_{1}$ to $E_{1}$ and from $B_{2}$ to
$E_{2}$. Then the first inequality follows because $I(Y;B_{1}B_{2}|X)_{\sigma
}-I(Y;E_{1}E_{2}|X)_{\sigma}\geq0$. The fourth equality follows by expanding
the mutual informations, and the fifth equality follows because the state
$\sigma$ on systems $B_{1}B_{2}E_{1}E_{2}$ is pure when conditioning on the
classical systems $X$ and $Y$. The sixth equality follows from algebra, and
the seventh follows by rewriting the entropies. It holds that $I(B_{1}%
;B_{2})_{\sigma}\geq I(E_{1};E_{2})_{\sigma}$ because there is a degrading
channel from $B_{1}$ to $E_{1}$ and from $B_{2}$ to $E_{2}$. Then the
inequality follows because $I(B_{1};B_{2})_{\sigma}-I(E_{1};E_{2})_{\sigma
}\geq0$. The third inequality follows because the entropy difference
$H(B_{i})-H(E_{i})$ is always less than the coherent information of the
channel, and the final equality follows because the coherent information of a
channel is equal to its private information when the channel is degradable
(Theorem~\ref{thm-ie:degradable-priv-coh}).
\end{proof}

\begin{corollary}
Suppose that a quantum channel $\mathcal{N}$ is
\index{degradable channel}%
degradable. Then the regularized private information $P_{\operatorname{reg}%
}(\mathcal{N})$ of the channel is equal to its private information
$P(\mathcal{N})$:%
\begin{equation}
P_{\operatorname{reg}}(\mathcal{N})=P(\mathcal{N}).
\end{equation}

\end{corollary}

\begin{proof}
A proof follows by the same induction argument as in
Corollary~\ref{cor-ie:class-mut=reg-class-mut}\ and by exploiting the result
of Theorem~\ref{thm:private-additivity-degradable} and the fact that the
tensor power channel $\mathcal{N}^{\otimes n}$ is degradable if the original
channel $\mathcal{N}$ is.
\end{proof}

\section{Summary}

We conclude this chapter with a table that summarizes the main results
regarding the mutual information of a classical channel~$I(p_{Y|X})$, the
private information of a classical wiretap channel~$P(p_{Y,Z|X})$, the Holevo
information of a quantum channel~$\chi(\mathcal{N})$, the mutual information
of a quantum channel~$I(\mathcal{N})$, the coherent information of a quantum
channel~$Q(\mathcal{N})$, and the private information of a quantum
channel~$P(\mathcal{N})$. The table exploits the following definitions:%
\begin{align}
\rho_{XA^{\prime}}  &  \equiv\sum p_{X}(x)|x\rangle\langle x|_{X}\otimes
\phi_{A^{\prime}}^{x},\\
\sigma_{XA^{\prime}}  &  \equiv\sum p_{X}(x)|x\rangle\langle x|_{X}\otimes
\rho_{A^{\prime}}^{x}.
\end{align}

\begin{tabular}
[c]{c|c|c|c|c}\hline\hline
\textbf{Quantity} & \textbf{Input} & \textbf{Output} & \textbf{Formula} &
\textbf{Single-letter}\\\hline\hline
$I(p_{Y|X})$ & $p_{X}$ & $p_{X}p_{Y|X}$ & $\max_{p_{X}}I(X;Y)$ & all
channels\\\hline
$P(p_{Y,Z|X})$ & $p_{X}$ & $p_{X}p_{Y,Z|X}$ & $\max_{p_{U,X}}I(U;Y)-I(U;Z)$ &
all channels\\\hline
$\chi(\mathcal{N})$ & $\rho_{XA^{\prime}}$ & $\mathcal{N}_{A^{\prime
}\rightarrow B}(\rho_{XA^{\prime}})$ & $\max_{\rho}I(X;B)$ & some
channels\\\hline
$I(\mathcal{N})$ & $\phi_{AA^{\prime}}$ & $\mathcal{N}_{A^{\prime}\rightarrow
B}(\phi_{AA^{\prime}})$ & $\max_{\phi}I(A;B)$ & all channels\\\hline
$Q(\mathcal{N})$ & $\phi_{AA^{\prime}}$ & $\mathcal{N}_{A^{\prime}\rightarrow
B}(\phi_{AA^{\prime}})$ & $\max_{\phi}I(A\rangle B)$ & degradable%
\index{degradable channel}%
\\\hline
$P(\mathcal{N})$ & $\sigma_{XA^{\prime}}$ & $U_{A^{\prime}\rightarrow
BE}^{\mathcal{N}}(\sigma_{XA^{\prime}})$ & $\max_{\sigma}I(X;B)-I(X;E)$ &
degradable\\\hline\hline
\end{tabular}

\section{History and Further Reading}

\cite{BV04} is a good reference for the theory and practice of convex
optimization, which is helpful for computing capacity formulas. \cite{W75}
introduced the classical wiretap channel and proved that the private
information is additive for degraded wiretap channels. \cite{CK78}\ proved
that the private information is additive for general wiretap
channels.\ \cite{Hol98} and \cite{PhysRevA.56.131} provided an operational
interpretation of the Holevo information of a quantum channel. \cite{S02}
showed the additivity of the Holevo information for
\index{entanglement-breaking channel}%
entanglement-breaking channels. \cite{PhysRevA.56.3470} introduced the mutual
information of a quantum channel, and they proved several of its important
properties that appear in this chapter:\ non-negativity, additivity, and
concavity. \cite{PhysRevLett.83.3081,ieee2002bennett} later gave an
operational interpretation for this information quantity as the
entanglement-assisted classical capacity of a quantum channel.
\cite{PhysRevA.55.1613}, \cite{capacity2002shor}, and \cite{ieee2005dev} gave
increasingly rigorous proofs that the coherent information of a quantum
channel is an achievable rate for quantum communication. \cite{cmp2005dev}
showed that the coherent information of a quantum channel is additive for
degradable%
\index{degradable channel}
channels. \cite{YHD05MQAC} proved that the coherent information of a quantum
channel is a concave function of the input state whenever the channel is
degradable. \cite{DJKR06,GPLS09} discussed the reverse coherent information of
a quantum channel and showed that it is additive for all quantum channels.
\cite{ieee2005dev}\ and \cite{1050633} independently introduced the private
classical capacity of a quantum channel, and both papers proved that it is an
achievable rate for private classical communication over a quantum channel.
\cite{S08} showed that the private classical information is additive and equal
to the coherent information for degradable quantum channels.

\chapter{Classical Typicality}

\label{chap:classical-typicality}This chapter begins our first technical foray
into the asymptotic theory of information. We start with the classical setting
in an effort to build up our intuition of asymptotic behavior before delving
into the asymptotic theory of quantum information.

The central concept of this chapter is the asymptotic equipartition property.
The name of this property may sound somewhat technical at first, but it is
merely an application of the law of large numbers to a sequence drawn%
\index{law of large numbers}
independently and identically from a distribution~$p_{X}( x) $ for some random
variable $X$. The asymptotic equipartition property reveals that we can divide
sequences into two classes when their length becomes large:\ those that are
overwhelmingly likely to occur and those that are overwhelmingly likely not to
occur. The sequences that are likely to occur are%
\index{typical sequence}
the \textit{typical} sequences, and the ones that are not likely to occur are
the \textit{atypical} sequences. Additionally, the size of the set of typical
sequences is exponentially smaller than the size of the set of all sequences
whenever the random variable generating the sequences is not uniform. These
properties are an example of a more general mathematical phenomenon known as
\textquotedblleft measure concentration,\textquotedblright\ in which a smooth
function over a high-dimensional space or over a large number of random
variables tends to concentrate around a constant value with high probability.

The asymptotic equipartition%
\index{asymptotic equipartition theorem}
property immediately leads to the intuition behind Shannon's scheme for
compressing classical information.\ The scheme first generates a realization
of a random sequence and asks the question:\ Is the produced sequence typical
or atypical? If it is typical, compress it. Otherwise, throw it away. The
error probability of this compression scheme is non-zero for any fixed length
of a sequence, but it vanishes in the asymptotic limit because the probability
of the sequence being in the typical set converges to one, while the
probability that it is in the atypical set converges to zero. This compression
scheme has a straightforward generalization to the quantum setting, in which
we wish to compress qubits instead of classical bits.

The bulk of this chapter is here to present the many technical details needed
to make rigorous statements in the asymptotic theory of information. We begin
with an example, follow with the formal definition of a typical sequence and a
typical set, and prove the three important properties of a typical set. We
then discuss other forms of typicality such as joint typicality and
conditional typicality. These other notions turn out to be useful for proving
Shannon's classical capacity theorem as well (recall that Shannon's theorem
gives the ultimate rate at which a sender can transmit classical information
over a classical channel to a receiver). We also introduce the method of
types, which is a powerful technique in classical information theory, and
apply this method in order to develop a stronger notion of typicality. The
chapter then features a development of the strong notions of joint and
conditional typicality and ends with a concise proof of Shannon's channel
capacity theorem.

\section{An Example of Typicality}%

\begin{figure}
[ptb]
\begin{center}
\includegraphics[
width=4.4434in
]%
{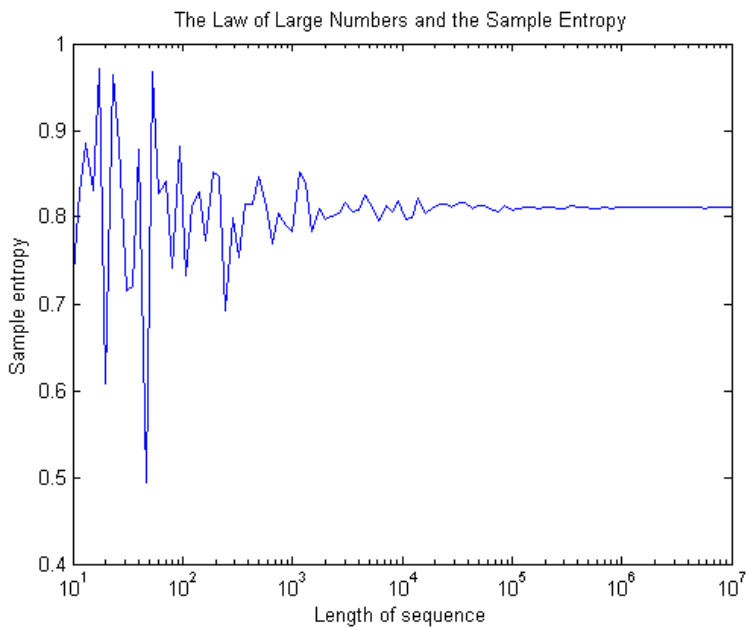}%
\caption{This figure depicts the sample entropy of a realization of a random
binary sequence as a function of its length. The source is a binary random
variable with distribution $(\frac{3}{4},\frac{1}{4})$. For the realizations
generated, the sample entropy of the sequences is converging to the true
entropy of the source.}%
\label{fig-ct:typicality}%
\end{center}
\end{figure}
Suppose that Alice possesses a binary random variable $X$ that takes the value
zero with probability $\frac{3}{4}$ and the value one with probability
$\frac{1}{4}$. Such a random source might produce the following sequence:%
\begin{equation}
0110001101,
\end{equation}
if we generate ten independent realizations. The probability that such a
sequence occurs is%
\begin{equation}
\left(  \frac{1}{4}\right)  ^{5}\left(  \frac{3}{4}\right)  ^{5},
\end{equation}
determined simply by counting the number of ones and zeros in the above
sequence and by applying the assumption that the source is i.i.d.

The \textit{information content} of the above sequence is the negative
logarithm of its probability divided by its length:%
\begin{equation}
-\frac{1}{10} \log\left(  \left(  \frac{1}{4}\right)  ^{5}\left(  \frac{3}%
{4}\right)  ^{5} \right)  = -\frac{5}{10}\log\left(  \frac{1}{4}\right)
-\frac{5}{10}\log\left(  \frac{3}{4}\right)  \approx1.207.
\end{equation}
We also refer to this quantity as the \textit{sample entropy}. The true
entropy of the source is%
\begin{equation}
-\frac{1}{4}\log\left(  \frac{1}{4}\right)  -\frac{3}{4}\log\left(  \frac
{3}{4}\right)  \approx0.8113. \label{eq-ct:true-entropy-ex}%
\end{equation}
We would expect that the sample entropy of a random sequence tends to approach
the true entropy as its length increases because the number of zeros should be
approximately $n\left(  3/4\right)  $ and the number of ones should be
approximately $n\left(  1/4\right)  $ according to the law of large numbers.

Another sequence of length 100 might be as follows:%
\begin{align}
&  00000000100010001000000000000110011010000000100000\nonumber\\
&  00000110101001000000010000001000000010000100010000,
\end{align}
featuring 81 zeros and 19 ones. Its sample entropy is%
\begin{equation}
-\frac{1}{100} \log\left(  \left(  \frac{1}{4}\right)  ^{19}\left(  \frac
{3}{4}\right)  ^{81} \right)  = -\frac{19}{100}\log\left(  \frac{1}{4}\right)
-\frac{81}{100}\log\left(  \frac{3}{4}\right)  \approx0.7162.
\end{equation}
The above sample entropy is closer to the true entropy in
\eqref{eq-ct:true-entropy-ex} than the sample entropy of the previous
sequence, but it still deviates significantly from it.

Figure~\ref{fig-ct:typicality}\ continues this game by generating random
sequences according to the distribution $\left(  \frac{3}{4},\frac{1}%
{4}\right)  $, and the result is that a concentration around the true entropy
begins to occur around $n\approx10^{6}$. That is, it becomes highly likely
that the sample entropy of a random sequence is close to the true entropy if
we increase the length of the sequence, and this holds for the realizations
generated in Figure~\ref{fig-ct:typicality}.

\section{Weak Typicality}

\label{sec-ct:weak-typ}This first section generalizes the example from the
introduction to an arbitrary discrete, finite-cardinality random variable. Our
first notion of typicality is the same as discussed in the example---we define
a sequence to be typical if its sample entropy%
\index{sample entropy}
is
\index{typicality!weak}%
close to the true entropy of the random variable that generates it. This
notion of typicality is known as \textit{weak typicality}.
Section~\ref{sec-ct:strong-typ}\ introduces another notion of typicality that
implies weak typicality, but the implication does not hold in the other
direction. For this reason, we distinguish the two different notions of
typicality as weak typicality and strong typicality.

Suppose that a random variable $X$ takes values in an alphabet $\mathcal{X}$
with cardinality $\left\vert \mathcal{X}\right\vert $. Let us label the
symbols in the alphabet as $a_{1}$, $a_{2}$, \ldots, $a_{\left\vert
\mathcal{X}\right\vert }$. An i.i.d.~information source samples
\textit{independently} from the distribution of random variable $X$\ and emits
$n$\ realizations $x_{1}$, \ldots, $x_{n}$. Let $X^{n}\equiv X_{1}\cdots
X_{n}$ denote the $n$ random variables that describe the information source,
and let $x^{n}\equiv x_{1}\cdots x_{n}$ denote an emitted realization of
$X^{n}$. The probability $p_{X^{n}}( x^{n}) $\ of a particular string $x^{n}$
is as follows:%
\begin{equation}
p_{X^{n}}( x^{n}) \equiv p_{X_{1},\ldots,X_{n}}( x_{1},\ldots,x_{n}) ,
\end{equation}
and $p_{X^{n}}( x^{n}) $\ factors as follows because the source is i.i.d.:%
\begin{equation}
p_{X^{n}}( x^{n}) =p_{X_{1}}( x_{1}) \cdots p_{X_{n}} (x_{n}) =p_{X}( x_{1})
\cdots p_{X}( x_{n}) =%
{\textstyle\prod\limits_{i=1}^{n}}
p_{X}( x_{i}) .
\end{equation}

Roughly speaking, we expect a long string $x^{n}$\ to contain about $np_{X}(
a_{1}) $ occurrences of symbol$~a_{1}$, $np_{X}( a_{2}) $ occurrences of
symbol $a_{2}$, etc., when $n$ is large, due to the law of large numbers. If
this is occurring, the probability that the source emits a particular string
$x^{n}$ is approximately%
\begin{equation}
p_{X^{n}}( x^{n}) =p_{X}( x_{1}) \cdots p_{X}( x_{n}) \approx p_{X}( a_{1})
^{np_{X}( a_{1}) }\cdots p_{X}( a_{\left\vert \mathcal{X}\right\vert })
^{np_{X}( a_{\left\vert \mathcal{X}\right\vert }) },
\end{equation}
and the information content of a given string is thus roughly%
\begin{equation}
-\frac{1}{n}\log( p_{X^{n}}( x^{n}) ) \approx-\sum_{i=1}^{\left\vert
\mathcal{X}\right\vert }p_{X}( a_{i}) \log( p_{X}( a_{i}) ) =H( X) .
\end{equation}
The above intuitive argument shows that the information content divided by the
length of the sequence is roughly equal to the entropy in the limit of large
$n$. It then makes sense to think of this quantity as the \textit{sample
entropy} of the sequence $x^{n}$.

\begin{definition}
[Sample Entropy]The sample entropy%
\index{sample entropy}
$\overline{H}( x^{n}) $\ of a sequence $x^{n}$ with respect to a probability
distribution $p_{X}(x)$ is defined as follows:%
\begin{equation}
\overline{H}( x^{n}) \equiv-\frac{1}{n}\log( p_{X^{n}}( x^{n}) ) ,
\end{equation}
where $p_{X^{n}}( x^{n}) = \prod_{i=1}^{n} p_{X}(x_{i})$.
\end{definition}

This definition of sample entropy leads us to our first important definitions
in asymptotic information theory.

\begin{definition}
[Typical Sequence]%
\index{typical sequence}%
A sequence $x^{n}$\ is $\delta$-\textit{typical} if its sample entropy
$\overline{H}( x^{n}) $ is $\delta$-close to the entropy $H( X) $\ of random
variable $X$, where this random variable is the source of the sequence.
\end{definition}

\begin{definition}
[Typical Set]\label{def-ct:weak-typ}The $\delta$-\textit{typical set}
\index{typical set}%
$T_{\delta}^{X^{n}}$ is the set of all $\delta$-typical sequences$~x^{n}$:%
\begin{equation}
T_{\delta}^{X^{n}}\equiv\left\{  x^{n}:\left\vert \overline{H}( x^{n}) -H( X)
\right\vert \leq\delta\right\}  .
\end{equation}

\end{definition}

\section{Properties of the Typical Set}

\begin{figure}[ptb]
\begin{center}
\includegraphics[
width=4.8456in
]{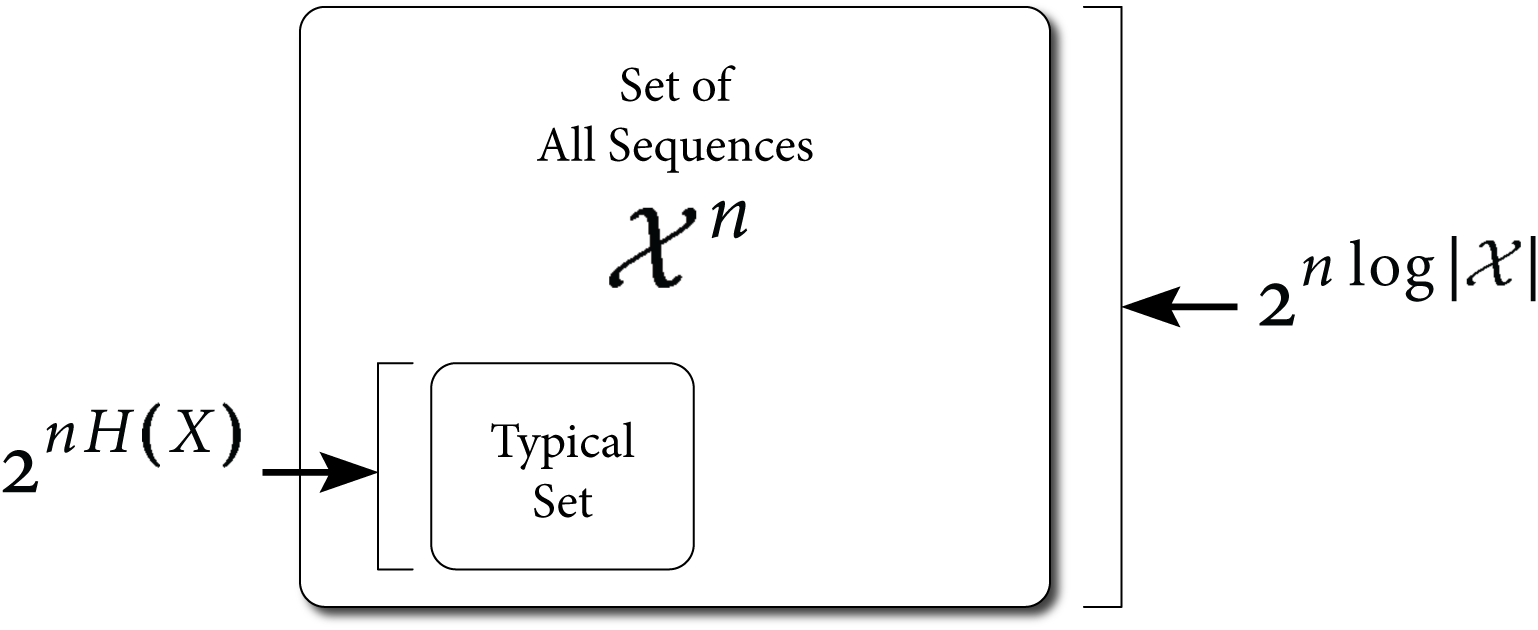}
\end{center}
\caption{This figure depicts the idea that the typical set is exponentially
smaller than the set of all sequences because $\left\vert \mathcal{X}%
\right\vert ^{n}=2^{n\log\left\vert \mathcal{X}\right\vert }>2^{nH( X) }$
whenever $X$ is not a uniform random variable. Yet, this exponentially smaller
set contains nearly all of the probability.}%
\label{fig-ct:typical-set}%
\end{figure}

The set of typical sequences enjoys three useful and beautifully surprising
properties
\index{typical set}%
that occur when we step into the \textquotedblleft land of large
numbers.\textquotedblright\ We can summarize these properties as follows: the
typical set contains almost all the probability, yet it is exponentially
smaller than the set of all sequences, and each typical sequence has almost
uniform probability. Figure~\ref{fig-ct:typical-set} attempts to depict the
main idea of the typical set.

\begin{property}
[Unit Probability]\label{prop-ct:unit-prob}The typical set asymptotically has
probability one. So as $n$ becomes large, it is highly likely that a source
emits a typical sequence. We formally state this property as follows:%
\begin{equation}
\Pr\left\{  X^{n}\in T_{\delta}^{X^{n}}\right\}  =\sum_{x^{n}\in T_{\delta
}^{X^{n}}}p_{X^{n}}( x^{n}) \geq1-\varepsilon,
\end{equation}
for all $\varepsilon\in(0,1)$, $\delta>0$, and sufficiently large $n$.
\end{property}

\begin{property}
[Exponentially Smaller Cardinality]\label{prop-ct:exp-small}The number
$\left\vert T_{\delta}^{X^{n}}\right\vert $\ of $\delta$-typical sequences is
exponentially smaller than the total number $\left\vert \mathcal{X}\right\vert
^{n}$\ of sequences for every random variable $X$ besides the uniform random
variable. We formally state this property as follows:%
\begin{equation}
\left\vert T_{\delta}^{X^{n}}\right\vert \leq2^{n\left(  H( X) +\delta\right)
}.
\end{equation}
We can also bound the size of the $\delta$-typical set from below:%
\begin{equation}
\left\vert T_{\delta}^{X^{n}}\right\vert \geq\left(  1-\varepsilon\right)
2^{n\left(  H( X) -\delta\right)  },
\end{equation}
for all $\varepsilon\in(0,1)$, $\delta>0$, and sufficiently large $n$.
\end{property}

\begin{property}
[Equipartition]\label{prop-ct:equi}The probability of a particular $\delta
$-typical sequence $x^{n}$\ is approximately uniform:%
\begin{equation}
2^{-n\left(  H( X) +\delta\right)  }\leq p_{X^{n}}( x^{n}) \leq2^{-n\left(  H(
X) -\delta\right)  }.
\end{equation}
This last property represents the \textquotedblleft
equipartition\textquotedblright\ in \textquotedblleft asymptotic equipartition
property\textquotedblright\ because all typical sequences occur with nearly
the same probability when $n$ is large.
\end{property}

The size $\left\vert T_{\delta}^{X^{n}}\right\vert $\ of the $\delta$-typical
set is approximately equal to the total number $\left\vert \mathcal{X}%
\right\vert ^{n}$ of sequences only when random variable $X$\ is uniform
because $H( X) =\log\left\vert \mathcal{X}\right\vert $ in such a case and
thus%
\begin{equation}
\left\vert T_{\delta}^{X^{n}}\right\vert \leq2^{n\left(  H( X) +\delta\right)
}=2^{n\left(  \log\left\vert \mathcal{X}\right\vert +\delta\right)
}=\left\vert \mathcal{X}\right\vert ^{n}\cdot2^{n\delta}\approx\left\vert
\mathcal{X}\right\vert ^{n}.
\end{equation}

\subsection{Proofs of Typical Set Properties}

\begin{proof}
[Proof of the Unit Probability Property (Property~\ref{prop-ct:unit-prob})]The
weak law of large numbers states that the sample mean converges in probability
to the expectation. More precisely, consider a sequence of i.i.d.~random
variables $Z_{1}$, \ldots, $Z_{n}$ that each have expectation~$\mu$. The
sample average of this sequence is as follows:%
\begin{equation}
\overline{Z}=\frac{1}{n}\sum_{i=1}^{n}Z_{i}.
\end{equation}
The formal statement of the law of large numbers is that $\forall
\varepsilon\in(0,1),\delta>0\ \ \exists n_{0}:\forall n>n_{0}$
\begin{equation}
\Pr\left\{  \left\vert \overline{Z}-\mu\right\vert \leq\delta\right\}  \geq1
-\varepsilon.
\end{equation}
We can now consider the sequence of random variables $-\log( p_{X}(X_{1})) $,
\ldots, $-\log( p_{X}(X_{n})) $. The sample average of this sequence is equal
to the sample entropy of $X^{n}$:%
\begin{align}
-\frac{1}{n}\sum_{i=1}^{n}\log( p_{X}(X_{i}))  &  =-\frac{1}{n}\log( p_{X^{n}%
}(X^{n}))\\
&  =\overline{H}(X^{n}).
\end{align}
Recall from \eqref{eq-cie:expected-info-content} that the expectation of the
random variable $-\log( p_{X}(X)) $ is equal to the Shannon entropy:%
\begin{equation}
\mathbb{E}_{X}\left\{  -\log( p_{X}(X)) \right\}  =H(X).
\end{equation}
Then we can apply the law of large numbers and find that $\forall
\varepsilon\in(0,1),\delta>0\ \ \exists n_{0}:\forall n>n_{0}$ such that
\begin{equation}
\Pr\left\{  \left\vert \overline{H}(X^{n})-H(X)\right\vert \leq\delta\right\}
\geq1-\varepsilon.
\end{equation}
The event $\left\{  \left\vert \overline{H}\left(  X^{n}\right)
-H(X)\right\vert \leq\delta\right\}  $ is precisely the condition for a random
sequence $X^{n}$ to be in the typical set $T_{\delta}^{X^{n}}$,\ and the
probability of this event goes to one as $n$ becomes large.
\end{proof}

\bigskip

\begin{proof}
[Proof of the Exponentially Smaller Cardinality Property
(Property~\ref{prop-ct:exp-small})]Consider the following chain of
inequalities:%
\begin{multline}
1=\sum_{x^{n}\in\mathcal{X}^{n}}p_{X^{n}}( x^{n}) \geq\sum_{x^{n}\in
T_{\delta}^{X^{n}}}p_{X^{n}}( x^{n})\label{eq-ct:proof-typ}\\
\geq\sum_{x^{n}\in T_{\delta}^{X^{n}}}2^{-n\left(  H( X) +\delta\right)
}=2^{-n\left(  H( X) +\delta\right)  }\left\vert T_{\delta}^{X^{n}}\right\vert
.
\end{multline}
The first inequality uses the fact that the probability of the typical set is
smaller than the probability of the set of all sequences. The second
inequality uses the equipartition property of typical sets\ (proved below).
After rearranging the leftmost side of \eqref{eq-ct:proof-typ} with its
rightmost side, we find that%
\begin{equation}
\left\vert T_{\delta}^{X^{n}}\right\vert \leq2^{n\left(  H( X) +\delta\right)
}.
\end{equation}
The second part of the property follows because the \textquotedblleft unit
probability\textquotedblright\ property holds for sufficiently large $n$. Then
the following chain of inequalities holds:%
\begin{multline}
1-\varepsilon\leq\Pr\left\{  X^{n}\in T_{\delta}^{X^{n}}\right\}  =\sum
_{x^{n}\in T_{\delta}^{X^{n}}}p_{X^{n}}( x^{n})\\
\leq\sum_{x^{n}\in T_{\delta}^{X^{n}}}2^{-n\left(  H( X) -\delta\right)
}=2^{-n\left(  H( X) -\delta\right)  }\left\vert T_{\delta}^{X^{n}}\right\vert
.
\end{multline}
We can then bound the size of the typical set as follows:%
\begin{equation}
\left\vert T_{\delta}^{X^{n}}\right\vert \geq2^{n\left(  H( X) -\delta\right)
}\left(  1-\varepsilon\right)  ,
\end{equation}
for all $\varepsilon\in(0,1)$, $\delta>0$, and sufficiently large $n$.
\end{proof}

\bigskip

\begin{proof}
[Proof of the Equipartition Property (Property~\ref{prop-ct:equi})]The
property follows immediately by manipulating the definition of a typical set.
\end{proof}

\section{Application: Data Compression}

\label{sec:shannon-compression}%
\index{Shannon compression}%
\index{data compression}
The above three properties of typical sequences immediately give our first
application in asymptotic information theory. It is Shannon's compression
protocol, which is a scheme for compressing the output of an
i.i.d.~information source.

We begin by defining the information-processing task and a corresponding
$\left(  n,R,\varepsilon\right)  $ source code. It is helpful to recall the
picture in Figure~\ref{fig-intro:classical-source-code}. An information source
outputs a sequence $x^{n}$ drawn independently according to the distribution
of some random variable $X$. A sender Alice encodes this sequence according to
some encoding map $E$ where%
\begin{equation}
E:\mathcal{X}^{n}\rightarrow\left\{  0,1\right\}  ^{nR}.
\end{equation}
The encoding takes elements from the set $\mathcal{X}^{n}$ of all sequences to
a set $\left\{  0,1\right\}  ^{nR}$\ of size $2^{nR}$. She then transmits the
codewords over $nR$ uses of a noiseless classical bit channel. Bob decodes
according to some decoding map $D:\left\{  0,1\right\}  ^{nR}\rightarrow
\mathcal{X}^{n}$. The probability of error for an $\left(  n,R,\varepsilon
\right)  $ source code is%
\begin{equation}
p( e) \equiv\Pr\left\{  \left(  D\circ E\right)  \left(  X^{n}\right)  \neq
X^{n}\right\}  \leq\varepsilon,
\end{equation}
where $\varepsilon\in(0,1)$. The rate of the source code is equal to the
number of channel uses divided by the length of the sequence, and it is equal
to $R$ for the above scheme. A particular compression rate $R$\ is
\textit{achievable} for $X$ if there exists an $\left(  n,R+\delta
,\varepsilon\right)  $ source code for all $\varepsilon\in(0,1) ,\delta>0$,
and sufficiently large $n$. We can now state Shannon's lossless compression theorem.

\begin{theorem}
[Shannon Compression]The entropy of an information source specified by a
discrete random variable $X$ is the smallest achievable rate for compression:%
\begin{equation}
\inf\left\{  R:R\text{ is achievable for }X\right\}  =H( X) .
\end{equation}

\end{theorem}

A proof of this theorem consists of two parts,\ traditionally called the
direct coding theorem and the converse theorem. The direct coding theorem is
the direction LHS $\leq$ RHS---the proof exhibits a coding scheme with an
achievable rate and demonstrates that its rate converges to the entropy in the
asymptotic limit. The converse theorem is the direction LHS $\geq$ RHS and is
a statement of optimality---it establishes that any coding scheme with rate
below the entropy is not achievable. The proofs of each part are usually
completely different. We employ typical sequences and their properties for
proving a direct coding theorem, while the converse part resorts to entropy
inequalities from Chapter~\ref{chap:info-entropy}.\footnote{The direct part of
a quantum coding theorem can employ the properties of typical subspaces
(discussed in Chapter~\ref{chap:quantum-typicality}), and the proof of a
converse theorem for quantum information usually employs the quantum entropy
inequalities from Chapter~\ref{chap:q-info-entropy}.} For now, we prove the
direct coding theorem and hold off on the converse part until we reach
Schumacher compression for quantum information in Chapter~\ref{chap:schumach}.
Our main goal here is to illustrate a simple application of typical sequences,
and we can wait on the converse part because Shannon compression is in some
sense a special case of Schumacher compression.%
\begin{figure}
[ptb]
\begin{center}
\includegraphics[
width=4.3431in
]%
{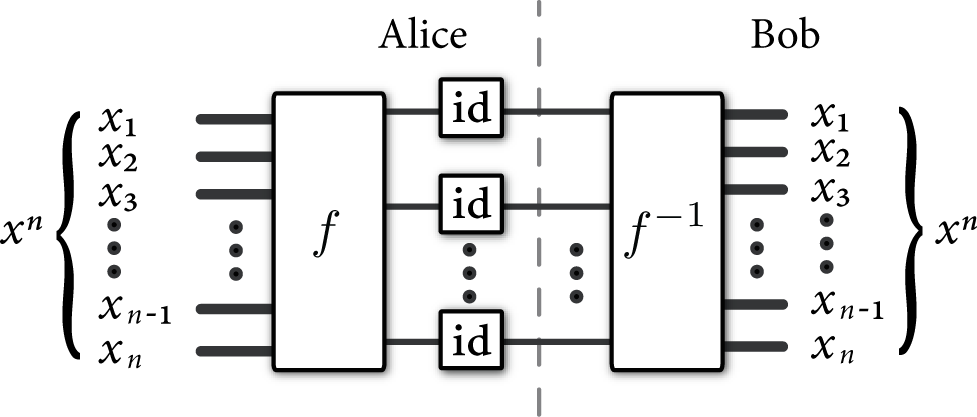}%
\caption{Shannon's scheme for the compression of classical data. The encoder
$f$ is a map from the typical set to a set of binary sequences of size
$\approx2^{nH( X) }$ where $H( X) $ is the entropy of the information source.
The map $f$\ is invertible on the typical set but maps an atypical sequence to
a constant. Alice then transmits the compressed data over $\approx nH( X) $
uses of a noiseless classical channel. The inverse map $f^{-1}$ (the decoder)
is the inverse of $f$ on the typical set and decodes to some error sequence
otherwise.}%
\label{fig-ct:source-coding-proof}%
\end{center}
\end{figure}

The idea behind the proof of the direct coding theorem is simple:\ just keep
the typical sequences and throw away the rest. This coding strategy succeeds
with asymptotically vanishing probability of error because the typical set
asymptotically has all of the probability. Since we are only concerned with
error probabilities in communication protocols, it makes sense that we should
only be keeping track of a set where all of the probability concentrates. We
can formally state a proof as follows. Pick an $\varepsilon\in(0,1)$, a
$\delta>0$, and a sufficiently large $n$ such that
Property~\ref{prop-ct:unit-prob} holds. Consider that
Property~\ref{prop-ct:exp-small} then holds so that the size of the typical
set is no larger than $2^{n\left[  H(X)+\delta\right]  }$. We choose the
encoding to be a one-to-one function $f$ that maps a typical sequence to a
binary sequence in $\left\{  0,1\right\}  ^{nR}$, where $R=H(X)+\delta$. We
define $f$ to map any atypical sequence to $0^{n}$. This scheme gives up on
encoding the atypical sequences because they have vanishingly small
probability. We define the decoding operation to be the inverse of $f$. This
scheme has probability of error less than $\varepsilon$, by considering
Property~\ref{prop-ct:unit-prob}. Figure~\ref{fig-ct:source-coding-proof}%
\ depicts this coding scheme.

Shannon's scheme for compression suffers from a problem that plagues all
results in classical and quantum information theory. The proof guarantees that
there exists a scheme that can compress at the rate of entropy in the
asymptotic limit. But the complexity of encoding and decoding is far from
practical---without any further specification of the encoding, it could
require resources that are prohibitively exponential in the length of the sequence.

The above scheme certainly gives an achievable rate for compression of
classical information, but how can we know that it is optimal?\ The converse
theorem addresses this point (recall that a converse theorem gives a sense of
optimality for a particular protocol) and completes the operational
interpretation of the entropy as the fundamental limit on the compressibility
of classical information. For now, we do not prove a converse theorem and
instead choose to wait until we cover Schumacher compression because its
converse proof applies to Shannon compression as well.

\section{Weak Joint Typicality}

\label{sec-ct:weak-joint-typ}Joint typicality
\index{typicality!weak!joint}%
is a concept similar to typicality, but the difference is that it applies to
any two random variables $X$ and $Y$. That is, there are analogous notions of
typicality for the joint random variable$~\left(  X,Y\right)  $.

\begin{definition}
[Joint Sample Entropy]Consider $n$ independent realizations $x^{n}=x_{1}\cdots
x_{n}$ and $y^{n}=y_{1}\cdots y_{n}$ of respective random variables $X$ and
$Y$. The sample joint entropy $\overline{H}(x^{n},y^{n})$ of these two
sequences is%
\begin{equation}
\overline{H}(x^{n},y^{n})\equiv-\frac{1}{n}\log\left(  p_{X^{n},Y^{n}}%
(x^{n},y^{n})\right)  ,
\end{equation}
where we assume that the joint distribution $p_{X^{n},Y^{n}}(x^{n},y^{n})$ has
the i.i.d.~property:%
\begin{equation}
p_{X^{n},Y^{n}}(x^{n},y^{n})\equiv p_{X,Y}(x_{1},y_{1})\cdots p_{X,Y}%
(x_{n},y_{n}).
\end{equation}

\end{definition}

\begin{figure}
[ptb]
\begin{center}
\includegraphics[
width=4.8456in
]%
{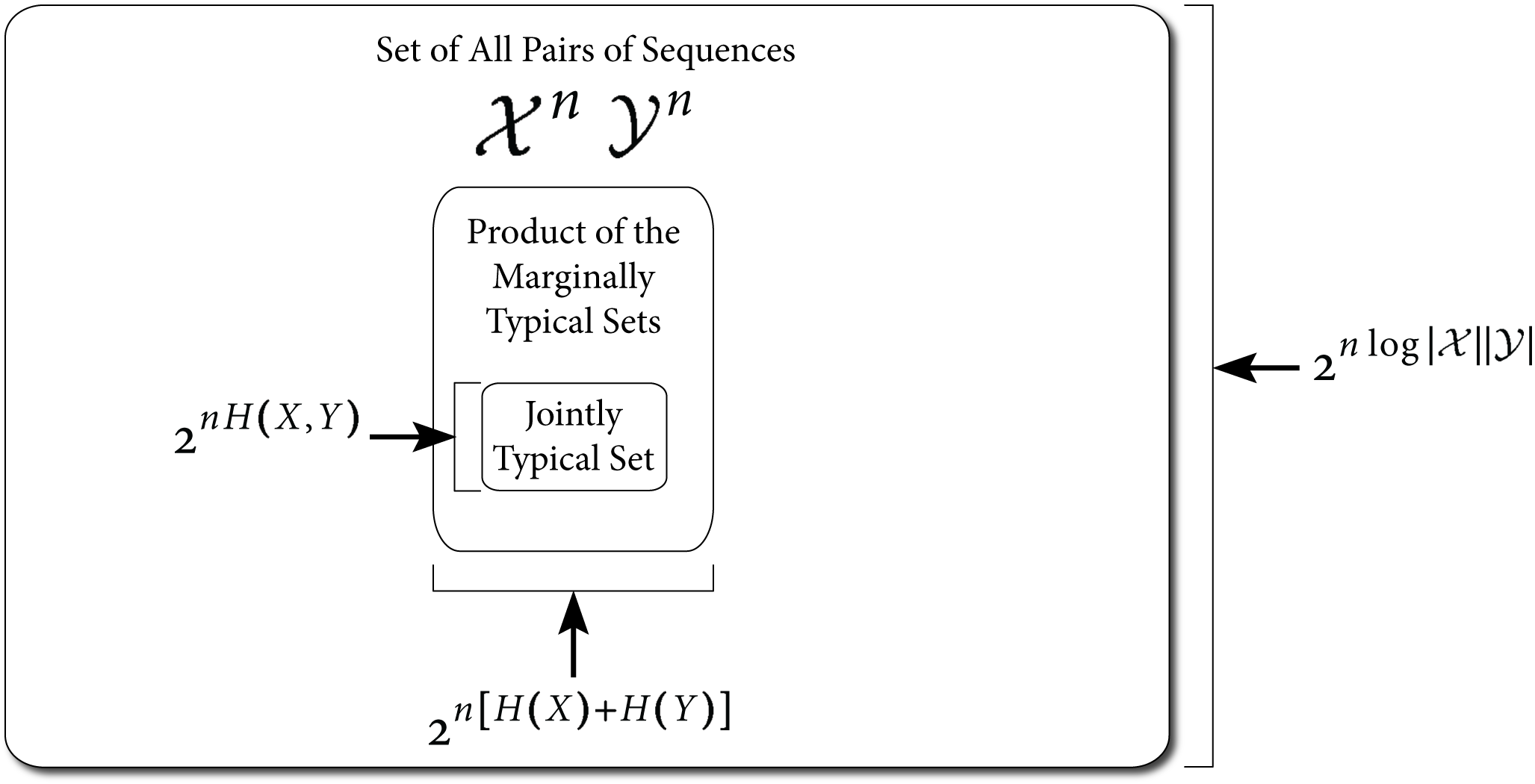}%
\caption{A depiction of the jointly typical set. Some sequence pairs $(
x^{n},y^{n}) $\ are such that $x^{n}$ is typical or such that $y^{n}$ is
typical, but fewer are such that the pair is jointly typical. The jointly
typical set has size roughly equal to $2^{nH( X,Y) }$, which is smaller than
the Cartesian product of the marginally typical sets if random variables $X$
and $Y$ are not independent.}%
\label{fig-ct:joint-typ}%
\end{center}
\end{figure}

This notion of joint sample entropy%
\index{sample entropy!joint}
immediately leads to the following definition of joint typicality.
Figure~\ref{fig-ct:joint-typ}\ attempts to depict the notion of joint typicality.

\begin{definition}
[Jointly Typical Sequence]Two sequences $x^{n},y^{n}$\ are $\delta$-jointly
\textit{typical} if
\index{typical sequence!jointly}%
their sample joint entropy $\overline{H}( x^{n},y^{n}) $ is $\delta$-close to
the joint entropy $H( X,Y) $ of random variables $X$ and $Y$ and if both
$x^{n}$ and $y^{n}$ are marginally typical.
\end{definition}

\begin{definition}
[Jointly Typical Set]\label{def-ct:weak-joint-typ-set}The $\delta$-jointly
\textit{typical set} $T_{\delta}^{X^{n}Y^{n}}$
\index{typical set!jointly}%
consists of all $\delta$-jointly typical sequences:%
\begin{equation}
T_{\delta}^{X^{n}Y^{n}}\equiv\left\{  (x^{n},y^{n}):\left\vert \overline{H}(
x^{n},y^{n}) -H( X,Y) \right\vert \leq\delta,\ \ x^{n}\in T_{\delta}^{X^{n}%
},\ \ y^{n}\in T_{\delta}^{Y^{n}}\right\}  .
\end{equation}

\end{definition}

The extra conditions on the marginal sample entropies are necessary to have a
sensible definition of joint typicality. That is, it does not necessarily
follow that the marginal sample entropies are close to the marginal true
entropies if the joint ones are close, but it intuitively makes sense that
this condition should hold. Thus, we add these extra conditions to the
definition of jointly typical sequences. Later, we find in
Section~\ref{sec-ct:strong-typ}\ that the intuitive implication holds (it is
not necessary to include the marginals) when we introduce a different
definition of typicality.%

\subsection{Properties of the Jointly Typical Set}

The set $T_{\delta}^{X^{n}Y^{n}}$\ of jointly typical sequences enjoys three
properties similar to what we have seen in Section~\ref{sec-ct:weak-typ}, and
the proofs of these properties are nearly identical to those in
Section~\ref{sec-ct:weak-typ}.

\begin{property}
[Unit Probability]The jointly typical set asymptotically has probability one.
So as $n$ becomes large, it is highly likely that a source emits a jointly
typical sequence. We formally state this property as follows:%
\begin{equation}
\Pr\left\{  (X^{n},Y^{n})\in T_{\delta}^{X^{n}Y^{n}}\right\}  \geq
1-\varepsilon,
\end{equation}
for all $\varepsilon\in(0,1)$, $\delta>0$, and sufficiently large $n$.
\end{property}

\begin{property}
[Exponentially Smaller Cardinality]The number $\left\vert T_{\delta}%
^{X^{n}Y^{n}}\right\vert $\ of $\delta$-jointly typical sequences is
exponentially smaller than the total number $\left(  \left\vert \mathcal{X}%
\right\vert \left\vert \mathcal{Y}\right\vert \right)  ^{n}$\ of sequences for
any joint random variable $\left(  X,Y\right)  $ that is not uniform. We
formally state this property as follows:%
\begin{equation}
\left\vert T_{\delta}^{X^{n}Y^{n}}\right\vert \leq2^{n\left(  H\left(
X,Y\right)  +\delta\right)  }.
\end{equation}
We can also bound the size of the $\delta$-jointly typical set from below:%
\begin{equation}
\left\vert T_{\delta}^{X^{n}Y^{n}}\right\vert \geq\left(  1-\varepsilon
\right)  2^{n\left(  H( X,Y) -\delta\right)  },
\end{equation}
for all $\varepsilon\in(0,1)$, $\delta>0$, and sufficiently large $n$.
\end{property}

\begin{property}
[Equipartition]\label{prop-ct:equi-joint}The probability of a particular
$\delta$-jointly typical sequence $x^{n}y^{n}$\ is approximately uniform:%
\begin{equation}
2^{-n\left(  H(X,Y)+\delta\right)  }\leq p_{X^{n},Y^{n}}(x^{n},y^{n}%
)\leq2^{-n\left(  H(X,Y)-\delta\right)  }.
\end{equation}

\end{property}

\begin{exercise}
Prove the above three properties of the jointly typical set.
\end{exercise}

The above three properties may be similar to what we have seen before, but
there is another interesting property of jointly typical sequences that we
give below. It states that two sequences drawn independently according to the
marginal distributions $p_{X}( x) $ and $p_{Y}( y) $ are jointly typical
according to the joint distribution $p_{X,Y}( x,y) $ with probability
$\approx2^{-nI( X;Y) }$. This property gives a simple interpretation of the
mutual information that is related to its most important operational
interpretation as the classical channel capacity discussed briefly in
Section~\ref{sec-ccs:channel-cap}.

\begin{property}
[Probability of Joint Typicality]\label{prop:joint-independent}Consider two
independent random variables $\tilde{X}^{n}$ and $\tilde{Y}^{n}$ whose
respective probability density functions $p_{\tilde{X}^{n}}( x^{n}) $\ and
$p_{\tilde{Y}^{n}}( y^{n}) $\ are equal to the marginal densities of the joint
density $p_{X^{n},Y^{n}}( x^{n},y^{n}) $:%
\begin{equation}
( \tilde{X}^{n},\tilde{Y}^{n}) \sim p_{X^{n}}( x^{n}) p_{Y^{n}}( y^{n}) .
\end{equation}
Then we can bound the probability that two random sequences $\tilde{X}^{n}$
and $\tilde{Y}^{n}$ are in the jointly typical set $T_{\delta}^{X^{n}Y^{n}}$:%
\begin{equation}
\Pr\left\{  ( \tilde{X}^{n},\tilde{Y}^{n}) \in T_{\delta}^{X^{n}Y^{n}%
}\right\}  \leq2^{-n\left(  I( X;Y) -3\delta\right)  }.
\end{equation}

\end{property}

\begin{exercise}
Prove Property~\ref{prop:joint-independent}. (Hint:\ Consider that%
\begin{equation}
\Pr\left\{  ( \tilde{X}^{n},\tilde{Y}^{n}) \in T_{\delta}^{X^{n}Y^{n}%
}\right\}  =\sum_{x^{n},y^{n}\in T_{\delta}^{X^{n}Y^{n}}}p_{X^{n}}( x^{n})
p_{Y^{n}}( y^{n}) ,
\end{equation}
and use the properties of typical and jointly typical sets to bound this probability.)
\end{exercise}

\section{Weak Conditional Typicality}

\label{sec-ct:weak-cond-typ}Conditional typicality%
\index{typicality!weak!conditional}
is a property that we expect to hold for any two random sequences---it is also
a useful tool in the proofs of coding theorems. Suppose two random variables
$X$ and $Y$ have respective alphabets $\mathcal{X}$ and $\mathcal{Y}$ and a
joint distribution $p_{X,Y}( x,y) $. We can factor the joint distribution
$p_{X,Y}( x,y) $ as the product of a marginal distribution $p_{X}( x) $ and a
conditional distribution $p_{Y|X}( y|x) $, and this factoring leads to a
particular way that we can think about generating realizations of the joint
random variable. We can consider random variable $Y$ to be a noisy version of
$X$, where we first generate a realization $x$\ of the random variable $X$
according to the distribution $p_{X}( x) $ and follow by generating a
realization $y$ of the random variable $Y$ according to the conditional
distribution $p_{Y|X}( y|x) $.

Suppose that we generate $n$ independent realizations of random variable $X$
to obtain the sequence $x^{n}=x_{1}\cdots x_{n}$. We then record these values
and use the conditional distribution $p_{Y|X}( y|x) $ $n$ times to generate
$n$ independent realizations of random variable $Y$. Let $y^{n}=y_{1}\cdots
y_{n}$ denote the resulting sequence.%
\begin{figure}
[ptb]
\begin{center}
\includegraphics[
width=4.8456in
]%
{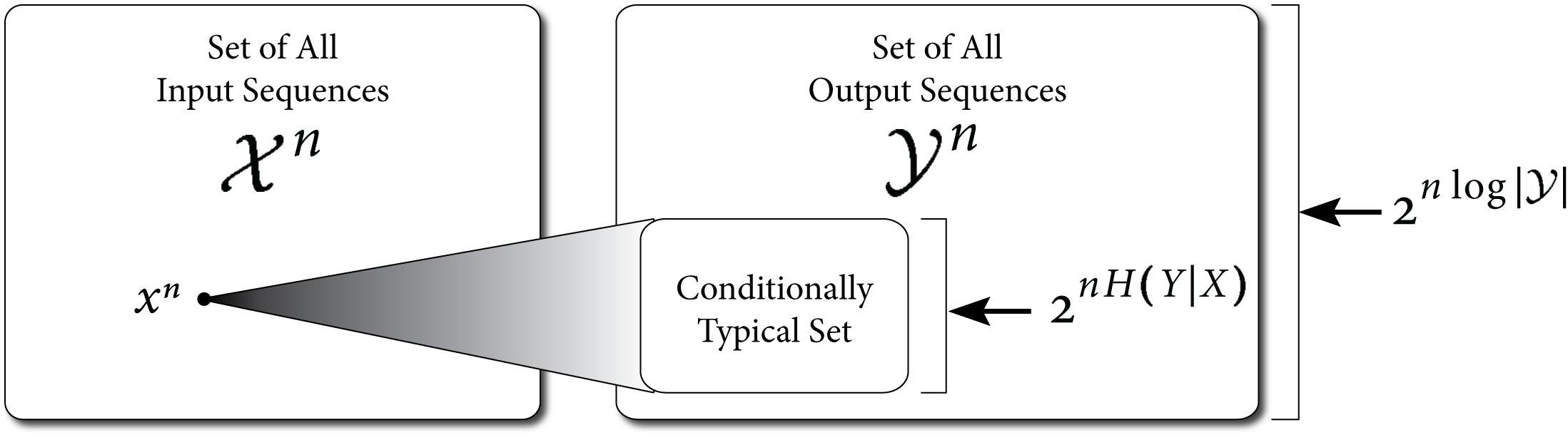}%
\caption{The notion of the conditionally typical set. A typical sequence
$x^{n}$\ in $T_{\delta}^{X^{n}}$ maps stochastically through many
instantiations of a conditional distribution $p_{Y|X}( y|x) $ to some sequence
$y^{n}$. It is overwhelmingly likely that $y^{n}$ is in a conditionally
typical set $T_{\delta}^{Y^{n}|x^{n}}$\ when $n$ becomes large. This
conditionally typical set has size around $2^{nH( Y|X) }$. It contains nearly
all of the probability but is exponentially smaller than the set of all
sequences $\mathcal{Y}^{n}$.}%
\label{fig-ct:conditional-typical}%
\end{center}
\end{figure}

\begin{definition}
[Conditional Sample Entropy]\label{def:cond-typ-def}The conditional sample
entropy%
\index{sample entropy!conditional}
$\overline{H}( y^{n}|x^{n}) $ of two sequences $x^{n}$ and $y^{n}$ with
respect to $p_{X,Y}( x,y) = p_{X}( x) p_{Y|X}( y|x)$ is%
\begin{equation}
\overline{H}( y^{n}|x^{n}) =-\frac{1}{n}\log p_{Y^{n}|X^{n}}( y^{n}|x^{n}) ,
\end{equation}
where%
\begin{equation}
p_{Y^{n}|X^{n}}( y^{n}|x^{n}) \equiv p_{Y|X}( y_{1}|x_{1}) \cdots p_{Y|X}(
y_{n}|x_{n}) .
\end{equation}

\end{definition}

\begin{definition}
[Conditionally Typical Set]Let $x^{n}\in\mathcal{X}^{n}$. The $\delta
$-conditionally typical set $T_{\delta}^{Y^{n}|x^{n}}$ consists of all
sequences whose conditional sample entropy is $\delta$-close to the true
\index{typical set!conditionally}%
conditional entropy:%
\begin{equation}
T_{\delta}^{Y^{n}|x^{n}}\equiv\left\{  y^{n}:\left\vert \overline{H}%
(y^{n}|x^{n})-H(Y|X)\right\vert \leq\delta\right\}  .
\end{equation}

\end{definition}

Figure~\ref{fig-ct:conditional-typical} provides an intuitive picture of the
notion of conditional typicality.

\subsection{Properties of the Conditionally Typical Set}

The set $T_{\delta}^{Y^{n}|x^{n}}$ of conditionally typical sequences enjoys
properties similar to what we have seen before, and we list them for completeness.

\begin{property}
[Unit Probability]\label{prop-ct:unit-prob-weak-cond-typ}The set $T_{\delta
}^{Y^{n}|x^{n}}$ asymptotically has probability one on average with respect to
a random sequence $X^{n}$. So as $n$ becomes large, it is highly likely that
random sequences $Y^{n}$\ and $X^{n}$ are such that $Y^{n}$ is a conditionally
typical sequence (on average with respect to $X^{n}$). We formally state this
property as follows:%
\begin{equation}
\mathbb{E}_{X^{n}}\left\{  \Pr_{Y^{n}|X^{n}}\left\{  Y^{n}\in T_{\delta
}^{Y^{n}|X^{n}}\right\}  \right\}  \geq1-\varepsilon,
\end{equation}
for all $\varepsilon\in(0,1)$, $\delta>0$, and sufficiently large $n$.
\end{property}

\begin{property}
[Exponentially Smaller Cardinality]\label{eq-ct:exp-smaller-weak-cond-typ}The
number $\left\vert T_{\delta}^{Y^{n}|x^{n}}\right\vert $\ of $\delta
$-conditionally typical sequences is exponentially smaller than the total
number $\left\vert \mathcal{Y}\right\vert ^{n}$\ of sequences for any
conditional random variable $Y|X$ that is not uniform. We formally state this
property as follows:%
\begin{equation}
\left\vert T_{\delta}^{Y^{n}|x^{n}}\right\vert \leq2^{n\left(  H(Y|X)+\delta
\right)  }.
\end{equation}
We can also bound the expected size of the $\delta$-conditionally typical set
from below when $x^{n}$ is a random sequence:%
\begin{equation}
\mathbb{E}_{X^{n}}\left\{  \left\vert T_{\delta}^{Y^{n}|X^{n}}\right\vert
\right\}  \geq\left(  1-\varepsilon\right)  2^{n\left(  H(Y|X)-\delta\right)
},
\end{equation}
for all $\varepsilon\in(0,1)$, $\delta>0$, and sufficiently large $n$.
\end{property}

\begin{property}
[Equipartition]\label{eq-ct:w-cond-typ-equi}The probability of a given
$\delta$-conditionally typical sequence $y^{n}$\ (corresponding to the
sequence $x^{n}$)\ is approximately uniform:%
\begin{equation}
2^{-n\left(  H( Y|X) +\delta\right)  }\leq p_{Y^{n}|X^{n}}( y^{n}|x^{n})
\leq2^{-n\left(  H( Y|X) -\delta\right)  }.
\end{equation}

\end{property}

In summary, averaged over realizations of the random variable $X^{n}$, the
conditionally typical set $T_{\delta}^{Y^{n}|X^{n}}$ has almost all the
probability, and its size is exponentially smaller than the size of the set of
all sequences. For each realization of $X^{n}$, each $\delta$-conditionally
typical sequence has an approximate uniform probability of occurring.

Our last note on the weak conditionally typical set is that there is a
subtlety in the statement of Property~\ref{prop-ct:unit-prob-weak-cond-typ}
that allows for a relatively straightforward proof. This subtlety is that we
average over the sequence $X^{n}$ as well, and this allows one to exploit the
extra randomness in order to establish a proof. We do not impose such a
constraint later on in Section~\ref{sec-ct:strong-cond-typ}\ when we introduce
the notion of a strong conditionally typical sequence. We instead impose the
constraint that the sequence $x^{n}$ is a strongly typical sequence, and this
property is sufficient to prove that similar properties hold for a strong
conditionally typical set.

We now prove the first property. This is just again another application of the
law of large numbers. Consider that%
\begin{align}
&  \mathbb{E}_{X^{n}}\left\{  \Pr_{Y^{n}|X^{n}}\left\{  Y^{n}\in T_{\delta
}^{Y^{n}|X^{n}}\right\}  \right\} \nonumber\\
&  =\mathbb{E}_{X^{n}}\left\{  \mathbb{E}_{Y^{n}|X^{n}}\left\{  {I}%
_{T_{\delta}^{Y^{n}|X^{n}}}(Y^{n})\right\}  \right\} \\
&  =\mathbb{E}_{X^{n},Y^{n}}\left\{  {I}_{T_{\delta}^{Y^{n}|X^{n}}}%
(Y^{n})\right\} \\
&  =\sum_{x^{n}\in\mathcal{X}^{n},y^{n}\in\mathcal{Y}^{n}}p_{X^{n},Y^{n}%
}(x^{n},y^{n}){I}_{T_{\delta}^{Y^{n}|x^{n}}}(y^{n}),
\label{eq-ct:cond-typ-rewrite-high-prob}%
\end{align}
where ${I}$ denotes an indicator function. Given random variables $X$ and $Y$,
let us define the random variable $g(X,Y)=-\log p_{Y|X}(Y|X)$. The expectation
of this random variable is%
\begin{align}
\mathbb{E}_{X,Y}\{g(X,Y)\}  &  =\mathbb{E}_{X,Y}\{\left[  -\log p_{Y|X}%
(Y|X)\right]  \}\\
&  =\sum_{x,y}p_{X,Y}(x,y)\left[  -\log p_{Y|X}(y|x)\right] \\
&  =H(Y|X).
\end{align}
Consider that the conditional sample entropy $\overline{H}(Y^{n}|X^{n})$ for
the random sequences $X^{n}$ and $Y^{n}$ factors as follows:%
\begin{align}
\overline{H}(Y^{n}|X^{n})  &  =-\frac{1}{n}\log p_{Y^{n}|X^{n}}(Y^{n}|X^{n})\\
&  =\frac{1}{n}\sum_{i=1}^{n}\left[  -\log p_{Y|X}(Y_{i}|X_{i})\right] \\
&  =\frac{1}{n}\sum_{i=1}^{n}g( X_{i},Y_{i}) .
\end{align}
This is the sample average of the random variables $g(X_{i},Y_{i})$ for all
$i\in\{1, \ldots, n\}$. Given all of the above and the definition of weak
conditional typicality, we can rewrite
\eqref{eq-ct:cond-typ-rewrite-high-prob} as follows:%
\begin{equation}
\Pr_{X^{n}Y^{n}}\left\{  \left\vert \frac{1}{n}\sum_{i=1}^{n}g\left(
X_{i},Y_{i}\right)  -\mathbb{E}_{X,Y}\{g(X,Y)\}\right\vert \leq\delta\right\}
.
\end{equation}
By applying the law of large numbers, this is $\geq1-\varepsilon$ for all
$\varepsilon\in(0,1)$, $\delta>0$, and sufficiently large $n$.

\begin{exercise}
Prove Properties~\ref{eq-ct:exp-smaller-weak-cond-typ} and
\ref{eq-ct:w-cond-typ-equi} for the weak conditionally typical set.
\end{exercise}

\section{Strong Typicality}

\label{sec-ct:strong-typ}In the development
\index{typicality!strong}
in the previous sections, we showed how the law of large numbers is the
underpinning method to prove many of the interesting results regarding typical
sequences. These results are satisfactory and provide an intuitive notion of
typicality through the idea of the sample entropy approaching the true entropy
for sufficiently long sequences.

It is possible to develop a stronger notion of typicality with a different
definition. Instead of requiring that the sample entropy of a random sequence
is close to the true entropy of a distribution for sufficiently long
sequences, strong typicality requires that the empirical distribution or
relative frequency of symbols of a random sequence has a small deviation from
the true probability distribution for sufficiently large sequence length.

We begin with a simple example to help illustrate this stronger notion of
typicality. Suppose that we generate a binary sequence i.i.d.~according to the
distribution $p( 0) =1/4$ and $p( 1) =3/4$. Such a random generation process
could lead to the following sequence:%
\begin{equation}
0110111010. \label{eq-ct:example-sequence}%
\end{equation}
Rather than computing the sample entropy of this sequence and comparing it
with the true entropy, we can count the number of zeros or ones that appear in
the sequence and compare their normalizations with the true distribution of
the information source. For the above example, the number of zeros is equal to
4, and the number of ones (the Hamming weight of the sequence) is equal to 6:%
\begin{equation}
N( 0\ |\ 0110111010) =4,\ \ \ \ \ \ \ \ \ \ \ N( 1\ |\ 0110111010) =6.
\end{equation}
We can compute the empirical distribution of this sequence by normalizing the
above numbers by the length of the sequence:%
\begin{equation}
\frac{1}{10}N( 0\ |\ 0110111010) =\frac{2}{5},\ \ \ \ \ \ \ \ \ \ \ \frac
{1}{10}N( 1\ |\ 0110111010) =\frac{3}{5}.
\end{equation}
This empirical distribution deviates from the true distribution by the
following amount:%
\begin{equation}
\max\left\{  \left\vert \frac{1}{4}-\frac{2}{5}\right\vert ,\left\vert
\frac{3}{4}-\frac{3}{5}\right\vert \right\}  =\frac{3}{20},
\end{equation}
which is a fairly significant deviation. However, suppose that the length of
the sequence grows large enough so that the law of large numbers comes into
play. We would then expect it to be highly likely that the empirical
distribution of a random sequence does not deviate much from the true
distribution, and the law of large numbers again gives a theoretical
underpinning for this intuition. This example gives the essence of strong typicality.

We wish to highlight another important aspect of the above example. The
particular sequence in \eqref{eq-ct:example-sequence} has a Hamming weight of
six, but this sequence is not the only one with this Hamming weight. By a
simple counting argument, there are $\binom{10}{6}-1=209$ other sequences with
the same length and Hamming weight. That is, all these other sequences have
the same empirical distribution and thus have the same deviation from the true
distribution as the original sequence in \eqref{eq-ct:example-sequence}. We
say that all these sequences are in the same \textquotedblleft type
class,\textquotedblright\ which simply means that they have the same empirical
distribution. The type class is thus an equivalence class on sequences where
the equivalence relation is the empirical distribution of the sequence.

We mention a few interesting properties of the type class before giving more
formal definitions. We can partition the set of all possible sequences
according to type classes. Consider that the set of all binary sequences of
length ten has size $2^{10}$. There is one sequence with all zeros,
$\binom{10}{1}$ sequences with Hamming weight one, $\binom{10}{2}$ sequences
with Hamming weight two, etc. The binomial theorem guarantees that the total
number of sequences is equal to the number of sequences in all of the type
classes:%
\begin{equation}
2^{10}=\sum_{i=0}^{10}\binom{10}{i}.
\end{equation}

Suppose now that we generate ten i.i.d.~realizations of the Bernoulli
distribution $p( 0) =1/4$ and $p( 1) =3/4$. Without knowing anything else, our
best description of the distribution of the random sequence is%
\begin{equation}
p( x_{1},\ldots,x_{10}) =p( x_{1}) \cdots p( x_{10}) ,
\end{equation}
where $x_{1},\ldots,x_{10}$ are different realizations of the binary random
variable. But suppose that a third party tells us the Hamming weight $w_{0}%
$\ of the generated sequence. This information allows us to update our
knowledge of the distribution of the sequence, and we can say that any
sequence with Hamming weight not equal to $w_{0}$ has zero probability. All
the sequences with the same Hamming weight have the same distribution because
we generated the sequence in an i.i.d.~way, and each sequence with Hamming
weight $w_{0}$ has a uniform distribution after renormalizing. Thus,
conditioned on the Hamming weight $w_{0}$, our best description of the
distribution of the random sequence is%
\begin{equation}
p( x_{1},\ldots,x_{10}|w_{0}) =\left\{
\begin{array}
[c]{ccc}%
0 & : & w\left(  x_{1},\ldots,x_{10}\right)  \neq w_{0}\\
\binom{10}{w_{0}}^{-1} & : & w\left(  x_{1},\ldots,x_{10}\right)  =w_{0}%
\end{array}
\right.  ,
\end{equation}
where $w$ is a function that gives the Hamming weight of a binary sequence.
This property has important consequences for asymptotic information processing
because it gives us a way to extract uniform randomness from an
i.i.d.~distribution, and we later see that it has applications in several
quantum information-processing protocols as well.

\subsection{Types and Strong Typicality}

\label{sec-ct:method-of-types}We now formally develop the notion of a
\index{type}%
type and strong typicality. Let $x^{n}$ denote a sequence $x_{1}x_{2}\dots
x_{n}$, where each $x_{i}$ belongs to the alphabet $\mathcal{X}$. Let
$|\mathcal{X}|$ be the cardinality of $\mathcal{X}$. Let $N(x|x^{n})$ be the
number of occurrences of the symbol $x\in\mathcal{X}$ in the sequence $x^{n}$.

\begin{definition}
[Type]The \textit{type} or empirical distribution $t_{x^{n}}$ of a sequence
$x^{n}$ is a probability mass function whose elements are $t_{x^{n}}( x) $
where%
\begin{equation}
t_{x^{n}}( x) \equiv\frac{1}{n}N(x|x^{n}).
\end{equation}

\end{definition}

\begin{definition}
[Strongly Typical Set]\label{def-ct:strong-typ}The $\delta$-strongly typical
set $T_{\delta}^{X^{n}}$\ is the set of all sequences
\index{typical set!strongly}
with an empirical distribution $\frac{1}{n}N(x|x^{n})$\ that has maximum
deviation $\delta$ from the true distribution $p_{X}( x) $. Furthermore, the
empirical distribution $\frac{1}{n}N(x|x^{n})$\ of any sequence in $T_{\delta
}^{X^{n}}$\ vanishes for any letter $x$ for which $p_{X}( x) =0$:%
\begin{equation}
T_{\delta}^{X^{n}}\equiv\left\{  x^{n}:\forall x\in\mathcal{X},\ \left\vert
\frac{1}{n}N(x|x^{n})-p_{X}( x) \right\vert \leq\delta\text{ if }p_{X}( x)
>0,\text{ else }\frac{1}{n}N(x|x^{n})=0\right\}  .
\end{equation}

\end{definition}

The extra condition where $\frac{1}{n}N(x|x^{n})=0$ when $p_{X}(x)=0$ is a
somewhat technical condition, nevertheless intuitive, that is necessary to
prove the three desired properties for the strongly typical set. Also, we are
using the same notation $T_{\delta}^{X^{n}}$ to indicate both the weakly and
strongly typical set, but which one is appropriate should be clear from the
context, or we will explicitly indicate which one we are using.

The notion of type class becomes useful for us in our later developments---it
is simply a way for grouping together all the sequences with the same
empirical distribution. Its most important use is as a way for obtaining a
uniform distribution from an arbitrary i.i.d.~distribution (recall that we can
do this by conditioning on a particular type).

\begin{definition}
[Type Class]\label{def-ct:type-class}Let $T_{t}^{X^{n}}$ denote the
\textit{type class} of a particular
\index{type class}%
type $t$. The type class $T_{t}^{X^{n}}$ is the set of all sequences with
length$~n$ and type $t$:%
\begin{equation}
T_{t}^{X^{n}}\equiv\{x^{n}\in\mathcal{X}^{n}:t_{x^{n}}=t\}.
\end{equation}

\end{definition}

\begin{property}
[Bound on the Number of Types]\label{prop-typ:bound-num-types}The number
\index{type!bound on number of types}%
of types for a given sequence of length$~n$ containing symbols from an alphabet
$\mathcal{X}$\ is exactly equal to%
\begin{equation}
\binom{n+\left\vert \mathcal{X}\right\vert -1}{\left\vert \mathcal{X}%
\right\vert -1}. \label{eq-ct:number-of-types}%
\end{equation}
A good upper bound on the number of types is $\left(  n+1\right)  ^{\left\vert
\mathcal{X}\right\vert }$.
\end{property}

\begin{proof}
The number of types is equal to the number of ways that the symbols in a
sequence of length $n$\ can form $\left\vert \mathcal{X}\right\vert $ distinct
groups. Consider the following visual aid:%
\begin{equation}
\bullet\bullet\bullet\bullet\bullet|\bullet\bullet\bullet\bullet\bullet
\bullet|\bullet\bullet\bullet\bullet\bullet\bullet\bullet\bullet
\bullet|\bullet\bullet\bullet\bullet\ .
\end{equation}
We can think of the number of types as the number of different ways of
arranging $\left\vert \mathcal{X}\right\vert -1$ vertical bars to group the
$n$\ dots into $\left\vert \mathcal{X}\right\vert $ distinct groups, which
gives \eqref{eq-ct:number-of-types}. The upper bound follows from a simple
argument. The number of types is the number of different ways that $\left\vert
\mathcal{X}\right\vert $ positive numbers can sum to $n$. Overestimating the
count, we can choose the first number in $n+1$ different ways (it can be any
number from $0$ to $n$), and we can choose the $\left\vert \mathcal{X}%
\right\vert -1$ other numbers in $n+1$ different ways. Multiplying all of
these possibilities together gives an upper bound $\left(  n+1\right)
^{\left\vert \mathcal{X}\right\vert }$ on the number of types. This bound
illustrates that the number of types is only \textit{polynomial} in the length
$n$\ of the sequence (compare with the total number $\left\vert \mathcal{X}%
\right\vert ^{n}$\ of sequences of length $n$ being exponential in the length
of the sequence).
\end{proof}

\begin{definition}
[Typical Type]\label{def-ct:typical-type}Let $p_{X}( x) $ denote the true
probability distribution of symbols $x$\ in the alphabet $\mathcal{X}$. For
$\delta>0$, let $\tau_{\delta}$ denote the set of all \textit{typical} types
that have maximum deviation $\delta$ from the true distribution $p_{X}( x) $:%
\begin{equation}
\tau_{\delta}\equiv\{t:\forall x\in\mathcal{X},\ \left\vert t( x) -p_{X}( x)
\right\vert \leq\delta\text{ if }p_{X}( x) >0\text{ else }t( x) =0\}.
\end{equation}

\end{definition}

We can then equivalently define the set of strongly $\delta$-typical sequences
of length $n$ as a union over all the type classes of the typical types in
$\tau_{\delta}$:%
\begin{equation}
T_{\delta}^{X^{n}}=%
{\displaystyle\bigcup\limits_{t\in\tau_{\delta}}}
T_{t}^{X^{n}}.
\end{equation}

\subsection{Properties of the Strongly Typical Set}

The strongly typical set enjoys many useful properties (similar to the weakly
typical set).

\begin{property}
[Unit Probability]\label{prop-ct:strong-unit}The strongly typical set
asymptotically has probability one. So as $n$ becomes large, it is highly
likely that a source emits a strongly typical sequence. We formally state this
property as follows:%
\begin{equation}
\Pr\left\{  X^{n}\in T_{\delta}^{X^{n}}\right\}  \geq1-\varepsilon,
\end{equation}
for all $\varepsilon\in(0,1)$, $\delta>0$, and sufficiently large $n$.
\end{property}

\begin{property}
[Exponentially Smaller Cardinality]\label{prop-ct:strong-exp-small}The number
$\left\vert T_{\delta}^{X^{n}}\right\vert $\ of $\delta$-typical sequences is
exponentially smaller than the total number $\left\vert \mathcal{X}\right\vert
^{n}$\ of sequences for most random variables $X$. We formally state this
property as follows:%
\begin{equation}
\left\vert T_{\delta}^{X^{n}}\right\vert \leq2^{n\left(  H( X) +c\delta
\right)  },
\end{equation}
where $c$ is some positive constant. We can also bound the size of the
$\delta$-typical set from below:%
\begin{equation}
\left\vert T_{\delta}^{X^{n}}\right\vert \geq\left(  1-\varepsilon\right)
2^{n\left(  H( X) -c\delta\right)  },
\end{equation}
for all $\varepsilon\in(0,1)$, $\delta>0$, and sufficiently large $n$.
\end{property}

\begin{property}
[Equipartition]\label{prop-ct:strong-equi}The probability of a given $\delta
$-typical sequence $x^{n}$\ occurring is approximately uniform:%
\begin{equation}
2^{-n\left(  H( X) +c\delta\right)  }\leq p_{X^{n}}( x^{n}) \leq2^{-n\left(
H( X) -c\delta\right)  }.
\end{equation}

\end{property}

This last property of strong typicality demonstrates that it implies weak
typicality up to a constant $c$.

\subsection{Proofs of the Properties of the Strongly Typical Set}

\label{sec-ct:proof-weak-typ}

\begin{proof}
[Proof of the Unit Probability Property (Property~\ref{prop-ct:strong-unit})]A
proof proceeds similarly to the proof of the unit probability property for the
weakly typical set. The law of large numbers states that the sample mean of a
random sequence converges in probability to the expectation of the random
variable from which we generate the sequence. So consider a sequence of
i.i.d.~random variables $Z_{1}$, \ldots, $Z_{n}$ where each random variable in
the sequence has expectation $\mu$. The sample average of this sequence is as
follows:%
\begin{equation}
\overline{Z}\equiv\frac{1}{n}\sum_{i=1}^{n}Z_{i}.
\end{equation}
The precise statement of the weak law of large numbers is that $\forall
\varepsilon\in(0,1),\delta>0\ \ \exists n_{0}:\forall n>n_{0}$ such that
\begin{equation}
\Pr\left\{  \left\vert \overline{Z}-\mu\right\vert >\delta\right\}
<\varepsilon.
\end{equation}
We can now consider the indicator random variables $I( X_{1}=a) $, \ldots, $I(
X_{n}=a) $. The sample mean of a random sequence of indicator variables is
equal to the empirical distribution $N( a|X^{n}) /n$:%
\begin{equation}
\frac{1}{n}\sum_{i=1}^{n}I( X_{i}=a) =\frac{1}{n}N( a|X^{n}) ,
\end{equation}
and the expectation of the indicator random variable $I( X=a) $ is equal to
the probability of the symbol $a$:%
\begin{equation}
\mathbb{E}_{X}\left\{  I( X=a) \right\}  =p_{X}( a) .
\end{equation}
Also, any random sequence $X^{n}$ has probability zero if one of its symbols
$x_{i}$ is such that $p_{X}( x_{i}) =0$. Thus, the probability that $\frac
{1}{n}N( a|X^{n}) =0$ is equal to one whenever $p_{X}( a) =0$:%
\begin{equation}
\Pr\left\{  \frac{1}{n}N( a|X^{n}) =0:p_{X}( a) =0\right\}  =1,
\end{equation}
and we can consider the cases when $p_{X}( a) >0$. We apply the law of large
numbers to find that%
\begin{equation}
\forall\varepsilon\in(0,1),\delta>0\ \ \exists n_{0,a}:\forall n>n_{0,a}%
\ \ \ \ \Pr\left\{  \left\vert \frac{1}{n}N( a|X^{n}) -p_{X}( a) \right\vert
>\delta\right\}  <\frac{\varepsilon}{\left\vert \mathcal{X}\right\vert }.
\end{equation}
Choosing $n_{0}=\max_{a\in\mathcal{X}}\left\{  n_{0,a}\right\}  $, the
following condition holds by the union bound of probability theory:%
\begin{multline}
\forall\varepsilon\in(0,1),\delta>0\ \ \exists n_{0}:\forall n>n_{0}\\
\Pr\left\{
{\displaystyle\bigcup\limits_{a\in\mathcal{X}}}
\left\vert \frac{1}{n}N( a|X^{n}) -p_{X}( a) \right\vert >\delta\right\} \\
\leq\sum_{a\in\mathcal{X}}\Pr\left\{  \left\vert \frac{1}{n}N\left(
a|X^{n}\right)  -p_{X}( a) \right\vert >\delta\right\}  <\varepsilon.
\end{multline}
Thus the complement of the above event on the left has high probability. That
is, $\forall\varepsilon\in(0,1),\delta>0\ \ \exists n_{0}:\forall n>n_{0}$
such that
\begin{equation}
\Pr\left\{  \forall a\in\mathcal{X},\ \left\vert \frac{1}{n}N\left(
a|X^{n}\right)  -p_{X}( a) \right\vert \leq\delta\right\}  \geq1-\varepsilon.
\end{equation}
The event $\left\{  \forall a\in\mathcal{X},\ \left\vert \frac{1}{n}N(
a|X^{n}) -p_{X}( a) \right\vert \leq\delta\right\}  $ is the condition for a
random sequence $X^{n}$ to be in the strongly typical set $T_{\delta}^{X^{n}}%
$,\ and the probability of this event goes to one as $n$ becomes sufficiently large.
\end{proof}

\bigskip

\begin{proof}
[Proof of the Exponentially Smaller Cardinality Property
(Property~\ref{prop-ct:strong-exp-small})]By the proof of
Property~\ref{prop-ct:strong-equi} (proved below), we know that the following
relation holds for any sequence $x^{n}$ in the strongly typical set:%
\begin{equation}
2^{-n\left(  H(X)+c\delta\right)  }\leq p_{X^{n}}(x^{n})\leq2^{-n\left(
H(X)-c\delta\right)  },
\end{equation}
where $c$ is some positive constant that we define when we prove
Property~\ref{prop-ct:strong-equi}. Summing over all sequences in the typical
set, we get the following inequalities:%
\begin{align}
\sum_{x^{n}\in T_{\delta}^{X^{n}}}2^{-n\left(  H(X)+c\delta\right)  }  &
\leq\Pr\left\{  X^{n}\in T_{\delta}^{X^{n}}\right\}  \leq\sum_{x^{n}\in
T_{\delta}^{X^{n}}}2^{-n\left(  H(X)-c\delta\right)  },\\
\Rightarrow2^{-n\left(  H(X)+c\delta\right)  }\left\vert T_{\delta}^{X^{n}%
}\right\vert  &  \leq\Pr\left\{  X^{n}\in T_{\delta}^{X^{n}}\right\}
\leq2^{-n\left(  H(X)-c\delta\right)  }\left\vert T_{\delta}^{X^{n}%
}\right\vert .
\end{align}
By the unit probability property of the strongly typical set, we know that the
following relation holds for sufficiently large $n$:%
\begin{equation}
1\geq\Pr\left\{  X^{n}\in T_{\delta}^{X^{n}}\right\}  \geq1-\varepsilon.
\end{equation}
Then the following inequalities result by combining the above inequalities:%
\begin{equation}
2^{n\left(  H(X)-c\delta\right)  }\left(  1-\varepsilon\right)  \leq\left\vert
T_{\delta}^{X^{n}}\right\vert \leq2^{n\left(  H(X)+c\delta\right)  },
\end{equation}
concluding the proof.
\end{proof}

\bigskip

\begin{proof}
[Proof of the Equipartition Property (Property~\ref{prop-ct:strong-equi})]The
following relation holds from the i.i.d.~property of the distribution
$p_{X^{n}}(x^{n})$ and because the sequence $x^{n}$ is strongly typical:%
\begin{equation}
p_{X^{n}}(x^{n})=%
{\displaystyle\prod\limits_{x\in\mathcal{X}^{+}}}
p_{X}(x)^{N(x|x^{n})}%
\end{equation}
where $\mathcal{X}^{+}$ denotes all the letters $x$\ in $\mathcal{X}$ with
$p_{X}(x)>0$. (The fact that the sequence $x^{n}$ is strongly typical
according to Definition~\ref{def-ct:strong-typ}\ allows us to employ this
modified alphabet.) Take the logarithm of the above expression:%
\begin{equation}
\log\left(  p_{X^{n}}(x^{n})\right)  =\sum_{x\in\mathcal{X}^{+}}N(x|x^{n}%
)\log\left(  p_{X}(x)\right)  .
\end{equation}
Multiply both sides by $-\frac{1}{n}$:%
\begin{equation}
-\frac{1}{n}\log\left(  p_{X^{n}}(x^{n})\right)  =-\sum_{x\in\mathcal{X}^{+}%
}\frac{1}{n}N(x|x^{n})\log\left(  p_{X}(x)\right)  .
\label{eq-ct:info-content-sub}%
\end{equation}
The following relation holds because the sequence $x^{n}$ is strongly typical:%
\begin{equation}
\forall x\in\mathcal{X}^{+}:\left\vert \frac{1}{n}N(x|x^{n})-p_{X}%
(x)\right\vert \leq\delta,
\end{equation}
and it implies that%
\begin{equation}
\Rightarrow\forall x\in\mathcal{X}^{+}:-\delta+p_{X}(x)\leq\frac{1}%
{n}N(x|x^{n})\leq\delta+p_{X}(x). \label{eq-ct:strongly-typical-expansion1}%
\end{equation}
Now multiply \eqref{eq-ct:strongly-typical-expansion1} by $-\log\left(
p_{X}(x)\right)  >0$, sum over all letters in the alphabet $\mathcal{X}^{+}$,
and apply the substitution in \eqref{eq-ct:info-content-sub}. This procedure
gives the following set of inequalities:%
\begin{multline}
-\sum_{x\in\mathcal{X}^{+}}\left(  -\delta+p_{X}(x)\right)  \log\left(
p_{X}(x)\right)  \leq-\frac{1}{n}\log\left(  p_{X^{n}}(x^{n})\right) \\
\leq-\sum_{x\in\mathcal{X}^{+}}\left(  \delta+p_{X}(x)\right)  \log\left(
p_{X}(x)\right)  ,
\end{multline}%
\begin{align}
\Rightarrow-c\delta+H(X)  &  \leq-\frac{1}{n}\log\left(  p_{X^{n}}%
(x^{n})\right)  \leq c\delta+H(X),\\
\Rightarrow2^{-n\left(  H(X)+c\delta\right)  }  &  \leq p_{X^{n}}(x^{n}%
)\leq2^{-n\left(  H(X)-c\delta\right)  },
\end{align}
where%
\begin{equation}
c\equiv-\sum_{x\in\mathcal{X}^{+}}\log\left(  p_{X}(x)\right)  \geq0.
\end{equation}
It now becomes apparent why we require the technical condition in the
definition of strong typicality (Definition~\ref{def-ct:strong-typ}). Were it
not there, then the constant $c$ would not be finite, and we would not be able
to obtain a reasonable bound on the probability of a strongly typical sequence.
\end{proof}

\subsection{Cardinality of a Typical Type Class}

\label{sec-ct:typical-type-class-card}Recall that a typical type class
\index{type class!typical}%
is defined to be the set of all sequences with the same empirical
distribution, and the empirical distribution happens to have maximum deviation
$\delta$ from the true distribution. It might seem that the size $\left\vert
T_{t}^{X^{n}}\right\vert $ of a typical type class $T_{t}^{X^{n}}$ should be
smaller than the size of the strongly typical set. But the following property
overrides this intuition and shows that a given typical type class
$T_{t}^{X^{n}}$ in some sense has almost as many sequences in it as the
strongly typical set $T_{\delta}^{X^{n}}$\ for sufficiently large $n$.

\begin{property}
[Minimal Cardinality of a Typical Type Class]%
\label{prop-ct:min-card-typical-type}Let $X$ be a random variable with
alphabet $\mathcal{X}$. Fix $\delta\in(0,2/|\mathcal{X}|]$. For $t\in
\tau_{\delta}$, the size $\left\vert T_{t}^{X^{n}}\right\vert $ of the typical
type class $T_{t}^{X^{n}}$\ is bounded from below as follows:%
\begin{equation}
\left\vert T_{t}^{X^{n}}\right\vert \geq\frac{1}{\left(  n+1\right)
^{\left\vert \mathcal{X}\right\vert }}2^{n\left[  H(X)-\eta(\left\vert
\mathcal{X}\right\vert \delta)\right]  }=2^{n\left[  H(X)-\eta(\left\vert
\mathcal{X}\right\vert \delta)-\left\vert \mathcal{X}\right\vert \frac{1}%
{n}\log\left(  n+1\right)  \right]  },
\end{equation}
where $\eta(\delta)$ is some function such that $\eta(\delta)\rightarrow0$ as
$\delta\rightarrow0$. Thus, a typical type class is of size roughly
$2^{nH(X)}$ when $n\rightarrow\infty$ and $\delta\rightarrow0$ (it is about as
large as the typical set when$~n$ becomes large).
\end{property}

\begin{proof}
We first show that if $X_{1}$, \ldots, $X_{n}$ are random variables drawn
i.i.d.~from a distribution $q(x)$, then the probability $q^{n}(x^{n})$ of a
particular sequence $x^{n}$ depends only on its type:%
\begin{equation}
q^{n}(x^{n})=2^{-n\left(  H\left(  t_{x^{n}}\right)  +D(t_{x^{n}}\Vert
q)\right)  },
\end{equation}
where $D(t_{x^{n}}\Vert q)$ is the relative entropy between $t_{x^{n}}$ and
$q$. To see this, consider the following chain of equalities:%
\begin{align}
q^{n}(x^{n})  &  =\prod\limits_{i=1}^{n}q(x_{i})=\prod\limits_{x\in
\mathcal{X}}q(x)^{N(x|x^{n})}=\prod\limits_{x\in\mathcal{X}}q(x)^{nt_{x^{n}%
}(x)}\\
&  =\prod\limits_{x\in\mathcal{X}}2^{nt_{x^{n}}(x)\log q(x)}=2^{n\sum
_{x\in\mathcal{X}}t_{x^{n}}(x)\log q(x)}\\
&  =2^{n\sum_{x\in\mathcal{X}}t_{x^{n}}(x)\log q(x)-t_{x^{n}}(x)\log t_{x^{n}%
}(x)+t_{x^{n}}(x)\log t_{x^{n}}(x)}\\
&  =2^{-n\left(  D(t_{x^{n}}\Vert q)+H\left(  t_{x^{n}}\right)  \right)  }.
\label{eq-cie:probability-type-equality}%
\end{align}
It then follows that the probability of the sequence $x^{n}$\ is
$2^{-nH\left(  t_{x^{n}}\right)  }$ if the distribution $q(x)=t_{x^{n}}(x)$.
Now consider that each type class $T_{t}^{X^{n}}$ has size%
\begin{equation}
\binom{n}{nt_{x^{n}}(x_{1}),\ldots,nt_{x^{n}}(x_{\left\vert \mathcal{X}%
\right\vert })},
\end{equation}
where the distribution $t=\left(  t_{x^{n}}(x_{1}),\ldots,t_{x^{n}%
}(x_{\left\vert \mathcal{X}\right\vert })\right)  $ and the letters of
$\mathcal{X}$ are $x_{1},\ldots,x_{\left\vert \mathcal{X}\right\vert }$. This
result follows because the size of a type class is just the number of ways of
arranging $nt_{x^{n}}(x_{1}),\ldots,nt_{x^{n}}(x_{\left\vert \mathcal{X}%
\right\vert })$ in a sequence of length $n$. We now prove that the type class
$T_{t}^{X^{n}}$ has the highest probability among all type classes when the
probability distribution is $t$:%
\begin{equation}
t^{n}( T_{t}^{X^{n}}) \geq t^{n}( T_{t^{\prime}}^{X^{n}}) \text{ for all
}t^{\prime}\in\mathcal{P}_{n},
\end{equation}
where $t^{n}$ is the i.i.d.~distribution induced by the type $t$ and
$\mathcal{P}_{n}$ is the set of all types. Consider the following equalities:%
\begin{align}
\frac{t^{n}( T_{t}^{X^{n}}) }{t^{n}( T_{t^{\prime}}^{X^{n}}) }  &
=\frac{\left\vert T_{t}^{X^{n}}\right\vert \prod\limits_{x\in\mathcal{X}%
}t_{x^{n}}(x)^{nt_{x^{n}}(x)}}{\left\vert T_{t^{\prime}}^{X^{n}}\right\vert
\prod\limits_{x\in\mathcal{X}}t_{x^{n}}(x)^{nt_{x^{n}}^{\prime}(x)}}\\
&  =\frac{\binom{n}{nt_{x^{n}}(x_{1}),\ldots,nt_{x^{n}}(x_{\left\vert
\mathcal{X}\right\vert })}\prod\limits_{x\in\mathcal{X}}t_{x^{n}%
}(x)^{nt_{x^{n}}(x)}}{\binom{n}{nt_{x^{n}}^{\prime}\left(  x_{1}\right)
,\ldots,nt_{x^{n}}^{\prime}(x_{\left\vert \mathcal{X}\right\vert })}%
\prod\limits_{x\in\mathcal{X}}t_{x^{n}}(x)^{nt_{x^{n}}^{\prime}(x)}}\\
&  =\prod\limits_{x\in\mathcal{X}}\frac{nt_{x^{n}}^{\prime}(x)!}{nt_{x^{n}%
}(x)!}t_{x^{n}}(x)^{n\left(  t_{x^{n}}(x)-t_{x^{n}}^{\prime}(x)\right)  }.
\end{align}
Now apply the bound $\frac{m!}{n!}\geq n^{m-n}$ (that holds for any positive
integers $m$ and $n$) to get%
\begin{align}
\frac{t^{n}( T_{t}^{X^{n}}) }{t^{n}( T_{t^{\prime}}^{X^{n}}) }  &  \geq
\prod\limits_{x\in\mathcal{X}}\left[  nt_{x^{n}}(x)\right]  ^{n\left(
t_{x^{n}}^{\prime}(x)-t_{x^{n}}(x)\right)  }t_{x^{n}}(x)^{n\left(  t_{x^{n}%
}(x)-t_{x^{n}}^{\prime}(x)\right)  }\\
&  =\prod\limits_{x\in\mathcal{X}}n^{n\left(  t_{x^{n}}^{\prime}(x)-t_{x^{n}%
}(x)\right)  }\\
&  =n^{n\sum_{x\in\mathcal{X}}t_{x^{n}}^{\prime}(x)-t_{x^{n}}(x)}\\
&  =n^{n\left(  1-1\right)  }\\
&  =1.
\end{align}
Thus, it holds that $t^{n}( T_{t}^{X^{n}}) \geq t^{n}( T_{t^{\prime}}^{X^{n}})
$ for all $t^{\prime}$. Now we are close to obtaining the desired bound in
Property~\ref{prop-ct:min-card-typical-type}. Consider the following chain of
inequalities:%
\begin{align}
1  &  =\sum_{t^{\prime}\in\mathcal{P}_{n}}t^{n}( T_{t^{\prime}}^{X^{n}})
\leq\sum_{t^{\prime}\in\mathcal{P}_{n}}\max_{t^{\prime}}t^{n}( T_{t^{\prime}%
}^{X^{n}}) =\sum_{t^{\prime}\in\mathcal{P}_{n}}t^{n}( T_{t}^{X^{n}%
})\label{eq-ct:typical-type-proof-1}\\
&  \leq\left(  n+1\right)  ^{\left\vert \mathcal{X}\right\vert }t^{n}(
T_{t}^{X^{n}}) =\left(  n+1\right)  ^{\left\vert \mathcal{X}\right\vert }%
\sum_{x^{n}\in T_{t}^{X^{n}}}t^{n}(x^{n})\\
&  =\left(  n+1\right)  ^{\left\vert \mathcal{X}\right\vert }\sum_{x^{n}\in
T_{t}^{X^{n}}}2^{-nH(t)}\\
&  =\left(  n+1\right)  ^{\left\vert \mathcal{X}\right\vert }2^{-nH(t)}%
\left\vert T_{t}^{X^{n}}\right\vert . \label{eq-ct:typical-type-proof-last}%
\end{align}
Recall that $t$ is a typical type, implying that $\left\vert
t(x)-p(x)\right\vert \leq\delta$ for all $x$. This then implies that the
variational distance between the distributions is small:%
\begin{equation}
\sum_{x}\left\vert t(x)-p(x)\right\vert \leq\left\vert \mathcal{X}\right\vert
\delta.
\end{equation}
We can apply the continuity of entropy (Theorem~\ref{thm-ie:fannes-actual}) to
get a bound on the difference of entropies:%
\begin{equation}
\left\vert H(t)-H(X)\right\vert \leq\frac{1}{2}\left\vert \mathcal{X}%
\right\vert \delta\log\left\vert \mathcal{X}\right\vert +h_{2}(\left\vert
\mathcal{X}\right\vert \delta/2). \label{eq-ct:last-step-card-typ-type}%
\end{equation}
The desired bound then follows with $\eta(\left\vert \mathcal{X}\right\vert
\delta)\equiv\left\vert \mathcal{X}\right\vert \delta/2\log\left\vert
\mathcal{X}\right\vert +h_{2}(\left\vert \mathcal{X}\right\vert \delta/2)$.
\end{proof}

\begin{exercise}
\label{ex-cie:prob-type-class-bound}Prove that $2^{nH( t) }$ is an upper bound
on the number of sequences $x^{n}$ of type$~t$:%
\begin{equation}
\left\vert T_{t}^{X^{n}}\right\vert \leq2^{nH( t) }.
\end{equation}
Use this bound and \eqref{eq-cie:probability-type-equality} to prove the
following upper bound on the probability of a type class where each sequence
is generated i.i.d.~according to a distribution $q( x) $:%
\begin{equation}
\Pr\left\{  T_{t}^{X^{n}}\right\}  \leq2^{-nD\left(  t\Vert q\right)  }.
\end{equation}

\end{exercise}

\section{Strong Joint Typicality}

It is possible to extend the above notions of
\index{typicality!strong!joint}%
strong typicality to jointly typical sequences. In a marked difference with
the weakly typical case, we can show that strong joint typicality implies
marginal typicality. Thus there is no need to impose this constraint in the definition.

Let $N(x,y|x^{n},y^{n})$ be the number of occurrences of the symbol
$x\in\mathcal{X},y\in\mathcal{Y}$ in the respective sequences $x^{n}$ and
$y^{n}$. The \emph{type }or empirical distribution $t_{x^{n}y^{n}}$ of
sequences $x^{n}$ and $y^{n}$ is a probability mass function whose elements
are $t_{x^{n}y^{n}}( x,y) $ where%
\begin{equation}
t_{x^{n}y^{n}}( x,y) \equiv\frac{1}{n}N(x,y|x^{n},y^{n}).
\end{equation}

\begin{definition}
[Strong Jointly Typical Sequence]Two sequences $x^{n},y^{n}$\ are $\delta
$-strongly jointly \textit{typical} if their empirical distribution has
maximum deviation $\delta$ from the true distribution and vanishes for any two
symbols $x$ and $y$ for which $p_{X,Y}( x,y) =0$.
\end{definition}

\begin{definition}
[Strong Jointly Typical Set]The $\delta$-jointly \textit{typical set}
$T_{\delta}^{X^{n}Y^{n}}$ is the set of all $\delta$-jointly typical
sequences:%
\begin{multline}
T_{\delta}^{X^{n}Y^{n}}\equiv\\
\left\{
\begin{array}
[c]{c}%
x^{n},y^{n}:\forall( x,y) \in\mathcal{X}\times\mathcal{Y}%
\begin{array}
[c]{cc}%
\left\vert \frac{1}{n}N(x,y|x^{n},y^{n})-p_{X,Y}( x,y) \right\vert \leq\delta
& \operatorname{if\ }p_{X,Y}( x,y) >0\\
\frac{1}{n}N(x,y|x^{n},y^{n})=0 & \operatorname{otherwise}%
\end{array}
\end{array}
\right\}  .
\end{multline}

\end{definition}

It follows from the above definitions that strong joint typicality implies
marginal typicality for both sequences$~x^{n}$ and$~y^{n}$. We leave
justifying this statement as the following exercise.

\begin{exercise}
\label{ex-ct:strong-joint-typ->marg}Prove that strong joint typicality implies
marginal typicality for either the sequence$~x^{n}$ or$~y^{n}$.
\end{exercise}

\subsection{Properties of the Strong Jointly Typical Set}

The set $T_{\delta}^{X^{n}Y^{n}}$\ of strong jointly typical sequences enjoys
properties similar to what we have seen before.

\begin{property}
[Unit Probability]The strong jointly typical set $T_{\delta}^{X^{n}Y^{n}}$
asymptotically has probability one. So as $n$ becomes large, it is highly
likely that a source emits a strong jointly typical sequence. We formally
state this property as follows:%
\begin{equation}
\Pr\left\{  X^{n}Y^{n}\in T_{\delta}^{X^{n}Y^{n}}\right\}  \geq1-\varepsilon,
\end{equation}
for all $\varepsilon\in(0,1)$, $\delta>0$, and sufficiently large $n$.
\end{property}

\begin{property}
[Exponentially Smaller Cardinality]The number $\left\vert T_{\delta}%
^{X^{n}Y^{n}}\right\vert $\ of $\delta$-jointly typical sequences is
exponentially smaller than the total number $\left(  \left\vert \mathcal{X}%
\right\vert \left\vert \mathcal{Y}\right\vert \right)  ^{n}$\ of sequences for
any joint random variable $\left(  X,Y\right)  $ that is not uniform. We
formally state this property as follows:%
\begin{equation}
\left\vert T_{\delta}^{X^{n}Y^{n}}\right\vert \leq2^{n\left(  H\left(
X,Y\right)  +c\delta\right)  },
\end{equation}
where $c$ is a positive constant. We can also bound the size of the $\delta
$-jointly typical set from below:%
\begin{equation}
\left\vert T_{\delta}^{X^{n}Y^{n}}\right\vert \geq\left(  1-\varepsilon
\right)  2^{n\left(  H(X,Y)-c\delta\right)  },
\end{equation}
for all $\varepsilon\in(0,1)$, $\delta>0$, and sufficiently large $n$.
\end{property}

\begin{property}
[Equipartition]The probability of a given $\delta$-jointly typical sequence
$x^{n}y^{n}$\ occurring is approximately uniform:%
\begin{equation}
2^{-n\left(  H(X,Y)+c\delta\right)  }\leq p_{X^{n},Y^{n}}(x^{n},y^{n}%
)\leq2^{-n\left(  H(X,Y)-c\delta\right)  },
\end{equation}
where $c$ is a positive constant.
\end{property}

\begin{property}
[Probability of Strong Joint Typicality]Consider two independent random
variables $\tilde{X}^{n}$ and $\tilde{Y}^{n}$ whose respective probability
density functions $p_{\tilde{X}^{n}}(x^{n})$\ and $p_{\tilde{Y}^{n}}(y^{n}%
)$\ are equal to the marginal densities of the joint density $p_{X^{n},Y^{n}%
}(x^{n},y^{n})$:%
\begin{equation}
(\tilde{X}^{n},\tilde{Y}^{n})\sim p_{X^{n}}(x^{n})p_{Y^{n}}(y^{n}).
\end{equation}
Then we can bound the probability that two random sequences $\tilde{X}^{n}$
and $\tilde{Y}^{n}$ are in the jointly typical set $T_{\delta}^{X^{n}Y^{n}}$:%
\begin{equation}
\Pr\left\{  (\tilde{X}^{n},\tilde{Y}^{n})\in T_{\delta}^{X^{n}Y^{n}}\right\}
\leq2^{-n\left(  I(X;Y)-3c\delta\right)  }.
\end{equation}

\end{property}

Proofs of the first three properties are the same as in the previous section,
and the proof of the last property is the same as that for the weakly typical case.

\section{Strong Conditional Typicality}

\label{sec-ct:strong-cond-typ}Strong conditional
\index{typicality!strong!conditional}%
typicality bears some similarities to strong typicality, but it is
sufficiently different for us to provide a discussion of it. We first
introduce it with a simple example.

Suppose that we draw a sequence from an alphabet $\left\{  0,1,2\right\}  $
according to the distribution:%
\begin{equation}
p_{X}(0)=\frac{1}{4},\ \ \ \ \ p_{X}(1)=\frac{1}{4},\ \ \ \ \ p_{X}%
(2)=\frac{1}{2}.
\end{equation}
A particular realization sequence could be as follows:%
\begin{equation}
2010201020120212122220202222.
\end{equation}
We count up the occurrences of each symbol and find them to be%
\begin{align}
N(0\ |\ 2010201020120212122220202222)  &  =8,\\
N(1\ |\ 2010201020120212122220202222)  &  =5,\\
N(2\ |\ 2010201020120212122220202222)  &  =15.
\end{align}
The maximum deviation of the sequence's empirical distribution from the true
distribution of the source is as follows:%
\begin{equation}
\max\left\{  \left\vert \frac{1}{4}-\frac{8}{28}\right\vert ,\left\vert
\frac{1}{4}-\frac{5}{28}\right\vert ,\left\vert \frac{1}{2}-\frac{15}%
{28}\right\vert \right\}  =\max\left\{  \frac{1}{28},\frac{2}{28},\frac{1}%
{28}\right\}  =\frac{1}{14}.
\end{equation}
We now consider generating a different sequence from an alphabet $\left\{
a,b,c\right\}  $. However, we generate it according to the following
\textit{conditional} probability distribution:%
\begin{equation}
\left[
\begin{array}
[c]{ccc}%
p_{Y|X}(a|0)=\frac{1}{5} & p_{Y|X}(a|1)=\frac{1}{6} & p_{Y|X}(a|2)=\frac{2}%
{4}\\
p_{Y|X}(b|0)=\frac{2}{5} & p_{Y|X}(b|1)=\frac{3}{6} & p_{Y|X}(b|2)=\frac{1}%
{4}\\
p_{Y|X}(c|0)=\frac{2}{5} & p_{Y|X}(c|1)=\frac{2}{6} & p_{Y|X}(c|2)=\frac{1}{4}%
\end{array}
\right]  . \label{eq-ct:stochastic-matrix}%
\end{equation}
The second generated sequence should thus have correlations with the original
sequence. A possible realization of the second sequence could be as follows:%
\begin{equation}
abbcbccabcabcabcabcbcbabacba.
\end{equation}
We would now like to analyze how close the empirical conditional distribution
is to the true conditional distribution for all input and output sequences. A
useful conceptual first step is to apply a permutation to the first sequence
so that all of its symbols appear in lexicographic order, and we then apply
the same permutation to the second sequence:%
\begin{gather}%
\begin{array}
[c]{cccccccccccccccccccccccccccc}%
2 & 0 & 1 & 0 & 2 & 0 & 1 & 0 & 2 & 0 & 1 & 2 & 0 & 2 & 1 & 2 & 1 & 2 & 2 &
2 & 2 & 0 & 2 & 0 & 2 & 2 & 2 & 2\\
a & b & b & c & b & c & c & a & b & c & a & b & c & a & b & c & a & b & c &
b & c & b & a & b & a & c & b & a
\end{array}
\nonumber\\
\underrightarrow{\operatorname{permute}}\nonumber\\%
\begin{array}
[c]{cccccccccccccccccccccccccccc}%
0 & 0 & 0 & 0 & 0 & 0 & 0 & 0 & 1 & 1 & 1 & 1 & 1 & 2 & 2 & 2 & 2 & 2 & 2 &
2 & 2 & 2 & 2 & 2 & 2 & 2 & 2 & 2\\
b & c & c & a & c & c & b & b & b & c & a & b & a & a & b & b & b & a & c &
b & c & b & c & a & a & c & b & a
\end{array}
.
\end{gather}
This rearrangement makes it easy to count up the empirical conditional
distribution of the second sequence. We first place the joint occurrences of
the symbols into the following matrix:%
\begin{equation}
\left[
\begin{array}
[c]{ccc}%
N(0,a)=1 & N(1,a)=2 & N(2,a)=5\\
N(0,b)=3 & N(1,b)=2 & N(2,b)=6\\
N(0,c)=4 & N(1,c)=1 & N(2,c)=4
\end{array}
\right]  ,
\end{equation}
and we obtain the empirical conditional distribution matrix by dividing these
entries by the marginal distribution of the first sequence:%
\begin{equation}
\left[
\begin{array}
[c]{ccc}%
\frac{N\left(  0,a\right)  }{N(0)}=\frac{1}{8} & \frac{N\left(  1,a\right)
}{N(1)}=\frac{2}{5} & \frac{N\left(  2,a\right)  }{N\left(  2\right)  }%
=\frac{5}{15}\\
\frac{N\left(  0,b\right)  }{N(0)}=\frac{3}{8} & \frac{N\left(  1,b\right)
}{N(1)}=\frac{2}{5} & \frac{N\left(  2,b\right)  }{N(2)}=\frac{6}{15}\\
\frac{N\left(  0,c\right)  }{N(0)}=\frac{4}{8} & \frac{N\left(  1,c\right)
}{N(1)}=\frac{1}{5} & \frac{N\left(  2,c\right)  }{N(2)}=\frac{4}{15}%
\end{array}
\right]  .
\end{equation}
We then compare the maximal deviation of the elements in this matrix with the
elements in the stochastic matrix in \eqref{eq-ct:stochastic-matrix}:%
\begin{multline}
\max\left\{  \left\vert \frac{1}{5}-\frac{1}{8}\right\vert ,\left\vert
\frac{2}{5}-\frac{3}{8}\right\vert ,\left\vert \frac{2}{5}-\frac{4}%
{8}\right\vert ,\left\vert \frac{1}{6}-\frac{2}{5}\right\vert ,\left\vert
\frac{3}{6}-\frac{2}{5}\right\vert ,\left\vert \frac{2}{6}-\frac{1}%
{5}\right\vert ,\left\vert \frac{2}{4}-\frac{5}{15}\right\vert ,\left\vert
\frac{1}{4}-\frac{6}{15}\right\vert ,\left\vert \frac{1}{4}-\frac{4}%
{15}\right\vert \right\} \\
=\max\left\{  \frac{3}{40},\frac{1}{40},\frac{1}{10},\frac{7}{30},\frac{1}%
{10},\frac{2}{15},\frac{1}{6},\frac{3}{20},\frac{1}{60}\right\}  =\frac{7}%
{30}.
\end{multline}
The above analysis applies to a finite realization to illustrate the notion of
conditional typicality, and there is a large deviation from the true
distribution in this case. We would again expect this deviation to vanish for
a random sequence in the limit as the length of the sequence becomes  large.

\subsection{Definition of Strong Conditional Typicality}

We now give a formal definition of strong conditional typicality.

\begin{definition}
[Conditional Empirical Distribution]The conditional empirical distribution
$t_{y^{n}|x^{n}}( y|x) $ is as follows:%
\begin{equation}
t_{y^{n}|x^{n}}( y|x) =\frac{t_{x^{n}y^{n}}( x,y) }{t_{x^{n}}( x) }.
\end{equation}

\end{definition}

\begin{definition}
[Strong Conditional Typicality]\label{def-ct:strong-cond-typ}Suppose that a
sequence $x^{n}$ is a strongly typical sequence in$~T_{\delta}^{X^{n}}$. Then
the $\delta$-strong conditionally typical set$~T_{\delta}^{Y^{n}|x^{n}}$
corresponding to the sequence $x^{n}$ consists of all sequences whose joint
empirical distribution$~\frac{1}{n}N(x,y|x^{n},y^{n})$\ is $\delta$-close to
the product of the true conditional distribution$~p_{Y|X}(y|x)$ with the
marginal empirical distribution~$\frac{1}{n}N(x|x^{n})$:%
\begin{multline}
T_{\delta}^{Y^{n}|x^{n}}\equiv\\
\left\{  y^{n}:\forall(x,y)\in\mathcal{X}\times\mathcal{Y}%
\begin{array}
[c]{cc}%
\left\vert N(x,y|x^{n},y^{n})-p(y|x)N(x|x^{n})\right\vert \leq n\delta &
\operatorname{if}p(y|x)>0\\
N(x,y|x^{n},y^{n})=0 & \operatorname{otherwise}%
\end{array}
\right\}  ,
\end{multline}
where we abbreviate $p_{Y|X}(y|x)$ as $p(y|x)$.
\end{definition}

The above definition of strong conditional typicality implies that the
conditional empirical distribution is close to the true conditional
distribution, in the sense that%
\begin{equation}
\left\vert \frac{t_{x^{n}y^{n}}( x,y) }{t_{x^{n}}( x) }-p_{Y|X}( y|x)
\right\vert \leq\frac{1}{t_{x^{n}}( x) }\delta.
\end{equation}
Of course, such a relation only makes sense if the marginal empirical
distribution $t_{x^{n}}( x) $\ is non-zero.

The extra technical condition ($N(x,y|x^{n},y^{n})=0$ if $p_{Y|X}(y|x)=0$) in
Definition~\ref{def-ct:strong-cond-typ}\ is present again for a reason that we
found in the proof of the Equipartition Property for Strong Typicality
(Property~\ref{prop-ct:strong-equi}).

\subsection{Properties of the Strong Conditionally Typical Set}

The set $T_{\delta}^{Y^{n}|x^{n}}$ of conditionally typical sequences enjoys a
few useful properties that are similar to what we have for the weak
conditionally typical set, but the initial sequence $x^{n}$ can be
deterministic. However, we do impose the constraint that it has to be strongly
typical so that we can prove useful properties for the corresponding strong
conditionally typical set. So first suppose that a given sequence $x^{n}\in
T_{\delta^{\prime}}^{X^{n}}$ for some $\delta^{\prime} > 0 $.

\begin{property}
[Unit Probability]\label{prop-ct:strong-cond-typ-unit}The set $T_{\delta
}^{Y^{n}|x^{n}}$ asymptotically has probability one. So as $n$ becomes large,
it is highly likely that a random sequence $Y^{n}$\ corresponding to a given
typical sequence $x^{n}$ is a conditionally typical sequence. We formally
state this property as follows:%
\begin{equation}
\Pr\left\{  Y^{n}\in T_{\delta}^{Y^{n}|x^{n}}\right\}  \geq1-\varepsilon,
\end{equation}
for all $\varepsilon\in(0,1)$, $\delta>0$, and sufficiently large $n$.
\end{property}

\begin{property}
[Exponentially Smaller Cardinality]The number $\left\vert T_{\delta}%
^{Y^{n}|x^{n}}\right\vert $\ of $\delta$-conditionally typical sequences is
exponentially smaller than the total number $\left\vert \mathcal{Y}\right\vert
^{n}$\ of sequences for any conditional random variable $Y$ that is not
uniform. We formally state this property as follows:%
\begin{equation}
\left\vert T_{\delta}^{Y^{n}|x^{n}}\right\vert \leq2^{n\left(  H(Y|X)+c(\delta
+\delta^{\prime})\right)  }.
\end{equation}
We can also bound the size of the $\delta$-conditionally typical set from
below:%
\begin{equation}
\left\vert T_{\delta}^{Y^{n}|x^{n}}\right\vert \geq\left(  1-\varepsilon
\right)  2^{n\left(  H(Y|X)-c(\delta+\delta^{\prime})\right)  },
\end{equation}
for all $\varepsilon\in(0,1)$, $\delta>0$, and sufficiently large $n$.
\end{property}

\begin{property}
[Equipartition]\label{prop-ct:equi-cond}The probability of a particular
$\delta$-conditionally typical sequence $y^{n}$\ is approximately uniform:%
\begin{equation}
2^{-n\left(  H(Y|X)+c(\delta+\delta^{\prime})\right)  }\leq p_{Y^{n}|X^{n}%
}(y^{n}|x^{n})\leq2^{-n\left(  H(Y|X)-c(\delta+\delta^{\prime})\right)  }.
\end{equation}

\end{property}

In summary, given a realization $x^{n}$ of the random variable $X^{n}$, the
conditionally typical set $T_{\delta}^{Y^{n}|x^{n}}$ has almost all the
probability, its size is exponentially smaller than the size of the set of all
sequences, and each $\delta$-conditionally typical sequence has an
approximately uniform probability of occurring.

\subsection{Proofs of the Properties of the Strong Conditionally Typical Set}

\label{sec-ct:proof-strong-typ}

\begin{proof}
[Proof of the Unit Probability Property
(Property~\ref{prop-ct:strong-cond-typ-unit})]A proof of this property is
somewhat more complicated for strong conditional typicality. Since we are
dealing with an i.i.d.~distribution, we can assume that the sequence $x^{n}$
is lexicographically ordered with an order on the alphabet $\mathcal{X}$. We
write the elements of $\mathcal{X}$ as $x_{1}$, \ldots, $x_{\left\vert
\mathcal{X}\right\vert }$. Then the lexicographic ordering means that we can
write the sequence $x^{n}$ as follows:%
\begin{equation}
x^{n}=\underbrace{x_{1}\cdots x_{1}}_{N\left(  x_{1}|x^{n}\right)
}\underbrace{x_{2}\cdots x_{2}}_{N\left(  x_{2}|x^{n}\right)  }\cdots
\underbrace{x_{\left\vert \mathcal{X}\right\vert }\cdots x_{\left\vert
\mathcal{X}\right\vert }}_{N(x_{\left\vert \mathcal{X}\right\vert }|x^{n})}.
\end{equation}
It follows that $N(x|x^{n})\geq n\left(  p_{X}(x)-\delta^{\prime}\right)  $
from the typicality of $x^{n}$, and the law of large numbers comes into play
for each block $x_{i}\cdots x_{i}$ with length $N(x_{i}|x^{n})$ when this
length is large enough. Let $p_{Y|X=x}(y)$ be the distribution for the
conditional random variable $Y|\left(  X=x\right)  $. Then the following is an
equivalent way to write the notion of conditional typicality:%
\begin{equation}
\left\{  y^{n}\in T_{\delta}^{Y^{n}|x^{n}}\right\}  \Leftrightarrow
\bigwedge\limits_{x\in\mathcal{X}}\left\{  y^{N(x|x^{n})}\in T_{\delta
}^{\left(  Y|\left(  X=x\right)  \right)  ^{N(x|x^{n})}}\right\}  ,
\end{equation}
where the symbol $\wedge$ denotes concatenation (note that the lexicographic
ordering of $x^{n}$ applies to the ordering of the sequence $y^{n}$ as well).
Also, $T_{\delta}^{\left(  Y|\left(  X=x\right)  \right)  ^{N(x|x^{n})}}$ is
the typical set for a sequence of conditional random variables $Y|\left(
X=x\right)  $ with length $N(x|x^{n})$:%
\begin{equation}
T_{\delta}^{\left(  Y|\left(  X=x\right)  \right)  ^{N(x|x^{n})}}%
\equiv\left\{  y^{N(x|x^{n})}:\forall y\in\mathcal{Y},\ \ \left\vert
\frac{N(y|y^{N(x|x^{n})})}{N(x|x^{n})}-p_{Y|X=x}(y)\right\vert \leq
\delta\right\}  .
\end{equation}
We can apply the law of large numbers to each of these typical sets
$T_{\delta}^{\left(  Y|\left(  X=x\right)  \right)  ^{N(x|x^{n})}}$ where the
length $N(x|x^{n})$\ becomes large. It then follows that%
\begin{align}
\Pr\left\{  Y^{n}\in T_{\delta}^{Y^{n}|x^{n}}\right\}   &  =%
{\displaystyle\prod\limits_{x\in\mathcal{X}}}
\Pr\left\{  Y^{N(x|x^{n})}\in T_{\delta}^{\left(  Y|\left(  X=x\right)
\right)  ^{N(x|x^{n})}}\right\} \\
&  \geq\left(  1-\varepsilon\right)  ^{\left\vert \mathcal{X}\right\vert }\\
&  \geq1-\left\vert \mathcal{X}\right\vert \varepsilon,
\end{align}
concluding the proof.
\end{proof}

\bigskip

\begin{proof}
[Proof of the Equipartition Property (Property~\ref{prop-ct:equi-cond})]The
following relation holds from the i.i.d.~property of the conditional
distribution $p_{Y^{n}|X^{n}}(y^{n}|x^{n})$ and because the sequence $y^{n}$
is strong conditionally typical according to
Definition~\ref{def-ct:strong-cond-typ}:%
\begin{equation}
p_{Y^{n}|X^{n}}(y^{n}|x^{n})=%
{\displaystyle\prod\limits_{\left(  \mathcal{X},\mathcal{Y}\right)  ^{+}}}
p_{Y|X}(y|x)^{N(x,y|x^{n},y^{n})}%
\end{equation}
where $\left(  \mathcal{X},\mathcal{Y}\right)  ^{+}$ denotes all the letters
$x,y$\ in $\mathcal{X},\mathcal{Y}$ with $p_{Y|X}(y|x)>0$. Take the logarithm
of the above expression:%
\begin{equation}
\log\left(  p_{Y^{n}|X^{n}}(y^{n}|x^{n})\right)  =\sum_{x,y\in\left(
\mathcal{X},\mathcal{Y}\right)  ^{+}}N(x,y|x^{n},y^{n})\log\left(
p_{Y|X}(y|x)\right)  .
\end{equation}
Multiply both sides by $-\frac{1}{n}$:%
\begin{equation}
-\frac{1}{n}\log\left(  p_{Y^{n}|X^{n}}(y^{n}|x^{n})\right)  =-\sum
_{x,y\in\left(  \mathcal{X},\mathcal{Y}\right)  ^{+}}\frac{1}{n}%
N(x,y|x^{n},y^{n})\log\left(  p_{Y|X}(y|x)\right)  .
\label{eq-ct:info-content-sub-cond}%
\end{equation}
The following relations hold because the sequence $x^{n}$ is strongly typical
and $y^{n}$ is strong conditionally typical:%
\begin{gather}
\forall x\in\mathcal{X}^{+}:\left\vert \frac{1}{n}N(x|x^{n})-p_{X}%
(x)\right\vert \leq\delta^{\prime},\\
\Rightarrow\forall x\in\mathcal{X}^{+}:-\delta^{\prime}+p_{X}(x)\leq\frac
{1}{n}N(x|x^{n})\leq\delta^{\prime}+p_{X}(x),\label{eq-ct:x-is-strong-typ}\\
\forall x,y\in\left(  \mathcal{X},\mathcal{Y}\right)  ^{+}:\left\vert \frac
{1}{n}N(x,y|x^{n},y^{n})-p_{Y|X}(y|x)\frac{1}{n}N(x|x^{n})\right\vert
\leq\delta
\end{gather}%
\begin{multline}
\Rightarrow\forall x,y\in\left(  \mathcal{X},\mathcal{Y}\right)  ^{+}%
:-\delta+p_{Y|X}(y|x)\frac{1}{n}N(x|x^{n})\leq\frac{1}{n}N(x,y|x^{n}%
,y^{n})\label{eq-ct:strongly-typ-expan-cond}\\
\leq\delta+p_{Y|X}(y|x)\frac{1}{n}N(x|x^{n}).
\end{multline}
Now multiply \eqref{eq-ct:strongly-typ-expan-cond} by $-\log\left(
p_{Y|X}(y|x)\right)  >0$, sum over all letters in the alphabet $\left(
\mathcal{X},\mathcal{Y}\right)  ^{+}$, and apply the substitution in
\eqref{eq-ct:info-content-sub-cond}. This procedure gives the following set of
inequalities:%
\begin{multline}
-\sum_{x,y\in\left(  \mathcal{X},\mathcal{Y}\right)  ^{+}}\left(
-\delta+p_{Y|X}(y|x)\frac{1}{n}N(x|x^{n})\right)  \log\left(  p_{Y|X}%
(y|x)\right) \\
\leq-\frac{1}{n}\log\left(  p_{Y^{n}|X^{n}}(y^{n}|x^{n})\right) \\
\leq-\sum_{x,y\in\left(  \mathcal{X},\mathcal{Y}\right)  ^{+}}\left(
\delta+p_{Y|X}(y|x)\frac{1}{n}N(x|x^{n})\right)  \log\left(  p_{Y|X}%
(y|x)\right)  .
\end{multline}
Now apply the inequalities in \eqref{eq-ct:x-is-strong-typ} (assuming that
$p_{X}(x)\geq\delta^{\prime}$ for $x\in\mathcal{X}^{+}$) to get that%
\begin{gather}
\Rightarrow-\sum_{x,y\in\left(  \mathcal{X},\mathcal{Y}\right)  ^{+}}\left(
-\delta+p_{Y|X}(y|x)\left(  -\delta^{\prime}+p_{X}(x)\right)  \right)
\log\left(  p_{Y|X}(y|x)\right) \\
\leq-\frac{1}{n}\log\left(  p_{Y^{n}|X^{n}}(y^{n}|x^{n})\right) \\
\leq-\sum_{x,y\in\left(  \mathcal{X},\mathcal{Y}\right)  ^{+}}\left(
\delta+p_{Y|X}(y|x)\left(  \delta^{\prime}+p_{X}(x)\right)  \right)
\log\left(  p_{Y|X}(y|x)\right) \\
\Rightarrow-c(\delta+\delta^{\prime})+H(Y|X)\leq-\frac{1}{n}\log\left(
p_{Y^{n}|X^{n}}(y^{n}|x^{n})\right)  \leq c(\delta+\delta^{\prime})+H(Y|X),\\
\Rightarrow2^{-n\left(  H(Y|X)+c(\delta+\delta^{\prime})\right)  }\leq
p_{Y^{n}|X^{n}}(y^{n}|x^{n})\leq2^{-n\left(  H(Y|X)-c(\delta+\delta^{\prime
})\right)  },
\end{gather}
where%
\begin{equation}
c\equiv-\sum_{x,y\in\left(  \mathcal{X},\mathcal{Y}\right)  ^{+}}\log\left(
p_{Y|X}(y|x)\right)  \geq0.
\end{equation}
It again becomes apparent why we require the technical condition in the
definition of strong conditional typicality
(Definition~\ref{def-ct:strong-cond-typ}). Were it not there, then the
constant $c$ would not be finite, and we would not be able to obtain a
reasonable bound on the probability of a strong conditionally typical sequence.
\end{proof}

We close this section with a lemma that relates strong conditional, marginal,
and joint typicality.

\begin{lemma}
\label{lem-ct:cond-marg-joint-typ}Suppose that $y^{n}$ is a conditionally
typical sequence in $T_{\delta}^{Y^{n}|x^{n}}$\ and its conditioning sequence
$x^{n}$ is a typical sequence in $T_{\delta^{\prime}}^{X^{n}}$. Then $x^{n}$
and $y^{n}$ are jointly typical in the set $T_{\delta+\delta^{\prime}}%
^{X^{n}Y^{n}}$, and $y^{n}$ is a typical sequence in $T_{\left\vert
\mathcal{X}\right\vert (\delta+\delta^{\prime})}^{Y^{n}}$.
\end{lemma}

\begin{proof}
It follows from the above that $\forall x\in\mathcal{X},y\in\mathcal{Y}:$%
\begin{align}
p_{X}(x)-\delta^{\prime}  &  \leq\frac{1}{n}N(x|x^{n})\leq\delta^{\prime
}+p_{X}(x),\\
p_{Y|X}(y|x)\frac{1}{n}N(x|x^{n})-\delta &  \leq\frac{1}{n}N(x,y|x^{n}%
,y^{n})\leq\delta+p_{Y|X}(y|x)\frac{1}{n}N(x|x^{n}).
\end{align}
Substituting the upper bound on $\frac{1}{n}N(x|x^{n})$ gives%
\begin{align}
\frac{1}{n}N(x,y|x^{n},y^{n})  &  \leq\delta+p_{Y|X}(y|x)\left(
\delta^{\prime}+p_{X}(x)\right) \\
&  =\delta+p_{Y|X}(y|x)\delta^{\prime}+p_{X}(x)p_{Y|X}(y|x)\\
&  \leq\delta+\delta^{\prime}+p_{X,Y}(x,y).
\end{align}
Similarly, substituting the lower bound on $\frac{1}{n}N(x|x^{n})$ gives%
\begin{equation}
\frac{1}{n}N(x,y|x^{n},y^{n})\geq p_{X,Y}(x,y)-\delta-\delta^{\prime}.
\end{equation}
Putting both of the above bounds together, we get the following bound:%
\begin{equation}
\left\vert \frac{1}{n}N(x,y|x^{n},y^{n})-p_{X,Y}(x,y)\right\vert \leq
\delta+\delta^{\prime}.
\end{equation}
This then implies that the sequences $x^{n}$ and $y^{n}$ lie in the strong
jointly typical set $T_{\delta+\delta^{\prime}}^{X^{n}Y^{n}}$. It follows from
the result of Exercise~\ref{ex-ct:strong-joint-typ->marg} that $y^{n}\in
T_{\left\vert \mathcal{X}\right\vert (\delta+\delta^{\prime})}^{Y^{n}}$.
\end{proof}

\section{Application:\ Channel Capacity Theorem}

\label{sec-ct:capacity}We close the technical content of this chapter with a
remarkable application of conditional typicality: Shannon's channel capacity
theorem. As discussed in Section~\ref{sec-cst:channel-code}, this theorem is
one of the central results of classical information theory, appearing in
Shannon's seminal paper. The theorem establishes that the highest achievable
rate for communication over many independent uses of a classical channel is
equal to a simple function of the channel.

We begin by defining the information-processing task and a corresponding
$\left(  n,R,\varepsilon\right)  $ channel code. It is helpful to recall
Figure~\ref{fig-intro:classical-channel-coding} depicting a general protocol
for communication over a classical channel $\mathcal{N}\equiv p_{Y|X}(y|x)$.
Before communication begins, the sender Alice and receiver Bob have already
established a codebook $\left\{  x^{n}(m)\right\}  _{m\in\mathcal{M}}$, where
each codeword $x^{n}(m)$ corresponds to a message $m$ that Alice might wish to
send to Bob. If Alice wishes to send message $m$, she inputs the codeword
$x^{n}(m)$ to the i.i.d.~channel $\mathcal{N}^{n}\equiv p_{Y^{n}|X^{n}}%
(y^{n}|x^{n})$. More formally, her encoding is some map $E^{n}:\mathcal{M}%
\rightarrow\mathcal{X}^{n}$. She then exploits $n$ uses of the channel to send
$x^{n}(m)$. Bob receives some sequence $y^{n}$ from the output of the channel,
and he performs a decoding $D^{n}:\mathcal{Y}^{n}\rightarrow\mathcal{M}$ in
order to recover the message $m$ that Alice transmits. The rate $R$\ of the
code is equal to $\left[  \log\left\vert \mathcal{M}\right\vert \right]  /n$,
measured in bits per channel use. The probability of error $p_{e}$ for an
$\left(  n,R,\varepsilon\right)  $ channel code is bounded from above as%
\begin{equation}
p_{e}\equiv\max_{m}\Pr\left\{  D^{n}\left(  \mathcal{N}^{n}\left(
E^{n}(m)\right)  \right)  \neq m\right\}  \leq\varepsilon.
\end{equation}
A communication rate $R$ is \textit{achievable} for $\mathcal{N}$ if there
exists an $\left(  n,R-\delta,\varepsilon\right)  $ channel code for all
$\varepsilon\in(0,1),\delta>0$ and sufficiently large $n$. The channel
capacity $C(\mathcal{N})$ of $\mathcal{N}$\ is the supremum of all achievable
rates for $\mathcal{N}$. We can now state Shannon's channel capacity theorem:

\begin{theorem}
[Shannon Channel Capacity]The maximum mutual information $I(\mathcal{N})$\ is
equal to the capacity $C(\mathcal{N})$\ of a channel $\mathcal{N}\equiv
p_{Y|X}(y|x)$:%
\begin{equation}
C(\mathcal{N})=I(\mathcal{N})\equiv\max_{p_{X}(x)}I(X;Y).
\end{equation}

\end{theorem}

\begin{proof}
A proof consists of two parts. The first part, known as the direct coding
theorem, demonstrates that the RHS $\leq$ LHS. That is, there is a sequence of
channel codes with rate $I( \mathcal{N}) $, demonstrating that this rate is
achievable. The second part, known as the converse part, demonstrates that the
LHS\ $\leq$ RHS. That is, it demonstrates that the rate on the RHS\ is
optimal, and it is impossible to have achievable rates exceeding it. Here, we
prove the direct coding theorem and hold off on proving the converse part
until we reach the HSW\ theorem in Chapter~\ref{chap:classical-comm-HSW}%
\ because the converse theorem there suffices as the converse part for this
classical theorem. We have already outlined the proof of the direct coding
theorem in Section~\ref{sec:random-code-idea}, and it might be helpful at this
point to review this section. In particular, the proof breaks down into three
parts: random coding to establish the encoding, the decoding algorithm for the
receiver, and the error analysis. We now give all of the technical details of
the proof because this chapter has established all the tools that we need.

\textbf{Code Construction.} Before communication begins, Alice and Bob agree
upon a code by the following random selection procedure. For every message
$m\in\mathcal{M}$, generate a codeword $x^{n}(m)$ i.i.d.~according to the
product distribution $p_{X^{n}}(x^{n})$, where $p_{X}(x)$ is a distribution
that maximizes $I(\mathcal{N})$. Importantly, this random construction is such
that every codeword is generated independently of the other codewords.

\textbf{Encoding.} If Alice wishes to send message $m$, she inputs the
codeword $x^{n}( m) $ to the channels.

\textbf{Decoding Algorithm.} After receiving the sequence $y^{n}$ from the
channel outputs, Bob tests whether $y^{n}$ is in the typical set $T_{\delta
}^{Y^{n}}$ corresponding to the distribution $p_{Y}(y)\equiv\sum_{x}%
p_{Y|X}(y|x)p_{X}(x)$. If it is not, then he reports an error. He then tests
if there is some message $m$ such that the sequence $y^{n}$ is in the
conditionally typical set $T_{\delta}^{Y^{n}|x^{n}(m)}$. If $m$ is the unique
message such that $y^{n}\in T_{\delta}^{Y^{n}|x^{n}(m)}$, then he declares $m$
to be the transmitted message. If there is no message $m$ such that $y^{n}\in
T_{\delta}^{Y^{n}|x^{n}(m)}$ or multiple messages $m^{\prime}$\ such that
$y^{n}\in T_{\delta}^{Y^{n}|x^{n}(m^{\prime})}$, then he reports an error.
Observe that the decoder is a function of the channel, so that we might say
that we construct channel codes \textquotedblleft from the
channel.\textquotedblright

\textbf{Error Analysis.} As discussed in the above decoding algorithm, there
are three kinds of errors that can occur in this communication scheme when
Alice sends the codeword $x^{n}( m) $ over the channels:

\begin{description}
\item[$\mathcal{E}_{0}( m) $\textbf{:}] The event that the channel output
$y^{n}$ is not in the typical set $T_{\delta}^{Y^{n}}$.

\item[$\mathcal{E}_{1}( m) $\textbf{:}] The event that the channel output
$y^{n}$ is in $T_{\delta}^{Y^{n}}$ but not in the conditionally typical set
$T_{\delta}^{Y^{n}|x^{n}( m) }$.

\item[$\mathcal{E}_{2}(m)$\textbf{:}] The event that the channel output
$y^{n}$ is in $T_{\delta}^{Y^{n}}$ but it is in the conditionally typical set
for some other message:%
\begin{equation}
\left\{  y^{n}\in T_{\delta}^{Y^{n}}\right\}  \text{ and }\left\{  \exists
m^{\prime}\neq m:y^{n}\in T_{\delta}^{Y^{n}|x^{n}(m^{\prime})}\right\}  .
\end{equation}

\end{description}

Recall from Section~\ref{sec:random-code-idea} that it is helpful to analyze
the expectation of the average error probability, where the expectation is
with respect to the random selection of the code and the average is with
respect to a uniformly random choice of the message $m$. Let $\mathcal{C}
\equiv\{ X^{n}(1), X^{n}(2), \ldots, X^{n}(|\mathcal{M}|)\}$ denote the random
variable corresponding to the random selection of a code. The expectation of
the average error probability of a randomly selected code is as follows:%
\begin{equation}
\mathbb{E}_{\mathcal{C}}\left\{  \frac{1}{\left\vert \mathcal{M}\right\vert
}\sum_{m}\Pr\left\{  \mathcal{E}_{0}( m) \cup\mathcal{E}_{1}( m)
\cup\mathcal{E}_{2}( m) \right\}  \right\}  .
\label{eq-ct:error-bound-capacity}%
\end{equation}
Our first \textquotedblleft move\textquotedblright\ is to exchange the
expectation and the sum, following from linearity of the expectation:%
\begin{equation}
\frac{1}{\left\vert \mathcal{M}\right\vert }\sum_{m}\mathbb{E}_{\mathcal{C}
}\left\{  \Pr\left\{  \mathcal{E}_{0}( m) \cup\mathcal{E}_{1}( m)
\cup\mathcal{E}_{2}( m) \right\}  \right\}  .
\end{equation}
Since all codewords are selected in the same way (randomly and independently
of the message $m$ and according to the same distribution $p_{X^{n}}(x^{n}))$,
the following equality holds for all $m,m^{\prime}\in\mathcal{M}$:
\begin{equation}
\mathbb{E}_{\mathcal{C} }\left\{  \Pr\left\{  \mathcal{E}_{0}( m)
\cup\mathcal{E}_{1}( m) \cup\mathcal{E}_{2}( m) \right\}  \right\}  =
\mathbb{E}_{\mathcal{C} }\left\{  \Pr\left\{  \mathcal{E}_{0}( m^{\prime})
\cup\mathcal{E}_{1}( m^{\prime}) \cup\mathcal{E}_{2}( m^{\prime}) \right\}
\right\}  ,
\end{equation}
implying that it suffices to analyze $\mathbb{E}_{\mathcal{C} }\left\{
\Pr\left\{  \mathcal{E}_{0}( m) \cup\mathcal{E}_{1}( m) \cup\mathcal{E}_{2}(
m) \right\}  \right\}  $ for just a single message $m$. Without loss of
generality, we can pick $m=1$ (the first message). Using the above, we find
that the expectation of the average error probability simplifies as follows:
\begin{equation}
\frac{1}{\left\vert \mathcal{M}\right\vert }\sum_{m}\mathbb{E}_{\mathcal{C}
}\left\{  \Pr\left\{  \mathcal{E}_{0}( m) \cup\mathcal{E}_{1}( m)
\cup\mathcal{E}_{2}( m) \right\}  \right\}  = \mathbb{E}_{\mathcal{C}
}\left\{  \Pr\left\{  \mathcal{E}_{0}( 1) \cup\mathcal{E}_{1}( 1)
\cup\mathcal{E}_{2}( 1) \right\}  \right\}  .
\end{equation}
So we can then apply the union bound:%
\begin{multline}
\mathbb{E}_{\mathcal{C} }\left\{  \Pr\left\{  \mathcal{E}_{0}( 1)
\cup\mathcal{E}_{1}( 1) \cup\mathcal{E}_{2}( 1) \right\}  \right\} \\
\leq\mathbb{E}_{\mathcal{C} }\left\{  \Pr\left\{  \mathcal{E}_{0}\left(
1\right)  \right\}  \right\}  +\mathbb{E}_{\mathcal{C} }\left\{  \Pr\left\{
\mathcal{E}_{1}( 1) \right\}  \right\}  +\mathbb{E}_{\mathcal{C} }\left\{
\Pr\left\{  \mathcal{E}_{2}( 1) \right\}  \right\}  .
\end{multline}

We now analyze each error individually. For each of the above events, we can
exploit indicator functions in order to simplify the error analysis (we are
also doing this to help build a bridge between this classical proof and the
packing lemma approach for the quantum case in Chapter~\ref{chap:packing}%
---projectors in some sense replace indicator functions later on). Recall that
an indicator function $I_{\mathcal{A}}(x)$ is equal to one if $x\in
\mathcal{A}$ and equal to zero otherwise. So the following three functions
being equal to one or larger then corresponds to error events $\mathcal{E}%
_{0}(1)$, $\mathcal{E}_{1}(1)$, and $\mathcal{E}_{2}(1)$, respectively:%
\begin{align}
&  1-I_{T_{\delta}^{Y^{n}}}(y^{n}),\label{eq-ct:indicator-1}\\
&  I_{T_{\delta}^{Y^{n}}}(y^{n})\left(  1-I_{T_{\delta}^{Y^{n}|x^{n}(1)}%
}(y^{n})\right)  ,\label{eq-ct:indicator-2}\\
&  \sum_{m^{\prime}\neq1}I_{T_{\delta}^{Y^{n}}}(y^{n})I_{T_{\delta}%
^{Y^{n}|x^{n}(m^{\prime})}}(y^{n}). \label{eq-ct:indicator-3}%
\end{align}
(The last sum of indicators is a consequence of applying the union bound again
to the error $\mathcal{E}_{2}(1)$, which itself is a union of events.)

By exploiting the indicator function from \eqref{eq-ct:indicator-1}, we have
that%
\begin{align}
&  \mathbb{E}_{\mathcal{C}}\left\{  \Pr\left\{  \mathcal{E}_{0}( 1) \right\}
\right\} \nonumber\\
&  =\mathbb{E}_{X^{n}(1)}\left\{  \mathbb{E}_{Y^{n}|X^{n}(1)}\left\{
1-I_{T_{\delta}^{Y^{n}}}(Y^{n})\right\}  \right\} \\
&  =1-\mathbb{E}_{X^{n}(1),Y^{n}}\left\{  I_{T_{\delta}^{Y^{n}}}%
(Y^{n})\right\} \\
&  =1-\mathbb{E}_{Y^{n}}\left\{  I_{T_{\delta}^{Y^{n}}}(Y^{n})\right\} \\
&  =\Pr\left\{  Y^{n}\notin T_{\delta}^{Y^{n}}\right\}  \leq\varepsilon,
\end{align}
where the first line follows because $Y^{n}$ is generated according to the
conditional distribution $p_{Y^{n}|X^{n}}$ and from $X^{n}(1)$ (since the
first message was transmitted) and all other codewords have no role in the
test, so that we marginalize over them. In the last line we have exploited the
high probability property of the typical set $T_{\delta}^{Y^{n}}$. In the
above, we are also exploiting the fact that $\mathbb{E}\left\{  I_{\mathcal{A}%
}\right\}  =\Pr\left\{  \mathcal{A}\right\}  $. By exploiting the indicator
function from \eqref{eq-ct:indicator-2}, we have that%
\begin{align}
&  \mathbb{E}_{\mathcal{C}}\left\{  \Pr\left\{  \mathcal{E}_{1}( 1) \right\}
\right\} \nonumber\\
&  =\mathbb{E}_{X^{n}(1)}\left\{  \mathbb{E}_{Y^{n}|X^{n}(1)}\left\{
I_{T_{\delta}^{Y^{n}}}(Y^{n})\left(  1-I_{T_{\delta}^{Y^{n}|X^{n}(1)}}%
(Y^{n})\right)  \right\}  \right\} \\
&  \leq\mathbb{E}_{X^{n}(1)}\left\{  \mathbb{E}_{Y^{n}|X^{n}(1)}\left\{
1-I_{T_{\delta}^{Y^{n}|X^{n}(1)}}(Y^{n})\right\}  \right\} \\
&  =1-\mathbb{E}_{X^{n}(1)}\left\{  \mathbb{E}_{Y^{n}|X^{n}(1)}\left\{
I_{T_{\delta}^{Y^{n}|X^{n}(1)}}(Y^{n})\right\}  \right\} \\
&  =\mathbb{E}_{X^{n}(1)}\left\{  \Pr_{Y^{n}|X^{n}(1)}\left\{  Y^{n}\notin
T_{\delta}^{Y^{n}|X^{n}(1)}\right\}  \right\}  \leq\varepsilon,
\end{align}
where in the last line we have exploited the high probability property of the
conditionally typical set $T_{\delta}^{Y^{n}|X^{n}(1)}$. We finally consider
the probability of the last kind of error by exploiting the indicator function
in \eqref{eq-ct:indicator-3}:%
\begin{align}
&  \mathbb{E}_{\mathcal{C}}\left\{  \Pr\left\{  \mathcal{E}_{2}( 1) \right\}
\right\} \nonumber\\
&  \leq\mathbb{E}_{\mathcal{C}}\left\{  \sum_{m^{\prime}\neq1}I_{T_{\delta
}^{Y^{n}}}(y^{n})I_{T_{\delta}^{Y^{n}|X^{n}(m^{\prime})}}(y^{n})\right\} \\
&  =\sum_{m^{\prime}\neq1}\mathbb{E}_{\mathcal{C}}\left\{  I_{T_{\delta
}^{Y^{n}}}(y^{n})I_{T_{\delta}^{Y^{n}|X^{n}(m^{\prime})}}(y^{n})\right\} \\
&  =\sum_{m^{\prime}\neq1}\mathbb{E}_{X^{n}(1),X^{n}(m^{\prime}),Y^{n}%
}\left\{  I_{T_{\delta}^{Y^{n}}}(y^{n})I_{T_{\delta}^{Y^{n}|X^{n}(m^{\prime}%
)}}(y^{n})\right\} \\
&  =\sum_{m^{\prime}\neq1}\sum_{x^{n}(1),x^{n}(m^{\prime}),y^{n}}p_{X^{n}%
}(x^{n}(1))p_{X^{n}}(x^{n}(m^{\prime}))\times\nonumber\\
&  \ \ \ \ \ \ \ \ \ \ \ \ p_{Y^{n}|X^{n}}(y^{n}|x^{n}(1))I_{T_{\delta}%
^{Y^{n}}}(y^{n})I_{T_{\delta}^{Y^{n}|x^{n}(m^{\prime})}}(y^{n})\\
&  =\sum_{m^{\prime}\neq1}\sum_{x^{n}(m^{\prime}),y^{n}}p_{X^{n}}%
(x^{n}(m^{\prime}))p_{Y^{n}}(y^{n})I_{T_{\delta}^{Y^{n}}}(y^{n})I_{T_{\delta
}^{Y^{n}|x^{n}(m^{\prime})}}(y^{n}).
\end{align}
The first inequality is from the union bound, and the first equality follows
from the way that we select the random code: for every message $m$, the
codewords are selected independently and randomly according to $p_{X^{n}}$ so
that the distribution for the joint random variable $X^{n}(1)X^{n}(m^{\prime
})Y^{n}$ is%
\begin{equation}
p_{X^{n}}(x^{n}(1))\,p_{X^{n}}(x^{n}(m^{\prime}))\,p_{Y^{n}|X^{n}}(y^{n}%
|x^{n}(1)).
\end{equation}
The second equality follows from marginalizing over $X^{n}(1)$. Continuing, we
have%
\begin{align}
&  \leq2^{-n\left[  H(Y)-\delta\right]  }\sum_{m^{\prime}\neq1}\sum
_{x^{n}(m^{\prime}),y^{n}}p_{X^{n}}(x^{n}(m^{\prime}))I_{T_{\delta}%
^{Y^{n}|x^{n}(m^{\prime})}}(y^{n})\\
&  =2^{-n\left[  H(Y)-\delta\right]  }\sum_{m^{\prime}\neq1}\sum
_{x^{n}(m^{\prime})}p_{X^{n}}(x^{n}(m^{\prime}))\sum_{y^{n}}I_{T_{\delta
}^{Y^{n}|x^{n}(m^{\prime})}}(y^{n})\\
&  \leq2^{-n\left[  H(Y)-\delta\right]  }2^{n\left[  H(Y|X)+\delta\right]
}\sum_{m^{\prime}\neq1}\sum_{x^{n}(m^{\prime})}p_{X^{n}}(x^{n}(m^{\prime}))\\
&  \leq\left\vert \mathcal{M}\right\vert 2^{-n\left[  I(X;Y)-2\delta\right]
}.
\end{align}
The first inequality follows from the bound $p_{Y^{n}}(y^{n})I_{T_{\delta
}^{Y^{n}}}(y^{n})\leq2^{-n\left[  H\left(  Y\right)  -\delta\right]  }$ that
holds for typical sequences. The second inequality follows from the
cardinality bound $\left\vert T_{\delta}^{Y^{n}|x^{n}(m^{\prime})}\right\vert
\leq2^{n\left[  H(Y|X)+\delta\right]  }$ on the conditionally typical set. The
last inequality follows because%
\begin{equation}
\sum_{x^{n}(m^{\prime})}p_{X^{n}}(x^{n}(m^{\prime}))=1,
\end{equation}
$\left\vert \mathcal{M}\right\vert $ is an upper bound on $\sum_{m^{\prime
}\neq1}1=|\mathcal{M}|-1$, and by the identity $I(X;Y)=H(Y)-H(Y|X)$. Thus, we
can make this error arbitrarily small by choosing the message set size
$\left\vert \mathcal{M}\right\vert =2^{n\left[  I(X;Y)-3\delta\right]  }$.
Putting everything together, we have the following bound on
\eqref{eq-ct:error-bound-capacity}:%
\begin{equation}
\varepsilon^{\prime}\equiv2\varepsilon+2^{-n\delta},
\end{equation}
as long as we choose the message set size as given above. It follows that
there exists a particular code with the same error bound on its average error
probability. We can then exploit an expurgation argument as discussed in
Section~\ref{sec:random-code-idea}\ to convert an average error bound into a
maximal one (the expurgation step throws away the worse half of the codewords,
guaranteeing a bound of $2\varepsilon^{\prime}$ on the maximum error
probability).
Thus, we have shown the achievability of an $\left(  n,C(\mathcal{N}%
)-\delta^{\prime},2\varepsilon^{\prime}\right)  $ channel code for all
$\delta^{\prime}>0,\varepsilon^{\prime}\in(0,1/2)$ and sufficiently large $n$
(where $\delta^{\prime}=3\delta+1/n$). Finally, as a simple observation, our
proof above does not rely on whether the definition of conditional typicality
employed is weak or strong.
\end{proof}

\section{Concluding Remarks}

This chapter deals with many different definitions and flavors of typicality
in the classical world, but the essential theme is Shannon's central
insight---the application of the law of large numbers in information theory.
Our main goal in information theory is to analyze the probability of error in
the transmission or compression of information. Thus, we deal with
probabilities and we do not care much what happens for all sequences, but we
instead only care what happens for the likely sequences. This frame of mind
immediately leads to the definition of a typical sequence and to a simple
scheme for the compression of information---keep only the typical sequences
and performance is optimal in the asymptotic limit. Despite the seemingly
different nature of quantum information when compared to its classical
counterpart, the intuition developed in this chapter carries over to the
quantum world in the next chapter where we define several different notions of
quantum typicality.

\section{History and Further Reading}

\cite{book1991cover} contains a great presentation of typicality in the
classical case. The proof of Property~\ref{prop-ct:min-card-typical-type} is
directly from the Cover and Thomas book. \cite{B77} introduced strong
typicality, and \cite{CK97} systematically developed it. Other helpful books
on information theory are those of \cite{B71} and \cite{Y02}. There are other
notions of typicality which are useful, including those presented in
\cite{el2010lecture}\ and \cite{Wolf78}. Our proof of Shannon's channel
capacity theorem is similar to that in \cite{Savov12}.

\chapter{Quantum Typicality}

\label{chap:quantum-typicality}This chapter marks the beginning of our study
of the asymptotic theory of quantum information, where we develop the
technical tools underpinning this theory. The intuition for it is similar to
the intuition we developed in the previous chapter on typical sequences, but
we will find some important differences between the classical and quantum cases.

So far, there is not a single known information-processing task%
\index{quantum typicality}
in quantum Shannon theory where the tools from this chapter are not helpful in
proving the achievability part of a coding theorem. For the most part, we can
straightforwardly import many of the ideas from the previous chapter about
typical sequences for use in the asymptotic theory of quantum information.
However, one might initially think that there are some obstacles to doing so.
For example, what is the analogy of a quantum information source? Once we have
established this notion, how would we determine if a state emitted from a
quantum information source is a typical state? In the classical case, a simple
way of determining typicality is to inspect all of the bits in the sequence.
But there is a problem with this approach in the quantum
domain---\textquotedblleft looking at quantum bits\textquotedblright\ is
equivalent to performing a measurement and doing so destroys delicate
superpositions that we would want to preserve in any subsequent quantum
information-processing task.

So how can we get around the aforementioned problem and construct a useful
notion of quantum typicality? Well, we should not be so destructive in
determining the answer to a question when it has only two possible answers.
After all, we are only asking \textquotedblleft Is the state typical or
not?\textquotedblright, and we can be a bit more delicate in the way that we
ask this question. As an analogy, suppose Bob is curious to determine whether
Alice could join him for dinner at a nice restaurant on the beach. He would
likely just phone her and politely ask, \textquotedblleft Sweet Alice, are you
available for a lovely oceanside dinner?\textquotedblright, as opposed to
barging into her apartment, probing through all of her belongings in search of
her calendar, and demanding that she join him if she is available. This latter
infraction would likely disturb her so much that she would never speak to him
again (and what would become of quantum Shannon theory without these two
communicating!). It is the same with quantum information---we must be gentle
when handling quantum states. Otherwise, we will disturb the state so much
that it will not be useful in any future quantum information-processing task.

We can gently ask a binary question of a quantum system by constructing an
incomplete measurement with only two outcomes. If one outcome has a high
probability of occurring, then we do not learn much about the state after
learning this outcome, and thus we would expect that this inquiry does not
disturb the state very much. For the case above, we can formulate the
question, \textquotedblleft Is the state typical or not?\textquotedblright\ as
a binary measurement that returns only the answer to this question and no more
information. Since it is highly likely that the state is indeed a typical
state, we would expect this inquiry not to disturb the state very much, and we
could use it for further quantum information-processing tasks. This is the
essential content of this chapter, and there are several technicalities
necessary to provide a rigorous underpinning.

We structure this chapter as follows. We first discuss the notion of a typical
subspace (the quantum analogy of the typical set). We can employ weak or
strong notions of typicality in the definition of quantum typicality.
Section~\ref{sec-qt:cond-typ}\ then discusses conditional quantum typicality,
a form of quantum typicality that applies to quantum states chosen randomly
according to a classical sequence. We end this chapter with a brief discussion
of the method of types for quantum systems. All of these developments are
important for understanding the asymptotic nature of quantum information and
for determining the ultimate limits of storage and transmission with quantum media.

\section{The Typical Subspace}

Our first task is to establish the notion of a
\index{quantum information source}%
quantum information source. It is analogous to the notion of a classical
information source, in the sense that the source randomly outputs a quantum
state according to some probability distribution, but the states that it
outputs do not necessarily have to be distinguishable as in the classical case.

\begin{definition}
[Quantum Information Source]\label{def-qt:q-info-source}A quantum information
source is some device that randomly emits pure qudit states in a Hilbert space
$\mathcal{H}_{A}$ of finite dimension.
\end{definition}

We use the symbol $A$ to denote the quantum system for the quantum information
source. Suppose that the source outputs states $|\psi_{y}\rangle$ randomly
according to some probability distribution $p_{Y}( y) $. Note that the states
$|\psi_{y}\rangle$ do not necessarily have to form an orthonormal set. Then
the density operator $\rho_{A}$\ of the source is the expected state emitted:%
\begin{equation}
\rho_{A}\equiv\mathbb{E}_{Y}\left\{  |\psi_{Y}\rangle\langle\psi_{Y}%
|_{A}\right\}  =\sum_{y}p_{Y}( y) |\psi_{y}\rangle\langle\psi_{y}|_{A}.
\end{equation}
There are many decompositions of a density operator as a convex sum of
rank-one projectors (and the above decomposition is one such example), but
perhaps the most important such decomposition is a spectral decomposition of the
density operator $\rho$:%
\begin{equation}
\rho_{A}=\sum_{x\in\mathcal{X}}p_{X}( x) \vert x\rangle\langle x\vert_{A}.
\label{eq-qt:spec-dec}%
\end{equation}
The above states $\vert x\rangle_{A}$ are eigenvectors of $\rho_{A}$ and form
a complete orthonormal basis for Hilbert space $\mathcal{H}_{A}$, and the
non-negative, convex real numbers $p_{X}( x) $ are the eigenvalues
of$~\rho_{A}$.

We have written the states $\vert x\rangle_{A}$ and the eigenvalues $p_{X}( x)
$ in a suggestive notation because it is actually possible to think of our
quantum source as a classical information source---the emitted states $\{\vert
x\rangle_{A}\}_{x\in\mathcal{X}}$ are orthonormal and each corresponding
eigenvalue $p_{X}( x) $ acts as a probability for choosing $\vert x\rangle
_{A}$. We can say that our source is classical because it is emitting the
orthogonal, and thus distinguishable, states $\vert x\rangle_{A}$ with
probability $p_{X}( x) $. This description is equivalent to the ensemble
$\left\{  p_{Y}( y) ,|\psi_{y}\rangle\right\}  _{y}$ because the two ensembles
lead to the same density operator (recall that two ensembles that have the
same density operator are essentially equivalent because they lead to the same
probabilities for outcomes of any measurement performed on the system). Our
quantum information source then corresponds to the pure-state ensemble:%
\begin{equation}
\left\{  p_{X}( x) ,\vert x\rangle_{A}\right\}  _{x\in\mathcal{X}}.
\end{equation}

Recall that the quantum entropy $H(A)_{\rho}$\ of the density operator
$\rho_{A}$\ is as follows (Definition~\ref{def:quantum-entropy}):%
\begin{equation}
H(A)_{\rho}\equiv-\operatorname{Tr}\left\{  \rho_{A}\log\rho_{A}\right\}  .
\end{equation}
It is straightforward to show that the quantum entropy $H(A)_{\rho}$\ is equal
to the Shannon entropy$~H(X)$ of a random variable $X$\ with distribution
$p_{X}(x)$\ because the basis states $|x\rangle_{A}$ are orthonormal.

Suppose now that the quantum information source emits a large number $n$ of
random quantum states so that the density operator describing the emitted
state is as follows:%
\begin{equation}
\rho_{A^{n}}\equiv\left.
\begin{array}
[c]{c}%
\underbrace{\rho_{A_{1}}\otimes\cdots\otimes\rho_{A_{n}}}\\
n
\end{array}
\right.  =(\rho_{A})^{\otimes n}. \label{eq-qt:n-copies}%
\end{equation}
The labels $A_{1}$, \ldots, $A_{n}$ denote the Hilbert spaces corresponding to
the different quantum systems, but the density operator is the same for each
quantum system $A_{1}$, \ldots, $A_{n}$\ and is equal to $\rho_{A}$. The above
description of a quantum source is within the i.i.d.~setting for the quantum
domain. A spectral decomposition of the state in \eqref{eq-qt:n-copies} is as
follows:%
\begin{align}
\rho_{A^{n}}  &  =\sum_{x_{1}\in\mathcal{X}}p_{X}(x_{1})|x_{1}\rangle\langle
x_{1}|_{A_{1}}\otimes\cdots\otimes\sum_{x_{n}\in\mathcal{X}}p_{X}(x_{n}%
)|x_{n}\rangle\langle x_{n}|_{A_{n}}\\
&  =\sum_{x_{1},\cdots,x_{n}\in\mathcal{X}}p_{X}(x_{1})\cdots p_{X}%
(x_{n})\left(  |x_{1}\rangle\cdots|x_{n}\rangle\right)  \left(  \langle
x_{1}|\cdots\langle x_{n}|\right)  _{A_{1},\ldots,A_{n}}%
\label{eq:quant-source}\\
&  =\sum_{x^{n}\in\mathcal{X}^{n}}p_{X^{n}}(x^{n})|x^{n}\rangle\langle
x^{n}|_{A^{n}},
\end{align}
where we employ the shorthand:%
\begin{equation}
p_{X^{n}}(x^{n})\equiv p_{X}(x_{1})\cdots p_{X}(x_{n}),\ \ \ \ \ \ \ \ |x^{n}%
\rangle_{A^{n}}\equiv|x_{1}\rangle_{A_{1}}\cdots|x_{n}\rangle_{A_{n}}.
\label{eq-qt:nth-basis-state-1}%
\end{equation}
The above quantum description of the density operator is essentially
equivalent to the classical picture of $n$ realizations of random variable $X$
with each eigenvalue $p_{X_{1}}(x_{1})\cdots p_{X_{n}}(x_{n})$ acting as a
probability because the set of states $\{|x_{1}\rangle\cdots|x_{n}%
\rangle_{A_{1},\ldots,A_{n}}\}_{x_{1},\cdots,x_{n}\in\mathcal{X}}$ is an
orthonormal set.

We can now \textquotedblleft quantize\textquotedblright\ or extend the notion
of typicality to the quantum information source. The definitions follow
directly from the classical definitions in
Chapter~\ref{chap:classical-typicality}. The quantum definition of typicality
can employ either the weak notion as in Definition~\ref{def-ct:weak-typ}\ or
the strong notion as in Definition~\ref{def-ct:strong-typ}. We do not
distinguish the notation for a typical subspace%
\index{typical subspace}
and a typical set because it should be clear from the context which kind of
typicality we are employing.

\begin{definition}
[Typical Subspace]\label{def-qt:typ-subspace}The $\delta$-\textit{typical
subspace} $T_{A^{n}}^{\delta}$ is a subspace of the full Hilbert space
$\mathcal{H}_{A^{n}}=\mathcal{H}_{A_{1}}\otimes\cdots\otimes\mathcal{H}%
_{A_{n}}$, associated with many copies of a density operator, such as the one
in \eqref{eq-qt:spec-dec}. It is spanned by states $|x^{n}\rangle_{A^{n}}%
$\ whose corresponding classical sequences $x^{n}$\ are $\delta$-typical:%
\begin{equation}
T_{A^{n}}^{\delta}\equiv\operatorname{span}\left\{  |x^{n}\rangle_{A^{n}%
}:x^{n}\in T_{\delta}^{X^{n}}\right\}  ,
\end{equation}
where it is implicit that the typical subspace $T_{A^{n}}^{\delta}$ on the
left-hand side\ is with respect to a density operator $\rho$ and the typical
set $T_{\delta}^{X^{n}}$ on the right-hand side\ is with respect to the
distribution $p_{X}(x)$ from the spectral decomposition of $\rho$ in
\eqref{eq-qt:spec-dec}. We could also denote the typical subspace as
$T_{A^{n}}^{\rho,\delta}$ if we would like to make the dependence of the space
on $\rho$ more explicit.
\end{definition}

\subsection{The Typical Subspace Measurement}

The definition of the typical subspace (Definition~\ref{def-qt:typ-subspace})
gives a way to divide up the Hilbert space of $n$ qudits into two subspaces:
the typical subspace and the atypical subspace. The properties of the typical
subspace are similar to what we found for typical sequences.
That is, the typical subspace is exponentially smaller than the full Hilbert
space of $n$ qudits, yet it contains nearly all of the probability (in a sense
that we show below). The intuition for these properties of the typical
subspace is the same as it is classically, as depicted in
Figure~\ref{fig-ct:typical-set}, once we have a spectral decomposition of a
density operator.

The \textit{typical projector}%
\index{typical projector}
is a projector onto the typical subspace, and the complementary projector
projects onto the atypical subspace. These projectors play an important
operational role in quantum Shannon theory because we can construct a quantum
measurement from them. That is, this measurement is the best way of asking the
question, \textquotedblleft Is the state typical or not?\textquotedblright%
\ because it minimally disturbs the state while still retrieving this one bit
of information.

\begin{definition}
[Typical Projector]\label{def-qt:typ-proj}Let $\Pi_{A^{n}}^{\delta}$ denote
the typical projector for the typical subspace of a density operator $\rho
_{A}$ with spectral decomposition in \eqref{eq-qt:spec-dec}. It is a projector
onto the typical subspace:%
\begin{equation}
\Pi_{A^{n}}^{\delta}\equiv\sum_{x^{n}\in T_{\delta}^{X^{n}}}|x^{n}%
\rangle\langle x^{n}|_{A^{n}},
\end{equation}
where it is implicit that the $x^{n}$ below the summation is a classical
sequence in the typical set $T_{\delta}^{X^{n}}$, and the state $|x^{n}%
\rangle$ is a quantum state given in \eqref{eq-qt:nth-basis-state-1} and
associated with the classical sequence $x^{n}$ via the spectral decomposition
of $\rho$ in \eqref{eq-qt:spec-dec}. We can also denote the typical projector
as $\Pi_{A^{n}}^{\rho,\delta}$ if we would like to make its dependence on
$\rho$ explicit.
\end{definition}

The action of multiplying the density operator $\rho_{A^{n}}$ by the typical
projector $\Pi_{A^{n}}^{\delta}$ is to select out all the basis states of
$\rho_{A^{n}}$ that are in the typical subspace and form a \textquotedblleft
sliced\textquotedblright\ operator $\tilde{\rho}_{A^{n}}$\ that is close to
the original density operator $\rho_{A^{n}}$:%
\begin{equation}
\tilde{\rho}_{A^{n}}\equiv\Pi_{A^{n}}^{\delta}\rho_{A^{n}}\Pi_{A^{n}}^{\delta
}=\sum_{x^{n}\in T_{\delta}^{X^{n}}}p_{X^{n}}(x^{n})|x^{n}\rangle\langle
x^{n}|_{A^{n}}.
\end{equation}
That is, the effect of projecting a state onto the typical subspace $T_{A^{n}%
}^{\delta}$ is to \textquotedblleft slice\textquotedblright\ out any component
of the state $\rho_{A^{n}}$ that does not lie in the typical subspace
$T_{A^{n}}^{\delta}$.

\begin{exercise}
\label{ex:typ-commute}Show that the typical projector $\Pi_{A^{n}}^{\delta}$
commutes with the density operator$~\rho_{A^{n}}$:%
\begin{equation}
\rho_{A^{n}}\Pi_{A^{n}}^{\delta}=\Pi_{A^{n}}^{\delta}\rho_{A^{n}}.
\end{equation}

\end{exercise}

The typical projector allows us to formulate an operational method for
delicately asking the question:\ \textquotedblleft Is the state typical or
not?\textquotedblright\ We can construct a quantum measurement that consists
of two outcomes:\ the outcome \textquotedblleft1\textquotedblright\ reveals
that the state is in the typical subspace, and \textquotedblleft%
0\textquotedblright\ reveals that it is not. This typical subspace measurement
is often one of the first important steps in most protocols in quantum Shannon theory.

\begin{definition}
[Typical Subspace Measurement]\label{def:typ-sub-meas}The following map is a
quantum instrument (see Section~\ref{sec-pqt:instrument}) that realizes the
\index{typical subspace!measurement}%
typical subspace measurement:%
\begin{equation}
\sigma\rightarrow\left(  I-\Pi_{A^{n}}^{\delta}\right)  \sigma\left(
I-\Pi_{A^{n}}^{\delta}\right)  \otimes|0\rangle\langle0|+\Pi_{A^{n}}^{\delta
}\sigma\Pi_{A^{n}}^{\delta}\otimes|1\rangle\langle1|,
\end{equation}
where $\sigma$ is some density operator acting on the Hilbert space
$\mathcal{H}_{A^{n}}$. It associates a classical register with the outcome of
the measurement---the value of the classical register is $|0\rangle$ for the
support of the state $\sigma$\ that is not in the typical subspace, and it is
equal to $|1\rangle$ for the support of the state $\sigma$ that is in the
typical subspace.
\end{definition}

The implementation of a typical subspace measurement is currently far from the
reality of what is experimentally accessible if we would like to have the
measure concentration effects necessary for proving many of the results in
quantum Shannon theory. Recall from Figure~\ref{fig-ct:typicality}\ that we
required a sequence of about a million bits in order to have the needed
measure concentration effects. We would need a similar number of qubits
emitted from a quantum information source, and furthermore, we would require
the ability to perform noiseless coherent operations over about a million or
more qubits in order to implement the typical subspace measurement. Such a
daunting requirement firmly places quantum Shannon theory as a
\textquotedblleft highly theoretical theory,\textquotedblright\ rather than
being a theory that can make close connection to current experimental
practice.\footnote{We should note that this was certainly the case as well for
information theory when Claude Shannon developed it in 1948, but in the many
years since then, there has been much progress in the development of practical
classical codes for achieving the classical capacity of a classical channel.}

\subsection{The Difference between the Typical Set and the Typical Subspace}

We now offer a simple example to discuss the difference between the classical
viewpoint associated with the typical set and the quantum viewpoint associated
with the typical subspace. Suppose that a quantum information source emits the
state $\vert+\rangle$ with probability $1/2$ and it emits the state
$\vert0\rangle$ with probability $1/2$. For the moment, let us ignore the fact
that the two states $\vert+\rangle$ and $\vert0\rangle$ are not perfectly
distinguishable and instead suppose that they are. Then it would turn out that
nearly every sequence emitted from this source is a typical sequence because
the distribution of the source is uniform. Recall that the typical set has
size roughly equal to $2^{nH( X) }$, and in this case, the entropy of the
distribution $\left(  \frac{1}{2},\frac{1}{2}\right)  $ is equal to one bit.
Thus the size of the typical set is roughly the same as the size of the set of
all sequences for this distribution because $2^{nH( X) }=2^{n}$.

Now let us take into account the fact that the states $\left\vert
+\right\rangle $ and $|0\rangle$ are not perfectly distinguishable and use the
prescription given in Definition~\ref{def-qt:typ-subspace}\ for the typical
subspace. The density operator of the above ensemble is as follows:%
\begin{equation}
\frac{1}{2}|+\rangle\langle+|+\frac{1}{2}|0\rangle\langle0|=%
\begin{bmatrix}
\frac{3}{4} & \frac{1}{4}\\
\frac{1}{4} & \frac{1}{4}%
\end{bmatrix}
,
\end{equation}
where its matrix representation is with respect to the computational basis.
The spectral decomposition of the density operator is%
\begin{equation}
\cos^{2}(\pi/8)|\psi_{0}\rangle\langle\psi_{0}|+\sin^{2}(\pi/8)|\psi
_{1}\rangle\langle\psi_{1}|,
\end{equation}
where the states $|\psi_{0}\rangle$ and $|\psi_{1}\rangle$\ are orthogonal,
and thus distinguishable from one another. The quantum information source that
outputs $|0\rangle$ and $|+\rangle$ with equal probability is thus equivalent
to a source that outputs $|\psi_{0}\rangle$ with probability $\cos^{2}(\pi/8)$
and $|\psi_{1}\rangle$ with probability $\sin^{2}(\pi/8)$.

We construct the projector onto the typical subspace by taking sums of typical
strings of the states $|\psi_{0}\rangle\langle\psi_{0}|$ and $|\psi_{1}%
\rangle\langle\psi_{1}|$ rather than the states $|0\rangle\langle0|$ and
$|+\rangle\langle+|$, where typicality is with respect to the distribution
$\left(  \cos^{2}(\pi/8),\sin^{2}(\pi/8)\right)  $. The dimension of the
typical subspace corresponding to the quantum information source is far
different from the size of the aforementioned typical set corresponding to the
distribution $\left(  1/2,1/2\right)  $. It is roughly equal to $2^{0.6n}$
because the entropy of the distribution $\left(  \cos^{2}(\pi/8),\sin^{2}%
(\pi/8)\right)  $ is about 0.6 bits. This stark contrast in the sizes has to
do with the non-orthogonality of the states from the original description of
the ensemble. That is, non-orthogonality of states in an ensemble implies that
the size of the typical subspace can potentially be dramatically smaller than
the size of the typical set corresponding to the distribution of the states in
the ensemble. This result has implications for the compressibility of quantum
information, and we will discuss these ideas in more detail in
Chapter~\ref{chap:schumach}. For now, we continue with the technical details
of typical subspaces.

\subsection{Properties of the Typical Subspace}

The typical subspace $T_{A^{n}}^{\delta}$ enjoys several useful properties
that are \textquotedblleft quantized\textquotedblright\ versions of the
typical sequence properties:

\begin{property}
[Unit Probability]\label{prop-qt:unit}Suppose that we perform a typical
subspace measurement of a state $\rho_{A^{n}}$. Then the probability that the
quantum state $\rho_{A^{n}}$ is in the typical subspace $T_{A^{n}}^{\delta}$
approaches one as $n$ becomes large. That is,
\begin{equation}
\operatorname{Tr}\left\{  \Pi_{A^{n}}^{\delta}\rho_{A^{n}}\right\}
\geq1-\varepsilon,
\end{equation}
for all $\varepsilon\in(0,1)$, $\delta>0$, and sufficiently large $n$, where
$\Pi_{A^{n}}^{\delta}$ is the typical subspace projector from
Definition~\ref{def-qt:typ-proj}.
\end{property}

\begin{property}
[Exponentially Smaller Dimension]\label{prop-qt:exp-small}The dimension
$\dim(T_{A^{n}}^{\delta})$\ of the $\delta$-typical subspace is exponentially
smaller than the dimension $\left\vert A\right\vert ^{n}$\ of the
entire space of quantum states when the output of the quantum information
source is not maximally mixed. We formally state this property as follows:%
\begin{equation}
\operatorname{Tr}\left\{  \Pi_{A^{n}}^{\delta}\right\}  \leq2^{n\left(
H(A)+c\delta\right)  },
\end{equation}
where $c$ is some positive constant that depends on whether we employ the weak
or strong notion of typicality. We can also bound the dimension $\dim
(T_{A^{n}}^{\delta})$\ of the $\delta$-typical subspace from below:%
\begin{equation}
\operatorname{Tr}\left\{  \Pi_{A^{n}}^{\delta}\right\}  \geq\left(
1-\varepsilon\right)  2^{n\left(  H(A)-c\delta\right)  },
\end{equation}
for all $\varepsilon\in(0,1)$, $\delta>0$, and sufficiently large $n$.
\end{property}

\begin{property}
[Equipartition]\label{prop-qt:equi}The operator $\Pi_{A^{n}}^{\delta}%
\rho_{A^{n}}\Pi_{A^{n}}^{\delta}$ corresponds to a \textquotedblleft
slicing\textquotedblright\ of the density operator $\rho_{A^{n}}$ where we
slice out and keep only the part with support in the typical subspace. We can
then bound all of the eigenvalues of the sliced operator $\Pi_{A^{n}}^{\delta
}\rho_{A^{n}}\Pi_{A^{n}}^{\delta}$ as follows:%
\begin{equation}
2^{-n\left(  H(A)+c\delta\right)  }\Pi_{A^{n}}^{\delta}\leq\Pi_{A^{n}}%
^{\delta}\rho_{A^{n}}\Pi_{A^{n}}^{\delta}\leq2^{-n\left(  H(A)-c\delta\right)
}\Pi_{A^{n}}^{\delta}.
\end{equation}
The above inequality is an operator inequality. It is a statement about the
eigenvalues of the operators $\Pi_{A^{n}}^{\delta}\rho_{A^{n}}\Pi_{A^{n}%
}^{\delta}$ and $\Pi_{A^{n}}^{\delta}$, and these operators have the same
eigenvectors because they commute. Therefore, the above inequality is
equivalent to the following inequality that applies in the classical case:%
\begin{equation}
\forall x^{n}\in T_{\delta}^{X^{n}}:2^{-n\left(  H(A)+c\delta\right)  }\leq
p_{X^{n}}(x^{n})\leq2^{-n\left(  H(A)-c\delta\right)  }.
\end{equation}
This equivalence holds because each probability $p_{X^{n}}(x^{n})$ is an
eigenvalue of $\Pi_{A^{n}}^{\delta}\rho_{A^{n}}\Pi_{A^{n}}^{\delta}$.
\end{property}

The dimension $\dim(T_{A^{n}}^{\delta})$\ of the $\delta$-typical subspace is
approximately equal to the dimension $\left\vert \mathcal{X}\right\vert ^{n}$
of the entire space only when the density operator of the quantum information
source is maximally mixed because%
\begin{equation}
\operatorname{Tr}\left\{  \Pi_{A^{n}}^{\delta}\right\}  \leq\left\vert
A\right\vert ^{n}\cdot2^{n\delta}\simeq\left\vert A
\right\vert ^{n}.
\end{equation}

Proofs of the above properties are essentially identical to those from the
classical case in Sections~\ref{sec-ct:proof-weak-typ} and
\ref{sec-ct:proof-strong-typ}, regardless of whether we employ a weak or
strong notion of quantum typicality. We leave the proofs as the three
exercises below.

\begin{exercise}
Prove the unit probability property of the $\delta$-typical subspace
(Property~\ref{prop-qt:unit}). First show that the probability that many
copies of a density operator is in the $\delta$-typical subspace is equal to
the probability that a random sequence is $\delta$-typical:%
\begin{equation}
\operatorname{Tr}\left\{  \Pi_{A^{n}}^{\delta}\rho_{A^{n}}\right\}
=\Pr\left\{  X^{n}\in T_{\delta}^{X^{n}}\right\}  .
\end{equation}

\end{exercise}

\begin{exercise}
\label{ex:dim-typ}Prove the exponentially smaller dimension property of the
$\delta$-typical subspace (Property~\ref{prop-qt:exp-small}). First show that
the trace of the typical projector $\Pi_{A^{n}}^{\delta}$ is equal to the
dimension of the typical subspace $T_{A^{n}}^{\delta}$:%
\begin{equation}
\dim(T_{A^{n}}^{\delta})=\operatorname{Tr}\left\{  \Pi_{A^{n}}^{\delta
}\right\}  .
\end{equation}
Then prove the property.
\end{exercise}

\begin{exercise}
Prove the equipartition property of the $\delta$-typical subspace
(Property~\ref{prop-qt:equi}). First show that%
\begin{equation}
\Pi_{A^{n}}^{\delta}\rho_{A^{n}}\Pi_{A^{n}}^{\delta}=\sum_{x^{n}\in T_{\delta
}^{X^{n}}}p_{X^{n}}(x^{n})|x^{n}\rangle\langle x^{n}|_{A^{n}},
\end{equation}
and then argue the proof.
\end{exercise}

The result of the following exercise shows that the sliced operator
$\tilde{\rho}_{A^{n}}\equiv\Pi_{A^{n}}^{\delta}\rho_{A^{n}}\Pi_{A^{n}}%
^{\delta}$ is a good approximation to the original state $\rho_{A^{n}}$ in the
limit of many copies of the states, and it effectively gives a scheme for
quantum data compression (more on this in Chapter~\ref{chap:schumach}).

\begin{exercise}
\label{ex:gentle-typ-close}Use the gentle operator lemma
(Lemma~\ref{lem-dm:gentle-operator}) to show that $\rho_{A^{n}}$ is
$2\sqrt{\varepsilon}$-close to the sliced operator $\tilde{\rho}_{A^{n}}$ when
$n$ is large:%
\begin{equation}
\left\Vert \rho_{A^{n}}-\tilde{\rho}_{A^{n}}\right\Vert _{1}\leq
2\sqrt{\varepsilon}.
\end{equation}
Use the gentle measurement lemma (Lemma~\ref{lem-dm:gentle-measurement}) to
show that the sliced state%
\begin{equation}
\left[  \operatorname{Tr}\left\{  \Pi_{A^{n}}^{\delta}\rho_{A^{n}}\right\}
\right]  ^{-1}\tilde{\rho}_{A^{n}}%
\end{equation}
is $2\sqrt{\varepsilon}$-close in trace distance to $\tilde{\rho}_{A^{n}}$.
\end{exercise}

\begin{exercise}
Show that the purity $\operatorname{Tr}\left\{  \left(  \tilde{\rho}_{A^{n}%
}\right)  ^{2}\right\}  $ of the sliced state $\tilde{\rho}_{A^{n}}%
$\ satisfies the following bound for sufficiently large $n$ and any
$\varepsilon\in(0,1)$ (use weak quantum typicality):%
\begin{equation}
\left(  1-\varepsilon\right)  2^{-n\left(  H(A)+\delta\right)  }%
\leq\operatorname{Tr}\left\{  \left(  \tilde{\rho}_{A^{n}}\right)
^{2}\right\}  \leq2^{-n\left(  H(A)-\delta\right)  }.
\end{equation}

\end{exercise}

\begin{exercise}
Show that the following bounds hold for the rank and the $\infty$-norm of the
sliced state $\tilde{\rho}_{A^{n}}$ for any $\varepsilon\in(0,1)$ and
sufficiently large $n$:%
\begin{align}
\left(  1-\varepsilon\right)  2^{n\left(  H(A)-\delta\right)  }  &
\leq\operatorname{rank}(\tilde{\rho}_{A^{n}})\leq2^{n\left(  H(A)+\delta
\right)  },\\
2^{-n\left(  H(A)+\delta\right)  }  &  \leq\left\Vert \tilde{\rho}_{A^{n}%
}\right\Vert _{\infty}\leq2^{-n\left(  H(A)-\delta\right)  }.
\end{align}
(Recall that the rank of an operator is equal to the size of its support and
that the infinity norm is equal to its maximum eigenvalue. Again use weak
quantum typicality.)
\end{exercise}

\subsection{The Typical Subspace for Bipartite or Multipartite States}

Recall from Section~\ref{sec-ct:weak-joint-typ}\ that two classical sequences
$x^{n}$ and $y^{n}$ are weak jointly typical if the joint sample entropy of
$x^{n}y^{n}$ is close to the joint entropy $H( X,Y) $ and if the sample
entropies of the individual sequences are close to their respective marginal
entropies $H( X) $ and $H( Y) $ (where the entropies are with respect to some
joint distribution $p_{X,Y}( x,y) $). How would we then actually check that
these conditions hold?\ The most obvious way is simply to look at the sequence
$x^{n}y^{n}$, compute its joint sample entropy, compare this quantity to the
true joint entropy, determine if the difference is under the threshold
$\delta$, and do the same for the marginal sequences. These two operations
both commute in the sense that we can determine first if the marginals are
typical and then if the joint sequence is typical or vice versa without any
difference in which one we do first.

But such a commutation does not necessarily hold in the quantum world. The way
that we determine whether a quantum state is typical is by performing a
typical subspace measurement. If we perform a typical subspace measurement of
the whole system followed by such a measurement on the marginals, the
resulting state is not necessarily the same as if we performed the marginal
measurements followed by the joint measurements. For this reason, the notion
of weak joint typicality as given in
Definition~\ref{def-ct:weak-joint-typ-set}\ does not really exist in general
for the quantum case. Nevertheless, we still briefly overview how one would
handle such a case and later give an example of a restricted class of states
for which weak joint typicality holds.

Suppose that we have a quantum system in the mixed state $\rho_{AB}$\ shared
between two parties $A$ and $B$. We can decompose the mixed state with the
spectral theorem:%
\begin{equation}
\rho_{AB}=\sum_{z\in\mathcal{Z}}p_{Z}(z)|\psi_{z}\rangle\langle\psi_{z}|_{AB},
\end{equation}
where the states $\{|\psi_{z}\rangle_{AB}\}_{z\in\mathcal{Z}}$ form an
orthonormal basis for the joint quantum system $AB$ and each of the states
$|\psi_{z}\rangle_{AB}$ can be entangled in general.

We can consider the $n$th extension $\rho_{A^{n}B^{n}}$ of the above state and
abbreviate its spectral decomposition as follows:%
\begin{equation}
\rho_{A^{n}B^{n}}\equiv(\rho_{AB})^{\otimes n}=\sum_{z^{n}\in\mathcal{Z}^{n}%
}p_{Z^{n}}(z^{n})|\psi_{z^{n}}\rangle\langle\psi_{z^{n}}|_{A^{n}B^{n}},
\end{equation}
where%
\begin{align}
p_{Z^{n}}(z^{n})  &  \equiv p_{Z}(z_{1})\cdots p_{Z}(z_{n}),\\
|\psi_{z^{n}}\rangle_{A^{n}B^{n}}  &  \equiv\left\vert \psi_{z_{1}%
}\right\rangle _{A_{1}B_{1}}\cdots\left\vert \psi_{z_{n}}\right\rangle
_{A_{n}B_{n}}.
\end{align}
This development immediately leads to the definition of the typical subspace
for a bipartite state.

\begin{definition}
[Typical Subspace of a Bipartite State]The $\delta$\textit{-typical subspace}
$T_{A^{n}B^{n}}^{\delta}$ of $\rho_{AB}$ is the space spanned by states
$|\psi_{z^{n}}\rangle_{A^{n}B^{n}}$ whose corresponding classical sequence
$z^{n}$\ is in the typical set $T_{\delta}^{Z^{n}}$:%
\begin{equation}
T_{A^{n}B^{n}}^{\delta}\equiv\operatorname{span}\left\{  |\psi_{z^{n}}%
\rangle_{A^{n}B^{n}}:z^{n}\in T_{\delta}^{Z^{n}}\right\}  .
\end{equation}
The states $|\psi_{z^{n}}\rangle_{A^{n}B^{n}}$ are from a spectral
decomposition of $\rho_{AB}$, and the distribution to consider for typicality
of the classical sequence $z^{n}$ is $p_{Z}(z)$ from the spectral decomposition.
\end{definition}

\begin{definition}
[Typical Projector of a Bipartite State]Let $\Pi_{A^{n}B^{n}}^{\delta}$ denote
the projector onto the typical subspace of $\rho_{AB}$:%
\begin{equation}
\Pi_{A^{n}B^{n}}^{\delta}\equiv\sum_{z^{n}\in T_{\delta}^{Z^{n}}}|\psi_{z^{n}%
}\rangle\langle\psi_{z^{n}}|_{A^{n}B^{n}}.
\end{equation}

\end{definition}

Thus, there is ultimately no difference between the typical subspace for a
bipartite state and the typical subspace for a single-party state because the
spectral decomposition gives a way for determining the typical subspace and
the typical projector in both cases. Perhaps the only difference is a cosmetic
one because $AB$ denotes the bipartite system while $Z$ indicates a random
variable with a distribution given from a spectral decomposition. Finally,
Properties~\ref{prop-qt:unit}--\ref{prop-qt:equi}\ hold for quantum typicality
of a bipartite state.

\subsection{The Jointly Typical Subspace for Classical States}

The notion of weak joint typicality may not hold in the general case, but it
does hold for a special class of states that are completely classical. Suppose
now that the mixed state $\rho_{AB}$\ shared between two parties $A$ and $B$
has the following special form:%
\begin{align}
\rho_{AB}  &  =\sum_{x\in\mathcal{X}}\sum_{y\in\mathcal{Y}}p_{X,Y}(x,y)\left(
|x\rangle\otimes|y\rangle\right)  \left(  \langle x|\otimes\langle y|\right)
_{AB}\label{eq-qt:special-joint-form}\\
&  =\sum_{x\in\mathcal{X}}\sum_{y\in\mathcal{Y}}p_{X,Y}(x,y)|x\rangle\langle
x|_{A}\otimes|y\rangle\langle y|_{B},
\end{align}
where the states $\{|x\rangle_{A}\}_{x\in\mathcal{X}}$ and $\{|y\rangle
_{B}\}_{y\in\mathcal{Y}}$\ form an orthonormal basis for the respective
systems $\mathcal{X}$ and $\mathcal{Y}$. This state has only classical
correlations because Alice and Bob can prepare it simply by local operations
and classical communication. That is, Alice can sample from the distribution
$p_{X,Y}(x,y)$ in her laboratory and send Bob the variable $y$. Furthermore,
the states on $A$ and $B$ locally form a distinguishable set.

We can consider the $n$th extension $\rho_{A^{n}B^{n}}$ of the above state:%
\begin{align}
\rho_{A^{n}B^{n}}  &  \equiv(\rho_{AB})^{\otimes n}%
\label{eq-qt:classical-joint-state-1}\\
&  =\sum_{x^{n}\in\mathcal{X}^{n},y^{n}\in\mathcal{Y}^{n}}p_{X^{n},Y^{n}%
}(x^{n},y^{n})\left(  |x^{n}\rangle\otimes|y^{n}\rangle\right)  \left(
\langle x^{n}|\otimes\langle y^{n}|\right)  _{A^{n}B^{n}}\\
&  =\sum_{x^{n}\in\mathcal{X}^{n},y^{n}\in\mathcal{Y}^{n}}p_{X^{n},Y^{n}%
}(x^{n},y^{n})|x^{n}\rangle\langle x^{n}|_{A^{n}}\otimes|y^{n}\rangle\langle
y^{n}|_{B^{n}}.
\end{align}
This development immediately leads to the definition of the weak jointly
typical subspace for this special case.

\begin{definition}
[Jointly Typical Subspace]The weak $\delta$-\textit{jointly typical subspace}
$T_{A^{n}B^{n}}^{\delta}$ is the space spanned by states $|x^{n}\rangle
|y^{n}\rangle_{A^{n}B^{n}}$ whose corresponding classical sequence $x^{n}%
y^{n}$\ is in the \textit{jointly }typical set:%
\begin{equation}
T_{A^{n}B^{n}}^{\delta}\equiv\operatorname{span}\left\{  |x^{n}\rangle_{A^{n}%
}|y^{n}\rangle_{B^{n}}:x^{n}y^{n}\in T_{\delta}^{X^{n}Y^{n}}\right\}  .
\end{equation}

\end{definition}

\begin{definition}
[Jointly Typical Projector]Let $\Pi_{A^{n}B^{n}}^{\delta}$ denote the jointly
typical projector. It is the projector onto the jointly typical subspace:%
\begin{equation}
\Pi_{A^{n}B^{n}}^{\delta}\equiv\sum_{x^{n},y^{n}\in T_{\delta}^{X^{n}Y^{n}}%
}|x^{n}\rangle\langle x^{n}|_{A^{n}}\otimes|y^{n}\rangle\langle y^{n}|_{B^{n}%
}.
\end{equation}

\end{definition}

\subsubsection{Properties of the Jointly Typical Projector for Classical
States}

Properties~\ref{prop-qt:unit}--\ref{prop-qt:equi} apply to the jointly typical
subspace $T_{A^{n}B^{n}}^{\delta}$ because it is a typical subspace. The
following property, analogous to Property~\ref{prop:joint-independent}\ for
classical joint typicality, holds whenever the state $\rho_{AB}$ has the
special form in \eqref{eq-qt:special-joint-form}:

\begin{property}
[Probability of Joint Typicality]\label{prop-qt:prob-joint-typ}Let
$\rho_{A^{n}B^{n}}$ be a classical state as given in
\eqref{eq-qt:classical-joint-state-1}. Consider the following marginal density
operators:%
\begin{equation}
\rho_{A^{n}}\equiv\operatorname{Tr}_{B^{n}}\left\{  \rho_{A^{n}B^{n}}\right\}
,\ \ \ \ \ \ \ \ \rho_{B^{n}}\equiv\operatorname{Tr}_{A^{n}}\left\{
\rho_{A^{n}B^{n}}\right\}  .
\end{equation}
Let us define $\rho_{\tilde{A}^{n}\tilde{B}^{n}}$ as the following density
operator:%
\begin{equation}
\rho_{\tilde{A}^{n}\tilde{B}^{n}}\equiv\rho_{A^{n}}\otimes\rho_{B^{n}}\neq
\rho_{A^{n}B^{n}}\operatorname{.}%
\end{equation}
The marginal density operators of $\rho_{\tilde{A}^{n}\tilde{B}^{n}}$\ are
therefore equivalent to the marginal density operators of $\rho_{A^{n}B^{n}}$.
Then we can bound the probability that the state $\rho_{\tilde{A}^{n}\tilde
{B}^{n}}$ lies in the typical subspace $T_{A^{n}B^{n}}^{\delta}$:%
\begin{equation}
\operatorname{Tr}\left\{  \Pi_{A^{n}B^{n}}^{\delta}\rho_{\tilde{A}^{n}%
\tilde{B}^{n}}\right\}  \leq2^{-n\left(  I( A;B) -3\delta\right)  }.
\end{equation}

\end{property}

\begin{exercise}
Prove the bound in Property~\ref{prop-qt:prob-joint-typ}:%
\begin{equation}
\operatorname{Tr}\left\{  \Pi_{A^{n}B^{n}}^{\delta}\rho_{\tilde{A}^{n}%
\tilde{B}^{n}}\right\}  \leq2^{-n\left(  I( A;B) -3\delta\right)  }.
\end{equation}

\end{exercise}

\section{Conditional Quantum Typicality}

\label{sec-qt:cond-typ}The notion of conditional quantum typicality%
\index{quantum typicality!conditional}
is somewhat similar to the notion of conditional typicality in the classical
domain, but we again quickly notice some departures because different quantum
states do not have to be perfectly distinguishable. The technical tools for
conditional quantum typicality developed in this section are important for
developing schemes that send public or private classical information over a
quantum channel (topics discussed in Chapters~\ref{chap:classical-comm-HSW}
and \ref{chap:private-cap}).

We first develop the notion of a conditional quantum information source.
Consider a random variable $X$ with probability distribution $p_{X}(x)$. Let
$\mathcal{X}$ be the alphabet of the random variable, and let $\left\vert
\mathcal{X}\right\vert $ denote its cardinality. We also associate a quantum
system $X$\ with the random variable $X$ and use an orthonormal set $\left\{
|x\rangle\right\}  _{x\in\mathcal{X}}$\ to represent its realizations. We
again label the elements of the alphabet $\mathcal{X}$ as $\left\{  x\right\}
_{x\in\mathcal{X}}$.

Suppose we generate a realization $x$ of random variable $X$ according to its
distribution $p_{X}(x)$, and we follow by generating a random quantum state
according to some conditional distribution. This procedure then gives us a set
of $\left\vert \mathcal{X}\right\vert $ quantum information sources (each of
them are as in Definition~\ref{def-qt:q-info-source}). We index them by the
classical index $x$, and the quantum information source has expected density
operator $\rho_{B}^{x}$ if the emitted classical index is $x$. Furthermore, we
impose the constraint that each $\rho_{B}^{x}$ has the same dimension (one
could achieve this by embedding the lower dimensional states into a larger
Hilbert space). This quantum information source is therefore a
\textquotedblleft conditional quantum information source.\textquotedblright%
\ Let $\mathcal{H}_{B}$ and $B$ denote the respective Hilbert space and system
label corresponding to the quantum output of the conditional quantum
information source. Let us call the resulting ensemble the \textquotedblleft
classical--quantum ensemble\textquotedblright\ and say that a
\textquotedblleft classical--quantum information source\textquotedblright%
\ generates it. The classical--quantum ensemble is as follows:%
\begin{equation}
\left\{  p_{X}(x),|x\rangle\langle x|_{X}\otimes\rho_{B}^{x}\right\}
_{x\in\mathcal{X}},
\end{equation}
where we correlate the classical state $|x\rangle_{X}$ with the density
operator $\rho_{B}^{x}$\ of the conditional quantum information source. The
expected density operator of the above classical--quantum ensemble is the
following classical--quantum state (discussed in
Section~\ref{sec-nqt:classical-quantum}):%
\begin{equation}
\rho_{XB}\equiv\sum_{x\in\mathcal{X}}p_{X}(x)|x\rangle\langle x|_{X}%
\otimes\rho_{B}^{x}. \label{eq-qt:cq-state}%
\end{equation}

The conditional quantum entropy $H(B|X)_{\rho}$\ of the classical--quantum
state $\rho_{XB}$\ is as follows:%
\begin{equation}
H(B|X)_{\rho}=\sum_{x\in\mathcal{X}}p_{X}(x)H(\rho_{B}^{x}).
\end{equation}
We can write a spectral decomposition of each conditional density operator
$\rho_{B}^{x}$ as follows:%
\begin{equation}
\sum_{y\in\mathcal{Y}}p_{Y|X}(y|x)|y_{x}\rangle\langle y_{x}|_{B},
\label{eq-qt:spec-dec-cond-ops}%
\end{equation}
where the elements of the set $\left\{  y\right\}  _{y\in\mathcal{Y}}$ label
the elements of an alphabet $\mathcal{Y}$, the orthonormal set $\{|y_{x}%
\rangle_{B}\}_{y\in\mathcal{Y}}$ is the set of eigenvectors of $\rho_{B}^{x}$,
and the corresponding eigenvalues are $\left\{  p_{Y|X}(y|x)\right\}
_{y\in\mathcal{Y}}$. We need the $x$ label for the orthonormal set
$\{|y_{x}\rangle_{B}\}_{y\in\mathcal{Y}}$ because the decomposition may be
different for different density operators $\rho_{B}^{x}$. The above notation
is again suggestive because the eigenvalues $p_{Y|X}(y|x)$ correspond to
conditional probabilities, and the set $\{|y_{x}\rangle_{B}\}$ of eigenvectors
corresponds to an orthonormal set of quantum states conditioned on label $x$.
With this respresentation, the conditional entropy $H(B|X)$ reduces to a
formula that looks like that for the classical conditional entropy:%
\begin{align}
H(B|X)  &  =\sum_{x\in\mathcal{X}}p_{X}(x)H(\rho_{B}^{x})\\
&  =\sum_{x\in\mathcal{X},y\in\mathcal{Y}}p_{X}(x)p_{Y|X}(y|x)\log\frac
{1}{p_{Y|X}(y|x)}.
\end{align}

We now consider when the classical--quantum information source emits a large
number $n$\ of states. The density operator for the output state $\rho
_{X^{n}B^{n}}$ is as follows:%
\begin{align}
&  \rho_{X^{n}B^{n}}\nonumber\\
&  \equiv\left(  \rho_{XB}\right)  ^{\otimes n}\\
&  =\left(  \sum_{x_{1}\in\mathcal{X}}p_{X}(x_{1})|x_{1}\rangle\langle
x_{1}|_{X_{1}}\otimes\rho_{B_{1}}^{x_{1}}\right)  \otimes\cdots\otimes\left(
\sum_{x_{n}\in\mathcal{X}}p_{X}(x_{n})|x_{n}\rangle\langle x_{n}|_{X_{n}%
}\otimes\rho_{B_{n}}^{x_{n}}\right) \\
&  =\sum_{x_{1},\ldots,x_{n}\in\mathcal{X}}p_{X}(x_{1})\cdots p_{X}%
(x_{n})|x_{1}\rangle\cdots|x_{n}\rangle\langle x_{1}|\cdots\langle
x_{n}|_{X^{n}}\otimes\left(  \rho_{B_{1}}^{x_{1}}\otimes\cdots\otimes
\rho_{B_{n}}^{x_{n}}\right)  .
\end{align}
We can abbreviate the above state as%
\begin{equation}
\sum_{x^{n}\in\mathcal{X}^{n}}p_{X^{n}}(x^{n})|x^{n}\rangle\langle
x^{n}|_{X^{n}}\otimes\rho_{B^{n}}^{x^{n}},
\end{equation}
where
\begin{align}
p_{X^{n}}(x^{n})  &  \equiv p_{X}(x_{1})\cdots p_{X}(x_{n}),\\
|x^{n}\rangle_{X^{n}}  &  \equiv|x_{1}\rangle_{X_{1}}\cdots|x_{n}%
\rangle_{X_{n}},\ \ \ \ \ \ \ \rho_{B^{n}}^{x^{n}}\equiv\rho_{B_{1}}^{x_{1}%
}\otimes\cdots\otimes\rho_{B_{n}}^{x_{n}},
\end{align}
and a spectral decomposition for the state $\rho_{B^{n}}^{x^{n}}$\ is%
\begin{equation}
\rho_{B^{n}}^{x^{n}}=\sum_{y^{n}\in\mathcal{Y}^{n}}p_{Y^{n}|X^{n}}(y^{n}%
|x^{n})|y_{x^{n}}^{n}\rangle\langle y_{x^{n}}^{n}|_{B^{n}},
\end{equation}
where%
\begin{align}
p_{Y^{n}|X^{n}}(y^{n}|x^{n})  &  \equiv p_{Y_{1}|X_{1}}(y_{1}|x_{1})\cdots
p_{Y_{n}|X_{n}}(y_{n}|x_{n}),\\
|y_{x^{n}}^{n}\rangle_{B^{n}}  &  \equiv|\left.  y_{1}\right.  _{x_{1}}%
\rangle_{B_{1}}\cdots|\left.  y_{n}\right.  _{x_{n}}\rangle_{B_{n}}.
\label{eq-qt:conditional-eigenstates}%
\end{align}
The above developments are a step along the way for formulating the
definitions of weak and strong conditional quantum typicality.

\subsection{Weak Conditional Quantum Typicality}

We can \textquotedblleft quantize\textquotedblright\ the notion of weak
classical conditional typicality so that it applies to a classical--quantum
information source.%
\index{quantum typicality!conditional!weak}%

\begin{definition}
[Weak Conditionally Typical Subspace]The conditionally typical subspace
$T_{B^{n}|x^{n}}^{\delta}$ corresponds to a particular sequence $x^{n}$ and an
ensemble $\left\{  p_{X}(x),\rho_{B}^{x}\right\}  $. It is the subspace
spanned by the states $|y_{x^{n}}^{n}\rangle_{B^{n}}$ whose conditional sample
entropy is $\delta$-close to the true conditional quantum entropy:%
\begin{equation}
T_{B^{n}|x^{n}}^{\delta}\equiv\operatorname{span}\left\{  |y_{x^{n}}%
^{n}\rangle_{B^{n}}:\left\vert \overline{H}(y^{n}|x^{n})-H(B|X)\right\vert
\leq\delta\right\}  ,
\end{equation}
where the states $|y_{x^{n}}^{n}\rangle_{B^{n}}$ are formed from the
eigenstates of the density operators $\rho_{B}^{x}$ (they are of the form in
\eqref{eq-qt:conditional-eigenstates}) and the sample entropy is with respect
to the distribution $p_{Y|X}(y|x)$ from \eqref{eq-qt:spec-dec-cond-ops}.
\end{definition}

\begin{definition}
[Weak Conditionally Typical Projector]The projector $\Pi_{B^{n}|x^{n}}%
^{\delta}$ onto the conditionally typical subspace $T_{B^{n}|x^{n}}^{\delta}%
$\ is as follows:%
\begin{equation}
\Pi_{B^{n}|x^{n}}^{\delta}\equiv\sum_{y_{x^{n}}^{n}\in T_{\delta
}^{Y^{n}|x^{n}}}|y_{x^{n}}^{n}\rangle\langle y_{x^{n}}^{n}|_{B^{n}}.
\end{equation}

\end{definition}

\subsection{Properties of the Weak Conditionally Typical Subspace}

The weak conditionally typical subspace%
\index{typical subspace!conditionally!weak}
$T_{B^{n}|x^{n}}^{\delta}$ enjoys several useful properties that are
\textquotedblleft quantized\textquotedblright\ versions of the properties for
weak conditionally typical sequences discussed in
Section~\ref{sec-ct:weak-cond-typ}. We should point out that we cannot really
say much for several of the properties for a particular sequence $x^{n}$, but
we can do so on average for a random sequence $X^{n}$. Thus, several of the
properties give expected behavior for a random sequence $X^{n}$. This
convention for quantum weak conditional typicality is the same as we had for
classical weak conditional typicality in Section~\ref{sec-ct:weak-cond-typ}.

\begin{property}
[Unit Probability]The expectation of the probability that we measure a random
quantum state $\rho_{B^{n}}^{X^{n}}$ to be in the conditionally typical
subspace $T_{B^{n}|X^{n}}^{\delta}$ approaches one as $n$ becomes large:%
\begin{equation}
\mathbb{E}_{X^{n}}\left\{  \operatorname{Tr}\left\{  \Pi_{B^{n}|X^{n}}%
^{\delta}\rho_{B^{n}}^{X^{n}}\right\}  \right\}  \geq1-\varepsilon,
\end{equation}
for all $\varepsilon\in(0,1)$, $\delta>0$, and sufficiently large $n$.
\end{property}

\begin{property}
[Exponentially Smaller Dimension]The dimension $\dim(T_{B^{n}|x^{n}}^{\delta
})$\ of the $\delta$-conditionally typical subspace is exponentially smaller
than the dimension $\left\vert \mathcal{Y}\right\vert ^{n}$\ of the entire
space of quantum states for most classical--quantum sources. We formally state
this property as follows:%
\begin{equation}
\operatorname{Tr}\left\{  \Pi_{B^{n}|x^{n}}^{\delta}\right\}  \leq2^{n\left(
H(B|X)+\delta\right)  }.
\end{equation}
We can also bound the dimension $\dim(T_{B^{n}|x^{n}}^{\delta})$\ of the
$\delta$-conditionally typical subspace from below:%
\begin{equation}
\mathbb{E}_{X^{n}}\left\{  \operatorname{Tr}\left\{  \Pi_{B^{n}|X^{n}}%
^{\delta}\right\}  \right\}  \geq\left(  1-\varepsilon\right)  2^{n\left(
H(B|X)-\delta\right)  },
\end{equation}
for all $\varepsilon\in(0,1)$, $\delta>0$, and sufficiently large $n$.
\end{property}

\begin{property}
[Equipartition]The density operator $\rho_{B^{n}}^{x^{n}}$ looks approximately
maximally mixed when projected to the conditionally typical subspace:%
\begin{equation}
2^{-n\left(  H( B|X) +\delta\right)  }\Pi_{B^{n}|x^{n}}^{\delta}\leq\Pi
_{B^{n}|x^{n}}^{\delta}\rho_{B^{n}}^{x^{n}}\Pi_{B^{n}|x^{n}}^{\delta}%
\leq2^{-n\left(  H( B|X) -\delta\right)  }\Pi_{B^{n}|x^{n}}^{\delta}.
\end{equation}

\end{property}

\begin{exercise}
Prove all three of the above properties for weak conditional quantum typicality.
\end{exercise}

\subsection{Strong Conditional Quantum Typicality}

\label{sec-qt:strong-quant-cond-typ}We now develop the notion of strong
conditional quantum
\index{quantum typicality!conditional!strong}%
typicality. This notion again applies to an ensemble or to a
classical--quantum state such as that given in \eqref{eq-qt:cq-state}.
However, it differs from weak conditional quantum typicality because we can
prove stronger statements about the asymptotic behavior of conditional quantum
systems (just as we could for the classical case in
Section~\ref{sec-ct:strong-cond-typ}). We begin this section with an example
to build up our intuition. We then follow with the formal definition of strong
conditional quantum typicality, and we end by proving some properties of the
strong conditionally typical subspace.

Recall the example from Section~\ref{sec-ct:strong-typ}. In a similar way to
this example, we can draw a sequence from an alphabet $\left\{  0,1,2\right\}
$ according to the following distribution:%
\begin{equation}
p_{X}(0)=\frac{1}{4},\ \ \ \ \ p_{X}(1)=\frac{1}{4},\ \ \ \ \ p_{X}%
(2)=\frac{1}{2}. \label{eq-qt:example-cond-typ}%
\end{equation}
One potential realization sequence is as follows:%
\begin{equation}
201020102212.
\end{equation}
The above sequence has four \textquotedblleft zeros,\textquotedblright\ three
\textquotedblleft ones,\textquotedblright\ and five \textquotedblleft
twos,\textquotedblright\ so that the empirical distribution of this sequence
is $\left(  1/3,1/4,5/12\right)  $ and has maximum deviation $1/12$ from the
true distribution in~\eqref{eq-qt:example-cond-typ}.

For each symbol in the above sequence, we could then draw from one of three
quantum information sources based on whether the classical index is $0$, $1$,
or $2$. Suppose that the expected density operator of the first quantum
information source is $\rho^{0}$, that of the second is $\rho^{1}$, and that
of the third is $\rho^{2}$. Then the density operator for the resulting
sequence of quantum states is as follows:%
\begin{equation}
\rho_{B_{1}}^{2}\otimes\rho_{B_{2}}^{0}\otimes\rho_{B_{3}}^{1}\otimes
\rho_{B_{4}}^{0}\otimes\rho_{B_{5}}^{2}\otimes\rho_{B_{6}}^{0}\otimes
\rho_{B_{7}}^{1}\otimes\rho_{B_{8}}^{0}\otimes\rho_{B_{9}}^{2}\otimes
\rho_{B_{10}}^{2}\otimes\rho_{B_{11}}^{1}\otimes\rho_{B_{12}}^{2},
\label{eq-qt:cond-states}%
\end{equation}
where the subscripts label the systems as usual. So, the state of systems
$B_{1}$, $B_{5}$, $B_{9}$, $B_{10}$, and $B_{12}$ is equal to five copies of
$\rho^{2}$, the state of systems $B_{2}$, $B_{4}$, $B_{6}$, and $B_{8}$ is
equal to four copies of $\rho^{0}$, and the state of systems $B_{3}$, $B_{7}$,
and $B_{11}$ is equal to three copies of $\rho^{1}$. Let $I_{x}$ be an
indicator set for each $x\in\left\{  0,1,2\right\}  $, so that $I_{x}$
consists of all the indices in the sequence for which a symbol is equal to
$x$. For the above example,%
\begin{equation}
I_{0}=\left\{  2,4,6,8\right\}  ,\ \ \ \ \ \ \ \ I_{1}=\left\{
3,7,11\right\}  ,\ \ \ \ \ \ \ \ I_{2}=\left\{  1,5,9,10,12\right\}  .
\end{equation}
These sets serve as a way of grouping all of the density operators that are
the same because they correspond to the same classical symbol, and it is
important to do so if we would like to consider concentration of measure
effects when we go to the asymptotic setting. As a visual aid, we could
permute the sequence of density operators in \eqref{eq-qt:cond-states} if we
would like to see systems with the same density operator grouped together:%
\begin{equation}
\rho_{B_{2}}^{0}\otimes\rho_{B_{4}}^{0}\otimes\rho_{B_{6}}^{0}\otimes
\rho_{B_{8}}^{0}\otimes\rho_{B_{3}}^{1}\otimes\rho_{B_{7}}^{1}\otimes
\rho_{B_{11}}^{1}\otimes\rho_{B_{1}}^{2}\otimes\rho_{B_{5}}^{2}\otimes
\rho_{B_{9}}^{2}\otimes\rho_{B_{10}}^{2}\otimes\rho_{B_{12}}^{2}.
\end{equation}
There is then a typical projector for the first four systems with density
operator $\rho^{0}$, a different typical projector for the next three systems
with density operator $\rho^{1}$, and an even different typical projector for
the last five systems with density operator $\rho^{2}$ (however, the length of
the above quantum sequence is certainly not large enough to observe any
measure concentration effects!). Thus, the indicator sets $I_{x}$ serve to
identify which systems have the same density operator so that we can know upon
which systems a particular typical projector should act.

This example helps build our intuition of strong conditional quantum
typicality, and we can now begin to state what we would expect in the
asymptotic setting. Suppose that the original classical sequence is large and
strongly typical, so that it has roughly $n/4$ occurrences of
\textquotedblleft zero,\textquotedblright\ $n/4$ occurrences of
\textquotedblleft one,\textquotedblright\ and $n/2$ occurrences of
\textquotedblleft two.\textquotedblright\ We would then expect the law of
large numbers to come into play for $n/4$ and $n/2$ when $n$ is large enough.
Thus, we can use the classical sequence to identify which quantum systems have
the same density operator, and apply a typical projector to each of these
subsets of quantum systems. Then all of the useful asymptotic properties of
typical subspaces apply whenever $n$ is large enough.

We can now state the definition of the
\index{typical subspace!conditionally!strong}
strong conditionally typical subspace and the
\index{typical projector!conditionally!strong}
strong conditionally typical projector, and we prove some of their asymptotic
properties by exploiting the properties of typical subspaces.

\begin{definition}
[Strong Conditionally Typical Subspace]\label{def-qt:cond-typ-sub}The strong
conditionally typical subspace corresponds to a sequence $x^{n}$ and an
ensemble $\left\{  p_{X}(x),\rho_{B}^{x}\right\}  $. Let a spectral
decomposition of each state $\rho_{B}^{x}$ be as in
\eqref{eq-qt:spec-dec-cond-ops} with distribution $p_{Y|X}(y|x)$ and
corresponding eigenstates $|y_{x}\rangle$. The strong conditionally typical
subspace $T_{B^{n}|x^{n}}^{\delta}$\ is then as follows:%
\begin{equation}
T_{B^{n}|x^{n}}^{\delta}\equiv\operatorname{span}\left\{  \bigotimes
\limits_{x\in\mathcal{X}}|y_{x}^{I_{x}}\rangle_{B^{I_{x}}}:\forall
x,\ \ \ \ y^{I_{x}}\in T_{\delta}^{\left(  Y|x\right)  ^{\left\vert
I_{x}\right\vert }}\right\}  ,
\end{equation}
where $I_{x}\equiv\left\{  i:x_{i}=x\right\}  $ is an indicator set that
selects the indices $i$\ in the sequence $x^{n}$ for which the $i$th symbol
$x_{i}$\ is equal to $x\in\mathcal{X}$, $B^{I_{x}}$ selects the systems from
$B^{n}$ where the classical sequence $x^{n}$ is equal to the symbol $x$,
$|y_{x}^{I_{x}}\rangle$ is some string of states from the set $\left\{
|y_{x}\rangle\right\}  $, $y^{I_{x}}$ is a classical string corresponding to
this string of states, $Y|x$ is a random variable with distribution
$p_{Y|X}(y|x)$, and $\left\vert I_{x}\right\vert $ is the cardinality of the
indicator set $I_{x}$.
\end{definition}

\begin{definition}
[Strong Conditionally Typical Projector]The strong conditionally typical
projector again corresponds to a sequence $x^{n}$ and an ensemble $\left\{
p_{X}( x) ,\rho_{B}^{x}\right\}  $. It is a tensor product of typical
projectors for each state $\rho_{B}^{x}$ in the ensemble:%
\begin{equation}
\Pi_{B^{n}|x^{n}}^{\delta}\equiv\bigotimes\limits_{x\in\mathcal{X}}%
\Pi_{B^{I_{x}}}^{\rho_{x},\delta},
\end{equation}
where $I_{x}$ is defined in Definition$~$\ref{def-qt:cond-typ-sub}, and
$B^{I_{x}}$ indicates the systems onto which a particular typical projector
for $\rho_{x}$ projects.\footnote{Having the conditional density operators in
the subscript breaks somewhat from our convention throughout this chapter, but
it is useful here to indicate explicitly which density operator corresponds to
a typical projector.}
\end{definition}

\subsection{Properties of the Strong Conditionally Typical Subspace}

The strong conditionally typical subspace admits several useful asymptotic
properties similar to what we have seen before, and the proof strategy for
proving all of them is similar to the way that we proved the analogous
properties for the strong conditionally typical set in
Section~\ref{sec-ct:proof-strong-typ}. Suppose that we draw a sequence $x^{n}$
from a probability distribution $p_{X}(x)$, and we are able to draw as many
samples as we wish so that it is very likely that the sequence $x^{n}$ is
strongly typical and the occurrences $N(x|x^{n})$ of each symbol $x$ are as
large as we wish. Then the following properties hold for $x^{n}$ strongly
typical and each $N(x|x^{n})$ large.

\begin{property}
[Unit Probability]\label{prop-qt:strong-cond-typ-unit}The probability that we
measure a quantum state $\rho_{B^{n}}^{x^{n}}$ to be in the conditionally
typical subspace $T_{B^{n}|x^{n}}^{\delta}$ has the following lower bound:%
\begin{equation}
\operatorname{Tr}\left\{  \Pi_{B^{n}|x^{n}}^{\delta}\rho_{B^{n}}^{x^{n}%
}\right\}  \geq1-\varepsilon,
\end{equation}
for all $\varepsilon\in(0,1)$, $\delta>0$, and sufficiently large $n$.
\end{property}

\begin{property}
[Exponentially Smaller Dimension]\label{prop-qt:strong-cond-typ-exp-small}The
dimension $\dim(T_{B^{n}|x^{n}}^{\delta})$\ of the $\delta$-conditionally
typical subspace is exponentially smaller than the dimension $\left\vert
B\right\vert ^{n}$\ of the entire space of quantum states for all
classical--quantum information sources besides ones where all their density
operators are maximally mixed. We formally state this property as follows:%
\begin{equation}
\operatorname{Tr}\left\{  \Pi_{B^{n}|x^{n}}^{\delta}\right\}  \leq2^{n\left(
H(B|X)+\delta^{\prime\prime}\right)  },
\end{equation}
where $\delta^{\prime\prime}$ is given in \eqref{eq-qt:typical-delta-double}.
We can also bound the dimension $\dim(T_{\delta}^{Y^{n}|x^{n}})$\ of the
$\delta$-conditionally typical subspace from below:%
\begin{equation}
\operatorname{Tr}\left\{  \Pi_{B^{n}|x^{n}}^{\delta}\right\}  \geq\left(
1-\varepsilon\right)  2^{n\left(  H(B|X)-\delta^{\prime\prime}\right)  },
\end{equation}
for all $\varepsilon\in(0,1)$, $\delta>0$, and sufficiently large $n$.
\end{property}

\begin{property}
[Equipartition]\label{prop-qt:strong-cond-typ-equi}The state $\rho_{B^{n}%
}^{x^{n}}$ is approximately maximally mixed when projected onto the strong
conditionally typical subspace:%
\begin{equation}
2^{-n\left(  H( B|X) +\delta^{\prime\prime}\right)  }\Pi_{B^{n}|x^{n}}%
^{\delta}\leq\Pi_{B^{n}|x^{n}}^{\delta}\rho_{B^{n}}^{x^{n}}\Pi_{B^{n}|x^{n}%
}^{\delta}\leq2^{-n\left(  H( B|X) -\delta^{\prime\prime}\right)  }\Pi
_{B^{n}|x^{n}}^{\delta},
\end{equation}
where $\delta^{\prime\prime}$ is given in \eqref{eq-qt:typical-delta-double}.
\end{property}

\subsection{Proofs of the Properties of the Strong Conditionally Typical
Subspace}

\label{sec-qt:proof-strong-typ}

\begin{proof}
[Proof of the Unit Probability Property
(Property~\ref{prop-qt:strong-cond-typ-unit})]A proof of this property is
similar to the proof of Property~\ref{prop-ct:strong-cond-typ-unit} for the
strong conditionally typical set. Since we are dealing with an
i.i.d.~distribution, we can assume without loss of generality that the
sequence $x^{n}$ is lexicographically ordered with an order on the alphabet
$\mathcal{X}$. We write the elements of $\mathcal{X}$ as $a_{1}$, \ldots,
$a_{\left\vert \mathcal{X}\right\vert }$. Then the lexicographic ordering
means that we can write the sequence of quantum states $\rho_{x^{n}}$ as
follows:%
\begin{equation}
\rho_{x^{n}}=\underbrace{\rho_{a_{1}}\otimes\cdots\otimes\rho_{a_{1}}%
}_{N(a_{1}|x^{n})}\otimes\underbrace{\rho_{a_{2}}\otimes\cdots\otimes
\rho_{a_{2}}}_{N(a_{2}|x^{n})}\otimes\cdots\otimes\underbrace{\rho
_{a_{\left\vert \mathcal{X}\right\vert }}\otimes\cdots\otimes\rho
_{a_{\left\vert \mathcal{X}\right\vert }}}_{N(a_{\left\vert \mathcal{X}%
\right\vert }|x^{n})}. \label{eq-qt:lexico-quantum}%
\end{equation}
It follows that $N(a_{i}|x^{n})\geq n\left(  p_{X}(a_{i})-\delta^{\prime
}\right)  $ from the typicality of $x^{n}$, and thus the law of large numbers
comes into play for each block $a_{i}\cdots a_{i}$ with length $N(a_{i}%
|x^{n})$. The strong conditionally typical projector $\Pi_{B^{n}|x^{n}%
}^{\delta}$ for this system is as follows:%
\begin{equation}
\Pi_{B^{n}|x^{n}}^{\delta}\equiv\bigotimes\limits_{x\in\mathcal{X}}%
\Pi_{B^{N(x|x^{n})}}^{\rho_{x},\delta}, \label{eq-qt:lexico-typ-proj}%
\end{equation}
because we assumed the lexicographic ordering of the symbols in the sequence
$x^{n}$. Each projector $\Pi_{B^{N(x|x^{n})}}^{\rho_{x},\delta}$ in the above
tensor product is a typical projector for the density operator $\rho_{x}$ when
$N(x|x^{n})\approx np_{X}(x)$ becomes very large. Then we can apply the unit
probability property (Property~\ref{prop-qt:unit})\ for each of these typical
projectors, and it follows that%
\begin{align}
\operatorname{Tr}\left\{ \Pi_{B^{n}|x^{n}}^{\delta}
 \rho_{B^{n}}^{x^{n}}\right\}   &  =\operatorname{Tr}\left\{  \bigotimes\limits_{x\in\mathcal{X}%
}\Pi_{B^{N(x|x^{n})}}^{\rho_{x},\delta}\rho_{x}^{\otimes N(x|x^{n})}\right\}
\\
&  =%
{\displaystyle\prod\limits_{x\in\mathcal{X}}}
\operatorname{Tr}\left\{  \Pi_{B^{N(x|x^{n})}}^{\rho_{x},\delta}\rho
_{x}^{\otimes N(x|x^{n})}\right\} \\
&  \geq\left(  1-\varepsilon\right)  ^{\left\vert \mathcal{X}\right\vert }\\
&  \geq1-\left\vert \mathcal{X}\right\vert \varepsilon,
\end{align}
concluding the proof.
\end{proof}

\bigskip

\begin{proof}
[Proof of the Equipartition Property
(Property~\ref{prop-qt:strong-cond-typ-equi})]We first assume without loss of
generality that we can write the state $\rho_{B^{n}}^{x^{n}}$ in lexicographic
order as in \eqref{eq-qt:lexico-quantum}. Then the strong conditionally
typical projector is again as in \eqref{eq-qt:lexico-typ-proj}. It follows
that%
\begin{equation}
\Pi_{B^{n}|x^{n}}^{\delta}\rho_{x^{n}}\Pi_{B^{n}|x^{n}}^{\delta}%
=\bigotimes\limits_{x\in\mathcal{X}}\Pi_{B^{N(x|x^{n})}}^{\rho_{x},\delta}%
\rho_{x}^{\otimes N(x|x^{n})}\Pi_{B^{N(x|x^{n})}}^{\rho_{x},\delta}.
\end{equation}
We can apply the equipartition property of the typical subspace for each
typical projector $\Pi_{B^{N(x|x^{n})}}^{\rho_{x},\delta}$%
\ (Property~\ref{prop-qt:equi}):%
\begin{multline}
\bigotimes\limits_{x\in\mathcal{X}}\Pi_{B^{N(x|x^{n})}}^{\rho_{x},\delta
}2^{-N(x|x^{n})\left(  H\left(  \rho_{x}\right)  +c\delta\right)  }%
\leq\bigotimes\limits_{x\in\mathcal{X}}\Pi_{B^{N(x|x^{n})}}^{\rho_{x},\delta
}\rho_{x}^{\otimes N(x|x^{n})}\Pi_{B^{N(x|x^{n})}}^{\rho_{x},\delta}\\
\leq\bigotimes\limits_{x\in\mathcal{X}}\Pi_{B^{N(x|x^{n})}}^{\rho_{x},\delta
}2^{-N(x|x^{n})\left(  H\left(  \rho_{x}\right)  -c\delta\right)  }.
\end{multline}
The following inequalities hold because the sequence $x^{n}$ is strongly
typical as defined in Definition~\ref{def-ct:strong-typ}:%
\begin{multline}
\bigotimes\limits_{x\in\mathcal{X}}\Pi_{B^{N(x|x^{n})}}^{\rho_{x},\delta
}2^{-n\left(  p_{X}(x)+\delta^{\prime}\right)  \left(  H(\rho_{x}%
)+c\delta\right)  }\leq\Pi_{B^{n}|x^{n}}^{\delta}\rho_{x^{n}}\Pi_{B^{n}|x^{n}%
}^{\delta}\\
\leq\bigotimes\limits_{x\in\mathcal{X}}\Pi_{B^{N(x|x^{n})}}^{\rho_{x},\delta
}2^{-n\left(  p_{X}(x)-\delta^{\prime}\right)  \left(  H(\rho_{x}%
)-c\delta\right)  }.
\end{multline}
We can factor out each term $2^{-n\left(  p_{X}(x)+\delta^{\prime}\right)
\left(  H(\rho_{x})+c\delta\right)  }$ from the tensor products:%
\begin{multline}
\prod\limits_{x\in\mathcal{X}}2^{-n\left(  p_{X}(x)+\delta^{\prime}\right)
\left(  H(\rho_{x})+c\delta\right)  }\bigotimes\limits_{x\in\mathcal{X}}%
\Pi_{B^{N(x|x^{n})}}^{\rho_{x},\delta}\leq\Pi_{B^{n}|x^{n}}^{\delta}%
\rho_{x^{n}}\Pi_{B^{n}|x^{n}}^{\delta}\\
\leq\prod\limits_{x\in\mathcal{X}}2^{-n\left(  p_{X}(x)-\delta^{\prime
}\right)  \left(  H(\rho_{x})-c\delta\right)  }\bigotimes\limits_{x\in
\mathcal{X}}\Pi_{B^{N(x|x^{n})}}^{\rho_{x},\delta}.
\end{multline}
We then multiply out the $\left\vert \mathcal{X}\right\vert $ terms
$2^{-n\left(  p_{X}(x)+\delta^{\prime}\right)  \left(  H\left(  \rho
_{x}\right)  +c\delta\right)  }$:%
\begin{multline}
2^{-n\left(  H(B|X)+\sum_{x}\left(  H(\rho_{x})\delta^{\prime}+cp_{X}%
(x)\delta+c\delta\delta^{\prime}\right)  \right)  }\Pi_{B^{n}|x^{n}}^{\delta
}\leq\Pi_{B^{n}|x^{n}}^{\delta}\rho_{x^{n}}\Pi_{B^{n}|x^{n}}^{\delta}\\
\leq2^{-n\left(  H(B|X)+\sum_{x}c\delta\delta^{\prime}-H\left(  \rho
_{x}\right)  \delta^{\prime}-cp_{X}(x)\delta\right)  }\Pi_{B^{n}|x^{n}%
}^{\delta}.
\end{multline}
The final step below follows because $\sum_{x}p_{X}(x)=1$ and because the
bound $\sum H(\rho_{x})\leq\left\vert \mathcal{X}\right\vert \log d$ applies
where $d$ is the dimension of the density operator $\rho_{x}$:%
\begin{equation}
2^{-n\left(  H(B|X)+\delta^{\prime\prime}\right)  }\Pi_{B^{n}|x^{n}}^{\delta
}\leq\Pi_{B^{n}|x^{n}}^{\delta}\rho_{x^{n}}\Pi_{B^{n}|x^{n}}^{\delta}%
\leq2^{-n\left(  H(B|X)-\delta^{\prime\prime}\right)  }\Pi_{B^{n}|x^{n}%
}^{\delta},
\end{equation}
where%
\begin{equation}
\delta^{\prime\prime}\equiv\delta^{\prime}\left\vert \mathcal{X}\right\vert
\log d+c\delta+\left\vert \mathcal{X}\right\vert c\delta\delta^{\prime}.
\label{eq-qt:typical-delta-double}%
\end{equation}
This concludes the proof.
\end{proof}

\begin{exercise}
Prove Property~\ref{prop-qt:strong-cond-typ-exp-small}.
\end{exercise}

\subsection{Strong Conditional and Marginal Quantum Typicality}

We end this section on strong conditional quantum typicality by proving a
final property that applies to a state drawn from an ensemble and the typical
subspace of the expected density operator of the ensemble.

\begin{property}
\label{prop-qt:cond-state-with-uncond-proj}Consider an ensemble of the form
$\left\{  p_{X}(x),\rho_{x}\right\}  $ with expected density operator
$\rho\equiv\sum_{x}p_{X}(x)\rho_{x}$. Suppose that $x^{n}$ is a strongly
typical sequence with respect to the distribution $p_{X}(x)$ and leads to a
conditional density operator $\rho_{x^{n}}$. Then the probability of measuring
$\rho_{x^{n}}$ in the strongly typical subspace of $\rho$ is high:%
\begin{equation}
\operatorname{Tr}\left\{  \Pi_{\rho,\delta}^{n}\ \rho_{x^{n}}\right\}
\geq1-\varepsilon,
\end{equation}
for all $\varepsilon\in(0,1)$, $\delta>0$, and sufficiently large $n$, where
the typical projector $\Pi_{\rho,\delta}^{n}$ is with respect to the density
operator $\rho$.
\end{property}

\begin{proof}
Let the expected density operator have the following spectral decomposition:%
\begin{equation}
\rho=\sum_{z}p_{Z}(z)|z\rangle\langle z|.
\end{equation}
We define the \textquotedblleft pinching\textquotedblright\ operation as a
dephasing with respect to the basis $\left\{  |z\rangle\right\}  $:%
\begin{equation}
\sigma\rightarrow\Delta(\sigma)\equiv\sum_{z}|z\rangle\langle z|\sigma
|z\rangle\langle z|.
\end{equation}
Let $\overline{\rho}_{x}$ denote the pinched version of the conditional
density operators $\rho_{x}$:%
\begin{equation}
\overline{\rho}_{x}\equiv\Delta(\rho_{x})=\sum_{z}|z\rangle\langle z|\rho
_{x}|z\rangle\langle z|=\sum_{z}p_{Z|X}(z|x)|z\rangle\langle z|,
\end{equation}
where $p_{Z|X}(z|x)\equiv\langle z|\rho_{x}|z\rangle$. This pinching is the
crucial insight for the proof because all of the pinched density operators
$\overline{\rho}_{x}$ have a common eigenbasis and the analysis reduces from a
quantum one to a classical one that exploits the properties of strong
marginal, conditional, and joint typicality. The following chain of
inequalities then holds by exploiting the above definitions:%
\begin{align}
\operatorname{Tr}\left\{  \Pi_{\rho,\delta}^{n} \rho_{x^{n}}\right\}   &
=\operatorname{Tr}\left\{  \sum_{z^{n}\in T_{\delta}^{Z^{n}}%
}|z^{n}\rangle\langle z^{n}|\rho_{x^{n}}\right\} \\
&  =\operatorname{Tr}\left\{  \sum_{z^{n}\in T_{\delta}^{Z^{n}}%
}|z^{n}\rangle\left\langle z^{n}|z^{n}\right\rangle \langle z^{n}|\rho_{x^{n}}\right\} \\
&  =\operatorname{Tr}\left\{  \sum_{z^{n}\in T_{\delta}^{Z^{n}}}|z^{n}%
\rangle\langle z^{n}|\rho_{x^{n}}|z^{n}\rangle\langle z^{n}|\right\} \\
&  =\operatorname{Tr}\left\{  \sum_{z^{n}\in T_{\delta}^{Z^{n}}}p_{Z^{n}%
|X^{n}}(z^{n}|x^{n})|z^{n}\rangle\langle z^{n}|\right\} \\
&  =\sum_{z^{n}\in T_{\delta}^{Z^{n}}}p_{Z^{n}|X^{n}}(z^{n}|x^{n}).
\end{align}
The first equality follows from the definition of the typical projector
$\Pi_{\rho,\delta}^{n}$. The second equality follows because $|z^{n}%
\rangle\langle z^{n}|$ is a projector, and the third follows from linearity
and cyclicity of the trace. The fourth equality follows because%
\begin{equation}
\langle z^{n}|\rho_{x^{n}}|z^{n}\rangle=\prod\limits_{i=1}^{n}\left\langle
z_{i}\right\vert \rho_{x_{i}}\left\vert z_{i}\right\rangle =\prod
\limits_{i=1}^{n}p_{Z|X}\left(  z_{i}|x_{i}\right)  \equiv p_{Z^{n}|X^{n}%
}(z^{n}|x^{n}).
\end{equation}
Now consider this final expression $\sum_{z^{n}\in T_{\delta}^{Z^{n}}}%
p_{Z^{n}|X^{n}}(z^{n}|x^{n})$. It is equal to the probability that a
random conditional sequence $Z^{n}|x^{n}$ is in the typical set for $p_{Z}%
(z)$:%
\begin{equation}
\Pr\left\{  Z^{n}|x^{n}\in T_{\delta}^{Z^{n}}\right\}  .
\end{equation}
By taking $n$ large enough, the law of large numbers guarantees that it is
highly likely (with probability greater than $1-\varepsilon$ for any
$\varepsilon>0$) that this random conditional sequence $Z^{n}|x^{n}$ is in the
conditionally typical set $T_{\delta^{\prime}}^{Z^{n}|x^{n}}$ for some
$\delta^{\prime}$. It then follows that this conditional sequence has a high
probability of being in the unconditionally typical set $T_{\delta}^{Z^{n}}$
because we assumed that the sequence $x^{n}$ is strongly typical and
Lemma~\ref{lem-ct:cond-marg-joint-typ}\ states that a sequence $z^{n}$ is
unconditionally typical if $x^{n}$ is strongly typical and $z^{n}$ is strong
conditionally typical.
\end{proof}

\section{The Method of Types for Quantum Systems}

\label{sec-qt:method-of-types-q}Our final development in this chapter is to
establish the method of types in the quantum domain, and the classical tools
from Section~\ref{sec-ct:strong-typ}\ have a straightforward generalization.

We can partition the Hilbert space of $n$ qudits into different type class
subspaces, just as we can partition the set of all sequences into different
type classes. For example, consider the Hilbert space of three qubits. The
computational basis is an orthonormal basis for the entire Hilbert space of
three qubits:%
\begin{equation}
\left\{  |000\rangle,|001\rangle,|010\rangle,|011\rangle,|100\rangle
,|101\rangle,|110\rangle,|111\rangle\right\}  .
\end{equation}
Then the computational basis states with the same Hamming weight form a basis
for each type class subspace. So, for the above example, the type class
subspaces are as follows:%
\begin{align}
T_{0}  &  \equiv\left\{  |000\rangle\right\}  ,\\
T_{1}  &  \equiv\left\{  |001\rangle,|010\rangle,|100\rangle\right\}  ,\\
T_{2}  &  \equiv\left\{  |011\rangle,|101\rangle,|110\rangle\right\}  ,\\
T_{3}  &  \equiv\left\{  |111\rangle\right\}  ,
\end{align}
and the projectors onto the different type class subspaces are as follows:%
\begin{align}
\Pi_{0}  &  \equiv|000\rangle\langle000|,\\
\Pi_{1}  &  \equiv|001\rangle\langle001|+|010\rangle\langle010|+|100\rangle
\langle100|,\\
\Pi_{2}  &  \equiv|011\rangle\langle011|+|101\rangle\langle101|+|110\rangle
\langle110|,\\
\Pi_{3}  &  \equiv|111\rangle\langle111|.
\end{align}

We can generalize the above example to an $n$-fold tensor product of qudit
systems using the method of types.

\begin{definition}
[Type Class Subspace]The%
\index{type class!subspace}
type class subspace is the subspace spanned by all states with the same type:%
\begin{equation}
T_{A^{n}}^{t}\equiv\operatorname{span}\left\{  |x^{n}\rangle_{A^{n}}:x^{n}\in
T_{t}^{X^{n}}\right\}  ,
\end{equation}
where the notation $T_{A^{n}}^{t}$ on the left-hand side\ indicates the type
class subspace, and the notation $T_{t}^{X^{n}}$ on the right-hand
side\ indicates the type class of the classical sequence $x^{n}$.
\end{definition}

\begin{definition}
[Type Class Projector]Let $\Pi_{A^{n}}^{t}$ denote the type class subspace%
\index{type class!projector}
projector:%
\begin{equation}
\Pi_{A^{n}}^{t}\equiv\sum_{x^{n}\in T_{t}^{X^{n}}}|x^{n}\rangle\langle
x^{n}|_{A^{n}}.
\end{equation}

\end{definition}

\begin{property}
[Resolution of the Identity with Type Class Projectors]The sum of all type
class projectors forms a resolution of the identity on the full Hilbert space
$\mathcal{H}_{A^{n}}$ of $n$ qudits:%
\begin{equation}
I_{A^{n}}=\sum_{t}\Pi_{A^{n}}^{t},
\end{equation}
where $I_{A^{n}}$ is the identity operator on $\mathcal{H}_{A^{n}}$.
\end{property}

\begin{definition}
[Maximally Mixed Type Class State]The maximally mixed density operator
proportional to the type class subspace projector is%
\begin{equation}
\pi_{A^{n}}^{t}\equiv{D_{t}}^{-1}{\Pi_{A^{n}}^{t},}%
\end{equation}
where $D_{t}$ is the dimension of the type class:%
\begin{equation}
D_{t}\equiv\operatorname{Tr}\left\{  {\Pi_{A^{n}}^{t}}\right\}  .
\end{equation}

\end{definition}

Recall from Definition~\ref{def-ct:typical-type} that a $\delta$-typical type
is one for which the empirical distribution has maximum deviation $\delta$
from the true distribution, and $\tau_{\delta}$ is the set of all $\delta
$-typical types. For the quantum case, we determine the maximum deviation
$\delta$\ of a type from the true distribution $p_{X}( x) $ (this is the
distribution from a spectral decomposition of a density operator $\rho$). This
definition allows us to write the strongly $\delta$-typical subspace projector
$\Pi_{A^{n}}^{\delta}$\ of $\rho$ as a sum over all of the $\delta$-typical
type class projectors $\Pi_{A^{n}}^{t}$:%
\begin{equation}
\Pi_{A^{n}}^{\delta}=\sum_{t\in\tau_{\delta}}\Pi_{A^{n}}^{t}.
\label{eq-qt:typ-sub-decomp-types}%
\end{equation}

Some protocols in quantum Shannon theory such as entanglement concentration in
Chapter~\ref{chap:ent-conc}\ employ the above decomposition of the typical
subspace projector into types. The way that such a protocol works is first to
perform a typical subspace measurement on many copies of a state, and this
measurement succeeds with high probability. One party involved in the protocol
then performs a type class measurement $\left\{  \Pi_{A^{n}}^{t}\right\}
_{t}$. We perform this latter measurement in a protocol if we would like the
state to have a uniform distribution over states in the type class. One might
initially think that the dimension of the remaining state would not be
particularly large, but it actually holds that the dimension is large because
we can obtain the following useful lower bound on the dimension of any typical
type class projector.

\begin{property}
[Minimal Dimension of a Typical Type Class Projector]%
\label{prop-qt:min-dim-typical-type}Suppose that $p_{X}(x)$ is the
distribution from a spectral decomposition of a density operator $\rho$, and
$\tau_{\delta}$ collects all the type class subspaces with maximum deviation
$\delta$ from the distribution $p_{X}(x)$. Then for any type $t\in\tau
_{\delta}$ and for sufficiently large $n$, we can bound the dimension of the
type class projector $\Pi_{A^{n}}^{t}$ from below as follows:%
\begin{equation}
\operatorname{Tr}\left\{  \Pi_{A^{n}}^{t}\right\}  \geq2^{n[H(\rho
)-\eta(d\delta)-d\frac{1}{n}\log\left(  n+1\right)  ]},
\end{equation}
where $d$ is the dimension of the Hilbert space on which $\rho$ acts and the
function $\eta(d\delta)\rightarrow0$ as $\delta\rightarrow0$.
\end{property}

\begin{proof}
A proof follows directly by exploiting
Property~\ref{prop-ct:min-card-typical-type}\ from the previous chapter.
\end{proof}

\section{Concluding Remarks}

This chapter is about the asymptotic nature of quantum information in the
i.i.d.~setting. The main technical development is the notion of the typical
subspace, and our approach here is simply to \textquotedblleft
quantize\textquotedblright\ the definition of the typical set from the
previous chapter. The typical subspace enjoys properties similar to those of
the typical set---the probability that many copies of a density operator lie
in the typical subspace approaches one as the number of copies approaches
infinity, the dimension of the typical subspace is exponentially smaller than
the dimension of the full Hilbert space, and many copies of a density operator
look approximately maximally mixed on the typical subspace. The rest of the
content in this chapter involves an extension of these ideas to conditional
quantum typicality.

The content in this chapter is here to provide a rigorous underpinning that we
can quickly cite later on, and after having mastered the results in this
chapter along with the tools in the next two chapters, we will be ready to
prove many of the important results in quantum Shannon theory.

\section{History and Further Reading}

\cite{OP93} devised the notion of a typical subspace, and later
\cite{PhysRevA.51.2738} independently devised it when he proved the quantum
data-compression theorem bearing his name. \cite{Hol98}
and\ \cite{PhysRevA.56.131} introduced the conditionally typical subspace in
order to prove the HSW~coding theorem. Winter's thesis is a good source for
proofs of several properties of quantum typicality~\citep{thesis1999winter}.
\cite{book2000mikeandike} use weak conditional quantum typicality to prove
the HSW\ theorem. \cite{ieee2002bennett} and \cite{Hol01a} introduced
frequency typical (or strongly typical) subspaces to quantum information
theory in order to prove the entanglement-assisted classical capacity theorem.
Devetak used strong typicality to prove the HSW\ coding theorem in Appendix~B
of \citep{ieee2005dev}.

\chapter{The Packing Lemma}

\label{chap:packing}The packing lemma\ is a general method for one party to
\textquotedblleft pack\textquotedblright\ or encode classical messages into a
Hilbert space so that another party can distinguish the encoded messages. The
first party can prepare an ensemble of quantum states, and the other party has
access to a set of projectors using which he can form a quantum measurement. If
the ensemble and the projectors satisfy the conditions of the
\index{packing lemma}%
packing lemma, then it guarantees the existence of a scheme by which the
second party can distinguish the classical messages that the first party prepares.

The statement of the packing lemma is quite general, and this approach has a
great advantage because we can use it as a primitive for many coding theorems.
Examples of coding theorems that we can prove using the packing lemma are the
Holevo--Schumacher--Westmoreland (HSW) theorem%
\index{HSW theorem}
for the transmission of classical information over a quantum channel and the
entanglement-assisted classical capacity theorem for the transmission of
classical information over an entanglement-assisted quantum channel
(furthermore, Chapter~\ref{chap:coh-comm-noisy} shows that these two protocols
are sufficient to generate most known protocols in quantum Shannon theory).
Combined with the covering lemma of the next chapter, the packing lemma gives
a method for transmitting private classical information over a quantum
channel, and this technique in turn gives a way to communicate quantum
information over a quantum channel. As long as we can determine an ensemble
and a set of projectors satisfying the conditions of the packing lemma, we can
apply it in a straightforward way. For example, we prove the HSW\ coding
theorem in Chapter~\ref{chap:classical-comm-HSW}\ largely by relying on the
properties of typical and conditionally typical subspaces that we proved in
the previous chapter, and some of these properties are equivalent to the
conditions of the packing lemma.

The packing lemma is a \textquotedblleft one-shot\textquotedblright\ lemma
because it applies to a general scenario that is not limited only to
i.i.d.~uses of a quantum channel. This \textquotedblleft
one-shot\textquotedblright\ approach is part of the reason that we can apply
it to a variety of situations. The technique of proving a \textquotedblleft
one-shot\textquotedblright\ result and applying it to the i.i.d.~scenario is a
common method of attack in quantum Shannon theory
(Chapter~\ref{chap:covering-lemma} does this also by establishing a covering
lemma, which helps in determining a method for sending private classical
information over a quantum channel).

We begin in the next section with a simple example that illustrates the main
ideas of the packing lemma. We then generalize this setting and give the
statement of the packing lemma. We dissect its proof in several sections that
explain the random selection of a code, the construction of a quantum
measurement (called the \textquotedblleft square-root
measurement\textquotedblright), and the error analysis. We then show how to
derandomize the packing lemma so that there exists some scheme for packing
classical messages into Hilbert space with negligible probability of error for
determining each classical message. Finally, we show how a different quantum
measurement, called a sequential decoder, can also be used by a receiver to
decode the messages transmitted.

\section{Introductory Example}

Suppose that Alice would like to communicate classical information to Bob, and
suppose further that she can prepare a message for Bob using the following
BB84 ensemble:%
\begin{equation}
\left\{  \vert0\rangle,\vert1\rangle,\vert+\rangle,\vert-\rangle\right\}  ,
\end{equation}
where each state occurs with equal probability. Let us label each of the above
states by the classical indices $a$, $b$, $c$, and $d$ so that $a$ labels
$\vert0\rangle$, $b$ labels $\vert1\rangle$, etc. She cannot use all of the
states for transmitting classical information because, for example,
$\vert0\rangle$ and $\vert+\rangle$ are non-orthogonal states and there is no
measurement that can distinguish them with high probability.

How can Alice communicate to Bob using this ensemble? She can choose a subset
of the states in the BB84 ensemble for transmitting classical information. She
can choose the states $|0\rangle$ and $|1\rangle$ for encoding one classical
bit of information. Bob can then perform a complete projective measurement in
the basis $\left\{  |0\rangle,|1\rangle\right\}  $ to determine the message
that Alice encodes. Alternatively, Alice and Bob can use the states
$|+\rangle$ and $|-\rangle$ in a similar fashion for encoding one classical
bit of information.

In the above example, Alice can send two messages by using the labels $a$ and
$b$ only. We say that the labels $a$ and $b$ constitute the \textit{code}. The
states $\vert0\rangle$ and $\vert1\rangle$ are the \textit{codewords}, the
projectors $\vert0\rangle\langle0\vert$ and $\vert1\rangle\langle1\vert$ are
each a \textit{codeword projector}, and the projector $\vert0\rangle
\langle0\vert+\vert1\rangle\langle1\vert$ is the \textit{code projector} (in
this case, the code projector projects onto the whole Hilbert space).

The construction in the above example gives a way to use a certain ensemble
for \textquotedblleft packing\textquotedblright\ classical information into
Hilbert space, but there is only so much room for packing. For example, it is
impossible to encode more than one bit of classical information into a qubit
such that someone else can access this classical information reliably---this
is the statement of the%
\index{Holevo bound}
Holevo bound\ (Exercise~\ref{ex-qie:holevo-bound}).

\section{The Setting of the Packing Lemma}

We generalize the above example to show how Alice can effectively pack
classical information into a Hilbert space such that Bob can retrieve it with
high probability. Suppose that Alice's resource for communication is an
ensemble $\left\{  p_{X}(x),\sigma_{x}\right\}  _{x\in\mathcal{X}}$ of quantum
states that she can prepare for Bob, where the states $\sigma_{x}$ are not
necessarily perfectly distinguishable. We define the ensemble as follows:

\begin{definition}
[Ensemble]\label{def:packing-lemma-ensemble}Suppose $\mathcal{X}$ is a set of
size $\left\vert \mathcal{X}\right\vert $ with elements $x$, and suppose $X$
is a random variable with probability density function $p_{X}(x)$. Suppose we
have an ensemble $\left\{  p_{X}(x),\sigma_{x}\right\}  _{x\in\mathcal{X}}%
$\ of quantum states where we encode each realization $x$ into a quantum state
$\sigma_{x}\in\mathcal{D}(\mathcal{H})$. The expected density operator of the
ensemble is%
\begin{equation}
\sigma\equiv\sum_{x\in\mathcal{X}}p_{X}(x)\sigma_{x}.
\end{equation}

\end{definition}

How can Alice transmit classical information reliably to Bob by making use of
this ensemble? As suggested in the example from the previous section, Alice
can select a subset of messages from the set $\mathcal{X}$, and Bob's task is
to distinguish this subset of states as best he can. We equip him with certain
tools: a \textit{code} subspace projector $\Pi$ and a set of \textit{codeword}
subspace projectors $\left\{  \Pi_{x}\right\}  _{x\in\mathcal{X}}$ with
certain desirable properties (we explain these terms in more detail below). As
a rough description, he can use these projectors to construct a quantum
measurement that determines the message Alice sends. He would like to be
almost certain that the received state lies in the subspace onto which the
code subspace projector $\Pi$ projects. He would also like to use the codeword
subspace projectors $\left\{  \Pi_{x}\right\}  _{x\in\mathcal{X}}$ to
determine the classical message that Alice sends. If the ensemble and the
projectors satisfy certain conditions, the four conditions of the packing
lemma, then it is possible for Bob to build up a measurement such that Alice
can communicate reliably with him.

Suppose that Alice chooses some subset $\mathcal{C}$ of $\mathcal{X}$ for
encoding classical information. The subset $\mathcal{C}$ that Alice chooses
constitutes a \textit{code}. Let us index the code $\mathcal{C}$\ by a message
set $\mathcal{M}$ with elements labeled by $m$. The set $\mathcal{M}$ contains
messages\ that Alice would like to transmit to Bob, and we assume that she
chooses each message $m$ with equal probability. The subensemble that Alice
uses for transmitting classical information is thus as follows:%
\begin{equation}
\left\{  \frac{1}{\left\vert \mathcal{M}\right\vert },\sigma_{c_{m}}\right\}
,
\end{equation}
where each $c_{m}$ is a \textit{codeword} that depends on the message $m$ and
takes a value in $\mathcal{X}$.

Bob needs a way to determine the classical message that Alice transmits. The
most general way that quantum mechanics offers for retrieving classical
information is a POVM. Thus, Bob performs some measurement described by a
POVM\ $\{\Lambda_{m}\}_{m\in\mathcal{M}}$. Bob constructs this POVM\ by using
the codeword subspace projectors $\left\{  \Pi_{x}\right\}  _{x\in\mathcal{X}%
}$ and the code subspace projector $\Pi$ (we give an explicit construction in
the proof of the packing lemma, called the \textquotedblleft
square-root\textquotedblright\ measurement---later we give a different
construction called sequential decoding). If Alice transmits a message $m$,
the probability that Bob correctly retrieves the message $m$ is as follows:%
\begin{equation}
\operatorname{Tr}\left\{  \Lambda_{m}\sigma_{c_{m}}\right\}  .
\end{equation}
Thus, the probability of error for a given message $m$ while using the code
$\mathcal{C}$ is as follows:%
\begin{align}
p_{e}(m,\mathcal{C})  &  \equiv1-\operatorname{Tr}\left\{  \Lambda_{m}%
\sigma_{c_{m}}\right\} \\
&  =\operatorname{Tr}\left\{  \left(  I-\Lambda_{m}\right)  \sigma_{c_{m}%
}\right\}  .
\end{align}

We are interested in the performance of the code $\mathcal{C}$\ that Alice and
Bob choose, and we consider three different measures of performance.

\begin{enumerate}
\item The first and strongest measure of performance is the \textit{maximal
probability of error of the code }$\mathcal{C}$. A code $\mathcal{C}$\ has
maximum probability of error $\varepsilon$ if%
\begin{equation}
\varepsilon=\max_{m\in\mathcal{M}}p_{e}(m,\mathcal{C}).
\end{equation}

\item A weaker measure of performance is the \textit{average probability of
error }$\bar{p}_{e}(\mathcal{C})$\textit{ of the code} $\mathcal{C}$, where%
\begin{equation}
\bar{p}_{e}(\mathcal{C})\equiv\frac{1}{\left\vert \mathcal{M}\right\vert }%
\sum_{m=1}^{\left\vert \mathcal{M}\right\vert }p_{e}(m,\mathcal{C}).
\end{equation}

\item The third measure of performance is even weaker than the previous two
but turns out to be the most useful in the mathematical proofs. It uses a
conceptually different notion of code called a \textit{random code}. Suppose
that Alice and Bob choose a code $\mathcal{C}$ randomly from the set of all
possible codes according to some probability density $p_{\mathcal{C}}$ (the
code $\mathcal{C}$ itself therefore becomes a random variable!). The third
measure of performance is the \textit{expectation of the average probability
of error of a random code }$\mathcal{C}$ where the expectation is with respect
to the set of all possible codes, with each code chosen according to some
density $p_{\mathcal{C}}$:%
\begin{align}
\mathbb{E}_{\mathcal{C}}\left\{  \bar{p}_{e}(\mathcal{C})\right\}   &
\equiv\mathbb{E}_{\mathcal{C}}\left\{  \frac{1}{\left\vert \mathcal{M}%
\right\vert }\sum_{m=1}^{\left\vert \mathcal{M}\right\vert }p_{e}%
(m,\mathcal{C})\right\} \\
&  =\sum_{\mathcal{C}}p_{\mathcal{C}}\left(  \frac{1}{\left\vert
\mathcal{M}\right\vert }\sum_{m=1}^{\left\vert \mathcal{M}\right\vert }%
p_{e}(m,\mathcal{C})\right)  .
\end{align}
We will see that considering this performance criterion simplifies the
mathematics in the proof of the packing lemma. Then we will employ a series of
arguments to strengthen the result from this weakest performance criterion to
the first and strongest performance criterion.
\end{enumerate}

\section{Statement of the Packing Lemma}

\begin{lemma}
[Packing Lemma]\label{lem-pack:pack}Suppose that we have an ensemble as in
Definition~\ref{def:packing-lemma-ensemble}. Suppose that a code subspace
projector $\Pi$ and codeword subspace projectors $\left\{  \Pi_{x}\right\}
_{x\in\mathcal{X}}$ exist, they project onto subspaces of $\mathcal{H}$, and
these projectors and the ensemble satisfy the following conditions:%
\begin{align}
\operatorname{Tr}\left\{  \Pi\sigma_{x}\right\}   &  \geq1-\varepsilon
,\label{eq:pack-1}\\
\operatorname{Tr}\left\{  \Pi_{x}\sigma_{x}\right\}   &  \geq1-\varepsilon
,\label{eq:pack-2}\\
\operatorname{Tr}\left\{  \Pi_{x}\right\}   &  \leq d,\label{eq:pack-3}\\
\Pi\sigma\Pi &  \leq\frac{1}{D}\Pi, \label{eq:pack-4}%
\end{align}
where $\varepsilon\in(0,1)$, $D>0$, and $d\in(0,D)$. Suppose that
$\mathcal{M}$ is a set of size $\left\vert \mathcal{M}\right\vert $ with
elements $m$. We generate a set $\mathcal{C}=\left\{  C_{m}\right\}
_{m\in\mathcal{M}}$ of random variables $C_{m}$ independently at random
according to $p_{X}(x)$, so that each random variable $C_{m}$ takes a value in
$\mathcal{X}$ and corresponds to the message $m$, but its distribution is
independent of the particular message $m$. The set $\mathcal{C}$\ constitutes
a random code. Then there exists a corresponding POVM\ $\{\Lambda_{m}%
\}_{m\in\mathcal{M}}$ that reliably distinguishes the states $\{\sigma_{C_{m}%
}\}_{m\in\mathcal{M}}$, in the sense that the expectation of the average
probability of detecting the correct state is high:%
\begin{equation}
\mathbb{E}_{\mathcal{C}}\left\{  \frac{1}{\left\vert \mathcal{M}\right\vert
}\sum_{m\in\mathcal{M}}\operatorname{Tr}\left\{  \Lambda_{m}\sigma_{C_{m}%
}\right\}  \right\}  \geq1-2\left(  \varepsilon+2\sqrt{\varepsilon}\right)
-4\left\vert \mathcal{M}\right\vert \frac{d}{D},
\end{equation}
when $D/d$ is large, $\left\vert \mathcal{M}\right\vert \ll D/d$, and
$\varepsilon$ is small.
\end{lemma}

Condition \eqref{eq:pack-1} states that the code subspace with projector $\Pi$
contains each message $\sigma_{x}$ with high probability. Condition
\eqref{eq:pack-2}\ states that each codeword subspace projector $\Pi_{x}$
contains its corresponding state $\sigma_{x}$ with high probability. Condition
\eqref{eq:pack-3}\ states that the dimension of each codeword subspace
projector $\Pi_{x}$\ is less than some positive number $d\in(0,D)$. Condition
\eqref{eq:pack-4}\ states that the expected density operator $\sigma$\ of the
ensemble is approximately maximally mixed when projecting it onto the subspace
with projector $\Pi$. Conditions \eqref{eq:pack-1} and
\eqref{eq:pack-4}\ imply that%
\begin{equation}
\operatorname{Tr}\left\{  \Pi\right\}  \geq D\left(  1-\varepsilon\right)  ,
\end{equation}
so that the dimension of the code subspace projector $\Pi$ is approximately
$D$ if $\varepsilon$ is small. We show how to construct a code with messages
that Alice wants to send. These four conditions are crucial for constructing a
decoding POVM with the desirable property that it can distinguish the encoded
messages with high probability.

The main idea of the packing lemma is that we can pack $\left\vert
\mathcal{M}\right\vert $ classical messages into a subspace with corresponding
projector $\Pi.$ There is then a small probability of error when trying to
detect the classical messages using codeword subspace projectors $\Pi_{x}$. The
intuition is the same as that depicted in
Figure~\ref{fig-intro:Shannon-packing}. We are trying to pack as many
subspaces of size $d$ into a larger space of size $D$. In the proof of the HSW
coding theorem in Chapter~\ref{chap:classical-comm-HSW}, $D$ will be of size
$\approx2^{nH( B) }$ and $d$ will be of size $\approx2^{nH( B|X) }$,
suggesting that we can pack in $\approx2^{n\left[  H( B) -H( B|X) \right]
}=2^{nI( X;B) }$ messages while still being able to distinguish them reliably.

\section{Proof of the Packing Lemma}

The proof technique employs a Shannon-like argument in which we generate a
code at random. We first show how to construct a POVM, the \textquotedblleft
pretty good\textquotedblright\ or \textquotedblleft
square-root\textquotedblright\ measurement, that can decode a classical
message with high probability. We then prove that the expectation of the
average error probability is small (where the expectation is with respect to all random
codes). In a corollary in the next section, we finally use standard
Shannon-like arguments to show that a code exists whose maximal probability of
error for all messages is small.

\subsection{Code Construction}

\label{sec-pack:code-constr}We present a Shannon-like random coding argument
to simplify the mathematics that follow. We construct a code $\mathcal{C}$ at
random by independently generating $\left\vert \mathcal{M}\right\vert $
codewords according to the distribution $p_{X}(x)$. Let $\mathcal{C}%
\equiv\left\{  c_{m}\right\}  _{m\in\mathcal{M}}$\ be a collection of the
realizations $c_{m}$ of $\left\vert \mathcal{M}\right\vert $\ independent
random variables $C_{m}$. Each $C_{m}$ takes a value $c_{m}$ in $\mathcal{X}$
with probability $p_{X}(c_{m})$ and represents a classical codeword in the
random code $\mathcal{C}$. The probability $p(\mathcal{C})$ of choosing a
particular code $\mathcal{C}$ is equal to the following:%
\begin{equation}
p(\mathcal{C})=\prod_{m=1}^{\left\vert \mathcal{M}\right\vert }p_{X}(c_{m}).
\end{equation}
There is a great advantage to choosing the code in this way. The expectation
of any product $f(C_{m})g(C_{m^{\prime}})$\ of two functions $f$ and $g$ of
two different random codewords $C_{m}$ and $C_{m^{\prime}}$, where the
expectation is with respect to the random choice of code, factors as follows:%
\begin{align}
\mathbb{E}_{\mathcal{C}}\left\{  f(C_{m})g(C_{m^{\prime}})\right\}   &
=\sum_{c}p(c)f(c_{m})g(c_{m^{\prime}})\\
&  =\sum_{c_{1}\in\mathcal{X}}p_{X}(c_{1})\cdots\sum_{c_{\left\vert
\mathcal{M}\right\vert }\in\mathcal{X}}p_{X}(c_{\left\vert \mathcal{M}%
\right\vert })f(c_{m})g(c_{m^{\prime}})\\
&  =\sum_{c_{m}\in\mathcal{X}}p_{X}(c_{m})f(c_{m})\sum_{c_{m^{\prime}}%
\in\mathcal{X}}p_{X}(c_{m^{\prime}})g(c_{m^{\prime}})\\
&  =\mathbb{E}_{X}\left\{  f(X)\right\}  \mathbb{E}_{X}\left\{  g(X)\right\}
.
\end{align}
This factoring happens because of the random way in which we choose the code,
and we exploit this fact in the proof of the packing lemma. We employ the
following events in sequence:

\begin{enumerate}
\item We choose a random code as described above.

\item We reveal the code to the sender and receiver (i.e., they are allowed to
meet beforehand to agree on a strategy before communication begins).

\item The sender chooses a message $m$\ at random (with uniform probability
according to some random variable~$M$)\ from $\mathcal{M}$ and encodes it in
the codeword $c_{m}$. The quantum state that the sender transmits is then
equal to $\sigma_{c_{m}}$.

\item The receiver performs a POVM $\{\Lambda_{m}\}_{m\in\mathcal{M}}$ to
determine the message that the sender transmits, and each POVM\ element
$\Lambda_{m}$ corresponds to a message $m$ in the code. The receiver obtains a
classical result from the measurement, and we model it with the random
variable~$M^{\prime}$. The conditional probability $\Pr\left\{  M^{\prime
}=m|M=m\right\}  $ of obtaining the correct result from the measurement is
equal to%
\begin{equation}
\Pr\left\{  M^{\prime}=m|M=m\right\}  =\operatorname{Tr}\left\{  \Lambda
_{m}\sigma_{c_{m}}\right\}  .
\end{equation}

\item The receiver decodes correctly if $M^{\prime}=M$ and decodes incorrectly
if $M^{\prime}\neq M$.
\end{enumerate}

\subsection{POVM\ Construction}

We cannot directly use the projectors $\Pi_{x}$ in a POVM\ because they do not
satisfy the conditions for being a POVM. Namely, it is not necessarily true
that $\sum_{x\in\mathcal{X}}\Pi_{x}=I$. Furthermore, the codeword subspace
projectors $\Pi_{x}$ may have support outside that of the code subspace
projector $\Pi$. It is necessary for us to have the code subspace projector
involved in the analysis because we will need to invoke condition
\eqref{eq:pack-4}\ of the packing lemma.

To remedy these issues, first consider the following set of operators:%
\begin{equation}
\forall x\ \ \ \Upsilon_{x}\equiv\Pi\Pi_{x}\Pi. \label{eq-pack:coated-ops}%
\end{equation}
The operator $\Upsilon_{x}$\ is a positive semi-definite operator, and the
effect of \textquotedblleft coating\textquotedblright\ the codeword subspace
projector $\Pi_{x}$ with the code subspace projector $\Pi$ is to slice out any
part of the support of $\Pi_{x}$ that is not in the support of $\Pi$. From the
conditions \eqref{eq:pack-1}--\eqref{eq:pack-2} of the packing lemma, there
should be little probability for our states of interest to lie in the part of
the support of $\Pi_{x}$ outside the support of $\Pi$. The operators
$\Upsilon_{x}$ have the desirable property that they only have support inside
the subspace corresponding to the code subspace projector $\Pi$. So we have
remedied the second issue stated above.
Exercise~\ref{ex-pack:no-code-subspace-proj}\ explores an alternative way of
resolving this issue.

We now remedy the first problem stated above by constructing a set\ $\left\{
\Lambda_{m}\right\}  _{m\in\mathcal{M}}$ with the following elements:%
\begin{equation}
\Lambda_{m}\equiv\left(  \sum_{m^{\prime}=1}^{\left\vert \mathcal{M}%
\right\vert }\Upsilon_{c_{m^{\prime}}}\right)  ^{-1/2}\Upsilon_{c_{m}}\left(
\sum_{m^{\prime}=1}^{\left\vert \mathcal{M}\right\vert }\Upsilon
_{c_{m^{\prime}}}\right)  ^{-1/2}. \label{eq-pack:packing-POVM}%
\end{equation}
The elements of the set $\left\{  \Lambda_{m}\right\}  _{m\in\mathcal{M}}$
constitute
\index{square-root measurement}%
the \textquotedblleft pretty good\textquotedblright\ or \textquotedblleft
square-root\textquotedblright\ measurement. These\ elements generally have the
property that $\sum_{m=1}^{\left\vert \mathcal{M}\right\vert }\Lambda_{m}\leq
I$, and we can \textquotedblleft complete\textquotedblright\ the set to be a
full POVM\ by inserting the element $\Lambda_{0}\equiv I-\sum_{m=1}%
^{\left\vert \mathcal{M}\right\vert }\Lambda_{m}$ into the set $\left\{
\Lambda_{m}\right\}  _{m\in\mathcal{M}}$. The extra element $\Lambda_{0}$
corresponds to a failed decoding. Note that the inverse square root $A^{-1/2}%
$\ of a positive semi-definite operator $A$ is defined as the inverse square
root operation only on the support of $A$, per the usual convention from
Definition~\ref{def-qt:hermitian-op-function}. The idea of the pretty good
measurement is that the POVM elements $\left\{  \Lambda_{m}\right\}
_{m=1}^{\left\vert \mathcal{M}\right\vert }$ correspond to the messages sent
and the element $\Lambda_{0}$ corresponds to an error result (an inability to
identify any of the messages).

\subsection{Error Analysis}

Before proceeding with the error analysis, we need the following operator
inequality, which will be helpful in analyzing the error probability:

\begin{lemma}
[Hayashi--Nagaoka]\label{lem-pack:hayashi-nag}Let $S,T\in\mathcal{L}%
(\mathcal{H})$ be positive semi-definite operators such that $I-S$ is positive
semi-definite also.
\index{Hayashi--Nagaoka operator inequality}%
Then for a strictly positive constant $c$, the following operator inequality
holds%
\begin{equation}
I-\left(  S+T\right)  ^{-1/2}S\left(  S+T\right)  ^{-1/2}\leq\left(
1+c\right)  \left(  I-S\right)  +\left(  2+c+c^{-1}\right)  T.
\label{eq:nagoaka}%
\end{equation}

\end{lemma}

\begin{proof}
For any two operators $A,B\in\mathcal{L}(\mathcal{H})$, the following operator
inequality holds%
\begin{equation}
\left(  A-cB\right)  ^{\dag}\left(  A-cB\right)  \geq0,
\end{equation}
which is equivalent to%
\begin{equation}
c^{-1}A^{\dag}A+cB^{\dag}B\geq A^{\dag}B+B^{\dag}A.
\end{equation}
Now pick $A=\sqrt{T}R$ and $B=\sqrt{T}\left(  I-R\right)  $, where
$R\in\mathcal{L}(\mathcal{H})$, and plug into the above to find that%
\begin{equation}
c^{-1}R^{\dag}TR+c\left(  I-R\right)  ^{\dag}T\left(  I-R\right)  \geq
R^{\dag}T\left(  I-R\right)  +\left(  I-R\right)  ^{\dag}TR.
\end{equation}
Thus, we find that%
\begin{align}
T  &  =R^{\dag}TR+R^{\dag}T\left(  I-R\right)  +\left(  I-R\right)  ^{\dag
}TR+\left(  I-R\right)  ^{\dag}T\left(  I-R\right) \\
&  \leq\left(  1+c^{-1}\right)  R^{\dag}TR+\left(  1+c\right)  \left(
I-R\right)  ^{\dag}T\left(  I-R\right)  .
\end{align}
Setting $R=\left(  S+T\right)  ^{1/2}$, we find that%
\begin{multline}
T\leq\left(  1+c^{-1}\right)  \left(  S+T\right)  ^{1/2}T\left(  S+T\right)
^{1/2}\\
+\left(  1+c\right)  \left(  I-\left(  S+T\right)  ^{1/2}\right)  T\left(
I-\left(  S+T\right)  ^{1/2}\right)  .
\end{multline}
Since $T\leq S+T$, we can conclude that%
\begin{align}
T  &  \leq\left(  1+c^{-1}\right)  \left(  S+T\right)  ^{1/2}T\left(
S+T\right)  ^{1/2}\nonumber\\
&  \ \ \ \ \ \ \ +\left(  1+c\right)  \left(  I-\left(  S+T\right)
^{1/2}\right)  \left(  S+T\right)  \left(  I-\left(  S+T\right)  ^{1/2}\right)
\\
&  =\left(  S+T\right)  ^{1/2}\left[  \left(  1+c^{-1}\right)  T+\left(
1+c\right)  \left(  I+S+T-2\left(  S+T\right)  ^{1/2}\right)  \right]  \left(
S+T\right)  ^{1/2}\\
&  =\left(  S+T\right)  ^{1/2}\left[  \left(  2+c+c^{-1}\right)  T+\left(
1+c\right)  \left(  I+S-2\left(  S+T\right)  ^{1/2}\right)  \right]  \left(
S+T\right)  ^{1/2}\\
&  \leq\left(  S+T\right)  ^{1/2}\left[  \left(  2+c+c^{-1}\right)  T+\left(
1+c\right)  \left(  I+S-2S\right)  \right]  \left(  S+T\right)  ^{1/2}\\
&  =\left(  S+T\right)  ^{1/2}\left[  \left(  2+c+c^{-1}\right)  T+\left(
1+c\right)  \left(  I-S\right)  \right]  \left(  S+T\right)  ^{1/2}.
\end{align}
The last inequality follows because $S\leq S^{1/2}\leq\left(  S+T\right)
^{1/2}$. This makes use of the assumption that $S\leq I$ and that fact that
the square root function is operator monotone, meaning that if $X\leq Y$ for
positive semi-definite $X$ and $Y$, then $X^{1/2}\leq Y^{1/2}$. Multiplying
both sides of this operator inequality by $\left(  S+T\right)  ^{-1/2}$, we
find that%
\begin{equation}
\left(  S+T\right)  ^{-1/2}T\left(  S+T\right)  ^{-1/2}\leq\left(
2+c+c^{-1}\right)  T+\left(  1+c\right)  \left(  \Pi_{S+T}-S\right)  ,
\end{equation}
where $\Pi_{S+T}$ denotes the projector onto the support of $S+T$, so that
$\Pi_{S+T}T\Pi_{S+T}=T$ and $\Pi_{S+T}S\Pi_{S+T}=S$. Then consider that%
\begin{align}
&  I-\left(  S+T\right)  ^{-1/2}S\left(  S+T\right)  ^{-1/2}\nonumber\\
&  =I-\left(  S+T\right)  ^{-1/2}\left(  S+T\right)  \left(  S+T\right)
^{-1/2}+\left(  S+T\right)  ^{-1/2}T\left(  S+T\right)  ^{-1/2}\\
&  =I-\Pi_{S+T}+\left(  S+T\right)  ^{-1/2}T\left(  S+T\right)  ^{-1/2}\\
&  \leq\left(  1+c\right)  \left(  I-\Pi_{S+T}\right)  +\left(  2+c+c^{-1}%
\right)  T+\left(  1+c\right)  \left(  \Pi_{S+T}-S\right) \\
&  =\left(  2+c+c^{-1}\right)  T+\left(  1+c\right)  \left(  I-S\right)  .
\end{align}
This concludes the proof.
\end{proof}

\bigskip

Suppose we have chosen a particular code $\mathcal{C}$. Let $p_{e}%
(m,\mathcal{C})$ be the probability of decoding incorrectly given that message
$m$ was sent while using the code $\mathcal{C}$:%
\begin{equation}
p_{e}(m,\mathcal{C})\equiv\operatorname{Tr}\left\{  \left(  I-\Lambda
_{m}\right)  \sigma_{c_{m}}\right\}  .
\end{equation}
In Lemma~\ref{lem-pack:hayashi-nag}, we can set%
\begin{equation}
T=\sum_{m^{\prime}\neq m}^{\left\vert \mathcal{M}\right\vert }\Upsilon
_{c_{m^{\prime}}},\ \ \ \ \ \ \ \ S=\Upsilon_{c_{m}},
\end{equation}
and pick $c=1$,\footnote{For our purposes here, it suffices to pick the
parameter $c=1$, but for more fine-tuned error analyses, it is essential to
choose the parameter $c$ to vary with other code parameters.} so that the
bound in \eqref{eq:nagoaka} becomes%
\begin{equation}
I-\Lambda_{m}\leq2\left(  I-\Upsilon_{c_{m}}\right)  +4\sum_{m^{\prime}\neq
m}^{\left\vert \mathcal{M}\right\vert }\Upsilon_{c_{m^{\prime}}}.
\label{eq-pack:nagaoka-applied}%
\end{equation}
Then using \eqref{eq-pack:nagaoka-applied} and linearity of trace, we obtain
the following upper bound on the error probability:%
\begin{equation}
p_{e}(m,\mathcal{C})\leq2\operatorname{Tr}\left\{  \left(  I-\Upsilon_{c_{m}%
}\right)  \sigma_{c_{m}}\right\}  +4\sum_{m^{\prime}\neq m}^{\left\vert
\mathcal{M}\right\vert }\operatorname{Tr}\left\{  \Upsilon_{c_{m^{\prime}}%
}\sigma_{c_{m}}\right\}  . \label{eq:pack-bound}%
\end{equation}
The above bound on the message error probability for code $\mathcal{C}$ has a
similar interpretation as that in classical Shannon-like proofs. We bound the
error probability by the probability of incorrectly decoding the message $m$
with the message operator $\Upsilon_{c_{m}}$ (the first term in
\eqref{eq:pack-bound}) summed with the probability of confusing the
transmitted message with a message $c_{m^{\prime}}$ different from the
transmitted one (the second term in \eqref{eq:pack-bound}).

Consider the first term $\operatorname{Tr}\left\{  \left(  I-\Upsilon_{c_{m}%
}\right)  \sigma_{c_{m}}\right\}  $ on the right-hand side\ of
\eqref{eq:pack-bound}. We can bound it from above by a small number, simply by
applying \eqref{eq:pack-1}--\eqref{eq:pack-2} and the gentle operator lemma
(Lemma~\ref{lem-dm:gentle-operator}). Consider the following chain of
inequalities:%
\begin{align}
\operatorname{Tr}\left\{  \Upsilon_{c_{m}}\sigma_{c_{m}}\right\}   &
=\operatorname{Tr}\left\{  \Pi\Pi_{c_{m}}\Pi\sigma_{c_{m}}\right\} \\
&  =\operatorname{Tr}\left\{  \Pi_{c_{m}}\Pi\sigma_{c_{m}}\Pi\right\} \\
&  \geq\operatorname{Tr}\left\{  \Pi_{c_{m}}\sigma_{c_{m}}\right\}
-\left\Vert \Pi\sigma_{c_{m}}\Pi-\sigma_{c_{m}}\right\Vert _{1}\\
&  \geq1-\varepsilon-2\sqrt{\varepsilon}.
\end{align}
The first equality follows from the definition of $\Upsilon_{c_{m}}$ in
\eqref{eq-pack:coated-ops}. The second equality follows from cyclicity of the
trace. The first inequality follows from applying
Exercise~\ref{ex-dm:trace-ineq-herm}. The last inequality follows from
applying \eqref{eq:pack-1} to $\operatorname{Tr}\left\{  \Pi_{c_{m}}%
\sigma_{c_{m}}\right\}  $ and from applying \eqref{eq:pack-2} and the gentle
operator lemma (Lemma~\ref{lem-dm:gentle-operator}) to $\left\Vert \Pi
\sigma_{c_{m}}\Pi-\sigma_{c_{m}}\right\Vert _{1}$. The above bound then
implies the following one:%
\begin{equation}
\operatorname{Tr}\left\{  \left(  I-\Upsilon_{c_{m}}\right)  \sigma_{c_{m}%
}\right\}  =1-\operatorname{Tr}\left\{  \Upsilon_{c_{m}}\sigma_{c_{m}%
}\right\}  \leq\varepsilon+2\sqrt{\varepsilon},
\end{equation}
and by substituting into \eqref{eq:pack-bound}, we get the following bound on
the probability of error:%
\begin{equation}
p_{e}(m,\mathcal{C})\leq2\left(  \varepsilon+2\sqrt{\varepsilon}\right)
+4\sum_{m^{\prime}\neq m}^{\left\vert \mathcal{M}\right\vert }%
\operatorname{Tr}\left\{  \Upsilon_{c_{m^{\prime}}}\sigma_{c_{m}}\right\}  .
\label{eq-pack:packbnd-10}%
\end{equation}

The average error probability $\bar{p}_{e}(\mathcal{C})$\ over all transmitted
messages for code $\mathcal{C}$\ is%
\begin{equation}
\bar{p}_{e}(\mathcal{C})=\frac{1}{\left\vert \mathcal{M}\right\vert }%
\sum_{m=1}^{\left\vert \mathcal{M}\right\vert }p_{e}(m,\mathcal{C}),
\end{equation}
if we assume that Alice chooses the message $m$ according to the uniform
distribution. From \eqref{eq-pack:packbnd-10}, we see that the average error
probability $\bar{p}_{e}(\mathcal{C})$\ obeys the following bound:%
\begin{equation}
\bar{p}_{e}(\mathcal{C})\leq2\left(  \varepsilon+2\sqrt{\varepsilon}\right)
+\frac{4}{\left\vert \mathcal{M}\right\vert }\sum_{m=1}^{\left\vert
\mathcal{M}\right\vert }\sum_{m^{\prime}\neq m}^{\left\vert \mathcal{M}%
\right\vert }\operatorname{Tr}\left\{  \Upsilon_{c_{m^{\prime}}}\sigma_{c_{m}%
}\right\}  . \label{eq-pack:first-bound}%
\end{equation}

At this point, bounding the average error probability further is a bit
difficult, given the sheer number of combinations of terms $\operatorname{Tr}%
\{\Upsilon_{c_{m^{\prime}}}\sigma_{c_{m}}\}$ that we would have to consider to
do so. Thus, we now invoke the classic Shannon argument in order to simplify
the mathematics. Instead of considering the average probability of error, we
consider the expectation of the average error probability $\mathbb{E}%
_{\mathcal{C}}\left\{  \bar{p}_{e}(\mathcal{C})\right\}  $ with respect to all
possible random codes $\mathcal{C}$. Considering this error quantity
significantly simplifies the mathematics because of the way in which we
constructed the code. We can use the probability distribution $p_{X}(x)$ to
compute the expectation $\mathbb{E}_{\mathcal{C}}$ because we constructed our
code according to this distribution. The bound above becomes as follows, now
denoting each codeword as a random variable $C_{m}$:%
\begin{align}
\mathbb{E}_{\mathcal{C}}\left\{  \bar{p}_{e}(\mathcal{C})\right\}   &
\leq\mathbb{E}_{\mathcal{C}}\left\{  2\left(  \varepsilon+2\sqrt{\varepsilon
}\right)  +\frac{4}{\left\vert \mathcal{M}\right\vert }\sum_{m=1}^{\left\vert
\mathcal{M}\right\vert }\sum_{m^{\prime}\neq m}^{\left\vert \mathcal{M}%
\right\vert }\operatorname{Tr}\left\{  \Upsilon_{C_{m^{\prime}}}\sigma_{C_{m}%
}\right\}  \right\} \\
&  =2\left(  \varepsilon+2\sqrt{\varepsilon}\right)  +\frac{4}{\left\vert
\mathcal{M}\right\vert }\sum_{m=1}^{\left\vert \mathcal{M}\right\vert }%
\sum_{m^{\prime}\neq m}^{\left\vert \mathcal{M}\right\vert }\mathbb{E}%
_{\mathcal{C}}\left\{  \operatorname{Tr}\left\{  \Upsilon_{C_{m^{\prime}}%
}\sigma_{C_{m}}\right\}  \right\}  , \label{eq-pack:random-exp-err-1}%
\end{align}
where the equality follows from the linearity of expectation.

We now calculate the expectation of $\operatorname{Tr}\{\Upsilon
_{C_{m^{\prime}}}\sigma_{C_{m}}\}$ over all random codes $\mathcal{C}$:%
\begin{align}
\mathbb{E}_{\mathcal{C}}\left\{  \operatorname{Tr}\left\{  \Upsilon
_{C_{m^{\prime}}}\sigma_{C_{m}}\right\}  \right\}   &  =\mathbb{E}%
_{\mathcal{C}}\left\{  \operatorname{Tr}\left\{  \Pi\Pi_{C_{m^{\prime}}}%
\Pi\sigma_{C_{m}}\right\}  \right\} \label{eq-pack:other-bound-1}\\
&  =\mathbb{E}_{\mathcal{C}}\left\{  \operatorname{Tr}\left\{  \Pi
_{C_{m^{\prime}}}\Pi\sigma_{C_{m}}\Pi\right\}  \right\} \\
&  =\mathbb{E}_{C_{m},C_{m^{\prime}}}\left\{  \operatorname{Tr}\left\{
\Pi_{C_{m^{\prime}}}\Pi\sigma_{C_{m}}\Pi\right\}  \right\}  .
\end{align}
The first equality follows from the definition in \eqref{eq-pack:coated-ops},
and the second equality follows from cyclicity of trace. The third equality
follows because only the codewords for $m$ and $m^{\prime}$ are involved.
Independence of random variables $C_{m}$ and $C_{m^{\prime}}$ (from the code
construction) gives that the above expression equals%
\begin{align}
\operatorname{Tr}\left\{  \mathbb{E}_{C_{m^{\prime}}}\{\Pi_{C_{m^{\prime}}%
}\}\Pi\mathbb{E}_{C_{m}}\{\sigma_{C_{m}}\}\Pi\right\}   &  =\operatorname{Tr}%
\left\{  \mathbb{E}_{C_{m^{\prime}}}\{\Pi_{C_{m^{\prime}}}\}\Pi\sigma
\Pi\right\} \\
&  \leq\operatorname{Tr}\left\{  \mathbb{E}_{C_{m^{\prime}}}\{\Pi
_{C_{m^{\prime}}}\}\frac{1}{D}\Pi\right\} \\
&  =\frac{1}{D}\operatorname{Tr}\left\{  \mathbb{E}_{C_{m^{\prime}}}%
\{\Pi_{C_{m^{\prime}}}\}\Pi\right\}  ,
\end{align}
where the first equality uses the fact that $\mathbb{E}_{C_{m}}\left\{
\sigma_{C_{m}}\right\}  =\sum_{x\in\mathcal{X}}p(x)\sigma_{x}=\sigma$ and
$\Pi$ is a constant with respect to the expectation. The first inequality uses
the fourth condition \eqref{eq:pack-4}\ of the packing lemma, the fact that
$\Pi\sigma\Pi$, $\Pi$, and $\Pi_{C_{m^{\prime}}}$ are all positive
semi-definite operators, and $\operatorname{Tr}\left\{  CA\right\}
\geq\operatorname{Tr}\left\{  CB\right\}  $ for $C\geq0$ and $A\geq B$.
Continuing, we have%
\begin{align}
\frac{1}{D}\operatorname{Tr}\left\{  \mathbb{E}_{C_{m^{\prime}}}%
\{\Pi_{C_{m^{\prime}}}\}\Pi\right\}   &  \leq\frac{1}{D}\operatorname{Tr}%
\left\{  \mathbb{E}_{C_{m^{\prime}}}\{\Pi_{C_{m^{\prime}}}\}\right\} \\
&  =\frac{1}{D}\mathbb{E}_{C_{m^{\prime}}}\left\{  \operatorname{Tr}%
\{\Pi_{C_{m^{\prime}}}\}\right\} \\
&  \leq\frac{d}{D}. \label{eq-pack:other-bound-last}%
\end{align}
The first inequality follows from the fact that $\Pi\leq I$ and $\Pi
_{C_{m^{\prime}}}$ is a positive semi-definite operator. The last inequality
follows from \eqref{eq:pack-3}. The following inequality then holds by
considering the development from~\eqref{eq-pack:other-bound-1}
to~\eqref{eq-pack:other-bound-last}:%
\begin{equation}
\mathbb{E}_{\mathcal{C}}\left\{  \operatorname{Tr}\left\{  \Upsilon
_{C_{m^{\prime}}}\sigma_{C_{m}}\right\}  \right\}  \leq\frac{d}{D}.
\end{equation}
We substitute into \eqref{eq-pack:random-exp-err-1} to show that the
expectation $\mathbb{E}_{\mathcal{C}}\left\{  \bar{p}_{e}(\mathcal{C}%
)\right\}  $ of the average error probability $\bar{p}_{e}(\mathcal{C})$ over
all codes obeys the bound stated in the packing lemma:%
\begin{align}
\mathbb{E}_{\mathcal{C}}\left\{  \bar{p}_{e}(\mathcal{C})\right\}   &
\leq2\left(  \varepsilon+2\sqrt{\varepsilon}\right)  +\frac{4}{\left\vert
\mathcal{M}\right\vert }\sum_{m=1}^{\left\vert \mathcal{M}\right\vert }%
\sum_{m^{\prime}\neq m}^{\left\vert \mathcal{M}\right\vert }\mathbb{E}%
_{\mathcal{C}}\left\{  \operatorname{Tr}\left\{  \sigma_{c_{m}}\Upsilon
_{c_{m^{\prime}}}\right\}  \right\} \\
&  \leq2\left(  \varepsilon+2\sqrt{\varepsilon}\right)  +\frac{4}{\left\vert
\mathcal{M}\right\vert }\sum_{m=1}^{\left\vert \mathcal{M}\right\vert }%
\sum_{m^{\prime}\neq m}^{\left\vert \mathcal{M}\right\vert }\frac{d}{D}\\
&  \leq2\left(  \varepsilon+2\sqrt{\varepsilon}\right)  +4\left(  \left\vert
\mathcal{M}\right\vert -1\right)  \frac{d}{D}\\
&  \leq2\left(  \varepsilon+2\sqrt{\varepsilon}\right)  +4\left\vert
\mathcal{M}\right\vert \frac{d}{D}.
\end{align}

\section{Derandomization and Expurgation}

The above version of the packing lemma is a randomized version that shows how
the expectation of the average probability of error is small. We now prove a
derandomized version that guarantees the existence of a code with small
maximal error probability for each message. The last two arguments are
traditionally called \textit{derandomization} and \textit{expurgation}.

\begin{corollary}
\label{cor-pack:derandomized}Suppose we have the ensemble as in
Definition~\ref{def:packing-lemma-ensemble}. Suppose that a code subspace
projector $\Pi$ and codeword subspace projectors $\left\{  \Pi_{x}\right\}
_{x\in\mathcal{X}}$ exist, they project onto subspaces of $\mathcal{H}$, and
these projectors and the ensemble have the following properties:%
\begin{align}
\operatorname{Tr}\left\{  \Pi\sigma_{x}\right\}   &  \geq1-\varepsilon,\\
\operatorname{Tr}\left\{  \Pi_{x}\sigma_{x}\right\}   &  \geq1-\varepsilon,\\
\operatorname{Tr}\left\{  \Pi_{x}\right\}   &  \leq d,\\
\Pi\sigma\Pi &  \leq\frac{1}{D}\Pi,
\end{align}
where $\varepsilon\in(0,1)$ and $d\in(0,D)$. Suppose that $\mathcal{M}$ is a
set of size $\left\vert \mathcal{M}\right\vert $ with elements $m$. Then there
exists a code $\mathcal{C}_{0}=\left\{  c_{m}\right\}  _{m\in\mathcal{M}}$
with codewords $c_{m}$ depending on the message $m$\ and taking values in
$\mathcal{X}$, and there exists a corresponding POVM\ $\{\Lambda_{m}%
\}_{m\in\mathcal{M}}$ that reliably distinguishes the states $\{\sigma_{c_{m}%
}\}_{m\in\mathcal{M}}$ in the sense that the probability of detecting the
correct state is high:%
\begin{equation}
\forall m\in\mathcal{M\ \ \ \ \ }\operatorname{Tr}\left\{  \Lambda_{m}%
\sigma_{c_{m}}\right\}  \geq1-4\left(  \varepsilon+2\sqrt{\varepsilon}\right)
-16\left\vert \mathcal{M}\right\vert \frac{d}{D},
\end{equation}
if $\varepsilon$ is small and $\left\vert \mathcal{M}\right\vert \ll D/d$. We
can then use the code $\mathcal{C}_{0}$ and the POVM $\{\Lambda_{m}%
\}_{m\in\mathcal{M}}$ to encode and decode, respectively, $\left\vert
\mathcal{M}\right\vert $ classical messages with high success probability.
\end{corollary}

\begin{proof}
Generate a random code according to the construction in the previous lemma.
The expectation of the average error probability then satisfies the bound in
the statement of the packing lemma. We now make a few standard Shannon-like
arguments to strengthen the result of the previous lemma.

\textbf{Derandomization.} The expectation of the average error probability
$\mathbb{E}_{\mathcal{C}}\left\{  \bar{p}_{e}(\mathcal{C})\right\}  $
satisfies the following bound:%
\begin{equation}
\mathbb{E}_{\mathcal{C}}\left\{  \bar{p}_{e}(\mathcal{C})\right\}
\leq2\left(  \varepsilon+2\sqrt{\varepsilon}\right)  +4\left\vert
\mathcal{M}\right\vert \frac{d}{D}.
\end{equation}
It then follows that the average error probability of at least one code
$\mathcal{C}_{0}=\left\{  c_{m}\right\}  _{m\in\mathcal{M}}$ satisfies the
same bound:%
\begin{equation}
\bar{p}_{e}(\mathcal{C}_{0})\leq2\left(  \varepsilon+2\sqrt{\varepsilon
}\right)  +4\left\vert \mathcal{M}\right\vert \frac{d}{D}.
\end{equation}
Choose this code $\mathcal{C}_{0}$ as the code, and it is possible to find
this code $\mathcal{C}_{0}$ in practice by exhaustive search. This process is
known as \textit{derandomization}.

\begin{exercise}
\label{ex-pack:strong-derandomization}Use Markov's inequality to prove the
following strong statement: if $\varepsilon^{\prime}\equiv2\left(
\varepsilon+2\sqrt{\varepsilon}\right)  +4\left\vert \mathcal{M}\right\vert
\frac{d}{D}$ is small, then the overwhelming fraction $1-\sqrt{\varepsilon
^{\prime}}$ of codes contructed randomly have average error probability less
than $\sqrt{\varepsilon^{\prime}}$.
\end{exercise}

\textbf{Expurgation.}
\index{expurgation}%
We now go from average error probability to maximal error probability instead
by employing an expurgation argument. We know from
Exercise~\ref{ex-intro:expurgation} that $p_{e}(m)\leq2\left[  2\left(
\varepsilon+2\sqrt{\varepsilon}\right)  +4\left\vert \mathcal{M}\right\vert
\frac{d}{D}\right]  $ for at least half of the indices (if it were not true,
then these indices would contribute more than $\varepsilon^{\prime}$ to the
average error probability $\bar{p}_{e}$). Throw out the half of the codewords
with the worst decoding probability and redefine the code according to the new
set of indices. If we redefine the message set $\mathcal{M}^{\prime}$\ such
that the message size $\left\vert \mathcal{M}^{\prime}\right\vert =\left\vert
\mathcal{M}\right\vert /2$, then the error bound becomes $p_{e}(m)\leq2\left[
2\left(  \varepsilon+2\sqrt{\varepsilon}\right)  +8\left\vert \mathcal{M}%
^{\prime}\right\vert \frac{d}{D}\right]  $. We could then use the decoding
POVM\ from the original code and be guaranteed that every codeword has an
error probability no larger than this amount (alternatively, we could also use
a modified square-root decoding POVM that is built from the codeword subspace
projectors remaining after expurgation). These steps have a negligible effect
on the parameters of the code when we later consider a large number of uses of
a noisy quantum channel.
\end{proof}

\begin{exercise}
Use Markov's inequality%
\index{Markov's inequality}
to prove an even stronger expurgation argument (following on the result of
Exercise~\ref{ex-pack:strong-derandomization}). Prove that we can retain a
large fraction $1-\sqrt[4]{\varepsilon^{\prime}}$ of the codewords
(expurgating $\sqrt[4]{\varepsilon^{\prime}}$ of them) so that each remaining
codeword has error probability less than $\sqrt[4]{\varepsilon^{\prime}}$.
\end{exercise}

\begin{exercise}
\label{ex-pack:averaged-conditions}Prove that the packing lemma and its
corollary hold for the same ensemble and a set of projectors for which the
following conditions hold:%
\begin{align}
\sum_{x\in\mathcal{X}}p_{X}(x)\operatorname{Tr}\left\{  \sigma_{x}\Pi\right\}
&  \geq1-\varepsilon,\\
\sum_{x\in\mathcal{X}}p_{X}(x)\operatorname{Tr}\left\{  \sigma_{x}\Pi
_{x}\right\}   &  \geq1-\varepsilon,\\
\operatorname{Tr}\left\{  \Pi_{x}\right\}   &  \leq d,\\
\Pi\sigma\Pi &  \leq\frac{1}{D}\Pi.
\end{align}

\end{exercise}

\begin{exercise}
\label{ex-pack:no-code-subspace-proj}Prove that a variation of the packing
lemma holds in which the POVM\ is of the following form:%
\begin{equation}
\Lambda_{m}\equiv\left(  \sum_{m^{\prime}=1}^{\left\vert \mathcal{M}%
\right\vert }\Pi_{c_{m^{\prime}}}\right)  ^{-1/2}\Pi_{c_{m}}\left(
\sum_{m^{\prime}=1}^{\left\vert \mathcal{M}\right\vert }\Pi_{c_{m^{\prime}}%
}\right)  ^{-1/2}.
\end{equation}
That is, it is not actually necessary to \textquotedblleft
coat\textquotedblright\ each operator in the square-root measurement with the
overall message subspace projector.
\end{exercise}

\section{Sequential Decoding}

\label{sec-pack:sequential}We now prove a variation of the packing lemma by
making use of a completely different decoding scheme, called%
\index{sequential decoding}
\textit{sequential decoding}. This scheme has the receiver perform sequential
tests using the codeword subspace projectors to \textquotedblleft
ask\textquotedblright\ sequentially, \textquotedblleft Was the first codeword
sent?\textquotedblright, \textquotedblleft The second?\textquotedblright,
\textquotedblleft The third?\textquotedblright, etc., until the outcome of one
of these measurements is \textquotedblleft yes.\textquotedblright

It is perhaps unintuitive that such a measurement strategy should work. After
all, we are well aware by this point that measurements can disturb the state
of a quantum system. However, what the packing lemma demonstrates is that if
good codeword projectors are available and we do not try to pack in too many
messages, then reliable decoding is possible. This also means that the
codewords selected are approximately orthogonal. In this sense, it is perhaps
less surprising that such a sequential decoding strategy should work if we
know that the states to be distinguished are approximately orthogonal. The
main tool for analyzing the performance of a sequential decoder is the
\textquotedblleft non-commutative union bound\textquotedblright%
\ (Lemma~\ref{thm-pack:nc-union}) which is a generalization of the union bound
from probability theory.

Here we divide up the proof into several parts, and some of the analysis is
similar to that presented in the previous section. The parts of the
proof\ consist of codebook generation, POVM\ construction, and the error
analysis. We begin with all of the objects and premises of the Packing Lemma
(in Lemma~\ref{lem-pack:pack}).

\textbf{Codebook Generation.} This part is exactly the same as before, so we
instead point to the discussion of this part in
Section~\ref{sec-pack:code-constr}.

\textbf{POVM Construction}. The method for Bob to decode the state that Alice
transmits is as follows: Bob should first ask \textquotedblleft Is the
received state in the code subspace?\textquotedblright\ He can do this
operationally by performing the measurement $\left\{  \Pi,I-\Pi\right\}  $.
Next, he asks in sequential order, \textquotedblleft Is the received codeword
in the $i$th codeword subspace?\textquotedblright\ This is in some sense
equivalent to the question, \textquotedblleft Is the received codeword the
$i$th transmitted codeword?\textquotedblright\ He can ask these questions
operationally by performing the codeword subspace measurements $\left\{
\Pi_{c_{i}},I-\Pi_{c_{i}}\right\}  $.

Why should this sequential decoding scheme work well? Supposing that Alice
transmits message $m$, one reason is that the encoded state $\sigma_{c_{m}}$
lies in the code subspace with high probability, as indicated by
\eqref{eq:pack-1}. Also, the projector $\Pi_{c_{m}}$ is a \textquotedblleft
good detector\textquotedblright\ for the encoded state $\sigma_{c_{m}}$, due
to \eqref{eq:pack-2}.

\textbf{Error Analysis}. The probability of detecting the $m$th codeword
correctly under our sequential decoding scheme is equal to%
\begin{equation}
\text{Tr}\left\{  \Pi_{c_{m}}\hat{\Pi}_{c_{m-1}}\cdots\hat{\Pi}_{c_{1}}%
\Pi\sigma_{c_{m}}\Pi\hat{\Pi}_{c_{1}}\cdots\hat{\Pi}_{c_{m-1}}\Pi_{c_{m}%
}\right\}  ,
\end{equation}
where we make the abbreviation $\hat{\Pi}_{c_{m}}\equiv I-\Pi_{c_{m}}$. That
is, the receiver should get an initial \textquotedblleft yes\textquotedblright%
\ for projecting into the code subspace, $m-1$ \textquotedblleft
no's,\textquotedblright\ and finally a \textquotedblleft yes\textquotedblright%
\ for the $m$th test. Thus, the probability of an incorrect detection for the
$m$th codeword is given by%
\begin{equation}
1-\text{Tr}\left\{  \Pi_{c_{m}}\hat{\Pi}_{c_{m-1}}\cdots\hat{\Pi}_{c_{1}}%
\Pi\sigma_{c_{m}}\Pi\hat{\Pi}_{c_{1}}\cdots\hat{\Pi}_{c_{m-1}}\Pi_{c_{m}%
}\right\}  ,
\end{equation}
and the average error probability of this scheme is equal to%
\begin{equation}
1-\frac{1}{\left\vert \mathcal{M}\right\vert }\sum_{m}\text{Tr}\left\{
\Pi_{c_{m}}\hat{\Pi}_{c_{m-1}}\cdots\hat{\Pi}_{c_{1}}\Pi\sigma_{c_{m}}\Pi
\hat{\Pi}_{c_{1}}\cdots\hat{\Pi}_{c_{m-1}}\Pi_{c_{m}}\right\}  .
\end{equation}
Instead of analyzing the average error probability, we analyze the expectation
of the average error probability, where the expectation is with respect to the
random choice of code:%
\begin{equation}
1-\mathbb{E}_{\mathcal{C}}\left\{  \frac{1}{\left\vert \mathcal{M}\right\vert
}\sum_{m}\operatorname{Tr}\left\{  \Pi_{C_{m}}\hat{\Pi}_{C_{m-1}}\cdots
\hat{\Pi}_{C_{1}}\Pi\sigma_{C_{m}}\Pi\hat{\Pi}_{C_{1}}\cdots\hat{\Pi}%
_{C_{m-1}}\Pi_{C_{m}}\right\}  \right\}  . \label{eq-pack:seq-error-term}%
\end{equation}

We rewrite the above expression just slightly, by observing that%
\begin{align}
1  &  =\mathbb{E}_{\mathcal{C}}\left\{  \frac{1}{\left\vert \mathcal{M}%
\right\vert }\sum_{m}\operatorname{Tr}\left\{  \sigma_{C_{m}}\right\}
\right\} \\
&  =\mathbb{E}_{\mathcal{C}}\left\{  \frac{1}{\left\vert \mathcal{M}%
\right\vert }\sum_{m}\operatorname{Tr}\left\{  \Pi\sigma_{C_{m}}\right\}
+\text{Tr}\left\{  \hat{\Pi}\sigma_{C_{m}}\right\}  \right\} \\
&  \leq\mathbb{E}_{\mathcal{C}}\left\{  \frac{1}{\left\vert \mathcal{M}%
\right\vert }\sum_{m}\operatorname{Tr}\left\{  \Pi\sigma_{C_{m}}\Pi\right\}
\right\}  +\varepsilon,
\end{align}
where we have used \eqref{eq:pack-1}\ in the last line. Substituting into
\eqref{eq-pack:seq-error-term} (and forgetting about the small $\varepsilon$
term for now) gives an upper bound of
\begin{equation}
\mathbb{E}_{\mathcal{C}}\left\{  \frac{1}{\left\vert \mathcal{M}\right\vert
}\sum_{m}\operatorname{Tr}\left\{  \Pi\sigma_{C_{m}}\Pi\right\}
-\operatorname{Tr}\left\{  \Pi_{C_{m}}\hat{\Pi}_{C_{m-1}}\cdots\hat{\Pi
}_{C_{1}}\Pi\sigma_{C_{m}}\Pi\hat{\Pi}_{C_{1}}\cdots\hat{\Pi}_{C_{m-1}}%
\Pi_{C_{m}}\right\}  \right\}  . \label{eq-pack:seq-err-term-2}%
\end{equation}

We now need a tool for analyzing this error probability, which is known as the
non-commutative union bound:

\begin{lemma}
[Non-Commutative Union Bound]\label{thm-pack:nc-union}
\index{non-commutative union bound}
Let $\omega$ be such that $\omega\geq0$ and $\operatorname{Tr}\{\omega\}\leq
1$. Let $P_{1}$, \ldots, $P_{L}$ be Hermitian projectors. Then%
\begin{equation}
\operatorname{Tr}\{\omega\}-\operatorname{Tr}\{P_{L}\cdots P_{1}\omega
P_{1}\cdots P_{L}\}\leq2\sqrt{\sum_{i=1}^{L}\operatorname{Tr}\{(I-P_{i}%
)\omega\}}. \label{eq:non-commutative-bound}%
\end{equation}

\end{lemma}

\begin{proof}
It suffices to prove the following bound for a vector $|\psi\rangle$ such that
$\left\Vert |\psi\rangle\right\Vert _{2}^{2}\leq1$:%
\begin{equation}
\left\Vert |\psi\rangle\right\Vert _{2}^{2}-\left\Vert P_{L}\cdots P_{1}%
|\psi\rangle\right\Vert _{2}^{2}\leq2\sqrt{\sum_{i=1}^{L}\left\Vert \left(
I-P_{i}\right)  |\psi\rangle\right\Vert _{2}^{2}}.
\label{eq:simpler-non-commutative-bound}%
\end{equation}
This is because%
\begin{align}
\left\Vert |\psi\rangle\right\Vert _{2}^{2}  &  =\operatorname{Tr}%
\{|\psi\rangle\langle\psi|\},\\
\left\Vert P_{L}\cdots P_{1}|\psi\rangle\right\Vert _{2}^{2}  &
=\operatorname{Tr}\{P_{L}\cdots P_{1}|\psi\rangle\langle\psi|P_{1}\cdots
P_{L}\},\\
\left\Vert \left(  I-P_{i}\right)  |\psi\rangle\right\Vert _{2}^{2}  &
=\operatorname{Tr}\{(I-P_{i})|\psi\rangle\langle\psi|\},
\end{align}
and any $\omega$ satisfying the conditions given can be written as a convex
combination $\omega=\sum_{z}p(z)|\psi_{z}\rangle\langle\psi_{z}|$ where $p(z)$
is a probability distribution and each $|\psi_{z}\rangle$ satisfies
$\left\Vert |\psi_{z}\rangle\right\Vert _{2}^{2}\leq1$. Then
\eqref{eq:non-commutative-bound} follows from
\eqref{eq:simpler-non-commutative-bound} by concavity of the square root
function. So we now focus on proving \eqref{eq:simpler-non-commutative-bound}.
We begin by showing that%
\begin{equation}
\left\Vert |\psi\rangle-P_{L}\cdots P_{1}|\psi\rangle\right\Vert _{2}^{2}%
\leq\sum_{i=1}^{L}\left\Vert \left(  I-P_{i}\right)  |\psi\rangle\right\Vert
_{2}^{2}. \label{eq:non-comm-first-step}%
\end{equation}
To see this, consider that%
\begin{align}
\left\Vert |\psi\rangle-P_{L}\cdots P_{1}|\psi\rangle\right\Vert _{2}^{2}  &
=\left\Vert \left(  I-P_{L}\right)  |\psi\rangle\right\Vert _{2}%
^{2}+\left\Vert P_{L}\left(  |\psi\rangle-P_{L-1}\cdots P_{1}|\psi
\rangle\right)  \right\Vert _{2}^{2}\\
&  \leq\left\Vert \left(  I-P_{L}\right)  |\psi\rangle\right\Vert _{2}%
^{2}+\left\Vert |\psi\rangle-P_{L-1}\cdots P_{1}|\psi\rangle\right\Vert
_{2}^{2}\\
&  \leq\sum_{i=1}^{L}\left\Vert \left(  I-P_{i}\right)  |\psi\rangle
\right\Vert _{2}^{2}.
\end{align}
The first equality follows from the Pythagorean theorem. The first inequality
follows because a projection cannot increase the norm of a vector. The last
inequality is by induction. Now we take the square root of
\eqref{eq:non-comm-first-step}:%
\begin{equation}
\left\Vert |\psi\rangle-P_{L}\cdots P_{1}|\psi\rangle\right\Vert _{2}\leq
\sqrt{\sum_{i=1}^{L}\left\Vert \left(  I-P_{i}\right)  |\psi\rangle\right\Vert
_{2}^{2}},
\end{equation}
from which we can conclude the following by the triangle inequality:%
\begin{equation}
\left\Vert |\psi\rangle\right\Vert _{2}-\left\Vert P_{L}\cdots P_{1}%
|\psi\rangle\right\Vert _{2}\leq\sqrt{\sum_{i=1}^{L}\left\Vert \left(
I-P_{i}\right)  |\psi\rangle\right\Vert _{2}^{2}}.
\end{equation}
Then rearrange this as follows:%
\begin{equation}
\left\Vert |\psi\rangle\right\Vert _{2}-\sqrt{\sum_{i=1}^{L}\left\Vert \left(
I-P_{i}\right)  |\psi\rangle\right\Vert _{2}^{2}}\leq\left\Vert P_{L}\cdots
P_{1}|\psi\rangle\right\Vert _{2}%
\end{equation}
and square both sides to get%
\begin{align}
&  \left(  \left\Vert |\psi\rangle\right\Vert _{2}-\sqrt{\sum_{i=1}%
^{L}\left\Vert \left(  I-P_{i}\right)  |\psi\rangle\right\Vert _{2}^{2}%
}\right)  ^{2}\nonumber\\
&  =\left\Vert |\psi\rangle\right\Vert _{2}^{2}-2\left\Vert |\psi\rangle\right\Vert _{2}\sqrt{\sum_{i=1}%
^{L}\left\Vert \left(  I-P_{i}\right)  |\psi\rangle\right\Vert _{2}^{2}}%
+\sum_{i=1}^{L}\left\Vert \left(  I-P_{i}\right)  |\psi\rangle\right\Vert
_{2}^{2}\\
&  \leq\left\Vert P_{L}\cdots P_{1}|\psi\rangle\right\Vert _{2}^{2}.
\end{align}
This then implies \eqref{eq:simpler-non-commutative-bound} by dropping the
non-negative term $\sum_{i=1}^{L}\left\Vert \left(  I-P_{i}\right)
|\psi\rangle\right\Vert _{2}^{2}$.
\end{proof}

\begin{remark}
We can think of the bound in \eqref{eq:non-commutative-bound} as a
\textquotedblleft non-commutative union bound\textquotedblright\ because it is
analogous to the following union bound from probability theory:%
\begin{equation}
\Pr\left\{  \left(  A_{1}\cap\cdots\cap A_{N}\right)  ^{c}\right\}
=\Pr\left\{  A_{1}^{c}\cup\cdots\cup A_{N}^{c}\right\}  \leq\sum_{i=1}^{N}%
\Pr\left\{  A_{i}^{c}\right\}  ,
\end{equation}
where $A_{1}$, \ldots, $A_{N}$ are events. The analogous bound for projector
logic would be%
\begin{equation}
\operatorname{Tr}\left\{  \left(  I-P_{1}\cdots P_{N}\cdots P_{1}\right)
\rho\right\}  \leq\sum_{i=1}^{N}\operatorname{Tr}\left\{  \left(
I-P_{i}\right)  \rho\right\}  ,
\end{equation}
if we think of $P_{1}\cdots P_{N}$ as a projector onto the intersection of
subspaces. However, the above bound only holds if the projectors $P_{1}$,
\ldots, $P_{N}$ are commuting (choosing $P_{1}=|+\rangle\langle+|$,
$P_{2}=|0\rangle\langle0|$, and $\rho=|0\rangle\langle0|$ gives a
counterexample). If the projectors are non-commuting, then the non-commutative
union bound seems to be the next best thing and suffices for our purposes here.
\end{remark}

Continuing, we then apply the non-commutative union bound to the expression in
\eqref{eq-pack:seq-err-term-2}\ with $\omega=\Pi\sigma_{C_{m}}\Pi$ and the
sequential projectors as $\Pi_{C_{m}}$, $\hat{\Pi}_{C_{m-1}}$, \ldots,
$\hat{\Pi}_{C_{1}}$. This gives the following upper bound on
\eqref{eq-pack:seq-err-term-2}:%
\begin{multline}
\mathbb{E}_{\mathcal{C}}\left\{  \frac{1}{\left\vert \mathcal{M}\right\vert
}\sum_{m}2\left[  \text{Tr}\left\{  \left(  I-\Pi_{C_{m}}\right)  \Pi
\sigma_{C_{m}}\Pi\right\}  +\sum_{i=1}^{m-1}\text{Tr}\left\{  \Pi_{C_{i}}%
\Pi\sigma_{C_{m}}\Pi\right\}  \right]  ^{1/2}\right\} \\
\leq2\left[  \mathbb{E}_{\mathcal{C}}\left\{  \frac{1}{\left\vert
\mathcal{M}\right\vert }\sum_{m}\text{Tr}\left\{  \left(  I-\Pi_{C_{m}%
}\right)  \Pi\sigma_{C_{m}}\Pi\right\}  +\sum_{i\neq m}\text{Tr}\left\{
\Pi_{C_{i}}\Pi\sigma_{C_{m}}\Pi\right\}  \right\}  \right]  ^{1/2},
\end{multline}
where the inequality follows from concavity of the square root function and by
summing over all of the codewords not equal to the $m$th codeword (these terms
can only increase the expression). At this point, we have two error terms to
analyze that are essentially the same as those in \eqref{eq:pack-bound}. Thus,
we can invoke the analysis from before to conclude that%
\begin{align}
\text{Tr}\left\{  \left(  I-\Pi_{c_{m}}\right)  \Pi\sigma_{c_{m}}\Pi\right\}
&  \leq\varepsilon+2\sqrt{\varepsilon},\\
\mathbb{E}_{\mathcal{C}}\left\{  \frac{1}{\left\vert \mathcal{M}\right\vert
}\sum_{m}\sum_{i\neq m}\text{Tr}\left\{  \Pi_{C_{i}}\Pi\sigma_{C_{m}}%
\Pi\right\}  \right\}   &  \leq\left\vert \mathcal{M}\right\vert \frac{d}{D},
\end{align}
which leads to a final bound of%
\begin{multline}
1-\mathbb{E}_{\mathcal{C}}\left\{  \frac{1}{\left\vert \mathcal{M}\right\vert
}\sum_{m}\operatorname{Tr}\left\{  \Pi_{C_{m}}\hat{\Pi}_{C_{m-1}}\cdots
\hat{\Pi}_{C_{1}}\Pi\sigma_{C_{m}}\Pi\hat{\Pi}_{C_{1}}\cdots\hat{\Pi}%
_{C_{m-1}}\Pi_{C_{m}}\right\}  \right\} \\
\leq2\sqrt{\varepsilon+2\sqrt{\varepsilon}+\left\vert \mathcal{M}\right\vert
\frac{d}{D}}+\varepsilon.
\end{multline}
We can then derandomize and expurgate codewords as before to establish the
existence of a code and a sequential decoder with maximum error probability no
larger than $4(\varepsilon+2\sqrt{\varepsilon}+2\left\vert \mathcal{M}%
\right\vert d/D)^{1/2}+2\varepsilon$.

\begin{exercise}
Following Exercise~\ref{ex-pack:no-code-subspace-proj}, show that a variation
of the sequential decoder, in which there is no initial code subspace
projection, works well for decoding the messages transmitted.
\end{exercise}

\begin{exercise}
Show that the sequential decoding strategy works when assuming only the
conditions in Exercise~\ref{ex-pack:averaged-conditions}.
\end{exercise}

\section{History and Further Reading}

\cite{Hol98} and \cite{PhysRevA.56.131} did not prove the classical coding
theorem using the packing lemma, but they instead used other arguments to bound
the probability of error. The operator inequality in \eqref{eq:nagoaka} is at
the heart of the packing lemma.\ \cite{HN03} proved this operator inequality
in order to develop the more general setting of the quantum information
spectrum method, where there is no i.i.d.~constraint and essentially no
structure to a channel. \cite{itit2008hsieh} later exploited this operator
inequality in the context of entanglement-assisted classical coding and
followed the approach in \cite{HN03} to prove the packing lemma.

\cite{PhysRevA.85.012302,PhysRevLett.106.250501} introduced the method of
sequential decoding to quantum information theory. \cite{S11} proved the
non-commutative union bound and applied it to various problems in quantum
information theory. \cite{Wilde20130259} extended the non-commutative union
bound to non-projective tests (ones of the form $\{\Lambda,I-\Lambda\}$ for
$\Lambda\geq0$) and established a \textquotedblleft one-shot\textquotedblright%
\ characterization of sequential decoding for classical communication.

\chapter{The Covering Lemma}

\label{chap:covering-lemma}The goal of the covering lemma
\index{covering lemma}%
is perhaps opposite to that of the packing lemma because it applies to a
setting in which one party wishes to make messages \textit{indistinguishable}
to another party (instead of trying to make them distinguishable, as in the
packing lemma of the previous chapter). That is, the covering lemma is helpful
when one party is trying to simulate a noisy channel to another party, rather
than trying to simulate a noiseless channel. One party can accomplish this
task by randomly \textquotedblleft covering\textquotedblright\ the Hilbert
space of the other party (this viewpoint gives the covering lemma its name).

One can certainly simulate noise by choosing a quantum state uniformly at
random from a large set of quantum states and passing along the chosen quantum
state to a third party without indicating which state was chosen. But the problem
with this approach is that it could potentially be expensive if the set from
which we choose a random state is large, and we would really like to use as
few resources as possible in order to simulate noise. That is, we would like
the set from which we choose a quantum state uniformly at random to be as
small as possible when simulating noise. The covering lemma is similar to the
packing lemma in the sense that its conditions for application are general
(involving bounds on projectors and an ensemble), and it gives an effective
scheme for simulating noise when we apply it in an i.i.d.~setting.

One application of the covering lemma in quantum Shannon theory is in the
construction of a code for transmitting private classical information over a
quantum channel (discussed in Chapter~\ref{chap:private-cap}). The method of
proof for private classical transmission involves a clever combination of
packing messages so that Bob can distinguish them, while covering Eve's space
in such a way that Eve cannot distinguish the messages intended for Bob. A few
other applications of the covering lemma are in secret key distillation,
determining the amount of noise needed to destroy correlations in a bipartite
state, and compressing the outcomes of an i.i.d.~measurement on an
i.i.d.~quantum state.

We begin this chapter with a simple example to explain the main idea behind
the covering lemma. Section~\ref{sec-cl:setting-statement}\ then discusses its
general setting and gives its statement. We dissect its proof into several
different parts: the construction of a \textquotedblleft Chernoff
ensemble,\textquotedblright\ the construction of a \textquotedblleft Chernoff
code,\textquotedblright\ the application of the operator Chernoff bound, and
the error analysis.

The main tool that we use to prove the covering lemma is the operator Chernoff
bound. This bound is a generalization of the standard Chernoff bound from
probability theory, which states that the sample mean of a sequence of
i.i.d.~random variables converges exponentially fast to its true mean. A proof
of the operator version of the Chernoff bound appears in
Section~\ref{sec-cov:proof-op-chernoff}. The exponential convergence in the
Chernoff bound is much stronger than the polynomial convergence from
Chebyshev's inequality and is helpful for establishing the existence of good
private classical codes in Chapter~\ref{chap:private-cap}.

\section{Introductory Example}

Suppose that Alice is trying to communicate with Bob as before, but now there
is an eavesdropper Eve listening in on their communication. Alice wants the
messages that she is sending to Bob to be \textit{private} so that Eve does
not gain any information about the message that she is sending.

How can Alice make the information that she is sending private? The strongest
criterion for security is to ensure that whatever Eve receives is independent
of what Alice is sending. Alice may have to sacrifice the amount of
information she can communicate to Bob in order to have privacy, but this
sacrifice is worth it to her because she really does not want Eve to know
anything about the intended message for Bob.

We first give an example to motivate a general method that Alice can use to
make her information private. Suppose Alice can transmit one of four messages
$\left\{  a,b,c,d\right\}  $ to Bob, and suppose he receives them perfectly as
distinguishable quantum states. She chooses from these messages with equal
probability. Suppose further that Alice and Eve know that Eve receives one of
the following four states corresponding to each of Alice's messages:%
\begin{equation}
a\rightarrow|0\rangle,\ \ \ \ \ \ \ \ b\rightarrow|1\rangle
,\ \ \ \ \ \ \ \ c\rightarrow|+\rangle,\ \ \ \ \ \ \ \ d\rightarrow|-\rangle.
\label{eq:cover-orig-ensemble}%
\end{equation}
Observe that each of Eve's states lies in the two-dimensional Hilbert space of
a qubit. We refer to the quantum states in the above ensemble as
\textquotedblleft Eve's ensemble.\textquotedblright

We are not so much concerned for what Bob receives for the purposes of this
example, but we just make the assumption that he can distinguish the four
messages that Alice sends. Without loss of generality, let us just assume that
he receives the messages unaltered in some preferred orthonormal basis such as
$\left\{  \left\vert a\right\rangle ,\left\vert b\right\rangle ,\left\vert
c\right\rangle ,\left\vert d\right\rangle \right\}  $ so that he can
distinguish the four messages, and let us call this ensemble \textquotedblleft
Bob's ensemble.\textquotedblright

Both Alice and Eve then know that the expected density operator of Eve's
ensemble is the maximally mixed state if Eve does not know which message Alice
chooses:%
\begin{equation}
\frac{1}{4}\vert0\rangle\langle0\vert+\frac{1}{4}\vert1\rangle\langle
1\vert+\frac{1}{4}\vert+\rangle\langle+\vert+\frac{1}{4}\left\vert
-\right\rangle \langle-\vert=\pi,
\end{equation}
where $\pi\equiv I/2$ is the maximally mixed state of a qubit. How can Alice
ensure that Eve's information is independent of the message Alice is sending?
Alice can choose subsets or subensembles of the states in Eve's ensemble to
simulate the expected density operator of Eve's ensemble. Let us call these
new simulating ensembles the \textquotedblleft fake
ensembles.\textquotedblright\ Alice chooses the member states of the fake
ensembles according to the uniform distribution in order to randomize Eve's
knowledge. The density operator for each new fake ensemble is its
\textquotedblleft fake expected density operator.\textquotedblright

Which states work well for being members of the fake ensembles?\ An
equiprobable mixture of the states $|0\rangle$ and $|1\rangle$ suffices to
simulate the expected density operator of Eve's ensemble because the fake
expected density operator of this new ensemble is as follows: $\left[
|0\rangle\langle0|+|1\rangle\langle1|\right]  /2=\pi$. An equiprobable mixture
of the states $|+\rangle$ and $\left\vert -\right\rangle $ also works because
the fake expected density operator of this other fake ensemble is as follows:
$\left[  |+\rangle\langle+|+|-\rangle\langle-|\right]  /2=\pi$.

So it is possible for Alice to encode a private bit in this way. She first
generates a random bit that selects a particular message within each fake
ensemble. So she selects $a$ or $b$ according to the random bit if she wants
to transmit a \textquotedblleft0\textquotedblright\ privately to Bob, and she
selects $c$ or $d$ according to the random bit if she wants to transmit a
\textquotedblleft1\textquotedblright\ privately to Bob. In each of these
cases, Eve's resulting expected density operator is the maximally mixed state.
Thus, there is no measurement that Eve can perform to distinguish the original
message that Alice transmits, in the sense that she cannot do better than a
random guessing strategy. Bob, on the other hand, can perform a measurement in
the basis $\left\{  \left\vert a\right\rangle ,\left\vert b\right\rangle
,\left\vert c\right\rangle ,\left\vert d\right\rangle \right\}  $ to determine
Alice's private bit. In the case in which one private bit is being
transmitted, Eve can guess its value correctly with probability $1/2$, but
Alice and Bob can make this probability exponentially small if Alice sends
more private bits using this technique (the guessing probability is $2^{-n}$
if $n$ private bits are transmitted in this way).

We can explicitly calculate Eve's information about the private bit. Consider
Eve's description of the state if she does not know which message Alice
transmits---it is an equal mixture of the following states: $\left\{
|0\rangle,|1\rangle,|+\rangle,|-\rangle\right\}  $ (equal to the maximally
mixed state $\pi$). Eve's description of the state \textquotedblleft
improves\textquotedblright\ to an equal mixture of the states $\left\{
|0\rangle,|1\rangle\right\}  $ or $\left\{  |+\rangle,|-\rangle\right\}  $,
each having density operator $\pi$, if she does know which message Alice
transmits. The following classical--quantum state describes this setting:%
\begin{multline}
\rho_{MKE}\equiv\frac{1}{4}|0\rangle\langle0|_{M}\otimes|0\rangle\langle
0|_{K}\otimes|0\rangle\langle0|_{E}+\frac{1}{4}|0\rangle\langle0|_{M}%
\otimes|1\rangle\langle1|_{K}\otimes|1\rangle\langle1|_{E}\\
+\frac{1}{4}|1\rangle\langle1|_{M}\otimes|0\rangle\langle0|_{K}\otimes
|+\rangle\langle+|_{E}+\frac{1}{4}|1\rangle\langle1|_{M}\otimes|1\rangle
\langle1|_{K}\otimes|-\rangle\langle-|_{E},
\end{multline}
where we suppose that Eve never has access to the $K$ register. Tracing over
the register $K$ gives the reduced state $\rho_{ME}$:%
\begin{align}
\rho_{ME}  &  =\frac{1}{4}|0\rangle\langle0|_{M}\otimes|0\rangle\langle
0|_{E}+\frac{1}{4}|0\rangle\langle0|_{M}\otimes|1\rangle\langle1|_{E}%
\nonumber\\
&  \ \ \ \ +\frac{1}{4}|1\rangle\langle1|_{M}\otimes|+\rangle\langle
+|_{E}+\frac{1}{4}|1\rangle\langle1|_{M}\otimes|-\rangle\langle-|_{E}\\
&  =\frac{1}{2}|0\rangle\langle0|_{M}\otimes\frac{1}{2}\left[  |0\rangle
\langle0|_{E}+|1\rangle\langle1|_{E}\right]  +\frac{1}{2}|1\rangle
\langle1|_{M}\otimes\frac{1}{2}\left[  |+\rangle\langle+|_{E}+|-\rangle
\langle-|_{E}\right] \\
&  =\frac{1}{2}|0\rangle\langle0|_{M}\otimes\pi_{E}+\frac{1}{2}|1\rangle
\langle1|_{M}\otimes\pi_{E}\\
&  =\pi_{M}\otimes\pi_{E}.
\end{align}
Then Eve's register is completely independent of the private bit in $M$ and
her information about the private bit in $M$ is given by evaluating the mutual
information of the reduced state $\rho_{ME}$: $I(M;E)_{\rho}=0$, because the
state is product. Thus, using this scheme, Eve has no information about the
private bit as we argued before.

We are interested in making this scheme use as little noise as possible
because Alice would like to transmit as much information as she can to Bob
while still retaining privacy. Therefore, Alice should try to make the fake
ensembles use as little randomness as possible. In the above example, Alice
cannot make the fake ensembles any smaller because a smaller size would leak
information to Eve.

\section{Setting and Statement of the Covering Lemma}

\label{sec-cl:setting-statement}The setting of the covering lemma is a
generalization of the setting in the above example. It essentially uses the
same strategy for making information private, but the mathematical analysis
becomes more involved in the more general setting. In general, we cannot have
perfect privacy as in the above example, but instead we ask only for
approximate privacy. Approximate privacy then becomes perfect in the
asymptotic limit in the i.i.d.~setting.

We first define the relevant ensemble for the covering lemma. We call it the
\textquotedblleft true ensemble\textquotedblright\ in order to distinguish it
from the \textquotedblleft fake ensemble.\textquotedblright

\begin{definition}
[True Ensemble]\label{def:true-ensemble}Suppose $\mathcal{X}$ is a set of size
$\left\vert \mathcal{X}\right\vert $ with elements $x$. Suppose we have an
ensemble $\left\{  p_{X}(x),\sigma_{x}\right\}  _{x\in\mathcal{X}}$\ of
quantum states where each value $x$ occurs with probability $p_{X}(x)$
according to some random variable $X$, and suppose we encode each value $x$
into a quantum state $\sigma_{x}\in\mathcal{D}(\mathcal{H})$. The expected
density operator of the ensemble is $\sigma\equiv\sum_{x\in\mathcal{X}}%
p_{X}(x)\sigma_{x}$.
\end{definition}

\noindent The definition for a fake ensemble is similar to the way that we
constructed the fake ensembles in the example. It is merely a subset of the
states in the true ensemble chosen according to a uniform distribution.

\begin{definition}
[Fake Ensemble]Let $\mathcal{S}$ be a set such that $\mathcal{S}%
\subseteq\mathcal{X}$. The fake ensemble is defined as follows:%
\begin{equation}
\left\{  1/\left\vert \mathcal{S}\right\vert ,\sigma_{s}\right\}
_{s\in\mathcal{S}}.
\end{equation}
Let $\overline{\sigma}$ denote the \textquotedblleft fake expected density
operator\textquotedblright\ of the fake ensemble:%
\begin{equation}
\overline{\sigma}(\mathcal{S})\equiv\frac{1}{\left\vert \mathcal{S}\right\vert
}\sum_{s\in\mathcal{S}}\sigma_{s}.
\end{equation}

\end{definition}

In the example, Alice was able to obtain perfect privacy from Eve. We need a
good measure of privacy because it is not possible in general to obtain
perfect privacy, but Alice can instead obtain only approximate privacy. We
call this measure the \textquotedblleft obfuscation error\textquotedblright%
\ because it determines how well Alice can obfuscate the state that Eve receives.

\begin{definition}
[Obfuscation Error]The obfuscation error $o_{e}(\mathcal{S})$ of set
$\mathcal{S}$ is a measure of how close the fake expected density operator
$\overline{\sigma}(\mathcal{S})$ is to the actual expected density operator:%
\begin{equation}
o_{e}(\mathcal{S})=\left\Vert \overline{\sigma}(\mathcal{S})-\sigma\right\Vert
_{1}.
\end{equation}

\end{definition}

The goal for Alice is to make the size of her fake ensembles as small as
possible while still having privacy from Eve. The covering lemma quantifies
this trade-off by determining how small each fake ensemble can be in
order to obtain a certain obfuscation error.

The hypotheses of the covering lemma are somewhat similar to those of the
packing lemma. But as stated in the introduction to this chapter, the goal of
the covering lemma is much different.

\begin{lemma}
[Covering Lemma]\label{lemma-cov:covering}Let $\left\{  p_{X}(x),\sigma
_{x}\right\}  _{x\in\mathcal{X}}$ be an ensemble as in
Definition~\ref{def:true-ensemble}. Suppose a total subspace projector $\Pi$
and codeword subspace projectors $\left\{  \Pi_{x}\right\}  _{x\in\mathcal{X}%
}$ are given, they project onto subspaces of $\mathcal{H}$, and these projectors
and each state $\sigma_{x}$ satisfy the following conditions:%
\begin{align}
\operatorname{Tr}\left\{  \sigma_{x}\Pi\right\}   &  \geq1-\varepsilon
,\label{eq-cov:cov-1}\\
\operatorname{Tr}\left\{  \sigma_{x}\Pi_{x}\right\}   &  \geq1-\varepsilon
,\label{eq-cov:cov-2}\\
\operatorname{Tr}\left\{  \Pi\right\}   &  \leq D,\label{eq-cov:cov-3}\\
\Pi_{x}\sigma_{x}\Pi_{x}  &  \leq\frac{1}{d}\Pi_{x}, \label{eq-cov:cov-4}%
\end{align}
where $\varepsilon\in(0,1)$, $D>0$, and $d\in(0,D)$. Suppose that
$\mathcal{M}$ is a set of size $\left\vert \mathcal{M}\right\vert $ with
elements $m$. Let a random covering code $\mathcal{C}\equiv\left\{
C_{m}\right\}  _{m\in\mathcal{M}}$ consist of random codewords $C_{m}$
where\ the codewords $C_{m}$ are chosen independently according to the distribution
$p_{X}(x)$ and give rise to a fake ensemble $\left\{  1/\left\vert
\mathcal{M}\right\vert ,\sigma_{C_{m}}\right\}  _{m\in\mathcal{M}}$. Then
there is a high probability that the obfuscation error $o_{e}(\mathcal{C})$ of
the random covering code $\mathcal{C}$ is small:%
\begin{equation}
\Pr_{\mathcal{C}}\left\{  o_{e}(\mathcal{C})\leq\varepsilon+4\sqrt
{\varepsilon}+24\sqrt[4]{\varepsilon}\right\}  \geq1-2D\exp\left(
-\frac{\varepsilon^{3}}{4}\frac{\left\vert \mathcal{M}\right\vert d}%
{D}\right)  ,
\end{equation}
when $\varepsilon$ is small and $\left\vert \mathcal{M}\right\vert
\gg\varepsilon^{3}d/D$. Thus, it is highly likely that a given fake ensemble
$\left\{  1/\left\vert \mathcal{M}\right\vert ,\sigma_{c_{m}}\right\}
_{m\in\mathcal{M}}$\ has its expected density operator indistinguishable from
the expected density operator of the original ensemble $\left\{
p_{X}(x),\sigma_{x}\right\}  _{x\in\mathcal{X}}$. It is in this sense that the
fake ensemble $\left\{  1/\left\vert \mathcal{M}\right\vert ,\sigma_{c_{m}%
}\right\}  _{m\in\mathcal{M}}$ \textquotedblleft covers\textquotedblright\ the
original ensemble $\left\{  p_{X}(x),\sigma_{x}\right\}  _{x\in\mathcal{X}}$.
\end{lemma}

\section{Operator Chernoff Bound}

\label{sec-cov:proof-op-chernoff}Before giving the proof of the covering
lemma, we first state and prove the operator Chernoff bound, which is the most
critical tool for establishing the covering lemma. The operator Chernoff bound
is a theorem from the theory of large deviations and essentially states that
the sample average of a large number of i.i.d.~operator-valued random
variables is close to their expectation (with some constraints on the
operator-valued random variables).

\begin{lemma}
[Operator Chernoff Bound]\label{lemma:operator-chernoff}Let $\xi_{1}%
,\ldots,\xi_{K}\in\mathcal{L}(\mathcal{H})$ be $K$\ i.i.d.~positive
semi-definite operator-valued random%
\index{Operator Chernoff Bound}
variables. Suppose that each $\xi_{k}$ has all of its eigenvalues between zero
and one:%
\begin{equation}
\forall k\in\left[  K\right]  :0\leq\xi_{k}\leq I.
\end{equation}
Let $\overline{\xi}$ denote the sample average of the $K$ operator-valued
random variables:%
\begin{equation}
\overline{\xi}=\frac{1}{K}\sum_{k=1}^{K}\xi_{k}.
\end{equation}
Suppose that the expectation $\mathbb{E}_{\xi}\left\{  \xi_{k}\right\}
\equiv\mu$\ of each operator $\xi_{k}$ is positive definite, so that $\mu$
exceeds the identity operator scaled by a number $a\in\left(  0,1\right)  $: $
\mu\geq aI.
$
Then for every $\eta$ where $0<\eta<1/2$ and $\left(  1+\eta\right)  a\leq1$,
we can bound the probability that the sample average $\overline{\xi}$\ lies
inside the operator interval $\left[  \left(  1\pm\eta\right)  \mu\right]  $:%
\begin{equation}
\Pr_{\xi}\left\{  \left(  1-\eta\right)  \mu\leq\overline{\xi}\leq\left(
1+\eta\right)  \mu\right\}  \geq1-2\dim(\mathcal{H})\exp\left(  -\frac
{K\eta^{2}a}{4}\right)  .
\end{equation}
Thus it is highly likely that the sample average operator $\overline{\xi}$
becomes close to the true expected operator $\mu$ as $K$ becomes large.
\end{lemma}

We prove the above lemma by making a progression through the operator Markov
inequality all the way to the proof of the operator Chernoff bound. Recall
that we write $A\geq B$ if $A-B$ is a positive semi-definite operator, and we
write $A\ngeq B$ otherwise.
In what follows, we take the convention 
$\exp\{A\} = \sum_{i} \exp(a_i) \vert i \rangle \langle i \vert$ for a Hermitian operator $A$ with spectral decomposition $A = \sum_{i} a_i \vert i \rangle \langle i \vert$ (this differs from our usual convention $\exp\{A\} = \sum_{i:a_i\neq 0} \exp(a_i) \vert i \rangle \langle i \vert$).

\begin{lemma}
[Operator Markov Inequality]\label{lem-cover:op-Markov}Let $X\in
\mathcal{L}(\mathcal{H})$ be a positive semi-definite operator-valued random
variable. Let $\mathbb{E}\left\{  X\right\}  $ denote its expectation. Let $A$
be a fixed positive definite operator in $\mathcal{L}(\mathcal{H})$. Then%
\begin{equation}
\Pr\left\{  X\nleq A\right\}  \leq\operatorname{Tr}\left\{  \mathbb{E}\left\{
X\right\}  A^{-1}\right\}  .
\end{equation}

\end{lemma}

\begin{proof}
Observe that if $X\nleq A$ then $A^{-1/2}XA^{-1/2}\nleq I$. This then implies
that the largest eigenvalue of $A^{-1/2}XA^{-1/2}$ exceeds one:\ $\left\Vert
A^{-1/2}XA^{-1/2}\right\Vert _{\infty}>1$. Let $I_{X\nleq A}$ denote an
indicator function for the event $X\nleq A$. We then have that%
\begin{equation}
I_{X\nleq A}\leq\operatorname{Tr}\left\{  A^{-1/2}XA^{-1/2}\right\}  .
\label{eq-app:indicator-inequality}%
\end{equation}
The above inequality follows because the right-hand side\ is non-negative if
the indicator is zero. If the indicator is one, then the right-hand
side\ exceeds one because its largest eigenvalue is greater than one and the
trace exceeds the largest eigenvalue for a positive semi-definite operator. We
then have the following inequalities:%
\begin{align}
\Pr\left\{  X\nleq A\right\}   &  =\mathbb{E}\left\{  I_{X\nleq A}\right\}
\leq\mathbb{E}\left\{  \operatorname{Tr}\left\{  A^{-1/2}XA^{-1/2}\right\}
\right\} \\
&  =\mathbb{E}\left\{  \operatorname{Tr}\left\{  XA^{-1}\right\}  \right\}
=\operatorname{Tr}\left\{  \mathbb{E}\left\{  X\right\}  A^{-1}\right\}  ,
\end{align}
where the inequality follows from \eqref{eq-app:indicator-inequality} and the
second equality from cyclicity of trace.
\end{proof}

\begin{lemma}
[Bernstein Trick]\label{lem-app:bernstein-trick}Let $X$, $X_{1}$, \ldots,
$X_{K}\in\mathcal{L}(\mathcal{H})$ be i.i.d.~Hermitian operator-valued random
variables, and let $A$ be a fixed Hermitian operator. Then for any invertible
operator $T$, the following bound holds%
\begin{equation}
\Pr\left\{  \sum_{k=1}^{K}X_{k}\nleq KA\right\}  \leq\dim(\mathcal{H}%
)\left\Vert \mathbb{E}\left\{  \exp\left\{  T\left(  X-A\right)  T^{\dag
}\right\}  \right\}  \right\Vert _{\infty}^{K}.
\end{equation}

\end{lemma}

\begin{proof}
The proof of this lemma relies on the Golden--Thompson trace inequality from
statistical mechanics, which holds for any two Hermitian operators $A$ and $B$
(we state it without proof):%
\begin{equation}
\operatorname{Tr}\left\{  \exp\left\{  A+B\right\}  \right\}  \leq
\operatorname{Tr}\left\{  \exp\left\{  A\right\}  \exp\left\{  B\right\}
\right\}  .
\end{equation}
Consider the following chain of inequalities:%
\begin{align}
\Pr\left\{  \sum_{k=1}^{K}X_{k}\nleq KA\right\}   &  =\Pr\left\{  \sum
_{k=1}^{K}\left(  X_{k}-A\right)  \nleq0\right\} \\
&  =\Pr\left\{  \sum_{k=1}^{K}T\left(  X_{k}-A\right)  T^{\dag}\nleq0\right\}
\\
&  =\Pr\left\{  \exp\left\{  \sum_{k=1}^{K}T\left(  X_{k}-A\right)  T^{\dag
}\right\}  \nleq I\right\} \\
&  \leq\operatorname{Tr}\left\{  \mathbb{E}\left\{  \exp\left\{  \sum
_{k=1}^{K}T\left(  X_{k}-A\right)  T^{\dag}\right\}  \right\}  \right\}  .
\end{align}
The first two equalities are straightforward and the third follows because
$A\leq B$ is equivalent to $\exp\left\{  A\right\}  \leq\exp\left\{
B\right\}  $ for commuting operators $A$ and $B$. The inequality follows by
applying the operator Markov inequality (Lemma~\ref{lem-cover:op-Markov}).
Continuing, we have%
\begin{align}
&  =\mathbb{E}\left\{  \operatorname{Tr}\left\{  \exp\left\{  \sum_{k=1}%
^{K}T\left(  X_{k}-A\right)  T^{\dag}\right\}  \right\}  \right\} \\
&  \leq\mathbb{E}\left\{  \operatorname{Tr}\left\{  \exp\left\{  \sum
_{k=1}^{K-1}T\left(  X_{k}-A\right)  T^{\dag}\right\}  \exp\left\{  T\left(
X_{K}-A\right)  T^{\dag}\right\}  \right\}  \right\} \\
&  =\mathbb{E}_{X_{1},\ldots,X_{K-1}}\left\{  \operatorname{Tr}\left\{
\exp\left\{  \sum_{k=1}^{K-1}T\left(  X_{k}-A\right)  T^{\dag}\right\}
\mathbb{E}_{X_{K}}\left\{  \exp\left\{  T\left(  X_{K}-A\right)  T^{\dag
}\right\}  \right\}  \right\}  \right\} \\
&  =\mathbb{E}_{X_{1},\ldots,X_{K-1}}\left\{  \operatorname{Tr}\left\{
\exp\left\{  \sum_{k=1}^{K-1}T\left(  X_{k}-A\right)  T^{\dag}\right\}
\mathbb{E}_{X}\left\{  \exp\left\{  T\left(  X-A\right)  T^{\dag}\right\}
\right\}  \right\}  \right\}  .
\end{align}
The first equality follows from exchanging the expectation and the trace. The
inequality follows from applying the Golden--Thompson trace inequality. The second
and third equalities follow from the i.i.d.~assumption. Continuing,%
\begin{align}
&  \leq\mathbb{E}_{X_{1},\ldots,X_{K-1}}\left\{  \operatorname{Tr}\left\{
\exp\left\{  \sum_{k=1}^{K-1}T\left(  X_{k}-A\right)  T^{\dag}\right\}
\right\}  \right\}  \left\Vert \mathbb{E}_{X}\left\{  \exp\left\{  T\left(
X-A\right)  T^{\dag}\right\}  \right\}  \right\Vert _{\infty}\\
&  \leq\operatorname{Tr}\left\{  I\right\}  \left\Vert \mathbb{E}_{X}\left\{
\exp\left\{  T\left(  X-A\right)  T^{\dag}\right\}  \right\}  \right\Vert
_{\infty}^{K}\\
&  =\dim(\mathcal{H})\left\Vert \mathbb{E}_{X}\left\{  \exp\left\{  T\left(
X-A\right)  T^{\dag}\right\}  \right\}  \right\Vert _{\infty}^{K}.
\end{align}
The first inequality follows from $\operatorname{Tr}\left\{  AB\right\}
\leq\operatorname{Tr}\left\{  A\right\}  \left\Vert B\right\Vert _{\infty}$
for $A$ positive semi-definite. The second inequality follows from a repeated
application of the same steps. The final equality follows because the trace of
the identity operator is equal to the dimension of the Hilbert space. This
proves the \textquotedblleft Bernstein trick\textquotedblright\ lemma.
\end{proof}

\bigskip

\begin{proof}
[Proof of the Operator Chernoff Bound (Lemma~\ref{lemma:operator-chernoff})]We
first prove that the following inequality holds for i.i.d.~Hermitian
operator-valued random variables $X$, $X_{1}$, \ldots, $X_{K}$ such that
$\mathbb{E}\left\{  X\right\}  \leq mI$, $A\geq aI$, and $1\geq a>m\geq0$:%
\begin{equation}
\Pr\left\{  \sum_{k=1}^{K}X_{k}\nleq KA\right\}  \leq\dim(\mathcal{H}%
)\exp\left\{  -KD(a\Vert m)\right\}  , \label{eq-app:oper-cher-1}%
\end{equation}
where $D(a\Vert m)$ is the binary relative entropy:%
\begin{equation}
D(a\Vert m)=a\ln a-a\ln m+\left(  1-a\right)  \ln\left(  1-a\right)  -\left(
1-a\right)  \ln\left(  1-m\right)  .
\end{equation}
We first apply the Bernstein Trick (Lemma~\ref{lem-app:bernstein-trick}) with
$T=\sqrt{t}I$, for $t>0$:%
\begin{align}
\Pr\left\{  \sum_{k=1}^{K}X_{k}\nleq KA\right\}   &  \leq\Pr\left\{
\sum_{k=1}^{K}X_{k}\nleq KaI\right\} \\
&  \leq\dim(\mathcal{H})\left\Vert \mathbb{E}\left\{  \exp\left\{  tX\right\}
\exp\left\{  -ta\right\}  \right\}  \right\Vert _{\infty}^{K}.
\end{align}
So it is clear that it is best to optimize $t$ in such a way that%
\begin{equation}
\left\Vert \mathbb{E}\left\{  \exp\left\{  tX\right\}  \exp\left\{
-ta\right\}  \right\}  \right\Vert _{\infty}<1,
\end{equation}
so that we have exponential decay with increasing $K$. Now consider the
following inequality:%
\begin{equation}
\exp\left\{  tX\right\}  -I\leq X\left(  \exp\left\{  t\right\}  -1\right)  ,
\end{equation}
which holds because a similar one holds for all real $x\in\left[  0,1\right]
$:%
\begin{equation}
\left(  \exp\left\{  tx\right\}  -1\right)  \leq x (\exp\left\{
t\right\}  -1).
\end{equation}
Applying this inequality gives%
\begin{align}
\mathbb{E}\left\{  \exp\left\{  tX\right\}  \right\}   &  \leq\mathbb{E}%
\left\{  X\right\}  \left(  \exp\left\{  t\right\}  -1\right)  +I\\
&  \leq mI\left(  \exp\left\{  t\right\}  -1\right)  +I\\
&  =\left(  m\exp\left\{  t\right\}  +1-m\right)  I.
\end{align}
which in turn implies%
\begin{equation}
\left\Vert \mathbb{E}\left\{  \exp\left\{  tX\right\}  \exp\left\{
-ta\right\}  \right\}  \right\Vert _{\infty}\leq\left(  m\exp\left\{
t\right\}  +1-m\right)  \exp\left\{  -ta\right\}  .
\end{equation}
Choosing%
\begin{equation}
t=\ln\left(  \frac{a}{m}\cdot\frac{1-m}{1-a}\right)  >0,
\end{equation}
which follows from the assumption that $a>m$, gives%
\begin{align}
&  \left(  m\exp\left\{  t\right\}  +1-m\right)  \exp\left\{  -ta\right\}
\nonumber\\
&  =\left(  m\left(  \frac{a}{m}\cdot\frac{1-m}{1-a}\right)  +1-m\right)
\exp\left\{  -\ln\left(  \frac{a}{m}\cdot\frac{1-m}{1-a}\right)  a\right\} \\
&  =\left(  a\cdot\frac{1-m}{1-a}+1-m\right)  \exp\left\{  -a\ln\left(
\frac{a}{m}\right)  -a\ln\left(  \frac{1-m}{1-a}\right)  \right\} \\
&  =\left(  \frac{1-m}{1-a}\right)  \exp\left\{  -a\ln\left(  \frac{a}%
{m}\right)  -a\ln\left(  \frac{1-m}{1-a}\right)  \right\} \\
&  =\exp\left\{  -a\ln\left(  \frac{a}{m}\right)  -\left(  1-a\right)
\ln\left(  \frac{1-a}{1-m}\right)  \right\} \\
&  =\exp\left\{  -D(a\Vert m)\right\}  ,
\end{align}
proving the desired bound in \eqref{eq-app:oper-cher-1}.

By substituting $Y_{k}=I-X_{k}$ and $B=I-A$ into \eqref{eq-app:oper-cher-1}
and having the opposite conditions $\mathbb{E}\left\{  X\right\}  \geq mI$,
$A\leq aI$, and $0\leq a<m\leq1$, we can show that the following inequality
holds for i.i.d.~operators $X$, $X_{1}$, \ldots, $X_{K}$:%
\begin{equation}
\Pr\left\{  \sum_{k=1}^{K}X_{k}\ngeq KA\right\}  \leq\dim(\mathcal{H}%
)\exp\left\{  -KD(a\Vert m)\right\}  . \label{eq-app:oper-cher-2}%
\end{equation}

To finish off the proof of the operator Chernoff bound, consider the variables
$Z_{k}=L\mu^{-1/2}X_{k}\mu^{-1/2}$ with $\mu\equiv\mathbb{E}\left\{
X\right\}  \geq LI$. Then $\mathbb{E}\left\{  Z_{k}\right\}  =LI$ and $0\leq
Z_{i}\leq I$. The following events are thus equivalent%
\begin{equation}
\left(  1-\eta\right)  \mu\leq\frac{1}{K}\sum_{k=1}^{K}X_{k}\leq\left(
1+\eta\right)  \mu\Longleftrightarrow\left(  1-\eta\right)  LI\leq\frac{1}%
{K}\sum_{k=1}^{K}Z_{k}\leq\left(  1+\eta\right)  LI,
\end{equation}
and we can apply \eqref{eq-app:oper-cher-1}, \eqref{eq-app:oper-cher-2}, and
the union bound to obtain%
\begin{align}
&  \Pr\left\{  \left(  \left(  1-\eta\right)  \mu\nleq\frac{1}{K}\sum
_{k=1}^{K}X_{k}\right)  \bigcup\left(  \frac{1}{K}\sum_{k=1}^{K}X_{k}%
\nleq\left(  1+\eta\right)  \mu\right)  \right\} \nonumber\\
&  \leq\dim(\mathcal{H})\exp\left\{  -KD(\left(  1-\eta\right)  L\Vert
L)\right\}  +\dim(\mathcal{H})\exp\left\{  -KD(\left(  1+\eta\right)  L\Vert
L)\right\} \\
&  \leq2\dim(\mathcal{H})\exp\left\{  -K\frac{\eta^{2}L}{4}\right\}  ,
\end{align}
where the last line exploits the following inequality valid for $-1/2\leq
\eta\leq1/2$ and $\left(  1+\eta\right)  L\leq1$:%
\begin{equation}
D(  \left(  1+\eta\right)  L\Vert L)  \geq\frac{\eta^{2}L}{4}.
\label{eq:calc-arg-Op-Chern-end}
\end{equation}
This last inequality follows from a calculus argument. To see it, define
$f(\eta)\equiv D(  \left(  1+\eta\right)  L\Vert L)$, and note that
$f'(\eta) = L \ln (1+\eta) - L \ln (1- \eta L / (1-L))$ and 
$f''(\eta) = L / [ (1+\eta)(1-  (1+\eta)L)]\geq L /  (1+\eta)$. By Taylor's theorem,
it follows that $f(\eta) = f(0) + \eta f'(0) + \tfrac{\eta^2}{2} f''(\xi)$, for some
$\xi \in [0,\eta]$. Since $f(0) = f'(0) = 0$, it follows that $f(\eta) = \tfrac{\eta^2}{2} f''(\xi)
\geq \tfrac{\eta^2 L}{2(1+\xi)}\geq \tfrac{\eta^2 L}{2(1+\eta)}$. Since $\eta < 1$, then $1/(1+\eta)\geq 1/2$,
implying that $\tfrac{\eta^2 L}{2(1+\eta)}\geq \tfrac{\eta^2 L}{4}$ and finally gives \eqref{eq:calc-arg-Op-Chern-end}.
This concludes the proof of Lemma~\ref{lemma:operator-chernoff}.
\end{proof}

\section{Proof of the Covering Lemma}

The first step of the proof of the covering lemma is to construct an alternate
ensemble that is close to the original ensemble yet satisfies the conditions
of the operator Chernoff bound (Lemma~\ref{lemma:operator-chernoff}). We call
this alternate ensemble the \textquotedblleft Chernoff
ensemble.\textquotedblright\ We then generate a random code, a set of
$M$\ i.i.d.~random variables, using the Chernoff ensemble. Call this random
code the \textquotedblleft Chernoff code.\textquotedblright\ We apply the
operator Chernoff bound to the Chernoff code to obtain a good bound on the
obfuscation error of the Chernoff code. We finally show that the bound holds
for a covering code generated by the original ensemble because the original
ensemble is close to the Chernoff ensemble in trace distance.

\subsection{Construction of the Chernoff Ensemble}

We first establish a few definitions to construct intermediary ensembles. We
then use these intermediary ensembles to construct the Chernoff ensemble. We
construct the first \textquotedblleft primed\textquotedblright\ ensemble
$\left\{  p_{X}(x),\sigma_{x}^{\prime}\right\}  $ by using the projection
operators $\Pi_{x}$ to slice out some of the support of the states $\sigma
_{x}$:%
\begin{equation}
\forall x\ \ \ \ \sigma_{x}^{\prime}\equiv\Pi_{x}\sigma_{x}\Pi_{x}.
\end{equation}
The above \textquotedblleft slicing\textquotedblright\ operation cuts outs any
part of the support of $\sigma_{x}$ that is not in the support of $\Pi_{x}$.
The expected operator $\sigma^{\prime}$ for the first primed ensemble is as
follows:%
\begin{equation}
\sigma^{\prime}\equiv\sum_{x\in\mathcal{X}}p_{X}(x)\sigma_{x}^{\prime}.
\end{equation}
We then continue slicing with the projector $\Pi$ and form the second primed
ensemble $\left\{  p_{X}(x),\sigma_{x}^{\prime\prime}\right\}  $ as follows:%
\begin{equation}
\forall x\ \ \ \ \sigma_{x}^{\prime\prime}\equiv\Pi\sigma_{x}^{\prime}\Pi.
\end{equation}
The expected operator for the second primed ensemble is as follows:%
\begin{equation}
\sigma^{\prime\prime}\equiv\sum_{x\in\mathcal{X}}p_{X}(x)\sigma_{x}%
^{\prime\prime}.
\end{equation}
Let $\hat{\Pi}$ be the projector onto the subspace spanned by the eigenvectors
of $\sigma^{\prime\prime}$ whose corresponding eigenvalues are greater than
$\varepsilon/D$. We would expect that this extra slicing does not change the
state very much when $D$ is large and $\varepsilon$ is small. We construct states $\omega_{x}$ in the
Chernoff ensemble by using the projector $\hat{\Pi}$ to slice out some more
elements of the support of the original ensemble:%
\begin{equation}
\forall x\ \ \ \ \omega_{x}\equiv\hat{\Pi}\sigma_{x}^{\prime\prime}\hat{\Pi}.
\end{equation}
The expected operator $\omega$\ for the Chernoff ensemble is then as follows:%
\begin{equation}
\omega\equiv\sum_{x\in\mathcal{X}}p_{X}(x)\omega_{x}.
\end{equation}
The Chernoff ensemble satisfies the conditions necessary to apply the operator
Chernoff bound. We wait to apply the operator Chernoff bound and for now show
how to construct a random covering code.

\subsection{Chernoff Code Construction}

We present a Shannon-like random coding argument. We construct a covering code
$\mathcal{C}$ at random by independently generating $\left\vert \mathcal{M}%
\right\vert $ codewords according to the distribution $p_{X}(x)$. Let
$\mathcal{C}=\left\{  c_{m}\right\}  _{m\in\mathcal{M}}$\ be a collection of
the realizations $c_{m}$ of $\left\vert \mathcal{M}\right\vert $\ independent
random variables $C_{m}$. Each $C_{m}$ takes a value $c_{m}$ in $\mathcal{X}$
with probability $p_{X}(c_{m})$ and represents a codeword in the random code
$\mathcal{C}$. This process generates the Chernoff code $\mathcal{C}%
$\ consisting of $\left\vert \mathcal{M}\right\vert $ quantum states $\left\{
\omega_{c_{m}}\right\}  _{m\in\mathcal{M}}$. The fake expected operator
$\overline{\omega}(\mathcal{C})$\ of the states in the Chernoff code is as
follows:%
\begin{equation}
\overline{\omega}(\mathcal{C})\equiv\frac{1}{\left\vert \mathcal{M}\right\vert
}\sum_{m=1}^{\left\vert \mathcal{M}\right\vert }\omega_{c_{m}},
\end{equation}
because we assume that Alice randomizes codewords in the Chernoff code
according to a uniform distribution (notice that there is a difference in the
distribution that we use to choose the code and the distribution that Alice
uses to randomize the selected codewords). The expectation $\mathbb{E}%
_{\mathcal{C}}\left\{  \omega_{C_{m}}\right\}  $ of each operator
$\omega_{C_{m}}$ is equal to the expected operator $\omega$ because of the way
that we constructed the covering code. We can also define codes with respect
to the primed ensembles as follows: $\left\{  \sigma_{c_{m}}\right\}
_{m\in\mathcal{M}}$, $\left\{  \sigma_{c_{m}}^{\prime}\right\}  _{m\in
\mathcal{M}}$, $\left\{  \sigma_{c_{m}}^{\prime\prime}\right\}  _{m\in
\mathcal{M}}$. These codes respectively have fake expected operators of the
following form:%
\begin{align}
\overline{\sigma}(\mathcal{C})  &  \equiv\frac{1}{\left\vert \mathcal{M}%
\right\vert }\sum_{m=1}^{\left\vert \mathcal{M}\right\vert }\sigma_{c_{m}},\\
\overline{\sigma}^{\prime}(\mathcal{C})  &  \equiv\frac{1}{\left\vert
\mathcal{M}\right\vert }\sum_{m=1}^{\left\vert \mathcal{M}\right\vert }%
\sigma_{c_{m}}^{\prime},\\
\overline{\sigma}^{\prime\prime}(\mathcal{C})  &  \equiv\frac{1}{\left\vert
\mathcal{M}\right\vert }\sum_{m=1}^{\left\vert \mathcal{M}\right\vert }%
\sigma_{c_{m}}^{\prime\prime}.
\end{align}

\textbf{Applying the Operator Chernoff Bound.} We make one final modification
before applying the operator Chernoff bound. The operators $\omega_{c_{m}}$
are in the operator interval between the zero operator$~0$ and$~\frac{1}%
{d}\hat{\Pi}$:%
\begin{equation}
\forall m\in\mathcal{M}:0\leq\omega_{c_{m}}\leq\frac{1}{d}\hat{\Pi}.
\end{equation}
The above statement holds because the operators $\sigma_{x}^{\prime}$ satisfy
$\sigma_{x}^{\prime}=\Pi_{x}\sigma_{x}\Pi_{x}\leq\frac{1}{d}\Pi_{x}$ (the
fourth condition of the covering lemma) and this condition implies the
following inequalities:%
\begin{align}
\sigma_{x}^{\prime}=\Pi_{x}\sigma_{x}\Pi_{x}  &  \leq\frac{1}{d}\Pi_{x}\\
\Rightarrow\Pi\sigma_{x}^{\prime}\Pi=\sigma_{x}^{\prime\prime}  &  \leq
\frac{1}{d}\Pi\Pi_{x}\Pi\leq\frac{1}{d}\Pi\\
\Rightarrow\omega_{x}=\hat{\Pi}\sigma_{x}^{\prime\prime}\hat{\Pi}  &
\leq\frac{1}{d}\hat{\Pi}\Pi\hat{\Pi}\leq\frac{1}{d}\hat{\Pi}.
\end{align}
Therefore, we consider another set of operators (not necessarily density
operators)\ where we scale each $\omega_{c_{m}}$\ by $d$ so that%
\begin{equation}
\forall m\in\mathcal{M}:0\leq d\omega_{c_{m}}\leq\hat{\Pi}.
\end{equation}
This code satisfies the conditions of the operator Chernoff bound with
$a=\varepsilon d/D$ and with $\hat{\Pi}$ acting as the identity on the
subspace onto which it projects. We can now apply the operator Chernoff bound
to bound the probability that the sample average\ $\overline{\omega}%
\equiv\left\vert \mathcal{M}\right\vert ^{-1}\sum_{m\in\mathcal{M}}%
\omega_{c_{m}}$\ falls in the operator interval $\left[  \left(
1\pm\varepsilon\right)  \omega\right]  $:%
\begin{align}
\Pr\left\{  \left(  1-\varepsilon\right)  \omega\leq\overline{\omega}%
\leq\left(  1+\varepsilon\right)  \omega\right\}   &  =\Pr\left\{  d\left(
1-\varepsilon\right)  \omega\leq d\overline{\omega}\leq d\left(
1+\varepsilon\right)  \omega\right\} \\
&  \geq1-2\operatorname{Tr}\left\{  \hat{\Pi}\right\}  \exp\left(
-\frac{\left\vert \mathcal{M}\right\vert \varepsilon^{2}\left(  \varepsilon
d/D\right)  }{4}\right) \\
&  \geq1-2D\exp\left(  -\frac{\varepsilon^{3}}{4}\frac{\left\vert
\mathcal{M}\right\vert d}{D}\right)  .
\end{align}

\subsection{Obfuscation Error of the Covering Code}

The random covering code is a set\ of $\left\vert \mathcal{M}\right\vert $
quantum states $\left\{  \sigma_{C_{m}}\right\}  _{m\in\mathcal{M}}$ where the
quantum states arise from the original ensemble. Recall that our goal is to
show that the obfuscation error of the random covering code $\mathcal{C}$,%
\begin{equation}
o_{e}(\mathcal{C})=\left\Vert \overline{\sigma}(\mathcal{C})-\sigma\right\Vert
_{1},
\end{equation}
has a high probability of being small.

We now show that the obfuscation error of this random covering code is highly
likely to be small, by relating it to the Chernoff ensemble. Our method of
proof is simply to exploit the triangle inequality, the gentle operator lemma
(Lemma~\ref{lem-dm:gentle-operator}), and Exercise~\ref{ex-dm:trace-ineq-herm}
several times. The triangle inequality gives the following bound for the
obfuscation error:%
\begin{align}
o_{e}(\mathcal{C})  &  =\left\Vert \overline{\sigma}(\mathcal{C}%
)-\sigma\right\Vert _{1}\nonumber\\
&  =\left\Vert \overline{\sigma}(\mathcal{C})-\overline{\sigma}^{\prime\prime
}(\mathcal{C})-\left(  \overline{\omega}(\mathcal{C})-\overline{\sigma
}^{\prime\prime}(\mathcal{C})\right)  +\left(  \overline{\omega}%
(\mathcal{C})-\omega\right)  +\left(  \omega-\sigma^{\prime\prime}\right)
-\left(  \sigma-\sigma^{\prime\prime}\right)  \right\Vert _{1}\\
&  \leq\left\Vert \overline{\sigma}(\mathcal{C})-\overline{\sigma}%
^{\prime\prime}(\mathcal{C})\right\Vert _{1}+\left\Vert \overline{\omega
}(\mathcal{C})-\overline{\sigma}^{\prime\prime}(\mathcal{C})\right\Vert
_{1}\nonumber\\
&  \ \ \ \ \ \ \ +\left\Vert \overline{\omega}(\mathcal{C})-\omega\right\Vert
_{1}+\left\Vert \omega-\sigma^{\prime\prime}\right\Vert _{1}+\left\Vert
\sigma-\sigma^{\prime\prime}\right\Vert _{1}.
\label{eq-cover:obfus-error-total}%
\end{align}
We show how to obtain a good bound for each of the above five terms.

First consider the rightmost term $\left\Vert \sigma-\sigma^{\prime\prime
}\right\Vert _{1}$ in \eqref{eq-cover:obfus-error-total}. Consider that the
projected state $\sigma_{x}^{\prime}=\Pi_{x}\sigma_{x}\Pi_{x}$ is close to the
original state $\sigma_{x}$ by applying \eqref{eq-cov:cov-2} and the gentle
operator lemma:%
\begin{equation}
\left\Vert \sigma_{x}-\sigma_{x}^{\prime}\right\Vert _{1}\leq2\sqrt
{\varepsilon}. \label{eq:first-coat-bound}%
\end{equation}
Consider that%
\begin{equation}
\left\Vert \sigma_{x}^{\prime}-\sigma_{x}^{\prime\prime}\right\Vert _{1}%
\leq2\sqrt{\varepsilon+2\sqrt{\varepsilon}} \label{eq-cov:cov-bound-1}%
\end{equation}
because $\sigma_{x}^{\prime\prime}=\Pi\sigma_{x}^{\prime}\Pi$ and from
applying the gentle operator lemma to%
\begin{align}
\operatorname{Tr}\left\{  \Pi\sigma_{x}^{\prime}\right\}   &  \geq
\operatorname{Tr}\left\{  \Pi\sigma_{x}\right\}  -\left\Vert \sigma_{x}%
-\sigma_{x}^{\prime}\right\Vert _{1}\\
&  \geq1-\varepsilon-2\sqrt{\varepsilon}, \label{eq-cov:sigma-double-bound}%
\end{align}
where the first inequality follows from Exercise~\ref{ex-dm:trace-ineq-herm}
and the second from \eqref{eq-cov:cov-1} and \eqref{eq:first-coat-bound}. Then
the state $\sigma_{x}^{\prime\prime}$ is close to the original state
$\sigma_{x}$ for all $x$ because%
\begin{align}
\left\Vert \sigma_{x}-\sigma_{x}^{\prime\prime}\right\Vert _{1}  &
\leq\left\Vert \sigma_{x}-\sigma_{x}^{\prime}\right\Vert _{1}+\left\Vert
\sigma_{x}^{\prime}-\sigma_{x}^{\prime\prime}\right\Vert _{1}\\
&  \leq2\sqrt{\varepsilon}+2\sqrt{\varepsilon+2\sqrt{\varepsilon}},
\label{eq:second-coat-bound}%
\end{align}
where we first applied the triangle inequality and the bounds from
\eqref{eq:first-coat-bound} and \eqref{eq-cov:cov-bound-1}. Convexity of the
trace distance then gives a bound on $\left\Vert \sigma-\sigma^{\prime\prime
}\right\Vert _{1}$:%
\begin{align}
\left\Vert \sigma-\sigma^{\prime\prime}\right\Vert _{1}  &  =\left\Vert
\sum_{x\in\mathcal{X}}p_{X}(x)\sigma_{x}-\sum_{x\in\mathcal{X}}p_{X}%
(x)\sigma_{x}^{\prime\prime}\right\Vert _{1}\\
&  =\left\Vert \sum_{x\in\mathcal{X}}p_{X}(x)\left(  \sigma_{x}-\sigma
_{x}^{\prime\prime}\right)  \right\Vert _{1}\\
&  \leq\sum_{x\in\mathcal{X}}p_{X}(x)\left\Vert \sigma_{x}-\sigma_{x}%
^{\prime\prime}\right\Vert _{1}\\
&  \leq\sum_{x\in\mathcal{X}}p_{X}(x)\left(  2\sqrt{\varepsilon}%
+2\sqrt{\varepsilon+2\sqrt{\varepsilon}}\right) \\
&  =2\sqrt{\varepsilon}+2\sqrt{\varepsilon+2\sqrt{\varepsilon}}.
\end{align}

We now consider the second rightmost term $\left\Vert \omega-\sigma
^{\prime\prime}\right\Vert _{1}$ in \eqref{eq-cover:obfus-error-total}. The
support of $\sigma^{\prime\prime}$ has dimension less than $D$ by
\eqref{eq-cov:cov-3}, the third condition in the covering lemma. Therefore,
eigenvalues smaller than $\varepsilon/D$ contribute at most $\varepsilon$ to
$\operatorname{Tr}\left\{  \sigma^{\prime\prime}\right\}  $. In particular, if
$\sum_{i}\lambda_{i}|i\rangle\langle i|$ is a spectral decomposition of
$\sigma^{\prime\prime}$, with $\lambda_{i}\geq0$ for all $i$ and $\sum
_{i}\lambda_{i}\leq1$, we have that%
\begin{align}
\operatorname{Tr}\left\{  \sigma^{\prime\prime}\right\}  -\operatorname{Tr}%
\left\{  \omega\right\}   &  =\sum_{i}\lambda_{i}-\sum_{i:\lambda_{i}%
\geq\varepsilon/D}\lambda_{i}\\
&  =\sum_{i:\lambda_{i}<\varepsilon/D}\lambda_{i}\leq\frac{\varepsilon}%
{D}\cdot D=\varepsilon.
\end{align}
We can use this to bound the trace of $\omega$ as follows:%
\begin{align}
\operatorname{Tr}\left\{  \omega\right\}   &  \geq\operatorname{Tr}\left\{
\sigma^{\prime\prime}\right\}  -\varepsilon\\
&  =\sum_{x\in\mathcal{X}}p_{X}(x)\operatorname{Tr}\left\{  \sigma_{x}%
^{\prime\prime}\right\}  -\varepsilon\\
&  \geq\left(  \sum_{x\in\mathcal{X}}p_{X}(x)\right)  \left(  1-\varepsilon
-2\sqrt{\varepsilon}\right)  -\varepsilon\\
&  =1-2\left(  \varepsilon+\sqrt{\varepsilon}\right)  , \label{eq:omega-bound}%
\end{align}
where the first inequality applies the above \textquotedblleft
eigenvalue-bounding\textquotedblright\ argument and the second inequality
employs the bound in \eqref{eq-cov:sigma-double-bound}. This argument shows
that the average operator of the Chernoff ensemble has trace almost equal to
one. We can then apply the gentle operator lemma to $\operatorname{Tr}\left\{
\omega\right\}  \geq1-2\left(  \varepsilon+\sqrt{\varepsilon}\right)  $ to
give%
\begin{equation}
\left\Vert \omega-\sigma^{\prime\prime}\right\Vert _{1}\leq2\sqrt{2\left(
\varepsilon+\sqrt{\varepsilon}\right)  }.
\end{equation}

We now consider the middle term $\left\Vert \overline{\omega}-\omega
\right\Vert _{1}$ in \eqref{eq-cover:obfus-error-total}. The Chernoff bound
gives us a probabilistic estimate and not a deterministic estimate like the
other two bounds we have shown above. So we suppose for now that the fake
operator $\overline{\omega}$ of the Chernoff code is close to the average
operator $\omega$\ of the Chernoff ensemble:%
\begin{equation}
\overline{\omega}(\mathcal{C})\equiv\frac{1}{\left\vert \mathcal{M}\right\vert
}\sum_{m\in\mathcal{M}}\omega_{c_{m}}\in\left[  \left(  1\pm\varepsilon
\right)  \omega\right]  .
\end{equation}
With this assumption, it holds that%
\begin{equation}
\left\Vert \overline{\omega}(\mathcal{C})-\omega\right\Vert _{1}%
\leq\varepsilon,
\end{equation}
by employing Lemma~\ref{lemma:op-interval} from
Appendix~\ref{chap:appendix-math}\ and $\operatorname{Tr}\left\{
\omega\right\}  \leq1$.

We consider the second leftmost term $\left\Vert \overline{\omega}%
(\mathcal{C})-\overline{\sigma}^{\prime\prime}(\mathcal{C})\right\Vert _{1}$
in \eqref{eq-cover:obfus-error-total}. The following inequality holds:%
\begin{equation}
\operatorname{Tr}\left\{  \overline{\omega}(\mathcal{C})\right\}
\geq1-3\varepsilon-2\sqrt{\varepsilon},
\end{equation}
because in \eqref{eq:omega-bound} we showed that $\operatorname{Tr}\left\{
\omega\right\}  \geq1-2\left(  \varepsilon+\sqrt{\varepsilon}\right)  $, and
the triangle inequality implies that%
\begin{align}
\operatorname{Tr}\left\{  \overline{\omega}(\mathcal{C})\right\}   &
=\left\Vert \overline{\omega}(\mathcal{C})\right\Vert _{1}\\
&  =\left\Vert \omega-\left(  \omega-\overline{\omega}(\mathcal{C})\right)
\right\Vert _{1}\\
&  \geq\left\Vert \omega\right\Vert _{1}-\left\Vert \omega-\overline{\omega
}(\mathcal{C})\right\Vert _{1}\\
&  =\operatorname{Tr}\left\{  \omega\right\}  -\left\Vert \omega
-\overline{\omega}(\mathcal{C})\right\Vert _{1}\\
&  \geq\left(  1-2\left(  \varepsilon+\sqrt{\varepsilon}\right)  \right)
-\varepsilon\\
&  =1-3\varepsilon-2\sqrt{\varepsilon}.
\end{align}
Applying the gentle operator lemma to $\operatorname{Tr}\left\{
\overline{\omega}(\mathcal{C})\right\}  \geq1-3\varepsilon-2\sqrt{\varepsilon
}$ gives that%
\begin{equation}
\left\Vert \overline{\omega}(\mathcal{C})-\overline{\sigma}^{\prime\prime
}(\mathcal{C})\right\Vert _{1}\leq2\sqrt{3\varepsilon+2\sqrt{\varepsilon}}.
\end{equation}

We finally bound the leftmost term $\left\Vert \overline{\sigma}%
(\mathcal{C})-\overline{\sigma}^{\prime\prime}(\mathcal{C})\right\Vert _{1}$
in \eqref{eq-cover:obfus-error-total}. We can use convexity of trace distance
and \eqref{eq:second-coat-bound} to obtain the following bounds:%
\begin{align}
\left\Vert \overline{\sigma}(\mathcal{C})-\overline{\sigma}^{\prime\prime
}(\mathcal{C})\right\Vert _{1}  &  \leq\frac{1}{\left\vert \mathcal{M}%
\right\vert }\sum_{m\in\mathcal{M}}\left\Vert \sigma_{C_{m}}-\sigma_{C_{m}%
}^{\prime\prime}\right\Vert _{1}\\
&  \leq2\sqrt{\varepsilon}+2\sqrt{\varepsilon+2\sqrt{\varepsilon}}.
\end{align}

We now combine all of the above bounds with the triangle inequality in order
to bound the obfuscation error of the covering code $\mathcal{C}$:%
\begin{align}
&  o_{e}(\mathcal{C})\nonumber\\
&  =\left\Vert \overline{\sigma}(\mathcal{C})-\sigma\right\Vert _{1}\\
&  =\left\Vert \overline{\sigma}(\mathcal{C})-\overline{\sigma}^{\prime\prime
}(\mathcal{C})-\left(  \overline{\omega}(\mathcal{C})-\overline{\sigma
}^{\prime\prime}(\mathcal{C})\right)  +\left(  \overline{\omega}%
(\mathcal{C})-\omega\right)  +\left(  \omega-\sigma^{\prime\prime}\right)
-\left(  \sigma-\sigma^{\prime\prime}\right)  \right\Vert _{1}\\
&  \leq\left\Vert \overline{\sigma}(\mathcal{C})-\overline{\sigma}%
^{\prime\prime}(\mathcal{C})\right\Vert _{1}+\left\Vert \overline{\omega
}(\mathcal{C})-\overline{\sigma}^{\prime\prime}(\mathcal{C})\right\Vert
_{1}\nonumber\\
&  \ \ \ \ \ \ \ \ +\left\Vert \overline{\omega}(\mathcal{C})-\omega
\right\Vert _{1}+\left\Vert \omega-\sigma^{\prime\prime}\right\Vert
_{1}+\left\Vert \sigma-\sigma^{\prime\prime}\right\Vert _{1}\\
&  \leq\left(  2\sqrt{\varepsilon}+2\sqrt{\varepsilon+2\sqrt{\varepsilon}%
}\right)  +\left(  2\sqrt{3\varepsilon+2\sqrt{\varepsilon}}\right)
+\varepsilon\nonumber\\
&  \ \ \ \ \ \ \ \ +\left(  2\sqrt{2\left(  \varepsilon+\sqrt{\varepsilon
}\right)  }\right)  +\left(  2\sqrt{\varepsilon}+2\sqrt{\varepsilon
+2\sqrt{\varepsilon}}\right) \\
&  =\varepsilon+4\sqrt{\varepsilon}+4\sqrt{\varepsilon+2\sqrt{\varepsilon}%
}+2\sqrt{3\varepsilon+2\sqrt{\varepsilon}}+2\sqrt{2\left(  \varepsilon
+\sqrt{\varepsilon}\right)  }\\
&  \leq\varepsilon+4\sqrt{\varepsilon}+24\sqrt[4]{\varepsilon}.
\end{align}
Observe from the above that the event that the quantity $\varepsilon$ bounds
the obfuscation error $o_{e}(\mathcal{C})$ of the Chernoff code with states
$\omega_{C_{m}}$ implies the event when the quantity $\varepsilon
+4\sqrt{\varepsilon}+24\sqrt[4]{\varepsilon}$ bounds the obfuscation error
$o_{e}(\mathcal{C})$ of the original code with states $\sigma_{C_{m}}$. Thus,
we can bound the probability of obfuscation error of the covering code by
applying the Chernoff bound:%
\begin{align}
\Pr\left\{  o_{e}(\mathcal{C},\left\{  \sigma_{C_{m}}\right\}  )\leq
\varepsilon+4\sqrt{\varepsilon}+24\sqrt[4]{\varepsilon}\right\}   &  \geq
\Pr\left\{  o_{e}(\mathcal{C},\left\{  \omega_{C_{m}}\right\}  )\leq
\varepsilon\right\} \\
&  \geq1-2D\exp\left(  -\frac{\varepsilon^{3}}{4}\frac{\left\vert
\mathcal{M}\right\vert d}{D}\right)  .
\end{align}
This argument shows that it is highly likely that a random covering code is
good in the sense that it has a low obfuscation error.

\begin{exercise}
Prove that the covering lemma holds for the same ensemble and a set of
projectors for which the following conditions hold:%
\begin{align}
\sum_{x\in\mathcal{X}}p_{X}( x) \operatorname{Tr}\left\{  \sigma_{x}%
\Pi\right\}   &  \geq1-\varepsilon,\\
\sum_{x\in\mathcal{X}}p_{X}( x) \operatorname{Tr}\left\{  \sigma_{x}\Pi
_{x}\right\}   &  \geq1-\varepsilon,\\
\operatorname{Tr}\left\{  \Pi\right\}   &  \leq D,\\
\Pi_{x}\sigma_{x}\Pi_{x}  &  \leq\frac{1}{d}\Pi_{x}.
\end{align}

\end{exercise}

\begin{exercise}
Show that there exists a particular covering code with the property that the
obfuscation error is small.
\end{exercise}

\section{History and Further Reading}

\cite{AW02} introduced the operator Chernoff bound in the context of quantum
identification. \cite{WM01,Winter01a} later applied it to quantum measurement
compression. \cite{DW03} applied the covering lemma to classical compression
with quantum side information and to distilling secret key from quantum
states~\citep{DW05}. \cite{ieee2005dev}\ and \cite{1050633} applied it to
private classical communication over a quantum channel, and \cite{GPW05}
applied it to study the destruction of correlations in a bipartite state.

\part{Noiseless Quantum Shannon Theory}

\chapter{Schumacher Compression}

\label{chap:schumach}One of the fundamental%
\index{Schumacher compression}
tasks in classical information theory is the compression of information. Given
access to many uses of a noiseless classical channel, what is the best that a
sender and receiver can make of this resource for compressed data
transmission?\ Shannon's compression theorem demonstrates that the Shannon
entropy is the fundamental limit for the compression rate in the
i.i.d.~setting (recall the development in
Section~\ref{sec:shannon-compression}). That is, if one compresses at a rate
above the Shannon entropy, then it is possible to recover the compressed data
perfectly in the asymptotic limit, and otherwise, it is not possible to do
so.\footnote{Technically, we did not prove the converse part of Shannon's
data-compression theorem, but the converse of this chapter suffices for
Shannon's classical theorem as well.} This theorem establishes the prominent
role of the entropy in Shannon's theory of information.

In the quantum world, it very well could be that one day a sender and a
receiver would have many uses of a noiseless quantum channel
available,\footnote{How we hope so! If working, coherent fault-tolerant
quantum computers come along one day, they stand to benefit from quantum
compression protocols.} and the sender could use this resource to transmit
compressed quantum information. But what exactly does this mean in the quantum
setting? A simple model of a quantum information source%
\index{quantum information source}
is an ensemble of quantum states $\left\{  p_{X}(x),|\psi_{x}\rangle\right\}
$; i.e., the source outputs the state $|\psi_{x}\rangle$ with probability
$p_{X}(x)$, and the states $\left\{  |\psi_{x}\rangle\right\}  $ do not
necessarily have to form an orthonormal basis. Let us suppose for the moment
that the classical data $x$ is available as well, even though this might not
necessarily be the case in practice. A naive strategy for compressing this
quantum information source would be to ignore the quantum states coming out,
handle the classical data instead, and exploit Shannon's compression protocol
from Section~\ref{sec:shannon-compression}. That is, the sender compresses the
sequence $x^{n}$ emitted from the quantum information source at a rate equal
to the Shannon entropy $H(X)$, sends the compressed classical bits over the
noiseless quantum channels, the receiver reproduces the classical sequence
$x^{n}$ at his end, and finally reconstructs the sequence $|\psi_{x^{n}%
}\rangle$ of quantum states corresponding to the classical sequence$~x^{n}$.

The above strategy will certainly work, but it makes no use of the fact that
the noiseless quantum channels are quantum! It is clear that noiseless quantum
channels will be expensive in practice, and the above strategy is wasteful in
this sense because it could have merely exploited classical channels (channels
that cannot preserve superpositions) to achieve the same goals. Schumacher
compression is a strategy that makes effective use of noiseless quantum
channels to compress a quantum information source down to a rate equal to the
quantum entropy. This has a great benefit from a practical standpoint---recall
from Exercise~\ref{ex-qie:shannon-vs-von-neumann}\ that the quantum entropy of
a quantum information source is strictly lower than the source's Shannon
entropy if the states in the ensemble are non-orthogonal. In order to execute
the protocol, the sender and receiver simply need to know the density operator
$\rho\equiv\sum_{x}p_{X}(x)|\psi_{x}\rangle\langle\psi_{x}|$ of the source.
Furthermore, Schumacher compression is provably optimal in the sense that any
protocol that compresses a quantum information source of the above form at a
rate below the quantum entropy cannot have a vanishing error in the asymptotic limit.

Schumacher compression thus gives an operational interpretation of the quantum
entropy%
\index{von Neumann entropy!operational interpretation}
as the fundamental limit on the rate of quantum data compression. Also, it
sets the term \textquotedblleft qubit\textquotedblright\ on a firm foundation
in an information-theoretic sense as a measure of the amount of quantum
information \textquotedblleft contained\textquotedblright\ in a quantum
information source.

We begin this chapter by giving the details of the general
information-processing task corresponding to quantum data compression. We then
prove that the quantum entropy is an achievable rate of compression and follow
by showing that it is optimal (these two respective parts are the direct
coding theorem and the converse theorem for quantum data compression). We
illustrate how much savings one can gain in quantum data compression by
detailing a specific example. The final section of the chapter closes with a
presentation of more general forms of Schumacher compression.

\section{The Information-Processing Task}

\begin{figure}
[ptb]
\begin{center}
\includegraphics[
width=4.8456in
]%
{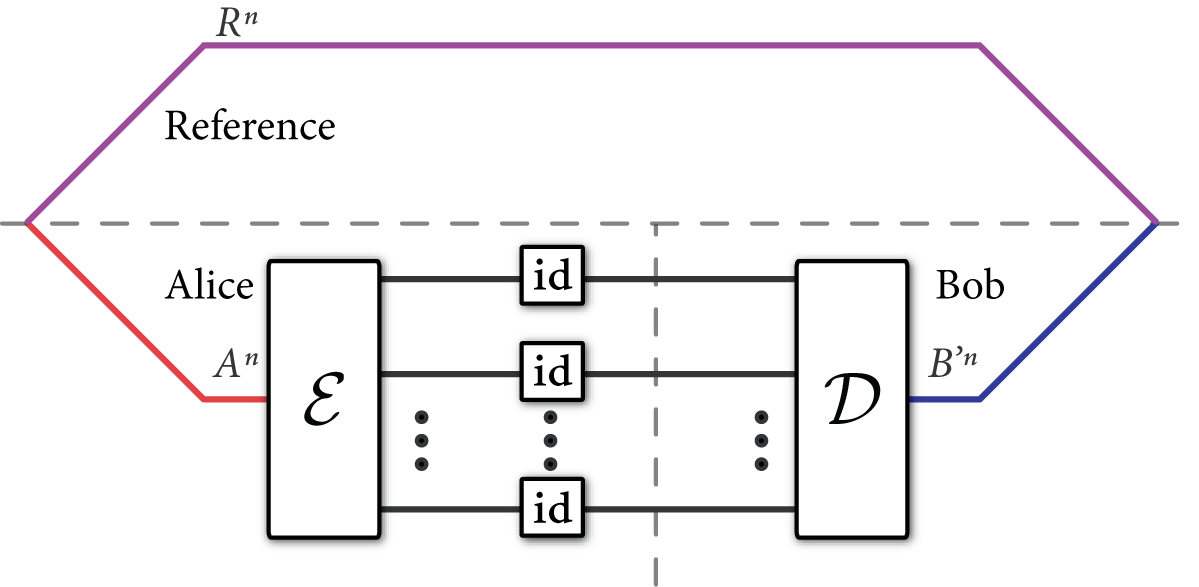}%
\caption{The most general protocol for quantum compression. Alice begins with
the output of some quantum information source whose density operator is
$\rho^{\otimes n}$ on some system $A^{n}$. The inaccessible reference system
holds the purification of this density operator. She performs some
CPTP\ encoding map $\mathcal{E}$, sends the compressed qubits through $nR$
uses of a noiseless qubit channel, and Bob performs some CPTP\ decoding map
$\mathcal{D}$ to decompress the qubits. The scheme is successful if the
initial state and the final state are indistinguishable in the asymptotic
limit $n\rightarrow\infty$.}%
\label{fig:general-protocol-comp}%
\end{center}
\end{figure}

We first discuss the general task that any quantum compression protocol
attempts to accomplish. Three parameters $n$, $R$, and $\varepsilon$,
corresponding to the length of the original quantum data sequence, the rate,
and the error, respectively, characterize any such protocol. An $\left(
n,R,\varepsilon\right)  $ quantum compression code consists of four
steps:\ state preparation, encoding, transmission, and decoding.
Figure~\ref{fig:general-protocol-comp} depicts a general protocol for quantum
compression.%

\textbf{State Preparation.} The quantum information source outputs a sequence
$|\psi_{x^{n}}\rangle_{A^{n}}$\ of quantum states according to the ensemble
$\left\{  p_{X}(x),|\psi_{x}\rangle\right\}  $ where%
\begin{equation}
|\psi_{x^{n}}\rangle_{A^{n}}\equiv\left\vert \psi_{x_{1}}\right\rangle
_{A_{1}}\otimes\cdots\otimes\left\vert \psi_{x_{n}}\right\rangle _{A_{n}}.
\end{equation}
The density operator, from the perspective of someone ignorant of the
classical sequence $x^{n}$, is equal to the tensor power state $\rho^{\otimes
n}$ where%
\begin{equation}
\rho\equiv\sum_{x}p_{X}(x)|\psi_{x}\rangle\langle\psi_{x}|.
\end{equation}
Also, we can think about the purification of the above density operator. That
is, a related picture is to imagine that the quantum
information source produces states of the form%
\begin{equation}
|\varphi_{\rho}\rangle_{RA}\equiv\sum_{x}\sqrt{p_{X}(x)}|x\rangle_{R}|\psi
_{x}\rangle_{A},
\end{equation}
where $R$ is the label for an inaccessible reference system (not to be
confused with the rate $R$!). The resulting i.i.d.~state produced is
$(|\varphi_{\rho}\rangle_{RA})^{\otimes n}$.

\textbf{Encoding.} Alice encodes the systems $A^{n}$ according to some
 compression channel $\mathcal{E}_{A^{n}\rightarrow W}$ where $W$ is a
quantum system of size $2^{nR}$. Recall that $R$ is the rate of compression:%
\begin{equation}
R=\frac{1}{n}\log\dim(\mathcal{H}_{W}).
\end{equation}

\textbf{Transmission.} Alice transmits the system $W$ to Bob using $nR$
noiseless qubit channels.

\textbf{Decoding.} Bob sends the system $W$ through a decompression channel
$\mathcal{D}_{W\rightarrow\hat{A}^{n}}$.

The protocol has $\varepsilon\in\left[  0,1\right]  $ error if the compressed
and decompressed state is $\varepsilon$-close in normalized trace distance to the
original state $(|\varphi_{\rho}\rangle_{RA})^{\otimes n}$:%
\begin{equation}
\frac{1}{2}\left\Vert \left(  \varphi_{RA}^{\rho}\right)  ^{\otimes
n}-(\mathcal{D}_{W\rightarrow\hat{A}^{n}}\circ\mathcal{E}_{A^{n}\rightarrow
W})(\left(  \varphi_{RA}^{\rho}\right)  ^{\otimes n})\right\Vert _{1}%
\leq\varepsilon. \label{eq:schu-perf-crit}%
\end{equation}
We say that a quantum compression rate $R$ is \textit{achievable} if there
exists an $\left(  n,R+\delta,\varepsilon\right)  $ quantum compression code
for all $\delta>0,\varepsilon\in(0,1)$, and sufficiently large $n$. The
\textit{quantum data compression limit} of $\rho$ is equal to the infimum of
all achievable quantum compression rates.

\begin{exercise}
\label{ex-sc:error-criterion}An alternate figure of merit for the performance
of Schumacher compression is the average ensemble trace distance, given by%
\begin{equation}
\frac{1}{2}\sum_{x^{n}}p_{X^{n}}(x^{n})\left\Vert \psi_{x^{n}}-(\mathcal{D}%
_{W\rightarrow\hat{A}^{n}}\circ\mathcal{E}_{A^{n}\rightarrow W})(\psi_{x^{n}%
})\right\Vert _{1}. \label{eq-sc:less-stringent-error}%
\end{equation}
Prove that a quantum compression protocol satisfying \eqref{eq:schu-perf-crit}
also has an average ensemble trace distance no larger than $\varepsilon$.
(Hint:\ Consider using the monotonicity of trace distance with respect to
quantum channels and acting on the reference systems with a particular channel.)
\end{exercise}

\section{The Quantum Data-Compression Theorem}

Schumacher's compression theorem establishes the quantum entropy as the
fundamental limit on quantum data compression.

\begin{theorem}
[Quantum Data Compression]Suppose that $\rho_{A}$ is the density operator corresponding to a quantum information source. Then the quantum entropy $H(A)_{\rho}$ is
equal to the quantum data compression limit of $\rho$.
\end{theorem}

\subsection{The Direct Coding Theorem}

\begin{figure}
[ptb]
\begin{center}
\includegraphics[
width=4.8456in
]%
{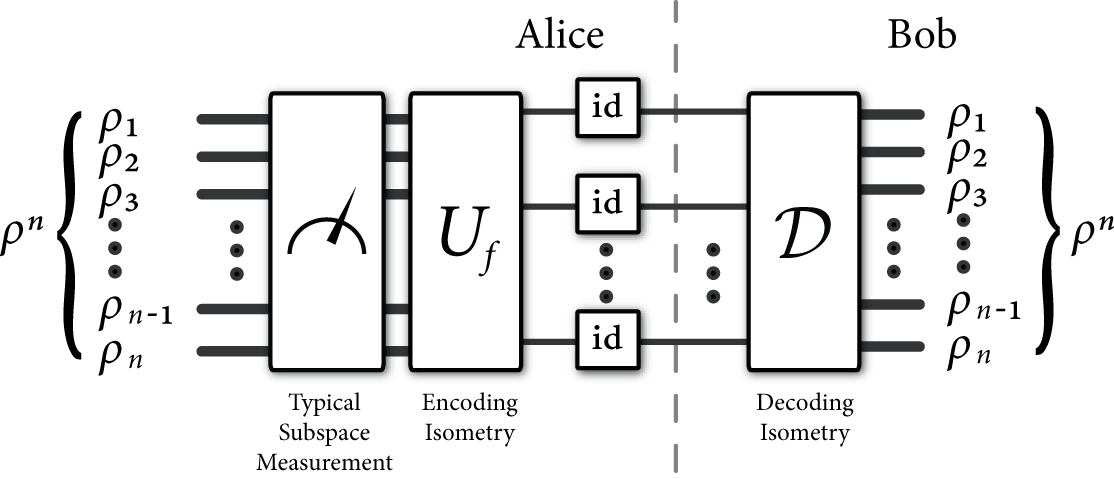}%
\caption{Schumacher's compression protocol. Alice begins with many copies of
the output of the quantum information source. She performs a measurement onto
the typical subspace corresponding to the state $\rho$ and then performs a
compression isometry of the typical subspace to a space of dimension
$2^{n\left[  H(\rho)+\delta\right]  }$, corresponding to $n\left[
H(\rho)+\delta\right]  $ qubits. She transmits these compressed qubits over
$n\left[  H(\rho)+\delta\right]  $\ uses of a noiseless qubit channel. Bob
performs the inverse of the isometry to uncompress the qubits. The protocol is
successful in the asymptotic limit due to the properties of typical
subspaces.}%
\label{fig:schu-comp}%
\end{center}
\end{figure}

Schumacher's compression protocol demonstrates that the quantum entropy
$H(A)_{\rho}$ is an achievable rate for quantum data compression. It is
remarkably similar to Shannon's compression protocol from
Section~\ref{sec:shannon-compression}, but it has some subtle differences that
are necessary for the quantum setting. The basic steps of the encoding are to
perform a typical subspace measurement and an isometry that compresses the
typical subspace. The decoder then performs the inverse of the isometry to
decompress the state. The protocol is successful if the typical subspace
measurement successfully projects onto the typical subspace, and it fails
otherwise. Just like in the classical case, the law of large numbers
guarantees that the protocol is successful in the asymptotic limit as
$n\rightarrow\infty$. Figure~\ref{fig:schu-comp}\ provides an illustration of
the protocol, and we now provide a rigorous argument.%

Alice begins with $n$ copies of the state $\left(  \varphi_{RA}^{\rho}\right)
^{\otimes n}$. Suppose that a spectral decomposition of $\rho$ is as follows:%
\begin{equation}
\rho=\sum_{z}p_{Z}(z)|z\rangle\langle z|,
\end{equation}
where $p_{Z}(z)$ is some probability distribution, and $\left\{
|z\rangle\right\}  $ is some orthonormal basis. Her first step is to perform a
typical subspace measurement onto the typical subspace of $A^{n}$, where the
typical projector is with respect to the density operator $\rho$. Recall from
the Shannon compression protocol in Section~\ref{sec:shannon-compression} that
we exploited a one-to-one function $f$ that mapped from the set of typical
sequences\ to a set of binary sequences $\left\{  0,1\right\}  ^{n\left[
H(\rho)+\delta\right]  }$. Now, we can construct a linear map $U_{f}$ that is
a coherent version of this classical function$~f$. It simply maps the
orthonormal basis $\{|z^{n}\rangle_{A^{n}}\}$ to the basis $\{\left\vert
f(z^{n})\right\rangle _{W}\}$:%
\begin{equation}
U_{f}\equiv\sum_{z^{n}\in T_{\delta}^{Z^{n}}}\left\vert f(z^{n})\right\rangle
_{W}\langle z^{n}|_{A^{n}},
\end{equation}
where $Z$ is a random variable corresponding to the distribution $p_{Z}(z)$,
so that $T_{\delta}^{Z^{n}}$ is its typical set. The inverse of the above
operator is an isometry because the input space span$\left\{  |z^{n}%
\rangle_{A^{n}}:z^{n}\in T_{\delta}^{Z^{n}}\right\}  $\ is a subspace of size
at most $2^{n\left[  H(\rho)+\delta\right]  }$ (recall
Property~\ref{prop-qt:exp-small}) embedded in a larger space of size $\left[
\dim(\mathcal{H}_{A})\right]  ^{n}$ and the output space is of size at most
$2^{n\left[  H(\rho)+\delta\right]  }$. So her next step is to perform the
compression conditioned on the typical subspace measurement being successful.
We can thus write the encoding as a single quantum channel as follows:%
\begin{equation}
\mathcal{E}_{A^{n}\rightarrow W}(X_{A^{n}})\equiv U_{f}\Pi_{A^{n}}^{\delta
}X_{A^{n}}\Pi_{A^{n}}^{\delta}U_{f}^{\dag}+\operatorname{Tr}\{(I_{A^{n}}%
-\Pi_{A^{n}}^{\delta})X_{A^{n}}\}\sigma_{W}, \label{eq:first-step-schu}%
\end{equation}
where $X_{A^{n}}\in\mathcal{L}(\mathcal{H}_{A^{n}})$ and $\sigma_{W}$ is any
density operator with support in $\operatorname{span}\{\left\vert
f(z^{n})\right\rangle _{W}:z^{n}\in T_{\delta}^{Z^{n}}\}$. Alice then sends
the $W$ system of $\mathcal{E}_{A^{n}\rightarrow W}(\left(  \varphi_{RA}%
^{\rho}\right)  ^{\otimes n})$ (the compressed qubits) over $n\left[
H(\rho)+\delta\right]  $ uses of the noiseless qubit channel.

Bob's decoding $\mathcal{D}_{W\rightarrow A^{n}}$ essentially performs the
inverse of the linear map $U_{f}$, implemented as the following quantum
channel:%
\begin{equation}
\mathcal{D}_{W\rightarrow A^{n}}(Y_{W})\equiv U_{f}^{\dag}Y_{W}U_{f}%
+\operatorname{Tr}\{(I-U_{f}U_{f}^{\dag})Y_{W}\}\tau_{A^{n}},
\end{equation}
where $Y_{W}\in\mathcal{L}(\mathcal{H}_{W})$ and $\tau_{A^{n}}\in
\mathcal{D}(\mathcal{H}_{A^{n}})$.

We now can analyze how this protocol performs with respect to our performance
criterion in~\eqref{eq:schu-perf-crit}. Consider that%
\begin{align}
&  \left(  \mathcal{D}_{W\rightarrow A^{n}}\circ\mathcal{E}_{A^{n}\rightarrow
W}\right)  (\left(  \varphi_{RA}^{\rho}\right)  ^{\otimes n})\nonumber\\
&  =\mathcal{D}_{W\rightarrow A^{n}}\left(  U_{f}\Pi_{A^{n}}^{\delta}\left(
\varphi_{RA}^{\rho}\right)  ^{\otimes n}\Pi_{A^{n}}^{\delta}U_{f}^{\dag
}+\operatorname{Tr}_{A^{n}}\{(I_{A^{n}}-\Pi_{A^{n}}^{\delta})\left(
\varphi_{RA}^{\rho}\right)  ^{\otimes n}\}\otimes\sigma_{W}\right) \\
&  =\mathcal{D}_{W\rightarrow A^{n}}\left(  U_{f}\Pi_{A^{n}}^{\delta}\left(
\varphi_{RA}^{\rho}\right)  ^{\otimes n}\Pi_{A^{n}}^{\delta}U_{f}^{\dag
}\right) \nonumber\\
&  \ \ \ \ +\mathcal{D}_{W\rightarrow A^{n}}\left(  \operatorname{Tr}_{A^{n}%
}\{(I_{A^{n}}-\Pi_{A^{n}}^{\delta})\left(  \varphi_{RA}^{\rho}\right)
^{\otimes n}\}\otimes\sigma_{W}\right) \\
&  =\Pi_{A^{n}}^{\delta}\left(  \varphi_{RA}^{\rho}\right)  ^{\otimes n}%
\Pi_{A^{n}}^{\delta}+\operatorname{Tr}_{A^{n}}\{(I_{A^{n}}-\Pi_{A^{n}}%
^{\delta})\left(  \varphi_{RA}^{\rho}\right)  ^{\otimes n}\}\otimes
\mathcal{D}_{W\rightarrow A^{n}}(\sigma_{W}).
\end{align}
Then%
\begin{align}
&  \left\Vert \left(  \varphi_{RA}^{\rho}\right)  ^{\otimes n}-\left(
\mathcal{D}_{W\rightarrow A^{n}}\circ\mathcal{E}_{A^{n}\rightarrow W}\right)
(\left(  \varphi_{RA}^{\rho}\right)  ^{\otimes n})\right\Vert _{1}\nonumber\\
&  \leq\left\Vert \left(  \varphi_{RA}^{\rho}\right)  ^{\otimes n}-\Pi_{A^{n}%
}^{\delta}\left(  \varphi_{RA}^{\rho}\right)  ^{\otimes n}\Pi_{A^{n}}^{\delta
}\right\Vert _{1}\nonumber\\
&  \ \ \ \ +\left\Vert \operatorname{Tr}_{A^{n}}\{(I_{A^{n}}-\Pi_{A^{n}%
}^{\delta})\left(  \varphi_{RA}^{\rho}\right)  ^{\otimes n}\}\otimes
\mathcal{D}_{W\rightarrow A^{n}}(\sigma_{W})\right\Vert _{1}\\
&  \leq2\sqrt{\varepsilon}+\varepsilon.
\end{align}
The first inequality follows from the triangle inequality. The second
inequality follows from the first property of typical subspaces:%
\begin{equation}
\operatorname{Tr}\left\{  \Pi_{A^{n}}^{\delta}\left(  \varphi_{RA}^{\rho
}\right)  ^{\otimes n}\right\}  =\operatorname{Tr}\left\{  \Pi_{A^{n}}%
^{\delta}\rho^{\otimes n}\right\}  \geq1-\varepsilon,
\end{equation}
the gentle operator lemma (Lemma~\ref{lem-dm:gentle-operator}), and the fact
that%
\begin{align}
&  \left\Vert \operatorname{Tr}_{A^{n}}\{(I_{A^{n}}-\Pi_{A^{n}}^{\delta
})\left(  \varphi_{RA}^{\rho}\right)  ^{\otimes n}\}\otimes\mathcal{D}%
_{W\rightarrow A^{n}}(\sigma_{W})\right\Vert _{1}\nonumber\\
&  =\left\Vert \operatorname{Tr}_{A^{n}}\{(I_{A^{n}}-\Pi_{A^{n}}^{\delta
})\left(  \varphi_{RA}^{\rho}\right)  ^{\otimes n}\}\right\Vert _{1}\left\Vert
\mathcal{D}_{W\rightarrow A^{n}}(\sigma_{W})\right\Vert _{1}\\
&  \leq\operatorname{Tr}\{(I_{A^{n}}-\Pi_{A^{n}}^{\delta})\left(  \varphi
_{RA}^{\rho}\right)  ^{\otimes n}\}\leq\varepsilon.
\end{align}

We remark that it is important for the typical subspace measurement in
\eqref{eq:first-step-schu}\ to be implemented (as indicated) as a
non-destructive quantum measurement. That is, the only information that this
measurement should learn is whether the state is typical or not. Otherwise,
there would be too much disturbance to the quantum information, and the
protocol would fail at the desired task of compression. Such precise control
on so many qubits is possible in principle, but it is rather daunting to
implement in practice!

\subsection{The Converse Theorem}

We now prove the converse theorem for quantum data compression by considering
the most general compression protocol that meets the success criterion in
\eqref{eq:schu-perf-crit} and demonstrating that a sequence of such protocols
with error approaching zero should have their rate of compression above the
quantum entropy of the source. Alice would like to compress a state
$\rho^{\otimes n}$ that acts on a Hilbert space $A^{n}$. A purification
$\phi_{R^{n}A^{n}}\equiv\left(  \varphi_{RA}^{\rho}\right)  ^{\otimes n}$ of
this state represents the state of the joint systems$~A^{n}$ and $R^{n}$ where
$R^{n}$ is the purifying system (again, we should not confuse reference system
$R^{n}$ with rate $R$). If she can compress any system on$~A^{n}$ and recover
it faithfully, then she should be able to do so for the purification of the
state. An $\left(  n,R+\delta,\varepsilon\right)  $ compression code has the
property that it can compress at a rate $R+\delta\equiv\left[  \log
\dim(\mathcal{H}_{W})\right]  /n$ with only error $\varepsilon$. The quantum
data processing is%
\begin{equation}
A^{n}\ \ \ \underrightarrow{\mathcal{E}_{A^{n}\rightarrow W}}%
\ \ \ W\ \ \ \underrightarrow{\mathcal{D}_{W\rightarrow\hat{A}^{n}}}%
\ \ \ \hat{A}^{n}%
\end{equation}
and the following inequality holds for a quantum compression protocol with
error $\varepsilon$:%
\begin{equation}
\frac{1}{2}\left\Vert \omega_{R^{n}\hat{A}^{n}}-\left(  \varphi_{RA}^{\rho
}\right)  ^{\otimes n}\right\Vert _{1}\leq\varepsilon,
\label{eq:schu-converse-perf-crit}%
\end{equation}
where%
\begin{equation}
\omega_{R^{n}\hat{A}^{n}}\equiv\mathcal{D}_{W\rightarrow\hat{A}^{n}%
}(\mathcal{E}_{A^{n}\rightarrow W}(\left(  \varphi_{RA}^{\rho}\right)
^{\otimes n})).
\end{equation}
Let $\tau_{R^{n}W}\equiv\mathcal{E}_{A^{n}\rightarrow W}(\left(  \varphi
_{RA}^{\rho}\right)  ^{\otimes n})$. Consider the following chain of
inequalities:%
\begin{align}
2\log\dim(\mathcal{H}_{W})  &  \geq I(W;R^{n})_{\tau} \label{eq-sc:converse-1}%
\\
&  \geq I(\hat{A}^{n};R^{n})_{\omega}\\
&  \geq I(A^{n};R^{n})_{\varphi^{\otimes n}}-f(n,\varepsilon)\\
&  =nI(A;R)_{\varphi}-f(n,\varepsilon)\label{eq-sc:converse-last-2nd}\\
&  =2nH(A)_{\varphi}-f(n,\varepsilon). \label{eq-sc:converse-last}%
\end{align}
The first inequality is a consequence of a dimension bound for the quantum
mutual information $I(E;F)\leq2\log\left(  \min\left\{  \left\vert
E\right\vert ,\left\vert F\right\vert \right\}  \right)  $ (see
Exercise~\ref{ex-qie:dim-bound-MI}). The second inequality follows from the
quantum data-processing inequality (Bob processes $W$ using the decoder to get
$\hat{A}^{n}$). The third inequality follows because $H(R^{n})_{\omega
}=H(R^{n})_{\varphi^{\otimes n}}$, so that%
\begin{align}
&  \left\vert I(\hat{A}^{n};R^{n})_{\omega}-I(A^{n};R^{n})_{\varphi^{\otimes
n}}\right\vert \nonumber\\
&  =\left\vert H(R^{n})_{\omega}-H(R^{n}|\hat{A}^{n})_{\omega}-\left[
H(A^{n})_{\varphi^{\otimes n}}-H(R^{n}|A^{n})_{\varphi^{\otimes n}}\right]
\right\vert \\
&  =\left\vert H(R^{n}|A^{n})_{\varphi^{\otimes n}}-H(R^{n}|\hat{A}%
^{n})_{\omega}\right\vert \\
&  \leq f(n,\varepsilon)\equiv2\varepsilon n\log\dim(\mathcal{H}%
_{R})+g_2(\varepsilon).
\end{align}
The inequality directly above follows from the AFW inequality
(Theorem~\ref{thm-qie:AFW-cont-ent}) applied to the success criterion in
\eqref{eq:schu-converse-perf-crit}. The function $f(n,\varepsilon)$ has the
property that $\lim_{\varepsilon\rightarrow0}\lim_{n\rightarrow\infty}\frac
{1}{n}f(n,\varepsilon)=0$. The equality in
\eqref{eq-sc:converse-last-2nd}\ follows because the quantum mutual
information is additive for tensor-product states. The equality in
\eqref{eq-sc:converse-last}\ follows because the quantum mutual information of
a pure, bipartite state is equal to twice the marginal entropy. Putting
everything together, we find that%
\begin{equation}
R+\delta=\frac{1}{n}\log\dim(\mathcal{H}_{W})\geq H(A)_{\varphi^{\rho}}%
-\frac{1}{2n}f(n,\varepsilon).
\end{equation}
Taking the limit as $n\rightarrow\infty$ and $\varepsilon,\delta\rightarrow0$
allows us to conclude that an achievable rate $R$ of quantum data compression
necessarily satisfies $R\geq H(A)_{\varphi^{\rho}}$.

\begin{exercise}
\label{ex-sc:alt-converse-bound}We proved the converse theorem for Schumacher
compression with respect to the error criterion in \eqref{eq:schu-perf-crit}.
However, it might be the case that if we use the less stringent error
criterion in \eqref{eq-sc:less-stringent-error} that we could achieve a
smaller rate of quantum data compression. Show that this is not the case, by
establishing that the converse theorem holds with this less stringent error
criterion. (Hint:\ The development is essentially the same as in
\eqref{eq-sc:converse-1}--\eqref{eq-sc:converse-last}, except you should have
the reference system be classical, containing a classical label for which
state in the i.i.d.~ensemble was chosen.)
\end{exercise}

\section{Quantum Compression Example}

We now highlight a particular example where Schumacher compression gives a big
savings in compression rates if noiseless qubit channels are available.
Suppose that the ensemble is of the following form:%
\begin{equation}
\left\{  \left(  \frac{1}{2},|0\rangle\right)  ,\left(  \frac{1}{2}%
,|+\rangle\right)  \right\}  .
\end{equation}
This ensemble is known as the Bennett-92 ensemble because it is useful in
Bennett's protocol for quantum key distribution. The naive strategy would be
for Alice and Bob to exploit Shannon's compression protocol. That is, Alice
would ignore the quantum nature of the states, and supposing that the
classical label for them were available, she would encode the classical label.
However, the entropy of the uniform distribution on two states is equal to one
bit, and she would have to transmit classical messages at a rate of one bit
per channel use.

A different strategy is to employ Schumacher compression. The density operator
of the above ensemble is%
\begin{equation}
\frac{1}{2}|0\rangle\langle0|+\frac{1}{2}|+\rangle\langle+|,
\end{equation}
which has the following spectral decomposition:%
\begin{equation}
\cos^{2}(\pi/8)\left\vert +^{\prime}\right\rangle \left\langle +^{\prime
}\right\vert +\sin^{2}(\pi/8)\left\vert -^{\prime}\right\rangle \left\langle
-^{\prime}\right\vert ,
\end{equation}
where%
\begin{align}
\left\vert +^{\prime}\right\rangle  &  \equiv\cos(\pi/8)|0\rangle+\sin
(\pi/8)|1\rangle,\\
\left\vert -^{\prime}\right\rangle  &  \equiv\sin(\pi/8)|0\rangle-\cos
(\pi/8)|1\rangle.
\end{align}
The binary entropy $h_{2}(\cos^{2}(\pi/8))$ of the distribution $\left[
\cos^{2}(\pi/8),\sin^{2}(\pi/8)\right]  $ is approximately equal to%
\begin{equation}
0.6009\text{ qubits,}%
\end{equation}
and thus they can save a significant amount in terms of compression rate by
employing Schumacher compression. This type of savings will always occur
whenever the ensemble includes non-orthogonal quantum states.

\begin{exercise}
In the above example, suppose that Alice associates a classical label with the
states, so that the ensemble instead is%
\begin{equation}
\left\{  \left(  \frac{1}{2},\vert0\rangle\langle0\vert\otimes\vert
0\rangle\langle0\vert\right)  ,\left(  \frac{1}{2},\vert1\rangle\langle
1\vert\otimes\vert+\rangle\langle+\vert\right)  \right\}  .
\end{equation}
Does this help in reducing the amount of qubits she has to transmit to Bob?
\end{exercise}

\section{Variations on the Schumacher Theme}

\label{sec-schu-comp:variations}We can propose several variations on the
Schumacher compression theme. For example, suppose that the quantum
information source corresponds to the following ensemble instead:%
\begin{equation}
\left\{  p_{X}(x),\rho_{A}^{x}\right\}  , \label{eq-sc:mixed-state-source}%
\end{equation}
where each $\rho_{x}$ is a mixed state. Then the situation is not as
\textquotedblleft clear-cut\textquotedblright\ as in the simpler model for a
quantum information source, and the entropy of the source does not necessarily
serve as a lower bound on the ultimate compressibility rate. This depends on
the figure of merit that we choose for mixed-state compression, and there are
at least three interesting figures of merit that we could consider. To see
these different ones, observe that the following state serves as a
purification of the mixed-state source in \eqref{eq-sc:mixed-state-source}:%
\begin{equation}
|\phi\rangle_{XX^{\prime}RA}\equiv\sum_{x}\sqrt{p_{X}(x)}|x\rangle
_{X}|x\rangle_{X^{\prime}}\left\vert \phi^{\rho^{x}}\right\rangle _{RA},
\end{equation}
where $|\phi^{\rho^{x}}\rangle_{RA}$ is a purification of $\rho_{A}^{x}$,
so that the purifying system of $A$ is the joint system $XX^{\prime}R$. The
three figures of merit to consider for any encoding-decoding pair
$(\mathcal{E}_{A^{n}\rightarrow W},\mathcal{D}_{W\rightarrow\hat{A}^{n}})$ are
as follows:%
\begin{align}
&  \frac{1}{2}\left\Vert \phi_{XX^{\prime}RA}^{\otimes n}-(\mathcal{D}%
_{W\rightarrow\hat{A}^{n}}\circ\mathcal{E}_{A^{n}\rightarrow W})(\phi
_{XX^{\prime}RA}^{\otimes n})\right\Vert _{1}, \label{eq-sc:mixed-error-crit}%
\\
&  \frac{1}{2}\sum_{x^{n}}p_{X^{n}}(x^{n})\left\Vert \phi_{R^{n}A^{n}}%
^{\rho^{x^{n}}}-(\mathcal{D}_{W\rightarrow\hat{A}^{n}}\circ\mathcal{E}%
_{A^{n}\rightarrow W})(\phi_{R^{n}A^{n}}^{\rho^{x^{n}}})\right\Vert
_{1},\label{eq-sc:mixed-error-crit-2}\\
&  \frac{1}{2}\sum_{x^{n}}p_{X^{n}}(x^{n})\left\Vert \rho_{A^{n}}^{x^{n}%
}-(\mathcal{D}_{W\rightarrow\hat{A}^{n}}\circ\mathcal{E}_{A^{n}\rightarrow
W})(\rho_{A^{n}}^{x^{n}})\right\Vert _{1}. \label{eq-sc:mixed-error-crit-3}%
\end{align}
Consider that satisfying the first error criterion up to some $\varepsilon
\in\left[  0,1\right]  $ implies that second criterion is satisfied as well,
which in turn implies that the third is satisfied\ (this follows from a
reasoning similar to that in the hint for Exercise~\ref{ex-sc:error-criterion}).

How should we handle the mixed source case in general? Let us consider the
direct coding theorem and the converse theorem. The direct coding theorem for
this case is essentially equivalent to Schumacher's protocol for quantum
compression---there does not appear to be a better approach in the general
case. The density operator of the source is equal to%
\begin{equation}
\rho_{A}=\sum_{x}p_{X}(x)\rho_{A}^{x}.
\end{equation}
A compression rate $R\geq H(A)_{\rho}$ is achievable if we form the typical
subspace measurement from the typical subspace projector $\Pi_{A^{n}}^{\delta
}$\ onto the state $(\rho_{A})^{\otimes n}$, and the error analysis from
before shows that this holds for the error criterion in
\eqref{eq-sc:mixed-error-crit}, which implies that it holds for the other two
error criteria mentioned above.

Although the direct coding theorem stays the same, the converse theorem
changes somewhat. If we demand that the converse hold for the error criterion
in \eqref{eq-sc:mixed-error-crit}, then the method of proof in
\eqref{eq-sc:converse-1}--\eqref{eq-sc:converse-last} demonstrates that the
entropy $H(A)_{\rho}$ serves as a converse bound. Thus, in this case, we have
a statement of optimality. However, if we demand that the converse hold for
the error criterion in \eqref{eq-sc:mixed-error-crit-2}, then the same method
of proof gives a converse bound of $\frac{1}{2}I(XR;A)_{\phi}$. If we demand
that the converse hold for the error criterion in
\eqref{eq-sc:mixed-error-crit-3}, then the method of proof from
Exercise~\ref{ex-sc:alt-converse-bound}\ gives a converse bound of
$I(X;A)_{\phi}$. In general, these latter two lower bounds are incomparable,
but we can deduce that a sequence of compression schemes each meeting the
error criterion in \eqref{eq-sc:mixed-error-crit-2} should have a compression
rate larger than or equal to $\max\{\frac{1}{2}I(XR;A)_{\phi},I(X;A)_{\phi}%
\}$, given that the error criterion in \eqref{eq-sc:mixed-error-crit-2} is
more stringent than that in \eqref{eq-sc:mixed-error-crit-3}.

Let us consider a special example of the above situation, which allows for
comparing the two different error criteria in \eqref{eq-sc:mixed-error-crit}
and \eqref{eq-sc:mixed-error-crit-3} and the optimal rates. Suppose that the
mixed states $\rho_{x}$ act on orthogonal subspaces, and let $\rho_{A}%
=\sum_{x}p_{X}(x)\rho_{x}$ denote the expected density operator of the
ensemble. The states in $\{\rho_{x}\}$ are then perfectly distinguishable by a
measurement whose projectors project onto the different orthogonal subspaces.
As a consequence, Alice could perform this measurement and associate classical
labels with each of the states, leading to the following classical--quantum
state:%
\begin{equation}
\sigma_{XA}\equiv\sum_{x}p_{X}(x)|x\rangle\langle x|_{X}\otimes\rho_{A}^{x}.
\end{equation}
Furthermore, she can do this in principle without disturbing the state in any
way, and therefore the entropy of the state $\sigma_{XA}$ is equivalent to the
original entropy of the state $\rho_{A}$:%
\begin{equation}
H(A)_{\rho}=H(XA)_{\sigma}.
\end{equation}

Applying Schumacher compression to such a source meets the error criterion in
\eqref{eq-sc:mixed-error-crit} at an optimal rate equal to $H(A)_{\rho}$. This
compression rate is equal to%
\begin{equation}
H(XA)_{\sigma}=H(X)_{\sigma}+H(A|X)_{\sigma},
\end{equation}
and in this case, $H(A|X)_{\sigma}\geq0$ because the conditioning system is
classical. Furthermore, if at least one $\rho^{x}$ is truly mixed then we have
a strict inequality $H(A|X)_{\sigma}>0$. However, if we are only interested in
a scheme which meets the error criterion in \eqref{eq-sc:mixed-error-crit-3},
then a much better strategy than Schumacher compression is for Alice to
measure the classical variable $X$ directly, compress it with Shannon
compression, and transmit to Bob so that he can reconstruct the quantum states
at his end of the channel. The rate of compression here is equal to the
Shannon entropy $H(X)_{\sigma}$, which is provably lower than $H(XA)_{\rho}$
for this example. Since $I(X;A)_{\sigma}=H(X)_{\sigma}$ (see exercise below),
this example has an optimal rate for the error criterion in
\eqref{eq-sc:mixed-error-crit-3} and we see that there can be a strict
difference in optimal rates if we consider different error criteria in
mixed-state compression.

The next exercise asks you to verify that ensembles of mixed states on
orthogonal subspaces saturate the lower bound of $I(X;A)_{\phi}$.

\begin{exercise}
Show that the Holevo information of an ensemble of mixed states on orthogonal
subspaces has its Shannon information equal to its Holevo information. Thus,
this is an example of a class of ensembles that meet the lower bound
$I(X;A)_{\phi}$\ on compressibility.
\end{exercise}

\section{Concluding Remarks}

Schumacher compression was the first quantum Shannon-theoretic result
discovered and is the simplest one that we encounter in this book. The proof
is remarkably similar to the proof of Shannon's noiseless coding theorem, with
the main difference being that we should be more careful in the quantum case
not to be learning any more information than necessary when performing
measurements. The intuition that we gain for future quantum protocols is that
it often suffices to consider only what happens to a high probability subspace
rather than the whole space itself if our primary goal is to have a small
probability of error in a communication task. In fact, this intuition is the
same needed for understanding information-processing tasks such as
entanglement concentration, classical communication, private classical
communication, and quantum communication.

The problem of characterizing the lower and upper bounds for the quantum
compression rate of a mixed state quantum information source still remains
open, despite considerable efforts in this direction. It is only in special
cases, such as the example mentioned in Section~\ref{sec-schu-comp:variations}%
, that we know of a matching lower and upper bound as in Schumacher's original theorem.

\section{History and Further Reading}

\cite{OP93} devised the notion of a typical subspace, and
\cite{PhysRevA.51.2738} independently introduced typical subspaces and
additionally proved the quantum data-compression theorem. \cite{JS94} later
generalized this proof, and \cite{L95} further generalized the theorem to
mixed state sources. There are other generalizations in \cite{H98,BCFJS01}.
Several schemes for universal quantum data compression
exist~\citep{RHHH98,JP03,BHL06}, in which the sender does not need to have a
description of the quantum information source in order to compress its output.
There are also practical schemes for quantum data compression discussed in
work about quantum Huffman codes~\citep{BFGL00}.

Going beyond the settings considered here, researchers have considered error
exponents, strong converses, and second-order characterizations for quantum
data compression (we explain what these terms mean in
Section~\ref{sec-cc:history}). \cite{thesis1999winter} established a strong
converse theorem, \cite{PhysRevA.66.032321} derived error exponents, and
\cite{DL15} established second-order characterizations.

\chapter{Entanglement Manipulation}

\label{chap:ent-conc}Entanglement is one of the most useful resources in
quantum information. If Alice and Bob share noiseless entanglement in the form
of maximally entangled states, then they can teleport quantum bits between
each other with the help of classical communication, or they can double the
capacity of a noiseless qubit channel for transmitting classical information.
We will see further applications in Chapter~\ref{chap:EA-classical} in which
they can exploit noiseless entanglement to assist in the transmission of
classical or quantum data over a quantum channel.

Given the utility of maximal entanglement, a reasonable question to ask is
what two spatially separated parties can accomplish if they share pure
entangled states that are not maximally entangled. In the quantum
Shannon-theoretic setting, we make the further assumption that the two parties
share many copies of the same pure entangled state. We find out in this
chapter that they can \textquotedblleft concentrate\textquotedblright\ these
non-maximally entangled states to maximally entangled ebits by performing
local operations on their systems, and the optimal rate at which they can do
so is equal to the \textquotedblleft entropy of entanglement\textquotedblright%
\ (the quantum entropy of the marginal density operator of the original
state). Entanglement concentration%
\index{entanglement concentration}
is thus another fundamental task in noiseless quantum Shannon theory, and it
gives a different operational interpretation to the quantum entropy.
Entanglement concentration is perhaps complementary to Schumacher compression
in the sense that it gives a firm quantum information-theoretic interpretation
of the term \textquotedblleft ebit\textquotedblright\ (just as Schumacher
compression did for the term \textquotedblleft qubit\textquotedblright), and
it plays a part in demonstrating how the entropy of entanglement is the unique
measure of entanglement for pure bipartite states.

More generally, Alice and Bob could try to convert a large number of copies of
a pure state $|\psi\rangle_{AB}$ into as many copies as possible of another
bipartite pure state $|\phi\rangle_{AB}$, by performing only local operations
and exchanging classical messages (called \textquotedblleft
LOCC,\textquotedblright\
\index{LOCC}%
an abbreviation for \textquotedblleft local operations and classical
communication\textquotedblright). It is important to place a constraint on
their allowed operations---if they could perform arbitrary global operations
on their systems, then the task becomes trivial. We call such a task
\textquotedblleft entanglement manipulation\textquotedblright%
\index{entanglement manipulation}%
\ as a generalization of the aforementioned entanglement concentration task.

The main result in this chapter is that the optimal rate of conversion for
such an entanglement manipulation task is equal to $H(A)_{\psi}/H(A)_{\phi}$.
That is, if the goal is to convert $|\psi\rangle_{AB}^{\otimes n}$ by LOCC\ to
a state that has very high fidelity with $|\phi\rangle_{AB}^{\otimes nE}$,
then this transformation is possible for large $n$ if and only if $E\leq
H(A)_{\psi}/H(A)_{\phi}$. This conversion rate $H(A)_{\psi}/H(A)_{\phi}$ is
known as the \textquotedblleft entanglement manipulation
limit.\textquotedblright\ The achievability part of this theorem follows by
breaking the task into two parts. In the first part, Alice and Bob perform
entanglement concentration to convert $|\psi\rangle_{AB}^{\otimes n}$ to
$\approx nH(A)_{\psi}$ ebits. The next part makes use of a protocol called
\textquotedblleft entanglement dilution,\textquotedblright\
\index{entanglement dilution}%
which converts $nH(A)_{\phi}$ ebits to $\approx n$ copies of $|\phi
\rangle_{AB}$ by means of LOCC. Scaling the conversion rate appropriately, it
follows that $nH(A)_{\psi}$ ebits can be converted to $\approx n\left[
H(A)_{\psi}/H(A)_{\phi}\right]  $ copies of $|\phi\rangle_{AB}$, concluding
the direct part. The converse part of this theorem follows by exploiting the
properties of an information quantity called the relative entropy of
entanglement (the main properties that we need are that it does not increase under
the action of an LOCC\ channel, it is never smaller than the coherent
information, and it is equal to the entropy of entanglement for pure states).

The entanglement manipulation theorem mentioned above is one of the most
important results in the resource theory of quantum entanglement. It
demonstrates that the conversion of pure-state entanglement from one form to
another is essentially reversible in the limit of many copies. That is, when
$n$ is very large, Alice and Bob could first concentrate $|\psi\rangle
_{AB}^{\otimes n}$ to $\approx nH(A)_{\psi}$ ebits. Then they could execute an
entanglement dilution protocol to transform these $\approx nH(A)_{\psi}$ ebits
back to the original state $|\psi\rangle_{AB}^{\otimes n(1-\delta)}$, where
$\delta\in(0,1)$ is a small number that can be made to go to zero as $n$
becomes large. Alternatively, they could take the original state to some other
\textquotedblleft in-between\textquotedblright\ state besides the ebit at a
rate given by the ratio of the entropies of entanglement, but the main
advantage of the ebit is that its entropy of entanglement is equal to one, so
that we can think of it as a unit resource.

The technique for proving that the quantum entropy
\index{von Neumann entropy!operational interpretation}
is an achievable rate for entanglement concentration exploits the method of
types outlined in Sections~\ref{sec-ct:strong-typ} and
\ref{sec-qt:method-of-types-q} for classical and quantum typicality,
respectively (the most important property is
Property~\ref{prop-ct:min-card-typical-type} which states that the
exponentiated entropy is a lower bound on the size of a typical type class).
In hindsight, it is perhaps surprising that a typical type class is
exponentially large in the large $n$ limit (on the same order as the typical
set itself), and we soon discover the quantum Shannon-theoretic consequences
of this~result. The protocol for entanglement dilution is in some sense just that
for entanglement concentration \textquotedblleft run
backwards,\textquotedblright\ additionally making use of quantum teleportation.

We begin this chapter by discussing a simple example of entanglement
concentration for three copies of a state, and then we sketch out how more
general entanglement concentration and dilution protocols operate.
Section~\ref{sec-em:LOCC-rel-ent-enta} gives a formal definition of LOCC and
introduces the relative entropy of entanglement (an LOCC monotone).
Section~\ref{sec-ent-con:info-task}\ then details the information-processing
task for entanglement manipulation, and Section~\ref{sec-ent-con:theorem}
states the entanglement manipulation theorem and proves both the direct coding
and converse parts.

\section{Sketch of Entanglement Manipulation}

\label{sec-ent-con:example}This section sketches the main ideas underlying
entanglement concentration and dilution.

\subsection{Three-Copy Entanglement Concentration Example}%

\index{entanglement concentration}%
A simple example illustrates the main idea underlying the concentration of
entanglement. Consider the following partially entangled state:%
\begin{equation}
\left\vert \Phi_{\theta}\right\rangle _{AB}\equiv\cos(\theta)|00\rangle
_{AB}+\sin(\theta)|11\rangle_{AB}, \label{eq-ent-conc:example-state}%
\end{equation}
where $\theta$ is some parameter such that $0<\theta<\pi/2$. The Schmidt
decomposition (Theorem~\ref{thm-qt:schmidt}) guarantees that the above state
is the most general form to consider for a pure bipartite entangled state on
qubits. Now suppose that Alice and Bob share three copies of the above state.
We can rewrite the three copies of the above state using simple algebra:%
\begin{multline}
\left\vert \Phi_{\theta}\right\rangle _{A_{1}B_{1}}\left\vert \Phi_{\theta
}\right\rangle _{A_{2}B_{2}}\left\vert \Phi_{\theta}\right\rangle _{A_{3}%
B_{3}}=\cos^{3}(\theta)|000\rangle_{A}|000\rangle_{B}+\sin^{3}(\theta
)|111\rangle_{A}|111\rangle_{B}\\
+\sqrt{3}\cos(\theta)\sin^{2}(\theta)\frac{1}{\sqrt{3}}\left(  |110\rangle
_{A}|110\rangle_{B}+|101\rangle_{A}|101\rangle_{B}+|011\rangle_{A}%
|011\rangle_{B}\right) \\
+\sqrt{3}\cos^{2}(\theta)\sin(\theta)\frac{1}{\sqrt{3}}\left(  |100\rangle
_{A}|100\rangle_{B}+|010\rangle_{A}|010\rangle_{B}+\left\vert 001\right\rangle
_{A}\left\vert 001\right\rangle _{B}\right)  ,
\end{multline}
where we have relabeled all of the systems on Alice and Bob's respective sides as
$A\equiv A_{1}A_{2}A_{3}$ and $B\equiv B_{1}B_{2}B_{3}$. Observe that the
subspace with coefficient $\cos^{3}(\theta)$ whose states have zero
\textquotedblleft ones\textquotedblright\ is one-dimensional. The subspace
whose states have three \textquotedblleft ones\textquotedblright\ is also
one-dimensional. But the subspace with coefficient $\cos(\theta)\sin
^{2}(\theta)$ whose states have two \textquotedblleft ones\textquotedblright%
\ is three-dimensional, and the same holds for the subspace whose states each
have one \textquotedblleft one.\textquotedblright

A protocol for entanglement concentration in this scenario is then
straightforward. Alice performs a projective measurement consisting of the
operators $\Pi_{0}$, $\Pi_{1}$, $\Pi_{2}$, $\Pi_{3}$ where%
\begin{align}
\Pi_{0}  &  \equiv|000\rangle\langle000|_{A},\\
\Pi_{1}  &  \equiv|001\rangle\langle001|_{A}+|010\rangle\langle010|_{A}%
+|100\rangle\langle100|_{A},\\
\Pi_{2}  &  \equiv|110\rangle\langle110|_{A}+|101\rangle\langle101|_{A}%
+|011\rangle\langle011|_{A},\\
\Pi_{3}  &  \equiv|111\rangle\langle111|_{A}.
\end{align}
The subscript $i$\ of the projection operator $\Pi_{i}$\ corresponds to the
Hamming weight of the basis states in the corresponding subspace. Bob can
perform the same \textquotedblleft Hamming weight\textquotedblright%
\ measurement on his side. With probability $\cos^{6}(\theta)+\sin^{6}%
(\theta)$, the procedure fails because it results in $|000\rangle
_{A}|000\rangle_{B}$ or $|111\rangle_{A}|111\rangle_{B}$ which are both
product states with no entanglement at all. But with probability $3\cos
^{2}(\theta)\sin^{4}(\theta)$, the state is in the subspace with Hamming
weight two, and it has the following form:%
\begin{equation}
\frac{1}{\sqrt{3}}\left(  |110\rangle_{A}|110\rangle_{B}+|101\rangle
_{A}|101\rangle_{B}+|011\rangle_{A}|011\rangle_{B}\right)  ,
\end{equation}
and with probability $3\cos^{4}(\theta)\sin^{2}(\theta)$, the state is in the
subspace with Hamming weight one, and it has the following form:%
\begin{equation}
\frac{1}{\sqrt{3}}\left(  |100\rangle_{A}|100\rangle_{B}+|010\rangle
_{A}|010\rangle_{B}+|001\rangle_{A}|001\rangle_{B}\right)  .
\end{equation}
Alice and Bob can then perform local isometric operations on their respective
systems to rotate either of these states to a maximally entangled state with
Schmidt rank three:%
\begin{equation}
\frac{1}{\sqrt{3}}\left(  |0\rangle_{A}|0\rangle_{B}+|1\rangle_{A}%
|1\rangle_{B}+|2\rangle_{A}|2\rangle_{B}\right)  .
\end{equation}

\subsection{Sketch of Entanglement Concentration}%

\index{entanglement concentration}%
The simple protocol outlined above is the basis for the entanglement
concentration protocol, but it unfortunately fails with a non-negligible
probability in this case. On the other hand, if we allow Alice and Bob to have
a large number of copies of a pure bipartite entangled state, the probability
of failing becomes negligible in the asymptotic limit due to the properties of
typicality, and each type class subspace contains an exponentially large
maximally entangled state. The proof of the direct coding theorem in
Section~\ref{sec-ent-con:direct-cod-th} makes this intuition precise.

Generalizing the procedure outlined above to an arbitrary number of copies is
straightforward. Suppose Alice and Bob share $n$ copies of the partially
entangled state $\left\vert \Phi_{\theta}\right\rangle $. We can then write
the state as follows:%
\begin{align}
\left\vert \Phi_{\theta}\right\rangle _{A^{n}B^{n}}  &  =\sum_{k=0}^{n}%
\cos^{n-k}(\theta)\sin^{k}(\theta)\sum_{x:w(x)=k}|x\rangle_{A^{n}}%
|x\rangle_{B^{n}}\\
&  =\sum_{k=0}^{n}\sqrt{\binom{n}{k}}\cos^{n-k}(\theta)\sin^{k}(\theta)\left(
\frac{1}{\sqrt{\binom{n}{k}}}\sum_{x:w(x)=k}|x\rangle_{A^{n}}|x\rangle_{B^{n}%
}\right)  ,
\end{align}
where $w(x)$ is the Hamming weight of the binary vector $x$. Alice performs a
\textquotedblleft Hamming weight\textquotedblright\ measurement whose
projective operators are as follows:%
\begin{equation}
\Pi_{k}=\sum_{x:w(x)=k}|x\rangle\langle x|_{A^{n}},
\end{equation}
and the Schmidt rank of the maximally entangled state that they then share is
$\binom{n}{k}$.

A rough analysis of the protocol's performance for large $n$
 follows by exploiting Stirling's approximation and typicality. Recalling that Stirling's approximation
is $n!\approx\sqrt{2\pi n}\left(  n/e\right)  ^{n}$ and thinking of $k/n$ as a constant (as is the case for typical sequences), this gives%
\begin{align}
\binom{n}{k}  &  =\frac{n!}{k!n-k!}\approx\frac{\sqrt{2\pi n}\left(
n/e\right)  ^{n}}{\sqrt{2\pi k}\left(  k/e\right)  ^{k}\sqrt{2\pi\left(
n-k\right)  }\left(  \left(  n-k\right)  /e\right)  ^{n-k}} \label{eq:Stirling-approx-ent-conc-1}\\
&  =\sqrt{\frac{n}{2\pi k\left(  n-k\right)  }}\frac{n^{n}}{\left(
n-k\right)  ^{n-k}k^{k}}\\
&  =\sqrt{\frac{1}{2\pi n \left(\tfrac{k}{n}\right)\left(  1-\tfrac{k}{n}\right)  }}\ \left(  \frac{n-k}{n}\right)  ^{-\left(
n-k\right)  }\left(  \frac{k}{n}\right)  ^{-k}\\
&  =[\operatorname{poly}(n)]^{-1}\cdot  2^{n\left[  -\left(  \left(  n-k\right)
/n\right)  \log\left(  \left(  n-k\right)  /n\right)  -(k/n)\log(k/n)\right]
}\\
&  =[\operatorname{poly}(n)]^{-1}\cdot  2^{nh_{2}(k/n)},
\label{eq:Stirling-approx-ent-conc-last}
\end{align}
where$~h_{2}$ is the binary entropy function in \eqref{eq-intro:bin-entropy}
and~poly$(n)$ indicates a term at most polynomial in$~n$. When $n$ is large,
the exponential term $2^{nh_{2}(k/n)}$ dominates the polynomial
$\left(2\pi n \left(\tfrac{k}{n}\right)\left(  1-\tfrac{k}{n}\right)  \right)^{-1/2}$, so that the polynomial term begins to behave merely
as a constant. So, the protocol is for Alice to perform a strongly typical
subspace measurement with respect to the distribution $\left(  \cos^{2}%
(\theta),\sin^{2}(\theta)\right)  $, and the state then reduces to the
following one with high probability:%
\begin{equation}
\frac{1}{\sqrt{\mathcal{N}}}\sum_{\substack{k=0\ :\\\left\vert k/n-\sin
^{2}(\theta)\right\vert \leq\delta}}^{n}\sqrt{\binom{n}{k}}\cos
^{n-k}(\theta)\sin^{k}(\theta)\left(  \frac{1}{\sqrt{\binom{n}{k}}}%
\sum_{x:w(x)=k}|x\rangle_{A^{n}}|x\rangle_{B^{n}}\right)  ,
\end{equation}
where $\mathcal{N}\geq1-\varepsilon$ is an appropriate normalization constant.
Alice and Bob then both perform a Hamming weight measurement and the state
reduces to a state of the form%
\begin{equation}
\frac{1}{\sqrt{[\operatorname{poly}(n)]^{-1} \cdot 2^{nh_{2}(k/n)}}}\sum_{x:w(x)=k}%
|x\rangle_{A^{n}}|x\rangle_{B^{n}},
\end{equation}
depending on the outcome$~k$ of the measurement. In the above, we have employed the approximation in
\eqref{eq:Stirling-approx-ent-conc-1}--\eqref{eq:Stirling-approx-ent-conc-last}, being valid since $k/n$ is approximately constant $k/n\approx \sin
^{2}(\theta)$, due to typicality.
The above state is a
maximally entangled state with Schmidt rank $[\operatorname{poly}(n)]^{-1} \cdot2^{nh_{2}(k/n)}$, and it
follows that%
\begin{equation}
h_{2}(k/n)\geq h_{2}(  \sin^{2}(\theta))  -\delta = h_{2}(  \cos^{2}(\theta))  -\delta,
\end{equation}
from the assumption that the state first projects into the typical subspace.
Alice and Bob can then perform local operations to rotate this state to
approximately $nh_{2}(  \cos^{2}(\theta))  $ ebits. Thus, this
procedure concentrates the original non-maximally entangled state to ebits at
a rate equal to the entropy of entanglement of the state $\left\vert
\Phi_{\theta}\right\rangle _{AB}$ in \eqref{eq-ent-conc:example-state}. The
above proof is a bit rough, and it applies only to entangled qubit systems in
a pure state. The proof of the direct coding theorem in
Section~\ref{sec-ent-con:direct-cod-th} generalizes this proof to pure
entangled states on $d$-dimensional systems.

\subsection{Sketch of Entanglement Dilution}

\label{sec-em:ent-dil-sketch}Entanglement dilution is easier to sketch out if
we are not concerned about the classical communication cost of the protocol.
Suppose that the goal is to create $n$ copies of $|\phi\rangle_{AB}$ from
$\approx nH(A)_{\phi}$ ebits. Then Alice can prepare $n$ copies of
$|\phi\rangle_{AB}$ in her laboratory. She performs Schumacher compression on
the $B$ systems, which compresses these systems to $\approx nH(A)_{\phi}$
qubits while causing only a small disturbance to the state (when $n$ is
large). If she and Bob share $\approx nH(A)_{\phi}$ ebits, then she can
teleport the compressed qubits to Bob (this costs $\approx n2H(A)_{\phi}$
classical bits). Bob receives the compressed qubits and decompresses them,
which is the end of the protocol. So this version of the entanglement dilution
protocol is rather straightforward. Later on, we see how the classical
communication cost can be much smaller---in fact a modified protocol requires
a number of classical bits required which is sublinear in $n$, so that the
rate of classical communication needed vanishes in the large $n$ limit.

\section{LOCC\ and Relative Entropy of Entanglement}

\label{sec-em:LOCC-rel-ent-enta}%
\index{local operations and classical communication}%
Before describing the information-processing task for entanglement
manipulation, we should formally define what we mean by LOCC, the allowed set
of operations. An LOCC\ channel consists of a finite number of compositions of
the following:

\begin{enumerate}
\item Alice performs a quantum instrument, which has both a quantum and
classical output. She forwards the classical output to Bob, who then performs
a quantum channel conditioned on the classical data received. This sequence of
actions corresponds to a channel of the following form:%
\begin{equation}
\sum_{x}\mathcal{F}_{A}^{x}\otimes\mathcal{G}_{B}^{x},
\label{eq-em:LOCC-channel}%
\end{equation}
where $\{\mathcal{F}_{A}^{x}\}$ is a collection of completely positive maps
such that $\sum_{x}\mathcal{F}_{A}^{x}$ is a quantum channel and
$\{\mathcal{G}_{B}^{x}\}$ is a collection of quantum channels.

\item The situation is reversed, with Bob performing the initial instrument,
who forwards the classical data to Alice, who then performs a quantum channel
conditioned on the classical data.\ This sequence of actions corresponds to a
channel of the form in \eqref{eq-em:LOCC-channel}, with the $A$ and $B$ labels switched.
\end{enumerate}

An information
\index{LOCC}%
measure is an \textit{LOCC\ monotone}
\index{LOCC monotone}%
if it is non-increasing with respect to an LOCC channel. One such information
measure is the relative entropy of entanglement, defined as follows:

\begin{definition}
[Relative Entropy of Entanglement]%
\index{relative entropy of entanglement}%
Let $\rho_{AB}\in\mathcal{D}(\mathcal{H}_{A}\otimes\mathcal{H}_{B})$. The
relative entropy of entanglement of $\rho_{AB}$ is equal to the
\textquotedblleft relative entropy distance\textquotedblright\ between
$\rho_{AB}$ and the closest separable state:%
\begin{equation}
E_{R}(A;B)_{\rho}\equiv\min_{\sigma_{AB}\in\operatorname{SEP}(A:B)}D(\rho
_{AB}\Vert\sigma_{AB}).
\end{equation}

\end{definition}

That $E_{R}(A;B)_{\rho}$ is an LOCC\ monotone follows from the monotonicity of
relative entropy with respect to channels (Theorem~\ref{thm-qie:mono-rel-ent}%
)\ and the fact that LOCC\ channels take separable states to separable states.
That is, from the definition of LOCC\ channels given above, we can see that
such operations have Kraus operators of the form $\{F_{A}^{z}\otimes G_{B}%
^{z}\}$, so that an LOCC\ channel $\Lambda_{AB}$ acts as follows on a
separable state $\sigma_{AB}=\sum_{y}p(y)\omega_{A}^{y}\otimes\tau_{B}^{y}$:%
\begin{align}
\Lambda_{AB}(\sigma_{AB})  &  =\sum_{z}\left(  F_{A}^{z}\otimes G_{B}%
^{z}\right)  \left[  \sum_{y}p(y)\omega_{A}^{y}\otimes\tau_{B}^{y}\right]
\left(  F_{A}^{z}\otimes G_{B}^{z}\right)  ^{\dag}\\
&  =\sum_{z,y}p(y)F_{A}^{z}\omega_{A}^{y}\left(  F_{A}^{z}\right)  ^{\dag
}\otimes G_{B}^{z}\tau_{B}^{y}\left(  G_{B}^{z}\right)  ^{\dag},
\end{align}
which is clearly a separable state. So this means that for any separable state
$\sigma_{AB}$ and LOCC\ channel $\Lambda_{AB\rightarrow A^{\prime}B^{\prime}}%
$, we find that%
\begin{equation}
D(\rho_{AB}\Vert\sigma_{AB})\geq D(\Lambda_{AB\rightarrow A^{\prime}B^{\prime
}}(\rho_{AB})\Vert\Lambda_{AB\rightarrow A^{\prime}B^{\prime}}(\sigma
_{AB}))\geq E_{R}(A^{\prime};B^{\prime})_{\Lambda(\rho)},
\end{equation}
where the first inequality follows from the monotonicity of relative entropy
(Theorem~\ref{thm-qie:mono-rel-ent}). Since the inequality holds for all
$\sigma_{AB}\in\operatorname{SEP}(A:B)$, we can conclude that%
\begin{equation}
E_{R}(A;B)_{\rho}\geq E_{R}(A^{\prime};B^{\prime})_{\Lambda(\rho)},
\end{equation}
which is equivalent to the following:

\begin{theorem}
\label{thm-em:rel-ent-LOCC-mono}The relative entropy of entanglement is an LOCC\ monotone.
\end{theorem}

\noindent We need two other properties of the relative entropy of entanglement:

\begin{proposition}
\label{prop-em:rel-ent-enta-bigger-coh-info}Let $\rho_{AB}\in\mathcal{D}%
(\mathcal{H}_{A}\otimes\mathcal{H}_{B})$. Then the relative entropy of
entanglement is never smaller than the coherent information:%
\begin{equation}
E_{R}(A;B)_{\rho}\geq\max\{I(A\rangle B)_{\rho},I(B\rangle A)_{\rho}\}.
\end{equation}

\end{proposition}

\begin{proof}
Let $\sigma_{AB}\in\operatorname{SEP}(A\!:\!B)$. Then%
\begin{equation}
\sigma_{AB}=\sum_{y}p(y)\omega_{A}^{y}\otimes\tau_{B}^{y}\leq\sum_{y}%
p(y)I_{A}\otimes\tau_{B}^{y}=I_{A}\otimes\sigma_{B}.
\end{equation}
The operator inequality follows because $\omega_{A}^{y}\leq I_{A}$, which
implies that $\sum_{y}p(y)\left(  I_{A}-\omega_{A}^{y}\right)  \otimes\tau
_{B}^{y}$ is positive semi-definite. We can then conclude that%
\begin{align}
D(\rho_{AB}\Vert\sigma_{AB})  &  \geq D(\rho_{AB}\Vert I_{A}\otimes\sigma
_{B})\\
&  \geq\min_{\sigma_{B}}D(\rho_{AB}\Vert I_{A}\otimes\sigma_{B})\\
&  =I(A\rangle B)_{\rho}.
\end{align}
The first inequality follows from Proposition~\ref{prop-qie:rel-ent-s-s'}. The
equality follows from Exercise~\ref{ex-qie:rel-ent-cond-ent}. Since the
inequality holds for all $\sigma_{AB}\in\operatorname{SEP}(A\!:\!B)$, we can
conclude that $E_{R}(A;B)_{\rho}\geq I(A\rangle B)_{\rho}$. The other
inequality follows by a symmetric proof.
\end{proof}

\begin{proposition}
\label{prop-em:rel-ent=ent-of-ent}Let $|\psi\rangle_{AB}\in\mathcal{H}%
_{A}\otimes\mathcal{H}_{B}$ be a pure bipartite state. Then the relative
entropy of entanglement is equal to the entropy of entanglement:%
\begin{equation}
H(A)_{\psi}=E_{R}(A;B)_{\psi}.
\end{equation}

\end{proposition}

\begin{proof}
From Proposition~\ref{prop-em:rel-ent-enta-bigger-coh-info}, we know that
$E_{R}(A;B)_{\psi}\geq I(A\rangle B)_{\psi}=H(A)_{\psi}$. So it remains to
prove the other inequality. Suppose that $|\psi\rangle_{AB}$ has a Schmidt
decomposition as follows:%
\begin{equation}
|\psi\rangle_{AB}=\sum_{x}\sqrt{p(x)}|x\rangle_{A}\otimes|x\rangle_{B}.
\end{equation}
Let $\Delta$ denote the following channel:%
\begin{equation}
\Delta(X)\equiv PXP+(I-P)X(I-P),
\end{equation}
where $P\equiv\sum_{x}|x\rangle\langle x|_{A}\otimes|x\rangle\langle x|_{B}$.
Let%
\begin{equation}
\overline{\psi}_{AB}\equiv\sum_{x}p(x)|x\rangle\langle x|_{A}\otimes
|x\rangle\langle x|_{B}.
\end{equation}
Note that $\overline{\psi}_{AB}$ is a separable state. Consider that%
\begin{align}
H(A)_{\psi}  &  =I(A\rangle B)_{\psi}\\
&  =D(\psi_{AB}\Vert I_{A}\otimes\psi_{B})\\
&  \geq D(\Delta(\psi_{AB})\Vert\Delta(I_{A}\otimes\psi_{B}))\\
&  =D(\psi_{AB}\Vert\overline{\psi}_{AB})\\
&  \geq\min_{\sigma_{AB}\in\operatorname{SEP}(A:B)}D(\psi_{AB}\Vert\sigma
_{AB})\\
&  =E_{R}(A;B)_{\psi}.
\end{align}
The second equality follows from Exercise~\ref{ex-qie:rel-ent-cond-ent}. The
first inequality follows from the monotonicity of quantum relative entropy
(Theorem~\ref{thm-qie:mono-rel-ent}). The third equality follows because
$\Delta(\psi_{AB})=\psi_{AB}$ and
\begin{align}
\Delta(I_{A}\otimes\psi_{B})  &  =P(I_{A}\otimes\psi_{B})P+(I-P)(I_{A}%
\otimes\psi_{B})(I-P).\\
&  =\overline{\psi}_{AB}+(I-P)(I_{A}\otimes\psi_{B})(I-P).
\end{align}
Since $\psi_{AB}$ does not have support outside of the subspace onto which $P$
projects, it follows that%
\begin{equation}
D(\psi_{AB}\Vert\overline{\psi}_{AB}+(I-P)(I_{A}\otimes\psi_{B})(I-P))=D(\psi
_{AB}\Vert\overline{\psi}_{AB}).
\end{equation}
This concludes the proof.
\end{proof}

\section{Entanglement Manipulation Task}

\label{sec-ent-con:info-task}We can now define an $\left(  n,m/n,\varepsilon
\right)  $ entanglement manipulation protocol. Alice and Bob begin with $n$
copies of a pure bipartite, entangled state $|\psi\rangle_{AB}$. They then
perform an LOCC\ channel $\Lambda_{A^{n}B^{n}\rightarrow A^{m}B^{m}}^{(n)}$ in
an attempt to convert the original state $(|\psi\rangle_{AB})^{\otimes n}$ to
$m$ copies of another bipartite pure state $|\phi\rangle_{AB}$. Let
$\omega_{A^{m}B^{m}}$\ denote the state after the LOCC\ channel:%
\begin{equation}
\omega_{A^{m}B^{m}}\equiv\Lambda_{A^{n}B^{n}\rightarrow A^{m}B^{m}}^{(n)}%
(\psi_{AB}^{\otimes n}).
\end{equation}
The protocol has $\varepsilon$ error if the final state $\omega_{A^{m}B^{m}}$
is $\varepsilon$-close to $\phi_{AB}^{\otimes m}$:%
\begin{equation}
\frac{1}{2}\left\Vert \omega_{A^{m}B^{m}}-\phi_{AB}^{\otimes m}\right\Vert
_{1}\leq\varepsilon,
\end{equation}
where $\varepsilon\in\left[  0,1\right]  $ and the rate of entanglement
conversion is $m/n$.

We say that a particular rate $E$ of entanglement manipulation is
\textit{achievable} if there exists an $\left(  n,E-\delta,\varepsilon\right)
$ entanglement manipulation protocol for all $\varepsilon\in(0,1)$, $\delta
>0$, and sufficiently large $n$. The \textit{entanglement manipulation limit}
$E(\psi\rightarrow\phi)$\ for the conversion $|\psi\rangle_{AB}\rightarrow
|\phi\rangle_{AB}$ is equal to the supremum of all achievable rates.

\section{The Entanglement Manipulation Theorem}

\label{sec-ent-con:theorem}We first state the entanglement manipulation
theorem and then prove it below in two parts (the converse theorem and the
direct coding theorem).

\begin{theorem}
[Entanglement Manipulation]\label{thm-ent-con:ent-con}Let $|\psi\rangle
_{AB},|\phi\rangle_{AB}\in\mathcal{H}_{A}\otimes\mathcal{H}_{B}$ be pure
bipartite states. The entanglement manipulation limit for the conversion
$|\psi\rangle_{AB}\rightarrow|\phi\rangle_{AB}$ is equal to the ratio
$H(A)_{\psi}/H(A)_{\phi}$:%
\begin{equation}
E(\psi\rightarrow\phi)=\frac{H(A)_{\psi}}{H(A)_{\phi}}.
\end{equation}

\end{theorem}

\begin{remark}
Theorem~\ref{thm-ent-con:ent-con} implies that the entanglement concentration
and dilution protocols are individually optimal. That is, if the goal is to
convert $n$ copies of $|\psi\rangle_{AB}$ to as many ebits as possible, then
the maximal rate of ebit generation is equal to $H(A)_{\psi}$. Furthermore, if
the goal is to convert $nR$ ebits to $n$ copies of $|\psi\rangle_{AB}$, then
the minimal rate $R$ of ebit consumption is equal to $H(A)_{\psi}$.
\end{remark}

\subsection{The Converse Theorem}

We now prove the converse theorem for entanglement manipulation, i.e., that
the entanglement manipulation limit for the conversion $|\psi\rangle
_{AB}\rightarrow|\phi\rangle_{AB}$ does not exceed $H(A)_{\psi}/H(A)_{\phi}$.
Suppose that there is a sequence of LOCC\ transformations $\left\{
\Lambda^{(n)}\right\}  $,\ each of which takes $n$ copies of a pure state
$|\psi\rangle_{AB}$ to $m_{n}$ approximate copies of a pure state
$|\phi\rangle_{AB}$. That is, $\Lambda^{(n)}$ is such that%
\begin{equation}
\frac{1}{2}\left\Vert \Lambda^{(n)}(\psi_{AB}^{\otimes n})-\phi_{AB}^{\otimes
m_{n}}\right\Vert _{1}\leq\varepsilon,
\end{equation}
where $\varepsilon\in(0,1)$. Let $\omega_{A^{m_{n}}B^{m_{n}}}\equiv
\Lambda^{(n)}(\psi_{AB}^{\otimes n})$. To bound the rate $m_{n}/n$, we use the
relative entropy of entanglement. Consider that%
\begin{align}
nH(A)_{\psi}  &  =H(A^{n})_{\psi^{\otimes n}}\\
&  =E_{R}(A^{n};B^{n})_{\psi^{\otimes n}}\\
&  \geq E_{R}(A^{m_{n}};B^{m_{n}})_{\omega}\\
&  \geq I(A^{m_{n}}\rangle B^{m_{n}})_{\omega}\\
&  \geq I(A^{m_{n}}\rangle B^{m_{n}})_{\phi^{\otimes m_{n}}}-f(m_{n}%
,\varepsilon)\\
&  =H(A^{m_{n}})_{\phi^{\otimes m_{n}}}-f(m_{n},\varepsilon)\\
&  =m_{n}H(A)_{\phi}-f(m_{n},\varepsilon).
\end{align}
The first equality follows because the entropy is additive for a
tensor-product state. The second equality follows from
Proposition~\ref{prop-em:rel-ent=ent-of-ent}. The first inequality follows
because the relative entropy of entanglement is an LOCC\ monotone
(Theorem~\ref{thm-em:rel-ent-LOCC-mono}). The second inequality follows from
Proposition~\ref{prop-em:rel-ent-enta-bigger-coh-info}. The third inequality
is a consequence of continuity of conditional entropy (the AFW\ inequality),
with $f(m_{n},\varepsilon)\equiv2\varepsilon m_{n}\log\dim(\mathcal{H}%
_{A})+g_2(\varepsilon)$. The third equality follows because the coherent information of a
pure state is equal to the marginal entropy. The last equality follows because
the entropy of a tensor-product state is additive. Putting everything
together, we find that%
\begin{equation}
\frac{m_{n}}{n}\left(  1-2\varepsilon\log\dim(\mathcal{H}_{A})/H(A)_{\phi
}\right)  \leq\frac{H(A)_{\psi}}{H(A)_{\phi}}+\frac{g_2(\varepsilon)}{nH(A)_{\phi}}.
\end{equation}
Thus, if we are considering a sequence of $(n,m_{n}/n,\varepsilon)$
entanglement manipulation protocols with rate $E-\delta_{n}=m_{n}/n$, such
that $\lim_{n\rightarrow\infty}\delta_{n}=0$, then the above bound becomes%
\begin{equation}
\left(  E-\delta_{n}\right)  \left(  1-2\varepsilon\log\dim(\mathcal{H}%
_{A})/H(A)_{\phi}\right)  \leq\frac{H(A)_{\psi}}{H(A)_{\phi}}+\frac{g_2(\varepsilon)}{nH(A)_{\phi}}.
\end{equation}
Taking the limit as $n\rightarrow\infty$ and $\varepsilon\rightarrow0$, we
find that any achievable rate $E$ of entanglement manipulation for
$\psi\rightarrow\phi$ necessarily satisfies the following bound:%
\begin{equation}
E\leq\frac{H(A)_{\psi}}{H(A)_{\phi}}.
\end{equation}

\subsection{The Direct Coding Theorem}

\label{sec-ent-con:direct-cod-th}As discussed in the introduction of this
chapter, we break the direct coding theorem into two parts:\ entanglement
concentration and entanglement dilution. We begin by discussing entanglement
concentration. To do so, it is helpful to discuss a related, exclusively
classical task known as randomness concentration (also known as randomness extraction).

\subsubsection{Randomness Concentration}%

\index{randomness concentration}%
In a randomness concentration protocol, the goal is to extract as many
approximately uniformly random bits as possible from a given distribution.
Since we are operating in a quantum Shannon theoretic regime, let us suppose
that a sequence $x^{n}$ is generated according to an i.i.d.~distribution%
\begin{equation}
p_{X^{n}}(x^{n})\equiv\prod\limits_{i=1}^{n}p_{X}(x_{i}).
\end{equation}
Recall the method of types from Section~\ref{sec-ct:method-of-types}. Suppose
that Alice performs a mapping%
\begin{equation}
x^{n}\rightarrow(t(x^{n}),f_{t}(x^{n})),
\end{equation}
where $t(x^{n})$ is the type (or empirical distribution) of the sequence
$x^{n}$ and the function $f_{t}(x^{n})$ is an index keeping track of the
ordering of the symbols in the sequence $x^{n}$ for a given type class
$T_{t}^{X^{n}}$. Note that this mapping is reversible (given an output, one
can determine the input uniquely).

For example, all three-bit sequences map in this way as follows:%
\begin{align}
000  &  \rightarrow\left(  0,0\right)  ,\ \ \ \ 001\rightarrow\left(
1,0\right)  ,\ \ \ \ 010\rightarrow\left(  1,1\right)  ,\\
011  &  \rightarrow\left(  2,0\right)  ,\ \ \ \ 100\rightarrow\left(
1,2\right)  ,\ \ \ \ 101\rightarrow\left(  2,1\right)  ,\\
110  &  \rightarrow\left(  2,2\right)  ,\ \ \ \ 111\rightarrow\left(
3,0\right)  .
\end{align}
For binary sequences, the type is the Hamming weight. Thus, $000$ has type $0$
and is the only sequence in this type class, thus receiving an index $0$.
Also, $011$ has type $2$ and we label it as the first sequence in this type
class, indexed with $0$. The sequence $101$ has type $2$ and it is the second
sequence in this type class, indexed with $1$, etc.

What is the use of performing this mapping?\ First consider that the joint
probability of observing a particular type $t$ and index $f$ is equal to%
\begin{equation}
p_{t(X^{n}),f_{t}(X^{n})}(t,f)=p_{X^{n}}(x^{n}),
\end{equation}
where $x^{n}\in T_{t}^{X^{n}}$. This is because the mapping $x^{n}%
\rightarrow(t(x^{n}),f_{t}(x^{n}))$ is reversible. Consider furthermore that
for $x^{n}\in T_{t}^{X^{n}}$, we have that%
\begin{equation}
p_{X^{n}}(x^{n})=\prod\limits_{x\in\mathcal{X}}p_{X}(x)^{N(x|x^{n})}%
=\prod\limits_{x\in\mathcal{X}}p_{X}(x)^{N(x|x_{t}^{n})}=p_{X^{n}}(x_{t}^{n}),
\end{equation}
where $x_{t}^{n}$ is a representative sequence of the type class $T_{t}%
^{X^{n}}$ (having type $t$ as well). Then the probability for observing a
particular type $t$ is equal to%
\begin{equation}
p_{t(X^{n})}(t)=\sum_{x^{n}\in T_{t}^{X^{n}}}p_{X^{n}}(x^{n})=\sum_{x^{n}\in
T_{t}^{X^{n}}}p_{X^{n}}(x_{t}^{n})=\left\vert T_{t}^{X^{n}}\right\vert
\prod\limits_{x\in\mathcal{X}}p_{X}(x)^{N(x|x_{t}^{n})}.
\end{equation}
That is, $p_{t(X^{n})}(t)$ just depends on the size of the type class
$\left\vert T_{t}^{X^{n}}\right\vert $ and the empirical distribution of the
type $t$. These considerations then imply that the conditional probability
distribution $p_{f_{t}(X^{n})|t(X^{n})}$ is uniform because%
\begin{equation}
p_{f_{t}(X^{n})|t(X^{n})}(f|t)=\frac{p_{t(X^{n}),f_{t}(X^{n})}(t,f)}%
{p_{t(X^{n})}(t)}=\frac{1}{\left\vert T_{t}^{X^{n}}\right\vert }.
\label{eq-em:uniform-RV-types}%
\end{equation}
That is, conditioned on observing a particular type $t$, all of the sequences
in the type class $T_{t}^{X^{n}}$ are uniformly distributed. Thus, the mapping
$x^{n}\rightarrow(t(x^{n}),f_{t}(x^{n}))$ \textquotedblleft
reshapes\textquotedblright\ the distribution of $x^{n}$ in the above way.

This \textquotedblleft distribution reshaping\textquotedblright\ leads to a
first idea for a randomness concentration protocol. Given a sequence $x^{n}$,
send it through the reversible mapping $x^{n}\rightarrow(t(x^{n}),f_{t}%
(x^{n}))$. Given the type $t$, the value $f=f_{t}(x^{n})$ is uniformly random
on a set of size $\left\vert T_{t}^{X^{n}}\right\vert $, so that we recover
$\log\left\vert T_{t}^{X^{n}}\right\vert $ uniformly random bits.

There are two main problems with the above method. First, some type classes
are very small and lead to little randomness or none at all. Second, the
amount of randomness that the above procedure yields is not consistent:\ it
varies from type to type. What we would prefer is to have a method which takes
a random sequence $X^{n}$ and maps it to exactly $nR$ bits, such that the
distribution of these $nR$ bits is nearly indistinguishable from a uniform distribution.

How can we accomplish this? To start, we should first have a preprocessing
step in which we only proceed with the above method if $x^{n}$ is a strongly
typical sequence and otherwise declare failure. Equivalently, we only proceed
if $t(x^{n})$ is a strongly typical type, such that the empirical distribution
$N(x|x^{n})$ satisfies $\max_{x}\left\vert N(x|x^{n})-p_{X}(x)\right\vert
\leq\delta$ for some $\delta>0$. For sufficiently large $n$, we are guaranteed
that this preprocessing step fails with probability no larger than an
arbitrarily small constant $\varepsilon\in(0,1)$, due to the law of large
numbers (or the \textquotedblleft high probability\textquotedblright\ property
of typicality). By Property~\ref{prop-ct:min-card-typical-type}, this
preprocessing step guarantees that every strongly typical type class has size
bounded as follows:%
\begin{equation}
\left\vert T_{t}^{X^{n}}\right\vert \geq2^{n\left[  H(X)-\eta(\left\vert
\mathcal{X}\right\vert \delta)-\left\vert \mathcal{X}\right\vert \frac{1}%
{n}\log (n+1)\right]  }.
\end{equation}
Note that we also have the following upper bound:%
\begin{equation}
\left\vert T_{t}^{X^{n}}\right\vert \leq2^{n\left[  H(X)+c\delta\right]  },
\end{equation}
because the size of a typical type class cannot exceed the size of the
strongly typical set, for some constant $c>0$. Thus, the preprocessing step
solves the first problem because it guarantees that we will have at least
$n\left[  H(X)-\eta(\left\vert \mathcal{X}\right\vert \delta)-\left\vert
\mathcal{X}\right\vert \frac{1}{n}\log (n+1)\right]  $ uniformly random bits if it
is successful.

From here, we can then solve the second problem by performing a hashing
function, in order to hash down every typical type class to a set of size
$2^{n\left[  H(X)-\eta(\left\vert \mathcal{X}\right\vert \delta)-\left\vert
\mathcal{X}\right\vert \frac{1}{n}\log (n+1)\right]  }2^{-n\delta}$, so that we
decrease the size of the set by a factor of $2^{n\delta}$. Even though the size
drops by an amount that is exponential in $n$, we only lose $\delta$ on the
\textit{rate} of randomness concentration, which is the main parameter of
interest in the large $n$ limit. That is, at the end of the protocol, we will
be left with $n\left[  H(X)-\eta(\left\vert \mathcal{X}\right\vert
\delta)-\left\vert \mathcal{X}\right\vert \frac{1}{n}\log (n+1)-\delta\right]  $
bits that are nearly indistinguishable from uniformly random bits. As
$n\rightarrow\infty$, the rate of randomness concentration is equal to $H(X)$
uniformly random bits per source symbol.

To have a complete proof, we need the following hashing lemma, which gives a
way for hashing down a uniform random variable on a larger set to one on a
much smaller set:

\begin{lemma}
[Hashing]\label{lem-em:hashing}Let $k$ and $l$ be positive integers such that
$k\geq l$ (in fact think of $k$ as being much larger than $l$). Let $W_{k}$ be
uniformly distributed on the set $\{1,\ldots,k\}$, $W_{l}$ be uniformly
distributed on the set $\{1,\ldots,l\}$, and $W_{r}$ be uniformly distributed
on the set $\left\{  1,\ldots,r=\left\lceil k/l\right\rceil \right\}  $. Then
there exists a one-to-one function $g:\left\{  1,\ldots,k\right\}
\rightarrow\left\{  1,\cdots,l\right\}  \times\left\{  1,\ldots,r\right\}  $
such that%
\begin{equation}
\frac{1}{2}\left\Vert p_{W_{l}}\times p_{W_{r}}-p_{g(W_{k})}\right\Vert
_{1}\leq\frac{l}{k}.
\end{equation}

\end{lemma}

\begin{proof}
Divide the set $\left\{  1,\ldots,k\right\}  $ into $l$ bins, each of which
has size no more than $r=\left\lceil k/l\right\rceil $. Then take $g(w_{k})$
to be the one-to-one mapping which outputs the bin index and the location
inside of that bin. The distribution of $g(W_{k})$ is uniformly random, equal
to $1/k$ for $k$ of the possible $l\cdot\left\lceil k/l\right\rceil $ output
values and zero for the others (the unfilled locations if $k/l$ is not an
integer). The distribution of the joint random variable $(W_{l},W_{r})$ is
uniformly random also, equal to $1/\left(  l\cdot\left\lceil k/l\right\rceil
\right)  $ for each of the possible $l\cdot\left\lceil k/l\right\rceil $
values. The distributions then overlap on exactly $k$ of the possible output
values, implying that the trace distance between the two distributions is as
follows:%
\begin{equation}
\left\Vert p_{g(W_{k})}-p_{W_{l}}\times p_{W_{r}}\right\Vert _{1}=k\left\vert
\frac{1}{k}-\frac{1}{l\cdot\left\lceil k/l\right\rceil }\right\vert +\left(
l\cdot\left\lceil k/l\right\rceil -k\right)  \left\vert \frac{1}%
{l\cdot\left\lceil k/l\right\rceil }\right\vert .
\end{equation}
Consider that%
\begin{align}
k\left\vert \frac{1}{k}-\frac{1}{l\cdot\left\lceil k/l\right\rceil
}\right\vert  &  =k\left\vert \frac{\left\lceil k/l\right\rceil }%
{k\cdot\left\lceil k/l\right\rceil }-\frac{k/l}{k\cdot\left\lceil
k/l\right\rceil }\right\vert \\
&  =\frac{1}{\left\lceil k/l\right\rceil }\left\vert \left\lceil
k/l\right\rceil -k/l\right\vert \leq\frac{1}{\left\lceil k/l\right\rceil }%
\leq\frac{l}{k}.
\end{align}
Furthermore,%
\begin{align}
\left(  l\cdot\left\lceil k/l\right\rceil -k\right)  \left\vert \frac
{1}{l\cdot\left\lceil k/l\right\rceil }\right\vert  &  =\left(  \left\lceil
k/l\right\rceil -k/l\right)  \left\vert \frac{1}{\left\lceil k/l\right\rceil
}\right\vert \\
&  \leq\frac{1}{\left\lceil k/l\right\rceil }\leq\frac{l}{k}.
\end{align}
This concludes the proof.
\end{proof}

We can now specify the complete protocol for randomness concentration. It
begins with a sequence $x^{n}$ being generated at random according to
$p_{X^{n}}(x^{n})$. Alice computes whether it is a strongly typical sequence,
declaring failure of the protocol in case it is not. Let $\widetilde{X}^{n}$
be a random variable with the following distribution:%
\begin{equation}
p_{\widetilde{X}^{n}}(x^{n})\equiv\left\{
\begin{array}
[c]{cc}%
p_{X^{n}}(x^{n})/\sum_{x^{n}\in T_{\delta}^{X^{n}}}p_{X^{n}}(x^{n}) & \text{if
}x^{n}\in T_{\delta}^{X^{n}}\\
0 & \text{else}%
\end{array}
\right.  , \label{eq-em:truncated-iid-dist}%
\end{equation}
where $T_{\delta}^{X^{n}}$ is the strongly typical set. This random variable
represents the distribution of the sequence conditioned on it being strongly
typical. Set the rate%
\begin{equation}
R=H(X)-\eta(\left\vert \mathcal{X}\right\vert \delta)-\left\vert
\mathcal{X}\right\vert \frac{1}{n}\log (n+1)-\delta.
\end{equation}
Let $\tau_{\delta}$ denote the set of all strongly typical types for the
distribution $p_{X}$, for sequences of length $n$ and tolerance $\delta$. Note
that $\left\vert \tau_{\delta}\right\vert \leq\left(  n+1\right)  ^{\left\vert
\mathcal{X}\right\vert }$ (see Property~\ref{prop-typ:bound-num-types}), so
that only $\left\vert \mathcal{X}\right\vert \log (n+1)$ bits are required to
record the type. If $x^{n}$ is strongly typical, then Alice applies the
one-to-one mapping%
\begin{equation}
x^{n}\rightarrow(t(x^{n}),g_{t}(f_{t}(x^{n}))), \label{eq-em:ent-conc-encoder}%
\end{equation}
where%
\begin{align}
t  &  :T_{\delta}^{X^{n}}\rightarrow\tau_{\delta},\\
g_{t}  &  :T_{t}^{X^{n}}\rightarrow\left\{  0,1\right\}  ^{nR}\times\left\{
0,1\right\}  ^{n\left[  (1+c)\delta+\eta(\left\vert \mathcal{X}\right\vert
\delta)+\left\vert \mathcal{X}\right\vert \frac{1}{n}\log (n+1)\right]  },
\end{align}
with $g_{t}$ the one-to-one function guaranteed by Lemma~\ref{lem-em:hashing},
with $k=\left\vert T_{t}^{X^{n}}\right\vert $ and $l=2^{nR}$. Let
$W_{\operatorname{out}}$ denote a uniform random variable over the set
$\left\{  0,1\right\}  ^{nR}$, and let $W_{\operatorname{rem}}|t(\widetilde
{X}^{n})=t$ denote a conditional random variable that is uniform over a subset
of $\left\{  0,1\right\}  ^{n\left[  (1+c)\delta+\eta(\left\vert
\mathcal{X}\right\vert \delta)+\left\vert \mathcal{X}\right\vert \frac{1}%
{n}\log (n+1)\right]  }$ of size $\left\lceil k/l\right\rceil $.\ Consider from
our discussion around \eqref{eq-em:uniform-RV-types} that the conditional
random variable $f_{t}(\widetilde{X}^{n})|t(\widetilde{X}^{n})=t$, for
$t\in\tau_{\delta}$, is a uniform random variable. Thus, we can apply
Lemma~\ref{lem-em:hashing}, taking $k=\left\vert T_{t}^{X^{n}}\right\vert $
and $l=2^{nR}$, and find that%
\begin{multline}
\frac{1}{2}\left\Vert p_{g_{t}(f_{t}(\widetilde{X}^{n}))|t(\widetilde{X}%
^{n})=t}-p_{W_{\operatorname{out}}}\times p_{W_{\operatorname{rem}%
}|t(\widetilde{X}^{n})=t}\right\Vert _{1}\leq\frac{2^{nR}}{\left\vert
T_{t}^{X^{n}}\right\vert }\\
\leq\frac{2^{n\left[  H(X)-\eta(\left\vert \mathcal{X}\right\vert
\delta)-\left\vert \mathcal{X}\right\vert \frac{1}{n}\log (n+1)-\delta\right]  }%
}{2^{n\left[  H(X)-\eta(\left\vert \mathcal{X}\right\vert \delta)-\left\vert
\mathcal{X}\right\vert \frac{1}{n}\log (n+1)\right]  }}=2^{-n\delta}.
\end{multline}
This bound is then sufficient for us to conclude that%
\begin{equation}
\frac{1}{2}\left\Vert p_{t(\widetilde{X}^{n}),g_{t}(f_{t}(\widetilde{X}^{n}%
))}-p_{t(\widetilde{X}^{n}),W_{\operatorname{rem}}}\times
p_{W_{\operatorname{out}}}\right\Vert _{1}\leq2^{-n\delta},
\label{eq-em:almost-final-bound-RC}%
\end{equation}
because%
\begin{multline}
\left\Vert p_{t(\widetilde{X}^{n}),g_{t}(f_{t}(\widetilde{X}^{n}%
))}-p_{t(\widetilde{X}^{n}),W_{\operatorname{rem}}}\times
p_{W_{\operatorname{out}}}\right\Vert _{1}\\
=\sum_{t\in\tau_{\delta}}p_{t(\widetilde{X}^{n})}(t)\left\Vert p_{g_{t}%
(f_{t}(\widetilde{X}^{n}))|t(\widetilde{X}^{n})=t}-p_{W_{\operatorname{out}}%
}\times p_{W_{\operatorname{rem}}|t(\widetilde{X}^{n})=t}\right\Vert _{1}.
\end{multline}

For all $\varepsilon\in(0,1)$ and sufficiently large $n$, we know that
$P\equiv\sum_{x^{n}\in T_{\delta}^{X^{n}}}p_{X^{n}}(x^{n})\geq1-\varepsilon$.
Then we can conclude that the distributions $p_{\widetilde{X}^{n}}$ and
$p_{X^{n}}$ are nearly indistinguishable because%
\begin{align}
\left\Vert p_{X^{n}}-p_{\widetilde{X}^{n}}\right\Vert _{1}  &  =\sum_{x^{n}\in
T_{\delta}^{X^{n}}}\left\vert p_{X^{n}}(x^{n})-p_{X^{n}}(x^{n})/P\right\vert
+\sum_{x^{n}\notin T_{\delta}^{X^{n}}}p_{X^{n}}(x^{n})\\
&  =\sum_{x^{n}\in T_{\delta}^{X^{n}}}\left\vert 1-P\right\vert \frac
{p_{X^{n}}(x^{n})}{P}+\sum_{x^{n}\notin T_{\delta}^{X^{n}}}p_{X^{n}}(x^{n})\\
&  \leq2\varepsilon. \label{eq-em:typ-proj-close}%
\end{align}
We can then finally conclude that%
\begin{equation}
\frac{1}{2}\left\Vert p_{t(X^{n}),g_{t}(f_{t}(X^{n}))}-p_{t(\widetilde{X}%
^{n}),W_{\operatorname{rem}}}\times p_{W_{\operatorname{out}}}\right\Vert
_{1}\leq\varepsilon+2^{-n\delta},
\end{equation}
by applying the triangle inequality to
\eqref{eq-em:almost-final-bound-RC}\ and \eqref{eq-em:typ-proj-close}, and
using the fact that%
\begin{equation}
\left\Vert p_{r(Z)}-p_{r(Y)}\right\Vert _{1}=\left\Vert p_{Z}-p_{Y}\right\Vert
_{1}%
\end{equation}
for random variables $Z$ and $Y$ and a one-to-one mapping $r$.

The final step of the randomness concentration protocol consists of discarding
the type register containing $t$ and the $W_{\operatorname{rem}}$ register.
Monotonicity of the trace distance with respect to discardings implies that%
\begin{equation}
\frac{1}{2}\left\Vert \operatorname{Tr}_{W_{\operatorname{rem}}}%
\{p_{g_{t}(f_{t}(X^{n}))}\}-p_{W_{\operatorname{out}}}\right\Vert _{1}%
\leq\varepsilon+2^{-n\delta}.
\end{equation}
Even though the final step of randomness concentration consists of this
discarding, we have developed the protocol using one-to-one functions because
it is essential for our development of the entanglement dilution protocol,
which follows just after we discuss entanglement concentration next.

\subsubsection{Entanglement Concentration}%

\index{entanglement concentration}%
We have actually done the bulk of the \textquotedblleft hard
work\textquotedblright\ in the previous section, when developing the method
for randomness concentration. We now just need to apply the one-to-one mapping
in \eqref{eq-em:ent-conc-encoder}\ in a coherent way in order to have a method
for entanglement concentration. In the setting of entanglement concentration,
Alice and Bob begin with $n$ copies of the state $|\psi\rangle_{AB}$. Suppose
that $|\psi\rangle_{AB}$\ has a Schmidt decomposition of the following form:%
\begin{equation}
|\psi\rangle_{AB}=\sum_{x\in\mathcal{X}}\sqrt{p_{X}(x)}|x\rangle_{A}%
|x\rangle_{B}. \label{eq-ent-con:Schmidt-form}%
\end{equation}
Then the state $|\psi\rangle_{AB}^{\otimes n}$ has the following form:%
\begin{equation}
|\psi\rangle_{AB}^{\otimes n}=\sum_{x^{n}\in\mathcal{X}^{n}}\sqrt{p_{X^{n}%
}(x^{n})}|x^{n}\rangle_{A^{n}}|x^{n}\rangle_{B^{n}}.
\end{equation}
So we need to figure out local quantum channels that Alice and Bob can each
perform in order to convert this state to as many ebits as possible. Before
doing so, we state the following lemma, which will allow us to quantify the
performance of a coherent version of a classical protocol:

\begin{lemma}
\label{lem-em:incoh-to-coh}Let $p_{X}$ and $q_{X}$ be probability
distributions such that%
\begin{equation}
\left\Vert p_{X}-q_{X}\right\Vert _{1}\leq\varepsilon.
\end{equation}
Then the quantum states $|\psi^{p}\rangle_{AB}\equiv\sum_{x}\sqrt{p_{X}%
(x)}|x\rangle_{A}|x\rangle_{B}$ and $|\psi^{q}\rangle_{AB}\equiv\sum_{x}%
\sqrt{q_{X}(x)}|x\rangle_{A}|x\rangle_{B}$ are such that%
\begin{equation}
\left\Vert \psi_{AB}^{p}-\psi_{AB}^{q}\right\Vert _{1}\leq2\sqrt{\varepsilon}.
\label{eq-em:coherent-from-classical}%
\end{equation}

\end{lemma}

\begin{proof}
Consider that the quantum fidelity between $|\psi^{p}\rangle_{AB}$ and
$|\psi^{q}\rangle_{AB}$ is equal to the classical fidelity of the
distributions $p_{X}$ and $q_{X}$:%
\begin{equation}
\sqrt{F}(\psi_{AB}^{p},\psi_{AB}^{q})=\left\vert \langle\psi^{q}|\psi
^{p}\rangle_{AB}\right\vert =\sum_{x}\sqrt{q_{X}(x)p_{X}(x)}.
\end{equation}
From Corollary~\ref{cor-dm:trace-imp-fid}, we can conclude that $F(\psi
_{AB}^{p},\psi_{AB}^{q})\geq1-\varepsilon$.
Corollary~\ref{cor-dm:fid-imp-trace} in turn implies \eqref{eq-em:coherent-from-classical}.
\end{proof}

Let $U_{A^{n}\rightarrow TW_{\operatorname{out}}W_{\operatorname{rem}}}$denote
the following isometric implementation of the one-to-one mapping in
\eqref{eq-em:ent-conc-encoder}:%
\begin{equation}
U_{A^{n}\rightarrow TW_{\operatorname{out}}W_{\operatorname{rem}}}\equiv
\sum_{x^{n}\in T_{\delta}^{X^{n}}}|t(x^{n}),g_{t}(f_{t}(x^{n}))\rangle
_{TW_{\operatorname{out}}W_{\operatorname{rem}}}\langle x^{n}|_{A^{n}}.
\label{eq-em:isometric-encoder}%
\end{equation}
Then we set Alice's encoding to be the following quantum channel:%
\begin{equation}
\mathcal{E}_{A^{n}\rightarrow TW_{\operatorname{out}}W_{\operatorname{rem}}%
}(Y_{A^{n}})\equiv U\Pi_{A^{n}}^{\delta}Y_{A^{n}}\Pi_{A^{n}}^{\delta}U^{\dag
}+\operatorname{Tr}\{(I_{A^{n}}-\Pi_{A^{n}}^{\delta})Y_{A^{n}}\}\sigma
_{TW_{\operatorname{out}}W_{\operatorname{rem}}},
\label{eq-em:ent-conc-encoding}%
\end{equation}
where $\Pi_{A^{n}}^{\delta}$ is the strongly typical projector for the density
operator $\sum_{x\in\mathcal{X}}p_{X}(x)|x\rangle\langle x|_{A}$ and
$\sigma_{TW_{\operatorname{out}}W_{\operatorname{rem}}}$ is some state of the
registers $TW_{\operatorname{out}}W_{\operatorname{rem}}$. This channel
isometrically embeds the typical subspace for Alice's system into the
registers $TW_{\operatorname{out}}W_{\operatorname{rem}}$. Let $\mathcal{E}%
_{B^{n}\rightarrow(TW_{\operatorname{out}}W_{\operatorname{rem}})_{B}}$ denote
the same encoding for Bob, and let $\mathcal{E}_{A^{n}}$ and $\mathcal{E}%
_{B^{n}}$ be a shorthand for Alice and Bob's encodings, respectively.

Let $|\widetilde{\psi}^{n}\rangle_{A^{n}B^{n}}$ denote the following state:%
\begin{equation}
|\widetilde{\psi}^{n}\rangle_{A^{n}B^{n}}\equiv\sum_{x^{n}\in\mathcal{X}^{n}%
}\sqrt{p_{\widetilde{X}^{n}}(x^{n})}|x^{n}\rangle_{A^{n}}|x^{n}\rangle_{B^{n}%
},
\end{equation}
where the distribution $p_{\widetilde{X}^{n}}$ is defined in
\eqref{eq-em:truncated-iid-dist}. By invoking
\eqref{eq-em:typ-proj-close}\ and Lemma~\ref{lem-em:incoh-to-coh}, we find
that%
\begin{equation}
\left\Vert \psi_{AB}^{\otimes n}-\widetilde{\psi}_{A^{n}B^{n}}^{n}\right\Vert
_{1}\leq2\sqrt{2\varepsilon}. \label{eq-em:ent-dilu-1}%
\end{equation}
The monotonicity of trace distance then implies that%
\begin{equation}
\left\Vert (\mathcal{E}_{A^{n}}\otimes\mathcal{E}_{B^{n}})(\psi_{AB}^{\otimes
n})-(\mathcal{E}_{A^{n}}\otimes\mathcal{E}_{B^{n}})(\widetilde{\psi}%
_{A^{n}B^{n}}^{n})\right\Vert _{1}\leq2\sqrt{2\varepsilon}.
\label{eq-em:ent-conc-1}%
\end{equation}
Consider that the coherent version of the distribution $p_{t(\widetilde{X}%
^{n}),W_{\operatorname{rem}}}\times p_{W_{\operatorname{out}}}$ is the
following state:%
\begin{equation}
\Upsilon_{T_{A}W_{A^{\prime}}T_{B}W_{B^{\prime}}}\otimes\left(  \Phi_{AB}%
^{+}\right)  ^{\otimes nR},
\end{equation}
where $\Upsilon_{T_{A}W_{A^{\prime}}T_{B}W_{B^{\prime}}}$ is a coherent
version of the distribution $p_{t(\widetilde{X}^{n}),W_{\operatorname{rem}}}$,
defined as%
\begin{equation}
|\Upsilon\rangle_{T_{A}W_{A^{\prime}}T_{B}W_{B^{\prime}}}\equiv\sum_{t\in
\tau_{\delta},w}\sqrt{p_{t(\widetilde{X}^{n}),W_{\operatorname{rem}}}%
(t,w)}|t,w\rangle_{T_{A}W_{A^{\prime}}}|t,w\rangle_{T_{B}W_{B^{\prime}}},
\end{equation}
and $|\Phi^{+}\rangle_{AB}\equiv\left[  |00\rangle_{AB}+|11\rangle
_{AB}\right]  /\sqrt{2}$. By invoking \eqref{eq-em:almost-final-bound-RC}\ and
Lemma~\ref{lem-em:incoh-to-coh}, we find that%
\begin{equation}
\left\Vert (\mathcal{E}_{A^{n}}\otimes\mathcal{E}_{B^{n}})(\widetilde{\psi
}_{A^{n}B^{n}}^{n})-\Upsilon_{T_{A}W_{A^{\prime}}T_{B}W_{B^{\prime}}}%
\otimes\left(  \Phi_{AB}^{+}\right)  ^{\otimes nR}\right\Vert _{1}\leq
2\sqrt{2\cdot2^{-n\delta}}. \label{eq-em:ent-conc-2}%
\end{equation}
Applying the triangle inequality to \eqref{eq-em:ent-conc-1} and
\eqref{eq-em:ent-conc-2}, we find that%
\begin{equation}
\left\Vert (\mathcal{E}_{A^{n}}\otimes\mathcal{E}_{B^{n}})(\psi_{AB}^{\otimes
n})-\Upsilon_{T_{A}W_{A^{\prime}}T_{B}W_{B^{\prime}}}\otimes\left(  \Phi
_{AB}^{+}\right)  ^{\otimes nR}\right\Vert _{1}\leq2\left[  \sqrt
{2\varepsilon}+\sqrt{2\cdot2^{-n\delta}}\right]  .
\end{equation}

The final step is for Alice and Bob to discard the registers $T_{A}%
T_{B}W_{A^{\prime}}W_{B^{\prime}}$, which implies that%
\begin{equation}
\left\Vert ((\operatorname{Tr}_{T_{A}W_{A^{\prime}}}\circ\mathcal{E}_{A^{n}%
})\otimes(\operatorname{Tr}_{T_{B}W_{B^{\prime}}}\circ\mathcal{E}_{B^{n}%
}))(\psi_{AB}^{\otimes n})-\left(  \Phi_{AB}^{+}\right)  ^{\otimes
nR}\right\Vert _{1}\leq2\left[  \sqrt{2\varepsilon}+\sqrt{2\cdot2^{-n\delta}%
}\right]  .
\end{equation}
The rate of ebit generation is equal to $R=H(A)_{\psi}-\eta(\left\vert
\mathcal{X}\right\vert \delta)-\left\vert \mathcal{X}\right\vert \frac{1}%
{n}\log (n+1)-\delta$. No classical communication is required. This concludes the
proof for entanglement concentration.

\subsubsection{Entanglement Dilution}%

\index{entanglement dilution}%
\label{sec-em:ent-dilution}We have already outlined an entanglement dilution
protocol in Section~\ref{sec-em:ent-dil-sketch}. However, the protocol
sketched there uses far more classical communication than is necessary. Here,
we show that entanglement dilution requires a rate of classical communication
that vanishes as $n$ becomes large. This result demonstrates that the resource
theory of entanglement for pure states is truly a reversible theory, in the
sense that the only resource we need to count is the rate of entanglement
conversion, given that the classical communication rate is negligible.

We now discuss such an entanglement dilution protocol. The main idea is really
just to take the entanglement concentration protocol from the previous section
and \textquotedblleft run it backwards.\textquotedblright\ So we keep the
system labels as they were in the previous section. Let Alice and Bob share
the following maximally entangled state at the beginning:%
\begin{equation}
\Phi_{T_{A}W_{A^{\prime}}T_{B}W_{B^{\prime}}}\otimes\left(  \Phi_{AB}%
^{+}\right)  ^{\otimes nR},
\end{equation}
Consider that%
\begin{align}
\log\dim(\mathcal{H}_{T_{A}})  &  =\left\vert \mathcal{X}\right\vert \log n,\\
\log\dim(\mathcal{H}_{W_{A^{\prime}}})  &  =n(1+c)\delta+n\eta(\left\vert
\mathcal{X}\right\vert \delta)+\left\vert \mathcal{X}\right\vert \log n,
\end{align}
implying that the total number of ebits in the state $\Phi_{T_{A}W_{A^{\prime
}}T_{B}W_{B^{\prime}}}$ is equal to $n(1+c)\delta+n\eta(\left\vert
\mathcal{X}\right\vert \delta)+2\left\vert \mathcal{X}\right\vert \log n$, and
the total number of ebits that they share overall is equal to $nH(A)_{\psi
}+nc\delta+\left\vert \mathcal{X}\right\vert \log n$. Alice prepares the state
$\Upsilon_{T_{A}W_{A^{\prime}}T_{B}W_{B^{\prime}}}$ locally in her lab and
uses the state $\Phi_{T_{A}W_{A^{\prime}}T_{B}W_{B^{\prime}}}$ to teleport the
$T_{B}W_{B^{\prime}}$\ systems of $\Upsilon$ to Bob. This requires%
\begin{equation}
2\left[  n(1+c)\delta+n\eta(\left\vert \mathcal{X}\right\vert \delta
)+2\left\vert \mathcal{X}\right\vert \log n\right]
\end{equation}
bits of classical communication. At this point, Alice and Bob share the
following state:%
\begin{equation}
\Upsilon_{T_{A}W_{A^{\prime}}T_{B}W_{B^{\prime}}}\otimes\left(  \Phi_{AB}%
^{+}\right)  ^{\otimes nR}.
\end{equation}
They each then perform the following quantum channel, which is essentially the
\textquotedblleft inverse\textquotedblright\ of the encoding in
\eqref{eq-em:ent-conc-encoding}:%
\begin{equation}
\mathcal{E}^{\left(  -1\right)  }(Z)=U^{\dag}ZU+\operatorname{Tr}%
\{(I-UU^{\dag})Z\}\omega,
\end{equation}
where $U$ is the isometry defined in \eqref{eq-em:isometric-encoder}$\ $%
and$\ \omega$ is any state having support in $\operatorname{span}%
\{|x^{n}\rangle:x^{n}\in T_{\delta}^{X^{n}}\}$. In fact, since $\widetilde
{\psi}_{A^{n}B^{n}}^{n}$ exclusively has support in $\operatorname{span}%
\{|x^{n}\rangle:x^{n}\in T_{\delta}^{X^{n}}\}$, it follows that%
\begin{equation}
(\mathcal{E}_{A^{n}}^{(-1)}\otimes\mathcal{E}_{B^{n}}^{(-1)})(\mathcal{E}%
_{A^{n}}\otimes\mathcal{E}_{B^{n}})(\widetilde{\psi}_{A^{n}B^{n}}%
^{n})=\widetilde{\psi}_{A^{n}B^{n}}^{n},
\end{equation}
which along with the monotonicity of trace distance applied to
\eqref{eq-em:ent-conc-2}, implies that%
\begin{equation}
\left\Vert \widetilde{\psi}_{A^{n}B^{n}}^{n}-(\mathcal{E}_{A^{n}}%
^{(-1)}\otimes\mathcal{E}_{B^{n}}^{(-1)})\left(  \Upsilon_{T_{A}W_{A^{\prime}%
}T_{B}W_{B^{\prime}}}\otimes\left(  \Phi_{AB}^{+}\right)  ^{\otimes
nR}\right)  \right\Vert _{1}\leq2\sqrt{2\cdot2^{-n\delta}}.
\label{eq-em:ent-dilu-2}%
\end{equation}
Applying the triangle inequality to \eqref{eq-em:ent-dilu-1} and
\eqref{eq-em:ent-dilu-2}, we find that%
\begin{equation}
\left\Vert \psi_{AB}^{\otimes n}-(\mathcal{E}_{A^{n}}^{(-1)}\otimes
\mathcal{E}_{B^{n}}^{(-1)})\left(  \Upsilon_{T_{A}W_{A^{\prime}}%
T_{B}W_{B^{\prime}}}\otimes\left(  \Phi_{AB}^{+}\right)  ^{\otimes nR}\right)
\right\Vert _{1}\leq2\left[  \sqrt{\varepsilon}+\sqrt{2\cdot2^{-n\delta}%
}\right]  ,
\end{equation}
which concludes the error analysis.

The rate of ebits needed to form $\psi_{AB}^{\otimes n}$\ is equal to
$H(A)_{\psi}+c\delta+\frac{\left\vert \mathcal{X}\right\vert }{n}\log n$ ebits
per copy of $\psi_{AB}$, and the rate of classical communication needed is
$2\left[  (1+c)\delta+\eta(\left\vert \mathcal{X}\right\vert \delta
)+2\left\vert \mathcal{X}\right\vert \frac{1}{n}\log n\right]  $ cbits per
copy of $\psi_{AB}$. For large $n$, we can take $\delta$ to be order $\sqrt
{n}$ and the central limit theorem implies that we can achieve any constant
error $\varepsilon\in(0,1)$ (this is a modified version of typicality in which
$\delta$ changes with $n$). At the same time, the ebit rate converges to
$H(A)_{\psi}$ and the classical communication rate vanishes.

\subsubsection{Entanglement Manipulation}

We can now put together entanglement concentration and dilution to give a
general achievable strategy for an entanglement manipulation protocol. The
goal here is to convert $n$ copies of $\psi_{AB}$ to as many copies of
$\phi_{AB}$ as possible. In order to do so, Alice and Bob first conduct an
entanglement concentration protocol, which takes $n$ copies of $\psi_{AB}$ to
$\approx nH(A)_{\psi}$ approximate ebits. Then, using entanglement dilution
and a negligible rate of classical communication, they can convert these
$\approx nH(A)_{\psi}$ approximate ebits to $\approx n\left[  H(A)_{\psi
}/H(A)_{\psi}\right]  $ approximate copies of $\phi_{AB}$. The accuracy of the
protocol becomes arbitrarily small and the rate of entanglement conversion
approaches $H(A)_{\psi}/H(A)_{\psi}$ in the limit as $n$ becomes large.

\section{Concluding Remarks}

Entanglement concentration was one of the earliest discovered protocols in
quantum Shannon theory. The protocol exploits one of the fundamental tools of
classical information theory (the method of types), but it applies the method
in a coherent fashion so that a type class measurement learns only the type
and nothing more. The protocol is similar to Schumacher compression in this
regard (in that it learns only the necessary information required to execute
the protocol and preserves coherent superpositions), and we will continue to
see this idea of applying classical techniques in a coherent way in future
quantum Shannon-theoretic protocols. For example, the protocol for quantum
communication over a quantum channel is a coherent version of a protocol to
transmit private classical information over a quantum channel.

\section{History and Further Reading}

\cite{E72} constructed a protocol for randomness concentration in an early
paper. \cite{BBPS96} offered two different protocols for entanglement
concentration (one of which we developed in this chapter). \cite{N99} later
connected entanglement concentration protocols to the theory of majorization.
\cite{PhysRevA.63.022301,PhysRevLett.83.1459} studied entanglement
concentration and the classical communication cost of the inverse protocol
(entanglement dilution). \cite{PhysRevA.67.012326} characterized the classical
communication cost of entanglement dilution, as did \cite{HL04}. \cite{KM01}
developed practical networks for entanglement concentration, and recently,
\cite{BCG09} took this line of research a step further by considering
streaming protocols for entanglement concentration. \cite{HM01} also developed
protocols for universal entanglement concentration.

\cite{VP98} introduced the relative entropy of entanglement as one of the
first LOCC\ monotones in quantum information theory.
Proposition~\ref{prop-em:rel-ent-enta-bigger-coh-info}\ is due to \cite{PVP00}.

Going beyond the settings considered here, researchers have considered error
exponents, strong converses, and second-order characterizations for
entanglement manipulation tasks (we explain what these terms mean in
Section~\ref{sec-cc:history}). \cite{PhysRevA.63.022301} established a strong
converse theorem for entanglement concentration. \cite{HKMMW03} derived error
exponents and an exact strong converse for entanglement concentration.
\cite{PhysRevLett.111.130407} established exact second-order characterizations
of entanglement concentration and dilution.

\part{Noisy Quantum Shannon Theory}

\noindent Before quantum information theory became an established discipline, John~R.~Pierce issued the following quip at the end of his
1973 retrospective article on the history of information theory~\citep{P73}:

\textquotedblleft I think that I have never met a physicist who understood
information theory. I wish that physicists would stop talking about
reformulating information theory and would give us a general expression for
the capacity of a channel with quantum effects taken into account rather than
a number of special cases.\textquotedblright

Since the publication of Pierce's article, we have learned much more about
quantum mechanics and information theory than he might have imagined at the
time, but we have also realized that there is much more to discover. In spite
of all that we have learned, we still unfortunately have not been able to
address Pierce's concern in the above quote in full generality.

The most basic question that we could ask in quantum Shannon theory (and the
one with which Pierce was concerned) is how much classical information can a
sender transmit to a receiver by exploiting a quantum channel. We have
determined many special cases of quantum channels for which we do know their
classical capacities, but we also now know that this most basic question is
still wide open in the general case.

What Pierce might not have imagined at the time is that a quantum channel has a
much larger variety of capacities than does a classical channel. For example,
we might wish to determine the classical capacity of a quantum channel
assisted by entanglement shared between the sender and receiver. We have seen
that in the simplest of cases, such as the noiseless qubit channel, shared
entanglement boosts the classical capacity up to two bits, and we now refer to
this phenomenon as the super-dense coding effect (see
Chapter~\ref{chap:three-noiseless}). Interestingly, the entanglement-assisted
capacity of a quantum channel is one of the few scenarios where we can claim
to have a complete understanding of the channel's transmission capabilities.
From the results regarding the entanglement-assisted capacity, we have learned
that shared entanglement is often a \textquotedblleft friend\textquotedblright%
\ because it tends to simplify results in both quantum Shannon theory and
other subfields of quantum information science.

Additionally, we might consider the capacity of a quantum channel for
transmitting quantum information. In 1973, it was not even clear what was
meant by \textquotedblleft quantum information,\textquotedblright\ but we have
since been able to formulate what it means, and we have been able to
characterize the quantum capacity of a quantum channel. The task of
transmitting quantum information over a quantum channel bears some
similarities with the task of transmitting private classical information over
that channel, where we are concerned with keeping the classical information
private from the environment of the channel. This connection has given insight
for achieving good rates of quantum communication over a noisy quantum
channel, and there is even a certain class of channels for which we already
have a good expression for the quantum capacity (the expression being the
coherent information of the channel). However, the problem of determining a
good expression for the quantum capacity in the general case is still wide open.

The remaining chapters of the book are an attempt to summarize many items the
quantum information community has learned in the past few decades, all of
which are an attempt to address Pierce's concern in various ways. The most
important open problem in quantum Shannon theory is to find better expressions
for these capacities so that we can actually compute them for an arbitrary
quantum channel.

\chapter{Classical Communication}

\label{chap:classical-comm-HSW}This chapter begins our exploration of
\textquotedblleft dynamic\textquotedblright\ information-processing tasks in
quantum Shannon theory, where the term \textquotedblleft
dynamic\textquotedblright\ indicates that a quantum channel connects a sender
to a receiver and their goal is to exploit this resource for communication. We
specifically consider the scenario in which a sender Alice would like to
communicate classical information to a receiver Bob, and the capacity theorem
that we prove here is one particular generalization of Shannon's noisy channel
coding theorem from classical information theory (reviewed in
Section~\ref{sec-ccs:channel-cap}). In later chapters, we will see other
generalizations of Shannon's theorem, depending on what resources are
available to assist their communication or whether they are trying to
communicate classical or quantum information. For this reason and others,
quantum Shannon theory is quite a bit richer than classical information theory.

The naive approach to communicate classical information over a quantum channel
is for Alice and Bob simply to mimic the approach used in Shannon's noisy
channel coding theorem. That is, they randomly select a classical code
according to some distribution $p_{X}(x)$, and Bob performs individual
measurements of the outputs of a quantum channel according to some POVM. The
POVM\ at the output induces some conditional probability distribution
$p_{Y|X}(y|x)$, which we can in turn think of as an induced classical channel.
The classical mutual information $I(X;Y)$ of this channel is an achievable
rate for communication, and the best strategy for Alice and Bob is to optimize
the mutual information over all of Alice's inputs to the channel and over all
measurements that Bob could perform at the output. The resulting quantity is
equal to Bob's optimized accessible information, which we previously discussed
in Section~\ref{sec-cie:acc-info}.

If the aforementioned coding strategy were optimal, then there would not be
anything much interesting to say for the information-processing task of
classical communication (in fact, there would not be any need for all of the
tools we developed in Chapters~\ref{chap:quantum-typicality} and
\ref{chap:packing}!). This is perhaps one first clue that the above strategy
is not necessarily optimal. Furthermore, we know from
Chapter~\ref{chap:q-info-entropy}\ that the Holevo information is an upper
bound on the accessible information, and this bound might prompt us to wonder
if it is also an achievable rate for classical communication, given that the
accessible information is achievable.

The main theorem of this chapter is the classical capacity theorem (also known
as the Holevo--Schumacher--Westmoreland theorem), and it states that the
Holevo information of a quantum channel is an achievable rate for classical
communication. The Holevo information is easier to manipulate mathematically
than is the accessible information. The proof of its achievability
demonstrates that the aforementioned strategy is not generally optimal, and
the proof also shows how performing a collective measurement on all of the
channel outputs allows the sender and receiver to achieve the Holevo
information as a rate for classical communication. Thus, this strategy
fundamentally makes use of quantum-mechanical effects at the decoder and
suggests that such an approach is necessary to achieve the Holevo information.
Although this strategy exploits a collective measurement at the decoder, it
does not make use of entangled states at the encoder. That is, the sender
could input quantum states that are entangled across all of the channel
inputs, and this encoder entanglement might potentially increase classical
communication rates.

One major drawback of the classical capacity theorem (also the case for many
other results in quantum Shannon theory) is that it only demonstrates that the
Holevo information is an achievable rate for classical communication---the
converse theorem is a \textquotedblleft multi-letter\textquotedblright%
\ converse, meaning that it might be necessary in the general case to evaluate
the Holevo information over a potentially infinite number of uses of the
channel. The multi-letter nature of the capacity theorem implies that the
optimization task for general channels is intractable and thus further implies
that we know very little about the actual classical capacity of general
quantum channels. Now, there are many natural quantum channels such as the
depolarizing channel and the dephasing channel for which the classical
capacity is known (the Holevo information becomes \textquotedblleft
single-letter\textquotedblright\ for these channels), and these results imply
that we have a complete understanding of the classical
information-transmission capabilities of these channels. All of these results
have to do with the additivity of the Holevo information of a quantum channel,
which we studied previously in Chapter~\ref{chap:additivity}.

We mentioned that the Holevo--Schumacher--Westmoreland coding strategy does
not make use of entangled inputs at the encoder. But a natural question is to
wonder whether entanglement at the encoder could boost classical
information-transmission rates, given that it is a resource for many quantum
protocols. This question was known as the additivity conjecture and went
unsolved for many years, but \cite{H09} offered a proof that entangled inputs
can increase communication rates for certain channels. Thus, for these
channels, the single-letter Holevo information is not the proper
characterization of classical capacity (however, this is not to say that there
could be some alternate characterization of the classical capacity other than
the Holevo information which would be single-letter). These results
demonstrate that we still know little about classical communication in the
general case and furthermore that quantum Shannon theory is an active area of research.

We structure this chapter as follows. We first discuss the aforementioned
naive strategy in detail, so that we can understand the difference between it
and the Holevo--Schumacher--Westmoreland strategy.
Section~\ref{sec-cc:info-proc-task} describes the steps needed in any protocol
for classical communication over a quantum channel.
Section~\ref{sec-cc:cc-theorem}\ provides a statement of the classical
capacity theorem, and its two subsections prove the corresponding direct
coding theorem and the converse theorem. The direct coding theorem exploits
two tools:\ quantum typicality from Chapter~\ref{chap:quantum-typicality}\ and
the packing lemma from Chapter~\ref{chap:packing}. The converse theorem
exploits two tools from Chapter~\ref{chap:q-info-entropy}:\ continuity of
entropies (the AFW inequality in Theorem~\ref{thm-qie:AFW-cont-ent}) and the
quantum data-processing inequality (Theorem~\ref{cor-qie:QDP}). We then detail
how to calculate the classical capacity of several exemplary channels such as
\index{entanglement-breaking channel}%
entanglement-breaking channels, quantum Hadamard channels, erasure channels,
and depolarizing channels---these are channels for which we have a complete
understanding of their classical capacity. Finally, we end with a discussion
of the proof that the Holevo information can be superadditive (that is,
entangled inputs at the encoder can enhance classical communication rates for
some channels).

\section{Naive Approach:\ Product Measurements}

\label{sec-cc:naive}%
\begin{figure}
[ptb]
\begin{center}
\includegraphics[
width=3.5397in
]%
{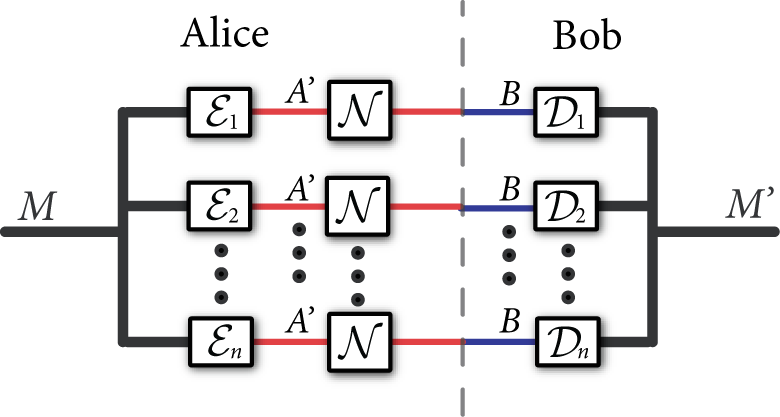}%
\caption{The most naive strategy for Alice and Bob to communicate classical
information over many independent uses of a quantum channel. Alice wishes to
send some message $M$ and selects some tensor product state to input to the
channel, conditioned on the message $M$. She transmits the codeword over the
channel, and Bob then receives a noisy version of it. He performs individual
measurements of his quantum systems and produces some estimate $M^{\prime}$ of
the original message $M$. This scheme is effectively a classical scheme
because it makes no use of quantum-mechanical features such as entanglement or a
collective measurement.}%
\label{fig-cc:naive-classical-capacity}%
\end{center}
\end{figure}
We begin by discussing in more detail the most naive strategy that a sender
and receiver can exploit for the transmission of classical information over
many uses of a quantum channel. Figure~\ref{fig-cc:naive-classical-capacity}%
\ depicts this naive approach. This first approach mimics certain features of
Shannon's classical approach without making any use of quantum-mechanical
effects. Alice and Bob agree on a codebook beforehand, such that each
classical codeword $x^{n}(m)$ in the codebook corresponds to some message $m$
that Alice wishes to transmit. Alice can exploit some set $\left\{  \rho
_{x}\right\}  $ of density operators to act as input to the quantum channel.
That is, the quantum codewords are of the form%
\begin{equation}
\rho_{x^{n}(m)}\equiv\rho_{x_{1}(m)}\otimes\rho_{x_{2}(m)}\otimes\cdots
\otimes\rho_{x_{n}(m)}.
\end{equation}
Bob then performs individual measurements of the outputs of the quantum
channel by exploiting some POVM $\left\{  \Lambda_{y}\right\}  $. This scheme
induces the following conditional probability distribution:%
\begin{align}
&  p_{Y_{1}\cdots Y_{n}|X_{1}\cdots X_{n}}(y_{1}\cdots y_{n}|x_{1}(m)\cdots
x_{n}(m))\nonumber\\
&  =\operatorname{Tr}\left\{  \Lambda_{y_{1}}\otimes\cdots\otimes
\Lambda_{y_{n}}\left(  \mathcal{N}\otimes\cdots\otimes\mathcal{N}\right)
\left(  \rho_{x_{1}(m)}\otimes\cdots\otimes\rho_{x_{n}(m)}\right)  \right\} \\
&  =\operatorname{Tr}\left\{  \left(  \Lambda_{y_{1}}\otimes\cdots
\otimes\Lambda_{y_{n}}\right)  \left(  \mathcal{N}(\rho_{x_{1}(m)}%
)\otimes\cdots\otimes\mathcal{N}(\rho_{x_{n}(m)})\right)  \right\} \\
&  =\prod\limits_{i=1}^{n}\operatorname{Tr}\left\{  \Lambda_{y_{i}}%
\mathcal{N}(\rho_{x_{i}(m)})\right\}  ,
\end{align}
which we immediately realize is equivalent to many i.i.d.~instances of the following
classical channel:%
\begin{equation}
p_{Y|X}(y|x)\equiv\operatorname{Tr}\left\{  \Lambda_{y}\mathcal{N}(\rho
_{x})\right\}  .
\end{equation}
Thus, if they exploit this scheme, the optimal rate at which they can
communicate is equal to the following expression:%
\begin{equation}
I_{\operatorname{acc}}(\mathcal{N})\equiv\max_{\left\{  p_{X}(x),\rho
_{x},\Lambda\right\}  }I(X;Y),
\end{equation}
where the maximization of the classical mutual information is with respect to
all input distributions, all input density operators, and all POVMs that Bob
could perform at the output of the channel. This information quantity is known
as the accessible information
\index{accessible information!of a channel}%
of the channel.%

\begin{figure}
[ptb]
\begin{center}
\includegraphics[
width=3.5405in
]%
{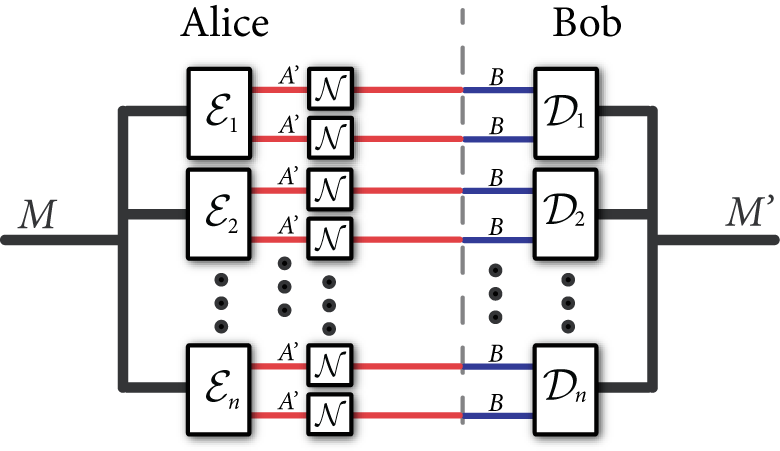}%
\caption{A coding strategy that can outperform the previous naive strategy,
simply by making use of entanglement at the encoder and decoder. }%
\label{fig-cc:cc-entangled-acc}%
\end{center}
\end{figure}
The above strategy is not necessarily an optimal strategy if the channel is
truly a quantum channel---it does not make use of any quantum effects such as
entanglement or collective measurements (an example of a collective
measurement is a Bell measurement, as in quantum teleportation). A first
simple modification of the protocol to allow for such effects would be to
consider coding for the tensor-product channel $\mathcal{N}\otimes\mathcal{N}$
rather than the original channel. The input states would be entangled across
two channel uses, and the output measurements would be over two channel
outputs at a time. In this way, they would be exploiting entangled states at
the encoder and collective measurements at the decoder.
Figure~\ref{fig-cc:cc-entangled-acc}\ illustrates the modified protocol, and
the rate of classical communication that they can achieve with such a strategy
is $\frac{1}{2}I_{\operatorname{acc}}(\mathcal{N}\otimes\mathcal{N})$. This
quantity is always at least as large as $I_{\operatorname{acc}}(\mathcal{N})$
because a special case of the strategy for the tensor-product channel
$\mathcal{N}\otimes\mathcal{N}$ is to choose the distribution $p_{X}(x)$, the
states $\rho_{x}$, and the POVM\ $\Lambda$ to be tensor products of the ones
that maximize $I_{\operatorname{acc}}(\mathcal{N})$. We can then extend this
construction inductively by forming codes for the tensor-product channel
$\mathcal{N}^{\otimes k}$ (where $k$ is a positive integer), and this extended
strategy achieves the classical communication rate of $\frac{1}{k}%
I_{\operatorname{acc}}(\mathcal{N}^{\otimes k})$ for any finite $k$. These
results then suggest that the ultimate classical capacity of the channel is
the regularization of the accessible information of the channel:%
\begin{equation}
I_{\operatorname{reg}}(\mathcal{N})\equiv\lim_{k\rightarrow\infty}\frac{1}%
{k}I_{\operatorname{acc}}(\mathcal{N}^{\otimes k}).
\end{equation}

The regularization of the accessible information is intractable for general
quantum channels, but the optimization task could simplify immensely if the
accessible information is additive (additive in the sense discussed in
Chapter~\ref{chap:additivity}). In this case, the regularized accessible
information $I_{\operatorname{reg}}(\mathcal{N})$ would be equal to the
accessible information $I_{\operatorname{acc}}(\mathcal{N})$. However, even if
the quantity is additive, the optimization could still be difficult to perform
in practice. A simple upper bound on the accessible information is the Holevo
information $\chi(\mathcal{N})$ of the channel, defined as%
\begin{equation}
\chi(\mathcal{N})\equiv\max_{\rho}I(X;B), \label{eq-cc:holevo-formula}%
\end{equation}
where the maximization is with respect to classical--quantum states $\rho
_{XB}$ of the following form:%
\begin{equation}
\rho_{XB}\equiv\sum_{x}p_{X}(x)|x\rangle\langle x|_{X}\otimes\mathcal{N}%
_{A^{\prime}\rightarrow B}(\psi_{A^{\prime}}^{x}).
\end{equation}
The Holevo information is a more desirable quantity to characterize classical
communication over a quantum channel because it is always an upper bound on
the accessible information and it does not involve an optimization over measurements.

Thus, a natural question to ask is whether Alice and Bob can achieve the
Holevo information rate, and the main theorem of this chapter states that it
is possible to do so. The resulting coding scheme bears some similarities with
the techniques in Shannon's channel coding theorem, but the main difference is
that the decoding POVM\ is a collective measurement over all of the channel outputs.

\section{The Information-Processing Task}

\label{sec-cc:info-proc-task}

\subsection{Classical Communication}

\begin{figure}
[ptb]
\begin{center}
\includegraphics[
width=3.1816in
]%
{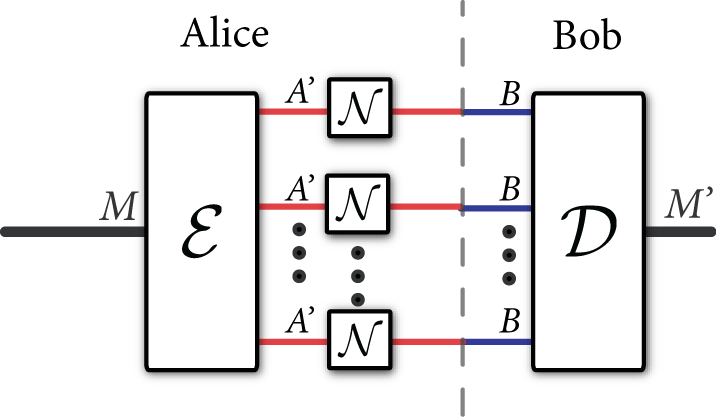}%
\caption{The most general protocol for classical communication over a quantum
channel. Alice selects some message $M$ and encodes it as a quantum codeword
for input to many independent uses of the noisy quantum channel. Bob performs
some POVM\ over all of the channel outputs to determine the message that Alice
transmits.}%
\label{fig-cc:info-proc-task}%
\end{center}
\end{figure}

We now discuss the most general form of the information-processing task and
give the criterion for a classical communication rate $C$ to be
achievable---i.e., we define an $\left(  n,C,\varepsilon\right)  $ code for
classical communication over a quantum channel. Alice begins by selecting some
classical message $m$ that she would like to transmit to Bob---she selects
from a set of messages $\left\{  1,\ldots,\left\vert \mathcal{M}\right\vert
\right\}  $. Let $M$ denote the random variable corresponding to Alice's
choice of message, and let $\left\vert \mathcal{M}\right\vert $ denote its
cardinality. She then prepares some state $\rho_{A^{\prime n}}^{m}$ as input
to the many independent uses of the channel---the input systems are $n$ copies
of the channel input system $A^{\prime}$. She transmits this state over $n$
independent uses of the channel $\mathcal{N}$, and the state at Bob's
receiving end is%
\begin{equation}
\mathcal{N}^{\otimes n}(\rho_{A^{\prime n}}^{m}).
\end{equation}
Bob has some decoding POVM\ $\left\{  \Lambda_{m}\right\}  $\ that he can
exploit to determine which message Alice transmits.
Figure~\ref{fig-cc:info-proc-task} depicts such a general protocol for
classical communication over a quantum channel.%

Let $M^{\prime}$ denote the random variable for Bob's estimate of the message.
The probability that he determines the correct message $m$ is as follows:%
\begin{equation}
\Pr\left\{  M^{\prime}=m|M=m\right\}  =\operatorname{Tr}\left\{  \Lambda
_{m}\mathcal{N}^{\otimes n}(\rho_{A^{\prime n}}^{m})\right\}  ,
\end{equation}
and thus the probability of error for a particular message $m$ is%
\begin{align}
p_{e}(m)  &  \equiv1-\Pr\left\{  M^{\prime}=m|M=m\right\} \\
&  =\operatorname{Tr}\left\{  \left(  I-\Lambda_{m}\right)  \mathcal{N}%
^{\otimes n}(\rho_{A^{\prime n}}^{m})\right\}  .
\end{align}
The maximal probability of error for any coding scheme is then%
\begin{equation}
p_{e}^{\ast}\equiv\max_{m\in\mathcal{M}}p_{e}(m).
\end{equation}
The rate $C$ of communication is%
\begin{equation}
C\equiv\frac{1}{n}\log\left\vert \mathcal{M}\right\vert ,
\end{equation}
and the code has $\varepsilon\in\left[  0,1\right]  $ error if $p_{e}^{\ast
}\leq\varepsilon$.

A rate $C$ of classical communication is \textit{achievable} for a channel
$\mathcal{N}$\ if there exists an $\left(  n,C-\delta,\varepsilon\right)  $
code for all $\delta>0$, $\varepsilon\in(0,1)$, and sufficiently large $n$.
The classical capacity $C(\mathcal{N})$\ of $\mathcal{N}$ is equal to the
supremum of all achievable rates for classical communication.

\subsection{Randomness Distribution}

A sender and receiver can exploit a quantum channel for the alternate but
related task of \textit{randomness distribution}. Here, Alice prepares a local
classical system in a uniformly random state, she makes a copy of it, and the
goal is for Bob to have the copy, such that Alice and Bob share a state of the
following form at the end of the protocol:
\begin{equation}
\overline{\Phi}_{MM^{\prime}}\equiv\sum_{m\in\mathcal{M}}\frac{1}{\left\vert
\mathcal{M}\right\vert }\vert m\rangle\langle m\vert_{M}\otimes\vert
m\rangle\langle m\vert_{M^{\prime}}. \label{eq-cc:common-randomness}%
\end{equation}
Such shared randomness is not particularly useful as a resource, but this
viewpoint is helpful for proving the converse theorem of this chapter and
later on when we encounter other information-processing tasks in quantum
Shannon theory. The main point to note is that a noiseless classical bit
channel can always generate one bit of noiseless shared randomness. Thus, if a
quantum channel has a particular capacity for classical communication, it can
always achieve the same capacity for randomness distribution. In fact, the
capacity for randomness distribution can only be larger than that for
classical communication because shared randomness is a weaker resource than
classical communication. This relationship gives a simple way to bound the
capacity for classical communication from above by the capacity for randomness distribution.

The most general protocol for randomness distribution is as follows. Alice
begins by locally preparing a state of the form in
\eqref{eq-cc:common-randomness}. She then performs an encoding channel that
transforms this state to the following one:%
\begin{equation}
\sum_{m\in\mathcal{M}}\frac{1}{\left\vert \mathcal{M}\right\vert }%
|m\rangle\langle m|_{M}\otimes\rho_{A^{\prime n}}^{m},
\end{equation}
and she transmits the $A^{\prime n}$ systems over $n$ independent uses of the
quantum channel $\mathcal{N}$, producing the following state:%
\begin{equation}
\omega_{MB^{n}}\equiv\sum_{m\in\mathcal{M}}\frac{1}{\left\vert \mathcal{M}%
\right\vert }|m\rangle\langle m|_{M}\otimes\mathcal{N}^{\otimes n}%
(\rho_{A^{\prime n}}^{m}). \label{eq-cc:cq-state-after-channel}%
\end{equation}
Bob then performs a quantum instrument on the received systems (exploiting
some POVM $\left\{  \Lambda_{m}\right\}  $), and the resulting state is%
\begin{equation}
\sum_{m,m^{\prime}\in\mathcal{M}}\frac{1}{\left\vert \mathcal{M}\right\vert
}|m\rangle\langle m|_{M}\otimes\sqrt{\Lambda_{m^{\prime}}}\mathcal{N}^{\otimes
n}(\rho_{A^{\prime n}}^{m})\sqrt{\Lambda_{m^{\prime}}}\otimes|m^{\prime
}\rangle\langle m^{\prime}|_{M^{\prime}}. \label{eq-cc:shared-state}%
\end{equation}
The state%
\begin{equation}
\omega_{MM^{\prime}}=\sum_{m,m^{\prime}\in\mathcal{M}}\frac{1}{\left\vert
\mathcal{M}\right\vert }\operatorname{Tr}\left\{  \Lambda_{m^{\prime}%
}\mathcal{N}^{\otimes n}(\rho_{A^{\prime n}}^{m})\right\}  |m\rangle\langle
m|_{M}\otimes|m^{\prime}\rangle\langle m^{\prime}|_{M^{\prime}}%
\end{equation}
should then be $\varepsilon$-close in trace distance to the original state in
\eqref{eq-cc:common-randomness} for an $(n,C,\varepsilon)$ protocol for
randomness distribution:%
\begin{equation}
\frac{1}{2}\left\Vert \overline{\Phi}_{MM^{\prime}}-\omega_{MM^{\prime}%
}\right\Vert _{1}\leq\varepsilon. \label{eq-cc:crg-err}%
\end{equation}

A rate $C$ for randomness distribution is achievable if there exists an
$\left(  n,C-\delta,\varepsilon\right)  $ randomness distribution code for all
$\delta>0$, $\varepsilon\in(0,1)$, and sufficiently large $n$. The
capacity\ for randomness distribution is equal to the supremum of all
achievable rates. Clearly, from the definitions, it follows that the classical
capacity of a channel can never exceed the capacity for randomness distribution.

\section{The Classical Capacity Theorem}

\label{sec-cc:cc-theorem}We now state the main theorem of this chapter, the
\index{classical capacity theorem}%
classical capacity theorem.

\begin{theorem}
[Holevo--Schumacher--Westmoreland]\label{thm-cc:HSW}The classical capacity of
a quantum channel
\index{HSW theorem}%
is equal to the regularization of the Holevo information of the channel:%
\begin{equation}
C(\mathcal{N})=\chi_{\operatorname{reg}}(\mathcal{N}),
\label{eq-cc:capacity-theorem}%
\end{equation}
where%
\begin{equation}
\chi_{\operatorname{reg}}(\mathcal{N})\equiv\lim_{k\rightarrow\infty}\frac
{1}{k}\chi(\mathcal{N}^{\otimes k}), \label{eq-cc:reg-holevo}%
\end{equation}
and the Holevo information $\chi(\mathcal{N})$ of a channel $\mathcal{N}$ is
defined in \eqref{eq-cc:holevo-formula}.
\end{theorem}

The regularization in the above characterization is a reflection of our
ignorance of a better formula for the classical capacity of a quantum channel.
The proof of the above theorem in the next two sections demonstrates that the
above quantity is indeed equal to the classical capacity, but the
regularization implies that the above characterization is intractable for
general quantum channels. However, if the Holevo information of a particular
channel is additive (in the sense discussed in Chapter~\ref{chap:additivity}),
then $\chi_{\operatorname{reg}}(\mathcal{N})=\chi(\mathcal{N})$, the classical
capacity formula simplifies for such a channel, and we can claim to have a
complete understanding of the channel's classical transmission capabilities.
This \textquotedblleft all-or-nothing\textquotedblright\ situation with
capacities is quite common in quantum Shannon theory, and it implies that we
still have much remaining to understand regarding classical information
transmission over quantum channels.

The next two sections prove the above capacity theorem in two parts:\ the
direct coding theorem and the converse theorem. The proof of the direct coding
theorem demonstrates the inequality LHS $\geq$ RHS in
\eqref{eq-cc:capacity-theorem}. That is, it shows that the regularized Holevo
information is an achievable rate for classical communication, and it exploits
typical and conditionally typical subspaces and the packing lemma to do so.
The proof of the converse theorem shows the inequality LHS\ $\leq$ RHS\ in
\eqref{eq-cc:capacity-theorem}. That is, it shows that any sequence of
protocols with achievable rate $C$ (with vanishing error in the large $n$
limit) should have this rate below the regularized Holevo information. The
proof of the converse theorem exploits the aforementioned idea of randomness
distribution, continuity of entropy, and the quantum data-processing inequality.

\subsection{The Direct Coding Theorem}

\label{sec-cc:direct-coding}We first prove the direct coding theorem.
\index{HSW theorem!direct part}%
Suppose that a quantum channel $\mathcal{N}$ connects Alice to Bob, and they
are allowed access to many independent uses of this quantum channel. Alice can
choose some ensemble $\left\{  p_{X}(x),\rho^{x}\right\}  $ of states which
she can exploit to make a random code for this channel. She selects
$\left\vert \mathcal{M}\right\vert $ codewords $\left\{  x^{n}(m)\right\}
_{m\in\left\{  1,\ldots,\left\vert \mathcal{M}\right\vert \right\}  }$
independently according to the following distribution:%
\begin{equation}
p_{X^{\prime n}}^{\prime}(x^{n})=\left\{
\begin{array}
[c]{cc}%
\left[  \sum_{x^{n}\in T_{\delta}^{X^{n}}}p_{X^{n}}(x^{n})\right]
^{-1}p_{X^{n}}(x^{n}) & :x^{n}\in T_{\delta}^{X^{n}}\\
0 & :x^{n}\notin T_{\delta}^{X^{n}}%
\end{array}
\right.  ,
\end{equation}
where $X^{\prime n}$ is a random variable with
distribution $p_{X^{\prime n}}^{\prime}(x^{n})$, $p_{X^{n}}(x^{n})=p_{X}%
(x_{1})\cdots p_{X}(x_{n})$, and $T_{\delta}^{X^{n}}$ denotes the set of
strongly typical sequences for the distribution $p_{X^{n}}(x^{n})$ (see
Section~\ref{sec-ct:strong-typ}). This \textquotedblleft
pruned\textquotedblright\ distribution is approximately close to the
i.i.d.~distribution $p_{X^{n}}(x^{n})$ because the probability mass of the
typical set is nearly one. In fact, from \eqref{eq-em:typ-proj-close}, we have
that the following holds if\ $\Pr\left\{  X^{n}\in T_{\delta}^{X^{n}}\right\}
\geq1-\varepsilon$:%
\begin{equation}
\sum_{x^{n}\in\mathcal{X}^{n}}\left\vert p_{X^{\prime n}}^{\prime}%
(x^{n})-p_{X^{n}}(x^{n})\right\vert \leq2\varepsilon.
\end{equation}
Indeed, we know from typicality that $\Pr\left\{  X^{n}\in T_{\delta}^{X^{n}%
}\right\}  \geq1-\varepsilon$\ for all $\varepsilon\in(0,1)$ and sufficiently
large $n$.

These classical codewords $\left\{  x^{n}(m)\right\}  _{m\in\left\{
1,\ldots,\left\vert \mathcal{M}\right\vert \right\}  }$ lead to quantum
codewords of the following form:%
\begin{equation}
\rho^{x^{n}(m)}\equiv\rho^{x_{1}(m)}\otimes\cdots\otimes\rho^{x_{n}(m)},
\end{equation}
by exploiting the quantum states in the ensemble $\left\{  p_{X}(x),\rho
^{x}\right\}  $. Alice then transmits these codewords through the channel,
leading to the following tensor-product density operators:%
\begin{align}
\sigma^{x^{n}(m)}  &  \equiv\sigma^{x_{1}(m)}\otimes\cdots\otimes\sigma
^{x_{n}(m)}\\
&  \equiv\mathcal{N}(\rho^{x_{1}(m)})\otimes\cdots\otimes\mathcal{N}%
(\rho^{x_{n}(m)}).
\end{align}
Bob then detects which codeword Alice transmits by exploiting some detection
POVM $\left\{  \Lambda_{m}\right\}  $ that acts on all of the channel outputs.

At this point, we would like to exploit the packing lemma
(Lemma~\ref{lem-pack:pack} from Chapter~\ref{chap:packing}). Recall that four
objects are needed to apply the packing lemma, and they should satisfy four
inequalities. The first object needed is an ensemble from which we can select
a code randomly, and the ensemble in our case is $\left\{  p_{X^{\prime n}%
}^{\prime}(x^{n}),\sigma^{x^{n}}\right\}  $. The next object is the expected
density operator of this ensemble:%
\begin{equation}
\mathbb{E}_{X^{\prime n}}\left\{  \sigma^{X^{\prime n}}\right\}  =\sum
_{x^{n}\in\mathcal{X}^{n}}p_{X^{\prime n}}^{\prime}(x^{n})\sigma^{x^{n}}.
\end{equation}
Finally, we need a message subspace projector and a total subspace projector,
and we let these respectively be the conditionally typical projector
$\Pi_{B^{n}|x^{n}}^{\delta}$ for the state $\sigma^{x^{n}}$ and the typical
projector $\Pi_{B^{n}}^{\delta}$ for the tensor product state $\sigma^{\otimes
n}$ where $\sigma\equiv\sum_{x}p_{X}(x)\sigma^{x}$. Intuitively, the tensor
product state $\sigma^{\otimes n}$ should be close to the expected state
$\mathbb{E}_{X^{\prime n}}\left\{  \sigma^{X^{\prime n}}\right\}  $, and the
next exercise asks you to verify this statement.

\begin{exercise}
Prove that the trace distance between the expected state $\mathbb{E}%
_{X^{\prime n}}\left\{  \sigma^{X^{\prime n}}\right\}  $ and the tensor
product state $\sigma^{\otimes n}$ is small for all sufficiently large $n$:%
\begin{equation}
\left\Vert \mathbb{E}_{X^{\prime n}}\left\{  \sigma^{X^{\prime n}}\right\}
-\sigma^{\otimes n}\right\Vert _{1}\leq2\varepsilon,
\end{equation}
where $\varepsilon$ is an arbitrarily small positive number such that
$\Pr\left\{  X^{n}\in T_{\delta}^{X^{n}}\right\}  \geq1-\varepsilon$.
\end{exercise}

If the four conditions of the packing lemma
\index{packing lemma}%
are satisfied (see \eqref{eq:pack-1}--\eqref{eq:pack-4}), then there exists a
coding scheme with a detection POVM\ that has an arbitrarily low maximal
probability of error as long as the number of messages in the code is not too
high. We now show how to satisfy these four conditions by exploiting the
properties of typical and conditionally typical projectors. The following
three conditions follow from the properties of typical subspaces:%
\begin{align}
\operatorname{Tr}\left\{  \Pi_{B^{n}}^{\delta}\sigma_{B^{n}}^{x^{n}}\right\}
&  \geq1-\varepsilon,\\
\operatorname{Tr}\left\{  \Pi_{B^{n}|x^{n}}^{\delta}\sigma_{B^{n}}^{x^{n}%
}\right\}   &  \geq1-\varepsilon,\\
\operatorname{Tr}\left\{  \Pi_{B^{n}|x^{n}}^{\delta}\right\}   &
\leq2^{n\left(  H(B|X)+c\delta\right)  },
\end{align}
where $c$ is a strictly positive constant. The first inequality follows from
Property~\ref{prop-qt:cond-state-with-uncond-proj}. The second inequality
follows from Property~\ref{prop-qt:strong-cond-typ-unit}, and the third from
Property~\ref{prop-qt:strong-cond-typ-exp-small}. We leave the proof of the
fourth inequality for the packing lemma as an exercise.

\begin{exercise}
Prove that the following inequality holds:%
\begin{equation}
\Pi_{B^{n}}^{\delta}\mathbb{E}_{X^{\prime n}}\left\{  \sigma_{B^{n}%
}^{X^{\prime n}}\right\}  \Pi_{B^{n}}^{\delta}\leq\left[  1-\varepsilon
\right]  ^{-1}2^{-n\left(  H(B)-c^{\prime}\delta\right)  }\Pi_{B^{n}}^{\delta
},
\end{equation}
where $c^{\prime}$ is a strictly positive constant. (Hint:\ First show that
$\mathbb{E}_{X^{\prime n}}\left\{  \sigma_{B^{n}}^{X^{\prime n}}\right\}
\leq\left[  1-\varepsilon\right]  ^{-1}\sigma_{B^{n}}$ and then apply the
third property of typical subspaces---Property~\ref{prop-qt:equi}.)
\end{exercise}

With these four conditions holding, it follows from
Corollary~\ref{cor-pack:derandomized} (the derandomized version of the packing
lemma) that there exists a deterministic code and a POVM\ $\left\{
\Lambda_{m}\right\}  $ that can detect the transmitted states with arbitrarily
low maximal probability of error as long as the size $\left\vert
\mathcal{M}\right\vert $\ of the message set is small enough:%
\begin{align}
p_{e}^{\ast}  &  \equiv\max_{m}\operatorname{Tr}\left\{  \left(  I-\Lambda
_{m}\right)  \mathcal{N}^{\otimes n}(\rho^{x^{n}(m)})\right\} \\
&  \leq4\left(  \varepsilon+2\sqrt{\varepsilon}\right)  +16\left[
1-\varepsilon\right]  ^{-1}2^{-n\left(  H(B)-H(B|X)-(c+c^{\prime}%
)\delta\right)  }\left\vert \mathcal{M}\right\vert \\
&  =4\left(  \varepsilon+2\sqrt{\varepsilon}\right)  +16\left[  1-\varepsilon
\right]  ^{-1}2^{-n\left(  I(X;B)-(c+c^{\prime})\delta\right)  }\left\vert
\mathcal{M}\right\vert .
\end{align}
Thus, we can choose the size of the message set to be $\left\vert
\mathcal{M}\right\vert =2^{n\left(  I(X;B)-(c+c^{\prime}+1)\delta\right)  }$
so that the rate of communication is the Holevo information$~I(X;B)$:%
\begin{equation}
\frac{1}{n}\log\left\vert \mathcal{M}\right\vert =I(X;B)-(c+c^{\prime
}+1)\delta,
\end{equation}
and the bound on the maximal probability of error becomes%
\begin{equation}
p_{e}^{\ast}\leq4\left(  \varepsilon+2\sqrt{\varepsilon}\right)  +16\left[
1-\varepsilon\right]  ^{-1}2^{-n\delta}.
\end{equation}
Let $\varepsilon^{\prime}\in(0,1)$ and $\delta^{\prime}>0$. By picking $n$
large enough, it is clear that we can have both $4\left(  \varepsilon
+2\sqrt{\varepsilon}\right)  +16\left[  1-\varepsilon\right]  ^{-1}%
2^{-n\delta}\leq\varepsilon^{\prime}$ and $(c+c^{\prime}+1)\delta\leq
\delta^{\prime}$. Thus, the Holevo information $I(X;B)_{\rho}$, with respect
to the following classical--quantum state:%
\begin{equation}
\rho_{XB}\equiv\sum_{x\in\mathcal{X}}p_{X}(x)|x\rangle\langle x|_{X}%
\otimes\mathcal{N}(\rho^{x}),
\end{equation}
is an achievable rate for the transmission of classical information over
$\mathcal{N}$.

Alice and Bob can achieve the Holevo information $\chi(\mathcal{N})$\ of the
channel $\mathcal{N}$ simply by selecting a random code according to the
ensemble $\left\{  p_{X}(x),\rho^{x}\right\}  $ that maximizes $I(X;B)_{\rho}%
$. Lastly, they can achieve the rate $\frac{1}{k}\chi(\mathcal{N}^{\otimes
k})$ by coding instead for the tensor-product channel $\mathcal{N}^{\otimes
k}$, and this last result implies that they can achieve the regularization
$\chi_{\operatorname{reg}}(\mathcal{N})$ by making the blocks for which they
are coding be arbitrarily large. This concludes the proof of the direct part
of the coding theorem.

We comment more on the role of entanglement at the encoder before moving on to
the proof of the converse theorem. First, the above coding scheme for the
channel $\mathcal{N}$ does not make use of entangled inputs at the encoder
because the codeword states $\rho^{x^{n}(m)}$ are separable across the channel
inputs. It is only when we code for the tensor-product channel $\mathcal{N}%
^{\otimes k}$ that entanglement comes into play. Here, the codeword states are
of the form%
\begin{equation}
\rho_{x^{n}(m)}\equiv\rho_{A^{\prime k}}^{x_{1}(m)}\otimes\cdots\otimes
\rho_{A^{\prime k}}^{x_{n}(m)}.
\end{equation}
That is, the states $\rho_{A^{\prime k}}^{x_{i}(m)}$\ act on the
tensor-product Hilbert space of $k$ channel inputs and can be entangled across
these $k$ systems. Whether entanglement at the encoder could increase
classical communication rates over general quantum channels was the subject of
much intense work over the past few years, but it is now known that there exists a channel for which exploiting entanglement at
the encoder is strictly better than not exploiting entanglement (see
Section~\ref{sec-cc:superadditivity}).

It is worth re-examining the proof of the packing lemma
(Lemma~\ref{lem-pack:pack}) in order to understand better the decoding POVM at
the receiving end. The particular decoding POVM elements employed in the
packing lemma have the following form:%
\begin{align}
\Lambda_{m}  &  \equiv\left(  \sum_{m^{\prime}\in\mathcal{M}}\Gamma
_{m^{\prime}}\right)  ^{-\frac{1}{2}}\Gamma_{m}\left(  \sum_{m^{\prime}%
\in\mathcal{M}}\Gamma_{m^{\prime}}\right)  ^{-\frac{1}{2}},\\
\Gamma_{m}  &  \equiv\Pi_{B^{n}}^{\delta}\Pi_{B^{n}|x^{n}(m)}^{\delta}%
\Pi_{B^{n}}^{\delta}.
\end{align}
(Simply substitute the conditionally typical projector $\Pi_{B^{n}|x^{n}%
(m)}^{\delta}$ and the typical projector $\Pi_{B^{n}}^{\delta}$ into
\eqref{eq-pack:packing-POVM}.) A POVM\ with the above elements is known as a
\textquotedblleft square-root\textquotedblright\ measurement because
\index{square-root measurement}%
of its particular form. We employ such a measurement at the decoder because it
has nice analytic properties that allow us to obtain a good bound on the
expectation of the average error probability (in particular, we can exploit
the operator inequality from Lemma~\ref{lem-pack:hayashi-nag}). This
measurement is a collective measurement because the conditionally typical
projector and the typical projector are both acting on all of the channel
outputs, and we construct the square-root measurement from these projectors.
Such a decoding POVM\ is far more exotic than the naive strategy overviewed in
Section~\ref{sec-cc:naive} where Bob measures the channel outputs
individually---it is for the construction of this decoding POVM\ and the proof
that it is asymptotically good that Holevo, Schumacher, and Westmoreland were
given much praise for their work. However, there is no known way to implement
this decoding POVM\ efficiently, and the original efficiency problems with the
decoder in the proof of Shannon's noisy classical channel coding theorem
plague the decoders in the quantum world as well.

Alternatively, Bob's decoding operation could take the form of a sequential
decoder%
\index{sequential decoding}%
, as discussed in Section~\ref{sec-pack:sequential}. That is, Bob would
perform the measurements $\{\Pi_{B^{n}|x^{n}(m)}^{\delta},I_{B^{n}}-\Pi
_{B^{n}|x^{n}(m)}^{\delta}\}$ one after another in an attempt to learn which
message was transmitted. This decoding strategy is also a collective
measurement strategy because each of the above measurements is a collective
measurement acting on all of the channel outputs. This scheme is also
inefficient:\ even if there is an efficient implementation for each of the
individual tests, there are an exponential number of them to perform in the
worst case since there are an exponential number of messages in a codebook
(exponential in $n$).

\begin{exercise}
Show that a measurement with POVM\ elements of the following form is
sufficient to achieve the Holevo information of a quantum channel:%
\begin{equation}
\Lambda_{m}\equiv\left(  \sum_{m^{\prime}\in\mathcal{M}}\Pi_{B^{n}%
|x^{n}(m^{\prime})}^{\delta}\right)  ^{-1/2}\Pi_{B^{n}|x^{n}(m)}^{\delta
}\left(  \sum_{m^{\prime}\in\mathcal{M}}\Pi_{B^{n}|x^{n}(m^{\prime})}^{\delta
}\right)  ^{-1/2}.
\end{equation}

\end{exercise}

\subsubsection{Alternate Proofs of the Direct Part}

\label{sec-cc:HSW-alternate-direct}There are at least two alternate proofs of
the direct part of the HSW\ theorem, which can be useful as building blocks
for other scenarios. The first is based on weak typicality and the second is
called the \textit{constant-composition coding} approach. We discuss these
briefly here.

Let $\{p_{X}(x),\rho_{A^{\prime}}^{x}\}$ be an ensemble of states for the
input of the channel $\mathcal{N}_{A^{\prime}\rightarrow B}$, which leads to
an ensemble $\{p_{X}(x),\sigma_{B}^{x}\}$ at the output, where $\sigma_{B}%
^{x}\equiv\mathcal{N}_{A^{\prime}\rightarrow B}(\rho_{A^{\prime}}^{x})$. Let
$\sigma_{B}\equiv\sum_{x}p_{X}(x)\sigma_{B}^{x}$\ be the expected density
operator of the output ensemble. In the weak typicality approach, the
codewords $\{x^{n}(m)\}$ are chosen independently at random according to the
product distribution $p_{X^{n}}(x^{n})=\prod\limits_{i=1}^{n}p_{X}(x_{i})$.
One then sets the total subspace projector to be the weakly typical projection
$\Pi_{B^{n}}^{\delta}$ for $\sigma_{B}^{\otimes n}$ and each message subspace
projection (for the codeword $x^{n}(m)$) to be the weak conditionally typical
projection $\Pi_{B^{n}|x^{n}(m)}^{\delta}$. From the properties of weak
typicality (see Chapter~\ref{chap:quantum-typicality}), the following
conditions for the average version of the packing lemma (see
Exercise~\ref{ex-pack:averaged-conditions}) are satisfied:%
\begin{align}
\sum_{x^{n}\in\mathcal{X}^{n}}p_{X^{n}}(x^{n})\operatorname{Tr}\left\{
\Pi_{B^{n}}^{\delta}\sigma_{B^{n}}^{x^{n}}\right\}   &  \geq1-\varepsilon,\\
\sum_{x^{n}\in\mathcal{X}^{n}}p_{X^{n}}(x^{n})\operatorname{Tr}\left\{
\Pi_{B^{n}|x^{n}}^{\delta}\sigma_{B^{n}}^{x^{n}}\right\}   &  \geq
1-\varepsilon,\\
\operatorname{Tr}\left\{  \Pi_{B^{n}|x^{n}}^{\delta}\right\}   &
\leq2^{n\left(  H(B|X)+\delta\right)  },\\
\Pi_{B^{n}}^{\delta}\sigma_{B}^{\otimes n}\Pi_{B^{n}}^{\delta}  &
\leq2^{-n(H(B)-\delta)}\Pi_{B^{n}}^{\delta}.
\end{align}
We can then conclude that the rate $I(X;B)$ is achievable for classical
communication over $\mathcal{N}_{A^{\prime}\rightarrow B}$, by making use of
this slightly different scheme. (The Holevo information is with respect to the
output ensemble $\{p_{X}(x),\sigma_{B}^{x}\}$.)

The next scheme that we mention is the constant-composition
\index{constant-composition coding}%
coding scheme. Consider again the output ensemble $\{p_{X}(x),\sigma_{B}%
^{x}\}$ with expectation $\sigma_{B}\equiv\sum_{x}p_{X}(x)\sigma_{B}^{x}$. Now
select a typical type class $T_{t}$, as discussed in
Definition~\ref{def-ct:typical-type}\ and
Section~\ref{sec-ct:typical-type-class-card}. This is a set of all the
sequences $x^{n}$ with empirical distribution $t(x)$\ that deviates from the
true distribution $p_{X}(x)$\ by no more than $\delta>0$. All the sequences in
the same type class are related to one another by a permutation, and all of
them are strongly typical. The idea for selecting a code at random is now to
pick all of the codewords independently and uniformly at random from the
typical type class $t$. Thus, the ensemble from which we are selecting
codewords is now $\{1/\left\vert T_{t}\right\vert ,\sigma_{B^{n}}^{x^{n}%
}\}_{x^{n}\in T_{t}}$ and we would like to show that in doing so, we can still
achieve a rate equal to $I(X;B)$. Let%
\begin{equation}
\widetilde{\sigma}_{B^{n}}\equiv\left\vert T_{t}\right\vert ^{-1}\sum
_{x^{n}\in T_{t}}\sigma_{B^{n}}^{x^{n}},\ \ \ \ \ \ \overline{\sigma}%
_{B}\equiv\sum_{x\in\mathcal{X}}t(x)\sigma_{B}^{x}.
\end{equation}
Observe that $\frac{1}{2}\left\Vert \overline{\sigma}_{B}-\sigma
_{B}\right\Vert _{1}\leq\left\vert \mathcal{X}\right\vert \delta/2$, and thus
$\left\vert H(B)_{\overline{\sigma}}-H(B)_{\sigma}\right\vert \leq
\eta(\left\vert \mathcal{X}\right\vert \delta)$, with $\eta(\cdot)$ defined
just after \eqref{eq-ct:last-step-card-typ-type} and such that $\lim
_{\delta\rightarrow0}\eta(\left\vert \mathcal{X}\right\vert \delta)=0$. We can
take the total subspace projector to be the strongly typical projection
$\Pi_{B^{n}}^{\delta}$\ for $\overline{\sigma}_{B}^{\otimes n}$, and we take
the message subspace projection (for the codeword $x^{n}(m)$) to be the strong
conditionally typical projector $\Pi_{B^{n}|x^{n}(m)}^{\delta}$. We then need
to verify that the conditions of the packing lemma
(Corollary~\ref{cor-pack:derandomized}) hold. Consider that%
\begin{align}
\operatorname{Tr}\left\{  \Pi_{B^{n}}^{\delta}\sigma_{B^{n}}^{x^{n}}\right\}
&  \geq1-\varepsilon,\\
\operatorname{Tr}\left\{  \Pi_{B^{n}|x^{n}}^{\delta}\sigma_{B^{n}}^{x^{n}%
}\right\}   &  \geq1-\varepsilon,\\
\operatorname{Tr}\left\{  \Pi_{B^{n}|x^{n}}^{\delta}\right\}   &
\leq2^{n\left(  H(B|X)+c\delta\right)  }.
\end{align}
The first inequality follows because each $x^{n}\in T_{t}$ is strongly typical
with respect to $t(x)$, so that we can apply
Property~\ref{prop-qt:cond-state-with-uncond-proj}. The second two
inequalities are properties of strong conditional typicality. So we need to
figure out something about the last condition. Let $t^{n}(x^{n})\equiv
\prod\limits_{i=1}^{n}t(x_{i})$, i.e., the product distribution realized by
$t(x)$. Consider that%
\begin{align}
\widetilde{\sigma}_{B^{n}}  &  =\frac{1}{\left\vert T_{t}\right\vert }%
\sum_{x^{n}\in T_{t}}\sigma_{B^{n}}^{x^{n}}=\sum_{x^{n}\in\mathcal{X}^{n}%
}\frac{I(x^{n}\in T_{t})}{\left\vert T_{t}\right\vert }\sigma_{B^{n}}^{x^{n}%
}\\
&  \leq\sum_{x^{n}\in\mathcal{X}^{n}}\left(  n+1\right)  ^{\left\vert
\mathcal{X}\right\vert }t^{n}(x^{n})\sigma_{B^{n}}^{x^{n}}=\left(  n+1\right)
^{\left\vert \mathcal{X}\right\vert }\overline{\sigma}_{B}^{\otimes n}.
\end{align}
The inequality follows from the development in
\eqref{eq-ct:typical-type-proof-1}--\eqref{eq-ct:typical-type-proof-last}.
Using this, we find that%
\begin{equation}
\Pi_{B^{n}}^{\delta}\widetilde{\sigma}_{B^{n}}\Pi_{B^{n}}^{\delta}%
\leq2^{-n(H(B)_{\overline{\sigma}}-\frac{1}{n}\left\vert \mathcal{X}%
\right\vert \log(n+1))}\Pi_{B^{n}}^{\delta},
\end{equation}
which is the last condition we need for the packing lemma
(Corollary~\ref{cor-pack:derandomized}). We can then conclude that the rate
$H(B)_{\overline{\sigma}}-H(B|X)$ is achievable for classical communication
using constant-composition codes. However, since $\left\vert H(B)_{\overline
{\sigma}}-H(B)_{\sigma}\right\vert \leq\eta(\left\vert \mathcal{X}\right\vert
\delta)$, we can also conclude that the rate $I(X;B)$ is achievable for
classical communication using constant-composition codes, where the Holevo
information is with respect to the original output ensemble $\{p_{X}%
(x),\sigma_{B}^{x}\}$.

\subsection{The Converse Theorem}

\label{sec-cc:converse}The second part of the classical capacity theorem is
\index{HSW theorem!converse}
the converse theorem, and we provide a simple proof of it in this section.
Suppose that Alice and Bob are trying to accomplish randomness distribution
rather than classical communication---the capacity for such a task can only be
larger than that for classical communication as we argued before in
Section~\ref{sec-cc:info-proc-task}. Recall that in such a task, Alice first
prepares a maximally correlated state $\overline{\Phi}_{MM^{\prime}}$ so that
the rate of randomness distribution is equal to $\frac{1}{n}\log\left\vert
\mathcal{M}\right\vert $. Alice and Bob share a state of the form in
\eqref{eq-cc:shared-state} after encoding, channel transmission, and decoding.
We now show that the regularized Holevo information in
\eqref{eq-cc:reg-holevo} bounds the capacity of randomness distribution. As a
result, the regularized Holevo information also bounds the capacity for
classical communication from above. Consider the following chain of
inequalities:%
\begin{align}
\log\left\vert \mathcal{M}\right\vert  &  =I(M;M^{\prime})_{\overline{\Phi}}\\
&  \leq I(M;M^{\prime})_{\omega}+f(\left\vert \mathcal{M}\right\vert
,\varepsilon)\label{eq-cc:first-ineq-converse}\\
&  \leq I(M;B^{n})_{\omega}+f(\left\vert \mathcal{M}\right\vert ,\varepsilon
)\\
&  \leq\chi(\mathcal{N}^{\otimes n})+f(\left\vert \mathcal{M}\right\vert
,\varepsilon). \label{eq-cc:final-upper-bound}%
\end{align}
The first equality follows because the mutual information of the shared
randomness state $\overline{\Phi}_{MM^{\prime}}$\ is equal to $\log\left\vert
\mathcal{M}\right\vert $ bits. The first inequality follows from the error
criterion in \eqref{eq-cc:crg-err} and by applying the AFW inequality
(Theorem~\ref{thm-qie:AFW-cont-ent}). That is, since $H(M)_{\overline{\Phi}%
}=H(M)_{\omega}$, we know that%
\begin{align}
&  \left\vert I(M;M^{\prime})_{\overline{\Phi}}-I(M;M^{\prime})_{\omega
}\right\vert \\
&  =\left\vert H(M)_{\overline{\Phi}}-H(M|M^{\prime})_{\overline{\Phi}%
}-\left[  H(M)_{\omega}-H(M|M^{\prime})_{\omega}\right]  \right\vert \\
&  =\left\vert H(M|M^{\prime})_{\omega}-H(M|M^{\prime})_{\overline{\Phi}%
}\right\vert \\
&  \leq f(\left\vert \mathcal{M}\right\vert ,\varepsilon)\equiv\varepsilon
\log\left\vert \mathcal{M}\right\vert +g_2(\varepsilon).
\end{align}
The second inequality results from the quantum data-processing inequality for
quantum mutual information (Theorem~\ref{cor-qie:QDP})---recall that Bob
processes the $B^{n}$ system with a quantum instrument to get the classical
system $M^{\prime}$. Also, the quantum mutual information is evaluated on a
classical--quantum state of the form in \eqref{eq-cc:cq-state-after-channel}.
The final inequality follows because the classical--quantum state in
\eqref{eq-cc:cq-state-after-channel} has a particular distribution and choice
of states, and this choice always leads to a value of the quantum mutual
information that cannot be greater than the Holevo information of the
tensor-product channel $\mathcal{N}^{\otimes n}$.\ Putting everything
together, we find that%
\begin{equation}
\frac{1}{n}\log\left\vert \mathcal{M}\right\vert \left(  1-\varepsilon\right)
\leq\frac{1}{n}\chi(\mathcal{N}^{\otimes n})+\frac{1}{n}g_2(\varepsilon).
\end{equation}
Thus, if we are considering a sequence of $(n,\left[  \log\left\vert
\mathcal{M}\right\vert \right]  /n,\varepsilon_{n})$\ classical communication
protocols with rate $C-\delta_{n}=\frac{1}{n}\log\left\vert \mathcal{M}%
\right\vert $, such that $\lim_{n\rightarrow\infty}\varepsilon_{n}%
=\lim_{n\rightarrow\infty}\delta_{n}=0$, then the above bound becomes%
\begin{equation}
\left(  C-\delta_{n}\right)  \left(  1-\varepsilon_{n}\right)  \leq\frac{1}%
{n}\chi(\mathcal{N}^{\otimes n})+\frac{1}{n}g_2(\varepsilon_n).
\end{equation}
Taking the limit as $n\rightarrow\infty$ then establishes that an achievable
rate $C$ necessarily satisfies $C\leq\chi_{\operatorname{reg}}(\mathcal{N})$,
where $\chi_{\operatorname{reg}}(\mathcal{N})$ is the regularized Holevo
formula given in \eqref{eq-cc:reg-holevo}.

\section{Examples of Channels}

Observe that the final upper bound in \eqref{eq-cc:final-upper-bound} on the
rate $C$ is the multi-letter Holevo information of the channel. It would be
more desirable to have $\chi(\mathcal{N})$ as the upper bound on $C$ rather
than $\frac{1}{n}\chi(\mathcal{N}^{\otimes n})$ because the former is simpler,
and the optimization problem set out in the latter quantity is simply
impossible to compute in general using finite computational resources (for
large $n$). However, the upper bound in \eqref{eq-cc:final-upper-bound} is the
best known upper bound if we do not know anything else about the structure of
the channel, and for this reason, the best known characterization of the
classical capacity is the one given in \eqref{eq-cc:capacity-theorem}.

If we know that the Holevo information of the tensor product of a certain
channel with an arbitrary number of copies of itself is additive, then there
is no need for the regularization $\chi_{\operatorname{reg}}(\mathcal{N})$,
and the characterization in Theorem~\ref{thm-cc:HSW}\ reduces to a very good
one:\ the Holevo information $\chi(\mathcal{N})$. There are many examples of
channels for which the classical capacity reduces to the Holevo information of
the channel, and we detail a few such classes of examples in this section:\
\index{entanglement-breaking channel}%
entanglement-breaking channels, quantum Hadamard channels, erasure channels,
and quantum depolarizing channels. The proof that demonstrates additivity of
the Holevo information for each of these channels depends explicitly on
structural properties of each one, and there is unfortunately not much to
learn from these proofs in order to say anything about additivity of the
Holevo information of general quantum channels. Nevertheless, it is good to
have some natural channels for which we can compute their classical capacity,
and it is instructive to examine these proofs in detail to understand what it
is about each channel that makes their Holevo information additive.

\subsection{Classical Capacity of Entanglement-Breaking Channels}

We have already seen in Section~\ref{sec-add:holevo-additivity-ent-break}%
\ that the Holevo information of an entanglement-breaking channel is additive.
As a consequence, we can conclude that the capacity of an
\index{entanglement-breaking channel}%
entanglement-breaking channel $\mathcal{N}$\ is given by $\chi(\mathcal{N})$.

We now focus our discussion on cq channels. Recall from
Section~\ref{sec-nqt:entanglement-breaking}\ that a quantum channel is a
particular kind of entanglement-breaking channel (cq channel) if the action of
the channel is equivalent to performing first a complete projective
measurement of the input and then preparing a quantum state conditioned on the
value of the classical variable resulting from the measurement. Additionally,
Theorem~\ref{thm-ie:holevo-concave-input} states that the Holevo information
is a concave function of the input distribution over which we are optimizing
for such channels. Thus, computing the classical capacity of cq channels can
be performed by optimization techniques because the Holevo information is
additive for them.

\subsubsection{The Relation to General Channels}

We can always exploit the above result regarding cq entanglement-breaking
channels to get a reasonable lower bound on the classical capacity of any
quantum channel $\mathcal{N}$. The sender Alice can simulate an
entanglement-breaking channel by modifying the processing at the input of an
arbitrary quantum channel. She can first measure the input to her simulated
channel in the basis $\left\{  |x\rangle\langle x|\right\}  $, prepare a state
$\rho_{x}$ conditioned on the outcome of the measurement, and subsequently
feed this state into the channel $\mathcal{N}$. These actions are equivalent
to the following channel:%
\begin{equation}
\sigma\rightarrow\sum_{x}\langle x|\sigma|x\rangle\mathcal{N}(\rho_{x}),
\end{equation}
and the capacity of this simulated channel is equal to%
\begin{equation}
I(X;B)_{\rho},
\end{equation}
where%
\begin{align}
\rho_{XB}  &  \equiv\sum_{x}p_{X}(x)|x\rangle\langle x|_{X}\otimes
\mathcal{N}(\rho_{x}),\\
p_{X}(x)  &  \equiv\langle x|\sigma|x\rangle.
\end{align}
Of course, Alice has the freedom to prepare whichever state $\sigma$ she would
like to be input to the simulated channel, and she also has the ability to
prepare whichever states $\rho_{x}$\ she would like to be conditioned on the
outcomes of the first measurement, so we should let her maximize the Holevo
information over all these inputs. Thus, the capacity of the
entanglement-breaking channel composed with the actual channel is equal to the
Holevo information of the original channel:%
\begin{equation}
\max_{p_{X}(x),\rho_{x}}I(X;B)_{\rho}.
\end{equation}
This capacity is also known as the product-state capacity of the channel
because it is the capacity achieved by inputting unentangled, separable states
at the encoder (Alice can in fact just input product states), and it can be a
good lower bound on the true classical capacity of a quantum channel, even if
it does not allow for entanglement at the encoder.

\subsection{Classical Capacity of Quantum Hadamard Channels}

Recall from Section~\ref{sec-nqt:hadamard-channel}\ that quantum Hadamard
channels
\index{Hadamard channel}%
are those with a complementary channel that is entanglement breaking, and this
property allows us to prove that the Holevo information of the original
channel is additive. Several important natural channels are quantum Hadamard
channels. A trivial example is the noiseless qubit channel because Bob could
perform a projective measurement of his system and send a constant state to
Eve. A less trivial example of a quantum Hadamard channel is a generalized
dephasing channel (see Section~\ref{sec-nqt:gen-deph}), but this channel
trivially has a maximal classical capacity of $\log d$ bits per channel use
because this channel transmits a preferred orthonormal basis without error. A
quantum Hadamard channel with a more interesting classical capacity is known
as a cloning channel, the channel induced by a universal cloning machine
(however we will not discuss this channel in any detail).

\begin{theorem}
\label{thm-cc:hadamard-additive}The Holevo information of a quantum Hadamard
channel $\mathcal{N}_{\operatorname{H}}$ and any other channel $\mathcal{N}$
is additive:%
\begin{equation}
\chi(\mathcal{N}_{\operatorname{H}}\otimes\mathcal{N})=\chi(\mathcal{N}%
_{\operatorname{H}})+\chi(\mathcal{N}).
\end{equation}

\end{theorem}

\begin{proof}
First, recall from Theorem~\ref{thm-add:pure-states-suff-holevo}\ that it is
sufficient to consider ensembles of pure states at the input of the channel
when maximizing its Holevo information. That is, we only need to consider
classical--quantum states of the following form:%
\begin{equation}
\sigma_{XA^{\prime}}\equiv\sum_{x}p_{X}(x)|x\rangle\langle x|_{X}\otimes
|\phi_{x}\rangle\langle\phi_{x}|_{A^{\prime}},
\label{eq-cc:HSW-c-q-state-pure}%
\end{equation}
where $A^{\prime}$ is the input to some channel $\mathcal{N}_{A^{\prime
}\rightarrow B}$. Let $\omega_{XBE}\equiv\mathcal{U}_{A^{\prime}\rightarrow
BE}^{\mathcal{N}}(\sigma_{XA^{\prime}})$ where $U_{A^{\prime}\rightarrow
BE}^{\mathcal{N}}$ is an isometric extension of the channel and $\mathcal{U}%
_{A^{\prime}\rightarrow BE}^{\mathcal{N}}$ denotes the corresponding channel.
Thus, the Holevo information of $\mathcal{N}_{A^{\prime}\rightarrow B}$ is
equal to a different expression:%
\begin{align}
\chi(\mathcal{N})  &  \equiv\max_{\sigma}I(X;B)_{\omega}\\
&  =\max_{\sigma}\left[  H(B)_{\omega}-H(B|X)_{\omega}\right] \\
&  =\max_{\sigma}\left[  H(B)_{\omega}-H(E|X)_{\omega}\right]  ,
\label{eq-cc:holevo-info}%
\end{align}
where the second equality follows from the definition of the quantum mutual
information, and the third equality follows because, conditioned on $X$, the
input to the channel is pure and the entropies $H(B|X)_{\omega}$ and
$H(E|X)_{\omega}$ are equal.

\begin{exercise}
\label{ex-cc:Bob-minus-cond-Eve-pure-suff}Prove that it is sufficient to
consider pure-state inputs when maximizing the following entropy difference
over classical--quantum states:%
\begin{equation}
\max_{\sigma}\left[  H( B) _{\omega}-H( E|X) _{\omega}\right]  .
\end{equation}

\end{exercise}

Suppose now that $\sigma$ is a state that maximizes the Holevo information of
the joint channel $\mathcal{N}_{\operatorname{H}}\otimes\mathcal{N}$, and
suppose it has the following form:%
\begin{equation}
\sigma_{XA_{1}^{\prime}A_{2}^{\prime}}\equiv\sum_{x}p_{X}(x)|x\rangle\langle
x|_{X}\otimes|\phi_{x}\rangle\langle\phi_{x}|_{A_{1}^{\prime}A_{2}^{\prime}}.
\label{eq-cc:additivity-state}%
\end{equation}
Let%
\begin{equation}
\omega_{XB_{1}B_{2}E_{1}E_{2}}\equiv(\mathcal{U}_{A_{1}^{\prime}\rightarrow
B_{1}E_{1}}^{\mathcal{N}_{\operatorname{H}}}\otimes\mathcal{U}_{A_{2}^{\prime
}\rightarrow B_{2}E_{2}}^{\mathcal{N}})(\sigma_{XA_{1}^{\prime}A_{2}^{\prime}%
}).
\end{equation}
The Hadamard channel is
\index{degradable channel}%
degradable, and the degrading channel from Bob to Eve takes a particular
form:\ it is a measurement that produces a classical variable $Y$, followed by
the preparation of a state conditioned on the outcome of the measurement. Let
$\mathcal{D}_{B_{1}\rightarrow Y}^{1}$ be the first part of the degrading
channel that produces the classical variable $Y$, and let $\theta
_{XYE_{1}B_{2}E_{2}}\equiv\mathcal{D}_{B_{1}\rightarrow Y}^{1}(\omega
_{XB_{1}B_{2}E_{1}E_{2}})$. Let $\mathcal{D}_{Y\rightarrow E_{1}^{\prime}}%
^{2}$ be the second part of the degrading channel that produces the state of
$E_{1}$ conditioned on the classical variable $Y$, and let $\tau
_{XE_{1}^{\prime}E_{1}B_{2}E_{2}}\equiv\mathcal{D}_{Y\rightarrow E_{1}%
^{\prime}}^{2}(\theta_{XYE_{1}B_{2}E_{2}})$.
Figure~\ref{fig-cc:additivity-hadamard}\ summarizes these structural
relationships.
\begin{figure}
[ptb]
\begin{center}
\includegraphics[
width=3.5405in
]%
{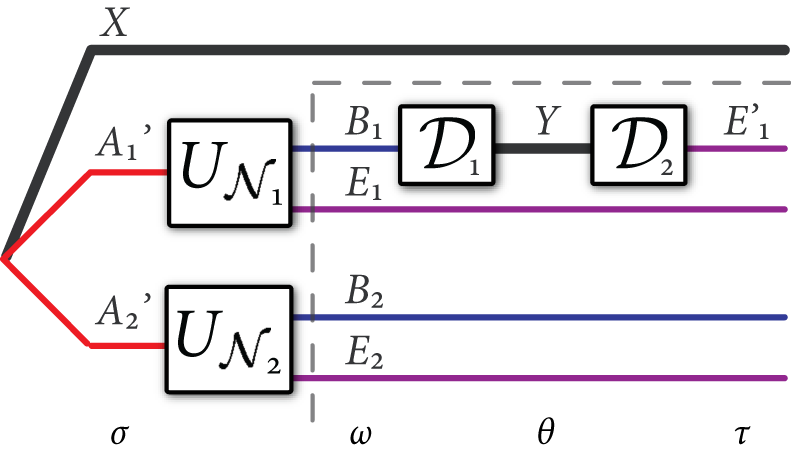}%
\caption{A summary of the structural relationships for the additivity question
if one channel is a quantum Hadamard channel. Alice first prepares a state of
the form in \eqref{eq-cc:additivity-state}. She transmits one system
$A_{1}^{\prime}$ through the quantum Hadamard channel and the other
$A_{2}^{\prime}$ through the other channel. The first Bob $B_{1}$ at the
output of the Hadamard channel can simulate the channel to the first Eve
$E_{1}$ because the first channel is a quantum Hadamard channel. He performs a
complete projective measurement of his system, leading to a classical variable
$Y$, followed by the preparation of some state conditioned on the value of the
classical variable $Y$. The bottom of the figure labels the state of the
systems at each step.}%
\label{fig-cc:additivity-hadamard}%
\end{center}
\end{figure}
Consider the following chain of inequalities:%
\begin{align}
I(X;B_{1}B_{2})_{\omega}  &  =H(B_{1}B_{2})_{\omega}-H(B_{1}B_{2}|X)_{\omega
}\\
&  =H(B_{1}B_{2})_{\omega}-H(E_{1}E_{2}|X)_{\omega}\\
&  \leq H(B_{1})_{\omega}+H(B_{2})_{\omega}-H(E_{1}|X)_{\omega}-H(E_{2}%
|E_{1}X)_{\omega}\\
&  =H(B_{1})_{\omega}-H(E_{1}|X)_{\omega}+H(B_{2})_{\omega}-H(E_{2}%
|E_{1}^{\prime}X)_{\tau}\\
&  \leq H(B_{1})_{\omega}-H(E_{1}|X)_{\omega}+H(B_{2})_{\theta}-H(E_{2}%
|YX)_{\theta}\\
&  \leq\chi(\mathcal{N}_{\operatorname{H}})+\chi(\mathcal{N}).
\end{align}
The first equality follows from the definition of the quantum mutual
information. The second equality follows because $H(B_{1}B_{2}|X)_{\omega
}=H(E_{1}E_{2}|X)_{\omega}$ when the conditional inputs $|\phi_{x}%
\rangle_{A_{1}^{\prime}A_{2}^{\prime}}$ to the channel are pure states. The
next inequality follows from subadditivity of entropy $H(B_{1}B_{2})_{\omega
}\leq H(B_{1})_{\omega}+H(B_{2})_{\omega}$ and from the chain rule for
entropy: $H(E_{1}E_{2}|X)_{\omega}=H(E_{1}|X)_{\omega}+H(E_{2}|E_{1}%
X)_{\omega}$. The third equality follows from a rearrangement of terms and
realizing that the state of $\tau$ on systems $E_{1}^{\prime}E_{2}X$ is equal
to the state of $\omega$ on the same systems. The second inequality follows
from the quantum data-processing inequality $I(E_{2};E_{1}^{\prime}|X)_{\tau
}\leq I(E_{2};Y|X)_{\theta}$. The final inequality follows because the state
$\omega$ is a state of the form in \eqref{eq-cc:holevo-info}, because the
entropy difference is never greater than the Holevo information of the first
channel, and from the result of
Exercise~\ref{ex-cc:Bob-minus-cond-Eve-pure-suff}. The same reasoning follows
for the other entropy difference and by noting that the classical system is
the composite system $XY$.
\end{proof}

\subsection{Classical Capacity of the Quantum Erasure Channel}%

\index{quantum erasure channel}%
The quantum erasure channel is one of the simplest channels for which we can
compute the classical capacity. Recall from Section~\ref{sec-nqt:erasure}%
\ that the qudit erasure channel is defined as follows:%
\begin{equation}
\rho\rightarrow(1-\varepsilon)\rho+\varepsilon|e\rangle\langle e|,
\end{equation}
where $\rho$ is a $d$-dimensional qudit input state, $\varepsilon\in\left[
0,1\right]  $, and $|e\rangle$ is an erasure symbol orthogonal to the input
space of the channel.

\begin{theorem}
[Classical Capacity of the Erasure Channel]The classical capacity of the
$d$-dimensional quantum erasure channel is equal to $\left(  1-\varepsilon
\right)  \log d$.
\end{theorem}

\begin{proof}
The rate $\left(  1-\varepsilon\right)  \log d$ is achievable by picking the
input ensemble to be $\left\{  1/d,|i\rangle\right\}  $ and evaluating the
Holevo information. So we need to show that it is not possible to achieve a
rate higher than this. Let $\mathcal{M}_{A_{1}\rightarrow B_{1}}$ be some
quantum channel and let $\mathcal{N}_{A_{2}\rightarrow B_{2}}^{\varepsilon}$
denote the quantum erasure channel. Let $\rho_{XA_{1}A_{2}}$ be the following
classical--quantum state%
\begin{equation}
\rho_{XA_{1}A_{2}}\equiv\sum_{x}p_{X}(x)|x\rangle\langle x|_{X}\otimes
\phi_{A_{1}A_{2}}^{x},
\end{equation}
and suppose that it achieves the Holevo information $\chi(\mathcal{M}%
\otimes\mathcal{N}^{\varepsilon})$. Consider that%
\begin{align}
\omega_{XB_{1}B_{2}}  &  \equiv(\mathcal{M}_{A_{1}\rightarrow B_{1}}%
\otimes\mathcal{N}_{A_{2}\rightarrow B_{2}}^{\varepsilon})(\rho_{XA_{1}A_{2}%
})\\
&  =\sum_{x}p_{X}(x)|x\rangle\langle x|_{X}\otimes\left[  (1-\varepsilon
)\mathcal{M}_{A_{1}\rightarrow B_{1}}(\phi_{A_{1}A_{2}}^{x})+\varepsilon
\mathcal{M}_{A_{1}\rightarrow B_{1}}(\phi_{A_{1}}^{x})\otimes|e\rangle\langle
e|_{B_{2}}\right]  .
\end{align}
Consider that the isometry $\left[  |0\rangle\langle0|_{B_{2}}+\ldots
|d-1\rangle\langle d-1|_{B_{2}}\right]  \otimes|0\rangle_{Y}+|e\rangle\langle
e|_{B_{2}}\otimes|1\rangle_{Y}$ takes the above state to the following one:%
\begin{multline}
\omega_{XB_{1}B_{2}Y}\equiv\sum_{x}p_{X}(x)|x\rangle\langle x|_{X}%
\otimes\mathcal{M}_{A_{1}\rightarrow B_{1}}(\phi_{A_{1}A_{2}}^{x}%
)\otimes(1-\varepsilon)|0\rangle\langle0|_{Y}\\
+\sum_{x}p_{X}(x)|x\rangle\langle x|_{X}\otimes\mathcal{M}_{A_{1}\rightarrow
B_{1}}(\phi_{A_{1}}^{x})\otimes|e\rangle\langle e|_{B_{2}}\otimes
\varepsilon|1\rangle\langle1|_{Y},
\end{multline}
so that the $Y$ register is a flag indicating whether an erasure occurred.
Then%
\begin{align}
\chi(\mathcal{M}\otimes\mathcal{N}^{\varepsilon})  &  =I(X;B_{1}B_{2}%
)_{\omega}\\
&  =I(X;B_{1}B_{2}Y)_{\omega}\\
&  =I(X;B_{1}B_{2}|Y)_{\omega}+I(X;Y)_{\omega}\\
&  =I(X;B_{1}B_{2}|Y)_{\omega}\\
&  =\left(  1-\varepsilon\right)  I(X;B_{1}A_{2})_{\mathcal{M}(\phi^{x}%
)}+\varepsilon I(X;B_{1})_{\mathcal{M}(\phi^{x})}\\
&  \leq\left(  1-\varepsilon\right)  \chi(\mathcal{M}\otimes\operatorname{id}%
)+\varepsilon\chi(\mathcal{M})\\
&  =\chi(\mathcal{M})+\left(  1-\varepsilon\right)  \chi(\operatorname{id})\\
&  =\chi(\mathcal{M})+\left(  1-\varepsilon\right)  \log d.
\end{align}
The second equality follows because the aforementioned isometry takes
$B_{1}B_{2}$ to $B_{1}B_{2}Y$. The third equality follows from the chain rule
for mutual information, and the fourth because $I(X;Y)_{\omega}=0$. The fifth
equality follows because $Y$ is a classical system, so that we can expand the
mutual information as a convex combination of individual mutual informations.
The inequality follows by maximizing the Holevo informations with respect to
all input ensembles. The second-to-last equality follows because the identity
channel is a Hadamard channel, so that Theorem~\ref{thm-cc:hadamard-additive}
implies that $\chi(\mathcal{M}\otimes\operatorname{id})=\chi(\mathcal{M}%
)+\chi(\operatorname{id})$.

By setting $\mathcal{M}=(\mathcal{N}^{\varepsilon})^{\otimes\left[
n-1\right]  }$ and iterating the above, we see that $\chi((\mathcal{N}%
^{\varepsilon})^{\otimes n})\leq n\left(  1-\varepsilon\right)  \log d$, so
that the regularized Holevo information cannot exceed $\left(  1-\varepsilon
\right)  \log d$. Since this rate is also achievable, this concludes the proof.
\end{proof}

\subsection{Classical Capacity of the Depolarizing Channel}

The qudit depolarizing channel%
\index{depolarizing channel}
is another example of a channel for which we can compute its classical
capacity. Additionally, we will see that achieving the classical capacity of
this channel requires a strategy which is very \textquotedblleft
classical\textquotedblright---it is sufficient to prepare classical states
$\left\{  \vert x\rangle\langle x\vert\right\}  $ at the input of the channel
and to measure each channel output in the same basis (see
Exercise~\ref{ex-cc:dep-ach}). However, we will later see in
Chapter~\ref{chap:quantum-capacity}\ that the depolarizing channel has some
rather bizarre, uniquely quantum features when considering its quantum
capacity, even though the features of its classical capacity are rather classical.

Recall from Section~\ref{sec-nqt:depolarizing}\ that the depolarizing channel
is the following map:%
\begin{equation}
\mathcal{N}_{\operatorname{D}}( \rho) =( 1-p) \rho+p\pi,
\end{equation}
where $\pi$ is the maximally mixed state.

\begin{theorem}
[Classical Capacity of the Depolarizing Channel]%
\label{thm-cc:cap-depolarizing}The classical capacity of the qudit
depolarizing channel $\mathcal{N}_{\operatorname{D}}$ is as follows:%
\begin{equation}
\chi(\mathcal{N}_{\operatorname{D}})=\log d+\left(  1-p+\frac{p}{d}\right)
\log\left(  1-p+\frac{p}{d}\right)  +\left(  d-1\right)  \frac{p}{d}%
\log\left(  \frac{p}{d}\right)  . \label{eq-cc:class-cap-dep}%
\end{equation}

\end{theorem}

\begin{figure}
[ptb]
\begin{center}
\includegraphics[
width=4.3439in
]%
{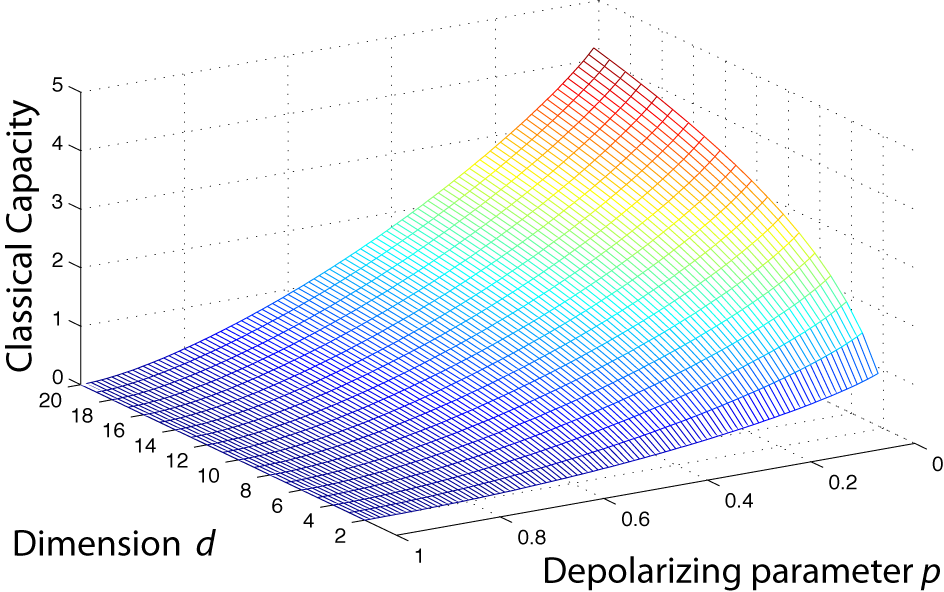}%
\caption{The classical capacity of the quantum depolarizing channel as a
function of the dimension $d$ of the channel and the depolarizing parameter
$p$. The classical capacity vanishes when $p=1$ because the channel replaces
the input with the maximally mixed state. The classical capacity is maximal at
$\log d$ when $p=0$ because there is no noise. In between these two extremes,
the classical capacity is a smooth function of $p$ and $d$ given by the
expression in \eqref{eq-cc:class-cap-dep}.}%
\label{fig-cc:class-cap-dep}%
\end{center}
\end{figure}

\begin{proof}
The first part of the proof of this theorem relies on a somewhat technical
result, namely, that the Holevo information of the tensor-product channel
$\mathcal{N}_{\operatorname{D}}\otimes\mathcal{N}$ is additive (where the
first channel is the depolarizing channel and the other is arbitrary):%
\begin{equation}
\chi(\mathcal{N}_{\operatorname{D}}\otimes\mathcal{N})=\chi(\mathcal{N}%
_{\operatorname{D}})+\chi(\mathcal{N}).
\end{equation}
This result is due to \cite{K03}, and it exploits a few properties of the
depolarizing channel. The result implies that the classical capacity of the
depolarizing channel is equal to its Holevo information. We now show how to
compute the Holevo information of the depolarizing channel. To do so, we first
determine the minimum output entropy of the channel.

\begin{definition}
[Minimum Output Entropy]\label{def-cc:min-out-entropy}The minimum output
\index{minimum output entropy}
entropy $H^{\min}(\mathcal{N})$ of a channel $\mathcal{N}$ is the minimum of
the entropy at the output of the channel:%
\begin{equation}
H^{\min}(\mathcal{N})\equiv\min_{\rho}H(\mathcal{N}(\rho)),
\end{equation}
where the minimization is over all states input to the channel.
\end{definition}

\begin{exercise}
Prove that it is sufficient to minimize over only pure-state inputs to the
channel when computing the minimum output entropy. That is,%
\begin{equation}
H^{\min}(\mathcal{N})=\min_{|\psi\rangle}H(\mathcal{N}(|\psi\rangle\langle
\psi|)).
\end{equation}

\end{exercise}

The depolarizing channel is a highly symmetric channel. For example, if we
input a pure state $|\psi\rangle$\ to the channel, the output is as follows:%
\begin{align}
(1-p)\psi+p\pi &  =(1-p)\psi+\frac{p}{d}I\\
&  =(1-p)\psi+\frac{p}{d}(\psi+I-\psi)\\
&  =\left(  1-p+\frac{p}{d}\right)  \psi+\frac{p}{d}(I-\psi).
\end{align}
Observe that the eigenvalues of the output state are the same for any pure
state and are equal to $1-p+\frac{p}{d}$ with multiplicity one and $\frac
{p}{d}$ with multiplicity $d-1$. Thus, the minimum output entropy of the
depolarizing channel is just%
\begin{equation}
H^{\min}(\mathcal{N}_{\operatorname{D}})=-\left(  1-p+\frac{p}{d}\right)
\log\left(  1-p+\frac{p}{d}\right)  -\left(  d-1\right)  \frac{p}{d}%
\log\left(  \frac{p}{d}\right)  .
\end{equation}
We now compute the Holevo information of the depolarizing channel. Recall from
Theorem~\ref{thm-add:pure-states-suff-holevo}\ that it is sufficient to
consider optimizing the Holevo information over a classical--quantum state
with conditional states that are pure (a state $\sigma_{XA^{\prime}}$ of the
form in \eqref{eq-cc:HSW-c-q-state-pure}). Also, the Holevo information has
the following form:%
\begin{equation}
\max_{\sigma}I(X;B)_{\omega}=\max_{\sigma}\left[  H(B)_{\omega}-H(B|X)_{\omega
}\right]  ,
\end{equation}
where $\omega_{XB}$ is the output state. Consider the following augmented
input ensemble:%
\begin{multline}
\rho_{XIJA^{\prime}}\equiv\\
\frac{1}{d^{2}}\sum_{x}\sum_{i,j=0}^{d-1}p_{X}(x)|x\rangle\langle
x|_{X}\otimes|i\rangle\langle i|_{I}\otimes|j\rangle\langle j|_{J}\otimes
X(i)Z(j)\psi_{A^{\prime}}^{x}Z^{\dag}(j)X^{\dag}(i),
\end{multline}
where $X(i)$ and $Z(j)$ are the generalized Pauli operators from
Section~\ref{sec-nqt:generalized-Pauli}. Suppose that we trace over the $IJ$
system. Then the state $\rho_{XA^{\prime}}$ is as follows:%
\begin{equation}
\rho_{XA^{\prime}}=\sum_{x}p_{X}(x)|x\rangle\langle x|_{X}\otimes
\pi_{A^{\prime}},
\end{equation}
by recalling the result of Exercise~\ref{ex-qt:uniformly-random-unitary}.
Also, note that inputting the maximally mixed state to the depolarizing
channel results in the maximally mixed state at its output. Consider the
following chain of inequalities:%
\begin{align}
I(X;B)_{\omega}  &  =H(B)_{\omega}-H(B|X)_{\omega}\\
&  \leq H(B)_{\rho}-H(B|X)_{\omega}\\
&  =\log d-H(B|XIJ)_{\rho}\\
&  =\log d-\sum_{x}p_{X}(x)H(B)_{\mathcal{N}_{D}(\psi^{x})}%
\label{eq-cc:dep-proof-1}\\
&  \leq\log d-\min_{x}H(B)_{\mathcal{N}_{D}(\psi^{x})}%
\label{eq-cc:dep-proof-2}\\
&  \leq\log d-H^{\min}(\mathcal{N}_{\operatorname{D}}).
\end{align}
The first equality follows by expanding the quantum mutual information. The
first inequality follows from concavity of entropy. The second equality
follows because the state of $\rho$ on system $B$ is the maximally mixed state
$\pi$ and from the following chain of equalities:%
\begin{align}
H(B|XIJ)_{\rho}  &  =\frac{1}{d^{2}}\sum_{x}\sum_{i,j=0}^{d-1}p_{X}%
(x)H(B)_{\mathcal{N}_{D}(X(i)Z(j)\psi^{x}Z^{\dag}(j)X^{\dag}(i))}\\
&  =\frac{1}{d^{2}}\sum_{x}\sum_{i,j=0}^{d-1}p_{X}(x)H(B)_{X(i)Z(j)\mathcal{N}%
_{D}(\psi^{x})Z^{\dag}(j)X^{\dag}(i)}\\
&  =\sum_{x}p_{X}(x)H(B)_{\mathcal{N}_{D}(\psi^{x})}\\
&  =H(B|X)_{\omega}.
\end{align}
The third equality in \eqref{eq-cc:dep-proof-1} follows from the above chain
of equalities. The second inequality in~\eqref{eq-cc:dep-proof-2} follows
because the expectation can never be smaller than the minimum (this step is
not strictly necessary for the depolarizing channel). The last inequality
follows because $\min_{x}H(B)_{\mathcal{N}_{D}(\psi^{x})}\geq H^{\min
}(\mathcal{N}_{\operatorname{D}})$ (though it is actually an equality for the
depolarizing channel).

An ensemble of the following form suffices to achieve
the classical capacity of the depolarizing channel:%
\begin{equation}
\frac{1}{d}\sum_{i=0}^{d-1}|i\rangle\langle i|_{I}\otimes|i\rangle\langle
i|_{A^{\prime}},
\end{equation}
because we only require that the reduced state on $A^{\prime}$ be equal to the
maximally mixed state. The final expression for the classical capacity of the
depolarizing channel is as stated in Theorem~\ref{thm-cc:cap-depolarizing},
which we plot in Figure~\ref{fig-cc:class-cap-dep}\ as a function of the
dimension $d$ and the depolarizing parameter $p$.
\end{proof}

\begin{exercise}
[Achieving the Classical Capacity of the Depolarizing Channel]%
\label{ex-cc:dep-ach}We actually know that even more is true regarding the
method for achieving the classical capacity of the depolarizing channel. Prove
that it is possible to achieve the classical capacity of the depolarizing
channel by choosing states from an ensemble $\left\{  \frac{1}{d}%
,|x\rangle\langle x|\right\}  $ and performing a complete projective
measurement in the same basis at the output of each channel. That is, the
naive scheme outlined in Section~\ref{sec-cc:naive}\ is sufficient to attain
the classical capacity of the depolarizing channel. (Hint:\ First show that
the classical channel $p_{Y|X}(y|x)$ induced by inputting a state $|x\rangle$
to the depolarizing channel and measuring $|y\rangle$ at the output is as
follows:%
\begin{equation}
p_{Y|X}(y|x)=(1-p)\delta_{x,y}+\frac{p}{d}.
\end{equation}
Then show that the distribution $p_{Y}(y)$ is uniform if $p_{X}(x)$ is
uniform. Finally, show that%
\begin{equation}
H(Y|X)=-\left(  1-p+\frac{p}{d}\right)  \log\left(  1-p+\frac{p}{d}\right)
-\left(  d-1\right)  \left(  \frac{p}{d}\right)  \log\left(  \frac{p}%
{d}\right)  .
\end{equation}
Conclude that the classical capacity of the induced channel $p_{Y|X}(y|x)$ is
the same as that for the quantum depolarizing channel.)
\end{exercise}

\begin{exercise}
A covariant channel $\mathcal{N}_{\operatorname{C}}$\ is one for which the
state resulting from a unitary $U$ acting on the input state before the
channel occurs is equivalent to one where there is a unitary representation
$R_{U}$ of the unitary $U$ acting on the output of the channel:%
\begin{equation}
\mathcal{N}_{\operatorname{C}}(U\rho U^{\dag})=R_{U}\mathcal{N}%
_{\operatorname{C}}(\rho)R_{U}^{\dag}.
\end{equation}
Show that the Holevo information $\chi(\mathcal{N}_{\operatorname{C}})$ of a
covariant channel is equal to%
\begin{equation}
\chi(\mathcal{N}_{\operatorname{C}})=\log d-H(\mathcal{N}_{\operatorname{C}%
}(\psi)),
\end{equation}
where $\psi$ is an arbitrary pure state.
\end{exercise}

\section{Superadditivity of the Holevo Information}

\label{sec-cc:superadditivity}Many researchers thought for some time that the
Holevo information would be additive for all quantum channels, implying that
it would be a good characterization
\index{Holevo information!of a channel!superadditivity}
of the classical capacity in the general case---this conjecture was known as
the additivity conjecture. Researchers thought that this conjecture would hold
because they discovered a few channels for which it did hold, but without any
common theme occurring in the proofs for the different channels, they soon
began looking in the other direction for a counterexample to disprove it.
After some time, \cite{H09} found the existence of a counterexample to the
additivity conjecture, demonstrating that it cannot hold in the general case.
This result demonstrates that even one of the most basic questions in quantum
Shannon theory still remains wide open and that entanglement at the encoder
can help increase classical communication rates over a quantum channel.

We first review a relation between the Holevo information and the minimum
output entropy of a tensor-product channel. Suppose that we have two channels
$\mathcal{N}$ and $\mathcal{M}$. The Holevo information of the tensor-product
channel is additive if%
\begin{equation}
\chi(\mathcal{N}\otimes\mathcal{M})=\chi(\mathcal{N})+\chi(\mathcal{M}).
\end{equation}
Since the Holevo information is always superadditive for any two channels:%
\begin{equation}
\chi(\mathcal{N}\otimes\mathcal{M})\geq\chi(\mathcal{N})+\chi(\mathcal{M})
\end{equation}
(recall the statement at the beginning of the proof of
Theorem~\ref{thm-add:holevo-EB-add}), we say that it is non-additive if it is
strictly superadditive:%
\begin{equation}
\chi(\mathcal{N}\otimes\mathcal{M})>\chi(\mathcal{N})+\chi(\mathcal{M}).
\end{equation}
The minimum output entropy $H^{\min}(\mathcal{N}\otimes\mathcal{M})$ of the
tensor-product channel is a quantity related to the Holevo information (see
Definition~\ref{def-cc:min-out-entropy}). It is additive if%
\begin{equation}
H^{\min}(\mathcal{N}\otimes\mathcal{M})=H^{\min}(\mathcal{N})+H^{\min
}(\mathcal{M}).
\end{equation}
Since the minimum output entropy is always subadditive:%
\begin{equation}
H^{\min}(\mathcal{N}\otimes\mathcal{M})\leq H^{\min}(\mathcal{N})+H^{\min
}(\mathcal{M}),
\end{equation}
we say that it is non-additive if it is strictly subadditive:%
\begin{equation}
H^{\min}(\mathcal{N}\otimes\mathcal{M})<H^{\min}(\mathcal{N})+H^{\min
}(\mathcal{M}).
\end{equation}
Additivity of these two quantities is in fact related---it is possible to show
that additivity of the Holevo information implies additivity of the minimum
output entropy and vice versa (we leave one of these implications as an
exercise). Thus, researchers focused on additivity of minimum output entropy
rather than additivity of Holevo information because it is a simpler quantity
to manipulate.

\begin{exercise}
Prove that non-additivity of the minimum output entropy implies non-additivity
of the Holevo information:%
\begin{multline}
H^{\min}(\mathcal{N}_{1}\otimes\mathcal{N}_{2})<H^{\min}(\mathcal{N}%
_{1})+H^{\min}(\mathcal{N}_{2})\ \ \\
\Rightarrow\ \ \chi(\mathcal{N}_{1}\otimes\mathcal{N}_{2})>\chi(\mathcal{N}%
_{1})+\chi(\mathcal{N}_{2}).
\end{multline}
(\textit{Hint}:\ Consider an augmented version $\mathcal{N}_{i}^{\prime}$ of
each channel $\mathcal{N}_{i}$, that has its first input be the same as the
input to $\mathcal{N}_{i}$ and its second input be a control input, and the
action of the channel is equivalent to measuring the auxiliary input $\sigma$
and applying a generalized Pauli operator:%
\begin{equation}
\mathcal{N}_{i}^{\prime}(\rho\otimes\sigma)\equiv\sum_{k,l}X(k)Z(l)\mathcal{N}%
_{i}(\rho)Z^{\dag}(l)X^{\dag}(k)\ \langle k|\langle l|\sigma|k\rangle
\left\vert l\right\rangle .
\end{equation}
What is the Holevo information of the augmented channel $\mathcal{N}%
_{i}^{\prime}$? What is the Holevo information of the tensor product of the
augmented channels $\mathcal{N}_{1}^{\prime}\otimes\mathcal{N}_{2}^{\prime}$?)
After proving the above statement, we can also conclude that additivity of the
Holevo information implies additivity of the minimum output entropy.
\end{exercise}

We briefly overview the main ideas behind the construction of a channel for
which the Holevo information is not additive. Consider a random-unitary
channel of the following form:%
\begin{equation}
\mathcal{E}(\rho)\equiv\sum_{i=1}^{D}p_{i}U_{i}\rho U_{i}^{\dag},
\end{equation}
where the dimension of the input state is $N$ and the number of random
unitaries is $D$. This channel is \textquotedblleft random
unitary\textquotedblright\ because it applies a particular unitary $U_{i}$
with probability $p_{i}$ to the state $\rho$. The cleverness behind the
construction is not actually to provide a deterministic instance of this
channel, but rather, to provide a random instance of the channel where both
the distribution and the unitaries are chosen at random, and the dimension $N$
and the number $D$ of chosen unitaries satisfy the following relationships:%
\begin{equation}
1\ll D\ll N.
\end{equation}
The other channel to consider to disprove additivity is the conjugated channel%
\begin{equation}
\overline{\mathcal{E}}(\rho)\equiv\sum_{i=1}^{D}p_{i}\overline{U}_{i}%
\rho\overline{U}_{i}^{\dag},
\end{equation}
where $p_{i}$ and $U_{i}$ are the same respective probability distribution and
unitaries from the channel $\mathcal{E}$, and here $\overline{U}_{i}$ denotes the
complex conjugate of $U_{i}$. The goal is then to show that there is a
non-zero probability over all channels of these forms that the minimum output
entropy is non-additive:%
\begin{equation}
H^{\min}(\mathcal{E}\otimes\overline{\mathcal{E}})<H^{\min}(\mathcal{E}%
)+H^{\min}(\overline{\mathcal{E}}).
\end{equation}

A good candidate for a state that could saturate the minimum output entropy
$H^{\min}(\mathcal{E}\otimes\overline{\mathcal{E}})$ of the tensor-product
channel is the maximally entangled state $\left\vert \Phi\right\rangle $,
where%
\begin{equation}
\left\vert \Phi\right\rangle \equiv\frac{1}{\sqrt{N}}\sum_{i=0}^{N-1}%
|i\rangle|i\rangle.
\end{equation}
Consider the effect of the tensor-product channel $\mathcal{E}\otimes
\overline{\mathcal{E}}$ on the maximally entangled state $\Phi$:%
\begin{align}
&  (\mathcal{E}\otimes\overline{\mathcal{E}})(\Phi)\nonumber\\
&  =\sum_{i,j=1}^{D}p_{i}p_{j}(U_{i}\otimes\overline{U}_{j})\Phi(U_{i}^{\dag
}\otimes\overline{U}_{j}^{\dag})\\
&  =\sum_{i=j}p_{i}^{2}(U_{i}\otimes\overline{U}_{i})\Phi(U_{i}^{\dag}%
\otimes\overline{U}_{i}^{\dag})+\sum_{i\neq j}p_{i}p_{j}(U_{i}\otimes
\overline{U}_{j})\Phi(U_{i}^{\dag}\otimes\overline{U}_{i}^{\dag})\\
&  =\left(  \sum_{i=1}^{D}p_{i}^{2}\right)  \Phi+\sum_{i\neq j}p_{i}%
p_{j}(U_{i}\otimes\overline{U}_{j})\Phi(U_{i}^{\dag}\otimes\overline{U}%
_{i}^{\dag}),
\end{align}
where the last line uses the fact that $\left(  M\otimes I\right)  \left\vert
\Phi\right\rangle =\left(  I\otimes M^{T}\right)  \left\vert \Phi\right\rangle
$ for any operator $M$ (this implies that $\left(  U\otimes\overline
{U}\right)  \left\vert \Phi\right\rangle =\left\vert \Phi\right\rangle $).
When comparing the above state to one resulting from inputting a product state
to the channel, there is a sense in which the above state is less noisy than
the product state because $D$ of the combinations of the random unitaries (the
ones which have the same index) have no effect on the maximally entangled
state. Using techniques from \cite{H09}, we can make this intuition precise
and obtain the following upper bound on the minimum output entropy:%
\begin{align}
H^{\min}(\mathcal{E}\otimes\overline{\mathcal{E}})  &  \leq H((\mathcal{E}%
\otimes\overline{\mathcal{E}})(\Phi))\\
&  \leq2\ln D-\frac{\ln D}{D},
\end{align}
for $N$ and $D$ large enough. However, using techniques in the same paper, we
can also show that%
\begin{equation}
H^{\min}(\mathcal{E})\geq\ln D-\delta S^{\max},
\end{equation}
where%
\begin{equation}
\delta S^{\max}\equiv\frac{c}{D}+\operatorname{poly}(D)O\left(  \sqrt
{\frac{\ln N}{N}}\right)  ,
\end{equation}
$c$ is a constant, and poly$(D)$ indicates a term polynomial in $D$. Thus, for
large enough $D$ and $N$, it follows that%
\begin{equation}
2\delta S^{\max}<\frac{\ln D}{D},
\end{equation}
and we get the existence of a channel for which a violation of additivity
occurs, because%
\begin{align}
H^{\min}(\mathcal{E}\otimes\overline{\mathcal{E}})  &  \leq2\ln D-\frac{\ln
D}{D}\\
&  <2\ln D-2\delta S^{\max}\\
&  \leq H^{\min}(\mathcal{E})+H^{\min}(\overline{\mathcal{E}}).
\end{align}

\section{Concluding Remarks}

The HSW theorem offers a good characterization of the classical capacity of
certain classes of channels, but at the same time, it also demonstrates our
lack of understanding of classical transmission over general quantum channels.
To be more precise, the Holevo information is a useful characterization of the
classical capacity of a quantum channel whenever it is additive, but the
regularized Holevo information is not particularly useful as a
characterization of it because we cannot even compute this quantity. This
suggests that there could be some other formula that better characterizes the
classical capacity (if such a formula were additive). As of the writing of
this book, such a formula is unknown.

Despite the drawbacks of the HSW\ theorem, it is still interesting because it
at least offers a step beyond the most naive characterization of the classical
capacity of a quantum channel in terms of the regularized accessible
information. The major insight of HSW\ was the construction of an explicit
POVM\ (corresponding to a random choice of code) that allows the sender and
receiver to communicate at a rate equal to the Holevo information of the
channel. This theorem is also useful for determining achievable rates in
different communication scenarios: for example, when two senders are trying to
communicate over a noisy medium to a single receiver or when a single sender
is trying to transmit both classical and quantum information to a receiver.

The depolarizing channel is an example of a quantum channel for which there is
a simple expression for its classical capacity. Furthermore, the expression
reveals that the scheme needed to achieve the capacity of the channel is
rather classical---it is only necessary for the sender to select codewords
uniformly at random from some orthonormal basis, and it is only necessary for
the receiver to perform measurements of the individual channel outputs in the
same orthonormal basis. Thus, the coding scheme is classical because
entanglement plays no role at the encoder and the decoding measurements act on
the individual channel outputs.

Finally, we discussed Hastings' construction of a quantum channel for which
the heralded additivity conjecture does not hold. That is, there exists a
channel for which entanglement at the encoder can improve communication rates.
This superadditive effect is a uniquely quantum phenomenon (recall that
Theorem~\ref{thm-ie:additivity-classical} states that the classical mutual
information of a classical channel is additive, and thus correlations at the
input cannot increase capacity). This result implies that our best known
characterization of the classical capacity of a quantum channel in terms of
the channel's Holevo information is far from being a satisfactory
characterization of the true capacity, and we still have much more to discover here.

\section{History and Further Reading}

\label{sec-cc:history}\cite{Holevo73} was the first to prove the bound bearing
his name, regarding the transmission of classical information using a quantum
channel, and \cite{Hol98} and \cite{PhysRevA.56.131} many years later proved
that the Holevo information is an achievable rate for classical data
transmission. Just prior to these works, \cite{PhysRevA.54.1869} proved
achievability of the Holevo information for the special case of a channel that
accepts a classical input and outputs a pure state conditioned on the input.
They also published a preliminary article~\citep{HSWW95}\ in which they
answered the catchy question (for the special case of pure states),
\textquotedblleft How many bits can you fit into a quantum-mechanical
it?\textquotedblright

\cite{S02} established the additivity of the Holevo information for
\index{entanglement-breaking channel}%
entanglement-breaking channels. \cite{BN05} proved that the classical capacity
of an entanglement-breaking channel is not enhanced by a classical feedback
channel from receiver to sender. \cite{K02} first proved additivity of the
Holevo information for unital qubit channels and later showed it for the
depolarizing channel \citep{K03}. \cite{S04} later showed the equivalence of
several additivity conjectures (that they are either all true or all false).
\cite{H07}, \cite{W07}, and a joint paper between them \citep{HW08} proved
some results leading up to the work of \cite{H09}, who demonstrated a
counterexample to the additivity conjecture. Thus, by Shor's aforementioned
paper, all of the additivity conjectures are false in general. There has been
much follow-up work in an attempt to understand Hastings' result \citep{BH10,fukuda:042201,FKM10,ASW10}.

In a landmark result, \cite{GHG15} established the classical capacity of all
phase-insensitive quantum Gaussian channels, building on a long series of
works starting with that of \cite{HW01}.

Some other papers have tried to understand the HSW\ coding theorem from the
perspective of hypothesis testing. \cite{HN03} began much of this work, and
\cite{H06book} covers quite a bit of quantum hypothesis testing in his book.
\cite{mosonyi:072104} followed up with some work along these lines, as did
\cite{WR10} and \cite{Wilde20130259}.

Researchers have invested quite a bit of effort in refining the HSW\ theorem,
with regard to error exponents, strong converse exponents and second-order
characterizations. From the proof of the HSW\ theorem given in this chapter,
we see that if we pick the rate $R$ of classical communication to be smaller
than the capacity by an additive constant, then it is possible to make the
decoding error decay exponentially fast to zero with an increasing number $n$
of channel uses. The optimal exponential decay rate of the error for a given
communication rate is known as the error exponent. \cite{BH98} derived a lower
bound on the optimal error exponent for pure-state classical--quantum
channels, and \cite{H00}\ then derived a lower bound on the error exponent for
mixed-state classical--quantum channels. \cite{Hay07} later improved upon this
result for mixed-state classical--quantum channels. \cite{D13} derived an
upper bound on the optimal error exponent of classical--quantum channels,
called the sphere-packing bound.

One can also ask about the behavior of the error probability when the
communication rate exceeds the capacity by an additive constant. This regime
is known as the strong converse regime, and a channel obeys the strong
converse property if the error probability tends to one when the communication
rate exceeds the capacity. \cite{itit1999winter} and \cite{ON99} proved the
strong converse for the classical capacity of classical--quantum channels,
\cite{KW09} for channels with certain symmetry, \cite{WWY13} for
\index{entanglement-breaking channel}%
entanglement-breaking and Hadamard channels, \cite{BPWW14} for
phase-insensitive quantum Gaussian channels, and \cite{DingW15} for
entanglement-breaking channels assisted by a noiseless classical feedback channel.

A second-order characterization asks how fast the communication rate can
converge to capacity if the error probability is fixed to be a constant. One
of the main tools here is the Berry--Esseen refinement of the central limit
theorem (see, e.g., \cite{F71,tyurin10}). \cite{TT13} derived an optimal
second-order characterization for classical--quantum channels (even going
beyond and establishing it for \textquotedblleft
image-additive\textquotedblright\ channels). \cite{WRG15} characterized an
achievable second-order strategy for the pure-loss bosonic channel (with a
converse part remaining open).

\chapter{Entanglement-Assisted Classical Communication}

\label{chap:EA-classical}We have learned that shared entanglement is often
helpful in quantum
\index{entanglement-assisted!classical communication}
communication. This is certainly true for the case of a noiseless qubit
channel. Without shared entanglement, the most classical information that a
sender can reliably transmit over a noiseless qubit channel is just one
classical bit (recall Exercise~\ref{ex-nqt:nayak}\ and
\index{Holevo bound}%
the Holevo bound in Exercise~\ref{ex-qie:holevo-bound}). With shared
entanglement, they can achieve the super-dense coding resource inequality from
Chapter~\ref{chap:coherent-communication}:%
\begin{equation}
\left[  q\rightarrow q\right]  +\left[  qq\right]  \geq2\left[  c\rightarrow
c\right]  .
\end{equation}
That is, using one noiseless qubit channel and one shared noiseless ebit, the
sender can reliably transmit two classical bits.

A natural question then for us to consider is whether shared entanglement
could be helpful in transmitting classical information over a noisy quantum
channel $\mathcal{N}$. As a first simplifying assumption, we let Alice and Bob
have access to an infinite supply of entanglement, in whatever form they wish,
and we would like to know how much classical information Alice can reliably
transmit to Bob over such an entanglement-assisted quantum channel. That is,
we would like to determine the highest achievable rate$~C$ of classical
communication in the following resource inequality:%
\begin{equation}
\left\langle \mathcal{N}\right\rangle +\infty\left[  qq\right]  \geq C\left[
c\rightarrow c\right]  .
\end{equation}

The answer to this question is one of the strongest known results in quantum
Shannon theory, and it is given by the entanglement-assisted classical
capacity theorem. This theorem states that the mutual information
$I(\mathcal{N})$ of a quantum channel$~\mathcal{N}$ is equal to its
entanglement-assisted classical capacity, where%
\begin{equation}
I(\mathcal{N})\equiv\max_{\varphi_{AA^{\prime}}}I(A;B)_{\rho},
\label{eq-eac:EAC-formula}%
\end{equation}
$\rho_{AB}\equiv\mathcal{N}_{A^{\prime}\rightarrow B}(\varphi_{AA^{\prime}})$,
and the maximization is with respect to all pure bipartite states of the form
$\varphi_{AA^{\prime}}$. We should stress that there is no need to regularize
this formula in order to characterize the capacity (as done in the previous
chapter and as is so often needed in quantum Shannon theory). The value of
this formula \textit{is equal to} the capacity. Also, the optimization task that the
formula in \eqref{eq-eac:EAC-formula} sets out is a straightforward convex
optimization program. Any local maximum is a global maximum because the
quantum mutual information is concave with respect to the input state $\varphi_{A^{\prime}%
}$ (recall Theorem~\ref{thm-ie:mutual-concave-input}\ from
Chapter~\ref{chap:additivity}) and the set of density operators is convex.

From the perspective of an information theorist, we should only say that a
capacity theorem has been solved if there is a tractable formula equal to the
optimal rate for achieving a particular operational task. The formula should
apply to an arbitrary quantum channel, and it should be a function of that
channel. Otherwise, the capacity theorem is still unsolved. There are several
operative words in the above sentences that we should explain in more detail.
The formula should be tractable, meaning that it sets out an optimization task
which is efficient to solve in the dimension of the channel's input system.
The formula should give the optimal achievable rate for the given
information-processing task, meaning that if a rate exceeds the capacity of
the channel, then the probability of error for any such protocol should be
bounded away from zero as the number of channel uses grows large.\footnote{We
could strengthen this requirement even more by demanding that the probability
of error increases exponentially to one in the asymptotic limit. Fulfilling
such a demand would constitute a proof of a \textit{strong converse theorem}.}
Finally, perhaps the most stringent (though related) criterion is that the
formula itself (and \textit{not} its regularization) should give the capacity
of an arbitrary quantum channel. Despite the success of the HSW\ coding
theorem in demonstrating that the Holevo information of a channel is an
achievable rate for classical communication, the classical capacity of a
quantum channel is still unsolved because there is an example of a channel for
which the Holevo information is not equal to that channel's capacity (see
Section~\ref{sec-cc:superadditivity}). Thus, it is rather impressive that the
formula in \eqref{eq-eac:EAC-formula} is equal to the entanglement-assisted
classical capacity of an arbitrary channel, given the stringent requirements
that we have set out for a formula to give the capacity. In this sense,
shared entanglement simplifies quantum Shannon theory.

This chapter presents a comprehensive study of the entanglement-assisted
classical capacity theorem. We begin by defining the information-processing
task, consisting of all the steps in a general protocol for classical
communication over an entanglement-assisted quantum channel. We then present a
simple example of a strategy for entanglement-assisted classical coding that
is inspired by super-dense coding, and in turn, that inspires a strategy for the
general case. Section~\ref{sec-eac:BSST-theorem} states the
entanglement-assisted classical capacity theorem.
Section~\ref{sec-eac:direct-coding}\ gives a proof of the direct coding
theorem, making use of strong quantum typicality from
Chapter~\ref{chap:quantum-typicality} and the packing lemma from
Chapter~\ref{chap:packing}. It demonstrates that the rate in
\eqref{eq-eac:EAC-formula} is an achievable rate for entanglement-assisted
classical communication. After taking a step back from the protocol, we can
realize that it is merely a glorified super-dense coding applied to noisy
quantum channels. Section~\ref{sec-eac:converse}\ gives a proof of the
converse of the entanglement-assisted classical capacity theorem. It exploits
familiar tools such as the AFW inequality, the quantum data-processing
inequality, and the chain rule for quantum mutual information (all from
Chapter~\ref{chap:q-info-entropy}), and the last part of it exploits
additivity of the mutual information of a quantum channel (from
Chapter~\ref{chap:additivity}). The converse theorem establishes that the rate
in \eqref{eq-eac:EAC-formula} is optimal. With the proof of the capacity
theorem complete, we then show the interesting result that the classical
capacity of a quantum channel assisted by a quantum feedback channel is equal
to the entanglement-assisted classical capacity of that channel. We close the
chapter by computing the entanglement-assisted classical capacity of both a
quantum erasure channel and an amplitude damping channel, and we leave the
computation of the entanglement-assisted capacities of two other channels as exercises.

\section{The Information-Processing Task}

\label{sec-eac:info-task}
\begin{figure}
[ptb]
\begin{center}
\includegraphics[
width=3.9652in
]%
{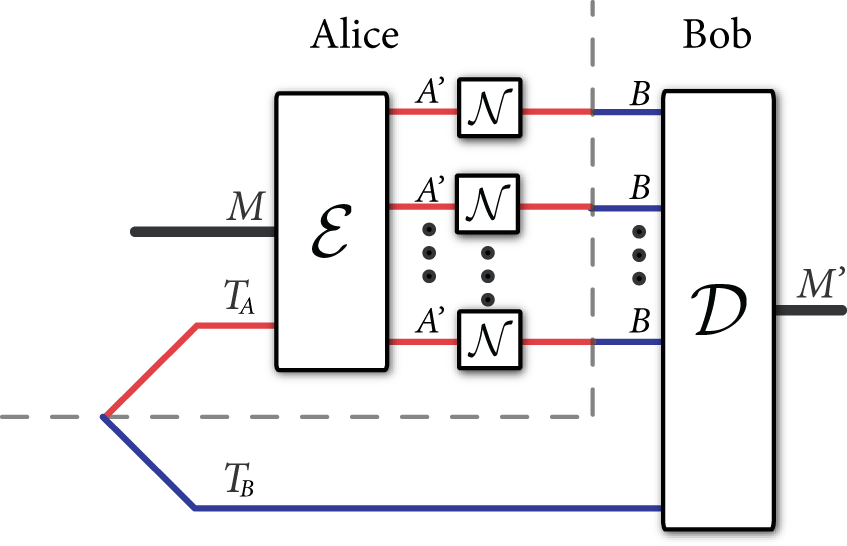}%
\caption{The most general protocol for entanglement-assisted classical
communication. Alice applies an encoder to her classical message $M$ and her
share $T_{A}$ of the entanglement, and she inputs the encoded systems
$A^{\prime n}$ to many uses of the channel. Bob receives the outputs of the
channels, combines them with his share of the entanglement, and performs some
decoding operation to estimate Alice's transmitted message.}%
\label{fig-eac:info-proc-task}%
\end{center}
\end{figure}

We begin by explicitly defining the
information-processing task of entanglement-assisted classical communication. That is,
 we define an $\left(  n,C,\varepsilon\right)  $ entanglement-assisted
classical code and what it means for a rate $C$ to be achievable. Prior to the
start of the protocol, we assume that Alice and Bob share pure-state
entanglement in whatever form they wish. Let $\Psi_{T_{A}T_{B}}$ denote this
state. Alice selects some message $m$ from a set $\mathcal{M}$ of messages.
Let $M$ denote the random variable corresponding to Alice's choice of message,
and let $\left\vert \mathcal{M}\right\vert $ be the cardinality of the
set~$\mathcal{M}$. She applies some encoding channel $\mathcal{E}%
_{T_{A}\rightarrow A^{\prime n}}^{m}$ to her share of the entangled state
$\Psi_{T_{A}T_{B}}$ depending on her choice of message~$m$. The global state
then becomes%
\begin{equation}
\mathcal{E}_{T_{A}\rightarrow A^{\prime n}}^{m}(\Psi_{T_{A}T_{B}}).
\end{equation}
Alice transmits the systems $A^{\prime n}$ over $n$ independent uses of a
noisy channel $\mathcal{N}_{A^{\prime}\rightarrow B}$, leading to the
following state:%
\begin{equation}
\mathcal{N}_{A^{\prime n}\rightarrow B^{n}}(\mathcal{E}_{T_{A}\rightarrow
A^{\prime n}}^{m}(\Psi_{T_{A}T_{B}})),
\end{equation}
where $\mathcal{N}_{A^{\prime n}\rightarrow B^{n}}\equiv(\mathcal{N}%
_{A^{\prime}\rightarrow B})^{\otimes n}$. Bob receives the systems $B^{n}$,
combines them with his share $T_{B}$ of the entanglement, and performs a
POVM\ $\{\Lambda_{B^{n}T_{B}}^{m}\}$ on the channel outputs $B^{n}$ and his
share $T_{B}$ of the entanglement in order to detect the message $m$ that
Alice transmits. Figure~\ref{fig-eac:info-proc-task}\ depicts such a general
protocol for entanglement-assisted classical communication.%

Let $M^{\prime}$ denote the random variable corresponding to the output of Bob's decoding
POVM (this represents Bob's estimate of the message). The probability of Bob
correctly decoding Alice's message is%
\begin{equation}
\Pr\{  M^{\prime}=m|M=m\}  =\operatorname{Tr}\{\Lambda_{B^{n}T_{B}%
}^{m}\mathcal{N}_{A^{\prime n}\rightarrow B^{n}}(\mathcal{E}_{T_{A}\rightarrow
A^{\prime n}}^{m}(\Psi_{T_{A}T_{B}}))\},
\end{equation}
and thus the probability of error $p_{e}(m)$\ for message $m$ is%
\begin{equation}
p_{e}(m)\equiv\operatorname{Tr}\{\left(  I-\Lambda_{B^{n}T_{B}}^{m}\right)
\mathcal{N}_{A^{\prime n}\rightarrow B^{n}}(\mathcal{E}_{T_{A}\rightarrow
A^{\prime n}}^{m}(\Psi_{T_{A}T_{B}}))\}.
\end{equation}
The maximal probability of error $p_{e}^{\ast}$ for the coding scheme is%
\begin{equation}
p_{e}^{\ast}\equiv\max_{m\in\mathcal{M}}p_{e}(m).
\end{equation}
The rate $C$ of communication is%
\begin{equation}
C\equiv\frac{1}{n}\log\left\vert \mathcal{M}\right\vert ,
\end{equation}
and the code has $\varepsilon$ error if $p_{e}^{\ast}\leq\varepsilon$.

A rate $C$ of entanglement-assisted classical communication is
\textit{achievable} if there exists an $\left(  n,C-\delta,\varepsilon\right)
$ entanglement-assisted classical code for all $\varepsilon\in(0,1)$,
$\delta>0$, and sufficiently large~$n$. The entanglement-assisted classical
capacity $C_{\operatorname{EA}}(\mathcal{N})$ of a quantum channel
$\mathcal{N}$ is equal to the supremum of all achievable rates.

\section{A Preliminary Example}

\label{sec-eac:prelim-example}Let us first recall a few items about qudits.
The maximally entangled qudit state is defined as $\left\vert \Phi
\right\rangle _{AB}\equiv d^{-1/2}\sum_{i=0}^{d-1}|i\rangle_{A}|i\rangle_{B}$.
Recall from Section~\ref{sec-nqt:generalized-Pauli} that the Heisenberg--Weyl
operators $X(x)$\ and $Z(z)$\ are an extension of the Pauli matrices to $d$
dimensions:%
\begin{equation}
X(x)\equiv\sum_{x^{\prime}=0}^{d-1}|x+x^{\prime}\rangle\langle x^{\prime
}|,\ \ \ \ \ \ \ \ Z(z)\equiv\sum_{z^{\prime}=0}^{d-1}e^{2\pi izz^{\prime}%
/d}|z^{\prime}\rangle\langle z^{\prime}|.
\end{equation}
Let $|\Phi^{x,z}\rangle_{AB}$ denote the state that results when Alice applies
the operator $X(x)Z(z)$ to her share of the maximally entangled state
$\left\vert \Phi\right\rangle _{AB}$:%
\begin{equation}
|\Phi^{x,z}\rangle_{AB}\equiv\left(  X_{A}(x)Z_{A}(z)\otimes I_{B}\right)
\left\vert \Phi\right\rangle _{AB}.
\end{equation}
Recall from Exercise~\ref{ex-qt:qudit-bell-states-ortho}\ that the set of
states $\left\{  |\Phi^{x,z}\rangle_{AB}\right\}  _{x,z=0}^{d-1}$ forms a
complete orthonormal basis:%
\begin{equation}
\langle\Phi^{x_{1},z_{1}}|\Phi^{x_{2},z_{2}}\rangle_{AB}=\delta_{x_{1},x_{2}%
}\delta_{z_{1},z_{2}},\ \ \ \ \ \ \ \ \ \ \sum_{x,z=0}^{d-1}|\Phi^{x,z}%
\rangle\langle\Phi^{x,z}|_{AB}=I_{AB}. \label{eq-eac:bell-ortho}%
\end{equation}
Let $\pi_{AB}$ denote the maximally mixed state on Alice and Bob's system:
$\pi_{AB}\equiv I_{AB}/d^{2}$, and let $\pi_{A}$ and $\pi_{B}$ denote the
respective maximally mixed states on Alice and Bob's systems: $\pi_{A}\equiv
I_{A}/d$ and $\pi_{B}\equiv I_{B}/d$. Observe that $\pi_{AB}=\pi_{A}\otimes
\pi_{B}$.%
\begin{figure}
[ptb]
\begin{center}
\includegraphics[
width=4.5653in
]%
{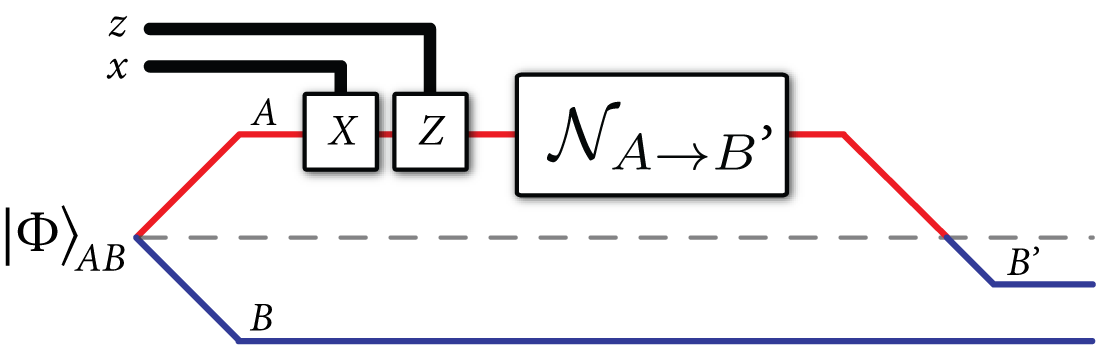}%
\caption{A simple scheme, inspired by super-dense coding, for Alice and Bob to
exploit shared entanglement and a noisy channel in order to establish an
ensemble at Bob's receiving end.}%
\label{fig-eac:eac-simple-protocol}%
\end{center}
\end{figure}

We now consider a simple strategy, inspired by super-dense coding and the
HSW\ coding scheme from Theorem~\ref{thm-cc:HSW}, that Alice and Bob can
employ for entanglement-assisted classical communication. That is, we show how
a strategy similar to super-dense coding induces a particular ensemble at
Bob's receiving end, to which we can then apply the HSW\ coding theorem in
order to establish the existence of a good code for entanglement-assisted
classical communication. Suppose that Alice and Bob possess a maximally
entangled qudit state $\left\vert \Phi\right\rangle _{AB}$. Alice chooses two
symbols $x$ and $z$\ uniformly at random, each in $\left\{  0,\ldots
,d-1\right\}  $. She applies the operators $X(x)Z(z)$ to her share of the
maximally entangled state $\left\vert \Phi\right\rangle _{AB}$, and the
resulting state is $|\Phi^{x,z}\rangle_{AB}$. She then sends her system $A$
over the noisy channel $\mathcal{N}_{A\rightarrow B^{\prime}}$, and Bob
receives the output $B^{\prime}$ from the channel. The noisy channel on the
whole system is $\mathcal{N}_{A\rightarrow B^{\prime}}\otimes\operatorname{id}%
_{B}$, and the ensemble that Bob receives is as follows:%
\begin{equation}
\left\{  d^{-2},\left(  \mathcal{N}_{A\rightarrow B^{\prime}}\otimes
\operatorname{id}_{B}\right)  (\Phi_{AB}^{x,z})\right\}  .
\label{eq-eac:hsw-eac-ensemble}%
\end{equation}
This constitutes an ensemble that they can prepare with one use of the channel
and one shared entangled state (Figure~\ref{fig-eac:eac-simple-protocol}%
\ depicts all of these steps). But, in general, we allow them to exploit many
uses of the channel and however much entanglement that they need. Bob can then
perform a collective measurement on both his share of the entanglement and the
channel outputs in order to determine a message that Alice is transmitting.

Consider that the above scenario is similar to HSW coding.
Theorem~\ref{thm-cc:HSW} from the previous chapter states that the Holevo
information of the above ensemble is an achievable rate for classical
communication over this entanglement-assisted quantum channel. Thus, we can
already state and prove the following corollary of Theorem~\ref{thm-cc:HSW},
simply by calculating the Holevo information of the ensemble in \eqref{eq-eac:hsw-eac-ensemble}.

\begin{corollary}
\label{cor-eac:HSW-eac-achievable}The quantum mutual information
$I(A;B)_{\sigma}$ of the state $\sigma_{AB}\equiv\mathcal{N}_{A^{\prime
}\rightarrow B}(\Phi_{AA^{\prime}})$ is an achievable rate for
entanglement-assisted classical communication over a quantum channel
$\mathcal{N}_{A^{\prime}\rightarrow B}$.
\end{corollary}

\begin{proof}
Observe that the ensemble in \eqref{eq-eac:hsw-eac-ensemble} maps to the
following classical--quantum state:%
\begin{equation}
\rho_{XZB^{\prime}B}\equiv\sum_{x,z=0}^{d-1}\frac{1}{d^{2}}|x\rangle\langle
x|_{X}\otimes|z\rangle\langle z|_{Z}\otimes\left(  \mathcal{N}_{A\rightarrow
B^{\prime}}\otimes\operatorname{id}_{B}\right)  (\Phi_{AB}^{x,z}).
\label{eq-eac:cq-state-simple-example}%
\end{equation}
The Holevo information of this classical--quantum state is $I(XZ;B^{\prime
}B)_{\rho}=H(B^{\prime}B)_{\rho}-H(B^{\prime}B|XZ)_{\rho}$, and it is an
achievable rate for entanglement-assisted classical communication over the
channel $\mathcal{N}_{A^{\prime}\rightarrow B}$ by the direct part of
Theorem~\ref{thm-cc:HSW}. We now proceed to calculate it. First, we determine
the entropy $H(B^{\prime}B)_{\rho}$ by tracing over the classical registers
$XZ$:%
\begin{align}
\operatorname{Tr}_{XZ}\{  \rho_{XZB^{\prime}B}\}   &  =\sum
_{x,z=0}^{d-1}\frac{1}{d^{2}}\left(  \mathcal{N}_{A\rightarrow B^{\prime}%
}\otimes\operatorname{id}_{B}\right)  (\Phi_{AB}^{x,z})\\
&  =\left(  \mathcal{N}_{A\rightarrow B^{\prime}}\otimes\operatorname{id}%
_{B}\right)  \left(  \sum_{x,z=0}^{d-1}\frac{1}{d^{2}}\Phi_{AB}^{x,z}\right)
\\
&  =\left(  \mathcal{N}_{A\rightarrow B^{\prime}}\otimes\operatorname{id}%
_{B}\right)  (\pi_{AB})\\
&  =\mathcal{N}_{A\rightarrow B^{\prime}}(\pi_{A})\otimes\pi_{B},
\end{align}
where the third equality follows from \eqref{eq-eac:bell-ortho}. Thus, the
entropy $H(B^{\prime}B)$ is as follows:%
\begin{equation}
H(B^{\prime}B)=H(\mathcal{N}_{A\rightarrow B^{\prime}}(\pi_{A}))+H(\pi_{B}).
\end{equation}
We now determine the conditional quantum entropy $H(B^{\prime}B|XZ)_{\rho}$:%
\begin{align}
&  H(B^{\prime}B|XZ)_{\rho}\nonumber\\
&  =\sum_{x,z=0}^{d-1}\frac{1}{d^{2}}H\left(  \left(  \mathcal{N}%
_{A\rightarrow B^{\prime}}\otimes\operatorname{id}_{B}\right)  (\Phi
_{AB}^{x,z})\right) \\
&  =\frac{1}{d^{2}}\sum_{x,z=0}^{d-1}H\left(  \mathcal{N}_{A\rightarrow
B^{\prime}}\left[  \left(  X_{A}(x)Z_{A}(z)\right)  (\Phi_{AB})\left(
Z_{A}^{\dag}(z)X_{A}^{\dag}(x)\right)  \right]  \right) \\
&  =\frac{1}{d^{2}}\sum_{x,z=0}^{d-1}H\left(  \mathcal{N}_{A\rightarrow
B^{\prime}}\left[  Z_{B}^{T}(z)X_{B}^{T}(x)(\Phi_{AB})X_{B}^{\ast}%
(x)Z_{B}^{\ast}(z)\right]  \right) \\
&  =\frac{1}{d^{2}}\sum_{x,z=0}^{d-1}H\left(  Z_{B}^{T}(z)X_{B}^{T}(x)\left[
\left(  \mathcal{N}_{A\rightarrow B^{\prime}}\right)  (\Phi_{AB})\right]
\left(  X_{B}^{\ast}(x)Z_{B}^{\ast}(z)\right)  \right) \\
&  =H(\mathcal{N}_{A\rightarrow B^{\prime}}(\Phi_{AB})).
\end{align}
The first equality follows because the system $XZ$ is classical (recall the
result in Section~\ref{sec-qie:cond-ent-cq}). The second equality follows from
the definition of the state $\Phi_{AB}^{x,z}$. The third equality follows by
exploiting the \textquotedblleft transpose trick\textquotedblright\ in
Exercise~\ref{ex-qt:bell-state-matrix-identity}. The fourth equality follows
because the transposed unitaries acting on Bob's system commute with the
action of the channel. Finally, the entropy of a state is invariant with
respect to unitaries. So the Holevo information $I(XZ;B^{\prime}B)_{\rho}$ of
the state $\rho_{XZB^{\prime}B}$ in \eqref{eq-eac:cq-state-simple-example} is
equal to%
\begin{equation}
I(XZ;B^{\prime}B)_{\rho}=H(\mathcal{N}(\pi_{A}))+H(\pi_{B})-H(\left(
\mathcal{N}_{A\rightarrow B^{\prime}}\otimes\operatorname{id}_{B}\right)
(\Phi_{AB})).
\end{equation}
Equivalently, we can write it as the quantum mutual information
$I(A;B)_{\sigma}$, evaluated with respect to the state $\sigma_{AB}%
\equiv\mathcal{N}_{A^{\prime}\rightarrow B}(\Phi_{AA^{\prime}})$.
\end{proof}

For some channels, the quantum mutual information in
Corollary~\ref{cor-eac:HSW-eac-achievable}\ is equal to that channel's
entanglement-assisted classical capacity. This occurs for the depolarizing
channel, a dephasing channel, and an erasure channel to name a few. But there
are examples of channels, such as the amplitude damping channel, where the
quantum mutual information in Corollary~\ref{cor-eac:HSW-eac-achievable} is
not equal to the entanglement-assisted capacity. In the general case, it might
perhaps be intuitive that the quantum mutual information of the channel in
\eqref{eq-eac:EAC-formula} is equal to the entanglement-assisted capacity of
the channel, and it is the goal of the next sections to prove this result.

\begin{exercise}
Consider the following strategy for transmitting and detecting classical
information over an entanglement-assisted depolarizing channel. Alice selects
a state $\left\vert \Phi^{x_{1},z_{1}}\right\rangle _{AB}$ uniformly at random
and sends the $A$ system over the quantum depolarizing channel $\mathcal{N}%
_{A\rightarrow B^{\prime}}^{\operatorname{D}}$, where%
\begin{equation}
\mathcal{N}_{A\rightarrow B^{\prime}}^{\operatorname{D}}(\rho)\equiv(1-p)\rho+p\pi.
\end{equation}
Bob receives the output $B^{\prime}$ of the channel and combines it with his
share $B$ of the entanglement. He then performs a measurement of these systems
in the Bell basis $\left\{  \vert \Phi^{x_{2},z_{2}}\rangle
\langle \Phi^{x_{2},z_{2}}\vert _{B^{\prime}B}\right\}  $.
Determine a simplified expression for the induced classical channel
$p_{Z_{2}X_{2}|Z_{1}X_{1}}(z_{2},x_{2}|z_{1},x_{1})$ where%
\begin{equation}
p_{Z_{2}X_{2}|Z_{1}X_{1}}(z_{2},x_{2}|z_{1},x_{1})\equiv\left\langle
\Phi^{x_{2},z_{2}}\right\vert \left(  \mathcal{N}_{A\rightarrow B^{\prime}%
}^{\operatorname{D}}\otimes\operatorname{id}_{B}\right)  \left(  \vert \Phi^{x_{1}%
,z_{1}}\rangle \langle \Phi^{x_{1},z_{1}}\vert _{AB}\right)
\left\vert \Phi^{x_{2},z_{2}}\right\rangle .
\end{equation}
Show that the classical capacity of the channel $p_{Z_{2}X_{2}|Z_{1}X_{1}%
}(z_{2},x_{2}|z_{1},x_{1})$ is equal to the entanglement-assisted classical
capacity of the depolarizing channel (you can take it for granted that the
entanglement-assisted classical capacity of the depolarizing channel is given
by Corollary~\ref{cor-eac:HSW-eac-achievable}). Thus, there is no need for the
receiver to perform a collective measurement on many channel outputs in order
to achieve capacity---it suffices to perform single-channel Bell measurements
at the receiving end.
\end{exercise}

\section{Entanglement-Assisted Capacity Theorem}

\label{sec-eac:BSST-theorem}We now state the entanglement-assisted classical
capacity theorem. Section~\ref{sec-eac:direct-coding}\ proves the direct part
of this theorem, and Section~\ref{sec-eac:converse}\ proves its converse part.%
\index{entanglement-assisted capacity theorem}%

\begin{theorem}
[Bennett--Shor--Smolin--Thapliyal]\label{thm-eac:BSST}The
entanglement-assisted classical capacity of a quantum channel is equal to the
channel's mutual information:%
\begin{equation}
C_{\operatorname{EA}}(\mathcal{N})=I(\mathcal{N}),
\end{equation}
where the mutual information $I(\mathcal{N})$ of a channel $\mathcal{N}$ is
defined as $I(\mathcal{N})\equiv\max_{\varphi_{AA^{\prime}}}I(A;B)_{\rho}$,
$\rho_{AB}\equiv\mathcal{N}_{A^{\prime}\rightarrow B}(\varphi_{AA^{\prime}})$,
and $\varphi_{AA^{\prime}}$ is a pure bipartite state.
\end{theorem}

\section{The Direct Coding Theorem}

\label{sec-eac:direct-coding}The direct coding theorem is a statement of
achievability:
\index{entanglement-assisted capacity theorem!direct part}%

\begin{theorem}
[Direct Coding]\label{thm-eac:direct-coding}The following resource inequality
corresponds to an achievable protocol for entanglement-assisted classical
communication over a quantum channel $\mathcal{N}_{A^{\prime}\rightarrow B}$:%
\begin{equation}
\left\langle \mathcal{N}\right\rangle +H(A)_{\rho}\left[  qq\right]  \geq
I(A;B)_{\rho}\left[  c\rightarrow c\right]  ,
\end{equation}
where $\rho_{AB}\equiv\mathcal{N}_{A^{\prime}\rightarrow B}(\varphi
_{AA^{\prime}})$.
\end{theorem}

\begin{proof}
We suppose that Alice and Bob share $n$ copies of an arbitrary pure, bipartite
entangled state $|\varphi\rangle_{AB}$. This is allowed in the setting of
entanglement-assisted communication as discussed in
Section~\ref{sec-eac:info-task}. Alternatively, they could convert $nH(A)$
shared ebits to $\approx n$ copies of $|\varphi\rangle_{AB}$ by making use of
the entanglement dilution protocol discussed in
Section~\ref{sec-em:ent-dilution} (they would need a sublinear amount of
classical communication to do so, which has negligible rate). We would like to
apply a similar coding technique as outlined in
Section~\ref{sec-eac:prelim-example}. For example, it would be useful to
exploit the transpose trick from
Exercise~\ref{ex-qt:bell-state-matrix-identity}, but we cannot do so directly
because this trick only applies to maximally entangled states. However, we can
instead exploit the fact that Alice and Bob share many copies of the state
$|\varphi\rangle_{AB}$ that decompose into a direct sum of maximally entangled
states. First, recall that every pure, bipartite state has a Schmidt
decomposition%
\index{Schmidt decomposition}
(see Theorem~\ref{thm-qt:schmidt}):%
\begin{equation}
|\varphi\rangle_{AB}\equiv\sum_{x}\sqrt{p_{X}(x)}|x\rangle_{A}|x\rangle_{B},
\end{equation}
where $p_{X}(x)>0$ for all $x$, $\sum_{x}p_{X}(x)=1$, and $\{|x\rangle_{A}\}$
and $\{|x\rangle_{B}\}$ are orthonormal bases for Alice and Bob's respective
systems. Let us take $n$ copies of the above state, giving a state of the
following form:%
\begin{equation}
|\varphi\rangle_{A^{n}B^{n}}\equiv\sum_{x^{n}}\sqrt{p_{X^{n}}(x^{n})}%
|x^{n}\rangle_{A^{n}}|x^{n}\rangle_{B^{n}},
\end{equation}
where $x^{n}\equiv x_{1}\cdots x_{n}$, $p_{X^{n}}(x^{n})\equiv p_{X}%
(x_{1})\cdots p_{X}(x_{n})$, and $|x^{n}\rangle\equiv|x_{1}\rangle\cdots
|x_{n}\rangle$. We can write the above state in terms of its
\index{type class!subspace}%
type decomposition (see Section~\ref{sec-ct:method-of-types}):%
\begin{align}
|\varphi\rangle_{A^{n}B^{n}}  &  =\sum_{t}\sum_{x^{n}\in T_{t}}\sqrt{p_{X^{n}%
}(x^{n})}|x^{n}\rangle_{A^{n}}|x^{n}\rangle_{B^{n}}%
\label{eq-eac:type-decomp-1}\\
&  =\sum_{t}\sqrt{p_{X^{n}}(x_{t}^{n})}\sum_{x^{n}\in T_{t}}|x^{n}%
\rangle_{A^{n}}|x^{n}\rangle_{B^{n}}\\
&  =\sum_{t}\sqrt{p_{X^{n}}(x_{t}^{n})d_{t}}\frac{1}{\sqrt{d_{t}}}\sum
_{x^{n}\in T_{t}}|x^{n}\rangle_{A^{n}}|x^{n}\rangle_{B^{n}}\\
&  =\sum_{t}\sqrt{p(t)}|\Phi_{t}\rangle_{A^{n}B^{n}},
\label{eq-eac:type-decomp-4}%
\end{align}
with the following definitions:%
\begin{align}
p(t)  &  \equiv p_{X^{n}}(x_{t}^{n})d_{t},\label{eq-eac:pt-dist}\\
|\Phi_{t}\rangle_{A^{n}B^{n}}  &  \equiv\frac{1}{\sqrt{d_{t}}}\sum_{x^{n}\in
T_{t}}|x^{n}\rangle_{A^{n}}|x^{n}\rangle_{B^{n}}.
\end{align}
The first equality in \eqref{eq-eac:type-decomp-1}\ follows by decomposing the
state into its different type class subspaces. The next equality follows
because $p_{X^{n}}(x^{n})$ is the same for all sequences $x^{n}$ in the same
type class and because the distribution is i.i.d.~(let $x_{t}^{n}$ be some
representative sequence of all sequences in the type class $T_{t}$). The third
equality follows by introducing $d_{t}$ as the dimension of a type class
subspace $T_{t}$, and the final equality in \eqref{eq-eac:type-decomp-4}
follows from the definitions. Observe that the state $|\Phi_{t}\rangle
_{A^{n}B^{n}}$ is maximally entangled.

Each state $|\Phi_{t}\rangle_{A^{n}B^{n}}$ is maximally entangled with Schmidt
rank $d_{t}$, and we can thus apply the transpose trick\ for operators acting
on the type class subspaces. Inspired by the dense-coding-like strategy from
Section~\ref{sec-eac:prelim-example}, we allow Alice to choose unitary
operators from the Heisenberg--Weyl%
\index{Heisenberg-Weyl operators}
set of $d_{t}^{2}$ operators that act on the $A^{n}$ share of $|\Phi
_{t}\rangle_{A^{n}B^{n}}$. We denote one of these operators as $V(x_{t}%
,z_{t})\equiv X(x_{t})Z(z_{t})$ where $x_{t},z_{t}\in\left\{  0,\ldots
,d_{t}-1\right\}  $. If she does this for every type class subspace and
applies a phase $\left(  -1\right)  ^{b_{t}}$ in each subspace, then the
resulting unitary operator $U(s)$ acting on all of her $A^{n}$ systems is a
direct sum of all of these unitaries:%
\begin{equation}
U(s)\equiv\bigoplus\limits_{t}\left(  -1\right)  ^{b_{t}}V(x_{t},z_{t}),
\label{eq-eac:unitary-encoding}%
\end{equation}
where $s$ is a vector containing all of the indices needed to specify the
unitary $U(s)$:%
\begin{equation}
s\equiv\left(  \left(  x_{t},z_{t},b_{t}\right)  \right)  _{t}.
\label{eq-eac:vector-s}%
\end{equation}
Let $\mathcal{S}$ denote the set of all possible vectors $s$. The transpose
trick holds for these particular unitary operators:%
\begin{equation}
\left(  U_{A^{n}}(s)\otimes I_{B^{n}}\right)  |\varphi\rangle_{A^{n}B^{n}%
}=\left(  I_{A^{n}}\otimes U_{B^{n}}^{T}(s)\right)  |\varphi\rangle
_{A^{n}B^{n}} \label{eq-eac:transpose-trick}%
\end{equation}
because it applies in each type class subspace:%
\begin{align}
&  \left(  U_{A^{n}}(s)\otimes I_{B^{n}}\right)  |\varphi\rangle_{A^{n}B^{n}%
}\nonumber\\
&  =\left(  \bigoplus\limits_{t}\left(  -1\right)  ^{b_{t}}V_{A^{n}}%
(x_{t},z_{t})\right)  \sum_{t}\sqrt{p(t)}|\Phi_{t}\rangle_{A^{n}B^{n}}\\
&  =\sum_{t}\sqrt{p(t)}\left(  -1\right)  ^{b_{t}}V_{A^{n}}(x_{t},z_{t}%
)|\Phi_{t}\rangle_{A^{n}B^{n}}\\
&  =\sum_{t}\sqrt{p(t)}\left(  -1\right)  ^{b_{t}}V_{B^{n}}^{T}(x_{t}%
,z_{t})|\Phi_{t}\rangle_{A^{n}B^{n}}\\
&  =\left(  \bigoplus\limits_{t}\left(  -1\right)  ^{b_{t}}V_{B^{n}}^{T}%
(x_{t},z_{t})\right)  \sum_{t}\sqrt{p(t)}|\Phi_{t}\rangle_{A^{n}B^{n}}\\
&  =\left(  I_{A^{n}}\otimes U_{B^{n}}^{T}(s)\right)  |\varphi\rangle
_{A^{n}B^{n}}.
\end{align}

Now we need to establish a means by which Alice can select a random code. For
every message $m\in\mathcal{M}$ that Alice would like to transmit, she chooses
the elements of the vector $s\in\mathcal{S}$ uniformly at random, leading to a
particular unitary operator $U(s)$. We can write $s(m)$ instead of just $s$ to
denote the explicit association of the vector $s$ with the message~$m$---we
can think of each chosen vector $s(m)$ as a classical codeword, with the
codebook being $\left\{  s(m)\right\}  _{m\in\left\{  1,\ldots,\left\vert
\mathcal{M}\right\vert \right\}  }$. This random selection procedure leads to
entanglement-assisted quantum codewords of the following form:%
\begin{equation}
|\varphi_{m}\rangle_{A^{n}B^{n}}\equiv\left(  U_{A^{n}}(s(m))\otimes I_{B^{n}%
}\right)  |\varphi\rangle_{A^{n}B^{n}}.
\end{equation}
Alice then transmits her systems $A^{n}$ through many uses of the quantum
channel $\mathcal{N}_{A\rightarrow B^{\prime}}$, leading to the following
state that is entirely in Bob's control:%
\begin{equation}
\mathcal{N}_{A^{n}\rightarrow B^{\prime n}}(|\varphi_{m}\rangle\langle
\varphi_{m}|_{A^{n}B^{n}}).
\end{equation}
Interestingly, the above state is equal to the state in
\eqref{eq-eac:transpose-state-many-copies} below, by exploiting the transpose
trick from \eqref{eq-eac:transpose-trick}:%
\begin{align}
&  \mathcal{N}_{A^{n}\rightarrow B^{\prime n}}(|\varphi_{m}\rangle
\langle\varphi_{m}|_{A^{n}B^{n}})\nonumber\\
&  =\mathcal{N}_{A^{n}\rightarrow B^{\prime n}}(U_{A^{n}}(s(m))|\varphi
\rangle\langle\varphi|_{A^{n}B^{n}}U_{A^{n}}^{\dag}(s(m)))\\
&  =\mathcal{N}_{A^{n}\rightarrow B^{\prime n}}(U_{B^{n}}^{T}(s(m))|\varphi
\rangle\langle\varphi|_{A^{n}B^{n}}U_{B^{n}}^{\ast}(s(m)))\\
&  =U_{B^{n}}^{T}(s(m))\mathcal{N}_{A^{n}\rightarrow B^{\prime n}}%
(|\varphi\rangle\langle\varphi|_{A^{n}B^{n}})U_{B^{n}}^{\ast}(s(m)).
\label{eq-eac:transpose-state-many-copies}%
\end{align}
Observe that the transpose trick allows us to commute the action of the
channel with Alice's encoding unitary $U(s(m))$, just as we did in the example
in the proof of Corollary~\ref{cor-eac:HSW-eac-achievable}. Let%
\begin{equation}
\rho_{B^{\prime n}B^{n}}\equiv\mathcal{N}_{A^{n}\rightarrow B^{\prime n}%
}(|\varphi\rangle\langle\varphi|_{A^{n}B^{n}})
\end{equation}
so that%
\begin{equation}
\mathcal{N}_{A^{n}\rightarrow B^{\prime n}}(|\varphi_{m}\rangle\langle
\varphi_{m}|_{A^{n}B^{n}})=U_{B^{n}}^{T}(s(m))\rho_{B^{\prime n}B^{n}}%
U_{B^{n}}^{\ast}(s(m)). \label{eq-eac:channel-output-codewords}%
\end{equation}

\begin{remark}
[Tensor-Power Channel Output States]\label{rem-eac:tensor-power-eac}When using
the coding scheme given above, the reduced state on the channel output
(obtained by ignoring Bob's share of the entanglement in $B^{n}$) is the same
tensor-power state, independent of the unitary that Alice applies at the
channel input:%
\begin{align}
\operatorname{Tr}_{B^{n}}\{  \mathcal{N}_{A^{n}\rightarrow B^{\prime n}%
}(|\varphi_{m}\rangle\langle\varphi_{m}|_{A^{n}B^{n}})\}   &
=\rho_{B^{\prime n}}\\
&  =\mathcal{N}_{A^{n}\rightarrow B^{\prime n}}(\varphi_{A^{n}}),
\end{align}
where $\varphi_{A^{n}}=\left(  \operatorname{Tr}_{B}\{  \varphi
_{AB}\}  \right)  ^{\otimes n}$. This follows directly from
\eqref{eq-eac:channel-output-codewords} and taking the partial trace over
$B^{n}$. We exploit this feature in the next chapter, where we construct codes
for transmitting both classical and quantum information with the help of
shared entanglement.
\end{remark}

After Alice has transmitted her entanglement-assisted quantum codewords over
the channel, it becomes Bob's task to determine which message $m$ Alice
transmitted, and he should do so with some POVM\ $\left\{  \Lambda
^{m}\right\}  $ that depends on the random choice of code.
Figure~\ref{fig-eac:eac-protocol}\ depicts the protocol.
\begin{figure}
[ptb]
\begin{center}
\includegraphics[
width=4.8456in
]%
{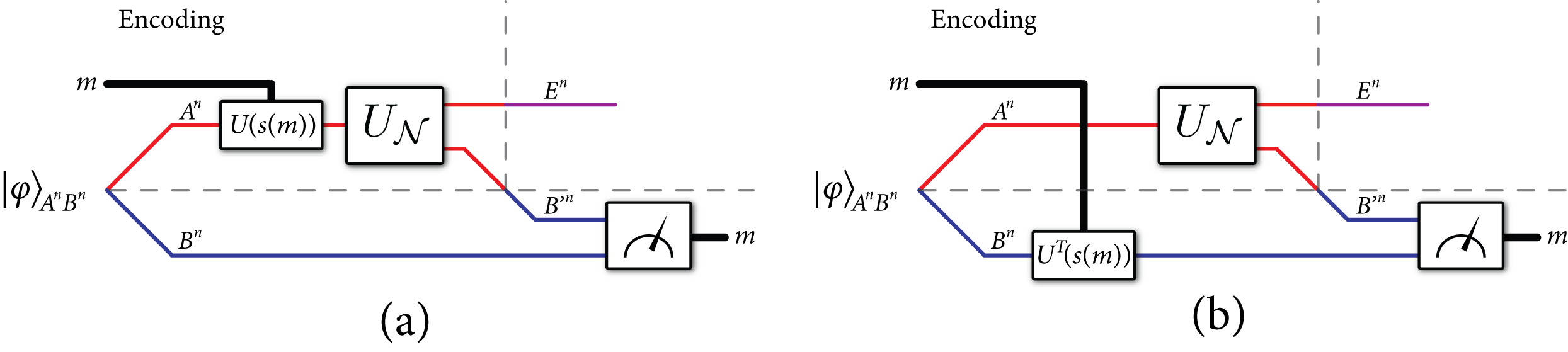}%
\caption{(a) Alice shares many copies of a pure, bipartite state $\left\vert
\varphi\right\rangle ^{\otimes n}$ with Bob. She encodes a message~$m$
according to some unitary of the form in \eqref{eq-eac:unitary-encoding}. She
transmits her share of the entanglement-assisted quantum codeword over many
uses of the quantum channel, and it is Bob's task to determine which message
she transmits. (b) Alice acting locally with the unitary $U(s(m))$ on her
share $A^{n}$ of the entanglement $\left\vert \varphi\right\rangle ^{\otimes
n}$ is the same as her acting non-locally with $U^{T}(s(m))$ on Bob's share
$B^{n}$ of the entanglement. This follows because of the particular structure
of the unitaries in \eqref{eq-eac:unitary-encoding}.}%
\label{fig-eac:eac-protocol}%
\end{center}
\end{figure}

At this point, we would like to exploit the
\index{packing lemma}%
packing lemma from Chapter~\ref{chap:packing} in order to establish the
existence of a reliable decoding POVM for Bob. Recall that the packing lemma
requires four objects, and these four objects should satisfy the four
inequalities in\ \eqref{eq:pack-1}--\eqref{eq:pack-4}. The first object
required is an ensemble from which Alice and Bob can select a code randomly,
and in our case, the ensemble is%
\begin{equation}
\left\{  \frac{1}{\left\vert \mathcal{S}\right\vert },\ U_{B^{n}}^{T}%
(s)\rho_{B^{\prime n}B^{n}}U_{B^{n}}^{\ast}(s)\right\}  _{s\in\mathcal{S}}.
\end{equation}
The next object required is the expected density operator of this ensemble:%
\begin{align}
\overline{\rho}_{B^{\prime n}B^{n}}  &  \equiv\mathbb{E}_{S}\left\{
U^{T}(S)_{B^{n}}\rho_{B^{\prime n}B^{n}}U^{\ast}(S)_{B^{n}}\right\} \\
&  =\frac{1}{\left\vert \mathcal{S}\right\vert }\sum_{s\in\mathcal{S}}%
U_{B^{n}}^{T}(s)\rho_{B^{\prime n}B^{n}}U_{B^{n}}^{\ast}(s).
\end{align}
We later prove that this expected density operator has the following simpler
form:%
\begin{equation}
\overline{\rho}_{B^{\prime n}B^{n}}=\sum_{t}p(t)\ \mathcal{N}_{A^{n}%
\rightarrow B^{\prime n}}(\pi_{A^{n}}^{t})\otimes\pi_{B^{n}}^{t},
\label{eq-eac:expected-dens-op-form}%
\end{equation}
where $p(t)$ is the distribution from \eqref{eq-eac:pt-dist} and $\pi_{A^{n}%
}^{t}$ is the maximally mixed state on a type class subspace: $\pi_{A^{n}}%
^{t}\equiv I_{t}/d_{t}$. The final two objects that we require for the packing
lemma are the message subspace projectors and the total subspace projector. We
assign these respectively as%
\begin{align}
&  U_{B^{n}}^{T}(s)\Pi_{B^{\prime n}B^{n}}^{\rho,\delta}U_{B^{n}}^{\ast}(s),\\
&  \Pi_{B^{\prime n}}^{\rho,\delta}\otimes\Pi_{B^{n}}^{\rho,\delta},
\end{align}
where $\Pi_{B^{\prime n}B^{n}}^{\rho,\delta}$, $\Pi_{B^{\prime n}}%
^{\rho,\delta}$, and $\Pi_{B^{n}}^{\rho,\delta}$ are the typical projectors
for many copies of the states $\rho_{B^{\prime}B}\equiv\mathcal{N}%
_{A\rightarrow B^{\prime}}(\varphi_{AB})$, $\ \rho_{B^{\prime}}%
=\operatorname{Tr}_{B}\{\rho_{B^{\prime}B}\}$, and $\rho_{B}=\operatorname{Tr}%
_{B^{\prime}}\{\rho_{B^{\prime}B}\}$, respectively. Observe that the size of
each message subspace projector is $\approx2^{nH(B^{\prime}B)}$, and the size
of the total subspace projector is $\approx2^{n\left[  H(B^{\prime
})+H(B)\right]  }$. By dimension counting, this is suggesting that we can pack
in $\approx2^{n\left[  H(B^{\prime})+H(B)\right]  }/2^{nH(B^{\prime}%
B)}=2^{nI(B^{\prime};B)}$ messages using this coding technique.

If the four conditions of the packing lemma are satisfied (see
\eqref{eq:pack-1}--\eqref{eq:pack-4}), then there exists a detection
POVM\ that can reliably decode Alice's transmitted messages as long as the
number of messages in the code is not too high. The four conditions in
\eqref{eq:pack-1}--\eqref{eq:pack-4} translate to the following four
conditions for our case:%
\begin{align}
\operatorname{Tr}\left\{  \left(  \Pi_{B^{\prime n}}^{\rho,\delta}\otimes
\Pi_{B^{n}}^{\rho,\delta}\right)  \left(  U_{B^{n}}^{T}(s)\rho_{B^{\prime
n}B^{n}}U_{B^{n}}^{\ast}(s)\right)  \right\}   &  \geq1-\varepsilon
,\label{eq-eac:pack-2}\\
\operatorname{Tr}\left\{  \left(  U_{B^{n}}^{T}(s)\Pi_{B^{\prime n}B^{n}%
}^{\rho,\delta}U_{B^{n}}^{\ast}(s)\right)  \left(  U_{B^{n}}^{T}%
(s)\rho_{B^{\prime n}B^{n}}U_{B^{n}}^{\ast}(s)\right)  \right\}   &
\geq1-\varepsilon,\label{eq-eac:pack-1}\\
\operatorname{Tr}\left\{  U_{B^{n}}^{T}(s)\Pi_{B^{\prime n}B^{n}}^{\rho
,\delta}U_{B^{n}}^{\ast}(s)\right\}   &  \leq2^{n\left[  H(B^{\prime}B)_{\rho
}+c\delta\right]  }, \label{eq-eac:pack-3}%
\end{align}%
\begin{multline}
\left(  \Pi_{B^{\prime n}}^{\rho,\delta}\otimes\Pi_{B^{n}}^{\rho,\delta
}\right)  \overline{\rho}_{B^{\prime n}B^{n}}\left(  \Pi_{B^{\prime n}}%
^{\rho,\delta}\otimes\Pi_{B^{n}}^{\rho,\delta}\right) \label{eq-eac:pack-4}\\
\leq2^{-n\left[  H(B^{\prime})_{\rho}+H(B)_{\rho}-\eta(n,\delta)-c\delta
\right]  }\left(  \Pi_{B^{\prime n}}^{\rho,\delta}\otimes\Pi_{B^{n}}%
^{\rho,\delta}\right)  ,
\end{multline}
where $c$ is some positive constant and $\eta(n,\delta)$ is a function that
approaches zero as $n\rightarrow\infty$ and $\delta\rightarrow0$.

We now prove the four inequalities in
\eqref{eq-eac:pack-2}--\eqref{eq-eac:pack-4}, attacking them in the order of
increasing difficulty. The condition in \eqref{eq-eac:pack-1} holds because%
\begin{align}
&  \operatorname{Tr}\left\{  \left(  U_{B^{n}}^{T}(s)\Pi_{B^{\prime n}B^{n}%
}^{\rho,\delta}U_{B^{n}}^{\ast}(s)\right)  \left(  U_{B^{n}}^{T}%
(s)\rho_{B^{\prime n}B^{n}}U_{B^{n}}^{\ast}(s)\right)  \right\} \nonumber\\
&  =\operatorname{Tr}\left\{  \Pi_{B^{\prime n}B^{n}}^{\rho,\delta}%
\rho_{B^{\prime n}B^{n}}\right\} \\
&  \geq1-\varepsilon.
\end{align}
The equality holds by cyclicity of the trace and because $U^{\ast}U^{T}=I$.
The inequality holds by exploiting the unit probability property of typical
projectors (Property~\ref{prop-qt:unit}). From this inequality, observe that
we choose each message subspace projector so that it is exactly the one that
should identify the entanglement-assisted quantum codeword $U_{B^{n}}%
^{T}(s)\rho_{B^{\prime n}B^{n}}U_{B^{n}}^{\ast}(s)$\ with high probability.

We next consider the condition in \eqref{eq-eac:pack-3}:%
\begin{align}
\operatorname{Tr}\left\{  U_{B^{n}}^{T}(s)\Pi_{B^{\prime n}B^{n}}^{\rho
,\delta}U_{B^{n}}^{\ast}(s)\right\}   &  =\operatorname{Tr}\left\{
\Pi_{B^{\prime n}B^{n}}^{\rho,\delta}\right\} \\
&  \leq2^{n\left[  H(B^{\prime}B)_{\rho}+c\delta\right]  }.
\end{align}
The equality holds again by cyclicity of trace, and the inequality follows
from the \textquotedblleft exponentially smaller cardinality\textquotedblright%
\ property of the typical subspace (Property~\ref{prop-qt:exp-small}).

Consider the condition in \eqref{eq-eac:pack-2}. First, define $\hat{P}=I-P$.
Then%
\begin{align}
\Pi_{B^{\prime n}}^{\rho,\delta}\otimes\Pi_{B^{n}}^{\rho,\delta}  &
=(I-\hat{\Pi}_{B^{\prime n}}^{\rho,\delta})\otimes(I-\hat{\Pi}_{B^{n}}%
^{\rho,\delta})\label{eq-eac:projector-complements-1}\\
&  =\left(  I_{B^{\prime n}}\otimes I_{B^{n}}\right)  -(\hat{\Pi}_{B^{\prime
n}}^{\rho,\delta}\otimes I_{B^{n}})\nonumber\\
&  \ \ \ \ \ \ \ -(I_{B^{\prime n}}\otimes\hat{\Pi}_{B^{n}}^{\rho,\delta
})+(\hat{\Pi}_{B^{\prime n}}^{\rho,\delta}\otimes\hat{\Pi}_{B^{n}}%
^{\rho,\delta})\\
&  \geq\left(  I_{B^{\prime n}}\otimes I_{B^{n}}\right)  -(\hat{\Pi
}_{B^{\prime n}}^{\rho,\delta}\otimes I_{B^{n}})-(I_{B^{\prime n}}\otimes
\hat{\Pi}_{B^{n}}^{\rho,\delta}). \label{eq-eac:projector-complements-4}%
\end{align}
Consider the following chain of inequalities:%
\begin{align}
&  \operatorname{Tr}\left\{  (\Pi_{B^{\prime n}}^{\rho,\delta}\otimes
\Pi_{B^{n}}^{\rho,\delta})\left(  U_{B^{n}}^{T}(s)\rho_{B^{\prime n}B^{n}%
}U_{B^{n}}^{\ast}(s)\right)  \right\} \nonumber\\
&  \geq\operatorname{Tr}\left\{  U_{B^{n}}^{T}(s)\rho_{B^{\prime n}B^{n}%
}U_{B^{n}}^{\ast}(s)\right\} \nonumber\\
&  \qquad -\operatorname{Tr}\left\{  (\hat{\Pi}_{B^{\prime n}%
}^{\rho,\delta}\otimes I_{B^{n}})\left(  U_{B^{n}}^{T}(s)\rho_{B^{\prime
n}B^{n}}U_{B^{n}}^{\ast}(s)\right)  \right\} \nonumber\\
&  \qquad -\operatorname{Tr}\left\{  (I_{B^{\prime n}}%
\otimes\hat{\Pi}_{B^{n}}^{\rho,\delta})\left(  U_{B^{n}}^{T}(s)\rho_{B^{\prime
n}B^{n}}U_{B^{n}}^{\ast}(s)\right)  \right\} \\
&  =1-\operatorname{Tr}\left\{  \hat{\Pi}_{B^{\prime n}}^{\rho,\delta}%
\rho_{B^{\prime n}}\right\}  -\operatorname{Tr}\left\{  \hat{\Pi}_{B^{n}%
}^{\rho,\delta}\rho_{B^{n}}\right\} \\
&  \geq1-2\varepsilon.
\end{align}
The first inequality follows from the development in
\eqref{eq-eac:projector-complements-1}--\eqref{eq-eac:projector-complements-4}.
The first equality follows because $\operatorname{Tr}\left\{  U_{B^{n}}%
^{T}(s)\rho_{B^{\prime n}B^{n}}U_{B^{n}}^{\ast}(s)\right\}  =1$ and from
performing a partial trace on $B^{n}$ and $B^{\prime n}$, respectively (while
noting that we can apply the transpose trick for the second one). The final
inequality follows from the unit probability property of the typical
projectors $\Pi_{B^{\prime n}}^{\rho,\delta}$ and $\Pi_{B^{n}}^{\rho,\delta}$
(Property~\ref{prop-qt:unit}).

The last inequality in \eqref{eq-eac:pack-4} requires the most effort to
prove. We first need to prove that the expected density operator
$\overline{\rho}_{B^{\prime n}B^{n}}$ takes the form given in
\eqref{eq-eac:expected-dens-op-form}. To simplify the development, we evaluate
the expectation without the channel applied, and we then apply the channel to
the state at the end of the development. Consider that%
\begin{align}
\overline{\rho}_{A^{n}B^{n}}  &  =\frac{1}{\left\vert \mathcal{S}\right\vert
}\sum_{s\in\mathcal{S}}U_{B^{n}}^{T}(s)|\varphi\rangle\langle\varphi
|_{A^{n}B^{n}}U_{B^{n}}^{\ast}(s)\label{eq-eac:average-state-1}\\
&  =\frac{1}{\left\vert \mathcal{S}\right\vert }\sum_{s\in\mathcal{S}}%
U_{B^{n}}^{T}(s)\left(  \sum_{t}\sqrt{p(t)}|\Phi_{t}\rangle_{A^{n}B^{n}%
}\right)  \left(  \sum_{t^{\prime}}\langle\Phi_{t^{\prime}}|_{A^{n}B^{n}}%
\sqrt{p(t^{\prime})}\right)  U_{B^{n}}^{\ast}(s)\\
&  =\frac{1}{\left\vert \mathcal{S}\right\vert }\sum_{s\in\mathcal{S}}\left(
\sum_{t}\sqrt{p(t)}\left(  -1\right)  ^{b_{t}(s)}\left(  V_{B^{n}}^{T}%
((z_{t},x_{t})(s))\right)  |\Phi_{t}\rangle_{A^{n}B^{n}}\right) \nonumber\\
&  \ \ \ \ \ \ \ \ \ \ \ \left(  \sum_{t^{\prime}}\langle\Phi_{t^{\prime}%
}|_{A^{n}B^{n}}\left(  -1\right)  ^{b_{t^{\prime}}(s)}\left(  V_{B^{n}}^{\ast
}((z_{t^{\prime}},x_{t^{\prime}})(s))\right)  \sqrt{p(t^{\prime})}\right)  .
\label{eq-eac:average-state-2}%
\end{align}
Let us first consider the case when $t=t^{\prime}$. Then the expression in
\eqref{eq-eac:average-state-2} becomes%
\begin{align}
&  \frac{1}{\left\vert \mathcal{S}\right\vert }\sum_{s\in\mathcal{S}}\sum
_{t}p(t)\left(  V_{B^{n}}^{T}((z_{t},x_{t})(s))\right)  |\Phi_{t}%
\rangle\langle\Phi_{t}|_{A^{n}B^{n}}\left(  V_{B^{n}}^{\ast}((z_{t}%
,x_{t})(s))\right) \\
&  =\sum_{t}p(t)\left[  \frac{1}{d_{t}^{2}}\sum_{x_{t},z_{t}}V_{B^{n}}%
^{T}(z_{t},x_{t})|\Phi_{t}\rangle\langle\Phi_{t}|_{A^{n}B^{n}}V_{B^{n}}^{\ast
}(z_{t},x_{t})\right] \\
&  =\sum_{t}p(t)\pi_{A^{n}}^{t}\otimes\pi_{B^{n}}^{t}.
\end{align}
These equalities hold because the sum over all the elements in $\mathcal{S}$
implies that we are uniformly mixing the maximally entangled states $|\Phi
_{t}\rangle_{A^{n}B^{n}}$ on the type class subspaces and
Exercise~\ref{ex-qt:uniformly-random-unitary}\ gives us that the resulting
state on each type class subspace is equal to $\operatorname{Tr}_{B^{n}%
}\left\{  \left[  \Phi_{t}\right]  _{A^{n}B^{n}}\right\}  \otimes\pi_{B^{n}%
}^{t}=\pi_{A^{n}}^{t}\otimes\pi_{B^{n}}^{t}$. Let us now consider the case
when $t\neq t^{\prime}$. Then the expression in \eqref{eq-eac:average-state-2}
becomes%
\begin{align}
&  \frac{1}{\left\vert \mathcal{S}\right\vert }\sum_{s\in\mathcal{S}}%
\sum_{t^{\prime},t\neq t^{\prime}}\sqrt{p(t)p(t^{\prime})}\left(  -1\right)
^{b_{t}(s)+b_{t^{\prime}}(s)}\times\nonumber\\
&  \ \ \ \ \ \ \ \left(  V_{B^{n}}^{T}\left(  (z_{t},x_{t})(s)\right)
\right)  |\Phi_{t}\rangle\langle\Phi_{t^{\prime}}|_{A^{n}B^{n}}\left(
V_{B^{n}}^{\ast}\left(  (z_{t^{\prime}},x_{t^{\prime}})(s)\right)  \right)
\nonumber\\
&  =\sum_{t^{\prime},t\neq t^{\prime}}\frac{1}{d_{t}^{2}d_{t^{\prime}}^{2}%
4}\sum_{b_{t},b_{t^{\prime}},x_{t},z_{t},x_{t^{\prime}},z_{t^{\prime}}}%
\sqrt{p(t)p(t^{\prime})}\left(  -1\right)  ^{b_{t}+b_{t^{\prime}}}%
\times\nonumber\\
&  \ \ \ \ \ \ \ V_{B^{n}}^{T}(z_{t},x_{t})|\Phi_{t}\rangle\langle
\Phi_{t^{\prime}}|_{A^{n}B^{n}}V_{B^{n}}^{\ast}(z_{t^{\prime}},x_{t^{\prime}%
})\\
&  =\sum_{t^{\prime},t\neq t^{\prime}}\frac{1}{d_{t}^{2}d_{t^{\prime}}^{2}%
}\sum_{b_{t},b_{t^{\prime}}}\frac{\left(  -1\right)  ^{b_{t}+b_{t^{\prime}}}%
}{4}\times\nonumber\\
&  \ \ \ \ \ \ \ \left(  \sum_{x_{t},z_{t},x_{t^{\prime}},z_{t^{\prime}}}%
\sqrt{p(t)p(t^{\prime})}V_{B^{n}}^{T}(z_{t},x_{t})|\Phi_{t}\rangle\langle
\Phi_{t^{\prime}}|_{A^{n}B^{n}}V_{B^{n}}^{\ast}(z_{t^{\prime}},x_{t^{\prime}%
})\right) \\
&  =0.
\end{align}
It then follows that%
\begin{equation}
\frac{1}{\left\vert \mathcal{S}\right\vert }\sum_{s\in\mathcal{S}}U_{B^{n}%
}^{T}(s)|\varphi\rangle\langle\varphi|_{A^{n}B^{n}}U_{B^{n}}^{\ast}%
(s)=\sum_{t}p(t)\pi_{A^{n}}^{t}\otimes\pi_{B^{n}}^{t},
\end{equation}
and by linearity, that%
\begin{multline}
\frac{1}{\left\vert \mathcal{S}\right\vert }\sum_{s\in\mathcal{S}}U_{B^{n}%
}^{T}(s)\mathcal{N}_{A^{n}\rightarrow B^{\prime n}}(|\varphi\rangle
\langle\varphi|_{A^{n}B^{n}})U_{B^{n}}^{\ast}%
(s)\label{eq-eac:expected-state-result}\\
=\sum_{t}p(t)\mathcal{N}_{A^{n}\rightarrow B^{\prime n}}(\pi_{A^{n}}%
^{t})\otimes\pi_{B^{n}}^{t}.
\end{multline}

We now prove the final condition in \eqref{eq-eac:pack-4} for the packing
lemma. Consider the following chain of inequalities:%
\begin{align}
&  \left(  \Pi_{B^{\prime n}}^{\rho,\delta}\otimes\Pi_{B^{n}}^{\rho,\delta
}\right)  \overline{\rho}_{B^{\prime n}B^{n}}\left(  \Pi_{B^{\prime n}}%
^{\rho,\delta}\otimes\Pi_{B^{n}}^{\rho,\delta}\right) \nonumber\\
&  =\left(  \Pi_{B^{\prime n}}^{\rho,\delta}\otimes\Pi_{B^{n}}^{\rho,\delta
}\right)  \left(  \sum_{t}p(t)\mathcal{N}_{A^{n}\rightarrow B^{\prime n}}%
(\pi_{A^{n}}^{t})\otimes\pi_{B^{n}}^{t}\right)  \left(  \Pi_{B^{\prime n}%
}^{\rho,\delta}\otimes\Pi_{B^{n}}^{\rho,\delta}\right) \\
&  =\sum_{t}p(t)\left(  \Pi_{B^{\prime n}}^{\rho,\delta}\mathcal{N}%
_{A^{n}\rightarrow B^{\prime n}}(\pi_{A^{n}}^{t})\Pi_{B^{\prime n}}%
^{\rho,\delta}\otimes\Pi_{B^{n}}^{\rho,\delta}\pi_{B^{n}}^{t}\Pi_{B^{n}}%
^{\rho,\delta}\right) \\
&  =\sum_{t}p(t)\left(  \Pi_{B^{\prime n}}^{\rho,\delta}\mathcal{N}%
_{A^{n}\rightarrow B^{\prime n}}(\pi_{A^{n}}^{t})\Pi_{B^{\prime n}}%
^{\rho,\delta}\otimes\Pi_{B^{n}}^{\rho,\delta}\frac{\Pi_{B^{n}}^{t}%
}{\operatorname{Tr}\left\{  \Pi_{B^{n}}^{t}\right\}  }\Pi_{B^{n}}^{\rho
,\delta}\right) \\
&  \leq\sum_{t}p(t)\left(  \Pi_{B^{\prime n}}^{\rho,\delta}\mathcal{N}%
_{A^{n}\rightarrow B^{\prime n}}(\pi_{A^{n}}^{t})\Pi_{B^{\prime n}}%
^{\rho,\delta}\otimes2^{-n\left[  H(B)_{\rho}-\eta(n,\delta)\right]  }%
\Pi_{B^{n}}^{\rho,\delta}\right)  .
\end{align}
The first equality follows from \eqref{eq-eac:expected-state-result}. The
second equality follows by a simple manipulation. The third equality follows
because the maximally mixed state $\pi_{B^{n}}^{t}$ is equal to the normalized
type class projection operator $\Pi_{B^{n}}^{t}$. The inequality follows from
Property~\ref{prop-qt:min-dim-typical-type} and $\Pi_{B^{n}}^{\rho,\delta}%
\Pi_{B^{n}}^{t}\Pi_{B^{n}}^{\rho,\delta}\leq\Pi_{B^{n}}^{\rho,\delta}$ (the
support of a typical type projector is always in the support of the typical
projector and the intersection of the support of an atypical type with the
typical projector is null). Continuing, by linearity, the last line above is
equal to%
\begin{align}
&  \Pi_{B^{\prime n}}^{\rho,\delta}\mathcal{N}_{A^{n}\rightarrow B^{\prime n}%
}\left(  \sum_{t}p(t)\pi_{A^{n}}^{t}\right)  \Pi_{B^{\prime n}}^{\rho,\delta
}\otimes2^{-n\left[  H(B)_{\rho}-\eta(n,\delta)\right]  }\Pi_{B^{n}}%
^{\rho,\delta}\nonumber\\
&  =\Pi_{B^{\prime n}}^{\rho,\delta}\mathcal{N}_{A^{n}\rightarrow B^{\prime
n}}(\varphi_{A^{n}})\Pi_{B^{\prime n}}^{\rho,\delta}\otimes2^{-n\left[
H(B)_{\rho}-\eta(n,\delta)\right]  }\Pi_{B^{n}}^{\rho,\delta}\\
&  \leq2^{-n\left[  H(B^{\prime})_{\rho}-c\delta\right]  }\Pi_{B^{\prime n}%
}^{\rho,\delta}\otimes2^{-n\left[  H(B)_{\rho}-\eta(n,\delta)\right]  }%
\Pi_{B^{n}}^{\rho,\delta}\\
&  =2^{-n\left[  H(B^{\prime})_{\rho}+H(B)_{\rho}-\eta(n,\delta)-c\delta
\right]  }\Pi_{B^{\prime n}}^{\rho,\delta}\otimes\Pi_{B^{n}}^{\rho,\delta}.
\end{align}
The first equality follows because $\varphi_{A^{n}}=\sum_{t}p(t)\pi_{A^{n}%
}^{t}$. The inequality follows from the equipartition property of typical
projectors (Property~\ref{prop-qt:equi}). The final equality follows by
rearranging terms.

With the four conditions in \eqref{eq-eac:pack-2}--\eqref{eq-eac:pack-4}
holding, it follows from Corollary~\ref{cor-pack:derandomized} (the
derandomized version of the packing lemma) that there exists a deterministic
code and a POVM\ $\left\{  \Lambda_{B^{\prime n}B^{n}}^{m}\right\}  $ that can
detect the transmitted states with arbitrarily low maximal probability of
error as long as the size $\left\vert \mathcal{M}\right\vert $\ of the message
set is small enough:%
\begin{align}
p_{e}^{\ast}  &  \equiv\max_{m}\operatorname{Tr}\left\{  \left(
I-\Lambda_{B^{\prime n}B^{n}}^{m}\right)  U_{B^{n}}^{T}(s(m))\rho_{B^{\prime
n}B^{n}}U_{B^{n}}^{\ast}(s(m))\right\} \\
&  \leq4\left(  2\varepsilon+2\sqrt{2\varepsilon}\right)  +16\cdot2^{-n\left[
H(B^{\prime})_{\rho}+H(B)_{\rho}-\eta(n,\delta)-c\delta\right]  }2^{n\left[
H(B^{\prime}B)_{\rho}+c\delta\right]  }\left\vert \mathcal{M}\right\vert \\
&  =4\left(  2\varepsilon+2\sqrt{2\varepsilon}\right)  +16\cdot2^{-n\left[
I(B^{\prime};B)_{\rho}-\eta(n,\delta)-2c\delta\right]  }\left\vert
\mathcal{M}\right\vert .
\end{align}
We can choose the size of the message set to be $\left\vert \mathcal{M}%
\right\vert =2^{n\left[  I(B^{\prime};B)-\eta(n,\delta)-3c\delta\right]  }$ so
that the rate of communication is%
\begin{equation}
\frac{1}{n}\log\left\vert \mathcal{M}\right\vert =I(B^{\prime};B)-\eta
(n,\delta)-3c\delta,
\end{equation}
and the bound on the maximal probability of error becomes%
\begin{equation}
p_{e}^{\ast}\leq4\left(  2\varepsilon+2\sqrt{2\varepsilon}\right)
+16\cdot2^{-nc\delta}.
\end{equation}
Let $\varepsilon^{\prime}\in(0,1)$ and $\delta^{\prime}>0$. It is then clear
that by picking $n$ large enough and $\delta$ small enough, we can have both
$4\left(  2\varepsilon+2\sqrt{2\varepsilon}\right)  +16\cdot2^{-nc\delta}%
\leq\varepsilon^{\prime}$ and $\eta(n,\delta)+3c\delta\leq\delta^{\prime}$.
Thus, the quantum mutual information $I(B^{\prime};B)_{\rho}$, with respect to
the state $\rho_{B^{\prime}B}\equiv\mathcal{N}_{A\rightarrow B^{\prime}%
}(\varphi_{AB})$ is an achievable rate for the entanglement-assisted
transmission of classical information over the channel~$\mathcal{N}$. To obtain the
precise statement in Theorem~\ref{thm-eac:BSST}, we can simply rewrite the
quantum mutual information as $I(A;B)_{\rho}$ with respect to the state
$\rho_{AB}\equiv\mathcal{N}_{A^{\prime}\rightarrow B}\left(  \varphi
_{AA^{\prime}}\right)  $. Alice and Bob can achieve the maximum rate of
communication simply by determining a state $\varphi_{AA^{\prime}}$ that
maximizes the quantum mutual information $I(A;B)_{\rho}$ and by generating
entanglement-assisted classical codes from the state $\rho_{AB}$.
\end{proof}

\section{The Converse Theorem}

\label{sec-eac:converse}This section contains a proof of the converse part of%
\index{entanglement-assisted capacity theorem!converse}
the entanglement-assisted classical capacity theorem. Let us begin by
supposing that Alice and Bob are trying to use the entanglement-assisted
channel many times to accomplish the task of randomness distribution (recall
that we took this approach for the converse of the classical capacity theorem
in Section~\ref{sec-cc:converse}).
An upper bound on the rate at which Alice can distribute randomness to Bob
also serves as an upper bound on the rate at which they can communicate
because a noiseless classical channel can distribute randomness. In such a
task, Alice and Bob share entanglement in some state $\Psi_{T_{A}T_{B}}$.
Alice first prepares the maximally correlated state $\overline{\Phi
}_{MM^{\prime}}$, as defined in \eqref{eq-cc:common-randomness}, and the rate
of randomness in this state is $\frac{1}{n}\log\left\vert \mathcal{M}%
\right\vert $. Alice then applies some encoding channel $\mathcal{E}%
_{M^{\prime}T_{A}\rightarrow A^{n}}$ to the classical system $M^{\prime}$ and
her share $T_{A}$ of $\Psi_{T_{A}T_{B}}$. The resulting state is%
\begin{equation}
\theta_{MA^{n}T_{B}}\equiv\mathcal{E}_{M^{\prime}T_{A}\rightarrow A^{n}%
}(\overline{\Phi}_{MM^{\prime}}\otimes\Psi_{T_{A}T_{B}}).
\end{equation}
She sends her $A^{n}$ systems through many uses $\mathcal{N}_{A^{n}\rightarrow
B^{n}}$\ of the channel $\mathcal{N}_{A\rightarrow B}$, and Bob receives the
systems $B^{n}$, producing the state%
\begin{equation}
\omega_{MT_{B}B^{n}}\equiv\mathcal{N}_{A^{n}\rightarrow B^{n}}(\mathcal{E}%
_{M^{\prime}T_{A}\rightarrow A^{n}}(\overline{\Phi}_{MM^{\prime}}\otimes
\Psi_{T_{A}T_{B}})).
\end{equation}
Finally, Bob performs some decoding channel $\mathcal{D}_{B^{n}T_{B}%
\rightarrow\hat{M}}$ on the above state to give%
\begin{equation}
\omega_{M\hat{M}}^{\prime}\equiv\mathcal{D}_{B^{n}T_{B}\rightarrow\hat{M}%
}(\omega_{MT_{B}B^{n}}).
\end{equation}
An $(n,\left[  \log\left\vert \mathcal{M}\right\vert \right]  /n,\varepsilon)$
protocol for randomness distribution is such that the actual state
$\omega_{M\hat{M}}^{\prime}$ resulting from the protocol is $\varepsilon
$-close in trace distance to the ideal shared randomness state:%
\begin{equation}
\frac{1}{2}\left\Vert \omega_{M\hat{M}}^{\prime}-\overline{\Phi}_{M\hat{M}%
}\right\Vert _{1}\leq\varepsilon. \label{eq-eac:converse-error-crit}%
\end{equation}

We now show that the quantum mutual information of the channel serves as an
upper bound on the rate of any reliable protocol for entanglement-assisted
randomness distribution (a protocol meeting the error criterion in
\eqref{eq-eac:converse-error-crit}). Consider the following:%
\begin{align}
\log\left\vert \mathcal{M}\right\vert  &  =I(M;\hat{M})_{\overline{\Phi}}\\
&  \leq I(M;\hat{M})_{\omega^{\prime}}+f(\left\vert \mathcal{M}\right\vert
,\varepsilon)\\
&  \leq I(M;B^{n}T_{B})_{\omega}+f(\left\vert \mathcal{M}\right\vert
,\varepsilon)\\
&  =I(T_{B}M;B^{n})_{\omega}+I(M;T_{B})_{\omega}-I(B^{n};T_{B})_{\omega
}+f(\left\vert \mathcal{M}\right\vert ,\varepsilon)\\
&  =I(T_{B}M;B^{n})_{\omega}-I(B^{n};T_{B})_{\omega}+f(\left\vert
\mathcal{M}\right\vert ,\varepsilon)\\
&  \leq I(T_{B}M;B^{n})_{\omega}+f(\left\vert \mathcal{M}\right\vert
,\varepsilon)\\
&  \leq I(\mathcal{N}^{\otimes n})+f(\left\vert \mathcal{M}\right\vert
,\varepsilon)\\
&  =nI(\mathcal{N})+f(\left\vert \mathcal{M}\right\vert ,\varepsilon).
\end{align}
The first equality follows by evaluating the quantum mutual information of the
shared randomness state $\overline{\Phi}_{M\hat{M}}$. The first inequality
follows for the exact same reasons as \eqref{eq-cc:first-ineq-converse} does
(the first inequality of the converse for the HSW\ theorem), with
$f(\left\vert \mathcal{M}\right\vert ,\varepsilon)\equiv\varepsilon
\log\left\vert \mathcal{M}\right\vert +\left(  1+\varepsilon\right)
h_{2}(\varepsilon/\left[  1+\varepsilon\right]  )$. The second inequality
follows from quantum data processing (Theorem~\ref{cor-qie:QDP})---Bob
processes the state $\omega_{MT_{B}B^{n}}$ with the decoder $\mathcal{D}%
_{B^{n}T_{B}\rightarrow\hat{M}}$ to get the state $\omega_{M\hat{M}}^{\prime}%
$. The second equality follows from the chain rule for quantum mutual
information (see Exercise~\ref{ex-qie:chain-rule-mut-info}). The third
equality follows because the systems $M$ and $T_{B}$ are in a product state,
so that $I(M;T_{B})_{\omega}=0$. The third inequality follows because
$I(B^{n};T_{B})_{\omega}\geq0$. Observe that the state $\omega_{MT_{B}B^{n}%
}=\mathcal{N}_{A^{n}\rightarrow B^{n}}(\theta_{MA^{n}T_{B}})$, so that the
systems $M$ and $T_{B}$ extend the $A^{n}$ system that is input to
$\mathcal{N}^{\otimes n}$. Thus, the mutual information between $MT_{B}$ and
$B^{n}$ can never exceed the maximum mutual information of the channel, where
we need to apply the result of Exercise~\ref{ex-add:mut-info-pure-suffice}.
Finally, the mutual information of a quantum channel is additive
(Theorem~\ref{thm-add:mut-info-additive}), and
Corollary~\ref{cor-add:mutual-info-single-letter} implies that $I(\mathcal{N}%
^{\otimes n})=nI(\mathcal{N})$. We can then rewrite the above bound as
follows:%
\begin{equation}
\frac{1}{n}\log\left\vert \mathcal{M}\right\vert (1-\varepsilon)\leq
I(\mathcal{N})+\frac{1+\varepsilon}{n}h_{2}(\varepsilon/\left[  1+\varepsilon
\right]  ).
\end{equation}

Thus, if we are considering a sequence of $(n,\left[  \log\left\vert
\mathcal{M}\right\vert \right]  /n,\varepsilon_{n})$\ entanglement-assisted
classical communication protocols with rate $C-\delta_{n}=\frac{1}{n}%
\log\left\vert \mathcal{M}\right\vert $, such that $\lim_{n\rightarrow\infty
}\varepsilon_{n}=\lim_{n\rightarrow\infty}\delta_{n}=0$, then the above bound
becomes%
\begin{equation}
\left(  C-\delta_{n}\right)  \left(  1-\varepsilon_{n}\right)  \leq
I(\mathcal{N})+\frac{1+\varepsilon_{n}}{n}h_{2}(\varepsilon_{n}/\left[
1+\varepsilon_{n}\right]  ).
\end{equation}
Taking the limit as $n\rightarrow\infty$ then establishes that an achievable
rate $C$ necessarily satisfies $C\leq I(\mathcal{N})$. This demonstrates a
single-letter upper bound on the entanglement-assisted classical capacity of a
quantum channel and completes the proof of Theorem~\ref{thm-eac:BSST}.

\subsection{Feedback Does Not Increase Capacity}

The entanglement-assisted classical capacity
\index{entanglement-assisted!classical communication!with feedback}%
formula is the closest formal analogy to Shannon's capacity formula for a
classical channel. The mutual information~$I(\mathcal{N})$ of a quantum
channel~$\mathcal{N}$ is the optimum of the quantum mutual information over
all bipartite input states:%
\begin{equation}
I(\mathcal{N})=\max_{\phi_{AA^{\prime}}}I(A;B), \label{eq-eac:EAC-cap}%
\end{equation}
and it is equal to the channel's entanglement-assisted classical capacity by
Theorem~\ref{thm-eac:direct-coding}. The mutual information~$I(p_{Y|X})$ of a
classical channel$~p_{Y|X}$ is the optimum of the classical mutual information
over all correlated inputs to the channel:%
\begin{equation}
I(p_{Y|X})=\max_{XX^{\prime}}I(X;Y), \label{eq-eac:class-cap}%
\end{equation}
where $XX^{\prime}$ are correlated random variables with the distribution
$p_{X,X^{\prime}}(x,x^{\prime})=p_{X}(x)\delta_{x,x^{\prime}}$. The formula is
equal to the classical capacity of a classical channel by Shannon's noisy
coding theorem. Both formulas not only appear similar in form, but they also
have the important property of being \textquotedblleft
single-letter,\textquotedblright\ meaning that the above formulas are equal to
the capacity (this was not generally the case for the Holevo information from the
previous chapter).

We now consider another way in which the entanglement-assisted classical
capacity is a good candidate for being the fully quantum generalization of
Shannon's formula to the quantum world. Though it might be surprising, it is
well known that free access to a classical feedback channel from receiver to
sender does not increase the capacity of a classical channel. We state this
result as the following theorem.

\begin{theorem}
[Feedback Does Not Increase Classical Capacity]\label{thm-eac:feed-class}The
feedback capacity of a classical channel$~p_{Y|X}(y|x)$\ is equal to the
mutual information of that channel:%
\begin{equation}
\sup\left\{  C:C\text{ is achievable for }p_{Y|X}\text{ with feedback
}\right\}  =I(p_{Y|X}),
\end{equation}
where$~I(p_{Y|X})$ is defined in \eqref{eq-eac:class-cap}.
\end{theorem}

\begin{proof}
We first define an $\left(  n,C-\delta,\varepsilon\right)  $\ classical
feedback code as one in which every symbol $x_{i}(m,Y^{i-1})$ of a
codeword~$x^{n}(m)$\ is a function of the message $m\in\mathcal{M}$ and all of
the previously received values $Y_{1}$, \ldots, $Y_{i-1}$ from the receiver.
The decoder consists of the decoding function $g:\mathcal{Y}^{n}%
\rightarrow\left\{  1,2,\ldots,\left\vert \mathcal{M}\right\vert \right\}  $
such that%
\begin{equation}
\Pr\left\{  M^{\prime}\neq M\right\}  \leq\varepsilon,
\label{eq-eac:cl-feed-good}%
\end{equation}
where $M^{\prime}\equiv g(Y^{n})$. The lower bound LHS$\ \geq\ $RHS follows
because we can always avoid the use of the feedback channel and achieve the
mutual information of the classical channel by employing Shannon's noisy
coding theorem. The upper bound LHS $\leq\ $RHS is less obvious, but it
follows from the memoryless structure of the channel and the structure of a
feedback code. Consider the following chain of inequalities:%
\begin{align}
\log\left\vert \mathcal{M}\right\vert  &  =H(M)=I(M;M^{\prime})+H(M|M^{\prime
})\label{eq-eac:class-feed-con-1}\\
&  \leq I(M;M^{\prime})+1+\varepsilon\log\left\vert \mathcal{M}\right\vert \\
&  \leq I(M;Y^{n})+1+\varepsilon\log\left\vert \mathcal{M}\right\vert .
\label{eq-eac:class-feed-con-4}%
\end{align}
The first equality follows because we assume that the message $M$ is uniformly
distributed. The first inequality follows from Fano's inequality (see
Theorem~\ref{thm-cie:fano}) and the assumption in \eqref{eq-eac:cl-feed-good}
that the protocol is good up to error~$\varepsilon$. The last inequality
follows from classical data processing. Continuing, we can bound $I(M;Y^{n})$
from above:%
\begin{align}
I(M;Y^{n})  &  =H(Y^{n})-H(Y^{n}|M)=H(Y^{n})-\sum_{k=1}^{n}H(Y_{k}|Y^{k-1}M)\\
&  =H(Y^{n})-\sum_{k=1}^{n}H(Y_{k}|Y^{k-1}MX_{k})=H(Y^{n})-\sum_{k=1}%
^{n}H(Y_{k}|X_{k})\\
&  \leq\sum_{k=1}^{n}H(Y_{k})-H(Y_{k}|X_{k})=\sum_{k=1}^{n}I(X_{k};Y_{k})\\
&  \leq n\max_{XX^{\prime}}I(X;Y).
\end{align}
The first equality follows from the definition of mutual information. The
second equality follows from the chain rule for entropy (see
Exercise~\ref{ex-ie:ent-chain-rule}). The third equality follows because
$X_{k}$ is a function of $Y^{k-1}$ and $M$. The fourth equality follows
because $Y_{k}$ is conditionally independent of $Y^{k-1}$ and $M$ through
$X_{k}$ ($Y^{k-1}M\rightarrow X_{k}\rightarrow Y_{k}$ forms a Markov chain).
The first inequality follows from subadditivity of entropy. The fifth equality
follows by definition, and the final inequality follows because the individual
mutual informations in the sum can never exceed the maximum over all inputs.
Putting everything together, our final bound on the feedback-assisted capacity
of a classical channel is%
\begin{equation}
C-\delta\leq I(p_{Y|X})+\frac{1}{n}+\frac{\varepsilon}{n}\log\left\vert
\mathcal{M}\right\vert ,
\end{equation}
which becomes $C\leq I(p_{Y|X})$ as $n\rightarrow\infty$ and $\varepsilon
,\delta\rightarrow0$.
\end{proof}

Given the above result, we might wonder if a similar result could hold for the
entanglement-assisted classical capacity. Such a result would more firmly
place the entanglement-assisted classical capacity as a good generalization of
Shannon's coding theorem. Indeed, the following theorem states that this
result holds.

\begin{theorem}
[Quantum Feedback Does Not Increase EAC\ Capacity]The classical capacity of a
quantum channel assisted by a quantum feedback channel is equal to that
channel's entanglement-assisted classical capacity:%
\begin{equation}
\sup\left\{  C\ |\ C\text{ is achievable for }\mathcal{N}\text{\ with quantum
feedback}\right\}  =I(\mathcal{N}),
\end{equation}
where $I(\mathcal{N})$ is defined in \eqref{eq-eac:EAC-cap}.
\end{theorem}

\begin{proof}
We define free access to a quantum feedback channel to mean that there is a
noiseless quantum channel of arbitrarily large dimension connecting the
receiver Bob to the sender Alice. The bound LHS $\geq$ RHS follows because Bob
can use the quantum feedback channel to establish an arbitrarily large amount
of entanglement with Alice. They then just execute the protocol from
Section~\ref{sec-eac:direct-coding}\ to achieve the entanglement-assisted
classical capacity.

The bound LHS $\leq$ RHS is much less obvious, and it requires a proof that is
different from the proof of Theorem~\ref{thm-eac:feed-class}. We first need to
determine the most general protocol for classical communication with the
assistance of a quantum feedback channel.
Figure~\ref{fig-eac:feedback-protocol}\ depicts such a protocol. %
\begin{figure}
[ptb]
\begin{center}
\includegraphics[
width=4.9156in
]%
{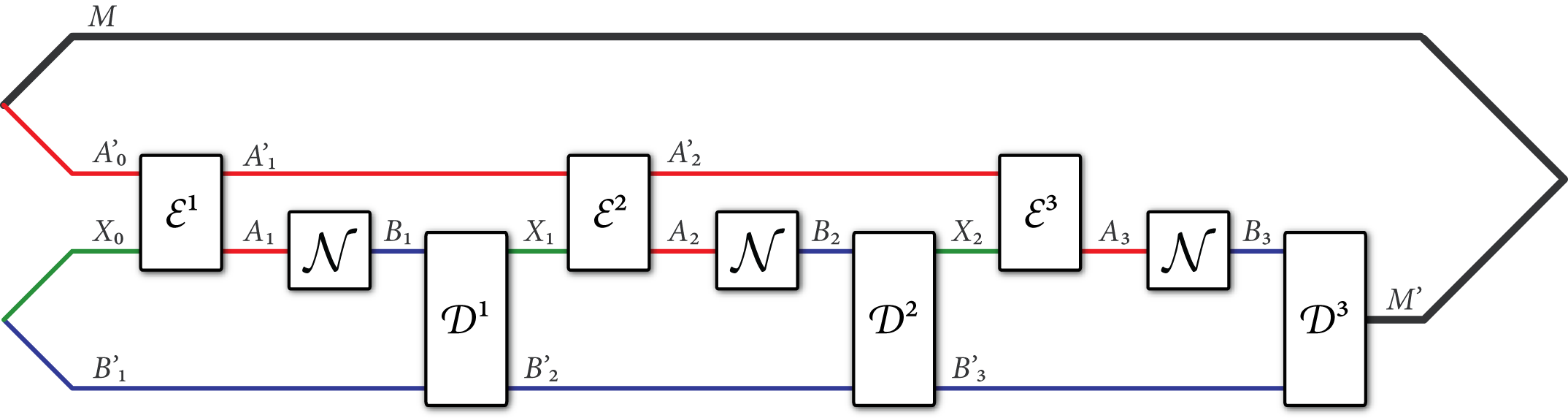}%
\caption{Three rounds of the most general protocol for classical communication
with a quantum feedback channel.}%
\label{fig-eac:feedback-protocol}%
\end{center}
\end{figure}
The protocol begins with Alice preparing a classical register $M$ with a
uniformly random message to be sent, which is correlated with some system
$A_{0}^{\prime}$. Bob uses the quantum feedback channel to send a quantum
system $X_{0}$ to Alice, which is correlated with some quantum system
$B_{1}^{\prime}$. Alice performs an encoding $\mathcal{E}_{A_{0}^{\prime}%
X_{0}\rightarrow A_{1}^{\prime}A_{1}}^{1}$. Alice sends system $A_{1}$ through
the first use of the channel $\mathcal{N}$. Bob now applies the decoding
channel $\mathcal{D}_{B_{1}B_{1}^{\prime}\rightarrow X_{1}B_{2}^{\prime}}^{1}%
$. The next encoder of Alice occurs, and the procedure repeats. The last
decoding channel of Bob outputs a classical system $M^{\prime}$ which contains
Bob's estimate of the message that Alice transmitted. The state of registers
$MB_{n}B_{n}^{\prime}$ after the $n$th channel $\mathcal{N}_{A_{n}\rightarrow
B_{n}}$\ has been applied has the following form:%
\begin{equation}
\omega_{MB_{n}B_{n}^{\prime}}^{\left(  n\right)  }\equiv\mathcal{N}%
_{A_{n}\rightarrow B_{n}}(\rho_{MB_{n}^{\prime}A_{n}}^{\left(  n\right)  }),
\label{eq-eac:feedback-state}%
\end{equation}
where $\rho_{MB_{n}^{\prime}A_{n}}^{\left(  n\right)  }$ is the state of
registers $MB_{n}^{\prime}A_{n}$ after the $n$th encoding channel has been
applied. Let $\psi_{R^{\left(  n\right)  }MB_{n}^{\prime}A_{n}}^{\left(
n\right)  }$ be a purification of $\rho_{MB_{n}^{\prime}A_{n}}^{\left(
n\right)  }$, and let%
\begin{equation}
\omega_{R^{\left(  n\right)  }MB_{n}B_{n}^{\prime}}^{\left(  n\right)  }%
\equiv\mathcal{N}_{A_{n}\rightarrow B_{n}}(\psi_{R^{\left(  n\right)  }%
MB_{n}^{\prime}A_{n}}^{\left(  n\right)  }).
\end{equation}

This protocol is the most general for classical communication with quantum
feedback. We can now proceed with proving the upper bound LHS $\leq$ RHS. To
do so, we assume that the random variable $M$ modeling Alice's message
selection is a uniform random variable, and Bob obtains a random variable
$M^{\prime}$ by measuring all of his systems $B_{n}$ and $B_{n}^{\prime}$ at
the end of the protocol. For any $\varepsilon$-good protocol for classical communication,
the bound $\Pr\left\{  M^{\prime}\neq M\right\}  \leq\varepsilon$ applies.
Consider the following chain of inequalities (these steps are essentially the
same as those
in~\eqref{eq-eac:class-feed-con-1}--\eqref{eq-eac:class-feed-con-4}):%
\begin{align}
\log\left\vert \mathcal{M}\right\vert  &  =H(M)=I(M;M^{\prime})+H(M|M^{\prime
})\\
&  \leq I(M;M^{\prime})+1+\varepsilon\log\left\vert \mathcal{M}\right\vert \\
&  \leq I(M;B_{n}B_{n}^{\prime})_{\omega^{\left(  n\right)  }}+1+\varepsilon
\log\left\vert \mathcal{M}\right\vert ,
\end{align}
where the last mutual information is with respect to the state in
\eqref{eq-eac:feedback-state}. This chain of inequalities follows for the same
reason as those in
\eqref{eq-eac:class-feed-con-1}--\eqref{eq-eac:class-feed-con-4}, with the
last step following from quantum data processing. Continuing, we have%
\begin{align}
I(M;B_{n}B_{n}^{\prime})_{\omega^{\left(  n\right)  }}  &  =I(M;B_{n}%
|B_{n}^{\prime})_{\omega^{\left(  n\right)  }}+I(M;B_{n}^{\prime}%
)_{\omega^{\left(  n\right)  }}\label{eq-eac:1st-feedback-step}\\
&  \leq I(MB_{n}^{\prime};B_{n})_{\omega^{\left(  n\right)  }}+I(M;B_{n}%
^{\prime})_{\omega^{\left(  n\right)  }}\\
&  \leq I(R^{\left(  n\right)  }MB_{n}^{\prime};B_{n})_{\omega^{\left(
n\right)  }}+I(M;B_{n}^{\prime})_{\omega^{\left(  n\right)  }}.
\end{align}
The first equality is the chain rule for mutual information. The first
inequality follows because $I(M;B_{n}|B_{n}^{\prime})=I(MB_{n}^{\prime}%
;B_{n})-I(B_{n}^{\prime};B_{n})\leq I(MB_{n}^{\prime};B_{n})$. The second
inequality follows from quantum data processing. Now, given that the mutual
information $I(R^{\left(  n\right)  }MB_{n}^{\prime};B_{n})$\ is with respect
to the state in \eqref{eq-eac:feedback-state} and this state has the following
form%
\begin{equation}
\mathcal{N}_{A_{n}\rightarrow B_{n}}(\phi_{RA_{n}}),
\end{equation}
where $\phi_{RA_{n}}$ is some pure state and $R$ is some system not going into
the channel (here identified with $R^{\left(  n\right)  }MB_{n}^{\prime}$), we
can optimize over all such inputs to find that%
\begin{equation}
I(R^{\left(  n\right)  }MB_{n}^{\prime};B_{n})_{\omega^{\left(  n\right)  }%
}\leq I(\mathcal{N}),
\end{equation}
where $I(\mathcal{N})$ is the quantum mutual information of the channel. So
this means that%
\begin{align}
I(M;B_{n}B_{n}^{\prime})_{\omega^{\left(  n\right)  }}  &  \leq I(\mathcal{N}%
)+I(M;B_{n}^{\prime})_{\omega^{\left(  n\right)  }}\\
&  \leq I(\mathcal{N})+I(M;B_{n-1}B_{n-1}^{\prime})_{\omega^{\left(
n-1\right)  }}. \label{eq-eac:last-feedback-step}%
\end{align}
where the last inequality follows from quantum data processing (the system
$B_{n}^{\prime}$ results from applying the $n-1$ decoder to the systems
$B_{n-1}B_{n-1}^{\prime}$). At this point, we realize that the above chain of
steps \eqref{eq-eac:1st-feedback-step}--\eqref{eq-eac:last-feedback-step} can
be applied to $I(M;B_{n-1}B_{n-1}^{\prime})_{\omega^{\left(  n-1\right)  }}$,
so we iterate this sequence until we go all the way back to the beginning of
the protocol. Putting everything together, we get the following upper bound on
any achievable rate~$C$ for classical communication with quantum feedback:%
\begin{equation}
C-\delta\leq I(\mathcal{N})+\frac{1}{n}+\frac{\varepsilon}{n}\log\left\vert
\mathcal{M}\right\vert ,
\end{equation}
which becomes $C\leq I(\mathcal{N})$ as $n\rightarrow\infty$ and
$\varepsilon,\delta\rightarrow0$.
\end{proof}

\begin{corollary}
\label{cor-eac:feedback}The capacity of a quantum channel with unlimited
entanglement and classical feedback is equal to the entanglement-assisted
classical capacity of $\mathcal{N}$.
\end{corollary}

\begin{proof}
This result follows because $I(\mathcal{N})$ is a lower bound on this capacity
(simply by avoiding use of the classical feedback channel). Also,
$I(\mathcal{N})$ is an upper bound on this capacity because the entanglement
and classical feedback channel can simulate an arbitrarily large quantum
feedback channel via teleportation, and the above theorem gives an upper bound
of $I(\mathcal{N})$ for this setting.
\end{proof}

\section{Examples of Channels}

This section shows how to compute the entanglement-assisted classical capacity
of both the quantum erasure channel and the amplitude damping channel, while
leaving the capacity of the quantum depolarizing channel and the dephasing
channel as exercises. For three of these channels (erasure, depolarizing, and
dephasing), a super-dense-coding-like strategy suffices to achieve capacity.
This strategy involves Alice locally rotating an ebit shared with Bob, sending
one share of it through the noisy channel, and Bob performing measurements in
the Bell basis to determine what Alice sent. This process induces a classical
channel from Alice to Bob, for which its capacity is equal to the
entanglement-assisted capacity of the original quantum channel (in the case of
depolarizing, dephasing, and erasure channels). For the amplitude damping
channel, this super-dense-coding-like strategy does not achieve capacity---in
general, it is necessary for Bob to perform a large, collective measurement on
all of the channel outputs in order for him to determine Alice's message.

\begin{figure}
[ptb]
\begin{center}
\includegraphics[
width=3.5405in
]%
{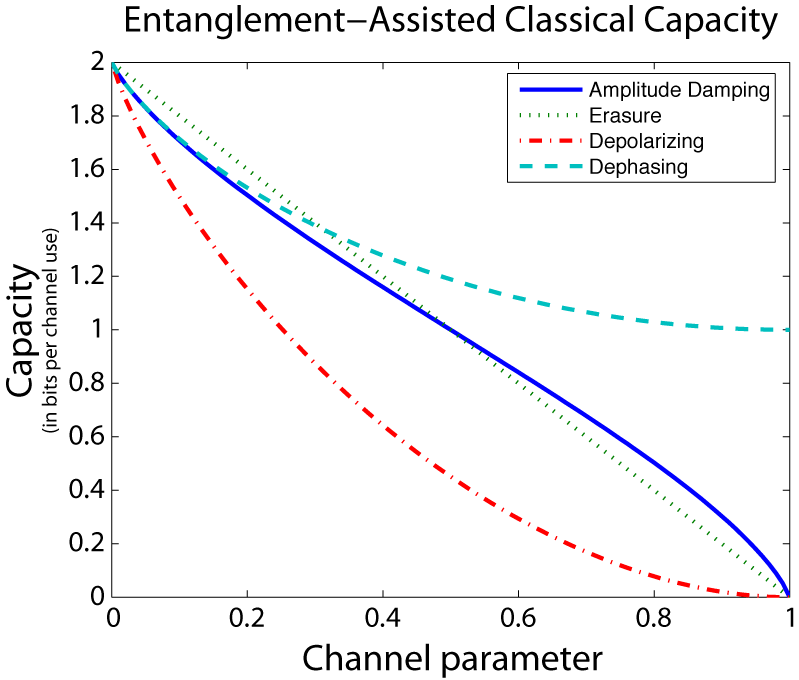}%
\caption{The entanglement-assisted classical capacity of the amplitude damping
channel, the erasure channel, the depolarizing channel, and the dephasing
channel as a function of each channel's noise parameter.}%
\label{fig-eac:EAC-capacities}%
\end{center}
\end{figure}

Figure~\ref{fig-eac:EAC-capacities}\ plots the entanglement-assisted
capacities of these four channels as a function of their noise parameters. As
expected, the depolarizing channel has the worst performance because it is a
\textquotedblleft worst-case scenario\textquotedblright\ channel---it either
sends the state through or replaces it with a completely random state. The
erasure channel's capacity is just a line of constant slope down to
zero---this is because the receiver can easily determine the fraction of the
time that he receives something from the channel. The dephasing channel
eventually becomes a completely classical channel, for which entanglement
cannot increase capacity beyond one bit per channel use. Finally, perhaps the
most interesting curve is for the amplitude damping channel. This channel's
capacity is convex when its noise parameter is less than $1/2$ and concave
when it is greater than $1/2$.%

\subsection{The Quantum Erasure Channel}

Recall that the quantum erasure channel%
\index{erasure channel}
acts as follows on an input density operator $\rho_{A^{\prime}}$:%
\begin{equation}
\rho_{A^{\prime}}\rightarrow\left(  1-\varepsilon\right)  \rho_{B}%
+\varepsilon|e\rangle\langle e|_{B},
\end{equation}
where $\varepsilon\in\left[  0,1\right]  $ is the erasure probability and
$|e\rangle_{B}$ is an erasure state that is orthogonal to the support of the
input state $\rho$.

\begin{proposition}
\label{prop-eac:eac-erasure}The entanglement-assisted classical capacity of a
quantum erasure channel with erasure probability $\varepsilon$\ is equal to
$2\left(  1-\varepsilon\right)  \log d_{A}$, where $d_{A}$ is the dimension of
the input system.
\end{proposition}

\begin{proof}
To determine the entanglement-assisted classical capacity of this channel, we
need to compute its mutual information. So, consider that sending one share of
a bipartite state $\phi_{AA^{\prime}}$ through the channel produces the output%
\begin{equation}
\sigma_{AB}\equiv\left(  1-\varepsilon\right)  \phi_{AB}+\varepsilon\phi
_{A}\otimes|e\rangle\langle e|_{B}.
\end{equation}
We could now attempt to calculate and optimize the quantum mutual information
$I(A;B)_{\sigma}$. However, observe that Bob can apply the following isometry
$U_{B\rightarrow BX}$ to his state:%
\begin{equation}
U_{B\rightarrow BX}\equiv\Pi_{B}\otimes|0\rangle_{X}+|e\rangle\langle
e|_{B}\otimes|1\rangle_{X},
\end{equation}
where $\Pi_{B}$ is a projector onto the support of the input state (for
qubits, it would be just $|0\rangle\langle0|+|1\rangle\langle1|$). Applying
this isometry leads to a state $\sigma_{ABX}$ where%
\begin{align}
\sigma_{ABX}  &  \equiv U_{B\rightarrow BX}\sigma_{AB}U_{B\rightarrow
BX}^{\dag}\\
&  =\left(  1-\varepsilon\right)  \phi_{AB}\otimes|0\rangle\langle
0|_{X}+\varepsilon\phi_{A}\otimes|e\rangle\langle e|_{B}\otimes|1\rangle
\langle1|_{X}.
\end{align}
The quantum mutual information $I(A;BX)_{\sigma}$ is equal to $I(A;B)_{\sigma
}$ because entropies do not change under the isometry $U_{B\rightarrow BX}$.
We now calculate $I(A;BX)_{\sigma}$:%
\begin{align}
I(A;BX)_{\sigma}  &  =H(A)_{\sigma}+H(BX)_{\sigma}-H(ABX)_{\sigma}\\
&  =H(A)_{\phi}+H(B|X)_{\sigma}-H(AB|X)_{\sigma}\\
&  =H(A)_{\phi}+\left(  1-\varepsilon\right)  \left[  H(B)_{\phi}-H(AB)_{\phi
}\right] \nonumber\\
&  \ \ \ \ \ \ \ +\varepsilon\left[  H(B)_{|e\rangle\langle e|}-H(AB)_{\phi_{A}%
\otimes|e\rangle\langle e|}\right] \\
&  =H(A)_{\phi}+\left(  1-\varepsilon\right)  H(B)_{\phi}-\varepsilon\left[
H(A)_{\phi}+H(B)_{|e\rangle}\right] \\
&  =\left(  1-\varepsilon\right)  \left[  H(A)_{\phi}+H(B)_{\phi}\right] \\
&  =2\left(  1-\varepsilon\right)  H(A)_{\phi}\\
&  \leq2\left(  1-\varepsilon\right)  \log d_{A}.
\end{align}
The first equality follows by the definition of quantum mutual information.
The second equality follows from $\phi_{A}=\operatorname{Tr}_{BX}\left\{
\sigma_{ABX}\right\}  $, from the chain rule of entropy, and by canceling
$H(X)$ on both sides. The third equality follows because the $X$ register is a
classical register, indicating whether the erasure occurs. The fourth equality
follows because $H(AB)_{\phi}=0$, $H(B)_{|e\rangle\langle e|}=0$, and $H(AB)_{\phi
_{A}\otimes|e\rangle\langle e|}=H(A)_{\phi}+H(B)_{|e\rangle\langle e|}$. The fifth
equality follows again because $H(B)_{|e\rangle\langle e|}=0$ and by collecting terms.
The final equality follows because $H(A)_{\phi}=H(B)_{\phi}$ ($\phi_{AB}$ is a
pure bipartite state). The final inequality follows because the entropy of a
state on system $A$ is never greater than the logarithm of the dimension of
$A$. We can conclude that the maximally entangled state $\Phi_{AA^{\prime}}$
achieves the entanglement-assisted classical capacity of the quantum erasure
channel because $H(A)_{\Phi}=\log d_{A}$.
\end{proof}

The strategy for achieving the entanglement-assisted classical capacity of the
quantum erasure channel is straightforward. Alice and Bob simply employ a
super-dense coding strategy on all of the channel uses (this means that Bob
performs measurements on each channel output with his share of the
entanglement---there is no need for a large, collective measurement on all of
the channel outputs). For a good fraction $1-\varepsilon$ of the time, this
strategy works and Alice can communicate $2\log d_{A}$ bits to Bob. For the
other fraction $\varepsilon$, all is lost to the environment. In order for
this to work, Alice and Bob need to make use of a feedback channel from Bob to
Alice so that Bob can report which messages come through and which do not, but
Corollary~\ref{cor-eac:feedback} states that this feedback cannot improve the
capacity. Thus, the rate of communication they can achieve is equal to the
capacity $2\left(  1-\varepsilon\right)  \log d_{A}$.

\subsection{The Amplitude Damping Channel}

We now compute the entanglement-assisted classical capacity of the amplitude
damping channel%
\index{amplitude damping channel}
$\mathcal{N}_{\operatorname{AD}}$. Recall that this channel acts as follows on
an input qubit in state $\rho$:%
\begin{equation}
\mathcal{N}_{\operatorname{AD}}(\rho)=A_{0}\rho A_{0}^{\dag}+A_{1}\rho
A_{1}^{\dag},
\end{equation}
where, for $\gamma\in\left[  0,1\right]  $,%
\begin{equation}
A_{0}\equiv|0\rangle\langle0|+\sqrt{1-\gamma}|1\rangle\langle
1|,\ \ \ \ \ \ A_{1}\equiv\sqrt{\gamma}|0\rangle\langle1|.
\end{equation}

\begin{proposition}
\label{thm-eac:eac-amp-damp}The entanglement-assisted classical capacity of an
amplitude damping channel with damping parameter $\gamma\in\left[  0,1\right]
$ is equal to%
\begin{equation}
I(\mathcal{N}_{\operatorname{AD}})=\max_{p\in\left[  0,1\right]  }%
h_{2}(p)+h_{2}(\left(  1-\gamma\right)  p)-h_{2}(\gamma p),
\end{equation}
where $h_{2}(x)\equiv-x\log x-(1-x)\log(1-x)$ is the binary entropy function.
\end{proposition}

\begin{proof}
Suppose that a matrix representation of the input qubit density operator
$\rho$ in the computational basis is%
\begin{equation}
\rho=%
\begin{bmatrix}
1-p & \eta^{\ast}\\
\eta & p
\end{bmatrix}
. \label{eq-eac:input-dens-amp}%
\end{equation}
One can readily verify that the density operator for Bob has the following
matrix representation:%
\begin{equation}
\mathcal{N}_{\operatorname{AD}}(\rho)=%
\begin{bmatrix}
1-\left(  1-\gamma\right)  p & \sqrt{1-\gamma}\eta^{\ast}\\
\sqrt{1-\gamma}\eta & \left(  1-\gamma\right)  p
\end{bmatrix}
, \label{eq-eac:amp-to-bob}%
\end{equation}
and by calculating the elements $\operatorname{Tr}\{A_{i}\rho A_{j}^{\dag
}\}|i\rangle\langle j|$, we can obtain a matrix representation for Eve's
density operator:%
\begin{equation}
\mathcal{N}_{\operatorname{AD}}^{c}(\rho)=%
\begin{bmatrix}
1-\gamma p & \sqrt{\gamma}\eta^{\ast}\\
\sqrt{\gamma}\eta & \gamma p
\end{bmatrix}
, \label{eq-eac:amp-to-eve}%
\end{equation}
where $\mathcal{N}_{\operatorname{AD}}^{c}$ is the complementary channel to
Eve. By comparing \eqref{eq-eac:amp-to-bob} and \eqref{eq-eac:amp-to-eve}, we
can see that the channel to Eve is an amplitude damping channel with damping
parameter $1-\gamma$. The entanglement-assisted classical capacity of
$\mathcal{N}_{\operatorname{AD}}$ is equal to its mutual information:%
\begin{equation}
I(\mathcal{N}_{\operatorname{AD}})=\max_{\phi_{AA^{\prime}}}I(A;B)_{\sigma},
\label{eq-eac:eac-general-formula-amp-channel}%
\end{equation}
where $\phi_{AA^{\prime}}$ is some pure bipartite input state and $\sigma
_{AB}=\mathcal{N}_{\operatorname{AD}}(\phi_{AA^{\prime}})$. We need to
determine the input density operator that maximizes the above formula as a
function of $\gamma$. As it stands now, the optimization depends on three
parameters:\ $p$, $\operatorname{Re}\left\{  \eta\right\}  $, and
$\operatorname{Im}\left\{  \eta\right\}  $. We can show that it is sufficient
to consider an optimization over only $p$ with $\eta=0$. The formula in
\eqref{eq-eac:eac-general-formula-amp-channel} also has the following form:%
\begin{equation}
I(\mathcal{N}_{\operatorname{AD}})=\max_{\rho}\left[  H(\rho)+H(\mathcal{N}%
_{\operatorname{AD}}(\rho))-H(\mathcal{N}_{\operatorname{AD}}^{c}%
(\rho))\right]  , \label{eq-eac:ea-formula-amp}%
\end{equation}
because%
\begin{align}
I(A;B)_{\sigma}  &  =H(A)_{\phi}+H(B)_{\sigma}-H(AB)_{\sigma}\\
&  =H(A^{\prime})_{\phi}+H(\mathcal{N}_{\operatorname{AD}}(\rho))-H(E)_{\sigma
}\\
&  =H(\rho)+H(\mathcal{N}_{\operatorname{AD}}(\rho))-H(\mathcal{N}%
_{\operatorname{AD}}^{c}(\rho))\\
&  \equiv I_{\operatorname{mut}}(\rho,\mathcal{N}_{\operatorname{AD}}).
\end{align}
The three entropies in\ \eqref{eq-eac:ea-formula-amp} depend only on the
eigenvalues of the three density operators in
\eqref{eq-eac:input-dens-amp}--\eqref{eq-eac:amp-to-eve}, respectively, which
are as follows:%
\begin{align}
&  \frac{1}{2}\left(  1\pm\sqrt{(1-2p)^{2}+4\left\vert \eta\right\vert ^{2}%
}\right)  ,\label{eq-eac:eig-alice-amp}\\
&  \frac{1}{2}\left(  1\pm\sqrt{\left(  1-2\left(  1-\gamma\right)  p\right)
^{2}+4\left\vert \eta\right\vert ^{2}\left(  1-\gamma\right)  }\right)
,\label{eq-eac:eig-bob-amp}\\
&  \frac{1}{2}\left(  1\pm\sqrt{\left(  1-2\gamma p\right)  ^{2}+4\left\vert
\eta\right\vert ^{2}\gamma}\right)  . \label{eq-eac:eig-eve-amp}%
\end{align}
The above eigenvalues are in the order of Alice, Bob, and Eve. All of the
above eigenvalues have a similar form, and their dependence on $\eta$ is only
through its magnitude. Thus, it suffices to consider $\eta\in\mathbb{R}$ (this
eliminates one parameter). Next, the eigenvalues do not change if we flip the
sign of $\eta$ (this is equivalent to rotating the original state $\rho$ by
$Z$, to $Z\rho Z$), and thus, the mutual information does not change as well:%
\begin{equation}
I_{\operatorname{mut}}(\rho,\mathcal{N}_{\operatorname{AD}}%
)=I_{\operatorname{mut}}(Z\rho Z,\mathcal{N}_{\operatorname{AD}}).
\end{equation}
By the above relation and concavity of quantum mutual information in the input
density operator (Theorem$~$\ref{thm-ie:mutual-concave-input}), the following
inequality holds:%
\begin{align}
I_{\operatorname{mut}}(\rho,\mathcal{N}_{\operatorname{AD}})  &  =\frac{1}%
{2}\left[  I_{\operatorname{mut}}(\rho,\mathcal{N}_{\operatorname{AD}%
})+I_{\operatorname{mut}}(Z\rho Z,\mathcal{N}_{\operatorname{AD}})\right] \\
&  \leq I_{\operatorname{mut}}(\left(  \rho+Z\rho Z\right)  /2,\mathcal{N}%
_{\operatorname{AD}})\\
&  =I_{\operatorname{mut}}(\overline{\Delta}(\rho),\mathcal{N}%
_{\operatorname{AD}}),
\end{align}
where $\overline{\Delta}$ is a completely dephasing channel in the
computational basis. This demonstrates that it is sufficient to consider
diagonal density operators $\rho$ when optimizing the quantum mutual
information. Thus, the eigenvalues in
\eqref{eq-eac:eig-alice-amp}--\eqref{eq-eac:eig-eve-amp} respectively become%
\begin{align}
&  \left\{  p,1-p\right\}  ,\\
&  \left\{  \left(  1-\gamma\right)  p,1-\left(  1-\gamma\right)  p\right\}
,\\
&  \left\{  \gamma p,1-\gamma p\right\}  ,
\end{align}
giving our final expression in the statement of the proposition.
\end{proof}

\begin{exercise}
Consider the qubit
\index{depolarizing channel}%
depolarizing channel:\ $\rho\rightarrow(1-p)\rho+p\pi$. Prove that its
entanglement-assisted classical capacity is equal to%
\begin{equation}
2+\left(  1-3p/4\right)  \log(1-3p/4)+\left(  3p/4\right)  \log(p/4).
\end{equation}

\end{exercise}

\begin{exercise}
Consider the
\index{dephasing channel}%
dephasing channel:\ $\rho\rightarrow\left(  1-p/2\right)  \rho+(p/2)Z\rho Z$.
Prove that its entanglement-assisted classical capacity is equal to
$2-h_{2}(p/2)$, where $p$ is the dephasing parameter.
\end{exercise}

\section{Concluding Remarks}

Shared entanglement has the desirable property of simplifying quantum Shannon
theory. The entanglement-assisted capacity theorem is one of the strongest
known results in quantum Shannon theory because it states that the quantum
mutual information of a channel is equal to its entanglement-assisted
capacity. This function of the channel is concave in the input state and the
set of input states is convex, implying that finding a local maximum is
equivalent to finding a global one. The converse theorem and additivity of the
channel mutual information demonstrates that there is no need to take the
regularization of the formula. Furthermore, quantum feedback does not improve
this capacity, just as it does not for the classical case of Shannon's
setting. In these senses, the entanglement-assisted classical capacity is the
most natural generalization of Shannon's capacity formula to the quantum setting.

The direct coding part of the capacity theorem exploits a strategy similar to
super-dense coding---effectively the technique is to perform super-dense
coding in the type class subspaces of many copies of a shared entangled state.
This strategy is \textit{equivalent} to super-dense coding if the initial
shared state is a maximally entangled state. The particular protocol that we
outlined in this chapter has the appealing feature that we can easily make it
coherent, similar to the way that coherent dense coding is a coherent version
of the super-dense coding protocol. We take this approach in the next chapter
and show that we can produce a whole host of other protocols using this
technique, eventually leading to a proof of the direct coding part of the
quantum capacity theorem.

This chapter features the calculation of the entanglement-assisted classical
capacity of certain channels of practical interest:\ the depolarizing channel,
the dephasing channel, the amplitude damping channel, and the erasure channel.
Each one of these channels has a single parameter that governs its noisiness,
and the capacity in each case is a straightforward function of this parameter.
One could carry out a similar type of analysis to determine the
entanglement-assisted capacity of any channel, although it generally will be
necessary to employ techniques from convex optimization.

Unfortunately, quantum Shannon theory only gets more complicated from here
onward.\footnote{We could also view this \textquotedblleft
unfortunate\textquotedblright\ situation as being fortunate for conducting
open-ended research in quantum Shannon theory.} For the other capacity
theorems that we will study, such as the private classical capacity or the
quantum capacity, the best expressions that we have for them are good only up
to regularization of the formulas. In certain cases, these formulas completely
characterize the capabilities of the channel for these particular operational
tasks, but these formulas are not particularly useful in the general case. One
important goal for future research in quantum Shannon theory would be to
improve upon these formulas, in the hopes that we could further our
understanding of the best strategy for achieving the information-processing
tasks corresponding to these other capacity questions.

\section{History and Further Reading}

\cite{PhysRevA.56.3470} figured that the mutual information of a quantum
channel would play an important role in quantum Shannon theory, and they
proved several of its most important properties.
\cite{PhysRevLett.83.3081,ieee2002bennett} later demonstrated that the quantum
mutual information of a channel has an operational interpretation as its
entanglement-assisted classical capacity. Our proof of the direct part of the
entanglement-assisted classical capacity theorem is the same as that in
\cite{itit2008hsieh}. We exploit this approach because it leads to all of the
results in the next chapter, implying that this protocol is sufficient to
generate all of the known protocols in quantum Shannon theory (with the
exception of private classical communication). \cite{PhysRevA.71.032314}
determined several capacities of the amplitude damping channel, and
\cite{PhysRevA.75.012303} made some further observations regarding it.
\cite{B04} proved that the classical capacity of a channel assisted by
unbounded quantum feedback is equal to its entanglement-assisted classical capacity.

There has also been work on the strong converse and second-order
characterization for entanglement-assisted capacity (see
Section~\ref{sec-cc:history}\ for a discussion of these terms). \cite{BDHSW09}
proved the
\index{quantum reverse Shannon theorem}%
quantum reverse Shannon theorem, which quantifies the rate of classical
communication needed to simulate a quantum channel in the presence of
unlimited entanglement shared between sender and receiver. \cite{BCR09}
provided an alternate proof of the quantum reverse Shannon theorem. By a
simulation argument, the
\index{quantum reverse Shannon theorem}%
quantum reverse Shannon theorem implies a strong converse for
entanglement-assisted capacity \citep{BDHSW09}. \cite{GW15} gave a direct
proof for the strong converse by making use of R\'{e}nyi entropies.
\cite{CMW14} later showed that the same strong converse bound still holds in
the presence of a quantum feedback channel, strengthening the result of
\cite{B04}. \cite{DTW14} established a second-order achievability result for
entanglement-assisted classical communication and proved that this
characterization is tight for some channels, by making use of prior results of
\cite{MW12}.

\chapter{Coherent Communication with Noisy Resources}

\label{chap:coh-comm-noisy}This chapter demonstrates the power of both
coherent communication%
\index{coherent communication}
from Chapter~\ref{chap:coherent-communication}\ and the particular protocol
for entanglement-assisted classical coding from the previous chapter. Recall
that
\index{coherent dense coding}%
coherent dense coding is a version of the dense coding protocol in which the
sender and receiver perform all of its steps coherently.\footnote{Performing a
protocol coherently means that we replace conditional unitaries with
controlled unitaries and measurements with controlled gates (e.g., compare
Figures~\ref{fig:dense-coding} and~\ref{fig:coherent-dense}).} Since our
protocol for entanglement-assisted classical coding from the previous chapter
is really just a glorified dense coding protocol, the sender and receiver can
perform each of its steps coherently, generating a protocol for
entanglement-assisted coherent coding. Then, by exploiting the fact that two
coherent bits are equivalent to a qubit and an ebit, we obtain a protocol for
entanglement-assisted quantum coding that consumes far less entanglement than
a naive strategy would in order to accomplish this task. We next combine this
entanglement-assisted quantum coding protocol with entanglement distribution
(Section~\ref{sec-3np:ent-dist}) and obtain a protocol for quantum
communication at a rate equal to the channel's coherent information
(Section~\ref{sec-add:coh-info}). This sequence of steps demonstrates an
alternate proof of the direct part of the quantum capacity theorem stated in
Chapter~\ref{chap:quantum-capacity}.

Entanglement-assisted classical communication is one generalization of
super-dense coding, in which the noiseless qubit channel becomes an arbitrary
noisy quantum channel while the noiseless ebits remain noiseless. Another
generalization of super-dense coding is a protocol named
\index{noisy super-dense coding}
\textit{noisy super-dense coding}, in which the shared entanglement becomes a
shared noisy state $\rho_{AB}$ and the noiseless qubit channels remain
noiseless. Interestingly, the protocol that we employ in this chapter for
noisy super-dense coding is essentially equivalent to the protocol from the
previous chapter for entanglement-assisted classical communication, with some
slight modifications to account for the different setting. We can also
construct a coherent version of noisy super-dense coding, leading to a
protocol that we name \textit{coherent state transfer}. Coherent state
transfer accomplishes not only the task of generating coherent communication
between Alice and Bob, but it also allows Alice to transfer her share of the
state $\rho_{AB}$ to Bob. By combining coherent state transfer with both the
coherent communication identity and teleportation, we obtain protocols for
quantum-assisted state transfer and classical-assisted state transfer,
respectively. The latter protocol gives an operational interpretation to the
conditional quantum entropy $H(A|B)_{\rho}$---if it is positive, then the
protocol consumes entanglement at the rate $H(A|B)_{\rho}$, and if it is
negative, the protocol generates entanglement at the rate $|H(A|B)_{\rho}|$.

The final part of this chapter shows that our particular protocol for
entanglement-assisted classical communication is even more powerful than
suggested in the first paragraph. It allows for a sender to communicate both
coherent bits and incoherent classical bits to a receiver, and they can trade
off these two resources against one another. The structure of the
entanglement-assisted protocol allows for this possibility, by taking
advantage of Remark~\ref{rem-eac:tensor-power-eac}\ and by combining it with
the HSW\ classical communication protocol from
Chapter~\ref{chap:classical-comm-HSW}. Then, by exploiting the coherent
communication identity, we obtain a protocol for entanglement-assisted
communication of classical and quantum information.
Chapter~\ref{chap:trade-off} demonstrates that this protocol, teleportation,
super-dense coding, and entanglement distribution are sufficient to accomplish
any task in dynamic quantum Shannon theory involving the three unit resources
of classical bits, qubits, and ebits. These four protocols give a
three-dimensional achievable rate region that is the best known
characterization for any information-processing task that a sender and
receiver would like to accomplish using a quantum channel and the three unit
resources. Chapter~\ref{chap:trade-off}\ discusses this triple trade-off
scenario in full detail.

\section{Entanglement-Assisted Quantum Communication}

The entanglement-assisted classical capacity theorem states that the quantum
mutual information of a channel is equal to its capacity for transmitting
classical information with the help of shared entanglement, and the direct
coding theorem from Section~\ref{sec-eac:direct-coding}\ provides a protocol
that achieves the capacity. We were not much concerned with the rate at which
this protocol consumes entanglement, but a direct calculation reveals that it
consumes $H(A)_{\varphi}$ ebits per channel use, where $|\varphi\rangle_{AB}$
is the bipartite state that they share before the protocol begins. This result
follows because they can concentrate $n$ copies of the state $|\varphi
\rangle_{AB}$ to $\approx nH(A)_{\varphi}$ ebits, as we learned in
Chapter~\ref{chap:ent-conc}. Also, they can \textquotedblleft
dilute\textquotedblright\ $nH(A)_{\varphi}$ ebits to $\approx n$ copies of
$|\varphi\rangle_{AB}$ with the help of a sublinear amount of classical
communication that does not factor into the resource count, as also discussed
in Chapter~\ref{chap:ent-conc}.

Suppose now that Alice is interested in exploiting the channel and shared
entanglement in order to transmit quantum information to Bob. There is a
simple (and as we will see, naive) way that we can convert the protocol in
Section~\ref{sec-eac:direct-coding}\ to one that transmits quantum
information: they can just combine it with teleportation. This naive strategy
requires consuming ebits at an additional rate of $\frac{1}{2}I(A;B)_{\rho}$
in order to have enough entanglement to combine with teleportation, where
$\rho_{AB}\equiv\mathcal{N}_{A^{\prime}\rightarrow B}(\varphi_{AA^{\prime}})$.
To see this, consider the following resource inequalities:%
\begin{align}
\left\langle \mathcal{N}\right\rangle +\left(  H(A)_{\rho}+\frac{1}%
{2}I(A;B)_{\rho}\right)  \left[  qq\right]   &  \geq I(A;B)_{\rho}\left[
c\rightarrow c\right]  +\frac{1}{2}I(A;B)_{\rho}\left[  qq\right]
\label{eq-eac:bad-eaq-1}\\
&  \geq\frac{1}{2}I(A;B)_{\rho}\left[  q\rightarrow q\right]  .
\label{eq-eac:bad-eaq-2}%
\end{align}
The first inequality follows by having them exploit the channel and the
$nH(A)_{\rho}$ ebits to generate classical communication at a rate
$I(A;B)_{\rho}$ (while doing nothing with the extra $n\frac{1}{2}I(A;B)_{\rho
}$ ebits). Alice then exploits the ebits and the classical communication in a
teleportation protocol to send $n\frac{1}{2}I(A;B)_{\rho}$ qubits to Bob. This
rate of quantum communication is provably optimal---were it not so, it would
be possible to combine the protocol in
\eqref{eq-eac:bad-eaq-1}--\eqref{eq-eac:bad-eaq-2} with super-dense coding and
beat the optimal rate for classical communication given by the
entanglement-assisted classical capacity theorem.

Although the above protocol achieves the entanglement-assisted quantum
capacity, we are left thinking that the entanglement consumption rate of $H(
A) _{\rho}+\frac{1}{2}I( A;B) _{\rho}$ ebits per channel use might be a bit
more than necessary because teleportation and super-dense coding are not dual
under resource reversal. That is, if we combine the protocol with super-dense
coding and teleportation \textit{ad infinitum}, then it consumes an infinite
amount of entanglement. In practice, this \textquotedblleft back and
forth\textquotedblright\ with teleportation and super-dense coding would be a
poor way to consume the precious resource of entanglement.

How might we make more judicious use of shared entanglement?\ Recall that
coherent communication from Chapter~\ref{chap:coherent-communication}\ was
helpful for doing so, at least in the noiseless case. A sender and receiver
can combine coherent teleportation and coherent dense coding \textit{ad
infinitum} without any net loss in entanglement, essentially because these two
protocols are dual under resource reversal. The following theorem shows how we
can upgrade the protocol in Section~\ref{sec-eac:direct-coding} to one that
generates coherent communication instead of just classical communication. The
resulting protocol is one way to have a version of coherent dense coding in
which one noiseless resource is replaced by a noisy one.

\begin{theorem}
[Entanglement-Assisted Coherent Communication]\label{thm-eac:ea-coh}The
following resource inequality corresponds to an achievable protocol for
entanglement-assisted coherent communication%
\index{entanglement-assisted!coherent communication}
over a quantum channel $\mathcal{N}_{A^{\prime}\rightarrow B}$:%
\begin{equation}
\left\langle \mathcal{N}\right\rangle +H(A)_{\rho}\left[  qq\right]  \geq
I(A;B)_{\rho}\left[  q\rightarrow qq\right]  ,
\end{equation}
where $\rho_{AB}\equiv\mathcal{N}_{A^{\prime}\rightarrow B}(\varphi
_{AA^{\prime}})$.
\end{theorem}

\begin{proof}
Suppose that Alice and Bob share many copies of some pure, bipartite entangled
state $|\varphi\rangle_{AB}$. Consider the code from the direct coding theorem
in Section~\ref{sec-eac:direct-coding}. We can say that it is a set of
$D^{2}\approx2^{nI(A;B)_{\rho}}$ unitaries $U(s(m))$, from which Alice can
select, and she applies a particular unitary $U(s(m))$ to her share $A^{n}%
$\ of the entanglement in order to encode message $m$. Also, Bob has a
detection POVM\ $\left\{  \Lambda_{B^{\prime n}B^{n}}^{m}\right\}  $ acting on
his share of the entanglement and the channel outputs that he can exploit to
detect message $m$. Just as we were able to construct a coherent super-dense
coding protocol in Chapter~\ref{chap:coherent-communication}\ by performing
all the steps in dense coding coherently, we can do so for the
entanglement-assisted classical coding protocol in
Section~\ref{sec-eac:direct-coding}. We track the steps in such a protocol.
Suppose Alice shares a state with a reference system $R$ to which she does not
have access:%
\begin{equation}
|\psi\rangle_{RA_{1}}\equiv\sum_{l,m=1}^{D^{2}}\alpha_{l,m}\left\vert
l\right\rangle _{R}|m\rangle_{A_{1}},
\end{equation}
where $\left\{  \left\vert l\right\rangle \right\}  $ and $\left\{  \left\vert
m\right\rangle \right\}  $ are some orthonormal bases for $R$ and $A_{1}$,
respectively. We say that Alice and Bob have implemented a coherent channel if
they execute the map $|m\rangle_{A_{1}}\rightarrow\left\vert m\right\rangle
_{A_{1}}|m\rangle_{B_{1}}$, which transforms the above state to%
\begin{equation}
\sum_{l,m=1}^{D^{2}}\alpha_{l,m}\left\vert l\right\rangle _{R}\left\vert
m\right\rangle _{A_{1}}|m\rangle_{B_{1}}.
\end{equation}
We say that they have implemented a coherent channel \textit{approximately} if
the state resulting from the protocol is $\varepsilon$-close in trace distance
to the above state. If we can show that $\varepsilon\in(0,1)$ approaches zero
in the limit of many channel uses, then the simulation of an approximate
coherent channel becomes perfect in the asymptotic limit. Alice's first step
is to append her shares of the entangled state $|\varphi\rangle_{A^{n}B^{n}}$
to $|\psi\rangle_{RA_{1}}$ and apply the following controlled unitary from her
system $A_{1}$ to her system $A^{n}$:%
\begin{equation}
\sum_{m}|m\rangle\langle m|_{A_{1}}\otimes U_{A^{n}}(s(m)).
\label{eq-eac:coherent-controlled-unitary}%
\end{equation}
The resulting global state is as follows:%
\begin{equation}
\sum_{l,m}\alpha_{l,m}\left\vert l\right\rangle _{R}|m\rangle_{A_{1}}U_{A^{n}%
}(s(m))|\varphi\rangle_{A^{n}B^{n}}.
\end{equation}
By the structure of the unitaries $U(s(m))$ (see
\eqref{eq-eac:unitary-encoding} and \eqref{eq-eac:transpose-trick}), the above
state is equivalent to the following one:%
\begin{equation}
\sum_{l,m}\alpha_{l,m}\left\vert l\right\rangle _{R}|m\rangle_{A_{1}}U_{B^{n}%
}^{T}(s(m))|\varphi\rangle_{A^{n}B^{n}}.
\end{equation}
Interestingly, observe that Alice applying the controlled gate in
\eqref{eq-eac:coherent-controlled-unitary} is the same as her applying the
non-local controlled gate $\sum_{m}|m\rangle\langle m|_{A_{1}}\otimes
U_{B^{n}}^{T}(s(m))$, due to the non-local (and perhaps spooky!) properties of
the entangled state $|\varphi\rangle_{A^{n}B^{n}}$. Alice then sends her
systems $A^{n}$ through many uses of the quantum channel $\mathcal{N}%
_{A\rightarrow B^{\prime}}$, whose isometric extension is $U_{A\rightarrow
B^{\prime}E}^{\mathcal{N}}$. Let $|\varphi\rangle_{B^{\prime n}E^{n}B^{n}}$
denote the state resulting from $n$ instances of the isometric extension
$U_{A\rightarrow B^{\prime}E}^{\mathcal{N}}$ of the channel acting on the
state $|\varphi\rangle_{A^{n}B^{n}}$:%
\begin{equation}
|\varphi\rangle_{B^{\prime n}E^{n}B^{n}}\equiv U_{A^{n}\rightarrow B^{\prime
n}E^{n}}^{\mathcal{N}}|\varphi\rangle_{A^{n}B^{n}},
\end{equation}
where $U_{A^{n}\rightarrow B^{\prime n}E^{n}}^{\mathcal{N}}\equiv
(U_{A\rightarrow B^{\prime}E}^{\mathcal{N}})^{\otimes n}$. After Alice
transmits through the isometric extension, the state becomes%
\begin{equation}
\sum_{l,m}\alpha_{l,m}\left\vert l\right\rangle _{R}|m\rangle_{A_{1}}U_{B^{n}%
}^{T}(s(m))|\varphi\rangle_{B^{\prime n}E^{n}B^{n}},
\end{equation}
where Bob now holds his shares $B^{n}$ of the entanglement and the channel
outputs $B^{\prime n}$. (Observe that the action of the controlled unitary in
\eqref{eq-eac:coherent-controlled-unitary} commutes with the action of the
channel.) Rather than perform an incoherent measurement with the
POVM$\ \left\{  \Lambda_{B^{\prime n}B^{n}}^{m}\right\}  $, Bob applies a
coherent gentle measurement (see Section~\ref{sec-pt:coherent-measurement}),
an isometry of the following form:%
\begin{equation}
\sum_{m}\sqrt{\Lambda_{B^{\prime n}B^{n}}^{m}}\otimes|m\rangle_{B_{1}}.
\end{equation}
Using the result of Exercise~\ref{ex-pt:coherent-measurement}, we can readily
check that the resulting state is $2\sqrt{2\varepsilon}$-close in trace
distance to the following state:%
\begin{equation}
\sum_{l,m}\alpha_{l,m}\left\vert l\right\rangle _{R}|m\rangle_{A_{1}}U_{B^{n}%
}^{T}(s(m))|\varphi\rangle_{B^{\prime n}E^{n}B^{n}}|m\rangle_{B_{1}}.
\end{equation}
Thus, for the rest of the protocol, we pretend that they are acting on the
above state. Alice and Bob would like to coherently remove the coupling of
their index $m$ to the environment, so Bob performs the following controlled
unitary:%
\begin{equation}
\sum_{m}|m\rangle\langle m|_{B_{1}}\otimes U_{B^{n}}^{\ast}(s(m)),
\end{equation}
and the final state is%
\begin{multline}
\sum_{l,m=1}^{D^{2}}\alpha_{l,m}\left\vert l\right\rangle _{R}\left\vert
m\right\rangle _{A_{1}}|\varphi\rangle_{B^{\prime n}E^{n}B^{n}}|m\rangle
_{B_{1}}\\
=\left(  \sum_{l,m=1}^{D^{2}}\alpha_{l,m}\left\vert l\right\rangle
_{R}|m\rangle_{A_{1}}|m\rangle_{B_{1}}\right)  \otimes|\varphi\rangle
_{B^{\prime n}E^{n}B^{n}}.
\end{multline}
Thus, this protocol implements a $D^{2}$-dimensional coherent channel up to an
arbitrarily small error, which implies that the resource inequality in the
statement of the theorem holds. Figure~\ref{fig-eac:eac-coherent}\ depicts the
entanglement-assisted coherent coding protocol.
\end{proof}

\begin{figure}
[ptb]
\begin{center}
\includegraphics[
width=4.8456in
]%
{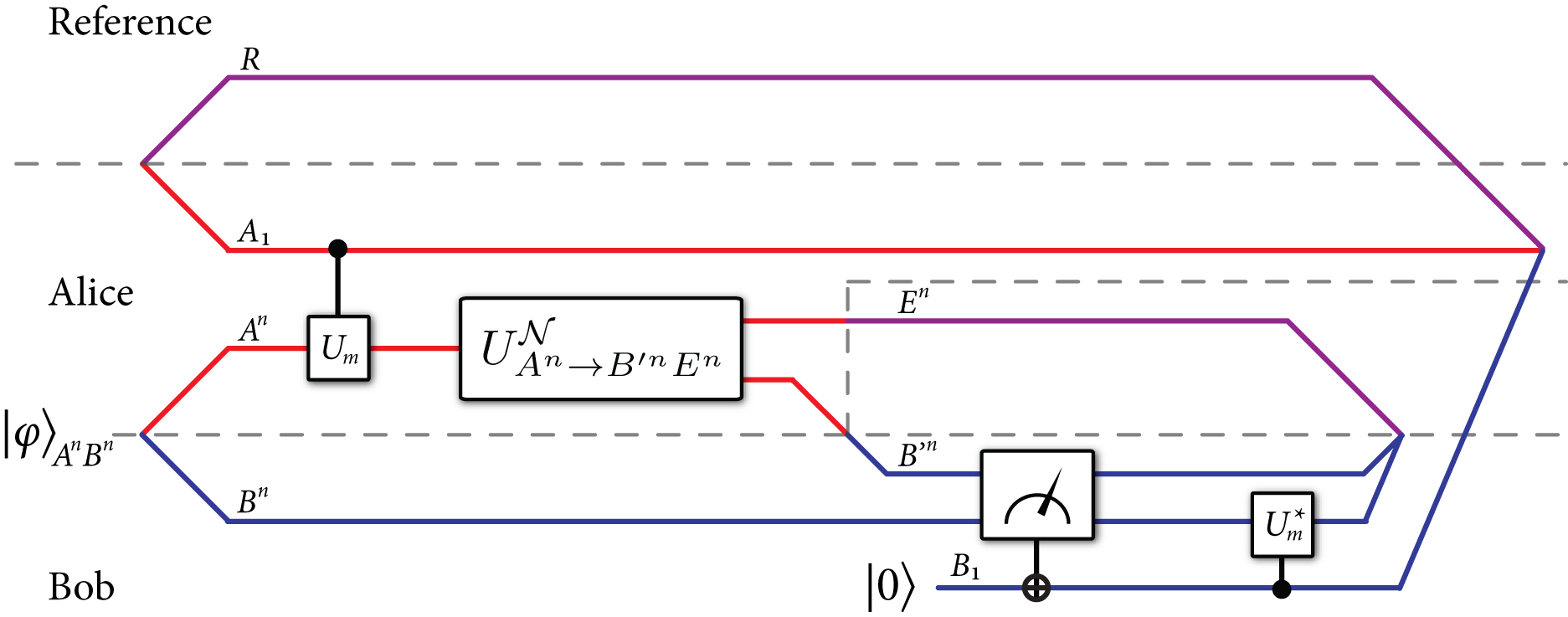}%
\caption{The protocol for entanglement-assisted coherent communication.
Observe that it is the coherent version of the protocol for
entanglement-assisted classical communication, just as coherent dense coding
is the coherent version of super-dense coding (compare this figure and
Figure~\ref{fig-eac:eac-protocol}\ with Figures~\ref{fig:dense-coding}
and~\ref{fig:coherent-dense}). Instead of applying conditional unitaries,
Alice applies a controlled unitary from her system $A_{1}$ to her share of the
entanglement and sends the encoded state through many uses of the noisy
channel. Rather than performing a POVM, Bob performs a coherent gentle
measurement from his systems $B^{\prime n}$ and $B^{n}$ to an ancilla $B_{1}$.
Finally, he applies a similar controlled unitary in order to decouple the
environment from the state of his ancilla $B_{1}$.}%
\label{fig-eac:eac-coherent}%
\end{center}
\end{figure}

It is now a straightforward task to convert the protocol from
Theorem~\ref{thm-eac:ea-coh}\ into one for entanglement-assisted quantum
communication, by exploiting the coherent communication identity from
Section~\ref{sec-cc:coh-comm-identity}.

\begin{corollary}
[Entanglement-Assisted Quantum Communication]\label{thm-eac:eaq-rate}The
following resource inequality corresponds to an achievable protocol for
entanglement-assisted quantum
\index{entanglement-assisted!quantum communication}
communication over a quantum channel $\mathcal{N}_{A^{\prime}\rightarrow B}$:%
\begin{equation}
\left\langle \mathcal{N}\right\rangle +\frac{1}{2}I(A;E)_{\varphi}\left[
qq\right]  \geq\frac{1}{2}I(A;B)_{\varphi}\left[  q\rightarrow q\right]  ,
\label{eq-ccn:RI-father}%
\end{equation}
where $|\varphi\rangle_{ABE}\equiv U_{A^{\prime}\rightarrow BE}^{\mathcal{N}%
}|\varphi\rangle_{AA^{\prime}}$ and $U_{A^{\prime}\rightarrow BE}%
^{\mathcal{N}}$ is an isometric extension of the channel $\mathcal{N}%
_{A^{\prime}\rightarrow B}$.
\end{corollary}

\begin{proof}
Consider the coherent communication identity%
\index{coherent communication identity}
from Section~\ref{sec-cc:coh-comm-identity}. This identity states that a
$D^{2}$-dimensional coherent channel can perfectly simulate a $D$-dimensional
quantum channel and a maximally entangled state $\left\vert \Phi\right\rangle
_{AB}$ with Schmidt rank $D$. In terms of cobits, qubits, and ebits, the
coherent communication identity is the following resource equality for
$D$-dimensional systems:%
\begin{equation}
2\log D\left[  q\rightarrow qq\right]  =\log D\left[  q\rightarrow q\right]
+\log D\left[  qq\right]  .
\end{equation}
Consider the following chain of resource inequalities:%
\begin{align}
\left\langle \mathcal{N}\right\rangle +H(A)_{\varphi}\left[  qq\right]   &
\geq I(A;B)_{\varphi}\left[  q\rightarrow qq\right] \\
&  \geq\frac{1}{2}I(A;B)_{\varphi}\left[  q\rightarrow q\right]  +\frac{1}%
{2}I(A;B)_{\varphi}\left[  qq\right]  .
\end{align}
The first resource inequality is the statement of Theorem~\ref{thm-eac:ea-coh}%
, and the second resource inequality follows from an application of coherent
teleportation. If we then allow for catalytic protocols, in which we allow for
some use of a resource with the demand that it be returned at the end of the
protocol, we have a protocol for entanglement-assisted quantum communication:%
\begin{equation}
\left\langle \mathcal{N}\right\rangle +\frac{1}{2}I(A;E)_{\varphi}\left[
qq\right]  \geq\frac{1}{2}I(A;B)_{\varphi}\left[  q\rightarrow q\right]  ,
\end{equation}
because $H(A)_{\varphi}-\frac{1}{2}I(A;B)_{\varphi}=\frac{1}{2}I(A;E)_{\varphi
}$ (see Exercise~\ref{ex-qie:entropy-games}).
\end{proof}

When comparing the entanglement consumption rate of the naive protocol in
\eqref{eq-eac:bad-eaq-1}--\eqref{eq-eac:bad-eaq-2} with that of the protocol
in Corollary~\ref{thm-eac:eaq-rate}, we see that the former requires an
additional $I(A;B)_{\rho}$ ebits per channel use. Also,
Corollary~\ref{thm-eac:eaq-rate} leads to a simple proof of the achievability
part of the quantum capacity theorem, as we see in the next section.

\begin{exercise}
\label{ex-ccn:feedback-father}Suppose that Alice can obtain the environment
$E$\ of the channel $U_{A^{\prime}\rightarrow BE}^{\mathcal{N}}$. Such a
channel is known as a \textit{coherent feedback isometry}. Show how they can
achieve the following resource inequality with the coherent feedback isometry
$U_{A^{\prime}\rightarrow BE}^{\mathcal{N}}$:%
\begin{equation}
\langle U_{A^{\prime}\rightarrow BE}^{\mathcal{N}}\rangle\geq\frac{1}%
{2}I(A;B)_{\varphi}\left[  q\rightarrow q\right]  +\frac{1}{2}I(E;B)_{\varphi
}\left[  qq\right]  ,
\end{equation}
where $|\varphi\rangle_{ABE}=U_{A^{\prime}\rightarrow BE}^{\mathcal{N}%
}|\varphi\rangle_{AA^{\prime}}$ and $\rho_{A^{\prime}}=\operatorname{Tr}%
_{A}\{\varphi_{AA^{\prime}}\}$. This protocol is a generalization of coherent
teleportation from Section~\ref{sec-coh:coh-tele} because it reduces to
coherent teleportation in the case that $U_{A^{\prime}\rightarrow
BE}^{\mathcal{N}}$ is equivalent to two coherent channels.
\end{exercise}

\section{Quantum Communication}

\label{sec-ccn:q-cap}We can obtain a protocol for quantum communication simply
by combining the protocol from Theorem~\ref{thm-eac:eaq-rate} further with
entanglement distribution. The resulting protocol again makes catalytic use of
entanglement, in the sense that it exploits some amount of entanglement shared
between Alice and Bob at the beginning of the protocol, but it generates the
same amount of entanglement at the end, so that the net entanglement
consumption rate of the protocol is zero. The resulting rate of quantum
communication turns out to be the same as we find for the quantum channel
coding theorem in Chapter~\ref{chap:quantum-capacity} (though the protocol
given there does not make catalytic use of shared entanglement).

\begin{corollary}
[Quantum Communication]\label{cor-ccn:q-comm}The coherent information
$Q(\mathcal{N})$\ is an achievable rate for quantum communication over a
quantum channel $\mathcal{N}$. That is, the following resource inequality
holds:%
\begin{equation}
\left\langle \mathcal{N}\right\rangle \geq Q(\mathcal{N})\left[  q\rightarrow
q\right]  ,
\end{equation}
where $Q(\mathcal{N})\equiv\max_{\varphi}I(A\rangle B)_{\rho}$ and $\rho
_{AB}\equiv\mathcal{N}_{A^{\prime}\rightarrow B}(\varphi_{AA^{\prime}})$.
\end{corollary}

\begin{proof}
If we further combine the entanglement-assisted quantum communication protocol
from Theorem~\ref{thm-eac:eaq-rate} with entanglement distribution at a rate
$\frac{1}{2}I(A;E)_{\rho}$, we obtain the following resource inequalities:%
\begin{align}
&  \left\langle \mathcal{N}\right\rangle +\frac{1}{2}I(A;E)_{\rho}\left[
qq\right] \nonumber\\
&  \geq\frac{1}{2}\left[  I(A;B)_{\rho}-I(A;E)_{\rho}\right]  \left[
q\rightarrow q\right]  +\frac{1}{2}I(A;E)_{\rho}\left[  q\rightarrow q\right]
\\
&  \geq\frac{1}{2}\left[  I(A;B)_{\rho}-I(A;E)_{\rho}\right]  \left[
q\rightarrow q\right]  +\frac{1}{2}I(A;E)_{\rho}\left[  qq\right]  ,
\end{align}
which after resource cancelation, becomes%
\begin{equation}
\left\langle \mathcal{N}\right\rangle \geq I(A\rangle B)_{\rho}\left[
q\rightarrow q\right]  ,
\end{equation}
because $I(A\rangle B)_{\rho}=\frac{1}{2}\left[  I(A;B)_{\rho}-I(A;E)_{\rho
}\right]  $ (see Exercise~\ref{ex-qie:entropy-games}). They can achieve the
coherent information of the channel simply by generating codes from the state
$\varphi_{AA^{\prime}}$ that maximizes the channel's coherent information.
\end{proof}

\section{Noisy Super-Dense Coding}

Recall that the resource inequality for super-dense coding is%
\begin{equation}
\left[  q\rightarrow q\right]  +\left[  qq\right]  \geq2\left[  c\rightarrow
c\right]  . \label{eq-ccn:SD-RI}%
\end{equation}
The entanglement-assisted classical communication protocol from the previous
chapter is one way to generalize this protocol to a noisy setting, simply by
replacing the noiseless qubit channels in \eqref{eq-ccn:SD-RI}\ with many uses
of a noisy quantum channel. This replacement leads to the setting of
entanglement-assisted classical communication presented in the previous chapter.

Another way to generalize super-dense coding%
\index{super-dense coding!noisy}
is to let the entanglement be noisy while keeping the quantum channels
noiseless. We allow Alice and Bob access to many copies of some shared noisy
state $\rho_{AB}$ and to many uses of a noiseless qubit channel with the goal
of generating noiseless classical communication. One might expect the
resulting protocol to be similar to that for entanglement-assisted classical
communication, and this is indeed the case. The resulting protocol is known as
\textit{noisy super-dense coding}:

\begin{theorem}
[Noisy Super-Dense Coding]\label{thm-ccn:noisy-SD}The following resource
inequality corresponds to an achievable protocol for quantum-assisted
classical communication with a shared quantum state:%
\begin{equation}
\langle\rho_{AB}\rangle+H(A)_{\rho}\left[  q\rightarrow q\right]  \geq
I(A;B)_{\rho}\left[  c\rightarrow c\right]  ,
\end{equation}
where $\rho_{AB}$ is some bipartite state that Alice and Bob share at the
beginning of the protocol.
\end{theorem}

\begin{proof}
The proof of the existence of a protocol proceeds similarly to the proof of
Theorem~\ref{thm-eac:direct-coding}, with a few modifications to account for
our different setting here. We simply need to establish a way for Alice and
Bob to select a code randomly, and then we can invoke the packing
lemma~(Lemma~\ref{lem-pack:pack})\ to establish the existence of a detection
POVM\ that Bob can employ to detect Alice's messages. The method by which they
select a random code is exactly the same as they do in the proof of
Theorem~\ref{thm-eac:direct-coding}, and for this reason, we only highlight
the key aspects of the proof. First consider the state $\rho_{AB}$, and
suppose that $|\varphi\rangle_{ABR}$ is a purification of this state, with $R$
a reference system to which Alice and Bob do not have access. We can say that
the state $|\varphi\rangle_{ABR}$\ arises from some isometry $U_{A^{\prime
}\rightarrow BR}^{\mathcal{N}}$ acting on system $A^{\prime}$ of a pure state
$|\varphi\rangle_{AA^{\prime}}$, so that $|\varphi\rangle_{AA^{\prime}}$ is
defined by $|\varphi\rangle_{ABR}=U_{A^{\prime}\rightarrow BR}^{\mathcal{N}%
}|\varphi\rangle_{AA^{\prime}}$. We can also then think that the state
$\rho_{AB}$ arises from sending the state $|\varphi\rangle_{AA^{\prime}}$
through a channel $\mathcal{N}_{A^{\prime}\rightarrow B}$, obtained by tracing
out the environment $R$ of $U_{A^{\prime}\rightarrow BR}^{\mathcal{N}}$. Our
setting here is becoming closer to the setting in the proof of
Theorem~\ref{thm-eac:direct-coding}, and we now show how it becomes nearly
identical. Observe that the state $|\varphi\rangle_{AA^{\prime}}^{\otimes n}$
admits a type decomposition, similar to the type decomposition in
\eqref{eq-eac:type-decomp-1}--\eqref{eq-eac:type-decomp-4}:%
\begin{equation}
|\varphi\rangle_{AA^{\prime}}^{\otimes n}=\sum_{t}\sqrt{p(t)}|\Phi_{t}%
\rangle_{A^{n}A^{\prime n}}.
\end{equation}
Similarly, we can write $|\varphi\rangle_{ABR}^{\otimes n}$ as%
\begin{equation}
|\varphi\rangle_{ABR}^{\otimes n}=\sum_{t}\sqrt{p(t)}|\Phi_{t}\rangle
_{A^{n}|B^{n}R^{n}},
\end{equation}
where the vertical line in $A^{n}|B^{n}R^{n}$ indicates the bipartite cut
between systems $A^{n}$ and $B^{n}R^{n}$. Alice can select a unitary
$U_{A^{n}}(s)$ of the form in \eqref{eq-eac:unitary-encoding} uniformly at
random, and the expected density operator with respect to this random choice
of unitary is%
\begin{align}
\overline{\rho}_{A^{n}B^{n}}  &  \equiv\mathbb{E}_{S}\left\{  U_{A^{n}}%
(S)\rho_{A^{n}B^{n}}U_{A^{n}}^{\dag}(S)\right\} \\
&  =\sum_{t}p(t)\pi_{A^{n}}^{t}\otimes\mathcal{N}_{A^{\prime n}\rightarrow
B^{n}}(\pi_{A^{\prime n}}^{t}),
\end{align}
by exploiting the development in
\eqref{eq-eac:average-state-1}--\eqref{eq-eac:expected-state-result}. For each
message $m$ that Alice would like to send, she selects a vector $s$ of the
form in \eqref{eq-eac:vector-s} uniformly at random, and we can write $s(m)$
to denote the explicit association of the vector $s$ with the message $m$
after Alice makes the assignment. This leads to quantum-assisted
codewords\footnote{We say that the codewords are \textquotedblleft
quantum-assisted\textquotedblright\ because we will allow the assistance of
quantum communication in transmitting them to Bob.} of the following form:%
\begin{equation}
U_{A^{n}}(s(m))\rho_{A^{n}B^{n}}U_{A^{n}}^{\dag}(s(m)).
\label{eq-ccn:QCA-codewords}%
\end{equation}
We would now like to exploit the
\index{packing lemma}%
packing lemma\ (Lemma~\ref{lem-pack:pack}), and we require message subspace
projectors and a total subspace projector in order to do so. We choose them
respectively as%
\begin{align}
&  U_{A^{n}}(s)\Pi_{A^{n}B^{n}}^{\rho,\delta}U_{A^{n}}^{\dag}(s),\\
&  \Pi_{A^{n}}^{\rho,\delta}\otimes\Pi_{B^{n}}^{\rho,\delta},
\end{align}
where $\Pi_{A^{n}B^{n}}^{\rho,\delta}$, $\Pi_{A^{n}}^{\rho,\delta}$, and
$\Pi_{B^{n}}^{\rho,\delta}$ are typical projectors for $\rho_{A^{n}B^{n}}$,
$\rho_{A^{n}}$, and $\rho_{B^{n}}$, respectively. The following four
conditions for the packing lemma hold, for the same reasons that they hold in
\eqref{eq-eac:pack-2}--\eqref{eq-eac:pack-4}:%
\begin{align}
\operatorname{Tr}\left\{  \left(  \Pi_{A^{n}}^{\rho,\delta}\otimes\Pi_{B^{n}%
}^{\rho,\delta}\right)  \left(  U_{A^{n}}(s)\rho_{A^{n}B^{n}}U_{A^{n}}^{\dag
}(s)\right)  \right\}   &  \geq1-\varepsilon,\label{eq-ccn:NSD-pack-1}\\
\operatorname{Tr}\left\{  \left(  U_{A^{n}}(s)\Pi_{A^{n}B^{n}}^{\rho,\delta
}U_{A^{n}}^{\dag}(s)\right)  \left(  U_{A^{n}}(s)\rho_{A^{n}B^{n}}U_{A^{n}%
}^{\dag}(s)\right)  \right\}   &  \geq1-\varepsilon,\\
\operatorname{Tr}\left\{  U_{A^{n}}(s)\Pi_{A^{n}B^{n}}^{\rho,\delta}U_{A^{n}%
}^{\dag}(s)\right\}   &  \leq2^{n\left[  H(AB)_{\rho}+c\delta\right]  },
\end{align}%
\begin{multline}
\left(  \Pi_{A^{n}}^{\rho,\delta}\otimes\Pi_{B^{n}}^{\rho,\delta}\right)
\overline{\rho}_{A^{n}B^{n}}\left(  \Pi_{A^{n}}^{\rho,\delta}\otimes\Pi
_{B^{n}}^{\rho,\delta}\right) \label{eq-ccn:NSD-pack-4}\\
\leq2^{-n\left[  H(A)_{\rho}+H(B)_{\rho}-\eta(n,\delta)-c\delta\right]
}\left(  \Pi_{A^{n}}^{\rho,\delta}\otimes\Pi_{B^{n}}^{\rho,\delta}\right)  ,
\end{multline}
where $c$ is some positive constant and $\eta(n,\delta)$ is a function that
approaches zero as $n\rightarrow\infty$ and $\delta\rightarrow0$. Let us
assume for the moment that Alice simply sends her $A^{n}$ systems to Bob with
many uses of a noiseless qubit channel. It then follows from
Corollary~\ref{cor-pack:derandomized} (the derandomized version of the packing
lemma) that there exists a code and a POVM\ $\left\{  \Lambda_{A^{n}B^{n}}%
^{m}\right\}  $ that can detect the transmitted codewords of the form in
\eqref{eq-ccn:QCA-codewords} with arbitrarily low maximal probability of
error, as long as the size $\left\vert \mathcal{M}\right\vert $\ of Alice's
message set is small enough:%
\begin{align}
p_{e}^{\ast}  &  \equiv\max_{m}\operatorname{Tr}\left\{  \left(
I-\Lambda_{A^{n}B^{n}}^{m}\right)  U(s(m))_{B^{n}}\rho_{A^{n}B^{n}}U_{B^{n}%
}^{\ast}(s(m))\right\} \\
&  \leq4\left(  \varepsilon+2\sqrt{\varepsilon}\right)  +16\cdot2^{-n\left[
H(A)_{\rho}+H(B)_{\rho}-\eta(n,\delta)-c\delta\right]  }2^{n\left[
H(AB)_{\rho}+c\delta\right]  }\left\vert \mathcal{M}\right\vert \\
&  =4\left(  \varepsilon+2\sqrt{\varepsilon}\right)  +16\cdot2^{-n\left[
I(A;B)_{\rho}-\eta(n,\delta)-2c\delta\right]  }\left\vert \mathcal{M}%
\right\vert .
\end{align}
Thus, we can choose the size of the message set to be $\left\vert
\mathcal{M}\right\vert =2^{n\left[  I(A;B)-\eta(n,\delta)-3c\delta\right]  }$
so that the rate of classical communication is%
\begin{equation}
\frac{1}{n}\log\left\vert \mathcal{M}\right\vert =I(A;B)_{\rho}-\eta
(n,\delta)-3c\delta,
\end{equation}
and the bound on the maximal probability of error becomes%
\begin{equation}
p_{e}^{\ast}\leq4\left(  \varepsilon+2\sqrt{\varepsilon}\right)
+16\cdot2^{-nc\delta}.
\end{equation}
Let $\varepsilon^{\prime}\in(0,1)$ and $\delta^{\prime}>0$. By picking $n$
large enough and $\delta$ small enough, we can make $4\left(  \varepsilon
+2\sqrt{\varepsilon}\right)  +16\cdot2^{-nc\delta}\leq\varepsilon^{\prime}$
and $\eta(n,\delta)+3c\delta\leq\delta^{\prime}$. Thus, the quantum mutual
information $I(A;B)_{\rho}$, with respect to the state $\rho_{AB}$ is an
achievable rate for noisy super-dense coding with $\rho$.%
\begin{figure}
[ptb]
\begin{center}
\includegraphics[
width=4.8456in
]%
{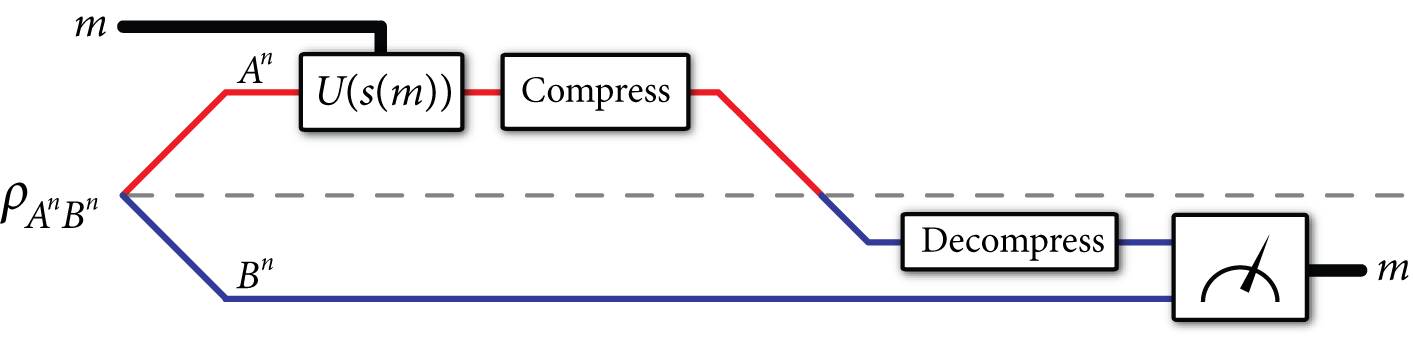}%
\caption{The protocol for noisy super-dense coding that corresponds to the
resource inequality in Theorem~\ref{thm-ccn:noisy-SD}. Alice first projects
her share into its typical subspace (not depicted). She then applies a unitary
encoding $U(s(m))$, based on her message $m$, to her share of the state
$\rho_{A^{n}B^{n}}$. She compresses her state to approximately $nH(A)_{\rho}$
qubits and transmits these qubits over noiseless qubit channels. Bob
decompresses the state and performs a decoding POVM\ that gives Alice's
message $m$ with high probability.}%
\label{fig-ccn:NSD-protocol}%
\end{center}
\end{figure}

We now summarize the protocol (with a final modification). Alice and Bob begin
with the state $\rho_{A^{n}B^{n}}$. Alice first performs a typical subspace
measurement of her system $A^{n}$. This measurement succeeds with high
probability and reduces the size of her system $A^{n}$ to a subspace with size
approximately equal to $nH(A)_{\rho}$ qubits. If Alice wishes to send message
$m$, she applies the unitary $U_{A^{n}}(s(m))$ to her share of the state. She
then performs a compression isometry from her subspace of $A^{n}$ to
$nH(A)_{\rho}$ qubits. She transmits her qubits over $nH(A)_{\rho}$ noiseless
qubit channels, and Bob receives them. Bob performs the decompression isometry
from the space of $nH(A)_{\rho}$ noiseless qubits to a space isomorphic to
Alice's original systems $A^{n}$. He then performs the decoding POVM $\left\{
\Lambda_{A^{n}B^{n}}^{m}\right\}  $ and determines Alice's message $m$ with
vanishingly small error probability. \textit{Note}:\ The only modification to
the protocol is the typical subspace measurement at the beginning, and one can
readily check that this measurement does not affect any of the conditions in
\eqref{eq-ccn:NSD-pack-1}--\eqref{eq-ccn:NSD-pack-4}.
Figure~\ref{fig-ccn:NSD-protocol}\ depicts the protocol.
\end{proof}

\section{State Transfer}

We can also construct a
\index{state transfer!coherent}%
coherent version of the noisy super-dense coding protocol, in a manner similar
to the way in which the proof of Theorem~\ref{thm-eac:ea-coh}\ constructs a
coherent version of entanglement-assisted classical communication. However,
the coherent version of noisy super-dense coding achieves an additional
task:\ the transfer of Alice's share of the state $\rho_{AB}^{\otimes n}$ to
Bob. The resulting protocol is known as coherent state transfer, and from this
protocol, we can derive a protocol for quantum-communication-assisted state
transfer, or quantum-assisted state transfer\footnote{This protocol goes by
several other names in the quantum Shannon theory literature:\ state transfer,
fully quantum Slepian--Wolf, state merging, and the merging mother.} for short.

\begin{theorem}
[Coherent State Transfer]\label{thm-ccn:cst}The following resource inequality
corresponds to an achievable protocol for coherent state transfer using a
\index{coherent state transfer}
state $\rho_{AB}$:%
\begin{equation}
\left\langle W_{S\rightarrow AB}:\rho_{S}\right\rangle +H(A)_{\rho}\left[
q\rightarrow q\right]  \geq I(A;B)_{\rho}\left[  q\rightarrow qq\right]
+\langle\operatorname{id}_{S\rightarrow\hat{B}B}:\rho_{S}\rangle,
\label{eq-ccn:cst}%
\end{equation}
where $\rho_{AB}$ is a bipartite state that Alice and Bob share at the
beginning of the protocol.
\end{theorem}

The resource inequality in \eqref{eq-ccn:cst} features some notation that we
have not seen yet. The expression $\left\langle W_{S\rightarrow AB}:\rho
_{S}\right\rangle $ means that a source party $S$ distributes many copies of
the state $\rho_{S}$ to Alice and Bob, by applying some isometry
$W_{S\rightarrow AB}$ to the state $\rho_{S}$. This resource is effectively
equivalent to Alice and Bob sharing many copies of the state $\rho_{AB}$, a
resource we expressed in Theorem~\ref{thm-ccn:noisy-SD}\ as $\left\langle
\rho_{AB}\right\rangle $. The expression $\langle\operatorname{id}%
_{S\rightarrow\hat{B}B}:\rho_{S}\rangle$ means that a source party applies the
identity map to $\rho_{S}$ and gives the full state to Bob. We can now state
the meaning of the resource inequality in \eqref{eq-ccn:cst}: Using
$n$\ copies of the state $\rho_{AB}$ and $nH(A)_{\rho}$ noiseless qubit
channels, Alice can simulate $nI(A;B)_{\rho}$ noiseless coherent channels to
Bob while at the same time transferring her share of the state $\rho
_{AB}^{\otimes n}$ to him.

\bigskip

\begin{figure}
[ptb]
\begin{center}
\includegraphics[
width=4.8456in
]%
{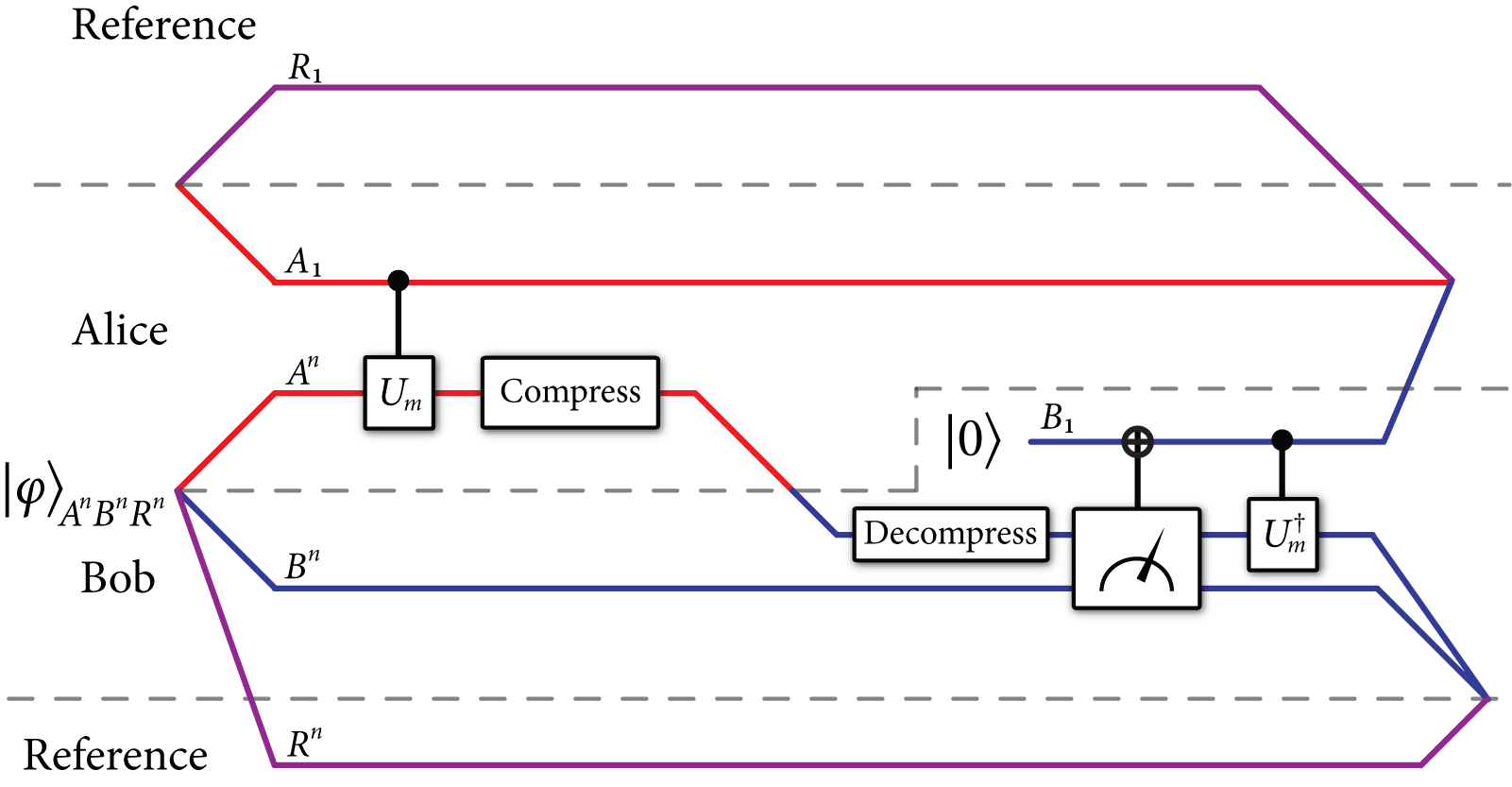}%
\caption{The protocol for coherent state transfer, a coherent version of the
noisy super-dense coding protocol that accomplishes the task of state transfer
in addition to coherent communication.}%
\label{fig-ccn:NSD-coherent}%
\end{center}
\end{figure}

\begin{proof}
A proof proceeds similarly to the proof of Theorem~\ref{thm-eac:ea-coh}. Let
$|\varphi\rangle_{ABR}$ be a purification of $\rho_{AB}$. Alice begins with a
state that she shares with a reference system $R_{1}$, on which she would like
to simulate coherent channels:%
\begin{equation}
|\psi\rangle_{R_{1}A_{1}}\equiv\sum_{l,m=1}^{D^{2}}\alpha_{l,m}\left\vert
l\right\rangle _{R_{1}}|m\rangle_{A_{1}},
\end{equation}
where $D^{2}\approx2^{nI(A;B)_{\rho}}$. She appends $|\psi\rangle_{R_{1}A_{1}%
}$ to $|\varphi\rangle_{A^{n}B^{n}R^{n}}\equiv|\varphi\rangle_{ABR}^{\otimes
n}$ and applies a typical subspace measurement to her system $A^{n}$. (In what
follows, we use the same notation for the typical projected state because the
states are the same up to a vanishingly small error.) She applies the
following controlled unitary to her systems $A_{1}A^{n}$:%
\begin{equation}
\sum_{m}|m\rangle\langle m|_{A_{1}}\otimes U_{A^{n}}(s(m)),
\end{equation}
resulting in the overall state%
\begin{equation}
\sum_{l,m}\alpha_{l,m}\left\vert l\right\rangle _{R_{1}}\left\vert
m\right\rangle _{A_{1}}U_{A^{n}}(s(m))|\varphi\rangle_{A^{n}B^{n}R^{n}}.
\end{equation}
Alice compresses her $A^{n}$ systems, sends them over $nH(A)_{\rho}$ noiseless
qubit channels, and Bob receives them. He decompresses them and places them in
systems $\hat{B}^{n}$ isomorphic to $A^{n}$. The resulting state is the same
as $|\varphi\rangle_{A^{n}B^{n}R^{n}}$, with the systems $A^{n}$ replaced by
$\hat{B}^{n}$. Bob performs a coherent gentle measurement of the following
form:%
\begin{equation}
\sum_{m}\sqrt{\Lambda_{\hat{B}^{n}B^{n}}^{m}}\otimes|m\rangle_{B_{1}},
\end{equation}
resulting in a state that is close in trace distance to%
\begin{equation}
\sum_{l,m}\alpha_{l,m}\left\vert l\right\rangle _{R_{1}}\left\vert
m\right\rangle _{A_{1}}|m\rangle_{B_{1}}U_{\hat{B}^{n}}(s(m))|\varphi
\rangle_{\hat{B}^{n}B^{n}R^{n}}.
\end{equation}
He finally performs the controlled unitary%
\begin{equation}
\sum_{m}|m\rangle\langle m|_{B_{1}}\otimes U_{\hat{B}^{n}}^{\dag}(s(m)),
\end{equation}
resulting in the state%
\begin{equation}
\left(  \sum_{l,m}\alpha_{l,m}\left\vert l\right\rangle _{R_{1}}\left\vert
m\right\rangle _{A_{1}}|m\rangle_{B_{1}}\right)  \otimes|\varphi\rangle
_{\hat{B}^{n}B^{n}R^{n}}.
\end{equation}
Thus, Alice has simulated $nI(A;B)_{\rho}$ coherent channels to Bob with
arbitrarily small error, while also transferring her share of the state
$|\varphi\rangle_{A^{n}B^{n}R^{n}}$ to him. Figure~\ref{fig-ccn:NSD-coherent}%
\ depicts the protocol.
\end{proof}

We obtain the following resource inequality for
\index{state transfer!quantum-assisted}%
quantum-assisted state transfer, by combining the above protocol with the
coherent communication identity:

\begin{corollary}
[Quantum-Assisted State Transfer]\label{thm-ccn:qast}The following resource
inequality corresponds to an achievable protocol for quantum-assisted state
transfer using a shared state $\rho_{AB}$:%
\begin{equation}
\langle W_{S\rightarrow AB}:\rho_{S}\rangle+\frac{1}{2}I(A;R)_{\varphi}\left[
q\rightarrow q\right]  \geq\frac{1}{2}I(A;B)_{\varphi}\left[  qq\right]
+\langle\operatorname{id}_{S\rightarrow\hat{B}B}:\rho_{S}\rangle,
\label{eq-ccn:QA-state-transfer}%
\end{equation}
where $\rho_{AB}$ is a bipartite state that Alice and Bob share at the
beginning of the protocol, and $|\varphi\rangle_{ABR}$ is a purification of it.
\end{corollary}

\begin{proof}
Consider the following chain of resource inequalities:%
\begin{align}
\langle W_{S\rightarrow AB}  &  :\rho_{S}\rangle+H(A)_{\varphi}\left[
q\rightarrow q\right] \nonumber\\
&  \geq I(A;B)_{\varphi}\left[  q\rightarrow qq\right]  +\langle
\operatorname{id}_{S\rightarrow\hat{B}B}:\rho_{S}\rangle\\
&  \geq\frac{1}{2}I(A;B)_{\varphi}\left[  q\rightarrow q\right]  +\frac{1}%
{2}I(A;B)_{\varphi}\left[  qq\right]  +\langle\operatorname{id}_{S\rightarrow
\hat{B}B}:\rho_{S}\rangle,
\end{align}
where the first follows from coherent state transfer and the second follows
from the coherent communication identity. By resource cancelation, we obtain
the resource inequality in the statement of the theorem because $\frac{1}%
{2}I(A;R)_{\varphi}=H(A)_{\rho}-\frac{1}{2}I(A;B)_{\rho}$.
\end{proof}

\begin{corollary}
[Classical-Assisted State Transfer]\label{thm-ccn:cast}The following resource
inequality corresponds to an achievable protocol
\index{state transfer!classical-assisted}
for classical-assisted state transfer using a shared state $\rho_{AB}$:%
\begin{equation}
\langle W_{S\rightarrow AB}:\rho_{S}\rangle+I(A;R)_{\varphi}\left[
c\rightarrow c\right]  \geq I(A\rangle B)_{\varphi}\left[  qq\right]
+\langle\operatorname{id}_{S\rightarrow\hat{B}B}:\rho_{S}\rangle,
\end{equation}
where $\rho_{AB}$ is a bipartite state that Alice and Bob share at the
beginning of the protocol, and $|\varphi\rangle_{ABR}$ is a purification of it.
\end{corollary}

\begin{proof}
We simply combine the protocol above with teleportation:%
\begin{align}
\langle W_{S\rightarrow AB}  &  :\rho_{S}\rangle+I(A;R)_{\varphi}\left[
c\rightarrow c\right]  +\frac{1}{2}I(A;R)_{\varphi}\left[  qq\right]
\nonumber\\
&  \geq\langle W_{S\rightarrow AB}:\rho_{S}\rangle+\frac{1}{2}I(A;R)_{\varphi
}\left[  q\rightarrow q\right] \\
&  \geq\frac{1}{2}I(A;B)_{\varphi}\left[  qq\right]  +\langle\operatorname{id}%
_{S\rightarrow\hat{B}B}:\rho_{S}\rangle.
\end{align}
Using the fact that $\frac{1}{2}I(A;B)_{\varphi}-\frac{1}{2}I(A;R)_{\varphi
}=I(A\rangle B)_{\varphi}$, we obtain the resource inequality in the statement
of the corollary.
\end{proof}

The above protocol gives a wonderful operational interpretation to the
coherent information (or negative conditional entropy $-H( A|B) _{\rho}$).
When the coherent information is positive, Alice and Bob share that rate of
entanglement at the end of the protocol (and thus the ability to teleport if
extra classical communication is available). When the coherent information is
negative, they need to consume entanglement at a rate of $H( A|B) _{\rho}$
ebits per copy in order for the state transfer process to complete.

\begin{exercise}
Suppose that Alice actually possesses the reference $R$ in the above
protocols. Show that Alice and Bob can achieve the following resource
inequality:%
\begin{equation}
\left\langle \psi_{ABR}\right\rangle +\frac{1}{2}I(A;R)_{\psi}\left[
q\rightarrow q\right]  \geq\frac{1}{2}\left(  H(A)_{\psi}+H(B)_{\psi
}+H(R)_{\psi}\right)  \left[  qq\right]  ,
\end{equation}
where $|\psi\rangle_{ABR}$ is some pure state.
\end{exercise}

\subsection{The Dual Roles of Quantum Mutual Information}

The resource inequality for entanglement-assisted quantum communication in
\eqref{eq-ccn:RI-father} and that for quantum-assisted state transfer in
\eqref{eq-ccn:QA-state-transfer} appear to be strikingly similar. Both contain
a noisy resource and both consume a noiseless quantum resource in order to
generate another noiseless quantum resource. We say that these two protocols
are related by \textit{source--channel duality} because we obtain one protocol
from another by changing channels to states and vice versa.

Also, both protocols require the consumed rate of the noiseless quantum
resource to be equal to half the quantum mutual information between the system
$A$ for which we are trying to preserve quantum coherence and the environment
to which we do not have access. In both cases, our goal is to break the
correlations between the system $A$ and the environment, and the quantum
mutual information is quantifying how much quantum coherence is required to
break these correlations. Both protocols in \eqref{eq-ccn:RI-father} and
\eqref{eq-ccn:QA-state-transfer} have their rates for the generated noiseless
quantum resource equal to half the quantum mutual information between the
system $A$ and the system $B$. Thus, the quantum mutual information is also
quantifying how much quantum correlations we can establish between two
systems---it plays the dual role of quantifying both the destruction and
creation of correlations.

\section{Trade-off Coding}

Suppose
\index{trade-off coding}%
that you are a communication engineer working at a quantum communication
company. Suppose further that your company has made
quite a profit from entanglement-assisted classical communication, beating out
the communication rates that other companies can achieve simply because your
company has been able to generate high-quality noiseless entanglement between
several nodes in its network, while the competitors have not been able to do
so. But now suppose that your customer base has become so large that there is
not enough entanglement to support protocols that achieve the rates given in
the entanglement-assisted classical capacity theorem
(Theorem~\ref{thm-eac:BSST}). Your boss would like you to make the best of
this situation, by determining the optimal rates of classical communication
for a fixed entanglement budget. He is hoping that you will be able to design
a protocol such that there will only be a slight decrease in communication
rates. You tell him that you will do your best.

What should you do in this situation? Your first thought might be that we have
already determined unassisted classical codes with a communication rate equal
to the channel Holevo information $\chi(\mathcal{N})$ and we have also
determined entanglement-assisted codes with a communication rate equal to the
channel mutual information $I(\mathcal{N})$. It might seem that a reasonable
strategy is to mix these two strategies, using some fraction $\lambda$\ of the
channel uses for the unassisted classical code and the other fraction
$1-\lambda$ of the channel uses for the entanglement-assisted code. This
strategy achieves a rate of%
\begin{equation}
\lambda\chi(\mathcal{N})+\left(  1-\lambda\right)  I(\mathcal{N}),
\label{eq-ccn:time-sharing}%
\end{equation}
and it has an error no larger than the sum of the errors of the individual
codes (thus, this error vanishes asymptotically). Meanwhile, it consumes
entanglement at a lower rate of $\left(  1-\lambda\right)  E$ ebits per
channel use, if $E$ is the amount of entanglement that the original protocol
for entanglement-assisted classical communication consumes. This simple mixing
strategy is known
\index{time-sharing}%
as \textit{time sharing}. You figure this strategy might perform well, and you
suggest it to your boss. After your boss reviews your proposal, he sends it
back to you, telling you that he already thought of this solution and suggests
that you are going to have to be a bit more clever---otherwise, he suspects
that the existing customer base will notice the drop in communication rates.

Another strategy for communication is known as \textit{trade-off coding}. We
explore this strategy in the forthcoming section and in a broader context in
Chapter~\ref{chap:trade-off}. Trade-off coding beats time sharing for many
channels of interest, but for other channels, it just reduces to time sharing.
It is not clear \textit{a priori} how to determine which channels benefit from
trade-off coding, but it certainly depends on the channel for which Alice and
Bob are coding. Chapter~\ref{chap:trade-off} follows up on the development
here by demonstrating that this trade-off coding strategy is provably optimal
for certain channels, and for general channels, it is optimal in the sense of
regularized formulas. Trade-off coding is our best known way to deal with the
above situation with a fixed entanglement budget, and your boss should be
pleased with these results. Furthermore, we can upgrade the protocol outlined
below to one that achieves entanglement-assisted communication of both
classical and quantum information.

\subsection{Classical Communication with Limited Entanglement}

We first show that the resource inequality given in the following theorem is
achievable, and we follow up with an interpretation of it in the context of
trade-off coding. We name the protocol \textit{CE\ trade-off coding} because
it captures the trade-off between classical communication and entanglement consumption.

\begin{theorem}
[CE\ Trade-off Coding]\label{thm-ccn:ce-trading}The following resource
inequality corresponds to an achievable protocol for entanglement-assisted
classical communication over a quantum channel $\mathcal{N}_{A^{\prime
}\rightarrow B}$:%
\begin{equation}
\left\langle \mathcal{N}\right\rangle +H(A|X)_{\rho}\left[  qq\right]  \geq
I(AX;B)_{\rho}\left[  c\rightarrow c\right]  ,
\end{equation}
where $\rho_{XAB}$ is a state of the following form:%
\begin{equation}
\rho_{XAB}\equiv\sum_{x}p_{X}(x)|x\rangle\langle x|_{X}\otimes\mathcal{N}%
_{A^{\prime}\rightarrow B}(\varphi_{AA^{\prime}}^{x}),
\label{eq-ccn:ce-trade-off-code-state}%
\end{equation}
and the states $\varphi_{AA^{\prime}}^{x}$ are pure.
\end{theorem}

\begin{proof}
The proof of the above trade-off coding theorem exploits the direct parts of
both the HSW\ coding theorem (Theorem~\ref{thm-cc:HSW}) and the
entanglement-assisted classical capacity theorem
(Theorem~\ref{thm-eac:direct-coding}). In particular, we exploit the
constant-composition coding variant of the HSW\ theorem, described in
Section~\ref{sec-cc:HSW-alternate-direct}, and that the entanglement-assisted
quantum codewords from Theorem~\ref{thm-eac:direct-coding}\ are tensor-power
states after tracing over Bob's shares of the entanglement\ (this is the
observation mentioned in Remark~\ref{rem-eac:tensor-power-eac}). Suppose that
Alice and Bob exploit a constant-composition HSW\ code for the channel
$\mathcal{N}_{A^{\prime}\rightarrow B}$. Such a code consists of a codebook
$\left\{  \rho^{x^{n}(m)}\right\}  _{m}$ with $\approx2^{nI(X;B)_{\rho}}$
quantum codewords. The Holevo information $I(X;B)_{\rho}$ is with respect to
some classical--quantum state $\rho_{XB}$ where%
\begin{equation}
\rho_{XB}\equiv\sum_{x}p_{X}(x)|x\rangle\langle x|_{X}\otimes\mathcal{N}%
_{A^{\prime}\rightarrow B}(\rho_{A^{\prime}}^{x}),
\end{equation}
and each codeword $\rho^{x^{n}(m)}$ is a tensor-product state of the form%
\begin{equation}
\rho_{x^{n}(m)}=\rho^{x_{1}(m)}\otimes\rho^{x_{2}(m)}\otimes\cdots\otimes
\rho^{x_{n}(m)}.
\end{equation}
Corresponding to the codebook is some decoding POVM\ $\left\{  \Lambda_{B^{n}%
}^{m}\right\}  $, which Bob can employ to decode each codeword transmitted
through the channel with arbitrarily high probability for all $\varepsilon
\in(0,1)$:%
\begin{equation}
\forall m\ \ \ \operatorname{Tr}\left\{  \Lambda_{B^{n}}^{m}\mathcal{N}%
_{A^{\prime n}\rightarrow B^{n}}(\rho_{A^{\prime n}}^{x^{n}(m)})\right\}
\geq1-\varepsilon.
\end{equation}
Recall from the constant-composition HSW\ coding variant described in
Section~\ref{sec-cc:HSW-alternate-direct} that we select each codeword
$x^{n}(m)$ from a typical type class, typical with respect to the distribution
$p_{X}(x)$. Let $t(x)$ denote the empirical distribution for the typical type
class, and it is such that $\max_{x}\left\vert t(x)-p_{X}(x)\right\vert
\leq\delta$ for some $\delta>0$. This implies that each classical codeword
$x^{n}(m)$ has approximately $np_{X}(a_{1})$ occurrences of the symbol
$a_{1}\in\mathcal{X}$, $np_{X}(a_{2})$ occurrences of the symbol $a_{2}%
\in\mathcal{X}$, and so on, for all letters in the alphabet $\mathcal{X}$.
However, for a typical type class, all sequences have exactly the same
empirical distribution, so that there exists some permutation $\pi_{m}$\ that
rearranges each sequence $x^{n}(m)$ in lexicographical order according to the
alphabet $\mathcal{X}$. That is, this permutation $\pi_{m}$\ arranges the
sequence $x^{n}(m)$ into $\left\vert \mathcal{X}\right\vert $ blocks, each of
length $nt(a_{1})$, \ldots, $nt(a_{\left\vert \mathcal{X}\right\vert })$:%
\begin{equation}
\pi_{m}(x^{n}(m))=\underbrace{a_{1}\cdots a_{1}}_{nt(a_{1})}\ \underbrace
{a_{2}\cdots a_{2}}_{nt(a_{2})}\ \cdots\ \underbrace{a_{\left\vert
\mathcal{X}\right\vert }\cdots a_{\left\vert \mathcal{X}\right\vert }%
}_{nt(a_{\left\vert \mathcal{X}\right\vert })}.
\label{eq-ccn:lexico-classical-seq}%
\end{equation}
The same holds true for the corresponding permutation operator $\pi_{m}%
$\ applied to a quantum state $\rho^{x^{n}(m)}$ corresponding to the sequence
$x^{n}(m)$:%
\begin{equation}
\pi_{m}(\rho^{x^{n}(m)})=\underbrace{\rho^{a_{1}}\otimes\cdots\otimes
\rho^{a_{1}}}_{nt(a_{1})}\otimes\underbrace{\rho^{a_{2}}\otimes\cdots
\otimes\rho^{a_{2}}}_{nt(a_{2})}\otimes\ \cdots\ \otimes\underbrace
{\rho^{a_{\left\vert \mathcal{X}\right\vert }}\otimes\cdots\otimes
\rho^{a_{\left\vert \mathcal{X}\right\vert }}}_{nt(a_{\left\vert
\mathcal{X}\right\vert })}.
\end{equation}
Now, we assume that $n$ is quite large, so large that each of $nt(a_{1})$,
\ldots, $nt(a_{\left\vert \mathcal{X}\right\vert })$ are large enough for the
law of large numbers to come into play for each block in the permuted sequence
$\pi_{m}(x^{n}(m))$ and tensor-product state $\pi_{m}(\rho^{x^{n}(m)})$. Let
$\varphi_{AA^{\prime}}^{x}$ be a purification of each $\rho_{A^{\prime}}^{x}$
in the ensemble $\left\{  p_{X}(x),\rho_{A^{\prime}}^{x}\right\}  $, where we
assume that Alice has access to system $A^{\prime}$ and Bob has access to $A$.
Then, for every HSW\ quantum codeword $\rho_{A^{\prime n}}^{x^{n}(m)}$, there
is some purification $\varphi_{A^{n}A^{\prime n}}^{x^{n}(m)}$, where%
\begin{equation}
\varphi_{A^{n}A^{\prime n}}^{x^{n}(m)}\equiv\varphi_{A_{1}A_{1}^{\prime}%
}^{x_{1}(m)}\otimes\varphi_{A_{2}A_{2}^{\prime}}^{x_{2}(m)}\otimes
\cdots\otimes\varphi_{A_{n}A_{n}^{\prime}}^{x_{n}(m)},
\end{equation}
Alice has access to the systems $A^{\prime n}\equiv A_{1}^{\prime}\cdots
A_{n}^{\prime}$, and Bob has access to $A^{n}\equiv A_{1}\cdots A_{n}$.
Applying the permutation $\pi_{m}$ to any purified tensor-product state
$\varphi^{x^{n}}$ gives%
\begin{equation}
\pi_{m}(\varphi^{x^{n}(m)})=\underbrace{\varphi^{a_{1}}\otimes\cdots
\otimes\varphi^{a_{1}}}_{nt(a_{1})}\otimes\underbrace{\varphi^{a_{2}}%
\otimes\cdots\otimes\varphi^{a_{2}}}_{nt(a_{2})}\otimes\cdots\otimes
\underbrace{\varphi^{a_{\left\vert \mathcal{X}\right\vert }}\otimes
\cdots\otimes\varphi^{a_{\left\vert \mathcal{X}\right\vert }}}%
_{nt(a_{\left\vert \mathcal{X}\right\vert })},
\label{eq-ccn:entangled-states-permuted}%
\end{equation}
where we have assumed that the permutation applies on both the purification
systems $A^{n}$ and the systems $A^{\prime n}$.

We can now formulate a strategy for trade-off coding. Alice begins with a
standard classical sequence $\hat{x}^{n}$ that is in lexicographical order, of
the form in \eqref{eq-ccn:lexico-classical-seq}. According to this sequence,
she arranges the states $\{\varphi_{AA^{\prime}}^{a_{i}}\}$ to be in
$\left\vert \mathcal{X}\right\vert $ blocks, each of length $n_{i}\equiv
nt(a_{i})\approx np_{X}(a_{i})$---the resulting state is of the same form as
in \eqref{eq-ccn:entangled-states-permuted}. Since $nt(a_{i})$ is large enough
for the law of large numbers to come into play, for each block, there exists
an entanglement-assisted classical code with $\approx2^{n_{i}%
I(A;B)_{\mathcal{N}(\varphi^{a_{i}})}}$ entanglement-assisted quantum
codewords, where the quantum mutual information $I(A;B)_{\mathcal{N}%
(\varphi^{a_{i}})}$ is with respect to the state $\mathcal{N}_{A^{\prime
}\rightarrow B}(\varphi_{AA^{\prime}}^{a_{i}})$. Then each of these
$\left\vert \mathcal{X}\right\vert $ entanglement-assisted classical codes
consumes $\approx n_{i}H(A)_{\varphi_{A}^{a_{i}}}$ ebits (i.e., each state
$\left(  \varphi^{a_{i}}\right)  ^{\otimes n_{i}}$ is produced from $\approx
n_{i}H(A)_{\varphi_{A}^{a_{i}}}$ ebits via entanglement dilution and a
negligible rate of classical communication). The entanglement-assisted quantum
codewords for each block are of the form%
\begin{equation}
U_{A^{n_{i}}}(s(l_{i}))(\varphi_{A^{n_{i}}A^{\prime n_{i}}}^{a_{i}%
})U_{A^{n_{i}}}^{\dag}(s(l_{i})), \label{eq-ccn:EA-codewords}%
\end{equation}
where $l_{i}$ is a message in the message set of size $\approx2^{n_{i}%
I(A;B)_{\varphi^{a_{i}}}}$, the state $\varphi_{A^{n_{i}}A^{\prime n_{i}}%
}^{a_{i}}=\varphi_{A_{1}A_{1}^{\prime}}^{a_{i}}\otimes\cdots\otimes
\varphi_{A_{n_{i}}A_{n_{i}}^{\prime}}^{a_{i}}$, and the unitaries
$U_{A^{n_{i}}}(s(l_{i}))$ are of the form in \eqref{eq-eac:unitary-encoding}.
Observe that the codewords in \eqref{eq-ccn:EA-codewords} are all equal to
$\rho_{A^{\prime n_{i}}}^{a_{i}}$ after tracing over Bob's systems $A^{n_{i}}%
$, regardless of the particular unitary that Alice applies (this is the
content of Remark~\ref{rem-eac:tensor-power-eac}). Alice then determines the
permutation $\pi_{m}^{-1}$ needed to permute the standard sequence $\hat
{x}^{n}$ to a codeword sequence $x^{n}(m)$, and she applies the permutation
operator $\pi_{m}^{-1}$ to her systems $A^{\prime n}$ so that her channel
input density operator is the HSW\ quantum codeword $\rho_{A^{\prime n}%
}^{x^{n}(m)}$ (we are tracing over Bob's systems $A^{n}$ and applying
Remark~\ref{rem-eac:tensor-power-eac} to obtain this result). She transmits
her systems $A^{\prime n}$ over the channel to Bob. If Bob ignores his share
of the entanglement in $A^{n}$, the state that he receives from the channel is
$\mathcal{N}_{A^{\prime n}\rightarrow B^{n}}(\rho_{A^{\prime n}}^{x^{n}(m)})$.
He then applies his HSW\ measurement $\{\Lambda_{B^{n}}^{m}\}$ to the systems
$B^{n}$ received from the channel, and he determines the sequence $x^{n}(m)$,
and hence the message $m$, with nearly unit probability. Also, this
measurement has negligible disturbance on the state, so that the
post-measurement state is $2\sqrt{\varepsilon}$-close in trace distance to the
state that Alice transmitted through the channel (in what follows, we assume
that the measurement does not change the state, and we collect error terms at
the end of the proof). Now that he knows $m$, he applies the permutation
operator $\pi_{m}$ to his systems $B^{n}$, and we are assuming that he already
has his share $A^{n}$ of the entanglement arranged in lexicographical order
according to the standard sequence $\hat{x}^{n}$. His state is then as
follows:%
\begin{equation}
\bigotimes\limits_{i=1}^{\left\vert \mathcal{X}\right\vert }U_{A^{n_{i}}%
}(s(l_{i}))\left(  \varphi_{A^{n_{i}}A^{\prime n_{i}}}^{a_{i}}\right)
U_{A^{n_{i}}}^{\dag}(s(l_{i})).
\end{equation}
At this point, he can decode the message $l_{i}$ in the $i$th block by
performing a collective measurement on the systems $A^{n_{i}}A^{\prime n_{i}}%
$. He does this for each of the $\left\vert \mathcal{X}\right\vert $
entanglement-assisted classical codes, and this completes the protocol for
trade-off coding. The total error accumulated is no larger than the
entanglement dilution error, the sum of $\varepsilon$ for the first
measurement, $2\sqrt{\varepsilon}$ for the disturbance of the state, and
$\left\vert \mathcal{X}\right\vert \varepsilon$ for the error from the final
measurement of the $\left\vert \mathcal{X}\right\vert $ blocks.
Figure~\ref{fig-ccn:eac-trading}\ depicts this protocol for an example.%
\begin{figure}
[ptb]
\begin{center}
\includegraphics[
width=4.8456in
]%
{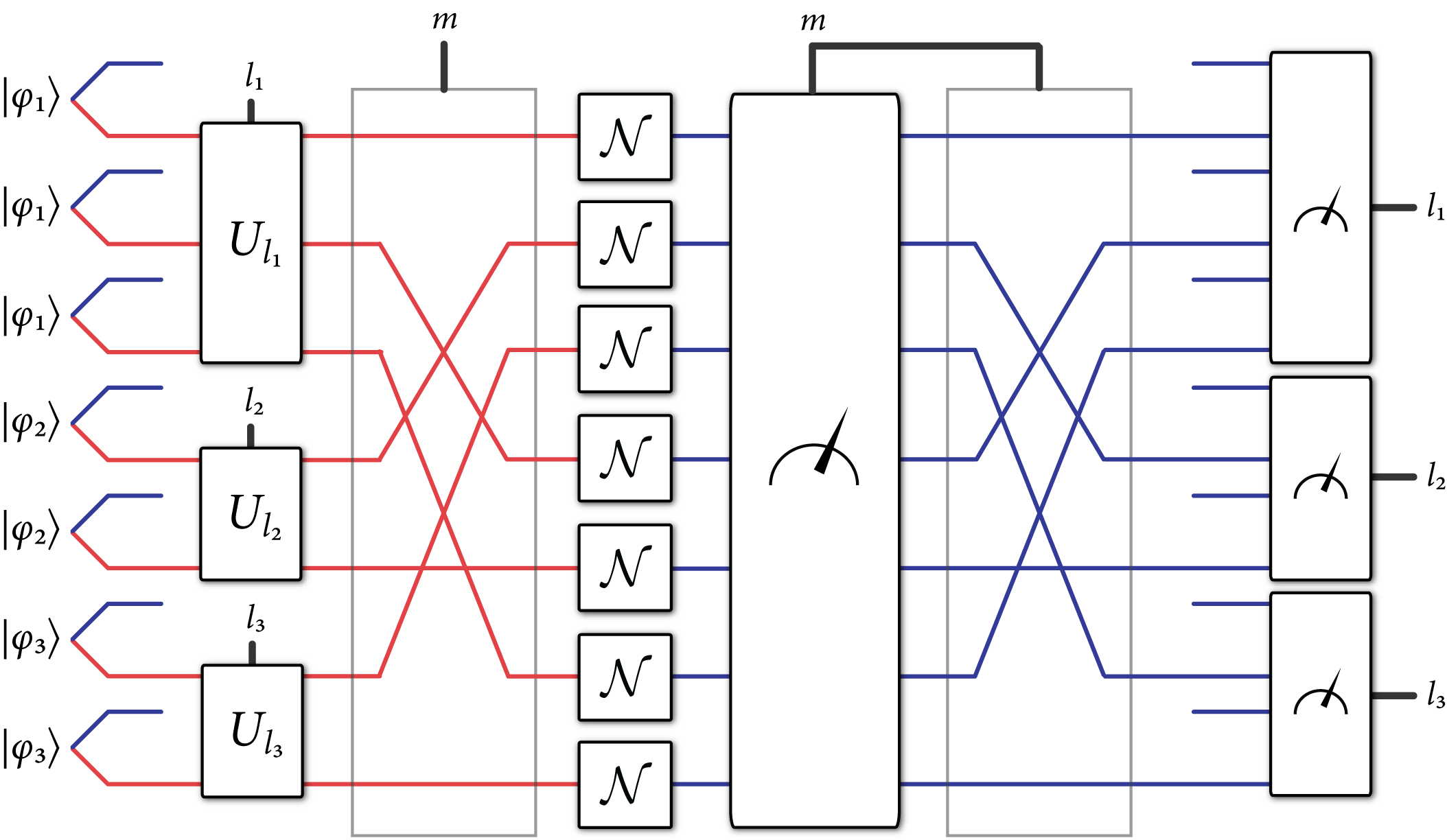}%
\caption{A simple protocol for trade-off coding between assisted and
unassisted classical communication. Alice wishes to send the classical message
$m$ while also sending the messages $l_{1}$, $l_{2}$, and $l_{3}$. Her
HSW\ codebook has the message $m$ map to the sequence $1231213$, which in turn
gives the HSW\ quantum codeword $\rho^{1}\otimes\rho^{2}\otimes\rho^{3}%
\otimes\rho^{1}\otimes\rho^{2}\otimes\rho^{1}\otimes\rho^{3}$. A purification
of these states is the following tensor product of pure states: $\varphi
^{1}\otimes\varphi^{2}\otimes\varphi^{3}\otimes\varphi^{1}\otimes\varphi
^{2}\otimes\varphi^{1}\otimes\varphi^{3}$, where Bob possesses the
purification of each state in the tensor product. She begins with these states
arranged in lexicographic order in three blocks (there are three letters in
this alphabet). For each block $i$, she encodes the message $l_{i}$ with the
local unitaries for an entanglement-assisted classical code. She then permutes
her shares of the entangled states according to the permutation associated
with the message $m$. She inputs her systems to many uses of the channel, and
Bob receives the outputs. His first action is to ignore his shares of the
entanglement and perform a collective HSW\ measurement on all of the channel
outputs. With high probability, he can determine the message $m$ while causing
a negligible disturbance to the state of the channel outputs. Based on the
message $m$, he performs the inverse of the permutation that Alice used at the
encoder. He combines his shares of the entanglement with the permuted channel
outputs. His final three measurements are those given by the three
entanglement-assisted codes Alice used at the encoder, and they detect the
messages $l_{1}$, $l_{2}$, and $l_{3}$ with high probability.}%
\label{fig-ccn:eac-trading}%
\end{center}
\end{figure}

We now show how the total rate of classical communication adds up to
$I(AX;B)_{\rho}$ where $\rho_{XAB}$ is a state of the form in
\eqref{eq-ccn:ce-trade-off-code-state}. First, we can apply the chain rule for
quantum mutual information to observe that the total rate $I(AX;B)_{\rho}$ is
the sum of a Holevo information $I(X;B)_{\rho}$ and a conditional quantum
mutual information $I(A;B|X)_{\rho}$:%
\begin{equation}
I(AX;B)_{\rho}=I(X;B)_{\rho}+I(A;B|X)_{\rho}.
\end{equation}
They achieve the rate $I(X;B)_{\rho}$ because Bob first decodes the
HSW\ quantum codeword, of which there can be $\approx2^{nI(X;B)}$. His next
step is to permute and decode the $\left\vert \mathcal{X}\right\vert $ blocks,
each consisting of an entanglement-assisted classical code on $\approx
np_{X}(x)$ channel uses. Each entanglement-assisted classical code
communicates $\approx np_{X}(x)I(A;B)_{\mathcal{N}(\varphi^{x})}$ bits and consumes
$\approx  np_{X}(x)H(A)_{\varphi^{x}}$ ebits. Thus, the total rate of classical
communication for this last part is%
\begin{align}
\frac{\#\text{ of bits generated}}{\#\text{ of channel uses}}  &  \approx
\frac{\sum_{x}n\ p_{X}(x)I(A;B)_{\mathcal{N}(\varphi^{x})}}{\sum_{x}%
n\ p_{X}(x)}\\
&  =\sum_{x}p_{X}(x)I(A;B)_{\mathcal{N}(\varphi^{x})}\\
&  =I(A;B|X)_{\rho},
\end{align}
and similarly, the total rate of entanglement consumption is%
\begin{align}
\frac{\#\text{ of ebit consumed}}{\#\text{ of channel uses}}  &  \approx
\frac{\sum_{x}n\ p_{X}(x)H(A)_{\varphi^{x}}}{\sum_{x}n\ p_{X}(x)}\\
&  =\sum_{x}p_{X}(x)H(A)_{\varphi^{x}}\\
&  =H(A|X)_{\rho}.
\end{align}
This gives the resource inequality in the statement of the theorem.
\end{proof}

\subsection{Trade-off Coding Subsumes Time Sharing}

\label{sec-ccn:trade-off-vs-TS}Before proceeding to other trade-off coding
settings, we show how time sharing emerges as a special case of a trade-off
coding strategy. Recall from \eqref{eq-ccn:time-sharing} that time sharing can
achieve the rate $\lambda\chi(\mathcal{N})+\left(  1-\lambda\right)
I(\mathcal{N})$ for any $\lambda$ such that $0\leq\lambda\leq1$. Suppose that
$\phi_{AA^{\prime}}$ is the pure state that maximizes the channel mutual
information $I(\mathcal{N})$, and suppose that $\{p_{X}(x),\psi_{A^{\prime}%
}^{x}\}$ is an ensemble of pure states that maximizes the channel Holevo
information $\chi(\mathcal{N})$ (recall from
Theorem~\ref{thm-add:pure-states-suff-holevo} that it is sufficient to
consider pure states for maximizing the Holevo information of a channel). Time
sharing simply mixes between these two strategies, and we can construct a
classical--quantum state of the form in
\eqref{eq-ccn:ce-trade-off-code-state}, for which time sharing turns out to be
the strategy executed by the constructed trade-off code:%
\begin{multline}
\sigma_{UXAB}\equiv\left(  1-\lambda\right)  |0\rangle\langle0|_{U}%
\otimes|0\rangle\langle0|_{X}\otimes\mathcal{N}_{A^{\prime}\rightarrow B}%
(\phi_{AA^{\prime}})\\
+\lambda|1\rangle\langle1|_{U}\otimes\sum_{x}p_{X}(x)|x\rangle\langle
x|_{X}\otimes|0\rangle\langle0|_{A}\otimes\mathcal{N}_{A^{\prime}\rightarrow
B}(\psi_{A^{\prime}}^{x}).
\end{multline}
In the above, the register $U$ is acting as a classical binary flag to
indicate whether the code should be an entanglement-assisted classical
capacity-achieving code or a code that achieves the channel's Holevo
information. The amount of classical bits that Alice can communicate to Bob
using a trade-off code is $I(AUX;B)_{\sigma}$, where we have assumed that $U$
and $X$ together form the classical register. We can then evaluate this mutual
information by applying the chain rule:%
\begin{align}
I(AUX;B)_{\sigma}  &  =I(A;B|XU)_{\sigma}+I(X;B|U)_{\sigma}+I(U;B)_{\sigma}\\
&  =\left(  1-\lambda\right)  I(A;B)_{\mathcal{N}(\phi)}+\lambda\left[
\sum_{x}p_{X}(x)I(A;B)_{|0\rangle\langle0|\otimes\mathcal{N}(\psi^{x}%
)}\right]  +\nonumber\\
&  \ \ \ \ \ \ \ \left(  1-\lambda\right)  I(X;B)_{|0\rangle\langle
0|\otimes\mathcal{N}(\phi)}+\lambda I(X;B)_{\left\{  p(x),\psi^{x}\right\}
}+I(U;B)_{\sigma}\\
&  \geq\left(  1-\lambda\right)  I(\mathcal{N})+\lambda\chi(\mathcal{N}).
\end{align}
The second equality follows by evaluating the first two conditional mutual
informations. The inequality follows from the assumptions that $I(\mathcal{N}%
)=I(A;B)_{\mathcal{N}(\phi)}$ and $\chi(\mathcal{N})=I(X;B)_{\left\{
p(x),\psi^{x}\right\}  }$, the fact that quantum mutual information vanishes
on product states, and $I(U;B)_{\sigma}\geq0$. Thus, in certain cases, this
strategy might do slightly better than time sharing, but for channels for
which $\phi_{A^{\prime}}=\sum_{x}p(x)\psi_{A^{\prime}}^{x}$, this strategy is
equivalent to time sharing because $I(U;B)_{\sigma}=0$ in this latter case.

Thus, time sharing emerges as a special case of trade-off coding. In general,
we can try to see if trade-off coding beats time sharing for certain channels
by optimizing the rates in Theorem~\ref{thm-ccn:ce-trading}\ over all possible
choices of states of the form in \eqref{eq-ccn:ce-trade-off-code-state}.

\subsection{Trading between Coherent and Classical Communication}

We obtain the following corollary of Theorem~\ref{thm-ccn:ce-trading}, simply
by upgrading the $\left\vert \mathcal{X}\right\vert $ entanglement-assisted
classical codes to entanglement-assisted coherent codes. The upgrading is
along the same lines as that in the proof of Theorem~\ref{thm-eac:ea-coh}, and
for this reason, we omit the proof.

\begin{corollary}
\label{thm-ccn:coherent-e-trading}The following resource inequality
corresponds to an achievable protocol for entanglement-assisted coherent
communication over a quantum channel $\mathcal{N}$:%
\begin{equation}
\left\langle \mathcal{N}\right\rangle +H(A|X)_{\rho}\left[  qq\right]  \geq
I(A;B|X)_{\rho}\left[  q\rightarrow qq\right]  +I(X;B)_{\rho}\left[
c\rightarrow c\right]  ,
\end{equation}
where $\rho_{XAB}$ is a state of the following form:%
\begin{equation}
\rho_{XAB}\equiv\sum_{x}p_{X}(x)|x\rangle\langle x|_{X}\otimes\mathcal{N}%
_{A^{\prime}\rightarrow B}(\varphi_{AA^{\prime}}^{x}),
\end{equation}
and the states $\varphi_{AA^{\prime}}^{x}$ are pure.
\end{corollary}

\subsection{Trading between Classical Communication and Entanglement-Assisted
Quantum Communication}

\label{sec-ccn:trade-off-CQE}We end this section with a protocol that achieves
entanglement-assisted communication of both classical and quantum information.
It is essential to the trade-off between a quantum channel and the three
resources of noiseless classical communication, noiseless quantum
communication, and noiseless entanglement. We study this trade-off in full
detail in Chapter~\ref{chap:trade-off}, where we show that combining this
protocol with teleportation, super-dense coding, and entanglement distribution
is sufficient to achieve any task in dynamic quantum Shannon theory involving
the three unit resources.

\begin{corollary}
[CQE\ Trade-off Coding]\label{cor-ccn:CQE-trading}The following resource
inequality corresponds to an achievable protocol for entanglement-assisted
communication of classical and quantum information over a quantum channel
$\mathcal{N}_{A^{\prime}\rightarrow B}$:%
\begin{equation}
\left\langle \mathcal{N}\right\rangle +\frac{1}{2}I(A;E|X)_{\rho}\left[
qq\right]  \geq\frac{1}{2}I(A;B|X)_{\rho}\left[  q\rightarrow q\right]
+I(X;B)_{\rho}\left[  c\rightarrow c\right]  , \label{eq-ccn:CQE-protocol}%
\end{equation}
where $\rho_{XABE}$ is a state of the following form:%
\begin{equation}
\rho_{XABE}\equiv\sum_{x}p_{X}(x)|x\rangle\langle x|_{X}\otimes\mathcal{U}%
_{A^{\prime}\rightarrow BE}^{\mathcal{N}}(\varphi_{AA^{\prime}}^{x}),
\end{equation}
the states $\varphi_{AA^{\prime}}^{x}$ are pure, and $U_{A^{\prime}\rightarrow
BE}^{\mathcal{N}}$ is an isometric extension of the channel $\mathcal{N}%
_{A^{\prime}\rightarrow B}$.
\end{corollary}

\begin{proof}
Consider the following chain of resource inequalities:%
\begin{align}
&  \left\langle \mathcal{N}\right\rangle +H(A|X)_{\rho}\left[  qq\right]
\nonumber\\
&  \geq I(A;B|X)_{\rho}\left[  q\rightarrow qq\right]  +I(X;B)_{\rho}\left[
c\rightarrow c\right] \\
&  \geq\frac{1}{2}I(A;B|X)_{\rho}\left[  qq\right]  +\frac{1}{2}%
I(A;B|X)_{\rho}\left[  q\rightarrow q\right]  +I(X;B)_{\rho}\left[
c\rightarrow c\right]  .
\end{align}
The first inequality is the statement in
Corollary~\ref{thm-ccn:coherent-e-trading}, and the second inequality follows
from the coherent communication identity. After resource cancelation and
noting that $H(A|X)_{\rho}-\frac{1}{2}I(A;B|X)_{\rho}=\frac{1}{2}%
I(A;E|X)_{\rho}$, the resulting resource inequality is equivalent to the one
in \eqref{eq-ccn:CQE-protocol}.
\end{proof}

\subsection{Trading between Classical and Quantum Communication}

Our final trade-off coding protocol that we consider is that between classical
and quantum communication. The proof of the  resource inequality below follows
by combining the protocol in Corollary~\ref{cor-ccn:CQE-trading}\ with
entanglement distribution, in much the same way as we did in
Corollary~\ref{cor-ccn:q-comm}. Thus, we omit the proof.

\begin{corollary}
[CQ Trade-off Coding]\label{thm-ccn:CQ-trading}The following resource
inequality corresponds to an achievable protocol for simultaneous classical
and quantum communication over a quantum channel $\mathcal{N}_{A^{\prime
}\rightarrow B}$:%
\begin{equation}
\left\langle \mathcal{N}\right\rangle \geq I(A\rangle BX)_{\rho}\left[
q\rightarrow q\right]  +I(X;B)_{\rho}\left[  c\rightarrow c\right]  ,
\end{equation}
where $\rho_{XAB}$ is a state of the following form:%
\begin{equation}
\rho_{XAB}\equiv\sum_{x}p_{X}(x)|x\rangle\langle x|_{X}\otimes\mathcal{N}%
_{A^{\prime}\rightarrow B}(\varphi_{AA^{\prime}}^{x}),
\end{equation}
and the states $\varphi_{AA^{\prime}}^{x}$ are pure.
\end{corollary}

\section{Concluding Remarks}

The maintainence of quantum coherence is the theme of this chapter. Alice and
Bob can execute powerful protocols if they perform encoding and decoding in
superposition. In both entanglement-assisted coherent communication and
coherent state transfer, Alice performs controlled gates instead of
conditional gates and Bob performs coherent measurements that place
measurement outcomes in an ancilla register without destroying superpositions.
Also, Bob's final action in both of these protocols is to perform a controlled
decoupling unitary, ensuring that the state of the environment is independent
of Alice and Bob's final state. Thus, the same protocol accomplishes the
different tasks of entanglement-assisted coherent communication and coherent
state transfer, and these in turn can generate a whole host of other protocols
by combining them with entanglement distribution and the coherent and
incoherent versions of teleportation and super-dense coding. Among these other
generated protocols are entanglement-assisted quantum communication, quantum
communication, quantum-assisted state transfer, and classical-assisted state
transfer. The exercises in this chapter explore further possibilities if Alice
has access to the environments of the different protocols---the most general
version of coherent teleportation arises in such a case.

Trade-off coding is the theme of the last part of this chapter. Here, we are
addressing the question:\ Given a fixed amount of a certain resource, how much
of another resource can Alice and Bob generate? Noisy quantum channels are the
most fundamental description of a medium over which information can propagate,
and it is thus important to understand the best ways to make effective use of
such a resource for a variety of purposes. We determined a protocol that
achieves the task of entanglement-assisted communication of classical and
quantum information, simply by combining the protocols we have already found
for classical communication and entanglement-assisted coherent communication.
Chapter~\ref{chap:trade-off}\ continues this theme of trade-off coding in a
much broader context and demonstrates that the protocol given here, when
combined with teleportation, super-dense coding, and entanglement
distribution, is optimal for some channels of interest and essentially optimal
in the general case.

\section{History and Further Reading}

\cite{DHW03,DHW05RI} showed that it is possible to make the protocols for
entanglement-assisted classical communication and noisy super-dense coding
coherent, leading to Theorems~\ref{thm-eac:eaq-rate} and \ref{thm-ccn:qast}.
They called these protocols the \textquotedblleft father\textquotedblright%
\ and \textquotedblleft mother,\textquotedblright\ respectively, because they
generated many other protocols in quantum Shannon theory by combining them
with entanglement distribution, teleportation, and super-dense coding.
\cite{H3LT01} formulated a protocol for noisy super-dense coding, but our
protocol here makes use of the coding technique in \cite{itit2008hsieh}.
\cite{Shor_CE} first proved a coding theorem for trading between assisted and
unassisted classical communication, and \cite{cmp2005dev} followed up on this
result by finding a scheme for trade-off coding between classical and quantum
communication. Some time later, \cite{HW08GFP}\ generalized these two coding
schemes to produce the result of Theorem~\ref{cor-ccn:CQE-trading}.\ The
proofs given in this chapter for these trade-off results are different from
those which appeared in \citep{Shor_CE,cmp2005dev,HW08GFP}.

\chapter{Private Classical Communication}

\label{chap:private-cap}We have now seen in
Chapters~\ref{chap:classical-comm-HSW}--\ref{chap:coh-comm-noisy} how Alice
can communicate classical or quantum information to Bob, perhaps even with the
help of shared entanglement. One might argue that these communication tasks
are the most fundamental tasks in quantum Shannon theory, given that they have
furthered our understanding of the nature of information transmission over
quantum channels. However, when discussing the communication of classical
information, we made no stipulation as to whether this classical information
should be public, so that any third party might have partial or full access to
it, or private, so that no third party has access.

This chapter establishes the
\index{private classical communication}
private classical capacity theorem, which gives the maximum rate at which
Alice can communicate classical information privately to Bob without anyone
else in the universe knowing what she sent to him. A variation of the
information-processing task corresponding to this theorem was one of the
earliest studied in quantum information theory, with the Bennett--Brassard-84
\index{quantum key distribution}%
quantum key distribution protocol being the first proposed protocol for
exploiting quantum mechanics to establish a shared secret key between two
parties. The private classical capacity theorem is important for quantum key
distribution because it establishes the maximum rate at which two parties can
generate a shared secret key.

Another equally important, but less obvious utility of private classical
communication is in establishing a protocol for quantum communication at the
coherent information rate. Section~\ref{sec-ccn:q-cap}\ demonstrated a
somewhat roundabout way of arriving at the conclusion that it is possible to
communicate quantum information reliably at the coherent information
rate---recall that we \textquotedblleft coherified\textquotedblright\ the
entanglement-assisted classical capacity theorem and then exploited the
coherent communication identity and catalytic use of entanglement.
Establishing achievability of the coherent information rate via private
classical coding is another way of arriving at the same result, with the added
benefit that the resulting protocol does not require the catalytic use of entanglement.

The intuition for quantum communication via privacy arises from the%
\index{no-cloning theorem}
no-cloning theorem. Suppose that Alice is able to communicate private
classical messages to Bob, so that the channel's environment (Eve) is not able
to distinguish which message Alice is transmitting to Bob. That is, Eve's
state is completely independent of Alice's message if the transmitted message
is private. Then we might expect it to be possible to make a coherent version
of this private classical code by exploiting superpositions of the private
classical codewords. Since Eve's states are independent of the quantum message
that Alice is sending through the channel, she is not able to
\textquotedblleft steal\textquotedblright\ any of the coherence in Alice's
superposed states. Given that the overall evolution of the channel to Bob and
Eve is unitary and the fact that Eve does not receive any quantum information
with this scheme, we should expect that the quantum information appears at the
receiving end of the channel so that Bob can decode it. Were Eve able to
obtain any information about the private classical messages, then Bob would
not be able to decode all of the quantum information when they construct a
coherent version of this private classical code. Otherwise, they would violate
the no-cloning theorem. We discuss this important application of private
classical communication in the next chapter.

This chapter follows a similar structure as previous chapters. We first detail
the information-processing task for private classical communication.
Section~\ref{sec-pcc:theorem}\ then states the private classical capacity
theorem, with the following two sections proving the achievability part and
the converse part. We end with a general discussion of the private classical
capacity and a brief overview of the secret-key-assisted private classical capacity.

\section{The Information-Processing Task}

We begin by describing the information-processing task for private classical
communication (we define an $\left(  n,P,\varepsilon\right)  $ private
classical code). Alice selects a message $m$ from a set $\mathcal{M}$ of
messages. Alice prepares some state $\rho_{A^{\prime n}}^{m}$ as input to many
uses of the quantum channel~$\mathcal{N}_{A^{\prime}\rightarrow B}$ and
transmits it, producing the following state at Bob's receiving end:%
\begin{equation}
\mathcal{N}_{A^{\prime n}\rightarrow B^{n}}(\rho_{A^{\prime n}}^{m}),
\end{equation}
where $\mathcal{N}_{A^{\prime n}\rightarrow B^{n}}\equiv(\mathcal{N}%
_{A^{\prime}\rightarrow B})^{\otimes n}$.

Bob employs a decoding POVM $\left\{  \Lambda_{m}\right\}  $ in order to
detect Alice's transmitted message $m$. The probability of error for a
particular message $m$ is as follows:%
\begin{equation}
p_{e}(m)=\operatorname{Tr}\left\{  \left(  I-\Lambda_{m}\right)
\mathcal{N}_{A^{\prime n}\rightarrow B^{n}}(\rho_{A^{\prime n}}^{m})\right\}
,
\end{equation}
so that the maximal probability of error is%
\begin{equation}
p_{e}^{\ast}\equiv\max_{m\in\mathcal{M}}p_{e}(m),
\end{equation}
where $p_{e}^{\ast}\leq\varepsilon\in\left[  0,1\right]  $ for an
$(n,P,\varepsilon)$ code. The rate $P$ of the code is%
\begin{equation}
P\equiv\frac{1}{n}\log\left\vert \mathcal{M}\right\vert .
\end{equation}

\begin{figure}
[ptb]
\begin{center}
\includegraphics[
width=4.5455in
]%
{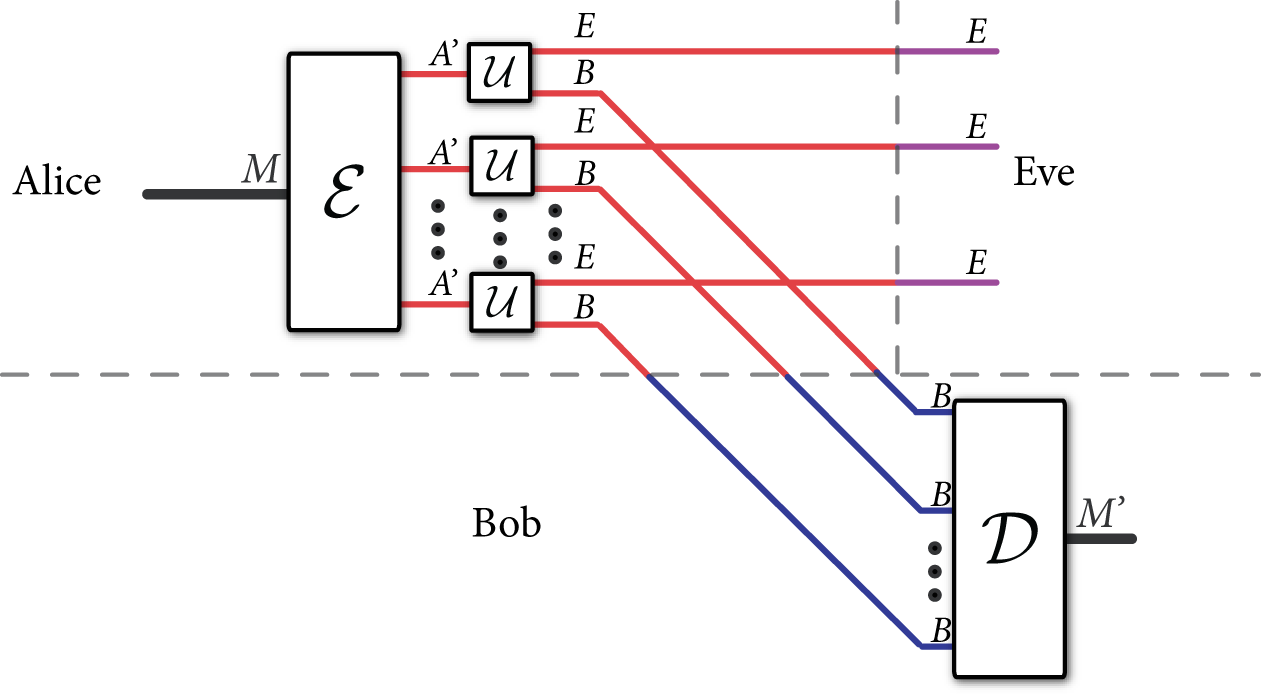}%
\caption{The information-processing task for private classical communication.
Alice encodes some private message~$m$ into a quantum codeword $\rho
_{A^{\prime n}}^{m}$ and transmits it over many uses of a quantum channel. The
goal of such a protocol is for Bob to be able to reliably distinguish the
message, while the channel's environment Eve should not be able to learn
anything about it.}%
\label{fig-pcc:private-info-task}%
\end{center}
\end{figure}

So far, the above specification of a private classical code is nearly
identical to that for the transmission of classical information outlined in
Section~\ref{sec-cc:info-proc-task}. What distinguishes a private classical
code from a public one is the following extra condition for privacy. Let
$U_{A^{\prime}\rightarrow BE}^{\mathcal{N}}$ be an isometric extension of the
channel $\mathcal{N}_{A^{\prime}\rightarrow B}$, so that the complementary
channel~$\widehat{\mathcal{N}}_{A^{\prime}\rightarrow E}$ to the environment
Eve is as follows:%
\begin{equation}
\widehat{\mathcal{N}}_{A^{\prime}\rightarrow E}(\sigma)\equiv\operatorname{Tr}%
_{B}\{\mathcal{U}_{A^{\prime}\rightarrow BE}^{\mathcal{N}}(\sigma)\}.
\end{equation}
If Alice transmits a message $m$, then the state for Eve is as follows:%
\begin{equation}
\omega_{E^{n}}^{m}\equiv\widehat{\mathcal{N}}_{A^{\prime n}\rightarrow E^{n}%
}(\rho_{A^{\prime n}}^{m}).
\end{equation}
Our condition for $\varepsilon$-privacy is that Eve's state is always close to
a constant state $\sigma_{E^{n}}$, regardless of which message~$m$ Alice
transmits through the channel:%
\begin{equation}
\forall m\in\mathcal{M}:\frac{1}{2}\left\Vert \omega_{E^{n}}^{m}-\sigma
_{E^{n}}\right\Vert _{1}\leq\varepsilon. \label{eq-pcc:security-crit}%
\end{equation}
This definition is the strongest definition of privacy because it implies that
Eve cannot learn anything about the message$~m$ that Alice transmits through
the channel. Figure~\ref{fig-pcc:private-info-task}\ depicts the
information-processing task for private classical communication.%

A rate$~P$ of private classical communication is achievable for $\mathcal{N}%
_{A^{\prime}\rightarrow B}$\ if there exists an $\left(  n,P-\delta
,\varepsilon\right)  $ private classical code for all $\varepsilon\in(0,1)$,
$\delta>0$, and sufficiently large$~n$, where $\varepsilon$ characterizes both
the reliability and the privacy of the code. The private classical capacity
$C_{P}(\mathcal{N})$\ of a channel $\mathcal{N}_{A^{\prime}\rightarrow B}$ is
equal to the supremum of all achievable rates for private classical communication.

\subsection{Mutual Information of Eve}

\label{sec-pcc:mut-info-Eve}We comment briefly how the condition in
\eqref{eq-pcc:security-crit}\ implies that Eve has little mutual information
about the transmitted message. It follows from
\eqref{eq-pcc:security-crit}\ that%
\begin{equation}
\varepsilon\geq\frac{1}{2}\sum_{m\in\mathcal{M}}\frac{1}{\left\vert
\mathcal{M}\right\vert }\left\Vert \omega_{E^{n}}^{m}-\sigma_{E^{n}%
}\right\Vert _{1}=\frac{1}{2}\left\Vert \omega_{ME^{n}}-\pi_{M}\otimes
\sigma_{E^{n}}\right\Vert _{1}, \label{eq-pcc:average-security}%
\end{equation}
where%
\begin{equation}
\omega_{ME^{n}}\equiv\sum_{m\in\mathcal{M}}\frac{1}{\left\vert \mathcal{M}%
\right\vert }|m\rangle\langle m|_{M}\otimes\omega_{E^{n}}^{m}.
\end{equation}
The criterion in \eqref{eq-pcc:average-security} implies that Eve's Holevo
information with $M$ is small:%
\begin{align}
I(M;E^{n})_{\omega}  &  =H(M)_{\omega}-H(M|E^{n})_{\omega}\\
&  =H(M|E^{n})_{\pi\otimes\sigma}-H(M|E^{n})_{\omega}\\
&  \leq\varepsilon\log\left\vert \mathcal{M}\right\vert +g_2(\varepsilon).
\end{align}
The inequality follows from applying the AFW inequality
(Theorem~\ref{thm-qie:AFW-cont-ent}) to both entropies. Thus, if $\varepsilon$
is exponentially small in $n$ (which will be the case for our codes), then it
is possible to make Eve's information about the message become arbitrarily
small in the asymptotic limit.

\section{The Private Classical Capacity Theorem}

\label{sec-pcc:theorem}We now state the main theorem of this chapter, the
\index{private classical capacity theorem}%
private classical capacity theorem.

\begin{theorem}
[Devetak--Cai--Winter--Yeung]\label{thm-pcc:private-capacity}The private
classical capacity $C_{P}(\mathcal{N})$ of a quantum channel$~\mathcal{N}%
_{A^{\prime}\rightarrow B}$ is equal to the regularized private information of
the channel:%
\begin{equation}
C_{P}(\mathcal{N})=P_{\operatorname{reg}}(\mathcal{N}),
\end{equation}
where%
\begin{equation}
P_{\operatorname{reg}}(\mathcal{N})\equiv\lim_{k\rightarrow\infty}\frac{1}%
{k}P(\mathcal{N}^{\otimes k}). \label{eq-pcc:regularization-private}%
\end{equation}
The private information $P(\mathcal{N})$ is defined as%
\begin{equation}
P(\mathcal{N})\equiv\max_{\rho}\left[  I(X;B)_{\sigma}-I(X;E)_{\sigma}\right]
, \label{eq-pcc:private-information}%
\end{equation}
where $\rho_{XA^{\prime}}$ is a classical--quantum state of the following
form:%
\begin{equation}
\rho_{XA^{\prime}}\equiv\sum_{x}p_{X}(x)|x\rangle\langle x|_{X}\otimes
\rho_{A^{\prime}}^{x},
\end{equation}
and $\sigma_{XBE}\equiv\mathcal{U}_{A^{\prime}\rightarrow BE}^{\mathcal{N}%
}(\rho_{XA^{\prime}})$, with $U_{A^{\prime}\rightarrow BE}^{\mathcal{N}}$ an
isometric extension of the channel $\mathcal{N}_{A^{\prime}\rightarrow B}$.
\index{private classical capacity theorem}%

\end{theorem}

We first prove the achievability part of the coding theorem and follow with
the converse proof. Recall that the private information is additive whenever
the channel is
\index{degradable channel}%
degradable (Theorem~\ref{thm:private-additivity-degradable}). Thus, for this
class of channels, the regularization in \eqref{eq-pcc:regularization-private}
is not necessary and the private information of the channel is equal to the
private classical capacity (in fact, the results from
Theorem~\ref{thm-ie:degradable-priv-coh}\ and the next chapter demonstrate
that the private information of a degradable channel is also equal to its
quantum capacity). The regularization of the private information seems to be
necessary in general in order to characterize the private capacity because
there is an example of a channel for which the private information is superadditive.

\section{The Direct Coding Theorem}

\label{sec-pcc:direct-coding}This section gives a proof that the private
information in \eqref{eq-pcc:private-information} is an achievable rate for
\index{private classical capacity theorem!direct part}
private classical communication over a quantum channel~$\mathcal{N}%
_{A^{\prime}\rightarrow B}$. We first give the intuition behind the protocol.
Alice's goal is to build a doubly indexed codebook $\left\{  x^{n}%
(m,k)\right\}  _{m\in\mathcal{M},k\in\mathcal{K}}$ that satisfies two properties:

\begin{enumerate}
\item Bob should be able to detect the message$~m$ and the \textquotedblleft
junk\textquotedblright\ variable$~k$ with high probability. From the classical
coding theorem of Chapter~\ref{chap:classical-comm-HSW}, our intuition is that
he should be able to do so as long as $\left\vert \mathcal{M}\right\vert
\left\vert \mathcal{K}\right\vert \approx2^{nI( X;B) }$.

\item Randomizing over the \textquotedblleft junk\textquotedblright%
\ variable$~k$ should approximately cover the typical subspace of Eve's
system, so that every state of Eve depending on the message~$m$ looks like a
constant, independent of the message$~m$ Alice sends (we would like the code
to satisfy \eqref{eq-pcc:security-crit}). Our intuition from the covering
lemma (Chapter~\ref{chap:covering-lemma}) is that the size of the
\textquotedblleft junk\textquotedblright\ variable set~$\mathcal{K}$ needs to
be at least $\left\vert \mathcal{K}\right\vert \approx2^{nI(X;E)}$ in order
for Alice to approximately cover Eve's typical subspace.
\end{enumerate}

\begin{figure}
[ptb]
\begin{center}
\includegraphics[
width=4.8456in
]%
{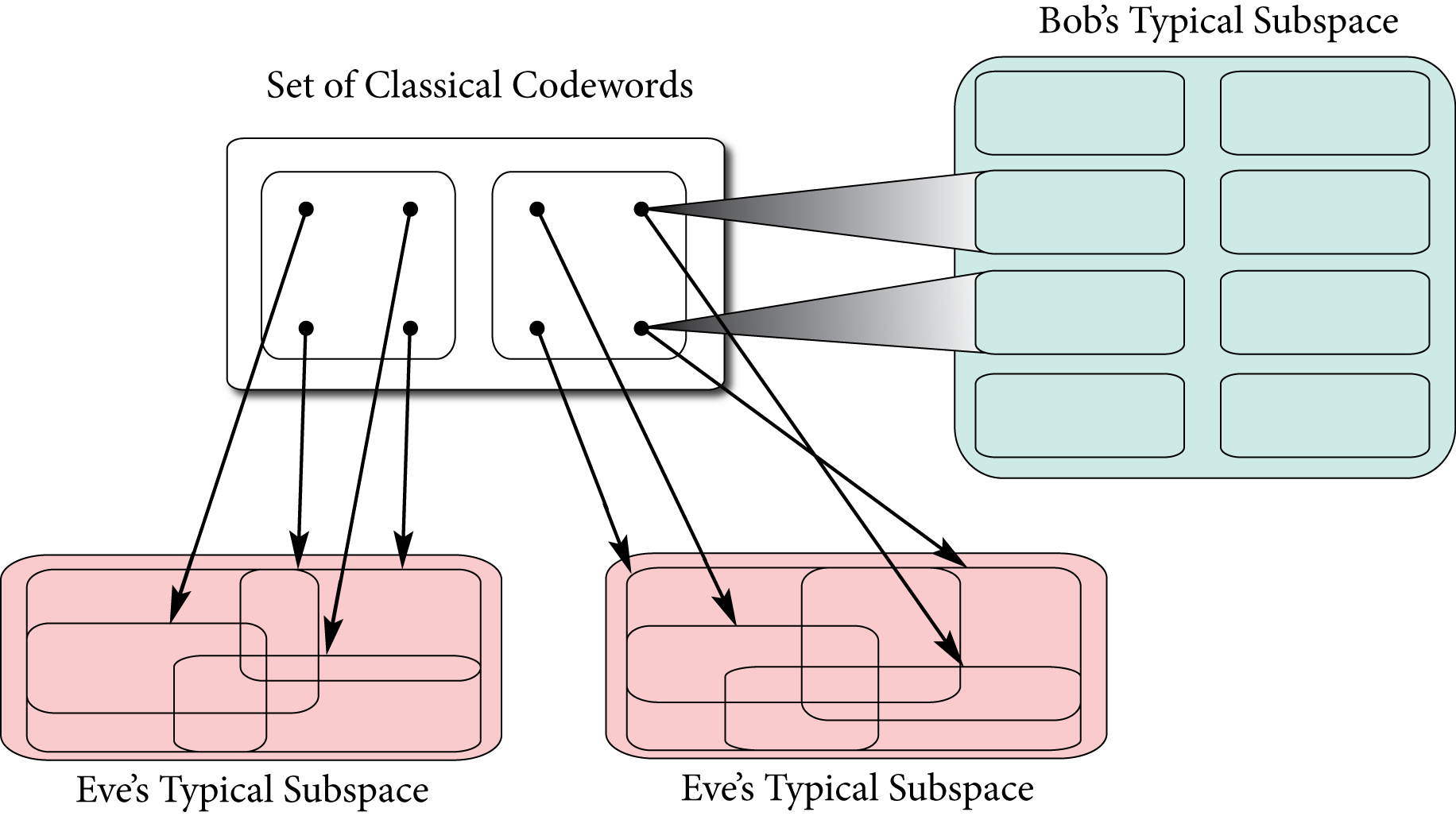}%
\caption{The anatomy of a code for private classical communication. In this
illustrative example, Alice has eight codewords, with each depicted as a
$\bullet$ and indexed by $m\in\left\{  1,2\right\}  $ and $k\in\left\{
1,2,3,4\right\}  $. Thus, she is interested in sending one of two messages and
has the \textquotedblleft junk\textquotedblright\ variable$~k$ available for
randomizing Eve's state. Each classical codeword $x^{n}(m,k)$ maps to a
distinguishable subspace on Bob's typical subspace (we show two of the
mappings in the figure, while displaying eight distinguishable subspaces).
From the packing lemma, our intuition is that Alice can reliably send about
$2^{nI(X;B)}$ distinguishable signals. The codewords $\left\{  x^{n}%
(1,k)\right\}  _{k\in\left\{  1,2,3,4\right\}  }$ and $\left\{  x^{n}%
(2,k)\right\}  _{k\in\left\{  1,2,3,4\right\}  }$ are each grouped in a box to
indicate that they form a privacy amplification set. When randomizing $k$, the
codewords $\left\{  x^{n}(1,k)\right\}  _{k\in\left\{  1,2,3,4\right\}  }$
uniformly cover Eve's typical subspace (and so does the set $\left\{
x^{n}(2,k)\right\}  _{k\in\left\{  1,2,3,4\right\}  }$), so that it becomes
nearly impossible in the asymptotic limit for Eve to distinguish whether Alice
is sending a codeword in $\left\{  x^{n}(1,k)\right\}  _{k\in\left\{
1,2,3,4\right\}  }$ or $\left\{  x^{n}(2,k)\right\}  _{k\in\left\{
1,2,3,4\right\}  }$. In this way, Eve cannot determine which message Alice is
transmitting. The minimum size for each privacy amplification set in the
asymptotic limit is $\approx2^{nI(X;E)}$.}%
\label{fig-pcc:private-classical-intuition}%
\end{center}
\end{figure}

Our method for generating a code is again random because we can invoke the
typicality properties that hold in the asymptotic limit of many channel uses.
Thus, if Alice chooses a code that satisfies the above criteria, she can send
approximately $\left\vert \mathcal{M}\right\vert \approx2^{n\left[
I(X;B)-I(X;E)\right]  }$ distinguishable signals to Bob such that they are
indistinguishable to Eve. We devote the remainder of this section to proving
that the above intuition is correct.
Figure~\ref{fig-pcc:private-classical-intuition}\ displays the anatomy of a
private classical code.%

\subsection{Dimensionality Arguments}

Before giving the proof of achievability, we confirm the above intuition with
some dimensionality arguments and show how to satisfy the conditions of both
the packing and covering lemmas. Suppose that Alice has some ensemble
$\{p_{X}(x),\rho_{A^{\prime}}^{x}\}$ from which she can generate random codes.
Let~$U_{A^{\prime}\rightarrow BE}^{\mathcal{N}}$ denote an isometric extension
of the channel~$\mathcal{N}_{A^{\prime}\rightarrow B}$, and let $\rho_{BE}%
^{x}$ denote the joint state of Bob and Eve after Alice inputs $\rho
_{A^{\prime}}^{x}$:%
\begin{equation}
\rho_{BE}^{x}\equiv\mathcal{U}_{A^{\prime}\rightarrow BE}^{\mathcal{N}}%
(\rho_{A^{\prime}}^{x}).
\end{equation}
The local respective density operators for Bob and Eve given a letter $x$ are
as follows:%
\begin{equation}
\rho_{B}^{x}\equiv\operatorname{Tr}_{E}\left\{  \rho_{BE}^{x}\right\}
,\ \ \ \ \ \ \ \ \rho_{E}^{x}\equiv\operatorname{Tr}_{B}\left\{  \rho_{BE}%
^{x}\right\}  .
\end{equation}
The expected respective density operators for Bob and Eve are as follows:%
\begin{equation}
\rho_{B}=\sum_{x}p_{X}(x)\rho_{B}^{x},\ \ \ \ \ \ \ \ \rho_{E}=\sum_{x}%
p_{X}(x)\rho_{E}^{x}.
\end{equation}
Given a particular input sequence $x^{n}$, we define the $n$th extensions of
the above states as follows:%
\begin{align}
\rho_{B^{n}}^{x^{n}}  &  \equiv\operatorname{Tr}_{E^{n}}\left\{  \rho
_{B^{n}E^{n}}^{x^{n}}\right\}  ,\\
\rho_{E^{n}}^{x^{n}}  &  \equiv\operatorname{Tr}_{B^{n}}\left\{  \rho
_{B^{n}E^{n}}^{x^{n}}\right\}  ,\\
\rho_{B^{n}}  &  =\sum_{x^{n}\in\mathcal{X}^{n}}p_{X^{n}}(x^{n})\rho_{B^{n}%
}^{x^{n}},\\
\rho_{E^{n}}  &  =\sum_{x^{n}\in\mathcal{X}^{n}}p_{X^{n}}(x^{n})\rho_{E^{n}%
}^{x^{n}}. \label{eq-pcc:eve-average-state}%
\end{align}
Table~\ref{TableKey}\ organizes these various density operators and their
corresponding typical subspaces and projectors.%
\begin{table}[tbp] \centering
\begin{tabular}
[c]{l|l|l|l}\hline\hline
\textbf{Party} & \textbf{Quantity} & \textbf{Typical Set/Subspace} &
\textbf{Projector}\\\hline\hline
Alice & $X$ & $T_{\delta}^{X^{n}}$ & N/A\\
Bob & $\rho_{B^{n}}$ & $T_{B^{n}}^{\delta}$ & $\Pi_{B^{n}}^{\delta}$\\
Bob conditioned on $x^{n}$ & $\rho_{B^{n}}^{x^{n}}$ & $T_{B^{n}|x^{n}}%
^{\delta}$ & $\Pi_{B^{n}|x^{n}}^{\delta}$\\
Eve & $\rho_{E^{n}}$ & $T_{E^{n}}^{\delta}$ & $\Pi_{E^{n}}^{\delta}$\\
Eve conditioned on $x^{n}$ & $\rho_{E^{n}}^{x^{n}}$ & $T_{E^{n}|x^{n}}%
^{\delta}$ & $\Pi_{E^{n}|x^{n}}^{\delta}$\\\hline\hline
\end{tabular}
\caption{This table lists several mathematical quantities
involved in the construction of a random private code. The first column
lists the party to whom the quantities belong. The second column lists the random
classical or quantum states. The third column gives the
appropriate typical set or subspace. The final column lists the appropriate
projector onto the typical subspace for the quantum states.}\label{TableKey}%
\end{table}%

The following four conditions corresponding to the packing lemma hold for
Bob's states $\left\{  \rho_{B^{n}}^{x^{n}}\right\}  $, Bob's average density
operator $\rho_{B^{n}}$, Bob's typical subspace $T_{B^{n}}^{\delta}$, and
Bob's conditionally typical subspace $T_{B^{n}|x^{n}}^{\delta}$:%
\begin{align}
\operatorname{Tr}\left\{  \Pi_{B^{n}}^{\delta}\rho_{B^{n}}^{x^{n}}\right\}
&  \geq1-\varepsilon,\label{eq-pcc:bob-pack-1}\\
\operatorname{Tr}\left\{  \Pi_{B^{n}|x^{n}}^{\delta}\rho_{B^{n}}^{x^{n}%
}\right\}   &  \geq1-\varepsilon,\\
\operatorname{Tr}\left\{  \Pi_{B^{n}|x^{n}}^{\delta}\right\}   &
\leq2^{n\left(  H(B|X)+c\delta\right)  },\\
\Pi_{B^{n}}^{\delta}\rho_{B^{n}}\Pi_{B^{n}}^{\delta}  &  \leq2^{-n\left(
H(B)-c\delta\right)  }\Pi_{B^{n}}^{\delta}, \label{eq-pcc:bob-pack-4}%
\end{align}
where $c$ is some positive constant (see
Properties~\ref{prop-qt:cond-state-with-uncond-proj}, \ref{prop-qt:unit},
\ref{prop-qt:exp-small}, and \ref{prop-qt:equi}).

The following four conditions corresponding to the covering lemma hold for
Eve's states $\left\{  \rho_{E^{n}}^{x^{n}}\right\}  $, Eve's typical subspace
$T_{E^{n}}^{\delta}$, and Eve's conditionally typical subspace $T_{E^{n}%
|x^{n}}^{\delta}$:%
\begin{align}
\operatorname{Tr}\left\{  \Pi_{E^{n}}^{\delta}\rho_{E^{n}}^{x^{n}}\right\}
&  \geq1-\varepsilon,\\
\operatorname{Tr}\left\{  \Pi_{E^{n}|x^{n}}^{\delta}\rho_{E^{n}}^{x^{n}%
}\right\}   &  \geq1-\varepsilon,\\
\operatorname{Tr}\left\{  \Pi_{E^{n}}^{\delta}\right\}   &  \leq2^{n\left(
H(E)+c\delta\right)  },\\
\Pi_{E^{n}|x^{n}}^{\delta}\rho_{E^{n}}^{x^{n}}\Pi_{E^{n}|x^{n}}^{\delta}  &
\leq2^{-n\left(  H(E|X)-c\delta\right)  }\Pi_{E^{n}|x^{n}}^{\delta}.
\end{align}

The above properties suggest that we can use the methods of both the packing
lemma and the covering lemma for constructing a private code. Consider two
sets $\mathcal{M}$ and $\mathcal{K}$ with the following respective sizes:%
\begin{align}
\left\vert \mathcal{M}\right\vert  &  =2^{n\left[  I(X;B)-I(X;E)-6c\delta
\right]  },\\
\left\vert \mathcal{K}\right\vert  &  =2^{n\left[  I(X;E)+3c\delta\right]  },
\label{eq-pcc:k-size}%
\end{align}
so that the product set $\mathcal{M}\times\mathcal{K}$ indexed by the ordered
pairs $(m,k)$ is of size%
\begin{equation}
\left\vert \mathcal{M}\times\mathcal{K}\right\vert =\left\vert \mathcal{M}%
\right\vert \left\vert \mathcal{K}\right\vert =2^{n\left[  I(X;B)-3c\delta
\right]  }. \label{eq-pcc:private-code-size}%
\end{equation}
The sizes of these sets suggest that we can use the product set $\mathcal{M}%
\times\mathcal{K}$\ for sending classical information, but we can use
$\left\vert \mathcal{M}\right\vert $ \textquotedblleft privacy
amplification\textquotedblright\ sets each of size $\left\vert \mathcal{K}%
\right\vert $ for reducing Eve's knowledge of the message$~m$ (see
Figure~\ref{fig-pcc:private-classical-intuition}).

\subsection{Random Code Construction}

We now argue for the existence of a good private classical code with rate
$P\approx I(X;B)-I(X;E)$ if Alice selects it randomly according to the
ensemble%
\begin{equation}
\{p_{X^{\prime n}}^{\prime}(x^{n}),\rho_{A^{\prime n}}^{x^{n}}\},
\end{equation}
where $p_{X^{\prime n}}^{\prime}(x^{n})$ is the pruned distribution (see
Section~\ref{sec-cc:direct-coding}---recall that this distribution is close to
the i.i.d.~distribution). Let us choose $\left\vert \mathcal{M}\right\vert
\left\vert \mathcal{K}\right\vert $ random variables $X^{n}(m,k)$ according to
the distribution $p_{X^{\prime n}}^{\prime}(x^{n})$ where the realizations of
the random variables $X^{n}(m,k)$ take values in $\mathcal{X}^{n}$. After
selecting these codewords randomly, the code $\mathcal{C}=\left\{
x^{n}(m,k)\right\}  _{m\in\mathcal{M},k\in\mathcal{K}}$ is then a fixed set of
codewords $x^{n}(m,k)$ depending on the message $m$ and the randomization
variable~$k$.

We first consider how well Bob can distinguish the pair $(m,k)$ and argue that
the random code is a good code in the sense that the expectation of the
average error probability over all codes is low. The packing lemma is the
basis of our argument. By applying the packing lemma
(Lemma~\ref{lem-pack:pack}) to
\eqref{eq-pcc:bob-pack-1}--\eqref{eq-pcc:bob-pack-4}, there exists a
POVM\ $\{\Lambda_{m,k}\}_{(m,k)\in\mathcal{M}\times\mathcal{K}}$ corresponding
to the random choice of code that reliably distinguishes the states
$\{\rho_{B^{n}}^{X^{n}(m,k)}\}_{m\in\mathcal{M},k\in\mathcal{K}}$ in the
following sense:%
\begin{align}
\mathbb{E}_{\mathcal{C}}\left\{  \bar{p}_{e}(\mathcal{C})\right\}   &
=1-\mathbb{E}_{\mathcal{C}}\left\{  \frac{1}{\left\vert \mathcal{M}\right\vert
\left\vert \mathcal{K}\right\vert }\sum_{m\in\mathcal{M}}\sum_{k\in
\mathcal{K}}\operatorname{Tr}\left\{  \Lambda_{m,k}\rho_{B^{n}}^{X^{n}%
(m,k)}\right\}  \right\} \\
&  \leq2\left(  \varepsilon+2\sqrt{\varepsilon}\right)  +4\left(
\frac{2^{n\left(  H(B|X)+c\delta\right)  }\left\vert \mathcal{M}%
\times\mathcal{K}\right\vert }{2^{n\left(  H(B)-c\delta\right)  }}\right) \\
&  =2\left(  \varepsilon+2\sqrt{\varepsilon}\right)  +4\left(  \frac
{2^{n\left(  H(B|X)+c\delta\right)  }2^{n\left[  I(X;B)-3c\delta\right]  }%
}{2^{n\left(  H(B)-c\delta\right)  }}\right) \\
&  =2\left(  \varepsilon+2\sqrt{\varepsilon}\right)  +4\cdot2^{-nc\delta
}\equiv\varepsilon^{\prime},
\end{align}
where the first equality follows by definition, the first inequality follows
by application of the packing lemma to the conditions in
\eqref{eq-pcc:bob-pack-1}--\eqref{eq-pcc:bob-pack-4}, the second equality
follows by substitution of \eqref{eq-pcc:private-code-size}, and the last
equality follows by a straightforward calculation. We can make $\varepsilon
^{\prime}$ arbitrarily small by choosing $n$ large enough.

Let us now consider the corresponding density operators $\rho_{E^{n}}%
^{X^{n}(m,k)}$ for Eve. Consider dividing the random code $\mathcal{C}$ into
$\left\vert \mathcal{M}\right\vert $ privacy amplification sets, each of size
$\left\vert \mathcal{K}\right\vert $. Each privacy amplification set
$\mathcal{C}_{m}\equiv\{\rho_{E^{n}}^{X^{n}(m,k)}\}_{k\in\mathcal{K}}$ of
density operators forms a good covering code according to the covering lemma
(Lemma~\ref{lemma-cov:covering}). The fake density operator of each privacy
amplification set\ $\mathcal{C}_{m}$\ is as follows:%
\begin{equation}
\hat{\rho}_{E^{n}}^{m}\equiv\frac{1}{\left\vert \mathcal{K}\right\vert }%
\sum_{k\in\mathcal{K}}\rho_{E^{n}}^{X^{n}(m,k)},
\end{equation}
because Alice chooses the randomizing variable $k$ uniformly at random. The
obfuscation error $o_{e}(\mathcal{C}_{m})$ of each privacy amplification set
$\mathcal{C}_{m}$ is as follows:%
\begin{equation}
o_{e}(\mathcal{C}_{m})\equiv\left\Vert \hat{\rho}_{E^{n}}^{m}-\rho_{E^{n}%
}\right\Vert _{1},
\end{equation}
where $\rho_{E^{n}}$ is defined in \eqref{eq-pcc:eve-average-state}. The
covering lemma (Lemma~\ref{lemma-cov:covering})\ states that the obfuscation
error for each random privacy amplification set $\mathcal{C}_{m}$\ has a high
probability of being small if $n$ is sufficiently large and $\left\vert
\mathcal{K}\right\vert $ is chosen as in \eqref{eq-pcc:k-size}:%
\begin{align}
&  \Pr\left\{  o_{e}(\mathcal{C}_{m})\leq\varepsilon+4\sqrt{\varepsilon
}+24\sqrt[4]{\varepsilon}\right\} \nonumber\\
&  \geq1-2d_{E}^{n}\exp\left\{  \frac{-\varepsilon^{3}}{4}\frac{\left\vert
\mathcal{K}\right\vert 2^{n\left(  H(E|X)-c\delta\right)  }}{2^{n\left(
H(E)+c\delta\right)  }}\right\} \\
&  =1-2d_{E}^{n}\exp\left\{  \frac{-\varepsilon^{3}}{4}\frac{2^{n\left[
I(X;E)+3c\delta\right]  }2^{n\left(  H(E|X)-c\delta\right)  }}{2^{n\left(
H(E)+c\delta\right)  }}\right\} \\
&  =1-2d_{E}^{n}\exp\left\{  \frac{-\varepsilon^{3}}{4}2^{nc\delta
}\right\}  .
\end{align}
In particular, let us choose $n$ large enough so that the following bound
holds:%
\begin{equation}
\Pr\left\{  o_{e}(\mathcal{C}_{m})\leq\varepsilon+4\sqrt{\varepsilon
}+24\sqrt[4]{\varepsilon}\right\}  \geq1-\frac{\varepsilon}{\left\vert
\mathcal{M}\right\vert }.
\end{equation}
That we can do so follows from the important fact that $\exp\left\{
-\varepsilon^{3}2^{nc\delta}/\left(  4\right)  \right\}  $ is doubly
exponentially decreasing in $n$. (We also see here why it is 
necessary to have the \textquotedblleft wiggle room\textquotedblright\ given
by an arbitrarily small, yet strictly positive $\delta$.)

This random construction already has some of the desirable features that we
are looking for in a private code just by choosing $n$ to be sufficiently
large. The expectation of Bob's average error probability for detecting the
pair $m,k$ is small, and the obfuscation error of each privacy amplification
set has a high probability of being small. Our hope is that there exists some
code for which Bob can retrieve the message $m$ with the guarantee that Eve's
state is independent of this message $m$. We argue in the next two sections
that such a good private code exists.

\subsection{Derandomization}

We now apply a derandomization argument similar to the one that is needed in
the proof of the HSW\ coding theorem. The argument in this case is more subtle
because we would like to find a code that has good classical communication
with the guarantee that it also has good privacy. We need to determine the
probability over all codes that there exists a good private code. If this
probability is non-zero, then we are sure that a good private code exists.

As we have said at the beginning of this section, a good private code has two
qualities: the code is $\varepsilon$-good for classical communication and it
is $\varepsilon$-private as well. Let $E_{0}$ denote the event that the random
code $\mathcal{C}$ is $\varepsilon$-good for classical communication:%
\begin{equation}
E_{0}=\left\{  \bar{p}_{e}(\mathcal{C})\leq\varepsilon\right\}  ,
\end{equation}
where we restrict the performance criterion to the average probability of
error for now. Let $E_{m}$ denote the event that the $m$th message in the
random code is $\varepsilon$-private:%
\begin{equation}
E_{m}=\left\{  o_{e}(\mathcal{C}_{m})\leq\varepsilon\right\}  .
\end{equation}
We would like all of the above events to be true, or, equivalently, we would
like the intersection of the above events to occur:%
\begin{equation}
E_{\operatorname{priv}}\equiv E_{0}\cap%
{\displaystyle\bigcap\limits_{m\in\mathcal{M}}}
E_{m}.
\end{equation}
If there is a positive probability over all codes that the above event is
true, then there exists a particular code that satisfies the above conditions.
Let us instead consider the complement of the above event (the event that a
good private code does not exist):%
\begin{equation}
E_{\operatorname{priv}}^{c}=E_{0}^{c}\cup\bigcup\limits_{m\in\mathcal{M}}%
E_{m}^{c}.
\end{equation}
We can then exploit the union bound from probability theory to bound the
probability of the complementary event$~E_{\operatorname{priv}}^{c}$ as
follows:%
\begin{equation}
\Pr\left\{  E_{0}^{c}\cup\bigcup\limits_{m\in\mathcal{M}}E_{m}^{c}\right\}
\leq\Pr\left\{  E_{0}^{c}\right\}  +\sum_{m\in\mathcal{M}}\Pr\left\{
E_{m}^{c}\right\}  .
\end{equation}
So if we can make the probability of the event $E_{\operatorname{priv}}^{c}$
small, then the probability of the event $E_{\operatorname{priv}}$, that there
exists a good private code, is high.

Let us first bound the probability of the event $E_{0}^{c}$. Markov's
inequality states that the following holds for a non-negative random variable
$Y$ and strictly positive $\alpha$:%
\begin{equation}
\Pr\left\{  Y\geq\alpha\right\}  \leq\frac{\mathbb{E}\left\{  Y\right\}
}{\alpha}.
\end{equation}
We can apply Markov's inequality because the random average error probability
$\bar{p}_{e}(\mathcal{C})$ is always non-negative:%
\begin{equation}
\Pr\left\{  E_{0}^{c}\right\}  =\Pr\left\{  \bar{p}_{e}(\mathcal{C}%
)\geq\left(  \varepsilon^{\prime}\right)  ^{3/4}\right\}  \leq\frac
{\mathbb{E}_{\mathcal{C}}\left\{  \bar{p}_{e}(\mathcal{C})\right\}  }{\left(
\varepsilon^{\prime}\right)  ^{3/4}}\leq\frac{\varepsilon^{\prime}}{\left(
\varepsilon^{\prime}\right)  ^{3/4}}=\sqrt[4]{\varepsilon^{\prime}}.
\label{eq:aver-error-prob-private}%
\end{equation}
So we now have a good bound on the probability of the complementary
event~$E_{0}^{c}$.

Let us now bound the probability of the events $E_{m}^{c}$. The bounds in the
previous section already give us what we need:%
\begin{align}
\Pr\left\{  E_{m}^{c}\right\}   &  =\Pr\left\{  o_{e}(\mathcal{C}%
_{m})>\varepsilon+4\sqrt{\varepsilon}+24\sqrt[4]{\varepsilon}\right\} \\
&  <\frac{\varepsilon}{\left\vert \mathcal{M}\right\vert },
\end{align}
implying that%
\begin{equation}
\sum_{m\in\mathcal{M}}\Pr\left\{  E_{m}^{c}\right\}  <\left\vert
\mathcal{M}\right\vert \frac{\varepsilon}{\left\vert \mathcal{M}\right\vert
}=\varepsilon.
\end{equation}

So it now follows that the probability of the complementary event is small:%
\begin{equation}
\Pr\left\{  E_{\operatorname{priv}}^{c}\right\}  \leq\sqrt[4]{\varepsilon
^{\prime}}+\varepsilon,
\end{equation}
and there is a high probability that there is a good code:%
\begin{equation}
\Pr\left\{  E_{\operatorname{priv}}\right\}  \geq1-\left(  \sqrt[4]%
{\varepsilon^{\prime}}+\varepsilon\right)  . \label{eq:prob-good-private-code}%
\end{equation}

Thus, there exists a particular code $\mathcal{C}$\ such that its average
probability of error is small for decoding the classical information:%
\begin{equation}
\bar{p}_{e}(\mathcal{C})\leq\left(  \varepsilon^{\prime}\right)  ^{3/4},
\end{equation}
and the obfuscation error of each privacy amplification set is small:%
\begin{equation}
\forall m:o_{e}(\mathcal{C}_{m})\leq\varepsilon+4\sqrt{\varepsilon}%
+24\sqrt[4]{\varepsilon}. \label{eq-pcc:deterministic-obfuscation-errors}%
\end{equation}
The derandomized code $\mathcal{C}$ is as follows:%
\begin{equation}
\mathcal{C}\equiv\left\{  x^{n}(m,k)\right\}  _{m\in\mathcal{M},k\in
\mathcal{K}},
\end{equation}
so that each codeword $x^{n}(m,k)$ is a deterministic variable. Each privacy
amplification set for the derandomized code is as follows:%
\begin{equation}
\mathcal{C}_{m}\equiv\left\{  x^{n}(m,k)\right\}  _{k\in\mathcal{K}}.
\end{equation}

The result in \eqref{eq:prob-good-private-code} is perhaps astonishing in
hindsight. By choosing a private code in a random way and choosing the block
length $n$ of the private code to be sufficiently large, the overwhelming
majority of codes constructed in this fashion are good private codes!

\subsection{Expurgation}

\label{sec-pcc:expurgate}We would like to
\index{expurgation}%
strengthen the above result even more, so that the code has a low maximal
probability of error, not just a low average error probability. We expurgate
codewords from the code as before, but we have to be careful with the
expurgation argument because we need to make sure that the code still has good
privacy after expurgation.

We can apply Markov's inequality for the expurgation in a way similar as in
Exercise~\ref{ex-intro:expurgation}. It is possible to apply Markov's
inequality to the bound on the average error probability in
\eqref{eq:aver-error-prob-private} to show that at most a fraction
$\sqrt{\varepsilon^{\prime}}$ of the codewords have error probability greater
than $\sqrt[4]{\varepsilon^{\prime}}$. We could merely expurgate the worst
$\sqrt{\varepsilon^{\prime}}$ codewords from the private code. But expurgating
in this fashion does not guarantee that each privacy amplification set has the
same number of codewords. Therefore, we expurgate the worst fraction
$\sqrt{\varepsilon^{\prime}}$ of the codewords in each privacy amplification
set. We then expurgate the worst fraction $\sqrt{\varepsilon^{\prime}}$ of the
privacy amplification sets. The expurgated sets $\mathcal{M}^{\prime}$ and
$\mathcal{K}^{\prime}$ both become a fraction $1-\sqrt{\varepsilon^{\prime}}$
of their original size. We denote the expurgated code as follows:%
\begin{equation}
\mathcal{C}^{\prime}\equiv\left\{  x^{n}(m,k)\right\}  _{m\in\mathcal{M}%
^{\prime},k\in\mathcal{K}^{\prime}},
\end{equation}
and the expurgated code has the following privacy amplification sets:%
\begin{equation}
\mathcal{C}_{m}^{\prime}\equiv\left\{  x^{n}(m,k)\right\}  _{k\in
\mathcal{K}^{\prime}}.
\end{equation}
The expurgation has a negligible impact on the rate of the private code when
$n$ is large.

Does each privacy amplification set still have good privacy properties after
performing the above expurgation? The fake density operator for each
expurgated privacy amplification set is as follows:%
\begin{equation}
\hat{\rho}_{E^{n}}^{m\prime}\equiv\frac{1}{\left\vert \mathcal{C}_{m}^{\prime
}\right\vert }\sum_{k\in\mathcal{K}^{\prime}}\rho_{E^{n}}^{x^{n}(m,k)}.
\end{equation}
It is possible to show that the fake density operators in the derandomized
code are $2\sqrt{\varepsilon^{\prime}}$-close in trace distance to the fake
density operators in the expurgated code:%
\begin{equation}
\forall m\in\mathcal{M}^{\prime}\ \ \ \ \ \left\Vert \hat{\rho}_{E^{n}%
}^{m\prime}-\hat{\rho}_{E^{n}}^{m}\right\Vert _{1}\leq2\sqrt{\varepsilon
^{\prime}}, \label{eq-pcc:expurgated-distance}%
\end{equation}
because these operators only lose a small fraction of their mass after expurgation.

We now drop the primed notation to denote the expurgated code. It follows that
the expurgated code$~\mathcal{C}$\ has good privacy:%
\begin{equation}
\forall m\in\mathcal{M}\ \ \ \ \ \left\Vert \hat{\rho}_{E^{n}}^{m}-\rho
_{E^{n}}\right\Vert _{1}\leq\varepsilon+4\sqrt{\varepsilon}+24\sqrt[4]%
{\varepsilon}+2\sqrt{\varepsilon^{\prime}},
\end{equation}
and reliable communication:%
\begin{equation}
\forall m\in\mathcal{M},\ k\in\mathcal{K}\ \ \ \ \ p_{e}(\mathcal{C}%
,m,k)\leq\sqrt[4]{\varepsilon^{\prime}}.
\end{equation}
The first expression follows by application of the triangle inequality to
\eqref{eq-pcc:deterministic-obfuscation-errors} and \eqref{eq-pcc:expurgated-distance}.

We end the proof by summarizing the operation of the private code. Alice
chooses a message$~m$ from the message set $\mathcal{M}$ and the randomization
variable $k$ uniformly at random from $\mathcal{K}$. She encodes these as
$x^{n}(m,k)$ and inputs the quantum codeword $\rho_{A^{\prime n}}^{x^{n}%
(m,k)}$ to the channel. Bob receives the state $\rho_{B^{n}}^{x^{n}(m,k)}$ and
performs a POVM $\{\Lambda_{m,k}\}_{(m,k)\in\mathcal{M}\times\mathcal{K}}$
that determines the pair $m$ and $k$ correctly with probability $1-\sqrt[4]%
{\varepsilon^{\prime}}$. The code guarantees that Eve has almost no knowledge
about the message$~m$. The private communication rate$~P$\ of the private code
is equal to the following expression:%
\begin{equation}
P\equiv\frac{1}{n}\log\left\vert \mathcal{M}\right\vert
=I(X;B)-I(X;E)-6c\delta.
\end{equation}
This concludes the proof of the direct coding theorem.

We remark that the above proof applies even in the scenario in which Eve does not
get the full purification of the channel. That is, suppose that the channel
has one input~$A^{\prime}$ for Alice and two outputs $B$ and $E$ for Bob and
Eve, respectively. Then the channel has an isometric extension to some
environment~$F$. In this scenario, the private information $I(X;B)-I(X;E)$ is
still achievable for some classical--quantum state input such that the Holevo
information difference is non-negative. However, one could always give both
outputs $E$ and $F$ to an eavesdropper (this is the setting that we proved in
the above theorem). Giving the full purification of the channel to the
environment ensures that the transmitted information is private from the
\textquotedblleft rest of the universe\textquotedblright\ (anyone other than
the intended receiver), and it thus yields the highest standard of security in
any protocol for private information transmission.

\section{The Converse Theorem}%

\index{private classical capacity theorem!converse}%
We now prove the converse part of the private classical capacity theorem,
which demonstrates that the regularization of the private information is an
upper bound on the private classical capacity. We suppose instead that Alice
and Bob are trying to accomplish the task of secret key generation. As we have
argued in other converse proofs (see Sections~\ref{sec-cc:converse} and
\ref{sec-eac:converse}), the capacity for generating this static resource can
only be larger than the capacity for private classical communication because
Alice and Bob can always use a noiseless private channel to establish a shared
secret key. In such a task, Alice first prepares a maximally correlated state
$\overline{\Phi}_{MM^{\prime}}$ and encodes the $M^{\prime}$ variable as a
codeword $\rho_{A^{\prime n}}^{m}$. This encoding leads to a state of the
following form, after Alice transmits her systems~$A^{\prime n}$ over many
independent uses of the channel:%
\begin{equation}
\omega_{MB^{n}E^{n}}\equiv\frac{1}{\left\vert \mathcal{M}\right\vert }%
\sum_{m\in\mathcal{M}}|m\rangle\langle m|_{M}\otimes\mathcal{U}_{A^{\prime
n}\rightarrow B^{n}E^{n}}^{\mathcal{N}}(\rho_{A^{\prime n}}^{m}).
\label{eq-pcc:channel-cq-state}%
\end{equation}
Bob finally applies a decoding channel~$\mathcal{D}_{B^{n}\rightarrow
M^{\prime}}$ to recover his share of the secret key:%
\begin{equation}
\omega_{MM^{\prime}E^{n}}\equiv\mathcal{D}_{B^{n}\rightarrow M^{\prime}%
}(\omega_{MB^{n}E^{n}}).
\end{equation}
The following condition holds for an $(n,\left[  \log\left\vert \mathcal{M}%
\right\vert \right]  /n,\varepsilon)$ protocol for secret key generation:%
\begin{equation}
\frac{1}{2}\left\Vert \omega_{MM^{\prime}E^{n}}-\overline{\Phi}_{MM^{\prime}%
}\otimes\sigma_{E^{n}}\right\Vert _{1}\leq\varepsilon,
\label{eq-pcc:secret-key-security}%
\end{equation}
so that Eve's state$~\sigma_{E^{n}}$ is a constant state independent of the
secret key~$\overline{\Phi}_{MM^{\prime}}$. In particular, the above condition
implies that Eve's information about $M$ is small:%
\begin{equation}
I(M;E^{n})_{\omega}\leq f(\left\vert \mathcal{M}\right\vert ,\varepsilon),
\label{eq-pcc:eve-info-sneaky}%
\end{equation}
where we apply the reasoning in Section~\ref{sec-pcc:mut-info-Eve},
with$~f(\left\vert \mathcal{M}\right\vert ,\varepsilon)\equiv\varepsilon
\log\left\vert \mathcal{M}\right\vert +g_2(\varepsilon)$. The rate of secret key
generation is equal to $\frac{1}{n}\log\left\vert \mathcal{M}\right\vert $.
Consider the following chain of inequalities:%
\begin{align}
\log\left\vert \mathcal{M}\right\vert  &  =I(M;M^{\prime})_{\overline{\Phi}}\\
&  \leq I(M;M^{\prime})_{\omega}+f(\left\vert \mathcal{M}\right\vert
,\varepsilon)\\
&  \leq I(M;B^{n})_{\omega}+f(\left\vert \mathcal{M}\right\vert ,\varepsilon
)\\
&  \leq I(M;B^{n})_{\omega}-I(M;E^{n})_{\omega}+2f(\left\vert \mathcal{M}%
\right\vert ,\varepsilon)\\
&  \leq P(\mathcal{N}^{\otimes n})+2f(\left\vert \mathcal{M}\right\vert
,\varepsilon).
\end{align}
The first equality follows because the mutual information of the shared
randomness state $\overline{\Phi}_{MM^{\prime}}$ is equal to $\log\left\vert
\mathcal{M}\right\vert $. The first inequality follows from applying the AFW
inequality to \eqref{eq-pcc:secret-key-security}. The second inequality
follows from quantum data processing. The third inequality follows from
\eqref{eq-pcc:eve-info-sneaky}, and the final inequality follows because the
classical--quantum state in \eqref{eq-pcc:channel-cq-state} has a particular
distribution and choice of states, and this choice always leads to a value of
the private information that cannot be larger than the private information of
the tensor-product channel $\mathcal{N}^{\otimes n}$. Putting everything
together, we find that%
\begin{equation}
\frac{1}{n}\log\left\vert \mathcal{M}\right\vert \left(  1-2\varepsilon
\right)  \leq\frac{1}{n}P(\mathcal{N}^{\otimes n})+\frac{2}{n}g_2(\varepsilon).
\end{equation}
Thus, if we are considering a sequence of $(n,\left[  \log\left\vert
\mathcal{M}\right\vert \right]  /n,\varepsilon_{n})$\ private classical
communication protocols with rate $P-\delta_{n}=\frac{1}{n}\log\left\vert
\mathcal{M}\right\vert $, such that $\lim_{n\rightarrow\infty}\varepsilon
_{n}=\lim_{n\rightarrow\infty}\delta_{n}=0$, then the above bound becomes%
\begin{equation}
\left(  P-\delta_{n}\right)  \left(  1-2\varepsilon_{n}\right)  \leq\frac
{1}{n}P(\mathcal{N}^{\otimes n})+\frac{2}{n}g_2(\varepsilon_n).
\end{equation}
Taking the limit as $n\rightarrow\infty$ then establishes that an achievable
rate $P$ necessarily satisfies $P\leq P_{\operatorname{reg}}(\mathcal{N})$,
where $P_{\operatorname{reg}}(\mathcal{N})$ is the regularized private
information given in \eqref{eq-pcc:regularization-private}.

\begin{exercise}
Prove that free access to a forward public classical channel from Alice to Bob
cannot improve the private classical capacity of a quantum channel.
\end{exercise}

\section{Discussion of Private Classical Capacity}

This last section discusses some important aspects of the private classical
capacity. Two of these results have to do with the fact that
Theorem~\ref{thm-pcc:private-capacity}\ only provides a regularized
characterization of the private classical capacity, and the last asks what
rates of private classical communication are achievable if the sender and
receiver share a secret key before communication begins. For full details, we
refer the reader to the original papers in the quantum Shannon theory literature.

\subsection{Superadditivity of the Private Information}

Theorem~\ref{thm-pcc:private-capacity}\ states that the private classical
capacity of a quantum channel
\index{private information!suparadditivity}
is equal to the regularized private information of the channel. As we have
said before (at the beginning of Chapter~\ref{chap:EA-classical}), a
regularized formula is not particularly useful from a practical perspective
because it is impossible to perform the optimization task that it sets out,
and it is not desirable from an information-theoretical perspective because
such a regularization does not identify a formula as a unique measure of capacity.

In light of the unsatisfactory nature of a regularized formula, is it really
necessary to have the regularization in Theorem~\ref{thm-pcc:private-capacity}
for arbitrary quantum channels? Interestingly, the answer seems to be
\textquotedblleft yes\textquotedblright\ in the general case (however, we know
it is not necessary if the channel is
\index{degradable channel}%
degradable). The reason is that there exists an example of a
channel~$\mathcal{N}$ for which the private information is strictly
superadditive:%
\begin{equation}
mP(\mathcal{N})<P\left(  \mathcal{N}^{\otimes m}\right)  ,
\label{eq-pcc:superadditivity-private}%
\end{equation}
for some positive integer$~m$. Specifically, Smith \textit{et al}.~showed that
the private information of a particular Pauli channel exhibits this
superadditivity~\citep{smith:170502}. To do so, they calculated the private
information~$P(\mathcal{N})$ for such a channel. Next, they consider
performing an $m$-qubit \textquotedblleft repetition code\textquotedblright%
\ before transmitting qubits into the channel. A repetition code is a quantum
code that performs the following encoding:%
\begin{equation}
\alpha|0\rangle+\beta|1\rangle\rightarrow\alpha|0\rangle^{\otimes m}%
+\beta|1\rangle^{\otimes m}.
\end{equation}
Evaluating the private information when sending a particular state through the
repetition code and then through $m$ instances of the channel leads to a
higher value than $mP(\mathcal{N})$, implying the strict inequality in
\eqref{eq-pcc:superadditivity-private}. Thus, additivity of the private
information formula~$P(\mathcal{N})$ cannot hold in the general case.

The implications of this result are that we really do not understand the best
way of transmitting information privately over a quantum channel that is not
degradable, and it is thus the subject of ongoing research.

\subsection{Superadditivity of Private Classical Capacity}

\label{sec-pcc:super-private-capacity}The private information of a particular
channel can be superadditive (as discussed in the previous section), and so
the regularized
\index{private classical capacity!superadditivity}
private information is our best characterization of the capacity for this
information-processing task. In spite of this, we might hope that some
eventual formula for the private classical capacity would be additive (some
formula other than the private information~$P( \mathcal{N}) $). Interestingly,
this is also not the case.

To clarify this point, suppose that $P^{?}(\mathcal{N})$ is some formula for
the private classical capacity. If it were an additive formula, then it should
be additive as a function of channels:%
\begin{equation}
P^{?}(\mathcal{N}\otimes\mathcal{M})=P^{?}(\mathcal{N})+P^{?}(\mathcal{M}).
\label{eq-pcc:super-additivity-capacity}%
\end{equation}
Li \textit{et al}.~have shown that this cannot be the case for any proposed
private capacity formula, by making a clever argument with a construction of
channels~\citep{LWZG09}. Specifically, they constructed a particular channel
$\mathcal{N}$ which has a single-letter \textit{classical} capacity. The fact
that the channel's classical capacity is sharply upper bounded implies that
its private classical capacity is as well. Let $D$ be the upper bound so that
$P^{?}(\mathcal{N})\leq D$. Also, they considered a 50\% erasure channel, one
which gives the input state to Bob and an erasure symbol to Eve with
probability$~1/2$ and gives the input state to Eve and an erasure symbol to
Bob with probability$~1/2$. Such a channel has zero capacity for sending
private classical information essentially because Eve is getting the same
amount of information as Bob does on average. Thus, $P^{?}(\mathcal{M})=0$. In
spite of this, Li \textit{et al}.~show that the tensor-product channel
$\mathcal{N}\otimes\mathcal{M}$ has a private classical capacity that
exceeds~$D$. We can then make the conclusion that these two channels allow for
superadditivity of private classical capacity:%
\begin{equation}
P^{?}(\mathcal{N}\otimes\mathcal{M})>P^{?}(\mathcal{N})+P^{?}(\mathcal{M}),
\end{equation}
and that \eqref{eq-pcc:super-additivity-capacity} cannot hold in the general
case. More profoundly, their results demonstrate that the private classical
capacity itself is non-additive, even if a characterization of it is found
that is more desirable than that with the formula in
Theorem~\ref{thm-pcc:private-capacity}. Thus, it will likely be difficult to
obtain\ a desirable characterization of the private classical capacity for
general quantum channels.

\subsection{Secret-key Assisted Private Classical Communication}

\label{sec-pcc:ska}The direct coding part of
Theorem~\ref{thm-pcc:private-capacity}\ demonstrates how to send private
classical
\index{private classical communication!secret-key assisted}
information over a quantum channel~$\mathcal{N}$ at the private information
rate~$P(\mathcal{N})$. A natural extension to consider is the scenario in
which Alice and Bob share a secret key before communication begins. A secret
key shared between Alice and Bob and secure from Eve is a tripartite state of
the following form:%
\begin{equation}
\overline{\Phi}_{AB}\otimes\sigma_{E},
\end{equation}
where $\overline{\Phi}_{AB}$ is the maximally correlated state and $\sigma
_{E}$ is a state on Eve's system that is independent of the key shared between
Alice and Bob. Like the entanglement-assisted capacity theorem, we assume that
they obtain this secret key from some third party, and the third party ensures
that the key remains secure.

The resulting capacity theorem is known as the secret-key-assisted private
classical capacity theorem, and it characterizes the trade-off between secret
key consumption and private classical communication. The main idea for this
setting is to show the existence of a protocol that transmits private
classical information at a rate of $I( X;B) $ private bits per channel use
while consuming secret key at a rate of $I( X;E) $ secret key bits per channel
use, where the information quantities are with respect to the state in
Theorem~\ref{thm-pcc:private-capacity}. The protocol for achieving these rates
is almost identical to the one we gave in the proof of the direct coding
theorem, though with one difference. Instead of sacrificing classical bits at
a rate of $I( X;E) $ in order to randomize Eve's knowledge of the message
(recall that our randomization variable had to be chosen uniformly at random
from a set of size $\approx2^{nI( X;E) }$), the sender exploits the secret key
to do so. The converse proof shows that this strategy is optimal (with a
multi-letter characterization). Thus, we have the following capacity theorem.

\begin{theorem}
[Secret-key-assisted capacity theorem]The secret-key-assisted private
classical capacity region $C_{\operatorname{SKA}}(\mathcal{N})$ of a quantum
channel $\mathcal{N}$ is given by%
\begin{equation}
C_{\operatorname{SKA}}(\mathcal{N})=\overline{\bigcup\limits_{k=1}^{\infty
}\frac{1}{k}\widetilde{C}_{\operatorname{SKA}}^{(1)}(\mathcal{N}^{\otimes k}%
)},
\end{equation}
where the overbar indicates the closure of a set. $\widetilde{C}%
_{\operatorname{SKA}}^{(1)}(\mathcal{N})$ is the set of all $P,S\geq0$ such
that%
\begin{align}
P  &  \leq I(X;B)_{\sigma}-I(X;E)_{\sigma}+S,\\
P  &  \leq I(X;B)_{\sigma}.
\end{align}
where $P$ is the rate of private classical communication, $S$ is the rate of
secret key consumption, the state $\sigma_{XBE}$ is of the following form:%
\begin{equation}
\sigma_{XBE}\equiv\sum_{x}p_{X}(x)|x\rangle\langle x|_{X}\otimes
\mathcal{U}_{A^{\prime}\rightarrow BE}^{\mathcal{N}}\left(  \rho_{A^{\prime}%
}^{x}\right)  ,
\end{equation}
and $U_{A^{\prime}\rightarrow BE}^{\mathcal{N}}$ is an isometric extension of
the channel.
\end{theorem}

Showing that the above inequalities are achievable follows by time sharing
between the protocol from the direct coding part of
Theorem~\ref{thm-pcc:private-capacity}\ and the aforementioned protocol for
secret-key-assisted private classical communication.

\section{History and Further Reading}

\cite{bb84} devised the first protocol for sending private classical data over
a quantum channel. The protocol given there became known as quantum key
distribution, which has now become a thriving field in its own
right~\citep{SBCDLP09}. \cite{ieee2005dev} and \cite{1050633}\ proved the
characterization of the private classical capacity given in this chapter (both
using the techniques that we reviewed in this chapter). \cite{HLB08} proved
achievability of the secret-key-assisted protocol given in
Section~\ref{sec-pcc:ska}, and \cite{PhysRevA.83.046303} proved the converse
and stated the secret-key-assisted capacity theorem. Later work characterized
the full trade-off between public classical communication, private classical
communication, and secret key~\citep{PhysRevA.80.022306,WH10}.
\cite{smith:170502} showed that the private information can exhibit
superadditivity, and \cite{LWZG09} showed that the private classical capacity
is generally non-additive. \cite{PhysRevLett.115.040501} demonstrated a
striking superadditivity effect, which suggests that a regularized expression
is necessary to determine the private capacity of an arbitrary channel.
\cite{S08} later showed that the symmetric-side-channel-assisted private
classical capacity is additive. \cite{datta:122202} demonstrated universal
private codes for quantum channels.

A different, weaker notion of security is based on the eavesdropper not being
able to learn much from any measurement performed on her system. Although this
notion might seem similar to the notion of security that we discussed in this
chapter, it is in fact much different \citep{KRBM07}\ and is the basis for
the
\index{information locking}
information locking effect \citep{DHLST04}. Impressive information locking
schemes exist \citep{HLSW04,Dupuis20130289,FHS13}, and recently, the locking
\index{locking capacity}
capacity of a channel was introduced and bounded \citep{PhysRevX.4.011016},
calculated for particular channels \citep{W15lock}, and developed further \citep{PhysRevLett.113.160502,LL15}.

\chapter{Quantum Communication}

\label{chap:quantum-capacity}The
\index{quantum capacity theorem}
quantum capacity theorem is one of the most important theorems in quantum
Shannon theory. It is a fundamentally \textquotedblleft
quantum\textquotedblright\ theorem in that it demonstrates that a
fundamentally quantum information quantity, the coherent information, is an
achievable rate for quantum communication over a quantum channel. The fact
that the coherent information does not have a strong analog in classical
Shannon theory truly separates the quantum and classical theories of information.

The no-cloning
\index{no-cloning theorem}%
theorem (Section~\ref{sec-qt:no-cloning}) provides the intuition behind the
quantum capacity theorem. The goal of any quantum communication protocol is
for Alice to establish quantum correlations with the receiver Bob. We know
well now that every quantum channel has an isometric extension, so that we can
think of another receiver, the environment Eve, who is at a second output port
of a larger unitary evolution. Were Eve able to learn anything about the
quantum information that Alice is attempting to transmit to Bob, then Bob
could not be retrieving this information---otherwise, they would violate the
no-cloning theorem. Thus, Alice should figure out some subspace of the channel
input where she can place her quantum information such that only Bob has
access to it, while Eve does not. That the dimensionality of this subspace is
exponential in the coherent information is perhaps then unsurprising in light
of the above no-cloning reasoning. The
\index{coherent information}%
coherent information is an entropy difference $H( B) -H( E) $---a measure of
the amount of quantum correlations that Alice can establish with Bob less the
amount that Eve can gain.\footnote{Recall from
Exercise~\ref{ex-qie:entropy-games}\ that we can also write the coherent
information as half the difference of Bob's mutual information with Alice less
Eve's: $I( A\rangle B) =1/2\left[  I( A;B) -I( A;E) \right]  $.}

We proved achievability of the coherent information for quantum data
transmission in Corollary~\ref{cor-ccn:q-comm}, but the roundabout path that
we followed to prove achievability there perhaps does not give much insight
into the structure of a quantum code that achieves the coherent information.
Our approach in this chapter is different and should shed more light on this
structure. Specifically, we show how to make coherent versions of the private
classical codes from the previous chapter. By exploiting the privacy
properties of these codes, we can form subspaces where Alice can store her
quantum information such that Eve does not have access to it. Thus, this
approach follows the above \textquotedblleft no-cloning
intuition\textquotedblright\ more closely.

The best characterization that we have for the quantum capacity of a general
quantum channel is the regularized coherent information. It turns out that the
regularization is not necessary for the class of
\index{degradable channel}%
degradable channels, implying that we have a complete understanding of the
quantum data transmission capabilities of these channels. However, if a
channel is not degradable, there can be some startling consequences, and these
results imply that we have an incomplete understanding of quantum data
transmission in the general case. First, the coherent information can be
strictly superadditive for the depolarizing channel. This means that the best
strategy for achieving the quantum capacity is not necessarily the familiar
one where we generate random quantum codes from a single instance of a
channel. This result is also in marked contrast with the \textquotedblleft
classical\textquotedblright\ strategies that achieve the unassisted and
entanglement-assisted classical capacities of the depolarizing channel.
Second, perhaps the most surprising result in quantum Shannon theory is that
it is possible to \textquotedblleft superactivate\textquotedblright\
\index{superactivation}
the quantum capacity. That is, suppose that two channels on their own have
zero capacity for transmitting quantum information (for the phenomenon to
occur, these channels are specific channels). Then it is possible for the
joint channel (the tensor product of the individual channels) to have a
non-zero quantum capacity, in spite of them being individually useless for
quantum data transmission. This latter result implies that we are rather
distant from having a complete quantum theory of information, in spite of the
many successes reviewed in this book.

We structure this chapter as follows. We first overview the
information-processing task relevant for quantum communication. Next, we
discuss the no-cloning intuition for quantum capacity in some more detail,
presenting the specific example of a quantum erasure channel.
Section~\ref{sec-q-cap:theorem}\ states the quantum capacity theorem, and the
following two sections prove the direct coding and converse theorems
corresponding to it.
Section~\ref{sec-q-cap:examples} computes the quantum capacity of two
\index{degradable channel}%
degradable channels:\ the quantum erasure channel and the amplitude damping
channel. We then discuss superadditivity of coherent information and
superactivation of quantum capacity in Section~\ref{sec-q-cap:strangeness}.
Finally, we prove the existence of an entanglement distillation protocol,
whose proof bears some similarities to the proof of the direct coding part of
the quantum capacity theorem.

\section{The Information-Processing Task}

\begin{figure}
[ptb]
\begin{center}
\includegraphics[
width=3.531in
]%
{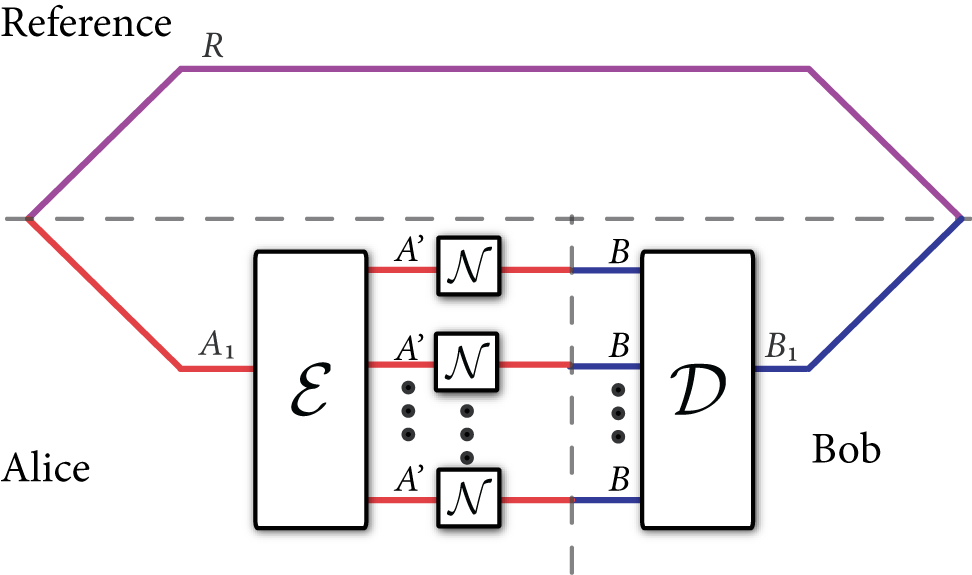}%
\caption{The information-processing task for entanglement transmission. Alice
is trying to preserve the entanglement with some inaccessible reference system
by encoding her system and transmitting the encoded quantum data over many
independent uses of a noisy quantum channel. Bob performs a decoding of the
systems he receives, and the state at the end of the protocol is close to the
original state shared between Alice and the reference if the protocol is any
good for entanglement transmission.}%
\label{fig-q-cap:info-task}%
\end{center}
\end{figure}

We begin the technical development in this chapter by describing the
information-processing task for quantum communication (we define an $\left(
n,Q,\varepsilon\right)  $ quantum communication code). First, there are
several different tasks that we can consider as quantum communication, but the
strongest definition of quantum capacity corresponds to a task known as
\textit{entanglement transmission}. Suppose that Alice shares entanglement
with a reference system to which she does not have access. Then their goal is
to devise a quantum coding scheme such that Alice can transfer this
entanglement to Bob. To this end, suppose that Alice and the reference share
an arbitrary state $|\varphi\rangle_{RA_{1}}$, where the systems $R$ and
$A_{1}$ have the same dimension. Alice then performs some encoder on system
$A_{1}$ to prepare it for input to many instances of a quantum channel
$\mathcal{N}_{A^{\prime}\rightarrow B}$. The resulting state is as follows:
$\mathcal{E}_{A_{1}\rightarrow A^{\prime n}}(\varphi_{RA_{1}})$. Alice
transmits the systems $A^{\prime n}$ through many independent uses of the
channel $\mathcal{N}_{A^{\prime}\rightarrow B}$, resulting in the following
state: $\mathcal{N}_{A^{\prime n}\rightarrow B^{n}}(\mathcal{E}_{A_{1}%
\rightarrow A^{\prime n}}(\varphi_{RA_{1}}))$, where $\mathcal{N}_{A^{\prime
n}\rightarrow B^{n}}\equiv(\mathcal{N}_{A^{\prime}\rightarrow B})^{\otimes n}%
$. After Bob receives the systems $B^{n}$ from the channel outputs, he
performs some decoding channel $\mathcal{D}_{B^{n}\rightarrow B_{1}}$, where
$B_{1}$ is some system of the same dimension as $A_{1}$. The final state after
Bob decodes is as follows:%
\begin{equation}
\omega_{RB_{1}}\equiv\mathcal{D}_{B^{n}\rightarrow B_{1}}(\mathcal{N}%
_{A^{\prime n}\rightarrow B^{n}}(\mathcal{E}_{A_{1}\rightarrow A^{\prime n}%
}(\varphi_{RA_{1}}))).
\end{equation}
Figure~\ref{fig-q-cap:info-task}\ depicts all of the above steps. For an
$(n,Q,\varepsilon)$ protocol, the following condition should hold for all
states $|\varphi\rangle_{RA_{1}}$:%
\begin{equation}
\frac{1}{2}\left\Vert \varphi_{RA_{1}}-\omega_{RB_{1}}\right\Vert _{1}%
\leq\varepsilon.
\end{equation}
The rate~$Q$ of this scheme is equal to the number of qubits transmitted per
channel use:%
\begin{equation}
Q\equiv\frac{1}{n}\log\dim(\mathcal{H}_{A_{1}}).
\end{equation}

A rate~$Q$ is achievable for $\mathcal{N}$\ if there exists an $\left(
n,Q-\delta,\varepsilon\right)  $ quantum communication code for all
$\varepsilon\in(0,1)$, $\delta>0$, and sufficiently large~$n$. The quantum
capacity $C_{Q}(\mathcal{N})$ is defined as the supremum of all achievable
rates for $\mathcal{N}$.%

The above notion of quantum communication encompasses other quantum
information-processing tasks such as mixed-state transmission, pure-state
transmission, and entanglement generation. Alice can transmit any mixed or
pure state if she can preserve the entanglement with a reference system. Also,
she can generate entanglement with Bob if she can preserve entanglement with a
reference system---she just needs to create an entangled state locally and
apply the above protocol to one system of the entangled state.

\begin{exercise}
Show that an $(n,Q,\varepsilon)$ protocol for quantum communication satisfies
the following:%
\begin{equation}
\frac{1}{2}\left\Vert \operatorname{id}_{A_{1}\rightarrow B_{1}}%
-\mathcal{D}_{B^{n}\rightarrow B_{1}}\circ\mathcal{N}_{A^{\prime}\rightarrow
B}^{\otimes n}\circ\mathcal{E}_{A_{1}\rightarrow A^{\prime n}}\right\Vert
_{\Diamond}\leq\varepsilon,
\end{equation}
where $\left\Vert \cdot\right\Vert _{\Diamond}$ denotes the diamond norm,
defined in Section~\ref{sec-dm:diamond-norm}.
\end{exercise}

\section{No-Cloning and Quantum Communication}

We first discuss quantum communication over a quantum erasure channel before
stating and proving the quantum capacity theorem. Consider the quantum erasure
channel that gives Alice's input state to Bob with probability $1-\varepsilon$
and an erasure flag to Bob with probability~$\varepsilon$:%
\begin{equation}
\rho\rightarrow\left(  1-\varepsilon\right)  \rho+\varepsilon|e\rangle\langle
e|,
\end{equation}
where $\langle e|\rho|e\rangle=0$ for all inputs $\rho$. Recall that an
isometric extension of this channel is as follows (see
Exercise~\ref{ex:iso-extension-erasure}):%
\begin{equation}
|\psi\rangle_{RA}\rightarrow\sqrt{1-\varepsilon}|\psi\rangle_{RB}|e\rangle
_{E}+\sqrt{\varepsilon}|\psi\rangle_{RE}|e\rangle_{B},
\label{eq-q-cap:iso-extend-erasure}%
\end{equation}
so that the channel now has the other interpretation that Eve gets the state
with probability~$\varepsilon$ while giving her the erasure flag with
probability $1-\varepsilon$.

Now suppose
\index{erasure channel}
that the erasure parameter is set to $1/2$.
\index{no-cloning theorem}
In such a scenario, the channel to Eve is the \textit{same} as the channel to
Bob, namely, both have the channel $\rho\rightarrow1/2\left(  \rho
+|e\rangle\langle e|\right)  $. We can argue that the quantum capacity of such
a channel should be equal to zero, by invoking the no-cloning theorem. More
specifically, suppose there is a scheme (an encoder and decoder as given in
Figure~\ref{fig-q-cap:info-task}) for Alice and Bob to communicate quantum
information reliably at a non-zero rate over such a channel. If so, Eve could
simply use the same decoder that Bob does, and she should also be able to
obtain the quantum information that Alice is sending. But the ability for both
Bob and Eve to decode the quantum information that Alice is transmitting
violates the no-cloning theorem. Thus, the quantum capacity of such a channel
should vanish.

\begin{exercise}
Argue that the quantum capacity of an amplitude damping channel vanishes if
its damping parameter is equal to $1/2$.
\end{exercise}

The no-cloning theorem plays a more general role in the analysis of quantum
communication over quantum channels. In the construction of a quantum code, we
are trying to find a \textquotedblleft no-cloning\textquotedblright\ subspace
of the input Hilbert space that is protected from Eve. If Eve is able to
obtain any of the quantum information in this subspace, then this information
cannot be going to Bob by the same no-cloning argument featured in the
previous paragraph. Thus, we might then suspect that the codes from the
previous chapter for private classical communication might play a role for
quantum communication because we constructed them in such a way that Eve would
not be able to obtain any information about the private message that Alice is
transmitting to Eve. The main insight needed is to make a coherent version of
these private classical codes, so that Alice and Bob conduct every step in
superposition (much like we did in Chapter~\ref{chap:coh-comm-noisy}).

\section{The Quantum Capacity Theorem}

\label{sec-q-cap:theorem}The main theorem of this chapter is the following
quantum capacity theorem.

\begin{theorem}
[Quantum Capacity]\label{thm-q-cap:q-cap-theorem}The quantum capacity
\index{quantum capacity theorem}
$C_{Q}(\mathcal{N})$\ of a quantum channel $\mathcal{N}_{A^{\prime}\rightarrow
B}$ is equal to the regularized coherent information of the channel:%
\begin{equation}
C_{Q}(\mathcal{N})=Q_{\operatorname{reg}}(\mathcal{N}),
\end{equation}
where%
\begin{equation}
Q_{\operatorname{reg}}(\mathcal{N})\equiv\lim_{k\rightarrow\infty}\frac{1}%
{k}Q(\mathcal{N}^{\otimes k}). \label{eq-qc:reg-coherent-info}%
\end{equation}
The channel coherent information $Q(\mathcal{N})$ is defined as%
\begin{equation}
Q(\mathcal{N})\equiv\max_{\phi}I(A\rangle B)_{\sigma},
\end{equation}
where the optimization is with respect to all pure, bipartite states $\phi_{AA^{\prime}}$
and $\sigma_{AB}\equiv\mathcal{N}_{A^{\prime}\rightarrow B}(\phi_{AA^{\prime}%
})$.
\end{theorem}

We prove this theorem in two parts:\ the direct coding theorem and the
converse theorem. The proof of the direct coding theorem proceeds by
exploiting the private classical codes from the previous chapter. The proof of
the converse theorem is similar to approaches from previous chapters---we
exploit the AFW inequality and quantum data processing in order to obtain an
upper bound on the quantum capacity. In general, the regularized coherent
information is our best characterization of the quantum capacity, but the
regularization is not necessary for the class of
\index{degradable channel}%
degradable channels. Since many channels of interest are degradable (including
dephasing, amplitude damping, and erasure channels), we can calculate their
quantum capacities.

\section{The Direct Coding Theorem}%

\index{quantum capacity theorem!direct part}
The proof of the direct coding part of the quantum capacity theorem follows by
taking advantage of the properties of the private classical codes constructed
in the previous chapter (see Section~\ref{sec-pcc:direct-coding}). We briefly
recall this construction. Suppose that a classical--quantum--quantum channel
connects Alice to Bob and Eve. Specifically, if Alice inputs a classical
letter~$x$ to the channel, then Bob receives a density operator $\rho_{B}^{x}$
and Eve receives a density operator~$\omega_{E}^{x}$. The direct coding part
of Theorem~\ref{thm-pcc:private-capacity} establishes the existence of a
codebook $\left\{  x^{n}(m,k)\right\}  _{m\in\mathcal{M},k\in\mathcal{K}}$
selected from a distribution $p_{X}(x)$ and a corresponding decoding
POVM\ $\{\Lambda_{B^{n}}^{m,k}\}$ such that Bob can detect Alice's message$~m$
and randomizing variable~$k$ with high probability:%
\begin{equation}
\forall m\in\mathcal{M},k\in\mathcal{K}:\operatorname{Tr}\left\{
\Lambda_{B^{n}}^{m,k}\rho_{B^{n}}^{x^{n}(m,k)}\right\}  \geq1-\varepsilon,
\label{eq-q-cap:good-detection-cond}%
\end{equation}
while Eve obtains a negligible amount of information about Alice's
message~$m$:%
\begin{equation}
\forall m\in\mathcal{M}:\left\Vert \frac{1}{\left\vert \mathcal{K}\right\vert
}\sum_{k\in\mathcal{K}}\omega_{E^{n}}^{x^{n}(m,k)}-\omega^{\otimes
n}\right\Vert _{1}\leq\varepsilon, \label{eq-q-cap:good-privacy-cond}%
\end{equation}
where $\omega$ is Eve's expected density operator:%
\begin{equation}
\omega\equiv\sum_{x}p_{X}(x)\omega_{E}^{x}.
\end{equation}
The above statements hold true for all $\varepsilon\in(0,1)$ and sufficiently
large~$n$ as long as%
\begin{align}
\left\vert \mathcal{M}\right\vert  &  \approx2^{n\left[  I(X;B)-I(X;E)\right]
},\\
\left\vert \mathcal{K}\right\vert  &  \approx2^{nI(X;E)}.
\end{align}

We can now construct a coherent version of the above code that is good for
quantum data transmission. First, suppose that there is some density operator
with the following spectral decomposition:%
\begin{equation}
\rho_{A^{\prime}}\equiv\sum_{x}p_{X}(x)|\psi_{x}\rangle\langle\psi
_{x}|_{A^{\prime}}. \label{eq-q-cap:code-density-operator}%
\end{equation}
Now suppose that the channel$~\mathcal{N}_{A^{\prime}\rightarrow B}$ has an
isometric extension $U_{A^{\prime}\rightarrow BE}^{\mathcal{N}}$, so that
inputting $|\psi_{x}\rangle_{A^{\prime}}$ leads to the following state shared
between Bob and Eve:%
\begin{equation}
|\psi_{x}\rangle_{BE}\equiv U_{A^{\prime}\rightarrow BE}^{\mathcal{N}}%
|\psi_{x}\rangle_{A^{\prime}}. \label{eq-q-cap:output-states}%
\end{equation}
From the direct coding part of Theorem~\ref{thm-pcc:private-capacity}, we know
that there exists a private classical code $\left\{  x^{n}(m,k)\right\}
_{m\in\mathcal{M},k\in\mathcal{K}}$ with the properties in
\eqref{eq-q-cap:good-detection-cond}--\eqref{eq-q-cap:good-privacy-cond} and
with rate%
\begin{equation}
I(X;B)_{\sigma}-I(X;E)_{\sigma}, \label{eq-q-cap:private-info}%
\end{equation}
where $\sigma_{XBE}$ is a classical--quantum state of the following form:%
\begin{equation}
\sigma_{XBE}\equiv\sum_{x}p_{X}(x)|x\rangle\langle x|_{X}\otimes|\psi
_{x}\rangle\langle\psi_{x}|_{BE}.
\end{equation}
The following identity demonstrates that the private information in
\eqref{eq-q-cap:private-info} is equal to the coherent information for the
particular state $\sigma_{XBE}$ above:%
\begin{align}
I(X;B)_{\sigma}-I(X;E)_{\sigma}  &  =H(B)_{\sigma}-H(B|X)_{\sigma
}-H(E)_{\sigma}+H(E|X)_{\sigma}\\
&  =H(B)_{\sigma}-H(B|X)_{\sigma}-H(E)_{\sigma}+H(B|X)_{\sigma}\\
&  =H(B)_{\sigma}-H(E)_{\sigma}.
\end{align}
The first equality follows from the identity $I(C;D)=H(D)-H(D|C)$. The second
equality follows because the state on systems $BE$ is pure when conditioned on
the classical variable$~X$. Observe that the last expression is a function
solely of the density operator $\rho$ in
\eqref{eq-q-cap:code-density-operator}, and it is also equal to the coherent
information of the channel for the particular input state~$\rho$ (see
Exercise~\ref{ex-qie:alternate-coh-info}).

Now we show how to construct a quantum code achieving the coherent information
rate $H(B)_{\sigma}-H(E)_{\sigma}$ by making a coherent version of the above
private classical code. Suppose that Alice shares a state~$|\varphi
\rangle_{RA_{1}}$ with a reference system~$R$, where%
\begin{equation}
|\varphi\rangle_{RA_{1}}\equiv\sum_{l,m\in\mathcal{M}}\alpha_{l,m}\left\vert
l\right\rangle _{R}|m\rangle_{A_{1}}, \label{eq-q-cap:unencoded-state}%
\end{equation}
and $\{\left\vert l\right\rangle _{R}\}$ is some orthonormal basis for $R$
while $\{|m\rangle_{A_{1}}\}$ is some orthonormal basis for $A_{1}$. Also, we
set $\left\vert \mathcal{M}\right\vert \approx2^{n\left[  H(B)_{\sigma
}-H(E)_{\sigma}\right]  }$. We would like for Alice and Bob to execute a
quantum communication protocol such that Bob can reconstruct Alice's share of
the above state on his system with Alice no longer entangled with the
reference (we would like for the final state to be close to $|\varphi
\rangle_{RA_{1}}$ where Bob is holding the $A_{1}$ system). To this end, Alice
creates a quantum codebook $\{\left\vert \phi_{m}\right\rangle _{A^{\prime n}%
}\}_{m\in\mathcal{M}}$ with quantum codewords:%
\begin{equation}
\left\vert \phi_{m}\right\rangle _{A^{\prime n}}\equiv\frac{1}{\sqrt
{\left\vert \mathcal{K}\right\vert }}\sum_{k\in\mathcal{K}}e^{i\gamma_{m,k}%
}|\psi^{x^{n}(m,k)}\rangle_{A^{\prime n}}, \label{eq-q-cap:quantum-codewords}%
\end{equation}
where the states $|\psi^{x^{n}(m,k)}\rangle_{A^{\prime n}}$ are the $n$th
extensions of the states arising from the spectral decomposition in
\eqref{eq-q-cap:code-density-operator}, the classical sequences $x^{n}(m,k)$
are from the codebook for private classical communication, and we specify how
to choose the phases $\gamma_{m,k}$ later. All the states $|\psi^{x^{n}%
(m,k)}\rangle_{A^{\prime n}}$ are orthonormal because they are chosen from the
spectral decomposition in \eqref{eq-q-cap:code-density-operator} and the
expurgation from Section~\ref{sec-pcc:expurgate}\ guarantees that they are
distinct (otherwise, they would not be good codewords!). The fact that the
states $|\psi^{x^{n}(m,k)}\rangle_{A^{\prime n}}$ are orthonormal implies that
the quantum codewords $\left\vert \phi_{m}\right\rangle _{A^{\prime n}}$ are
also orthonormal.

Alice's first action is to coherently copy the value of $m$ in the $A_{1}$
register to another register~$A_{2}$, so that the state in
\eqref{eq-q-cap:unencoded-state} becomes%
\begin{equation}
\sum_{l,m}\alpha_{l,m}\left\vert l\right\rangle _{R}|m\rangle_{A_{1}}%
|m\rangle_{A_{2}}. \label{eq-q-cap:coherent-copy}%
\end{equation}
Alice then performs some isometric encoding from $A_{2}$ to $A^{\prime n}$
that takes the above unencoded state to the following encoded state:%
\begin{equation}
\sum_{l,m}\alpha_{l,m}\left\vert l\right\rangle _{R}|m\rangle_{A_{1}%
}\left\vert \phi_{m}\right\rangle _{A^{\prime n}},
\label{eq-q-cap:quantum-encoder}%
\end{equation}
where each $\left\vert \phi_{m}\right\rangle _{A^{\prime n}}$ is a quantum
codeword of the form in \eqref{eq-q-cap:quantum-codewords}. Alice transmits
the systems $A^{\prime n}$ through many uses of the quantum channel, leading
to the following state shared between the reference, Alice, Bob, and Eve:%
\begin{equation}
\sum_{l,m}\alpha_{l,m}\left\vert l\right\rangle _{R}|m\rangle_{A_{1}%
}\left\vert \phi_{m}\right\rangle _{B^{n}E^{n}},
\end{equation}
where $\left\vert \phi_{m}\right\rangle _{B^{n}E^{n}}$ is defined from
\eqref{eq-q-cap:output-states} and \eqref{eq-q-cap:quantum-codewords}. Recall
from \eqref{eq-q-cap:good-detection-cond} that Bob can detect the message $m$
and the variable $k$ in the private classical code with high probability:%
\begin{equation}
\forall m,k:\operatorname{Tr}\left\{  \Lambda_{B^{n}}^{m,k}\psi_{B^{n}}%
^{x^{n}(m,k)}\right\}  \geq1-\varepsilon. \label{eq-q-cap:identify-states}%
\end{equation}
So Bob instead constructs a coherent version of this POVM:%
\begin{equation}
\sum_{m\in\mathcal{M},k\in\mathcal{K}}\sqrt{\Lambda_{B^{n}}^{m,k}}%
\otimes|m\rangle_{B_{1}}|k\rangle_{B_{2}}. \label{eq-q-cap:bob-coherent-meas}%
\end{equation}
He then performs this coherent POVM, resulting in the state%
\begin{equation}
\sum_{\substack{m^{\prime}\in\mathcal{M},\\k^{\prime}\in\mathcal{K}}%
}\sum_{l,m}\sum_{k\in\mathcal{K}}\frac{1}{\sqrt{\left\vert \mathcal{K}%
\right\vert }}\alpha_{l,m}\left\vert l\right\rangle _{R}\left\vert
m\right\rangle _{A_{1}}\sqrt{\Lambda_{B^{n}}^{m^{\prime},k^{\prime}}%
}e^{i\gamma_{k,m}}|\psi^{x^{n}(m,k)}\rangle_{B^{n}E^{n}}\left\vert m^{\prime
},k^{\prime}\right\rangle _{B_{1}B_{2}}. \label{eq-q-cap:ideal-measured-state}%
\end{equation}
We would like for the above state to be close in trace distance to the
following state:%
\begin{equation}
\sum_{l,m}\sum_{k\in\mathcal{K}}\frac{1}{\sqrt{\left\vert \mathcal{K}%
\right\vert }}\alpha_{l,m}\left\vert l\right\rangle _{R}\left\vert
m\right\rangle _{A_{1}}e^{i\delta_{m,k}}|\psi^{x^{n}(m,k)}\rangle_{B^{n}E^{n}%
}|m\rangle_{B_{1}}|k\rangle_{B_{2}}, \label{eq-q-cap:actual-measured-state}%
\end{equation}
where $\delta_{m,k}$ are some phases that we will specify shortly. To this
end, consider that the sets $\{\left\vert \chi_{m,k}\right\rangle _{B^{n}%
E^{n}B_{1}B_{2}}\}_{m,k}$ and $\{|\varphi_{m,k}\rangle_{B^{n}E^{n}B_{1}B_{2}%
}\}_{m,k}$ form orthonormal bases, where%
\begin{align}
\left\vert \chi_{m,k}\right\rangle _{B^{n}E^{n}B_{1}B_{2}}  &  \equiv
|\psi^{x^{n}(m,k)}\rangle_{B^{n}E^{n}}|m\rangle_{B_{1}}|k\rangle_{B_{2}},\\
|\varphi_{m,k}\rangle_{B^{n}E^{n}B_{1}B_{2}}  &  \equiv\sum_{m^{\prime}%
\in\mathcal{M},k^{\prime}\in\mathcal{K}}\sqrt{\Lambda_{B^{n}}^{m^{\prime
},k^{\prime}}}|\psi^{x^{n}(m,k)}\rangle_{B^{n}E^{n}}|m^{\prime}\rangle_{B_{1}%
}|k^{\prime}\rangle_{B_{2}}.
\end{align}
Also, consider that the overlap between corresponding states in the different
bases is high:%
\begin{align}
&  \left\langle \chi_{m,k}|\varphi_{m,k}\right\rangle \nonumber\\
&  =\langle\psi^{x^{n}(m,k)}|_{B^{n}E^{n}}\langle m|_{B_{1}}\langle k|_{B_{2}%
}\sum_{m^{\prime}\in\mathcal{M},k^{\prime}\in\mathcal{K}}\sqrt{\Lambda_{B^{n}%
}^{m^{\prime},k^{\prime}}}|\psi^{x^{n}(m,k)}\rangle_{B^{n}E^{n}}|m^{\prime
}\rangle_{B_{1}}|k^{\prime}\rangle_{B_{2}}\\
&  =\sum_{m^{\prime}\in\mathcal{M},k^{\prime}\in\mathcal{K}}\langle\psi
^{x^{n}(m,k)}|_{B^{n}E^{n}}\sqrt{\Lambda_{B^{n}}^{m^{\prime},k^{\prime}}}%
|\psi^{x^{n}(m,k)}\rangle_{B^{n}E^{n}}\left\langle m|m^{\prime}\right\rangle
_{B_{1}}\left\langle k|k^{\prime}\right\rangle _{B_{2}}\\
&  =\langle\psi^{x^{n}(m,k)}|_{B^{n}E^{n}}\sqrt{\Lambda_{B^{n}}^{m,k}}%
|\psi^{x^{n}(m,k)}\rangle_{B^{n}E^{n}}\\
&  \geq\langle\psi^{x^{n}(m,k)}|_{B^{n}E^{n}}\Lambda_{B^{n}}^{m,k}|\psi
^{x^{n}(m,k)}\rangle_{B^{n}E^{n}}\\
&  =\operatorname{Tr}\left\{  \Lambda_{B^{n}}^{m,k}\psi_{B^{n}}^{x^{n}%
(m,k)}\right\} \\
&  \geq1-\varepsilon,
\end{align}
where the first inequality follows from the fact that $\sqrt{\Lambda_{B^{n}%
}^{m,k}}\geq\Lambda_{B^{n}}^{m,k}$ for $\Lambda_{B^{n}}^{m,k}\leq I$ and the
second inequality follows from \eqref{eq-q-cap:identify-states}. By applying
Lemma~\ref{lem-app:fourier-states}\ from Appendix~\ref{chap:appendix-math}, we
know that there exist phases $\gamma_{m,k}$ and $\delta_{m,k}$ such that%
\begin{equation}
\left\langle \chi_{m}|\varphi_{m}\right\rangle \geq1-\varepsilon,
\end{equation}
where%
\begin{align}
\left\vert \chi_{m}\right\rangle _{B^{n}E^{n}B_{1}B_{2}}  &  \equiv\frac
{1}{\sqrt{\left\vert \mathcal{K}\right\vert }}\sum_{k}e^{i\delta_{m,k}%
}\left\vert \chi_{m,k}\right\rangle _{B^{n}E^{n}B_{1}B_{2}},\\
|\varphi_{m}\rangle_{B^{n}E^{n}B_{1}B_{2}}  &  \equiv\frac{1}{\sqrt{\left\vert
\mathcal{K}\right\vert }}\sum_{k}e^{i\gamma_{m,k}}|\varphi_{m,k}\rangle
_{B^{n}E^{n}B_{1}B_{2}}.
\end{align}
So we choose the phases in a way such that the above inequality holds. We can
then apply the above result to show that the state in
\eqref{eq-q-cap:ideal-measured-state} has high fidelity with the state in
\eqref{eq-q-cap:actual-measured-state}:%
\begin{align}
&  \left(  \sum_{l,m}\alpha_{l,m}^{\ast}\langle l|_{R}\langle m|_{A_{1}%
}\left\langle \chi_{m}\right\vert _{B^{n}E^{n}B_{1}B_{2}}\right)  \left(
\sum_{l^{\prime},m^{\prime}}\alpha_{l^{\prime},m^{\prime}}\left\vert
l^{\prime}\right\rangle _{R}|m^{\prime}\rangle_{A_{1}}|\varphi_{m^{\prime}%
}\rangle_{B^{n}E^{n}B_{1}B_{2}}\right) \nonumber\\
&  =\sum_{l,m,l^{\prime},m^{\prime}}\alpha_{l,m}^{\ast}\alpha_{l^{\prime
},m^{\prime}}\left\langle l|l^{\prime}\right\rangle _{R}\left\langle
m|m^{\prime}\right\rangle _{A_{1}}\left\langle \chi_{m}|\varphi_{m^{\prime}%
}\right\rangle _{B^{n}E^{n}B_{1}B_{2}}\\
&  =\sum_{l,m}\left\vert \alpha_{l,m}\right\vert ^{2}\left\langle \chi
_{m}|\varphi_{m}\right\rangle _{B^{n}E^{n}B_{1}B_{2}}\\
&  \geq1-\varepsilon.
\end{align}
Thus, the state resulting after Bob performs the coherent POVM\ is close in
trace distance to the following state:%
\begin{multline}
\sum_{l,m\in\mathcal{M}}\alpha_{l,m}\left\vert l\right\rangle _{R}\left\vert
m\right\rangle _{A_{1}}\frac{1}{\sqrt{\left\vert \mathcal{K}\right\vert }}%
\sum_{k\in\mathcal{K}}e^{i\delta_{m,k}}|\psi^{x^{n}(m,k)}\rangle_{B^{n}E^{n}%
}|m\rangle_{B_{1}}|k\rangle_{B_{2}}\label{eq-q-cap:final-step-1}\\
=\sum_{l,m\in\mathcal{M}}\alpha_{l,m}\left\vert l\right\rangle _{R}\left\vert
m\right\rangle _{A_{1}}|\widetilde{\phi}_{m}\rangle_{B^{n}E^{n}B_{2}}%
|m\rangle_{B_{1}},
\end{multline}
where%
\begin{equation}
|\widetilde{\phi}_{m}\rangle_{B^{n}E^{n}B_{2}}\equiv\frac{1}{\sqrt{\left\vert
\mathcal{K}\right\vert }}\sum_{k\in\mathcal{K}}e^{i\delta_{m,k}}|\psi
^{x^{n}(m,k)}\rangle_{B^{n}E^{n}}|k\rangle_{B_{2}}.
\end{equation}
Consider the state of Eve for a particular value of $m$:%
\begin{align}
\left[  \widetilde{\phi}_{m}\right]  _{E^{n}}  &  =\operatorname{Tr}%
_{B^{n}B_{2}}\left\{  |\widetilde{\phi}_{m}\rangle\langle\widetilde{\phi}%
_{m}|_{B^{n}E^{n}B_{2}}\right\} \\
&  =\operatorname{Tr}_{B^{n}B_{2}}\left\{  \sum_{k,k^{\prime}\in\mathcal{K}%
}\frac{1}{\left\vert \mathcal{K}\right\vert }e^{i(\delta_{m,k^{\prime}}%
-\delta_{m,k})}|\psi^{x^{n}(m,k)}\rangle\langle\psi^{x^{n}\left(  m,k^{\prime
}\right)  }|_{B^{n}E^{n}}\otimes|k\rangle\langle k^{\prime}|_{B_{2}}\right\}
\\
&  =\frac{1}{\left\vert \mathcal{K}\right\vert }\sum_{k\in\mathcal{K}}%
\psi_{E^{n}}^{x^{n}(m,k)}.
\end{align}
We are now in a position to apply the second property of the private classical
code. Recall from the privacy condition in \eqref{eq-q-cap:good-privacy-cond}
that Eve's state is guaranteed to be $\varepsilon$-close in trace distance to
the tensor-power state $\left[  \mathcal{N}_{A^{\prime}\rightarrow E}^{c}%
(\rho)\right]  ^{\otimes n}$, where $\mathcal{N}_{A^{\prime}\rightarrow E}%
^{c}$ is the complementary channel and $\rho$ is the density operator in
\eqref{eq-q-cap:code-density-operator}. Let $|\theta_{\mathcal{N}^{c}(\rho
)}\rangle_{E^{n}B_{3}}$ be some purification of this tensor-power state. By
Uhlmann's theorem and the relation between trace distance and fidelity (see
Definition~\ref{def-dm:uhlmann} and
Theorem~\ref{thm-dm:fidelity-trace-relation}), there is some isometry
$U_{B^{n}B_{2}\rightarrow B_{3}}^{m}$ for each value of $m$ such that the
following states are $2\sqrt{\varepsilon}$-close in trace distance (see
Exercise~\ref{ex-dm:decouple}):%
\begin{equation}
U_{B^{n}B_{2}\rightarrow B_{3}}^{m}|\widetilde{\phi}_{m}\rangle_{B^{n}%
E^{n}B_{2}}\ \ \ \ \ \ \overset{2\sqrt{\varepsilon}}{\approx}%
\ \ \ \ \ \ |\theta_{\mathcal{N}^{c}(\rho)}\rangle_{E^{n}B_{3}}.
\end{equation}
Bob then performs the following controlled isometry on his systems$~B^{n}$,
$B_{1}$, and $B_{2}$:%
\begin{equation}
\sum_{m}|m\rangle\langle m|_{B_{1}}\otimes U_{B^{n}B_{2}\rightarrow B_{3}}%
^{m}, \label{eq-q-cap:decoupler}%
\end{equation}
leading to a state that is close in trace distance to the following state:%
\begin{equation}
\left(  \sum_{l,m\in\mathcal{M}}\alpha_{l,m}\left\vert l\right\rangle
_{R}|m\rangle_{A_{1}}|m\rangle_{B_{1}}\right)  \otimes|\theta_{\mathcal{N}%
^{c}(\rho)}\rangle_{E^{n}B_{3}}. \label{eq-q-cap:final-step-2}%
\end{equation}
At this point, the key observation is that the state on $E^{n}B_{3}$ is
effectively decoupled from the state on systems~$R$, $A_{1}$, and $B_{1}$, so
that Bob can just throw away his system $B_{3}$. Thus, they have successfully
implemented an approximate coherent channel from system $A_{1}$ to $A_{1}%
B_{1}$.

\begin{figure}
[ptb]
\begin{center}
\includegraphics[
width=4.8456in
]%
{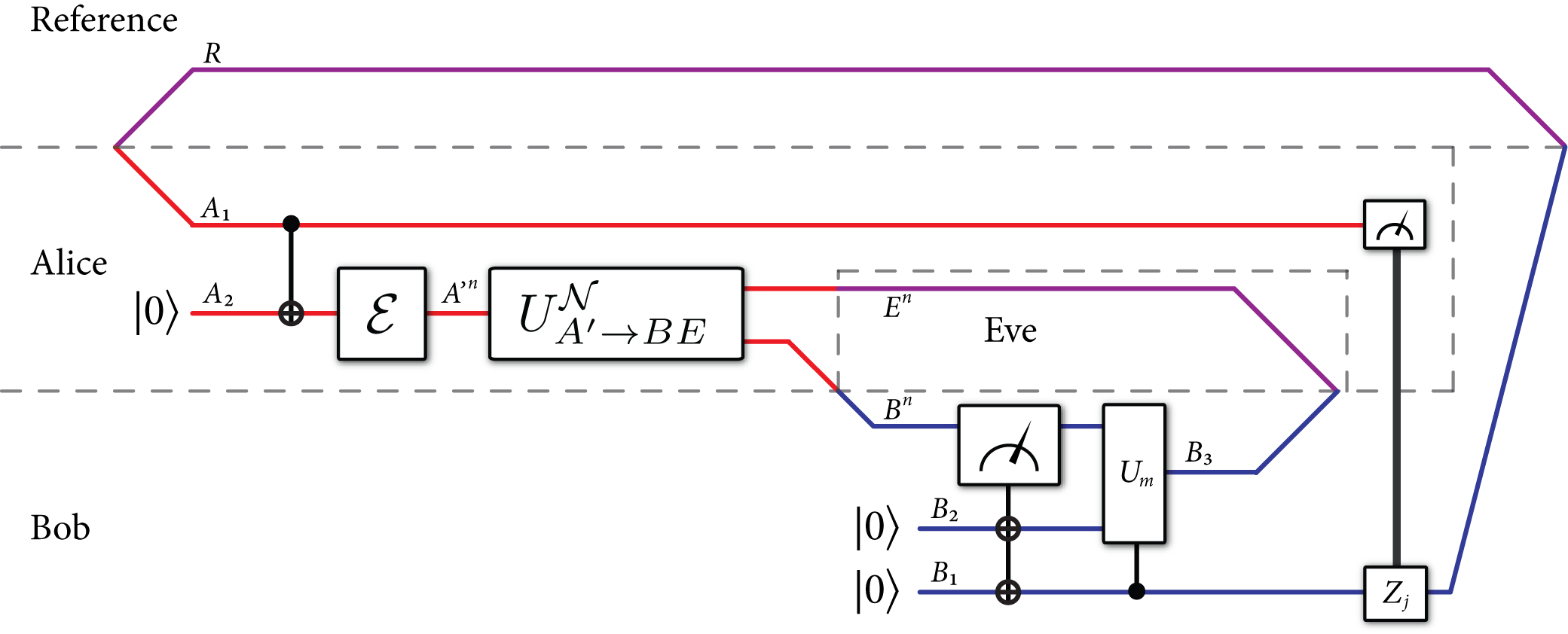}%
\caption{All of the steps in the protocol for quantum communication. Alice and
Bob's goal is to communicate as much quantum information as they can while
making sure that Eve's state is independent of what Alice is trying to
communicate to Bob. The figure depicts the series of controlled unitaries that
Alice and Bob perform and the final measurement and classical communication
that enables quantum communication from Alice to Bob at the coherent
information rate.}%
\label{fig-q-cap:direct-coding}%
\end{center}
\end{figure}

We now allow for Alice to communicate classical information to Bob in order
for them to implement a quantum communication channel rather than just a mere
coherent channel (in a moment we argue that this free forward classical
communication is not necessary). Alice performs a Fourier transform on the
register $A_{1}$, leading to the following state:%
\begin{equation}
\frac{1}{\sqrt{d_{A_{1}}}}\sum_{l,m,j\in\mathcal{M}}\alpha_{l,m}\exp\left\{
2\pi imj/d_{A_{1}}\right\}  \left\vert l\right\rangle _{R}|j\rangle_{A_{1}%
}|m\rangle_{B_{1}},
\end{equation}
where $d_{A_{1}}\equiv\dim(\mathcal{H}_{A_{1}})$. She then measures register
$A_{1}$ in the computational basis, leading to some outcome$~j$ and the
following post-measurement state:%
\begin{equation}
\left(  \sum_{l,m\in\mathcal{M}}\alpha_{l,m}\exp\left\{  2\pi imj/d_{A_{1}%
}\right\}  \left\vert l\right\rangle _{R}|m\rangle_{B_{1}}\right)
\otimes|j\rangle_{A_{1}}.
\end{equation}
She sends Bob the outcome~$j$ of her measurement over a classical channel, and
the protocol ends with Bob performing the following unitary:%
\begin{equation}
Z^{\dag}(j)|m\rangle_{B_{1}}=\exp\left\{  -2\pi imj/d_{A_{1}}\right\}
|m\rangle_{B_{1}}, \label{eq-q-cap:bob-phase-rotation}%
\end{equation}
leaving the desired state on the reference and Bob's system~$B_{1}$:%
\begin{equation}
\sum_{l,m\in\mathcal{M}}\alpha_{l,m}\left\vert l\right\rangle _{R}\left\vert
m\right\rangle _{B_{1}}.
\end{equation}
All of the errors accumulated in the above protocol are some finite sum of
$\varepsilon$ terms, and applying the triangle inequality several times
implies that the actual state is close to the desired state in the asymptotic
limit of large block length. Figure~\ref{fig-q-cap:direct-coding}\ depicts all
of the steps in this protocol for quantum communication.%

We now argue that the classical communication is not necessary---there exists
a scheme that does not require the use of this forward classical channel.
After reviewing the above protocol and glancing at
Figure~\ref{fig-q-cap:direct-coding}, we realize that Alice's encoder is a
quantum instrument of the following form:%
\begin{equation}
\mathcal{E}(\rho)\equiv\sum_{j}\mathcal{E}_{j}(\rho)\otimes|j\rangle\langle
j|.
\end{equation}
Each map $\mathcal{E}_{j}$ is trace-non-increasing and has the following
action on a pure-state input $|\varphi\rangle_{RA_{1}}$:%
\begin{equation}
\langle j|_{A_{1}}F_{A_{1}}\left(  \sum_{m^{\prime}}|\phi_{m^{\prime}}%
\rangle\langle m^{\prime}|_{A_{2}}\right)  \left(  \sum_{m}|m\rangle\langle
m|_{A_{1}}\otimes|m\rangle_{A_{2}}\right)  |\varphi\rangle_{RA_{1}},
\end{equation}
where $\sum_{m}|m\rangle\langle m|_{A_{1}}\otimes\left\vert m\right\rangle
_{A_{2}}$ is Alice's coherent copier in \eqref{eq-q-cap:coherent-copy},
$\sum_{m^{\prime}}|\phi_{m^{\prime}}\rangle\langle m^{\prime}|_{A_{2}}$ is her
quantum encoder in \eqref{eq-q-cap:quantum-encoder}, $F_{A_{1}}$ is the
Fourier transform, and $\langle j|_{A_{1}}$ represents the projection onto a
particular measurement outcome~$j$. We can simplify the above expression as
follows:%
\begin{align}
&  =\langle j|_{A_{1}}\sum_{m^{\prime}}|\widetilde{m}^{\prime}\rangle\langle
m^{\prime}|_{A_{1}}\left(  \sum_{m}|m\rangle\langle m|_{A_{1}}\otimes
\left\vert \phi_{m}\right\rangle _{A_{2}}\right)  |\varphi\rangle_{RA_{1}}\\
&  =\langle j|_{A_{1}}\sum_{m}|\widetilde{m}\rangle\langle m|_{A_{1}}%
\otimes\left\vert \phi_{m}\right\rangle _{A_{2}}|\varphi\rangle_{RA_{1}}\\
&  =\left(  \frac{1}{\sqrt{\left\vert \mathcal{M}\right\vert }}\sum
_{m}e^{i2\pi mj/\left\vert \mathcal{M}\right\vert }\left\vert \phi
_{m}\right\rangle _{A_{2}}\langle m|_{A_{1}}\right)  |\varphi\rangle_{RA_{1}}.
\label{eq-q-cap:encoding-isometries}%
\end{align}
It follows that the trace of each $\mathcal{E}_{j}$ is uniform and independent
of the input state $|\varphi\rangle_{RA_{1}}$:%
\begin{equation}
\operatorname{Tr}\left\{  \mathcal{E}_{j}(\varphi_{RA_{1}})\right\}  =\frac
{1}{\left\vert \mathcal{M}\right\vert }.
\end{equation}
Observe that multiplying the map in \eqref{eq-q-cap:encoding-isometries} by
$\sqrt{\left\vert \mathcal{M}\right\vert }$ gives a proper isometry that could
suffice as an encoding. Let $\mathcal{E}_{j}^{\prime}$ denote the rescaled
isometry. Corresponding to each encoder is a decoding map~$\mathcal{D}_{j}$
consisting of Bob's coherent measurement in
\eqref{eq-q-cap:bob-coherent-meas}, his decoupler in
\eqref{eq-q-cap:decoupler}, and his phase shifter in
\eqref{eq-q-cap:bob-phase-rotation}. We can thus represent the state output
from our classically coordinated protocol as follows:%
\begin{equation}
\sum_{j}\mathcal{D}_{j}\left(  \mathcal{N}^{\otimes n}\left(  \mathcal{E}%
_{j}\left(  \varphi_{RA_{1}}\right)  \right)  \right)  .
\end{equation}
From the analysis in the preceding paragraphs, we know that the trace distance
between the ideal state and the actual state is small for the classically
coordinated scheme:%
\begin{equation}
\left\Vert \sum_{j}\mathcal{D}_{j}\left(  \mathcal{N}^{\otimes n}\left(
\mathcal{E}_{j}\left(  \varphi_{RA_{1}}\right)  \right)  \right)
-\varphi_{RA_{1}}\right\Vert _{1}\leq\varepsilon^{\prime},
\end{equation}
where $\varepsilon^{\prime}$ is some arbitrarily small positive number. Thus,
the fidelity between these two states is high:%
\begin{equation}
F\left(  \sum_{j}\mathcal{D}_{j}\left(  \mathcal{N}^{\otimes n}\left(
\mathcal{E}_{j}\left(  \varphi_{RA_{1}}\right)  \right)  \right)
,\varphi_{RA_{1}}\right)  \geq1-\varepsilon^{\prime}.
\end{equation}
But we can rewrite the fidelity as follows:%
\begin{align}
&  F\left(  \sum_{j}\mathcal{D}_{j}\left(  \mathcal{N}^{\otimes n}\left(
\mathcal{E}_{j}\left(  \varphi_{RA_{1}}\right)  \right)  \right)
,\varphi_{RA_{1}}\right) \nonumber\\
&  =\langle\varphi|_{RA_{1}}\sum_{j}\mathcal{D}_{j}\left(  \mathcal{N}%
^{\otimes n}\left(  \mathcal{E}_{j}\left(  \varphi_{RA_{1}}\right)  \right)
\right)  |\varphi\rangle_{RA_{1}}\\
&  =\sum_{j}\langle\varphi|_{RA_{1}}\mathcal{D}_{j}\left(  \mathcal{N}%
^{\otimes n}\left(  \mathcal{E}_{j}\left(  \varphi_{RA_{1}}\right)  \right)
\right)  |\varphi\rangle_{RA_{1}}\\
&  =\sum_{j}\frac{1}{\left\vert \mathcal{M}\right\vert }\left[  \langle
\varphi|_{RA_{1}}\mathcal{D}_{j}\left(  \mathcal{N}^{\otimes n}\left(
\mathcal{E}_{j}^{\prime}\left(  \varphi_{RA_{1}}\right)  \right)  \right)
|\varphi\rangle_{RA_{1}}\right] \\
&  \geq1-\varepsilon^{\prime},
\end{align}
implying that at least one of the encoder--decoder pairs $(\mathcal{E}%
_{j}^{\prime},\mathcal{D}_{j})$ has arbitrarily high fidelity. Thus, Alice and
Bob simply agree beforehand to use a scheme $(\mathcal{E}_{j}^{\prime
},\mathcal{D}_{j})$ with high fidelity, obviating the need for the forward
classical communication channel.

The protocol given here achieves communication at the coherent information
rate. In order to achieve the regularized coherent information rate in the
statement of the theorem, Alice and Bob apply the same protocol to the
superchannel $(\mathcal{N}_{A^{\prime}\rightarrow B})^{\otimes k}$ instead of
the channel~$\mathcal{N}_{A^{\prime}\rightarrow B}$.

\section{Converse Theorem}%

\index{quantum capacity theorem!converse}
This section proves the converse part of the quantum capacity theorem,
demonstrating that the regularized coherent information is an upper bound on
the quantum capacity of any quantum channel. For the class of
\index{degradable channel}%
degradable channels, the coherent information itself is an upper bound on the
quantum capacity---this demonstrates that we completely understand the quantum
data transmission capabilities of these channels.

For this converse proof, we assume that Alice is trying to generate
entanglement with Bob. The capacity for this task is an upper bound on the
capacity for quantum data transmission because we can always use a noiseless
quantum channel to establish entanglement. We also allow Alice free forward
classical communication to Bob, and we demonstrate that this resource cannot
increase the quantum capacity (essentially because the coherent information is
convex). In a protocol for entanglement generation, Alice begins by preparing
the maximally entangled state $\Phi_{AA_{1}}$ of Schmidt rank $\left\vert
A\right\vert $ in her local laboratory, where $\frac{1}{n}\log\left\vert
A\right\vert $ is the rate of this entangled state. She performs some encoding
operation $\mathcal{E}_{A_{1}\rightarrow A^{\prime n}M}$ that outputs many
systems $A^{\prime n}$ and a classical register $M$. She then inputs the
systems $A^{\prime n}$ to many independent uses of a quantum
channel~$\mathcal{N}_{A^{\prime}\rightarrow B}$, resulting in the state%
\begin{equation}
\omega_{AMB^{n}}\equiv\mathcal{N}_{A^{\prime n}\rightarrow B^{n}}%
(\mathcal{E}_{A_{1}\rightarrow A^{\prime n}M}(\Phi_{AA_{1}})),
\end{equation}
where $\mathcal{N}_{A^{\prime n}\rightarrow B^{n}}\equiv(\mathcal{N}%
_{A^{\prime}\rightarrow B})^{\otimes n}$. Bob takes the outputs $B^{n}$ of the
channels and the classical register $M$ and performs some decoding operation
$\mathcal{D}_{B^{n}M\rightarrow B_{1}}$, resulting in the state%
\begin{equation}
\omega_{AB_{1}}^{\prime}\equiv\mathcal{D}_{B^{n}M\rightarrow B_{1}}%
(\omega_{AMB^{n}}).
\end{equation}
The following condition holds for an $(n,\left[  \log\left\vert A\right\vert
\right]  /n,\varepsilon)$ protocol for entanglement generation:%
\begin{equation}
\frac{1}{2}\left\Vert \omega_{AB_{1}}^{\prime}-\Phi_{AB_{1}}\right\Vert
_{1}\leq\varepsilon. \label{eq-q-cap:good-quantum-code}%
\end{equation}

The converse proof then proceeds in the following steps:%
\begin{align}
\log\left\vert A\right\vert  &  =I(A\rangle B_{1})_{\Phi}\\
&  \leq I(A\rangle B_{1})_{\omega^{\prime}}+f(\left\vert A\right\vert
,\varepsilon)\\
&  \leq I(A\rangle B^{n}M)_{\omega}+f(\left\vert A\right\vert ,\varepsilon).
\end{align}
The first equality follows because the coherent information of a maximally
entangled state is equal to the logarithm of the dimension of one of its
systems. The first inequality follows from an application of the AFW
inequality to the condition in \eqref{eq-q-cap:good-quantum-code}, with
$f(\left\vert A\right\vert ,\varepsilon)\equiv2\varepsilon\log\left\vert
A\right\vert +g_2(\varepsilon)  $. The second inequality follows from quantum
data processing. Now consider that the state $\omega_{MAB^{n}}$ is a
classical--quantum state of the following form:%
\begin{equation}
\omega_{MAB^{n}}\equiv\sum_{m}p_{M}(m)|m\rangle\langle m|_{M}\otimes
\mathcal{N}_{A^{\prime n}\rightarrow B^{n}}(\rho_{AA^{\prime n}}^{m}).
\end{equation}
We can then perform a spectral decomposition of each state $\rho_{m}$ as
follows:%
\begin{equation}
\rho_{AA^{\prime n}}^{m}=\sum_{l}p_{L|M}(l|m)|\phi_{l,m}\rangle\langle
\phi_{l,m}|_{AA^{\prime n}},
\end{equation}
and augment the above state as follows:%
\begin{equation}
\omega_{MLAB^{n}}\equiv\sum_{m,l}p_{M}(m)p_{L|M}(l|m)|m\rangle\langle
m|_{M}\otimes|l\rangle\langle l|_{L}\otimes\mathcal{N}_{A^{\prime
n}\rightarrow B^{n}}(\phi_{AA^{\prime n}}^{l,m}),
\label{eq-q-cap:augemented-state}%
\end{equation}
so that $\omega_{MAB^{n}}=\operatorname{Tr}_{L}\left\{  \omega_{MLAB^{n}%
}\right\}  $. We continue with bounding the relevant term:%
\begin{align}
I(A\rangle B^{n}M)_{\omega}  &  \leq I(A\rangle B^{n}ML)_{\omega}\\
&  =\sum_{m,l}p_{M}(m)p_{L|M}(l|m)I(A\rangle B^{n})_{\mathcal{N}(\phi^{l,m}%
)}\\
&  \leq I(A\rangle B^{n})_{\mathcal{N}(\phi_{l,m}^{\ast})}\\
&  \leq Q(\mathcal{N}^{\otimes n}).
\end{align}
The first inequality follows from the quantum data-processing inequality. The
first equality follows because the registers $M$ and $L$ are both classical,
and we can apply the result of Exercise~\ref{ex-qie:cond-coh-info}. The second
inequality follows because the expectation is always less than the maximal
value (where we define $\phi_{l,m}^{\ast}$ to be the state that achieves this
maximum). The final inequality follows from the definition of the channel
coherent information as the maximum of the coherent information over all pure,
bipartite inputs. Putting everything together, we find that%
\[
\frac{1}{n}\log\left\vert A\right\vert \left(  1-2\varepsilon\right)
\leq\frac{1}{n}Q(\mathcal{N}^{\otimes n})+\frac{g_2(\varepsilon)}{n}  .
\]
Thus, if we are considering a sequence of $(n,\left[  \log\left\vert
A\right\vert \right]  /n,\varepsilon_{n})$\ quantum communication protocols
with rate $Q-\delta_{n}=\frac{1}{n}\log\left\vert A\right\vert $, such that
$\lim_{n\rightarrow\infty}\varepsilon_{n}=\lim_{n\rightarrow\infty}\delta
_{n}=0$, then the above bound becomes%
\begin{equation}
\left(  Q-\delta_{n}\right)  \left(  1-2\varepsilon_{n}\right)  \leq\frac
{1}{n}Q(\mathcal{N}^{\otimes n})+\frac{g_(\varepsilon_{n})}{n}.
\end{equation}
Taking the limit as $n\rightarrow\infty$ then establishes that an achievable
rate $Q$ necessarily satisfies $Q\leq Q_{\operatorname{reg}}(\mathcal{N})$,
where $Q_{\operatorname{reg}}(\mathcal{N})$ is the regularized coherent
information given in \eqref{eq-qc:reg-coherent-info}. This concludes the proof
of the converse part of the quantum capacity theorem.

There are a few comments we should make regarding the converse theorem. First,
we see that classical communication cannot improve quantum capacity because
the coherent information is convex. We could obtain the same upper bound on
quantum capacity even if there were no classical communication. Second, it is
sufficient to consider isometric encoders for quantum communication---that is,
it is not necessary to exploit general noisy CPTP\ maps at the encoder. This
makes sense intuitively because it would seem odd if noisy encodings could
help in the noiseless transmission of quantum data. Our augmented state in
\eqref{eq-q-cap:augemented-state} and the subsequent development reveals that
this is so (again because the coherent information is convex).

We can significantly strengthen the statement of the quantum capacity theorem
for the class of
\index{degradable channel}%
degradable quantum channels because the following equality holds for them:%
\begin{equation}
Q(\mathcal{N}^{\otimes n})=nQ(\mathcal{N}).
\end{equation}
This inequality follows from the additivity of coherent information for
degradable channels (Theorem~\ref{thm:coh-info-additivity}). Also, the task of
optimizing the coherent information for these channels is straightforward
because it is a concave function of the input density operator
(Theorem~\ref{thm-add:coh-deg-concave-input}) and the set of density operators
is convex.

\section{An Interlude with Quantum Stabilizer Codes}

\label{qc-sec:stabilizer} We now describe a well-known class of quantum
\index{stabilizer codes}
error-correcting codes known as the stabilizer codes, and we prove that a
randomly chosen stabilizer code achieves a quantum communication rate known as
the hashing bound of a Pauli channel (the hashing bound is equal to the
coherent information of a Pauli channel when sending one share of a Bell state
through it). The proof of this theorem is different from our proof above that
the coherent information rate is achievable, and we consider it instructive to
see this other approach for the special case of stabilizer codes used for
protecting quantum information sent over many independent instances of a Pauli
channel. Before delving into the proof, we first briefly introduce the simple
repetition code and the more general stabilizer quantum codes.

\subsection{The Qubit Repetition Code}%

\index{repetition code!quantum}%
The simplest quantum error-correction code is the repetition code, which
encodes one qubit $|\psi\rangle\equiv\alpha|0\rangle+\beta|1\rangle$\ into
three physical qubits as follows:%
\begin{equation}
\alpha|0\rangle+\beta|1\rangle\rightarrow\alpha|000\rangle+\beta|111\rangle.
\label{qc-eq:rep-encoded-state}%
\end{equation}
A simple way to perform this encoding is to attach two ancilla qubits in the
state $|0\rangle$ to the original qubit and perform a CNOT\ gate from the
first qubit to the second and from the first to the last. This encoding
illustrates one of the fundamental principles of quantum error correction: the
quantum information is spread across the correlations between the three
physical qubits after the encoding takes place. (Of course, this was also the
case for the codes we constructed in the direct part of the quantum capacity theorem.)

The above encoding will protect the encoded qubit against an artificial noise
where either the first, second, or third qubit is subjected to a bit flip (and
no other errors occur). For example, if a bit flip occurs on the second qubit,
the encoded state changes as follows:%
\begin{equation}
X_{2}\left(  \alpha|000\rangle+\beta|111\rangle\right)  =\alpha|010\rangle
+\beta|101\rangle, \label{qc-eq:error-2nd-qubit}%
\end{equation}
where the notation $X_{2}$ indicates that a Pauli operator $X$ acts on the
second qubit. The procedure for the receiver to recover from such an error is
to perform collective measurements on all three qubits that learn only about
the error and nothing about the encoded quantum data. In this case, the
receiver can perform a measurement of the operators $Z_{1}Z_{2}$ and
$Z_{2}Z_{3}$ to learn only about the error, so that the coherent superposition
is preserved. One can easily verify that $Z_{1}Z_{2}$ and $Z_{2}Z_{3}$ are as
follows:%
\begin{align}
Z_{1}Z_{2}  &  \equiv Z\otimes Z\otimes I=\left[  \left(  |00\rangle
\langle00|+|11\rangle\langle11|\right)  -\left(  |01\rangle\langle
01|+|10\rangle\langle10|\right)  \right]  \otimes I,\\
Z_{2}Z_{3}  &  \equiv I\otimes Z\otimes Z=I\otimes\left[  \left(
|00\rangle\langle00|+|11\rangle\langle11|\right)  -\left(  |01\rangle
\langle01|+|10\rangle\langle10|\right)  \right]  ,
\end{align}
revealing that these measurements return a $+1$ if the parity of the basis
states is even and $-1$ if the parity is odd. So, for our example error in
(\ref{qc-eq:error-2nd-qubit}), the syndrome measurements will return $-1$ for
$Z_{1}Z_{2}$ and $-1$ for $Z_{2}Z_{3}$, which the receiver can use to identify
the error that occurs. He can then perform the bit-flip operator $X_{2}$ to
invert the action of the error. One can verify that the following syndrome
table identifies which type of error occurs:%
\begin{equation}%
\begin{tabular}
[c]{cc}\hline\hline
Measurement result & Error\\\hline\hline
$+1,+1$ & $I$\\\hline
$+1,-1$ & $X_{3}$\\\hline
$-1,+1$ & $X_{1}$\\\hline
$-1,-1$ & $X_{2}$\\\hline\hline
\end{tabular}
\end{equation}
Thus, if the only errors that occur are either no error or a single-qubit
bit-flip error, then it is possible to perfectly correct these. If errors
besides these ones occur, then it is not possible to correct them using this code.

\subsection{Stabilizer Codes}

We can generalize the main idea behind the above qubit repetition code to
formulate the class of quantum stabilizer codes. These stabilizer codes then
generalize the classical theory of linear error correction to the quantum case.

In the repetition code, observe that the encoded state in
(\ref{qc-eq:rep-encoded-state}) is a +1-eigenstate of the operators
$Z_{1}Z_{2}$ and $Z_{2}Z_{3}$, i.e., it holds that%
\begin{equation}
Z_{1}Z_{2}\left(  \alpha|000\rangle+\beta|111\rangle\right)  =\alpha
|000\rangle+\beta|111\rangle=Z_{2}Z_{3}\left(  \alpha|000\rangle
+\beta|111\rangle\right)  .
\end{equation}
We say that the operators $Z_{1}Z_{2}$ and $Z_{2}Z_{3}$ stabilize the encoded
state. The stabilizing operators form a group under multiplication because we
obtain another stabilizing operator if we multiply two of them:\ one can check
that the operator $Z_{1}Z_{3}$ stabilizes the encoded state and that
$Z_{1}Z_{3}=(Z_{1}Z_{2})(Z_{2}Z_{3})$. Also, the two operators $Z_{1}Z_{2}$
and $Z_{2}Z_{3}$ commute, implying that the encoded state is in the
simultaneous eigenspace of these operators, and that it is possible to measure
the operators $Z_{1}Z_{2}$ and $Z_{2}Z_{3}$ in any order, in order to learn
about errors that occur.

We now describe the theory of quantum stabilizer codes. Recall that the Pauli
matrices for one qubit are $I$, $X$, $Y$, and $Z$, whose action on the
computational basis is as follows:%
\begin{align}
I\vert0\rangle &  =\vert0\rangle, & I\vert1\rangle &  =\vert1\rangle,\\
X\vert0\rangle &  =\vert1\rangle, & X\vert1\rangle &  =\vert0\rangle,\\
Y\vert0\rangle &  =i\vert1\rangle, & Y\vert1\rangle &  =-i\vert0\rangle,\\
Z\vert0\rangle &  =\vert0\rangle, & Z\vert1\rangle &  =-\vert1\rangle.
\end{align}
The $X$ operator is known as the \textquotedblleft bit-flip\textquotedblright%
\ operator, $Z$ as the \textquotedblleft phase-flip\textquotedblright%
\ operator, and $Y$ as the \textquotedblleft bit and phase
flip\textquotedblright\ operator. The Pauli group $\mathcal{G}_{n}$\ acting on
$n$ qubits consists of $n$-fold tensor products of these operators along with
the phase factors $\pm1$ and $\pm i$:%
\begin{equation}
\mathcal{G}_{n}\equiv\left\{  \pm1,\pm i\right\}  \otimes\left\{
I,X,Y,Z\right\}  ^{\otimes n}.
\end{equation}
The inclusion of the phase factors, along with the relations $Y=iXZ$, $Z=iYX$,
and $X=iZY$ and the fact that any one of $X$, $Y$, and $Z$ anticommutes with
the other two ensures that the set $\mathcal{G}_{n}$ is closed under
multiplication. It is useful in the theory of quantum error correction to
consider the Pauli group quotiented out by its center:\ $\mathcal{G}%
_{n}/\left\{  \pm1,\pm i\right\}  $, essentially because global phases are not
physically observable. This reduced version of the Pauli group has $4^{n}$ elements.

Let $\mathcal{S}$ be an abelian subgroup of the Pauli group $\mathcal{G}_{n}$.
Any such subgroup $\mathcal{S}$ has size $2^{n-k}$ for some integer $k$ such
that $0\leq k\leq n$. This subgroup $\mathcal{S}$ can be generated by a set of
size $n-k$, so that $\mathcal{S}=\left\langle S_{1},\ldots,S_{n-k}%
\right\rangle $. A state $\vert\psi\rangle$ is stabilized by the subgroup
$\mathcal{S}$ if%
\begin{equation}
S\vert\psi\rangle=\vert\psi\rangle\ \ \ \ \ \ \ \forall S\in\mathcal{S}.
\end{equation}
The $2^{k}$-dimensional subspace of the full $2^{n}$-dimensional space for the
$n$ qubits that is stabilized by $\mathcal{S}$ is known as the codespace, or
equivalently, an $\left[  n,k\right]  $ stabilizer code that encodes $k$
logical qubits into $n$ physical qubits. The decoding operation that the
receiver performs is analogous to that for the repetition code---he just
measures the $n-k$ operators constituting some generating set of $\mathcal{S}$
and performs a recovery operation based on the results of these measurements.

We can define logical operations on the quantum information encoded inside an
$\left[  n,k\right]  $ stabilizer code. These are operations that manipulate
the quantum information inside the codespace without taking the encoded
information outside the codespace. These logical operations are part of the
normalizer of $\mathcal{S}$, defined as%
\begin{equation}
N(\mathcal{S})\equiv\left\{  U\in\mathbb{U}(2^{n}):U\mathcal{S}U^{\dag
}=\mathcal{S}\right\}  ,
\end{equation}
where $\mathbb{U}(2^{n})$ denotes the unitary group for $n$ qubits. We can
easily see that any $U\in N(\mathcal{S})$ does not take a state $|\psi\rangle$
in the codespace outside it. First, for all $U\in N(\mathcal{S})$, it follows
that $U^{\dag}\in N(\mathcal{S})$, so that for all $S\in\mathcal{S}$, we have%
\begin{equation}
SU|\psi\rangle=UU^{\dag}SU|\psi\rangle=US_{U}|\psi\rangle=U|\psi\rangle,
\end{equation}
where $S_{U}=U^{\dag}SU$ and $S_{U}\in\mathcal{S}$ from the definition of the
normalizer. From the above, we conclude that the state $U|\psi\rangle$ is in
the codespace since it is stabilized by all $S\in\mathcal{S}$: $SU|\psi
\rangle=U|\psi\rangle$. It also follows that $\mathcal{S}\subseteq
N(\mathcal{S})$ because $\mathcal{S}$ is abelian, implying that%
\begin{equation}
S_{1}S_{2}S_{1}^{\dag}=S_{2}S_{1}S_{1}^{\dag}=S_{2}\ \ \ \ \forall S_{1}%
,S_{2}\in\mathcal{S}.
\end{equation}

In quantum error correction, we are concerned with correcting a fixed set of
errors $\mathcal{E}\subseteq\mathcal{G}_{n}$ such that each element of
$\mathcal{E}$ acts on the $n$ physical qubits. In doing so, we might not be
able to correct all of the errors in a set $\mathcal{E}$ if there exists a
pair $E_{1},E_{2}\in\mathcal{E}$ such that%
\begin{equation}
E_{1}^{\dag}E_{2}\in N(\mathcal{S}).
\end{equation}
Consider that for all $S\in\mathcal{S}$, we have%
\begin{equation}
E_{1}^{\dag}E_{2}S=\left(  -1\right)  ^{g(S,E_{1})+g(S,E_{2})}SE_{1}^{\dag
}E_{2},
\end{equation}
where we define $g(P,Q)$ by $PQ=\left(  -1\right)  ^{g(P,Q)}QP$ for all
$P,Q\in\mathcal{G}_{n}$. The above relation then implies the following one for
all $S\in\mathcal{S}$:%
\begin{equation}
E_{1}^{\dag}E_{2}S(E_{1}^{\dag}E_{2})^{\dag}=\left(  -1\right)  ^{g(S,E_{1}%
)+g(S,E_{2})}S.
\end{equation}
Since we assumed that $E_{1}^{\dag}E_{2}\in N(\mathcal{S})$, the only way that
the above relation can be true for all $S\in\mathcal{S}$ is if $g(S,E_{1}%
)=g(S,E_{2})$. Thus, during the error correction procedure, Bob will measure a
set $\left\{  S_{j}\right\}  $\ of generators, and since the outcome of a
measurement of $S_{j}$ on $E|\psi\rangle$ is $g(S,E)$, the errors $E_{1}$ and
$E_{2}$ will be assigned the same syndrome. Since they have the same syndrome,
the receiver will have to reverse these errors with the same recovery
operation, and this is only possible if $E_{1}|\psi\rangle=E_{2}|\psi\rangle$
for all states $|\psi\rangle$ in the codespace. This latter condition is only
true if $E_{1}^{\dag}E_{2}\in\mathcal{S}$, leading us to the error-correcting
conditions for quantum stabilizer codes:

\begin{theorem}
It is possible to correct a set of errors $\mathcal{E}$ with a quantum
stabilizer code if every pair $E_{1},E_{2}\in\mathcal{E}$ satisfies%
\begin{equation}
E_{1}^{\dag}E_{2}\notin N(\mathcal{S})\backslash\mathcal{S}.
\end{equation}

\end{theorem}

A simple way to satisfy the error-correcting conditions is just to demand that
every pair of errors in $\mathcal{E}$ be such that $E_{1}^{\dag}E_{2}\notin N(
\mathcal{S}) $. In such a case, each error is assigned a unique syndrome, and
codes along with an error set satisfying this property are known as
non-degenerate codes. Codes with a corresponding error set not satisfying this
are known as degenerate codes.

\subsection{The Hashing Bound}%

\index{hashing bound}%
We now provide a proof that the hashing bound for a Pauli channel (coherent
information when sending one share of a Bell state through a Pauli channel) is
an achievable rate for quantum communication. Our proof of the direct part of
the quantum capacity theorem already suffices as a proof of this statement,
but we think it is instructive to provide a proof of this statement using the
theory of stabilizer codes. The main idea of the proof is to choose a
stabilizer code randomly from the set of all stabilizer codes and show that
such a code can correct the typical errors issued by a tensor-product Pauli channel.

\begin{theorem}
[Hashing Bound]There exists a stabilizer quantum error-correcting code that
achieves the hashing bound $R=1-H( \mathbf{p}) $\ for a Pauli channel of the
following form:%
\begin{equation}
\rho\rightarrow p_{I}\rho+p_{X}X\rho X+p_{Y}Y\rho Y+p_{Z}Z\rho Z,
\end{equation}
where $\mathbf{p}=\left(  p_{I},p_{X},p_{Y},p_{Z}\right)  $ and $H(
\mathbf{p}) $ is the entropy of this probability vector.
\end{theorem}

\begin{proof}
We consider a decoder that corrects only the typical errors. That is, consider
defining the typical error set as follows:%
\begin{equation}
T_{\delta}^{\mathbf{p}^{n}}\equiv\left\{  a^{n}:\left\vert -\frac{1}{n}%
\log\left(  \Pr\left\{  E_{a^{n}}\right\}  \right)  -H(\mathbf{p})\right\vert
\leq\delta\right\}  ,
\end{equation}
where $a^{n}$ is some sequence consisting of letters corresponding to the
Pauli operators $\left\{  I,X,Y,Z\right\}  $ and $\Pr\left\{  E_{a^{n}%
}\right\}  $ is the probability that an i.i.d.~Pauli channel issues some
tensor-product error $E_{a^{n}}\equiv E_{a_{1}}\otimes\cdots\otimes E_{a_{n}}%
$. This typical set consists of the likely errors in the sense that%
\begin{equation}
\sum_{a^{n}\in T_{\delta}^{\mathbf{p}^{n}}}\Pr\left\{  E_{a^{n}}\right\}
\geq1-\varepsilon, \label{eq:typical-errors}%
\end{equation}
for all $\varepsilon\in(0,1)$ and sufficiently large $n$. The error-correcting
conditions for a stabilizer code in this case are that $\{E_{a^{n}}:a^{n}\in
T_{\delta}^{\mathbf{p}^{n}}\}$ is a correctable set of errors if%
\begin{equation}
E_{a^{n}}^{\dag}E_{b^{n}}\notin N(\mathcal{S})\backslash\mathcal{S},
\end{equation}
for all error pairs $E_{a^{n}}$ and $E_{b^{n}}$ such that $a^{n},b^{n}\in
T_{\delta}^{\mathbf{p}^{n}}$. Also, we consider the expectation of the error
probability under a random choice of a stabilizer code. We proceed as follows:%
\begin{align}
\mathbb{E}_{\mathcal{S}}\left\{  p_{e}\right\}   &  =\mathbb{E}_{\mathcal{S}%
}\left\{  \sum_{a^{n}}\Pr\left\{  E_{a^{n}}\right\}  I(E_{a^{n}}\text{ is
uncorrectable using }\mathcal{S})\right\} \\
&  \leq\mathbb{E}_{\mathcal{S}}\left\{  \sum_{a^{n}\in T_{\delta}%
^{\mathbf{p}^{n}}}\Pr\left\{  E_{a^{n}}\right\}  I(E_{a^{n}}\text{ is
uncorrectable using }\mathcal{S})\right\}  +\varepsilon\\
&  =\sum_{a^{n}\in T_{\delta}^{\mathbf{p}^{n}}}\Pr\left\{  E_{a^{n}}\right\}
\mathbb{E}_{\mathcal{S}}\left\{  I(E_{a^{n}}\text{ is uncorrectable using
}\mathcal{S})\right\}  +\varepsilon\\
&  =\sum_{a^{n}\in T_{\delta}^{\mathbf{p}^{n}}}\Pr\left\{  E_{a^{n}}\right\}
\Pr_{\mathcal{S}}\left\{  E_{a^{n}}\text{ is uncorrectable using }%
\mathcal{S}\right\}  +\varepsilon. \label{eq-qc:random-stab-proof-1}%
\end{align}
The first equality follows by definition---$I$ is an indicator function equal
to one if $E_{a^{n}}$ is uncorrectable using $\mathcal{S}$ and equal to zero
otherwise. The first inequality follows from \eqref{eq:typical-errors}---we
correct only the typical errors because the atypical error set has negligible
probability mass. The second equality follows by exchanging the expectation
and the sum. The third equality follows because the expectation of an
indicator function is the probability that the event it selects occurs.
Continuing, we now bound the probability $\Pr_{\mathcal{S}}\{E_{a^{n}}$ is
uncorrectable using $\mathcal{S}\}$ when $a^{n}\in T_{\delta}^{\mathbf{p}^{n}%
}$:%
\begin{align}
&  \Pr_{\mathcal{S}}\left\{  E_{a^{n}}\text{ is uncorrectable using
}\mathcal{S}\right\} \nonumber \\
&  =\Pr_{\mathcal{S}}\left\{  \exists E_{b^{n}}:b^{n}\in T_{\delta
}^{\mathbf{p}^{n}},\ b^{n}\neq a^{n},\ E_{a^{n}}^{\dag}E_{b^{n}}\in
N(\mathcal{S})\backslash\mathcal{S}\right\} \\
&  \leq\Pr_{\mathcal{S}}\left\{  \exists E_{b^{n}}:b^{n}\in T_{\delta
}^{\mathbf{p}^{n}},\ b^{n}\neq a^{n},\ E_{a^{n}}^{\dag}E_{b^{n}}\in
N(\mathcal{S})\right\} \\
&  =\Pr_{\mathcal{S}}\left\{  \bigcup\limits_{b^{n}\in T_{\delta}%
^{\mathbf{p}^{n}},\ b^{n}\neq a^{n}}E_{a^{n}}^{\dag}E_{b^{n}}\in
N(\mathcal{S})\right\} \\
&  \leq\sum_{b^{n}\in T_{\delta}^{\mathbf{p}^{n}},\ b^{n}\neq a^{n}}%
\Pr_{\mathcal{S}}\left\{  E_{a^{n}}^{\dag}E_{b^{n}}\in N(\mathcal{S})\right\}
\\
&  \leq\sum_{b^{n}\in T_{\delta}^{\mathbf{p}^{n}},\ b^{n}\neq a^{n}%
}2^{-\left(  n-k\right)  }\\
&  \leq2^{n\left[  H(\mathbf{p})+\delta\right]  }2^{-\left(  n-k\right)  }\\
&  =2^{-n\left[  1-H(\mathbf{p})-k/n-\delta\right]  }.
\end{align}
The first equality follows from the error-correcting conditions for a quantum
stabilizer code, where $N(\mathcal{S})$ is the normalizer of $\mathcal{S}$.
The first inequality follows by ignoring any potential degeneracy in the
code---we consider an error uncorrectable if it lies in the normalizer
$N(\mathcal{S})$ and the probability can only be larger because $N(\mathcal{S}%
)\backslash\mathcal{S}\subseteq N(\mathcal{S})$. The second equality follows
by realizing that the probabilities for the existence criterion and the union
of events are equal. The second inequality follows by applying the union
bound. The third inequality follows from the fact that the probability for a
fixed operator $E_{a^{n}}^{\dag}E_{b^{n}}$ not equal to the identity commuting
with the stabilizer operators of a random stabilizer can be upper bounded as
follows:%
\begin{equation}
\Pr_{\mathcal{S}}\left\{  E_{a^{n}}^{\dag}E_{b^{n}}\in N(\mathcal{S})\right\}
=\frac{2^{n+k}-1}{2^{2n}-1}\leq2^{-\left(  n-k\right)  }.
\end{equation}
The random choice of a stabilizer code is equivalent to fixing operators
$Z_{1}$, \ldots, $Z_{n-k}$ and performing a uniformly random Clifford unitary
$U$. The probability that a fixed operator commutes with $UZ_{1}U^{\dag}$,
\ldots, $UZ_{n-k}U^{\dag}$ is then just the number of non-identity operators
in the normalizer ($2^{n+k}-1$) divided by the total number of non-identity
operators ($2^{2n}-1$). After applying the above bound, we then exploit the
typicality bound $|T_{\delta}^{\mathbf{p}^{n}}|\leq2^{n\left[  H(\mathbf{p}%
)+\delta\right]  }$. Plugging back into \eqref{eq-qc:random-stab-proof-1}, we
find that%
\begin{equation}
\mathbb{E}_{\mathcal{S}}\left\{  p_{e}\right\}  \leq2^{-n\left[
1-H(\mathbf{p})-k/n-\delta\right]  }+\varepsilon.
\end{equation}
We conclude that as long as the rate $k/n=1-H(\mathbf{p})-2\delta$, the
expectation of the error probability becomes arbitrarily small, so that there
exists at least one choice of a stabilizer code with the same bound on the
error probability.
\end{proof}

\section{Example Channels}

\label{sec-q-cap:examples}We now show how to calculate the quantum capacity
for two exemplary channels: the quantum erasure channel and the amplitude
damping channel. Both of these channels are
\index{degradable channel}%
degradable for particular channel parameters, simplifying the calculation of
their quantum capacities.

\subsection{The Quantum Erasure Channel}

Recall that the
\index{erasure channel}%
quantum erasure channel acts as follows on an input density operator
$\rho_{A^{\prime}}$:%
\begin{equation}
\rho_{A^{\prime}}\rightarrow\left(  1-\varepsilon\right)  \rho_{B}%
+\varepsilon|e\rangle\langle e|_{B},
\end{equation}
where $\varepsilon\in\left[  0,1\right]  $ is the erasure probability and
$|e\rangle_{B}$ is an erasure state that is orthogonal to the support of any
input state $\rho$.

\begin{proposition}
\label{prop-qc:erasure-q-cap}Let $d_{A}$ be the dimension of the input system
for the quantum erasure channel. The quantum capacity of a quantum erasure
channel with erasure probability $\varepsilon$\ is equal to $\left(
1-2\varepsilon\right)  \log d_{A}$ when $\varepsilon\in\lbrack0,1/2]$ and it
is equal to zero otherwise.
\end{proposition}

\begin{proof}
The quantum erasure channel is antidegradable%
\index{antidegradable channel}
for $\varepsilon\in\left[  1/2,1\right]  $. This follows from
Exercise~\ref{ex:iso-extension-erasure} and the fact that the erasure channel
is
\index{degradable channel}%
degradable for $\varepsilon\in\left[  0,1/2\right]  $ (see
Exercise~\ref{ex-add:erasure-degradable}). From
Exercise~\ref{ex-add:antidegrad-zero-coh-info} and the fact that
$\mathcal{N}^{\otimes n}$ is antidegradable if $\mathcal{N}$ is, we can then
conclude that the quantum capacity is equal to zero for $\varepsilon\in\left[
1/2,1\right]  $.

To determine the quantum capacity for $\varepsilon\in\left[  0,1/2\right]  $,
we know that it is degradable for this range (see
Exercise~\ref{ex-add:erasure-degradable}), so it suffices to compute its
coherent information. We can do so in a similar way as we did in
Proposition~\ref{prop-eac:eac-erasure}. Consider that sending one share of a
pure, bipartite state $\phi_{AA^{\prime}}$ through the channel produces the
output%
\begin{equation}
\sigma_{AB}\equiv\left(  1-\varepsilon\right)  \phi_{AB}+\varepsilon\phi
_{A}\otimes|e\rangle\langle e|_{B}.
\end{equation}
Recall that Bob can apply the following isometry $U_{B\rightarrow BX}$ to his
state:%
\begin{equation}
U_{B\rightarrow BX}\equiv\Pi_{B}\otimes|0\rangle_{X}+|e\rangle\langle
e|_{B}\otimes|1\rangle_{X}, \label{eq-q-cap:erasure-isometry-bob}%
\end{equation}
where $\Pi_{B}$ is a projector onto the support of the input state (for
qubits, it would be just $|0\rangle\langle0|+|1\rangle\langle1|$). Applying
this isometry leads to a state $\sigma_{ABX}$ where%
\begin{align}
\sigma_{ABX}  &  \equiv U_{B\rightarrow BX}\sigma_{AB}U_{B\rightarrow
BX}^{\dag}\\
&  =\left(  1-\varepsilon\right)  \phi_{AB}\otimes|0\rangle\langle
0|_{X}+\varepsilon\phi_{A}\otimes|e\rangle\langle e|_{B}\otimes|1\rangle
\langle1|_{X}.
\end{align}
The coherent information $I(A\rangle BX)_{\sigma}$ is equal to $I(A\rangle
B)_{\sigma}$ because entropies do not change under the isometry
$U_{B\rightarrow BX}$. We now calculate $I(A\rangle BX)_{\sigma}$:%
\begin{align}
I(A\rangle BX)_{\sigma}  &  =H(BX)_{\sigma}-H(ABX)_{\sigma}\\
&  =H(B|X)_{\sigma}-H(AB|X)_{\sigma}\\
&  =\left(  1-\varepsilon\right)  \left[  H(B)_{\phi}-H(AB)_{\phi}\right]
\nonumber\\
&  \ \ \ \ \ \ \ +\varepsilon\left[  H(B)_{|e\rangle\langle e|}-H(AB)_{\phi_{A}%
\otimes|e\rangle\langle e|}\right] \\
&  =\left(  1-\varepsilon\right)  H(B)_{\phi}-\varepsilon\left[  H(A)_{\phi
}+H(B)_{|e\rangle\langle e|}\right] \\
&  =\left(  1-2\varepsilon\right)  H(A)_{\phi}\\
&  \leq\left(  1-2\varepsilon\right)  \log d_{A}.
\end{align}
The first equality follows from the definition of coherent information. The
second equality follows from $\phi_{A}=\operatorname{Tr}_{BX}\left\{
\sigma_{ABX}\right\}  $, from the chain rule of entropy, and by canceling
$H(X)$ on both sides. The third equality follows because the $X$ register is a
classical register, indicating whether the erasure occurs. The fourth equality
follows because $H(AB)_{\phi}=0$, $H(B)_{|e\rangle\langle e|}=0$, and $H(AB)_{\phi
_{A}\otimes|e\rangle\langle e|}=H(A)_{\phi}+H(B)_{|e\rangle\langle e|}$. The fifth
equality follows again because $H(B)_{|e\rangle\langle e|}=0$, by collecting terms, and
because $H(A)_{\phi}=H(B)_{\phi}$ ($\phi_{AB}$ is a pure bipartite state). The
final inequality follows because the entropy of a state on system $A$ is never
greater than the logarithm of the dimension of $A$. We can conclude that the
maximally entangled state $\Phi_{AA^{\prime}}$ achieves the quantum capacity
of the quantum erasure channel for $\varepsilon\in\left[  0,1/2\right]  $
because $H(A)_{\Phi}=\log d_{A}$.
\end{proof}

\subsection{The Amplitude Damping Channel}

We now compute the
\index{amplitude damping channel}
quantum capacity of the amplitude damping channel $\mathcal{N}%
_{\operatorname{AD}}$. Recall that this channel acts as follows on an input
qubit in state $\rho$:%
\begin{equation}
\mathcal{N}_{\operatorname{AD}}(\rho)=A_{0}\rho A_{0}^{\dag}+A_{1}\rho
A_{1}^{\dag},
\end{equation}
where, for $\gamma\in\left[  0,1\right]  $,%
\begin{equation}
A_{0}\equiv|0\rangle\langle0|+\sqrt{1-\gamma}|1\rangle\langle
1|,\ \ \ \ \ \ A_{1}\equiv\sqrt{\gamma}|0\rangle\langle1|.
\end{equation}
The development here is similar to the development in the proof of
Proposition~\ref{thm-eac:eac-amp-damp}.

\begin{proposition}
The quantum capacity of an amplitude damping channel with damping parameter
$\gamma\in\left[  0,1\right]  $ is equal to the following%
\begin{equation}
\max_{p\in\left[  0,1\right]  }h_{2}((1-\gamma)p)-h_{2}(\gamma p),
\end{equation}
whenever $\gamma\in\lbrack0,1/2]$ (recall that $h_{2}(x)$ is the binary
entropy function). For $\gamma\in\left[  1/2,1\right]  $, the quantum capacity
is equal to zero.
\end{proposition}

\begin{proof}
Suppose that a matrix representation of the input qubit density operator
$\rho$ in the computational basis is%
\begin{equation}
\rho=%
\begin{bmatrix}
1-p & \eta^{\ast}\\
\eta & p
\end{bmatrix}
. \label{eq-q-cap:input-dens-amp}%
\end{equation}
One can readily verify that the density operator for Bob has the following
matrix representation:%
\begin{equation}
\mathcal{N}_{\operatorname{AD}}(\rho)=%
\begin{bmatrix}
1-\left(  1-\gamma\right)  p & \sqrt{1-\gamma}\eta^{\ast}\\
\sqrt{1-\gamma}\eta & \left(  1-\gamma\right)  p
\end{bmatrix}
, \label{eq-q-cap:amp-to-bob}%
\end{equation}
and by calculating the elements $\operatorname{Tr}\{A_{i}\rho A_{j}^{\dag
}\}|i\rangle\langle j|$, we can obtain a matrix representation for Eve's
density operator:%
\begin{equation}
\mathcal{N}_{\operatorname{AD}}^{c}(\rho)=%
\begin{bmatrix}
1-\gamma p & \sqrt{\gamma}\eta^{\ast}\\
\sqrt{\gamma}\eta & \gamma p
\end{bmatrix}
, \label{eq-q-cap:amp-to-eve}%
\end{equation}
where $\mathcal{N}_{\operatorname{AD}}^{c}$ is the complementary channel to
Eve. By comparing \eqref{eq-q-cap:amp-to-bob} and \eqref{eq-q-cap:amp-to-eve},
we can see that the channel to Eve is an amplitude damping channel with
damping parameter $1-\gamma$. One can verify that the channel is
antidegradable%
\index{antidegradable channel}
for $\gamma\in\left[  1/2,1\right]  $ and degradable for $\gamma\in\left[
0,1/2\right]  $ (\textit{Exercise}: find the degrading channel). By the same
reasoning in the previous proposition, the quantum capacity is equal to zero
for $\gamma\in\left[  1/2,1\right]  $ and it is equal to the optimized
coherent information for $\gamma\in\lbrack0,1/2]$. So we now focus on this
latter case. The quantum capacity of $\mathcal{N}_{\operatorname{AD}}$ is
equal to its coherent information:%
\begin{equation}
Q(\mathcal{N}_{\operatorname{AD}})=\max_{\phi_{AA^{\prime}}}I(A\rangle
B)_{\sigma}, \label{eq-q-cap:q-cap-general-formula-amp-channel}%
\end{equation}
where $\phi_{AA^{\prime}}$ is some pure bipartite input state and $\sigma
_{AB}=\mathcal{N}_{\operatorname{AD}}(\phi_{AA^{\prime}})$. We need to
determine the input density operator that maximizes the above formula as a
function of $\gamma$. So far, the optimization depends on three
parameters:\ $p$, $\operatorname{Re}\left\{  \eta\right\}  $, and
$\operatorname{Im}\left\{  \eta\right\}  $. We can show that it is sufficient
to consider an optimization over only $p$ with $\eta=0$. The formula in
\eqref{eq-q-cap:q-cap-general-formula-amp-channel} also has the following
form:%
\begin{equation}
Q(\mathcal{N}_{\operatorname{AD}})=\max_{\rho}\left[  H(\mathcal{N}%
_{\operatorname{AD}}(\rho))-H(\mathcal{N}_{\operatorname{AD}}^{c}%
(\rho))\right]  , \label{eq-q-cap:q-formula-amp}%
\end{equation}
because%
\begin{align}
I(A\rangle B)_{\sigma}  &  =H(B)_{\sigma}-H(AB)_{\sigma}\\
&  =H(\mathcal{N}_{\operatorname{AD}}(\rho))-H(E)_{\sigma}\\
&  =H(\mathcal{N}_{\operatorname{AD}}(\rho))-H(\mathcal{N}_{\operatorname{AD}%
}^{c}(\rho))\\
&  \equiv I_{\operatorname{coh}}(\rho,\mathcal{N}_{\operatorname{AD}}).
\end{align}
The two entropies in\ \eqref{eq-q-cap:q-formula-amp} depend only on the
eigenvalues of the two density operators in
\eqref{eq-q-cap:amp-to-bob}--\eqref{eq-q-cap:amp-to-eve}, respectively, which
are as follows:%
\begin{align}
&  \frac{1}{2}\left(  1\pm\sqrt{\left(  1-2\left(  1-\gamma\right)  p\right)
^{2}+4\left\vert \eta\right\vert ^{2}\left(  1-\gamma\right)  }\right)
,\label{eq-q-cap:eig-bob-amp}\\
&  \frac{1}{2}\left(  1\pm\sqrt{\left(  1-2\gamma p\right)  ^{2}+4\left\vert
\eta\right\vert ^{2}\gamma}\right)  . \label{eq-q-cap:eig-eve-amp}%
\end{align}
The above eigenvalues are in the order of Bob and Eve. All of the above
eigenvalues have a similar form, and their dependence on $\eta$ is only
through its magnitude. Thus, it suffices to consider $\eta\in\mathbb{R}$ (this
eliminates one parameter). Next, the eigenvalues do not change if we flip the
sign of $\eta$ (this is equivalent to rotating the original state $\rho$ by
$Z$, to $Z\rho Z$), and thus, the coherent information does not change as
well:%
\begin{equation}
I_{\operatorname{coh}}(\rho,\mathcal{N}_{\operatorname{AD}}%
)=I_{\operatorname{coh}}(Z\rho Z,\mathcal{N}_{\operatorname{AD}}).
\end{equation}
By the above relation and concavity of coherent information in the input
density operator for
\index{degradable channel}%
degradable channels (Theorem$~$\ref{thm-add:coh-deg-concave-input}), the
following inequality holds:%
\begin{align}
I_{\operatorname{coh}}(\rho,\mathcal{N}_{\operatorname{AD}})  &  =\frac{1}%
{2}\left[  I_{\operatorname{coh}}(\rho,\mathcal{N}_{\operatorname{AD}%
})+I_{\operatorname{coh}}(Z\rho Z,\mathcal{N}_{\operatorname{AD}})\right] \\
&  \leq I_{\operatorname{coh}}(\left(  \rho+Z\rho Z\right)  /2,\mathcal{N}%
_{\operatorname{AD}})\\
&  =I_{\operatorname{coh}}(\overline{\Delta}(\rho),\mathcal{N}%
_{\operatorname{AD}}),
\end{align}
where $\overline{\Delta}$ is a completely dephasing channel in the
computational basis. This demonstrates that it is sufficient to consider
diagonal density operators $\rho$ when optimizing the coherent information.
Thus, the eigenvalues in
\eqref{eq-q-cap:eig-bob-amp}--\eqref{eq-q-cap:eig-eve-amp} respectively become%
\begin{align}
&  \left\{  \left(  1-\gamma\right)  p,1-\left(  1-\gamma\right)  p\right\}
,\\
&  \left\{  \gamma p,1-\gamma p\right\}  ,
\end{align}
giving our final expression in the statement of the proposition.
\end{proof}

\begin{exercise}
Consider the dephasing channel:\ $\rho\rightarrow\left(  1-p/2\right)
\rho+(p/2)Z\rho Z$. Prove that its quantum capacity is equal to $1-h_{2}%
(p/2)$, where $p$ is the dephasing parameter.
\end{exercise}

\section{Discussion of Quantum Capacity}

\label{sec-q-cap:strangeness}The quantum capacity is particularly well-behaved
and understood for the class of
\index{degradable channel}%
degradable channels. Thus, we should not expect any surprises for this class
of channels. If a channel is not degradable, we currently cannot say much
about the exact value of its quantum capacity, but the study of non-degradable
channels has led to many surprises in quantum Shannon theory and this section
discusses two of these surprises. The first is the superadditivity of coherent
information for the depolarizing channel, and the second is a striking
phenomenon known as \textit{superactivation} of quantum capacity, where two
channels that individually have zero quantum capacity can combine to make a
channel with non-zero quantum capacity.

\subsection{Superadditivity of Coherent Information}

\begin{figure}[ptb]
\begin{center}
\includegraphics[
width=3.6876in
]{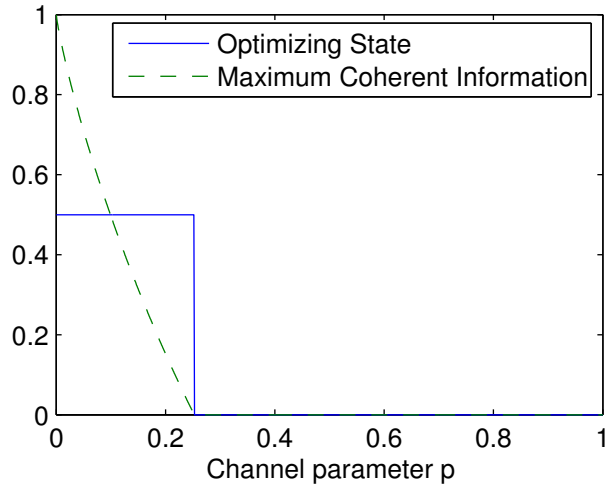}
\end{center}
\caption{This figure plots the maximum coherent information of the
depolarizing channel (dashed line) versus the depolarizing parameter
$p\in\left[  0,1\right]  $. It also plots the value $\mu\in\left[  0,1\right]
$ for the state which optimizes the value of the coherent information (solid
line) versus the depolarizing parameter $p$, considering states of the form
$\sqrt{\mu}|00\rangle_{AA^{\prime}}+\sqrt{1-\mu}|11\rangle_{AA^{\prime}}$. It
demonstrates that the maximum value is achieved for $\mu=1/2$ (the maximally
entangled state) for all values for which the coherent information is larger
than zero and for $\mu=0$ otherwise. }%
\label{fig-qc:coh-info-dep}%
\end{figure}

Recall that the
\index{depolarizing channel}
depolarizing
\index{coherent information!of a channel!superadditivity}
channel $\mathcal{N}^{\text{D}}$\ transmits its input with probability $1-p$
and replaces it with the maximally mixed state~$\pi$ with probability
$p\in\left[  0,1\right]  $:%
\begin{equation}
\mathcal{N}^{\text{D}}(\rho)=(1-p)\rho+p\pi.
\end{equation}
We focus on the case in which the input and output of this channel is a qubit.
The depolarizing channel is an example of a quantum channel that is not
degradable.\footnote{\cite{SS07}\ have given an explicit condition for
whether a channel is degradable.} As such, we might expect it to exhibit some
strange behavior with respect to its quantum capacity. Indeed, it is known
that its coherent information is strictly superadditive when the channel
becomes very noisy:%
\begin{equation}
5Q(\mathcal{N}^{\text{D}})<Q((\mathcal{N}^{\text{D}})^{\otimes5}).
\end{equation}
How can we show that this result is true? First, we can calculate the coherent
information of this channel with respect to one channel use. It is possible to
show that the maximally entangled state $\Phi_{AA^{\prime}}$ maximizes the
channel coherent information~$Q(\mathcal{N}^{\text{D}})$ for all values of the
coherent information for which it is non-negative. To see this, consider that
the depolarizing channel is unitarily covariant, so that $\mathcal{N}%
^{\text{D}}(U\rho U^{\dag})=U\mathcal{N}^{\text{D}}(\rho)U^{\dag}$ for any
unitary $U$ and any qubit input density operator $\rho$. Thus, for optimizing
the coherent information of $\mathcal{N}^{\text{D}}$, it suffices to consider
states of the form $\sqrt{\mu}|00\rangle_{AA^{\prime}}+\sqrt{1-\mu}%
|11\rangle_{AA^{\prime}}$ where $\mu\in\left[  0,1\right]  $. A numerical
optimization over all such states gives the plot in
Figure~\ref{fig-qc:coh-info-dep}, which demonstrates that the maximally
entangled state ($\mu=1/2$) is optimal for all values of the depolarizing
parameter $p$ for which the coherent information is non-negative, and for all
other values, a product state with $\mu=0$ is optimal.
Thus, we can calculate the coherent information as follows:%
\begin{equation}
Q(\mathcal{N}^{\text{D}})=H(B)_{\Phi}-H(AB)_{\mathcal{N}^{\text{D}}(\Phi
)}=1-H(AB)_{\mathcal{N}^{\text{D}}(\Phi)},
\end{equation}
where $H(B)_{\Phi}=1$ follows because the output state on Bob's system is the
maximally mixed state whenever the input to the channel is one share of a
maximally entangled state. In order to calculate $H(AB)_{\mathcal{N}%
^{\text{D}}(\Phi)}$, observe that the state on $AB$ is%
\begin{align}
(1-p)\Phi_{AB}+p\pi_{A}\otimes\pi_{B}  &  =(1-p)\Phi_{AB}+\frac{p}{4}I_{AB}\\
&  =(1-p)\Phi_{AB}+\frac{p}{4}\left(  \left[  I_{AB}-\Phi_{AB}\right]
+\Phi_{AB}\right) \\
&  =\left(  1-\frac{3p}{4}\right)  \Phi_{AB}+\frac{p}{4}\left(  I_{AB}%
-\Phi_{AB}\right)  .
\end{align}
Since $\Phi_{AB}$ and $I_{AB}-\Phi_{AB}$ are orthogonal, the eigenvalues of
this state are $1-3p/4$ with multiplicity one and $p/4$ with multiplicity
three. Thus, the entropy $H(AB)_{\mathcal{N}^{\text{D}}(\Phi)}$ is%
\begin{equation}
H(AB)_{\mathcal{N}^{\text{D}}(\phi)}=-\left(  1-\frac{3p}{4}\right)
\log\left(  1-\frac{3p}{4}\right)  -\frac{3p}{4}\log\left(  \frac{p}%
{4}\right)  ,
\end{equation}
and our final expression for the single-copy coherent information is%
\begin{equation}
Q(\mathcal{N}^{\text{D}})=1+\left(  1-\frac{3p}{4}\right)  \log\left(
1-\frac{3p}{4}\right)  +\frac{3p}{4}\log\left(  \frac{p}{4}\right)  .
\label{eq-q-cap:one-shot-dep-rate}%
\end{equation}

\begin{figure}
[ptb]
\begin{center}
\includegraphics[
width=4.8456in
]%
{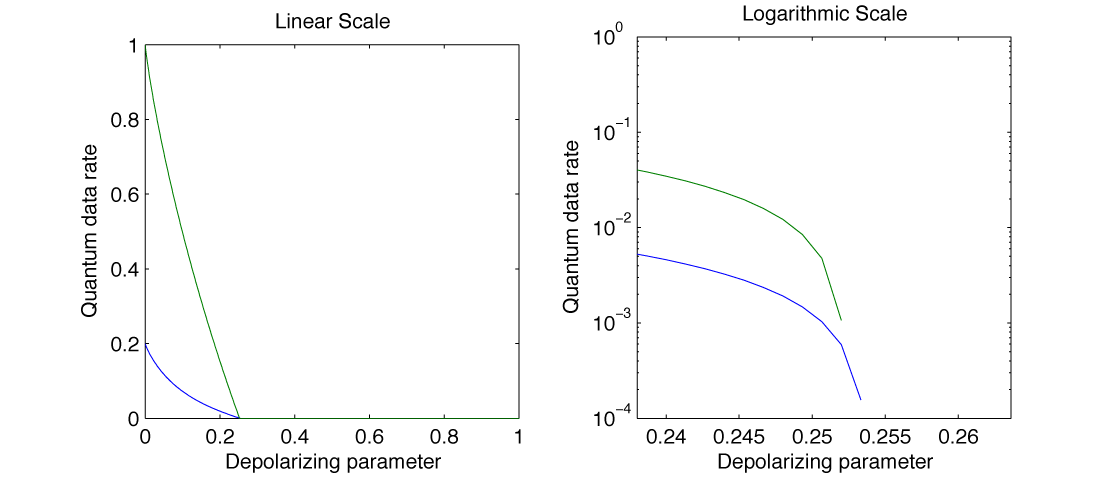}%
\caption{The figures plot the coherent information in
\eqref{eq-q-cap:one-shot-dep-rate} (bottom curves in both figures) and that in
\eqref{eq-q-cap:rep-strat} (top curves in both figures) versus the
depolarizing noise parameter~$p$. The figure on the left is on a linear scale,
and the one on the right is on a logarithmic scale. The notable features of
the figure on the left are that the quantum data rate of the top curve is
equal to one and the quantum data rate of the bottom curve is 1/5 when the
channel is noiseless (the latter rate is to be expected for a five-qubit
repetition code). Both data rates become small when $p$ is near 0.25, but the
figure on the right reveals that the repetition code concatenation strategy
still gets positive coherent information even when the rate of the random
coding strategy vanishes. This is an example of a channel for which the
coherent information can be superadditive.}%
\label{fig-q-cap:coh-superadd}%
\end{center}
\end{figure}

Another strategy for transmitting quantum data is to encode one share of the
maximally entangled state with a five-qubit repetition code:%
\begin{multline}
\frac{1}{\sqrt{2}}\left(  |00\rangle_{AA_{1}}+|11\rangle_{AA_{1}}\right) \\
\rightarrow\frac{1}{\sqrt{2}}\left(  |000000\rangle_{AA_{1}A_{2}A_{3}%
A_{4}A_{5}}+|111111\rangle_{AA_{1}A_{2}A_{3}A_{4}A_{5}}\right)  ,
\end{multline}
and calculate the following coherent information with respect to the state
resulting from sending the systems~$A_{1}\cdots A_{5}$ through the channel:%
\begin{equation}
\frac{1}{5}I(A\rangle B_{1}B_{2}B_{3}B_{4}B_{5}). \label{eq-q-cap:rep-strat}%
\end{equation}
(We normalize the above coherent information by five in order to make a fair
comparison between a code achieving this rate and one achieving the rate in
\eqref{eq-q-cap:one-shot-dep-rate}.) We know that the rate in
\eqref{eq-q-cap:rep-strat} is achievable by applying the direct part of the
quantum capacity theorem to the channel $(\mathcal{N}^{\text{D}})^{\otimes5}$,
and operationally, this strategy amounts to concatenating a random quantum
code with a five-qubit repetition code. The remarkable result is that this
concatenation strategy can beat the single-copy coherent information when the
channel becomes very noisy. Figure~\ref{fig-q-cap:coh-superadd}\ demonstrates
that the concatenation strategy has positive coherent information even when
the single-copy coherent information in \eqref{eq-q-cap:one-shot-dep-rate}
vanishes. This demonstrates superadditivity of coherent information.%

Why does this phenomenon occur?\ The simplest (though perhaps not completely
satisfying) explanation is that it results from a phenomenon known as
\textit{degeneracy}. Consider a qubit $\alpha|0\rangle+\beta|1\rangle$ encoded
in a repetition code:%
\begin{equation}
\alpha|00000\rangle+\beta|11111\rangle.
\end{equation}
If the \textquotedblleft error\textquotedblright\ $Z_{1}\otimes Z_{2}$ occurs,
then it actually has no effect on this state. The same holds for other
two-qubit combinations of $Z$ errors. When the channel noise is low,
degeneracy of the code with respect to these errors does not help very much
because these two-qubit error combinations are less likely to occur. However,
when the channel becomes really noisy, these errors are more likely to occur,
and the help from degeneracy of the repetition code offsets the loss in rate.

It is perhaps strange that the coherent information of a depolarizing channel
behaves in this way. The channel seems simple enough, and we could say that
the strategies for achieving the unassisted and entanglement-assisted
classical capacity of this channel are very \textquotedblleft
classical\textquotedblright\ strategies. Recall that the best strategy for
achieving the unassisted classical capacity is to generate random codes by
picking states uniformly at random from some orthonormal basis, and the
receiver measures each channel output in this same orthonormal basis. For
achieving the entanglement-assisted classical capacity, we choose a random
code by picking Bell states uniformly at random and the receiver measures each
channel output and his share of each entangled state in the Bell basis. Both
of these results follow from the additivity of the respective capacities. In
spite of these other results, the best strategy for achieving the quantum
capacity of the depolarizing channel remains very poorly understood.

\subsection{Superactivation of Quantum Capacity}

\label{sec-q-cap:superactivation}Perhaps the%
\index{superactivation}
most startling result in quantum communication is a phenomenon known as
\textit{superactivation}. Suppose that Alice is connected to Bob by a quantum
channel~$\mathcal{N}_{1}$ with zero capacity for transmitting quantum data.
Also, suppose that there is some other zero quantum capacity
channel~$\mathcal{N}_{2}$ connecting them. Intuitively, we would expect that
Alice should not be able to transmit quantum data reliably over the
tensor-product channel $\mathcal{N}_{1}\otimes\mathcal{N}_{2}$. That is, using
these channels in parallel seems like it should not give any advantage over
using the individual channels alone if they are both individually useless for
quantum data transmission (this is the intuition that we have whenever a
capacity formula is additive). But two examples of zero-capacity channels are
known that can \textit{superactivate} each other, such that the joint channel
has a non-zero quantum capacity. How is this possible?

First, consider a 50\% quantum erasure channel$~\mathcal{N}_{1}$ that
transmits its input state with probability 1/2 and replaces it with an
orthogonal erasure state with probability 1/2. As we have argued before with
the no-cloning theorem, such a channel has zero capacity for sending quantum
data reliably (see also Proposition~\ref{prop-qc:erasure-q-cap}). Now consider
some other channel $\mathcal{N}_{2}$. The following theorem states that the
coherent information of the joint channel $\mathcal{N}_{1}\otimes
\mathcal{N}_{2}$ is at least equal to half the private information of
$\mathcal{N}_{2}$ alone.

\begin{theorem}
Let $\{p_{X}(x),\rho_{A_{2}}^{x}\}$ be an ensemble of inputs for the
channel~$\mathcal{N}_{2}$, and let $\mathcal{N}_{1}$ be a 50\% erasure
channel. Then there exists a pure state $\varphi_{RA_{1}A_{2}}$ such that the
coherent information $H(B_{1}B_{2})-H(E_{1}E_{2})$ of the joint channel
$\mathcal{N}_{1}\otimes\mathcal{N}_{2}$\ is equal to half the private
information $I(X;B_{2})-I(X;E_{2})$ of the second channel:%
\begin{equation}
H(B_{1}B_{2})_{\omega}-H(E_{1}E_{2})_{\omega}=\frac{1}{2}\left[
I(X;B_{2})_{\rho}-I(X;E_{2})_{\rho}\right]  ,
\end{equation}
where%
\begin{align}
\omega_{RB_{1}B_{2}E_{1}E_{2}}  &  \equiv(\mathcal{U}_{A_{1}\rightarrow
B_{1}E_{1}}^{\mathcal{N}_{1}}\otimes\mathcal{U}_{A_{2}\rightarrow B_{2}E_{2}%
}^{\mathcal{N}_{2}})\left(  \varphi_{RA_{1}A_{2}}\right)  ,\\
\rho_{XB_{2}E_{2}}  &  \equiv\sum_{x}p_{X}(x)|x\rangle\langle x|_{X}%
\otimes\mathcal{U}_{A_{2}\rightarrow B_{2}E_{2}}^{\mathcal{N}_{2}}(\rho
_{A_{2}}^{x}),
\end{align}
and $U_{A_{1}\rightarrow B_{1}E_{1}}^{\mathcal{N}_{1}}$ and $U_{A_{2}%
\rightarrow B_{2}E_{2}}^{\mathcal{N}_{2}}$ are respective isometric extensions
of $\mathcal{N}_{1}$ and $\mathcal{N}_{2}$. This implies that%
\begin{equation}
Q(\mathcal{N}_{1}\otimes\mathcal{N}_{2})\geq P(\mathcal{N}_{2})/2.
\end{equation}

\end{theorem}

\begin{proof}
Consider the following classical--quantum state corresponding to the ensemble
$\{p_{X}(x),\rho_{A_{2}}^{x}\}$:%
\begin{equation}
\rho_{XA_{2}}\equiv\sum_{x}p_{X}(x)|x\rangle\langle x|_{X}\otimes\rho_{A_{2}%
}^{x}.
\end{equation}
Let $\rho_{XB_{2}E_{2}}\equiv\mathcal{U}_{A_{2}\rightarrow B_{2}E_{2}%
}^{\mathcal{N}_{2}}(\rho_{XA_{2}})$. A purification of this state is%
\begin{equation}
|\varphi\rangle_{XA_{1}A_{2}}\equiv\sum_{x}\sqrt{p_{X}(x)}|x\rangle
_{X}|x\rangle_{A_{1}^{\prime}}|\phi_{x}\rangle_{A_{1}^{\prime\prime}A_{2}},
\end{equation}
where we identify $A_{1}\equiv A_{1}^{\prime}A_{1}^{\prime\prime}$ and each
$|\phi_{x}\rangle_{A_{1}^{\prime\prime}A_{2}}$ is a purification of
$\rho_{A_{2}}^{x}$, so that the state $|\varphi\rangle_{XA_{1}A_{2}}$ is a
purification of $\rho_{XA_{2}}$. Let $|\varphi\rangle_{XB_{1}E_{1}B_{2}E_{2}}$
be the state resulting from sending $A_{1}$ and $A_{2}$ through the
tensor-product channel $\mathcal{U}_{A_{1}\rightarrow B_{1}E_{1}}%
^{\mathcal{N}_{1}}\otimes\mathcal{U}_{A_{2}\rightarrow B_{2}E_{2}%
}^{\mathcal{N}_{2}}$. Identifying the system $B_{1}\equiv B_{1}^{\prime}%
B_{1}^{\prime\prime}$ and $E_{1}\equiv E_{1}^{\prime}E_{1}^{\prime\prime}$, we
can write this state as follows by recalling the isometric extension of the
erasure channel in \eqref{eq-q-cap:iso-extend-erasure}:%
\begin{multline}
|\varphi\rangle_{XB_{1}E_{1}B_{2}E_{2}}\equiv\frac{1}{\sqrt{2}}\sum_{x}%
\sqrt{p_{X}(x)}|x\rangle_{X}|x\rangle_{B_{1}^{\prime}}|\phi_{x}\rangle
_{B_{1}^{\prime\prime}B_{2}E_{2}}|e\rangle_{E_{1}}\\
+\frac{1}{\sqrt{2}}\sum_{x}\sqrt{p_{X}(x)}|x\rangle_{X}|x\rangle
_{E_{1}^{\prime}}|\phi_{x}\rangle_{E_{1}^{\prime\prime}B_{2}E_{2}}%
|e\rangle_{B_{1}}.
\end{multline}
Recall that Bob can perform an isometry on $B_{1}$ of the form in
\eqref{eq-q-cap:erasure-isometry-bob} that identifies whether he receives the
state or the erasure symbol, and let $Z_{B}$ be a flag register indicating the
outcome. Eve can do the same, and let $Z_{E}$ indicate her flag register. The
resulting state is as follows:%
\begin{multline}
|\psi\rangle_{XB_{1}E_{1}B_{2}E_{2}Z_{B}Z_{E}}\equiv\frac{1}{\sqrt{2}}%
|\psi^{0}\rangle_{XB_{1}E_{1}B_{2}E_{2}}|0\rangle_{Z_{B}}|1\rangle_{Z_{E}}\\
+\frac{1}{\sqrt{2}}|\psi^{1}\rangle_{XB_{1}E_{1}B_{2}E_{2}}|1\rangle_{Z_{B}%
}|0\rangle_{Z_{E}},
\end{multline}
where%
\begin{align}
|\psi^{0}\rangle_{XB_{1}E_{1}B_{2}E_{2}}  &  \equiv\sum_{x}\sqrt{p_{X}%
(x)}|x\rangle_{X}|x\rangle_{B_{1}^{\prime}}|\phi_{x}\rangle_{B_{1}%
^{\prime\prime}B_{2}E_{2}}|e\rangle_{E_{1}},\\
|\psi^{1}\rangle_{XB_{1}E_{1}B_{2}E_{2}}  &  \equiv\sum_{x}\sqrt{p_{X}%
(x)}|x\rangle_{X}|x\rangle_{E_{1}^{\prime}}|\phi_{x}\rangle_{E_{1}%
^{\prime\prime}B_{2}E_{2}}|e\rangle_{B_{1}}.
\end{align}
Then we can evaluate the coherent information of the state resulting from
sending system $A_{1}$ through the erasure channel and $A_{2}$ through the
other channel~$\mathcal{N}_{2}$:%
\begin{align}
&  H(B_{1}B_{2})_{\varphi}-H(E_{1}E_{2})_{\varphi}\nonumber\\
&  =H(B_{1}Z_{B}B_{2})_{\psi}-H(E_{1}Z_{E}E_{2})_{\psi}\\
&  =H(B_{1}B_{2}|Z_{B})_{\psi}+H(Z_{B})_{\psi}-H(E_{1}E_{2}|Z_{E})_{\psi
}-H(Z_{E})_{\psi}\\
&  =H(B_{1}B_{2}|Z_{B})_{\psi}-H(E_{1}E_{2}|Z_{E})_{\psi}\\
&  =\frac{1}{2}\left[  H(B_{1}B_{2})_{\psi^{0}}+H(B_{2})_{\psi^{1}}\right]
-\frac{1}{2}\left[  H(E_{2})_{\psi^{0}}+H(E_{1}E_{2})_{\psi^{1}}\right] \\
&  =\frac{1}{2}\left[  H(XE_{2})_{\psi^{0}}+H(B_{2})_{\psi^{1}}\right]
-\frac{1}{2}\left[  H(E_{2})_{\psi^{0}}+H(XB_{2})_{\psi^{1}}\right] \\
&  =\frac{1}{2}\left[  H(XE_{2})_{\rho}+H(B_{2})_{\rho}\right]  -\frac{1}%
{2}\left[  H(E_{2})_{\rho}+H(XB_{2})_{\rho}\right] \\
&  =\frac{1}{2}\left[  I(X;B_{2})_{\rho}-I(X;E_{2})_{\rho}\right]  .
\end{align}
The first equality follows because Bob and Eve can perform the isometries that
identify whether they receive the state or the erasure flag. The second
equality follows from the chaining rule for entropy, and the third follows
because the entropies of the flag registers $Z_{B}$ and $Z_{E}$ are equal for
a 50\% erasure channel. The fourth equality follows because the registers
$Z_{B}$ and $Z_{E}$ are classical when tracing over the other $Z$ register, so
that we can evaluate the conditional entropies as a uniform convex sum of
different possibilities:\ Bob obtaining the state transmitted or not, and Eve
obtaining the state transmitted or not. The fifth equality follows because the
states $\psi^{0}$ and $\psi^{1}$ are pure. The sixth equality follows because
$\psi_{XE_{2}}^{0}=\rho_{XE_{2}}$, $\psi_{B_{2}}^{1}=\rho_{B_{2}}$,
$\psi_{E_{2}}^{0}=\rho_{E_{2}}$, and $\psi_{XB_{2}}^{1}=\rho_{XB_{2}}$.\ The
final equality follows from the definition of quantum mutual information.
\end{proof}

Armed with the above theorem, we need to find an example of a quantum channel
that has zero quantum capacity, but for which there exists an ensemble that
registers a non-zero private information. If such a channel were to exist, we
could combine it with a 50\% erasure channel in order to achieve a non-zero
coherent information (and thus a non-zero quantum capacity) for the joint
channel. Indeed, such a channel exists, and it is known as an
entanglement-binding channel. It has the ability to generate private classical
communication but no ability to transmit quantum information (we point the
reader to \cite{HHH96,H97}\ for further details on these channels). Thus, the
50\% erasure channel and the entanglement-binding channel can superactivate
each other.

The startling phenomenon of superactivation has important implications for
quantum data transmission. First, it implies that a quantum channel's ability
to transmit quantum information depends on the context in which it is used.
For example, if other seemingly useless channels are available, it could be
possible to transmit more quantum information than would be possible were the
channels used alone. Next, and more importantly for quantum Shannon theory, it
implies that whatever formula might eventually be found to characterize
quantum capacity (some characterization other than the regularized coherent
information in Theorem~\ref{thm-q-cap:q-cap-theorem}), it should be strongly
non-additive in some cases (strongly non-additive in the sense of
superactivation). That is, suppose that $Q^{?}(\mathcal{N})$ is some unknown
formula for the quantum capacity of $\mathcal{N}$ and $Q^{?}(\mathcal{M})$ is
the same formula characterizing the quantum capacity of $\mathcal{M}$. Then
this formula in general should be strongly non-additive in some cases:%
\begin{equation}
Q^{?}(\mathcal{N}\otimes\mathcal{M})>Q^{?}(\mathcal{N})+Q^{?}(\mathcal{M}).
\end{equation}
The discovery of superactivation has led us to realize that at present we are
much farther than we might have thought from understanding reliable quantum
communication rates over quantum channels.

\section{Entanglement Distillation}

We close out this chaper
\index{entanglement distillation}
with a final application of the techniques in the direct coding part of
Theorem~\ref{thm-q-cap:q-cap-theorem}\ to the task of entanglement
distillation. Entanglement distillation is a protocol where Alice and Bob
begin with many copies of some bipartite state $\rho_{AB}$. They attempt to
distill ebits from it at some positive rate by employing local operations and
forward classical communication from Alice to Bob. If the state is pure, then
Alice and Bob should simply perform the entanglement concentration protocol
from Chapter~\ref{chap:ent-conc}, and there is no need for forward classical
communication in this case. Otherwise, they can perform the protocol given in
the proof of the following theorem.

\begin{theorem}
[Devetak--Winter]Suppose that Alice and Bob share the state $\rho
_{AB}^{\otimes n}$ where $n$ is an arbitrarily large positive integer. Then it
is possible for them to distill ebits at the rate $I(A\rangle B)_{\rho}$ if
they are allowed forward classical communication from Alice to Bob.
\end{theorem}

We should mention that we have already proved the statement in the above
theorem with the protocol given in Corollary~\ref{thm-ccn:cast}. Nevertheless,
it is still instructive to exploit the techniques from this chapter in proving
the existence of an entanglement distillation protocol.

\bigskip

\begin{proof}
Suppose that Alice and Bob begin with a general bipartite state $\rho_{AB}$
with purification $\psi_{ABE}$. We can write the purification in Schmidt form
as follows:%
\begin{equation}
|\psi\rangle_{ABE}\equiv\sum_{x\in\mathcal{X}}\sqrt{p_{X}(x)}|x\rangle
_{A}\otimes|\psi_{x}\rangle_{BE}.
\end{equation}
The $n$th extension of the above state is%
\begin{equation}
|\psi\rangle_{A^{n}B^{n}E^{n}}\equiv\sum_{x^{n}\in\mathcal{X}^{n}}%
\sqrt{p_{X^{n}}(x^{n})}|x^{n}\rangle_{A^{n}}\otimes|\psi_{x^{n}}\rangle
_{B^{n}E^{n}}.
\end{equation}
The protocol begins with Alice performing a type class measurement given by
the type projectors (recall from \eqref{eq-qt:typ-sub-decomp-types} that the
typical projector decomposes into a sum of the type class projectors):%
\begin{equation}
\Pi_{t}^{n}\equiv\sum_{x^{n}\in T_{t}^{X^{n}}}|x^{n}\rangle\langle x^{n}|.
\end{equation}
If the type resulting from the measurement is not a typical type, then Alice
aborts the protocol (this result happens with arbitrarily small probability).
If it is a typical type, they can then consider a code over a particular type
class $t$ with the following structure:%
\begin{align}
LMK  &  \approx\left\vert T_{t}\right\vert \approx2^{nH(X)},\\
K  &  \approx2^{nI(X;E)},\\
MK  &  \approx2^{nI(X;B)},
\end{align}
where $t$ is the type class and the entropies are with respect to the
following dephased state:%
\begin{equation}
\sum_{x\in\mathcal{X}}p_{X}(x)|x\rangle\langle x|_{X}\otimes|\psi_{x}%
\rangle\langle\psi_{x}|_{BE}.
\end{equation}
It follows that $M\approx2^{n\left(  I(X;B)-I(X;E)\right)  }=2^{n\left[
H(B)-H(E)\right]  }$ and $L\approx2^{nH(X|B)}$. We label the codewords as
$x^{n}(l,m,k)$ where $x^{n}(l,m,k)\in T_{t}$. Thus, we instead operate on the
following state $|\tilde{\psi}_{t}\rangle_{A^{n}B^{n}E^{n}}$ resulting from
the type class measurement:%
\begin{equation}
|\tilde{\psi}_{t}\rangle_{A^{n}B^{n}E^{n}}\equiv\frac{1}{\sqrt{\left\vert
T_{t}\right\vert }}\sum_{x^{n}\in T_{t}}\ |x^{n}\rangle_{A^{n}}\otimes
|\psi_{x^{n}}\rangle_{B^{n}E^{n}}.
\end{equation}
The protocol proceeds as follows. Alice first performs an incomplete
measurement of the system $A^{n}$, with the following measurement operators:%
\begin{equation}
\left\{  \Gamma_{l}\equiv\sum_{m,k}|m,k\rangle\langle x^{n}(l,m,k)|_{A^{n}%
}\right\}  _{l}.
\end{equation}
This measurement collapses the above state as follows:%
\begin{equation}
\frac{1}{\sqrt{MK}}\sum_{m,k}|m,k\rangle_{A^{n}}\otimes\left\vert \psi
_{x^{n}(l,m,k)}\right\rangle _{B^{n}E^{n}}.
\end{equation}
Alice transmits the classical information in $l$ to Bob, using $nH(X|B)$ bits
of classical information. Bob needs to know $l$ so that he can know in which
code they are operating. Bob then constructs the following isometry, a
coherent POVM\ similar to that in \eqref{eq-q-cap:bob-coherent-meas}
(constructed from the POVM\ for a private classical communication code):%
\begin{equation}
\sum_{m,k}\sqrt{\Lambda_{B^{n}}^{m,k}}\otimes|m,k\rangle_{B}.
\end{equation}
After performing the above coherent POVM, the state is close to the following
one:%
\begin{equation}
\frac{1}{\sqrt{MK}}\sum_{m,k}|m,k\rangle_{A^{n}}\otimes|m,k\rangle
_{B}\left\vert \psi_{x^{n}(l,m,k)}\right\rangle _{B^{n}E^{n}}.
\end{equation}
Alice then performs a measurement of the $k$ register in the
Fourier-transformed basis:%
\begin{equation}
\left\{  \left\vert \hat{s}\right\rangle \equiv\frac{1}{\sqrt{K}}\sum
_{k}e^{i2\pi ks/K}|k\rangle\right\}  _{s\in\left\{  1,\ldots,K\right\}  }.
\end{equation}
Alice performs this particular measurement because she would like Bob and Eve
to maintain their entanglement in the $k$ variable. The state resulting from
this measurement is%
\begin{equation}
\frac{1}{\sqrt{MK}}\sum_{m,k}|m\rangle_{A^{n}}\otimes e^{i2\pi ks/K}%
|m,k\rangle_{B}\left\vert \psi_{x^{n}(l,m,k)}\right\rangle _{B^{n}E^{n}}.
\end{equation}
Alice then uses $nI(X;E)$ bits to communicate the $s$ variable to Bob. Bob
then applies the phase transformation $Z^{\dag}(s)$, where%
\begin{equation}
Z^{\dag}(s)=\sum_{k}e^{-i2\pi sk/K}|k\rangle\langle k|,
\end{equation}
to his $k$ variable in register $B$. The resulting state is
\begin{equation}
\frac{1}{\sqrt{MK}}\sum_{m,k}|m\rangle_{A^{n}}\otimes|m,k\rangle_{B}\left\vert
\psi_{x^{n}(l,m,k)}\right\rangle _{B^{n}E^{n}}.
\end{equation}
They then proceed as in the final steps
\eqref{eq-q-cap:final-step-1}--\eqref{eq-q-cap:final-step-2} of the protocol
from the direct coding part of Theorem~\ref{thm-q-cap:q-cap-theorem}, and they
extract a state close to a maximally entangled state of the following form:%
\begin{equation}
\frac{1}{\sqrt{M}}\sum_{m}|m\rangle_{A^{n}}\otimes\left\vert m\right\rangle
_{B},
\end{equation}
with rate equal to $\left(  \log M\right)  /n=H(B)-H(E)$.
\end{proof}

\begin{exercise}
Argue that the above protocol cannot perform the task of state transfer as can
the protocol in Corollary~\ref{thm-ccn:cast}.
\end{exercise}

\section{History and Further Reading}

The quantum capacity theorem has a long history that led to many important
discoveries in quantum information theory. \cite{PhysRevA.52.R2493} first
stated the problem of finding the quantum capacity of a quantum channel in his
seminal paper on quantum error correction. \cite{DSS98} demonstrated that the
coherent information of the depolarizing channel is superadditive by
concatenating a random code with a repetition code (this result in hindsight
was remarkable given that the coherent information was not even known at the
time). \cite{SS07} later extended this result to show that the coherent
information is strongly super-additive for several examples of Pauli channels.
\cite{PhysRevA.54.2629} demonstrated that the coherent information obeys a
quantum data-processing inequality, much like the classical data-processing
inequality for mutual information. \cite{PhysRevLett.80.5695} started making
connections between private communication and quantum communication.
\cite{BDSW96}\ and \cite{BKN98}\ demonstrated that forward classical
communication cannot increase the quantum capacity. In the same paper,
\cite{BDSW96} introduced the idea of entanglement distillation, which has
 connections with the quantum capacity.

\cite{PhysRevA.54.2614,PhysRevA.54.2629,BNS98,BKN98} made important progress
on the quantum capacity theorem in a series of papers that established the
coherent information upper bound on the quantum capacity.
\cite{PhysRevA.55.1613}, \cite{capacity2002shor}, and \cite{ieee2005dev} are
generally credited with proving the coherent information lower bound on the
quantum capacity, though an inspection of Lloyd's proof reveals that it is
perhaps not as rigorous as the latter two proofs. \cite{capacity2002shor}
delivered his proof of the lower bound in a lecture, though he never published
this proof in a journal. Later, \cite{qcap2008fourth} published a
paper\ detailing a proof of the quantum capacity theorem that they considered
to be close in spirit to the proof in \cite{capacity2002shor}. After Shor's
proof, \cite{ieee2005dev} provided a detailed proof of the lower bound on the
quantum capacity, by analyzing superpositions of the codewords from private
classical codes. This is the approach we have taken in this chapter. We should
also mention that \cite{H05} showed how to achieve the coherent information
for certain input states by using random stabilizer codes, and \cite{HP01}
showed how to achieve the coherent information rate for a very specific class
of channels.

\cite{thesis97gottesman} established the stabilizer formalism for quantum
error correction. Our discussion of stabilizer codes in this chapter follows
closely the development in the PhD thesis of \cite{Smith06}.

Another approach to
\index{decoupling approach}
proving the quantum capacity theorem is known as the decoupling
approach~\citep{qcap2008first}. This approach exploits a fundamental concept
introduced in \cite{qip2002schu}. Suppose that the reference, Bob, and Eve
share a tripartite pure entangled state~$\vert\psi\rangle_{RBE}$ after Alice
transmits her share of the entanglement with the reference through a noisy
channel. Then if the reduced state~$\psi_{RE}$ on the reference system and
Eve's system is approximately decoupled, meaning that%
\begin{equation}
\left\Vert \psi_{RE}-\psi_{R}\otimes\sigma_{E}\right\Vert _{1}\leq\varepsilon,
\end{equation}
where $\sigma_{E}$ is some arbitrary state, this implies that Bob can decode
the quantum information that Alice intended to send to him. Why is this so?
Let us suppose that the state is exactly decoupled. Then one purification of
the state $\psi_{RE}$ is the state $\vert\psi\rangle_{RBE}$ that they share
after the channel acts. Another purification of $\psi_{RE}=\psi_{R}%
\otimes\sigma_{E}$ is%
\begin{equation}
\vert\psi\rangle_{RB_{1}}\otimes\left\vert \sigma\right\rangle _{B_{2}E},
\end{equation}
where $\vert\psi\rangle_{RB_{1}}$ is the original state that Alice sent
through the channel and $\left\vert \sigma\right\rangle _{B_{2}E}$ is some
other state that purifies the state $\sigma_{E}$ of the environment. Since all
purifications are related by isometries and since Bob possesses the
purification of $R$ and $E$, there exists some unitary $U_{B\rightarrow
B_{1}B_{2}}$ such that%
\begin{equation}
U_{B\rightarrow B_{1}B_{2}}\vert\psi\rangle_{RBE}=\vert\psi\rangle_{RB_{1}%
}\otimes\left\vert \sigma\right\rangle _{B_{2}E}.
\end{equation}
This unitary is then Bob's decoder! Thus, the decoupling condition implies the
existence of a decoder for Bob, so that it is only necessary to show the
existence of an encoder that decouples the reference from the environment.
Simply put, the structure of quantum mechanics allows for this way of proving
the quantum capacity theorem.

Many researchers have now exploited the decoupling approach in a variety of
contexts. This approach is implicit in Devetak's proof of the quantum capacity
theorem~\citep{ieee2005dev}. \cite{Horodecki:2005:673,Horodecki:2007:107}
exploited it to prove the existence of a state-merging protocol.
\cite{YD09}\ and \cite{PhysRevA.78.030302} used it in their proofs of the
state redistribution protocol. \cite{DH2006}\ proved the best known
characterization of the entanglement-assisted quantum capacity of the
broadcast channel using this approach. The thesis of Dupuis and subsequent work
generalized this decoupling approach to settings beyond the traditional
i.i.d.~setting~\citep{D10,DBWR10}. Datta and coworkers have also applied this
approach in a variety of contexts~\citep{BD10,DH11a,DH11}, and \cite{WH10a}
used the approach to study quantum communication using a noisy channel and a
noisy state.

\cite{PhysRevLett.78.3217} found the quantum capacity of the erasure channel,
and \cite{PhysRevA.71.032314} computed the quantum capacity of the amplitude
damping channel. \cite{science2008smith} showed superactivation and later
showed superactivation for channels that can be realized more easily in the
laboratory~\citep{SSY11}. \cite{DW05} established that the coherent
information is achievable for entanglement distillation. \cite{CEMOGS15} and
\cite{PhysRevLett.115.040501} demonstrated a striking superadditivity effect,
which suggests that a regularized expression is necessary to determine the
quantum capacity of an arbitrary channel.

There have also been some results on error exponents, the strong converse, and
second-order characterizations of the quantum capacity (see
Section~\ref{sec-cc:history}\ for a discussion of the meaning of these terms).
\cite{BBCW13} proved that a quantity called the entanglement cost of a quantum
channel is a strong converse rate for quantum communication. \cite{MW13}
established what they called a \textquotedblleft pretty strong
converse\textquotedblright\ for the quantum capacity of degradable channels,
meaning that there is a sharp transition in the fidelity from one to $1/2$,
when the rate of communication goes from below to above the quantum capacity
(this is in the limit of many channel uses). \cite{WW14} demonstrated that
randomly selected codes with a communication rate exceeding the quantum
capacity of the quantum erasure channel lead to a fidelity that decreases
exponentially fast as the number of channel uses increases (a strong converse
would however demonstrate that this behavior occurs for all codes).
\cite{TWW14} proved that a quantity known as the Rains bound (defined in
\cite{R01}---see also the later work of \cite{AdMVW02}) is a strong converse rate
for quantum communication over any channel, which in turn establishes the
strong converse for any dephasing channel.

\cite{BDL15} and \cite{TBR15}\ established second-order achievability
characterizations for quantum capacity. \cite{BDL15} did so by making use of a
\textquotedblleft Petz recovery map\textquotedblright\ decoder and
\cite{TBR15} with a version of the decoupling theorem from \cite{MW13}.
\cite{TBR15} also gave a second-order converse for quantum communication by
making use of the Rains bound, and they obtained an exact second-order
characterization of quantum communication for dephasing channels.

\chapter{Trading Resources for Communication}

\label{chap:trade-off}This chapter unifies all of the channel coding theorems
that we have studied in this book. One of the most general
information-processing tasks that a sender and receiver can accomplish is to
transmit classical and quantum information and generate entanglement with many
independent uses of a quantum channel and with the assistance of classical
communication, quantum communication, and shared entanglement.\footnote{Recall
that Chapter~\ref{chap:unit-resource-cap} addressed a special case of this
task that applies to the scenario in which the sender and receiver do not have
access to many independent uses of a quantum channel.} The resulting rates for
communication are \textit{net} rates that give the generation rate of a
resource less its consumption rate. Since we have three resources, all
achievable rates are rate triples $\left(  C,Q,E\right)  $ that lie in a
three-dimensional capacity region, where $C$ is the net rate of classical
communication, $Q$ is the net rate of quantum communication, and $E$ is the
net rate of entanglement consumption/generation. The capacity theorem for this
general scenario is known as the quantum dynamic capacity theorem, and it is
the main theorem that we prove in this chapter. All of the rates given in the
channel coding theorems of previous chapters are special points in this
three-dimensional capacity region.

The proof of the quantum dynamic capacity theorem%
\index{quantum dynamic capacity theorem}
comes in two parts: the direct coding theorem and the converse theorem. The
direct coding theorem demonstrates that the strategy for achieving any point
in the three-dimensional capacity region is remarkably simple:\ we just
combine the protocol from Corollary~\ref{cor-ccn:CQE-trading}\ for
entanglement-assisted classical and quantum communication with the three unit
protocols of teleportation, super-dense coding, and entanglement distribution.
The interpretation of the achievable rate region is that it is the
\index{unit resource capacity region}%
unit resource capacity region from Chapter~\ref{chap:unit-resource-cap}%
\ translated along the points achievable with the protocol from
Corollary~\ref{cor-ccn:CQE-trading}. In the proof of the converse theorem, we
analyze the most general protocol that can consume and generate classical
communication, quantum communication, and entanglement along with the
consumption of many independent uses of a quantum channel, and we show that
the net rates for such a protocol are bounded by the regularization of the
achievable rate region. In the general case, our characterization is
multi-letter, meaning that the computation of the capacity region requires an
optimization over a potentially infinite number of channel uses and is thus
intractable. However, both the quantum Hadamard channels
\index{Hadamard channel}%
from Section~\ref{sec-nqt:hadamard-channel} and the quantum erasure channel
are special classes of channels for which the regularization is not necessary,
and we can compute their capacity regions with respect to a single instance of
the channel. Another important class of channels for which the capacity region
is known is the class of pure-loss bosonic channels (though the optimality
proof is only up to a long-standing conjecture which many researchers believe
to be true). These pure-loss bosonic channels model free-space communication
or loss in a fiber optic cable and thus have an elevated impetus for study
because of their importance in practical applications.

One of the most important questions for communication in this
three-dimensional setting is whether it is really necessary to exploit the
trade-off coding strategy given in Corollary~\ref{cor-ccn:CQE-trading}. That
is, would it be best simply to use a classical communication code for a
fraction of the channel uses, a quantum communication code for another
fraction, an entanglement-assisted code for another fraction, etc.? Such a
strategy is known as time sharing and allows the sender and receiver to
achieve convex combinations of any rate triples in the capacity region. The
answer to this question depends on the channel. For example, time sharing is
optimal for the quantum erasure channel, but it is not for a dephasing channel
or a pure-loss bosonic channel. In fact, trade-off coding for a pure-loss
bosonic channel can give tremendous performance gains over time sharing. How
can we know which one will perform better in the general case? It is hard to
say, but at the very least, we know that time sharing is a special case of
trade-off coding as we argued in Section~\ref{sec-ccn:trade-off-vs-TS}. Thus,
from this perspective, it might make sense simply to always use a trade-off strategy.

We organize this chapter as follows. We first review the
information-processing task corresponding to the quantum dynamic capacity
region. Section~\ref{sec-tr:capacity-theorem} states the quantum dynamic
capacity theorem and shows how many of the capacity theorems we studied
previously arise as special cases of it. The next two sections prove the
direct coding theorem and the converse theorem.
Section~\ref{sec-tr:dynamic-cap-formula}\ introduces the quantum dynamic
capacity formula, which is important for analyzing whether the quantum dynamic
capacity region is single-letter. In the final section of this chapter, we
compute and plot the quantum dynamic capacity region for the dephasing
channels, erasure channels, and the pure-loss bosonic channels.

\section{The Information-Processing Task}%

\begin{figure}
[ptb]
\begin{center}
\includegraphics[
width=4.8456in
]%
{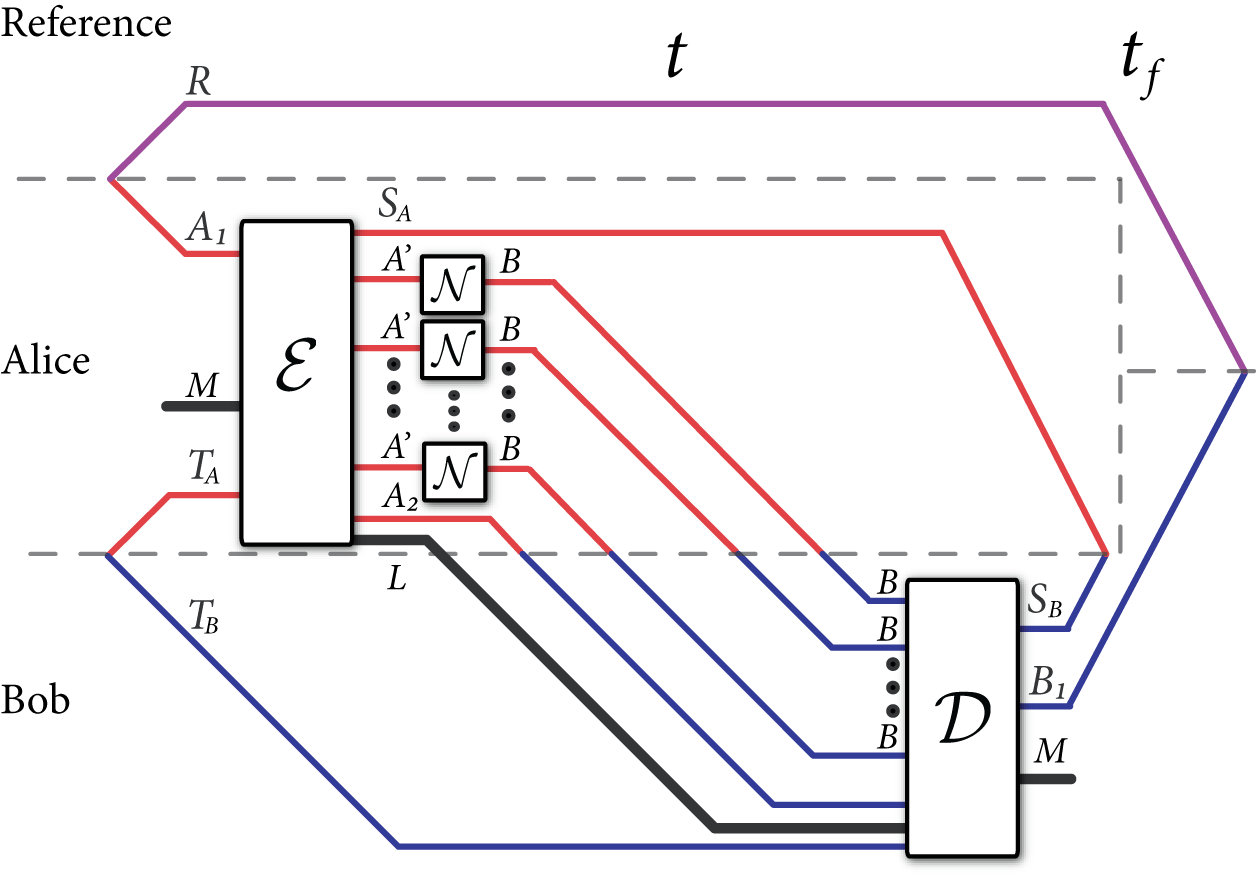}%
\caption{The most general protocol for generating classical communication,
quantum communication, and entanglement with the help of the same
respective resources and many uses of a quantum channel. Alice begins with her
classical register $M$, her quantum register $A_{1}$, and her share of the
entanglement in register $T_{A}$. She encodes according to some encoding
channel$\ \mathcal{E}$ that outputs a quantum register $S_{A}$, many registers
$A^{\prime n}$, a quantum register $A_{2}$, and a classical register~$L$. She
inputs $A^{\prime n}$ to many uses of the quantum channel $\mathcal{N}$ and
transmits $A_{2}$ over a noiseless quantum channel and $L$ over a noiseless
classical channel. Bob receives the channel outputs $B^{n}$, the quantum
register $A_{2}$, and the classical register $L$ and performs a decoding
$\mathcal{D}$ that recovers the quantum information and classical message. The
decoding also generates entanglement with system $S_{A}$. Many protocols are a
special case of the above one. For example, the protocol is
entanglement-assisted communication of classical and quantum information~if
the registers $L$, $S_{A}$, $S_{B}$, and $A_{2}$ are empty.}%
\label{fig-tr:catalytic-protocol}%
\end{center}
\end{figure}
Figure~\ref{fig-tr:catalytic-protocol} depicts the most general protocol for
generating classical communication, quantum communication, and entanglement
with the consumption of a noisy quantum channel $\mathcal{N}_{A^{\prime
}\rightarrow B}$ and the same respective resources. Alice possesses two
classical registers (each labeled by $M$ and of dimension
$2^{nC_{\operatorname{out}}}$), a quantum register $A_{1}$ of dimension
$2^{nQ_{\operatorname{out}}}$ entangled with a reference system $R$, and
another quantum register $T_{A}$ of dimension $2^{nE_{\operatorname{in}}}$
that contains her share of the state $\Phi_{T_{A}T_{B}}$\ maximally entangled
with Bob:%
\begin{equation}
\omega_{MMRA_{1}T_{A}T_{B}}\equiv\overline{\Phi}_{MM}\otimes\Psi_{RA_{1}%
}\otimes\Phi_{T_{A}T_{B}}.
\end{equation}
She passes one of the classical registers and the registers $A_{1}$ and
$T_{A}$ into an encoding channel $\mathcal{E}_{MA_{1}T_{A}\rightarrow
A^{\prime n}S_{A}LA_{2}}$ that outputs a quantum register $S_{A}$ of dimension
$2^{nE_{\operatorname{out}}}$ and a quantum register $A_{2}$ of dimension
$2^{nQ_{\operatorname{in}}}$, a classical register $L$ of dimension
$2^{nC_{\operatorname{in}}}$, and many quantum systems $A^{\prime n}$. The
register $S_{A}$\ is for creating entanglement with Bob. The state after the
encoding $\mathcal{E}$\ is as follows:%
\begin{equation}
\omega_{MA^{\prime n}S_{A}LA_{2}RT_{B}}\equiv\mathcal{E}_{MA_{1}%
T_{A}\rightarrow A^{\prime n}S_{A}LA_{2}}(\omega_{MMRA_{1}T_{A}T_{B}}).
\end{equation}
She sends the systems $A^{\prime n}$ through many uses $\mathcal{N}_{A^{\prime
n}\rightarrow B^{n}}\equiv(\mathcal{N}_{A^{\prime}\rightarrow B})^{\otimes n}$
of the quantum channel $\mathcal{N}_{A^{\prime}\rightarrow B}$, transmits $L$
over a noiseless classical channel, and transmits $A_{2}$ over a noiseless
quantum channel, producing the following state:%
\begin{equation}
\omega_{MB^{n}S_{A}LA_{2}RT_{B}}\equiv\mathcal{N}_{A^{\prime n}\rightarrow
B^{n}}(\omega_{MA^{\prime n}S_{A}LA_{2}RT_{B}}).
\label{eq-tr:state-after-channel}%
\end{equation}
The above state has the following form:%
\begin{equation}
\sum_{x}p_{X}(x)|x\rangle\langle x|_{X}\otimes\mathcal{N}_{A^{\prime
n}\rightarrow B^{n}}(\rho_{AA^{\prime n}}^{x}),
\end{equation}
with $A\equiv RT_{B}A_{2}S_{A}$ and $X\equiv ML$. Bob then applies a decoding
channel $\mathcal{D}_{B^{n}A_{2}T_{B}L\rightarrow B_{1}S_{B}\hat{M}}$ that
outputs a quantum system $B_{1}$, a quantum system $S_{B}$, and a classical
register $\hat{M}$. Let $\omega^{\prime}$ denote the final state. The
following condition holds for a protocol with error $\varepsilon\in(0,1)$:%
\begin{equation}
\frac{1}{2}\left\Vert \overline{\Phi}_{M\hat{M}}\otimes\Psi_{RB_{1}}%
\otimes\Phi_{S_{A}S_{B}}-\omega_{MB_{1}S_{B}\hat{M}S_{A}R}^{\prime}\right\Vert
_{1}\leq\varepsilon, \label{eq-tr:+++_good-code}%
\end{equation}
implying that Alice and Bob establish maximal classical correlations in $M$
and $\hat{M}$ and maximal entanglement between $S_{A}$ and $S_{B}$. The above
condition also implies that the coding scheme preserves the entanglement with
the reference system $R$. The net rate triple for the protocol is as follows:
$(C_{\operatorname{out}}-C_{\operatorname{in}},Q_{\operatorname{out}%
}-Q_{\operatorname{in}},E_{\operatorname{out}}-E_{\operatorname{in}})$. The
protocol generates a resource if its corresponding rate is positive, and it
consumes a resource if its corresponding rate is negative. A protocol of the
above form is an $(n,C_{\operatorname{out}}-C_{\operatorname{in}%
},Q_{\operatorname{out}}-Q_{\operatorname{in}},E_{\operatorname{out}%
}-E_{\operatorname{in}},\varepsilon)$ protocol.

We say that a rate triple $\left(  C,Q,E\right)  $ is achievable for
$\mathcal{N}$ if there exists a sequence of $\left(  n,C-\delta,Q-\delta
,E-\delta,\varepsilon\right)  $\ protocols for all $\delta>0$, $\varepsilon
\in(0,1)$ and sufficiently large$~n$. The quantum dynamic capacity region
$\mathcal{C}_{\operatorname{CQE}}(\mathcal{N})$\ is equal to the union of all
achievable rates.

\section{The Quantum Dynamic Capacity Theorem}

\label{sec-tr:capacity-theorem}The dynamic capacity theorem
\index{quantum dynamic capacity theorem}
gives bounds on the reliable communication rates of a noisy quantum channel
when combined with the noiseless resources of classical communication, quantum
communication, and shared entanglement. The theorem applies regardless of
whether a protocol consumes the noiseless resources or generates them.

\begin{theorem}
[Quantum Dynamic Capacity]\label{thm-tr:main-theorem}The dynamic capacity
region $\mathcal{C}_{\operatorname{CQE}}(\mathcal{N})$ of a quantum channel
$\mathcal{N}$ is equal to the following expression:%
\begin{equation}
\mathcal{C}_{\operatorname{CQE}}(\mathcal{N})=\overline{\bigcup_{k=1}^{\infty
}\frac{1}{k}\mathcal{C}_{\operatorname{CQE}}^{(1)}(\mathcal{N}^{\otimes k})},
\label{eq-tr:multi-letter}%
\end{equation}
where the overbar indicates the closure of a set. The region $\mathcal{C}%
_{\operatorname{CQE}}^{(1)}(\mathcal{N})$ is equal to the union of the
state-dependent\ regions $\mathcal{C}_{\operatorname{CQE},\sigma}%
^{(1)}(\mathcal{N})$:%
\begin{equation}
\mathcal{C}_{\operatorname{CQE}}^{(1)}(\mathcal{N})\equiv\bigcup_{\sigma
}\mathcal{C}_{\operatorname{CQE},\sigma}^{(1)}(\mathcal{N}).
\end{equation}
The state-dependent\ region $\mathcal{C}_{\operatorname{CQE},\sigma}%
^{(1)}(\mathcal{N})$ is the set of all rates $C$, $Q$, and $E$, such that%
\begin{align}
C+2Q  &  \leq I(AX;B)_{\sigma},\label{eq-tr:CQ-bound}\\
Q+E  &  \leq I(A\rangle BX)_{\sigma},\label{eq-tr:QE-bound}\\
C+Q+E  &  \leq I(X;B)_{\sigma}+I(A\rangle BX)_{\sigma}.
\label{eq-tr:CQE-bound}%
\end{align}
The above entropic quantities are with respect to a classical--quantum state
$\sigma_{XAB}$, where%
\begin{equation}
\sigma_{XAB}\equiv\sum_{x}p_{X}(x)\vert x\rangle\langle x\vert_{X}%
\otimes\mathcal{N}_{A^{\prime}\rightarrow B}(\phi_{AA^{\prime}}^{x}),
\label{eq-tr:main-theorem-state}%
\end{equation}
and the states $\phi_{AA^{\prime}}^{x}$ are pure. It is implicit that one
should consider states on $A^{\prime k}$ instead of $A^{\prime}$ when taking
the regularization in \eqref{eq-tr:multi-letter}.
\end{theorem}

The above theorem is a \textquotedblleft multi-letter\textquotedblright%
\ capacity theorem due to the regularization in
\eqref{eq-tr:multi-letter}. However, we show in
Section~\ref{sec-tr:single-letter-hadamard}\ that the regularization is not
necessary for the Hadamard class of channels. We prove the above theorem in
two parts:

\begin{enumerate}
\item The direct coding theorem in Section~\ref{sec-tr:direct-coding}\ shows
that combining the protocol from Corollary~\ref{cor-ccn:CQE-trading} with
teleportation, super-dense coding, and entanglement distribution achieves the
above region.

\item The converse theorem in Section~\ref{sec-tr:converse}\ demonstrates that
any coding scheme cannot do better than the regularization in
\eqref{eq-tr:multi-letter}, in the sense that a sequence of protocols with
decreasing error should have the communication rates below the above amounts.
\end{enumerate}

\begin{exercise}
\label{ex-tr:four-entropies}Show that it suffices to evaluate the
following four entropies in order to determine the state-dependent region in
Theorem~\ref{thm-tr:main-theorem}:%
\begin{align}
H(A|X)_{\sigma}  &  =\sum_{x}p_{X}(x)H(A)_{\phi_{x}},\\
H(B)_{\sigma}  &  =H\left(  \sum_{x}p_{X}(x)\mathcal{N}_{A^{\prime}\rightarrow
B}(\phi_{A^{\prime}}^{x})\right)  ,\\
H(B|X)_{\sigma}  &  =\sum_{x}p_{X}(x)H(\mathcal{N}_{A^{\prime}\rightarrow
B}(\phi_{A^{\prime}}^{x})),\\
H(E|X)_{\sigma}  &  =\sum_{x}p_{X}(x)H(\mathcal{N}_{A^{\prime}\rightarrow
E}^{c}(\phi_{A^{\prime}}^{x})),
\end{align}
where the state $\sigma_{XABE}$ extends the state in
\eqref{eq-tr:main-theorem-state} and is of the form
\begin{equation}
\sigma_{XABE}\equiv\sum_{x}p_{X}(x)\vert x\rangle\langle x\vert_{X}%
\otimes\mathcal{U}^{\mathcal{N}}_{A^{\prime}\rightarrow BE}(\phi_{AA^{\prime}%
}^{x}),
\end{equation}
where $U^{\mathcal{N}}_{A^{\prime}\rightarrow BE}$ is an isometric extension
of the channel ${\mathcal{N}}_{A^{\prime}\rightarrow BE}$.
\end{exercise}

\subsection{Special Cases of the Quantum Dynamic Capacity Theorem}

We first consider five special cases of the above capacity theorem that arise
when $Q$ and $E$ both vanish, $C$ and $E$ both vanish, or one of $C$, $Q$, or
$E$ vanishes. The first two cases correspond respectively to the classical
capacity theorem from Chapter~\ref{chap:classical-comm-HSW}\ and the quantum
capacity theorem from Chapter~\ref{chap:quantum-capacity}. Each of the other
special cases traces out a two-dimensional achievable rate region in the
three-dimensional capacity region. The five coding scenarios are as follows:

\begin{enumerate}
\item Classical communication (C)\ when there is no entanglement assistance or
quantum communication. The achievable rate region lies on the $(C,0,0)$ ray
extending from the origin.

\item Quantum communication (Q)\ when there is no entanglement assistance or
classical communication. The achievable rate region lies on the $(0,Q,0)$ ray
extending from the origin.

\item Entanglement-assisted quantum communication (QE) when there is no
classical communication. The achievable rate region lies in the $\left(
0,Q,-E\right)  $ quarter-plane of the three-dimensional region in
Theorem~\ref{thm-tr:main-theorem}.

\item Classically enhanced quantum communication (CQ)\ when there is no
entanglement assistance. The achievable rate region lies in the $\left(
C,Q,0\right)  $ quarter-plane of the three-dimensional region in
Theorem~\ref{thm-tr:main-theorem}.

\item Entanglement-assisted classical communication (CE) when there is no
quantum communication. The achievable rate region lies in the $\left(
C,0,-E\right)  $ quarter-plane of the three-dimensional region in
Theorem~\ref{thm-tr:main-theorem}.
\end{enumerate}

\subsubsection{Classical Capacity}

The following theorem gives the
\index{HSW theorem}%
one-dimensional capacity region $\mathcal{C}_{\operatorname{C}}(\mathcal{N})$
of a quantum channel $\mathcal{N}$ for classical communication.

\begin{theorem}
[Holevo--Schumacher--Westmoreland]The classical capacity region $\mathcal{C}%
_{\emph{C}}(\mathcal{N})$ is given by%
\begin{equation}
\mathcal{C}_{\emph{C}}(\mathcal{N})=\overline{\bigcup_{k=1}^{\infty}\frac
{1}{k}\mathcal{C}_{\emph{C}}^{(1)}(\mathcal{N}^{\otimes k})}.
\end{equation}
The region $\mathcal{C}_{\emph{C}}^{(1)}(\mathcal{N})$ is the union of the
state-dependent regions $\mathcal{C}_{\emph{C},\sigma}^{(1)}(\mathcal{N})$,
where $\mathcal{C}_{\emph{C},\sigma}^{(1)}(\mathcal{N})$ is the set of all
$C\geq0$, such that%
\begin{equation}
C\leq I(X;B)_{\sigma}+I(A\rangle BX)_{\sigma}. \label{eq-tr:HSW-bound}%
\end{equation}
The entropic quantity is with respect to the state $\sigma_{XAB}$ in \eqref{eq-tr:main-theorem-state}.
\end{theorem}

\noindent The bound in \eqref{eq-tr:HSW-bound} is never larger than the bound
in \eqref{eq-tr:CQE-bound} with $Q=0$ and $E=0$, given that $I(X;B)_{\sigma
}+I(A\rangle BX)_{\sigma}\leq I(AX;B)_{\sigma}$. The above characterization of
the classical capacity\ region may seem slightly different from the
characterization in Chapter~\ref{chap:classical-comm-HSW}, until we make a few
observations. First, we rewrite the coherent information $I(A\rangle
BX)_{\sigma}$ as $H(B|X)_{\sigma}-H(E|X)_{\sigma}$. Then $I(X;B)_{\sigma
}+I(A\rangle BX)_{\sigma}=H(B)_{\sigma}-H(E|X)_{\sigma}$. Next, pure states of
the form $|\varphi^{x}\rangle_{A^{\prime}}$ are sufficient to attain the
classical capacity of a quantum channel (see
Theorem~\ref{thm-add:pure-states-suff-holevo}). Then $H(E|X)_{\sigma
}=H(B|X)_{\sigma}$ so that $I(X;B)_{\sigma}+I(A\rangle BX)_{\sigma
}=H(B)_{\sigma}-H(B|X)_{\sigma}=I(X;B)_{\sigma}$ for states of this form.
Thus, the expression in \eqref{eq-tr:HSW-bound} can never exceed the classical
capacity and finds its maximum exactly at the Holevo information.

\subsubsection{Quantum Capacity}

The following theorem gives the
\index{quantum capacity theorem}
one-dimensional quantum capacity region $\mathcal{C}_{\operatorname{Q}%
}(\mathcal{N})$ of a quantum channel $\mathcal{N}$.

\begin{theorem}
[Quantum Capacity]The quantum capacity region $\mathcal{C}_{\emph{Q}%
}(\mathcal{N})$ is given by%
\begin{equation}
\mathcal{C}_{\emph{Q}}(\mathcal{N})=\overline{\bigcup_{k=1}^{\infty}\frac
{1}{k}\mathcal{C}_{\emph{Q}}^{(1)}(\mathcal{N}^{\otimes k})}.
\end{equation}
The region $\mathcal{C}_{\emph{Q}}^{(1)}(\mathcal{N})$ is the union of the
state-dependent regions $\mathcal{C}_{\emph{Q},\sigma}^{(1)}(\mathcal{N})$,
where $\mathcal{C}_{\emph{Q},\sigma}^{(1)}(\mathcal{N})$ is the set of all
$Q\geq0$, such that%
\begin{equation}
Q\leq I(A\rangle BX)_{\sigma}. \label{eq-tr:LSD-bound}%
\end{equation}
The entropic quantity is with respect to the state $\sigma_{XAB}$ in
\eqref{eq-tr:main-theorem-state} with the restriction that the density
$p_{X}(x)$ is degenerate.
\end{theorem}

\noindent The bound in \eqref{eq-tr:LSD-bound}\ is a special case of the bound
in \eqref{eq-tr:QE-bound} with $E=0$. The other bounds in
Theorem~\ref{thm-tr:main-theorem} are looser than the bound in
\eqref{eq-tr:QE-bound} when $C,E=0$.

\subsubsection{Entanglement-Assisted Quantum Capacity}

The following theorem gives the
\index{entanglement-assisted!quantum communication}
two-dimensional entanglement-assisted quantum capacity region $\mathcal{C}%
_{\operatorname{QE}}(\mathcal{N})$ of a quantum channel $\mathcal{N}$.

\begin{theorem}
[Devetak--Harrow--Winter]The entanglement-assisted quantum capacity region
$\mathcal{C}_{\emph{QE}}(\mathcal{N})$ is given by
\begin{equation}
\mathcal{C}_{\emph{QE}}(\mathcal{N})=\overline{\bigcup_{k=1}^{\infty}\frac
{1}{k}\mathcal{C}_{\emph{QE}}^{(1)}(\mathcal{N}^{\otimes k})}.
\label{eq-tr:eaQ}%
\end{equation}
The region $\mathcal{C}_{\emph{QE}}^{(1)}(\mathcal{N})$ is the union of the
state-dependent regions $\mathcal{C}_{\emph{QE},\sigma}^{(1)}(\mathcal{N})$,
where $\mathcal{C}_{\emph{QE},\sigma}^{(1)}(\mathcal{N})$ is the set of all
$Q,E\geq0$, such that%
\begin{align}
2Q  &  \leq I(AX;B)_{\sigma},\label{eq-tr:eaQ1}\\
Q  &  \leq I(A\rangle BX)_{\sigma}+\left\vert E\right\vert .
\label{eq-tr:eaQ2}%
\end{align}
The entropic quantities are with respect to the state $\sigma_{XAB}$ in
\eqref{eq-tr:main-theorem-state} with the restriction that the density
$p_{X}(x)$ is degenerate.
\end{theorem}

\noindent The bounds in \eqref{eq-tr:eaQ1} and \eqref{eq-tr:eaQ2} are a
special case of the respective bounds in \eqref{eq-tr:CQ-bound} and
\eqref{eq-tr:QE-bound} with $C=0$. The other bounds in
Theorem~\ref{thm-tr:main-theorem} are looser than the bounds in
\eqref{eq-tr:CQ-bound} and \eqref{eq-tr:QE-bound} when $C=0$. Observe that the
region is a union of general pentagons (see the $QE$-plane in
Figure~\ref{fig-tr:one-shot-region-state} for an example of one of these
general pentagons in the union).

\subsubsection{Classically-Enhanced Quantum Capacity}

The following theorem gives the two-dimensional capacity region $\mathcal{C}%
_{\operatorname{CQ}}(\mathcal{N})$ for classically enhanced quantum
communication over a quantum channel $\mathcal{N}$.

\begin{theorem}
[Devetak--Shor]The classically enhanced quantum capacity region $\mathcal{C}%
_{\emph{CQ}}(\mathcal{N})$ is given by
\begin{equation}
\mathcal{C}_{\emph{CQ}}(\mathcal{N})=\overline{\bigcup_{k=1}^{\infty}\frac
{1}{k}\mathcal{C}_{\emph{CQ}}^{(1)}(\mathcal{N}^{\otimes k})}.
\end{equation}
The region $\mathcal{C}_{\emph{CQ}}^{(1)}(\mathcal{N})$ is the union of the
state-dependent regions $\mathcal{C}_{\emph{CQ},\sigma}^{(1)}(\mathcal{N})$,
where $\mathcal{C}_{\emph{CQ},\sigma}^{(1)}(\mathcal{N})$ is the set of all
$C,Q\geq0$, such that%
\begin{align}
C+Q  &  \leq I(X;B)_{\sigma}+I(A\rangle BX)_{\sigma},\label{eq-tr:ceq-1}\\
Q  &  \leq I(A\rangle BX)_{\sigma}. \label{eq-tr:ceq-2}%
\end{align}
The entropic quantities are with respect to the state $\sigma_{XAB}$ in \eqref{eq-tr:main-theorem-state}.
\end{theorem}

\noindent The bounds in \eqref{eq-tr:ceq-1} and \eqref{eq-tr:ceq-2} are a
special case of the respective bounds in \eqref{eq-tr:QE-bound} and
\eqref{eq-tr:CQE-bound} with $E=0$. The first inequality in
\eqref{eq-tr:CQ-bound} is redundant because $Q\leq I(A\rangle BX)_{\sigma
}=H(A|EX)_{\sigma}\leq H(A|X)_{\sigma}$ and combining this with the inequality
in \eqref{eq-tr:ceq-1} gives \eqref{eq-tr:CQ-bound}. Observe that the region
is a union of trapezoids (see the $CQ$-plane in
Figure~\ref{fig-tr:one-shot-region-state} for an example of one of these
rectangles in the union).

\subsubsection{Entanglement-Assisted Classical Capacity with Limited
Entanglement}

\begin{theorem}
[Shor]\label{thm-tr:shors-theorem}The entanglement-assisted classical capacity
region $\mathcal{C}_{\emph{CE}}(\mathcal{N})$ of a quantum channel
$\mathcal{N}$\ is%
\begin{equation}
\mathcal{C}_{\emph{CE}}(\mathcal{N})=\overline{\bigcup_{k=1}^{\infty}\frac
{1}{k}\mathcal{C}_{\emph{CE}}^{(1)}(\mathcal{N}^{\otimes k})}.
\label{eq-tr:eaC}%
\end{equation}
The region $\mathcal{C}_{\emph{CE}}^{(1)}(\mathcal{N})$ is the union of the
state-dependent regions $\mathcal{C}_{\emph{CE},\sigma}^{(1)}(\mathcal{N})$,
where $\mathcal{C}_{\emph{CE},\sigma}^{(1)}(\mathcal{N})$ is the set of all
$C,E\geq0$, such that%
\begin{align}
C  &  \leq I(AX;B)_{\sigma},\label{eq-tr:eac1}\\
C  &  \leq I(X;B)_{\sigma}+I(A\rangle BX)_{\sigma}+\left\vert E\right\vert ,
\label{eq-tr:eac2}%
\end{align}
where the entropic quantities are with respect to the state $\sigma_{XAB}$ in \eqref{eq-tr:main-theorem-state}.
\end{theorem}

\noindent The bounds in \eqref{eq-tr:eac1} and \eqref{eq-tr:eac2} are a
special case of the respective bounds in \eqref{eq-tr:CQ-bound} and
\eqref{eq-tr:CQE-bound} with $Q=0$. Observe that the region is a union of
general polyhedra (see the CE-plane in
Figure~\ref{fig-tr:one-shot-region-state} for an example of one of these
general polyhedra in the union).

\section{The Direct Coding Theorem}

\label{sec-tr:direct-coding}The unit resource achievable region is what Alice%
\index{quantum dynamic capacity theorem!direct part}
and Bob can achieve using the protocols entanglement distribution,
teleportation, and super-dense coding (see
Chapter~\ref{chap:unit-resource-cap}). It is the cone of the rate triples
corresponding to these protocols:%
\begin{equation}
\left\{  \alpha\left(  0,-1,1\right)  +\beta\left(  2,-1,-1\right)
+\gamma\left(  -2,1,-1\right)  :\alpha,\beta,\gamma\geq0\right\}  .
\end{equation}
We can also write any rate triple $\left(  C,Q,E\right)  $ in the unit
resource capacity region with a matrix equation:%
\begin{equation}%
\begin{bmatrix}
C\\
Q\\
E
\end{bmatrix}
=%
\begin{bmatrix}
0 & 2 & -2\\
-1 & -1 & 1\\
1 & -1 & -1
\end{bmatrix}%
\begin{bmatrix}
\alpha\\
\beta\\
\gamma
\end{bmatrix}
. \label{eq-tr:unit-resource-achievable-region}%
\end{equation}
The inverse of the above matrix is as follows:%
\begin{equation}%
\begin{bmatrix}
-\frac{1}{2} & -1 & 0\\
0 & -\frac{1}{2} & -\frac{1}{2}\\
-\frac{1}{2} & -\frac{1}{2} & -\frac{1}{2}%
\end{bmatrix}
,
\end{equation}
and gives the following set of inequalities for the unit resource achievable
region:%
\begin{align}
C+2Q  &  \leq0,\\
Q+E  &  \leq0,\\
C+Q+E  &  \leq0,
\end{align}
by inverting the matrix equation in
\eqref{eq-tr:unit-resource-achievable-region} and applying the constraints
$\alpha,\beta,\gamma\geq0$.

Now, let us include the protocol from Corollary~\ref{cor-ccn:CQE-trading}\ for
entanglement-assisted communication of classical and quantum information.
Corollary~\ref{cor-ccn:CQE-trading}\ states that we can achieve the following
rate triple by channel coding for $\mathcal{N}_{A^{\prime}\rightarrow B}$:%
\begin{equation}
\left(  I(X;B)_{\sigma},\frac{1}{2}I(A;B|X)_{\sigma},-\frac{1}{2}%
I(A;E|X)_{\sigma}\right)  ,
\end{equation}
for any state $\sigma_{XABE}$\ of the form%
\begin{equation}
\sigma_{XABE}\equiv\sum_{x}p_{X}(x)|x\rangle\langle x|_{X}\otimes
\mathcal{U}_{A^{\prime}\rightarrow BE}^{\mathcal{N}}(\phi_{AA^{\prime}}^{x}),
\label{eq-tr:maximization-state}%
\end{equation}
where $U_{A^{\prime}\rightarrow BE}^{\mathcal{N}}$ is an isometric extension
of the quantum channel $\mathcal{N}_{A^{\prime}\rightarrow B}$. Specifically,
we showed in Corollary~\ref{cor-ccn:CQE-trading} that one can achieve the
above rates with vanishing error in the limit of large blocklength. Thus the
achievable rate region is the following translation of the unit resource
achievable region in \eqref{eq-tr:unit-resource-achievable-region}:%
\begin{equation}%
\begin{bmatrix}
C\\
Q\\
E
\end{bmatrix}
=%
\begin{bmatrix}
0 & 2 & -2\\
-1 & -1 & 1\\
1 & -1 & -1
\end{bmatrix}%
\begin{bmatrix}
\alpha\\
\beta\\
\gamma
\end{bmatrix}
+%
\begin{bmatrix}
I(X;B)_{\sigma}\\
\frac{1}{2}I(A;B|X)_{\sigma}\\
-\frac{1}{2}I(A;E|X)_{\sigma}%
\end{bmatrix}
.
\end{equation}
We can now determine bounds on an achievable rate region that employs the
above coding strategy. We apply the inverse of the matrix in
\eqref{eq-tr:unit-resource-achievable-region} to the left-hand side and
right-hand side, giving%
\begin{equation}%
\begin{bmatrix}
-\frac{1}{2} & -1 & 0\\
0 & -\frac{1}{2} & -\frac{1}{2}\\
-\frac{1}{2} & -\frac{1}{2} & -\frac{1}{2}%
\end{bmatrix}%
\begin{bmatrix}
C\\
Q\\
E
\end{bmatrix}
-%
\begin{bmatrix}
-\frac{1}{2} & -1 & 0\\
0 & -\frac{1}{2} & -\frac{1}{2}\\
-\frac{1}{2} & -\frac{1}{2} & -\frac{1}{2}%
\end{bmatrix}%
\begin{bmatrix}
I(X;B)_{\sigma}\\
\frac{1}{2}I(A;B|X)_{\sigma}\\
-\frac{1}{2}I(A;E|X)_{\sigma}%
\end{bmatrix}
=%
\begin{bmatrix}
\alpha\\
\beta\\
\gamma
\end{bmatrix}
.
\end{equation}
Then using the following identities:%
\begin{align}
I(X;B)_{\sigma}+I(A;B|X)_{\sigma}  &  =I(AX;B)_{\sigma},\\
\frac{1}{2}I(A;B|X)_{\sigma}-\frac{1}{2}I(A;E|X)_{\sigma}  &  =I(A\rangle
BX)_{\sigma},
\end{align}
and the constraints $\alpha,\beta,\gamma\geq0$, we obtain the inequalities in
\eqref{eq-tr:CQ-bound}--\eqref{eq-tr:CQE-bound}, corresponding exactly to the
state-dependent region in Theorem~\ref{thm-tr:main-theorem}. Taking the union
over all possible states $\sigma$ in \eqref{eq-tr:main-theorem-state} and
taking the regularization gives the full dynamic achievable rate region.

\begin{figure}
[ptb]
\begin{center}
\includegraphics[
width=4.8456in
]%
{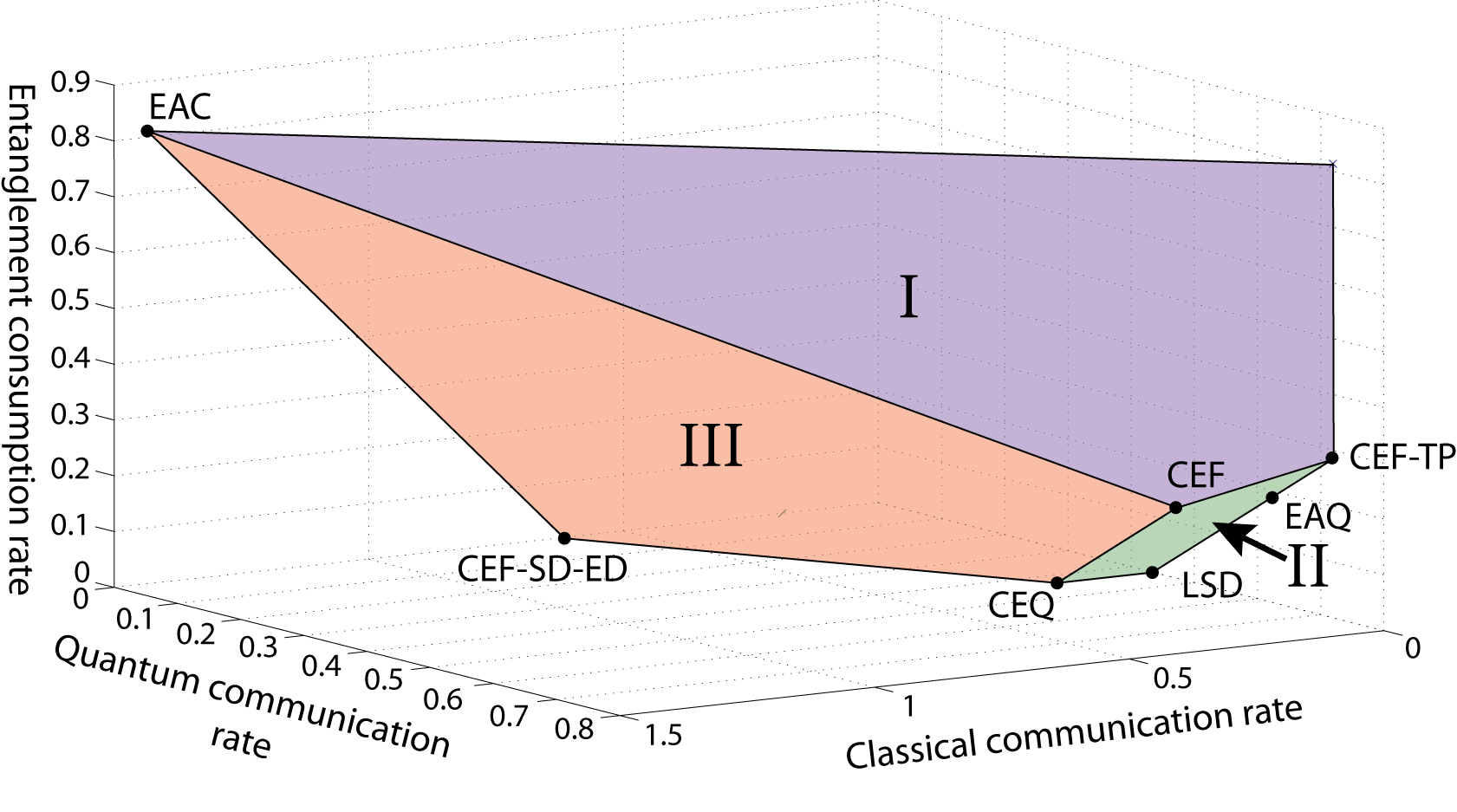}%
\caption{An example of the state-dependent achievable region $\mathcal{C}%
_{\operatorname{CQE}\sigma}^{(1)}(\mathcal{N})$ corresponding to a state
$\sigma_{XABE}$ that arises from a qubit dephasing channel with dephasing
parameter $p=0.2$. The figure depicts the octant corresponding to the
consumption of entanglement and the generation of classical and quantum
communication. The state input to the channel $\mathcal{N}$ is $\sigma
_{XAA^{\prime}}$, defined in \eqref{eq-tr:example-input-state}. The plot
features seven achievable corner points of the state-dependent region. We can
achieve the convex hull of these seven points by time sharing any two
different coding strategies. We can also achieve any point above an achievable
point by consuming more entanglement than necessary. The seven achievable
points correspond to entanglement-assisted quantum communication (EAQ), the
protocol from Corollary~\ref{thm-ccn:CQ-trading} for classically enhanced
quantum communication (CEQ), the protocol from
Theorem~\ref{thm-ccn:ce-trading}~for entanglement-assisted classical
communication with limited entanglement (EAC), quantum communication (LSD),
combining CEF\ with entanglement distribution and super-dense coding
(CEF-SD-ED), the protocol from Corollary~\ref{cor-ccn:CQE-trading} for
entanglement-assisted communication of classical and quantum information
(CEF), and combining CEF with teleportation (CEF-TP). Observe that we can
obtain EAC\ by combining CEF\ with super-dense coding, so that the points CEQ,
CEF, EAC, and CEF-SD-ED all lie in plane III. Observe that we can obtain CEQ
from CEF by entanglement distribution and we can obtain LSD from EAQ and EAQ
from CEF-TP,\ both by entanglement distribution. Thus, the points CEF, CEQ,
LSD, EAQ, and CEF-TP\ all lie in plane II. Finally, observe that we can obtain
all corner points by combining CEF with the unit protocols of teleportation,
super-dense coding, and entanglement distribution. The bounds in
\eqref{eq-tr:CQ-bound}--\eqref{eq-tr:CQE-bound} uniquely specify the
respective planes I-III. We obtain the full achievable region by taking the
union over all states $\sigma$ of the state-dependent regions $\mathcal{C}%
_{\sigma}^{(1)}(\mathcal{N})$ and taking the regularization, as outlined in
Theorem~\ref{thm-tr:main-theorem}. The above region is a translation of the
unit resource capacity region from Chapter~\ref{chap:unit-resource-cap}\ to
the protocol for entanglement-assisted communication of classical and quantum
information.}%
\label{fig-tr:one-shot-region-state}%
\end{center}
\end{figure}

Figure~\ref{fig-tr:one-shot-region-state}\ illustrates an example of the
general polyhedron specified by
\eqref{eq-tr:CQ-bound}--\eqref{eq-tr:CQE-bound}, where the channel is the
qubit dephasing channel\ $\rho\rightarrow(1-p)\rho+pZ\rho Z$ with dephasing
parameter $p=0.2$, and the input state is%
\begin{equation}
\sigma_{XAA^{\prime}}\equiv\frac{1}{2}(|0\rangle\langle0|_{X}\otimes
\phi_{AA^{\prime}}^{0}+|1\rangle\langle1|_{X}\otimes\phi_{AA^{\prime}}^{1}),
\label{eq-tr:example-input-state}%
\end{equation}
where%
\begin{align}
\left\vert \phi^{0}\right\rangle _{AA^{\prime}}  &  \equiv\sqrt{1/4}%
|00\rangle_{AA^{\prime}}+\sqrt{3/4}|11\rangle_{AA^{\prime}},\\
\left\vert \phi^{1}\right\rangle _{AA^{\prime}}  &  \equiv\sqrt{3/4}%
|00\rangle_{AA^{\prime}}+\sqrt{1/4}|11\rangle_{AA^{\prime}}.
\end{align}
The state $\sigma_{XABE}$ resulting from the channel is $\mathcal{U}%
_{A^{\prime}\rightarrow BE}^{\mathcal{N}}(\sigma_{XAA^{\prime}})$ where
$U_{A^{\prime}\rightarrow BE}^{\mathcal{N}}$ is an isometric extension of the
qubit dephasing channel. The figure caption provides a detailed explanation of
the state-dependent region $\mathcal{C}_{\operatorname{CQE},\sigma}^{(1)}$
(note that Figure~\ref{fig-tr:one-shot-region-state} displays the
state-dependent region and does not display the full capacity region).%

\section{The Converse Theorem}

\label{sec-tr:converse}We provide a catalytic,
\index{quantum dynamic capacity theorem!converse}%
information-theoretic converse proof of the dynamic capacity region, showing
that \eqref{eq-tr:multi-letter} gives a multi-letter characterization of it.
The catalytic approach means that we are considering the most general protocol
that \textit{consumes and generates} classical communication, quantum
communication, and entanglement in addition to the uses of the noisy quantum
channel. This approach has the advantage that we can prove the converse
theorem in \textquotedblleft one fell swoop.\textquotedblright\ We employ the
AFW inequality, the chain rule for quantum mutual information, elementary
properties of quantum entropy, and the quantum data-processing inequality to
prove the converse.

We show that the bounds in \eqref{eq-tr:CQ-bound}--\eqref{eq-tr:CQE-bound}
hold for randomness distribution instead of classical communication because a
capacity for generating shared randomness can only be better than that for
generating classical communication (classical communication can generate
shared randomness). We also consider a protocol that preserves maximal
entanglement with a reference system instead of one that generates quantum communication.

We prove that the converse theorem holds for a state of the following form:%
\begin{equation}
\sigma_{XAB^{n}}\equiv\sum_{x}p(x)|x\rangle\langle x|_{X}\otimes
\mathcal{N}_{A^{\prime}\rightarrow B}^{\otimes n}(\rho_{AA^{\prime n}}^{x}),
\label{eq-tr:converse-state}%
\end{equation}
where the states $\rho_{AA^{\prime n}}^{x}$ are mixed. We identify this state
with $\omega_{XAB^{n}}$ defined in \eqref{eq-tr:state-after-channel}, setting
$A\equiv RS_{A}A_{2}T_{B}$ and $X\equiv LM$. We do this rather than proving it
for a state of the form in \eqref{eq-tr:main-theorem-state}. Then we show in
Section~\ref{sec-tr:pure-suff}\ that it is not necessary to consider an
ensemble of mixed states---i.e., we can do just as well with an ensemble of
pure states, giving the statement of Theorem~\ref{thm-tr:main-theorem}.

We first prove the bound in \eqref{eq-tr:CQ-bound}. Consider the following
chain of inequalities:%
\begin{align}
&  n(C_{\operatorname{out}}+2Q_{\operatorname{out}})\nonumber\\
&  =I(M;\hat{M})_{\overline{\Phi}}+I(R;B_{1})_{\Phi}\\
&  =I(RM;B_{1}\hat{M})_{\Phi\otimes\overline{\Phi}}\\
&  \leq I(RM;B_{1}\hat{M})_{\omega^{\prime}}+n\delta^{\prime}\\
&  \leq I(RM;B^{n}A_{2}LT_{B})_{\omega}+n\delta^{\prime}\\
&  =I(RA_{2}T_{B}LM;B^{n})_{\omega}+I(RM;A_{2}T_{B}L)_{\omega}-I(B^{n}%
;A_{2}T_{B}L)_{\omega}+n\delta^{\prime}\\
&  \leq I(RS_{A}A_{2}T_{B}LM;B^{n})_{\omega}+I(RM;A_{2}T_{B}L)_{\omega
}+n\delta^{\prime}\\
&  =I(AX;B^{n})_{\omega}+I(RM;T_{B})_{\omega}+I(RM;L|T_{B})_{\omega
}+I(RM;A_{2}|T_{B}L)_{\omega}+n\delta^{\prime}\\
&  \leq I(AX;B^{n})_{\omega}+n(C_{\operatorname{in}}+2Q_{\operatorname{in}%
})+n\delta^{\prime}.
\end{align}
The first equality holds by evaluating the quantum mutual informations on the
respective states $\overline{\Phi}_{M\hat{M}}$ and $\Phi_{RB_{1}}$. The second
equality follows because the mutual information is additive with respect to
tensor-product states (see Exercise~\ref{ex-qie:add-mutual-info}). The first
inequality follows from the condition in \eqref{eq-tr:+++_good-code} and an
application of the AFW inequality where $\delta^{\prime}$ is a parameter that
vanishes as $n\to \infty$ and $\varepsilon\rightarrow0$. The second inequality follows from
quantum data processing. The third equality is a consequence of an identity
from Exercise~\ref{ex-qie:chain-rule-mut-info}. The third inequality follows
from quantum data processing and the fact that $I(B^{n};A_{2}T_{B}L)_{\omega
}\geq0$. The fourth equality follows by identifying $A\equiv RS_{A}A_{2}T_{B}$
and $X\equiv LM$, and also from the chain rule, which implies that
$I(RM;A_{2}T_{B}L)_{\omega}=I(RM;T_{B})_{\omega}+I(RM;L|T_{B})_{\omega
}+I(RM;A_{2}|T_{B}L)_{\omega}$. The final inequality follows because
$I(RM;T_{B})_{\omega}=0$ and from the dimension bounds $I(RM;L|T_{B})_{\omega
}\leq\log\dim(\mathcal{H}_{L})=nC_{\operatorname{in}}$ and $I(RM;A_{2}%
|T_{B}L)_{\omega}\leq2\log\dim(\mathcal{H}_{A_{2}})=n2Q_{\operatorname{in}}$
(see Exercise~\ref{ex-qie:CMI-dim-bound}). Thus, \eqref{eq-tr:CQ-bound} holds
for the net rates.

We now prove the second bound in \eqref{eq-tr:QE-bound}. Consider the
following chain of inequalities:%
\begin{align}
n(Q_{\operatorname{out}}+E_{\operatorname{out}})  &  =I(R\rangle B_{1})_{\Phi
}+I(S_{A}\rangle S_{B})_{\Phi}\\
&  =I(RS_{A}\rangle B_{1}S_{B})_{\Phi\otimes\Phi}\\
&  \leq I(RS_{A}\rangle B_{1}S_{B})_{\omega^{\prime}}+n\delta^{\prime}\\
&  \leq I(RS_{A}\rangle B^{n}A_{2}T_{B}LM)_{\omega}+n\delta^{\prime}\\
&  \leq I(RS_{A}A_{2}T_{B}\rangle B^{n}LM)_{\omega}+\log\dim(\mathcal{H}%
_{A_{2}}\otimes\mathcal{H}_{T_{B}})+n\delta^{\prime}\\
&  =I(A\rangle B^{n}X)_{\omega}+n(Q_{\operatorname{in}}+E_{\operatorname{in}%
})+n\delta^{\prime}.
\end{align}
The first equality follows by evaluating the coherent informations of the
respective states $\Phi_{RB_{1}}$ and $\Phi_{S_{A}S_{B}}$. The second equality
follows because $\Phi_{RB_{1}}\otimes\Phi_{S_{A}S_{B}}$ is a product state and
coherent information is additive with respect to tensor-product states. The
first inequality follows from the condition in \eqref{eq-tr:+++_good-code} and
an application of the AFW inequality with parameter $\delta^{\prime}$
vanishing when $n\to \infty$ and $\varepsilon\rightarrow0$. The second inequality follows from
quantum data processing. The third inequality follows from the dimension bound
in Exercise~\ref{ex-qie:coh-info-dim-bound-part-two}. The final equality
follows by identifying $A\equiv RS_{A}A_{2}T_{B}$ and $X\equiv LM$, and noting
that $\log\dim(\mathcal{H}_{A_{2}})=nQ_{\operatorname{in}}$ and $\log
\dim(\mathcal{H}_{T_{B}})=nE_{\operatorname{in}}$. Thus,
\eqref{eq-tr:QE-bound} holds for the net rates.

We prove the last bound in \eqref{eq-tr:CQE-bound}. Consider the following
chain of inequalities:%
\begin{align}
&  n(C_{\operatorname{out}}+Q_{\operatorname{out}}+E_{\operatorname{out}%
})\nonumber\\
&  =I(M;\hat{M})_{\overline{\Phi}}+I(RS_{A}\rangle B_{1}S_{B})_{\Phi
\otimes\Phi}\\
&  \leq I(M;\hat{M})_{\omega^{\prime}}+I(RS_{A}\rangle B_{1}S_{B}%
)_{\omega^{\prime}}+n\delta^{\prime}\\
&  \leq I(M;B^{n}A_{2}T_{B}L)_{\omega}+I(RS_{A}\rangle B^{n}A_{2}%
T_{B}LM)_{\omega}+n\delta^{\prime}\\
&  =I(ML;B^{n})_{\omega}+I(RS_{A}A_{2}T_{B}\rangle B^{n}LM)_{\omega
}+I(M;L)_{\omega}\nonumber\\
&  \ \ \ \ \ \ +H(A_{2}T_{B}|B^{n})_{\omega}-I(A_{2}B^{n}T_{B};L)_{\omega
}+n\delta^{\prime}\\
&  \leq I(X;B^{n})_{\omega}+I(A\rangle B^{n}X)_{\omega}+I(M;L)_{\omega
}+H(A_{2}T_{B}|B^{n})_{\omega}+n\delta^{\prime}\\
&  \leq I(X;B^{n})_{\omega}+I(A\rangle B^{n}X)_{\omega}+n(C_{\operatorname{in}%
}+Q_{\operatorname{in}}+E_{\operatorname{in}})+n\delta^{\prime}.
\end{align}
The first equality follows from evaluating the mutual information of the state
$\overline{\Phi}_{M\hat{M}}$ and the coherent information of the product state
$\Phi_{RB_{1}}\otimes\Phi_{S_{A}S_{B}}$. The first inequality follows from the
condition in \eqref{eq-tr:+++_good-code} and an application of the AFW
inequality with $\delta^{\prime}$ vanishing when
$n\to \infty$ and
$\varepsilon\rightarrow0$.
The second inequality follows from quantum data processing. The second
equality is an identity, verified by expanding all quantities as unconditional
entropies and seeing that both terms are equal to $H(M)_{\omega}+H(B^{n}%
A_{2}T_{B}L)_{\omega}-H(RS_{A}B^{n}A_{2}T_{B}LM)_{\omega}$. The third
inequality follows by identifying $A\equiv RT_{B}A_{2}S_{A}$ and $X\equiv ML$,
and also because $I(A_{2}B^{n}T_{B};L)_{\omega}\geq0$. The last inequality
follows from the dimension bounds $I(M;L)_{\omega}\leq\log\dim(\mathcal{H}%
_{L})=nC_{\operatorname{in}}$ and $H(A_{2}T_{B}|B^{n})_{\omega}\leq\log
\dim(\mathcal{H}_{A_{2}}\otimes\mathcal{H}_{T_{B}})=n(Q_{\operatorname{in}%
}+E_{\operatorname{in}})$. Thus, \eqref{eq-tr:CQE-bound} applies to the net
rates. This concludes the proof of the converse theorem.

\subsection{Pure-state Ensembles are Sufficient}

\label{sec-tr:pure-suff}We prove that it is sufficient to consider an ensemble
of pure states as in the statement of Theorem~\ref{thm-tr:main-theorem}%
\ rather than an ensemble of mixed states as in \eqref{eq-tr:converse-state}
in the proof of our converse theorem. We first determine a spectral
decomposition of the mixed-state ensemble, model the index of the pure states
in the decomposition as a classical variable $Y$, and then place this
classical variable $Y$ in a classical register. It follows that the
communication rates can only improve, and it is sufficient to consider an
ensemble of pure states.

Consider that each mixed state in the ensemble in \eqref{eq-tr:converse-state}
admits a spectral decomposition of the following form:%
\begin{equation}
\rho_{AA^{\prime}}^{x}=\sum_{y}p( y|x) \psi_{AA^{\prime}}^{x,y}.
\end{equation}
We can thus represent the ensemble as follows:%
\begin{equation}
\rho_{XAB}\equiv\sum_{x,y}p(x)p( y|x) \vert x\rangle\langle x\vert_{X}%
\otimes\mathcal{N}_{A^{\prime}\rightarrow B}(\psi_{AA^{\prime}}^{x,y}).
\label{eq-tr:non-isometric-state}%
\end{equation}
The inequalities in \eqref{eq-tr:CQ-bound}--\eqref{eq-tr:CQE-bound} for the
dynamic capacity region involve the mutual information $I(AX;B)_{\rho}$, the
Holevo information $I(X;B)_{\rho}$, and the coherent information $I(A\rangle
BX)_{\rho}$. As we show below, each of these entropic quantities can only
improve in each case if we make the variable $y$ be part of the classical
variable. This improvement then implies that it is only necessary to consider
pure states in the dynamic capacity theorem.

Let $\theta_{XYAB}$ denote an augmented state of the following form:%
\begin{equation}
\theta_{XYAB}\equiv\sum_{x}p(x)p(y|x)|x\rangle\langle x|_{X}\otimes
|y\rangle\langle y|_{Y}\otimes\mathcal{N}_{A^{\prime}\rightarrow B}%
(\psi_{AA^{\prime}}^{x,y}). \label{eq-tr:measured-state}%
\end{equation}
This state is actually a state of the form in \eqref{eq-tr:main-theorem-state}
if we subsume the classical variables $X$ and $Y$ into one classical variable.
The following three inequalities each follow from an application of the
quantum data-processing inequality:%
\begin{align}
I(X;B)_{\rho}  &  =I(X;B)_{\theta}\leq I(XY;B)_{\theta},\\
I(AX;B)_{\rho}  &  =I(AX;B)_{\theta}\leq I(AXY;B)_{\theta}\\
I(A\rangle BX)_{\rho}  &  =I(A\rangle BX)_{\theta}\leq I(A\rangle
BXY)_{\theta}.
\end{align}
Each of these inequalities proves the desired result for the respective Holevo
information, mutual information, and coherent information, and it suffices to
consider an ensemble of pure states in Theorem~\ref{thm-tr:main-theorem}. The
same argument holds with $\mathcal{N}_{A^{\prime}\rightarrow B}$ replaced by
$\mathcal{N}_{A^{\prime}\rightarrow B}^{\otimes n}$.

\subsection{The Quantum Dynamic Capacity Formula}

\label{sec-tr:dynamic-cap-formula}Here we introduce the quantum dynamic
capacity formula and show how its additivity implies that the computation of
the Pareto optimal trade-off surface of the capacity region requires an
optimization over a single copy of the channel, rather than a potentially
infinite number of them. The Pareto optimal trade-off surface consists of all
points in the capacity region that are Pareto optimal, in the sense that it is
not possible to make improvements in one resource without offsetting another
resource (these are essentially the boundary points of the region in our
case). We then show how several important capacity formulas discussed
previously in this book are special cases of the quantum dynamic capacity formula.

\begin{definition}
[Quantum Dynamic Capacity Formula]The quantum dynamic capacity formula of a
\index{quantum dynamic capacity formula}
quantum channel $\mathcal{N}$ is defined as follows:%
\begin{equation}
D_{\vec{\lambda}}(\mathcal{N})\equiv\max_{\sigma}\lambda_{1}I(AX;B)_{\sigma
}+\lambda_{2}I(A\rangle BX)_{\sigma}+\lambda_{3}\left[  I(X;B)_{\sigma
}+I(A\rangle BX)_{\sigma}\right]  , \label{eq-tr:objective}%
\end{equation}
where $\sigma$ is a state of the form in \eqref{eq-tr:main-theorem-state} and
$\vec{\lambda}\equiv(\lambda_{1},\lambda_{2},\lambda_{3})$ is a vector of
Lagrange multipliers such that $\lambda_{1},\lambda_{2},\lambda_{3}\geq0$.
\end{definition}

\begin{definition}
The regularized quantum dynamic capacity formula is defined as follows:%
\begin{equation}
D_{\vec{\lambda}}^{\operatorname{reg}}(\mathcal{N})\equiv\lim_{k\rightarrow
\infty}\frac{1}{k}D_{\vec{\lambda}}(\mathcal{N}^{\otimes k}).
\end{equation}

\end{definition}

\begin{lemma}
\label{thm-tr:CEQ-single-letter}Suppose the quantum dynamic capacity formula
is additive for a channel~$\mathcal{N}$ and any other arbitrary
channel~$\mathcal{M}$:%
\begin{equation}
D_{\vec{\lambda}}(\mathcal{N}\otimes\mathcal{M})=D_{\vec{\lambda}}%
(\mathcal{N})+D_{\vec{\lambda}}(\mathcal{M}). \label{eq-tr:q-dynamic-additive}%
\end{equation}
Then the regularized quantum dynamic capacity formula for $\mathcal{N}$ is
equal to the quantum dynamic capacity formula:%
\begin{equation}
D_{\vec{\lambda}}^{\operatorname{reg}}(\mathcal{N})=D_{\vec{\lambda}%
}(\mathcal{N}).
\end{equation}
In this sense, the regularized formula \textquotedblleft
single-letterizes.\textquotedblright
\end{lemma}

\begin{proof}
We prove the result using induction on $n$. The base case for $n=1$ is
trivial. Suppose the result holds for $n$: $D_{\vec{\lambda}}(\mathcal{N}%
^{\otimes n})=nD_{\vec{\lambda}}(\mathcal{N})$. Then the following chain of
equalities establishes the inductive step:%
\begin{equation}
D_{\vec{\lambda}}(\mathcal{N}^{\otimes n+1})=D_{\vec{\lambda}}(\mathcal{N}%
\otimes\mathcal{N}^{\otimes n})=D_{\vec{\lambda}}(\mathcal{N})+D_{\vec
{\lambda}}(\mathcal{N}^{\otimes n})=D_{\vec{\lambda}}(\mathcal{N}%
)+nD_{\vec{\lambda}}(\mathcal{N}).
\end{equation}
The first equality follows by expanding the tensor product. The second
critical equality follows from the assumption in
\eqref{eq-tr:q-dynamic-additive}, setting $\mathcal{M}=\mathcal{N}^{\otimes
n}$. The final equality follows from the induction hypothesis.
\end{proof}

\begin{theorem}
Single-letterization of the quantum dynamic capacity formula implies that the
computation of the Pareto optimal trade-off surface of the dynamic capacity
region requires an optimization over a single copy of the channel.
\end{theorem}

\begin{proof}
We employ ideas from optimization theory for the proof (see \cite{BV04}). We
would like to characterize all the points in the capacity region that are
Pareto optimal. Such a task is a standard vector optimization in the theory of
Pareto trade-off analysis (see Section~4.7 of \cite{BV04}).

Let $\vec{w}\equiv(w_{C},w_{Q},w_{E})\in\mathbb{R}^{3}$ be a weight vector,
$\vec{R}\equiv(C,Q,E)$ a rate vector, and $\mathcal{E}\equiv\{p_{X}%
(x),\phi_{AA^{\prime}}^{x}\}$ an ensemble. Our main goal here is to show that
the computational problem reduces to computing the quantum dynamic capacity
formula in \eqref{eq-tr:objective}\ for all values of $\vec{\lambda}$ such
that $\lambda_{1},\lambda_{2},\lambda_{3}\geq0$. As such, we can focus for now
on a single copy of the channel and deduce the statement of the theorem once
we are done. We can phrase the task of computing the boundary of the
single-copy capacity region as the following optimization problem:%
\begin{align}
P^{\ast}(\vec{w})  &  \equiv\sup_{\vec{R},\mathcal{E}}\vec{w}\cdot\vec
{R}\label{eq-tr:scalarization}\\
\text{subject to\ \ \ \ }C+2Q  &  \leq I(AX;B)_{\sigma},\label{eq-tr:cond-1}\\
Q+E  &  \leq I(A\rangle BX)_{\sigma},\label{eq-tr:cond-2}\\
C+Q+E  &  \leq I(X;B)_{\sigma}+I(A\rangle BX)_{\sigma}, \label{eq-tr:cond-3}%
\end{align}
where the optimization is with respect to all rate vectors $\vec{R}=(C,Q,E)$
and ensembles $\mathcal{E}=\{p_{X}(x),\phi_{AA^{\prime}}^{x}\}$, with
$\sigma_{XAB}$ a state of the form in \eqref{eq-tr:main-theorem-state}.

The geometric interpretation of the optimization task is that we are trying to
find a supporting plane of the dynamic capacity region such that the weight
vector $\vec{w}$ is the normal vector of the plane and the value of its inner
product with $\vec{R}$ characterizes the offset of the plane, so that
$P^{\ast}(\vec{w})$ for all $\vec{w}\in\mathbb{R}^{3}$\ characterizes the
boundary of the region. Note that the optimal value $P^{\ast}(\vec{w})$ can
sometimes be infinite. For example, if $\vec{w}=(-1,-1,-1)$, then the optimal
rates are $\vec{R}=(-\infty,-\infty,-\infty)$, which corresponds to simply
consuming all resources (the dynamic capacity theorem does not give any
constraint on all resources being consumed at the same time, but rather on the
generation of some resources while others are consumed).

For now, let us fix an ensemble $\mathcal{E}$ and let $P^{\ast}(\vec
{w},\mathcal{E})\equiv\sup_{\vec{R}}\vec{w}\cdot\vec{R}$, subject to the
constraints in \eqref{eq-tr:cond-1}--\eqref{eq-tr:cond-3}. The optimization
problem then becomes what is known as a linear program, given that the
objective function is linear in $\vec{R}$ and the constraints are linear
inequalities involving $\vec{R}$. We can then define the following Lagrangian,
which introduces Lagrange multipliers $\lambda_{1},\lambda_{2},\lambda_{3}%
$\ for the respective constraints in
\eqref{eq-tr:cond-1}--\eqref{eq-tr:cond-3}:%
\begin{align}
&  \mathcal{L}(\vec{w},\vec{R},\mathcal{E},\vec{\lambda})\nonumber\\
&  \equiv w_{C}C+w_{Q}Q+w_{E}E+\lambda_{1}\left[  I(AX;B)_{\sigma}-\left(
C+2Q\right)  \right] \nonumber\\
&  \ \ \ \ \ +\lambda_{2}\left[  I(A\rangle BX)_{\sigma}-\left(  Q+E\right)
\right] \nonumber\\
&  \ \ \ \ \ +\lambda_{3}\left[  I(X;B)_{\sigma}+I(A\rangle BX)_{\sigma
}-\left(  C+Q+E\right)  \right] \\
&  =\left(  w_{C}-\lambda_{1}-\lambda_{3}\right)  C+\left(  w_{Q}-2\lambda
_{1}-\lambda_{2}-\lambda_{3}\right)  Q+\left(  w_{E}-\lambda_{2}-\lambda
_{3}\right)  E\nonumber\\
&  \ \ \ \ \ +\lambda_{1}I(AX;B)_{\sigma}+\lambda_{2}I(A\rangle BX)_{\sigma
}+\lambda_{3}\left[  I(X;B)_{\sigma}+I(A\rangle BX)_{\sigma}\right]  .
\end{align}
The Lagrange dual function is defined as%
\begin{equation}
g(\vec{w},\mathcal{E},\vec{\lambda})\equiv\sup_{\vec{R}}\mathcal{L}(\vec
{w},\vec{R},\mathcal{E},\vec{\lambda}).
\end{equation}
By inspection, $g(\vec{w},\mathcal{E},\vec{\lambda})$ is infinite unless
$w_{C}=\lambda_{1}+\lambda_{3}$, $w_{Q}=2\lambda_{1}+\lambda_{2}+\lambda_{3}$,
$w_{E}=\lambda_{2}+\lambda_{3}$, or equivalently (by inverting these
equations), $g(\vec{w},\mathcal{E},\vec{\lambda})$ is infinite unless%
\begin{align}
\lambda_{1}  &  =\frac{1}{2}\left(  w_{Q}-w_{E}\right)
,\label{eq-tr:dual-func-finite-1}\\
\lambda_{2}  &  =\frac{1}{2}\left(  -2w_{C}+w_{Q}+w_{E}\right)  ,\\
\lambda_{3}  &  =\frac{1}{2}\left(  2w_{C}-w_{Q}+w_{E}\right)  .
\label{eq-tr:dual-func-finite-3}%
\end{align}
So we see that%
\begin{equation}
g(\vec{w},\mathcal{E},\vec{\lambda})=\lambda_{1}I(AX;B)_{\sigma}+\lambda
_{2}I(A\rangle BX)_{\sigma}+\lambda_{3}\left[  I(X;B)_{\sigma}+I(A\rangle
BX)_{\sigma}\right]  \label{eq-tr:Lagrange-dual-func-1}%
\end{equation}
if \eqref{eq-tr:dual-func-finite-1}--\eqref{eq-tr:dual-func-finite-3} hold and
$g(\vec{w},\mathcal{E},\vec{\lambda})=+\infty$ otherwise. Observe that
\eqref{eq-tr:Lagrange-dual-func-1} contains the expressions in the quantum
dynamic capacity formula in \eqref{eq-tr:objective}.

When $\lambda_{1},\lambda_{2},\lambda_{3}\geq0$, the Lagrange dual function
gives upper bounds on $P^{\ast}(\vec{w},\mathcal{E})$. To see this, let
$\vec{R}$ be a rate vector satisfying
\eqref{eq-tr:cond-1}--\eqref{eq-tr:cond-3} for fixed $\mathcal{E}$. Then%
\begin{equation}
\vec{w}\cdot\vec{R}\leq\mathcal{L}(\vec{w},\vec{R},\mathcal{E},\vec{\lambda
})\leq g(\vec{w},\mathcal{E},\vec{\lambda}).
\end{equation}
We can then obtain the tightest upper bound on $P^{\ast}(\vec{w},\mathcal{E})$
by taking an infimum over all $\vec{\lambda}$ satisfying $\lambda_{1}%
,\lambda_{2},\lambda_{3}\geq0$:%
\begin{equation}
P^{\ast}(\vec{w},\mathcal{E})\leq D^{\ast}(\vec{w},\mathcal{E})\equiv
\inf_{\lambda_{1},\lambda_{2},\lambda_{3}\geq0}g(\vec{w},\mathcal{E}%
,\vec{\lambda}).
\end{equation}
The optimization problem on the right-hand side is known as the \textit{dual
optimization problem} and the inequality $P^{\ast}(\vec{w},\mathcal{E})\leq
D^{\ast}(\vec{w},\mathcal{E})$ is known as \textit{weak duality}, which always
holds. Let%
\begin{align}
\lambda_{1}^{\ast}  &  =\frac{1}{2}\left(  w_{Q}-w_{E}\right)
,\label{eq-tr:dual-func-finite-1-opt}\\
\lambda_{2}^{\ast}  &  =\frac{1}{2}\left(  -2w_{C}+w_{Q}+w_{E}\right)  ,\\
\lambda_{3}^{\ast}  &  =\frac{1}{2}\left(  2w_{C}-w_{Q}+w_{E}\right)  .
\label{eq-tr:dual-func-finite-3-opt}%
\end{align}
Then by inspection, observe that%
\begin{equation}
D^{\ast}(\vec{w},\mathcal{E})=\lambda_{1}^{\ast}I(AX;B)_{\sigma}+\lambda
_{2}^{\ast}I(A\rangle BX)_{\sigma}+\lambda_{3}^{\ast}\left[  I(X;B)_{\sigma
}+I(A\rangle BX)_{\sigma}\right]  \label{eq-tr:dual-func-opt-formula}%
\end{equation}
if $\lambda_{1}^{\ast},\lambda_{2}^{\ast},\lambda_{3}^{\ast}\geq0$ and it is
equal to $+\infty$ otherwise.

In computing the dual optimal value $D^{\ast}(\vec{w},\mathcal{E})$, we first
maximized the Lagrangian $\mathcal{L}(\vec{w},\vec{R},\mathcal{E},\vec
{\lambda})$ with respect to the rate vector $\vec{R}$ and then we minimized
with respect to the Lagrange multipliers $\vec{\lambda}$, i.e.,%
\begin{equation}
D^{\ast}(\vec{w},\mathcal{E})=\inf_{\lambda_{1},\lambda_{2},\lambda_{3}\geq
0}\sup_{\vec{R}}\mathcal{L}(\vec{w},\vec{R},\mathcal{E},\vec{\lambda}).
\end{equation}
What happens if we conduct the optimizations in the opposite order? By
inspection, consider that%
\begin{equation}
\inf_{\lambda_{1},\lambda_{2},\lambda_{3}\geq0}\mathcal{L}(\vec{w},\vec
{R},\mathcal{E},\vec{\lambda})=\left\{
\begin{array}
[c]{cc}%
\vec{w}\cdot\vec{R} & \text{if
\eqref{eq-tr:cond-1}--\eqref{eq-tr:cond-3}\ hold}\\
-\infty & \text{else}%
\end{array}
\right.  .
\end{equation}
As a consequence, we find that%
\begin{equation}
\sup_{\vec{R}}\inf_{\lambda_{1},\lambda_{2},\lambda_{3}\geq0}\mathcal{L}%
(\vec{w},\vec{R},\mathcal{E},\vec{\lambda})=P^{\ast}(\vec{w},\mathcal{E}).
\end{equation}
The statement of \textit{strong duality} is that $P^{\ast}(\vec{w}%
,\mathcal{E})=D^{\ast}(\vec{w},\mathcal{E})$, which we see from the above
amounts to an exchange of a minimum and a maximum. Strong duality holds for
any linear program \cite[Exercise~5.23]{BV04}, provided that at least one of
the primal or dual problems is feasible. For our primal problem, this means
that there should exist a value $\vec{R}$\ meeting the constraints in
\eqref{eq-tr:cond-1}--\eqref{eq-tr:cond-3}. Since this is always the case, we
can conclude that strong duality holds:%
\begin{equation}
P^{\ast}(\vec{w},\mathcal{E})=D^{\ast}(\vec{w},\mathcal{E}),
\end{equation}
which in turn implies that the optimal value $P^{\ast}(\vec{w})$\ in
\eqref{eq-tr:scalarization}\ can be rewritten as%
\begin{equation}
P^{\ast}(\vec{w})=\sup_{\mathcal{E}}P^{\ast}(\vec{w},\mathcal{E}%
)=\sup_{\mathcal{E}}D^{\ast}(\vec{w},\mathcal{E}).
\end{equation}
By combining with \eqref{eq-tr:dual-func-opt-formula}, we see that the optimal
primal value is given by%
\begin{align}
P^{\ast}(\vec{w})  &  =\sup_{\mathcal{E}}D^{\ast}(\vec{w},\mathcal{E})\\
&  =\sup_{\mathcal{E}}\lambda_{1}^{\ast}I(AX;B)_{\sigma}+\lambda_{2}^{\ast
}I(A\rangle BX)_{\sigma}+\lambda_{3}^{\ast}\left[  I(X;B)_{\sigma}+I(A\rangle
BX)_{\sigma}\right]  , \label{eq-tr:final-QDC-formula}%
\end{align}
if $\lambda_{1}^{\ast},\lambda_{2}^{\ast},\lambda_{3}^{\ast}\geq0$ and it is
equal to $+\infty$ otherwise, with $\lambda_{1}^{\ast},\lambda_{2}^{\ast
},\lambda_{3}^{\ast}$ defined in
\eqref{eq-tr:dual-func-finite-1-opt}--\eqref{eq-tr:dual-func-finite-3-opt}.
Thus, it is now clear that the quantum dynamic capacity formula in
\eqref{eq-tr:objective} plays an essential role in computing the quantum
dynamic capacity region. If it is additive, then the computation of the Pareto
optimal trade-off surface requires an optimization with respect to a single
copy of the channel.
\end{proof}

\subsubsection{Special Cases of the Quantum Dynamic Capacity Formula}

We now show how several capacity formulas of a quantum channel, including the
entanglement-assisted classical capacity (Theorem~\ref{thm-eac:BSST}), the
quantum capacity formula (Theorem~\ref{thm-q-cap:q-cap-theorem}), and the
classical capacity formula (Theorem~\ref{thm-cc:HSW}) are special cases of the
quantum dynamic capacity formula.

We first give a geometric interpretation of these special cases before
proceeding to the proofs. Recall that the dynamic capacity region has the
simple interpretation as a translation of the three-faced unit resource
capacity region along the trade-off curve for entanglement-assisted classical
and quantum communication (see Figure~\ref{fig-tr:dephasing-plot} for an
example of the region for the dephasing channel). Any particular weight vector
$\left(  w_{C},w_{Q},w_{E}\right)  $ in \eqref{eq-tr:scalarization} gives a
set of parallel planes that slice through the $\left(  C,Q,E\right)  $\ space,
and the goal of the scalar optimization task is to find one of these planes
that is a supporting plane, intersecting a point (or a set of points) on the
trade-off surface of the dynamic capacity region. We consider three special planes:

\begin{enumerate}
\item The first corresponds to the plane containing the vectors of super-dense
coding and teleportation. The normal vector of this plane is $(1,2,0)$, and
suppose that we set the weight vector in \eqref{eq-tr:scalarization} to be
this vector. Then the optimization program finds the set of points on the
trade-off surface such that a plane with this normal vector is a supporting
plane for the region. The optimization program singles out the constraint in
\eqref{eq-tr:cond-1}, and by inspecting
\eqref{eq-tr:dual-func-finite-1-opt}--\eqref{eq-tr:dual-func-finite-3-opt},
this is equivalent to setting $\vec{\lambda}^{\ast}=(1,0,0)$ in the quantum
dynamic capacity formula in \eqref{eq-tr:final-QDC-formula}. We show below
that the optimization program becomes equivalent to finding the
entanglement-assisted capacity (Theorem~\ref{thm-eac:BSST}), in the sense that
the quantum dynamic capacity formula becomes the entanglement-assisted
capacity formula.

\item The next plane contains the vectors of teleportation and entanglement
distribution. The normal vector of this plane is $\left(  0,1,1\right)  $.
Setting the weight vector in \eqref{eq-tr:scalarization} to be this vector
makes the optimization program single out the constraint in
\eqref{eq-tr:cond-2}, and and by inspecting
\eqref{eq-tr:dual-func-finite-1-opt}--\eqref{eq-tr:dual-func-finite-3-opt},
this is equivalent to setting $\vec{\lambda}^{\ast}=(0,1,0)$ in the quantum
dynamic capacity formula in \eqref{eq-tr:final-QDC-formula}. We show below
that the optimization program becomes equivalent to finding the quantum
capacity (Theorem~\ref{thm-q-cap:q-cap-theorem}), in the sense that the
quantum dynamic capacity formula becomes the LSD formula for the quantum capacity.

\item A third plane contains the vectors of super-dense coding and
entanglement distribution. The normal vector of this plane is $\left(
1,1,1\right)  $. Setting the weight vector in \eqref{eq-tr:scalarization} to
be this vector makes the optimization program single out the constraint in
\eqref{eq-tr:cond-3}, and by inspecting
\eqref{eq-tr:dual-func-finite-1-opt}--\eqref{eq-tr:dual-func-finite-3-opt},
this is equivalent to setting $\vec{\lambda}^{\ast}=(0,0,1)$ in the quantum
dynamic capacity formula in \eqref{eq-tr:final-QDC-formula}. We show below
that the optimization becomes equivalent to finding the classical capacity
(Theorem~\ref{thm-cc:HSW}), in the sense that the quantum dynamic capacity
formula becomes the HSW\ formula for the classical capacity.
\end{enumerate}

\begin{corollary}
The quantum dynamic capacity formula is equal to the entanglement-assisted
classical capacity formula when $\lambda_{1}=1$, $\lambda_{2}=0$, and
$\lambda_{3}=0$:%
\begin{equation}
\max_{\sigma_{XAA^{\prime}}}I(AX;B)_{\sigma}=\max_{\phi_{AA^{\prime}}%
}I(A;B)_{\rho},
\end{equation}
where $\sigma_{XAA^{\prime}}\equiv\sum_{x}p_{X}(x)|x\rangle\langle
x|_{X}\otimes|\phi^{x}\rangle\langle\phi^{x}|_{AA^{\prime}}$, $\sigma
_{XAB}\equiv\mathcal{N}_{A^{\prime}\rightarrow B}(\sigma_{XAA^{\prime}})$, and
$\rho_{AB}\equiv\mathcal{N}_{A^{\prime}\rightarrow B}(\phi_{AA^{\prime}})$.
\end{corollary}

\begin{proof}
The inequality $\max_{\sigma}I(AX;B)_\sigma \geq\max_{\phi_{AA^{\prime}}}I(A;B)_{\rho}$
follows because the state $\sigma_{XAA'}$ is of the form in
\eqref{eq-tr:main-theorem-state} and we can always choose $p_{X}%
(x)=\delta_{x,x_{0}}$ and $\phi_{AA^{\prime}}^{x_{0}}$ to be the state that
maximizes $I(A;B)$.

We now show the other inequality $\max_{\sigma}I(AX;B)_{\sigma}\leq\max
_{\phi_{AA^{\prime}}}I(A;B)_{\rho}$. Consider that the following state purifies
$\sigma_{XAA^{\prime}}$:%
\begin{equation}
|\varphi\rangle_{XRAA^{\prime}}\equiv\sum_{x}\sqrt{p_{X}(x)}|x\rangle
_{X}|x\rangle_{R}|\phi^{x}\rangle_{AA^{\prime}}.
\end{equation}
Then by quantum data processing, we find that%
\begin{equation}
I(AX;B)_{\sigma}\leq I(RAX;B)_{\mathcal{N}(\varphi)}\leq\max_{\phi
_{AA^{\prime}}}I(A;B)_{\rho},
\end{equation}
where the second inequality follows because systems $RAX$ of $\varphi
_{XRAA^{\prime}}$ purify $A^{\prime}$ and the formula on the right involves an
optimization over all such purifications.
\end{proof}

\begin{corollary}
The quantum dynamic capacity formula is equal to the LSD\ quantum capacity
formula when $\lambda_{1}=0$, $\lambda_{2}=1$, and $\lambda_{3}=0$:%
\begin{equation}
\max_{\sigma}I(A\rangle BX)=\max_{\phi_{AA^{\prime}}}I(A\rangle B).
\end{equation}

\end{corollary}

\begin{proof}
The inequality $\max_{\sigma}I(A\rangle BX)\geq\max_{\phi_{AA^{\prime}}%
}I(A\rangle B)$ follows because the state $\sigma$ is of the form in
\eqref{eq-tr:main-theorem-state} and we can always choose $p_{X}%
(x)=\delta_{x,x_{0}}$ and $\phi_{AA^{\prime}}^{x_{0}}$ to be the state that
maximizes $I(A\rangle B)$. The inequality $\max_{\sigma}I(A\rangle BX)\leq
\max_{\phi_{AA^{\prime}}}I(A\rangle B)$ follows because $I(A\rangle
BX)=\sum_{x}p_{X}(x)I(A\rangle B)_{\mathcal{N}(\phi_{x})}$ and a maximum is never smaller
than the average.
\end{proof}

\begin{corollary}
The quantum dynamic capacity formula is equal to the HSW\ classical capacity
formula when $\lambda_{1}=0$, $\lambda_{2}=0$, and $\lambda_{3}=1$:%
\begin{equation}
\max_{\sigma}\left[  I(A\rangle BX)_{\sigma}+I(X;B)_{\sigma}\right]
=\max_{\left\{  p_{X}(x),\psi_{x}\right\}  }I(X;B).
\end{equation}

\end{corollary}

\begin{proof}
The inequality $\max_{\sigma}I(A\rangle BX)_{\sigma}+I(X;B)_{\sigma}\geq
\max_{\left\{  p_{X}(x),\psi_{x}\right\}  }I(X;B)$ follows by choosing
$\sigma$ to be the pure-state ensemble that maximizes $I(X;B)$ and noting that
$I(A\rangle BX)_{\sigma}$ vanishes for a pure-state ensemble. We now prove the
inequality $\max_{\sigma}I(A\rangle BX)_{\sigma}+I(X;B)_{\sigma}\leq
\max_{\left\{  p_{X}(x),\psi_{x}\right\}  }I(X;B)$. Consider a state
$\omega_{XYBE}$ obtained by performing a complete projective measurement on
the $A$ system of the state $\sigma_{XABE}$. Then%
\begin{align}
I(A\rangle BX)_{\sigma}+I(X;B)_{\sigma}  &  =H(B)_{\sigma}-H(E|X)_{\sigma}\\
&  =H(B)_{\omega}-H(E|X)_{\omega}\\
&  \leq H(B)_{\omega}-H(E|XY)_{\omega}\\
&  =H(B)_{\omega}-H(B|XY)_{\omega}\\
&  =I(XY;B)_{\omega}\\
&  \leq\max_{\left\{  p_{X}(x),\psi_{x}\right\}  }I(X;B).
\end{align}
The first equality follows by expanding the conditional coherent information
and the Holevo information. The second equality follows because the measured
$A$ system is not involved in the entropies. The first inequality follows
because conditioning does not increase entropy. The third equality follows
because the state $\omega$ is pure when conditioned on $X$ and $Y$. The fourth
equality follows by definition, and the last inequality follows for clear reasons.
\end{proof}

\section{Examples of Channels}

In this final section, we prove that a broad class of channels, known as the
Hadamard channels (see Section~\ref{sec-nqt:hadamard-channel}), have a
single-letter dynamic capacity region. We prove this result by analyzing the
quantum dynamic capacity formula for this class of channels. A dephasing
channel is a special case of a Hadamard channel, and so we can compute its
dynamic capacity region. We also show that the quantum erasure channel has a
single-letter dynamic capacity region, and in the process, we see that
time-sharing is an optimal coding strategy for this channel. We finally
overview the dynamic capacity region of a pure-loss bosonic channel, which is
a good model for free-space communication or loss in an optical fiber.
However, we only state the main results and do not get into too many details
of this channel (doing so requires the theory of quantum optics and
infinite-dimensional Hilbert spaces which is beyond the scope of this book).
The upshot for this channel is that trade-off coding can give remarkable gains
over time sharing.

\subsection{Quantum Hadamard Channels}

\label{sec-tr:single-letter-hadamard}Below we
\index{Hadamard channel}
show that the regularization in \eqref{eq-tr:multi-letter} is not necessary if
the quantum channel is a Hadamard channel. This result holds because a
Hadamard channel has a special structure (see
Section~\ref{sec-nqt:hadamard-channel}).

\begin{theorem}
\label{thm-tr:Hadamard-theorem}The dynamic capacity region $\mathcal{C}%
_{\mathrm{{CQE}}}(\mathcal{N}_{\operatorname{H}})$ of a quantum Hadamard
channel $\mathcal{N}_{\operatorname{H}}$ is equal to its single-letter region
$\mathcal{C}_{\mathrm{{CQE}}}^{(1)}(\mathcal{N}_{\operatorname{H}})$.
\end{theorem}

\noindent The proof of the above theorem follows in two parts:\ 1)\ the lemma
below states that the quantum dynamic capacity formula is additive when one of
the channels is Hadamard and 2)\ the induction argument in
Lemma~\ref{thm-tr:CEQ-single-letter}\ establishes single-letterization.

\begin{lemma}
\label{lem:CEQ-base-case}The following additivity relation holds for a
Hadamard channel $\mathcal{N}_{H}$, any other channel $\mathcal{N}$, and for
all $\vec{\lambda}$ such that $\lambda_{1}, \lambda_{2}, \lambda_{3} \geq0$:%
\begin{equation}
D_{\vec{\lambda}}(\mathcal{N}_{H}\otimes\mathcal{N})=D_{\vec{\lambda}%
}(\mathcal{N}_{H})+D_{\vec{\lambda}}(\mathcal{N}).
\end{equation}

\end{lemma}

\begin{proof}
We first note that the inequality $D_{\vec{\lambda}}(\mathcal{N}_{H}%
\otimes\mathcal{N})\geq D_{\vec{\lambda}}(\mathcal{N}_{H})+D_{\vec{\lambda}%
}(\mathcal{N})$ holds for any two channels simply by selecting the state
$\sigma$ in the maximization to be a tensor product of the ones that
individually maximize $D_{\vec{\lambda}}(\mathcal{N}_{H})$ and $D_{\vec
{\lambda}}(\mathcal{N})$.

So we prove that the non-trivial inequality $D_{\vec{\lambda}}(\mathcal{N}%
_{H}\otimes\mathcal{N})\leq D_{\vec{\lambda}}(\mathcal{N}_{H})+D_{\vec
{\lambda}}(\mathcal{N})$ holds when the first channel is a Hadamard channel.
Since the first channel is Hadamard, it is
\index{degradable channel}%
degradable and its degrading channel has a particular structure: there are
channels $\mathcal{D}_{B_{1}\rightarrow Y}^{1}$ and $\mathcal{D}_{Y\rightarrow
E_{1}}^{2}$ where $Y$ is a classical register and such that the degrading
channel is $\mathcal{D}_{Y\rightarrow E_{1}}^{2}\circ\mathcal{D}%
_{B_{1}\rightarrow Y}^{1}$. Suppose the state we are considering to input to
the tensor-product channel is%
\begin{equation}
\rho_{XAA_{1}^{\prime}A_{2}^{\prime}}\equiv\sum_{x}p_{X}(x)|x\rangle\langle
x|_{X}\otimes\phi_{AA_{1}^{\prime}A_{2}^{\prime}}^{x},
\end{equation}
and this state is the one that maximizes $D_{\vec{\lambda}}(\mathcal{N}%
_{H}\otimes\mathcal{N})$. Suppose that the output of the first channel is%
\begin{equation}
\theta_{XAB_{1}E_{1}A_{2}^{\prime}}\equiv\mathcal{U}_{A_{1}^{\prime
}\rightarrow B_{1}E_{1}}^{\mathcal{N}_{H}}(\rho_{XAA_{1}^{\prime}A_{2}%
^{\prime}}),
\end{equation}
and the output of the second channel is%
\begin{equation}
\omega_{XAB_{1}E_{1}B_{2}E_{2}}\equiv\mathcal{U}_{A_{2}^{\prime}\rightarrow
B_{2}E_{2}}^{\mathcal{N}}(\theta_{XAB_{1}E_{1}A_{2}^{\prime}}).
\end{equation}
Finally, we define the following state as the result of applying the first
part of the degrading channel (a complete projective measurement) to $\omega$:%
\begin{equation}
\sigma_{XYAE_{1}B_{2}E_{2}}\equiv\mathcal{D}_{B_{1}\rightarrow Y}^{1}%
(\omega_{XAB_{1}E_{1}B_{2}E_{2}}).
\end{equation}
In particular, the state $\sigma$ on systems $AE_{1}B_{2}E_{2}$ is pure when
conditioned on $X$ and $Y$. Then the following chain of inequalities holds:%
\begin{align}
&  D_{\vec{\lambda}}\left(  \mathcal{N}_{H}\otimes\mathcal{N}\right)
\nonumber\\
&  =\lambda_{1}I(AX;B_{1}B_{2})_{\omega}+\lambda_{2}I(A\rangle B_{1}%
B_{2}X)_{\omega}+\lambda_{3}\left[  I(X;B_{1}B_{2})_{\omega}+I(A\rangle
B_{1}B_{2}X)_{\omega}\right] \\
&  =\lambda_{1}H(B_{1}B_{2}E_{1}E_{2}|X)_{\omega}+\lambda_{2}H(B_{1}%
B_{2}|X)_{\omega}+\left(  \lambda_{1}+\lambda_{3}\right)  H(B_{1}%
B_{2})_{\omega}\nonumber\\
&  \qquad -\left(  \lambda_{1}+\lambda_{2}+\lambda_{3}\right)
H(E_{1}E_{2}|X)_{\omega}\\
&  =\lambda_{1}H(B_{1}E_{1}|X)_{\omega}+\lambda_{2}H(B_{1}|X)_{\omega}+\left(
\lambda_{1}+\lambda_{3}\right)  H(B_{1})_{\omega}\nonumber\\
&  \qquad -\left(  \lambda_{1}+\lambda_{2}+\lambda_{3}\right)
H(E_{1}|X)_{\omega}+\lambda_{1}H(B_{2}E_{2}|B_{1}E_{1}X)_{\omega}+\lambda
_{2}H(B_{2}|B_{1}X)_{\omega}\nonumber\\
&  \qquad +\left(  \lambda_{1}+\lambda_{3}\right)  H(B_{2}|B_{1})_{\omega
}-\left(  \lambda_{1}+\lambda_{2}+\lambda_{3}\right)  H(E_{2}|E_{1}X)_{\omega
}\\
&  \leq\lambda_{1}H(B_{1}E_{1}|X)_{\theta}+\lambda_{2}H(B_{1}|X)_{\theta
}+\left(  \lambda_{1}+\lambda_{3}\right)  H(B_{1})_{\theta}\nonumber\\
&  \qquad -\left(  \lambda_{1}+\lambda_{2}+\lambda_{3}\right)
H(E_{1}|X)_{\theta}+\lambda_{1}H(B_{2}E_{2}|YX)_{\sigma}+\lambda_{2}%
H(B_{2}|YX)_{\sigma}\nonumber\\
&  \qquad +\left(  \lambda_{1}+\lambda_{3}\right)  H(B_{2})_{\sigma
}-\left(  \lambda_{1}+\lambda_{2}+\lambda_{3}\right)  H(E_{2}|YX)_{\sigma}\\
&  =\lambda_{1}I(AA_{2}^{\prime}X;B_{1})_{\theta}+\lambda_{2}I(AA_{2}^{\prime
}\rangle B_{1}X)_{\theta}+\lambda_{3}\left[  I(X;B_{1})_{\theta}%
+I(AA_{2}^{\prime}\rangle B_{1}X)_{\theta}\right] \nonumber\\
&  \qquad +\lambda_{1}I(AE_{1}YX;B_{2})_{\sigma}+\lambda_{2}I(AE_{1}\rangle
B_{2}YX)_{\sigma}\nonumber\\
&  \qquad +\lambda_{3}\left[  I(YX;B_{2})_{\sigma}+I(AE_{1}\rangle
B_{2}YX)_{\sigma}\right] \\
&  \leq D_{\vec{\lambda}}\left(  \mathcal{N}_{H}\right)  +D_{\vec{\lambda}%
}(\mathcal{N}).
\end{align}
The first equality follows by evaluating the quantum dynamic capacity formula
$D_{\vec{\lambda}}\left(  \mathcal{N}_{H}\otimes\mathcal{N}\right)  $ on the
state $\rho$. The next two equalities follow by rearranging entropies and
because the state $\omega$ on systems $AB_{1}E_{1}B_{2}E_{2}$ is pure when
conditioned on $X$. The inequality in the middle is the crucial one and
follows from the Hadamard structure of the channel: we exploit monotonicity of
conditional entropy with respect to quantum channels so that
\begin{align}
H(B_{2}|B_{1}X)_{\omega}& \leq H(B_{2}|YX)_{\sigma}, \\
H(B_{2}E_{2}|B_{1}E_{1}X)_{\omega}
& \leq H(B_{2}E_{2}|YX)_{\sigma}, \\
H(E_{2}|YX)_{\sigma} & \leq
H(E_{2}|E_{1}X)_{\omega},\\
H(B_{2}|B_{1})_{\omega} & 
\leq H(B_{2})_{\omega}.
\end{align}
 The next equality follows by rearranging entropies
and the final inequality follows because~$\theta$ is a state of the form
\eqref{eq-tr:main-theorem-state} for the first channel while $\sigma$ is a
state of the form \eqref{eq-tr:main-theorem-state} for the second channel.
\end{proof}

\subsection{The Dephasing Channel}

\label{sec-tr:dephasing}
\begin{figure}
[ptb]
\begin{center}
\includegraphics[
width=4.0413in
]%
{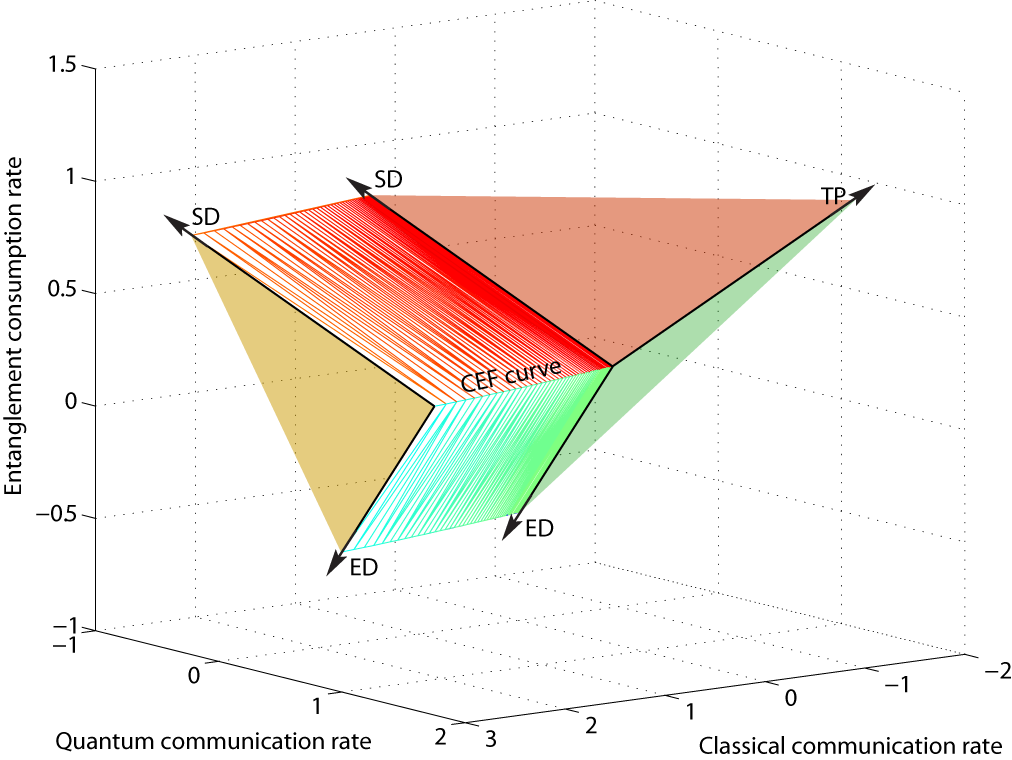}%
\caption{A plot of the dynamic capacity region for a qubit dephasing channel
with dephasing parameter $p=0.2$. The plot shows that the CEF\ trade-off curve
(the protocol from Corollary~\ref{cor-ccn:CQE-trading}) lies along the
boundary of the dynamic capacity region. The rest of the region is simply the
combination of the CEF points with the unit protocols teleportation (TP),
super-dense coding (SD), and entanglement distribution~(ED).}%
\label{fig-tr:full-plot-dephasing}%
\end{center}
\end{figure}

\begin{figure}
[ptb]
\begin{center}
\includegraphics[
width=4.8456in
]%
{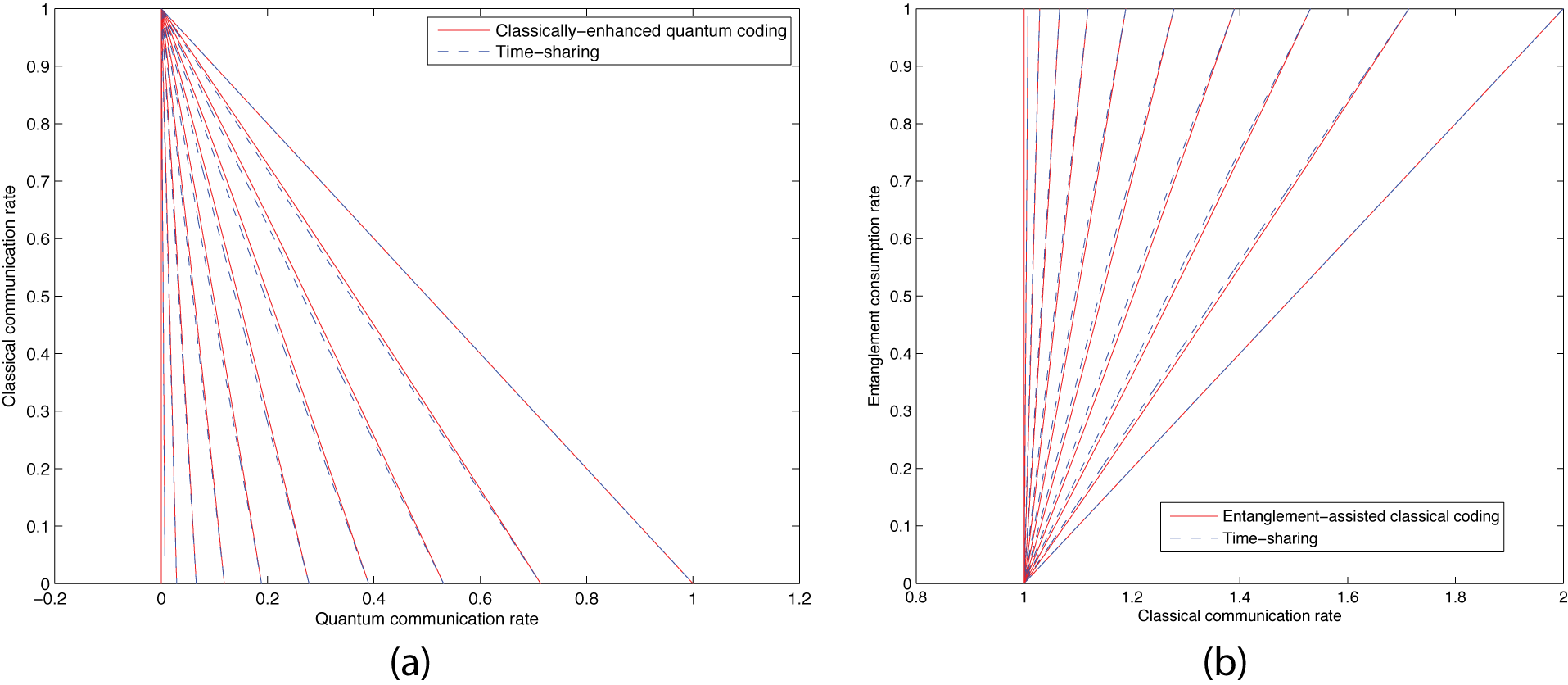}%
\caption{Plot of (a) the CQ trade-off curve and (b) the CE trade-off curve for
a $p$-dephasing qubit channel for $p=0$, $0.1$, $0.2$, \ldots, $0.9$, $1$. The
trade-off curves for $p=0$ correspond to those of a noiseless qubit channel
and are the rightmost trade-off curves in each plot. The trade-off curves for
$p=1$ correspond to those for a classical channel and are the leftmost
trade-off curves in each plot. Each trade-off curve between these two extremes
beats a time-sharing strategy, but these two extremes do not beat time
sharing.}%
\label{fig-tr:dephasing-plot}%
\end{center}
\end{figure}

The theorem below states
\index{dephasing channel}%
that the full dynamic capacity region admits a particularly simple form when
the noisy quantum channel is a qubit dephasing channel $\overline{\Delta}_{p}$
where%
\begin{align}
\overline{\Delta}_{p}(\rho)  &  \equiv(1-p)\rho+p\overline{\Delta}(\rho),\\
\overline{\Delta}(\rho)  &  \equiv\langle0|\rho|0\rangle|0\rangle
\langle0|+\langle1|\rho|1\rangle|1\rangle\langle1|,
\end{align}
and $p\in\left[  0,1\right]  $. Figure~\ref{fig-tr:full-plot-dephasing}\ plots
this region for the case of a dephasing channel with dephasing parameter
$p=0.2$. Figure~\ref{fig-tr:dephasing-plot}\ plots special two-dimensional
cases of the full region for various values of the dephasing parameter~$p$.
The figure demonstrates that trade-off coding just barely improves upon time
sharing.%

\begin{theorem}
\label{thm-tr:dephasing}The dynamic capacity region $\mathcal{C}%
_{\operatorname{CQE}}(\overline{\Delta}_{p})$ of a dephasing channel with
dephasing parameter $p\in[0,1]$\ is the set of all $C$, $Q$, and $E$ such that%
\begin{align}
C+2Q  &  \leq1+h_{2}(\nu)-h_{2}(\gamma(\nu,p)),\\
Q+E  &  \leq h_{2}(\nu)-h_{2}(\gamma(\nu,p)),\\
C+Q+E  &  \leq1-h_{2}(\gamma(\nu,p)),
\end{align}
where $\nu\in\left[  0,1/2\right]  $, $h_{2}$ is the binary entropy function,
and%
\begin{equation}
\gamma(\nu,p)\equiv\frac{1}{2}+\frac{1}{2}\sqrt{1-16\cdot\frac{p}{2}\left(
1-\frac{p}{2}\right)  \nu(1-\nu)}.
\end{equation}

\end{theorem}

\begin{proof}
The dephasing channel is a Hadamard channel, so that the region
single-letterizes and it suffices to simplify the quantum dynamic capacity
formula evaluated on a single copy of the channel. We then notice that it
suffices to consider an ensemble of pure states whose reductions to
$A^{\prime}$ are diagonal in the dephasing basis (see the following exercise).

\begin{exercise}
Prove that the following properties hold for a generalized dephasing channel
$\mathcal{N}_{D}$, its complement $\mathcal{N}_{D}^{c}$, the completely
dephasing channel~$\overline{\Delta}$, and all input states $\rho$:%
\begin{align}
\mathcal{N}_{D}(\overline{\Delta}(\rho))  &  =\overline{\Delta}(\mathcal{N}%
_{D}(\rho)),\\
\mathcal{N}_{D}^{c}(\overline{\Delta}(\rho))  &  =\mathcal{N}_{D}^{c}(\rho).
\end{align}
Conclude that%
\begin{align}
H(\rho)  &  \leq H(\overline{\Delta}(\rho)),\\
H(\mathcal{N}_{D}(\rho))  &  \leq H(\overline{\Delta}(\mathcal{N}_{D}%
(\rho)))=H(\mathcal{N}_{D}(\overline{\Delta}(\rho))),\\
H(\mathcal{N}_{D}^{c}(\rho))  &  =H(\mathcal{N}_{D}^{c}(\overline{\Delta}%
(\rho))),
\end{align}
so that it suffices to consider diagonal input states for the dephasing channel.
\end{exercise}

\noindent Next we prove below that it is sufficient to consider an ensemble of
the following form to characterize the boundary points of the region:%
\begin{equation}
\frac{1}{2}|0\rangle\langle0|_{X}\otimes\psi_{AA^{\prime}}^{0}+\frac{1}%
{2}|1\rangle\langle1|_{X}\otimes\psi_{AA^{\prime}}^{1},
\label{eq-tr:mu-cq-state-CEQ}%
\end{equation}
where $\psi_{AA^{\prime}}^{0}$ and $\psi_{AA^{\prime}}^{1}$ are pure states,
defined as follows for $\nu\in\left[  0,1/2\right]  $:%
\begin{align}
\operatorname{Tr}_{A}\left\{  \psi_{AA^{\prime}}^{0}\right\}   &
=\nu|0\rangle\langle0|_{A^{\prime}}+\left(  1-\nu\right)  |1\rangle
\langle1|_{A^{\prime}},\label{eq-tr:1st-mu-state-CEQ}\\
\operatorname{Tr}_{A}\left\{  \psi_{AA^{\prime}}^{1}\right\}   &  =\left(
1-\nu\right)  |0\rangle\langle0|_{A^{\prime}}+\nu|1\rangle\langle
1|_{A^{\prime}}. \label{eq-tr:2nd-mu-state-CEQ}%
\end{align}
We now prove the above claim. We assume without loss of generality that the
dephasing basis is the computational basis. Consider a classical--quantum
state with a finite number $N$ of conditional density operators $\phi
_{AA^{\prime}}^{x}$ whose reduction to $A^{\prime}$ is diagonal:%
\begin{equation}
\rho_{XAA^{\prime}}\equiv\sum_{x=0}^{N-1}p_{X}(x)|x\rangle\langle
x|_{X}\otimes\phi_{AA^{\prime}}^{x}.
\end{equation}
We can form a new classical--quantum state with double the number of
conditional density operators by \textquotedblleft
bit-flipping\textquotedblright\ the original conditional density operators:%
\begin{equation}
\sigma_{XAA^{\prime}}\equiv\frac{1}{2}\sum_{x=0}^{N-1}p_{X}(x)\left(
|x\rangle\langle x|_{X}\otimes\phi_{AA^{\prime}}^{x}+|x+N\rangle\langle
x+N|_{X}\otimes X_{A^{\prime}}\phi_{AA^{\prime}}^{x}X_{A^{\prime}}\right)  ,
\end{equation}
where $X$ is the $\sigma_{X}$ \textquotedblleft bit-flip\textquotedblright%
\ Pauli operator. Consider the following chain of inequalities that holds for
all $\lambda_{1},\lambda_{2},\lambda_{3}\geq0$:%
\begin{align}
&  \lambda_{1}I(AX;B)_{\rho}+\lambda_{2}I(A\rangle BX)_{\rho}+\lambda
_{3}\left[  I(X;B)_{\rho}+I(A\rangle BX)_{\rho}\right] \nonumber\\
&  =\lambda_{1}H(A|X)_{\rho}+\left(  \lambda_{1}+\lambda_{3}\right)
H(B)_{\rho}+\lambda_{2}H(B|X)_{\rho}-\left(  \lambda_{1}+\lambda_{2}%
+\lambda_{3}\right)  H(E|X)_{\rho}\\
&  \leq\left(  \lambda_{1}+\lambda_{3}\right)  H(B)_{\sigma}+\lambda
_{1}H(A|X)_{\sigma}+\lambda_{2}H(B|X)_{\sigma}-\left(  \lambda_{1}+\lambda
_{2}+\lambda_{3}\right)  H(E|X)_{\sigma}\\
&  =\left(  \lambda_{1}+\lambda_{3}\right)  +\lambda_{1}H(A|X)_{\sigma
}+\lambda_{2}H(B|X)_{\sigma}-\left(  \lambda_{1}+\lambda_{2}+\lambda
_{3}\right)  H(E|X)_{\sigma}\\
&  =\left(  \lambda_{1}+\lambda_{3}\right)  +\sum_{x}p_{X}(x)\left[
\lambda_{1}H(A)_{\phi_{x}}+\lambda_{2}H(B)_{\phi_{x}}-\left(  \lambda
_{1}+\lambda_{2}+\lambda_{3}\right)  H(E)_{\phi_{x}}\right] \\
&  \leq\left(  \lambda_{1}+\lambda_{3}\right)  +\max_{x}\left[  \lambda
_{1}H(A)_{\phi_{x}}+\lambda_{2}H(B)_{\phi_{x}}-\left(  \lambda_{1}+\lambda
_{2}+\lambda_{3}\right)  H(E)_{\phi_{x}}\right] \\
&  =\left(  \lambda_{1}+\lambda_{3}\right)  +\lambda_{1}H(A)_{\phi_{\ast}^{x}%
}+\lambda_{2}H(B)_{\phi_{\ast}^{x}}-\left(  \lambda_{1}+\lambda_{2}%
+\lambda_{3}\right)  H(E)_{\phi_{\ast}^{x}}.
\end{align}
The first equality follows by standard entropic manipulations. The second
equality follows because the conditional entropy $H(B|X)$ is invariant under a
bit-flipping unitary on the input state that commutes with the channel:
$H(B)_{X\rho_{B}^{x}X}=H(B)_{\rho_{B}^{x}}$. Furthermore, a bit flip on the
input state does not change the eigenvalues for the output of the dephasing
channel's complementary channel:%
\begin{equation}
H(E)_{\mathcal{N}^{c}(X\rho_{A^{\prime}}^{x}X)}=H(E)_{\mathcal{N}^{c}%
(\rho_{A^{\prime}}^{x})}.
\end{equation}
The first inequality follows because entropy is concave, i.e., the local state
$\sigma_{B}$ is a mixed version of $\rho_{B}$. The third equality follows
because%
\begin{equation}
H(B)_{\sigma_{B}}=H\left(  \sum_{x}\frac{1}{2}p_{X}(x)(\rho_{B}^{x}+X\rho
_{B}^{x}X)\right)  =H\left(  \frac{1}{2}\sum_{x}p_{X}(x)I\right)  =1.
\end{equation}
The fourth equality follows because the system $X$ is classical. The second
inequality follows because the maximum value of a realization of a random
variable is not less than its expectation. The final equality simply follows
by defining $\phi_{\ast}^{x}$ to be the conditional density operator on
systems $A$, $B$, and $E$ that arises from sending through the channel a state
whose reduction to $A^{\prime}$ is of the form $\nu|0\rangle\langle
0|_{A^{\prime}}+\left(  1-\nu\right)  |1\rangle\langle1|_{A^{\prime}}$. Thus,
an ensemble of the kind in \eqref{eq-tr:mu-cq-state-CEQ} is sufficient to
attain a point on the boundary of the region. Evaluating the entropic
quantities in Theorem~\ref{thm-tr:main-theorem}\ on a state of the above form
then gives the expression for the region in Theorem~\ref{thm-tr:dephasing}.
\end{proof}

\subsection{The Quantum Erasure Channel}

\begin{figure}
[ptb]
\begin{center}
\includegraphics[
width=4.5455in
]%
{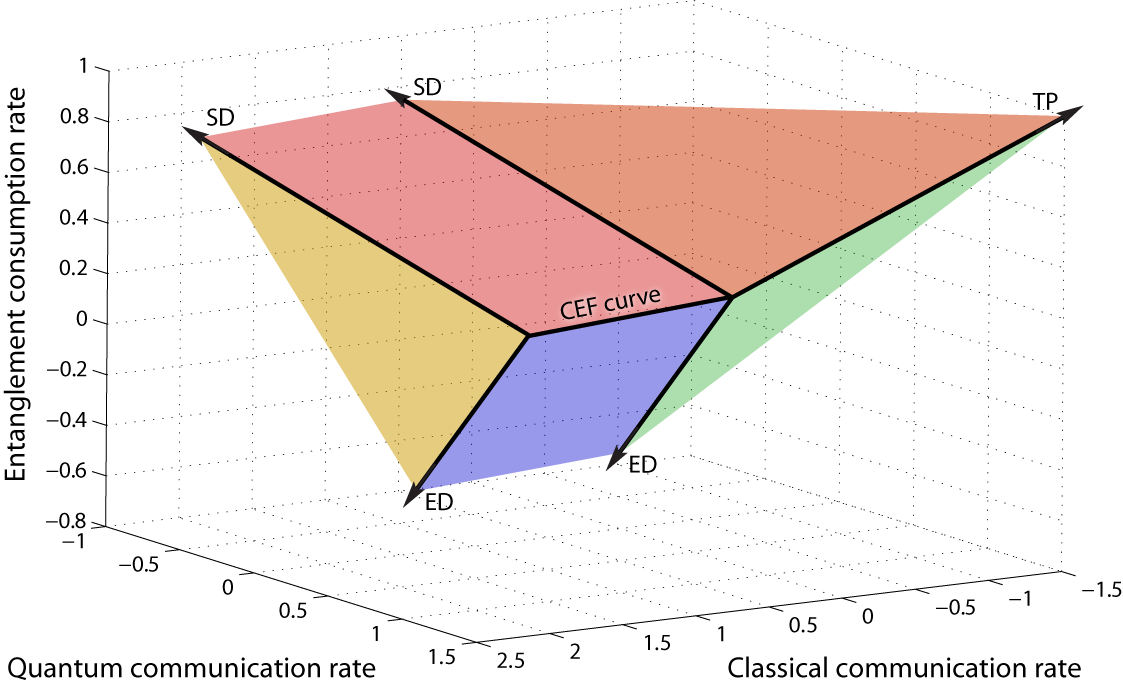}%
\caption{The quantum dynamic capacity region for the (qubit) quantum erasure
channel with $\varepsilon=1/4$. The plot demonstrates that time-sharing is
optimal.}%
\label{fig-tr:dynamic-erasure}%
\end{center}
\end{figure}

We can also obtain a simple characterization of the quantum dynamic capacity
region for a quantum erasure channel for $\varepsilon\in\left[  0,1/2\right]
$. Recall that the erasure channel is defined as follows:%
\begin{equation}
\mathcal{N}^{\varepsilon}(\rho)=\left(  1-\varepsilon\right)  \rho
+\varepsilon|e\rangle\langle e|,
\end{equation}
where $\rho$ is a $d$-dimensional input state, $|e\rangle$ is an erasure flag
state orthogonal to all inputs (so that the output space has dimension $d+1$),
and $\varepsilon\in\left[  0,1\right]  $ is the erasure probability.
Theorem~\ref{thm-tr:erasure-theorem} characterizes the quantum dynamic
capacity region for such a channel and establishes that time-sharing is an
optimal strategy. Figure~\ref{fig-tr:dynamic-erasure}\ plots an example of the
region for a qubit erasure channel with $\varepsilon=1/4$.%

\begin{theorem}
\label{thm-tr:erasure-theorem}Let $\mathcal{N}^{\varepsilon}$ be a quantum
erasure channel with $\varepsilon\in\left[  0,1/2\right]  $. Then the quantum
dynamic capacity region $\mathcal{C}_{\mathrm{{CQE}}}(\mathcal{N}%
^{\varepsilon})$ is equal to the union of the following regions, obtained by
varying $\lambda\in\left[  0,1\right]  $:%
\begin{align*}
C+2Q  &  \leq\left(  1-\varepsilon\right)  \left(  1+\lambda\right)  \log d,\\
Q+E  &  \leq\left(  1-2\varepsilon\right)  \lambda\log d,\\
C+Q+E  &  \leq\left(  1-\varepsilon-\varepsilon\lambda\right)  \log d.
\end{align*}
The region is achievable for $\varepsilon\in(1/2,1]$.
\end{theorem}

\begin{proof}
We need to show that the above region is achievable and that the regularized
quantum dynamic capacity formula simplifies to it as well. For the
achievability part, we choose a particular ensemble and show that the above
rate region is achievable using it. Let $|\phi\rangle_{AA^{\prime}}\equiv
\sum_{x}\sqrt{p_{X}(x)}|x\rangle_{A}|x\rangle_{A^{\prime}}$ where
$\{|x\rangle_{A}\}$ and $\{|x\rangle_{A^{\prime}}\}$ are the standard
computational bases for systems $A$ and $A^{\prime}$, respectively. Observe
that $H(p_{X})=H(A)_{\phi}$. We take the input ensemble as
$\{1/d,X(x)_{A^{\prime}}|\phi\rangle_{AA^{\prime}}\}$, where $X(x)_{A^{\prime
}}$ are the Heisenberg--Weyl shift operators. This ensemble has the property
that the expected density operator on system $A^{\prime}$ is the maximally
mixed state $\pi_{A^{\prime}}$. A classical--quantum state for evaluating the
rates in Theorem~\ref{thm-tr:main-theorem} is as follows:%
\begin{align}
\rho_{XAB}  &  \equiv\sum_{x}\frac{1}{d}|x\rangle\langle x|_{X}\otimes
\mathcal{N}_{A^{\prime}\rightarrow B}^{\varepsilon}[X(x)_{A^{\prime}}%
|\phi\rangle\langle\phi|_{AA^{\prime}}X^{\dag}(x)_{A^{\prime}}]\\
&  =\left(  1-\varepsilon\right)  \rho_{XAA^{\prime}}^{0}+\varepsilon
\rho_{XAB}^{1},
\end{align}
where%
\begin{align}
\rho_{XAA^{\prime}}^{0}  &  \equiv\sum_{x}\frac{1}{d}|x\rangle\langle
x|_{X}\otimes X(x)_{A^{\prime}}|\phi\rangle\langle\phi|_{AA^{\prime}}X^{\dag
}(x)_{A^{\prime}},\\
\rho_{XAB}^{1}  &  \equiv\sum_{x}\frac{1}{d}|x\rangle\langle x|_{X}\otimes
\phi_{A}\otimes|e\rangle\langle e|_{B}.
\end{align}
Then, from Theorem~\ref{thm-tr:main-theorem}, we see that it suffices to
compute the following three information quantities: $I(AX;B)_{\rho}$,
$I(A\rangle BX)_{\rho}$, and $I(X;B)_{\rho}$. Consider that we can perform the
following isometry on the $B$ system without affecting any of the information
quantities:%
\begin{equation}
\left[  |0\rangle\langle0|+\ldots+|d-1\rangle\langle d-1|\right]
\otimes|0\rangle_{Y}+|e\rangle\langle e|\otimes|1\rangle_{Y}.
\label{eq-tr:erasure-isometry}%
\end{equation}
Let $\omega_{XABY}$ denote the resulting state:%
\begin{equation}
\omega_{XABY}\equiv\left(  1-\varepsilon\right)  \rho_{XAA^{\prime}}%
^{0}\otimes|0\rangle\langle0|_{Y}+\varepsilon\rho_{XAB}^{1}\otimes
|1\rangle\langle1|_{Y}.
\end{equation}
Then, observing that $\omega_{XAY}=\pi_{X}\otimes\phi_{A}\otimes\left[
\left(  1-\varepsilon\right)  |0\rangle\langle0|_{Y}+\varepsilon
|1\rangle\langle1|_{Y}\right]  $, we find that%
\begin{align}
I(AX;B)_{\rho}  &  =I(AX;BY)_{\omega}=I(AX;B|Y)_{\omega},\\
I(A\rangle BX)_{\rho}  &  =I(A\rangle BXY)_{\omega}=I(A\rangle BX|Y)_{\omega
},\\
I(X;B)_{\rho}  &  =I(X;BY)_{\omega}=I(X;B|Y)_{\omega}.
\end{align}
We now evaluate each of these in turn:%
\begin{align}
I(AX;B|Y)_{\omega}  &  =\left(  1-\varepsilon\right)  I(AX;A^{\prime}%
)_{\rho^{0}}+\varepsilon I(AX;B)_{\rho^{1}}\nonumber\\
&  =\left(  1-\varepsilon\right)  \left(  \log d+H(p_{X})\right)  ,\\
I(A\rangle BX|Y)_{\omega}  &  =\left(  1-\varepsilon\right)  I(A\rangle
A^{\prime}X)_{\rho^{0}}+\varepsilon I(A\rangle BX)_{\rho^{1}}\nonumber\\
&  =\left(  1-\varepsilon\right)  H(p_{X})+\varepsilon\left[  -H(p_{X})\right]
\nonumber\\
&  =\left(  1-2\varepsilon\right)  H(p_{X}),\\
I(X;B|Y)_{\omega}  &  =\left(  1-\varepsilon\right)  I(X;A^{\prime})_{\rho
^{0}}+\varepsilon I(X;B)_{\rho^{1}}\nonumber\\
&  =\left(  1-\varepsilon\right)  \left(  \log d-H(p_{X})\right)  .
\end{align}
Then, by Theorem~\ref{thm-tr:main-theorem}, we can achieve the rate region in
the statement of the theorem. That is, we can pick distributions $p_{X}$ such
that $H(p_{X})=\lambda\log d$ for all $\lambda\in\left[  0,1\right]  $.

We now bound the quantum dynamic capacity formula for $(\mathcal{N}%
^{\varepsilon})^{\otimes n}$ from above for all positive integers $n$, and as
a consequence, establish that the region given in the statement of the theorem
is optimal for $\varepsilon\in\left[  0,1/2\right]  $. To begin with, let us
consider the quantum dynamic capacity formula for an erasure channel
$\mathcal{N}_{A_{1}\rightarrow B_{1}}^{\varepsilon}$ and an arbitrary
degradable channel $\mathcal{M}_{A_{2}\rightarrow B_{2}}$. We use the same
idea as above, applying the isometry in \eqref{eq-tr:erasure-isometry} to the
output $B_{1}$ of the erasure channel. The quantum dynamic capacity formula
for $\mathcal{N}_{A_{1}\rightarrow B_{1}}^{\varepsilon}\otimes\mathcal{M}%
_{A_{2}\rightarrow B_{2}}$ becomes%
\begin{multline}
\lambda_{1}I(AX;B_{1}YB_{2})_{\sigma}+\lambda_{2}I(A\rangle B_{1}%
YB_{2}X)_{\sigma}\\
+\lambda_{3}\left[  I(X;B_{1}YB_{2})_{\sigma}+I(A\rangle B_{1}YB_{2}%
X)_{\sigma}\right]  , \label{eq-tr:QDCF-erasure-dev}%
\end{multline}
where%
\begin{align}
\sigma_{XAB_{1}B_{2}Y}  &  \equiv\sigma_{XAA_{1}B_{2}}^{0}\otimes\left(
1-\varepsilon\right)  |0\rangle\langle0|_{Y}+\sigma_{XAB_{1}B_{2}}^{1}%
\otimes\varepsilon|1\rangle\langle1|_{Y},\\
\sigma_{XAA_{1}B_{2}}^{0}  &  \equiv\sum_{x}p_{X}(x)|x\rangle\langle
x|_{X}\otimes\mathcal{M}_{A_{2}\rightarrow B_{2}}(\phi_{AA_{1}A_{2}}^{x}),\\
\sigma_{XAB_{1}B_{2}}^{1}  &  \equiv\sum_{x}p_{X}(x)|x\rangle\langle
x|_{X}\otimes\mathcal{M}_{A_{2}\rightarrow B_{2}}(\phi_{AA_{2}}^{x}%
)\otimes|e\rangle\langle e|_{B_{1}}.
\end{align}
Observe that $\sigma_{XAB_{2}}^{0}=\sigma_{XAB_{2}}^{1}=\sigma_{XAB_{2}}$, so
that
\begin{equation}
\sigma_{XAB_{2}Y}=\sigma_{XAB_{2}}\otimes\left[  \left(  1-\varepsilon\right)
|0\rangle\langle0|_{Y}+\varepsilon|1\rangle\langle1|_{Y}\right]  .
\end{equation}
First we handle the quantum mutual information term:%
\begin{align}
I(AX;B_{1}YB_{2})_{\sigma}  &  =I(AX;B_{1}B_{2}|Y)_{\sigma}\\
&  =\left(  1-\varepsilon\right)  I(AX;A_{1}B_{2})_{\sigma^{0}}+\varepsilon
I(AX;B_{2})_{\sigma^{1}}\\
&  =\left(  1-\varepsilon\right)  \left[  I(AX;B_{2})_{\sigma^{0}}%
+I(AX;A_{1}|B_{2})_{\sigma^{0}}\right]  +\varepsilon I(AX;B_{2})_{\sigma^{1}%
}\\
&  =\left(  1-\varepsilon\right)  \left[  I(AX;B_{2})_{\sigma}+I(AX;A_{1}%
|B_{2})_{\sigma^{0}}\right]  +\varepsilon I(AX;B_{2})_{\sigma}\\
&  =\left(  1-\varepsilon\right)  I(AX;A_{1}|B_{2})_{\sigma^{0}}%
+I(AX;B_{2})_{\sigma}.
\end{align}
We can bound the term $I(AX;A_{1}|B_{2})_{\sigma^{0}}$\ from above as follows:%
\begin{align}
I(AX;A_{1}|B_{2})_{\sigma^{0}}  &  =H(A_{1}|B_{2})_{\sigma^{0}}-H(A_{1}%
|B_{2}AX)_{\sigma^{0}}\\
&  =H(A_{1}|B_{2})_{\sigma^{0}}+H(A_{1}|E_{2}X)_{\sigma^{0}}\\
&  \leq\log d+H(A_{1}|E_{2}X)_{\sigma^{0}},
\end{align}
where $\sigma^{0}$ is defined below in \eqref{eq-tr:sigma_0-erasure}. This
then gives the following bound:%
\begin{equation}
I(AX;B_{1}YB_{2})_{\sigma}\leq\left(  1-\varepsilon\right)  \left[  \log
d+H(A_{1}|E_{2}X)_{\sigma^{0}}\right]  +I(AX;B_{2})_{\sigma} \label{eq-tr:q-dyn-eras-EA-term} .
\end{equation}
Now we handle the Holevo information term, and the development is similar:%
\begin{align}
I(X;B_{1}YB_{2})_{\sigma}  &  =I(X;B_{1}B_{2}|Y)_{\sigma}\\
&  =\left(  1-\varepsilon\right)  I(X;A_{1}B_{2})_{\sigma^{0}}+\varepsilon
I(X;B_{2})_{\sigma^{1}}\\
&  =\left(  1-\varepsilon\right)  \left[  I(X;B_{2})_{\sigma^{0}}%
+I(X;A_{1}|B_{2})_{\sigma^{0}}\right]  +\varepsilon I(X;B_{2})_{\sigma^{1}}\\
&  =\left(  1-\varepsilon\right)  \left[  I(X;B_{2})_{\sigma}+I(X;A_{1}%
|B_{2})_{\sigma^{0}}\right]  +\varepsilon I(X;B_{2})_{\sigma}\\
&  =\left(  1-\varepsilon\right)  I(X;A_{1}|B_{2})_{\sigma^{0}}+I(X;B_{2}%
)_{\sigma}. \label{eq-tr:q-dyn-eras-holevo}
\end{align}
We finally handle the coherent information term. To this end, note that an
isometric extension of the erasure channel is as follows: $|\psi\rangle
_{A}\rightarrow\sqrt{1-\varepsilon}|\psi\rangle_{B}|e\rangle_{E}%
+\sqrt{\varepsilon}|e\rangle_{B}|\psi\rangle_{E}$. Both the receiver and the
environment can perform the isometry in \eqref{eq-tr:erasure-isometry} on
their systems, without affecting the information quantities. Let
$U_{A_{2}\rightarrow B_{2}E_{2}}^{\mathcal{M}}$ be an isometric extension of
$\mathcal{M}_{A_{2}\rightarrow B_{2}}$ and define%
\begin{multline}
|\sigma\rangle_{XX^{\prime}AB_{1}B_{2}E_{1}E_{2}YZ}\equiv|\sigma^{0}%
\rangle_{XX^{\prime}AA_{1}B_{2}E_{1}E_{2}}\otimes\sqrt{1-\varepsilon}%
|0\rangle_{Y}|1\rangle_{Z}\\
+|\sigma^{1}\rangle_{XX^{\prime}AB_{1}B_{2}A_{1}E_{2}}\otimes\sqrt
{\varepsilon}|1\rangle_{Y}|0\rangle_{Z},
\end{multline}
where%
\begin{align}
|\sigma^{0}\rangle_{XX^{\prime}AA_{1}B_{2}E_{1}E_{2}}  &  \equiv\sum_{x}%
\sqrt{p_{X}(x)}|x\rangle_{X}|x\rangle_{X^{\prime}}\otimes U_{A_{2}\rightarrow
B_{2}E_{2}}^{\mathcal{M}}|\phi^{x}\rangle_{AA_{1}A_{2}}|e\rangle_{E_{1}%
},\label{eq-tr:sigma_0-erasure}\\
|\sigma^{1}\rangle_{XX^{\prime}AB_{1}B_{2}A_{1}E_{2}}  &  \equiv\sum_{x}%
\sqrt{p_{X}(x)}|x\rangle_{X}|x\rangle_{X^{\prime}}\otimes U_{A_{2}\rightarrow
B_{2}E_{2}}^{\mathcal{M}}|\phi^{x}\rangle_{AA_{1}A_{2}}|e\rangle_{B_{1}}.
\end{align}
Then%
\begin{align}
&  I(A\rangle B_{1}YB_{2}X)_{\sigma}=I(A\rangle B_{1}B_{2}|XY)_{\sigma
}\nonumber\\
&  =H(B_{1}B_{2}|YX)_{\sigma}-H(E_{1}E_{2}|YX)_{\sigma}\\
&  =\left[  \left(  1-\varepsilon\right)  H(A_{1}B_{2}|X)_{\sigma^{0}%
}+\varepsilon H(B_{2}|X)_{\sigma^{1}}\right] \nonumber\\
&  \ \ \ \ -\left[  \left(  1-\varepsilon\right)  H(E_{2}|X)_{\sigma^{0}%
}+\varepsilon H(A_{1}E_{2}|X)_{\sigma^{1}}\right] \\
&  =\left[  \left(  1-\varepsilon\right)  \left(  H(B_{2}|X)_{\sigma^{0}%
}+H(A_{1}|B_{2}X)_{\sigma^{0}}\right)  +\varepsilon H(B_{2}|X)_{\sigma^{1}%
}\right] \nonumber\\
&  \ \ \ \ -\left[  \left(  1-\varepsilon\right)  H(E_{2}|X)_{\sigma^{0}%
}+\varepsilon\left(  H(E_{2}|X)_{\sigma^{1}}+H(A_{1}|E_{2}X)_{\sigma^{1}%
}\right)  \right] \\
&  =\left[  \left(  1-\varepsilon\right)  \left(  H(B_{2}|X)_{\sigma}%
+H(A_{1}|B_{2}X)_{\sigma^{0}}\right)  +\varepsilon H(B_{2}|X)_{\sigma}\right]
\nonumber\\
&  \ \ \ \ -\left[  \left(  1-\varepsilon\right)  H(E_{2}|X)_{\sigma
}+\varepsilon\left(  H(E_{2}|X)_{\sigma}+H(A_{1}|E_{2}X)_{\sigma^{1}}\right)
\right] \\
&  =H(B_{2}|X)_{\sigma}-H(E_{2}|X)_{\sigma}+\left(  1-\varepsilon\right)
H(A_{1}|B_{2}X)_{\sigma^{0}}-\varepsilon H(A_{1}|E_{2}X)_{\sigma^{1}}\\
&  =H(B_{2}|X)_{\sigma}-H(E_{2}|X)_{\sigma}+\left(  1-\varepsilon\right)
H(A_{1}|B_{2}X)_{\sigma^{0}}-\varepsilon H(A_{1}|E_{2}X)_{\sigma^{0}}. \label{eq-tr:q-dyn-eras-coh-info}
\end{align}
Applying the assumption that the channel $\mathcal{M}$ is degradable (so that
$H(A_{1}|B_{2}X)_{\sigma^{0}}\leq H(A_{1}|E_{2}X)_{\sigma^{0}}$), we find the
upper bound%
\begin{align}
&  \leq H(B_{2}|X)_{\sigma}-H(E_{2}|X)_{\sigma}+\left(  1-\varepsilon\right)
H(A_{1}|E_{2}X)_{\sigma^{0}}-\varepsilon H(A_{1}|E_{2}X)_{\sigma^{0}}\\
&  =H(B_{2}|X)_{\sigma}-H(E_{2}|X)_{\sigma}+\left(  1-2\varepsilon\right)
H(A_{1}|E_{2}X)_{\sigma^{0}}. \label{eq-tr:q-dyn-eras-coh-info-bnd}
\end{align}
It also follows from
\eqref{eq-tr:q-dyn-eras-holevo} and \eqref{eq-tr:q-dyn-eras-coh-info} that%
\begin{align}
&  I(X;B_{1}YB_{2})_{\sigma}+I(A\rangle B_{1}YB_{2}X)_{\sigma}\nonumber\\
&  =\left(  1-\varepsilon\right)  I(X;A_{1}|B_{2})_{\sigma^{0}}+I(X;B_{2}%
)_{\sigma}+H(B_{2}|X)_{\sigma}-H(E_{2}|X)_{\sigma}\nonumber\\
&  \ \ \ \ \ \ +\left(  1-\varepsilon\right)  H(A_{1}|B_{2}X)_{\sigma^{0}%
}-\varepsilon H(A_{1}|E_{2}X)_{\sigma^{0}}\\
&  =I(X;B_{2})_{\sigma}+H(B_{2}|X)_{\sigma}-H(E_{2}|X)_{\sigma}\nonumber\\
&  \ \ \ \ \ \ +\left(  1-\varepsilon\right)  H(A_{1}|B_{2})_{\sigma^{0}%
}-\left(  1-\varepsilon\right)  H(A_{1}|B_{2}X)_{\sigma^{0}}\nonumber\\
&  \ \ \ \ \ \ +\left(  1-\varepsilon\right)  H(A_{1}|B_{2}X)_{\sigma^{0}%
}-\varepsilon H(A_{1}|E_{2}X)_{\sigma^{0}}\\
&  =I(X;B_{2})_{\sigma}+H(B_{2}|X)_{\sigma}-H(E_{2}|X)_{\sigma}+\left(
1-\varepsilon\right)  H(A_{1}|B_{2})_{\sigma^{0}}-\varepsilon H(A_{1}%
|E_{2}X)_{\sigma^{0}}\\
&  \leq I(X;B_{2})_{\sigma}+H(B_{2}|X)_{\sigma}-H(E_{2}|X)_{\sigma}+\left(
1-\varepsilon\right)  \log d-\varepsilon H(A_{1}|E_{2}X)_{\sigma^{0}} \label{eq-tr:q-dyn-eras-Hol-info-bnd}
\end{align}
Now putting together \eqref{eq-tr:q-dyn-eras-EA-term},
 \eqref{eq-tr:q-dyn-eras-coh-info-bnd},
and \eqref{eq-tr:q-dyn-eras-Hol-info-bnd}, we find the following upper bound on the
quantum dynamic capacity formula in \eqref{eq-tr:QDCF-erasure-dev}:%
\begin{align}
&  \lambda_{1}I(AX;B_{1}YB_{2})_{\sigma}+\lambda_{2}I(A\rangle B_{1}%
YB_{2}X)_{\sigma}\nonumber\\
&  \ \ \ \ +\lambda_{3}\left[  I(X;B_{1}YB_{2})_{\sigma}+I(A\rangle
B_{1}YB_{2}X)_{\sigma}\right] \\
&  \leq\lambda_{1}\left(  1-\varepsilon\right)  \left[  \log d+H(A_{1}%
|E_{2}X)_{\sigma^{0}}\right]  +\lambda_{2}\left(  1-2\varepsilon\right)
H(A_{1}|E_{2}X)_{\sigma^{0}}\nonumber\\
&  \ \ \ +\lambda_{3}\left[  \left(  1-\varepsilon\right)  \log d-\varepsilon
H(A_{1}|E_{2}X)_{\sigma^{0}}\right] \nonumber\\
&  \ \ \ +\lambda_{1}I(AX;B_{2})_{\sigma}+\lambda_{2}\left[  H(B_{2}%
|X)_{\sigma}-H(E_{2}|X)_{\sigma}\right] \nonumber\\
&  \ \ \ +\lambda_{3}\left[  I(X;B_{2})_{\sigma}+H(B_{2}|X)_{\sigma}%
-H(E_{2}|X)_{\sigma}\right] \\
&  =\left(  \lambda_{1}+\lambda_{3}\right)  \left(  1-\varepsilon\right)  \log
d+\left[  \lambda_{1}\left(  1-\varepsilon\right)  +\lambda_{2}\left(
1-2\varepsilon\right)  -\lambda_{3}\varepsilon\right]  H(A_{1}|E_{2}%
X)_{\sigma^{0}}\nonumber\\
&  \ \ \ \ +\lambda_{1}I(AX;B_{2})_{\sigma}+\lambda_{2}\left[  H(B_{2}%
|X)_{\sigma}-H(E_{2}|X)_{\sigma}\right] \nonumber\\
&  \ \ \ \ +\lambda_{3}\left[  I(X;B_{2})_{\sigma}+H(B_{2}|X)_{\sigma}%
-H(E_{2}|X)_{\sigma}\right]  \label{eq-tr:branch-point-erasure} .
\end{align}
In the case that $\lambda_{1}\left(  1-\varepsilon\right)  +\lambda_{2}\left(
1-2\varepsilon\right)  -\lambda_{3}\varepsilon\geq0$, we can apply data
processing ($H(A_{1}|E_{2}X)_{\sigma^{0}}\leq H(A_{1}|X)_{\sigma^{0}}$) to
find that the last line is never larger than%
\begin{multline}
\left(  \lambda_{1}+\lambda_{3}\right)  \left(  1-\varepsilon\right)  \log
d+\left[  \lambda_{1}\left(  1-\varepsilon\right)  +\lambda_{2}\left(
1-2\varepsilon\right)  -\lambda_{3}\varepsilon\right]  H(A_{1}|X)_{\sigma^{0}%
}\\
+\lambda_{1}I(AX;B_{2})_{\sigma}+\lambda_{2}\left[  H(B_{2}|X)_{\sigma
}-H(E_{2}|X)_{\sigma}\right] \\
+\lambda_{3}\left[  I(X;B_{2})_{\sigma}+H(B_{2}|X)_{\sigma}-H(E_{2}%
|X)_{\sigma}\right]  .
\end{multline}
If we now take $\mathcal{M}_{A_{2}\rightarrow B_{2}}=(\mathcal{N}%
^{\varepsilon})^{\otimes n-1}$ and iterate this development $n-1$ more times,
we find that%
\begin{align}
\frac{1}{n}D_{\vec{\lambda}}((\mathcal{N}^{\varepsilon})^{\otimes n})  &
\leq\left(  \lambda_{1}+\lambda_{3}\right)  \left(  1-\varepsilon\right)  \log
d\nonumber\\
&  \ \ \ +\left[  \lambda_{1}\left(  1-\varepsilon\right)  +\lambda_{2}\left(
1-2\varepsilon\right)  -\lambda_{3}\varepsilon\right]  \left[  \frac{1}{n}%
\sum_{i=1}^{n}H(A_{i}|X)\right] \\
&  \leq\left(  \lambda_{1}+\lambda_{3}\right)  \left(  1-\varepsilon\right)
\log d\nonumber\\
&  \ \ \ +\left[  \lambda_{1}\left(  1-\varepsilon\right)  +\lambda_{2}\left(
1-2\varepsilon\right)  -\lambda_{3}\varepsilon\right]  \max_{i,x}%
H(A_{i})_{\phi^{x}}.
\end{align}
This establishes the optimality of the region whenever $\lambda_{1}\left(
1-\varepsilon\right)  +\lambda_{2}\left(  1-2\varepsilon\right)  -\lambda
_{3}\varepsilon\geq0$. For the case when $\lambda_{1}\left(  1-\varepsilon
\right)  +\lambda_{2}\left(  1-2\varepsilon\right)  -\lambda_{3}\varepsilon
<0$, we start from \eqref{eq-tr:branch-point-erasure} and use data processing
for the channel $\operatorname{Tr}_{B_{2}}\{\mathcal{U}_{A_{2}\rightarrow
B_{2}E_{2}}^{\mathcal{M}}(\cdot)\}$, giving $H(A_{1}|E_{2}X)_{\sigma^{0}}\geq
H(A_{1}|A_{2}X)_{\sigma^{0}}$, to find that%
\begin{multline}
\left(  \lambda_{1}+\lambda_{3}\right)  \left(  1-\varepsilon\right)  \log
d+\left[  \lambda_{1}\left(  1-\varepsilon\right)  +\lambda_{2}\left(
1-2\varepsilon\right)  -\lambda_{3}\varepsilon\right]  H(A_{1}|E_{2}%
X)_{\sigma^{0}}+\lambda_{1}I(AX;B_{2})_{\sigma}\\
+\lambda_{2}\left[  H(B_{2}|X)_{\sigma}-H(E_{2}|X)_{\sigma}\right]
+\lambda_{3}\left[  I(X;B_{2})_{\sigma}+H(B_{2}|X)_{\sigma}-H(E_{2}%
|X)_{\sigma}\right] \\
\leq\left(  \lambda_{1}+\lambda_{3}\right)  \left(  1-\varepsilon\right)  \log
d+\left[  \lambda_{1}\left(  1-\varepsilon\right)  +\lambda_{2}\left(
1-2\varepsilon\right)  -\lambda_{3}\varepsilon\right]  H(A_{1}|A_{2}%
X)_{\sigma^{0}}\\
+\lambda_{1}I(AX;B_{2})_{\sigma}+\lambda_{2}\left[  H(B_{2}|X)_{\sigma
}-H(E_{2}|X)_{\sigma}\right] \\
+\lambda_{3}\left[  I(X;B_{2})_{\sigma}+H(B_{2}|X)_{\sigma}-H(E_{2}%
|X)_{\sigma}\right] .
\end{multline}
If we now take $\mathcal{M}_{A_{2}\rightarrow B_{2}}=(\mathcal{N}%
^{\varepsilon})^{\otimes n-1}$ and iterate this development $n-1$ more times,
we find that%
\begin{align}
\frac{1}{n}D_{\vec{\lambda}}((\mathcal{N}^{\varepsilon})^{\otimes n})  &
\leq\left(  \lambda_{1}+\lambda_{3}\right)  \left(  1-\varepsilon\right)  \log
d\nonumber\\
&  \ \ \ \ +\left[  \lambda_{1}\left(  1-\varepsilon\right)  +\lambda
_{2}\left(  1-2\varepsilon\right)  -\lambda_{3}\varepsilon\right]  \left[
\frac{1}{n}\sum_{i=1}^{n}H(A_{i}|A^{i-1}X)\right] \\
&  =\left(  \lambda_{1}+\lambda_{3}\right)  \left(  1-\varepsilon\right)  \log
d\nonumber\\
&  \ \ \ \ +\left[  \lambda_{1}\left(  1-\varepsilon\right)  +\lambda
_{2}\left(  1-2\varepsilon\right)  -\lambda_{3}\varepsilon\right]  \left[
\frac{1}{n}H(A^{n}|X)\right] \\
&  \leq\left(  \lambda_{1}+\lambda_{3}\right)  \left(  1-\varepsilon\right)
\log d.
\end{align}
The last line follows because entropy is non-negative. This inequality shows
that the inequality $\lambda_{1}\left(  1-\varepsilon\right)  +\lambda
_{2}\left(  1-2\varepsilon\right)  -\lambda_{3}\varepsilon<0$ holding implies
that it is optimal to pick $\lambda=0$ in the statement of the theorem (which
just corresponds to the case in which we are maximizing the unassisted
classical capacity). This completes the proof.
\end{proof}

\subsection{The Pure-Loss Bosonic Channel}

One of the most important practical
\index{bosonic channel}
channels in quantum communication is known as the pure-loss bosonic channel.
This channel can model the communication of photons through free space or over
a fiber optic cable because the main source of noise in these settings is just
the loss of photons. The pure-loss bosonic channel has one parameter $\eta
\in\left[  0,1\right]  $ that characterizes the fraction of photons that make
it through the channel to the receiver on average. The environment Eve is able
to collect all of the photons that do not make it to the receiver---this
fraction is $1-\eta$. Usually, we also restrict the mean number of photons
that the sender is allowed to send through the channel (if we do not do so,
then there could be an infinite amount of energy available, which is
unphysical from a practical perspective, and furthermore, some of the
capacities become infinite, which is less interesting from an
information-theoretical perspective). So, we let $N_{S}\in\lbrack0,\infty)$ be
the mean number of photons available at the transmitter. Capacities of this
channel are then a function of these two parameters~$\eta$ and$~N_{S}$.

\begin{exercise}
Prove that the quantum capacity of a pure-loss bosonic channel vanishes when
$\eta=1/2$.
\end{exercise}

In this section, we show how trade-off coding for this channel can give a
remarkable gain over time sharing. Trade-off coding for this channel amounts
to a power-sharing strategy, in which the sender dedicates a fraction~$\lambda
\in[0,1] $ of the available photons to the quantum part of the code and the
other fraction~$1-\lambda$ to the classical part of the code. This
power-sharing strategy is provably optimal (up to a long-standing conjecture)
and can beat time sharing by significant margins (much more so than the
dephasing channel does, for example). Specifically, recall that a trade-off
coding strategy has the sender and receiver generate random codes from an
ensemble of the following form:%
\begin{equation}
\left\{  p_{X}(x),|\phi_{x}\rangle_{AA^{\prime}}\right\}  ,
\end{equation}
where $p_{X}(x)$ is some distribution and the states $|\phi_{x}\rangle
_{AA^{\prime}}$ are correlated with this distribution, with Alice feeding
system $A^{\prime}$ into the channel. For the pure-loss bosonic channel, it
turns out that the best ensemble to choose is of the following form:%
\begin{equation}
\left\{  p_{\left(  1-\lambda\right)  N_{S}}(\alpha),D_{A^{\prime}}%
(\alpha)|\psi_{\operatorname{TMS}}(\lambda)\rangle_{AA^{\prime}}\right\}  ,
\label{eq-tr:bosonic-ensemble}%
\end{equation}
where $\alpha$ is a complex variable. The distribution $p_{\left(
1-\lambda\right)  N_{S}}(\alpha)$ is an isotropic Gaussian distribution with
variance $\left(  1-\lambda\right)  N_{S}$:%
\begin{equation}
p_{\left(  1-\lambda\right)  N_{S}}(\alpha)\equiv\frac{1}{\pi\left(
1-\lambda\right)  N_{S}}\exp\left\{  -\left\vert \alpha\right\vert
^{2}/\left[  \left(  1-\lambda\right)  N_{S}\right]  \right\}  ,
\label{eq-tr:gaussian-prior}%
\end{equation}
where $\lambda\in\left[  0,1\right]  $ is the power-sharing or
photon-number-sharing parameter, indicating how many photons to dedicate to
the quantum part of the code, while $1-\lambda$ indicates how many photons to
dedicate to the classical part. In~\eqref{eq-tr:bosonic-ensemble},
$D_{A^{\prime}}(\alpha)$ is a \textquotedblleft displacement\textquotedblright%
\ unitary operator%
\index{displacement operator}
acting on system $A^{\prime}$ (more on this below), and $|\psi
_{\operatorname{TMS}}(\lambda)\rangle_{AA^{\prime}}$ is a \textquotedblleft
two-mode squeezed\textquotedblright\ (TMS) state
\index{two-mode squeezed state}%
of the following form:%
\begin{equation}
|\psi_{\operatorname{TMS}}(\lambda)\rangle_{AA^{\prime}}\equiv\sum
_{n=0}^{\infty}\sqrt{\frac{\left[  \lambda N_{S}\right]  ^{n}}{\left[  \lambda
N_{S}+1\right]  ^{n+1}}}|n\rangle_{A}|n\rangle_{A^{\prime}},
\label{eq-tr:two-mode-squeezed}%
\end{equation}
where $\vert n \rangle$ is a state of definite photon number $n$. Let
$\theta(\lambda)$ denote the state resulting from tracing over the mode $A$:%
\begin{align}
\theta(\lambda)  &  \equiv\operatorname{Tr}_{A}\left\{  |\psi
_{\operatorname{TMS}}(\lambda)\rangle\langle\psi_{\operatorname{TMS}}%
(\lambda)|_{AA^{\prime}}\right\} \\
&  =\sum_{n=0}^{\infty}\frac{\left[  \lambda N_{S}\right]  ^{n}}{\left[
\lambda N_{S}+1\right]  ^{n+1}}|n\rangle\langle n|_{A^{\prime}}.
\end{align}
The reduced state $\theta(\lambda)$ is known as a thermal
\index{thermal state}%
state with mean photon number $\lambda N_{S}$. We can readily check that its
mean photon number is $\lambda N_{S}$ simply by computing the expectation of
the photon number$~n$ with respect to the geometric distribution $\left[
\lambda N_{S}\right]  ^{n}/\left[  \lambda N_{S}+1\right]  ^{n+1}$:%
\begin{equation}
\sum_{n=0}^{\infty}n\frac{\left[  \lambda N_{S}\right]  ^{n}}{\left[  \lambda
N_{S}+1\right]  ^{n+1}}=\lambda N_{S}.
\end{equation}
The most important property of the displacement operators $D_{A^{\prime}%
}(\alpha)$ for our purposes is that averaging over a random choice of them
according to the Gaussian distribution $p_{\left(  1-\lambda\right)  N_{S}%
}(\alpha)$, where each operator acts on the state $\theta$, gives a thermal
state with mean photon number$~N_{S}$:%
\begin{align}
\overline{\theta}  &  \equiv\int d\alpha\ p_{\left(  1-\lambda\right)  N_{S}%
}(\alpha)\ D(\alpha)\theta(\lambda)D^{\dag}(\alpha)\\
&  =\sum_{n=0}^{\infty}\frac{\left[  N_{S}\right]  ^{n}}{\left[
N_{S}+1\right]  ^{n+1}}|n\rangle\langle n|_{A^{\prime}}.
\end{align}
Thus, the choice of ensemble in \eqref{eq-tr:bosonic-ensemble} meets the
constraint that the average number of photons input to the channel be equal to
$N_{S}$.

In order to calculate the quantum dynamic capacity region for this pure-loss
bosonic channel, it is helpful to observe that the entropy of a thermal state
with mean number of photons~$N_{S}$ is equal to%
\begin{equation}
g(N_{S})\equiv\left(  N_{S}+1\right)  \log\left(  N_{S}+1\right)  -N_{S}%
\log(N_{S}), \label{eq-tr:thermal-entropy}%
\end{equation}
because we will evaluate all of the relevant entropies on thermal states. From
Exercise~\ref{ex-tr:four-entropies}, we know that we should evaluate just the
following four entropies:%
\begin{align}
H(A|X)_{\sigma}  &  =\int d\alpha\ p_{\left(  1-\lambda\right)  N_{S}}%
(\alpha)\ H(D(\alpha)\theta(\lambda)D^{\dag}(\alpha)),\\
H(B)_{\sigma}  &  =H(\mathcal{N}(\overline{\theta})),\\
H(B|X)_{\sigma}  &  =\int d\alpha\ p_{\left(  1-\lambda\right)  N_{S}}%
(\alpha)\ H(\mathcal{N}(D(\alpha)\theta(\lambda)D^{\dag}(\alpha
))),\label{eq-tr:third-ent}\\
H(E|X)_{\sigma}  &  =\int d\alpha\ p_{\left(  1-\lambda\right)  N_{S}}%
(\alpha)\ H(\mathcal{N}^{c}(D(\alpha)\theta(\lambda)D^{\dag}(\alpha))),
\label{eq-tr:fourth-ent}%
\end{align}
where $\mathcal{N}$ is the pure-loss bosonic channel that transmits $\eta$ of
the input photons to the receiver and $\mathcal{N}^{c}$ is the complementary
channel that transmits $1-\eta$ of the input photons to the environment Eve.
We proceed with calculating the above four entropies:%
\begin{align}
\int d\alpha\ p_{\left(  1-\lambda\right)  N_{S}}(\alpha)\ H(D(\alpha
)\theta(\lambda)D^{\dag}(\alpha))  &  =\int d\alpha\ p_{\left(  1-\lambda
\right)  N_{S}}(\alpha)\ H(\theta(\lambda))\\
&  =H(\theta(\lambda))=g\left(  \lambda N_{S}\right)  .
\end{align}
The first equality follows because $D(\alpha)$ is a unitary operator, and the
third equality follows because $\theta$ is a thermal state with mean photon
number~$N_{S}$. Continuing, we have%
\begin{equation}
H(\mathcal{N}(\overline{\theta}))=g(\eta N_{S}),
\end{equation}
because $\overline{\theta}$ is a thermal state with mean photon number $N_{S}%
$, but the channel only lets a fraction~$\eta$ of the input photons through on
average. The third entropy in \eqref{eq-tr:third-ent} equals%
\begin{align}
&  \int d\alpha\ p_{\left(  1-\lambda\right)  N_{S}}(\alpha)\ H(\mathcal{N}%
(D(\alpha)\theta(\lambda)D^{\dag}(\alpha)))\nonumber\\
&  =\int d\alpha\ p_{\left(  1-\lambda\right)  N_{S}}(\alpha)\ H(D(\sqrt{\eta
}\alpha)\mathcal{N}(\theta(\lambda))D^{\dag}(\sqrt{\eta}\alpha))\\
&  =\int d\alpha\ p_{\left(  1-\lambda\right)  N_{S}}(\alpha)\ H(\mathcal{N}%
(\theta(\lambda)))\\
&  =H(\mathcal{N}(\theta(\lambda)))=g(\lambda\eta N_{S}).
\end{align}
The first equality follows because a displacement operator is covariant with
respect to the pure-loss channel (we do not justify this rigorously here). The
second equality follows because $D(\alpha)$ is a unitary operator. The final
equality follows because $\theta(\lambda)$ is a thermal state with mean photon
number $\lambda N_{S}$, but the channel only lets a fraction~$\eta$ of the
input photons through on average. By the same line of reasoning (except that
the complementary channel lets through only a fraction~$1-\eta$ of the input
photons), the fourth entropy in \eqref{eq-tr:fourth-ent} is equal to%
\begin{align}
&  \int d\alpha\ p_{\left(  1-\lambda\right)  N_{S}}(\alpha)\ H(\mathcal{N}%
^{c}(D(\alpha)\theta(\lambda)D^{\dag}(\alpha)))\nonumber\\
&  =\int d\alpha\ p_{\left(  1-\lambda\right)  N_{S}}(\alpha)\ H(D(\sqrt
{1-\eta}\alpha)\mathcal{N}^{c}(\theta(\lambda))D^{\dag}(\sqrt{1-\eta}%
\alpha))\\
&  =\int d\alpha\ p_{\left(  1-\lambda\right)  N_{S}}(\alpha)\ H(\mathcal{N}%
^{c}(\theta(\lambda)))\\
&  =H(\mathcal{N}^{c}(\theta(\lambda)))=g(\lambda(1-\eta)N_{S}).
\end{align}
Then, by the result of Exercise~\ref{ex-tr:four-entropies}\ and a matching
converse that holds whenever $\eta\geq1/2$,\footnote{We should clarify that
the converse holds only if a long-standing minimum-output entropy conjecture
is true (researchers have collected much evidence that it should be true).} we
have the following characterization of the quantum dynamic capacity region of
the pure-loss bosonic channel.%
\begin{figure}
[ptb]
\begin{center}
\includegraphics[
width=4.8456in
]%
{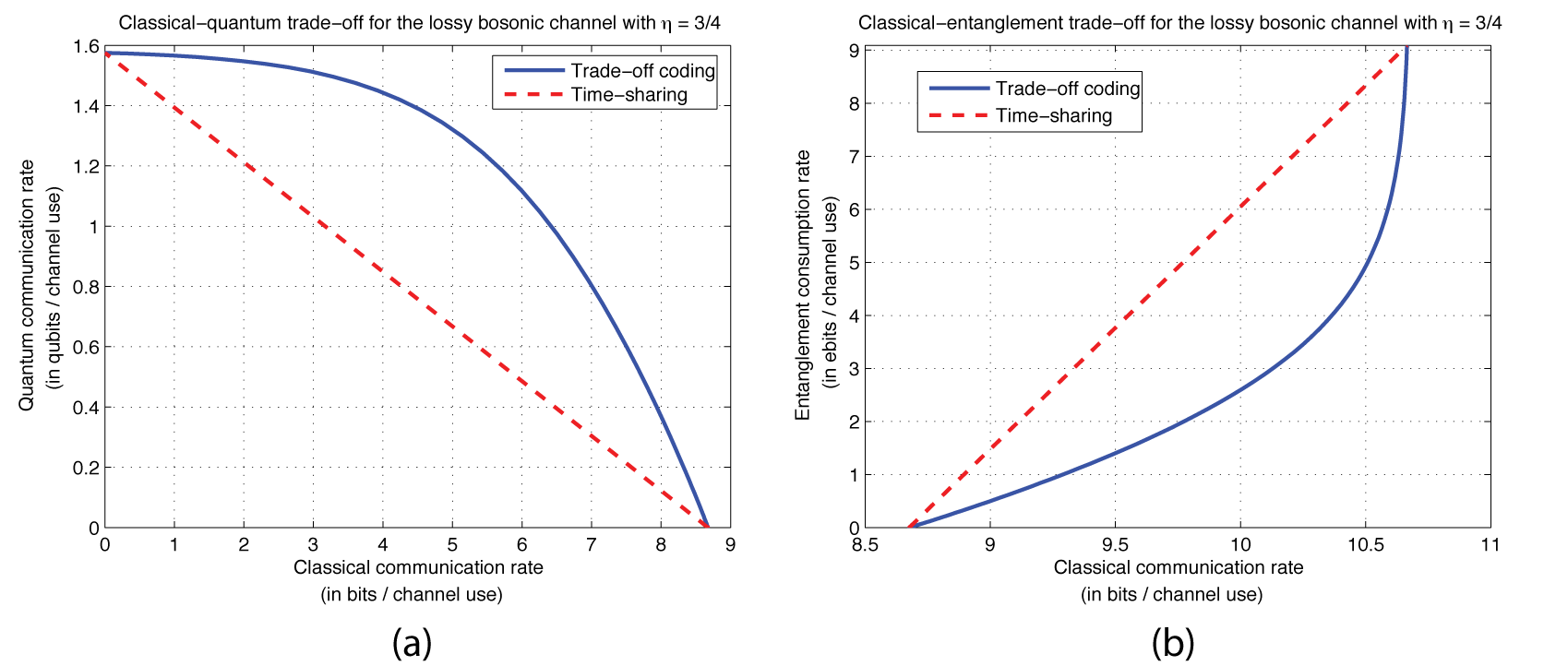}%
\caption{(a) Suppose a channel transmits on average $3/4$ of the photons to
the receiver, while losing the other $1/4$ en route. Such a channel can
reliably transmit a maximum of $\log\left(  3/4\right)  -\log\left(
1/4\right)  \approx1.58$ qubits per channel use, and a mean photon budget of
about $200$~photons per channel use at the transmitter is sufficient to nearly
achieve this quantum capacity. A trade-off coding strategy which lowers the
quantum data rate to about $1.4$~qubits per channel use while retaining the
same mean photon budget allows for a sender to reliably transmit an additional
$4.5$~classical bits per channel use, while time-sharing would only allow for
an additional $1$~classical bit per channel use with this photon budget. The
$6.5$~dB increase in the classical data rate that trade-off coding gives over
time-sharing for this example is strong enough to demand that quantum
communication engineers employ trade-off coding strategies in order to take
advantage of such theoretical performance gains. (b) The sender and the
receiver share entanglement, and the sender would like to transmit classical
information while minimizing the consumption of entanglement. With a mean
photon budget of $200$~photons per channel use over a channel that propagates
only $3/4$ of the photons input to it, the sender can reliably transmit a
maximum of about $10.7$~classical bits per channel use while consuming
entanglement at a rate of about $9.1$~entangled bits per channel use. With
trade-off coding, the sender can significantly reduce the entanglement
consumption rate to about $5$~entangled bits per channel use while still
transmitting about $10.5$~classical bits per channel use, only a $0.08$~dB
decrease in the rate of classical communication for a $2.6$~dB decrease in the
entanglement consumption rate. The savings in entanglement consumption could
be useful for them if they would like to have the extra entanglement for
future rounds of assisted communication.}%
\label{fig-tr:bosonic-trade-offs}%
\end{center}
\end{figure}

\begin{theorem}
Provided that Strong~Conjecture~2 of \citep{G08thesis} is true, the quantum dynamic capacity region for a pure-loss bosonic channel with
transmissivity $\eta\geq1/2$ is the union of regions of the form:%
\begin{align}
C+2Q  &  \leq g(\lambda N_{S})+g(\eta N_{S})-g(\left(  1-\eta\right)  \lambda
N_{S}),\label{eq:bosonic-region-1}\\
Q+E  &  \leq g(\eta\lambda N_{S})-g(\left(  1-\eta\right)  \lambda
N_{S}),\label{eq:bosonic-region-2}\\
C+Q+E  &  \leq g(\eta N_{S})-g(\left(  1-\eta\right)  \lambda N_{S}),
\label{eq:bosonic-region-3}%
\end{align}
where $\lambda\in\left[  0,1\right]  $ is a photon-number-sharing parameter
and $g(N)$ is the entropy 
defined in \eqref{eq-tr:thermal-entropy}. The region is achievable for all
$\eta \in [0,1]$.
\end{theorem}

\noindent Figure~\ref{fig-tr:bosonic-trade-offs} depicts two important special
cases of the region in the above theorem:\ (a) the trade-off between classical
and quantum communication without entanglement assistance and (b) the
trade-off between entanglement-assisted and unassisted classical
communication. The figure indicates the remarkable improvement over time
sharing that trade-off coding gives.

Other special cases of the above capacity region are the unassisted classical
capacity $g(\eta N_{S})$ when $\lambda_{1},\lambda_{2},Q,E=0$ and $\lambda
_{3}=1$, the quantum capacity $g(\eta N_{S})-g((1-\eta)N_{S})$ when
$\lambda_{2}=1$, $\lambda_{1},\lambda_{3},C,E=0$, and the
entanglement-assisted classical capacity $g(N_{S})+g(\eta N_{S})-g((1-\eta
)N_{S})$ when $\lambda_{1}=1$, $\lambda_{2},\lambda_{3},Q=0$, and $E=-\infty$.

\section{History and Further Reading}

\cite{Shor_CE} considered the classical capacity of a channel assisted by
rate-limited  entanglement. He calculated a trade-off curve
that determines how a sender can optimally trade the consumption of noiseless
entanglement with the generation of noiseless classical communication. This
trade-off curve also bounds a rate region consisting of rates of entanglement
consumption and generated classical communication. Shor's result then inspired
\cite{cmp2005dev} to consider a scenario in which a sender exploits a quantum
channel to simultaneously transmit both noiseless classical and quantum
information, a scenario later dubbed \textquotedblleft classically enhanced
quantum coding\textquotedblright\ in \citep{HW08GFP,HW09}\ after schemes
formulated in the theory of quantum error
correction~\citep{kremsky:012341,arx2008wildeUQCC}. \cite{cmp2005dev} provided
a multi-letter characterization of the classically enhanced quantum capacity
region for general channels, but they were able to show that both generalized
dephasing channels and erasure channels admit single-letter capacity regions.

The above scenarios are a part of the dynamic, double-resource quantum Shannon
theory, in which a sender can exploit a quantum channel to generate two
noiseless resources, or a sender can exploit a quantum channel in addition to
a noiseless resource to generate another noiseless resource. This theory
culminated with the work of \cite{DHW03,DHW05RI}, that provided a multi-letter
characterization for virtually every combination of two resources and a
quantum channel which one can consider. Other researchers concurrently
considered how noiseless resources might trade off against each other in tasks
outside of the dynamic, double-resource quantum Shannon theory, such as
quantum compression~\citep{KI01,BHJW01,hayden:4404}, remote state
preparation~\citep{PhysRevA.68.062319,BHLSW05}, and hybrid quantum memories~\citep{K03a}.

\cite{HW08GFP,HW09,WH10b} considered the dynamic, triple-resource quantum
Shannon theory by providing a multi-letter characterization of an
entanglement-assisted quantum channel's ability to transmit both classical and
quantum information. \cite{HW08GFP} also constructed a new protocol, dubbed
the \textquotedblleft classically enhanced father protocol,\textquotedblright%
\ that outperforms a time-sharing strategy for transmitting both classical and
quantum information over an entanglement-assisted quantum channel.
\cite{BHTW10} showed that the quantum Hadamard channels have a single-letter
capacity region. Later studies continued these efforts of exploring
information trade-offs~\citep{JBW11,WH10}.

\cite{WHG11,PhysRevA.86.062306} found the quantum dynamic capacity region of the
pure-loss bosonic channel (up to a long-standing minimum-output entropy
conjecture). The results there build on a tremendous body of literature for
bosonic channels. \cite{GGLMSY04} found the classical capacity of the
pure-loss bosonic channel. Others found the entanglement-assisted classical
and quantum capacities of the pure-loss bosonic
channel~\citep{ieee2002bennett,HW01,GLMS03,GLMS03a}\ and its quantum capacity
\citep{WPG07,GSE08}. The long-standing minimum-output entropy conjecture
(different from the one proved by \cite{GHG15}) is detailed in the papers of
\cite{GGLMS04,GSE08,GHLM10}.

\chapter{Summary and Outlook}

This brief final chapter serves as a compact summary of many results
presented in this book, it highlights information-processing tasks that we did
not cover, and it discusses new directions. We exploit the resource inequality
formalism in our summary.

A resource inequality is a statement of achievability:%
\begin{equation}
\sum_{k}\alpha_{k}\geq\sum_{j}\beta_{j},
\end{equation}
meaning that the resources $\left\{  \alpha_{k}\right\}  $ on the left-hand
side\ can simulate the resources $\left\{  \beta_{j}\right\}  $ on the
right-hand side. The simulation can be exact and finite or asymptotically
perfect. We can classify resources as follows:

\begin{enumerate}
\item Unit, noiseless, or noisy.

\item Dynamic or static. Moreover, dynamic resources can be \textit{relative}
(see below).

\item Classical, quantum, or hybrid.
\end{enumerate}

The unit resources are as follows: $\left[  c\rightarrow c\right]  $
represents one noiseless classical bit channel, $\left[  q\rightarrow
q\right]  $ represents one noiseless qubit channel, $\left[  qq\right]  $
represents one noiseless ebit, and $\left[  q\rightarrow qq\right]  $
represents one noiseless coherent bit channel. We also have $\left[
c\rightarrow c\right]  _{\operatorname{priv}}$ representing a noiseless private
classical bit channel and $\left[  cc\right]  _{\operatorname{priv}}$ representing a
noiseless bit of secret key. An example of a noiseless resource is a pure
bipartite state $\vert\phi\rangle_{AB}$ shared between Alice and Bob or an
identity channel $\operatorname{id}_{A\rightarrow B}$ from Alice to Bob. An
example of a noisy resource could be a mixed bipartite state $\rho_{AB}$ or a
noisy channel $\mathcal{N}_{A^{\prime}\rightarrow B}$. Unit resources are a
special case of noiseless resources, which are in turn a special case of noisy resources.

A shared state $\rho_{AB}$ is an example of a noisy static resource, and a
channel~$\mathcal{N}$\ is an example of a noisy dynamic resource. We indicate
these by $\left\langle \rho\right\rangle $ or $\left\langle \mathcal{N}%
\right\rangle $ in a resource inequality. We can be more precise if necessary
and write $\left\langle \mathcal{N}\right\rangle $ as a dynamic, relative
resource $\langle\mathcal{N}_{A^{\prime}\rightarrow B}:\sigma_{A^{\prime}%
}\rangle$, meaning that the protocol only works as it should if the state
input to the channel is $\sigma_{A^{\prime}}$.

It is obvious when a resource is classical or when it is quantum, and an
example of a hybrid resource is a classical--quantum state%
\begin{equation}
\rho_{XA}=\sum_{x}p_{X}( x) \vert x\rangle\langle x\vert_{X}\otimes\rho
_{A}^{x}.
\end{equation}

\section{Unit Protocols}

Chapter~\ref{chap:three-noiseless} discussed entanglement distribution%
\begin{equation}
\left[  q\rightarrow q\right]  \geq\left[  qq\right]  , \label{eq-sum:ED}%
\end{equation}
teleportation%
\begin{equation}
2\left[  c\rightarrow c\right]  +\left[  qq\right]  \geq\left[  q\rightarrow
q\right]  , \label{eq-sum:TP}%
\end{equation}
and super-dense coding%
\begin{equation}
\left[  q\rightarrow q\right]  +\left[  qq\right]  \geq2\left[  c\rightarrow
c\right]  . \label{eq-sum:SD}%
\end{equation}
Chapter~\ref{chap:coherent-communication} introduced coherent dense coding%
\begin{equation}
\left[  q\rightarrow q\right]  +\left[  qq\right]  \geq2\left[  q\rightarrow
qq\right]  ,
\end{equation}
and coherent teleportation%
\begin{equation}
2\left[  q\rightarrow qq\right]  \geq\left[  q\rightarrow q\right]  +\left[
qq\right]  .
\end{equation}
The fact that these two resource inequalities are dual under resource reversal
implies the coherent communication identity:%
\begin{equation}
2\left[  q\rightarrow qq\right]  =\left[  q\rightarrow q\right]  +\left[
qq\right]  .
\end{equation}
We also have the following resource inequalities:%
\begin{equation}
\left[  q\rightarrow q\right]  \geq\left[  q\rightarrow qq\right]  \geq\left[
qq\right]  .
\end{equation}

Other unit protocols not covered in this book are the one-time pad%
\begin{equation}
\left[  c\rightarrow c\right]  _{\operatorname{pub}}+\left[  cc\right]
_{\operatorname{priv}}\geq\left[  c\rightarrow c\right]  _{\operatorname{priv}%
},
\end{equation}
secret key distribution%
\begin{equation}
\left[  c\rightarrow c\right]  _{\operatorname{priv}}\geq\left[  cc\right]
_{\operatorname{priv}},
\end{equation}
and private-to-public transmission%
\begin{equation}
\left[  c\rightarrow c\right]  _{\operatorname{priv}}\geq\left[  c\rightarrow
c\right]  _{\operatorname{pub}}.
\end{equation}
The last protocol assumes a model where the receiver can locally copy
information and place it in a register to which Eve has access.

\section{Noiseless Quantum Shannon Theory}

Noiseless quantum Shannon theory consists of resource inequalities involving
unit resources and one non-unit, noiseless resource, such as an identity
channel or a pure bipartite state.

Schumacher compression%
\index{Schumacher compression}
from Chapter~\ref{chap:schumach}\ gives a way to simulate an identity channel
$\operatorname{id}_{A\rightarrow B}$ acting on a mixed state $\rho_{A}$ by
exploiting noiseless qubit channels at a rate equal to the entropy
$H(A)_{\rho}$:%
\begin{equation}
H(A)_{\rho}\left[  q\rightarrow q\right]  \geq\left\langle \operatorname{id}%
_{A\rightarrow B}:\rho_{A}\right\rangle . \label{eq-sum:schu}%
\end{equation}
We also know that if $n$ uses of an identity channel are available, then
achievability of the coherent information for quantum communication
(Chapter~\ref{chap:quantum-capacity}) implies that we can send quantum data
down this channel at a rate equal to $H(B)-H(E)$, where the entropies are with
respect to some input density operator $\rho_{A}$. But $H(E)=0$ because the
channel is the identity channel (the environment gets no information) and
$H(B)=H(A)_{\rho}$ because Alice's input goes directly to Bob. This gives us
the following resource inequality:%
\begin{equation}
\left\langle \operatorname{id}_{A\rightarrow B}:\rho_{A}\right\rangle \geq
H(A)_{\rho}\left[  q\rightarrow q\right]  , \label{eq-sum:schu-opp}%
\end{equation}
and combining \eqref{eq-sum:schu} and \eqref{eq-sum:schu-opp} gives the
following resource equality:%
\begin{equation}
\left\langle \operatorname{id}_{A\rightarrow B}:\rho_{A}\right\rangle
=H(A)_{\rho}\left[  q\rightarrow q\right]  .
\end{equation}

Entanglement concentration
\index{entanglement concentration}
from Chapter~\ref{chap:ent-conc}\ converts many copies of a pure, bipartite
state $|\phi\rangle_{AB}$ into ebits at a rate equal to the entropy of
entanglement:%
\begin{equation}
\left\langle \phi_{AB}\right\rangle \geq H(A)_{\phi}\left[  qq\right]  .
\end{equation}
Entanglement dilution
\index{entanglement dilution}
exploits a sublinear amount of classical communication to dilute ebits into
$n$ copies of a pure, bipartite state $|\phi\rangle_{AB}$. Ignoring the
sublinear rate of classical communication gives the following resource
inequality:%
\begin{equation}
H(A)_{\phi}\left[  qq\right]  \geq\left\langle \phi_{AB}\right\rangle .
\end{equation}
Combining entanglement concentration and entanglement dilution gives the
following resource equality:%
\begin{equation}
\left\langle \phi_{AB}\right\rangle =H(A)_{\phi}\left[  qq\right]  .
\end{equation}

The noiseless quantum Shannon theory is satisfactory in the sense that we can
obtain resource \textit{equalities}, illustrating the interconvertibility of
noiseless qubit channels with a relative identity channel and pure, bipartite
states with ebits.

\section{Noisy Quantum Shannon Theory}

Noisy quantum Shannon theory has resource inequalities with one noisy
resource, such as a noisy channel or a noisy state, interacting with other
unit resources. We can further classify a resource inequality as dynamic or
static, depending on whether the noisy resource involved is dynamic or static.

We first review the dynamic resource inequalities presented in this book.
These protocols involve a noisy channel interacting with the other unit
resources. Many of the protocols in noisy quantum Shannon theory generate
random codes from a state of the following form:%
\begin{equation}
\rho_{XABE}\equiv\sum_{x}p_{X}(x)|x\rangle\langle x|_{X}\otimes\mathcal{U}%
_{A^{\prime}\rightarrow BE}^{\mathcal{N}}(\phi_{AA^{\prime}}^{x}),
\end{equation}
where $\phi_{AA^{\prime}}^{x}$ is a pure, bipartite state and $U_{A^{\prime
}\rightarrow BE}^{\mathcal{N}}$ is an isometric extension of a channel
$\mathcal{N}_{A^{\prime}\rightarrow B}$. Also important is a special case of
the above form:%
\begin{equation}
\sigma_{ABE}\equiv\mathcal{U}_{A^{\prime}\rightarrow BE}^{\mathcal{N}}%
(\phi_{AA^{\prime}}),
\end{equation}
where $\phi_{AA^{\prime}}$ is a pure, bipartite state.
Holevo--Schumacher--Westmoreland
\index{HSW theorem}%
coding for classical communication over a quantum channel
(Chapter~\ref{chap:classical-comm-HSW}) is the following resource inequality:%
\begin{equation}
\left\langle \mathcal{N}\right\rangle \geq I(X;B)_{\rho}\left[  c\rightarrow
c\right]  .
\end{equation}
Devetak--Cai--Winter--Yeung coding for private classical communication
\index{private classical communication}%
over a quantum channel (Chapter~\ref{chap:private-cap}) is as follows:%
\begin{equation}
\left\langle \mathcal{N}\right\rangle \geq\left(  I(X;B)_{\rho}-I(X;E)_{\rho
}\right)  \left[  c\rightarrow c\right]  _{\operatorname{priv}}.
\end{equation}
Upgrading the private classical code to one that operates coherently gives
Devetak's method for coherent communication over a quantum channel
(Chapter~\ref{chap:quantum-capacity}):%
\begin{equation}
\left\langle \mathcal{N}\right\rangle \geq I(A\rangle B)_{\sigma}\left[
q\rightarrow qq\right]  ,
\end{equation}
which we showed can be converted
\index{LSD theorem}
asymptotically into a protocol for quantum communication:%
\begin{equation}
\left\langle \mathcal{N}\right\rangle \geq I(A\rangle B)_{\sigma}\left[
q\rightarrow q\right]  . \label{eq-sum:LSD}%
\end{equation}
Bennett--Shor--Smolin--Thapliyal coding for entanglement-assisted classical
communication over
\index{entanglement-assisted!classical communication}
a quantum channel (Chapter~\ref{chap:EA-classical}) is the following resource
inequality:%
\begin{equation}
\left\langle \mathcal{N}\right\rangle +H(A)_{\sigma}\left[  qq\right]  \geq
I(A;B)_{\sigma}\left[  c\rightarrow c\right]  .
\end{equation}
We showed how to upgrade this protocol to one for entanglement-assisted
coherent communication (Chapter~\ref{chap:coh-comm-noisy}):%
\begin{equation}
\left\langle \mathcal{N}\right\rangle +H(A)_{\sigma}\left[  qq\right]  \geq
I(A;B)_{\sigma}\left[  q\rightarrow qq\right]  ,
\label{eq-sum:coh-assisted-comm}%
\end{equation}
and combining with the coherent communication identity gives the following
protocol for
\index{entanglement-assisted!quantum communication}
entanglement-assisted quantum communication:%
\begin{equation}
\left\langle \mathcal{N}\right\rangle +\frac{1}{2}I(A;E)_{\sigma}\left[
qq\right]  \geq\frac{1}{2}I(A;B)_{\sigma}\left[  q\rightarrow q\right]  .
\end{equation}
Further combining with entanglement distribution gives the resource inequality
in \eqref{eq-sum:LSD} for quantum communication. By combining the HSW\ and
BSST\ protocols together (this needs to be done at the level of coding and not
at the level of resource inequalities---see Chapter~\ref{chap:coh-comm-noisy}%
), we recover a protocol
\index{trade-off coding}%
for entanglement-assisted communication of classical and quantum information:%
\begin{equation}
\left\langle \mathcal{N}\right\rangle +\frac{1}{2}I(A;E|X)_{\sigma}\left[
qq\right]  \geq\frac{1}{2}I(A;B|X)_{\sigma}\left[  q\rightarrow q\right]
+I(X;B)_{\sigma}\left[  c\rightarrow c\right]  .
\end{equation}
This protocol recovers any protocol in dynamic quantum Shannon theory that
involves a noisy channel and the three unit resources after combining it with
the three unit protocols in \eqref{eq-sum:ED}--\eqref{eq-sum:SD}. Important
special cases are entanglement-assisted classical communication with limited
entanglement:%
\begin{equation}
\left\langle \mathcal{N}\right\rangle +H(A|X)_{\sigma}\left[  qq\right]  \geq
I(AX;B)_{\sigma}\left[  c\rightarrow c\right]  ,
\end{equation}
and simultaneous classical and quantum communication:%
\begin{equation}
\left\langle \mathcal{N}\right\rangle \geq I(X;B)_{\sigma}\left[  c\rightarrow
c\right]  +I(A\rangle BX)_{\sigma}\left[  q\rightarrow q\right]  .
\end{equation}

Chapter~\ref{chap:coh-comm-noisy}\ touched on some important protocols in
static quantum Shannon theory. These protocols involve some noisy state
$\rho_{AB}$ interacting with the unit resources. The protocol for
coherent-assisted state transfer is the static couterpart to the protocol in
\eqref{eq-sum:coh-assisted-comm}:%
\begin{equation}
\langle W_{S\rightarrow AB}:\rho_{S}\rangle+H(A)_{\rho}\left[  q\rightarrow
q\right]  \geq I(A;B)_{\rho}\left[  q\rightarrow qq\right]  +\langle
\operatorname{id}_{S\rightarrow\hat{B}B}:\rho_{S}\rangle,
\label{eq-sum:assisted-coh-state-trans}%
\end{equation}
where $W$ is some isometry that distributes the state from a source $S$ to two
parties $A$ and $B$ and $\operatorname{id}_{S\rightarrow\hat{B}B}$ is the
identity. Ignoring the source and state transfer in the above protocol gives a
protocol for quantum-assisted coherent communication:%
\begin{equation}
\langle\rho\rangle+H(A)_{\rho}\left[  q\rightarrow q\right]  \geq
I(A;B)_{\rho}\left[  q\rightarrow qq\right]  . \label{eq-sum:q-assist-coh}%
\end{equation}
We can also combine \eqref{eq-sum:assisted-coh-state-trans} with the unit
protocols to obtain
\index{state transfer!quantum-assisted}
quantum-assisted state transfer:%
\begin{equation}
\langle W_{S\rightarrow AB}:\rho_{S}\rangle+\frac{1}{2}I(A;R)_{\varphi}\left[
q\rightarrow q\right]  \geq\frac{1}{2}I(A;B)_{\varphi}\left[  qq\right]
+\langle\operatorname{id}_{S\rightarrow\hat{B}B}:\rho_{S}\rangle,
\end{equation}
and
\index{state transfer!classical-assisted}%
classical-assisted state transfer:%
\begin{equation}
\langle W_{S\rightarrow AB}:\rho_{S}\rangle+I(A;R)_{\varphi}\left[
c\rightarrow c\right]  \geq I(A\rangle B)_{\varphi}\left[  qq\right]
+\langle\operatorname{id}_{S\rightarrow\hat{B}B}:\rho_{S}\rangle,
\end{equation}
where $|\varphi\rangle_{ABR}$ is a purification of $\rho_{AB}$. We also have
noisy super-dense coding%
\index{super-dense coding!noisy}%
\begin{equation}
\langle\rho\rangle+H(A)_{\rho}\left[  q\rightarrow q\right]  \geq
I(A;B)_{\rho}\left[  c\rightarrow c\right]  ,
\end{equation}
and noisy teleportation%
\index{quantum teleportation!noisy}%
\begin{equation}
\left\langle \rho\right\rangle +I(A;B)_{\rho}\left[  c\rightarrow c\right]
\geq I(A\rangle B)_{\rho}\left[  q\rightarrow q\right]  ,
\end{equation}
by combining \eqref{eq-sum:q-assist-coh} with the coherent communication
identity%
\index{coherent communication identity}
and the unit protocols.

\section{Protocols Not Covered In This Book}

There are many important protocols that we have not covered in this book because
our focus here was mostly on communication over quantum channels. One such
example is \textit{quantum state redistribution}. Suppose that Alice and Bob
share many copies of a tripartite state $\rho_{ACB}$ where Alice has the
shares $AC$ and Bob has the share $B$. The goal of state redistribution%
\index{state redistribution}
is for Alice to transfer the $C$ part of the state to Bob using the minimal
resources needed to do so. It is useful to identify a pure state
$\varphi_{RACB}$ as a purification of $\rho_{ACB}$, where $R$ is the purifying
system. \cite{DY08,YD09} showed the existence of the following state
redistribution protocol:%
\begin{multline}
\langle W_{S\rightarrow AC|B}:\rho_{S}\rangle+\frac{1}{2}I(  C;RB)
_{\varphi}\left[  q\rightarrow q\right]  +\frac{1}{2}I(  C;A)
_{\varphi}\left[  qq\right]  \geq\\
\langle W_{S\rightarrow A|CB}:\rho_{S}\rangle+\frac{1}{2}I(  C;B)
_{\varphi}\left[  q\rightarrow q\right]  +\frac{1}{2}I(  C;B)
_{\varphi}\left[  qq\right]  ,
\end{multline}
where $W_{S\rightarrow AC|B}$ is some isometry that distributes the system $S$
as $AC$ for Alice and $B$ for Bob and $W_{S\rightarrow A|CB}$ is defined
similarly. They also demonstrated that the above resource inequality gives an
optimal cost pair for the quantum communication rate $Q$ and the entanglement
consumption rate $E$, with%
\begin{align}
Q  &  =\frac{1}{2}I(  C;R|B)  _{\varphi},\\
E  &  =\frac{1}{2}\left[  I(  C;A)  _{\varphi}-I(  C;B)
_{\varphi}\right]  .
\end{align}
Thus, their protocol gives a direct operational interpretation to the
conditional quantum mutual information $\frac{1}{2}I(  C;R|B)
_{\varphi}$ as the net rate of quantum communication required in quantum state redistribution.

A simple version of the quantum reverse Shannon theorem%
\index{quantum reverse Shannon theorem}
gives a way to simulate the action of a channel $\mathcal{N}_{A^{\prime
}\rightarrow B}$ on some input state $\rho_{A^{\prime}}$ by exploiting
classical communication and
entanglement~\citep{ieee2002bennett,BDHSW09,BCR09}:%
\begin{equation}
H(B)_{\sigma}\left[  qq\right]  +I(R;B)_{\sigma}\left[  c\rightarrow c\right]
\geq\langle\mathcal{N}_{A^{\prime}\rightarrow B}:\rho_{A^{\prime}}\rangle,
\label{eq-sum:FQRS}%
\end{equation}
where $\sigma_{RB}\equiv\mathcal{N}_{A^{\prime}\rightarrow B}(\varphi
_{RA^{\prime}})$, with $\varphi_{RA^{\prime}}$ a purification of
$\rho_{A^{\prime}}$. One utility of the quantum reverse Shannon theorem is
that it gives an indication of how one channel might simulate another in the
presence of shared entanglement. In the simulation of the channel
$\mathcal{N}_{A^{\prime}\rightarrow B}$, the environment is also simulated and
ends up in Alice's possession. Thus, they end up simulating the quantum
feedback channel $\mathcal{U}_{A^{\prime}\rightarrow AB}^{\mathcal{N}}$, and
we can restate \eqref{eq-sum:FQRS} as follows:%
\begin{equation}
H(B)_{\sigma}\left[  qq\right]  +I(R;B)_{\sigma}\left[  c\rightarrow c\right]
\geq\langle\mathcal{U}_{A^{\prime}\rightarrow AB}^{\mathcal{N}}:\rho
_{A^{\prime}}\rangle.
\end{equation}
It is possible to upgrade the classical communication to coherent
communication~\citep{D06}, leading to the following coherent, fully-quantum
version of the quantum reverse Shannon theorem~\citep{ADHW06FQSW}:%
\begin{equation}
\frac{1}{2}I(A;B)_{\sigma}\left[  qq\right]  +\frac{1}{2}I(R;B)_{\sigma
}\left[  q\rightarrow q\right]  \geq\langle\mathcal{U}_{A^{\prime}\rightarrow
AB}^{\mathcal{N}}:\rho_{A^{\prime}}\rangle.
\end{equation}
Combining this resource inequality with the following one from
Exercise~\ref{ex-ccn:feedback-father}%
\begin{equation}
\langle\mathcal{U}_{A^{\prime}\rightarrow AB}^{\mathcal{N}}:\rho_{A^{\prime}%
}\rangle\geq\frac{1}{2}I(A;B)_{\sigma}\left[  qq\right]  +\frac{1}%
{2}I(R;B)_{\sigma}\left[  q\rightarrow q\right]
\end{equation}
gives the following satisfying resource equality:%
\begin{equation}
\langle\mathcal{U}_{A^{\prime}\rightarrow AB}^{\mathcal{N}}:\rho_{A^{\prime}%
}\rangle=\frac{1}{2}I(A;B)_{\sigma}\left[  qq\right]  +\frac{1}{2}%
I(R;B)_{\sigma}\left[  q\rightarrow q\right]  .
\end{equation}
The above resource equality is a generalization of the coherent communication
identity. A more general version of the quantum reverse Shannon theorem
quantifies the resources needed to simulate many independent instances of a
quantum channel on an arbitrary input state, and the proof in this case is
more involved \citep{BDHSW09,BCR09}.

Other protocols that we did not cover are
\index{remote state preparation}
remote state preparation~\citep{BDSSTW01,BHLSW05,PhysRevA.68.062319},
classical compression with quantum side information~\citep{DW03}, trade-offs
between public and private resources and channels~\citep{WH10}, trade-offs in
compression~\citep{hayden:4404}, a trade-off for a noisy state with the
three unit resources~\citep{HW09}, measurement compression \citep{Winter01a},
measurement compression with quantum side information
\citep{WHBH12}, and measurement channel simulation \citep{BRW14}. The resource inequality formalism is
helpful for devising new protocols in quantum Shannon theory by imagining some
resources to be unit and others to be noisy.

\section{Network Quantum Shannon Theory}

The field of network quantum Shannon theory%
\index{network quantum Shannon theory}
has arisen in recent years, motivated by the idea that one day we will be
dealing with a quantum Internet in which channels of increasing complexity can
connect a number of senders to a number of receivers. A quantum multiple
access channel has multiple senders and one receiver. Various authors have
considered classical communication over a multiple access channel
\citep{Winter01,FHSSW11,WS12,WG12,BN14}, quantum communication over multiple
access channels~\citep{Horodecki:2005:673,YHD05MQAC}, entanglement-assisted
protocols~\citep{itit2008hsieh}, and nonadditivity effects
\citep{CH08,PhysRevA.81.060305}. A quantum broadcast channel has one sender
and multiple receivers. Various authors have addressed similar scenarios in
this setting~\citep{YHD2006,DH2006,GS07,SW13,RSW14,HM15,STW15}. A quantum
interference channel has multiple senders and multiple receivers in which
certain sender-receiver pairs are interested in communicating. Recent progress
in this direction is in \citep{FHSSW11,S11,HMW14}. One could also consider
distributed compression tasks, and various authors have contributed to this
direction~\citep{ADHW06,ADHW06FQSW,Savov08}. We could imagine a future
textbook containing several chapters that summarize all of the progress in
network quantum Shannon theory and the novel techniques needed to handle
coding over such channels. \cite{Savov12} highlights much of this direction in
his PhD thesis (at least for classical communication).

\section{Future Directions}

Quantum Shannon theory has evolved from the first and simplest result
regarding Schumacher compression to a whole host of protocols that indicate
how much data we can transmit over noisy quantum channels or how much we can
compress information of varying types---the central question in any task is,
\textquotedblleft How many unit resources can we extract from a given non-unit
resource, perhaps with the help of other non-unit resources?\textquotedblright%
\ This book may give the impression that so much has been solved in the area
of quantum Shannon theory that little remains for the future, but this is
actually far from the truth! There remains much to do to improve our
understanding, and this final section briefly outlines just a few of these
important questions.

Find a better formula for the classical capacity other than the HSW\ formula.
Our best characterization of the classical capacity is with a regularized
version of the HSW\ formula, and this is unsatisfying in several ways that we
have mentioned before. In a similar vein, find a better formula for the
private classical capacity, the quantum capacity, and even for the trade-off
capacities. All of these formulas are unsatisfying because their
regularizations seem to be necessary in the general case. It could be the case
that an entropic expression evaluated on some finite tensor power of the
channels would be sufficient to characterize the capacity for different tasks,
but this is a difficult question to answer. Interestingly, recent work
suggests pursuing to find out whether this question is algorithmically
undecidable (see \cite{WCP11}). Effects such as superactivation of quantum
capacity (see Section~\ref{sec-q-cap:superactivation}) and non-additivity of
private capacity (see Section~\ref{sec-pcc:super-private-capacity}) have
highlighted how little we actually know about the corresponding
information-processing tasks in the general case. Also, it is important to
understand these effects more fully and to see if there is any way of
exploiting them in a practical communication scheme. Finally, a different
direction is to expand the number of channels that have additive capacities.
For example, finding the quantum capacity of a non-degradable quantum channel
would be a great result. Many questions remain open regarding second-order
characterizations, error exponents, and strong converses. Results in these
directions give much finer characterizations of communication tasks. We have
already highlighted progress in these directions at the end of relevant chapters.

Continue to explore network quantum Shannon theory. The single-sender,
single-receiver channel setting is a useful model for study and applies to
many practical scenarios, but eventually, we will be dealing with channels
connecting many inputs to many outputs. Having such an understanding for
information transmission in these scenarios could help guide the design of
practical communication schemes and might even shed light on the open problems
in the preceding paragraph.

\appendix

\chapter{Supplementary Results}

\label{chap:appendix-math}This section collects various useful definitions and
lemmas that we use throughout the proofs of certain theorems in this book.

\begin{lemma}
\label{lemma:positivity}Suppose that $M$ and $N$ are positive semi-definite
operators. Then the operators $M+N$, $MNM$, and $NMN$ are positive semi-definite.
\end{lemma}

\begin{lemma}
\label{lemma:op-interval}Suppose that the operators $\hat{\omega}$ and
$\omega$ have trace less than or equal to one. Suppose $\hat{\omega}$\ lies in
the operator interval $\left[  \left(  1-\varepsilon\right)  \omega,\left(
1+\varepsilon\right)  \omega\right]  $. Then%
\begin{equation}
\left\Vert \hat{\omega}-\omega\right\Vert _{1}\leq\varepsilon.
\end{equation}

\end{lemma}

\begin{proof}
The statement \textquotedblleft$\hat{\omega}$ lies in the operator interval
$\left[  \left(  1-\varepsilon\right)  \omega,\left(  1+\varepsilon\right)
\omega\right]  $\textquotedblright\ is equivalent to the following two
conditions:%
\begin{align}
\left(  1+\varepsilon\right)  \omega-\hat{\omega}  &  =\varepsilon
\omega-\left(  \hat{\omega}-\omega\right)  \geq0,\\
\hat{\omega}-\left(  1-\varepsilon\right)  \omega &  =\left(  \hat{\omega
}-\omega\right)  +\varepsilon\omega\geq0.
\end{align}
Let $\alpha\equiv\hat{\omega}-\omega$. Let us rewrite $\alpha$ in terms of the
positive semi-definite operators $P$ and $Q$:%
\begin{equation}
\alpha=P-Q,
\end{equation}
as we did in the proof of Lemma~\ref{lemma:trace-equiv}. The above conditions
become as follows:%
\begin{align}
\varepsilon\omega-\alpha &  \geq0,\\
\alpha+\varepsilon\omega &  \geq0.
\end{align}
Let the positive projectors $\Pi_{P}$ and $\Pi_{Q}$ project onto the
respective supports of $P$ and $Q$. We then apply the projector $\Pi_{P}$ to
the first condition:%
\begin{align}
\Pi_{P}\left(  \varepsilon\omega-\alpha\right)  \Pi_{P}  &  \geq0,\\
\Rightarrow\varepsilon\Pi_{P}\omega\Pi_{P}-\Pi_{P}\alpha\Pi_{P}  &  \geq0,\\
\Rightarrow\varepsilon\Pi_{P}\omega\Pi_{P}-P  &  \geq0,
\end{align}
where the first inequality follows from Lemma~\ref{lemma:positivity}. We apply
the projector $\Pi_{Q}$ to the second condition:%
\begin{align}
\Pi_{Q}\left(  \alpha+\varepsilon\omega\right)  \Pi_{Q}  &  \geq0,\\
\Rightarrow\Pi_{Q}\alpha\Pi_{Q}+\varepsilon\Pi_{Q}\omega\Pi_{Q}  &  \geq0,\\
\Rightarrow-Q+\varepsilon\Pi_{Q}\omega\Pi_{Q}  &  \geq0,
\end{align}
where the first inequality again follows from Lemma~\ref{lemma:positivity}.
Adding the two positive semi-definite operators together gives another
positive semi-definite operator by Lemma~\ref{lemma:positivity}:%
\begin{align}
\varepsilon\Pi_{P}\omega\Pi_{P}-P-Q+\varepsilon\Pi_{Q}\omega\Pi_{Q}  &
\geq0,\\
\Rightarrow\varepsilon\Pi_{P}\omega\Pi_{P}-\left\vert \hat{\omega}%
-\omega\right\vert +\varepsilon\Pi_{Q}\omega\Pi_{Q}  &  \geq0.
\end{align}
Apply the trace operation to get the following inequality:%
\begin{equation}
\varepsilon\operatorname{Tr}\left\{  \omega\right\}  \geq\operatorname{Tr}%
\left\{  \left\vert \hat{\omega}-\omega\right\vert \right\}  =\left\Vert
\hat{\omega}-\omega\right\Vert _{1}.
\end{equation}
Using the hypothesis that $\operatorname{Tr}\left\{  \omega\right\}  \leq1$
gives the desired result.
\end{proof}

\begin{theorem}
[Polar Decomposition]\label{thm-app:polar}Any operator
\index{polar decomposition}%
$A$ admits a left polar decomposition $A=U\sqrt{A^{\dag}A}$, and a right polar
decomposition $A=\sqrt{AA^{\dag}}V$.
\end{theorem}

\begin{proof}
We give a simple proof for just the right polar decomposition by appealing to
the singular value decomposition. Any operator $A$ admits a singular value
decomposition $A=U_{1}\Sigma U_{2}$, where $U_{1}$ and $U_{2}$ are unitary
operators and $\Sigma$ is an operator with positive singular values. Then
$AA^{\dag}=U_{1}\Sigma U_{2}U_{2}^{\dag}\Sigma U_{1}^{\dag}=U_{1}\Sigma
^{2}U_{1}^{\dag}$,\ and thus $\sqrt{AA^{\dag}}=U_{1}\Sigma U_{1}^{\dag}$. We
can take $V=U_{1}U_{2}$ and we obtain the right polar decomposition of $A$ as
$\sqrt{AA^{\dag}}V=U_{1}\Sigma U_{1}^{\dag}U_{1}U_{2}=U_{1}\Sigma U_{2}=A$.
\end{proof}

\begin{lemma}
\label{lem-app:fourier-states} Consider two collections of orthonormal states
$\{|\chi_{j}\rangle\}_{j\in\lbrack N]}$ and $\{|\zeta_{j}\rangle
\}_{j\in\lbrack N]}$ such that $\langle\chi_{j}|\zeta_{j}\rangle
\geq1-\varepsilon$ for all $j$. There exist phases $\gamma_{j}$ and
$\delta_{j}$ such that
\begin{equation}
{\langle\hat{\chi}|}\hat{\zeta}\rangle\geq1-\varepsilon,
\end{equation}
where%
\begin{equation}
|\hat{\chi}\rangle=\frac{1}{\sqrt{N}}\sum_{j=1}^{N}e^{i\gamma_{j}}|\chi
_{j}\rangle,\ \ \ \ \ \ \ \ \ \ |\hat{\zeta}\rangle=\frac{1}{\sqrt{N}}%
\sum_{j=1}^{N}e^{i\delta_{j}}|\zeta_{j}\rangle.
\end{equation}

\end{lemma}

\begin{proof}
Define the Fourier transformed states%
\begin{equation}
|\hat{\chi}_{s}\rangle\equiv\frac{1}{\sqrt{N}}\sum_{j=1}^{N}e^{2\pi
ijs/N}|\chi_{j}\rangle,
\end{equation}
and similarly define $|\hat{\zeta}_{s}\rangle$. By Parseval's relation, it
follows that%
\begin{equation}
\frac{1}{N}\sum_{s=1}^{N}{\langle\hat{\chi}_{s}|}\hat{\zeta}_{s}\rangle
=\frac{1}{N}\sum_{j=1}^{N}{\langle\chi_{j}|}\zeta_{j}\rangle\geq1-\varepsilon.
\end{equation}
Thus, at least one value of $s$ obeys the following inequality: $e^{i\theta
_{s}}{\langle\hat{\chi}_{s}|}\hat{\zeta}_{s}\rangle\geq1-\varepsilon$, for
some phase $\theta_{s}$. Setting $\gamma_{j}=2\pi js/N$ and $\delta_{j}%
=\gamma_{j}+\theta_{s}$ satisfies the statement of the lemma.
\end{proof}

The following ``support lemmas'' are taken directly from \cite[Appendix~B]%
{Renner2005}.

\begin{lemma}
\label{lem-app:support-1}Let $X_{AB}\in\mathcal{L}(\mathcal{H}_{A}%
\otimes\mathcal{H}_{B})$ be positive semi-definite, and let $X_{A}%
\equiv\operatorname{Tr}_{B}\{X_{AB}\}$ and $X_{B}\equiv\operatorname{Tr}%
_{A}\{X_{AB}\}$. Then $\operatorname{supp}(X_{AB})\subseteq\operatorname{supp}%
(X_{A})\otimes\operatorname{supp}(X_{B})$.
\end{lemma}

\begin{proof}
First suppose that $X_{AB}$ is rank one, so that $X_{AB}=|\Psi\rangle
\langle\Psi|_{AB}$ for some vector $|\Psi\rangle_{AB}\in\mathcal{H}_{A}%
\otimes\mathcal{H}_{B}$. Due to the Schmidt decomposition theorem
(Theorem~\ref{thm-qt:schmidt}), we have that%
\begin{equation}
|\Psi\rangle_{AB}=\sum_{z\in\mathcal{Z}}\gamma_{z}|\theta_{z}\rangle
_{A}\otimes|\xi_{z}\rangle_{B},
\end{equation}
where $\left\vert \mathcal{Z}\right\vert \leq\min\{\dim(\mathcal{H}_{A}%
),\dim(\mathcal{H}_{B})\}$, $\{\gamma_{z}\}$ is a set of strictly positive
numbers, and $\{|\theta_{z}\rangle_{A}\}$ and $\{|\xi_{z}\rangle_{B}\}$ are
orthonormal bases. Then%
\begin{align}
\operatorname{supp}(X_{AB})  &  =\operatorname{span}\{|\Psi\rangle_{AB}\}\\
&  \subseteq\operatorname{span}\{|\theta_{z}\rangle_{A}:z\in\mathcal{Z}%
\}\otimes\operatorname{span}\{|\xi_{z}\rangle_{B}:z\in\mathcal{Z}\}.
\end{align}
The statement then follows for this case because $\operatorname{supp}%
(X_{A})=\operatorname{span}\{|\theta_{z}\rangle_{A}:z\in\mathcal{Z}\}$ and
$\operatorname{supp}(X_{B})=\operatorname{span}\{|\xi_{z}\rangle_{B}%
:z\in\mathcal{Z}\}$.

Now suppose that $X_{AB}$ is not rank one. It admits a decomposition into
rank-one vectors of the following form:%
\begin{equation}
X_{AB}=\sum_{x\in\mathcal{X}}|\Psi^{x}\rangle\langle\Psi^{x}|_{AB},
\end{equation}
where $|\Psi^{x}\rangle_{AB}\in\mathcal{H}_{A}\otimes\mathcal{H}_{B}$ for all
$x\in\mathcal{X}$. Set $\Psi_{AB}^{x}=|\Psi^{x}\rangle\langle\Psi^{x}|_{AB}$,
and let $\Psi_{A}^{x}\equiv\operatorname{Tr}_{B}\{\Psi_{AB}^{x}\}$ and
$\Psi_{B}^{x}\equiv\operatorname{Tr}_{A}\{\Psi_{AB}^{x}\}$. Then%
\begin{align}
\operatorname{supp}(X_{AB})  &  =\operatorname{span}\{|\Psi^{x}\rangle
_{AB}:x\in\mathcal{X}\}\\
&  \subseteq\operatorname{span}\left[  \bigcup\limits_{x\in\mathcal{X}}\left[
\operatorname{supp}(\Psi_{A}^{x})\otimes\operatorname{supp}(\Psi_{B}%
^{x})\right]  \right] \\
&  \subseteq\operatorname{span}\left[  \bigcup\limits_{x\in\mathcal{X}%
}\operatorname{supp}(\Psi_{A}^{x})\right]  \otimes\operatorname{span}\left[
\bigcup\limits_{x\in\mathcal{X}}\operatorname{supp}(\Psi_{B}^{x})\right] \\
&  =\operatorname{supp}(X_{A})\otimes\operatorname{supp}(X_{B}),
\end{align}
concluding the proof.
\end{proof}

\begin{lemma}
\label{lem-app:support-2}Let $X_{AB},Y_{AB}\in\mathcal{L}(\mathcal{H}%
_{A}\otimes\mathcal{H}_{B})$ be positive semi-definite, and suppose that
$\operatorname{supp}(X_{AB})\subseteq\operatorname{supp}(Y_{AB})$. Then
$\operatorname{supp}(X_{A})\subseteq\operatorname{supp}(Y_{A})$, where
$X_{A}\equiv\operatorname{Tr}_{B}\{X_{AB}\}$ and $Y_{A}\equiv\operatorname{Tr}%
_{B}\{Y_{AB}\}$.
\end{lemma}

\begin{proof}
First suppose that $X_{AB}$ is rank one, as in the first part of the proof of
the previous lemma, and let us use the same notation as given there. Applying
the same lemma gives that%
\begin{equation}
\operatorname{supp}(X_{AB})\subseteq\operatorname{supp}(Y_{AB})\subseteq
\operatorname{supp}(Y_{A})\otimes\operatorname{supp}(Y_{B}),
\end{equation}
which in turn implies that $\operatorname{supp}(X_{AB})=\operatorname{span}%
\{|\Psi\rangle_{AB}\}\subseteq\operatorname{supp}(Y_{A})\otimes
\operatorname{supp}(Y_{B})$. This implies that $|\theta_{z}\rangle_{A}%
\in\operatorname{supp}(Y_{A})$ for all $z\in\mathcal{Z}$, and thus that
$\operatorname{span}\{|\theta_{z}\rangle_{A}\}\in\operatorname{supp}(Y_{A})$.
We can then conclude the statement in this case because $\operatorname{span}%
\{|\theta_{z}\rangle_{A}\}=\operatorname{supp}(X_{A})$.

Now suppose that $X_{AB}$ is not rank one. Then it admits a decomposition as
given in the proof of the previous lemma. Using the same notation, we have
that $\operatorname{supp}(\Psi_{AB}^{x})\subseteq\operatorname{supp}(Y_{AB}%
)$\ holds for all $x\in\mathcal{X}$. Since we have proven the lemma for
rank-one operators, we can conclude that $\operatorname{supp}(\Psi_{A}%
^{x})\subseteq\operatorname{supp}(Y_{A})$\ holds for all $x\in\mathcal{X}$. As
a consequence, we find that%
\begin{equation}
\operatorname{supp}(X_{A})=\operatorname{span}\left[  \bigcup\limits_{x\in
\mathcal{X}}\operatorname{supp}(\Psi_{A}^{x})\right]  \subseteq
\operatorname{supp}(Y_{A}),
\end{equation}
concluding the proof.
\end{proof}

\chapter{Unique Linear Extension of a Quantum Physical Evolution}

\label{app:unique-linear-ext}Recall in
Section~\ref{sec-nqt:axiomatic-approach}\ that we argued on physical grounds
how any quantum physical evolution $\mathcal{N}$\ should be convex linear when
acting on the space $\mathcal{D}(\mathcal{H}_{A})$\ of density operators:%
\begin{equation}
\mathcal{N}(\lambda\rho_{A}+(1-\lambda)\sigma_{A})=\lambda\mathcal{N}(\rho
_{A})+(1-\lambda)\mathcal{N}(\sigma_{A}),
\end{equation}
where $\rho_{A},\sigma_{A}\in\mathcal{D}(\mathcal{H}_{A})$ and $\lambda
\in\left[  0,1\right]  $. Here we show how to construct a unique linear
extension $\widetilde{\mathcal{N}}$\ of $\mathcal{N}$, whose action is well
defined on the space of all operators $X_{A}\in\mathcal{L}(\mathcal{H}_{A})$.
The development follows the approach given in Proposition~2.30 of \cite{HZ12}.

We first define $\widetilde{\mathcal{N}}(0)\equiv0$, where the inputs and
outputs are understood to be the zero operator. We next extend the action of
$\mathcal{N}$ to all positive semi-definite operators $P_{A}\neq0$ as follows:%
\begin{equation}
\widetilde{\mathcal{N}}(P_{A})\equiv\operatorname{Tr}\{P_{A}\}\mathcal{N}%
(\left[  \operatorname{Tr}\{P_{A}\}\right]  ^{-1}P_{A}),
\end{equation}
where it is clear that this is well defined from $\mathcal{N}$ because
$\left[  \operatorname{Tr}\{P_{A}\}\right]  ^{-1}P_{A}$ is a density operator.
Now consider for a constant $s>0$ that we have scale invariance:%
\begin{align}
\widetilde{\mathcal{N}}(sP_{A})  &  =\operatorname{Tr}\{sP_{A}\}\mathcal{N}%
(\left[  \operatorname{Tr}\{sP_{A}\}\right]  ^{-1}sP_{A})\\
&  =s\operatorname{Tr}\{P_{A}\}\mathcal{N}(\left[  \operatorname{Tr}%
\{P_{A}\}\right]  ^{-1}P_{A})\\
&  =s\widetilde{\mathcal{N}}(P_{A}).
\end{align}
Furthermore, for two non-zero positive semi-definite operators $P_{A}$ and
$Q_{A}$, we have the following additivity relation:%
\begin{equation}
\widetilde{\mathcal{N}}(P_{A}+Q_{A})=\widetilde{\mathcal{N}}(P_{A}%
)+\widetilde{\mathcal{N}}(Q_{A}), \label{eq-app-linear:additive-PSD}%
\end{equation}
which follows because%
\begin{align}
&  \widetilde{\mathcal{N}}(P_{A}+Q_{A})\nonumber\\
&  =\operatorname{Tr}\{P_{A}+Q_{A}\}\mathcal{N}(\left[  \operatorname{Tr}%
\{P_{A}+Q_{A}\}\right]  ^{-1}(P_{A}+Q_{A}))\\
&  =\operatorname{Tr}\{P_{A}+Q_{A}\}\mathcal{N}\left(  \frac{1}%
{\operatorname{Tr}\{P_{A}+Q_{A}\}}P_{A}+\frac{1}{\operatorname{Tr}%
\{P_{A}+Q_{A}\}}Q_{A}\right) \\
&  =\operatorname{Tr}\{P_{A}+Q_{A}\}\mathcal{N}\left(  \frac{\operatorname{Tr}%
\{P_{A}\}}{\operatorname{Tr}\{P_{A}+Q_{A}\}}\frac{P_{A}}{\operatorname{Tr}%
\{P_{A}\}}+\frac{\operatorname{Tr}\{Q_{A}\}}{\operatorname{Tr}\{P_{A}+Q_{A}%
\}}\frac{Q_{A}}{\operatorname{Tr}\{Q_{A}\}}\right) \\
&  =\operatorname{Tr}\{P_{A}\}\mathcal{N}\left(  \frac{P_{A}}%
{\operatorname{Tr}\{P_{A}\}}\right)  +\operatorname{Tr}\{Q_{A}\}\mathcal{N}%
\left(  \frac{Q_{A}}{\operatorname{Tr}\{Q_{A}\}}\right) \\
&  =\widetilde{\mathcal{N}}(P_{A})+\widetilde{\mathcal{N}}(Q_{A}),
\end{align}
where in the fourth equality, we exploited convex linearity of the quantum
physical evolution~$\mathcal{N}$.

For the next step, recall that any Hermitian operator $T_{A}$\ can be written
as a linear combination of a positive part and a negative part:\ $T_{A}%
=T_{A}^{+}-T_{A}^{-}$, where both $T_{A}^{+}$ and $T_{A}^{-}$ are positive
semi-definite operators. So we define the action of $\widetilde{\mathcal{N}}$
on any Hermitian operator $T_{A}$ as follows:%
\begin{equation}
\widetilde{\mathcal{N}}(T_{A})\equiv\widetilde{\mathcal{N}}(T_{A}%
^{+})-\widetilde{\mathcal{N}}(T_{A}^{-}).
\label{eq-app-linear:define-hermitian}%
\end{equation}
To see that the following additivity relation holds for all Hermitian $S_{A}$
and $T_{A}$%
\begin{equation}
\widetilde{\mathcal{N}}(S_{A}+T_{A})=\widetilde{\mathcal{N}}(S_{A}%
)+\widetilde{\mathcal{N}}(T_{A}), \label{eq-app-linear:additive-hermitian}%
\end{equation}
consider that%
\begin{equation}
S_{A}+T_{A}=(S_{A}+T_{A})^{+}-(S_{A}+T_{A})^{-},
\end{equation}
while also%
\begin{equation}
S_{A}+T_{A}=S_{A}^{+}+T_{A}^{+}-S_{A}^{-}-T_{A}^{-}.
\end{equation}
Equating both sides, we find that%
\begin{equation}
(S_{A}+T_{A})^{+}+S_{A}^{-}+T_{A}^{-}=(S_{A}+T_{A})^{-}+S_{A}^{+}+T_{A}^{+}.
\end{equation}
Now we exploit this equality, \eqref{eq-app-linear:additive-PSD}, and the
definition in \eqref{eq-app-linear:define-hermitian} to establish \eqref{eq-app-linear:additive-hermitian}.

The final step is to extend the action of $\widetilde{\mathcal{N}}$ to all
operators $X_{A}\in\mathcal{L}(\mathcal{H}_{A})$. Here, we recall that any
linear operator can be written in terms of a real and imaginary part as
follows:%
\begin{equation}
X_{A}^{R}\equiv\frac{1}{2}\left(  X_{A}+X_{A}^{\dag}\right)
,\ \ \ \ \ \ \ \ X_{A}^{I}\equiv\frac{1}{2i}\left(  X_{A}-X_{A}^{\dag}\right)
,
\end{equation}
where by inspection, $X_{A}^{R}$ and $X_{A}^{I}$ are Hermitian operators. So
we define%
\begin{equation}
\widetilde{\mathcal{N}}(X_{A})\equiv\widetilde{\mathcal{N}}(X_{A}%
^{R})+i\widetilde{\mathcal{N}}(X_{A}^{I}).
\end{equation}
This completes the development of a well defined linear extension
$\widetilde{\mathcal{N}}$\ of the quantum physical evolution $\mathcal{N}$.

To show that it is unique, recall that any operator $X_{A}$ can be expanded as
a linear combination of density operators from the basis $\{\rho_{A}^{x,y}\}$,
defined in \eqref{eq-nqt:density-op-basis}, as follows:%
\begin{equation}
X_{A}=\sum_{x,y}\alpha_{x,y}\rho_{A}^{x,y},
\end{equation}
where $\alpha_{x,y}\in\mathbb{C}$ for all $x$ and $y$. It is straightforward
to show from the above development that%
\begin{equation}
\widetilde{\mathcal{N}}(X_{A})=\sum_{x,y}\alpha_{x,y}\mathcal{N}(\rho
_{A}^{x,y}).
\end{equation}
Now suppose that $\mathcal{N}^{\prime}$ is some other linear map for which
$\mathcal{N}^{\prime}(\rho_{A})=\mathcal{N}(\rho_{A})$ for all $\rho_{A}%
\in\mathcal{D}(\mathcal{H}_{A})$. Then the following equality holds for all
$X_{A}\in\mathcal{L}(\mathcal{H}_{A})$:%
\begin{equation}
\mathcal{N}^{\prime}(X_{A})=\sum_{x,y}\alpha_{x,y}\mathcal{N}^{\prime}%
(\rho_{A}^{x,y})=\sum_{x,y}\alpha_{x,y}\mathcal{N}(\rho_{A}^{x,y}%
)=\widetilde{\mathcal{N}}(X_{A}).
\end{equation}
As a result, $\mathcal{N}^{\prime}=\widetilde{\mathcal{N}}$, given that they
have the same action on every operator $X_{A}\in\mathcal{L}(\mathcal{H}_{A})$.

{\footnotesize
\bibliographystyle{plainnat}
\bibliography{qit-notes}

\begin{thebibliography}{391}
\providecommand{\natexlab}[1]{#1}
\providecommand{\url}[1]{\texttt{#1}}
\expandafter\ifx\csname urlstyle\endcsname\relax
  \providecommand{\doi}[1]{doi: #1}\else
  \providecommand{\doi}{doi: \begingroup \urlstyle{rm}\Url}\fi

\bibitem[Abeyesinghe(2006)]{A06}
Anura Abeyesinghe.
\newblock \emph{Unification of Quantum Information Theory}.
\newblock PhD thesis, California Institute of Technology, 2006.

\bibitem[Abeyesinghe and Hayden(2003)]{PhysRevA.68.062319}
Anura Abeyesinghe and Patrick Hayden.
\newblock Generalized remote state preparation: Trading cbits, qubits, and
  ebits in quantum communication.
\newblock \emph{Physical Review A}, 68\penalty0 (6):\penalty0 062319, December
  2003.
\newblock \doi{10.1103/PhysRevA.68.062319}.
\newblock arXiv:quant-ph/0308143.

\bibitem[Abeyesinghe et~al.(2009)Abeyesinghe, Devetak, Hayden, and
  Winter]{ADHW06FQSW}
Anura Abeyesinghe, Igor Devetak, Patrick Hayden, and Andreas Winter.
\newblock The mother of all protocols: Restructuring quantum information's
  family tree.
\newblock \emph{Proceedings of the Royal Society A}, 465\penalty0
  (2108):\penalty0 2537--2563, August 2009.
\newblock arXiv:quant-ph/0606225.

\bibitem[Adami and Cerf(1997)]{PhysRevA.56.3470}
Christoph Adami and Nicolas~J. Cerf.
\newblock von {Neumann} capacity of noisy quantum channels.
\newblock \emph{Physical Review A}, 56\penalty0 (5):\penalty0 3470--3483,
  November 1997.
\newblock \doi{10.1103/PhysRevA.56.3470}.
\newblock arXiv:quant-ph/9609024.

\bibitem[Aharonov and Ben-Or(1997)]{258579}
Dorit Aharonov and Michael Ben-Or.
\newblock Fault-tolerant quantum computation with constant error.
\newblock In \emph{STOC '97: Proceedings of the Twenty-Ninth Annual ACM
  Symposium on Theory of Computing}, pages 176--188, New York, NY, USA, 1997.
  ACM.
\newblock ISBN 0-89791-888-6.
\newblock \doi{http://doi.acm.org/10.1145/258533.258579}.
\newblock arXiv:quant-ph/9906129.

\bibitem[Ahlswede and Winter(2002)]{AW02}
Rudolph Ahlswede and Andreas Winter.
\newblock Strong converse for identification via quantum channels.
\newblock \emph{IEEE Transactions on Information Theory}, 48\penalty0
  (3):\penalty0 569--579, March 2002.
\newblock arXiv:quant-ph/0012127.

\bibitem[Ahn et~al.(2006)Ahn, Doherty, Hayden, and Winter]{ADHW06}
Charlene Ahn, Andrew Doherty, Patrick Hayden, and Andreas Winter.
\newblock On the distributed compression of quantum information.
\newblock \emph{IEEE Transactions on Information Theory}, 52\penalty0
  (10):\penalty0 4349--4357, October 2006.
\newblock arXiv:quant-ph/0403042.

\bibitem[Alicki and Fannes(2004)]{AF04}
Robert Alicki and Mark Fannes.
\newblock Continuity of quantum conditional information.
\newblock \emph{Journal of Physics A: Mathematical and General}, 37\penalty0
  (5):\penalty0 L55--L57, February 2004.
\newblock arXiv:quant-ph/0312081.

\bibitem[Araki and Lieb(1970)]{araki1970}
Huzihiro Araki and Elliott~H. Lieb.
\newblock Entropy inequalities.
\newblock \emph{Communications in Mathematical Physics}, 18\penalty0
  (2):\penalty0 160--170, 1970.
\newblock URL \url{http://projecteuclid.org/euclid.cmp/1103842506}.

\bibitem[Aspect et~al.(1981)Aspect, Grangier, and Roger]{PhysRevLett.47.460}
Alain Aspect, Philippe Grangier, and G\'erard Roger.
\newblock Experimental tests of realistic local theories via {Bell's} theorem.
\newblock \emph{Physical Review Letters}, 47\penalty0 (7):\penalty0 460--463,
  August 1981.
\newblock \doi{10.1103/PhysRevLett.47.460}.

\bibitem[Aubrun et~al.(2011)Aubrun, Szarek, and Werner]{ASW10}
Guillaume Aubrun, Stanislaw Szarek, and Elisabeth Werner.
\newblock Hastings' additivity counterexample via {Dvoretzky's} theorem.
\newblock \emph{Communications in Mathematical Physics}, 305\penalty0
  (1):\penalty0 85--97, 2011.
\newblock arXiv:1003.4925.

\bibitem[Audenaert et~al.(2002)Audenaert, De~Moor, Vollbrecht, and
  Werner]{AdMVW02}
Koenraad Audenaert, Bart De~Moor, Karl Gerd~H. Vollbrecht, and Reinhard~F.
  Werner.
\newblock Asymptotic relative entropy of entanglement for orthogonally
  invariant states.
\newblock \emph{Physical Review A}, 66\penalty0 (3):\penalty0 032310, September
  2002.
\newblock \doi{10.1103/PhysRevA.66.032310}.
\newblock arXiv:quant-ph/0204143.

\bibitem[Audenaert(2007)]{A07}
Koenraad M.~R. Audenaert.
\newblock A sharp continuity estimate for the von {Neumann} entropy.
\newblock \emph{Journal of Physics A: Mathematical and Theoretical},
  40\penalty0 (28):\penalty0 8127, July 2007.
\newblock arXiv:quant-ph/0610146.

\bibitem[Bardhan et~al.(2015)Bardhan, Garcia-Patron, Wilde, and Winter]{BPWW14}
Bhaskar~Roy Bardhan, Raul Garcia-Patron, Mark~M. Wilde, and Andreas Winter.
\newblock Strong converse for the classical capacity of all phase-insensitive
  bosonic {Gaussian} channels.
\newblock \emph{IEEE Transactions on Information Theory}, 61\penalty0
  (4):\penalty0 1842--1850, April 2015.
\newblock arXiv:1401.4161.

\bibitem[Barnum and Knill(2002)]{BK02}
Howard Barnum and Emanuel Knill.
\newblock Reversing quantum dynamics with near-optimal quantum and classical
  fidelity.
\newblock \emph{Journal of Mathematical Physics}, 43\penalty0 (5):\penalty0
  2097--2106, May 2002.
\newblock arXiv:quant-ph/0004088.

\bibitem[Barnum et~al.(1998)Barnum, Nielsen, and Schumacher]{BNS98}
Howard Barnum, M.~A. Nielsen, and Benjamin Schumacher.
\newblock Information transmission through a noisy quantum channel.
\newblock \emph{Physical Review A}, 57\penalty0 (6):\penalty0 4153--4175, June
  1998.
\newblock \doi{10.1103/PhysRevA.57.4153}.

\bibitem[Barnum et~al.(2000)Barnum, Knill, and Nielsen]{BKN98}
Howard Barnum, Emanuel Knill, and Michael~A. Nielsen.
\newblock On quantum fidelities and channel capacities.
\newblock \emph{IEEE Transactions on Information Theory}, 46\penalty0
  (4):\penalty0 1317--1329, July 2000.
\newblock arXiv:quant-ph/9809010.

\bibitem[Barnum et~al.(2001{\natexlab{a}})Barnum, Caves, Fuchs, Jozsa, and
  Schumacher]{BCFJS01}
Howard Barnum, Carlton~M. Caves, Christopher~A. Fuchs, Richard Jozsa, and
  Benjamin Schumacher.
\newblock On quantum coding for ensembles of mixed states.
\newblock \emph{Journal of Physics A: Mathematical and General}, 34\penalty0
  (35):\penalty0 6767, September 2001{\natexlab{a}}.
\newblock arXiv:quant-ph/0008024.

\bibitem[Barnum et~al.(2001{\natexlab{b}})Barnum, Hayden, Jozsa, and
  Winter]{BHJW01}
Howard Barnum, Patrick Hayden, Richard Jozsa, and Andreas Winter.
\newblock On the reversible extraction of classical information from a quantum
  source.
\newblock \emph{Proceedings of the Royal Society A}, 457\penalty0
  (2012):\penalty0 2019--2039, July 2001{\natexlab{b}}.
\newblock arXiv:quant-ph/0011072.

\bibitem[Beigi et~al.(2015)Beigi, Datta, and Leditzky]{BDL15}
Salman Beigi, Nilanjana Datta, and Felix Leditzky.
\newblock Decoding quantum information via the {Petz} recovery map.
\newblock May 2015.
\newblock arXiv:1504.04449.

\bibitem[Bell(1964)]{bell1964}
John~Stewart Bell.
\newblock On the {Einstein-Podolsky-Rosen} paradox.
\newblock \emph{Physics}, 1:\penalty0 195--200, 1964.

\bibitem[Bennett(1992)]{Bennett:1992:3121}
Charles~H. Bennett.
\newblock Quantum cryptography using any two nonorthogonal states.
\newblock \emph{Physical Review Letters}, 68\penalty0 (21):\penalty0
  3121--3124, May 1992.
\newblock \doi{10.1103/PhysRevLett.68.3121}.

\bibitem[Bennett(1995)]{B95}
Charles~H. Bennett.
\newblock Quantum information and computation.
\newblock \emph{Physics Today}, 48\penalty0 (10):\penalty0 24--30, October
  1995.

\bibitem[Bennett(2004)]{Bennett04}
Charles~H. Bennett.
\newblock A resource-based view of quantum information.
\newblock \emph{Quantum Information and Computation}, 4:\penalty0 460--466,
  December 2004.
\newblock ISSN 1533-7146.

\bibitem[Bennett and Brassard(1984)]{bb84}
Charles~H. Bennett and Gilles Brassard.
\newblock Quantum cryptography: Public key distribution and coin tossing.
\newblock In \emph{Proceedings of IEEE International Conference on Computers
  Systems and Signal Processing}, pages 175--179, Bangalore, India, December
  1984.

\bibitem[Bennett and Wiesner(1992)]{PhysRevLett.69.2881}
Charles~H. Bennett and Stephen~J. Wiesner.
\newblock Communication via one- and two-particle operators on
  {Einstein-Podolsky-Rosen} states.
\newblock \emph{Physical Review Letters}, 69\penalty0 (20):\penalty0
  2881--2884, November 1992.
\newblock \doi{10.1103/PhysRevLett.69.2881}.

\bibitem[Bennett et~al.(1992{\natexlab{a}})Bennett, Brassard, and Ekert]{bbe92}
Charles~H. Bennett, Gilles Brassard, and Artur~K. Ekert.
\newblock Quantum cryptography.
\newblock \emph{Scientific American}, pages 50--57, October 1992{\natexlab{a}}.

\bibitem[Bennett et~al.(1992{\natexlab{b}})Bennett, Brassard, and
  Mermin]{Bennett:1992:557}
Charles~H. Bennett, Gilles Brassard, and N.~David Mermin.
\newblock Quantum cryptography without {Bell's} theorem.
\newblock \emph{Physical Review Letters}, 68\penalty0 (5):\penalty0 557--559,
  February 1992{\natexlab{b}}.
\newblock \doi{10.1103/PhysRevLett.68.557}.

\bibitem[Bennett et~al.(1993)Bennett, Brassard, Cr\'epeau, Jozsa, Peres, and
  Wootters]{PhysRevLett.70.1895}
Charles~H. Bennett, Gilles Brassard, Claude Cr\'epeau, Richard Jozsa, Asher
  Peres, and William~K. Wootters.
\newblock Teleporting an unknown quantum state via dual classical and
  {Einstein-Podolsky-Rosen} channels.
\newblock \emph{Physical Review Letters}, 70\penalty0 (13):\penalty0
  1895--1899, March 1993.
\newblock \doi{10.1103/PhysRevLett.70.1895}.

\bibitem[Bennett et~al.(1996{\natexlab{a}})Bennett, Bernstein, Popescu, and
  Schumacher]{BBPS96}
Charles~H. Bennett, Herbert~J. Bernstein, Sandu Popescu, and Benjamin
  Schumacher.
\newblock Concentrating partial entanglement by local operations.
\newblock \emph{Physical Review A}, 53\penalty0 (4):\penalty0 2046--2052, April
  1996{\natexlab{a}}.
\newblock \doi{10.1103/PhysRevA.53.2046}.
\newblock arXiv:quant-ph/9511030.

\bibitem[Bennett et~al.(1996{\natexlab{b}})Bennett, Brassard, Popescu,
  Schumacher, Smolin, and Wootters]{BBPSSW96EPP}
Charles~H. Bennett, Gilles Brassard, Sandu Popescu, Benjamin Schumacher,
  John~A. Smolin, and William~K. Wootters.
\newblock Purification of noisy entanglement and faithful teleportation via
  noisy channels.
\newblock \emph{Physical Review Letters}, 76\penalty0 (5):\penalty0 722--725,
  January 1996{\natexlab{b}}.
\newblock \doi{10.1103/PhysRevLett.76.722}.
\newblock arXiv:quant-ph/9511027.

\bibitem[Bennett et~al.(1996{\natexlab{c}})Bennett, DiVincenzo, Smolin, and
  Wootters]{BDSW96}
Charles~H. Bennett, David~P. DiVincenzo, John~A. Smolin, and William~K.
  Wootters.
\newblock Mixed-state entanglement and quantum error correction.
\newblock \emph{Physical Review A}, 54\penalty0 (5):\penalty0 3824--3851,
  November 1996{\natexlab{c}}.
\newblock \doi{10.1103/PhysRevA.54.3824}.
\newblock arXiv:quant-ph/9604024.

\bibitem[Bennett et~al.(1997)Bennett, DiVincenzo, and
  Smolin]{PhysRevLett.78.3217}
Charles~H. Bennett, David~P. DiVincenzo, and John~A. Smolin.
\newblock Capacities of quantum erasure channels.
\newblock \emph{Physical Review Letters}, 78\penalty0 (16):\penalty0
  3217--3220, April 1997.
\newblock \doi{10.1103/PhysRevLett.78.3217}.
\newblock arXiv:quant-ph/9701015.

\bibitem[Bennett et~al.(1999)Bennett, Shor, Smolin, and
  Thapliyal]{PhysRevLett.83.3081}
Charles~H. Bennett, Peter~W. Shor, John~A. Smolin, and Ashish~V. Thapliyal.
\newblock Entanglement-assisted classical capacity of noisy quantum channels.
\newblock \emph{Physical Review Letters}, 83\penalty0 (15):\penalty0
  3081--3084, October 1999.
\newblock \doi{10.1103/PhysRevLett.83.3081}.
\newblock arXiv:quant-ph/9904023.

\bibitem[Bennett et~al.(2001)Bennett, DiVincenzo, Shor, Smolin, Terhal, and
  Wootters]{BDSSTW01}
Charles~H. Bennett, David~P. DiVincenzo, Peter~W. Shor, John~A. Smolin,
  Barbara~M. Terhal, and William~K. Wootters.
\newblock Remote state preparation.
\newblock \emph{Physical Review Letters}, 87\penalty0 (7):\penalty0 077902,
  July 2001.
\newblock \doi{10.1103/PhysRevLett.87.077902}.

\bibitem[Bennett et~al.(2002)Bennett, Shor, Smolin, and
  Thapliyal]{ieee2002bennett}
Charles~H. Bennett, Peter~W. Shor, John~A. Smolin, and Ashish~V. Thapliyal.
\newblock Entanglement-assisted capacity of a quantum channel and the reverse
  {Shannon} theorem.
\newblock \emph{IEEE Transactions on Information Theory}, 48\penalty0
  (10):\penalty0 2637--2655, October 2002.
\newblock arXiv:quant-ph/0106052.

\bibitem[Bennett et~al.(2005)Bennett, Hayden, Leung, Shor, and Winter]{BHLSW05}
Charles~H. Bennett, Patrick Hayden, Debbie~W. Leung, Peter~W. Shor, and Andreas
  Winter.
\newblock Remote preparation of quantum states.
\newblock \emph{IEEE Transactions on Information Theory}, 51\penalty0
  (1):\penalty0 56--74, January 2005.
\newblock arXiv:quant-ph/0307100.

\bibitem[Bennett et~al.(2006)Bennett, Harrow, and Lloyd]{BHL06}
Charles~H. Bennett, Aram~W. Harrow, and Seth Lloyd.
\newblock Universal quantum data compression via nondestructive tomography.
\newblock \emph{Physical Review A}, 73\penalty0 (3):\penalty0 032336, March
  2006.
\newblock \doi{10.1103/PhysRevA.73.032336}.
\newblock arXiv:quant-ph/0403078.

\bibitem[Bennett et~al.(2014)Bennett, Devetak, Harrow, Shor, and
  Winter]{BDHSW09}
Charles~H. Bennett, Igor Devetak, Aram~W. Harrow, Peter~W. Shor, and Andreas
  Winter.
\newblock The quantum reverse {Shannon} theorem and resource tradeoffs for
  simulating quantum channels.
\newblock \emph{IEEE Transactions on Information Theory}, 60\penalty0
  (5):\penalty0 2926--2959, May 2014.
\newblock arXiv:0912.5537.

\bibitem[Berger(1971)]{B71}
Toby Berger.
\newblock \emph{Rate distortion theory: A mathematical basis for data
  compression}.
\newblock Prentice-Hall, Englewood Cliffs, New Jersey, USA, 1971.

\bibitem[Berger(1977)]{B77}
Toby Berger.
\newblock Multiterminal source coding.
\newblock \emph{The Information Theory Approach to Communications}, 1977.
\newblock Springer-Verlag, New York.

\bibitem[Bergh and L\"ofstr\"om(1976)]{BL76}
J.~Bergh and Jorgen L\"ofstr\"om.
\newblock \emph{Interpolation Spaces}.
\newblock Springer-Verlag Berlin Heidelberg, 1976.

\bibitem[Berta and Tomamichel(2016)]{BT15}
Mario Berta and Marco Tomamichel.
\newblock The fidelity of recovery is multiplicative.
\newblock \emph{IEEE Transactions on Information Theory}, 62\penalty0
  (4):\penalty0 1758--1763, April 2016.
\newblock ISSN 0018-9448.
\newblock \doi{10.1109/TIT.2016.2527683}.
\newblock arXiv:1502.07973.

\bibitem[Berta et~al.(2010)Berta, Christandl, Colbeck, Renes, and
  Renner]{BCCRR10}
Mario Berta, Matthias Christandl, Roger Colbeck, Joseph~M. Renes, and Renato
  Renner.
\newblock The uncertainty principle in the presence of quantum memory.
\newblock \emph{Nature Physics}, 6:\penalty0 659--662, 2010.
\newblock arXiv:0909.0950.

\bibitem[Berta et~al.(2011)Berta, Christandl, and Renner]{BCR09}
Mario Berta, Matthias Christandl, and Renato Renner.
\newblock The quantum reverse {Shannon} theorem based on one-shot information
  theory.
\newblock \emph{Communications in Mathematical Physics}, 306\penalty0
  (3):\penalty0 579--615, August 2011.
\newblock arXiv:0912.3805.

\bibitem[Berta et~al.(2013)Berta, Brandao, Christandl, and Wehner]{BBCW13}
Mario Berta, Fernando G. S.~L. Brandao, Matthias Christandl, and Stephanie
  Wehner.
\newblock Entanglement cost of quantum channels.
\newblock \emph{IEEE Transactions on Information Theory}, 59\penalty0
  (10):\penalty0 6779--6795, October 2013.
\newblock ISSN 0018-9448.
\newblock \doi{10.1109/TIT.2013.2268533}.
\newblock arXiv:1108.5357.

\bibitem[Berta et~al.(2014)Berta, Renes, and Wilde]{BRW14}
Mario Berta, Joseph~M. Renes, and Mark~M. Wilde.
\newblock Identifying the information gain of a quantum measurement.
\newblock \emph{IEEE Transactions on Information Theory}, 60\penalty0
  (12):\penalty0 7987--8006, December 2014.
\newblock ISSN 0018-9448.
\newblock \doi{10.1109/TIT.2014.2365207}.
\newblock arXiv:1301.1594.

\bibitem[Berta et~al.(2015{\natexlab{a}})Berta, Lemm, and Wilde]{BLW14}
Mario Berta, Marius Lemm, and Mark~M. Wilde.
\newblock Monotonicity of quantum relative entropy and recoverability.
\newblock \emph{Quantum Information and Computation}, 15\penalty0 (15 \&
  16):\penalty0 1333--1354, November 2015{\natexlab{a}}.
\newblock arXiv:1412.4067.

\bibitem[Berta et~al.(2015{\natexlab{b}})Berta, Seshadreesan, and Wilde]{BSW14}
Mario Berta, Kaushik Seshadreesan, and Mark~M. Wilde.
\newblock R\'enyi generalizations of the conditional quantum mutual
  information.
\newblock \emph{Journal of Mathematical Physics}, 56\penalty0 (2):\penalty0
  022205, February 2015{\natexlab{b}}.
\newblock arXiv:1403.6102.

\bibitem[Bhatia(1997)]{B97}
Rajendra Bhatia.
\newblock \emph{Matrix Analysis}.
\newblock Springer, Heidelberg, 1997.

\bibitem[Blume-Kohout et~al.(2014)Blume-Kohout, Croke, and Gottesman]{BCG09}
Robin Blume-Kohout, Sarah Croke, and Daniel Gottesman.
\newblock Streaming universal distortion-free entanglement concentration.
\newblock \emph{IEEE Transactions on Information Theory}, 60\penalty0
  (1):\penalty0 334--350, January 2014.
\newblock arXiv:0910.5952.

\bibitem[Boche and Notzel(2014)]{BN14}
Holger Boche and Janis Notzel.
\newblock The classical-quantum multiple access channel with conferencing
  encoders and with common messages.
\newblock \emph{Quantum Information Processing}, 13\penalty0 (12):\penalty0
  2595--2617, December 2014.
\newblock ISSN 1570-0755.
\newblock \doi{10.1007/s11128-014-0814-y}.
\newblock arXiv:1310.1970.

\bibitem[Bohm(1989)]{book1989bohm}
David Bohm.
\newblock \emph{Quantum Theory}.
\newblock {Courier Dover Publications}, 1989.

\bibitem[Bowen(2004)]{B04}
Garry Bowen.
\newblock Quantum feedback channels.
\newblock \emph{IEEE Transactions on Information Theory}, 50\penalty0
  (10):\penalty0 2429--2434, October 2004.
\newblock arXiv:quant-ph/0209076.

\bibitem[Bowen and Nagarajan(2005)]{BN05}
Garry Bowen and Rajagopal Nagarajan.
\newblock On feedback and the classical capacity of a noisy quantum channel.
\newblock \emph{IEEE Transactions on Information Theory}, 51\penalty0
  (1):\penalty0 320--324, January 2005.
\newblock arXiv:quant-ph/0305176.

\bibitem[Boyd and Vandenberghe(2004)]{BV04}
Stephen Boyd and Lieven Vandenberghe.
\newblock \emph{Convex Optimization}.
\newblock Cambridge University Press, The Edinburgh Building, Cambridge, CB2
  8RU, UK, 2004.

\bibitem[Br\'adler et~al.(2010)Br\'adler, Hayden, Touchette, and Wilde]{BHTW10}
Kamil Br\'adler, Patrick Hayden, Dave Touchette, and Mark~M. Wilde.
\newblock Trade-off capacities of the quantum {Hadamard} channels.
\newblock \emph{Physical Review A}, 81\penalty0 (6):\penalty0 062312, June
  2010.
\newblock \doi{10.1103/PhysRevA.81.062312}.
\newblock arXiv:1001.1732.

\bibitem[Brandao and Horodecki(2010)]{BH10}
Fernando G. S.~L. Brandao and Michal Horodecki.
\newblock On {Hastings'} counterexamples to the minimum output entropy
  additivity conjecture.
\newblock \emph{Open Systems \& Information Dynamics}, 17\penalty0
  (1):\penalty0 31--52, 2010.
\newblock arXiv:0907.3210.

\bibitem[Brandao et~al.(2011)Brandao, Christandl, and Yard]{BCY11}
Fernando G. S.~L. Brandao, Matthias Christandl, and Jon Yard.
\newblock Faithful squashed entanglement.
\newblock \emph{Communications in Mathematical Physics}, 306\penalty0
  (3):\penalty0 805--830, September 2011.
\newblock ISSN 0010-3616.
\newblock \doi{10.1007/s00220-011-1302-1}.
\newblock arXiv:1010.1750.

\bibitem[Brandao et~al.(2014)Brandao, Harrow, Oppenheim, and Strelchuk]{BHOS14}
Fernando G. S.~L. Brandao, Aram~W. Harrow, Jonathan Oppenheim, and Sergii
  Strelchuk.
\newblock Quantum conditional mutual information, reconstructed states, and
  state redistribution.
\newblock \emph{Physical Review Letters}, 115\penalty0 (5):\penalty0 050501,
  July 2014.
\newblock arXiv:1411.4921.

\bibitem[Braunstein et~al.(2000)Braunstein, Fuchs, Gottesman, and Lo]{BFGL00}
Samuel~L. Braunstein, Christopher~A. Fuchs, Daniel Gottesman, and Hoi-Kwong Lo.
\newblock A quantum analog of {Huffman} coding.
\newblock \emph{IEEE Transactions on Information Theory}, 46\penalty0
  (4):\penalty0 1644--1649, July 2000.
\newblock arXiv:quant-ph/9805080.

\bibitem[Brun()]{brun_lectures}
Todd~A. Brun.
\newblock Quantum information processing course lecture slides.
\newblock \verb+http://almaak.usc.edu/~tbrun/Course/+.

\bibitem[Burnashev and Holevo(1998)]{BH98}
M.~V. Burnashev and Alexander~S. Holevo.
\newblock On reliability function of quantum communication channel.
\newblock \emph{Probl. Peredachi Inform.}, 34\penalty0 (2):\penalty0 1--13,
  1998.
\newblock arXiv:quant-ph/9703013.

\bibitem[Buscemi and Datta(2010)]{BD10}
Francesco Buscemi and Nilanjana Datta.
\newblock The quantum capacity of channels with arbitrarily correlated noise.
\newblock \emph{IEEE Transactions on Information Theory}, 56\penalty0
  (3):\penalty0 1447--1460, March 2010.
\newblock arXiv:0902.0158.

\bibitem[Cai et~al.(2004)Cai, Winter, and Yeung]{1050633}
N.~Cai, Andreas Winter, and Raymond~W. Yeung.
\newblock Quantum privacy and quantum wiretap channels.
\newblock \emph{Problems of Information Transmission}, 40\penalty0
  (4):\penalty0 318--336, October 2004.
\newblock ISSN 0032-9460.
\newblock \doi{http://dx.doi.org/10.1007/s11122-005-0002-x}.

\bibitem[Calderbank and Shor(1996)]{PhysRevA.54.1098}
A.~Robert Calderbank and Peter~W. Shor.
\newblock Good quantum error-correcting codes exist.
\newblock \emph{Physical Review A}, 54\penalty0 (2):\penalty0 1098--1105,
  August 1996.
\newblock \doi{10.1103/PhysRevA.54.1098}.
\newblock arXiv:quant-ph/9512032.

\bibitem[Calderbank et~al.(1997)Calderbank, Rains, Shor, and
  Sloane]{PhysRevLett.78.405}
A.~Robert Calderbank, Eric~M. Rains, Peter~W. Shor, and N.~J.~A. Sloane.
\newblock Quantum error correction and orthogonal geometry.
\newblock \emph{Physical Review Letters}, 78\penalty0 (3):\penalty0 405--408,
  January 1997.
\newblock \doi{10.1103/PhysRevLett.78.405}.
\newblock arXiv:quant-ph/9605005.

\bibitem[Calderbank et~al.(1998)Calderbank, Rains, Shor, and
  Sloane]{ieee1998calderbank}
A.~Robert Calderbank, Eric~M. Rains, Peter~W. Shor, and N.~J.~A. Sloane.
\newblock Quantum error correction via codes over {GF(4)}.
\newblock \emph{IEEE Transactions on Information Theory}, 44\penalty0
  (4):\penalty0 1369--1387, July 1998.
\newblock arXiv:quant-ph/9608006.

\bibitem[Carlen and Lieb(2014)]{CL14}
Eric~A. Carlen and Elliott~H. Lieb.
\newblock Remainder terms for some quantum entropy inequalities.
\newblock \emph{Journal of Mathematical Physics}, 55\penalty0 (4):\penalty0
  042201, April 2014.
\newblock arXiv:1402.3840.

\bibitem[Cerf and Adami(1997)]{CA97}
Nicolas~J. Cerf and Christoph Adami.
\newblock Negative entropy and information in quantum mechanics.
\newblock \emph{Physical Review Letters}, 79\penalty0 (26):\penalty0
  5194--5197, December 1997.
\newblock arXiv:quant-ph/9512022.

\bibitem[Coles et~al.(2015)Coles, Berta, Tomamichel, and Wehner]{CBTW15}
Patrick Coles, Mario Berta, Marco Tomamichel, and Stephanie Wehner.
\newblock Entropic uncertainty relations and their applications.
\newblock November 2015.
\newblock arXiv:1511.04857.

\bibitem[Coles et~al.(2012)Coles, Colbeck, Yu, and Zwolak]{CCYZ11}
Patrick~J. Coles, Roger Colbeck, Li~Yu, and Michael Zwolak.
\newblock Uncertainty relations from simple entropic properties.
\newblock \emph{Physical Review Letters}, 108\penalty0 (21):\penalty0 210405,
  May 2012.
\newblock arXiv:1112.0543.

\bibitem[Cooney et~al.(2016)Cooney, Mosonyi, and Wilde]{CMW14}
Tom Cooney, Milan Mosonyi, and Mark~M. Wilde.
\newblock Strong converse exponents for a quantum channel discrimination
  problem and quantum-feedback-assisted communication.
\newblock \emph{Communications in Mathematical Physics}, 344\penalty0
  (3):\penalty0 797--829, June 2016.
\newblock arXiv:1408.3373.

\bibitem[Cover and Thomas(2006)]{book1991cover}
Thomas~M. Cover and Joy~A. Thomas.
\newblock \emph{Elements of Information Theory}.
\newblock Wiley-Interscience, second edition, 2006.

\bibitem[Csiszar(1967)]{C67}
Imre Csiszar.
\newblock Information-type measures of difference of probability distributions
  and indirect observations.
\newblock \emph{Studia Sci. Math. Hungar.}, 2:\penalty0 299--318, 1967.

\bibitem[Csisz\'{a}r and K\"orner(1978)]{CK78}
Imre Csisz\'{a}r and Janos K\"orner.
\newblock Broadcast channels with confidential messages.
\newblock \emph{IEEE Transactions on Information Theory}, 24\penalty0
  (3):\penalty0 339--348, May 1978.

\bibitem[Csisz\'{a}r and K\"orner(2011)]{CK97}
Imre Csisz\'{a}r and Janos K\"orner.
\newblock \emph{Information Theory: Coding Theorems for Discrete Memoryless
  Systems}.
\newblock Probability and mathematical statistics. Cambridge University Press,
  second edition, 2011.

\bibitem[Cubitt et~al.(2015)Cubitt, Elkouss, Matthews, Ozols, Perez-Garcia, and
  Strelchuk]{CEMOGS15}
Toby Cubitt, David Elkouss, William Matthews, Maris Ozols, David Perez-Garcia,
  and Sergii Strelchuk.
\newblock Unbounded number of channel uses may be required to detect quantum
  capacity.
\newblock \emph{Nature Communications}, 6:\penalty0 6739, March 2015.
\newblock arXiv:1408.5115.

\bibitem[Czekaj and Horodecki(2009)]{CH08}
Lukasz Czekaj and Pawel Horodecki.
\newblock Purely quantum superadditivity of classical capacities of quantum
  multiple access channels.
\newblock \emph{Physical Review Letters}, 102\penalty0 (11):\penalty0 110505,
  March 2009.
\newblock arXiv:0807.3977.

\bibitem[Dalai(2013)]{D13}
Marco Dalai.
\newblock Lower bounds on the probability of error for classical and
  classical-quantum channels.
\newblock \emph{IEEE Transactions on Information Theory}, 59\penalty0
  (12):\penalty0 8027--8056, December 2013.
\newblock ISSN 0018-9448.
\newblock \doi{10.1109/TIT.2013.2283794}.
\newblock arXiv:1201.5411.

\bibitem[Datta(2009)]{D09}
Nilanjana Datta.
\newblock Min- and max-relative entropies and a new entanglement monotone.
\newblock \emph{IEEE Transactions on Information Theory}, 55\penalty0
  (6):\penalty0 2816--2826, June 2009.
\newblock arXiv:0803.2770.

\bibitem[Datta and Hsieh(2010)]{datta:122202}
Nilanjana Datta and Min-Hsiu Hsieh.
\newblock Universal coding for transmission of private information.
\newblock \emph{Journal of Mathematical Physics}, 51\penalty0 (12):\penalty0
  122202, 2010.
\newblock \doi{10.1063/1.3521499}.
\newblock arXiv:1007.2629.

\bibitem[Datta and Hsieh(2011)]{DH11a}
Nilanjana Datta and Min-Hsiu Hsieh.
\newblock The apex of the family tree of protocols: Optimal rates and resource
  inequalities.
\newblock \emph{New Journal of Physics}, 13:\penalty0 093042, September 2011.
\newblock arXiv:1103.1135.

\bibitem[Datta and Hsieh(2013)]{DH11}
Nilanjana Datta and Min-Hsiu Hsieh.
\newblock One-shot entanglement-assisted quantum and classical communication.
\newblock \emph{IEEE Transactions on Information Theory}, 59\penalty0
  (3):\penalty0 1929--1939, March 2013.
\newblock arXiv:1105.3321.

\bibitem[Datta and Leditzky(2015)]{DL15}
Nilanjana Datta and Felix Leditzky.
\newblock Second-order asymptotics for source coding, dense coding, and
  pure-state entanglement conversions.
\newblock \emph{IEEE Transactions on Information Theory}, 61\penalty0
  (1):\penalty0 582--608, January 2015.
\newblock ISSN 0018-9448.
\newblock \doi{10.1109/TIT.2014.2366994}.
\newblock arXiv:1403.2543.

\bibitem[Datta and Renner(2009)]{DR09}
Nilanjana Datta and Renato Renner.
\newblock Smooth entropies and the quantum information spectrum.
\newblock \emph{IEEE Transactions on Information Theory}, 55\penalty0
  (6):\penalty0 2807--2815, June 2009.
\newblock arXiv:0801.0282.

\bibitem[Datta and Wilde(2015)]{DW15}
Nilanjana Datta and Mark~M. Wilde.
\newblock Quantum {Markov} chains, sufficiency of quantum channels, and
  {R\'enyi} information measures.
\newblock \emph{Journal of Physics A}, 48\penalty0 (50):\penalty0 505301,
  December 2015.
\newblock arXiv:1501.05636.

\bibitem[Datta et~al.(2016)Datta, Tomamichel, and Wilde]{DTW14}
Nilanjana Datta, Marco Tomamichel, and Mark~M. Wilde.
\newblock On the second-order asymptotics for entanglement-assisted
  communication.
\newblock \emph{Quantum Information Processing}, 15\penalty0 (6):\penalty0
  2569--2591, June 2016.
\newblock arXiv:1405.1797.

\bibitem[Davies and Lewis(1970)]{DL70}
E.~B. Davies and J.~T. Lewis.
\newblock An operational approach to quantum probability.
\newblock \emph{Communications in Mathematical Physics}, 17\penalty0
  (3):\penalty0 239--260, 1970.

\bibitem[de~Broglie(1924)]{debroglie1924}
Louis de~Broglie.
\newblock \emph{Recherches sur la th\'{e}orie des quanta}.
\newblock PhD thesis, Paris, 1924.

\bibitem[Deutsch(1985)]{prsla1985deutsch}
David Deutsch.
\newblock Quantum theory, the {Church-Turing} principle and the universal
  quantum computer.
\newblock \emph{Proceedings of the Royal Society of London A}, 400\penalty0
  (1818):\penalty0 97--117, July 1985.

\bibitem[Devetak(2005)]{ieee2005dev}
Igor Devetak.
\newblock The private classical capacity and quantum capacity of a quantum
  channel.
\newblock \emph{IEEE Transactions on Information Theory}, 51\penalty0
  (1):\penalty0 44--55, January 2005.
\newblock arXiv:quant-ph/0304127.

\bibitem[Devetak(2006)]{D06}
Igor Devetak.
\newblock Triangle of dualities between quantum communication protocols.
\newblock \emph{Physical Review Letters}, 97\penalty0 (14):\penalty0 140503,
  October 2006.
\newblock \doi{10.1103/PhysRevLett.97.140503}.

\bibitem[Devetak and Shor(2005)]{cmp2005dev}
Igor Devetak and Peter~W. Shor.
\newblock The capacity of a quantum channel for simultaneous transmission of
  classical and quantum information.
\newblock \emph{Communications in Mathematical Physics}, 256\penalty0
  (2):\penalty0 287--303, June 2005.
\newblock arXiv:quant-ph/0311131.

\bibitem[Devetak and Winter(2003)]{DW03}
Igor Devetak and Andreas Winter.
\newblock Classical data compression with quantum side information.
\newblock \emph{Physical Review A}, 68\penalty0 (4):\penalty0 042301, October
  2003.
\newblock \doi{10.1103/PhysRevA.68.042301}.
\newblock arXiv:quant-ph/0209029.

\bibitem[Devetak and Winter(2004)]{DW04}
Igor Devetak and Andreas Winter.
\newblock Relating quantum privacy and quantum coherence: An operational
  approach.
\newblock \emph{Physical Review Letters}, 93\penalty0 (8):\penalty0 080501,
  August 2004.
\newblock \doi{10.1103/PhysRevLett.93.080501}.
\newblock arXiv:quant-ph/0307053.

\bibitem[Devetak and Winter(2005)]{DW05}
Igor Devetak and Andreas Winter.
\newblock Distillation of secret key and entanglement from quantum states.
\newblock \emph{Proceedings of the Royal Society A}, 461\penalty0
  (2053):\penalty0 207--235, January 2005.
\newblock arXiv:quant-ph/0306078.

\bibitem[Devetak and Yard(2008)]{DY08}
Igor Devetak and Jon Yard.
\newblock Exact cost of redistributing multipartite quantum states.
\newblock \emph{Physical Review Letters}, 100\penalty0 (23):\penalty0 230501,
  June 2008.
\newblock \doi{10.1103/PhysRevLett.100.230501}.

\bibitem[Devetak et~al.(2004)Devetak, Harrow, and Winter]{DHW03}
Igor Devetak, Aram~W. Harrow, and Andreas Winter.
\newblock A family of quantum protocols.
\newblock \emph{Physical Review Letters}, 93\penalty0 (23):\penalty0 230504,
  December 2004.
\newblock arXiv:quant-ph/0308044.

\bibitem[Devetak et~al.(2006)Devetak, Junge, King, and Ruskai]{DJKR06}
Igor Devetak, Marius Junge, Christopher King, and Mary~Beth Ruskai.
\newblock Multiplicativity of completely bounded p-norms implies a new
  additivity result.
\newblock \emph{Communications in Mathematical Physics}, 266\penalty0
  (1):\penalty0 37--63, August 2006.
\newblock arXiv:quant-ph/0506196.

\bibitem[Devetak et~al.(2008)Devetak, Harrow, and Winter]{DHW05RI}
Igor Devetak, Aram~W. Harrow, and Andreas Winter.
\newblock A resource framework for quantum {Shannon} theory.
\newblock \emph{IEEE Transactions on Information Theory}, 54\penalty0
  (10):\penalty0 4587--4618, October 2008.
\newblock arXiv:quant-ph/0512015.

\bibitem[Dieks(1982)]{D82}
D.~Dieks.
\newblock Communication by {EPR} devices.
\newblock \emph{Physics Letters A}, 92:\penalty0 271, 1982.

\bibitem[Ding and Wilde(2015)]{DingW15}
David Ding and Mark~M. Wilde.
\newblock Strong converse exponents for the feedback-assisted classical
  capacity of entanglement-breaking channels.
\newblock June 2015.
\newblock arXiv:1506.02228.

\bibitem[Dirac(1982)]{citeulike:1280736}
P.~A.~M. Dirac.
\newblock \emph{The Principles of Quantum Mechanics (International Series of
  Monographs on Physics)}.
\newblock {Oxford University Press, USA}, February 1982.
\newblock ISBN 0198520115.

\bibitem[DiVincenzo et~al.(1998)DiVincenzo, Shor, and Smolin]{DSS98}
David~P. DiVincenzo, Peter~W. Shor, and John~A. Smolin.
\newblock Quantum-channel capacity of very noisy channels.
\newblock \emph{Physical Review A}, 57\penalty0 (2):\penalty0 830--839,
  February 1998.
\newblock \doi{10.1103/PhysRevA.57.830}.
\newblock arXiv:quant-ph/9706061.

\bibitem[DiVincenzo et~al.(2004)DiVincenzo, Horodecki, Leung, Smolin, and
  Terhal]{DHLST04}
David~P. DiVincenzo, Micha\l{} Horodecki, Debbie~W. Leung, John~A. Smolin, and
  Barbara~M. Terhal.
\newblock Locking classical correlations in quantum states.
\newblock \emph{Physical Review Letters}, 92\penalty0 (6):\penalty0 067902,
  February 2004.
\newblock \doi{10.1103/PhysRevLett.92.067902}.
\newblock arXiv:quant-ph/0303088.

\bibitem[Dowling and Milburn(2003)]{dowling2003}
Jonathan~P. Dowling and Gerard~J Milburn.
\newblock Quantum technology: The second quantum revolution.
\newblock \emph{Philosophical Transactions of The Royal Society of London
  Series A}, 361\penalty0 (1809):\penalty0 1655--1674, August 2003.
\newblock arXiv:quant-ph/0206091.

\bibitem[Dupuis(2010)]{D10}
Frederic Dupuis.
\newblock \emph{The decoupling approach to quantum information theory}.
\newblock PhD thesis, University of Montreal, April 2010.
\newblock arXiv:1004.1641.

\bibitem[Dupuis and Wilde(2016)]{DW15a}
Fr\'ed\'eric Dupuis and Mark~M. Wilde.
\newblock Swiveled {R\'enyi} entropies.
\newblock \emph{Quantum Information Processing}, 15\penalty0 (3):\penalty0
  1309--1345, March 2016.
\newblock \doi{10.1007/s11128-015-1211-x}.
\newblock arXiv:1506.00981.

\bibitem[Dupuis et~al.(2010)Dupuis, Hayden, and Li]{DH2006}
Frederic Dupuis, Patrick Hayden, and Ke~Li.
\newblock A father protocol for quantum broadcast channels.
\newblock \emph{IEEE Transactions on Information Theory}, 56\penalty0
  (6):\penalty0 2946--2956, June 2010.
\newblock arXiv:quant-ph/0612155.

\bibitem[Dupuis et~al.(2013)Dupuis, Florjanczyk, Hayden, and
  Leung]{Dupuis20130289}
Fr{\'e}d{\'e}ric Dupuis, Jan Florjanczyk, Patrick Hayden, and Debbie Leung.
\newblock The locking-decoding frontier for generic dynamics.
\newblock \emph{Proceedings of the Royal Society of London A: Mathematical,
  Physical and Engineering Sciences}, 469\penalty0 (2159), 2013.
\newblock ISSN 1364-5021.
\newblock \doi{10.1098/rspa.2013.0289}.
\newblock arXiv:1011.1612.

\bibitem[Dupuis et~al.(2014)Dupuis, Berta, Wullschleger, and Renner]{DBWR10}
Frederic Dupuis, Mario Berta, J\"{u}rg Wullschleger, and Renato Renner.
\newblock One-shot decoupling.
\newblock \emph{Communications in Mathematical Physics}, 328\penalty0
  (1):\penalty0 251--284, May 2014.
\newblock arXiv:1012.6044.

\bibitem[Dutil(2011)]{D11}
Nicolas Dutil.
\newblock \emph{Multiparty quantum protocols for assisted entanglement
  distillation}.
\newblock PhD thesis, McGill University, May 2011.
\newblock arXiv:1105.4657.

\bibitem[Einstein(1905)]{einstein1905}
Albert Einstein.
\newblock \"{U}ber einen die erzeugung und verwandlung des lichtes betreffenden
  heuristischen gesichtspunkt.
\newblock \emph{Annalen der Physik}, 17:\penalty0 132--148, 1905.

\bibitem[Einstein et~al.(1935)Einstein, Podolsky, and Rosen]{epr1935}
Albert Einstein, Boris Podolsky, and Nathan Rosen.
\newblock Can quantum-mechanical description of physical reality be considered
  complete?
\newblock \emph{Physical Review}, 47:\penalty0 777--780, 1935.

\bibitem[Ekert(1991)]{Ekert:1991:661}
Artur~K. Ekert.
\newblock Quantum cryptography based on {Bell's} theorem.
\newblock \emph{Physical Review Letters}, 67\penalty0 (6):\penalty0 661--663,
  August 1991.
\newblock \doi{10.1103/PhysRevLett.67.661}.

\bibitem[Elias(1972)]{E72}
Peter Elias.
\newblock The efficient construction of an unbiased random sequence.
\newblock \emph{Annals of Mathematical Statistics}, 43\penalty0 (3):\penalty0
  865--870, 1972.

\bibitem[Elkouss and Strelchuk(2015)]{PhysRevLett.115.040501}
David Elkouss and Sergii Strelchuk.
\newblock Superadditivity of private information for any number of uses of the
  channel.
\newblock \emph{Physical Review Letters}, 115\penalty0 (4):\penalty0 040501,
  July 2015.
\newblock \doi{10.1103/PhysRevLett.115.040501}.
\newblock arXiv:1502.05326.

\bibitem[Fannes(1973)]{Fannes73}
Mark Fannes.
\newblock A continuity property of the entropy density for spin lattices.
\newblock \emph{Communications~in~Mathematical~Physics}, 31:\penalty0 291,
  1973.

\bibitem[Fano(2008)]{F08}
Robert~Mario Fano.
\newblock Fano inequality.
\newblock \emph{Scholarpedia}, 3\penalty0 (10):\penalty0 6648, 2008.

\bibitem[Fawzi and Renner(2015)]{FR14}
Omar Fawzi and Renato Renner.
\newblock Quantum conditional mutual information and approximate {Markov}
  chains.
\newblock \emph{Communications in Mathematical Physics}, 340\penalty0
  (2):\penalty0 575--611, December 2015.
\newblock arXiv:1410.0664.

\bibitem[Fawzi et~al.(2012)Fawzi, Hayden, Savov, Sen, and Wilde]{FHSSW11}
Omar Fawzi, Patrick Hayden, Ivan Savov, Pranab Sen, and Mark~M. Wilde.
\newblock Classical communication over a quantum interference channel.
\newblock \emph{IEEE Transactions on Information Theory}, 58\penalty0
  (6):\penalty0 3670--3691, June 2012.
\newblock arXiv:1102.2624.

\bibitem[Fawzi et~al.(2013)Fawzi, Hayden, and Sen]{FHS13}
Omar Fawzi, Patrick Hayden, and Pranab Sen.
\newblock From low-distortion norm embeddings to explicit uncertainty relations
  and efficient information locking.
\newblock \emph{Journal of the ACM}, 60\penalty0 (6):\penalty0 44:1--44:61,
  November 2013.
\newblock ISSN 0004-5411.
\newblock \doi{10.1145/2518131}.
\newblock arXiv:1010.3007.

\bibitem[Feller(1971)]{F71}
William Feller.
\newblock \emph{An Introduction to Probability Theory and Its Applications}.
\newblock John Wiley and Sons, 2nd edition edition, 1971.

\bibitem[Feynman(1982)]{ijtp1982feynman}
Richard~P. Feynman.
\newblock Simulating physics with computers.
\newblock \emph{International Journal of Theoretical Physics}, 21:\penalty0
  467--488, 1982.

\bibitem[Feynman(1998)]{Feynman98}
Richard~P. Feynman.
\newblock \emph{Feynman Lectures On Physics (3 Volume Set)}.
\newblock {Addison Wesley Longman}, September 1998.
\newblock ISBN 0201021153.

\bibitem[Fuchs(1996)]{F96}
Christopher Fuchs.
\newblock \emph{Distinguishability and Accessible Information in Quantum
  Theory}.
\newblock PhD thesis, University of New Mexico, December 1996.
\newblock arXiv:quant-ph/9601020.

\bibitem[Fuchs and Caves(1995)]{FC95}
Christopher~A. Fuchs and Carlton~M. Caves.
\newblock Mathematical techniques for quantum communication theory.
\newblock \emph{Open Systems \& Information Dynamics}, 3\penalty0 (3):\penalty0
  345--356, 1995.
\newblock arXiv:quant-ph/9604001.

\bibitem[Fuchs and van~de Graaf(1998)]{FG98}
Christopher~A. Fuchs and Jeroen van~de Graaf.
\newblock Cryptographic distinguishability measures for quantum mechanical
  states.
\newblock \emph{IEEE Transactions on Information Theory}, 45\penalty0
  (4):\penalty0 1216--1227, May 1998.
\newblock arXiv:quant-ph/9712042.

\bibitem[Fukuda and King(2010)]{fukuda:042201}
Motohisa Fukuda and Christopher King.
\newblock Entanglement of random subspaces via the {Hastings} bound.
\newblock \emph{Journal of Mathematical Physics}, 51\penalty0 (4):\penalty0
  042201, 2010.
\newblock \doi{10.1063/1.3309418}.
\newblock arXiv:0907.5446.

\bibitem[Fukuda et~al.(2010)Fukuda, King, and Moser]{FKM10}
Motohisa Fukuda, Christopher King, and David~K. Moser.
\newblock Comments on {Hastings'} additivity counterexamples.
\newblock \emph{Communications in Mathematical Physics}, 296\penalty0
  (1):\penalty0 111--143, 2010.
\newblock arXiv:0905.3697.

\bibitem[Gamal and Kim(2012)]{el2010lecture}
Abbas~El Gamal and Young-Han Kim.
\newblock \emph{Network Information Theory}.
\newblock Cambridge University Press, January 2012.
\newblock arXiv:1001.3404.

\bibitem[Garc\'\i{}a-Patr\'on et~al.(2009)Garc\'\i{}a-Patr\'on, Pirandola,
  Lloyd, and Shapiro]{GPLS09}
Ra\'ul Garc\'\i{}a-Patr\'on, Stefano Pirandola, Seth Lloyd, and Jeffrey~H.
  Shapiro.
\newblock Reverse coherent information.
\newblock \emph{Physical Review Letters}, 102\penalty0 (21):\penalty0 210501,
  May 2009.
\newblock \doi{10.1103/PhysRevLett.102.210501}.
\newblock arXiv:0808.0210.

\bibitem[Gerlach and Stern(1922)]{sterngerlach}
Walther Gerlach and Otto Stern.
\newblock Das magnetische moment des silberatoms.
\newblock \emph{Zeitschrift f\"{u}r Physik}, 9:\penalty0 353--355, 1922.

\bibitem[Giovannetti and Fazio(2005)]{PhysRevA.71.032314}
Vittorio Giovannetti and Rosario Fazio.
\newblock Information-capacity description of spin-chain correlations.
\newblock \emph{Physical Review A}, 71\penalty0 (3):\penalty0 032314, March
  2005.
\newblock \doi{10.1103/PhysRevA.71.032314}.
\newblock arXiv:quant-ph/0405110.

\bibitem[Giovannetti et~al.(2003{\natexlab{a}})Giovannetti, Lloyd, Maccone, and
  Shor]{GLMS03}
Vittorio Giovannetti, Seth Lloyd, Lorenzo Maccone, and Peter~W. Shor.
\newblock Entanglement assisted capacity of the broadband lossy channel.
\newblock \emph{Physical Review Letters}, 91\penalty0 (4):\penalty0 047901,
  July 2003{\natexlab{a}}.
\newblock \doi{10.1103/PhysRevLett.91.047901}.
\newblock arXiv:quant-ph/0304020.

\bibitem[Giovannetti et~al.(2003{\natexlab{b}})Giovannetti, Lloyd, Maccone, and
  Shor]{GLMS03a}
Vittorio Giovannetti, Seth Lloyd, Lorenzo Maccone, and Peter~W. Shor.
\newblock Broadband channel capacities.
\newblock \emph{Physical Review A}, 68\penalty0 (6):\penalty0 062323, December
  2003{\natexlab{b}}.
\newblock \doi{10.1103/PhysRevA.68.062323}.
\newblock arXiv:quant-ph/0307098.

\bibitem[Giovannetti et~al.(2004{\natexlab{a}})Giovannetti, Guha, Lloyd,
  Maccone, and Shapiro]{GGLMS04}
Vittorio Giovannetti, Saikat Guha, Seth Lloyd, Lorenzo Maccone, and Jeffrey~H.
  Shapiro.
\newblock Minimum output entropy of bosonic channels: A conjecture.
\newblock \emph{Physical Review A}, 70\penalty0 (3):\penalty0 032315, September
  2004{\natexlab{a}}.
\newblock \doi{10.1103/PhysRevA.70.032315}.
\newblock arXiv:quant-ph/0404005.

\bibitem[Giovannetti et~al.(2004{\natexlab{b}})Giovannetti, Guha, Lloyd,
  Maccone, Shapiro, and Yuen]{GGLMSY04}
Vittorio Giovannetti, Saikat Guha, Seth Lloyd, Lorenzo Maccone, Jeffrey~H.
  Shapiro, and Horace~P. Yuen.
\newblock Classical capacity of the lossy bosonic channel: The exact solution.
\newblock \emph{Physical Review Letters}, 92\penalty0 (2):\penalty0 027902,
  January 2004{\natexlab{b}}.
\newblock \doi{10.1103/PhysRevLett.92.027902}.
\newblock arXiv:quant-ph/0308012.

\bibitem[Giovannetti et~al.(2010)Giovannetti, Holevo, Lloyd, and
  Maccone]{GHLM10}
Vittorio Giovannetti, Alexander~S. Holevo, Seth Lloyd, and Lorenzo Maccone.
\newblock Generalized minimal output entropy conjecture for one-mode {Gaussian}
  channels: definitions and some exact results.
\newblock \emph{Journal of Physics A: Mathematical and Theoretical},
  43\penalty0 (41):\penalty0 415305, 2010.
\newblock arXiv:1004.4787.

\bibitem[Giovannetti et~al.(2012)Giovannetti, Lloyd, and
  Maccone]{PhysRevA.85.012302}
Vittorio Giovannetti, Seth Lloyd, and Lorenzo Maccone.
\newblock Achieving the {Holevo} bound via sequential measurements.
\newblock \emph{Physical Review A}, 85\penalty0 (1):\penalty0 012302, January
  2012.
\newblock \doi{10.1103/PhysRevA.85.012302}.
\newblock arXiv:1012.0386.

\bibitem[Giovannetti et~al.(2015)Giovannetti, Holevo, and Garcia-Patron]{GHG15}
Vittorio Giovannetti, Alexander~S. Holevo, and Raul Garcia-Patron.
\newblock A solution of {Gaussian} optimizer conjecture for quantum channels.
\newblock \emph{Communications in Mathematical Physics}, 334\penalty0
  (3):\penalty0 1553--1571, March 2015.
\newblock ISSN 0010-3616.
\newblock \doi{10.1007/s00220-014-2150-6}.
\newblock arXiv:1312.2251.

\bibitem[Glauber(1963{\natexlab{a}})]{PhysRev.130.2529}
Roy~J. Glauber.
\newblock The quantum theory of optical coherence.
\newblock \emph{Physical Review}, 130\penalty0 (6):\penalty0 2529--2539, June
  1963{\natexlab{a}}.
\newblock \doi{10.1103/PhysRev.130.2529}.

\bibitem[Glauber(1963{\natexlab{b}})]{PhysRev.131.2766}
Roy~J. Glauber.
\newblock Coherent and incoherent states of the radiation field.
\newblock \emph{Physical Review}, 131\penalty0 (6):\penalty0 2766--2788,
  September 1963{\natexlab{b}}.
\newblock \doi{10.1103/PhysRev.131.2766}.

\bibitem[Glauber(2005)]{nobel2005glauber}
Roy~J. Glauber.
\newblock One hundred years of light quanta.
\newblock In Karl Grandin, editor, \emph{Les Prix Nobel. The Nobel Prizes
  2005}, pages 90--91. Nobel Foundation, 2005.

\bibitem[Gordon(1964)]{gordon1964}
James~P. Gordon.
\newblock Noise at optical frequencies; information theory.
\newblock In P.~A. Miles, editor, \emph{Quantum Electronics and Coherent Light;
  Proceedings of the International School of Physics Enrico Fermi, Course
  XXXI}, pages 156--181, Academic Press New York, 1964.

\bibitem[Gottesman(1996)]{PhysRevA.54.1862}
Daniel Gottesman.
\newblock Class of quantum error-correcting codes saturating the quantum
  {Hamming} bound.
\newblock \emph{Physical Review A}, 54\penalty0 (3):\penalty0 1862--1868,
  September 1996.
\newblock \doi{10.1103/PhysRevA.54.1862}.
\newblock arXiv:quant-ph/9604038.

\bibitem[Gottesman(1997)]{thesis97gottesman}
Daniel Gottesman.
\newblock \emph{Stabilizer Codes and Quantum Error Correction}.
\newblock PhD thesis, California Institute of Technology, 1997.
\newblock arXiv:quant-ph/9705052.

\bibitem[Grafakos(2008)]{G08}
Loukas Grafakos.
\newblock \emph{Classical Fourier Analysis}.
\newblock Springer, second edition, 2008.

\bibitem[Grassl et~al.(1997)Grassl, Beth, and Pellizzari]{GBP97}
Markus Grassl, Thomas Beth, and Thomas Pellizzari.
\newblock Codes for the quantum erasure channel.
\newblock \emph{Physical Review A}, 56\penalty0 (1):\penalty0 33--38, July
  1997.
\newblock \doi{10.1103/PhysRevA.56.33}.
\newblock arXiv:quant-ph/9610042.

\bibitem[Greene(1999)]{elegantuniverse}
Brian Greene.
\newblock \emph{The Elegant Universe: Superstrings, Hidden Dimensions, and the
  Quest for the Ultimate Theory}.
\newblock W. W. Norton \& Company, 1999.

\bibitem[Griffiths(1995)]{Grif95a}
David~J. Griffiths.
\newblock \emph{Introduction to Quantum Mechanics}.
\newblock {Prentice-Hall, Inc.}, 1995.

\bibitem[Groisman et~al.(2005)Groisman, Popescu, and Winter]{GPW05}
Berry Groisman, Sandu Popescu, and Andreas Winter.
\newblock Quantum, classical, and total amount of correlations in a quantum
  state.
\newblock \emph{Physical Review A}, 72\penalty0 (3):\penalty0 032317, September
  2005.
\newblock \doi{10.1103/PhysRevA.72.032317}.
\newblock arXiv:quant-ph/0410091.

\bibitem[Grudka and Horodecki(2010)]{PhysRevA.81.060305}
Andrzej Grudka and Pawe\l{} Horodecki.
\newblock Nonadditivity of quantum and classical capacities for entanglement
  breaking multiple-access channels and the butterfly network.
\newblock \emph{Physical Review A}, 81\penalty0 (6):\penalty0 060305, June
  2010.
\newblock \doi{10.1103/PhysRevA.81.060305}.
\newblock arXiv:0906.1305.

\bibitem[Guha(2008)]{G08thesis}
Saikat Guha.
\newblock \emph{Multiple-User Quantum Information Theory for Optical
  Communication Channels}.
\newblock PhD thesis, Massachusetts Institute of Technology, June 2008.

\bibitem[Guha and Shapiro(2007)]{GS07}
Saikat Guha and Jeffrey~H. Shapiro.
\newblock Classical information capacity of the bosonic broadcast channel.
\newblock In \emph{Proceedings of the IEEE International Symposium on
  Information Theory}, pages 1896--1900, Nice, France, June 2007.
\newblock arXiv:0704.1901.

\bibitem[Guha et~al.(2007)Guha, Shapiro, and Erkmen]{GSE07}
Saikat Guha, Jeffrey~H. Shapiro, and Baris~I. Erkmen.
\newblock Classical capacity of bosonic broadcast communication and a minimum
  output entropy conjecture.
\newblock \emph{Physical Review A}, 76\penalty0 (3):\penalty0 032303, September
  2007.
\newblock \doi{10.1103/PhysRevA.76.032303}.
\newblock arXiv:0706.3416.

\bibitem[Guha et~al.(2008)Guha, Shapiro, and Erkmen]{GSE08}
Saikat Guha, Jeffrey~H. Shapiro, and Baris~I. Erkmen.
\newblock Capacity of the bosonic wiretap channel and the entropy photon-number
  inequality.
\newblock In \emph{Proceedings of the IEEE International Symposium on
  Information Theory}, pages 91--95, Toronto, Ontario, Canada, July 2008.
\newblock arXiv:0801.0841.

\bibitem[Guha et~al.(2014)Guha, Hayden, Krovi, Lloyd, Lupo, Shapiro, Takeoka,
  and Wilde]{PhysRevX.4.011016}
Saikat Guha, Patrick Hayden, Hari Krovi, Seth Lloyd, Cosmo Lupo, Jeffrey~H.
  Shapiro, Masahiro Takeoka, and Mark~M. Wilde.
\newblock Quantum enigma machines and the locking capacity of a quantum
  channel.
\newblock \emph{Physical Review X}, 4\penalty0 (1):\penalty0 011016, January
  2014.
\newblock \doi{10.1103/PhysRevX.4.011016}.
\newblock arXiv:1307.5368.

\bibitem[Gupta and Wilde(2015)]{GW15}
Manish Gupta and Mark~M. Wilde.
\newblock Multiplicativity of completely bounded $p$-norms implies a strong
  converse for entanglement-assisted capacity.
\newblock \emph{Communications in Mathematical Physics}, 334\penalty0
  (2):\penalty0 867--887, March 2015.
\newblock arXiv:1310.7028.

\bibitem[Hamada(2005)]{H05}
Mitsuru Hamada.
\newblock Information rates achievable with algebraic codes on quantum discrete
  memoryless channels.
\newblock \emph{IEEE Transactions on Information Theory}, 51\penalty0
  (12):\penalty0 4263--4277, December 2005.
\newblock arXiv:quant-ph/0207113.

\bibitem[Harrington and Preskill(2001)]{HP01}
Jim Harrington and John Preskill.
\newblock Achievable rates for the {Gaussian} quantum channel.
\newblock \emph{Physical Review A}, 64\penalty0 (6):\penalty0 062301, November
  2001.
\newblock \doi{10.1103/PhysRevA.64.062301}.
\newblock arXiv:quant-ph/0105058.

\bibitem[Harrow(2004)]{prl2004harrow}
Aram Harrow.
\newblock Coherent communication of classical messages.
\newblock \emph{Physical Review Letters}, 92\penalty0 (9):\penalty0 097902,
  March 2004.
\newblock arXiv:quant-ph/0307091.

\bibitem[Harrow and Lo(2004)]{HL04}
Aram~W. Harrow and Hoi-Kwong Lo.
\newblock A tight lower bound on the classical communication cost of
  entanglement dilution.
\newblock \emph{IEEE Transactions on Information Theory}, 50\penalty0
  (2):\penalty0 319--327, February 2004.
\newblock arXiv:quant-ph/0204096.

\bibitem[Hastings(2009)]{H09}
Matthew~B. Hastings.
\newblock Superadditivity of communication capacity using entangled inputs.
\newblock \emph{Nature Physics}, 5:\penalty0 255--257, April 2009.
\newblock arXiv:0809.3972.

\bibitem[Hausladen et~al.(1995)Hausladen, Schumacher, Westmoreland, and
  Wootters]{HSWW95}
Paul Hausladen, Benjamin Schumacher, Michael Westmoreland, and William~K.
  Wootters.
\newblock Sending classical bits via quantum its.
\newblock \emph{Annals of the New York Academy of Sciences}, 755:\penalty0
  698--705, April 1995.

\bibitem[Hausladen et~al.(1996)Hausladen, Jozsa, Schumacher, Westmoreland, and
  Wootters]{PhysRevA.54.1869}
Paul Hausladen, Richard Jozsa, Benjamin Schumacher, Michael Westmoreland, and
  William~K. Wootters.
\newblock Classical information capacity of a quantum channel.
\newblock \emph{Physical Review A}, 54\penalty0 (3):\penalty0 1869--1876,
  September 1996.
\newblock \doi{10.1103/PhysRevA.54.1869}.

\bibitem[Hayashi(2002)]{PhysRevA.66.032321}
Masahito Hayashi.
\newblock Exponents of quantum fixed-length pure-state source coding.
\newblock \emph{Physical Review A}, 66\penalty0 (3):\penalty0 032321, September
  2002.
\newblock \doi{10.1103/PhysRevA.66.032321}.
\newblock arXiv:quant-ph/0202002.

\bibitem[Hayashi(2006)]{H06book}
Masahito Hayashi.
\newblock \emph{Quantum Information: An Introduction}.
\newblock Springer, 2006.

\bibitem[Hayashi(2007)]{Hay07}
Masahito Hayashi.
\newblock Error exponent in asymmetric quantum hypothesis testing and its
  application to classical-quantum channel coding.
\newblock \emph{Physical Review A}, 76\penalty0 (6):\penalty0 062301, December
  2007.
\newblock \doi{10.1103/PhysRevA.76.062301}.
\newblock arXiv:quant-ph/0611013.

\bibitem[Hayashi and Matsumoto(2001)]{HM01}
Masahito Hayashi and Keiji Matsumoto.
\newblock Variable length universal entanglement concentration by local
  operations and its application to teleportation and dense coding.
\newblock September 2001.
\newblock arXiv:quant-ph/0109028.

\bibitem[Hayashi and Nagaoka(2003)]{HN03}
Masahito Hayashi and Hiroshi Nagaoka.
\newblock General formulas for capacity of classical-quantum channels.
\newblock \emph{IEEE Transactions on Information Theory}, 49\penalty0
  (7):\penalty0 1753--1768, July 2003.
\newblock arXiv:quant-ph/0206186.

\bibitem[Hayashi et~al.(2003)Hayashi, Koashi, Matsumoto, Morikoshi, and
  Winter]{HKMMW03}
Masahito Hayashi, Masato Koashi, Keiji Matsumoto, Fumiaki Morikoshi, and
  Andreas Winter.
\newblock Error exponents for entanglement concentration.
\newblock \emph{Journal of Physics A: Mathematical and General}, 36\penalty0
  (2):\penalty0 527, January 2003.
\newblock arXiv:quant-ph/0206097.

\bibitem[Hayden(2007)]{H07}
Patrick Hayden.
\newblock The maximal p-norm multiplicativity conjecture is false.
\newblock July 2007.
\newblock arXiv:0707.3291.

\bibitem[Hayden and Winter(2003)]{PhysRevA.67.012326}
Patrick Hayden and Andreas Winter.
\newblock Communication cost of entanglement transformations.
\newblock \emph{Physical Review A}, 67\penalty0 (1):\penalty0 012326, January
  2003.
\newblock \doi{10.1103/PhysRevA.67.012326}.
\newblock arXiv:quant-ph/0204092.

\bibitem[Hayden and Winter(2008)]{HW08}
Patrick Hayden and Andreas Winter.
\newblock Counterexamples to the maximal p-norm multiplicativity conjecture for
  all p $>$ 1.
\newblock \emph{Communications in Mathematical Physics}, 284\penalty0
  (1):\penalty0 263--280, November 2008.
\newblock arXiv:0807.4753.

\bibitem[Hayden et~al.(2002)Hayden, Jozsa, and Winter]{hayden:4404}
Patrick Hayden, Richard Jozsa, and Andreas Winter.
\newblock Trading quantum for classical resources in quantum data compression.
\newblock \emph{Journal of Mathematical Physics}, 43\penalty0 (9):\penalty0
  4404--4444, September 2002.
\newblock \doi{10.1063/1.1497184}.
\newblock arXiv:quant-ph/0204038.

\bibitem[Hayden et~al.(2004{\natexlab{a}})Hayden, Jozsa, Petz, and
  Winter]{HJPW04}
Patrick Hayden, Richard Jozsa, Denes Petz, and Andreas Winter.
\newblock Structure of states which satisfy strong subadditivity of quantum
  entropy with equality.
\newblock \emph{Communications in Mathematical Physics}, 246\penalty0
  (2):\penalty0 359--374, April 2004{\natexlab{a}}.
\newblock arXiv:quant-ph/0304007.

\bibitem[Hayden et~al.(2004{\natexlab{b}})Hayden, Leung, Shor, and
  Winter]{HLSW04}
Patrick Hayden, Debbie Leung, Peter~W. Shor, and Andreas Winter.
\newblock Randomizing quantum states: Constructions and applications.
\newblock \emph{Communications in Mathematical Physics}, 250\penalty0
  (2):\penalty0 371--391, July 2004{\natexlab{b}}.
\newblock arXiv:quant-ph/0307104.

\bibitem[Hayden et~al.(2008{\natexlab{a}})Hayden, Horodecki, Winter, and
  Yard]{qcap2008first}
Patrick Hayden, Michal Horodecki, Andreas Winter, and Jon Yard.
\newblock A decoupling approach to the quantum capacity.
\newblock \emph{Open Systems \& Information Dynamics}, 15\penalty0
  (1):\penalty0 7--19, March 2008{\natexlab{a}}.
\newblock arXiv:quant-ph/0702005.

\bibitem[Hayden et~al.(2008{\natexlab{b}})Hayden, Shor, and
  Winter]{qcap2008fourth}
Patrick Hayden, Peter~W. Shor, and Andreas Winter.
\newblock Random quantum codes from {Gaussian} ensembles and an uncertainty
  relation.
\newblock \emph{Open Systems \& Information Dynamics}, 15\penalty0
  (1):\penalty0 71--89, March 2008{\natexlab{b}}.
\newblock arXiv:0712.0975.

\bibitem[Heinosaari and Ziman(2012)]{HZ12}
Teiko Heinosaari and M\'ario Ziman.
\newblock \emph{The Mathematical Language of Quantum Theory: From Uncertainty
  to Entanglement}.
\newblock Cambridge University Press, 2012.

\bibitem[Heisenberg(1925)]{1925heisenberg}
Werner Heisenberg.
\newblock \"{U}ber quantentheoretische umdeutung kinematischer und mechanischer
  beziehungen.
\newblock \emph{Zeitschrift f\"{u}r Physik}, 33:\penalty0 879--893, 1925.

\bibitem[Helstrom(1969)]{H69}
Carl~W. Helstrom.
\newblock Quantum detection and estimation theory.
\newblock \emph{Journal of Statistical Physics}, 1:\penalty0 231--252, 1969.
\newblock ISSN 0022-4715.

\bibitem[Helstrom(1976)]{Hel76}
Carl~W. Helstrom.
\newblock \emph{Quantum Detection and Estimation Theory}.
\newblock Academic, New York, 1976.

\bibitem[Hirche and Morgan(2015)]{HM15}
Christoph Hirche and Ciara Morgan.
\newblock An improved rate region for the classical-quantum broadcast channel.
\newblock \emph{Proceedings of the 2015 IEEE International Symposium on
  Information Theory}, pages 2782--2786, 2015.
\newblock arXiv:1501.07417.

\bibitem[Hirche et~al.(2016)Hirche, Morgan, and Wilde]{HMW14}
Christoph Hirche, Ciara Morgan, and Mark~M. Wilde.
\newblock Polar codes in network quantum information theory.
\newblock \emph{IEEE Transactions on Information Theory}, 62\penalty0
  (2):\penalty0 915--924, February 2016.
\newblock \doi{10.1109/TIT.2016.2514319}.
\newblock arXiv:1409.7246.

\bibitem[Hirschman(1952)]{H52}
Isidore~Isaac Hirschman.
\newblock A convexity theorem for certain groups of transformations.
\newblock \emph{Journal d'Analyse Math\'ematique}, 2\penalty0 (2):\penalty0
  209--218, December 1952.

\bibitem[Holevo(1972)]{Kholevo1972}
Alexander~S. Holevo.
\newblock On quasiequivalence of locally normal states.
\newblock \emph{Theoretical and Mathematical Physics}, 13\penalty0
  (2):\penalty0 1071--1082, November 1972.
\newblock ISSN 1573-9333.

\bibitem[Holevo(1973{\natexlab{a}})]{Holevo73}
Alexander~S. Holevo.
\newblock Bounds for the quantity of information transmitted by a quantum
  communication channel.
\newblock \emph{Problems of Information Transmission}, 9:\penalty0 177--183,
  1973{\natexlab{a}}.

\bibitem[Holevo(1973{\natexlab{b}})]{japan1973holevo}
Alexander~S. Holevo.
\newblock Statistical problems in quantum physics.
\newblock In \emph{Second Japan-USSR Symposium on Probability Theory}, volume
  330 of \emph{Lecture Notes in Mathematics}, pages 104--119. Springer Berlin /
  Heidelberg, 1973{\natexlab{b}}.
\newblock \doi{10.1007/BFb0061476}.

\bibitem[Holevo(1998)]{Hol98}
Alexander~S. Holevo.
\newblock The capacity of the quantum channel with general signal states.
\newblock \emph{IEEE Transactions on Information Theory}, 44\penalty0
  (1):\penalty0 269--273, January 1998.
\newblock arXiv:quant-ph/9611023.

\bibitem[Holevo(2000)]{H00}
Alexander~S. Holevo.
\newblock Reliability function of general classical-quantum channel.
\newblock \emph{IEEE Transactions on Information Theory}, 46\penalty0
  (6):\penalty0 2256--2261, September 2000.
\newblock ISSN 0018-9448.
\newblock \doi{10.1109/18.868501}.
\newblock arXiv:quant-ph/9907087.

\bibitem[Holevo(2002{\natexlab{a}})]{H02book}
Alexander~S. Holevo.
\newblock \emph{An Introduction to Quantum Information Theory}.
\newblock Moscow Center of Continuous Mathematical Education, Moscow,
  2002{\natexlab{a}}.
\newblock In Russian.

\bibitem[Holevo(2002{\natexlab{b}})]{Hol01a}
Alexander~S. Holevo.
\newblock On entanglement assisted classical capacity.
\newblock \emph{Journal of Mathematical Physics}, 43\penalty0 (9):\penalty0
  4326--4333, September 2002{\natexlab{b}}.
\newblock arXiv:quant-ph/0106075.

\bibitem[Holevo(2012)]{H12}
Alexander~S. Holevo.
\newblock \emph{Quantum Systems, Channels, Information}.
\newblock de Gruyter Studies in Mathematical Physics (Book 16). de Gruyter,
  November 2012.

\bibitem[Holevo and Werner(2001)]{HW01}
Alexander~S. Holevo and Reinhard~F. Werner.
\newblock Evaluating capacities of bosonic {Gaussian} channels.
\newblock \emph{Physical Review A}, 63\penalty0 (3):\penalty0 032312, February
  2001.
\newblock \doi{10.1103/PhysRevA.63.032312}.
\newblock arXiv:quant-ph/9912067.

\bibitem[Horodecki(1998)]{H98}
Michal Horodecki.
\newblock Limits for compression of quantum information carried by ensembles of
  mixed states.
\newblock \emph{Physical Review A}, 57\penalty0 (5):\penalty0 3364--3369, May
  1998.
\newblock \doi{10.1103/PhysRevA.57.3364}.
\newblock arXiv:quant-ph/9712035.

\bibitem[Horodecki et~al.(1996)Horodecki, Horodecki, and Horodecki]{HHH96}
Michal Horodecki, Pawel Horodecki, and Ryszard Horodecki.
\newblock Separability of mixed states: necessary and sufficient conditions.
\newblock \emph{Physics Letters A}, 223\penalty0 (1-2):\penalty0 1--8, November
  1996.
\newblock arXiv:quant-ph/9605038.

\bibitem[Horodecki et~al.(2001)Horodecki, Horodecki, Horodecki, Leung, and
  Terhal]{H3LT01}
Michal Horodecki, Pawel Horodecki, Ryszard Horodecki, Debbie Leung, and Barbara
  Terhal.
\newblock Classical capacity of a noiseless quantum channel assisted by noisy
  entanglement.
\newblock \emph{Quantum Information and Computation}, 1\penalty0 (3):\penalty0
  70--78, 2001.
\newblock arXiv:quant-ph/0106080.

\bibitem[Horodecki et~al.(2003)Horodecki, Shor, and Ruskai]{HSR03}
Michal Horodecki, Peter~W. Shor, and Mary~Beth Ruskai.
\newblock Entanglement breaking channels.
\newblock \emph{Reviews in Mathematical Physics}, 15\penalty0 (6):\penalty0
  629--641, 2003.
\newblock arXiv:quant-ph/0302031.

\bibitem[Horodecki et~al.(2005)Horodecki, Oppenheim, and
  Winter]{Horodecki:2005:673}
Michal Horodecki, Jonathan Oppenheim, and Andreas Winter.
\newblock Partial quantum information.
\newblock \emph{Nature}, 436:\penalty0 673--676, August 2005.

\bibitem[Horodecki et~al.(2007)Horodecki, Oppenheim, and
  Winter]{Horodecki:2007:107}
Michal Horodecki, Jonathan Oppenheim, and Andreas Winter.
\newblock Quantum state merging and negative information.
\newblock \emph{Communications in Mathematical Physics}, 269\penalty0
  (1):\penalty0 107--136, January 2007.
\newblock arXiv:quant-ph/0512247.

\bibitem[Horodecki(1997)]{H97}
Pawel Horodecki.
\newblock Separability criterion and inseparable mixed states with positive
  partial transposition.
\newblock \emph{Physics Letters A}, 232\penalty0 (5):\penalty0 333--339, August
  1997.
\newblock arXiv:quant-ph/9703004.

\bibitem[Horodecki and Horodecki(1994)]{HH94}
R.~Horodecki and P.~Horodecki.
\newblock Quantum redundancies and local realism.
\newblock \emph{Physics Letters A}, 194\penalty0 (3):\penalty0 147--152,
  October 1994.

\bibitem[Horodecki et~al.(2009)Horodecki, Horodecki, Horodecki, and
  Horodecki]{H42007}
Ryszard Horodecki, Pawel Horodecki, Michal Horodecki, and Karol Horodecki.
\newblock Quantum entanglement.
\newblock \emph{Reviews of Modern Physics}, 81\penalty0 (2):\penalty0 865--942,
  June 2009.
\newblock \doi{10.1103/RevModPhys.81.865}.
\newblock arXiv:quant-ph/0702225.

\bibitem[Hsieh and Wilde(2009)]{PhysRevA.80.022306}
Min-Hsiu Hsieh and Mark~M. Wilde.
\newblock Public and private communication with a quantum channel and a secret
  key.
\newblock \emph{Physical Review A}, 80\penalty0 (2):\penalty0 022306, August
  2009.
\newblock \doi{10.1103/PhysRevA.80.022306}.
\newblock arXiv:0903.3920.

\bibitem[Hsieh and Wilde(2010{\natexlab{a}})]{HW08GFP}
Min-Hsiu Hsieh and Mark~M. Wilde.
\newblock Entanglement-assisted communication of classical and quantum
  information.
\newblock \emph{IEEE Transactions on Information Theory}, 56\penalty0
  (9):\penalty0 4682--4704, September 2010{\natexlab{a}}.
\newblock arXiv:0811.4227.

\bibitem[Hsieh and Wilde(2010{\natexlab{b}})]{HW09}
Min-Hsiu Hsieh and Mark~M. Wilde.
\newblock Trading classical communication, quantum communication, and
  entanglement in quantum {Shannon} theory.
\newblock \emph{IEEE Transactions on Information Theory}, 56\penalty0
  (9):\penalty0 4705--4730, September 2010{\natexlab{b}}.
\newblock arXiv:0901.3038.

\bibitem[Hsieh et~al.(2008{\natexlab{a}})Hsieh, Devetak, and
  Winter]{itit2008hsieh}
Min-Hsiu Hsieh, Igor Devetak, and Andreas Winter.
\newblock Entanglement-assisted capacity of quantum multiple-access channels.
\newblock \emph{IEEE Transactions on Information Theory}, 54\penalty0
  (7):\penalty0 3078--3090, July 2008{\natexlab{a}}.
\newblock arXiv:quant-ph/0511228.

\bibitem[Hsieh et~al.(2008{\natexlab{b}})Hsieh, Luo, and Brun]{HLB08}
Min-Hsiu Hsieh, Zhicheng Luo, and Todd Brun.
\newblock Secret-key-assisted private classical communication capacity over
  quantum channels.
\newblock \emph{Physical Review A}, 78\penalty0 (4):\penalty0 042306, October
  2008{\natexlab{b}}.
\newblock \doi{10.1103/PhysRevA.78.042306}.
\newblock arXiv:0806.3525.

\bibitem[Jaynes(1957{\natexlab{a}})]{J57}
Edwin~T. Jaynes.
\newblock Information theory and statistical mechanics.
\newblock \emph{Physical Review}, 106:\penalty0 620, 1957{\natexlab{a}}.

\bibitem[Jaynes(1957{\natexlab{b}})]{J57a}
Edwin~T. Jaynes.
\newblock Information theory and statistical mechanics {II}.
\newblock \emph{Physical Review}, 108:\penalty0 171, 1957{\natexlab{b}}.

\bibitem[Jaynes(2003)]{J03}
Edwin~T. Jaynes.
\newblock \emph{Probability Theory: The Logic of Science}.
\newblock Cambridge University Press, 2003.

\bibitem[Jencova(2012)]{J12}
Anna Jencova.
\newblock Reversibility conditions for quantum operations.
\newblock \emph{Reviews in Mathematical Physics}, 24\penalty0 (7):\penalty0
  1250016, August 2012.
\newblock \doi{10.1142/S0129055X1250016X}.
\newblock arXiv:1107.0453.

\bibitem[Jochym-O'Connor et~al.(2011)Jochym-O'Connor, Br\'adler, and
  Wilde]{JBW11}
Tomas Jochym-O'Connor, Kamil Br\'adler, and Mark~M. Wilde.
\newblock Trade-off coding for universal qudit cloners motivated by the {Unruh}
  effect.
\newblock \emph{Journal of Physics A: Mathematical and Theoretical},
  44\penalty0 (41):\penalty0 415306, October 2011.
\newblock arXiv:1103.0286.

\bibitem[Jozsa(1994)]{J94}
Richard Jozsa.
\newblock Fidelity for mixed quantum states.
\newblock \emph{Journal of Modern Optics}, 41\penalty0 (12):\penalty0
  2315--2323, 1994.

\bibitem[Jozsa and Presnell(2003)]{JP03}
Richard Jozsa and Stuart Presnell.
\newblock Universal quantum information compression and degrees of prior
  knowledge.
\newblock \emph{Proceedings of the Royal Society A: Mathematical, Physical and
  Engineering Sciences}, 459\penalty0 (2040):\penalty0 3061--3077, December
  2003.
\newblock arXiv:quant-ph/0210196.

\bibitem[Jozsa and Schumacher(1994)]{JS94}
Richard Jozsa and Benjamin Schumacher.
\newblock A new proof of the quantum noiseless coding theorem.
\newblock \emph{Journal of Modern Optics}, 41\penalty0 (12):\penalty0
  2343--2349, December 1994.

\bibitem[Jozsa et~al.(1998)Jozsa, Horodecki, Horodecki, and Horodecki]{RHHH98}
Richard Jozsa, M.~Horodecki, Pawe\l{} Horodecki, and Ryszard Horodecki.
\newblock Universal quantum information compression.
\newblock \emph{Physical Review Letters}, 81\penalty0 (8):\penalty0 1714--1717,
  August 1998.
\newblock \doi{10.1103/PhysRevLett.81.1714}.
\newblock arXiv:quant-ph/9805017.

\bibitem[Junge et~al.(2015)Junge, Renner, Sutter, Wilde, and Winter]{Junge15}
Marius Junge, Renato Renner, David Sutter, Mark~M. Wilde, and Andreas Winter.
\newblock Universal recovery from a decrease of quantum relative entropy.
\newblock September 2015.
\newblock arXiv:1509.07127.

\bibitem[Kaye and Mosca(2001)]{KM01}
Phillip Kaye and Michele Mosca.
\newblock Quantum networks for concentrating entanglement.
\newblock \emph{Journal of Physics A: Mathematical and General}, 34\penalty0
  (35):\penalty0 6939, August 2001.
\newblock arXiv:quant-ph/0101009.

\bibitem[Kelvin(1901)]{kelvin1901}
William~Thomson Kelvin.
\newblock Nineteenth-century clouds over the dynamical theory of heat and
  light.
\newblock \emph{The London, Edinburgh and Dublin Philosophical Magazine and
  Journal of Science}, 2\penalty0 (6):\penalty0 1, 1901.

\bibitem[Kemperman(1969)]{K69}
Johannes H.~B. Kemperman.
\newblock On the optimum rate of transmitting information.
\newblock \emph{Lecture Notes in Mathematics}, 89:\penalty0 126--169, 1969.
\newblock In Probability and Information Theory.

\bibitem[Kim(2013)]{K13conj}
Isaac~H. Kim.
\newblock Application of conditional independence to gapped quantum many-body
  systems.
\newblock {http://www.physics.usyd.edu.au/quantum/Coogee2013}, January 2013.
\newblock Slide 43.

\bibitem[King(2002)]{K02}
Christopher King.
\newblock Additivity for unital qubit channels.
\newblock \emph{Journal of Mathematical Physics}, 43\penalty0 (10):\penalty0
  4641--4653, October 2002.
\newblock arXiv:quant-ph/0103156.

\bibitem[King(2003)]{K03}
Christopher King.
\newblock The capacity of the quantum depolarizing channel.
\newblock \emph{IEEE Transactions on Information Theory}, 49\penalty0
  (1):\penalty0 221--229, January 2003.
\newblock arXiv:quant-ph/0204172.

\bibitem[King et~al.(2007)King, Matsumoto, Nathanson, and Ruskai]{KMNR07}
Christopher King, Keiji Matsumoto, Michael Nathanson, and Mary~Beth Ruskai.
\newblock Properties of conjugate channels with applications to additivity and
  multiplicativity.
\newblock \emph{Markov Processes and Related Fields}, 13\penalty0 (2):\penalty0
  391--423, 2007.
\newblock J. T. Lewis memorial issue. arXiv:quant-ph/0509126.

\bibitem[Kitaev(1997)]{kitaev1997}
Alexei~Y. Kitaev.
\newblock \emph{Uspekhi Mat. Nauk.}, 52\penalty0 (53), 1997.

\bibitem[Klesse(2008)]{qcap2008second}
Rochus Klesse.
\newblock A random coding based proof for the quantum coding theorem.
\newblock \emph{Open Systems \& Information Dynamics}, 15\penalty0
  (1):\penalty0 21--45, March 2008.
\newblock arXiv:0712.2558.

\bibitem[Knill et~al.(1998)Knill, Laflamme, and Zurek]{KLZ98}
Emanuel~H. Knill, Raymond Laflamme, and Wojciech~H. Zurek.
\newblock Resilient quantum computation.
\newblock \emph{Science}, 279:\penalty0 342--345, 1998.
\newblock quant-ph/9610011.

\bibitem[Koashi and Imoto(2001)]{KI01}
Masato Koashi and Nobuyuki Imoto.
\newblock Teleportation cost and hybrid compression of quantum signals.
\newblock 2001.
\newblock arXiv:quant-ph/0104001.

\bibitem[Koenig and Wehner(2009)]{KW09}
Robert Koenig and Stephanie Wehner.
\newblock A strong converse for classical channel coding using entangled
  inputs.
\newblock \emph{Physical Review Letters}, 103\penalty0 (7):\penalty0 070504,
  August 2009.
\newblock arXiv:0903.2838.

\bibitem[Koenig et~al.(2009)Koenig, Renner, and Schaffner]{KRS09}
Robert Koenig, Renato Renner, and Christian Schaffner.
\newblock The operational meaning of min- and max-entropy.
\newblock \emph{IEEE Transactions on Information Theory}, 55\penalty0
  (9):\penalty0 4337--4347, September 2009.
\newblock arXiv:0807.1338.

\bibitem[K\"onig et~al.(2007)K\"onig, Renner, Bariska, and Maurer]{KRBM07}
Robert K\"onig, Renato Renner, Andor Bariska, and Ueli Maurer.
\newblock Small accessible quantum information does not imply security.
\newblock \emph{Physical Review Letters}, 98\penalty0 (14):\penalty0 140502,
  April 2007.
\newblock \doi{10.1103/PhysRevLett.98.140502}.
\newblock arXiv:quant-ph/0512021.

\bibitem[Kremsky et~al.(2008)Kremsky, Hsieh, and Brun]{kremsky:012341}
Isaac Kremsky, Min-Hsiu Hsieh, and Todd~A. Brun.
\newblock Classical enhancement of quantum-error-correcting codes.
\newblock \emph{Physical Review A}, 78\penalty0 (1):\penalty0 012341, 2008.
\newblock \doi{10.1103/PhysRevA.78.012341}.
\newblock arXiv:0802.2414.

\bibitem[Kullback(1967)]{K67}
Solomon Kullback.
\newblock A lower bound for discrimination in terms of variation.
\newblock \emph{IEEE-IT}, 13:\penalty0 126--127, 1967.

\bibitem[Kumagai and Hayashi(2013)]{PhysRevLett.111.130407}
Wataru Kumagai and Masahito Hayashi.
\newblock Entanglement concentration is irreversible.
\newblock \emph{Physical Review Letters}, 111\penalty0 (13):\penalty0 130407,
  September 2013.
\newblock \doi{10.1103/PhysRevLett.111.130407}.
\newblock arXiv:1305.6250.

\bibitem[Kuperberg(2003)]{K03a}
Greg Kuperberg.
\newblock The capacity of hybrid quantum memory.
\newblock \emph{IEEE Transactions on Information Theory}, 49\penalty0
  (6):\penalty0 1465--1473, June 2003.
\newblock arXiv:quant-ph/0203105.

\bibitem[Laflamme et~al.(1996)Laflamme, Miquel, Paz, and
  Zurek]{PhysRevLett.77.198}
Raymond Laflamme, Cesar Miquel, Juan~Pablo Paz, and Wojciech~Hubert Zurek.
\newblock Perfect quantum error correcting code.
\newblock \emph{Physical Review Letters}, 77\penalty0 (1):\penalty0 198--201,
  July 1996.
\newblock \doi{10.1103/PhysRevLett.77.198}.

\bibitem[Landauer(1995)]{landauer1995}
Rolf Landauer.
\newblock Is quantum mechanics useful?
\newblock \emph{Philosophical Transactions of the Royal Society: Physical and
  Engineering Sciences}, 353\penalty0 (1703):\penalty0 367--376, December 1995.

\bibitem[Lanford and Robinson(1968)]{LR68}
Oscar~E. Lanford and Derek~W. Robinson.
\newblock Mean entropy of states in quantum-statistical mechanics.
\newblock \emph{Journal of Mathematical Physics}, 9\penalty0 (7):\penalty0
  1120--1125, July 1968.
\newblock \doi{http://dx.doi.org/10.1063/1.1664685}.

\bibitem[Levitin(1969)]{levitin1969}
Lev~B. Levitin.
\newblock On the quantum measure of information.
\newblock In \emph{Proceedings of the Fourth All-Union Conference on
  Information and Coding Theory, Sec. II,}, Tashkent, 1969.

\bibitem[Li and Winter(2014)]{LW14}
Ke~Li and Andreas Winter.
\newblock Squashed entanglement, $k$-extendibility, quantum {M}arkov chains,
  and recovery maps, 2014.
\newblock arXiv:1410.4184.

\bibitem[Li et~al.(2009)Li, Winter, Zou, and Guo]{LWZG09}
Ke~Li, Andreas Winter, XuBo Zou, and Guang-Can Guo.
\newblock Private capacity of quantum channels is not additive.
\newblock \emph{Physical Review Letters}, 103\penalty0 (12):\penalty0 120501,
  September 2009.
\newblock \doi{10.1103/PhysRevLett.103.120501}.
\newblock arXiv:0903.4308.

\bibitem[Lieb(1973)]{L73}
Elliot~H. Lieb.
\newblock Convex trace functions and the {Wigner-Yanase-Dyson} conjecture.
\newblock \emph{Advances in Mathematics}, 11\penalty0 (3):\penalty0 267--288,
  December 1973.

\bibitem[Lieb and Ruskai(1973{\natexlab{a}})]{LR73}
Elliott~H. Lieb and Mary~Beth Ruskai.
\newblock Proof of the strong subadditivity of quantum-mechanical entropy.
\newblock \emph{Journal of Mathematical Physics}, 14\penalty0 (12):\penalty0
  1938--1941, December 1973{\natexlab{a}}.

\bibitem[Lieb and Ruskai(1973{\natexlab{b}})]{PhysRevLett.30.434}
Elliott~H. Lieb and Mary~Beth Ruskai.
\newblock A fundamental property of quantum-mechanical entropy.
\newblock \emph{Physical Review Letters}, 30\penalty0 (10):\penalty0 434--436,
  March 1973{\natexlab{b}}.
\newblock \doi{10.1103/PhysRevLett.30.434}.

\bibitem[Lindblad(1975)]{Lindblad1975}
G\"oran Lindblad.
\newblock Completely positive maps and entropy inequalities.
\newblock \emph{Communications in Mathematical Physics}, 40\penalty0
  (2):\penalty0 147--151, June 1975.
\newblock ISSN 1432-0916.
\newblock \doi{10.1007/bf01609396}.

\bibitem[Lloyd(1997)]{PhysRevA.55.1613}
Seth Lloyd.
\newblock Capacity of the noisy quantum channel.
\newblock \emph{Physical Review A}, 55\penalty0 (3):\penalty0 1613--1622, March
  1997.
\newblock \doi{10.1103/PhysRevA.55.1613}.
\newblock arXiv:quant-ph/9604015.

\bibitem[Lloyd et~al.(2011)Lloyd, Giovannetti, and
  Maccone]{PhysRevLett.106.250501}
Seth Lloyd, Vittorio Giovannetti, and Lorenzo Maccone.
\newblock Sequential projective measurements for channel decoding.
\newblock \emph{Physical Review Letters}, 106\penalty0 (25):\penalty0 250501,
  June 2011.
\newblock \doi{10.1103/PhysRevLett.106.250501}.
\newblock arXiv:1012.0106.

\bibitem[Lo(1995)]{L95}
Hoi-Kwong Lo.
\newblock Quantum coding theorem for mixed states.
\newblock \emph{Optics Communications}, 119\penalty0 (5-6):\penalty0 552--556,
  September 1995.
\newblock arXiv:quant-ph/9504004.

\bibitem[Lo and Popescu(1999)]{PhysRevLett.83.1459}
Hoi-Kwong Lo and Sandu Popescu.
\newblock Classical communication cost of entanglement manipulation: Is
  entanglement an interconvertible resource?
\newblock \emph{Physical Review Letters}, 83\penalty0 (7):\penalty0 1459--1462,
  August 1999.
\newblock \doi{10.1103/PhysRevLett.83.1459}.

\bibitem[Lo and Popescu(2001)]{PhysRevA.63.022301}
Hoi-Kwong Lo and Sandu Popescu.
\newblock Concentrating entanglement by local actions: Beyond mean values.
\newblock \emph{Physical Review A}, 63\penalty0 (2):\penalty0 022301, January
  2001.
\newblock \doi{10.1103/PhysRevA.63.022301}.
\newblock arXiv:quant-ph/9707038.

\bibitem[Luo and Zhang(2004)]{LZ04}
Shunlong Luo and Qiang Zhang.
\newblock Informational distance on quantum-state space.
\newblock \emph{Physical Review A}, 69\penalty0 (3):\penalty0 032106, March
  2004.

\bibitem[Lupo and Lloyd(2014)]{PhysRevLett.113.160502}
Cosmo Lupo and Seth Lloyd.
\newblock Quantum-locked key distribution at nearly the classical capacity
  rate.
\newblock \emph{Physical Review Letters}, 113\penalty0 (16):\penalty0 160502,
  October 2014.
\newblock \doi{10.1103/PhysRevLett.113.160502}.
\newblock arXiv:1406.4418.

\bibitem[Lupo and Lloyd(2015)]{LL15}
Cosmo Lupo and Seth Lloyd.
\newblock Quantum data locking for high-rate private communication.
\newblock \emph{New Journal of Physics}, 17\penalty0 (3):\penalty0 033022,
  2015.

\bibitem[MacKay(2003)]{M03}
David MacKay.
\newblock \emph{Information Theory, Inference, and Learning Algorithms}.
\newblock Cambridge University Press, September 2003.

\bibitem[Matthews and Wehner(2014)]{MW12}
William Matthews and Stephanie Wehner.
\newblock Finite blocklength converse bounds for quantum channels.
\newblock \emph{IEEE Transactions on Information Theory}, 60\penalty0
  (11):\penalty0 7317--7329, November 2014.
\newblock arXiv:1210.4722.

\bibitem[McEvoy and Zarate(2004)]{book2004mcevoy}
J.~P. McEvoy and Oscar Zarate.
\newblock \emph{Introducing Quantum Theory}.
\newblock Totem Books, third edition, October 2004.

\bibitem[Misner et~al.(2009)Misner, Thorne, and Zurek]{PT09}
Charles~W. Misner, Kip~S. Thorne, and Wojciech~H. Zurek.
\newblock John {Wheeler}, relativity, and quantum information.
\newblock \emph{Physics Today}, April 2009.

\bibitem[Morgan and Winter(2014)]{MW13}
Ciara Morgan and Andreas Winter.
\newblock {``Pretty strong''} converse for the quantum capacity of degradable
  channels.
\newblock \emph{IEEE Transactions on Information Theory}, 60\penalty0
  (1):\penalty0 317--333, January 2014.
\newblock arXiv:1301.4927.

\bibitem[Mosonyi(2005)]{M05}
Mil\'an Mosonyi.
\newblock \emph{Entropy, Information and Structure of Composite Quantum
  States}.
\newblock PhD thesis, Katholieke Universiteit Leuven, 2005.
\newblock Available at
  https://lirias.kuleuven.be/bitstream/1979/41/2/thesisbook9.pdf.

\bibitem[Mosonyi and Datta(2009)]{mosonyi:072104}
Mil\'{a}n Mosonyi and Nilanjana Datta.
\newblock Generalized relative entropies and the capacity of classical-quantum
  channels.
\newblock \emph{Journal of Mathematical Physics}, 50\penalty0 (7):\penalty0
  072104, July 2009.
\newblock \doi{10.1063/1.3167288}.
\newblock arXiv:0810.3478.

\bibitem[Mosonyi and Petz(2004)]{Mosonyi2004}
Mil\'an Mosonyi and D\'enes Petz.
\newblock Structure of sufficient quantum coarse-grainings.
\newblock \emph{Letters in Mathematical Physics}, 68\penalty0 (1):\penalty0
  19--30, April 2004.
\newblock ISSN 1573-0530.
\newblock \doi{10.1007/s11005-004-4072-2}.
\newblock arXiv:quant-ph/0312221.

\bibitem[Mullins(2001)]{ieeespec2001}
Justin Mullins.
\newblock The topsy turvy world of quantum computing.
\newblock \emph{IEEE Spectrum}, 38\penalty0 (2):\penalty0 42--49, February
  2001.

\bibitem[Nielsen(1998)]{N98}
Michael~A. Nielsen.
\newblock \emph{Quantum information theory}.
\newblock PhD thesis, University of New Mexico, 1998.
\newblock arXiv:quant-ph/0011036.

\bibitem[Nielsen(1999)]{N99}
Michael~A. Nielsen.
\newblock Conditions for a class of entanglement transformations.
\newblock \emph{Physical Review Letters}, 83\penalty0 (2):\penalty0 436--439,
  July 1999.
\newblock \doi{10.1103/PhysRevLett.83.436}.
\newblock arXiv:quant-ph/9811053.

\bibitem[Nielsen(2002)]{Nielsen2002249}
Michael~A. Nielsen.
\newblock A simple formula for the average gate fidelity of a quantum dynamical
  operation.
\newblock \emph{Physics Letters A}, 303\penalty0 (4):\penalty0 249 -- 252,
  2002.
\newblock ISSN 0375-9601.
\newblock \doi{DOI: 10.1016/S0375-9601(02)01272-0}.

\bibitem[Nielsen and Chuang(2000)]{book2000mikeandike}
Michael~A. Nielsen and Isaac~L. Chuang.
\newblock \emph{Quantum Computation and Quantum Information}.
\newblock Cambridge University Press, 2000.

\bibitem[Ogawa and Nagaoka(1999)]{ON99}
Tomohiro Ogawa and Hiroshi Nagaoka.
\newblock Strong converse to the quantum channel coding theorem.
\newblock \emph{IEEE Transactions on Information Theory}, 45\penalty0
  (7):\penalty0 2486--2489, November 1999.
\newblock arXiv:quant-ph/9808063.

\bibitem[Ogawa and Nagaoka(2007)]{ON07}
Tomohiro Ogawa and Hiroshi Nagaoka.
\newblock Making good codes for classical-quantum channel coding via quantum
  hypothesis testing.
\newblock \emph{IEEE Transactions on Information Theory}, 53\penalty0
  (6):\penalty0 2261--2266, June 2007.

\bibitem[Ohya and Petz(1993)]{OP93}
Masanori Ohya and Denes Petz.
\newblock \emph{Quantum Entropy and Its Use}.
\newblock Springer, 1993.

\bibitem[Ollivier and Zurek(2001)]{zurek}
Harold Ollivier and Wojciech~H. Zurek.
\newblock Quantum discord: A measure of the quantumness of correlations.
\newblock \emph{Physical Review Letters}, 88\penalty0 (1):\penalty0 017901,
  December 2001.
\newblock \doi{10.1103/PhysRevLett.88.017901}.
\newblock arXiv:quant-ph/0105072.

\bibitem[Ortigoso(2018)]{O18}
Juan Ortigoso.
\newblock Twelve years before the quantum no-cloning theorem.
\newblock \emph{American Journal of Physics}, 86\penalty0 (3):\penalty0
  201--205, March 2018.
\newblock \doi{10.1119/1.5021356}.
\newblock URL \url{https://doi.org/10.1119/1.5021356}.
\newblock arXiv:1707.06910.

\bibitem[Ozawa(1984)]{O79}
Masanao Ozawa.
\newblock Quantum measuring processes of continuous observables.
\newblock \emph{Journal of Mathematical Physics}, 25\penalty0 (1):\penalty0
  79--87, 1984.
\newblock \doi{10.1063/1.526000}.

\bibitem[Ozawa(2000)]{Ozawa2000158}
Masanao Ozawa.
\newblock Entanglement measures and the {Hilbert--Schmidt} distance.
\newblock \emph{Physics Letters A}, 268\penalty0 (3):\penalty0 158--160, April
  2000.
\newblock ISSN 0375-9601.
\newblock arXiv:quant-ph/0002036.

\bibitem[Park(1970)]{Park1970}
James~L. Park.
\newblock The concept of transition in quantum mechanics.
\newblock \emph{Foundations of Physics}, 1\penalty0 (1):\penalty0 23--33, Mar
  1970.
\newblock ISSN 1572-9516.
\newblock \doi{10.1007/BF00708652}.
\newblock URL \url{https://doi.org/10.1007/BF00708652}.

\bibitem[Pati and Braunstein(2000)]{PB00}
Arun~Kumar Pati and Samuel~L. Braunstein.
\newblock Impossibility of deleting an unknown quantum state.
\newblock \emph{Nature}, 404:\penalty0 164--165, March 2000.
\newblock arXiv:quant-ph/9911090.

\bibitem[Petz(1986)]{Petz1986}
D\'enes Petz.
\newblock Sufficient subalgebras and the relative entropy of states of a von
  {Neumann} algebra.
\newblock \emph{Communications in Mathematical Physics}, 105\penalty0
  (1):\penalty0 123--131, March 1986.
\newblock ISSN 1432-0916.
\newblock \doi{10.1007/bf01212345}.

\bibitem[Petz(1988)]{Petz1988}
D\'enes Petz.
\newblock Sufficiency of channels over von {Neumann} algebras.
\newblock \emph{Quarterly Journal of Mathematics}, 39\penalty0 (1):\penalty0
  97--108, 1988.
\newblock ISSN 1464-3847.
\newblock \doi{10.1093/qmath/39.1.97}.

\bibitem[Pierce(1973)]{P73}
John~R. Pierce.
\newblock The early days of information theory.
\newblock \emph{IEEE Transactions on Information Theory}, IT-19\penalty0
  (1):\penalty0 3--8, January 1973.

\bibitem[Pinsker(1960)]{P60}
Mark~Semenovich Pinsker.
\newblock Information and information stability of random variables and
  processes.
\newblock \emph{Problemy Peredaci Informacii}, 7, 1960.
\newblock AN SSSR, Moscow. English translation: Holden-Day, San Francisco, CA,
  1964.

\bibitem[Planck(1901)]{planck1901}
Max Planck.
\newblock Ueber das gesetz der energieverteilung im normalspectrum.
\newblock \emph{Annalen der Physik}, 4:\penalty0 553--563, 1901.

\bibitem[Plenio et~al.(2000)Plenio, Virmani, and Papadopoulos]{PVP00}
Martin~B. Plenio, Shashank Virmani, and P.~Papadopoulos.
\newblock Operator monotones, the reduction criterion and the relative entropy.
\newblock \emph{Journal of Physics A: Mathematical and General}, 33\penalty0
  (22):\penalty0 L193, June 2000.
\newblock arXiv:quant-ph/0002075.

\bibitem[Preskill(1998)]{preskill1998}
John Preskill.
\newblock Reliable quantum computers.
\newblock \emph{Proceedings of the Royal Society A: Mathematical, Physical and
  Engineering Sciences}, 454\penalty0 (1969):\penalty0 385--410, January 1998.
\newblock arXiv:quant-ph/9705031.

\bibitem[Radhakrishnan et~al.(2014)Radhakrishnan, Sen, and Warsi]{RSW14}
Jaikumar Radhakrishnan, Pranab Sen, and Naqueeb Warsi.
\newblock One-shot {Marton} inner bound for classical-quantum broadcast
  channel.
\newblock October 2014.
\newblock arXiv:1410.3248.

\bibitem[Rains(2001)]{R01}
Eric~M. Rains.
\newblock A semidefinite program for distillable entanglement.
\newblock \emph{IEEE Transactions on Information Theory}, 47\penalty0
  (7):\penalty0 2921--2933, November 2001.
\newblock arXiv:quant-ph/0008047.

\bibitem[Reed and Simon(1975)]{RS75}
Michael Reed and Barry Simon.
\newblock \emph{Methods of Modern Mathematical Physics II: Fourier Analysis,
  Self-Adjointness}.
\newblock Academic Press, 1975.

\bibitem[Renner(2005)]{Renner2005}
Renato Renner.
\newblock \emph{Security of Quantum Key Distribution}.
\newblock PhD thesis, ETH Zurich, September 2005.
\newblock arXiv:quant-ph/0512258.

\bibitem[Rivest et~al.(1978)Rivest, Shamir, and Adleman]{rsa}
Ronald Rivest, Adi Shamir, and Leonard Adleman.
\newblock A method for obtaining digital signatures and public-key
  cryptosystems.
\newblock \emph{Communications of the ACM}, 21\penalty0 (2):\penalty0 120--126,
  1978.

\bibitem[Sakurai(1994)]{book1994sakurai}
J.~J. Sakurai.
\newblock \emph{Modern Quantum Mechanics (2nd Edition)}.
\newblock {Addison Wesley}, January 1994.
\newblock ISBN 0201539292.

\bibitem[Sason(2013)]{Sason13}
Igal Sason.
\newblock Entropy bounds for discrete random variables via maximal coupling.
\newblock \emph{IEEE Transactions on Information Theory}, 59\penalty0
  (11):\penalty0 7118--7131, November 2013.
\newblock ISSN 0018-9448.
\newblock \doi{10.1109/TIT.2013.2274515}.
\newblock arXiv:1209.5259.

\bibitem[Savov(2008)]{Savov08}
Ivan Savov.
\newblock Distributed compression and squashed entanglement.
\newblock Master's thesis, McGill University, February 2008.
\newblock arXiv:0802.0694.

\bibitem[Savov(2012)]{Savov12}
Ivan Savov.
\newblock \emph{Network information theory for classical-quantum channels}.
\newblock PhD thesis, McGill University, School of Computer Science, July 2012.
\newblock arXiv:1208.4188.

\bibitem[Savov and Wilde(2015)]{SW13}
Ivan Savov and Mark~M. Wilde.
\newblock Classical codes for quantum broadcast channels.
\newblock \emph{IEEE Transactions on Information Theory}, 61\penalty0
  (12):\penalty0 7017--7028, December 2015.
\newblock ISSN 0018-9448.
\newblock \doi{10.1109/TIT.2015.2485998}.
\newblock arXiv:1111.3645.

\bibitem[Scarani(2013)]{S13}
Valerio Scarani.
\newblock The device-independent outlook on quantum physics (lecture notes on
  the power of {Bell}'s theorem).
\newblock 2013.
\newblock arXiv:1303.3081.

\bibitem[Scarani et~al.(2009)Scarani, Bechmann-Pasquinucci, Cerf,
  Du\ifmmode~\check{s}\else \v{s}\fi{}ek, L\"utkenhaus, and Peev]{SBCDLP09}
Valerio Scarani, Helle Bechmann-Pasquinucci, Nicolas~J. Cerf, Miloslav
  Du\ifmmode~\check{s}\else \v{s}\fi{}ek, Norbert L\"utkenhaus, and Momtchil
  Peev.
\newblock The security of practical quantum key distribution.
\newblock \emph{Reviews of Modern Physics}, 81\penalty0 (3):\penalty0
  1301--1350, September 2009.
\newblock \doi{10.1103/RevModPhys.81.1301}.
\newblock arXiv:0802.4155.

\bibitem[Schr\"{o}dinger(1926)]{schrodinger1926}
Erwin Schr\"{o}dinger.
\newblock Quantisierung als eigenwertproblem.
\newblock \emph{Annalen der Physik}, 79:\penalty0 361--376, 1926.

\bibitem[Schr\"{o}dinger(1935)]{S35}
Erwin Schr\"{o}dinger.
\newblock Discussion of probability relations between separated systems.
\newblock \emph{Proceedings of the Cambridge Philosophical Society},
  31:\penalty0 555--563, 1935.

\bibitem[Schumacher(1995)]{PhysRevA.51.2738}
Benjamin Schumacher.
\newblock Quantum coding.
\newblock \emph{Physical Review A}, 51\penalty0 (4):\penalty0 2738--2747, April
  1995.
\newblock \doi{10.1103/PhysRevA.51.2738}.

\bibitem[Schumacher(1996)]{PhysRevA.54.2614}
Benjamin Schumacher.
\newblock Sending entanglement through noisy quantum channels.
\newblock \emph{Physical Review A}, 54\penalty0 (4):\penalty0 2614--2628,
  October 1996.
\newblock \doi{10.1103/PhysRevA.54.2614}.

\bibitem[Schumacher and Nielsen(1996)]{PhysRevA.54.2629}
Benjamin Schumacher and Michael~A. Nielsen.
\newblock Quantum data processing and error correction.
\newblock \emph{Physical Review A}, 54\penalty0 (4):\penalty0 2629--2635,
  October 1996.
\newblock \doi{10.1103/PhysRevA.54.2629}.
\newblock arXiv:quant-ph/9604022.

\bibitem[Schumacher and Westmoreland(1997)]{PhysRevA.56.131}
Benjamin Schumacher and Michael~D. Westmoreland.
\newblock Sending classical information via noisy quantum channels.
\newblock \emph{Physical Review A}, 56\penalty0 (1):\penalty0 131--138, July
  1997.
\newblock \doi{10.1103/PhysRevA.56.131}.

\bibitem[Schumacher and Westmoreland(1998)]{PhysRevLett.80.5695}
Benjamin Schumacher and Michael~D. Westmoreland.
\newblock Quantum privacy and quantum coherence.
\newblock \emph{Physical Review Letters}, 80\penalty0 (25):\penalty0
  5695--5697, June 1998.
\newblock \doi{10.1103/PhysRevLett.80.5695}.
\newblock arXiv:quant-ph/9709058.

\bibitem[Schumacher and Westmoreland(2002)]{qip2002schu}
Benjamin Schumacher and Michael~D. Westmoreland.
\newblock Approximate quantum error correction.
\newblock \emph{Quantum Information Processing}, 1\penalty0 (1/2):\penalty0
  5--12, April 2002.
\newblock arXiv:quant-ph/0112106.

\bibitem[Sen(2011)]{S11}
Pranab Sen.
\newblock Achieving the {Han-Kobayashi} inner bound for the quantum
  interference channel by sequential decoding.
\newblock September 2011.
\newblock arXiv:1109.0802.

\bibitem[Seshadreesan and Wilde(2015)]{SW14}
Kaushik~P. Seshadreesan and Mark~M. Wilde.
\newblock Fidelity of recovery, squashed entanglement, and measurement
  recoverability.
\newblock \emph{Physical Review A}, 92\penalty0 (4):\penalty0 042321, October
  2015.
\newblock arXiv:1410.1441.

\bibitem[Seshadreesan et~al.(2015{\natexlab{a}})Seshadreesan, Berta, and
  Wilde]{SBW14}
Kaushik~P. Seshadreesan, Mario Berta, and Mark~M. Wilde.
\newblock R\'enyi squashed entanglement, discord, and relative entropy
  differences.
\newblock \emph{Journal of Physics A: Mathematical and Theoretical},
  48\penalty0 (39):\penalty0 395303, September 2015{\natexlab{a}}.
\newblock arXiv:1410.1443.

\bibitem[Seshadreesan et~al.(2015{\natexlab{b}})Seshadreesan, Takeoka, and
  Wilde]{STW15}
Kaushik~P. Seshadreesan, Masahiro Takeoka, and Mark~M. Wilde.
\newblock Bounds on entanglement distillation and secret key agreement for
  quantum broadcast channels.
\newblock March 2015{\natexlab{b}}.
\newblock arXiv:1503.08139.

\bibitem[Shannon(1948)]{bell1948shannon}
Claude~E. Shannon.
\newblock A mathematical theory of communication.
\newblock \emph{Bell System Technical Journal}, 27:\penalty0 379--423, 1948.

\bibitem[Shor(1994)]{Shor:1994:124}
Peter~W. Shor.
\newblock Algorithms for quantum computation: Discrete logarithms and
  factoring.
\newblock In \emph{Proceedings of the 35th Annual Symposium on Foundations of
  Computer Science}, pages 124--134, Los Alamitos, California, 1994. IEEE
  Computer Society Press.

\bibitem[Shor(1995)]{PhysRevA.52.R2493}
Peter~W. Shor.
\newblock Scheme for reducing decoherence in quantum computer memory.
\newblock \emph{Physical Review A}, 52\penalty0 (4):\penalty0 R2493--R2496,
  October 1995.
\newblock \doi{10.1103/PhysRevA.52.R2493}.

\bibitem[Shor(1996)]{10.1109/SFCS.1996.548464}
Peter~W. Shor.
\newblock Fault-tolerant quantum computation.
\newblock \emph{Annual IEEE Symposium on Foundations of Computer Science},
  page~56, 1996.
\newblock \doi{http://doi.ieeecomputersociety.org/10.1109/SFCS.1996.548464}.
\newblock arXiv:quant-ph/9605011.

\bibitem[Shor(2002{\natexlab{a}})]{S02}
Peter~W. Shor.
\newblock Additivity of the classical capacity of entanglement-breaking quantum
  channels.
\newblock \emph{Journal of Mathematical Physics}, 43\penalty0 (9):\penalty0
  4334--4340, 2002{\natexlab{a}}.
\newblock \doi{10.1063/1.1498000}.
\newblock arXiv:quant-ph/0201149.

\bibitem[Shor(2002{\natexlab{b}})]{capacity2002shor}
Peter~W. Shor.
\newblock The quantum channel capacity and coherent information.
\newblock In \emph{Lecture Notes, MSRI Workshop on Quantum Computation},
  2002{\natexlab{b}}.

\bibitem[Shor(2004{\natexlab{a}})]{S04}
Peter~W. Shor.
\newblock Equivalence of additivity questions in quantum information theory.
\newblock \emph{Communications in Mathematical Physics}, 246\penalty0
  (3):\penalty0 453--472, 2004{\natexlab{a}}.
\newblock arXiv:quant-ph/0305035.

\bibitem[Shor(2004{\natexlab{b}})]{Shor_CE}
Peter~W. Shor.
\newblock \emph{Quantum Information, Statistics, Probability (Dedicated to A.
  S. Holevo on the occasion of his 60th Birthday): The classical capacity
  achievable by a quantum channel assisted by limited entanglement}.
\newblock Rinton Press, Inc., 2004{\natexlab{b}}.
\newblock arXiv:quant-ph/0402129.

\bibitem[Smith(2006)]{Smith06}
Graeme Smith.
\newblock \emph{Upper and Lower Bounds on Quantum Codes}.
\newblock PhD thesis, California Institute of Technology, May 2006.

\bibitem[Smith(2008)]{S08}
Graeme Smith.
\newblock Private classical capacity with a symmetric side channel and its
  application to quantum cryptography.
\newblock \emph{Physical Review A}, 78\penalty0 (2):\penalty0 022306, August
  2008.
\newblock \doi{10.1103/PhysRevA.78.022306}.
\newblock arXiv:0705.3838.

\bibitem[Smith and Smolin(2007)]{SS07}
Graeme Smith and John~A. Smolin.
\newblock Degenerate quantum codes for {Pauli} channels.
\newblock \emph{Physical Review Letters}, 98\penalty0 (3):\penalty0 030501,
  January 2007.
\newblock \doi{10.1103/PhysRevLett.98.030501}.
\newblock arXiv:quant-ph/0604107.

\bibitem[Smith and Yard(2008)]{science2008smith}
Graeme Smith and Jon Yard.
\newblock Quantum communication with zero-capacity channels.
\newblock \emph{Science}, 321\penalty0 (5897):\penalty0 1812--1815, September
  2008.
\newblock arXiv:0807.4935.

\bibitem[Smith et~al.(2008)Smith, Renes, and Smolin]{smith:170502}
Graeme Smith, Joseph~M. Renes, and John~A. Smolin.
\newblock Structured codes improve the {Bennett-Brassard}-84 quantum key rate.
\newblock \emph{Physical Review Letters}, 100\penalty0 (17):\penalty0 170502,
  April 2008.
\newblock \doi{10.1103/PhysRevLett.100.170502}.
\newblock arXiv:quant-ph/0607018.

\bibitem[Smith et~al.(2011)Smith, Smolin, and Yard]{SSY11}
Graeme Smith, John~A. Smolin, and Jon Yard.
\newblock Quantum communication with {Gaussian} channels of zero quantum
  capacity.
\newblock \emph{Nature Photonics}, 5:\penalty0 624--627, August 2011.
\newblock arXiv:1102.4580.

\bibitem[Steane(1996)]{PhysRevLett.77.793}
Andrew~M. Steane.
\newblock Error correcting codes in quantum theory.
\newblock \emph{Physical Review Letters}, 77\penalty0 (5):\penalty0 793--797,
  July 1996.
\newblock \doi{10.1103/PhysRevLett.77.793}.

\bibitem[Stein(1956)]{S56}
Elias~M. Stein.
\newblock Interpolation of linear operators.
\newblock \emph{Transactions of the American Mathematical Society}, 83\penalty0
  (2):\penalty0 482--492, November 1956.

\bibitem[Stinespring(1955)]{S55}
William~F. Stinespring.
\newblock Positive functions on {C*}-algebras.
\newblock \emph{Proceedings of the American Mathematical Society}, 6:\penalty0
  211--216, 1955.

\bibitem[Sutter et~al.(2016{\natexlab{a}})Sutter, Fawzi, and Renner]{SOR15}
David Sutter, Omar Fawzi, and Renato Renner.
\newblock Universal recovery map for approximate markov chains.
\newblock \emph{Proceedings of the Royal Society A}, 472\penalty0 (2186),
  February 2016{\natexlab{a}}.
\newblock ISSN 1364-5021.
\newblock \doi{10.1098/rspa.2015.0623}.
\newblock arXiv:1504.07251.

\bibitem[Sutter et~al.(2016{\natexlab{b}})Sutter, Tomamichel, and
  Harrow]{Sutter15}
David Sutter, Marco Tomamichel, and Aram~W. Harrow.
\newblock Strengthened monotonicity of relative entropy via pinched {Petz}
  recovery map.
\newblock \emph{IEEE Transactions on Information Theory}, 62\penalty0
  (5):\penalty0 2907--2913, May 2016{\natexlab{b}}.
\newblock arXiv:1507.00303.

\bibitem[Tomamichel(2016)]{T15}
Marco Tomamichel.
\newblock \emph{Quantum Information Processing with Finite Resources ---
  Mathematical Foundations}, volume~5 of \emph{SpringerBriefs in Mathematical
  Physics}.
\newblock Springer, 2016.
\newblock \doi{10.1007/978-3-319-21891-5}.
\newblock arXiv:1504.00233.

\bibitem[Tomamichel and Renner(2011)]{TR11}
Marco Tomamichel and Renato Renner.
\newblock Uncertainty relation for smooth entropies.
\newblock \emph{Physical Review Letters}, 106\penalty0 (11):\penalty0 110506,
  March 2011.
\newblock \doi{10.1103/PhysRevLett.106.110506}.
\newblock arXiv:1009.2015.

\bibitem[Tomamichel and Tan(2015)]{TT13}
Marco Tomamichel and Vincent Y.~F. Tan.
\newblock Second-order asymptotics for the classical capacity of image-additive
  quantum channels.
\newblock \emph{Communications in Mathematical Physics}, 338\penalty0
  (1):\penalty0 103--137, August 2015.
\newblock arXiv:1308.6503.

\bibitem[Tomamichel et~al.(2015)Tomamichel, Berta, and Renes]{TBR15}
Marco Tomamichel, Mario Berta, and Joseph~M. Renes.
\newblock Quantum coding with finite resources.
\newblock 2015.
\newblock arXiv:1504.04617.

\bibitem[Tomamichel et~al.(2017)Tomamichel, Wilde, and Winter]{TWW14}
Marco Tomamichel, Mark~M. Wilde, and Andreas Winter.
\newblock Strong converse rates for quantum communication.
\newblock \emph{IEEE Transactions on Information Theory}, 63\penalty0
  (1):\penalty0 715--727, January 2017.
\newblock arXiv:1406.2946.

\bibitem[Tsirelson(1980)]{T80}
Boris~S. Tsirelson.
\newblock Quantum generalizations of {Bell}'s inequality.
\newblock \emph{Letters in Mathematical Physics}, 4\penalty0 (2):\penalty0
  93--100, March 1980.
\newblock ISSN 0377-9017.
\newblock \doi{10.1007/BF00417500}.

\bibitem[Tyurin(2010)]{tyurin10}
I.~S. Tyurin.
\newblock An improvement of upper estimates of the constants in the {Lyapunov}
  theorem.
\newblock \emph{Russian Mathematical Surveys}, 65\penalty0 (3):\penalty0
  201--202, 2010.

\bibitem[Uhlmann(1976)]{U73}
Armin Uhlmann.
\newblock The ``transition probability'' in the state space of a *-algebra.
\newblock \emph{Reports on Mathematical Physics}, 9\penalty0 (2):\penalty0
  273--279, 1976.

\bibitem[Uhlmann(1977)]{U77}
Armin Uhlmann.
\newblock Relative entropy and the {Wigner-Yanase-Dyson-Lieb} concavity in an
  interpolation theory.
\newblock \emph{Communications in Mathematical Physics}, 54\penalty0
  (1):\penalty0 21--32, 1977.
\newblock URL \url{http://projecteuclid.org/euclid.cmp/1103900757}.

\bibitem[Umegaki(1962)]{U62}
Hisaharu Umegaki.
\newblock Conditional expectations in an operator algebra {IV} (entropy and
  information).
\newblock \emph{Kodai Mathematical Seminar Reports}, 14\penalty0 (2):\penalty0
  59--85, 1962.

\bibitem[Unruh(1995)]{PhysRevA.51.992}
William~G. Unruh.
\newblock Maintaining coherence in quantum computers.
\newblock \emph{Physical Review A}, 51\penalty0 (2):\penalty0 992--997,
  February 1995.
\newblock \doi{10.1103/PhysRevA.51.992}.
\newblock arXiv:hep-th/9406058.

\bibitem[Vedral and Plenio(1998)]{VP98}
Vlatko Vedral and Martin~B. Plenio.
\newblock Entanglement measures and purification procedures.
\newblock \emph{Physical Review A}, 57\penalty0 (3):\penalty0 1619--1633, March
  1998.
\newblock \doi{10.1103/PhysRevA.57.1619}.
\newblock arXiv:quant-ph/9707035.

\bibitem[von Kretschmann(2007)]{K07}
Dennis von Kretschmann.
\newblock \emph{Information Transfer through Quantum Channels}.
\newblock PhD thesis, Technische Universit\"{a}t Braunschweig, 2007.

\bibitem[von Neumann(1996)]{book1996vonNeumann}
John von Neumann.
\newblock \emph{Mathematical Foundations of Quantum Mechanics}.
\newblock {Princeton University Press}, October 1996.
\newblock ISBN 0691028931.

\bibitem[Wang and Renner(2012)]{WR10}
Ligong Wang and Renato Renner.
\newblock One-shot classical-quantum capacity and hypothesis testing.
\newblock \emph{Physical Review Letters}, 108\penalty0 (20):\penalty0 200501,
  May 2012.
\newblock arXiv:1007.5456.

\bibitem[Watrous(2015)]{Wat15}
John Watrous.
\newblock \emph{Theory of Quantum Information}.
\newblock 2015.
\newblock Available at \verb+https://cs.uwaterloo.ca/~watrous/TQI/+.

\bibitem[Wehrl(1978)]{W78}
Alfred Wehrl.
\newblock General properties of entropy.
\newblock \emph{Reviews of Modern Physics}, 50\penalty0 (2):\penalty0 221--260,
  April 1978.

\bibitem[Werner(1989)]{W89}
Reinhard~F. Werner.
\newblock Quantum states with {Einstein-Podolsky-Rosen} correlations admitting
  a hidden-variable model.
\newblock \emph{Physical Review A}, 40\penalty0 (8):\penalty0 4277--4281,
  October 1989.
\newblock \doi{10.1103/PhysRevA.40.4277}.

\bibitem[Wiesner(1983)]{Wiesner:1983:78}
Stephen Wiesner.
\newblock Conjugate coding.
\newblock \emph{SIGACT News}, 15\penalty0 (1):\penalty0 78--88, 1983.
\newblock ISSN 0163-5700.
\newblock \doi{http://doi.acm.org/10.1145/1008908.1008920}.

\bibitem[Wilde(2011)]{PhysRevA.83.046303}
Mark~M. Wilde.
\newblock Comment on ``{Secret}-key-assisted private classical communication
  capacity over quantum channels''.
\newblock \emph{Physical Review A}, 83\penalty0 (4):\penalty0 046303, April
  2011.
\newblock \doi{10.1103/PhysRevA.83.046303}.

\bibitem[Wilde(2013)]{Wilde20130259}
Mark~M. Wilde.
\newblock Sequential decoding of a general classical-quantum channel.
\newblock \emph{Proceedings of the Royal Society of London A: Mathematical,
  Physical and Engineering Sciences}, 469\penalty0 (2157), September 2013.
\newblock ISSN 1364-5021.
\newblock \doi{10.1098/rspa.2013.0259}.
\newblock arXiv:1303.0808.

\bibitem[Wilde(2014)]{W14}
Mark~M. Wilde.
\newblock Multipartite quantum correlations and local recoverability.
\newblock \emph{Proceedings of the Royal Society A}, 471:\penalty0 20140941,
  March 2014.
\newblock arXiv:1412.0333.

\bibitem[Wilde(2015)]{W15}
Mark~M. Wilde.
\newblock Recoverability in quantum information theory.
\newblock \emph{Proceedings of the Royal Society A}, 471\penalty0
  (2182):\penalty0 20150338, October 2015.
\newblock ISSN 1364-5021.
\newblock \doi{10.1098/rspa.2015.0338}.
\newblock arXiv:1505.04661.

\bibitem[Wilde and Brun(2008)]{arx2008wildeUQCC}
Mark~M. Wilde and Todd~A. Brun.
\newblock Unified quantum convolutional coding.
\newblock In \emph{Proceedings of the IEEE International Symposium on
  Information Theory}, pages 359--363, Toronto, Ontario, Canada, July 2008.
\newblock arXiv:0801.0821.

\bibitem[Wilde and Guha(2012)]{WG12}
Mark~M. Wilde and Saikat Guha.
\newblock Explicit receivers for pure-interference bosonic multiple access
  channels.
\newblock \emph{Proceedings of the 2012 International Symposium on Information
  Theory and its Applications}, pages 303--307, October 2012.
\newblock arXiv:1204.0521.

\bibitem[Wilde and Hsieh(2010)]{WH10a}
Mark~M. Wilde and Min-Hsiu Hsieh.
\newblock Entanglement generation with a quantum channel and a shared state.
\newblock \emph{Proceedings of the 2010 IEEE International Symposium on
  Information Theory}, pages 2713--2717, June 2010.
\newblock arXiv:0904.1175.

\bibitem[Wilde and Hsieh(2012{\natexlab{a}})]{WH10}
Mark~M. Wilde and Min-Hsiu Hsieh.
\newblock Public and private resource trade-offs for a quantum channel.
\newblock \emph{Quantum Information Processing}, 11\penalty0 (6):\penalty0
  1465--1501, December 2012{\natexlab{a}}.
\newblock arXiv:1005.3818.

\bibitem[Wilde and Hsieh(2012{\natexlab{b}})]{WH10b}
Mark~M. Wilde and Min-Hsiu Hsieh.
\newblock The quantum dynamic capacity formula of a quantum channel.
\newblock \emph{Quantum Information Processing}, 11\penalty0 (6):\penalty0
  1431--1463, December 2012{\natexlab{b}}.
\newblock arXiv:1004.0458.

\bibitem[Wilde and Savov(2012)]{WS12}
Mark~M. Wilde and Ivan Savov.
\newblock Joint source-channel coding for a quantum multiple access channel.
\newblock \emph{Journal of Physics A: Mathematical and Theoretical},
  45\penalty0 (43):\penalty0 435302, November 2012.
\newblock arXiv:1202.3467.

\bibitem[Wilde and Winter(2014)]{WW14}
Mark~M. Wilde and Andreas Winter.
\newblock Strong converse for the quantum capacity of the erasure channel for
  almost all codes.
\newblock \emph{Proceedings of the 9th Conference on the Theory of Quantum
  Computation, Communication and Cryptography}, May 2014.
\newblock arXiv:1402.3626.

\bibitem[Wilde et~al.(2007)Wilde, Krovi, and Brun]{wilde:060303}
Mark~M. Wilde, Hari Krovi, and Todd~A. Brun.
\newblock Coherent communication with continuous quantum variables.
\newblock \emph{Physical Review A}, 75\penalty0 (6):\penalty0 060303, June
  2007.
\newblock \doi{10.1103/PhysRevA.75.060303}.
\newblock arXiv:quant-ph/0612170.

\bibitem[Wilde et~al.(2012{\natexlab{a}})Wilde, Hayden, Buscemi, and
  Hsieh]{WHBH12}
Mark~M. Wilde, Patrick Hayden, Francesco Buscemi, and Min-Hsiu Hsieh.
\newblock The information-theoretic costs of simulating quantum measurements.
\newblock \emph{Journal of Physics A: Mathematical and Theoretical},
  45\penalty0 (45):\penalty0 453001, November 2012{\natexlab{a}}.
\newblock arXiv:1206.4121.

\bibitem[Wilde et~al.(2012{\natexlab{b}})Wilde, Hayden, and
  Guha]{PhysRevA.86.062306}
Mark~M. Wilde, Patrick Hayden, and Saikat Guha.
\newblock Quantum trade-off coding for bosonic communication.
\newblock \emph{Physical Review A}, 86\penalty0 (6):\penalty0 062306, December
  2012{\natexlab{b}}.
\newblock \doi{10.1103/PhysRevA.86.062306}.
\newblock arXiv:1105.0119.

\bibitem[Wilde et~al.(2012{\natexlab{c}})Wilde, Hayden, and Guha]{WHG11}
Mark~M. Wilde, Patrick Hayden, and Saikat Guha.
\newblock Information trade-offs for optical quantum communication.
\newblock \emph{Physical Review Letters}, 108\penalty0 (14):\penalty0 140501,
  April 2012{\natexlab{c}}.
\newblock arXiv:1105.0119.

\bibitem[Wilde et~al.(2014)Wilde, Winter, and Yang]{WWY13}
Mark~M. Wilde, Andreas Winter, and Dong Yang.
\newblock Strong converse for the classical capacity of entanglement-breaking
  and {Hadamard} channels via a sandwiched {R\'enyi} relative entropy.
\newblock \emph{Communications in Mathematical Physics}, 331\penalty0
  (2):\penalty0 593--622, October 2014.
\newblock arXiv:1306.1586.

\bibitem[Wilde et~al.(2016)Wilde, Renes, and Guha]{WRG15}
Mark~M. Wilde, Joseph~M. Renes, and Saikat Guha.
\newblock Second-order coding rates for pure-loss bosonic channels.
\newblock \emph{Quantum Information Processing}, 15\penalty0 (3):\penalty0
  1289--1308, March 2016.
\newblock arXiv:1408.5328.

\bibitem[Winter(1999{\natexlab{a}})]{itit1999winter}
Andreas Winter.
\newblock Coding theorem and strong converse for quantum channels.
\newblock \emph{IEEE Transactions on Information Theory}, 45\penalty0
  (7):\penalty0 2481--2485, November 1999{\natexlab{a}}.
\newblock arXiv:1409.2536.

\bibitem[Winter(1999{\natexlab{b}})]{thesis1999winter}
Andreas Winter.
\newblock \emph{Coding Theorems of Quantum Information Theory}.
\newblock PhD thesis, Universit\"{a}t Bielefeld, July 1999{\natexlab{b}}.
\newblock arXiv:quant-ph/9907077.

\bibitem[Winter(2001)]{Winter01}
Andreas Winter.
\newblock The capacity of the quantum multiple access channel.
\newblock \emph{IEEE Transactions on Information Theory}, 47\penalty0
  (7):\penalty0 3059--3065, November 2001.
\newblock arXiv:quant-ph/9807019.

\bibitem[Winter(2004)]{Winter01a}
Andreas Winter.
\newblock ``{E}xtrinsic'' and ``intrinsic'' data in quantum measurements:
  asymptotic convex decomposition of positive operator valued measures.
\newblock \emph{Communications in Mathematical Physics}, 244\penalty0
  (1):\penalty0 157--185, January 2004.
\newblock arXiv:quant-ph/0109050.

\bibitem[Winter(2007)]{W07}
Andreas Winter.
\newblock The maximum output $p$-norm of quantum channels is not multiplicative
  for any $p > 2$.
\newblock July 2007.
\newblock arXiv:0707.0402.

\bibitem[Winter(2015{\natexlab{a}})]{W15lock}
Andreas Winter.
\newblock Weak locking capacity of quantum channels can be much larger than
  private capacity.
\newblock \emph{Journal of Cryptology}, pages 1--21, 2015{\natexlab{a}}.
\newblock ISSN 0933-2790.
\newblock \doi{10.1007/s00145-015-9215-3}.
\newblock arXiv:1403.6361.

\bibitem[Winter(2015{\natexlab{b}})]{Winter15}
Andreas Winter.
\newblock Tight uniform continuity bounds for quantum entropies: conditional
  entropy, relative entropy distance and energy constraints.
\newblock July 2015{\natexlab{b}}.
\newblock arXiv:1507.07775.

\bibitem[Winter and Li(2012)]{Winterconj}
Andreas Winter and Ke~Li.
\newblock A stronger subadditivity relation?
\newblock
  http://www.maths.bris.ac.uk/$\sim$csajw/stronger$\_$subadditivity.pdf, 2012.

\bibitem[Winter and Massar(2001)]{WM01}
Andreas Winter and Serge Massar.
\newblock Compression of quantum-measurement operations.
\newblock \emph{Physical Review A}, 64\penalty0 (1):\penalty0 012311, June
  2001.
\newblock \doi{10.1103/PhysRevA.64.012311}.
\newblock arXiv:quant-ph/0012128.

\bibitem[Wolf and P\'erez-Garc\'\i{}a(2007)]{PhysRevA.75.012303}
Michael~M. Wolf and David P\'erez-Garc\'\i{}a.
\newblock Quantum capacities of channels with small environment.
\newblock \emph{Physical Review A}, 75\penalty0 (1):\penalty0 012303, January
  2007.
\newblock \doi{10.1103/PhysRevA.75.012303}.
\newblock arXiv:quant-ph/0607070.

\bibitem[Wolf et~al.(2007)Wolf, P\'erez-Garc\'\i{}a, and Giedke]{WPG07}
Michael~M. Wolf, David P\'erez-Garc\'\i{}a, and Geza Giedke.
\newblock Quantum capacities of bosonic channels.
\newblock \emph{Physical Review Letters}, 98\penalty0 (13):\penalty0 130501,
  March 2007.
\newblock \doi{10.1103/PhysRevLett.98.130501}.
\newblock arXiv:quant-ph/0606132.

\bibitem[Wolf et~al.(2011)Wolf, Cubitt, and Perez-Garcia]{WCP11}
Michael~M. Wolf, Toby~S. Cubitt, and David Perez-Garcia.
\newblock Are problems in quantum information theory (un)decidable?
\newblock November 2011.
\newblock arXiv:1111.5425.

\bibitem[Wolfowitz(1978)]{Wolf78}
Jacob Wolfowitz.
\newblock \emph{Coding theorems of information theory}.
\newblock Springer-Verlag, 1978.

\bibitem[Wootters and Zurek(1982)]{nat1982}
William~K. Wootters and Wojciech~H. Zurek.
\newblock A single quantum cannot be cloned.
\newblock \emph{Nature}, 299:\penalty0 802--803, 1982.

\bibitem[Wyner(1975)]{W75}
Aaron~D. Wyner.
\newblock The wire-tap channel.
\newblock \emph{Bell System Technical Journal}, 54\penalty0 (8):\penalty0
  1355--1387, October 1975.

\bibitem[Yard(2005)]{Yard05a}
Jon Yard.
\newblock \emph{Simultaneous classical-quantum capacities of quantum multiple
  access channels}.
\newblock PhD thesis, Stanford University, Stanford, CA, 2005.
\newblock arXiv:quant-ph/0506050.

\bibitem[Yard and Devetak(2009)]{YD09}
Jon Yard and Igor Devetak.
\newblock Optimal quantum source coding with quantum side information at the
  encoder and decoder.
\newblock \emph{IEEE Transactions on Information Theory}, 55\penalty0
  (11):\penalty0 5339--5351, November 2009.
\newblock arXiv:0706.2907.

\bibitem[Yard et~al.(2005)Yard, Devetak, and Hayden]{YDH05}
Jon Yard, Igor Devetak, and Patrick Hayden.
\newblock Capacity theorems for quantum multiple access channels.
\newblock In \emph{Proceedings of the International Symposium on Information
  Theory}, pages 884--888, Adelaide, Australia, September 2005.
\newblock arXiv:cs/0508031.

\bibitem[Yard et~al.(2008)Yard, Hayden, and Devetak]{YHD05MQAC}
Jon Yard, Patrick Hayden, and Igor Devetak.
\newblock Capacity theorems for quantum multiple-access channels:
  Classical-quantum and quantum-quantum capacity regions.
\newblock \emph{IEEE Transactions on Information Theory}, 54\penalty0
  (7):\penalty0 3091--3113, July 2008.
\newblock arXiv:quant-ph/0501045.

\bibitem[Yard et~al.(2011)Yard, Hayden, and Devetak]{YHD2006}
Jon Yard, Patrick Hayden, and Igor Devetak.
\newblock Quantum broadcast channels.
\newblock \emph{IEEE Transactions on Information Theory}, 57\penalty0
  (10):\penalty0 7147--7162, October 2011.
\newblock arXiv:quant-ph/0603098.

\bibitem[Ye et~al.(2008)Ye, Bai, and Wang]{PhysRevA.78.030302}
Ming-Yong Ye, Yan-Kui Bai, and Z.~D. Wang.
\newblock Quantum state redistribution based on a generalized decoupling.
\newblock \emph{Physical Review A}, 78\penalty0 (3):\penalty0 030302, September
  2008.
\newblock \doi{10.1103/PhysRevA.78.030302}.
\newblock arXiv:0805.1542.

\bibitem[Yen and Shapiro(2005)]{PhysRevA.72.062312}
Brent~J. Yen and Jeffrey~H. Shapiro.
\newblock Multiple-access bosonic communications.
\newblock \emph{Physical Review A}, 72\penalty0 (6):\penalty0 062312, December
  2005.
\newblock \doi{10.1103/PhysRevA.72.062312}.
\newblock arXiv:quant-ph/0506171.

\bibitem[Yeung(2002)]{Y02}
Raymond~W. Yeung.
\newblock \emph{A First Course in Information Theory}.
\newblock Information Technology: Transmission, Processing, and Storage.
  Springer (Kluwer Academic/Plenum Publishers), New York, New York, USA, March
  2002.

\bibitem[Zhang(2014)]{Z14}
Lin Zhang.
\newblock A lower bound of quantum conditional mutual information.
\newblock March 2014.
\newblock arXiv:1403.1424.

\bibitem[Zhang(2007)]{Z07}
Zhengmin Zhang.
\newblock Estimating mutual information via {Kolmogorov} distance.
\newblock \emph{IEEE Transactions on Information Theory}, 53\penalty0
  (9):\penalty0 3280--3282, September 2007.
\newblock ISSN 0018-9448.
\newblock \doi{10.1109/TIT.2007.903122}.

\bibitem[Zurek(2000)]{Z00}
Wojciech~H. Zurek.
\newblock Einselection and decoherence from an information theory perspective.
\newblock \emph{Annalen der Physik}, 9\penalty0 (11-12):\penalty0 855--864,
  November 2000.
\newblock ISSN 1521-3889.
\newblock arXiv:quant-ph/0011039.

\end{thebibliography}
}

{\footnotesize \printindex
}

\end{document}